# THE NONLINEARITIES IN MICROWAVE SUPERCONDUCTIVITY

BY DIMITRI O. LEDENYOV AND VIKTOR O. LEDENYOV

# Modeling of Nonlinear Properties of HTS Thin Films, Using Bardeen, Cooper, Schrieffer and Lumped Element Circuit Theories, with Purpose to Enhance Microwave Power Handling Capabilities of Microwave Filters and Optimize Design of Microwave Circuits in Micro- and Nano-Electronics

**by Dimitri O. Ledenyov and Viktor O. Ledenyov**

*To all the scientists and engineers, who create the advancements in science and engineering for benefits of humanity.*

## Preface.

This brilliant book is a result of more than forty four years of research work in the field of superconductivity, which began by our father, Oleg P. Ledenyov, at Kharkov National University named after V. N. Karazin and at National Scientific Centre Kharkov Institute of Physics and Technology in Ukraine in 1970 and continues up to this date.

Dimitri O. Ledenyov first gained his knowledge of microwave superconductivity at Kharkov National University named after V. N. Karazin in 1994-1999 and at National Scientific Centre Kharkov Institute of Physics and Technology in 1999-2000, following by the visits to Kamerlingh Onnes Laboratory, Department of Physics, Leiden University, Leiden, The Netherlands in 1999 and to Department of Electronic, Electrical and Computing Engineering, University of Birmingham, Birmingham, U.K. in 2000, where his invited talk: "The Quantum Knots of Magnetic Vortices" was delivered at Marconi Seminar, making an important original theoretical proposition that the knot of Abricosov magnetic vortex is in an extreme quantum limit, and suggesting the design of quantum memory chips on the quantum knots of magnetic vortices, for the first time. The intensive innovative research in the field of microwave superconductivity has been conducted by Dimitri O. Ledenyov at Microwave and Electronic Material Research Group, Electrical and Computer Engineering Department, James Cook University, Queensland, Australia, during last decade. Prof. Janina E. Mazierska as a representative of Polish school of scientific thinking, created on the knowledge base accumulated by scientists at Warsaw University of Technology, Warsaw, Poland, had a profound influence on the Dimitri's research strategy formulation and implementation toward the accurate characterization of superconductors at microwaves, supervising the Dimitri's advanced research on the nonlinearities in microwave superconductivity at JCU in Australia for many years. The Australian knowledge base was extensively complemented by the theoretical and experimental data in the field of fundamental and applied microwave science gained at National Scientific Centre Kharkov Institute of Physics and Technology in Kharkov in Ukraine over the XX century and in the beginning of the XXI century.

The book starts with a remarkable fascinating story about the liquefaction of helium by Heike Kamerlingh Onnes in 1908, and the discovery of superconductivity by Heike Kamerlingh Onnes on April 8, 1911. In 1993, Viktor O. Ledenyov graduated from V. N. Karazin Kharkov

National University and conducted the advance research projects in the field of quantum computing, sublimated by his spirit of tireless innovation, at the National Scientific Centre Kharkov Institute of Physics and Technology in Kharkov, Ukraine. In 1998, Viktor O. Ledenyov accepted an invitation to make an invited highly innovative talk: "Measurements of Magnetic Flux Qubit by dc-SQUID" at Kamerling Onnes Laboratory Seminar at University of Leiden in Leiden in The Netherlands, inspiring the Dutch scientists with big names to begin their intensive research projects in the field of quantum computing with the superconductors. Dimitri O. Ledenyov visited the Kamerling Onnes Laboratory at University of Leiden in Leiden in The Netherlands in 1999, learning about the historical facts on the discovery of superconductivity during the active research efforts spearheaded by Heike Kamerlingh Onnes in 1911. Therefore, the authors are able to share their memories on some little known details about the superconductivity discovery with the readers, telling the fascinating story, which depicts the inspiring atmosphere of innovative scientific thinking at University of Leiden in Leiden in The Netherlands in the beginning of XX century.

Considering the microwave superconductivity topics, the authors accent reader's attention on the fact that very important experimental researches on properties of Low Temperature Superconductors (LTS) under application of high frequency alternate current flowing in the lead were firstly initiated and successfully completed by J. McLennan, A. C. Burton, A. Pitt, J. O. Wilhelm, at University of Toronto, Toronto, Canada in 1931-1932. One of the authors, Viktor O. Ledenyov, spent more than two years in 1998-1999 and 2005-2006, learning about the unnecessary forgotten historical developments in the field of microwave superconductivity at University of Toronto, Toronto, Canada, which took place in the beginning of the XX century. The authors think that a young generation of the European, Asian, North American scientists have no clue about the outstanding experimental research results, obtained by great scientist, J. McLennan, who, in our view, created the research foundation of microwave superconductivity.

Reading the excited story about the research on the superconductivity at ultra high frequency of $1.8\times10^{10}$Hz with the use of transmission line measurement method, made by Galkin and Lazarev in Kharkov, Ukraine in 1946, as well as the research on the frequency dependence of the surface resistance isotherms in Low Temperature Superconductors (LTS) in the frequency rage $(1.88 \div 4.5)\times10^{10}$Hz, completed by Galkin, Bezugly in Kharkov, Ukraine in 1954, the readers of our book will certainly learn some information about the world level research results in the field of microwave superconductivity, obtained in Ukraine on that time. Interestingly, the research achievements by Ukrainian scientists were partly described by an Australian scientist, John M. Blatt, in his well known book titled: “The Theory of Superconductivity.”

Discussing the theoretical foundations of high temperature superconductivity in HTS crystals and films, the authors make their emphasis on the clear identification of research contributions in the field of superconductivity made by different scientists in the World. In our opinion, many well known modern textbooks on the physics of superconductivity tend to be slightly biased, when describing the Bardeen, L. Cooper, and J. Schrieffer (BCS) theory to explain the nature of High Temperature Superconductivity, focusing on the groundbreaking research ideas by American physicists J. Bardeen, L. Cooper, and J. Schrieffer in 1957 only, and omitting the key theoretical propositions, made by Australian and European scientists early. Therefore, in our view, it may be of certain interest for the readers to learn about some little known interesting facts that, in 1946, Richard A. Ogg from Australia made a first ever suggestion of electron pairs and their Bose-Einstein condensation. In addition, the forgotten story on the creation of concept of the Fröhlich shell distribution of electrons in ***k***-space by Prof. H. Fröhlich of the University of Liverpool, U.K. is well described in this book.

Over the years, the understanding of nonlinearities in microwave superconductivity has been a formidable challenge for physicists, chemists, mathematicians and engineers. Most of the research attempts, directed to create the complete theories to explain the nonlinearities nature in microwave superconductivity, were not successful. The recent theoretical treatments of nonlinearities, including the Dahm-Scalapino theory and the Agassi-Oates theory, proved to be misleading and incorrect, because they were based on the incomplete knowledge, premises and assumptions about the main causes of nonlinearities origination in microwave superconductivity. For example, the Dahm-Scalapino theory implies a local electrodynamics for the nonlinear Meissner effect (NLME) in HTS thin films, concluding that the nonlinearities are caused by contributions from the current at the strip edges. On the other hand, the Agassi-Oates theory implies a nonlocal electrodynamics for the nonlinear Meissner effect (NLME), concluding that the nonlinearities are caused by contributions from the current at the strip midsection. However, the magnitude of experimentally observed nonlinear Meissner effect (NLME) is very small, hence its contribution to the nonlinearities origination is negligible, and it can not be considered a main source or cause of nonlinearities origination in microwave superconductivity, thus the DS and AO theories wrongly stipulate the assumption about a main role of the nonlinear Meissner effect (NLME) in the origination of nonlinearities in superconductors at high levels of applied microwave power. The author's critical opinion on the Dahm-Scalapino theory and the Agassi-Oates theory, which lack the physical grounds and have no derived Hamiltonian equations, is shared by many scientists at leading universities worldwide, because it is not possible to model the nonlinearities in the HTS thin films with the use of the DS and AO theories due to the fact that the computer modeling results can not approximate the obtained experimental curves closely. Therefore, the quest for a new theory, which can scientifically explain the nature of nonlinearities origination in superconductors at high levels of applied microwave power, and make it possible to model the nonlinearities, predicting their physical behaviour in superconductors at microwaves accurately, is going on now.

The Microwave and Electronic Material Research Group at Electrical and Computer Engineering Department, James Cook University, Townsville, Australia, established and led by Prof. Janina E. Mazierska, mainly focuses on the accurate characterisation of physical properties of HTS thin films at microwaves for application in HTS microwave filters in wireless communications. The recently published review by J. E. Mazierska and J. Krupka, presents some insights on the main research results obtained, describing the Resonant Techniques for the Complex Permittivity and Conductivity Measurements of Materials at Microwaves:

1. Principles and Limits of Loss Measurements of Microwave Resonance Techniques.
2. Measurements of Ferroelectrics and Conductive Materials Employing Composite Dielectric Resonators Technique.
3. Single Post Dielectric Resonators for Measurement of Metals and Semiconductors.
4. Split Post Dielectric Resonator Technique for Measurements of Thin Ferroelectric Films.

Among a big number of research results obtained by the Microwave and Electronic Material Research Group, it makes a lot of sense to highlight the important proposition by J. E. Mazierska to use the dielectric resonators as a possible standard for microwave characterisation of high temperature superconducting (HTS) films for microwave applications. All the gained knowledge by Microwave and Electronic Material Research Group at James Cook University was extensively used during the writing of the Dimitri O. Ledenyov's dissertation and this book.

Dimitri O. Ledenyov focused his main research interest on the advanced innovative research towards the modeling of nonlinear properties of High Temperature Superconductor (HTS) thin films, using the Bardeen, Cooper, Schrieffer and Lumped Element Circuit theories, with purpose to enhance microwave power handling capabilities of microwave filters and optimize design of microwave circuits in micro- and nano-electronics. The main experimental and theoretical researches are completed by Dimitri O. Ledenyov in the following fields:

1) precise characterization of microwave properties of MgO substrates in a split post dielectric resonator;

2) precise characterization of nonlinear surface resistance of $YBa_2Cu_3O_{7-\delta}$ thin films on MgO substrates in a dielectric resonator at ultra high frequencies;

3) precise characterization of nonlinear surface resistance of $YBa_2Cu_3O_{7-\delta}$ thin films on MgO substrates in superconducting microstrip resonators at ultra high frequencies.

4) Nature of noise in superconducting microwave resonators, including the microstrip resonators;

5) Design of High Temperature Superconducting (HTS) microwave filters with exceptional microwave power handling capabilities and low Intermodulation Distortions (IMD) for application in Cryogenic Transceiver Front End (CTFE) in wireless communication systems.

During the course of highly innovative research by Dimitri O. Ledenyov, the new lumped element models of a superconductor at microwaves are proposed. It makes sense to note that the accurate modeling and identification of lumped element model parameters for the superconducting Hakki-Coleman dielectric resonator is completed. The new formula for the accuracy calculation for completed experimental measurements is derived. The new original Ledenyov theory on the nonlinearities origination in superconductors at microwaves is proposed by Dimitri O. Ledenyov and Viktor O. Ledenyov. The Ledenyov theory is based on the theoretical assumption about a great influence by the Abricosov magnetic vortices on the physical mechanisms of nonlinearities origination in superconductors at high level of applied microwave power. The Ledenyov theory proposes the new formula for the surface resistance calculation $R_S(H_e)$, which takes to the consideration an influence by the Abricosov magnetic vortices on the physical mechanisms of nonlinearities origination in microwave superconductivity; and uses the modified formulas for the surface resistance calculation, derived from the Bardeen, Cooper, Schrieffer theory. The accurate modeling of nonlinear dependence of surface resistance on external magnetic field $R_S(H_e)$ in proximity to the critical magnetic fields $H_{c1}$ and $H_{c2}$ in HTS microstrip resonators at ultra high frequencies, using the Ledenyov theory with the modified formulas for the surface resistance calculation, derived from the Bardeen,

Cooper, Schrieffer theory, is performed. The obtained computer modeling results, using the Ledenyov theory, are in a good agreement with the experimental measurements data, obtained by Dimitri O. Ledenyov at JCU as well as with the experimental research data, obtained by other researchers at different universities. The research on newly discovered differential noise, including its modeling and suppression in HTS microwave filters is done. The definition of the magnetic dipole two-level systems (MTLS) is introduced firstly. The new quantum theory of *1/f* noise origin by the magnetic dipole two-level systems (MTLS) in superconducting microwave resonators is firstly formulated. The modern designs of HTS microstrip filters are reviewed with particular focus on the unique trimming and tuning techniques, used by different research groups. The extensive bibliography on the topic of research interest is also created.

Finally, Viktor O. Ledenyov's research work at Department of Physics, Technical University of Denmark, Lyngby, Denmark in 1995, 1996-1997 provided a useful insight on the modern research approaches to the understanding of nonlinear dynamics in complex systems, including the microwave resonators, extensively used by Chaos Group at Center for Modeling, Nonlinear Dynamics and Irreversible Thermodynamics (MIDIT) at Technical University of Denmark Lyngby, Denmark since early 1980's. The Danish period of innovative research thinking by Viktor O. Ledenyov resulted in the derivation of important research results on the "Measurements of Magnetic Flux Qubit by dc-SQUID", and it had a profound effect on the invention of the "Quantum Random Number Generator on Magnetic Flux Qubits" in 1996-97. At the same time, Viktor's frequent visits to participate in seminars at Niels Bohr Institute for Theoretical Physics and at University of Copenhagen in Copenhagen, Denmark in 1997, helped to better understand the Copenhagen interpretation of quantum theory by Niels Bohr, which is based on the analysis of the nature of measurement. This knowledge was mainly applied to model the nonlinear interactions between the electromagnetic waves in microwave resonators, using the parallel computing techniques, described in the book shortly.

Dimitri Olegovych Ledenyov, Townsville, Gold Coast and Brisbane in Queensland, Australia.
Viktor Olegovych Ledenyov, Kharkov, Ukraine.

March, 2014.

**Abstract of Dissertation.:** The aim of the thesis is to research the nonlinear properties of high temperature superconducting (*HTS*) thin films for application in electronic devices in information communication technologies at microwaves.

***Theoretical research consisted of:***

**1.** Analysis of present state of theoretical research on the superconductivity with focus on the electromagnetic and nonlinear properties of *HTS* materials at microwaves in Chapters 1, 2 and 3.

**2.** Creation of both a *lumped element model of a superconductor* and a *lumped element model of a microwave resonator* for accurate characterisation of nonlinear properties of *HTS* thin films in microwave resonators in Chapters 4. The consideration of limitations of *r*- parameter application for nonlinear models analysis in *HTS* microwave resonators is also described in Chapter 4.

**3.** Chapter 5 deals with modeling of nonlinear microwave responses of system with three types of *R* and *L* element dependences *(linear, quadratic & exponential*). The proposed lumped element model adequately describes the nonlinear properties of researched *HTS* thin films in ***Hakki-Coleman dielectric resonator*** **(*HCDR*).**

**4.** Chapter 6 focuses on the $R_s(T)$ and $R_s(P)$ dependences and theoretical consideration of measurement accuracy issues during accurate characterisation of *HTS* thin films in the *HCDR.*

**5.** Chapter 7 concentrates on the creation of a simplified microwave model, based on the Bardeen, Cooper, Schrieffer (*BCS)* theory, to determine nonlinear behaviour of *Rs(P)* of *HTS* thin films near $H_{c1}$ and $H_{c2}$ in ***microstrip resonator***. The simulated *S*-type dependences *Rs(P)* closely approximate the measured experimental results.

**6.** Chapter 8 presents research on noise in *HTS* microwave resonators with original proposals on:

a) New source of noise: *the magnetic dipole two-level systems (MTLS)* in *HTS* thin films;

b) New quantum theory of *1/f* noise in *HTS* microwave resonators;

c) Modeling of *differential noise* in $YBa_2Cu_3O_{7-\delta}$ thin films at microwaves.

***Experimental research included:***

**1.** Research on the microwave properties of *MgO* substrates in ***split post dielectric resonator*** (***SPDR***) at *f* = 10.48*GHz*. It is understood that the *MgO* substrate is not a source of nonlinearities in *HTS* thin films at microwaves in operational temperatures range in Chapter 6.

**2.** Research on the nonlinear surface resistance of $YBa_2Cu_3O_{7-\delta}$ thin films on *MgO* substrates in ***Hakki-Coleman dielectric resonator*** (***HCDR***) at $f = 25GHz$. It was observed that $YBa_2Cu_3O_{7-\delta}$ thin films have nonlinear characteristics in form of *S*-type dependence with increase of *Rs(P)* at elevated power levels, when $H_{rf}$ is higher than $H_{c1}$ in Chapter 6.

**3.** Research on nonlinear surface resistance of $YBa_2Cu_3O_{7-\delta}$ superconductor thin films on *MgO* substrates in ***microstrip resonators*** at $f_0$ = 1.985*GHz*. *Microstrip resonators* expressed *S*-type nonlinearity at the same power levels, when $H_{rf}$ is higher than $H_{c1}$ in Chapter 7.

***Thesis's conclusion*** is that the magnitude of critical magnetic field $H_{c1}$ of a superconductor needs to be firstly measured to predict the nonlinear behaviour of any *HTS* thin films, because the $H_{c1}$ is the level of magnitude of magnetic fields at which the nonlinear effects arise. The developed software program in *Matlab* can accurately model the nonlinearities in *HTS* thin films at microwaves, using the measured critical magnetic fields $H_{c1}$ and $H_{c2}$.

## Hypothesis of Dissertation.

The experimental and theoretical researches of nonlinear properties of high temperature superconductor (*HTS*) thin films at microwaves, including the modeling of nonlinearities using the *Bardeen, Cooper, Schrieffer (BCS)* and *Lumped Element Circuit* theories, to create the *HTS* passive/active microwave devices with advanced characteristics for application in microwave circuits in micro- and nano-electronics, constitute the core of dissertation. The main aim of the experimental research is to accurately measure the nonlinear dependences of surface resistance on microwave power $R_s(P)$ and on temperature of $R_s(T)$ of *HTS* thin films at elevated microwave powers at low temperatures. The second goal is to create a simple empirical physical model, which can closely describe the nonlinear properties of *HTS* thin films at microwaves in order to predict the nonlinear behaviour of superconducting materials for their possible use in wireless communication technologies.

To achieve these research objectives, $YBa_2Cu_3O_{7-\delta}$ thin films on *MgO* substrates were researched in the two types of microwave resonators. Firstly, the research on nonlinear surface resistance of $YBa_2Cu_3O_{7-\delta}$ thin films on *MgO* substrates was conducted in *Hakki-Coleman dielectric resonator (HCDR)* at the frequency of 25*GHz*. Secondly, the research on nonlinear surface resistance of $YBa_2Cu_3O_{7-\delta}$ thin films on *MgO* substrates was completed in *microstrip resonators* at the frequency of 1.985*GHz*. Microwave power *P* was changing from -30 *dBm* up to 25 *dBm*. Temperature *T* was from 15*K* to 90*K*. Thirdly, the research on microwave properties of *MgO* substrates in *split post dielectric resonator (SPDR)* was done at the frequency of 10.48*GHz*.

It was found that $YBa_2Cu_3O_{7-\delta}$ thin films have nonlinear characteristics in form of *S*-type dependence of surface resistance on microwave power *Rs(P)* at elevated microwave power levels, when $H_{rf}$ is higher than $H_{c1}$, during the measurements conducted in the *Hakki-Coleman dielectric resonator (HCDR)*. *Microstrip resonators* of $YBa_2Cu_3O_{7-\delta}$ also expressed *S*-type nonlinearity at elevated microwave power levels, when $H_{rf}$ is higher than $H_{c1}$. It was concluded that *MgO* substrates did not contribute to nonlinear properties of $YBa_2Cu_3O_{7-\delta}$ thin films on *MgO* substrates in the given microwave power and temperature ranges.

The theoretical research also concentrated on the creation of the two models using the theory of lumped element *RLC* circuits, and then the *BCS* theory to model microwave responses of the system. Simulation of nonlinear microwave responses of system with the three types of *R* and *L* element dependences (linear, quadratic & exponential) showed that the lumped element approach worked well for certain system parameters, but did not universally describe the nonlinear properties of *HTS* materials at ultra high frequencies (*UHF*). Simulation results demonstrated that the nonlinear behaviour of *Rs(P)* dependence near magnetic field $H_{c1}$ can be closely approximated by the *BCS* theory. Therefore, it is assumed that the high temperature superconductor undergoes the *Type II* phase transition, and turns into the mixed state, when the *Abricosov magnetic quantum vortices* begin to penetrate inside the *HTS* in magnetic field $H_{c1}$.

Thesis's main conclusion is that the magnitude of critical magnetic field $H_{c1}$ of a superconductor needs to be firstly measured to predict the nonlinear behaviour of *HTS* thin films with the *BCS* model, because the critical magnetic field $H_{c1}$ represents the level of magnitude of magnetic field at which the nonlinear effects begin to appear in $YBa_2Cu_3O_{7-\delta}$ thin films at microwaves. The developed software program in *Matlab*, *Maple*, and *Origin* can accurately model the nonlinearities in *HTS* thin films at microwaves, using the measured critical magnetic fields $H_{c1}$ and $H_{c2}$.

The successful application of *HTS* thin films in emerging microwave device technologies makes it possible to achieve the better technical characteristics of microwave device in linear regime of its operation, when the magnitude and phase changes are only imposed on complex input signals without the new signals generation. The problem is that microwave devices in nonlinear mode of operation can shift the complex input signals in frequency domain and/or generate new signals, including the harmonics and intermodulation products causing the severe signal distortion. Therefore, the determination of magnitude of critical magnetic field $H_{c1}$ of *HTS* thin film can provide the exact data on threshold magnitudes of microwave power at which the microwave device with *HTS* thin film will begin to exhibit the nonlinear behaviour, if driven with elevated microwave input signal power. The better linear specifications of microwave devices can be achieved by improving the quality of *HTS* thin film synthesis; by the use of improved *HTS* microwave filter designs with embedded ferroelectric; and by precisely controlling and maintaining the applied level of microwave input signal power and operational temperature of microwave component to gain the stable functional performance of microwave devices with *HTS* thin films in wireless communication systems and radars at ultra high frequencies.

# Book Contents.

**CHAPTER 5. DEVELOPMENT OF A LUMPED ELEMENT MODEL FOR ACCURATE MICROWAVE CHARACTERISATION OF SUPERCONDUCTORS.**

**CHAPTER 6. EXPERIMENTAL AND THEORETICAL RESEARCHES ON MICROWAVE PROPERTIES OF MgO SUBSTRATES IN A SPLIT POST DIELECTRIC RESONATOR AND NONLINEAR SURFACE RESISTANCE OF $YBa_2Cu_3O_{7-\delta}$ THIN FILMS ON MgO SUBSTRATES IN A DIELECTRIC RESONATOR AT ULTRA HIGH FREQUENCIES.**

# CHAPTER 1

## 1.1. INTRODUCTION.

The *analytic theory of heat*, which considers the *phenomena of conductive diffusion of heat*, had been created by Jean-Baptiste Joseph Fourier in Paris, France in 1822 [1, 2]. The *heat propagation in the solids* had been researched by Viktor Ya. Bunyakovsky and Augustin-Louis Cauchy in Paris, France in 1825 [3]. It was found that the *physical properties of solids* change with the *varying temperature* [3]. The *cooling down of* the mercury to the low temperatures of approximately 3*K* led to the discovery of the *superconductivity phenomena* in the *condensed matter* by Prof. Heike Kamerlingh Onnes and his research collaborators Cornelis Dorsman, Gerrit Jan Flim, and Gilles Holst, at Leiden University in Leiden in The Netherlands on April 8, 1911 [4]. The year 2011 marked 100$^{th}$ anniversary of the discovery of superconductivity, and this date was celebrated by many research institutions, including the Institute of Electrical and Electronics Engineering (*IEEE*) all around the world [5]. Professor Heike Kamerlingh Onnes won a *Nobel Prize* in the field of physics for his research on the liquefaction of *helium* and the discovery of *superconductivity* in 1913: "for his investigations on the properties of matter at low temperatures which led, inter alia, to the production of liquid helium" [22]. Fig. 1 presents some information on the discovery of superconductivity. The *IEEE* awarded the superconductivity discovery, the milestone in the electrical and computer engineering, a title that honours major technical innovations for benefits of humanity. Basic phenomenon behind the superconductivity lies in the disappearance of electrical resistance what enables materials to conduct the *dc* electrical current without any losses. Prof. Heike Kamerlingh Onnes discovered the effect for mercury, when cooling to very low temperatures of approximately 3*K*. During the next *75* years a number of other Low Temperature Superconductors (*LTS*) had been discovered with the element tungsten having the lowest critical temperature $T_C$ of 0.016*K*, and the inter-metallic compound $Nb_3Ge$ showing the highest $T_C$ of 23.2*K*. The main challenge for practical applications was to achieve the highest critical temperature possible to simplify the cryogenic systems needed to maintain superconductors in a temperatures range below the critical temperature $Tc$.

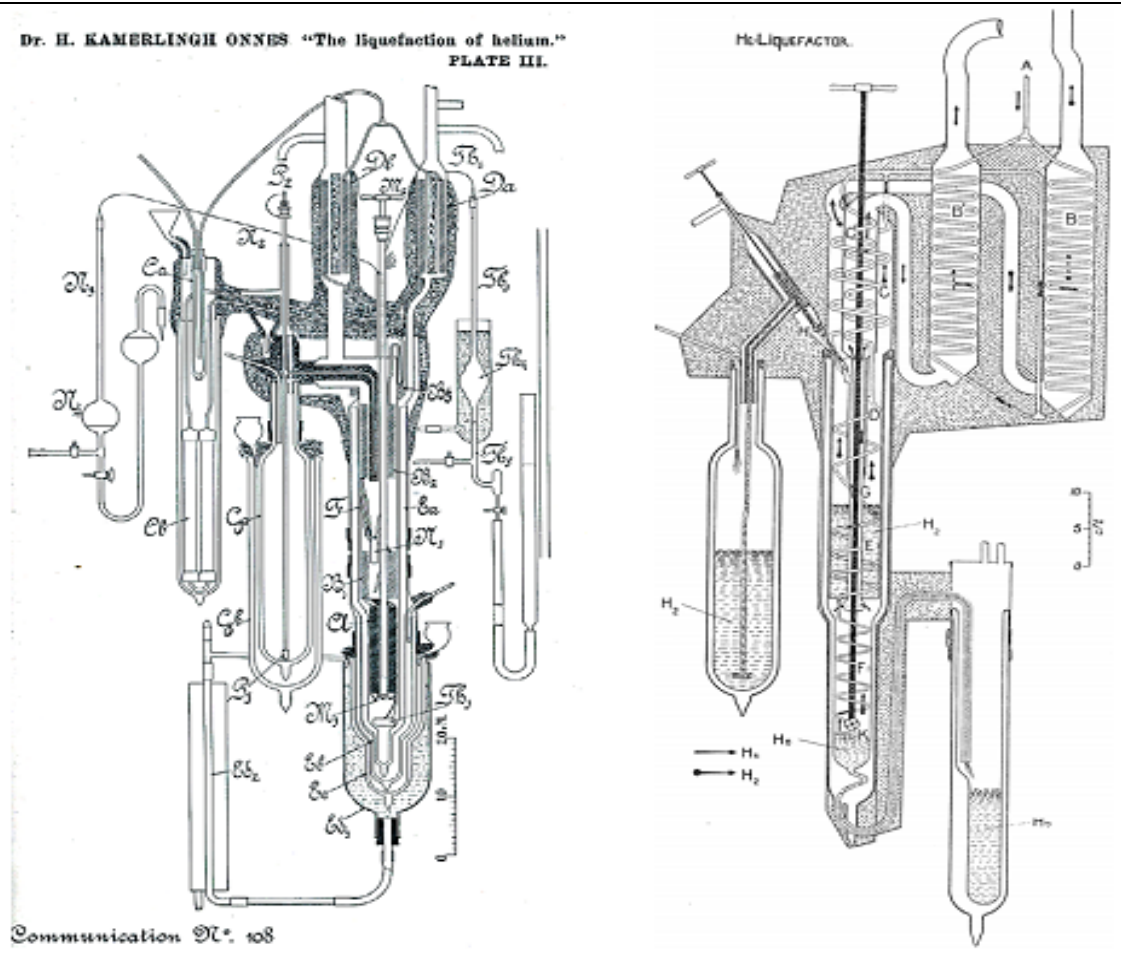

1 2

**Fig. 1. (1)** A schematic drawing of the first helium liquefier (1908) (after [22]). For full details see *Communications Physics Lab University Leiden* no 108. Production: 0.28 *l* liquid $h^{-1}$ (1908), in the improved version (1912): 0.5 *l* liquid $h^{-1}$. See **Fig. 2** with a photograph of the actual liquefier (after [24]).
**(2)** Schematic drawing of the first helium liquefier (1908) with the helium cryostat (1911) attached to the liquefier [24] (after [25]).

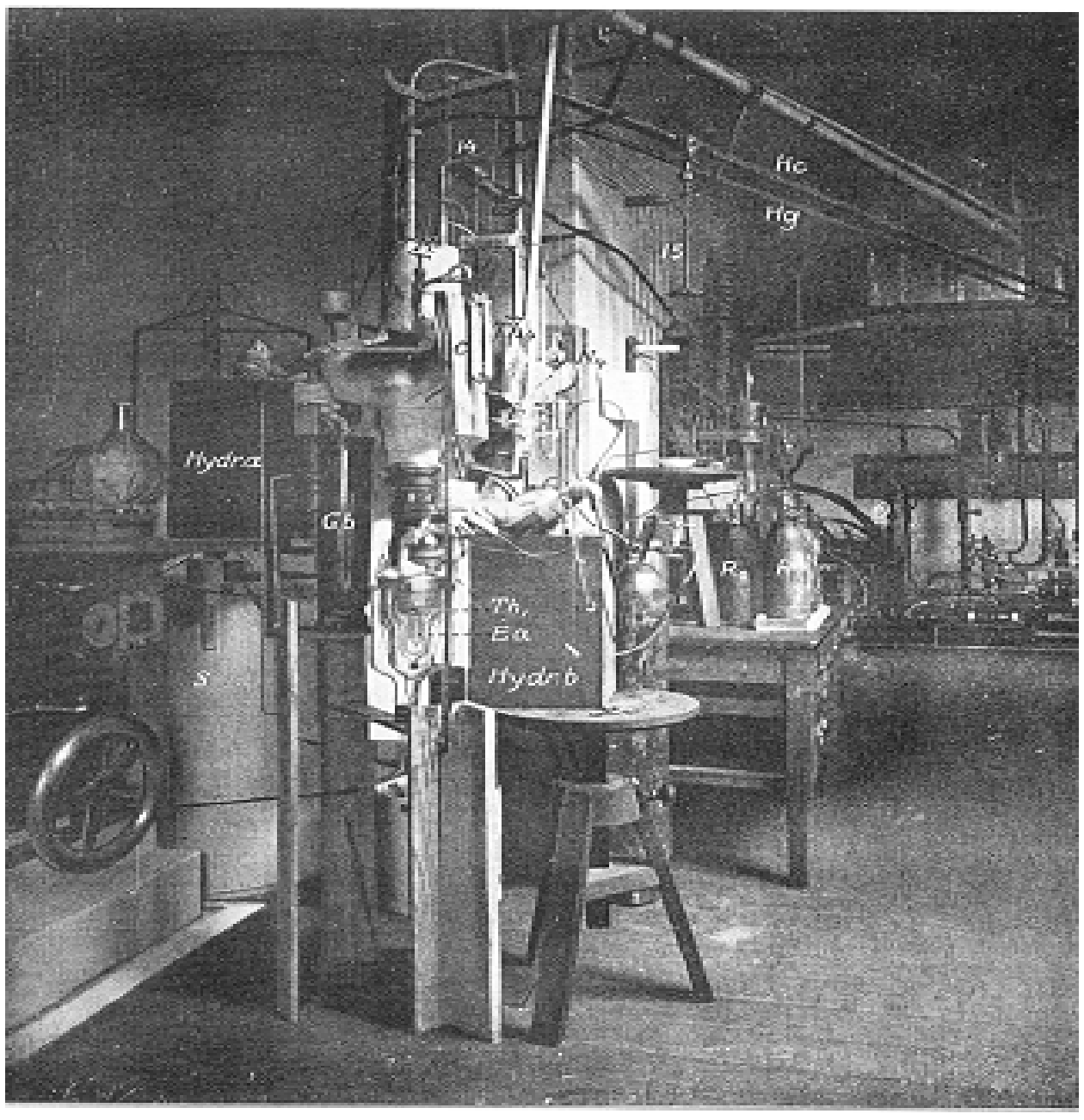

**Fig. 2.** The helium liquefactor (1908) (after [23, 25]). It has been arranged so that The liquid helium could be observed. Round the transparent bottom part of the vacuum glass Ea, a protection of liquid hydrogen (a second vacuum glass Eb, see **Fig. 1 (1)** has been applied. The hydrogen glass Eb is surrounded by a vacuum glass Ec with liquid air, which in its turn is surrounded by a glass Ed with alcohol (**Fig. 1 (1)**. 'By these contrivances and the extreme purity of helium we succeeded in keeping the apparatus perfectly transparent to the end of the experiment'. In **Fig. 1 (1)** is also drawn the German silver reservoir Th1 of the helium gas thermometer [25].

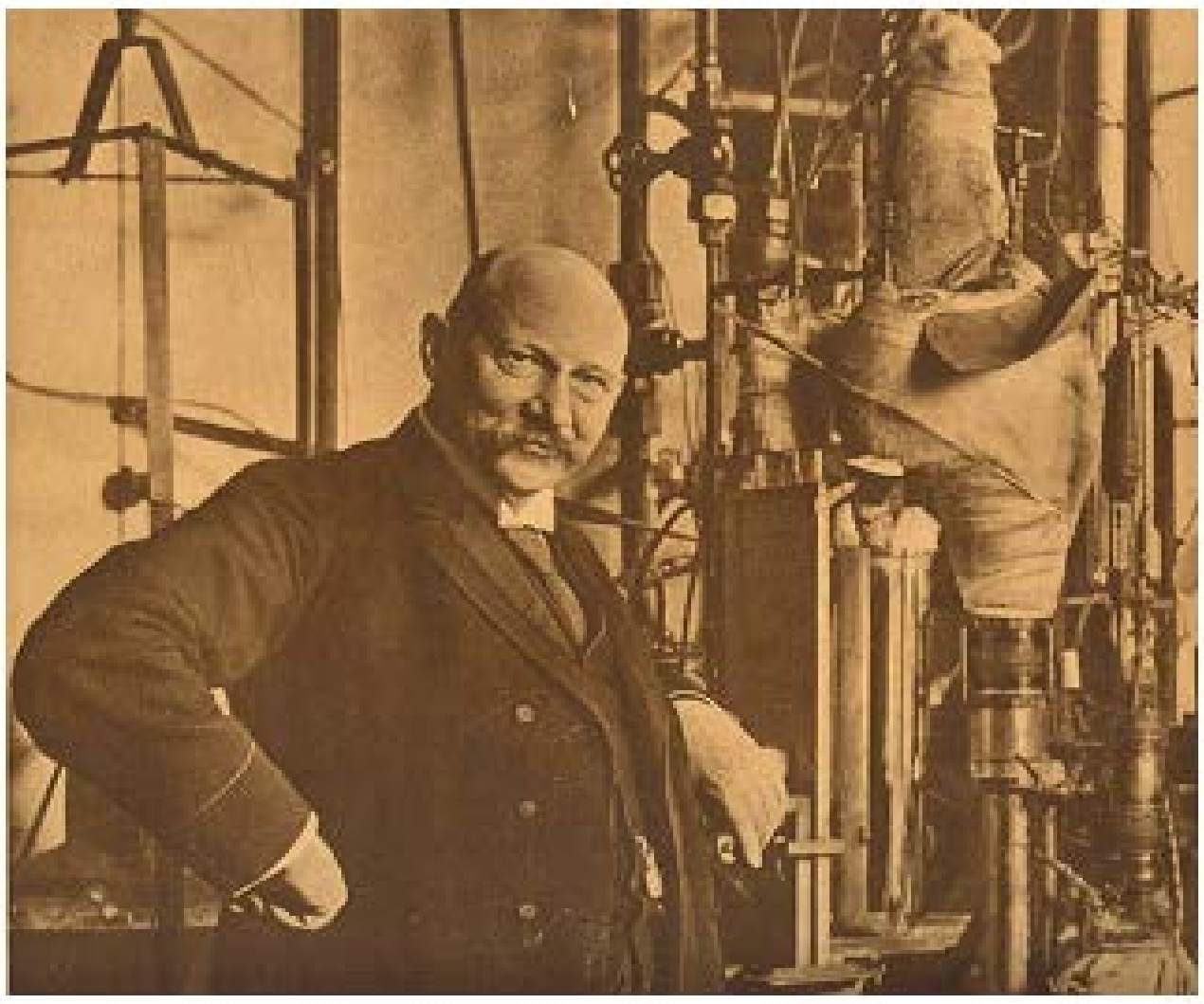

**Fig. 3.** Prof. Heike Kamerlingh Onnes (1853–1926), standing in front of the helium liquefier (Photograph is reproduced with permission, archives Leiden Institute of Physics, Leiden, The Netherlands) (after [25]).

| | |
|---|---|
| 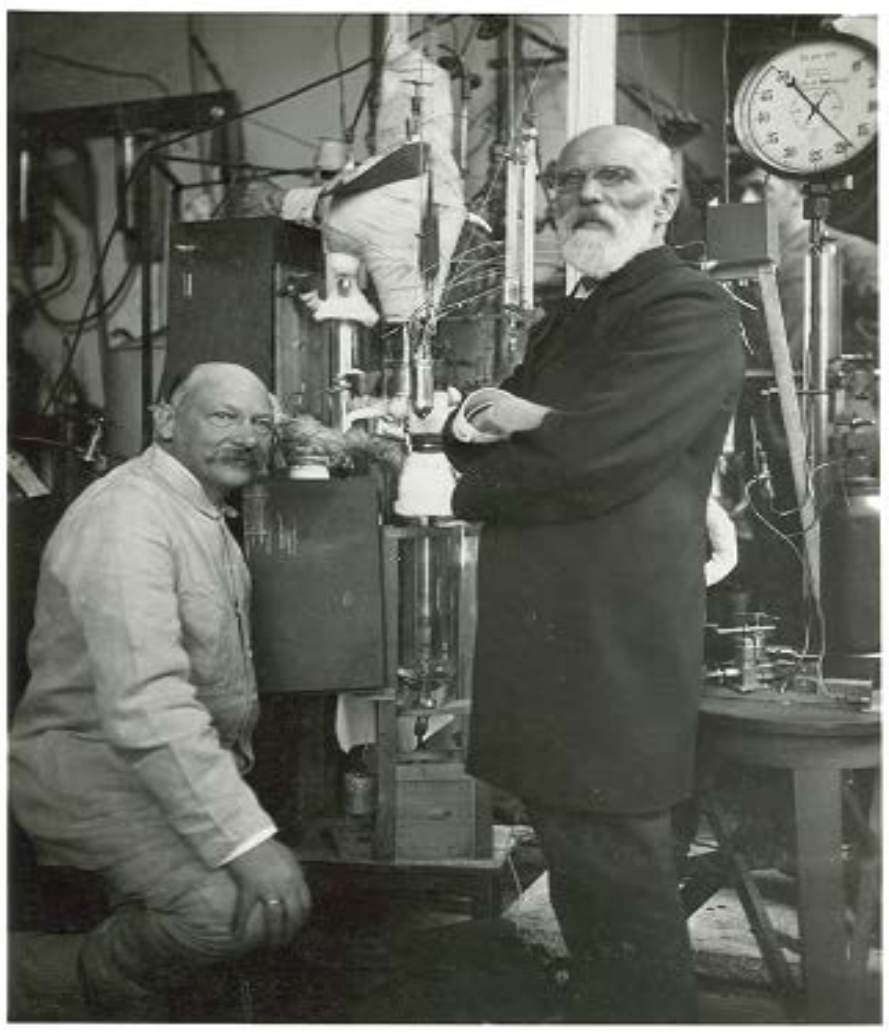 | **Fig. 4.** Profs. Heike Kamerlingh Onnes (left) and Van der Waals posing by the helium liquefier in 1911. (Photograph is courtesy of the Leiden Academisch Historisch Museum, Leiden, The Netherlands, and reprinted with permission Leiden Academisch Historisch Museum, Leiden, The Netherlands) (after [26]). |
| 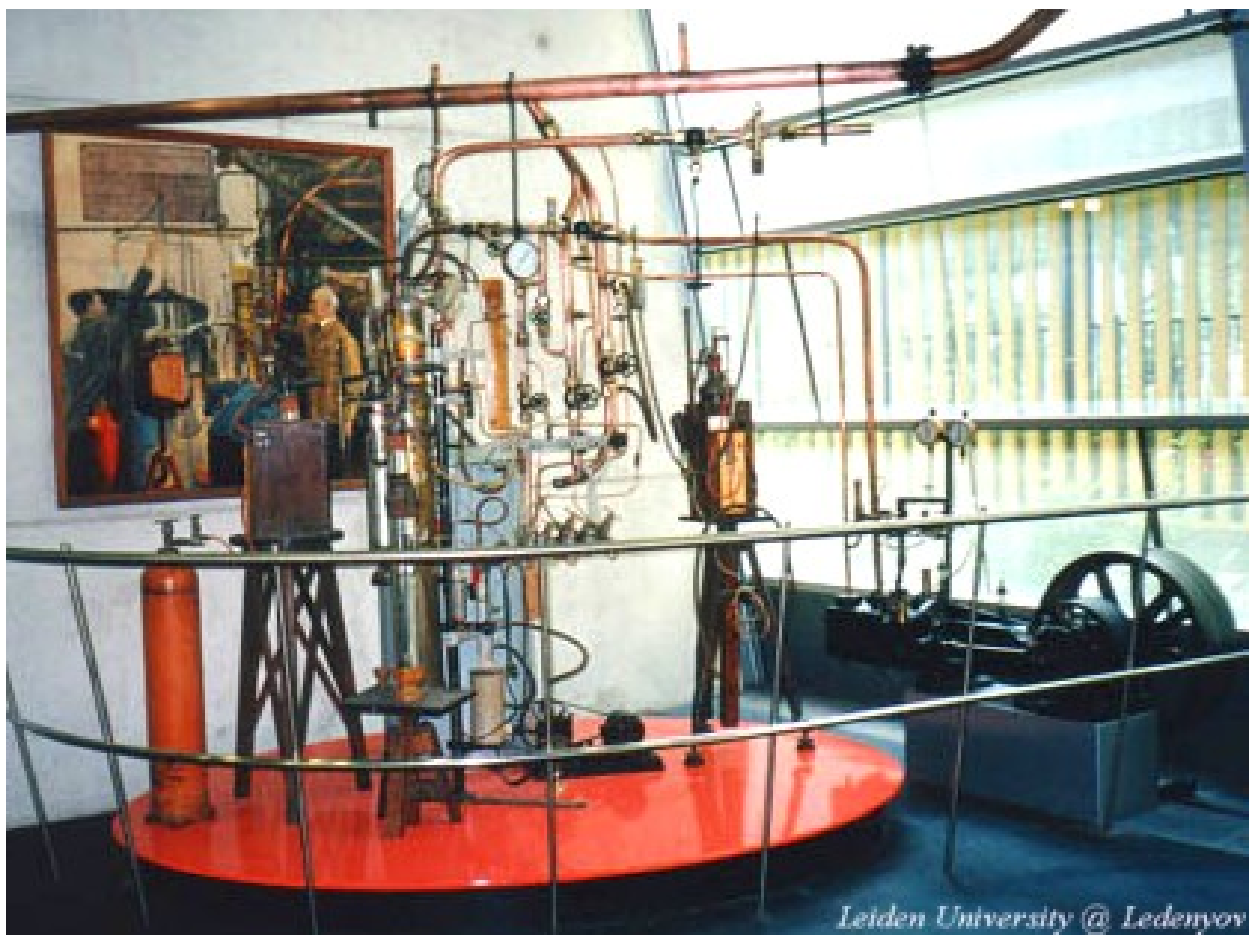  | **Fig. 5.** Experimental equipment set up used by Professor Heike Kamerlingh Onnes and his research collaborators, Cornelis Dorsman, Gerrit Jan Flim, and Gilles Holst to discover the superconductivity phenomena at Leiden University, Leiden, The Netherlands on April 8, 1911. The photo is taken by the author of dissertation during his invited visit to the New Building at Department of Physics, Leiden University, Leiden, The Netherlands in 1999. The visit was organized by Prof. Peter H. Kes, Dean of Department of Physics, Leiden University, Leiden, The Netherlands. |
| 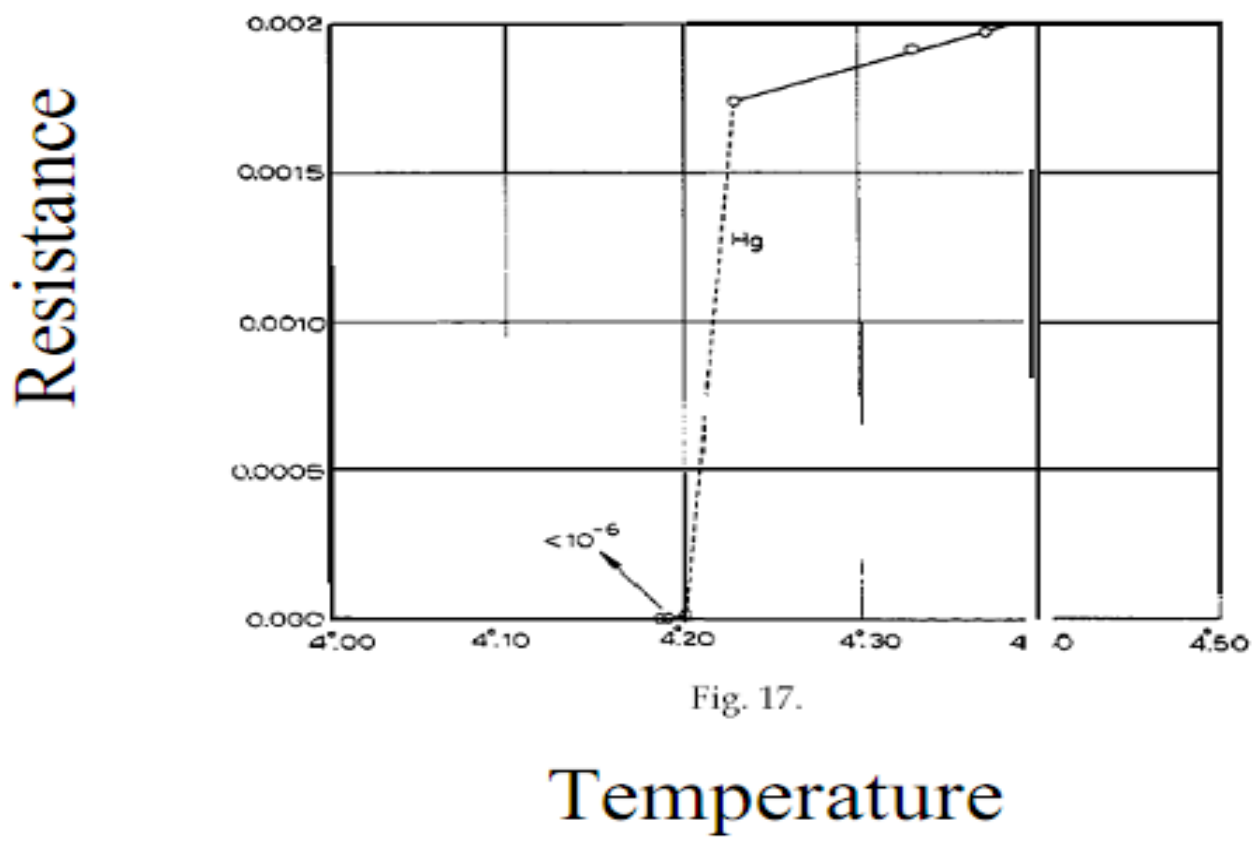  | **Fig. 6.** Discovery of superconductivity in mercury by Professor Heike Kamerlingh Onnes and his research collaborators, Cornelis Dorsman, Gerrit Jan Flim, and Gilles Holst, at Leiden University in Leiden in The Netherlands on April 8, 1911 (after [4]). |

**Tab. 1.** Historical information on the liquefaction helium and discovery of superconductivity phenomena by Prof. Heike Kamerlingh Onnes at Leiden University in Leiden, The Netherlands.

In April 1986, a new class of materials $Ba_xLa_{5-x}Cu_5O_{5(3-y)}$– the *complex layered copper oxide compounds* - was discovered by Bednorz and Muller at IBM Research Lab in Switzerland [6], using some research data on Ba-La-Cu-O system by Michel, Raveau [51]. The critical temperature *Tc* of $La_{2-x}Ba_xCuO_4$ compound was 35*K*, what was more than 10*K* higher than the highest critical temperature *Tc* known previously. This very significant increase in the critical temperature spurred widespread research that quickly led to the discovery of materials, known as High Temperature Superconductors (*HTS*), such as $YBa_2Cu_3O_{7-\delta}$ (*YBCO*) compound with $T_C$ of 93*K*, found by a group of scientists led by Wu of Alabama University in collaboration with Chu of Houston University in the U.S.A. in February, 1987 [7].

Since the discovery of High Temperature Superconductors (*HTS*), various scientific groups have been investigating microwave properties of *HTS* for application in microwave filters for cellular base station receivers and transmitters. The initial work concentrated on various aspects of measurements of the surface resistance *Rs* of superconductors and fabrication of *HTS* planar resonators [11-15]. Measurements of dependence of surface resistance on magnetic field $R_S(H_{rf})$ in *unpatterned HTS* films were performed using the cavities techniques with dielectric resonators [15], parallel plate resonators [16], or disk resonators [17]. Dielectric resonators were the preferred type for surface resistance *Rs* measurements, because there is no current *I* at the edges of the *HTS* film, and edge effects were eliminated. Dielectric resonator has been selected as a standard for *Rs* measurements in *HTS* thin films by International Electrochemical Commission (*IEC*). Many measurements of $R_S(H_{rf})$ were carried out with *patterned* films, implementing the techniques with common planar transmission-line geometries such as microstrip, coplanar waveguide, and stripline [18-19] due to the fact that many practical devices such as microwave filters and delay lines are fabricated from patterned films.

The distinct advantages of *HTS*-based planar microwave filters are in their very sharp bandpass, low insertion loss and noise as well as a compact size. These advanced technical characteristics of *HTS* microwave filters lead to a reduced adjacent band interference, increased coverage and better spectrum utilisation in comparison with other technologies of similar dimensions. Benefits of using the *HTS* technology to wireless communication can result in increased voice and data traffic, better coverage with fewer cells, better audio and video quality, less dropout calls, and hence financial benefits to telecommunications providers.

The Federal Government of Australia is committed to create a national broadband network, which will include the next generation optical and wireless networks with high data capacity and throughput, aiming to transform the Australian society [34]. Since the time of world's first commercially available mobile phone *Motorola Dynatac 8000X,* limited to voice applications in 1973-1983. The modern smart devices are emerging into the mini-computers (*Apple iPhone 6S+, Samsung Galaxy S6edge+ SM-G928F, Sony Xperia Z3, HTC One M8, LG G3*) that enable the end-users to browse high speed Internet with access to multimedia services and *GPS* navigation. The mobile data traffic is forecast to almost double annually over the coming years [35]. The development of telecommunication industry, which produces over five billion mobile phones around the world [10], creates a constant need for advanced electronic components with enhanced microwave properties. Superconductors exhibit much more superior microwave properties then the normal metals, and can be used to create the microwave devices with advanced characteristics [20]. The surface resistance of superconductors, and especially of *HTS* thin films, is much smaller than that of normal metals [8]. For example, $R_S$ of $YBa_2Cu_3O_{7-\delta}$ is around $Rs{\sim}1m\Omega$ at $f=10GHz$, $T=77K$, when the surface resistance $R_S$ of Copper is around $Rs{\sim}20m\Omega$ at $f=10GHz$ at $T=77K$. Due to the fact that $R_S$ of *HTS* thin films is small, the energy losses in superconductors in microwave resonators can be decreased, resulting in improved technical characteristics such as the better signal-to-noise ratio of microwave devices in communication systems [9].

The mass application of superconducting technology in telecommunication industry has not yet been achieved due to a number of factors. The Cryogenic Transceiver Front End (*CRFE*) systems with the *HTS* microwave filters and cooled Low Noise Amplifiers (*LNA*) seem to have large redundancy in terms of frequency selectivity and microwave signal amplification linearity [20]. Simpler technological solutions could be more economically feasible, including the microwave filters with dielectric resonators, in modern wireless communication transceivers. On the other hand, *CRFE* systems proved invaluable, when there is close adjacent channel interference, thus *CRFE* might be applicable with next generation wireless networks.

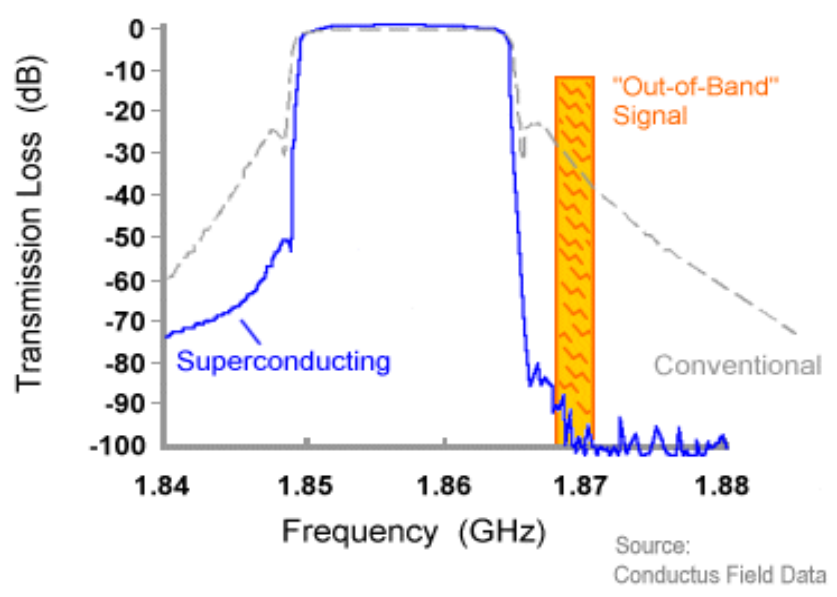

**Fig. 7.** Measured microwave characteristics of conventional metal microwave filters and High Temperature Superconducting (HTS) microwave filters in conditions of strong adjacent channel interference (after [34]).

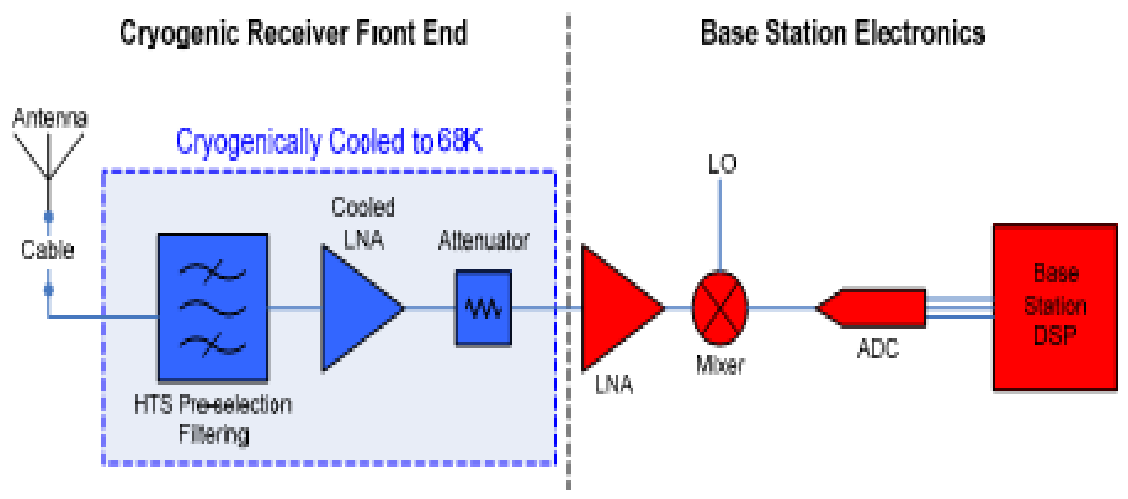

**Fig. 8.** Cryogenic Receiver Front Ends (*CRFE*) in base station (after [28, 29]).

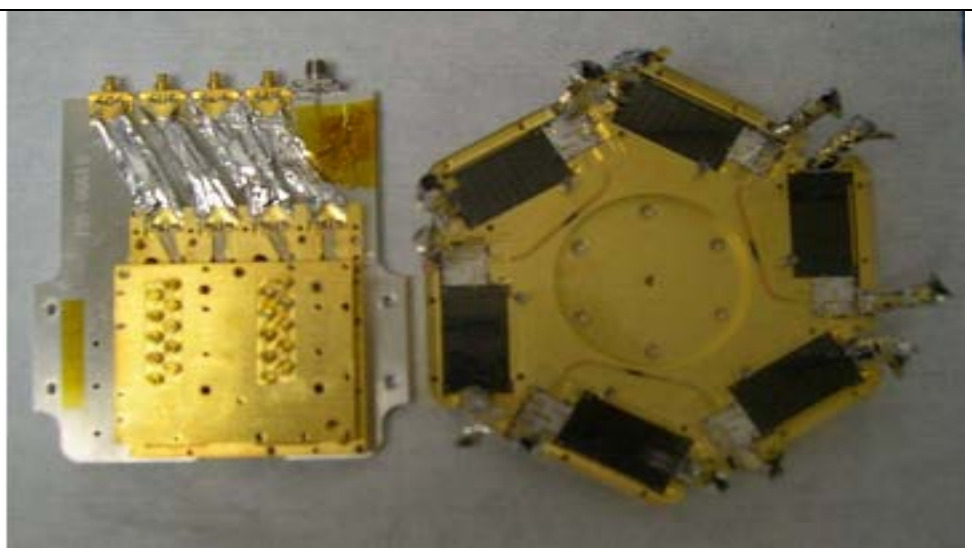

**Fig. 9.** Expanded views of one of the three microenclosures in a 2004 dewar (left) and a 2008 microenclosure [30].

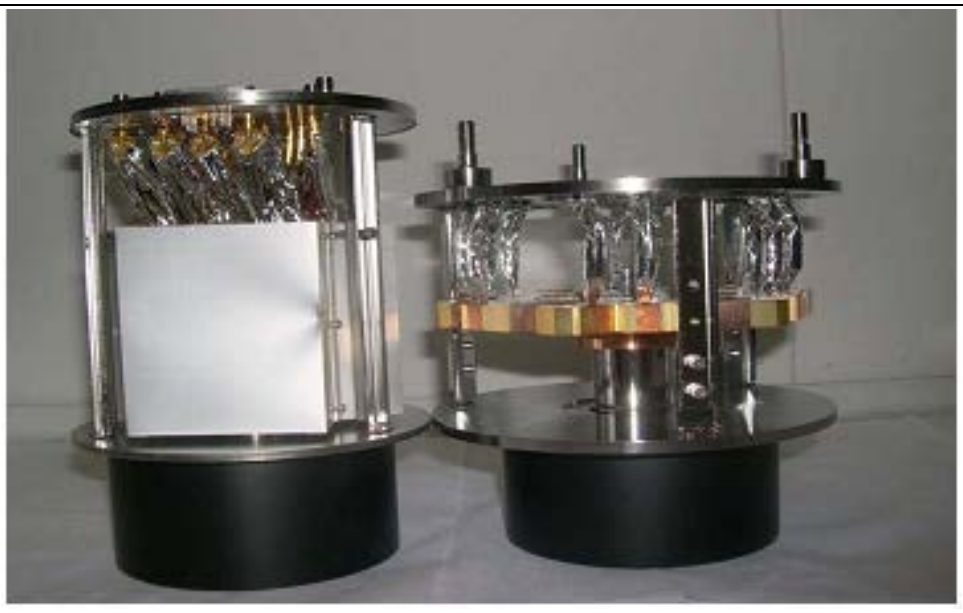

**Fig. 10.** Inside views of the cryogenic RF housings (microenclosures) inside 2004 dewar (left) and 2008 dewar (right). The 2008 dewar has been optimized to eliminate parts as compared to the 2004 design (after [30]).

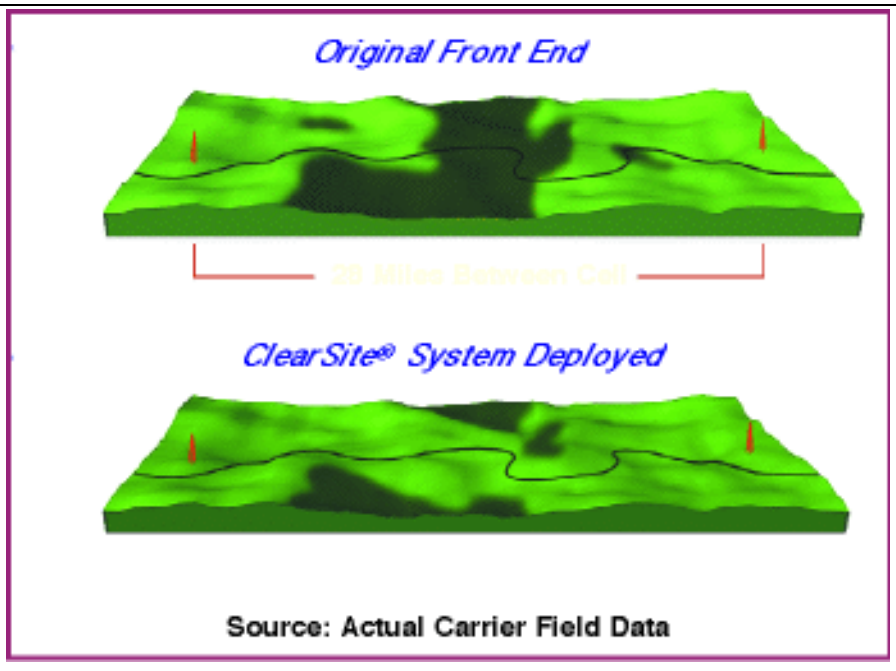

**Fig. 11.** Comparison of coverage before and after the HTS Cryogenic Transceiver Front End (CTFE) installation on a Californian highway in the U.S.A. (after [37]).

**Tab. 2.** Feasibility study of High Temperature Superconducting (*HTS*) Cryogenic Transceiver Front End (*CTFE*) application in modern basestations in 3G and 4G wireless communication networks.

In modern wireless communication systems with higher-data rates and/or longer-distance transmission requirements, the microwave filters in Cryogenic Transceiver Front Ends (*CTFE*) [28-30, 39-47] will be desired to satisfy the following characteristics in Yamanaka, Kurihara, Akasegawa [48]:

1. Higher power handling capability (*PHC*),
2. More compact dimensions, and
3. Wider tunability of resonance central frequency *fc*.

However, the *power handling capability* (*PHC*) of high temperature superconductors (*HTS*) in microwave applications is limited, because of the nonlinearities appearance at high levels of applied microwave power of modulated information carrier signal [21, 31] due to the following factors in Oates [21, 31]:

1. *Extrinsic effects* such as *grain boundaries, weak links,* or *impurities,* and
2. *Intrinsic effects* such as the *Cooper electron pair* breaking due to *Abricosov and Josephson magnetic vortices* generation.

The nonlinear increase of surface resistance with increase of microwave power of microwave signal at input port results in the appearance of:

1. Increased insertion loss of microwave filter,
2. Unwanted harmonics / sub-harmonics generation, and
3. Intermodulation distortions (*IMD*),

in measured spectrum of microwave signal at output port of *HTS* microwave filters in wireless communication systems and radars in Figs. 13, 14 [21, 31].

In Fig. 12, the transmitted microwave power vs. frequency for $YBa_2Cu_3O_{7-\delta}$ microwave resonator as a function of input microwave power is shown [21]. The maximum power is +30*dBm*, and the curves are in 5*dBm* steps. The frequency is 3*GHz*, and the temperature is 77*K* [21].

In Fig. 13, the surface resistance $R_S$ vs. peak magnetic field $H_{rf}$ in $YBa_2Cu_3O_{7-\delta}$ films at 1.5*GHz* at 75*K* is shown. Points are measured data. Solid and dashed lines are calculated from models created for the stripline and dielectric resonators respectively [21].

In Fig. 14, the surface resistance $R_S$ vs. temperature *T* in $YBa_2Cu_3O_{7-\delta}$ films on *MgO* at 2.3 *GHz* at two input power levels: −60*dBm*—*hatched squares*, and −20*dBm*—*dots* [50].

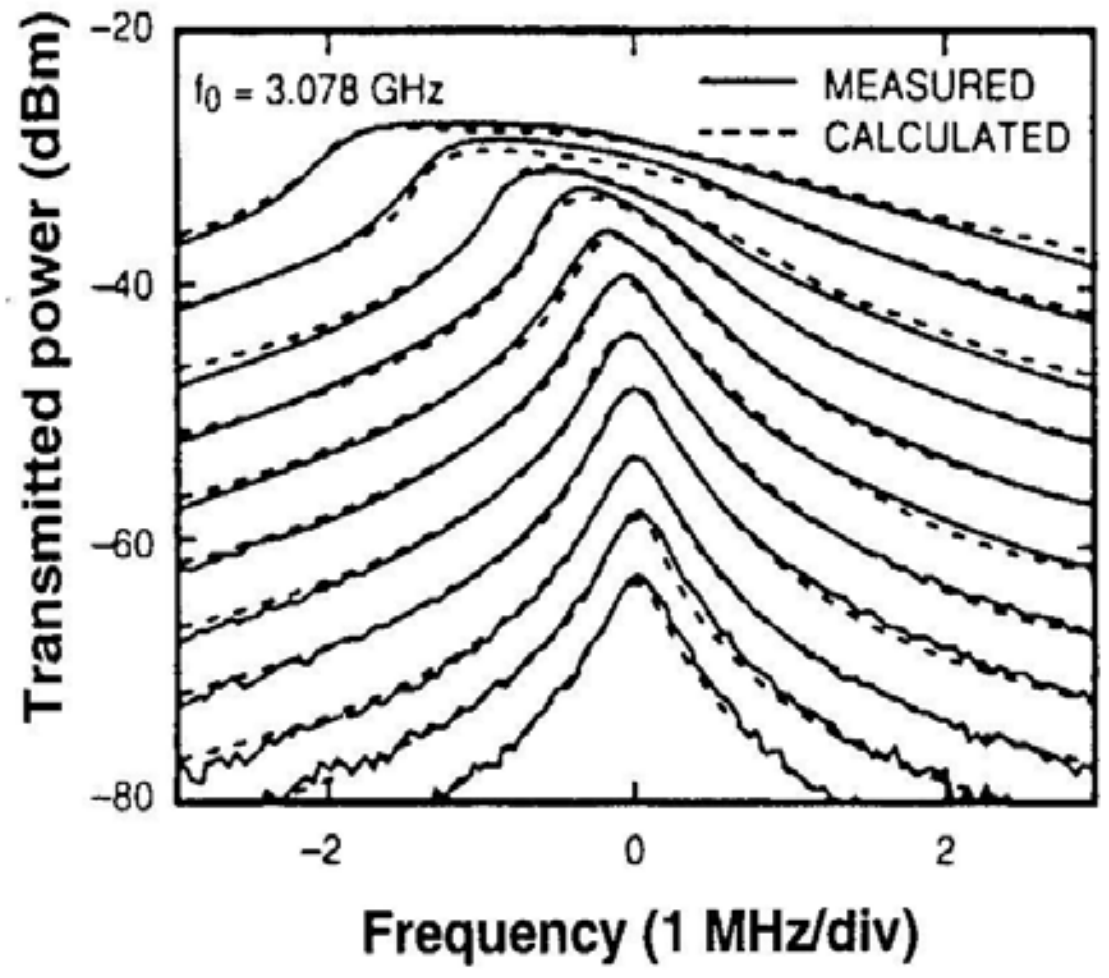

**Fig. 12.** Transmitted microwave power vs. frequency in $YBa_2Cu_3O_{7-\delta}$ microwave resonator as a function of input microwave power. The maximum power is +30dBm, and the curves are in 5*dBm* steps. The frequency is 3*GHz*, and the temperature is 77*K*. (after [21]).

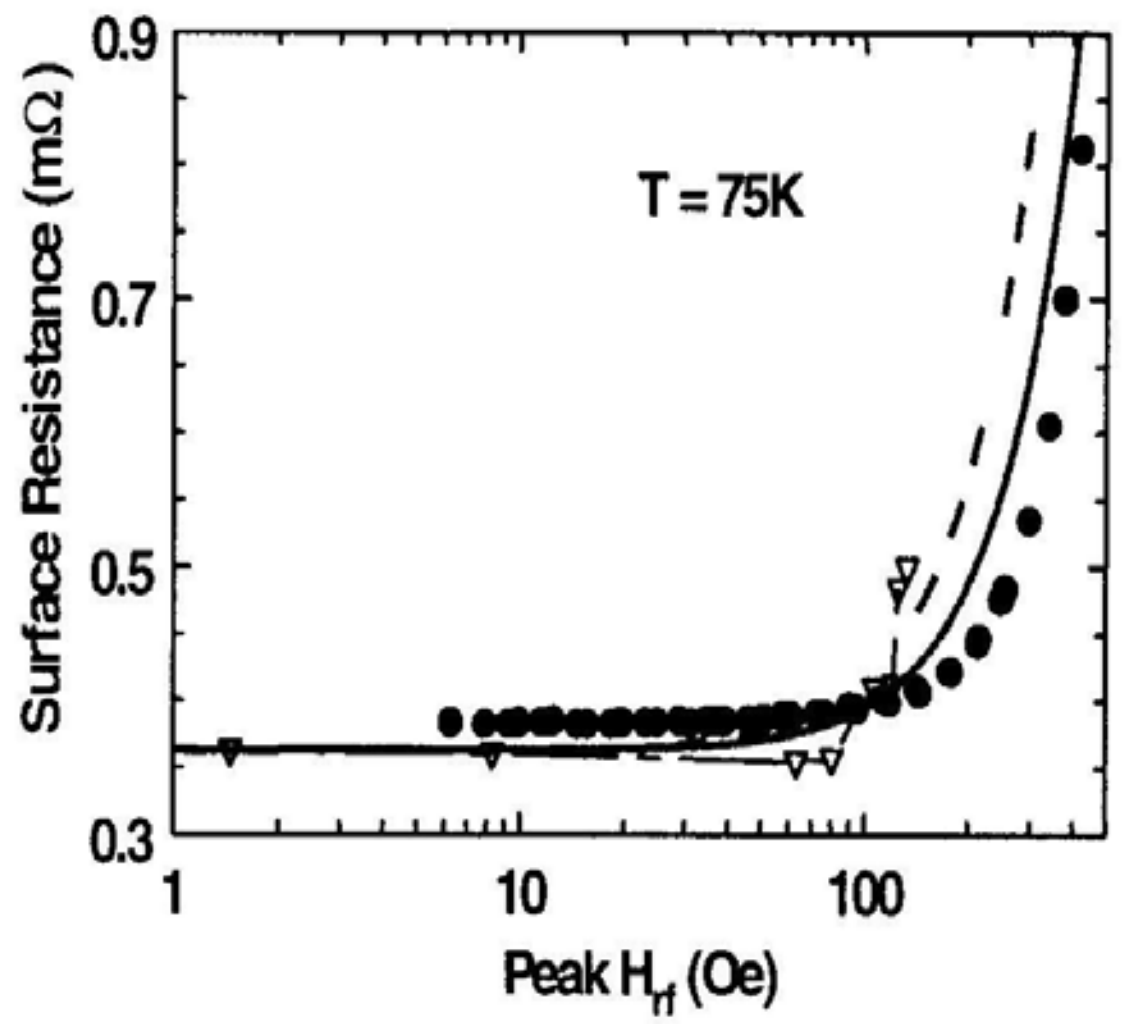

**Fig. 13.** Surface resistance $R_S$ vs. peak magnetic field $H_{rf}$ in $YBa_2Cu_3O_{7-\delta}$ films at 1.5*GHz* at 75*K*. Points are measured data. Solid and dashed lines are calculated from models created for the stripline and dielectric resonators respectively (after [21]).

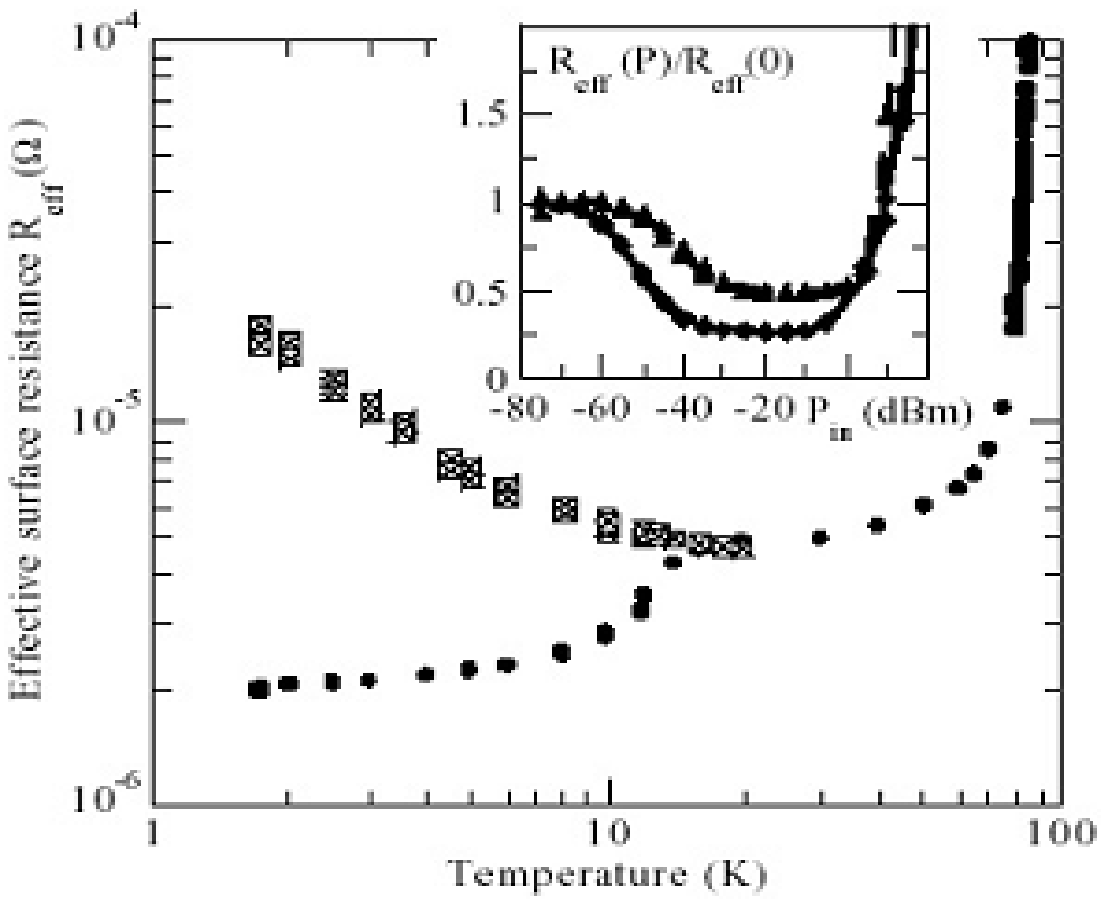

**Fig. 14.** Surface resistance $R_S$ vs. temperature $T$ in $YBa_2Cu_3O_{7-\delta}$ films on *MgO* at 2.3 *GHz* at two input power levels: −60 *dBm—hatched squares*, and −20 *dBm—dots*. (after [50]).

The main goal of many studies is to identify the sources of the nonlinearities so that the *HTS* materials for microwave applications could be improved. The extrinsic and intrinsic phenomena behind the nonlinear effects are quite complex, hence the efforts to develop *HTS* materials with minimized nonlinearities were hampered by a lack of theoretical understanding on the origin of nonlinear responses in *HTS* thin films [21, 31], and by varying results of experimental studies of many different device geometries, resulting in difficulties to compare the obtained research results quantitatively. Published theories and models are also complicated and difficult to use by non-physicists. Therefore, there is a clear need to develop a simplified model for accurate evaluation of nonlinear properties of high-*Tc* superconductors with aim to use it during microwave filters design by engineers.

In this dissertation, the empirical model to predict the nonlinear properties of superconductors at elevated microwave power levels has been created. The lumped element (*RLC)* circuits theory along with the Bardeen, Cooper, Schrieffer (*BCS)* theory of superconductors have been used to make accurate assessment and simulation of nonlinear behaviour of superconducting devices at microwaves. In order to identify parameters for the developed *RLC* model, the $YBa_2Cu_3O_{7-\delta}$ superconducting films on *MgO* substrates have been researched. The *MgO* substrates have also been separately characterised to determine their specific physical properties. The $YBa_2Cu_3O_{7-\delta}$ stripline resonators have been designed to research the *HTS* material properties for possible application in *HTS RF* filters in *4G*, *5G* wireless communication systems as schematically shown in the Wireless Communication Technology Roadmap in Fig. 15 [49].

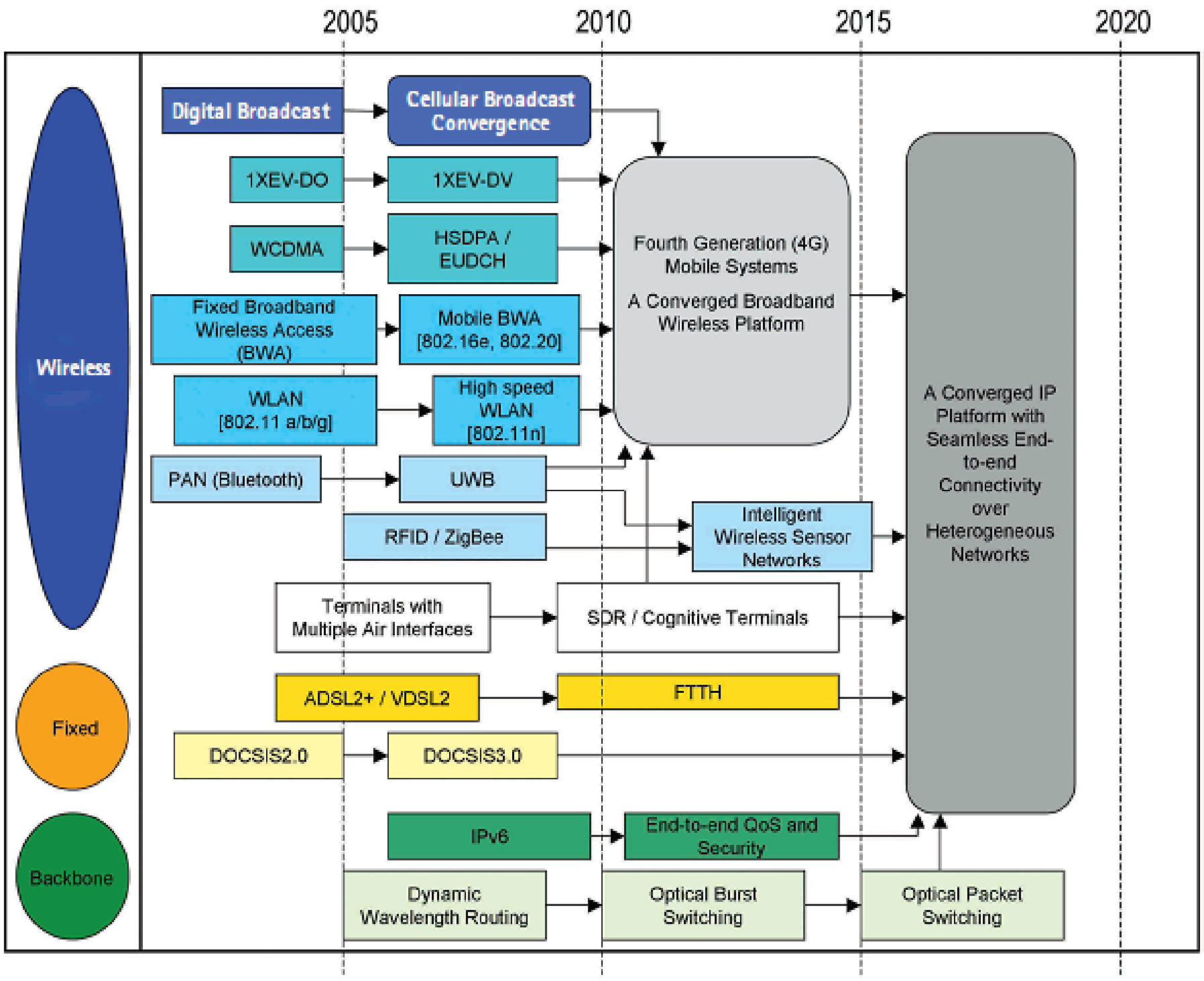

**Fig. 15.** Wireless Communication Technology Roadmap (after [49]).

The structure of the dissertation is as shown below:

***Chapter 1*** introduces to the research subject, and formulates the research task on the investigation of nonlinear phenomena in high-*Tc* superconducting materials at microwaves.

***Chapter 2*** considers the phenomena of superconductivity and technical applications of superconducting materials at microwaves. Special emphasis is made on the overview of theories towards understanding of superconductivity phenomenon, including the solid state physics and microwave properties of *HTS* superconductors. The basic concepts of microscopic electromagnetic theories developed towards better understanding of superconductivity phenomena are also presented. The two fluid approximation in terms of the complex conductivity approach is analyzed as a basis for the microscopic characterization of the superconductivity effect. The main technical applications of *HTS* materials, including sensors, magnets and microwave resonators are briefly outlined.

***Chapter 3*** focuses on the microwave properties of high-*Tc* superconducting materials, including the nonlinear effects, which are described in a comprehensive literature review on the researched dependencies of surface resistance $R_S$ as a function of the temperature *T*, frequency *f*, external magnetic field $H_e$ and *RF* magnetic field $H_{rf}$. It is shown that the surface resistance $R_S$ possesses *S-type* form dependences in the nonlinear case, which are observed in close proximity to both the critical magnetic field $H_{C1}$ as well as the critical magnetic field $H_{C2}$. Some aspects of accurate characterization of *Duffing like nonlinearities* in nonlinear dynamic resonant microwave systems are also considered.

***Chapter 4*** deals with the lumped element modelling of microwave properties of superconductors and superconducting resonators, starting with two-fluid representation for superconducting materials, introduced by Gorter and Casimir [31, 32]. This analytical review includes the description of complex conductivity of superconductor in terms of two-fluid approximation, and presents the equivalent circuits for calculation of conductivity, proposed by various authors. Special attention is paid to the parallel and series lumped element equivalent *RLC* circuits for modeling of microwave resonators. The role of *r-parameter* for nonlinear lumped element models analysis is also considered.

***Chapter 5*** describes main development steps undertaken to create a lumped element model to simulate the microwave characteristics of superconducting materials and devices, including the nonlinear effects. The *Hakki-Coleman type dielectric resonator* (*HCDR*) with incorporated superconducting plates is used for the modelling concept. Various mathematical models of *HCDR* resonator are analysed to determine the most suitable equivalent network. A theoretical mechanism, describing the nonlinear dependence of the surface resistance *Rs* in *Type II* superconductors under the action of external magnetic field $H_e$ and *RF* magnetic field $H_{rf}$, is introduced and analysed. Software for simulations of nonlinear surface resistance of superconductors near the critical magnetic fields $H_{c1}$, $H_{c2}$ and modeling of microwave power responses of networks, using the original algorithm, was developed in Matlab. Modeling of $Rs(H_e)$ and $Rs(H_{rf})$ dependences and identification of lumped element model parameters for superconducting *HCDR* are completed. It is shown that the nonlinearity occurs in a close proximity to the critical magnetic field $H_{c1}$ in agreement with the proposed model.

***Chapter 6*** presents the measurements of nonlinear resonance responses of $YBa_2Cu_3O_{7-\delta}$ thin films on *MgO* substrates in a dielectric resonator at different microwave powers and temperatures, selected as an example for the developed model. The microwave measurements procedure and practical measurements setup, including the *Hakki-Coleman resonator* with superconducting $YBa_2Cu_3O_{7-\delta}$ on *MgO* samples, microwave characterisation system and data processing network, are discussed. Precise microwave characterization of *MgO* substrates for *HTS* circuits with superconducting post dielectric resonator was completed firstly, and then the actual measurements of nonlinear resonance responses of $YBa_2Cu_3O_{7-\delta}$ thin films on MgO substrates in a dielectric resonator at microwaves were performed. An analysis of accuracy of measured results is also presented here.

***Chapter 7*** shows the results of experimental and theoretical researches on nonlinear surface resistance of $YBa_2Cu_3O_{7-\delta}$ superconductor thin films on *MgO* substrates in designed and fabricated microstrip resonators at microwaves, including the *S*-shape dependencies of $R_s(H_{rf})$. Influence of high frequency magnetic field $H_{rf}$ on the nonlinear dependence of surface resistance $R_s(H_{rf})$ of $YBa_2Cu_3O_{7-\delta}$ superconducting thin films in close proximity to the critical magnetic field $H_{c1}$ in microstrip resonators at microwaves is also researched. Microwave power dependences of transmission coefficient $S_{21}$ in microstrip resonator at different temperatures were obtained experimentally. Influence of geometrical form of researched sample on surface resistance $R_s$ in microstrip resonator is evaluated. Shift of resonant frequency $f_0$; change of magnitude of quality factor $Q$; and change of surface resistance $R_s$ depending on microwave input power in $YBa_2Cu_3O_{7-\delta}$ microstrip resonators were measured experimentally. Modeling of influence of external magnetic field $H_e$ on nonlinear dependence of surface resistance $R_s(H_e)$ in proximity to critical magnetic fields $H_{c1}$ and $H_{c2}$ was performed. Possible theoretical models, which explain the physical nature and characterize the nonlinear dependence of surface resistance $R_s$ of *HTS* thin films in close proximity to critical magnetic fields $H_{CJ1}$, $H_{C1}$ and $H_{C2}$ in $YBa_2Cu_3O_{7-\delta}$ thin films on *MgO* substrates at microwaves are proposed and discussed comprehensively.

***Chapter 8*** considers possible various sources of noise generation in linear regime of operation, and the influence by nonlinearities on noise generation in nonlinear regime of operation in *HTS* microwave filters. The new source of noise: *the magnetic dipole two-level systems (TLS) in HTS thin films,* based on the *Abricosov or Josephson magnetic vortices*, in superconducting microwave resonators is proposed firstly. Some theoretical aspects on the noise modeling in microwave filters are discussed, including the creation of new original quantum theory of *1/f noise* generation in *HTS* microwave resonators. The comprehensive analysis of physical characteristics of *differential noise* in the *HTS* microwave resonators is completed, using the comparative study of experimental and computer modeling results, obtained by author of dissertation at *ECE*, *JCU* in Australia. It is shown that it is possible to achieve the suppression of *differential noise,* using the nonlinear dependence $S_{21}(f)$ in *HTS* microstrip resonator. The proposed innovative method of suppression of differential noise in a wide range of technical applications in wireless communication systems and radars.

***Chapter 9*** covers the design of the High Temperature Superconducting (*HTS*) microwave filters for application in wireless communication systems. The considered modern designs of *HTS* microwave filters are described with the aim to show up the *HTS* microwave filters design layouts with frequency responses, and review a number of encountered design problems and limitations, because of nonlinearities appearance in *HTS* thin films at microwaves.

***Conclusion*** briefly summarises all the scientific results and outcomes, obtained during the work on this book. The main emphasis is made on the novelty of obtained results and their possible practical application in the *ICT* industry. Discussions about further progress towards the modeling and characterisation of microwave nonlinear properties in *HTS* microwave devices are also presented.

# CHAPTER 2

# THE SUPERCONDUCTIVITY: PRECISE CHARACTERISATION OF PHYSICAL PROPERTIES OF SUPERCONDUCTORS IN THEORIES AND EXPERIMENTS FOR ELECTRICAL, ELECTRONICS AND COMPUTER APPLICATIONS

## 2.1. Introduction.

The aim of this chapter is to make an introduction to the phenomenon of superconductivity and its applications. The microscopic electromagnetic theories developed towards better understanding of superconductivity is presented. The two fluid approximation in terms of the complex conductivity approach is selected as a basis for the microscopic characterisation of the superconductors in this thesis. It should be mentioned that this chapter is intended to serve as an introductory background for the upcoming chapters, and does in no way describe all the aspects of superconductivity.

## 2.2. The Superconductivity.

Visiting the Leiden University in Leiden, The Netherlands in 1999, the authors of book were quite interested in the science history in the field of physics made at Leiden University in Leiden in The Netherlands in the beginning of *XX* century. The main reason was very straightforward: the Department of Physics at Leiden University is a famous place, where the *helium* was firstly liquefied by Prof. Heike Kamerlingh Onnes in 1908, and the phenomenon of *superconductivity* was discovered by Prof. Heike Kamerlingh Onnes together with his colleagues in 1911. The original site of Department of Physics at Leiden University is located in the very heart of beautiful old city of Leiden, however the new modern building of Department of Physics is situated far away on other side of city of Leiden divided by the railway. Therefore, the main challenge for the author of dissertation was to find enough time, passion and energy to learn about the historical scientific events, which had place at the old building of Department of Physics at Leiden University in 1908-1911, when making the advanced research in the field of microwave superconductivity at the well equipped modern laboratories at new building of Department of Physics at Leiden University during the invited visit organised by prominent scientist, Prof. Peter H. Kes, Dean of Department of Physics, Leiden University, Leiden, The Netherlands. In Leiden, the authors of book learned a story that Prof. Hendrik Antoon Lorentz (1853–1928) was appointed to a chair of theoretical physics at the University of Leiden in 1878. Matricon, Waysand [156] write: “One man radically changed the way scientists work. He was responsible not only for the discovery of superconductivity but also for initiating developments that characterize modern scientific activity. The man is Heike Kamerling Onnes. Born to a family of rich merchants and industrialists, Kamerling Onnes studied at Groningen, the Netherlands, and spent several semesters in Heidelberg with G. R. Kirchoff and R. W. Bunsen. After his graduation, in 1879, he taught for several years in Delft before he accepted the chair of physics at the University of Leiden in 1882.”

The interesting fact is that the classical theory of electromagnetism obtained its final form due to the electron theory by Hendrik Antoon Lorentz, at the time often called the *Maxwell–Lorentz theory*. The Maxwell–Lorentz theory created a base for the clear explanation of interaction of the electromagnetic waves with the metals and superconductors. The Nobel Prize was awarded to Prof. Hendrik Antoon Lorentz in 1902. Prof. Hendrik Antoon Lorentz and Prof. Van Bemmelen favoured Prof. Heike Kamerlingh Onnes to become a chair of experimental physics at Leiden University, The Netherlands in 1882 [79]. Frits Berends, Emeritus Professor of Theoretical Physics at Leiden University, The Netherlands [79] provides an interesting story about the Prof. Heike Kamerlingh Onnes appointment at Leiden University in The Netherlands telling the story that a new higher education law was passed in 1876, with immediate implications for several physicists. Among other things, the Athenaeum Illustre school in Amsterdam gained university status and thus needed to appoint a professor of physics, while the existing chair of physics in Leiden was split in two. The incumbent Professor Rijke continued to occupy the Leiden chair of Experimental Physics, and a candidate was sought for the new chair of Mathematical Physics. Rijke thought in the first instance of his former student van der Waals, but the latter opted to go to Amsterdam. The choice thus fell on his other brilliant student, the younger Lorentz. He was appointed and, not yet 25 years old, held his inaugural lecture in early 1878 on the subject of “Molecular theories in physics.” The chair in theoretical physics, now occupied by Lorentz, was not only the first in the Netherlands but one of the first worldwide. Lorentz was to set an exemplary standard in the way he fulfilled his novel teaching and research remit. University chairs of this kind are now commonplace. The public mood was favourable in the Netherlands at the time, and the national budget for higher education was doubled in 1878. This meant more staff and better salaries, so giving the professors the necessary time and resources to conduct research. When Rijke retired, a new candidate had to be sought for the chair of experimental physics. The faculty was strongly divided over the shortlist. Van Bemmelen, the former high school teacher, who had already been a Leiden professor for several years, and Lorentz both favoured Kamerlingh Onnes. He was eventually appointed, and in 1882, Kamerlingh Onnes was able to start his programme of cryogenic research. ***The faculty now had a dream team: Kamerlingh Onnes set up a modern laboratory equipped for large-scale research, while Lorentz brought classical physics to its culmination.***

On the 10 July 1908, Prof. Heike Kamerlingh Onnes succeeded in the first liquefaction of *helium* at Leiden University, Leiden, The Netherlands [60-72]. The accumulated technological knowledge base from his predecessors was mainly formed by the ingenious discoveries of James Dewar, including the use of silvered vacuum glasses (1892), the liquefaction of *hydrogen* (1898), and the absorption of gases in charcoal at low temperatures used to purify helium (1905) [73, 74, 51]. Oxygen was liquefied in 1893 [75]. In 1898, Dewar liquefied the *hydrogen* for the first time in London, U.K. [73, 74, 51].

R. de Bruyn Ouboter [51] writes that, in 1898, Dewar beat Onnes in liquefying hydrogen by taking advantage of the Joule–Thomson effect, the temperature of a gas goes down as it expands through a valve when the gas is below its inversion temperature (the maximum inversion temperature is for hydrogen gas 204 K). In May 1898, Dewar could telegraph to Leiden: *Hydrogen* liquefied. Onnes reported: ‘A new glorious triumph was won by him’. On 10 May 1898 Dewar had produced 20 $cm^3$ of liquid hydrogen boiling quietly in a vacuum glass. This announcement was made by Dewar at a meeting of the Royal Society [73,74], two days later. Neither at that meeting, on 12 May 1898, nor at any future occasion did Dewar give a description of his liquefier. Although Dewar drove hydrogen down to its liquefaction temperature of 20 *K*, his apparatus produced only small amounts of liquid hydrogen. It was not until 1906 that the *hydrogen* liquefier of Onnes, based on the Linde principle was ready for use, but it was a real liquefier designed to supply as much as needed for the experiments and more importantly for the production of liquid helium. As it turned out the availability of a steady supply of liquid hydrogen was the key to the attempt to liquefy helium.

Sengers [50] states that Director Kamerlingh Onnes (1953–1926) had focused on the construction of a *hydrogen liquefier* that could produce several litres of *hydrogen* per hour. The design was modelled after that of Linde and Hampson: pressurized *hydrogen* gas is first pre-cooled by liquid air, cooled further in a recuperative heat exchanger, and finally expanded through a throttle valve [50]. After disappointing earlier attempts, in February, 1906, Kamerlingh Onnes and his staff triumphantly started operating the new liquefier [50]. The new hydrogen liquefier produced 4 $l$ $h^{-1}$, an unprecedented yield [50]. Thus, Prof. Heike Kamerlingh Onnes started with his cascade process of air liquefaction in 1892, and the large *hydrogen liquefier* of Onnes was ready for use in 1906 [50].

Sengers [50] notes that this achievement set the stage for the next goal: building a helium liquefier based on the same principle, but this time hydrogen is used to cool helium below its inversion temperature. The first hurdle was, of course, to obtain enough helium to liquefy not just a few drops, but an appreciable quantity. Although helium had been discovered in the spectrum of sunlight as early as 1868, only in the 1890s was it discovered on earth, in mines in the US. In the felicitous year 1906 Kamerlingh Onnes prepared his first few litres of pure helium gas from a load of monazite sand shipped from the US. The second hurdle was the lack of a reliable estimate of the critical temperature of helium. Kamerlingh Onnes’ plan was to use Van der Waals’ law of corresponding states, comparing pressure–volume isotherms of hydrogen and helium from 373 *K* down to 54 *K* (the lowest temperature reachable by reducing the pressure of a liquid-oxygen bath). In 1906, the programme was well under way for hydrogen, and it was initiated for helium as soon as the first sample of the gas became available. Although this phase of the work did not require a hydrogen cryostat, one was completed in 1906 anyway, in the expectation that it would be needed eventually [50].

Sengers [50] continues to retrospect the history of science, saying that, by the end of 1907, sufficient data had been obtained to produce an estimate of 5.2 *K* for the helium critical temperature, which turned out to be within 0.01 *K* from the current best value of 5.19 *K*. Although he did caution about possible departures from the law of corresponding states for these two cryogenic fluids, Kamerlingh Onnes was now convinced that liquefaction was feasible. In the middle of the following year, he and his staff produced the first ***liquid helium*** [50, 72, 76, 77, 78].

Next biggest hallmark of fundamental science in the XX century was the discovery of the phenomena of ***superconductivity*** by Professor Heike Kamerlingh Onnes and his research collaborators, Cornelis Dorsman, Gerrit Jan Flim, and Gilles Holst, at Leiden University in The Netherlands on April 8, 1911 [1, 52-69, 71]. The Dutch scientists observed that the resistance of mercury approached "practically zero" as its temperature was lowered to 3 *Kelvin* degrees [1, 52-69]. It had been known for years that the electrical resistance of metals should continuously decrease as they are cooled down below the room temperature, however the limiting values of the resistance of metals at the temperatures reduced toward the absolute zero were not known at that time. Professor Heike Kamerlingh Onnes, during his successful experiments with liquefying helium (see the experimental equipment setup shown in Fig. 1), was able to obtain the temperatures as low as 1 *Kelvin*, and latter observed that, instead of a smooth decrease of the resistance of mercury as the temperature was gradually reduced toward 0 *Kelvin*, the resistance of mercury sharply decreased to "practically zero" at about 4 *Kelvin*, and then completely disappeared below that temperature. Professor Heike Kamerlingh Onnes recognised that, at temperatures below 4 Kelvin, the mercury passes into a new physical state with the electrical properties unlike any known before, and this new state was called the ***superconducting state***.

Prof. Heike Kamerlingh Onnes explained in his Nobel lecture [81]: “As has been said, the experiment left no doubt that, as far as accuracy of measurement went, the resistance disappeared. At the same time, however, something unexpected occurred. The disappearance did not take place gradually but abruptly. From 1/500 the resistance at 4.2$^{o}K$ drop to a millionth part. At the lowest temperature, 1.5$^{o}K$, it could be established that the resistance had become less than a thousand-millionth part of that at normal temperature. Thus the mercury at 4.2$^{o}K$ has entered a new state, which, owing to its particular electrical properties, can be called the state of ***superconductivity***” in Fig. 1.

The threshold temperature below which any material begins to exhibit the superconducting properties, in other words: “losses it’s resistance,” is referred to as the ***critical temperature ($T_c$)*** or the ***transition temperature***.

Professor Heike Kamerlingh Onnes won a Nobel Prize in the field of physics for his research on the liquefaction of *helium* and the discovery of *superconductivity* in 1913: ***“for his investigations on the properties of matter at low temperatures which led, inter alia, to the production of liquid helium”*** [81].

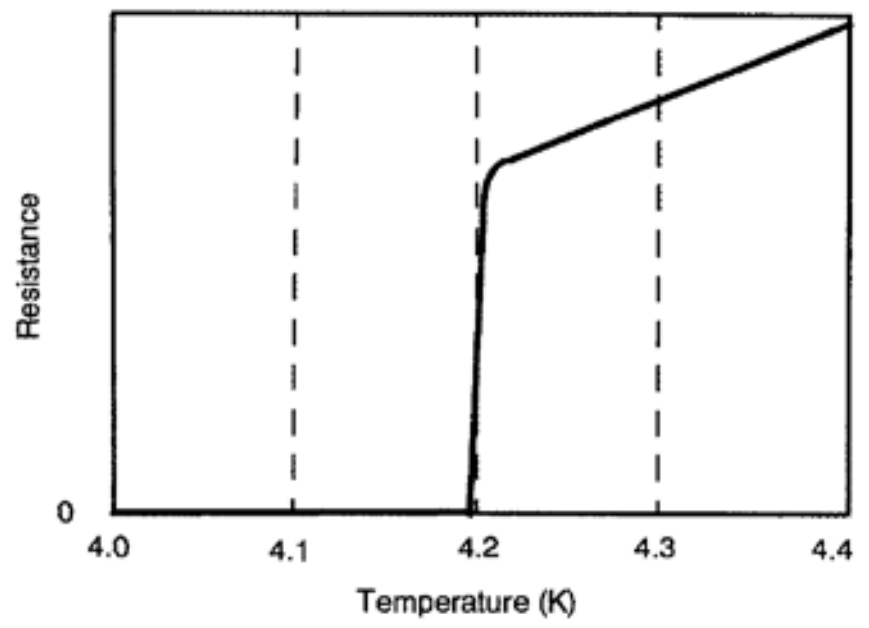

**Fig. 1.** Dependence of resistance on temperature *R(T)* of a mercury sample at low temperature (after [81, 91, 157]).

In 2010, the Board of Directors of the Institute of Electrical and Electronic Engineers (*IEEE*) has approved an *IEEE* Milestone in Electrical Engineering and Computing to commemorate the 100th anniversary of the discovery of superconductivity by Prof. Heike Kamerlingh Onnes and his collaborators in 1911. Milestone plaques was mounted on a wall near the entrance hall of the Kamerlingh Onnes Building of the Leiden University, Leiden, The Netherlands, where the discovery was made, and was dedicated on 8 April 2011, the precise 100th anniversary of the discovery [48].

Over the years, the different superconducting materials were discovered with varying critical temperatures. There were the metallic elements and alloys exhibiting superconducting properties below 30*K* with the element *Tungsten* having the lowest critical temperature of 0.016*K*, and the inter-metallic compound $Nb_3Ge$ showing the highest $T_c$ = 23,9*K* in Tab. 1 [402].

| Material | $T_c$, K | Year |
|---|---|---|
| Hg | 4,1 | 1911 |
| Pb | 7,2 | 1913 |
| Nb | 9,2 | 1930 |
| $Nb_3Sn$ | 18,1 | 1954 |
| $Nb_3Ge$ | 23,9 | 1973 |

**Tab. 1.** Low Temperature Superconductor (*LTS*) *critical temperature* records through the years as summarized by Ginzburg (after [46]).

The main challenge was to discover the superconductors with the highest *critical temperature* possible, as the higher the $T_c$ - the easier and cheaper are the cryogenic systems, needed to cool down the superconductors in electronic systems. Bednorz and Müller began to search the high-*Tc* superconductors among the ***perovskite oxides*** ceramics in 1983. Matricon, Waysand [156] explain: “The original perovskite is a calcium titanate, $CaTiO_3$; it was described in 1830 by Gustave Rose, a geologist who named it in honour of his colleague Count Lev Aleksevich von Perovski. The general formula for perovskites, natural or artificial, is $ABX_3$; clearly $CaTiO_3$ is an example.” Bednorz and Müller have been worked to research the $SrTiO_3$ superconductor at *IBM Research Laboratory* in Rüschlikon near Zurich in Switzerland for the long time. In 1985-86, Bednorz spent a big number of hours in the library in search for the literature on the topic of his research interest: the perovskite oxides, including the *Jahn-Teller ions* and *polarons*. One day, Bednorz found a research paper on the Ba-La-Cu-O by C. Michel, B. Raveau [137], and decided to synthesis Ba-La-Cu-O, using his own synthesis method: ***co-precipitation*** [156]. On January 27, 1986, Bednorz performed the measurements of Ba-La-Cu-O sample's parameters, observed the resistivity anomaly, measured the critical temperature *Tc*, and observed the *Meissner effect* (diamagnetism). It became clear to Bednorz and Müller that they discovered the *High Temperature Superconductivity* in Ba-La-Cu-O on January 27 in 1986. It is commonly accepted fact that Bednorz and Müller [2, 121, 138, 156] discovered a new class of materials: the ***complex layered copper oxide compounds***. The critical temperature of $La_{2-x}Ba_xCuO_4$ compound was $T_c$*=35 K*, that is more than 10 degrees higher, than the $T_c$*=23.2 K* in the *$Nb_3Ge$*, known previously [2]. This very significant increase in critical temperature *Tc* spurred widespread research [135, 136]. The name: "***High Temperature Superconductors** (HTS)*" was given to all the oxide superconductors, including $YBa_2Cu_3O_{7-\delta}$ superconductor with the critical temperature $T_C$ of 93*K*, discovered by M. K. Wu, J. R. Ashburn, C. J. Torng, P. H. Hor, R. L. Meng, L. Gao, Z. J. Huang, Y. Q. Wang, and C. W. Chu at Alabama University and at Houston University in Texas in the USA in February, 1987 [3, 134]. The highest $T_c$ detected at ambient pressure for a material formed stoichiometrically (by formula) stands at 138*K* for the thallium doped mercuric cuprate $(Hg_{0.8}Tl_{0.2})Ba_2Ca_2Cu_3O_{8.33}$ [4] in Tab. 2 and Fig. 2.

| Material | $T_c$, K | Year |
|---|---|---|
| $Ba_xLa_{5-x}Cu_5O_y$ | 30 - 35 | 1986 |
| $YBa_2Cu_3O_{7-\delta}$ | 91 - 93 | 1987 |
| $Bi_2Sr_2Ca_2Cu_3O_{10}$ | 106 - 110 | 1988 |
| $Tl_2Ba_2Ca_2Cu_3O_{10}$ | 125 | 1988 |
| $Tl_2Ba_2Ca_2Cu_3O_{10}$ (at 7 GPa) | 131 | 1993 |
| $HgBa_2Ca_2Cu_3O_{8+\delta}$ | 133 | 1993 |
| $HgBa_2Ca_2Cu_3O_{8+\delta}$ (at 25 GPa) | 155 | 1993 |
| $Hg_{0.8}Pb_{0.2}Ba_2Ca_2Cu_3O_x$ | 133 | 1994 |
| $HgBa_2Ca_2Cu_3O_{8+\delta}$ (at 30 GPa) | 164 | 1994 |

**Tab. 2.** High Temperature Superconductor (*HTS*) *critical temperature* records through the years as described by Vendik *et al*. (after [40]).

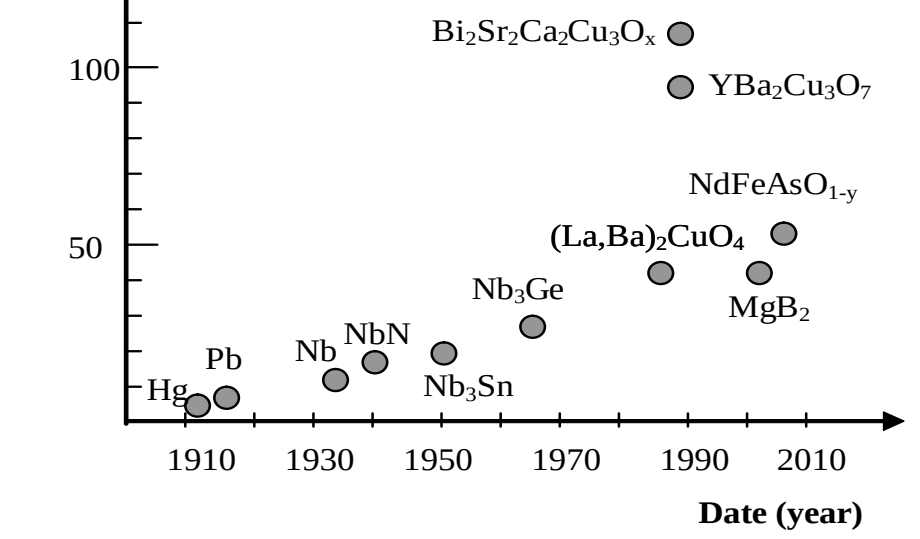

**Fig. 2.** Chronological progress in increase of critical temperature of superconductors.

Therefore, the *Low Temperature Superconductors* (*LTS*) and *High Temperature Superconductors* (*HTS*) are two different classes of superconducting materials with different underlying physical mechanisms responsible for the superconductivity phenomenon. Shin-ichi Uchida [124] points to the fact that the mechanism underlying the high-temperature (*Tc*) superconductivity (*HTS*) of copper oxides has remained unelucidated for some time, and it has not yet been completely clarified up to now, and it is one of the most difficult challenges of physics remaining in the 21st century. The *HTS* ceramics with unique physical properties [47] are of certain interest to the researchers, because of a possibility to achieve their operational temperatures, using easily available liquid *nitrogen* with costs of several orders of magnitude lower than in the case of *helium* for *LTS* materials, promising wider use of various *HTS* materials in electronic devices in different fields of science and technology.

In Fig. 3 (a) [86], $YBa_2Cu_3O_{6+x}$ compound has two $CuO_2$ planes per unit cell, approximately *3.2 Å* apart, and separated by *yttrium* ions [92]. Manske [92] mentions that the pairs of $CuO_2$ planes are themselves separated by atoms of *barium*, *oxygen*, and *copper*, forming the charge reservoir. The distance between adjacent pairs of these conducting planes is ~ *8.2 Å*. The number of carriers in the $CuO_2$ planes is controlled by the amount of charge transferred between the conducting layers and charge reservoir layers. In $YBa_2Cu_3O_{6+x}$, there are *Cu* atoms in the charge reservoir. In combination with oxygen, they form *Cu–O* chains along the *b* direction, which leads to an orthorhombic distortion. The *Cu–O* distance is about *1.9 Å*, as in the planes. For $YBa_2Cu_3O_7$, i.e. $x = 1$, the chains are well defined, but they are absent for the undoped parent compound $YBa_2Cu_3O_6$ [92]. It is usually believed that adding oxygen to the chains is equivalent to adding holes to the *CuO* planes [92].

Fig. 3 (b) shows the detailed crystal structure and atoms placement in an elementary cell of $YBa_2Cu_3O_{7-\delta}$ high temperature superconductor in V. O. Ledenyov, D. O. Ledenyov, O. P. Ledenyov [44]. The electrical conductivity and superconductivity are associated with the copper oxide planes while the other layers serve to chemically stabilise the structure of $YBa_2Cu_3O_{7-\delta}$ high temperature superconductor. It has been shown that the higher number of neighbouring copper oxide planes (up to a maximum of 3 or 4) for a given basic compound corresponds to the higher transition temperature of superconductor [5, 44].

Fig. 3. (c) shows the critical temperature $T_c$ vs. oxygen content $x$ in $YBa_2Cu_3O_{6+x}$ in *Cava et al.* (○) [87], *Jorgensen et al.* (Δ) [88, 89, and 86].

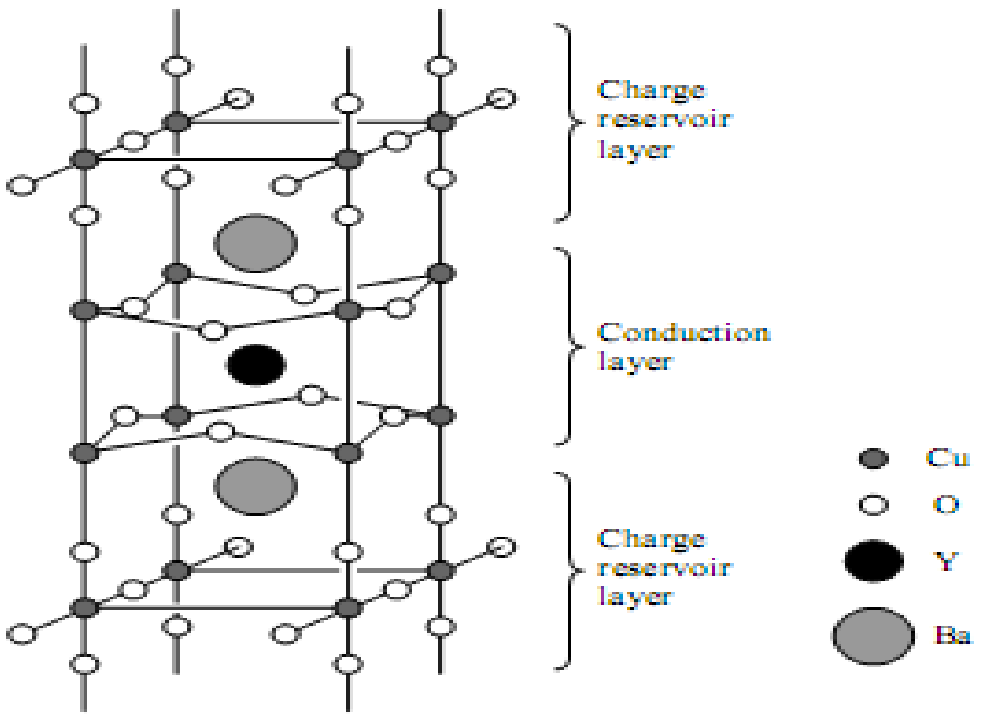

**a)**

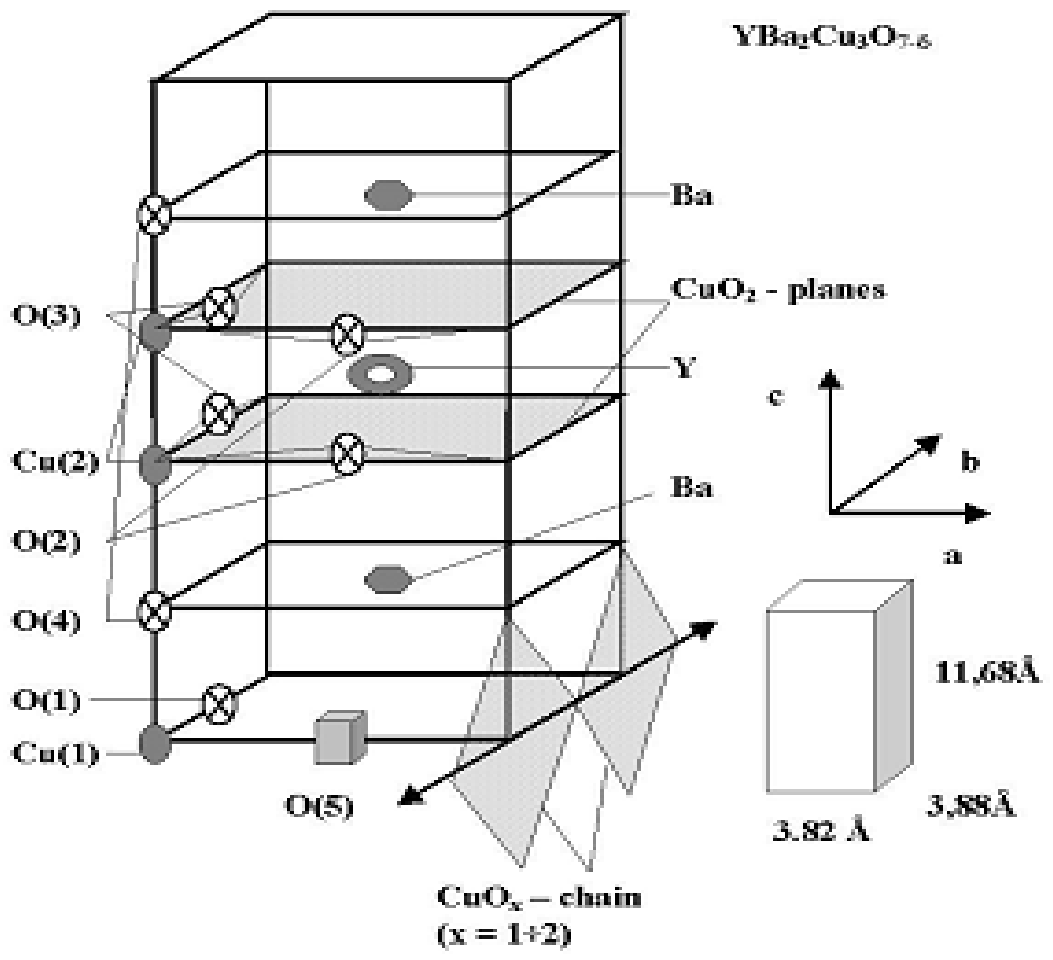

**b)**

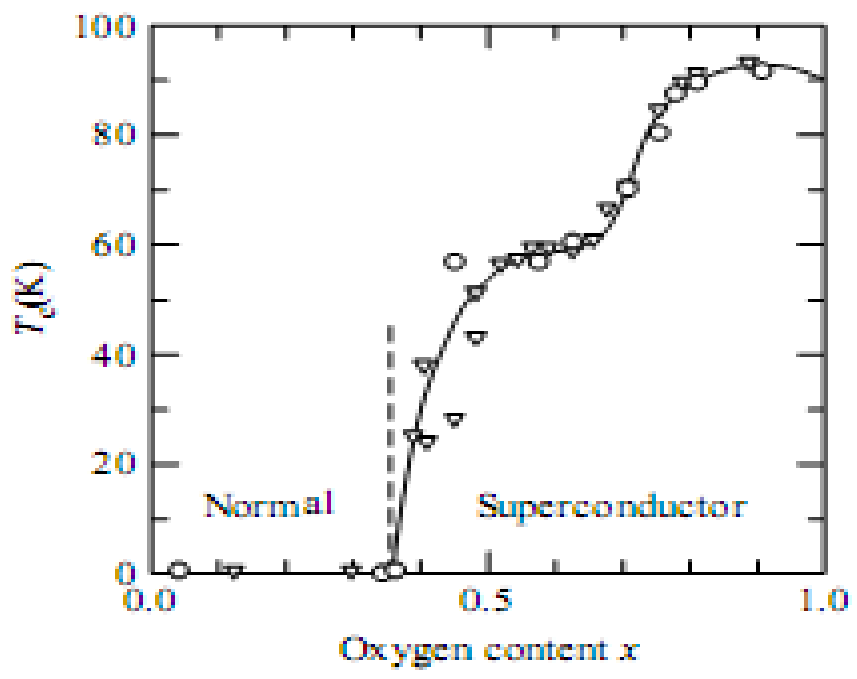

**c)**

**Fig. 3. (a)** Structure of layered orthorhombic $YBa_2Cu_3O_{7-\delta}$ (after [86]). **(b)** Crystal structure and atoms placement in elementary cell of $YBa_2Cu_3O_{7-\delta}$ (after [44]). **(c)** Critical temperature $T_c$ vs. oxygen content $x$ in $YBa_2Cu_3O_{6+x}$. *Cava et al.* (circles) [87], *Jorgensen et al.* (triangles) [88, 89] (after [86]).

Kitazawa [139] reviewed the main milestones in the development of new superconducting materials:

1. Kamerlingh Onnes first discovered superconductivity in mercury in 1911 [1];

2. Matthias, Geballe, Geller, Corenzwit discovered the superconductivity of a compound $Nb_3Sn$ in 1954 [140];

3. Bednorz and Muller discovered a cuprate superconductor in 1986 [2];

4. Wu, Ashburn, Torng, Hor, Meng, Gao, Huang, Wang, Chu discovered superconductivity with a critical temperature above the boiling point of liquid nitrogen for the first time in $YBa_2Cu_3O_{7-\delta}$ in 1987 [3];

5. Maeda, Tanaka, Fukutomi, Asano discovered superconductivity in $Bi_2Sr_2Ca_2Cu_3O_{10}$ at the highest critical temperature ever reported for potentially practically applicable superconductors in 1988 [141];

6. Nagamatsu, Nakagawa, Muranaka, Zenitani, Akimitsu discovered superconductivity in $MgB_2$ at a new highest reported critical temperature for metal superconductors in 2001 [142];

7. Kamihara, Watanabe, Hirano, Hosono discovered the Fe-based superconductor LaFeAsO in 2008 [143].

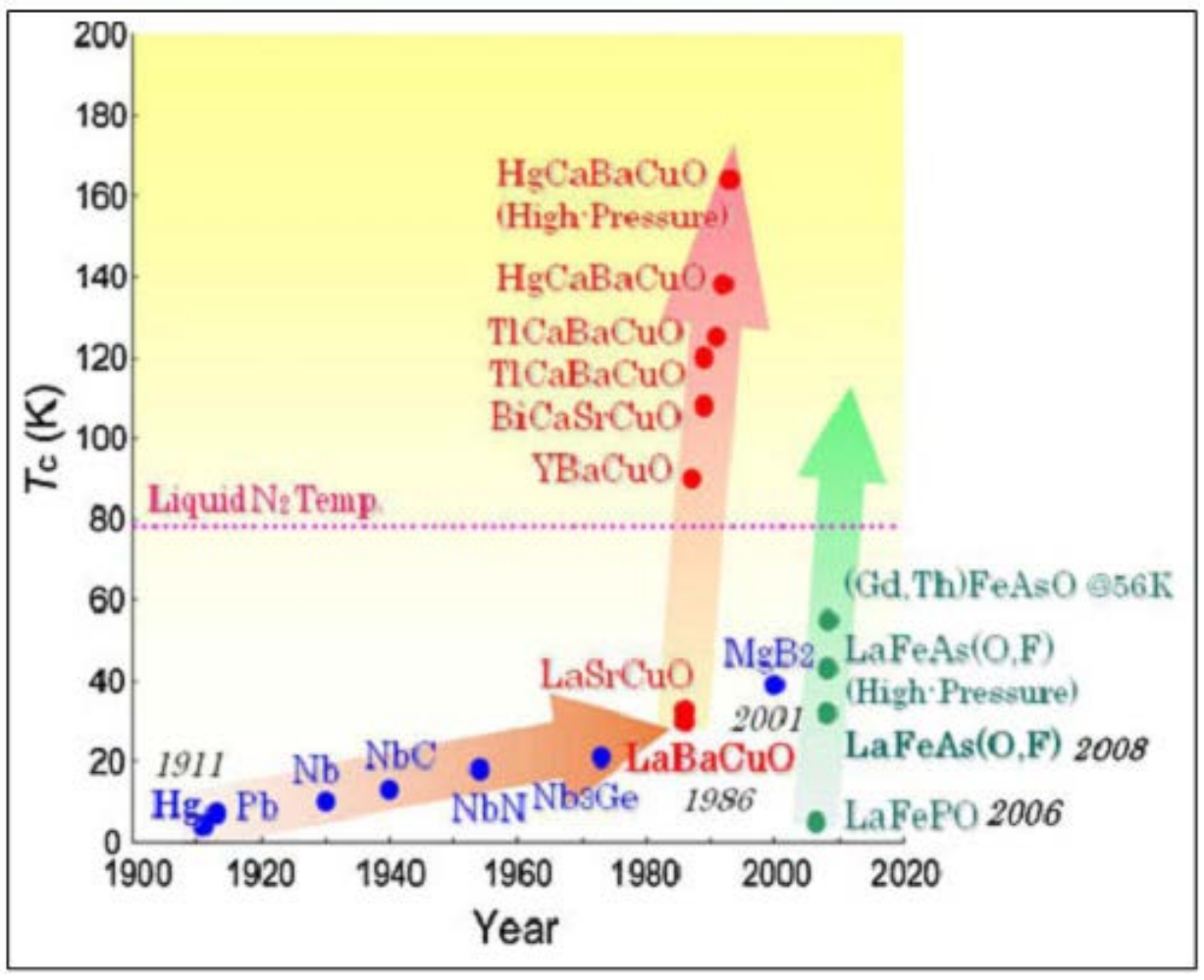

**Fig. 4.** Developing trend of superconducting critical temperature over last 100 years showing the three eras of discovery: the era of metal and intermetallic compound superconductors from 1911 to 1986, the era of cuprate superconductors from 1986 to 2000, and the subsequent era of superconductors based on new materials such as borides and Fe pnictides (after [139]).

## 2.3. Theories toward Understanding of Superconductivity.

Since the discovery of superconductivity phenomenon by Kamerlingh Onnes in 1911 [1], the quest for understanding the physics of superconductivity and formulation of theories has captivated the minds of scientists for almost half a century. The various fundamental theoretical approaches were proposed to better understand, accurately describe and predict the electromagnetic properties of superconducting materials. These phenomenological theories can be seen as developed using the existing knowledge on the phase transition, electromagnetism, and quantum mechanics rather than incorporating fundamental properties of superconductors. The *Two-Fluid phenomenological theory* was introduced by C. J. Gorter and H. Casimir with the goal to understand the nature of electrical losses in superconductors in 1934 [21, 22]. The *London theory* with electromagnetic *London equations* was proposed by brothers H. and F. London with the purpose to explain the electromagnetic properties of the superconductors in 1935 [23]. The *Ginzburg-Landau phenomenological quantum theory* was created by V. L. Ginzburg and L. D. Landau with the concentration entirely on the superconducting electrons rather than on the excitations in 1950 [24]. Schafroth [158, 159, 160] proposed that the superconductivity phenomena in the metals is due to an appearance of the charged bosons in the metal as described in Matricon, Waysand [156]. The first widely accepted theoretical microscopic understanding of superconductivity was advanced by American physicists J. Bardeen, L. Cooper, and J. Schrieffer in 1957 [15]. The *Theories of Superconductivity* by J. Bardeen, L. Cooper, and J. Schrieffer became know as the *BCS theory* [15], and the authors were awarded by a Nobel prize in 1972 [84, 85]. The mathematically-complex microscopic *BCS theory* explained superconductivity on the basis of electron-phonon interactions. The *BCS theory* relies on the fact that the energy of electrons, residing near the Fermi surfaces in the momentum space, can exchanged by virtual fluctuations of the crystalline lattice vibrations – by the phonons - that bring about an appearance of the attraction between any two electrons with opposite momentums or spins, and reduce the electrons energy in metal, and lead to an appearance of the *energy gap* $\Delta$. However, the *BCS theory* subsequently becomes inadequate theory to fully explain how the superconductivity happens in different superconductors at higher temperatures. Even now, almost two decades since the discovery of high temperature superconductors, there is no unified theory to explain the nature of high temperature superconductivity, because none of these theories explain the physical behaviour of High Temperature Superconductors (*HTS*) fully.

### 2.4. Gorter - Casimir Two-Fluid Phenomenological Theory and London Theory of Electrodynamics of Superconductors in AC Electromagnetic Fields.

In 1934, Gorter and Casimir [21, 22, 12] proposed the ***Two-Fluid* phenomenological theory** to understand electrical losses in superconductors, introducing the ***Two-Fluid model*** and assuming that the electrons in a superconductor may occupy either of two sets of states:

1. The superconducting state in which electrons are paired and resistance less, and
2. The normal state in which electrons behave like normal conduction electrons.

Therefore, in the *Two-Fluid model*, the current is considered to be flown by the two types of charge carriers in a superconductor:

1. The superconductive current is flown by *superconducting electrons* with carrier density $N_S$, and

2. The normal current is flown by *normal electrons* with carrier density $N_n$.

Thus, in the *classical Two-Fluid model*, it is postulated that the conduction electrons in a superconductor below its critical temperature *Tc* can be divided into two distinct groups of charge carriers. A fraction of the conduction electrons is condensed into a superconducting state, while the remainder is in the normal state. The superconducting electrons flow through the material without any resistance while the normal electrons encounter resistance due to collisions with other electrons or interactions with the crystal lattice similar to the conduction electrons in normal metal [21, 22, 12]. The total *electron density N* is the sum of the densities of the superelectrons *Ns* and the normal electrons $N_n$

$$N = N_s + N_n.$$

The densities of the superconducting and normal electrons in a superconductor are temperature dependent. Gorter and Casimir [21, 22, 12] found that, at temperatures below $T_C$, the equilibrium fractions of normal and superconducting electrons $N_n/N$ and $N_s/N$ vary with absolute temperature $T$ as described in eq. (2.1)

$$\frac{N_n}{N} = \left(\frac{T}{T_C}\right)^4, \qquad \frac{N_s}{N} = 1 - \left(\frac{T}{T_C}\right)^4 \tag{2.1}$$

At the absolute zero temperature, all the conduction electrons behave like superelectrons. At the temperatures above absolute zero, a few electrons start to behave as normal electrons, and during the further heating, the proportion of normal electrons increases. Eventually, all the electrons become normal ones and the metal loses its superconducting properties, when reaching the critical temperature.

The *Two-Fluid model* is a standard approach for understanding of electrical losses in superconductors, so that the energy dissipation can be expected and minimized in applications such as the microwave resonators. As this dissertation deals with the accurate characterization of physical properties of superconductors at microwaves, including the research on the nonlinear effects in superconductors in a microwave resonator of the *Hakki-Coleman type* and in *microstrip* resonators, hence the *Two-Fluid model* is considered as one of main theoretical foundations of electrodynamics of superconductor at microwaves. It is based on the superposition of responses of the "superconducting" and "normal" electron fluids to alternating electromagnetic fields as explained by Tinkham [12]. The validity of the model is restricted to the range of frequencies below the energy-gap frequency, since above that energy-gap frequency additional loss mechanisms set in and the dissipation approaches that in the normal state [12].

According to the *Two Fluids model*, the total current density $\bar{J}$ in a superconductor, is the sum of superconducting and normal currents (2.2) [12]:

$$\boldsymbol{J} = \boldsymbol{J}_n + \boldsymbol{J}_s = (\sigma_1 - i\sigma_2)\boldsymbol{E} \tag{2.2}$$

where $\sigma_1$ and $\sigma_2$ are the real and imaginary parts of the complex conductivity respectively (2.3)

$$\sigma_1 = \frac{N_n e^2 \tau}{m(1+\omega^2\tau^2)} \quad \text{and} \quad \sigma_2 = \frac{N_s e^2}{m\omega} + \frac{N_n e^2 (\omega\tau)^2}{m\omega(1+\omega^2\tau^2)}, \tag{2.3}$$

where $e$ is a fundamental unit of charge; $N_n$ and $N_s$ are the densities of normal and superconducting electrons respectively; $m$ is the mass of an electron; $\omega$ is the radian frequency; and $\tau$ is the momentum relaxation time [35].

In Fig. 5, the *Two Fluid lumped element model* of a superconductor with superconducting and normal channels is shown [97].

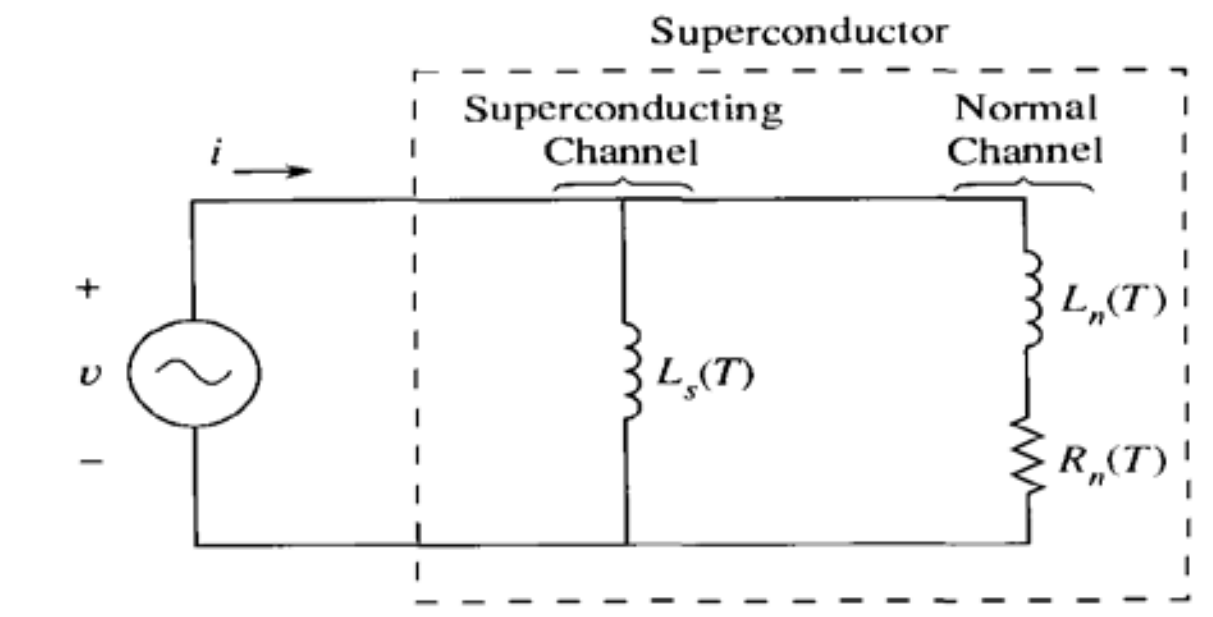

**Fig. 5.** Two Fluid lumped element model of a superconductor. In practice, the inductor $L_n$ is often neglected, which is similar to modeling the normal channel as nondispersive or independent of frequency (after [97]).

From the equation (2.2), it can be seen that the real part $\sigma_1$ of the complex conductivity involves only the normal state, while the imaginary part $\sigma_2$ includes the contributions from both the normal and superconducting fluids. The normal current $\boldsymbol{J}_n$ obeys only the Maxwell's equations [17, 18, 19] and the Ohm's law [8], while the super current $\boldsymbol{J}_s$ needs additionally to obey the *London equations*.

In 1933, Becker, Sauter, Heller [93] used a simple *free electron model* to analyse the electrodynamic behaviours of superconductors as perfect normal conductors, in which the electrons accelerate without any resistance under the exertion of an electric field $\boldsymbol{E}$. Becker, Sauter, Heller [478] argued that, if the electrons encountered no resistance, an applied electric field $\boldsymbol{E}$ would accelerate the electrons steadily.

In 1935, the brothers H. and F. London [23] proposed that, since the Becker, Sauter, Heller [93] macroscopic theory of a perfect conductor makes correct predictions about superconductors for the special case $B_o = 0$, it might be reasonable to suppose that the magnetic behavior of a superconductor may be correctly described, taking to the account the *Meissner effect*, and derived their famous equations. The ***London equations*** [23] describe the electrodynamic behaviour of superconductors on a macroscopic scale in weak electromagnetic fields and relate the microscopic strength ***E*** of electric field and induction ***B*** of magnetic field to the supercurrent density $\boldsymbol{J_S}$

$$\frac{\partial}{\partial t}\boldsymbol{J}_S = \frac{1}{\mu_0 \lambda_L^2}\boldsymbol{E} \tag{2.4}$$

$$\nabla \times \boldsymbol{J}_S = -\frac{1}{\mu_0 \lambda_L^2}\boldsymbol{B} \tag{2.5}$$

where $\lambda_L$ is the ***London penetration depth*** (2.6)

$$\lambda_L = \sqrt{\frac{m_s}{\mu_0 n_s e^2}} \tag{2.6}$$

where $m_S$ and $n_S$ are the superconducting electron mass and volume density respectively, *e* is the electron charge [23].

The first *London equation* (2.4) describes the perfect conductivity or resistance less property of a superconductor, showing that any applied electric field ***E*** *accelerates* the superconducting electrons rather than simply preserves their velocity against resistance as described in Ohm's law in a normal conductor. Therefore, any applied electric field ***E*** produces a change in the suppercurrent $\boldsymbol{J_S}$ flowing in a superconductor. From other side, there is no electric field ***E*** in a superconductor without the change of magnitude of supercurrent $\boldsymbol{J_S}$ [23, 12, 91].

The second *London equation* (2.5) describes the diamagnetic properties of a superconductor, namely the ***Meissner-Ochsenfeld effect*** [23, 12, 14, 91]. Magnetic field, which penetrates on a small distance $\lambda_L$ into superconductor, can be determined with the use of the second *London equation*. When a uniform magnetic field is applied in parallel to the surface of a superconductor, the magnetic flux density ***B*** can be obtained as a function of distance *x* inside the superconductor (2.7)

$$\boldsymbol{B}(x) = \boldsymbol{B}_0 \exp\left(-\frac{x}{\alpha^{1/2}}\right) \tag{2.7}$$

where $B_0$ is the density of the applied magnetic field at the surface, and the constant $\alpha=\lambda_L^2=m/\mu_0 n_s e^2$. The equation shows that the flux density ***B*** decays exponentially inside a superconductor or that a magnetic field ***B*** is exponentially screened from the interior of a superconductor, because of the ***Meissner-Ochsenfeld effect*** in Fig. 6 [14, 96, 97].

a)

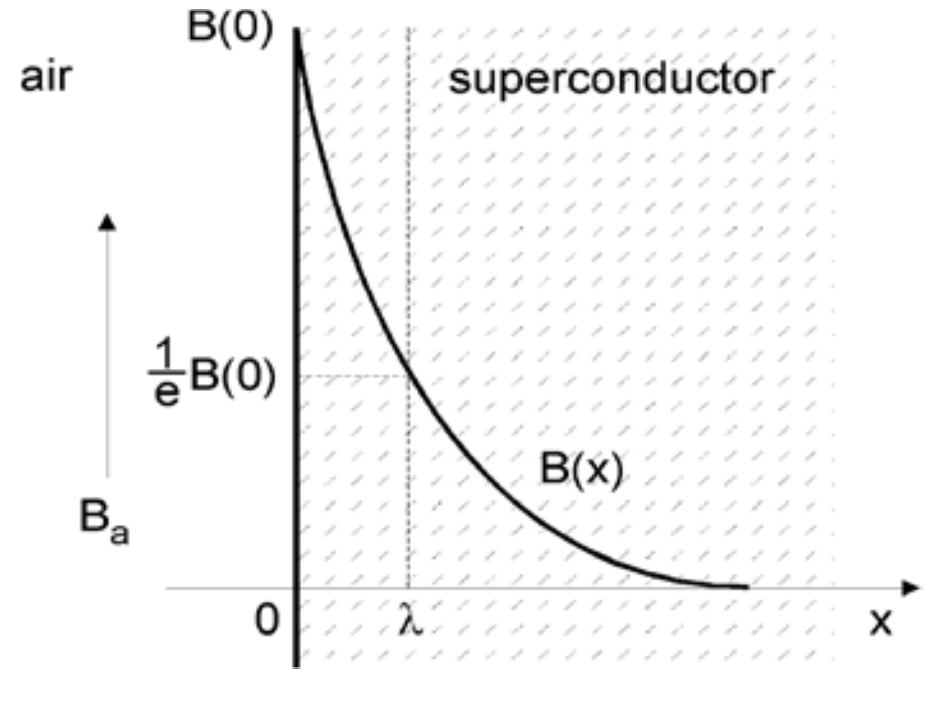

b)

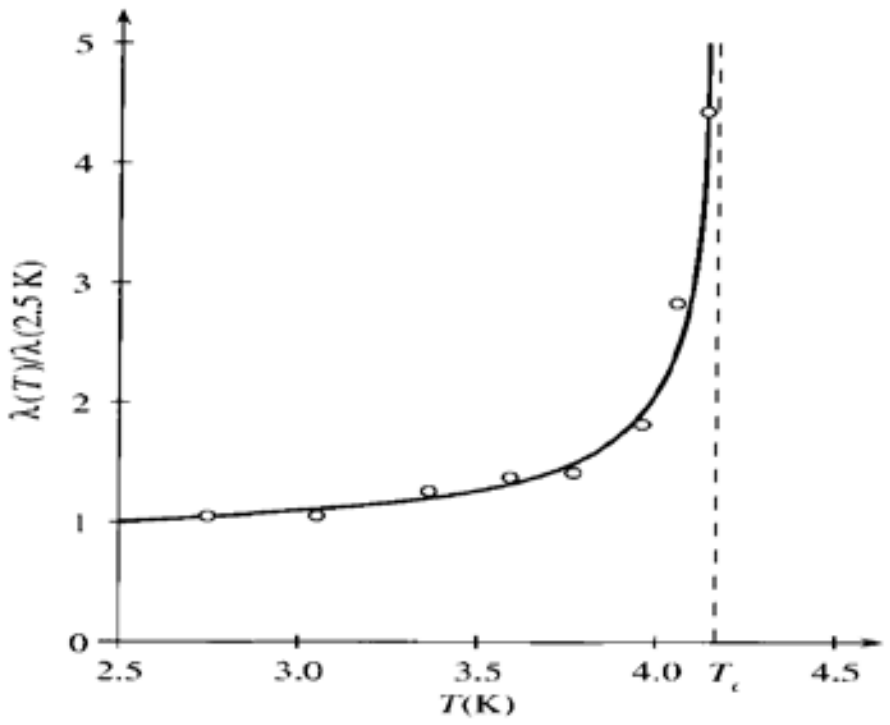

**Fig. 6. a)** Magnetic flux density as a function of distance ***B(x)*** inside the superconductor in eq. (2.7). **b)** Temperature dependence of the London penetration depth, *λ(T),* in mercury in eq. (2.8). Source: F. London, Superfluids: Macroscopic theory of superconductivity, *John Willey & Sons, Inc.*, vol. **1,** pp. 1-161, 1950 (after [96, 97]).

The distance, at which ***B(x)*** decreases to *1/e* $\approx$ *1/2,72* of its value at the surface of superconductor, is the ***London penetration depth*** $\lambda_L$ (2.6). The temperature dependence $\lambda_L(T)$ is found to be approximately described by (2.8)

$$\lambda(T) = \lambda(0) \Big/ \left[1 - (T / T_C)^4\right]^{1/2} \qquad (2.8)$$

The *London equations* do not replace the *Maxwell's equations* [17, 18, 19], which apply to all the currents and the fields in a superconductor [91]. The *London equations* are additional conditions obeyed by the suppercurrent in a superconductor [91]. The *London equations* are consequences of the phenomenon that superconductors do not allow a magnetic flux density to exists in its interior, because of the perfect diamagnetism (***Meissner-Ochsenfeld effect***) discovered in 1933 [14]. While a normal conductor develops a magnetic induction ***B*** equal to $\mu_r \mu_o \vec{H}_{ext}$, when exposed to an external magnetic field $\vec{H}_{ext}$, the magnetic induction inside superconducting materials is zero. In practice magnetic fields penetrate on a small distance into a superconductor. The penetration depth is very small, because the magnetic field decays exponentially with penetration distance.

The penetration depth $\lambda$ is related to the imaginary part of the conductivity

$$\lambda^2 = \frac{1}{\omega \mu_0 \sigma_2} \tag{2.9}$$

and hence is strongly temperature dependant, but does not vary with frequency. The lack of frequency dependence in $\lambda$. results in the *propagation constant* being constant for all frequencies as eq. (2.10) [35]

$$\gamma^2 = j\omega\mu_0 \ (\sigma_1 - i\sigma_2) = \frac{1}{\lambda^2} - i\frac{1}{\delta_1^2} \cong \frac{1}{\lambda^2} \tag{2.10}$$

As a result all the frequency components in a composite signal can travel through superconductors at the same velocity. In practice, the superconducting transmission lines exhibit some dispersion due to presence of supporting structures, but the level of dispersion is considered to be negligible in frequency ranges below 150 *GHz* [35].

The penetration depth $\lambda$ also determines the value of the surface reactance $X_s$ of superconductors as eq. (2.11)

$$X_s(T) = \omega \mu_o \lambda(T) \tag{2.11}$$

which is much bigger than the surface resistance $R_s$ ($X_S >> R_S$), unlike for conventional conductors.

The surface resistance of superconductors $R_s$ can be expressed as $\sigma_1$ and $\sigma_2$ dependent (2.12) [35]

$$R_s = \frac{\sigma_1}{2\sigma_2}\sqrt{\frac{\omega\mu_0}{\sigma_2}} = \frac{1}{2}\omega^2 \mu_o{}^2 \lambda^3 \sigma_1 \tag{2.12}$$

and hence is temperature dependence.

The ***surface resistance R<sub></sub>*** in *Type II* superconductors in *Two Fluid London theory* is given by eq. (2.13)

$$R_s = \omega^2 \mu_0^2 \lambda^3 \sigma_n \frac{N_n}{2N} \tag{2.13}$$

where $\sigma_n$ is the conductivity of normal electrons, $N = N_S + N_n$ is the full electrons density.

The temperature dependence of the surface resistance $R_s(T/Tc)$ can be written in form in eq. (2.14)

$$R_s \propto A(\omega)\frac{t^4}{(1-t^4)^{3/2}}, \qquad (2.14)$$

where $A(\omega)$ is the frequency factor and $t=T/T_C$.

The modified empirical expression for temperature dependence of surface resistance $R_s$ of pure superconductors was proposed by Pippard in eq. (2.15) [25, 9]

$$R_s \propto A(\omega)\frac{t^4(1-t^2)}{(1-t^4)^2}. \qquad (2.15)$$

## 2.5. Pippard Nonlocal Electrodynamics.

*London theory* is a *local theory* in the sense that the current densities are related to the electromagnetic potentials at the same point in space, shown, for instance. It was found by Pippard that the London theory had to be modified for certain superconductors in order to explain some experimental observations from his series of experiments for the measurements of the *penetration depth* of various types of superconductors, and its dependence on applied magnetic fields and on impurities in superconductors as well as its anisotropy [25]. In 1953, Pippard [25] conducted the experimental and theoretical study of the relation between magnetic field $H$ and current $I$ in a superconductor, considered the coherence concept in superconductivity, and introduced the *coherence length* $\xi_0$ while proposing a nonlocal generalization of the London equation (2.4). Pippard found that the *penetration depth* was noticeably dependent upon the impurity content, which could not be explained by the local theory of London since the density of superelectron and its effective mass could only be weak functions of the impurity concentration. Pippard [25] considered Reuter and Sondheimer [94] and Chambers [95] nonlocal generalization of *Ohm's law* $\boldsymbol{J(r)} = \boldsymbol{\sigma E(r)}$ for explaining anomalous skin effect [12]

$$\boldsymbol{J}_n(\boldsymbol{r}) = \frac{3\sigma}{4\pi\ell}\int\frac{\boldsymbol{R}[\boldsymbol{R}\bullet\boldsymbol{E}(\boldsymbol{r}')]e^{-R/\ell}}{R^4}dr',$$

where ***R*** = ***r*** – ***r***'; this formula takes into account the fact that the current at a point ***r*** depends on ***E*(*r*')** throughout a volume of radius ~$\ell$ about ***r***.

Pippard [25] argued that the superconducting wave-function should have a similar *characteristic dimension* $\xi_0$, which could be estimated by an uncertainty-principle argument, as follows: Only electrons within ~$kT_c$ of the *Fermi energy* can play a major role in a phenomenon, which sets in at $T_c$, and these electrons have a momentum range $\Delta p \approx kT_c/v_F$, where $v_F$ is the *Fermi velocity*. Thus, [12]

$$\Delta x \geq \hbar / \Delta p \approx \hbar v_F / kT_c,$$

leading to the definition of a *characteristic length* $\xi_0$ [12]

$$\xi_0 = a \frac{\hbar v_F}{kT_c},$$

where $\alpha$ is a numerical constant of order unity. Pippard found that $\alpha=0,15$ by fitting experimental data, the *BCS theory* gives $\alpha=0,18$. For typical elemental superconductors such as tin and aluminum, $\xi_0 >> \lambda_L(0)$. If $\xi_0$ represents the smallest size of a wave packet that the superconducting charge carriers can form, then one would expect a weakened supercurrent response to a vector potential ***A*(*r*)**, which did not maintain its full value over a volume of radius ~ $\xi_0$ about the point of interest. Thus, the $\xi_0$ plays a role analogous to the mean free path $\ell$ in the nonlocal electrodynamics of normal metals. Of course, if the ordinary mean free path is less than $\xi_0$, one might expect a further reduction in the response to an applied field [12]. Pippard proposed that a local relation should be replaced by a nonlocal relation of the form [12]

$$\boldsymbol{J}_s(\boldsymbol{r}) = -\frac{3}{4\pi\xi_0 \Lambda c} \int \frac{\mathbf{R}[\mathbf{R} \cdot \mathbf{A}(\mathbf{r}')]}{\mathrm{R}^4} e^{-R/\xi} d\mathbf{r}',$$

where ***R*** = ***r*** — ***r***' and the *effective coherence length* $\xi$ in the presence of scattering was assumed to be related to that of pure material *characteristic length* $\xi_0$ [12]

$$\frac{1}{\xi} = \frac{1}{\xi_0} + \frac{1}{\ell}.$$

Shu-Ang Zhou [91] concludes that the introduction of the concept of the coherence length in the nonlocal relation characterizes the fact that spatial variation of the density of superelectrons cannot occur over arbitrarily small distances and is only possible within a certain distance: *effective coherence length* $\xi$. Furthermore, the nonlocal model shows that the superconducting current density at a certain point depends on an average of the magnetic vector potential over a volume of radius about *effective coherence length* $\xi$ around the point of interest.

Tinkham [12] notes that Pippard's nonlocal electrodynamic equation $\boldsymbol{J_s(r)}$ not only fitted the experimental data, but it also anticipated the form of electrodynamics found several years later from the *BCS microscopic theory*.

Tab. 3 shows the material parameters such as the *penetration depth* $\lambda_0$ and *coherence length* $\xi$ for some *LTS* and *HTS* superconductors [91].

| Superconductor | $\lambda_0$ (nm) | $\xi_0$ (nm) | $T_c$ (K) |
|---|---|---|---|
| Al | 16 | 1500 | 1.2 |
| In | 25 | 400 | 3.3 |
| Sn | 28 | 300 | 3.7 |
| Pb | 28 | 110 | 7.2 |
| Nb | 32 | 39 | 8.95–9.3 |
| $Nb_3Sn$ | 50 | 6 | 18 |
| $YBa_2Cu_3O_x$ | 140 | 1.5 | 90 |

**Tab. 3.** Material parameters such as *penetration depth* $\lambda_0$ and *coherence length* $\xi$ for some *LTS* and *HTS* superconductors (after [91]).

## 2.6. Ginzburg-Landau (GL) Theory of Superconductivity.

In 1950, seven years before the *Bardeen, Cooper, Schrieffer (BCS) microscopic theory* [15] was developed, *Ginzburg and Landau (GL)* intuited a remarkable *phenomenological theory of the superconductivity* [24] that integrated the electrodynamics, quantum mechanical and thermodynamic properties of superconductors. In 1962, the Nobel Prize in physics was awarded to Leo D. Landau for “his pioneering theories for condensed matter, especially liquid helium” in [162, 163, 164, 165]. In 2003, the Nobel Prize in physics was awarded to Abrikosov and Ginzburg for “their pioneering contributions to the *theory of superconductors and superfluids*” in [162]. In Fig. 7, Prof. Lev D. Landau is pictured in Verkin B. I., Manzheliy V. G., Trapeznikova O. N., Gredeskul S. A., Pastur L. A., Freiman Yu. A., Khramov Yu. A. (editors) [154].

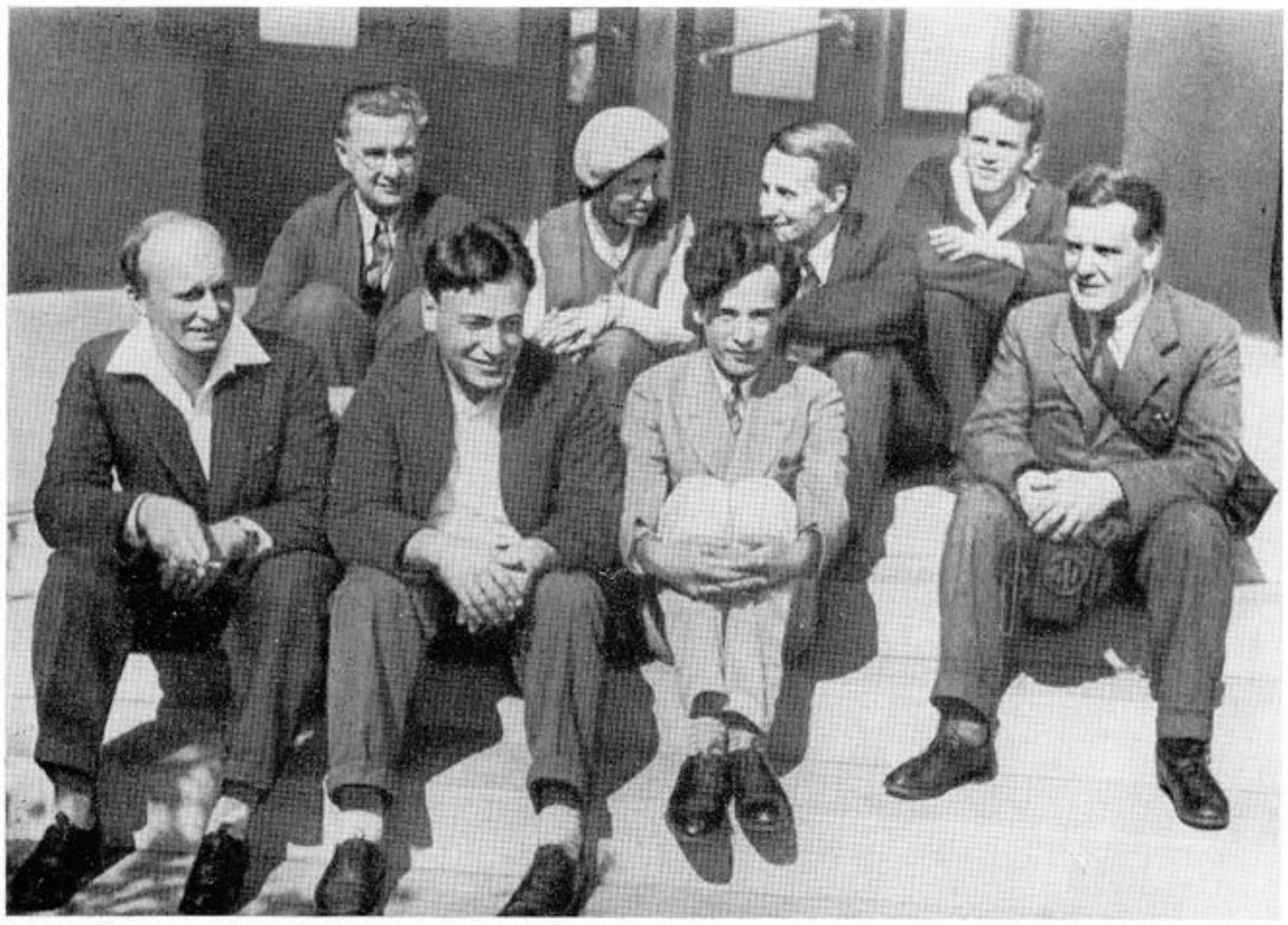

**Fig. 7.** Lev Schubnikov, Alexander I. Leipunski, Lev D. Landau, Peter Kapitza in the first raw; Boris Finkel'shtein, Olga Trapeznikova, Kyrill Leipunski, Yuri Rjabinin in the second raw are photographed at the front entry at the National Scientific Centre Kharkov Institute of Physics and Technology in Kharkov, Ukraine in 1935 (from left to right) (after Verkin B. I., Manzheliy V. G., Trapeznikova O. N., Gredeskul S. A., Pastur L. A., Freiman Yu. A., Khramov Yu. A (editors) [154]).

The ***Ginzburg-Landau (GL) theory*** states that the physical behaviour of superconducting electrons may be described by the *effective order parameter* or the *wave function*

$$\Psi(\mathbf{r}) = \Psi_0 \cdot \exp(i\theta(\mathbf{r})), \quad (2.16)$$

the *density of superconducting electrons* is

$$N_S = |\Psi|^2, \quad (2.17)$$

and the *superconducting current density* is

$$\boldsymbol{J}_S \propto \nabla\theta(\mathbf{r}),$$

where $\theta$ is the phase of the wave function, and $\nabla = d/d\mathbf{r}$, and $\boldsymbol{r}$ is the distance.

The wave function $\Psi$ is defined as zero for a normal metal, and must characterise the density of superconducting electrons in a superconductor. It has the quantum mechanical nature. The *GL theory* was developed by applying a variational method to an assumed expansion of the free-energy $F$ in powers of $|\Psi|^2$ and $|\nabla\Psi|^2$, leading to the pair of coupled differential equations for $\Psi$ and the vector potential $\boldsymbol{A}$ in eq. (2.18)

$$F = F_{n0} + \int \left\{ \frac{\boldsymbol{B}^2}{2\mu_0} + \frac{1}{2m^*} \left| (i\hbar\nabla + e^*\boldsymbol{A}) \right|^2 + \alpha |\Psi|^2 + \frac{\beta}{2} |\Psi|^4 \right\} dV \tag{2.18}$$

where $V$ is the volume of superconductor.

Minimizing expression for the free energy on $\Psi$ and vector potential $\boldsymbol{A}$ Ginzburg and Landau derived the two famous ***GL* equations** (2.19, 2.20) [24]

$$\alpha\Psi + \beta|\Psi|^2 \Psi + \frac{1}{2m^*}\left(i\hbar\nabla + e^*\boldsymbol{A}\right)^2 \Psi = 0 \tag{2.19}$$

$$\boldsymbol{J}_s = -i\frac{e^*\hbar}{2m^*}\left(\Psi^*\nabla\Psi - \Psi\nabla\Psi^*\right) + \frac{e^{*2}}{m^*}|\Psi|^2 \boldsymbol{A} \tag{2.20}$$

where $\boldsymbol{J}_S$ is the superconducting current density, $\alpha$ and $\beta$ are the expansion coefficients

$$\alpha = -\mu_0 H_C^2/|\Psi_0|^2$$

$$\beta = \mu_0 H_C^2/|\Psi_0|^4,$$

in which $H_C$ is the thermodynamic critical field, and $N_s=|\Psi_0|^2$ is the density of superconducting electrons in low magnetic field. The quantities $e^*$ and $m^*$ are the effective charge and mass of superelectrons respectively, and $\hbar$ is the reduced *Planck's constant.*

The first *GL* equation (2.20) establishes the interrelation between the kinetic energy of superconducting electrons, and the energy of their transition into a superconducting state, and the energy, which influences the vector-potential $\boldsymbol{A}$.

The second *GL* equation (2.20) describes the quantum interrelation of the superconducting current with the gradient of the wave function $\nabla\Psi$ and the vector-potential $\boldsymbol{A}$.

The result of the *GL* theory is a generalization of the *London theory* in application to the situations, in which the density of superconducting electrons $N_S$ changes in space, and also to deal with the nonlinear response to fields and currents, which are strong enough to change the density of superconducting electrons $N_S$. Tinkham [12] comments that the *GL theory* formalism made it possible to treat the following important features, which were beyond the scope of the *London theory*

a) the non-nonlinear effects caused by electromagnetic fields strong enough to change the density of superconducting electrons $N_S$ (or $|\Psi^2|$),
b) the spatial variation of density of superconducting electrons $N_S$,
c) the handling of the intermediate state of superconductors, in which superconducting and normal domains co-exist in the presence of $H << H_c$.

In 1957, on basis of the *GL theory*, Abrikosov developed the *theory of magnetic properties of the Type II superconductors* [16], and researched the special *magnetic lines* or the *quantum magnetic vortices*. Each of these vortices has the normal metal core with effective radius $\sim\xi$ and the circulating superconducting current in area with effective radius $\sim \lambda$.

In 1959, Gor'kov [26] showed that the *GL theory* is a direct consequence of the microscopic *BCS theory* [15] with $m^*=2m_e$ and $e^*=2e$. In the original theory, the analysis was restricted to the temperatures near $T_C$, although it gives a qualitative description of many phenomena, which seems to be valid in temperatures range away from $T_C$ [24]. The *GLAG* (***Ginzburg-Landau-Abrikosov-Gor'kov***) and *BCS* theories explained the main properties of the low temperatures superconductors. It is accepted to believe that, in some sense, the *Abrikosov–Gor'kov (AG)* theory was proposed as a bridge linking the macroscopic *GL* theory and the microscopic *BCS* theory, proving that these theories form a consistent theoretical framework, as explained in Kitazawa [139].

The *GLAG* theory introduces the characteristic temperature dependent length, usually called the ***GL coherence length*** $\xi_{GL}$ [24] (mentioned before in this chapter in identifications of the *Type I* and *Type II* superconductors) in eq. (2.21)

$$\xi_{GL}(T) = \frac{\hbar}{\left|2m^*\alpha(T)\right|^{1/2}}, \tag{2.21}$$

which characterises the distance over which the wave function $\Psi$ can change without the undue energy increase.

In a pure superconductor for temperatures much below $T_C$, $\xi_{GL}(T) \approx \xi_0$, which is the temperature independent *Pippard coherence length* [25]. Pippard was the first, who applied this concept in the theory of superconductivity. Pippard also proposed the empirical expression for the dependence of the coherence length on the electron mean free path *l* in eq. (2.22)

$$1/\xi(l)=1/\xi_0 + 1/al, \quad (2.22)$$

where *a* is the constant ≈1, *Pippard coherence length* $\xi_0$ is in eq. (2.23)

$$\xi_0 = 0.18\, \hbar v_F/k_B T_c, \quad (2.23)$$

where $v_F$ is the *Fermi velocity* of the electrons and $k_B$ is the *Boltzmann constant*. Near $T_C$, however, the *GL coherence length* follows the following rule in eq. (2.24)

$$\xi_{GL}(T) \propto (T_c - T)^{-1/2}, \quad (2.24)$$

since $\alpha$ vanishes as $(T - T_c)$. Thus, these two coherence lengths $\xi_{GL}$ and $\xi_0$ are related variables, but have different magnitudes.

The ratio between two characteristic lengths: $\lambda(T)$ and $\xi(T)$ defines the ***GL parameter k*** in eq. (2.25)

$$k = \frac{\lambda(T)}{\xi_{GL}(T)}, \quad (2.25)$$

where $\lambda(T)$ is the penetration depth for electromagnetic field and $\xi(T)$ is the coherence length.

Near $T_C$, the penetration depth $\lambda$ and the coherence length $\xi$ diverges as ~ $(T_C - T)^{-1/2}$, and the dimensionless ratio $\lambda(T)/\xi(T)$ is approximately independent on the temperature. For typical classic pure superconductors (*Type I*) $\lambda \approx 500$ *Å* and $\xi \approx 3000$ *Å*, i.e. the coherence length $\xi$ is larger than the penetration depth $\lambda$, $\xi > \lambda$. On the other hand, as Abrikosov remarkably identified in 1957 [16], for *Type II* superconductors, the coherence length $\xi$ is smaller than the penetration depth $\lambda$, $\xi < \lambda$. Abricosov also discovered that the exact break point between the Type I and *Type II* regimes was at $\kappa = 1/\sqrt{2}$. Abricosov showed that, for superconductor with $\kappa > 1/\sqrt{2}$, there is a continuous increase in flux penetration starting at a lower critical field $H_{C1}$ and reaching $B = H$ at an upper critical field $H_{C2}$. Because of the partial flux penetration, the diamagnetic energy cost of holding the field out is less, so $H_{C2}$ can be much greater than the thermodynamic critical field $H_C$. This property has made possible the creation of high magnetic field superconducting magnets.

Tab. 4 shows the data for critical temperatures *Tc*, coherence length $\xi$, penetration depth $\lambda$ and *GL* parameter $\boldsymbol{\kappa}$ for some *LTS* and *HTS* materials.

| Material | $T_c$ (K) | ξ (nm) | λ (nm) | κ (= λ/ξ) |
|---|---|---|---|---|
| Nb | 9.25 | 39 | 50 | 1.28 |
| Nb-Ti | 9.5 | 4 | 300 | 75 |
| $Nb_3Ge$ (A15) | 23.2 | 3 | 90 | 30 |
| $YBa_2Cu_3O_7$ | 89 | 1.8 | 170 | 95 |

**Tab. 4.** Data for critical temperatures *Tc*, coherence length $\xi$, penetration depth $\lambda$ and *GL* parameter $\kappa$ for some *LTS* and *HTS* superconductors (after [91]).

The Figs. 8 and 9 show the phase boundaries of *Type I* and *Type II* superconductors, when the ***GL parameter k*** is significantly less than 1: ***k*** << 1, or much more than 1: ***k*** >> 1 respectively in Eschrig [104].

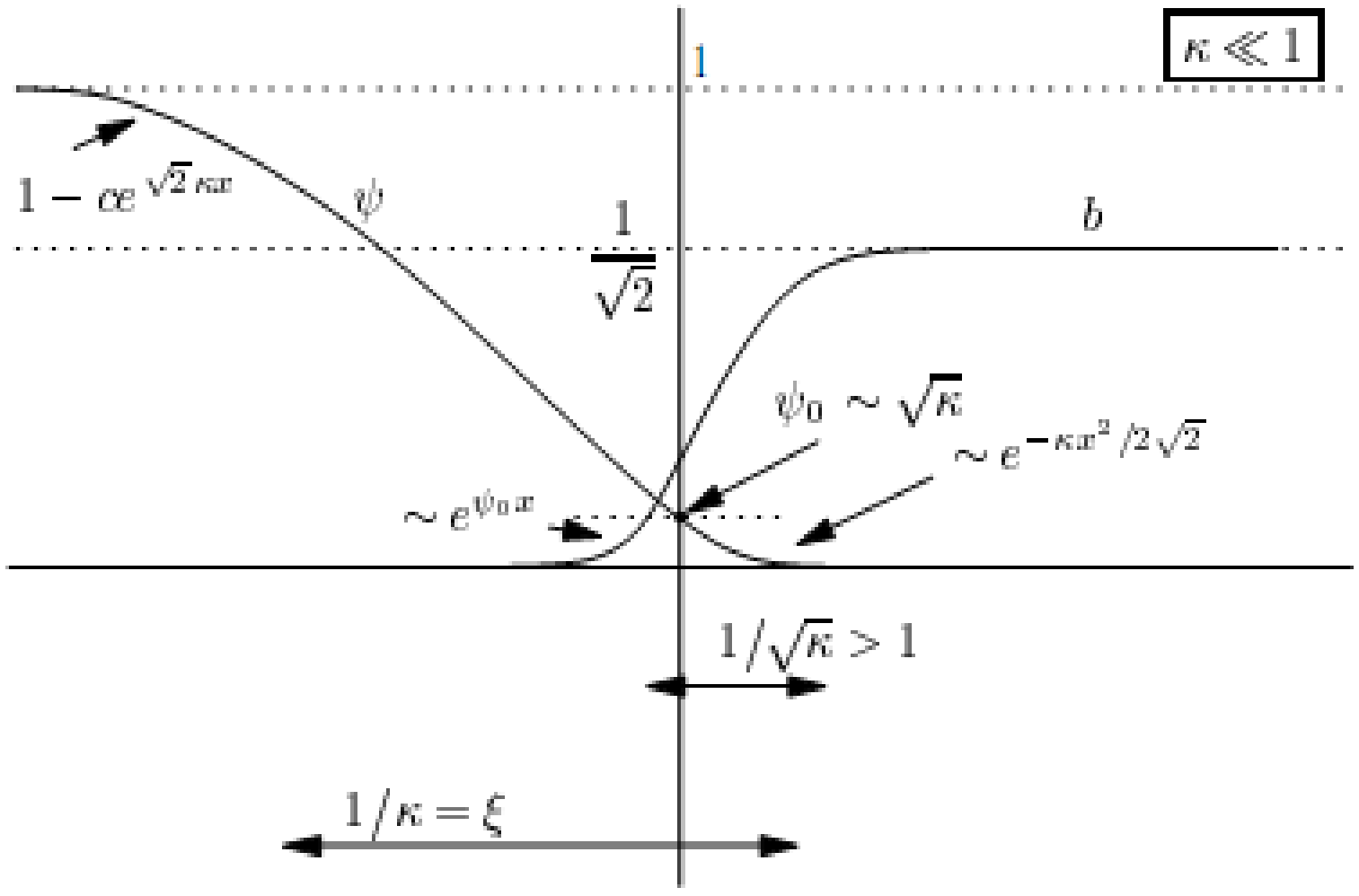

**Fig. 8.** The phase boundary of a *Type I* superconductor (after [104]).

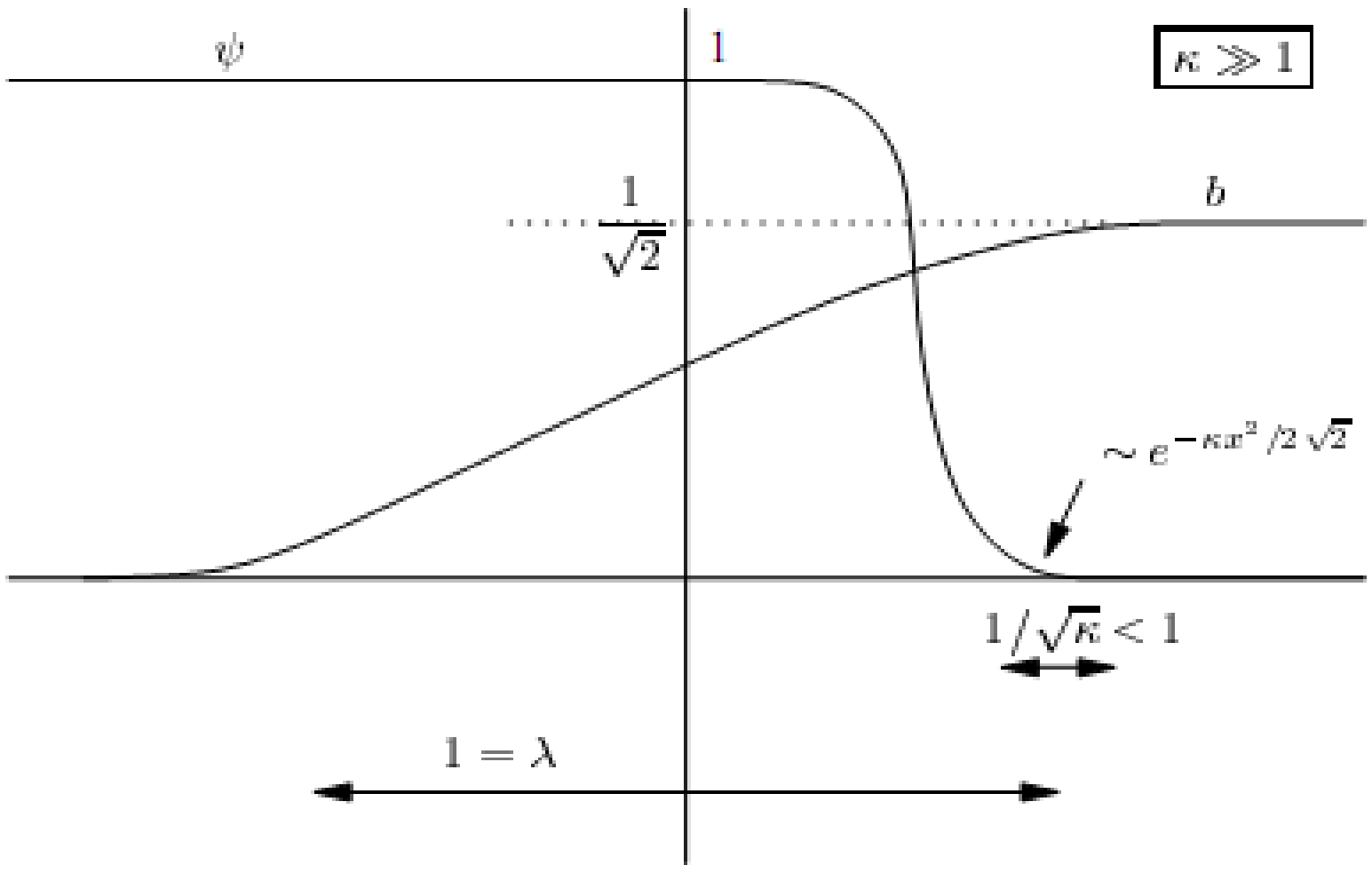

**Fig. 9.** The phase boundary of a *Type II* superconductor (after [104]).

## 2.7. Bardeen, Cooper and Schrieffer (BCS) Theory of Superconductivity.

In 1946, Richard A. Ogg [119] made a first ever suggestion of ***electron pairs*** and their ***Bose-Einstein condensation***, when describing his experiments on very dilute solutions of alkali metals in liquid ammonia. Blatt [120] writes that Ogg [119] claimed persistent ring currents at temperatures as high as -180 C, and ever higher. Ogg [119] obtained some indirect evidence for the existence of trapped *electron pairs* in his solutions, and estimated their *Bose-Einstein degeneracy temperature* on the basis of their approximately known concentration and a guess at their effective mass, to be in the right range. At that time, Gawov, a physicist from *Kharkov Institute of Physics and Technology* in Kharkov, Ukraine, who conducted his research in the field of physics at *Washington University* in the U.S.A. for many years, wrote a limerick on the *Ogg's bi-electron theory* [120]:

"In Ogg's theory it was his intent
That the current keep flowing, once sent;
So to save himself trouble,
He put them in double,
And instead of stopping, it went."

Blatt [120] concludes that there is no doubt about Ogg's priority of the suggestion of the ***electron pairs*** and their ***Bose-Einstein condensation***.

In 1950, Fröhlich [108] from University of Liverpool made his original proposition about the possible existence of weak attraction between the electrons as a result of electron-phonon interaction in crystal lattice of a superconductor. The Fröhlich theory is based on the *Hamiltonian*, which reproduces the results of the adiabatic approximation in the coupling between the electrons and phonons [115]

$$H = \sum \varepsilon_{\boldsymbol{k}} c^*_{\boldsymbol{k}\sigma} c_{\boldsymbol{k}\sigma} + \sum_{\boldsymbol{q}a} \hbar\omega_{a\boldsymbol{q}} (b^*_{a\boldsymbol{q}} b_{a\boldsymbol{q}} + \frac{1}{2}) + \sum_{\boldsymbol{k}_1,\boldsymbol{k}_2,\boldsymbol{q},K,a} c^*_{\boldsymbol{k}_1\sigma} c_{\boldsymbol{k}_2\sigma} (b^*_{a\boldsymbol{q}} + b_{a-\boldsymbol{q}}) g_{\boldsymbol{k}_1\boldsymbol{k}_2 qa} \delta_{\boldsymbol{k}_1+q,\boldsymbol{k}_2+K\bullet},$$

where $\varepsilon$ is the energy of electron of wave vector $q$ in normal state, $b^*$ and $b$ are the creation and destruction operators for phonons of wave vector $q$, $\omega$ is the frequency of a phonon of wave vector $q$, $\hbar$ is the *Planck's constant*, $\hbar\omega$ is the average phonon energy, $\boldsymbol{K}$ is the vector of reciprocal lattice, $\boldsymbol{k}$ is the wave vector, $\sigma$ is the spin of an electron, $g$ is the electron-phonon coupling constant, $c^*_{k\sigma}$ and $c_{k\sigma}$ are the creation and annihilation *Fermi operators*, $\delta$ is the delta-function. ***Fröhlich recognized that the interaction between the electrons and phonons could be represented by a Hamiltonian, and that this Hamiltonian implied an attractive interaction between the electrons, which might be responsible for superconductivity [116].*** Fig. 10 shows the Fröhlich shell distribution of electrons in ***k***-space [108, 113].

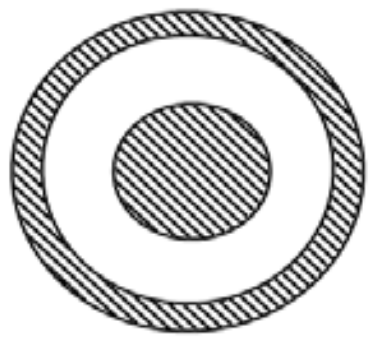

**Fig. 10.** Fröhlich shell distribution of electrons in ***k***-space. The shaded regions are occupied by electrons (after [108, 113]).

In 1953, during the Lorentz-Kamerling-Onnes conference, Niels Bohr [112] stated: "I most thoroughly appreciate the great importance of Fröhlich contribution to our understanding of the interaction between the electrons through their coupling with the ion lattice" [113].

In 1951, 1954, 1955, Schafroth [158, 159, 160] proposed that the superconductivity phenomena in the metals is due to an appearance of the ***charged bosons*** in the metal. Schafroth [158, 159, 160] made a research contribution to the electron pairing theory of superconductivity in Matricon, Waysand [156].

In 1956, Cooper [27] explored the dynamics of a pair of electrons restricted to *Bloch states* above the *Fermi sea*, and proposed that even the weak attraction potential $V$ can bind the pairs of electrons into a bound state or a resonant state [27]. The two electrons with opposite momentums $\boldsymbol{p_1} = -\boldsymbol{p_2}$ and opposite spins $s_1(\uparrow) = -s_2(\downarrow)$, which are in close proximity to the *Fermi surface* in the momentum space, can become paired, forming the so-called ***Cooper electron pair***. Cooper estimated ***the effective binding energy of a Cooper electron pair***. The energy of *Cooper pair* is less than the sum of energies of separate electrons

$$E \approx 2E_F - 2\hbar\omega_c exp[-1/N(0)V],$$

where $E_F$ is the *Fermi energy*, $\hbar\omega_c$ is the cutoff energy away from $E_F$, and $N(0)$ is the density of states at the *Fermi level* for electrons with one spin orientation. This result showed that ***the Fermi sea is unstable in presence of an attractive interaction between the electrons such as that mediated by phonons*** [114].

In 1957, Bardeen, Cooper and Schrieffer published the microscopic theory of superconductivity, known as the ***BCS theory*** [15, 161]. The *BCS theory* successfully explains the conventional *low temperature superconductivity*, and considers the superconductivity as a *microscopic quantum mechanical effect*, describing the microscopic ground state of superconductivity for the first time. In the *BCS theory* [15, 161], the physical behavior of a large number of electrons in a superconducting state is characterized on the basis of many-body interactions among individual electrons as explained in [139]. The *BCS theory* assumes that the energy of *Cooper electron pair* is $E<2E_F$, and the phonon exchange between electrons with opposite momentums and spins produces a weak interaction with potential $V$ in Fig. 11.

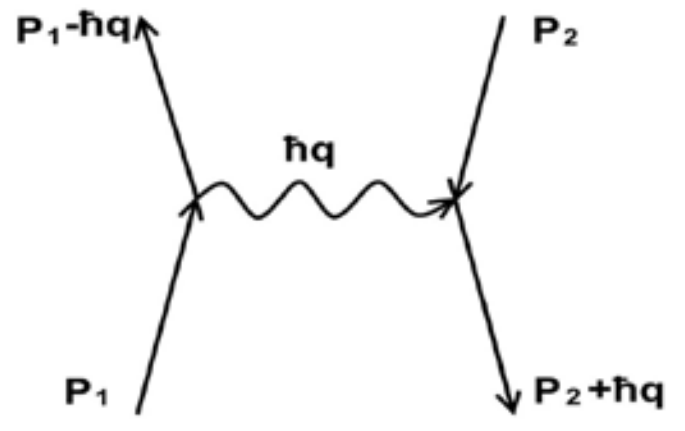

**Fig. 11.** Phonon exchange between two electrons with momentums $\boldsymbol{p_1} = -\boldsymbol{p_2}$ and phonon quasi momentum $\hbar\boldsymbol{q}$.

In superconductors, the attractive interaction between the electrons realized by the phonons exchange in vibrating crystal lattice. The electron, passing through the conductor, causes a slight increase in the density of positive charges around the electron in the crystal lattice. This increase of the density of positive charges, in turn, can attract another electron. In result, the two electrons are paired together with the certain binding energy. If this binding energy is higher than the energy of electrons collision with the oscillating atoms in superconductor, then the electron are paired, and not scattered anymore.

The *BCS theory* [15, 161] is based on the paired *Hamiltonian*

$$H = \sum_{k\sigma} \varepsilon_k n_{k\sigma} + \sum_{kl} V_{kl} c^*_{k\uparrow} c^*_{-k\downarrow} c_{-l\downarrow} c_{l\uparrow} ,$$

where $\hbar\boldsymbol{k} = \boldsymbol{p}$, and σ is the spin index, $n_{k\sigma} = c^*_{k\sigma} c_{k\sigma}$ is the particle number operator, $c^*_{k\sigma}$ and $c_{k\sigma}$ are the creation and annihilation *Fermi operators*. The first term in *Hamiltonian* relates to the electron states of normal metal, corresponding to the Fermi distribution. The second interaction term relates to the interaction energy of paired electrons, and $V_{kl}$ is the matrix element for the scattering from a state ($l\uparrow$,–$l\downarrow$) to one with ($k\uparrow$,–$k\downarrow$). Average energy of electron pair is

$$\langle V\rangle = \sum_{kl} V_{kl} c^*_{k\uparrow} c^*_{-k\downarrow} c_{-l\downarrow} c_{l\uparrow} = -\Delta_k ,$$

where $\Delta$ is the superconducting ***energy gap*** in the *BCS theory* [15, 161]. Comparing with the *Cooper theory*

$$\Delta \approx 2\hbar\omega_D \exp\left[-1/N(0)V\right], \qquad (2.26)$$

where $\omega_D$ is the highest limiting frequency of crystal lattice oscillations or the *Debye frequency* of phonons. The difference between the average energy of electrons in normal state and the average energy of electrons in superconducting state is

$$\langle E_n\rangle - \langle E_s\rangle = -N(0)\Delta^2/2,$$

where $\Delta=\Delta(T)$ is the temperature dependent term. In the *BCS theory,* at temperature of $T=0$, the energy gap is $\Delta(0)=1.76\cdot k_B T_c$. At temperature $T\to T_c$, the energy gap is approximately equal to $\Delta(T) \approx 1.74\Delta(0)(1-T/T_c)^{1/2}$ in Fig. 12 in Tinkham [12].

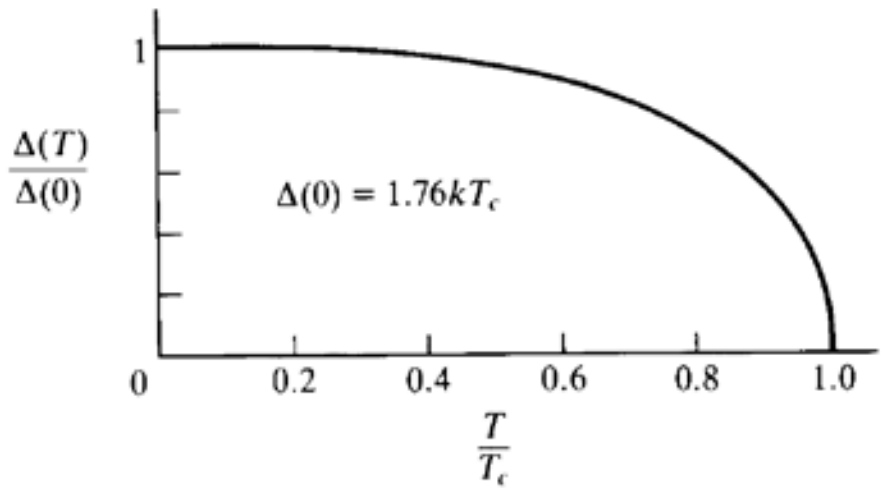

**Fig. 12**. Temperature dependence of the energy gap in the *BCS theory* (after [12]).

In the *BCS theory* [15, 161], the ***Cooper pair***s are the charge carriers in the superconductors. In the superconducting state, all the *Cooper pairs* have a single momentum and a coherent wave function in the momentum space in layer with the thickness $\delta\mathbf{p}\sim 2\Delta\cdot p_F/\varepsilon_F$ near the *Fermi surface* as shown by the author of dissertation in Fig. 13. The superconducting coherence length is $\xi_0 = 0.18\, \hbar v_F/k_B T_c \approx \hbar v_F/\pi\Delta(0)$, where $v_F$ is the *Fermi velocity* of the electrons. The wave functions of the Cooper electron pairs are in coherent *macroscopic quantum state*. The exact physical mechanisms of *Cooper pair* formation are continue to be researched by scientists [144].

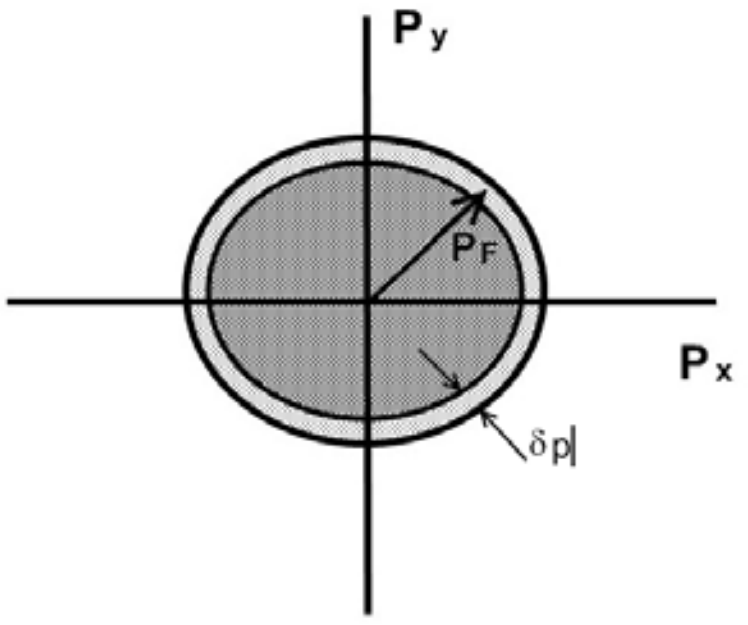

**Fig. 13.** Fermi surface and energy gap layer of superconducting electrons with thickness $\delta p$ in conventional superconductors (singlet *s-wave* wave function).

In the *BCS* theory, the origin of *ac* energy losses in a superconductor is explained as a co-existence of two types of charge carriers within the superconductor: the Cooper superconducting electron pairs and the normal electron excitations near the *Fermi surface*, which depend on the temperature *T*. The surface resistance of superconductors $R_s$ in the *BCS theory* is in eq. (2.27) [29]:

$$R_s \approx \frac{\Delta(T)}{k_B T}\omega^{3/2}\exp\left[-\Delta(T)/k_B T\right]. \qquad (2.27)$$

The *BCS theory* provides the satisfactory explanation of the phenomenon of superconductivity in *LTS* [117], and describes properties of superconductors with low critical temperature $T_C$ well [10].

Tohyama [125] writes that the *BCS* theory is a successful theory in condensed-matter physics, since a rigorous treatment of electron–phonon interaction is applicable, using the many-body perturbation theory. However, the *BCS* theory is inapplicable in unconventional superconductors, where electron-electron interaction plays an important role in electron pairing instead of electron-phonon interaction [125]. In these unconventional superconductors, the *superconducting energy gap function* may have the ***nodes*** to escape the effect of the Coulomb repulsion [125].

In the high temperature superconductors (*HTS*), the electron wave function can not only be in the even *s-wave* mode ($l$=0 and spin $S=0$) as in the conventional superconductors, but also be in the even *d-wave* mode with the orbital momentum number $l=2$ [30]. In these *d-pairing* superconductors, the *energy gap parameter* $\Delta$ has the ***nodes***, where it equals to zero on the *Fermi surface* shown by the author of dissertation in Fig. 14.

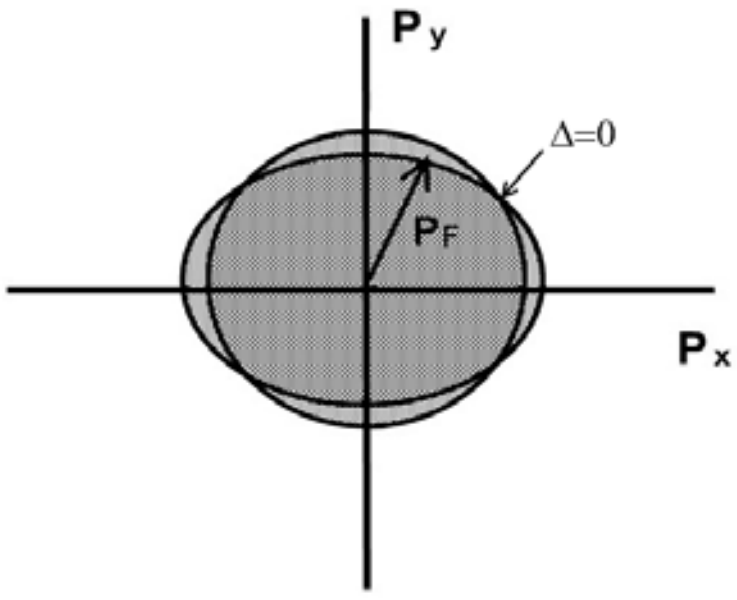

**Fig. 14.** Fermi surface and energy gap in *d-wave* high $T_c$ superconductors ($d_{x2-y2}$ wave function).

The pairing energy for electrons has the maximum magnitude, when the momentums of electrons are parallel to $P_X$ and $P_Y$ axes in Fig. 13. The energy gap is equal to zero $\Delta(p_x, p_y)=0$, when the momentum vector is positioned at 45° between the axes. In the *HTS* materials, such decrease of magnitude of energy gap in certain positions on the *Fermi surface* results in the change of dependence of the surface resistance on the temperature $R_S(T)$ (eq. 2.28) [30] and in some other effects [31]. These nodes give significant additional contribution to the density of state of normal electrons, and therefore they may change the temperature dependence of magnetic field penetration depth and the electronic properties of magnetic vortices in these superconductors [32]. The influence by these properties of electronic spectrum also appears in the change of impedance $Z_S$, especially its nonlinear part at microwaves, however presently, the theoretical research in this field was completed for the intermodulation effects only [37, 98].

The authors of book would like to note that, in the application of the *BCS* theory to *d-pairing* superconductors in case ξ<<λ (local approximation), the surface resistance $R_s$ can be expressed as an integral in eq. (2.28)

$$R_s \propto \frac{\omega^{3/2}}{k_B T} \int_S \Delta(p_x, p_y) \exp\left[-\Delta(p_x, p_y)/k_B T\right] dp_x dp_y \,, \qquad (2.28)$$

where the pairing energy gap is represented in every point $(p_x,\ p_y)$ of *Fermi surface* as $|\Delta(p_x,p_y)|=\Delta_0\cdot|p_x^2-p_y^2|$, and the integral is computed by all the *Fermi surface*.

Bardeen, Brattain, Shockley were awarded the Nobel Prize in the physics for the transistor invention in 1956; and Bardeen, Cooper, Schrieffer were awarded the Nobel prize in the physics for the theory of superconductivity formulation in 1972 [162]. John Bardeen is the only person to have received the Nobel Prize in the same research field for the two times [162].

## 2.8. Fundamental Properties of Superconducting Materials.

Superconducting materials have two fundamental phenomena associated with their physical properties such as the ***vanishing resistance phenomenon,*** and the ***diamagnetic behaviour phenomenon***. Certain conditions in terms of values of temperature, external magnetic field, and frequencies have to be satisfied to maintain the material superconducting state.

All the fundamental properties of superconductors are discussed in this section of dissertation. The principle behind the classification of superconductors, known as *Type-I* and *Type-II* is featured as well. The information about the prominent scientists, who made the significant contributions to the research on the *superconductivity* is shown in Tab. 1, and the world renowned scientist, who contributed to the research on the *He*, $^{3}He$, $^{4}He$ *and superfluidity* is presented in Tab. 2 in the *Appendix 2* [75-78]. More detailed information on the biographies of the Nobel laureates in the field of superconductivity is also presented in the *Appendix 2* in this dissertation [49, 86].

## 2.9. Ideal Conductivity of Superconductors.

As it is well known, the electrical resistivity of all metals and alloys decreases, when they are cooled down [8]. The nature of this effect lies on a microscopic level, where the current in a conductor is carried by “conduction electrons,” which are free to move through the material. As electrons have a wave-like nature, assuming a perfect crystal with perfectly periodic metal crystalline structure without any loss of momentum, then the current should experience no resistance. However, as we all know, the thermal vibrations and impurities or imperfections of microscopic structure can affect the perfect periodicity of a crystal lattice and so create resistance to current flow. Although, it is hypothetically possible to reach the zero resistance with a “perfect” sample of metal at temperature $T=0K$ such effect must not be considered as the phenomenon of superconductivity. In reality, any metal sample can not be perfectly pure and always contains some impurities. In this context, the nature of superconductivity is very remarkable, because the electrical resistance of superconductors decreases in usual way, but they exhibit the ideal conductivity until a certain critical temperature during the cooling process is reached in Fig. 15 [82, 3].

The fact that the superconductor has no resistance means that there is no voltage drop along the conductor, when the transport current passes through the sample. The phenomenon of “zero” resistance can be observed in the case of flow of direct current (*dc*) of constant magnitude in superconductor only. In 1931, the careful experiments by de Hass and Voogd [90] were conducted, and it was proposed that in "ideal' conditions the transition from the normal resistance state of the superconducting material to its superconducting state would be practically discontinuous. The temperature at which superconductivity first occurs in a material is thus termed the *superconducting transition temperature* or ***critical temperature*** of the material and is denoted by *Tc* [91].

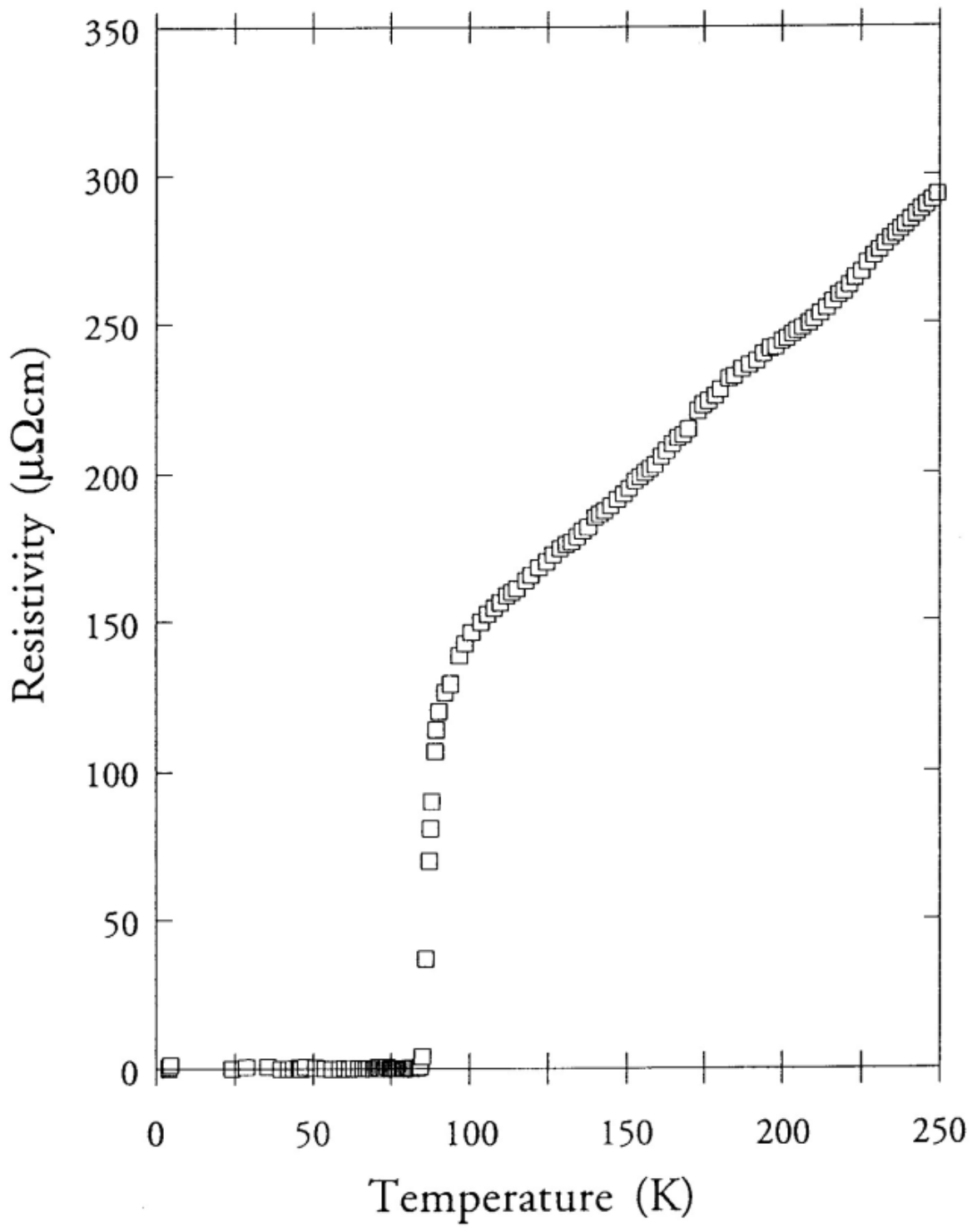

**Fig. 15.** Temperature dependence of *dc* resistivity in $YBa_2Cu_3O_{7-\delta}$ superconductor (after [3]).

*In the presence of a time varying magnetic field, the resistance of a superconductor is non-zero, although the magnitude of its resistance is still much lower than in the best conventional conductors.* The phenomenon behind the *ac* resistivity can be easily explained by “two fluid model” [6-9], considered in more details later in this chapter, in which the two types of conduction electrons are involved into the current distribution in a superconductor. The first type is “superelectrons,” which form so-called *Cooper pairs* and carry the transport current without any resistance, and the second type is “normal” electrons, they scatter and experience the resistance in the metal lattice [9-13].

## 2.10. Diamagnetic Behaviour - Meissner Effect.

Superconductors have a unique combination of electric and magnetic properties: *perfect conductivity* (zero resistance) and *perfect diamagnetism*. Although superconductors conduct current with no resistance, they do not simply behave as perfect conductors.

In a perfect conductor the imaginary path of resistance is zero, therefore the amount of magnetic flux trapped within this path cannot change: $dB/dt=0$. It means that, if the perfect conductor with no applied magnetic field is cooled down to very low temperature, and then the external magnetic field is applied, the magnetic flux will not be allowed to penetrate the superconductor sample due to the *screening currents* that are induced in order to have an unchanged magnetic field in the volume of material [14].

However, if the magnetic field is applied to the perfect conductor before the cooling process, the flux density inside of volume of material will be the same as that of the external field. Cooling down to the low temperatures will have no effect on the magnetisation, and the flux distribution will remain unchanged. If the applied field is then removed, the currents will be induced in the volume of perfect conductor in order to maintain the flux inside unchanged in Fig. 16 1) (a) [11].

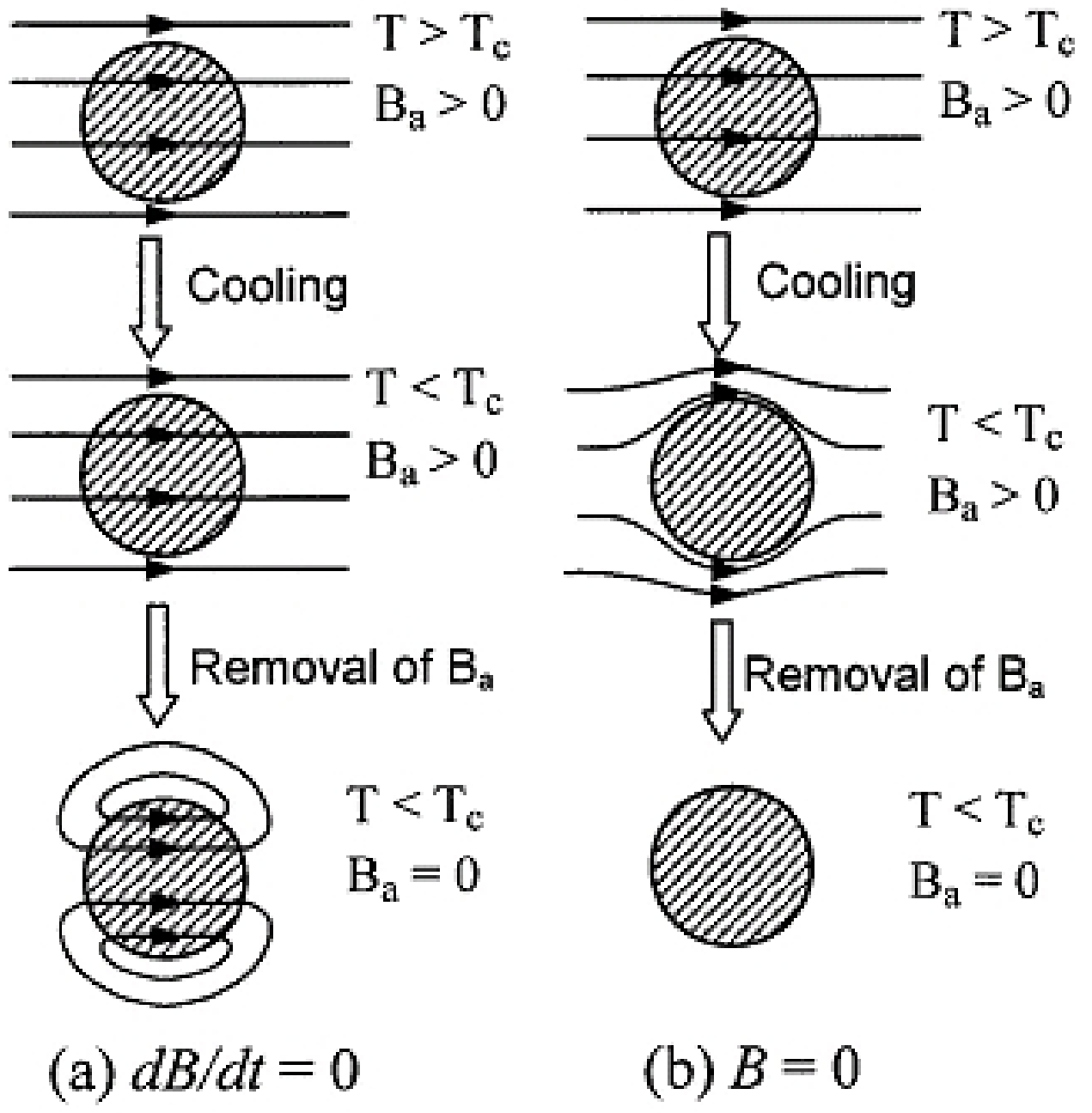

**Fig. 16. 1)** Magnetic behaviour of a "perfect" conductor **(a)**, and a superconductor **(b)** (2) (after [11]).

Superconductors behave in a different way than perfect conductors, when the magnetic field is applied before the cooling ($T > T_c$). The phenomenon, called the ***Meissner effect,*** was discovered by Meissner and Ochsenfeld in 1933 [14], when researchers measured the flux distribution in metallic superconductors cooled down to their critical temperatures, while in the magnetic field. Surprisingly, Meissner and Ochsenfeld observed that a metal in the superconducting state never allowed the magnetic flux density to exist in its inside volume, i.e. it exhibits the ***perfect diamagnetism*** in Fig. 15 1) (b) (2) [11, 12, 13]. The absence of any magnetic flux in a pure superconductor independent of the initial conditions is an additional fundamental property of the superconductor since it cannot be deduced from the perfect conductivity [91].

Fig. 17. 1) shows the schematic representation of the Meissner effect in a superconductor [131]. 2) shows the Meissner effect in *Type I* superconductors [82]. Fig. 17. 3) demonstrates the Meissner effect in *Type II* superconductors [82].

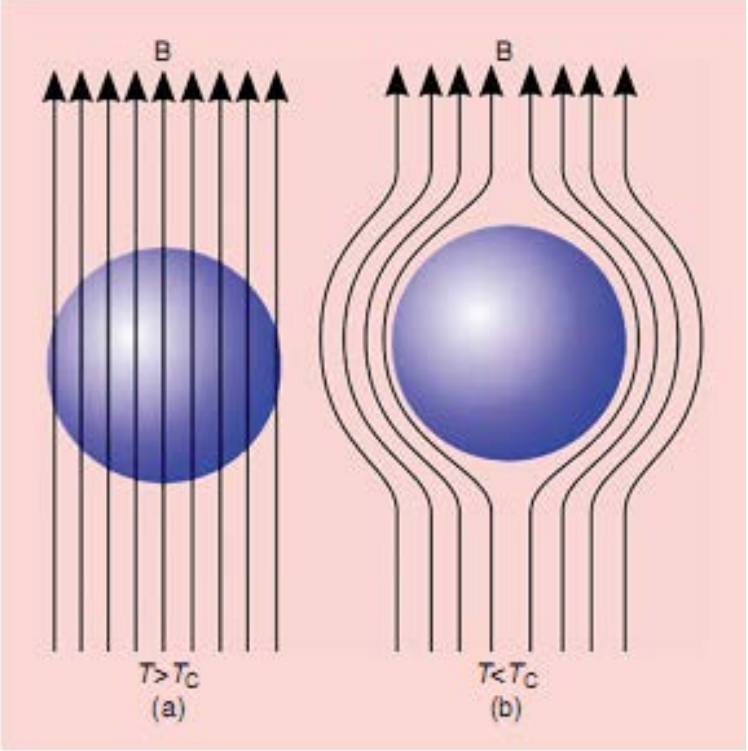

1)

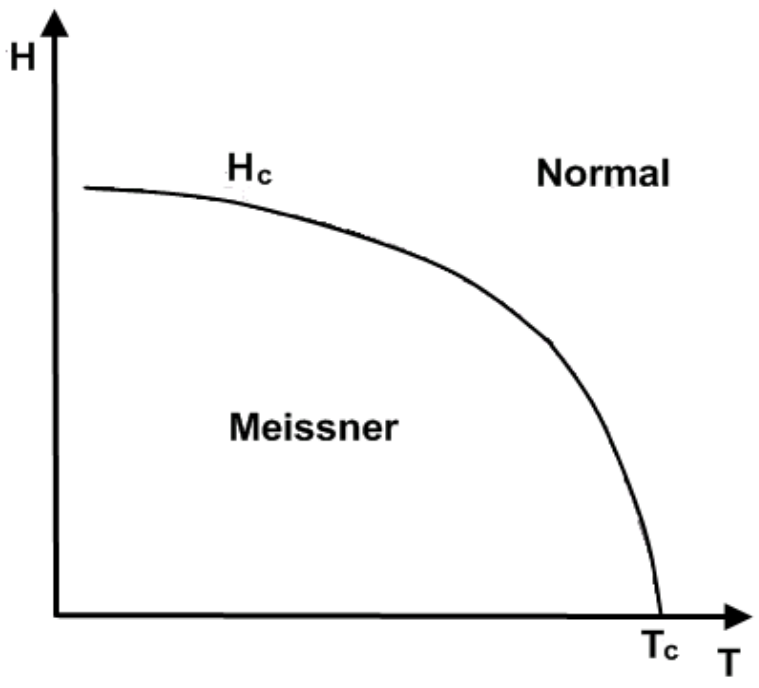

2)

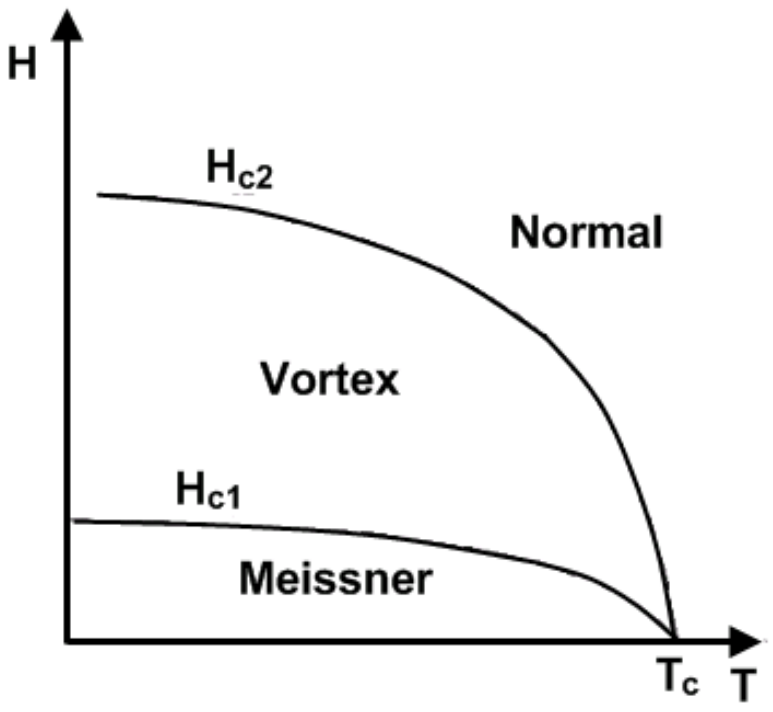

3)

**Fig. 17. 1)** Schematic representation of the Meissner effect in a superconductor: **(a)** The magnetic behavior when the specimen is in the normal state (i.e., at temperatures above Tc). **(b)** The magnetic behavior when the specimen is in the superconducting state (i.e., at temperatures below Tc) (after [131]).

**2)** Meissner Effect in *Type I* superconductors (after [82]).

**3)** Meissner Effect in *Type II* superconductors (after [82]).

On the other hand, it should be mentioned that currents, which circulate to cancel the applied magnetic flux inside a superconducting sample, cannot be confined entirely to the surface, therefore the applied magnetic field is not completely screened out from the superconductor. In fact, the field penetrates on a small distance into the material (approximately $10^{-5}$ *cm* [9]), enabling currents to flow within this very thin surface layer. The magnetic flux density decays exponentially as it penetrates into the superconductor. The depth at which the magnetic field falls to *1/e* of the field at the surface is called the *London penetration depth* $\lambda_L$, or simply the *penetration depth* $\lambda$ [9, 10, 11, 12].

## 2.11. Critical Parameters of Superconductors.

As mentioned earlier, just after the discovery of superconductivity by Kamerlingh Onnes, the superconductivity phenomenon, enabling the transport currents flow without any resistance, looked very promising for numerous practical applications. Further research, however, unveiled a number of physical pre-conditions that essentially slowed down the progress towards the industrial applications of superconductors for several decades. The discoveries of particular threshold magnitudes of applied magnetic field, current, and frequency, that need to be satisfied to keep the existence of superconductivity, were the main reasons for the limitations. In other words, the superconductor must not exceed one of certain critical magnitudes of the temperature, magnetic field, current and frequency in order to remain in superconducting state, because the superconductivity breaks down above the critical parameters magnitudes [9-13].

## 2.12. Critical Temperature.

The critical temperature ($T_c$), as mentioned earlier in this chapter, is the highest temperature at which the superconductor can maintain the superconducting state [9-13]. The higher $T_C$ of a superconductor is, the higher chances for its practical applications are, because the cheaper cryogenic refrigeration techniques is needed. Although, in real practical applications, the superconductor might have to satisfy some other criteria besides the temperature. The critical temperature for $YBa_2Cu_3O_{7-\delta}$ is *91-93K*. The evolution of the critical temperature of superconductors was shown in Fig. 2.

## 2.13. Critical Frequency.

The frequency at which the photons of electromagnetic waves excite the superconducting electrons with enough energy to drive them to the normal state is defined as the *critical frequency*. The magnitudes of the frequency $\omega_C$ , which force a material to lose its superconducting properties at $T=0$, are within the low optical frequency range of $10^{12}$ *Hz* [9-13] for low temperature superconductors (*LTS*), and in frequency range of $10^{13}$ *Hz* for high temperature superconductors (*HTS*) approximately. The critical frequency can be defined from equation

$$\hbar\omega_c(T)=2\Delta(T),$$

where $\hbar$ is the reduced Planck's constant, and $\Delta(T)$ is superconducting energy gap in the *BCS theory* [15].

The superconductors are mainly used in electronic devices at frequencies of well below *100GHz*, therefore the *critical frequency* parameter is not so important for many practical applications of superconductors [11]. Nisenoff [131] comments that for most practical high-frequency applications of superconducting materials, the operating frequency of the superconductor device should be about 10% or less of the frequency corresponding to the *energy gap* of the material.

## 2.14. Critical Magnetic Fields.

The *critical magnetic field* $H_c$ is defined as the highest value of external magnetic field strength applied to a superconductor without causing superconductivity to disappear [91]. In the case, when the strength of the applied magnetic field is increased, the screening currents must also increase and if the critical current density is reached by the screening currents the material will lose its superconductivity. This destruction of superconductivity by a sufficiently strong magnetic field is one of the most important properties of a superconductor. Magnitude of the critical magnetic field $H_c$ can be found from the equality densities of free energies of the normal metal and the superconductor in eq. (2.32)

$$F_N(T) = F_S(T) + \frac{1}{2}\mu_0 H_C^2(T). \qquad (2.32)$$

In terms of temperature dependence, the critical magnetic field of a superconductor, in general, has a zero value at $T_c$, and reaches its maximum at absolute zero temperature, as shown in Fig. 18 [91, 20], which illustrates the dependence of critical magnetic field as a function of temperature.

The critical field strength decreases as the square of temperature, so the critical field curves are closely approximated by parabolas of the form in eq. (2.33)

$$H_C = H_0\left[1-\left(\frac{T}{T_C}\right)^2\right], \tag{2.33}$$

where $H_0$ is the critical field at absolute zero and $T_c$ is the critical temperature [9-13].

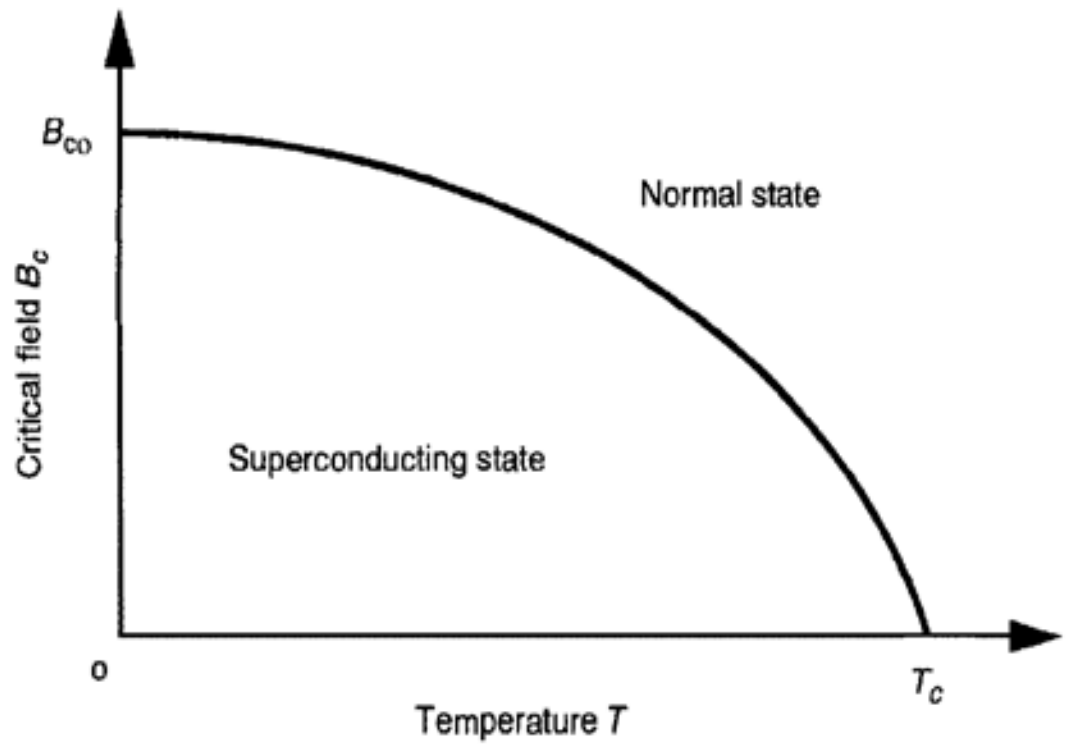

**Fig. 18.** The critical induction of magnetic field B=$\mu_0$H as a function of temperature (after [91]).

Tab. 5 provides information on the magnitudes of critical temperatures $T_c$ and critical magnetic fields $B_c=\mu_0 H_c$ of some *LTS* superconductors [91].

| Element | $T_c$ (K) | $B_{co}$ (Tesla) |
|---|---|---|
| Aluminum | 1.2 | $0.99 \times 10^{-2}$ |
| Gallium | 1.1 | $0.52 \times 10^{-2}$ |
| Indium | 3.4 | $2.76 \times 10^{-2}$ |
| Lead | 7.2 | $8.04 \times 10^{-2}$ |
| Mercury α<br>Mercury β | 4.2<br>4.0 | $4.15 \times 10^{-2}$<br>$3.39 \times 10^{-2}$ |
| Niobium | 9.3 | 0.2 |
| Rhenium | 1.7 | $2.01 \times 10^{-2}$ |
| Tantalum | 4.5 | $8.30 \times 10^{-2}$ |
| Thalium | 2.4 | $1.76 \times 10^{-2}$ |
| Tin | 3.7 | $3.02 \times 10^{-2}$ |
| Vanadium | 5.4 | 0.141 |
| Zinc | 0.9 | $0.53 \times 10^{-2}$ |
| Techneitium | 7.8 | 0.141 |
| Zirconium | 0.6 | $0.47 \times 10^{-2}$ |

**Tab. 5.** Critical temperatures $T_c$ and critical magnetic induction $B_c$ of some *LTS* superconductors (after [91]).

## 2.15. Classification of Superconductors in Type I and Type II.

In 1935, Rjabinin, Schubnikow [150, 151] experimentally discovered the *Type II* superconductors at the cryogenic laboratory at the National Scientific Center Kharkov Institute of Physics and Technology in Kharkov, Ukraine. Rjabinin, Schubnikow [150, 151] found that, at the certain temperature, the superconducting alloys (*Type II* superconductors) have the two critical magnetic fields $H_{c1}$ and $H_{c2}$ in distinction from the pure superconductors (*Type I* superconductors), which have the only critical magnetic field $H_c$.

In August, 1963, Bardeen and Schmitt on behalf of the members of International Conference on the Science of Superconductivity, conducted in Colgate University, Hamilton, N.Y., USA, addressed the *IUPAP* to "advise the use of the symbols $H_{c1}$ and $H_{c2}$ to express the limits of the mixed state in *Type II* superconductors. $H_{c2}$ is taken to be the upper field limit, while $H_{c1}$ is reserved for the lower field limit. This notation (in the Russian version, $H_{k1}$ and $H_{k2}$) is that introduced by Shubnikov in 1937, who was the first to give names to the critical fields." Bardeen and Schmitt referred to the experimental research by Schubnikov, Chotkevitsch, Schepelev, Rjabinin [153] in 1937; however the real fact is that the experimental discovery of the *Type II* superconductors was made by Rjabinin, Schubnikow [150, 151] in 1935 (see L. V. Schubnikov [154]).

In 1957, Abrikosov [16] explained that the original idea that the parameter $k = \xi/\lambda$ can be used to classify the *Type I* ($k<1/\sqrt{2}$) and *Type II* ($k>1/\sqrt{2}$) superconductors belongs to Lev D. Landau. Abrikosov [16] further developed the extensive theoretical grounds for the *Type I* and *Type II* superconductors classification at the external magnetic fields.

The physical properties of the *Type I* superconductors are well described earlier in this chapter. The magnetic field penetration depth $\lambda$ is smaller than the coherence length $\xi_0$ ($\lambda<\xi$), characterizing the spatial correlation between the pairs of superconducting electrons, in the *Type I* superconductors. This class of superconductors has a single critical field $H_C$ from $10^{-2}$ to $10^{-1}$ *Tesla*, at any fixed temperature [9, 10]. In the magnetic field $H<H_C$, which depends on the geometric form of superconductor, the superconductors can transit into an intermediate state. The intermediate state represents the layered structure of alternating normal and superconducting layers of superconductor in external magnetic fields (*Landau structure*) [10, 12]. The *quantum intermediate state* in high purity *Gallium* single crystal, which is *Type I* superconductor, was firstly observed and researched in O.P. Ledenyov [146]. All the pure superconducting metals, except *Nb*, are *Type I* superconductors.

Apart from the normal and superconducting states, associated with *Type I* superconductors, the *Type II* superconductors can be in a *mixed state*, where the *Meissner effect* is no longer present, and the magnetic flux may partially penetrate the superconductor [16]. *Type II* superconductors have the coherence length ξ, which is smaller than the penetration depth ($\xi < \lambda$). They are also unique in the way of having two critical magnetic fields, a ***lower critical magnetic*** $H_{c1}$ (2.34) and an ***upper critical magnetic*** $H_{c2}$ (2.35)

$$H_{C1} \approx \frac{H_C \ln k}{k\sqrt{2}}, \quad (2.34); \qquad H_{C2} = H_C k\sqrt{2}, \quad (2.35);$$

where $k = \xi/\lambda$ is the *GL parameter*, $H_C$ is the thermodynamic critical magnetic field.

Below $H_{c1}$, the superconductor is in the diamagnetic superconducting state with minor concentration of normal charge carriers, but when the external magnetic field exceeds the threshold magnitude $H_{c1}$, the superconductor transits into a mixed phase. This phase holds a large number of quantum magnetic vortices, where each vortex has the normal metal core with radius $\xi$ and traps the magnetic flux quantum $\phi_0 = 2{,}0679x10^{-15}$ *Weber* [12]. A unique condition exists within the mixed state, when the superconducting transport currents can still flow through the material even though significant domains of the superconductor material are in the normal state, but all the magnetic vortices are pinned by crystal structure defects and can not move [16]. The superconductivity disappears fully, when the external magnetic field exceeds the upper critical field $H_{c2}$ [16].

Two critical magnetic fields in the *Type II* superconductors have distinct temperature dependences as shown in Fig. 19 **(a)**. The ***upper critical magnetic field*** $H_{c2}$ is typically much higher than the critical magnetic field $H_c$ of the *Type I* superconductors. The *Type II* superconductor can have the $B_{c2} = \mu_0 H_{c2}$ magnitude above 100 *Tesla* at $T= 0$, while the critical magnetic induction $B_c$ is below 0.2 *Tesla* in the *Type I* superconductor. Therefore, the *Type II* superconductors have found their applications in strong magnets with superconductor wires for particle accelerators for generation of very strong magnetic fields.

In Fig. 19 **(b)**, when a current is applied to a *Type II* superconductor (blue rectangular box) in the mixed state, the *Abricosov magnetic vortices* (white cylinders) subjected to the *Lorentz force* action that pushes the *Abricosov magnetic vortices* at right angles toward the current flow with the dissipation of energy and appearance of resistance in superconductor. The *proximity effect* on the *NS* boundaries between the normal vortex core and superconductor can also be observed in a mixed state of superconductor [82, 83].

Fig. 19 **(c)** shows the detailed structure of an isolated *Abrikosov vortex line* in the *Type II* superconductor in Galperin [102], Eschrig [103], Ø. Fischer [126-130].

**a)**

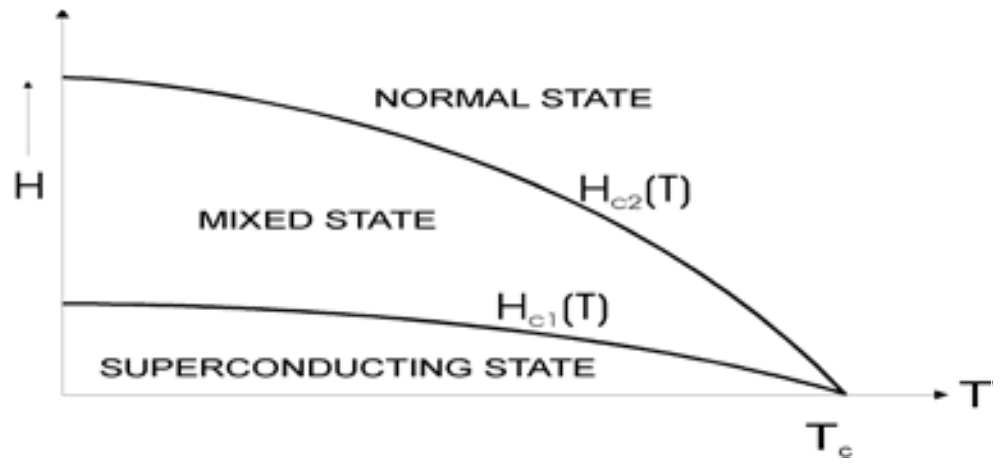

**b)**

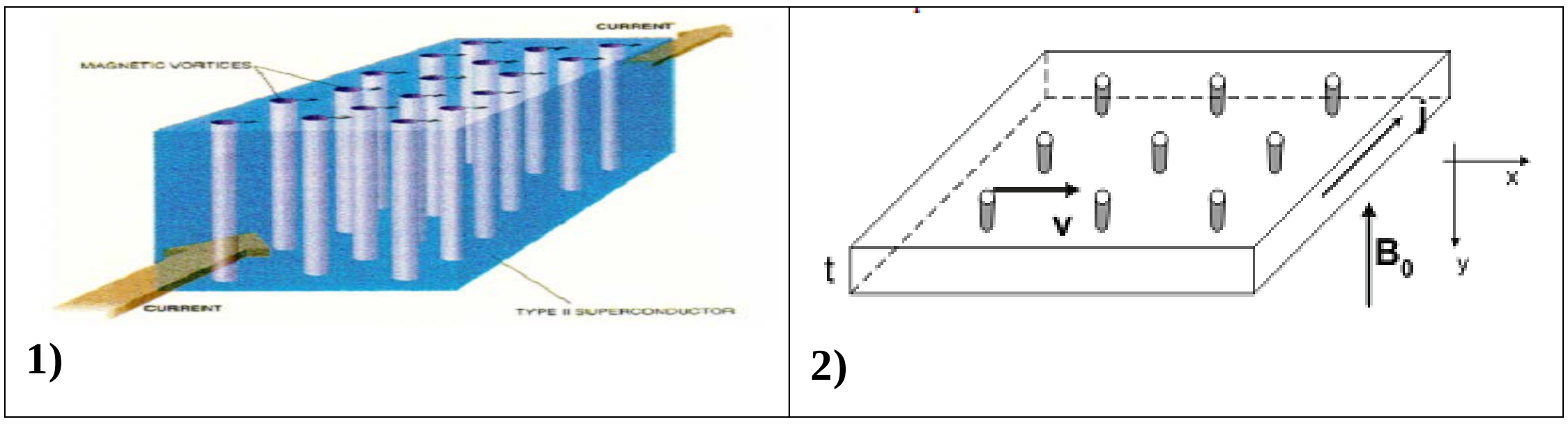

**c)**

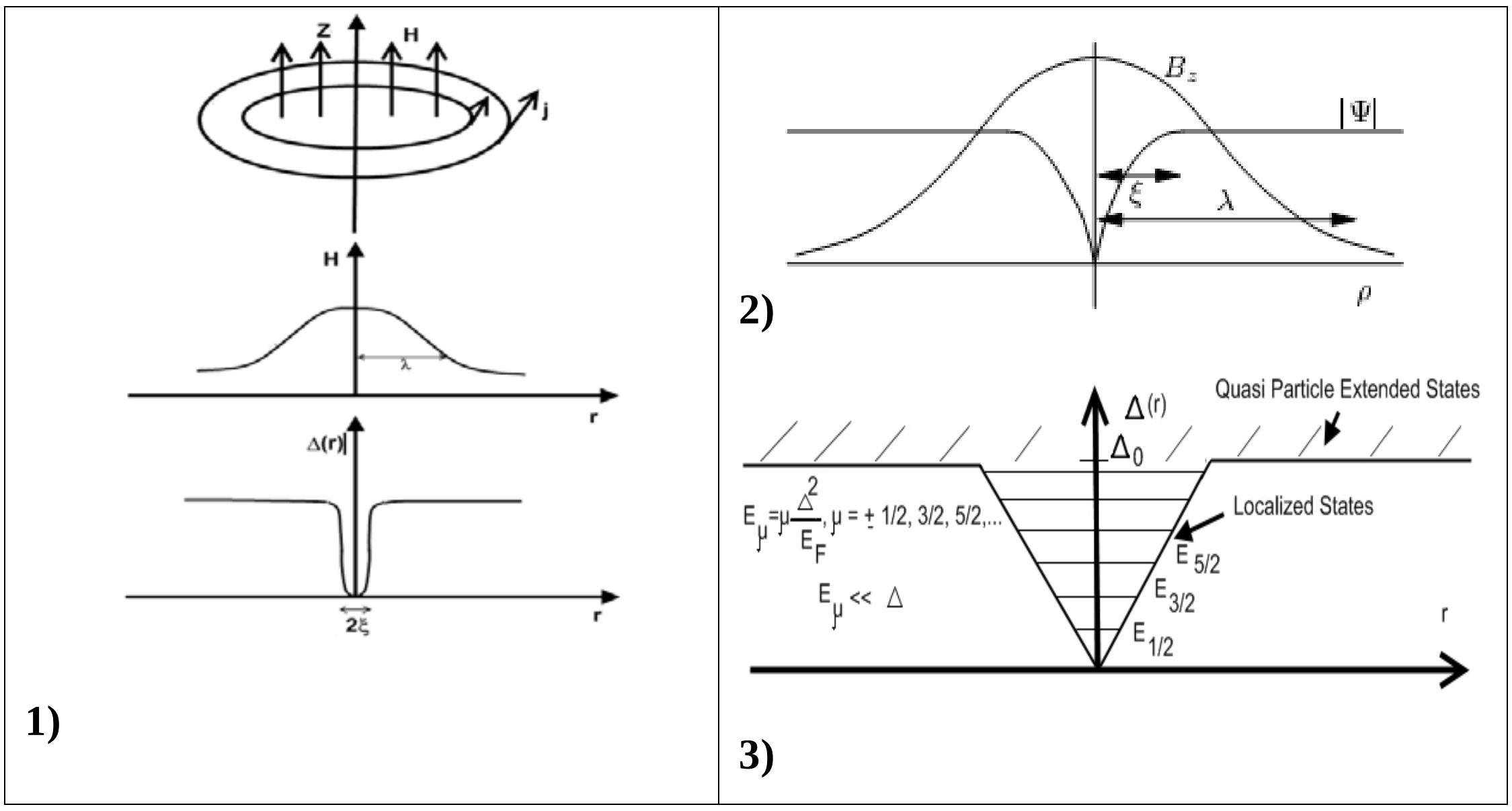

**Fig. 19. a)** Temperature dependences of critical magnetic fields $H_{C1}$ and $H_{C2}$ in *Type II* superconductors by Rose-Innes and Rhoderick (after [9]). **b) (1)** *Type II* superconductor in mixed state with *Abricosov magnetic vortices* and transport current direction shown by Orlando (after [82, 83, 132]), **(2)** *Abricosov magnetic vortex lattice* generation in *Type II HTS* thin film at application of external magnetic field ***B*** by Golovkina (after [109]). **c) (1)** Structure of an isolated *Abrikosov vortex line* in a *Type II* superconductor by Galperin (after [102, 133]), **(2)** by Eschrig (after [103]), and **(3)** electronic structure of a vortex core by Ø. Fischer (after [126-130]).

D. O. Ledenyov [149] made an innovative scientific discovery that the quantum knots of Abricosov and Josephson magnetic vortices are in an extreme quantum limit in 2000, which is important for the fundamental and applied physics.

Matsushita [145] has derived the force balance equation that determines the electric current flow in the superconductor by using the variation principle and researched the longitudinal magnetic field effect in superconductors in [146].

The capacity to conduct the current in *Type II* superconductor is sensitive to the presence of impurities in sample's volume. As known, the point-like impurities can cause the *Type II* superconductor to have small critical current [4, 16], meaning that the superconductivity phenomena may be observed, but the property to conduct the current is suppressed in *Type II* superconductors. However, on the other hand, the big critical currents can be observed in *Type II* superconductor with large concentration of inhomogeneities in sample, where the effective dimension of every inhomogeneity is $d \geq \xi$ [4, 16]. The critical currents can reach the big magnitudes, because the vortices can be pinned on the inhomogeneities, being in fixed positions until the moment, when the big magnitude of critical current is reached [126-130].

## 2.16. Critical Current Density.

Tinkham [12] considers a long superconducting wire of circular cross section with radius $a >> \lambda$, carrying a current *I*. This current produces a circumferential self-field at the surface of the wire of magnitude $H = I/2\pi a$. When this field reaches the critical field $H_c$, it will destroy the superconductivity (*Silsbee criterion*.) Thus, the critical current will be $I_c = 2\pi a H_c = 2\pi a B_c/\mu_0$, which scales with the perimeter, not the cross-sectional area, of the wire. This suggests that the current flows only in a surface layer of constant thickness of superconducting wire. It can be confirmed analytically by application of the *London and Maxwell equations* in this geometry that this is so, and that the thickness of the surface layer is $\lambda$. Since the cross-sectional area of this surface layer will be $2\pi a\lambda$, the critical current density $J_c$ will be $I_c/2\pi a\lambda$, namely,

$$J_C = \frac{B_c}{\mu_0 \lambda}$$

Shu-Ang Zhou [91] elucidates that the experimental evidence of that the current density in a superconducting wire could only be increased to a certain threshold value until a voltage occurred had led the introduction of the concept of *critical current density* $\boldsymbol{J_c}$. According to a hypothesis put forward by Silsbee [9-12] in 1916, the threshold values of the current and the magnetic field are simply related: The magnetic field produced by the critical current $\boldsymbol{J_c}$ at the surface of the superconductor equals the critical magnetic field, which gives

$$J_c(T) = B_c(T)/(\mu_0 \lambda(T)),$$

where $\lambda$ is the magnetic field penetration depth [91].

Author of dissertation would like to add that the maximum current that can be transported through a *Type I* superconductor ($\lambda < \xi$) without causing resistance to appear is called the *critical current* $\boldsymbol{I_c}$. This current is connected with the critical magnetic field $H_C$ and the diameter of superconductor $d$ by the *Silsbee rule* in eq. (2.29) [9-13]

$$I_c = \pi d H_c \text{ (Silsbee rule)}. \qquad (2.29)$$

In other words, the *Type I* superconductor loses its zero resistance, when, at any point on its surface, the magnitude of strength of total magnetic field $\boldsymbol{H}=\boldsymbol{H_I}+\boldsymbol{H_e}$,
which is created by the transport current and by the external magnetic field, reaches the critical field strength $H_c$ ($H= B/\mu_0$, where $B$ is the magnetic flux density).

In the case of *Type II* superconductors ($\lambda>\xi$), which are widely used in technology, the critical current density $J_C$ is an important parameter in eq. (2.30)

$$J_c = I_c/S, \qquad (2.30)$$

where $S$ is the cross-section area of superconductor and $I_c$ is the critical current, which in this instance is not connected with the *Sylsbee rule*. The critical current magnitude $J_c$ is defined by the pinning force of the *Abrikosov magnetic vortices* [16], which is present in *Type II* superconductors in range from the lower critical field $H_{c1}$ up to the upper critical field $H_{c2}$. These magnetic fields define the lower and upper boundaries of *mixed state of superconductor*. Superconducting current in the *Type II* superconductor in mixed state transports in the whole volume of superconductor. Each magnetic vortex traps a quantum of the magnetic flux $\phi_0$=2,0679 $10^{-15}$ *Wb*. Before that moment, while vortices are still pinned on defects of crystal lattice, the superconducting current transports in superconductor without dissipation. The vortices begin to move only, when the transport current reaches the critical density $J_C$ and the magnitude of the *Lorentz's force* in eq. (2.31)

$$\boldsymbol{F_L} = N\phi_0[\boldsymbol{n}\times\boldsymbol{J_C}], \qquad (2.31)$$

acting in unit of volume of superconductor on the part of current and on vortices, begins to exceed the pinning force $F_P$, which holds all the vortices in the unmoved state in this volume of superconductor ($N$ is a number of vortices in a unit of the volume, n is a single vector to the parallel magnetic field). As a result of vortices movement, the electric potentials difference appears in the sample, and its resistance becomes to be different from zero (the resistive state). In engineering applications, it is conditionally accepted to consider that the state is a resistive state, if the electric potentials difference on unit of length of superconductor, reaches $10^{-6}V/m$. The critical current density has a distinctive dependence on the temperature, it starts from $T_C$, and as temperature decreases, reaches its maximum at the absolute zero Kelvin.

The critical current density $J_c$ dependence on temperature for $YBa_2Cu_3O_{7-\delta}$ films on $LaMnO_3$-buffered biaxically textured Cu tapes for various magnitudes of the applied magnetic field is shown in Fig. 20 [116].

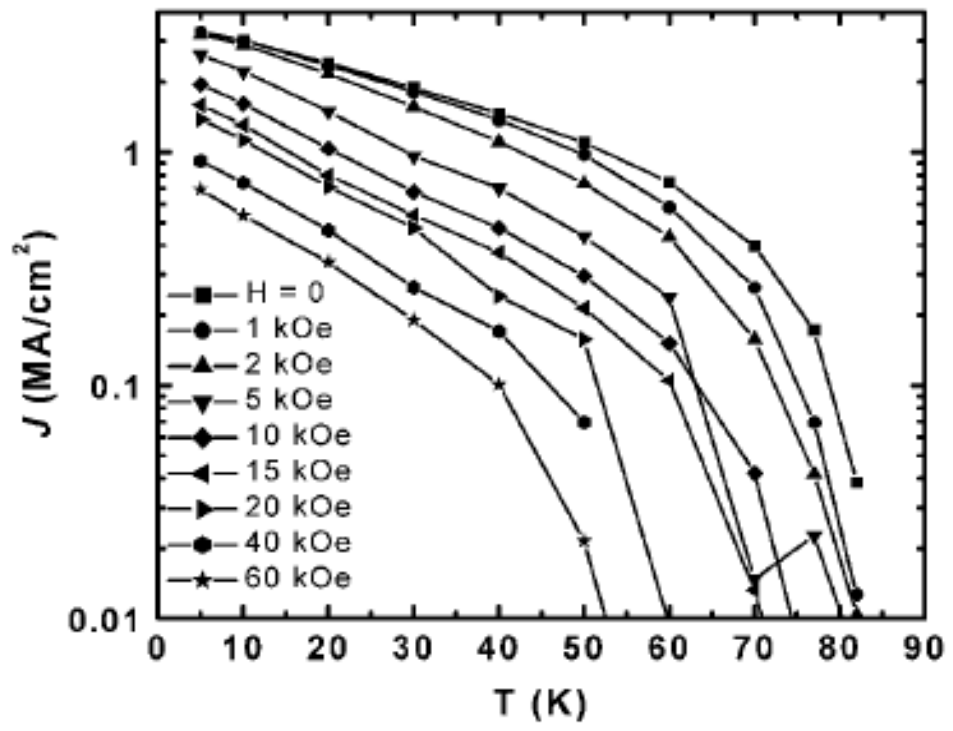

**Fig. 20.** Critical current density $J$ versus temperature $T$, with magnetic field $H$ applied normal to the $YBa_2Cu_3O_{7-\delta}$ superconducting tape (after [116]).

In different situation, when the superconductor is placed in variable electromagnetic field with the frequency $\omega$. The variable magnetic field

$$H_\omega = H_0\ exp(i\omega t)$$

penetrates in the surface layer of the superconductor, and generates the variable electric field

$$E_\omega = E_0 \cdot exp(i\omega t),$$

which connected with the magnetic field by *Maxwell equation* [17, 18, 19]

$$\nabla\times\mathbf{E}_{\omega} = -\mu_0\, d\mathbf{H}/dt = -i\omega\mu_0\mathbf{H}_{\omega}.$$

Under the action of field $E_{\omega}$, the variable currents (normal and superconducting) are

$$I_{\omega}=I_0\cdot exp(i\omega t),$$

which exist in close proximity to the surface of superconductor at the penetration depth $\lambda$. The variable electric field involves not only superconducting, but also normal electrons in motion, and total transport current is

$$I_0 = I_{S0}+I_{N0}.$$

The subsequent energy dissipation of normal electrons brings about partial dissipation of flowing current in superconductor, and in this sense, the superconductor is not an ideal conductor even in a variable field with frequency $\omega$, for instance, 50*Hz* or 60*Hz*, which are used in industry. However, the superconductor's surface resistance $R_S$ turns out to be in much smaller orders of magnitude than in comparison with any normal metal at low temperature, that property defines the superconductor's advantage in applications in the passive and active electronic devices at the ultra high frequencies (*UHF*) [13].

Fig. 21 shows the interdependence of the critical temperature $T_c$, critical magnetic field $H_c$, and critical current density $J_c$ in a superconductor [20, 131].

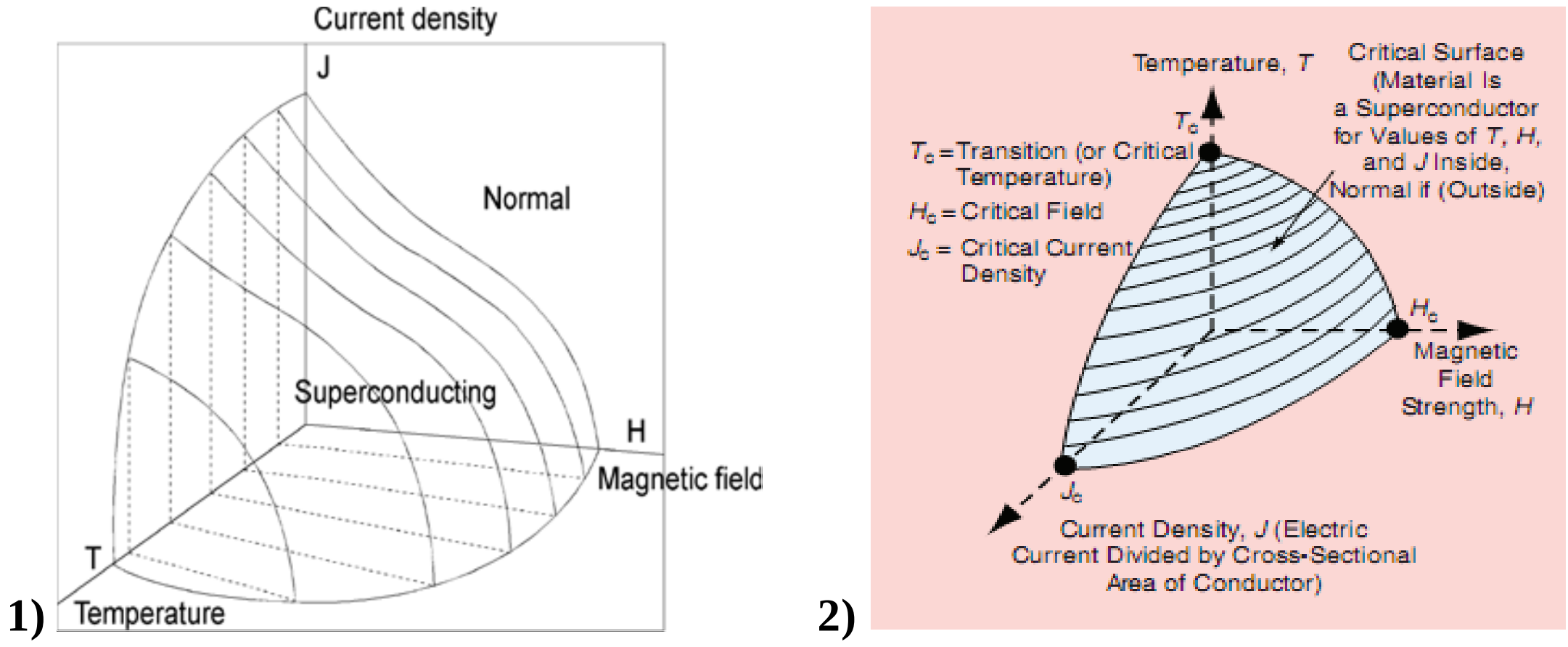

**Fig. 21. 1)** *THJ* phase diagram of interdependence of critical parameters of a superconductor (after [20]).**2)** A three-dimensional (3-D) plot of *Tc*, *Hc*, and *Jc* surface that defines superconductivity. Materials are superconducting, if all of these values are below the surface shown; whereas they are in normal state, if any of the three parameters fall outside the surface (after [131]).

## 2.17. Historical Overview on Technical Applications of Low Temperature Superconductors (LTS).

Discovery of superconductivity almost a century ago looked very promising for practical applications of superconductors. The research on material properties of superconductors led to the understanding of limitations imposed by the critical parameters of superconductors: *temperature*, *current*, and *magnetic field* on the use of superconductors in practical applications. The refrigeration systems presented an economical barrier for commercial applications of superconductors due to the need for costly liquid helium (4.2*K*). It has to be mentioned that the first real applications of superconductors were introduced in 1960's with the nearly coincident discoveries of *NbTi*, which made possible the development superconducting wires in magnets for the particle accelerators in nuclear physics, and then for the *Magnetic Resonance Imaging* (*MRI*) systems in medicine, where the costs and complexities associated with refrigeration were acceptable. The $Nb_3Sn$ tapes [148] were researched for similar large scale industrial applications listed in Tab. 6 [82].

**a)**

Small-Scale Applications

| Application | Techinical Points |
|---|---|
| Microwave filters in celluar stations | Low losses, smaller size, sharp filtering |
| Passive microwave devices, Resonators for oscillators | Lower surface losses, high quality factors, small size |
| Far-infrared bolometers | nonlinear tunneling SIS curves, high sensitivity |
| Microwave detectors | Uses nonlinear tunneling SIS curves, high conversion efficiency for mixing |
| X-ray detectors | High photon energy resolution |
| SQUID Magnetometers: Magneto-encephalography, NDT | Ultra-high sensitivity to magnetic fields |
| Voltage Standards | Quantum precision |
| Digital Circuits (SFQ) | Up to 750 GHz, ultra-fast, low-power |

Adapted from http://www.conectus.org/xxroadmap.html

**b)**

Large-Scale Applications

| Application | Techinical Points |
|---|---|
| Power cables | High current densities |
| Current Limiters | Uses highly nonlinear nature of transition |
| Transformers | High current densities and magentic fields, has lower losses |
| Motors/Generators | Smaller weight and size, lower losses |
| Energy Storage Magnets | Need high fields and currents<br>Smaller weight and size, lower losses |
| NMR magnets (MRI) | Ultra high field stability, large air gaps |
| Cavities for Accelerators | High microwave powers |
| Magnetic bearins | Low losses, self-controlled levitation |

Adapted from http://www.conectus.org/xxroadmap.html

**Tab. 6. a)** Small-scale applications of superconductors (after [82]).
**b)** Large scale applications of superconductors (after [82]).

The two types of *LTS* microwave resonators are distinguished:

***1. LTS microwave resonators with high intensity electromagnetic field***: *RF* transceivers, *GPS* navigation satellites, radio-astronomy, *RF* signal passive/active head-seekers in target acquisition/designation systems, and *RF* electronically scanned electronically steered phased array radars, *RF* synthetic aperture radars in remote sensing space applications in Tab. 7 **(a)**, **(b)** [42, 43].

***LTS microwave resonators with low intensity electromagnetic field*** processing elements (PE) for the quantum random number generators on magnetic flux qubits (*QRNG_MFQ*) and quantum computers (*QC*) in Tab. 8 **(1)**, **(2)** [45].

| | |
|---|---|
| 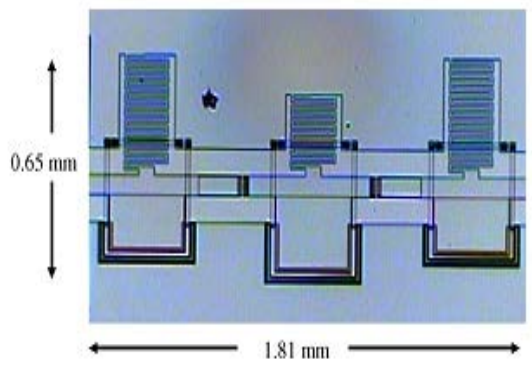  **a)** | 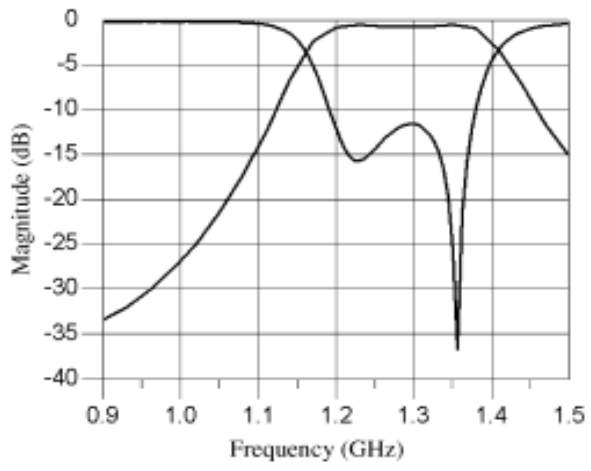  |
| 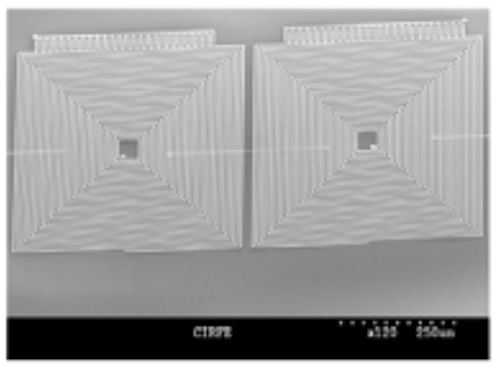 **b)** | 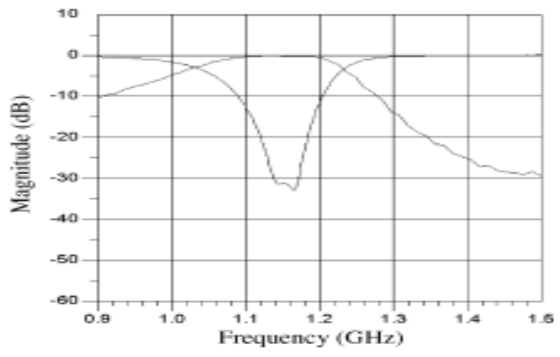  |
| 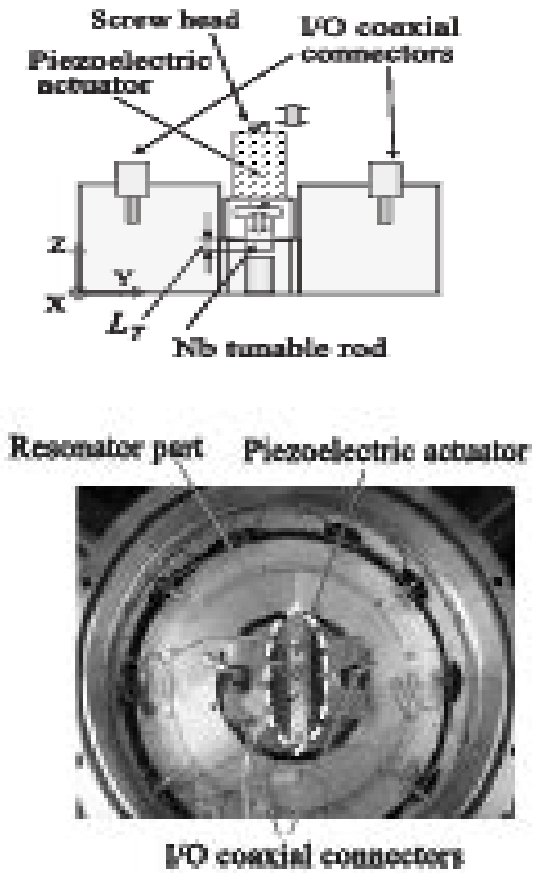  **c)** | 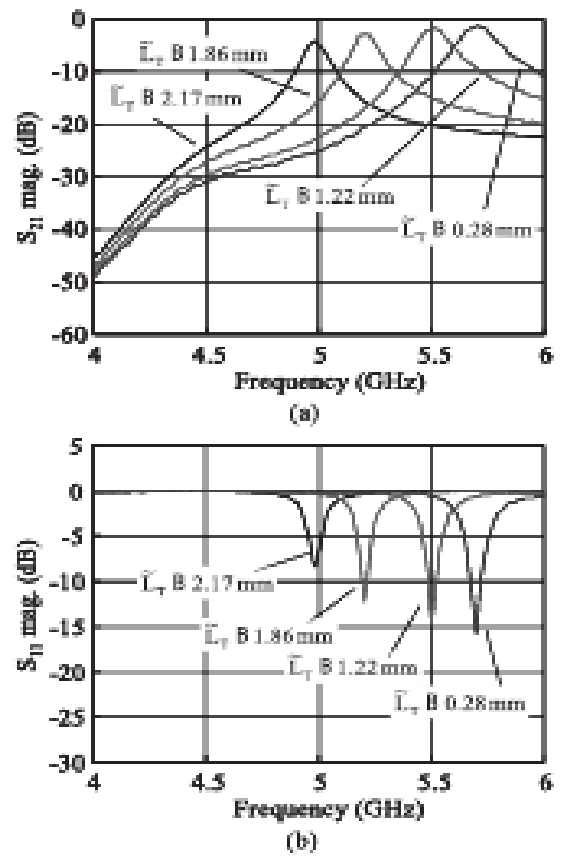  |

**Tab. 7. a)** Multilayer 3-pole coplanar waveguide lumped element *Nb* thin film filter and its measured $S_{21}(f)$ and $S_{11}(f)$ responses (after [42]); **b)** Scanning electron microscope image of the 2-pole microstrip *Nb* filter and its measured $S_{21}(f)$ and $S_{11}(f)$ responses (after [42]). **c)** A tunable *Nb* cavity resonator with a piezoelectric actuator for the model examination and its measured $S_{21}(f)$ and $S_{11}(f)$ responses (after [107]).

| | |
|---|---|
| **a), b), c), d)** 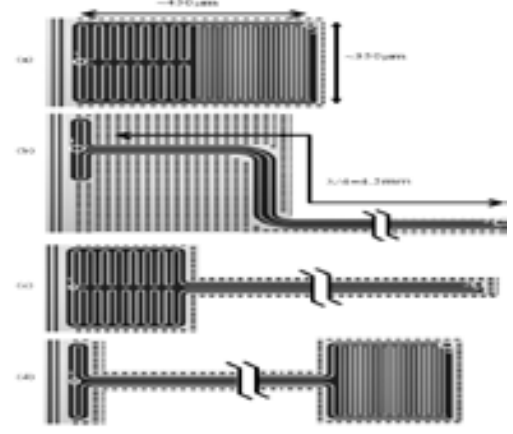 | 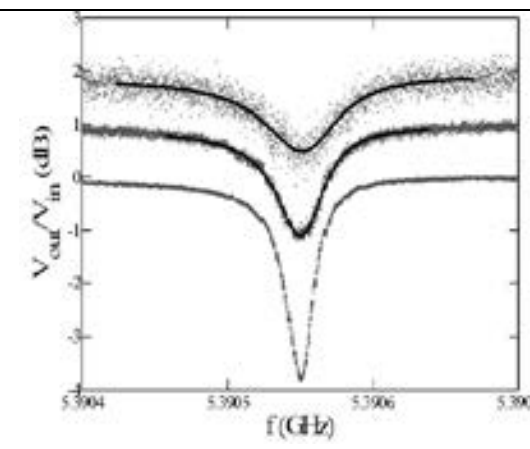  |

**Tab. 8.** Optical images of four resonators measured (gray/white is aluminium metal and black is sapphire substrate): **(a)** Quasi-lumped resonator (*QL*). **(b)** Coplanar strip resonator (*CPS*). **(c)** Quasi-lumped inductor with a *CPS* (*QLL-CPS*). **(d)** Quasi-lumped capacitor with a coplanar strip (*QLC-CPS*) (after [45]). **2)** The three Lorentzian curves are three different drive powers (after [45]).

## 2.18. Modern Industrial Applications of High Temperature Superconducting (HTS) Materials.

The discovery of High Temperature Superconductors (*HTS*) in the late 1980's with critical temperatures above 90 *Kelvin* degrees sparked tremendous interest towards various *HTS* applications [33]. The commercially available liquid nitrogen can be used to cool the *HTS* at 77*K*. The single-stage, close-cycle cryogenic refrigerators are available to provide refrigeration to 77*K* on a continuing basis with high reliability and at reasonable cost. The technological concepts of commercial applications of *HTS*, fabricated in five distinct forms, are [41, 122, 123, 156]:

1. ***Wire***: *HTS* wires are used in low-power magnets, power transmission lines, torque motors, compact electrical generators in Figs. 22 (a, b, c, e), and 21.

2. ***Tape*:** Superconducting tapes are most useful, where large lengths of material are required and must meet strict weight and size requirements.

3. ***Ribbon***: Ribbons are most suitable for space-borne and airborne superconducting devices, where implementation of monolithic microwave integrated circuit (*MMIC*) technology is highly desirable.

4. ***Thin films***: Thin films are generally employed in RF and microwave filters in Fig 22 (d) (f) [35, 36, 131, 139], patch antennas, optical detectors, and microelectronic circuits [106] with unique electronic devices such as the *SQUID*, based on the *Josephson junctions*. The digital integrated circuits [106], *Rapid Single Flux Quantum* (*RSFQ*) logics [38, 155], and *Analog-to-Digital Converters* (*ADC*) [131]. The microwave sensors, based on the *HTS SIS* bolometers in mm-wave imaging systems or *HTS* metamaterial with stacked *HTS* split-ring resonators (*SRR*) are designed with the aim to provide information about moisture content, density, structure and shape of different materials [110, 111, 131]. The 1024 *Quantum Random Number Generator on Magnetic Flux Qubits* (*QRNG_MFQ*), which uses the fundamental principles of quantum computing is the only fully operational quantum computing device as of 2013 [39].

5. ***Thick films***: Superconducting thick films are widely used in high-power microwave cavities in *RF* and filters, high-power lifting magnets, high power magnets for accelerators [80], and propulsion systems.

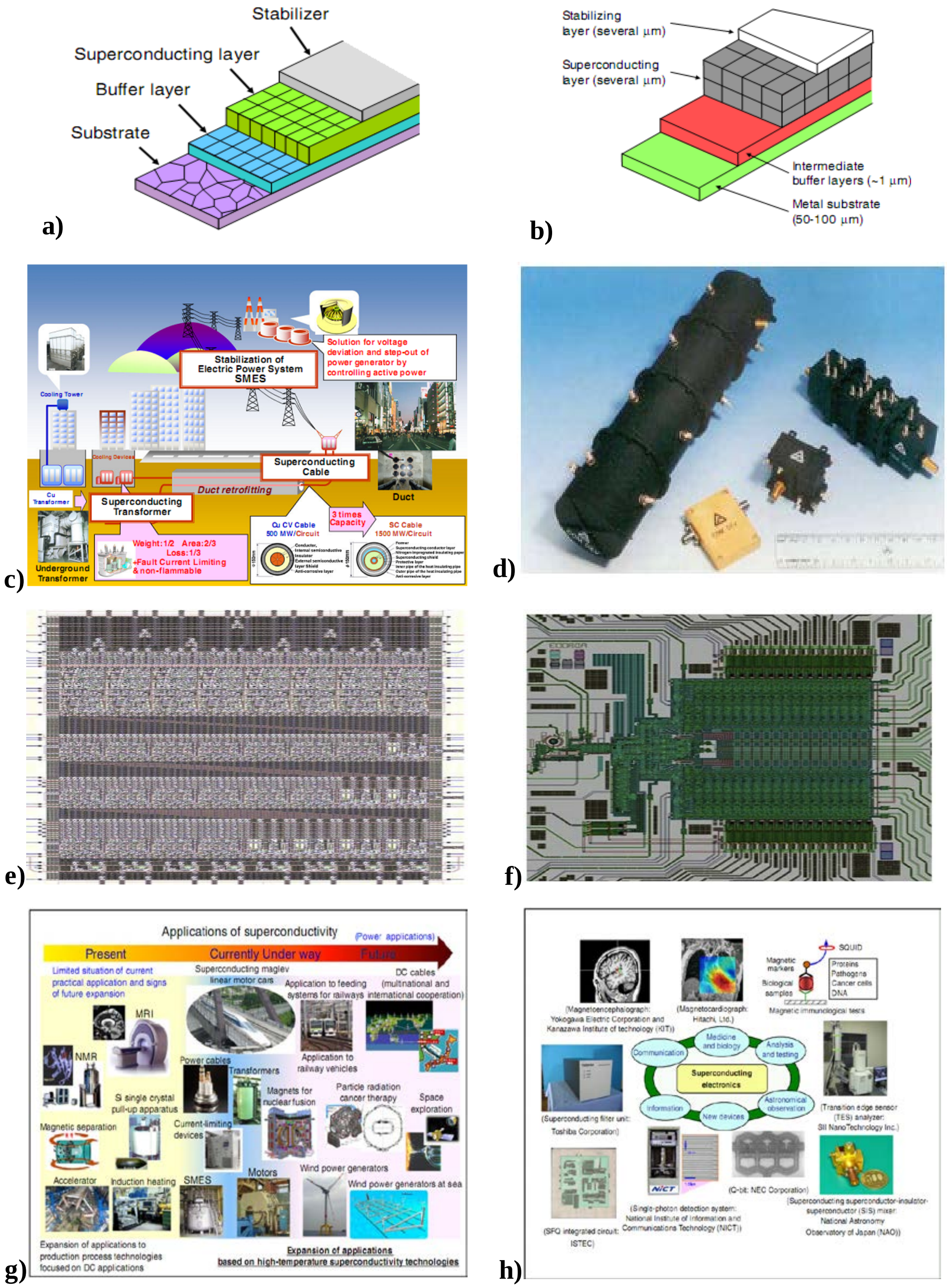

**Fig. 22. a)** Ion-beam-assisted deposition IBAD-processed $YBa_2Cu_3O_y$ (Y-123) superconducting wire (coated conductor) (after [122]). **b)** YBCO superconducting coated conductor (after [123]). **c)** Stable, large-capacity power supply system (after [123]). **d)** *HTS* microwave filters in the foreground (after [131]). **e)** RSFQ integrated circuit: an 8-bit asynchronous ALU with ~8,000 JJ, with latency below 400 ps (after [155]). **f)** RSFQ integrated circuit: X-band digital RF receiver operating with clock frequency up to 37 GHz (after [155]). **g)** Main applications of superconducting power and magnetic technologies (after [139]). **h)** Main applications of superconducting electronics (after [139]).

| Superconducting material | $T_c$ (K) | $H_{c2}$ at 4.2 K (T) |
|---|---|---|
| Oxide superconductor | | |
| $YBa_2Cu_3O_x$ | 92 | >100 |
| $Bi_2Sr_2CaCu_2O_y$ | 90 | |
| $Bi_2Sr_2Ca_2Cu_3O_z$ | 110 | |
| $Tl_2Ba_2Ca_2Cu_3O_w$ | 125 | |
| $HgBa_2Ca_2Cu_3O_v$ | 135 | |
| Metallic superconductor | | |
| Nb–Ti | 9.8 | 11.5 |
| Nb–Zr | 10.5 | 11 |
| $V_3Ga$ | 16 | 25 |
| $Nb_3Sn$ | 18 | 25 |
| $Nb_3Al$ | 18 | 32 |
| $Nb_3(Al,Ge)$ | 20 | 43 |
| $Nb_3Ga$ | 20 | 34 |
| $Nb_3Ge$ | 23 | 37 |
| $V_2(Hf,Zr)$ | 10.1 | 23 |
| NbCN | 17.8 | 12 |
| $MgB_2$ | 39 | 25 |
| Fe-based superconductor | | |
| $SmFeAsO_{1-x}F_x$ | 55 | >100 |
| $(Ba_{1-x}K_x)Fe_2As_2$ | 38 | 70 |

**Tab. 9.** Important superconducting materials with critical temperatures $T_c$ and critical magnetic fields $H_{c2}$ for development of superconducting wires (after [122]).

The modern applications of *HTS* superconductors are summarised by the author of dissertation in Fig. 23 [13, 41, 86, 99, 100, 101, 105, 110, 122, 123, 131].

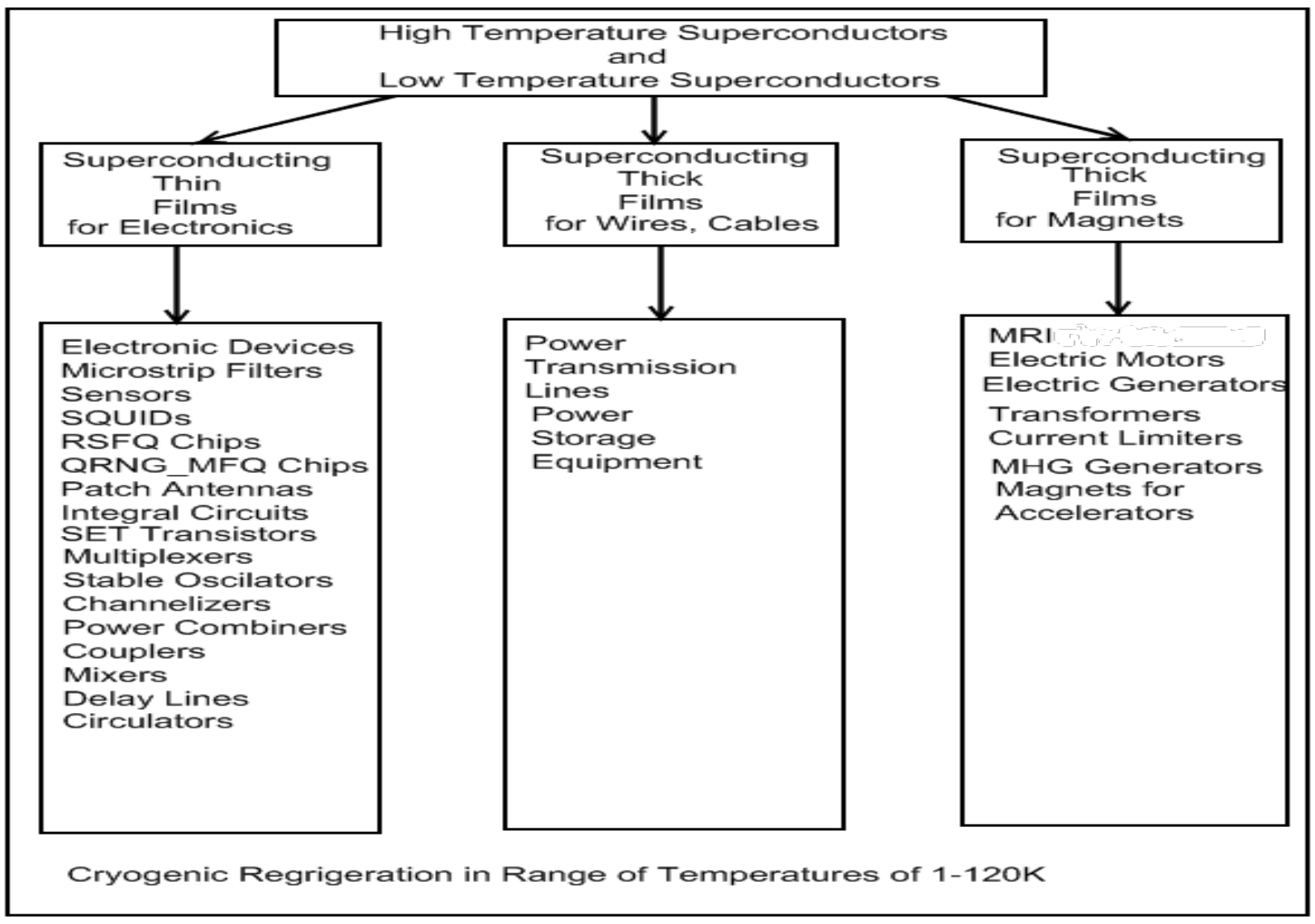

**Fig. 23.** General technical applications of High Temperature Superconductors (*HTS*) and Low Temperature Superconductors (*LTS*) summarized by author of dissertation, using the literature sources [13, 41, 86, 99, 100, 101, 105, 110, 122, 123, 131, 139].

The *HTS* thin films applications in microwave electronics, including the microwave filters in Cryogenic Transceiver Front Ends (*CTFE)* in wireless communications, driven by the *Internet of Things*, are described in Chapters 3, 9.

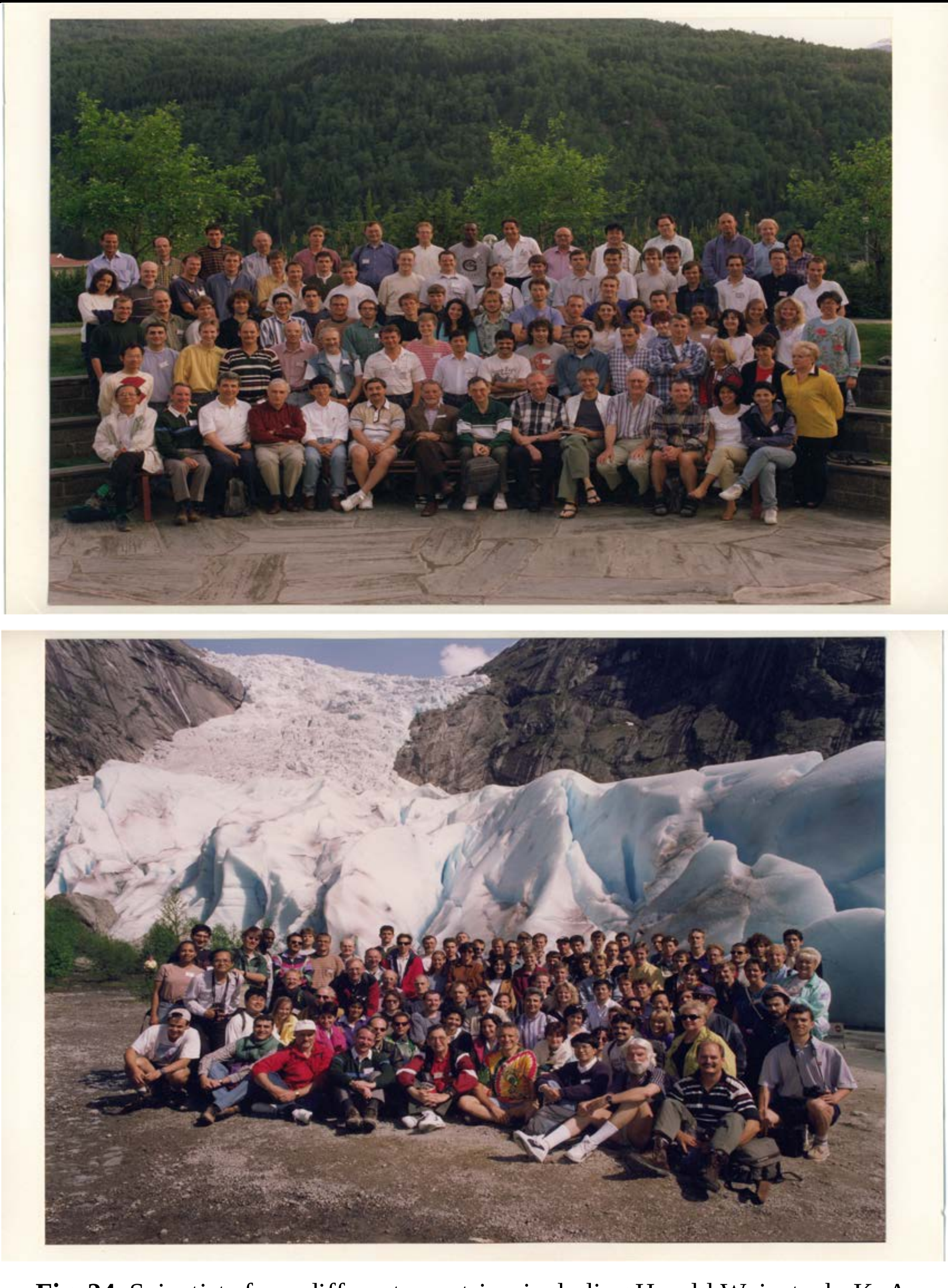

**Fig. 24.** Scientists from different countries, including Harold Weinstock, K. A. Muller, John Clarke, Konstantin K. Likharev, C. W. Chu, Viktor O. Ledenyov among other participants, who attended the *NATO* Advanced Study Institute (*ASI*) on Technical Applications of Superconductivity in Loen, Norway on June 1-10, 1997.

## Summary.

In this chapter, the introduction to the phenomenon of superconductivity and its technical applications are presented. The fundamental physical properties of superconductors, including the *ideal conductivity* of superconductors as well as the diamagnetic behaviour - *Meissner effect* are discussed. The considerable attention is focused on the theories toward understanding of superconductivity, including the *Two-Fluid thermodynamic theory* by Gorter and Casimir; *London theory of electrodynamics; Ginzburg-Landau-Abrikosov-Gor'kov phenomenological quantum theory*; *BCS microscopic theory of superconductivity* by Bardeen, Cooper, and Schrieffer. The critical parameters of superconductors such as the *critical temperature Tc*, *critical frequency*, *critical current density* $\boldsymbol{J_c}$, *critical magnetic fields* $H_{c1}$ and $H_{c2}$ are also described. It is explained that *Rjabinin, Schubnikow [150, 151]* experimentally discovered the *Type II* superconductors at the cryogenic laboratory at the National Scientific Center Kharkov Institute of Physics and Technology in Kharkov, Ukraine in 1935. The main characteristics of *Type I* and *Type II* superconductors are considered in details. In this book, the *Two Fluid model* in terms of the complex conductivity approach together with *London theory* were selected to form a basis for the microscopic characterisation of the superconductors at microwaves. In the course of research, the *Two Fluid phenomenological theory* and *London theory* were greatly complemented by the research findings from the *Bardeen, Cooper, Schrieffer* (*BCS*) theory with the purpose of accurate characterization of high temperature superconductors (*HTS)* at microwaves. Discussing the technical applications of superconductors, the authors of book concentrated on the *LTS/HTS* large and small scale applications [156]. The next Chapter 3 will specifically focus on the microwave superconductivity, including the research on the nature of nonlinearities in high temperature superconducting (*HTS*) thin films and applications of *HTS* thin films at the microwaves.

# CHAPTER 3

# MICROWAVE SUPERCONDUCTIVITY: ACCURATE CHARACTERISATION AND APPLICATIONS OF SUPERCONDUCTORS AT MICROWAVES

## 3.1. Introduction.

This chapter presents a theoretical description of microwave properties of superconductors together with main experimental findings. Characterisation of surface impedance $Z_s$ of *HTS* thin films within the temperature *T*, ultra high frequency *f* and microwave power *P* domains is described. Typical magnitudes of surface impedance *Zs* for classical and high-temperature superconductors are presented. The idea of *r-parameter* introduction for analysis of nonlinear properties of superconductors at microwaves is elucidated. Various possible sources of nonlinearities in *HTS* thin films at microwaves are also discussed. Most important theoretical and experimental research results in the field of microwave superconductivity are chronologically identified.

## 3.2.Microwave Superconductivity.

Microwave superconductivity as a research subject focuses on the accurate theoretical and experimental characterisation of physical properties of superconductors at microwaves. This research field is aimed at synthesis of superconductors with advanced microwave properties to be used in novel designs of microwave devices and circuits with improved technical characteristics for application in micro- and nano-electronics at microwaves [1-6].

Researches on properties of Low Temperature Superconductors (*LTS*) under application of high frequency alternate current flowing in the lead were firstly conducted by J. McLennan, A. C. Burton, A. Pitt, J. O. Wilhelm, in 1931-32 [7, 8]. In 1900–1932, Prof. John McLennan was Head of the Department of Physics at the University of Toronto [9-11]. McLennan was awarded the first *PhD* in science in Canada in 1900, based on his research at Cambridge University in the UK in 1898–1900. With enormous energy, McLennan built up a research lab that was one of the best in North America [661]. Then, the similar experimental research on the *LTS* superconductors at high frequency alternate current transferring in the tin was made by Silsbee [12] in 1932. It was discovered that there is a decrease of resistance of superconductor at critical temperature transition at high frequency $10^7$ *Hz*, however the resistance does not decrease to zero as it occurs under the conditions of direct current [12]. These experimental researches were continued by Shoenberg [650] in 1940, London [13] in 1941, and by Pippard [14] in period from 1947 to 1960. In a non-stationary process, the presence of normal conduction electrons requires the modification of London's electrodynamic equations. This has led to the early proposal of the Gorter–Casimir Two-Fluid phenomenological theory [15, 16]. The classical *Two-Fluid model* has been used widely and successfully to describe behaviors of superconductors at ultra high frequencies well below the energy gap frequency [15, 16, 1-6]. In particular, the surface resistance of superconductors has a frequency-dependence of $\omega^2$ according to the classical two-fluid model, while the surface resistance for normal conductors is proportional to $\omega^{1/2}$. The differences between the predicted data using the classical Two-Fluid model and experimental data at higher frequencies were observed by Pippard [17], Glover and Tinkham [18]. It was found by Pippard that the London theory had to be modified for certain superconductors in order to explain some experimental observations from his series of experiments for the measurements of the penetration depth of various types of superconductors, and its dependence on applied magnetic fields and on impurities in superconductors as well as its anisotropy [17]. In particular, Pippard found that the penetration depth was noticeably dependent upon the impurity content, which could not be explained by the local theory of London since the density of superelectron and its effective mass could only be weak functions of the impurity concentration [17]. Pippard's proposed the nonlocal theory, which is useful to describe the superconductors with relatively large coherence length at microwaves [17]. Tomonaga took an active part in the research toward the microwave circuits theory in Japan in 1947-1948 [19, 20], creating the knowledge base for the microwave superconductivity research in Japan.

Speaking about the microwave superconductivity development in Ukraine, Matricon, Waysand [649] write: “Kamerlingh Onnes, in spite of his cleverness and clear-sighted view of his goals, needed twenty years in Leiden to reach premier rank. Meissner made his leap much more quickly in Berlin, but there he profited from the close connection between the university and German industry. The Kharkov laboratory was built and became truly productive, at least in superconductivity, in less than four years. Twenty institutes were founded at the same time, of which physics was the most important. In fact, the laboratory was nominally created in 1929 before the arrival of Shubnikov. Ivan Obreimov, a scientist of the previous generation, the one who had introduced Shubnikov to crystallography, had been named director. Yet it was a young nuclear physicist named A. I. Leipunski who became the real leader. Most of the physicists came from Leningrad, and Shubnikov was named head of the low-temperature physics section in 1931. The lessons of Leiden were put into practice right away." The Shubnikov's scientific achievements in experimental physics are well known in [659]. The scientific school of thinking in the *experimental and theoretical physics of microwave superconductivity* was created by Boris Georgievich Lazarev at *National Scientific Centre Kharkov Institute of Physics and Technology (NSC KIPT)* in Kharkov in Ukraine in 1938 [651, 652, 656]. The *intensive research program* in the *microwave superconductivity* was started at *NSC KIPT* in Kharkov in Ukraine in 1938 for the first time in the former *USSR* in [660]. In the microwave superconductivity, the first experimental research on the properties of Low Temperature Superconductors (*LTS*) in the range of ultra high frequencies had been conducted by Lazarev, Galkin, Khotkevich [21] at *NSC KIPT* in Kharkov, Ukraine in 1941. The initially reported experimental results [21] had been discussed more comprehensively in Lazarev, Galkin, Khotkevich [653, 654] at *NSC KIPT* in Kharkov, Ukraine in 1947. The research on superconductivity at ultra high frequency of $1.8\times10^{10} Hz$ had been made by the *transmission line* measurement method by Galkin, Lazarev [22, 655] at *NSC KIPT* in Kharkov, Ukraine in 1948. The frequency dependence of the surface resistance isotherms in Low Temperature Superconductors (*LTS*) in the frequency rage $(1.88 \div 4.5) \times10^{10} Hz$ had been researched by Galkin, Bezugly [41] at *NSC KIPT* in Kharkov, Ukraine in 1954. Thus, Lazarev [651, 652, 656], Galkin [657], Khotkevich [658] completed the innovative experimental researches on the accurate characterization of physical properties of superconductors in a range of ultra high frequencies, using the *resonance and transmission line techniques* for the first time. These experimental measurement methodologies have been used by many scientists at a number of universities around the World during subsequent years.

Since the discovery of High Temperature Superconductors (*HTS*) by Bednorz and Muller [23] in 1986, the intensive experimental and theoretical researches on ultra high frequency properties of *HTS* thin films and crystals were conducted in a number of universities. The *HTS* materials [24] have a high critical temperature up to $T=92\ K$ and higher, and unique microwave properties, hence the *HTS* materials are of certain interest for technical applications in a variety of *RF* and microwave devices [25, 1-6].

The *HTS* materials have a number of features that need to be considered. The high-Tc oxide superconductors have very small coherence lengths $\xi$, and their surface resistance $R_s$ measurements do not always agree with the frequency-dependence $\omega^2$ behavior predicted by the classical two-fluid model in the *GHz* frequency range [3]. Therefore, a variety of modified forms of the two-fluid model have been proposed.

Withers [25] explains that the *HTS* superconductors have two physical properties, which make them attractive for application in microwave circuits:

**1.** The low surface resistance of superconductors results in very low propagation loss in transmission structures and very high quality factors *Q* in resonant structures;

**2.** The frequency-independent magnetic penetration depth of superconductors can be exploited to make dispersionless transmission structures.

Fossheim, Sudbo [26] express similar positive opinion on the advantages of application of *HTS* superconductors in microwave filters, because of two properties that differ greatly from those of normal metals at ultra high frequencies:

**1.** Much lower surface resistances is exhibited by superconductors, leading to much lower loss and much higher *Q*-values in *HTS* microwave system components.

**2.** Superconductors have a practically frequency independent penetration depth in the microwave frequency range.

These two properties have the important consequences, namely that superconductors introduce no energy dispersion into a microwave device up to 1*THz* frequency in low-*Tc* superconductors; and well above this frequency in high-*Tc* superconductors due to their larger gap frequency. The improvement over normal conducting performance is substantial. Superconductor passive devices may work quite well in cases where normal conductor devices would function very poorly or not at all. Normal metal components will, in many cases, have several orders of magnitude higher conductor loss than the superconducting ones. *HTS* passive microwave components like bandpass filters are becoming essential in mobile phone base stations [6]. Other applications of *HTS* are in chirp filters, which can provide improved resolution in radar images, and delay lines used in various contexts [26].

Zhao Xinjie *et al.* [27] explains that one of the most promising applications of this *HTS* material is the ***passive microwave devices***, for the *HTS* thin film surface resistance $R_S$ at liquid nitrogen temperature is extremely small, only one tenth of that for *Cu*. It has very low insertion loss and very high quality factor *Q*, when employed for passive microwave devices. ***The performance of the devices made of HTS thin films is several orders of magnitude higher compared with that of normal conductor, and the power loss, mass and weight are several orders of magnitude lower [196].*** The penetration depth of *HTS* is independent of frequency, therefore *HTS* thin films can be used for constructing transmission line with no dispersion, and for designing devices that is unattainable by using normal conductor. In ultra high frequencies range of electromagnetic waves up to $10^{11}$ *Hz*, the *HTS* microwave filters, delay lines, patch antennas with advanced characteristics, e.g. the high quality factor, narrow frequency selectivity were created [1-6, 25-28, 30].

## 3.3. Synthesis of HTS Thin Films for Application in Microwave Devices and Circuits.

Since the discovery of *HTS*, the *HTS* thin films have been synthesized, using the two techniques for the *HTS* thin films growth in Humphreys [32]: 1. ***ex situ** (two stages process)*: the *HTS* thin film is deposited on the substrate to form an amorphous layer, then it is annealed to form an epitaxial layer; 2. ***in situ** (one stage process)*: the *HTS* thin film is deposited to form the epitaxial layer. The high quality *HTS* thin films with thickness 400-600 nm have been fabricated by deposition of high-*Tc* superconductor on dielectric substrates with application of ***in situ*** synthesis technique by different methods such as in Withers [25], Braginski [29], Talisa [31], Humphreys [32]: 1. *Evaporation of superconductors* onto heated substrate in oxygen, 2. *Reactive co-evaporation* by cyclic deposition and reaction, 3. *Pulsed laser deposition (PLD)*, 4. *Metal-organic chemical vapor deposition* (*MOCVD*), 5. *Sputtering method* (*DC* and *RF* plasma discharge), 6. *Liquid phase* *epitaxy*. In Tab. 1, the examples of dielectric substrates for *HTS* thin films deposition are shown: *Sapphire* $\alpha$-$Al_2O_3$, *Lanthanum Aluminate* $LaAlO_3$ (*LAO*), *Strontium Lanthanum Aluminate* $LaSrAlO_4$, *Magnesium Oxide MgO*, and *Alumina* $Al_2O_3$ in Mazierska, Jacob [6]. Humphreys [32] writes that the main physical parameters of dielectric substrates are: 1. *lattice constant (a), 2. coefficient of linear expansion* ($\alpha$)*, 3. relative permittivity* ($\varepsilon$)*, 4. loss tangent* (tan $\delta$)*, 5. temperature of orthorhombic to tetragonal phase transition* ($T_t$).

| Substrate | Frequency | tanδ |
|---|---|---|
| *Sapphire* $\alpha$-$Al_2O_3$ | 9 GHz | $1.5x10^{-8}$ |
| *Lanthanum Aluminate* $LaAl_2O_3$ | 10 GHz | $7.6x10^{-6}$ |
| *Magnesium Oxide MgO* | 8 GHz | $2x10^{-6}$ |
| *Strontium Lanthanum Aluminate* $LaSrAlO_4$ | 12 GHz | $2x10^{-6}$ |
| *Alumina* $Al_2O_3$ | 7.7 GHz | $2x10^{-5}$ |

**Tab. 1.** Substrates for *HTS* thin film deposition and their loss tangent magnitude at specified frequencies (after [6]).

The first $YBa_2Cu_3O_{7-\delta}$ thin films on various substrates were prepared with application of different methods by the following research groups in 1987-1991:

**1.** $YBa_2Cu_3O_{7-\delta}$ thin films on $SrTiO_3$(100) by Laibowitz, Koch, Chaudhari, Gambino [33];

**2.** $YBa_2Cu_3O_{7-x}$ thin films on $SrTiO_3$(100) by co-evaporation by Mankiewich, Scofield, Skocpol, Howard, Dayem, Good [34];

**3.** $YBa_2Cu_3O_{7-x}$ thin films by Yoshizako, Tonouchi, Kobayashi [35];

**4.** $YBa_2Cu_3O_{7-x}$ thin films on $SrTiO_3$(100) and MgO(100) by activated reactive evaporation by Terashima, Iijima, Yamamoto, Hirata, Bando, Takada [36];

**5.** $YBa_2Cu_3O_{7-x}$ thin films on $SrTiO_3$(100) by laser ablation by Koren, Gupta, Baseman [142];

**6.** $YBa_2Cu_3O_{7-x}$ thin films on $SrTiO_3$(100) by pulse laser deposition by Koren, Gupta, Beserman, Lutwyche, Laibowitz [143];

**7.** $YBa_2Cu_3O_{7-x}$ thin films on $SrTiO_3$(100) by laser ablation by Koren, Gupta, Segmuller [144];

**8.** $YBa_2Cu_3O_{7-x}$ epitaxially grown thin films by Klein *et al.* [148];

**9.** $YBa_2Cu_3O_{7-x}$ thin films on MgO(100) by vapour deposition by Humphreys, Satchell, Chew, Edwards, Goodyear, Blenkinsop, Dosser, Cullis [37];

**10.** $YBa_2Cu_3O_7$ thin films on MgO(100) by laser ablation by Moeckly, Russek, Lathrop, Buhrman, Jian Li, Mayer [38, 39, 40];

**11.** $YBa_2Cu_3O_{7-x}$ thin films by off-axis magnetron sputtering by Newman, Char, Garisson, Barton, Taber, Eom, Geballe, Wilkens [41];

**12.** $YBa_2Cu_3O_{7-x}$ / Pr thin films by inverted cylindrical magnetron sputtering and pulsed laser deposition by Xi [42].

The aim of research of microstructure of the *HTS* thin films is to synthesis the ***high quality HTS thin films with good microwave properties.*** Considering the *HTS* thin films for microwave applications, Withers [25] states that the ***HTS thin films for microwave applications*** can be prepared by a variety of techniques, including: 1. *Pulsed-laser deposition;* 2. *Off-axis magnetron sputtering*; 3. *Co-evoparation*.

The review of methods of *HTS* thin films fabrication is presented in Braginski [29]. The methods exploiting the specific properties of *HTS* materials, including the inhibit patterning or controlled grain-boundary nucleation, appear more suitable than traditional approaches [29]. For example, the synthesis of $YBa_2Cu_3O_{7-\delta}$ thin films with ***low microwave surface resistance***, using the *Chemical Vapor Deposition* (*CVD*) techniques, was made by DeSisto, Henry, Newman, Osofsky, Cestone [43], and by other research groups.

Tab. 2 shows the *RF* device and circuit requirements for the fabrication of *HTS* thin films and multilayers on epitaxial substrates [29].

| | |
|---|---|
| **1.** | Growth at lowest possible temperature, $T_s$, insuring the exact stoichiometry of all constituents, formation of the desired single phase and full crystalline order. |
| **2.** | Adhesion to, but no inter-diffusion or reaction with, the substrate. |
| **3.** | Film thickness from monocell to >> $\lambda_L$ with narrow thickness tolerances (goal: << ±10%); $\lambda_L$ is the London penetration depth. |
| **4.** | Thermal expansion match over a broad temperature range. |
| **5.** | Smooth surfaces and interfaces (goal: roughness << 1 nm). |
| **6.** | Large, uniform area (goal: at least 10 cm in diameter). |
| **7.** | Epitaxial, *i.e.*, structural and lattice parameter match. |
| **8.** | Control of orientation. |
| **9.** | Epitaxial growth on *ex-situ* processed underlayer. |
| **10.** | No grain boundaries, except those controllably nucleated. |
| **11.** | High crystalline quality, low defect number density. |
| **12.** | No pinholes in insulators, barriers. |

**Tab. 2.** Device and circuit requirements for the fabrication of *HTS* films and multilayers on epitaxial substrates (after [29]).

Zhao Xinjie, Li Lin, Lei Chong, Tian Yongjun [27] elucidate, when discussing the high temperature superconducting thin films for microwave filters, that the synthesis of $YBa_2Cu_3O_{7-\delta}$ thin films is quite mature. Large-sized (≥8″) and double-sided $YBa_2Cu_3O_{7-\delta}$ thin films are commercialized in the developed countries, such as *Theva GmbH* in Germany, the films for microwave usage have $T_{C0}$ of 88 *K*, the critical current density $J_C \approx 3\times10^6\ A/cm^2$ (77 *K*, 0 *T*), and $R_S \approx 500$ $\mu\Omega$ (77 *K*, 10 *GHz*). By employing these $YBa_2Cu_3O_{7-\delta}$ thin films to make the microstrip filters, the filters can operate under liquid nitrogen temperature. Alternatively, much research work has also been carried out on the synthesis of $Tl_2Ba_2CaCu_2O_{8-\delta}$ thin films. It was found that two phases of $Tl_2Ba_2CaCu_2O_8$ thin films are suitable for passive microwave devices: $Tl_2Ba_2CaCu_2O_8$ (Tl-2212) and $Tl_2Ba_2CaCu_2O_{10}$ (*Tl-2223*) [44, 45]. The former has a $T_{C0}$ of 110 *K*, $J_C \approx 2\times106$ $A/cm^2$ and $R_S \approx 130\mu\Omega$ (77*K*,10*GHz*), and the latter has a $T_{C0}$ of *125 K*, $J_C$ *(77 K)*$\approx 105 A/cm^2$ ,and $R_S \approx 86$ $\mu\Omega$ (77*K*,10*GHz*) [45]. The synthesis of Tl-2212 is easier than *Tl-2223*, and because of the intergrowth of *Tl-2212* with *Tl-2223*, it is not possible to grow pure *Tl-2223* phase without doping, there is always some *Tl-2212* left in the *Tl-2223* film. Therefore for microwave applications it is better to use *Tl-2212*. The microstrip filter made by *Tl-2212* can be operated at 82 *K* [27]. Tab. 3 provides the information on the technical characteristics of *Type II* superconductors in Zhou [3].

| Material Properties | YBaCuO | TlBaCaCuO | Nb |
|---|---|---|---|
| Transition temperature $T_c$ | up to 95 K | up to 130 K | 9.2 K |
| Coherence length $\xi_{ab}$ | 1.5 nm | 3.0 nm | 39 nm |
| Coherence length $\xi_c$ | 0.2 nm | 0.1 nm | — |
| Penetration depth $\lambda_{ab}$ | 150 nm | 200 nm | 50 nm |
| Lower critical field $B_{c1}$ | 10–100 mT | 10 mT | 0.13 T |
| Upper critical field $B_{c2}$ | 100–200 T | 60 T | 0.3 T |

**Tab. 3.** Technical parameters of *Type II* superconductors (after [3]).

Microstructure of *HTS* thin films is studied by the atomic force microscopy (*AFM*), scanning electron microscopy (*SEM*), transmission electron microscopy (*TEM*), x-ray spectroscopy, modulated optical reflectance (*MOR*) and various microwave techniques in Tab. 4 [27, 46, 47, 48, 49] and in Tab. 5 [46, 50, 51, 52].

<table>
<tr><td>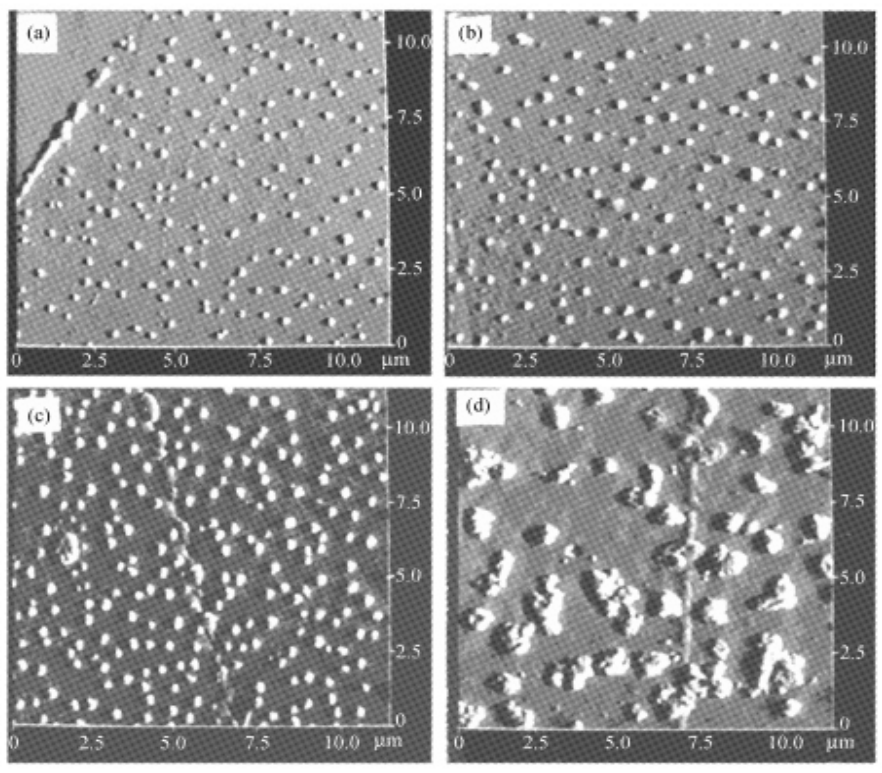
</td><td>Atomic Force Microscope (AFM) morphology of incipient $YBa_2Cu_3O_{7-\delta}$ film deposited at 800°C by PLD. The thicknesses of the films are 10nm (a), 25nm (b), 50 nm (c) and 150nm (d) (after [27]).</td></tr>
<tr><td>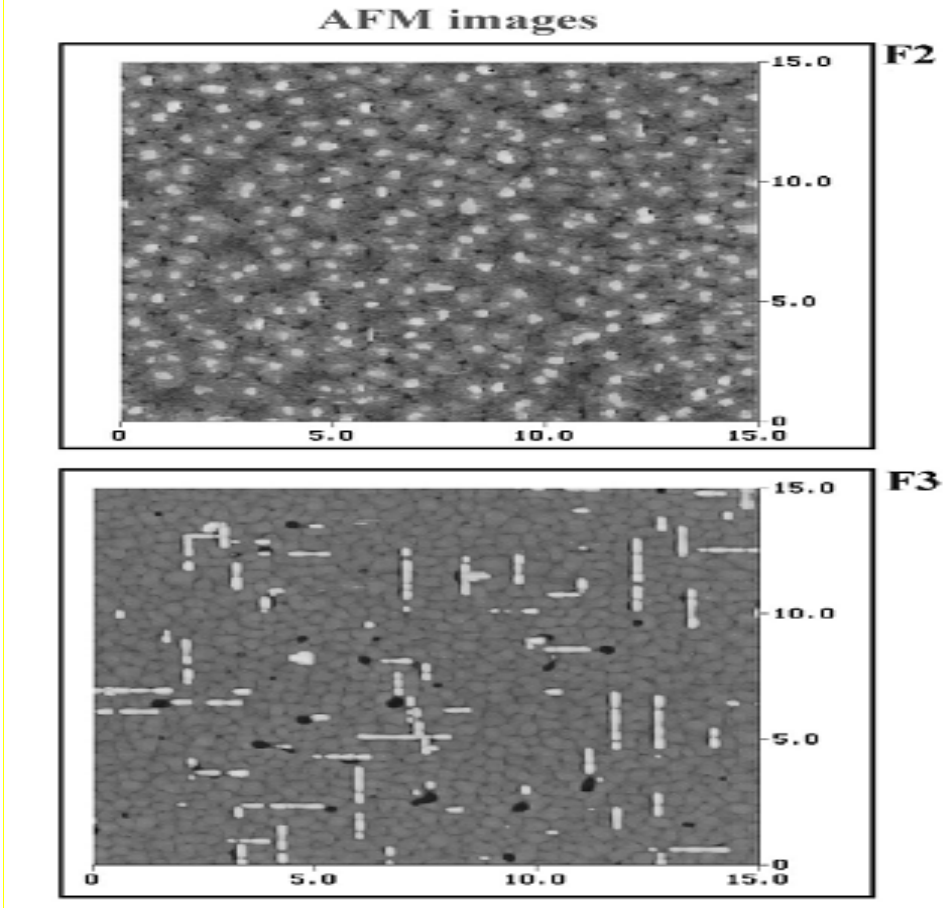
</td><td>Atomic Force Microscope (AFM) images of two $YBa_2Cu_3O_{7-\delta}$ films on MgO, F2 and F3. Reprinted by permission from D. W. Huish (after [46]).</td></tr>
<tr><td>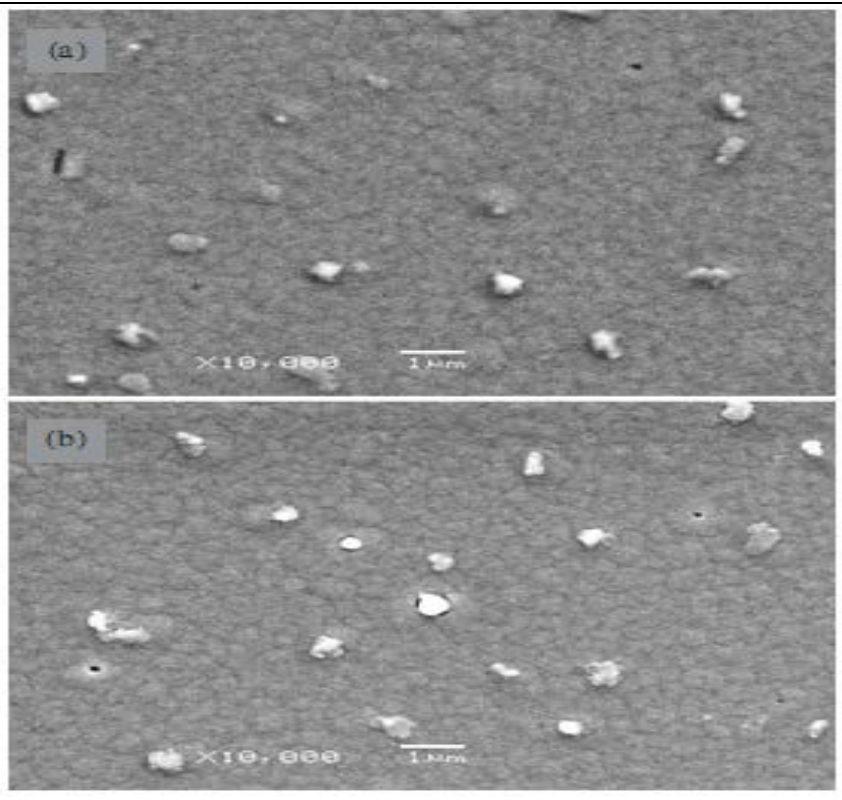
</td><td>Scanning Electron Microscope (SEM) images of (a) single-layer $YBa_2Cu_3O_{7-\delta}$ microwave filter and (b) tri-layer $YBa_2Cu_3O_{7-\delta}$ microwave filter (after [47]).</td></tr>
<tr><td>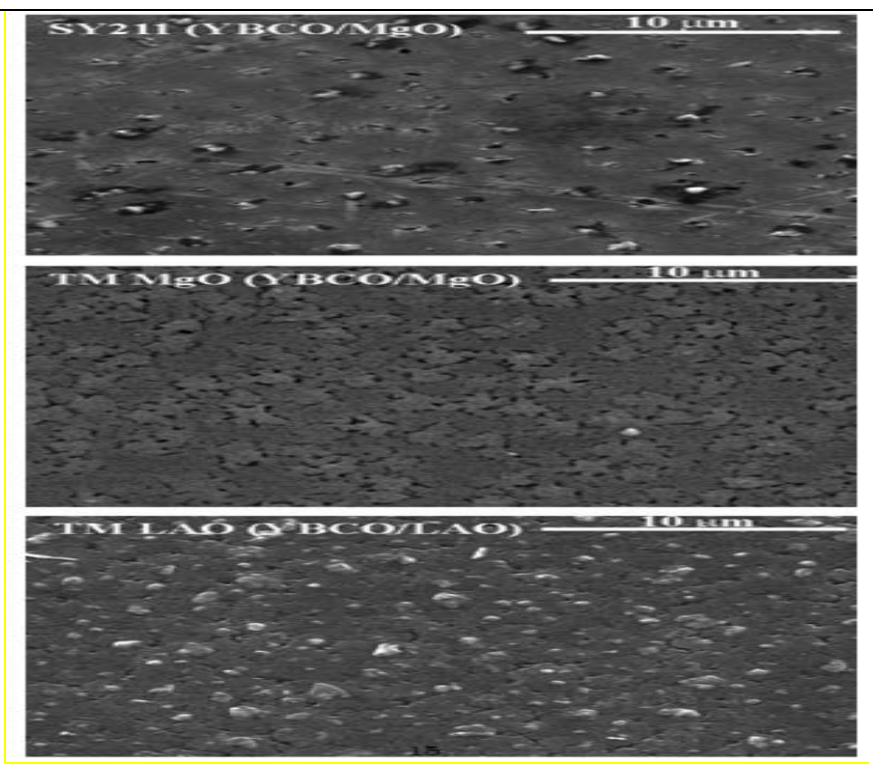
</td><td>Scanning Electron Microscope (SEM) images for the three studied $YBa_2Cu_3O_{7-\delta}$ thin films samples (after [48]).</td></tr>
</table>

| | |
|---|---|
| 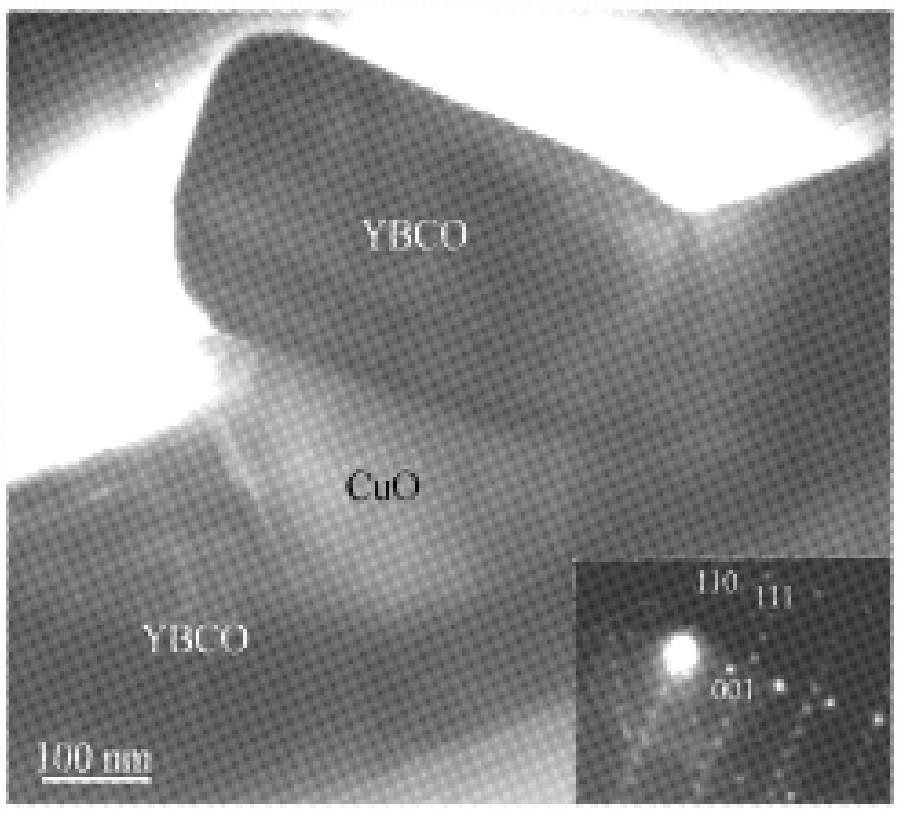 | Cross-sectional *TEM* of c-axis deviated $YBa_2Cu_3O_{7-\delta}$ and its *SAED*. The c-axis deviated $YBa_2Cu_3O_{7-\delta}$ is growing on the *CuO* and protrudes highly on the $YBa_2Cu_3O_{7-\delta}$ film surface (after [27]). |
| 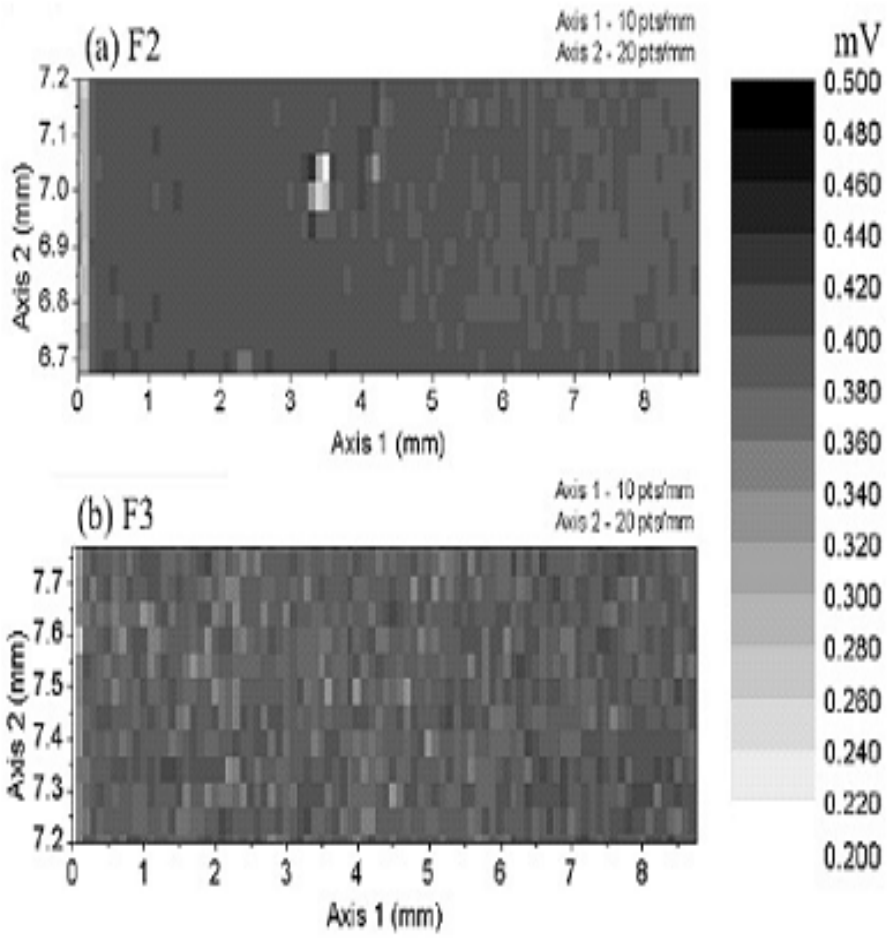 | Modulated Optical Reflection (*MOR*) scans of two $YBa_2Cu_3O_{7-\delta}$ films on MgO, F2 and F3. Reprinted by permission from D. W. Huish et al, *IEEE Trans. Appl. Supercond.*, vol. **13**, p. 3638, 2003. (© 2003 IEEE.) (after [46, 49]). |
| 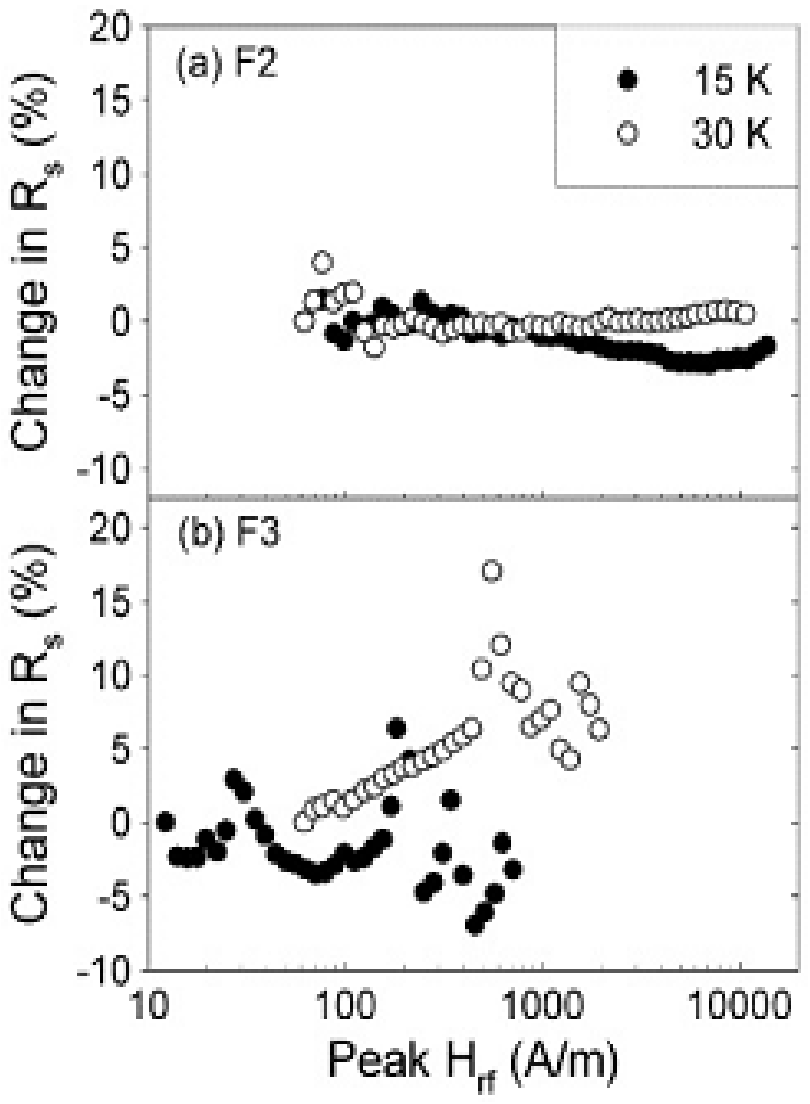 | $H_{rf}$ -dependences of $R_S$ for films F2 and F3 at frequency $f$=8*GHz*, the *MOR* scans for which are shown in Figure 22. Reprinted by permission from D. W. Huish et al, *IEEE Trans. Appl. Supercond.*, vol. **13**, p. 3638, 2003. (© 2003 IEEE.) (after [46, 49]). |

**Tab. 4.** Microstructures of $YBa_2Cu_3O_{7-\delta}$ thin films studied by atomic force microscopy (*AFM*), scanning electron microscopy (*SEM*), transmission electron microscopy (*TEM*), modulated optical reflectance (*MOR*) techniques (after [27, 46, 47, 48, 49]).

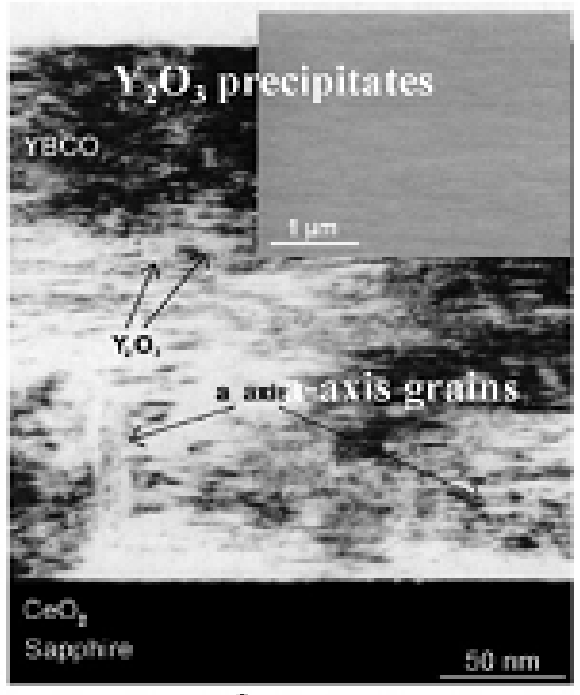

a)

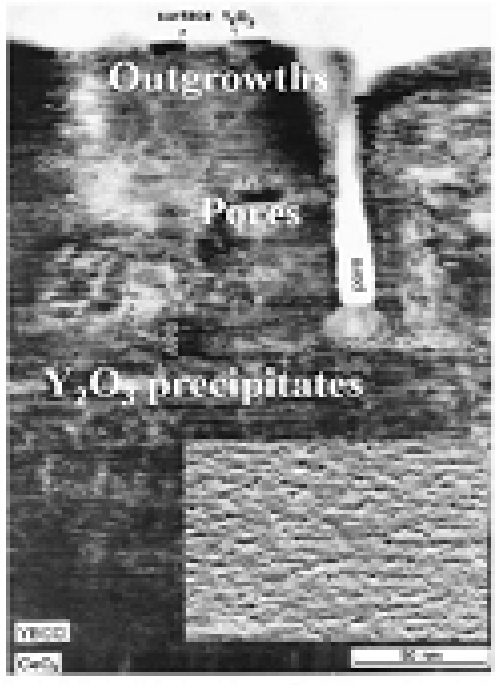

b)

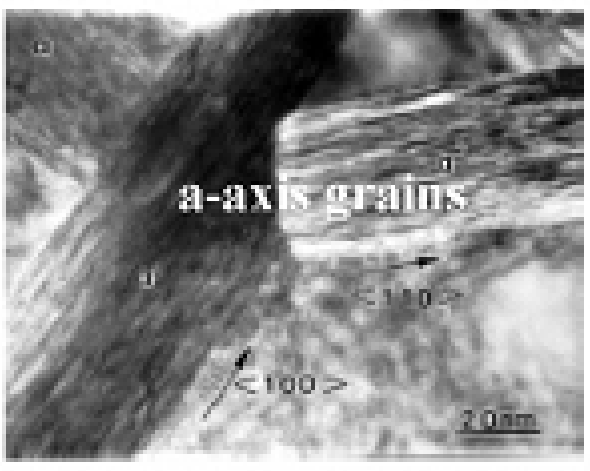

c)

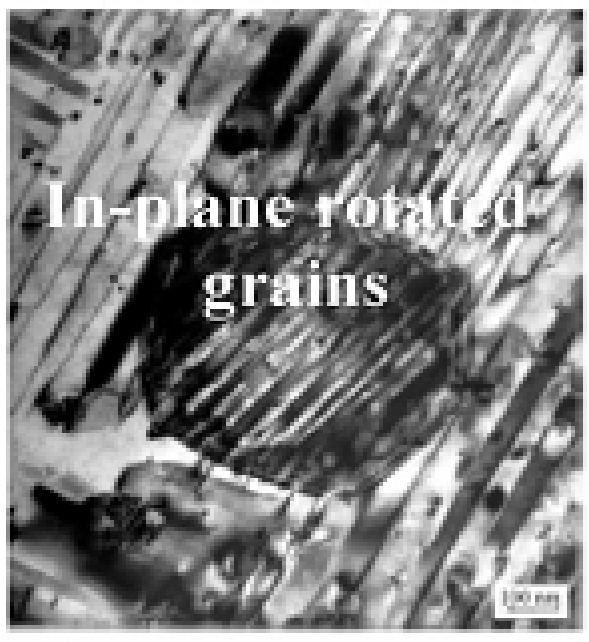

d)

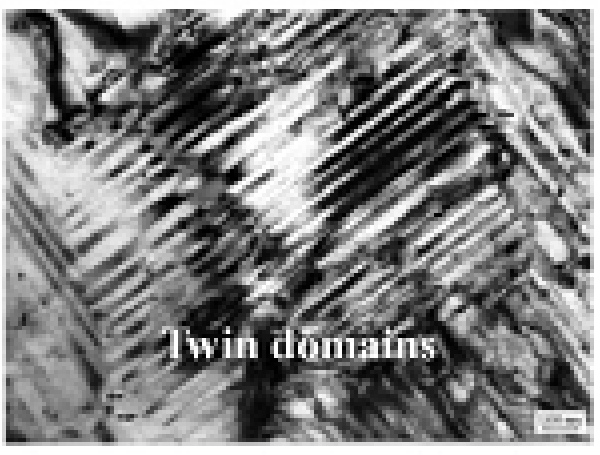

e)

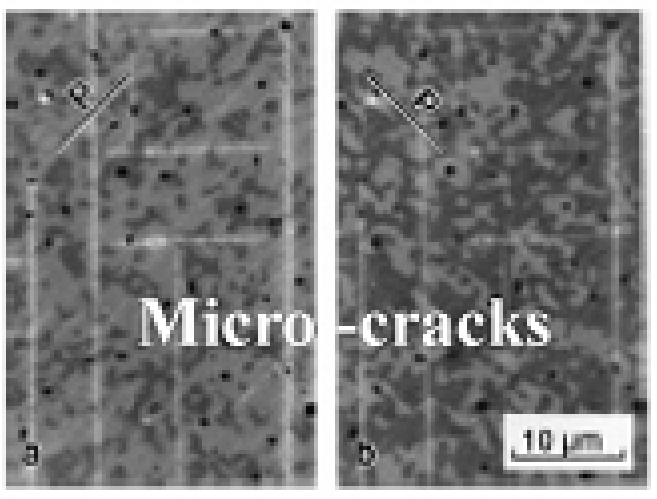

f)

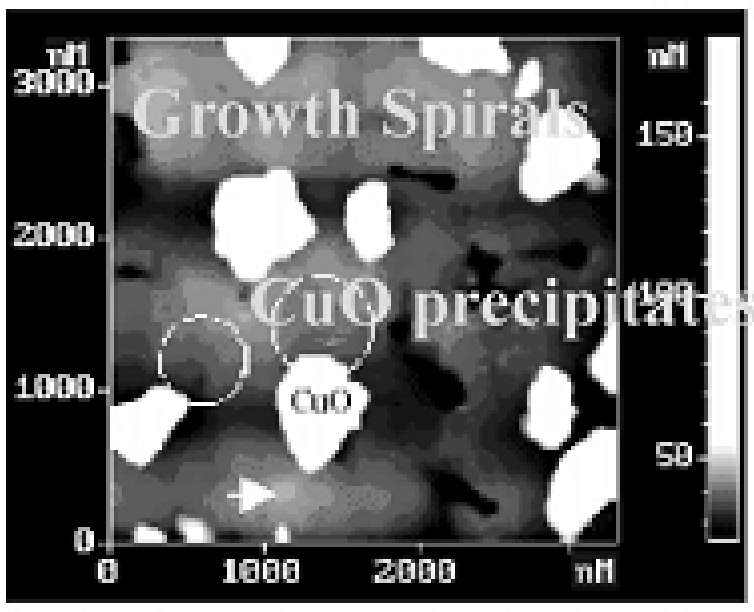

g)

**Tab. 5.** *TEM* images of magnetron-sputtered $YBa_2Cu_3O_{7-\delta}$ films on $CeO_2$-buffered sapphire substrate deposited with **(a)** low and **(b)** high ion beam energy respectively

[50] (reprinted from J. Einfeld, P. Lahl, R. Kutzner, R. Wornderweber, G. Kastner, *Physica C*, vol. **351**, p. 103, 2001., © 2001, with permission from Elsevier). Structural *TEM* images of a magnetron-sputtered $YBa_2Cu_3O_{7-\delta}$ film on $LaAlO_3$ (**(c)**, **(d)** and **(e)**), and optical micrograph image of a laser-ablated $YBa_2Cu_3O_{7-\delta}$ film on $CeO_2$-buffered sapphire substrates **(f)** [51], (reproduced by permission of *IOP Publishing Ltd.* from G. Kastner, C. Schafer, St. Senz, T. Kaiser, M. A. Hein, M. Lorenz, H. Hochmuth and D. Hesse, *Supercond. Sci. Technol.*, vol. **12**, p. 366, 1999.), *AFM* image of a $YBa_2Cu_3O_{7-\delta}$ film on *YSZ*-buffered sapphire substrate **(g)** [52], (reproduced by permission from A. K. Vorobiev, Y. N. Drozdov, S. A. Gusev, V. L. Mironov, N. V. Vostokov, E. B. Kluenkov, S. V. Gaponov and V. V. Talanov, *Supercond. Science Technology,* vol. **12**, p. 908, 1999.) (after [46]).

Velichko, Lancaster, Porch [46] conclude that the technology of fabricating *HTS* thin films has reached a very advanced stage, at which high quality films with excellent microwave properties, highly reproducible and with long lasting performance, can be routinely produced on a scale sufficiently large for wide-ranging applications. Authors [46] point to the fact that the major limitation for the successful microwave applications of *HTS* films in such devices as filters, delay lines, antennas, modulators, switches and power limiters still comes from ***microwave nonlinearity***.

The significant contributions to the research on the *LTS* films and crystals, which resulted in improvements of their microwave characteristics, were made by (in chronological order): McLennan [7, 8]; Silsbee [12]; London [13]; Pippard [14, 53-59]; Tomonaga [19, 20]; Lazarev, Galkin, Khotkevich [21]; Galkin, Lazarev [22]; Halbritter [60-73]; G. Dresselhaus, M. S. Dresselhaus [80, 81]; Mende, Bondarenko, Trubitsin [74]; Mende, Spitsin [75], Takken, Beasley, Pease [76].

The valuable research contributions to the *HTS* thin films and crystals synthesis, resulting in better microwave characteristics, were made by (in chronological order): Laibowitz, Koch, Chaudhari, Gambino [33], Mankiewich, Scofield, Skocpol, Howard, Dayem, Good [34], Yoshizako, Tonouchi, Kobayashi [35], Terashima, Iijima, Yamamoto, Hirata, Bando, Takada [36], Humphreys, Satchell, Chew, Edwards, Goodyear, Blenkinsop, Dosser, Cullis [37], Moeckly,

Russek, Lathrop, Buhrman, Jian Li, Mayer [38, 39, 40], Newman, Char, Garisson, Barton, Taber, Eom, Geballe, Wilkens [41], Xi [42], Kong [77, 78]; Xia, Kong, Shin [79]; G. Dresselhaus, M. S. Dresselhaus [82]; M. S. Dresselhaus, Oates, Shridhar [83]; Portis [84-87]; D. E. Oates [88-107]; Braginski [31]; Kobayashi [108-111]; Lancaster [112-120]; O. G. Vendik [121-141]; Koren [142-147], Klein [148-155]; Mazierska [156-184]; Krupka [185-199]; Piel, Müller [200]; Sridhar [201]; Porch [202-218]; Bonn [219-221]; I. S. Kim, Lee, Y. K. Park, J. Ch. Park [222]; Hein [223-242]; Golosovsky [243-245]; Belk [246]; Belk, Oates, Feld, Dresselhaus, Dresselhaus [247]; Willemsen [248-253]; Trunin [254-261]; Cherpak [262-269]; Velichko [46, 270-281]; Zaitsev [282-286]; Wosik [287-291]; Takken [292]; Booth [293-295]; Hashimoto, Kamijyou, Itamoto, Kobayashi [296]; Hashimoto, Kobayashi [297-300]; Berlinsky [301]; Kastner [302]; Vorobiev, Drozdov, Gusev, Mironov, Vostokov, Kluenkov, Gaponov, Talanov [303]; Talanov, Mercaldo, Anlage, Claassen [304]; Einfeld, Lahl, Kutzner, Wornderweber, Kastner [305]; Talisa [306]; Leong [307, 308]; Leong, Booth, Schima [309]; Gaganidze [310-312]; Zhao Xinjie, Li Lin, Lei Chong, Tian Yongjun [27]; Zhuravel [313, 314]; Xin [315]; Van der Beek [316]; Yingmin Zhang, Zhengxiang Luo, Kai Yang, Qishao Zhang [317]; D. O. Ledenyov, Mazierska, Allen, Jacob [318, 319]; Mazierska, D. O. Ledenyov, Jacob, Krupka [320-322]; Jacob, Mazierska, Krupka, D. O. Ledenyov, Takeuchi [323]; Jacob, Mazierska, Leong, D. O. Ledenyov, Krupka [324]; V. O. Ledenyov, D. O. Ledenyov, O. P. Ledenyov [325]; D. O. Ledenyov [326-328]; Jacob [329-333];. Jacob, Mazierska, Leong, Ledenyov, Krupka [330]; Jacob, Mazierska, Savvides, Ohshima, Oikawa [331]; Jacob, Mazierska, Srivastava, [332]; Lahl, Wördenweber [334]; Pan [335, 338]; Pan, Luzhbin, Kalenyuk, Kasatkin, Kamashko, Velichko, Lancaster, Storey [336]; Pan, Kalenyuk, Kasatkin, Komashko, Ivanyuta, Melkov [337]; Andreone [339]; Cifariello, Aurino, Di Gennaro, Lamura, Orgiani, Villegier, Xi, Andreone [340]; Cifariello [341]; Mateu, Collado, Menendez, O'Callaghan [342]; Mateu, Booth, Moeckly [343]; Kermorvant, van der Beek, Mage, Marcilhac, Lemaitre, Bernard, Briatico [344]; Kermorvant, van der Beek, Mage, Marcilhac, Lemaitre, Briatico, Bernard, Villegas [345-347]; Kermorvant [348, 349]; Wang [350]; Z. Kim [351]; Golovkina [352]; Weinstock [353-357], Nisenoff [356], and by many others.

## 3.4. Main Parameters for Accurate Characterization of High Temperature Superconducting Thin Films at Microwaves.

In 1940, the first accurate measurements of ***surface impedance*** $\boldsymbol{Z_S}$ of superconductors by London [13], using the microwave methods, provided the important information about the electromagnetic properties of superconductors at microwaves [358, 359]. It was understood that the surface impedance $Z_S$ is a measure of energy losses by an electromagnetic wave in superconductor at microwaves [358, 359]. The improved methods for the accurate measurements of surface impedance $Z_S$ were developed by Pippard [14, 17]. Blatt [360] mentions that the study of the surface impedance $Z_S$ of superconductors at microwaves led to the creation of the coherence length concept of Pippard [14, 17]. The significant initial scientific contributions to the experimental and theoretical researches on the surface impedance $Z_S$ of *LTS* superconductors at microwaves were made by Faber, Pippard [361], Galkin, Bezuglyi [362], Khaikin [363; 364], Spiewak [365, 366], Biondi, Garfunkel [367]; Kaplan, Nethercot, Boorse [368], Miller [369], Abrikosov, Gor'kov, Khalatnikov [370, 371], G. Dresselhaus, M. S. Dresselhaus [372], M. S. Dresselhaus, G. Dresselhaus [373], Williams [374], and by some others [358, 359].

The main parameter describing microwave properties of superconducting materials is the ***surface impedance*** $\boldsymbol{Z_S}$ in eq. (3.1) [1, 3, 4]:

$$Z_S = \left(\frac{E_x}{H_y}\right)_{surf} = R_S + jX_S = \left(\frac{j\omega\mu_0}{\sigma_S}\right)^{1/2} \quad (3.1)$$

where $\boldsymbol{R_S}$ is the ***surface resistance***

$$R_S = \frac{\sigma_1}{2\sigma_2}\left(\frac{\omega\mu}{\sigma_2}\right)^{1/2},$$

(*Note*: Biondi, Garfunkel [367] elaborate that the *surface resistance* $R_S$ is the real part of the *surface impedance* $Z$, defined as $Z=4\pi\,(E/H)_0$, where $(E/H)_0$ is the ratio of electric to magnetic fields at the surface of the metal. In general, the *surface impedance* $Z=R+iX$ is complex; the imaginary part is the *surface reactance*.)

and $\boldsymbol{X_S}$ is the ***surface reactance*** of a superconductor

$$X_S = \left(\frac{\omega\mu}{\sigma_2}\right)^{1/2},$$

$\boldsymbol{\sigma_S}$ is the ***complex surface conductivity*** of a superconductor

$$\sigma_S = \sigma_1 - j\sigma_2 = \frac{2\omega\mu_0 R_S X_S}{\left(R_S^2 + X_S^2\right)^2} - j\frac{\omega\mu_0\left(X_S^2 - R_S^2\right)}{\left(R_S^2 + X_S^2\right)^2},$$

$E_x$ and $H_y$ are the tangential electrical and magnetic fields components of the electromagnetic wave on the surface of a sample, $\omega=2\pi f$ is the circular frequency, $\mu_0$ is the magnetic permeability of free space.

In the simple two fluids model, all the properties of superconductors are temperature dependent. The real and imaginary parts of conductivity are

$$\sigma_1(T) = \frac{\sigma_n(T)}{1+\left(\omega\tau(T)\right)^2},$$

$$\sigma_2(T) = \frac{1}{\mu_0\omega\lambda^2(T)},$$

where $\sigma_n = n_n e^2\tau/m$, $n_n$ is the normal electrons density and $n_S$ is the superconducting electrons density and $n_n(T)+n_S(T)=1$, $\lambda$ is the magnetic penetration depth.

At low frequencies $\omega\tau<<1$ ($\omega < 10^{13}$ rad/s for $YBa_2Cu_3O_{7-\delta}$)

$$\lambda(T) \sim \lambda_L(T) = [m^*(T)/\mu_0 e^2 n_S(T)]^{1/2},$$

where $\lambda_L$ is the *London penetration depth* ($\lambda_L \approx 140$ *nm* for $YBa_2Cu_3O_{7-\delta}$ superconductor at $T=0K$ and coherence length $\xi_0 \approx 2.5$ *nm*), $m^*$ is the effective electron quasiparticle mass, $e$ is the electron charge.

The surface resistance and reactance may be re-written in the form in eqs. (3.2) and (3.3)

$$R_S(T) = \frac{\mu_0\omega^2\lambda^3(T)\sigma_n(T)}{2}, \quad (3.2)$$

$$X_S(T) = \omega L_S(T) = \omega\mu_0\lambda(T), \quad (3.3)$$

where $\sigma_n$ is the conductivity of normal electrons in superconductor, and $L_S = \mu_0\lambda$ is the surface inductance.

The temperature dependences of surface resistance $R_S$ and the surface reactance $X_S$ for high temperature superconductors were comprehensively researched [4, 375-378], and typical *Rs(T)* dependencies are demonstrated for $YBa_2Cu_3O_{7-\delta}$ single crystal in Fig. 1 [376]

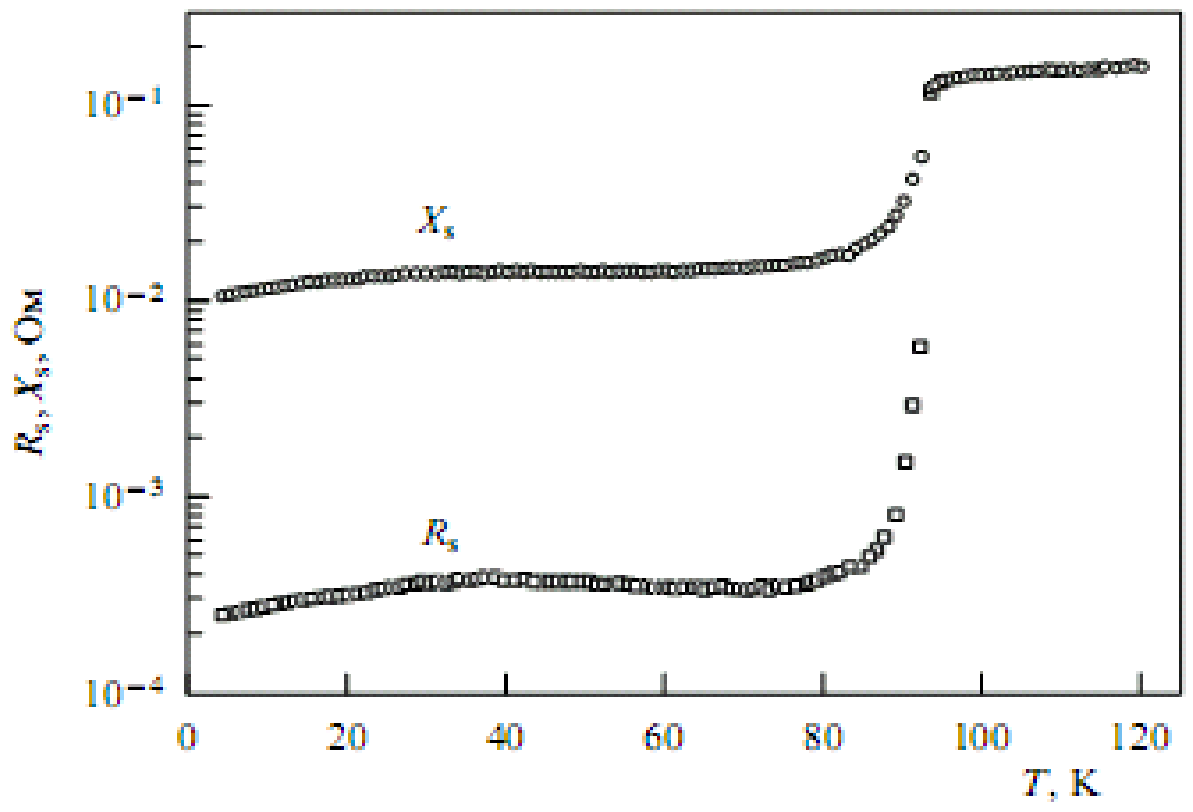

**Fig. 1.** Temperature dependences of surface resistance *Rs* and surface reactance *Xs* for superconducting state in $YBa_2Cu_3O_{7-\delta}$ single crystal at 9.4 *GHz* (after [376]).

Other *HTS* superconductors have a similar type of temperature dependence of surface resistance $R_S(T)$ as researched by Gaganidze, Heidenger, Halbritter, Shevchun, Trunin, Schneidewind [377]. Typical temperature dependences of surface resistance $R_S(T)$ of $YBa_2Cu_3O_{7-\delta}$ (*YBCO*) and $Tl_2Ba_2CaCu_2O_8$ (*TBCCO*) *HTS* thin films are presented in Fig. 2 [377]

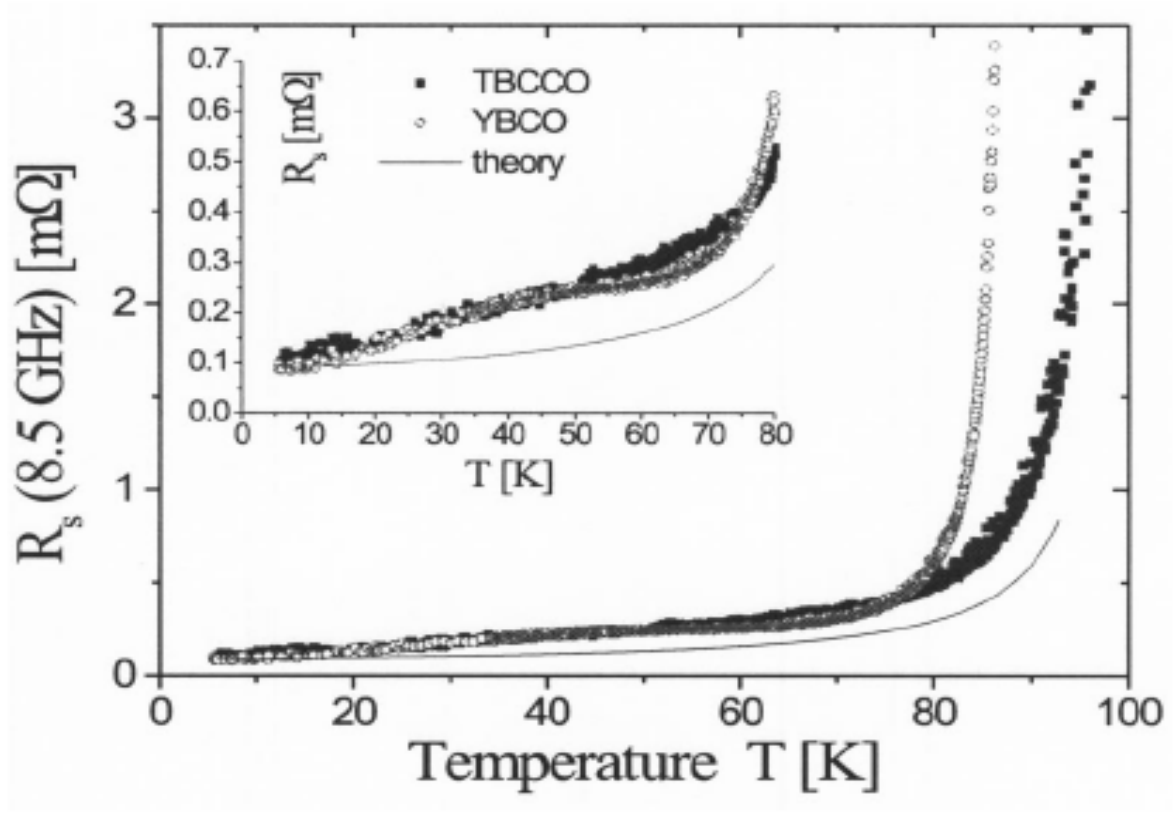

**Fig. 2.** Temperature dependence of surface resistance of $YBa_2Cu_3O_{7-\delta}$ and $Tl_2Ba_2CaCu_2O_8$ thin films at 8.5*GHz*. The inset shows the same data on a different scale. 50 $\mu\Omega$ were added to the theoretical curve to take into account the coupling losses (after [377]).

Authors of book would like to explain that, as it follows from theory and results from experimental data in Fig. 2, in the normal state at temperature $T>T_C$, the surface resistance $R_n$ is equal to the surface reactance $X_n$: $R_n = X_n$. However, in the superconducting state at low temperatures, the magnitudes of the surface resistance $R_S$ and surface reactance $X_S$ are significantly lower than their magnitudes in the case of normal metal: $R_S << R_n$, $X_S << X_n$ , and in this case, the magnitude of the surface reactance $X_S$ is significantly bigger than the magnitude of the surface resistance $R_S$: $X_S >> R_S$, as it is shown by the author of thesis in Fig. 3.

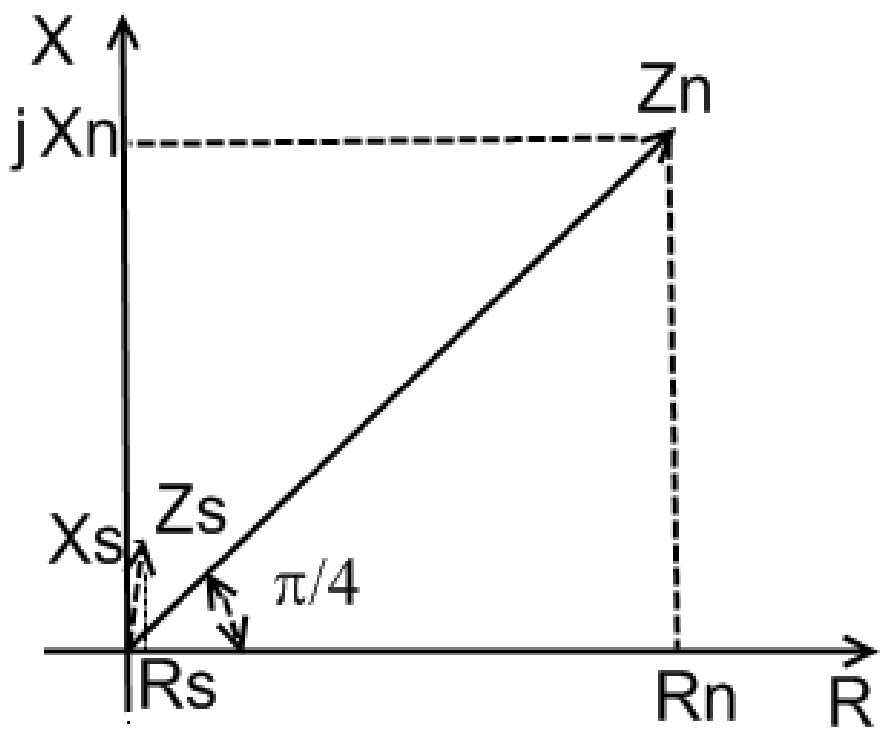

**Fig. 3.** Impedance vector for normal metal $Z_n$ and superconductor $Z_S$ in imaginary space.

In the normal metals: $X_n/R_n=1$, and in the superconductors: $X_S/R_S >> 1$.

The surface impedance of superconductor

$$Z_S = R_S + jX_S$$

can be represented as an equivalent lumped element model with the help of series electrical circuit, which has the corresponding characteristic magnitudes of surface resistance $R_S$ and reactance $X_S$ *as* shown by the author of dissertation in Fig. 4

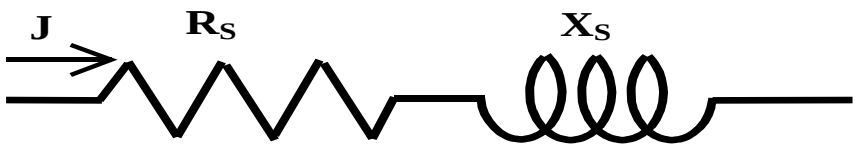

**Fig. 4.** Lumped element model of superconductor with application of series electrical circuit, which has an equivalent surface impedance $Z_S$ with superconductor.

The quality factor $Q$ of the circuit, when considering it as a part of a series resonator, may be written as

$$Q_S = X_S / R_S,$$

where $X_S = \omega \cdot L_S$.

The magnitude of quality factor is approximately $Q \approx 10^2$ in $YBa_2Cu_3O_{7-\delta}$ single crystal at low temperature $T<<T_C$ in the $GHz$ frequency range.

The quality factor can also be represented as a ratio of the dissipated and stored wave energy:

$$Q = W_d / W_S,$$

where $W_d$ is the dissipated and $W_S$ is the stored wave energy in surface layer of superconductor.

In superconductor, the main part of stored energy of electromagnetic wave is connected with its surface induction as

$$W_S \propto \omega L_S I^2,$$

where $I$ is the current, which is generated by the electromagnetic wave in the skin layer of superconductor. The dissipated energy in superconductor is

$$W_d \propto R_S I^2.$$

It should be noted that only a small part of the energy is dissipated in superconductor due to the surface resistance.

The superconductor has advanced microwave characteristics, because of high quality factor $Q$ in comparison with normal metal for which $Q_S = 1$.

The new ***r-parameter*** was introduced for analysis of nonlinear properties of superconductors by Halbritter [378, 379]. The ***r-parameter*** depends on the power $P_{rf}$ and magnetic field $H_{rf}$ of high frequency electromagnetic wave, as

$$r(H_{rf}) = \Delta X_S(H_{rf}) / \Delta R_S(H_{rf}).$$

Golosovsky, Snortland, Beasley [380] used the inversion form of this parameter: $r_G \sim 1/r$.

In the representation, proposed by the author of thesis, ***the r-parameter is the differential quality factor for superconducting thin film***

$$r(H_{rf}) = Q_{dif}(H_{rf}).$$

This parameter $r(H_{rf})$ will be discussed in the paragraph on the nonlinear properties of superconductors in this chapter later.

Dependences of surface resistance on frequency $R_S(f)$ for $YBa_2Cu_3O_{7-\delta}$ in the *GHz* frequency range, obtained by various research groups, are shown in Fig 5 [375]. It can be seen that the $YBa_2Cu_3O_{7-\delta}$ superconducting films exhibit much lower values of surface resistance than the pure normal metal *Cu* copper, which has one of smallest resistances $R_S$ among metals, in the frequency range of up to 100*GHz* at low temperature of 77 *K*. The measured results of $R_S(f)$ for the traditional low temperature superconductor niobium *Nb* are shown at the temperature of 7.7*K* [375].

Frequency dependence of surface resistance $R_S(f)$ for normal metal is close to expression $R_S \propto f^{1/2}$, while in the case of superconductors, this dependence is $R_S \propto f^{2}$, that is in good agreement with theories of normal metals and superconductors.

Magnitude of resistance of normal metal *Cu* at 77*K* corresponds to $R_S \approx 3.4 f^{1/2} m\Omega$, and value of resistance of superconductors at 77 *K* is $R_S \approx 5 \cdot f^{2} \mu\Omega$.

In superconductor $YBa_2Cu_3O_{7-\delta}$, the magnetic field penetration depth is $\lambda_L \approx 200nm$ at 77 *K*, and the London penetration depth is $\lambda_L \approx 150\ nm$ at 0 *K*.

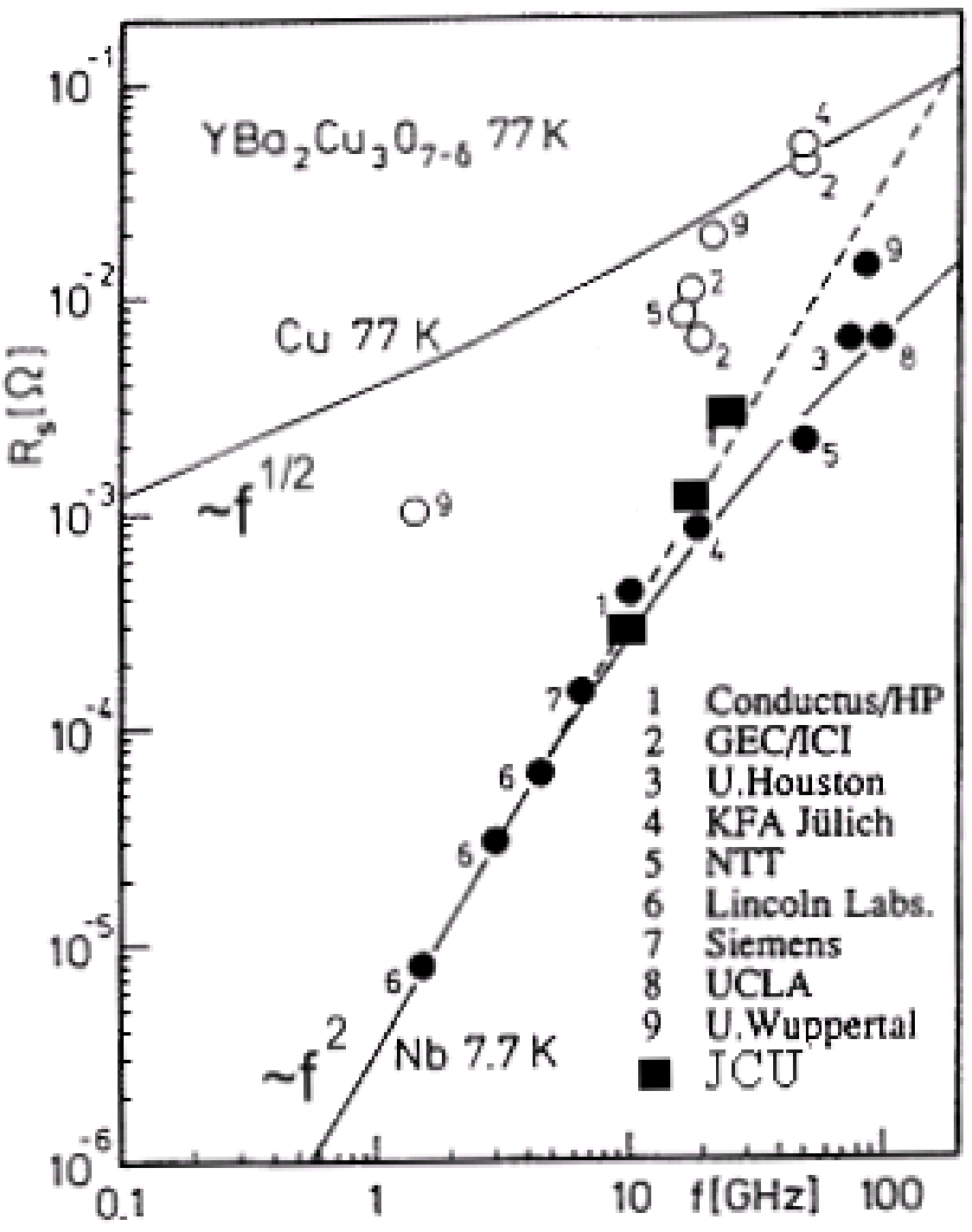

**Fig. 5.** Frequency dependence of surface resistance of $YBa_2Cu_3O_{7-\delta}$ thin films and copper Cu at 77*K*, as well as niobium *Nb* at 7.7*K*, as measured by various research groups (after [375]), and with *JCU* results presented by the author of thesis.

In Fig. 5, the superconductors have an advantage of smaller *Rs* in comparison with the normal metals up to the frequency range ~ 100 *GHz*. It means that the microwave devices with unique advanced parameters can be developed with the use of superconductors [1-6, 25-28, 30].

## 3.5. Experimental Methods and Types of Microwave Resonators for Precise Microwave Characterisation of Superconductors.

Microwave resonator techniques play an important role in analysis and precise characterisation of superconducting materials at microwaves in Mazierska, Jacob, Krupka [381]. The most accurate characterisation results are obtained, when the particular microwave resonator types are applied in Mazierska, Jacob, Leong, Ledenyov, Krupka [382], Mazierska [383]. To make the precise characterization of physical properties of *HTS* superconductors, the microstrip resonators are produced from superconductors, or the *HTS* thin films are researched with application of dielectric resonators [384]. Single crystal superconducting samples can also be researched in a dielectric resonator at ultra high frequencies (*UHF*) [376, 384-386].

Microwave resonators are well described by two main measurement parameters: the ***resonance frequency*** $\boldsymbol{f_0}$ and the ***quality factor Q***.

Fig. 6 shows the resonance curve of a simple *RLC* circuit in Zhou [3].

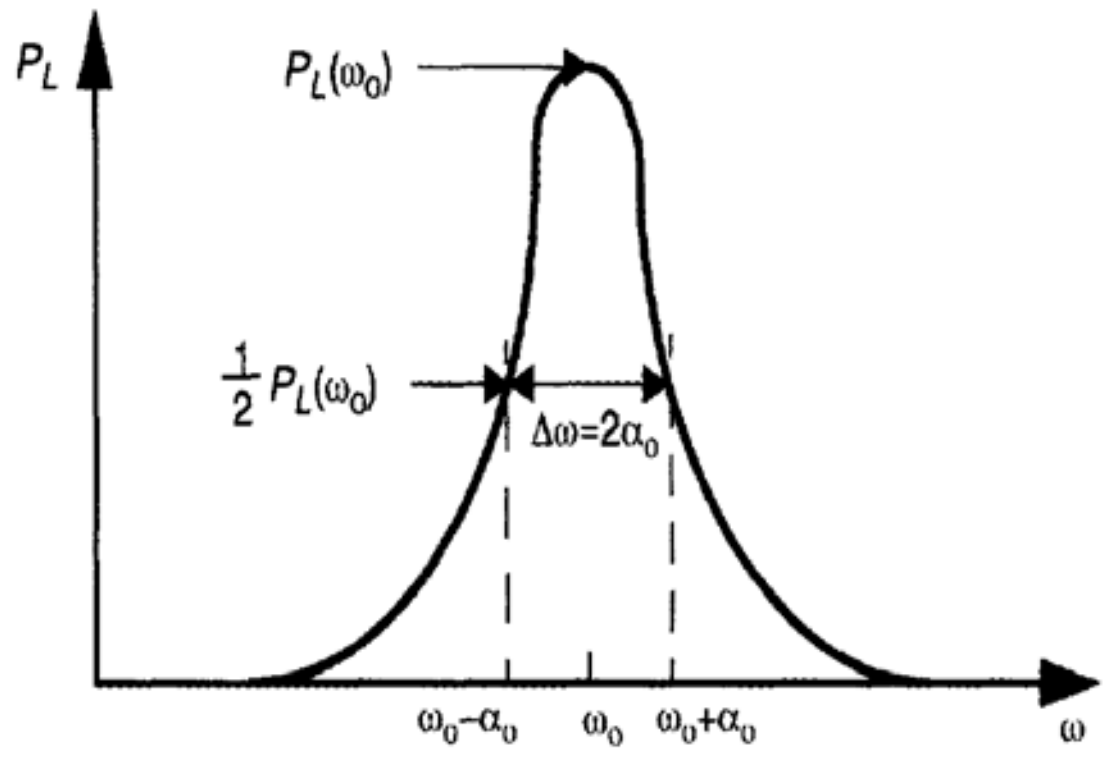

**Fig. 6.** Resonance curve of a simple *RLC* circuit (after [3]).

The resonance frequency $f_0$ or $\omega_0$ depends on the geometrical dimensions of a microwave resonator. The resonance frequency is lower, if volume of microwave resonator cavity is bigger. As a rule, the resonance frequency $f_0$ can be found by measuring of maximum amplitude of the electromagnetic signal in a microwave resonator, and by registering the maximum of the transmission coefficient $S_{21}(f)$ of the electromagnetic signal.

The *quality factor* of a microwave resonator is

$$Q = f_0 / \Delta f,$$

where $\Delta f$ is the width of the resonance curve at the -3*dB* level. The quality factor $Q$ of a microwave resonator is connected with the parameters of a superconducting sample and dielectric material as

$$\frac{1}{Q} = \frac{R_S}{A_S} + \frac{1}{Q_{par}} + \frac{\tan\delta}{A_{diel}},$$

where $A_S$ is the geometrical factor of a superconductor in a resonator, which is equal to the ratio between the volume of the resonator and the volume of the surface layer of a superconductor in which the magnetic field penetrates

$$A_S \approx V_{res} / V_s.$$

The second term with $Q_{par}$ is a ratio of the parasitic power dissipation in the housing of resonator to the circulating power. Third term denotes the losses in the dielectrics (in the dielectric resonators $A_{diel} \approx 1$), value of $\tan\delta$ for *sapphire* $\alpha$-$Al_2O_3$ is less than $\tan\delta < 10^{-6}$; and for magnesium *oxide MgO* and *lanthanum aluminate LAO* is $\tan\delta < 10^{-5}$ at temperature of *70 K* in Hein [375], Mazierska, D. O. Ledenyov, Jacob, Krupka [387].

Characterisation of superconducting materials in ultra high frequency electromagnetic fields is usually conducted with the use of different types of microwave resonators. The resonance systems can have various geometrical shapes such as rectangular, round, and some others with different technical characteristics. Microwave resonators can also be differentiated on either transmission or reflective types in relation to the electromagnetic wave propagation characteristics.

The two types of microwave resonators:

1. *Cavity resonators,* and
2. *Planar resonators,*

are being used for the accurate microwave characterization of superconductors.

The ***cavity type*** is represented by the bulk cavity resonators and cavity resonators with dielectrics, which may also have the two additional modifications: the open end dielectric resonators and the *Hakki-Colleman dielectric resonators* in Hakki, Coleman [388]. These types of resonators are being used for the microwave measurements of the unpatterned films. A high quality dielectric puck is usually placed at the centre of the resonator that enables to increase accumulated electromagnetic wave energy and concentrate the field within the dielectric area. The quality factor of the resonance system with a sample is determined as a ratio of accumulated wave energy within the free resonator volume to dissipated energy in resonator walls and researched sample for a period of oscillation. Therefore, the systems with high quality factors are more sensitive to a change of physical parameters of *HTS* samples, and widely used for accurate microwave characterisation of *HTS* thin films.

The ***planar resonators*** have several types such as the parallel plate resonator, the microstrip resonator, the stripline and coplanar resonators [46]. Planar resonators are used for the measurements of the patterned superconducting films. The microstrip resonators have smaller quality factors $Q$ than the cavity ones, although they produce higher fields at the same signal power levels. To some extend, this effect is an advantage in research on the superconducting nonlinear phenomena as the nonlinearities arise at elevated field strengths and currents in *HTS* samples.

Fig. 7 shows the eight main types of microwave resonators for the accurate *HTS* thin films characterization at microwaves in Velichko, Lancaster, Porch [46]:

a) *Open-ended dielectric resonator*;

b) *Hakki–Coleman dielectric resonator*;

c) *Parallel plate resonator*;

d) *Cavity resonator*;

e) *Coplanar resonator*;

f) *Disc resonator*;

g) *Microstrip resonator*;

h) *Stripline resonator*.

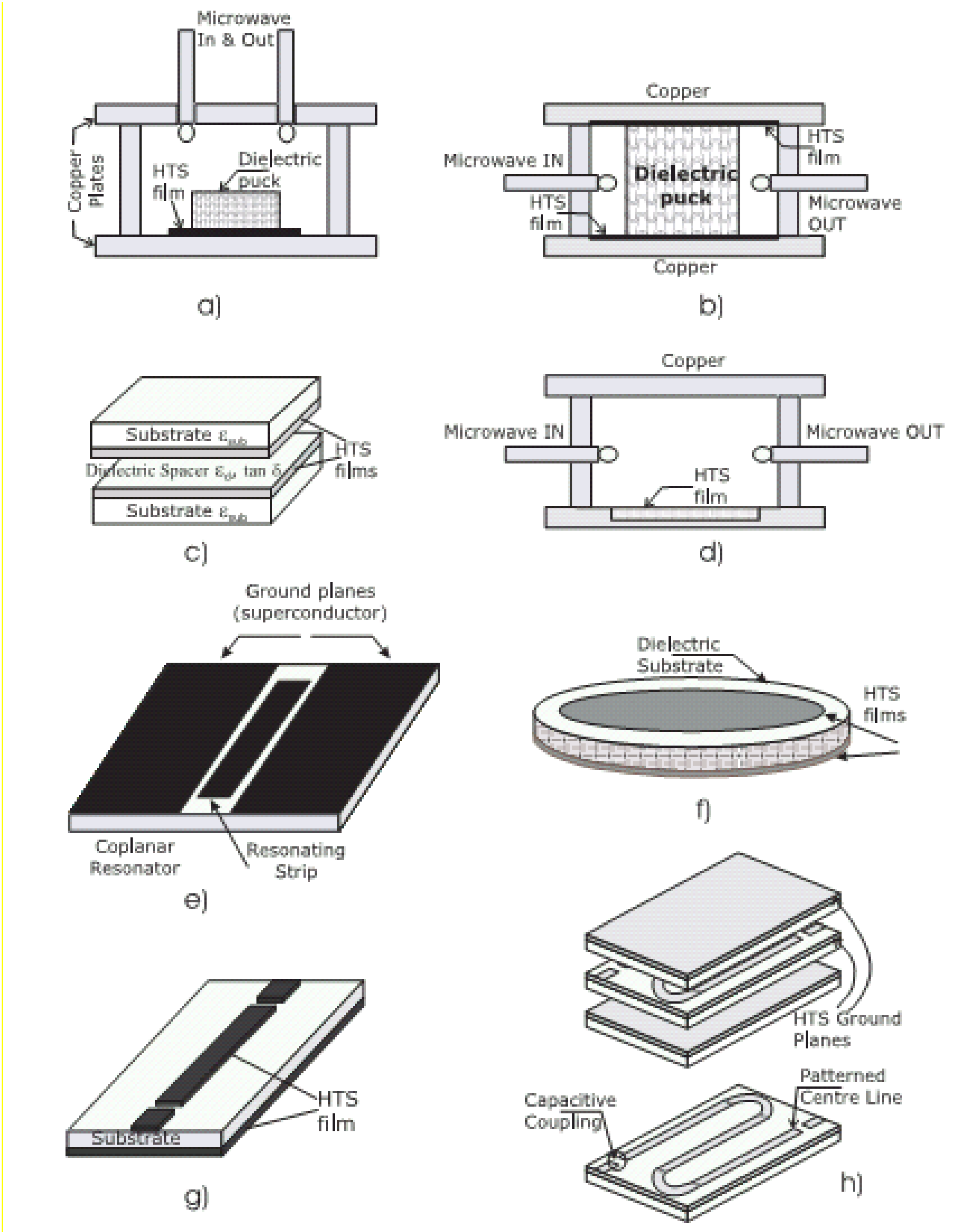

**Fig. 7.** Eight main types of microwave resonators used for thin film characterization: (a) open-ended dielectric resonator; (b) Hakki–Coleman dielectric resonator; (c) parallel plate resonator; (d) cavity resonator; (e) coplanar resonator; (f) disc resonator; (g) microstrip resonator; (h) stripline resonator (after [46]).

Tab. 6 provides the comprehensive review on the advantages and disadvantages of various types of microwave resonators employed for surface impedance measurements in Velichko, Lancaster, Porch [46].

| Type of resonator | Advantages | Disadvantages |
|---|---|---|
| Cavity | (a) Geometric factor can be calculated analytically; (b) if the cavity is made of low $T_c$ superconductor, it can have a very high quality factor $Q$ and therefore the sensitivity of the technique can be quite high; (c) easy sample mounting; (d) quite reproducible; (e) easy to realize a fixed temperature operation which allows the measurement of both $R_s$ and $\lambda$ (using a thermally isolated platform within the cavity) | (a) At typical microwave frequencies of interest (~10 GHz) the cavities are quite big ($\gtrsim$6 cm in diameter) and bulky; (b) they require large area films (several inches in diameter) or use diaphragms in the cavity end plate, which however bear potential problems due to current path interruption at the boundary between the plate and the film |
| Dielectric | (a) Quite sensitive (geometric factor is typically a few hundred ohms); (b) easy sample mounting; (c) does not require sample preparation or modification; (d) sufficiently reproducible; (e) relatively small size (at typical microwave frequencies of ~10 GHz for dielectric permittivity $\varepsilon \gtrsim$ 10 the pucks are usually $\lesssim$ 10 mm in diameter, which is acceptable for measuring both small (10 × 10 mm$^2$) and large (2″ in diameter or more) area HTS films) | (a) Penetration depth measurements are greatly encumbered because the change in frequency is mostly due to change in the dielectric permittivity of the puck, not change in $\lambda(T)$ of the film. The problem may be overcome by keeping the puck at fixed temperature, which is however quite a technical challenge. More details on the DR technique can be found in [6] |
| Parallel plate | (a) Low geometric factor and therefore high sensitivity; (b) many easily excitable modes | (a) Parasitic modes interfering; (b) two films required; (c) reproducibility is not great |
| Microstrip | (a) High sensitivity (low geometric factor); (b) large enhanced current crowding at the edges of the strip, which is useful for nonlinear measurements | (a) Possible influence of patterned edges; (b) loss of the substrate may noticeably contribute to the measurements; (c) requires double-sided films and measured $R_s$ will be weighted by the contributions of the two; (d) finite edge roughness due to patterning may enhance nonlinear effects and promote vortex penetration |
| Stripline | (a) Extremely low geometric factor; (b) possibly highest current crowding at the edges | (a) Possible influence of patterning; (b) three films required, although the major contribution ~90–95% comes from the film with the resonant centre strip; (c) impossible to perform DC field measurements due to screening of the superconducting groundplanes; (d) same as for the microstrip; (e) hard to analyse and requires numerical modelling |
| Coplanar | (a) Geometric factor comparable to stripline; (b) current crowding comparable to stripline; (c) only one single-sided film is required; (d) losses are nearly insensitive to the substrate; (e) allows application of DC magnetic field | (a) Same as for the stripline; (b) shape of resonance (especially higher order modes) highly sensitive to grounding; (c) finite edge roughness; (d) requires numerical modelling |

**Tab. 6.** Review of advantages and disadvantages of various types of microwave resonators employed for surface impedance measurements (after [46]).

The different resonant structures have been used by many researchers to perform the accurate measurements of surface resistance $R_S$ of the *LTS* and *HTS* thin films at microwaves. The author of dissertation would like to give a few examples:

1. ***Dielectric resonator*** by B. W. Hakki and P. D. Coleman [388]; Mazierska [389]; Ceremuga (Mazierska), Krupka, Modelski [390]; Ceremuga (Mazierska), Krupka, Geyer, Modelski [391]; Mazierska, Grabovickic [392]; Mazierska, Jacob, Leong, D. O. Ledenyov, Krupka [393]; Mazierska, Jacob, Krupka [394]; Mazierska, Jacob [395]; Mazierska, Jacob [396]; Mazierska, Krupka, Bialkowski, Jacob [397]; Jacob, Mazierska, D. O. Ledenyov, Krupka [398], Jacob, Mazierska, Leong, D. O. Ledenyov, Krupka [399]; D. O. Ledenyov, Mazierska, Allen, Jacob [400]; D. O. Ledenyov, Mazierska, Allen, Jacob [401]; Krupka, Klinger, Kuhn, Baranyak, Stiller, Hinken, Modelski [402]; Krupka, Mazierska [403]; Wosik, Xie, Mazierska, Grabovickic [404]; Misra, Kataria, Pinto, Tonouchi, Srivastava [405]; Zhang, Yan, Ji, Sun, Zhou, Fang, Zhao [406]; Bi Zhang, Fabbricatore, Gemme, Musenich, Parodi [407]; Trunin [408, 409]; Yingmin Zhang, Zhengxiang Luo, Kai Yang, Qishao Zhang [410]; Gaganidze, Heidinger, Halbritter, Schneidewind [411]; Gaganidze, Heidenger, Halbritter, Shevchun, Trunin, Schneidewind [412, 413]; Obara, Kosaka, Sawa, Yamasaki, Kobayashi, Hashimoto, Ohshima, Kusunoki, Inadomaru [414]; Obara, Kosaka [415]; DeGroot, Hogan, Kannewurf, Buchholz, Chang, Gao, Feng, Nordin [416]; Hashimoto, Kamijyou, Itamoto, Kobayashi [417]; Hashimoto, Kobayashi [418, 419]; Kobayashi, Katoh [420]; Kobayashi, Imai, Kayano [421, 422]; Kobayashi, Senju [423]; Kobayashi, Hashimoto [424]; Shen, Wilker, Pang, Holstein, Face, Kountz [425]; Shen [426]; Tellmann, Klein, Dahne, Scholen, Schiltz, Chaloupka [427]; Yoshikawa, Okajima, Kobayashi [428]; Mansour [429].

2. ***Quasi-optical dielectric resonator*** by Cherpak, Barannik, Filipov, Prokopenko, Vitusevich [430]; Cherpak, Barannik, Bunyaev, Prokopenko, Torokhtii, Vitusevich [431]; Barannik, Cherpak, Torokhtiy, Vitusevich, [432].

3. ***Confocal resonator*** by Martens [433].

4. ***Parallel plate dielectric resonator*** by Taber [434], Talanov, Mercaldo, Anlage, Claassen [435]; Mourachkine, Barel [436, 437].

5. ***Co-axial microwave resonator*** by Woodall [453].

5. ***Transmission line and microstrip line resonator*** by D. E. Oates, Anderson, Mankiewich [88]; D. E. Oates, Anderson, Alfredo, Sheen, Ali [89, 91]; Porch *et al.* [204]; Porch, Lancaster, Humphreys [206]; J. H. Oates, Shin, D. E. Oates, Tsuk, Nguyen [438]; Takken [292]; Booth, Beall, Rudman, Vale, Ono [293]; Pinto [439]; Golosovsky *et al.* [440]; O. G. Vendik, Kozyrev, Samoilova, Hollmann, Ockenfub, Wordenweber, Zaitsev [136]; Samoilova, O. G. Vendik, Hollmann, Kozyrev, Golovkov, Kalinikos [441]; D.O. Ledenyov [442], V.O. Ledenyov, D.O. Ledenyov [448], Sitnikova, I. B. Vendik, O. G. Vendik, Kholodnyak, Tural'chuk, Kolmakova, Belyavsky, Semenov [443]; Willemsen, King, Dahm, Scalipino [444]; Xin, D. E. Oates, Anderson, Slattery, G. Dresselhaus, M. S. Dresselhaus [445]; Schwab, Gaganidze, Halbritter, Heidinger, Aidam, Schneider [446], Chung, Yoo, Cheong, Kwak, Kang, Paek, Ryu, Kim, Kang [447], N. Zhao, J. Liu , H. Zhao, H. Li, T. Li, W. Chen [449]; Ohshima, Okuyama, Sawaya, Noguchi [450]; Okai, Ohshima, Kishida, Hatano [451], and by many others.

The author of dissertation would like to summarize that many types of microwave resonators have a great capacity for very accurate measurements of surface resistance $R_S$ in Misra, Kataria, Pinto, Tonouchi, Srivastava [452], but the best structures for the nonlinear measurements and simulations are ones for which the analytical solutions for electromagnetic fields can be properly obtained. Also, such factors as the technical convenience, reliability and accuracy in practical realisations should be considered as well in Mazierska, Krupka [181].

As the *Hakki-Coleman dielectric resonator* [388] satisfies to the above desirable requirements in Mazierska, Krupka [181], it has been chosen as a primary microwave device to perform the measurements toward the precise microwave characterisation of *HTS* thin films in this dissertation [318-330].

In this research, the three types of microwave resonators have been extensively used towards the accurate characterisations of superconducting thin films at microwaves:

1. ***Hakki-Colleman type dielectric resonator***,
2. ***Split post dielectric resonator***, and
3. ***Microstrip resonator*** [60, 61, 62, 66, 67, 68, 69].

The obtained experimental and computer modeling results are presented by the author of dissertation in details in Chapters 4, 5, 6, 7.

One of the most important tasks in the research of superconducting materials is a transition from the characteristics, obtained for the resonant system, to the characteristics of a researched superconductor in Mazierska, Krupka [181]. As the surface resistance $R_S$ of superconductors has a small value, the resonant systems with the high quality factors $Q$ are employed for the precise characterization of superconducting materials. The research problem to obtain the precise quantitative data about the surface resistance dependences on the temperature $R_S(T)$, magnetic field density $R_S(H)$, magnitudes of currents $R_S(I)$, and other parameters represents a formidable research task, in which the magnitude of the surface resistance $R_S$ needs to be retrieved on the basis of measured characteristics of a microwave resonator in Mazierska, Krupka [181].

Generally, knowing the structure and distribution of electromagnetic fields in a microwave resonator, along with the geometrical shape and position of the sample, it is possible to find the accurate characteristics of the superconductor by the direct calculations of the electromagnetic fields in Mazierska, Krupka [181]. However, the electromagnetic fields in the resonator can easily be calculated in the case, when the field frequency is either close or matches the resonance frequency only. Otherwise, the electromagnetic field distribution becomes essentially more complicated, and the calculations are approximate. In these conditions, another computational method for the characterisation of resonance system is more convenient. It is based on the possible representation of a resonator near any of its resonance frequencies as the lumped *RLC* element circuit in Zhou [3]. This method of modelling of resonator is less complex for calculations, and enables to extract the necessary superconductor parameters using the experimental data obtained for the resonator with the load. This approach has been employed in this dissertation, and is discussed in details in Chapters 4, 5.

## 3.6. Nonlinear Microwave Properties of Superconductors.

Main goal of dissertation is to develop an advanced model, using the microwave resonator techniques, for precise microwave characterisations of *HTS* thin films at microwaves. Trying to achieve the main research goal, the author of dissertation proposed the modeling of nonlinear properties of *HTS* thin films, using *Bardeen, Cooper, Schrieffer* and *Lumped Element Circuit* theories, with purposes to understand the nature of nonlinearities. As can be seen from the previous equations given for the ***linear case***, both the *surface resistance* $R_S$ and the *surface reactance* $X_S$ are functions of the *temperature T, RF magnetic field* $H_{rf}$, and *frequency* $\omega$. The *magnetic penetration depth* $\lambda$ and the *complex conductivity* $\sigma_n$ depend on the *temperature T* and magnitude of *magnetic field* $H_{rf}$. In the ***nonlinear case***, the surface impedance $Z_S$ is dependent on the ultra high frequency magnetic field $H_{rf}$ or current density $J_{rf}$

$$Z_S(H_{rf}) = R_S(H_{rf}) + jX_S(H_{rf}), \text{ also } \lambda \rightarrow \lambda(H_{rf}) \text{ and } \sigma_n \rightarrow \sigma_n(H_{rf}).$$

The nonlinear properties of *HTS* thin films at high microwave power levels are extensively researched by the author of dissertation in next sub-chapters. A great variety of research work has been performed on microwave power dependent nonlinear responses of *HTS* microwave resonators. The nonlinear effects have been observed and their modeling proposed in most researched cases.

Author's research incorporates the knowledge, obtained in the numerous modern completed research studies on the nonlinear modelling, involving the *transmission line - stripline resonator* techniques, which have been carried out by different research groups before. Especially, the research results on a *dielectric microwave resonator* by Wosik [454]; including the lumped element model of a dielectric cavity by Wosik [454] with the simple *RLC* equivalent circuit to simulate the nonlinear electromagnetic responses of a *dielectric resonator* with a linear function dependence of transmitted microwave power, proved to be very useful. However, in the author's opinion, the Wosik's model is not able to sufficiently represent the microwave behaviour of both the *HTS* thin films and the dielectric resonator at microwaves [454]. Therefore, a more comprehensive original approach for the modeling of superconductors embedded into a dielectric resonator, using the *RLC* lumped elements circuits, is engineered by the author of dissertation.

## 3.7. Nonlinear Surface Resistance $R_s$ in Low Temperature Superconductors at Ultra High Frequencies.

The initial research on nonlinear effects in conventional superconductors was started in the beginning of the 1950s, with research papers published by Faber, Pippard [361], Galkin, Bezuglyi [362], Khaikin [363; 364], Spiewak [365, 366], Biondi, Garfunkel [367]; Kaplan, Nethercot, Boorse [368], Miller [369], Abrikosov, Gor'kov, Khalatnikov [370, 371], G. Dresselhaus, M. S. Dresselhaus [372], M. S. Dresselhaus, G. Dresselhaus [373], Williams [374].

The author of dissertation created the Tab. 7, which presents the data, obtained by Faber, Pippard [361], Biondi, Garfunkel [367] during initial experimental research on the accurate measurements of nonlinear surface resistance $R_s$ in Low Temperature Superconductors (*LTS*) at microwaves [40, 46, 70].

Faber, Pippard [361] obtained the experimental dependence of surface resistance ratio $r=R_s/R_n$ as a function of reduced temperature $t=T/T_c$ for superconducting aluminum at wavelength of 25*cm* in London, U.K. in 1955 in Fig. 8 (a).

Biondi, Garfunkel [367] measured the temperature dependence of energy absorbed by superconducting aluminum, and found the surface resistance $R_S$ of superconducting aluminium in the range of wavelengths of 20-3*mm* in Pittsburgh, U.S.A. in 1959 in Fig 8 (b).

Considering the nonlinear surface resistance $R_s$ in *LTS* at microwaves in Fig. 8 (c), Abricosov [463] referred to the experimental results, obtained by Biondi, Garfunkel [367].

Over the years, the main contributions to the research on the nonlinear properties of *LTS* materials the transmission line techniques were made by a number of researchers: Faber, Pippard [361], Galkin, Bezuglyi [362], Khaikin [363; 364], Spiewak [365, 366], Biondi, Garfunkel [367]; Kaplan, Nethercot, Boorse [368], Miller [369], Abrikosov, Gor'kov, Khalatnikov [370, 371], Kaplan, Nethercot, Boorse [455], G. Dresselhaus, M. S. Dresselhaus [372], M. S. Dresselhaus, G. Dresselhaus [373], Williams [374], Nethercot, von Gutfeld [456]; Gittleman, Rosenblum, Seidel, Wicklund [457], Gittleman, Rosenblum [458]; Soderman, Rose [459], Turneaure, Weissman [460], and by some other researchers.

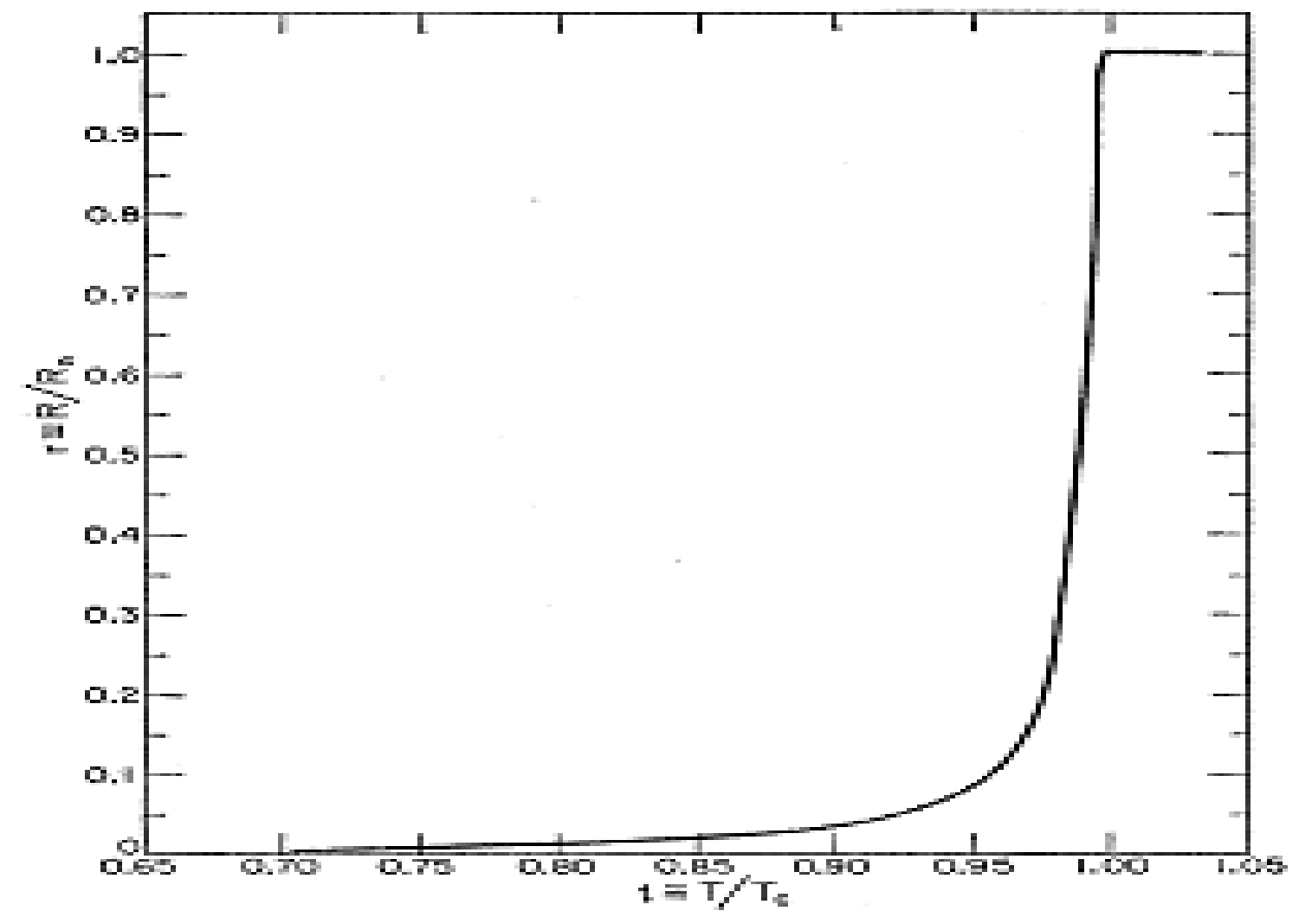

**Fig. 8. (a)** Surface resistance ratio $r=R_s/R_n$ as a function of reduced temperature $t=T/T_c$ for superconducting aluminum at wavelength of 25cm (after [361, 367]).

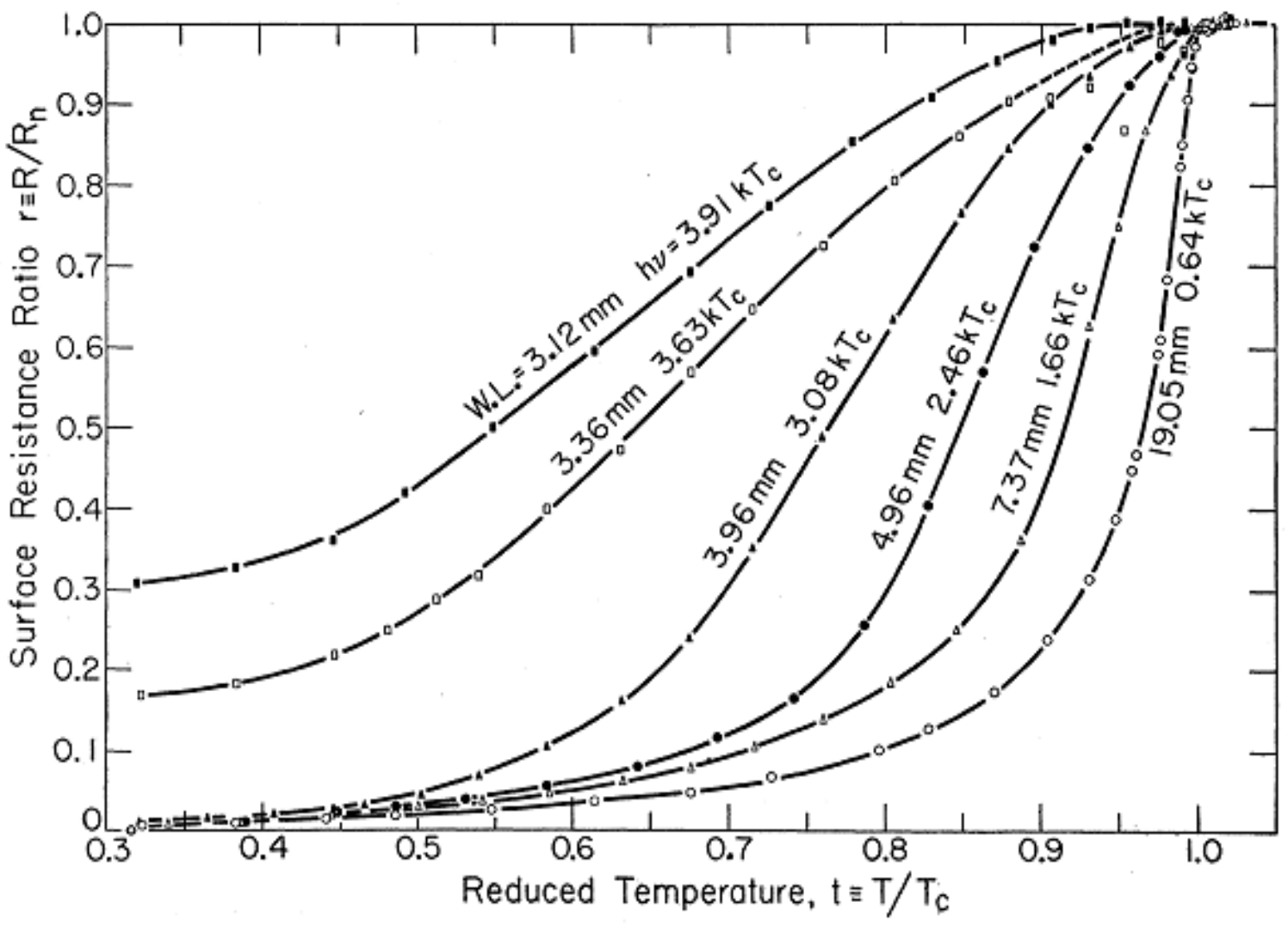

**Fig. 8. (b)** Measured values of the surface resistance ratio $r$ of superconducting aluminum as a function of the reduced temperature $t$ at several representative wavelengths. The wavelengths and corresponding photon energies are indicated on curves (after [367]).

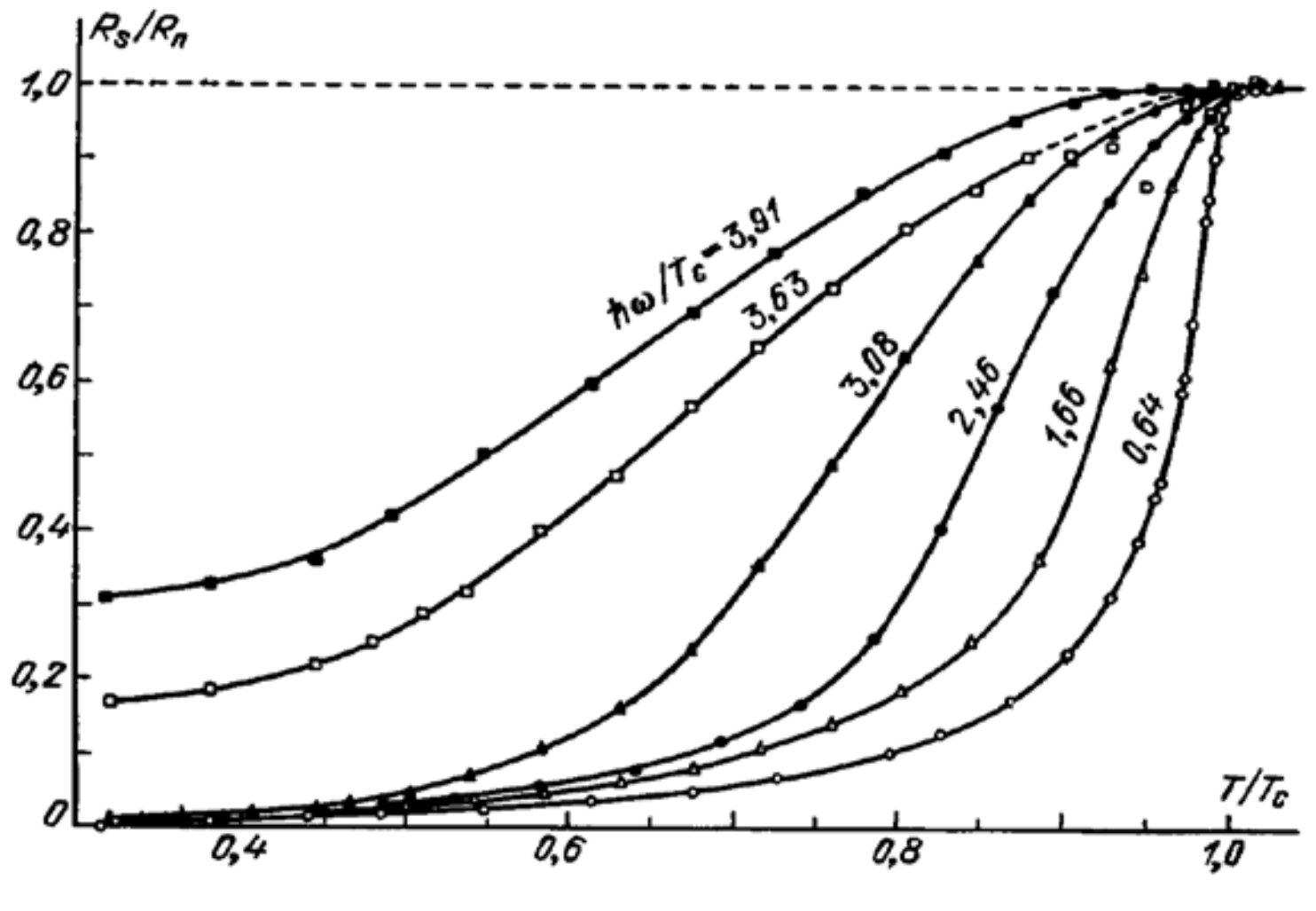

**Fig. 8. (c)** Dependence surface resistance ratio $r=R_s/R_n$ as a function of reduced temperature $t=T/T_c$ for superconducting aluminum at several wavelengths in Abrikosov's interpretation (after [367, 463]).

**Tab. 7.** Measurements of nonlinear surface resistance $R_S$ in Low Temperature Superconductors at microwaves (after [361, 367, 463]).

In 1959, Spiewak observed the nonlinear influence by the magnetic field on the impedance of a superconducting *tin* sample at the frequency of 1*GHz* [366]. Nonlinear microwave effects and switching speed of superconducting *tin* at frequency of 9.2 *GHz* were also researched by Nethercot, von Gutfeld in 1963 [456].

In 1965, Gittleman, Rosenblum, Seidel, Wicklund measured the nonlinear dependence of the imaginary part of conductivity on the transport current for the superconducting films of *tin*, *indium* and *tin-indium* alloys at various temperatures at 23 *GHz* [457]. The surface impedance of *Type-II* superconductors, using the different superconducting films, including $Nb_3Sn$, was researched by Gittleman, Rosenblum at the frequencies up to $\omega \approx 2 \cdot 10^{11}$ *rad/sec* in 1968 [458]. The experimental results are shown in Fig. 9 [458].

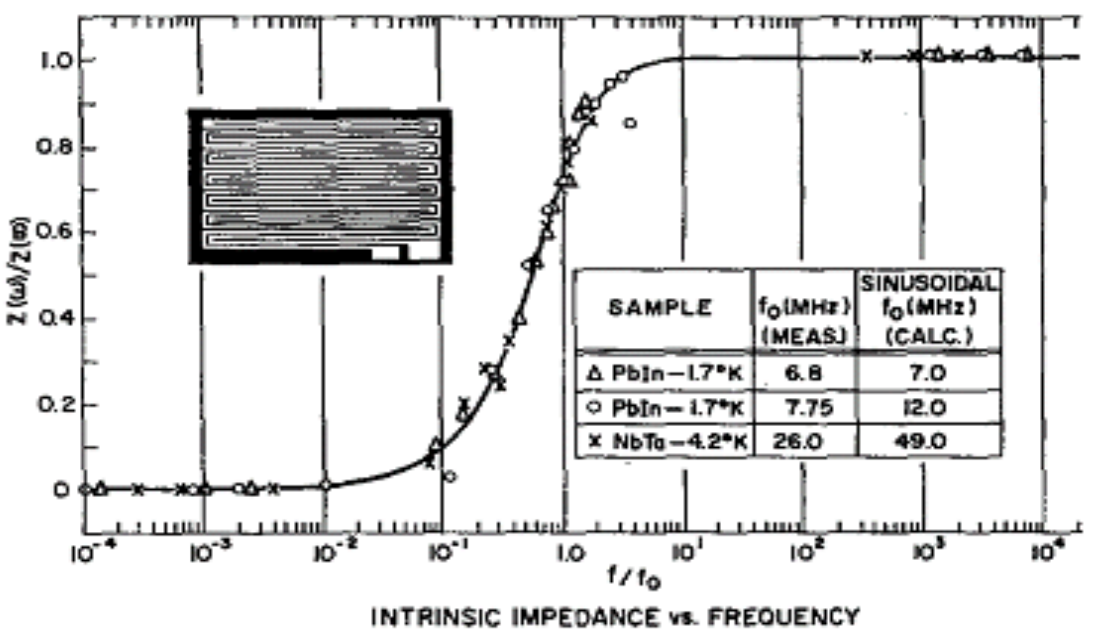

**Fig. 9.** Frequency dependence of intrinsic impedance of LTS films (after [458]).

In 1968, the nonlinear effects in *tin* superconducting films were researched by the pulse method in microwave power range of 100 $\mu W$ – 100 *W* at frequency of 9.360 *GHz* by Soderman, Rose [459]. The same year, Turneaure, Weissman measured the surface resistance of *Nb* superconductors in a cavity resonator at $TE_{011}$ mode at frequency of 11.2 *GHz*, and at $TM_{010}$ mode at 8.4 *GHz* in the strong *RF* magnetic fields up to 436 *Oe* [460].

In 1970, Halbritter presented the detailed analytical review on the microwave superconductivity researches conducted by various research groups in the World. In the perturbation theory of nonlinearities, Halbritter analysed the impedance $Z(h^2)$ in the power series, where $h^2 = H^2/H_C^2$, and used his theory to explain the impedance nonlinearities in [469].

The dependence of surface resistance $R_S$ on the external magnetic field $H_e$ in other *Type II* superconductors such as the *Ti-V* alloys at temperature 7*K* was investigated by Hackett, Maxwell, Kim [471] in Fig. 10 [471].

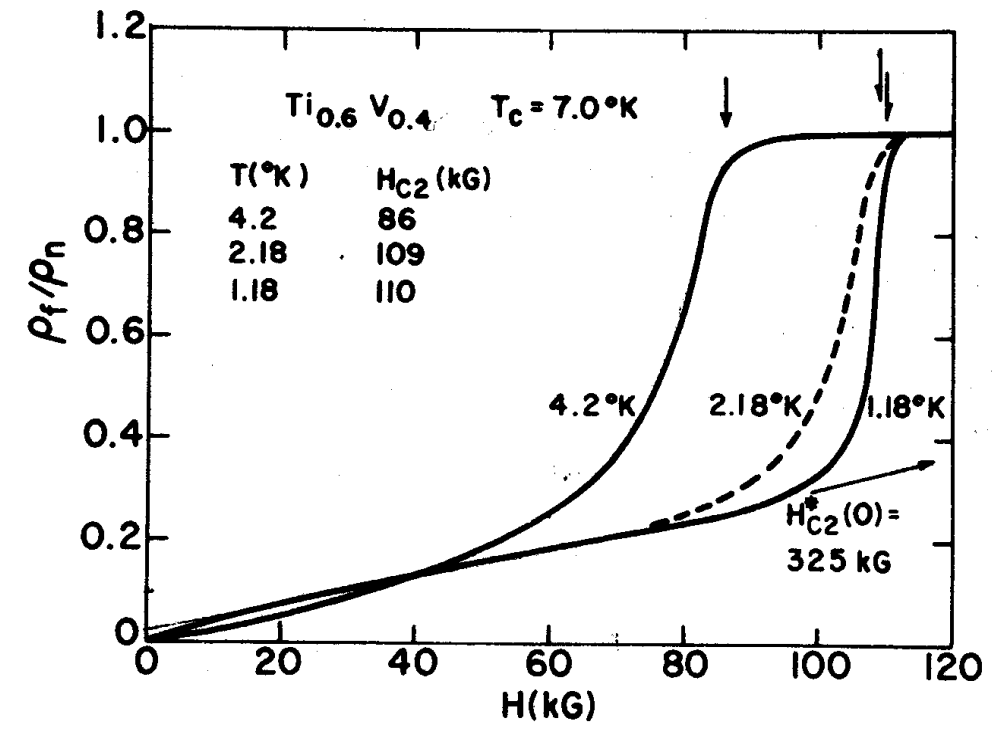

**Fig. 10.** Dependence of relative surface resistance (14.4 *GHz*) of the $Ti_{0.6}V_{0.4}$ superconductor at 7*K* on the external magnetic field (after [471]).

As it is shown, the strong nonlinear behaviour of $R_S(H_e)$ can be observed near the upper critical field $H_{c2}$ in the *TiV* superconductors. As the temperature decreases from 4.2*K* to 1.18*K,* the *S-type transition curve* gets sharper. When the temperature drops from 2.18*K* to 1.18*K,* the upper critical magnetic field $H_{c2}$ increases only by 1% while the transition sharpens considerably. In this paper the magnetic field $H_{rf}$ of electromagnetic wave was substantially smaller than the external field $H_{rf} << H_e$, and $H_{rf}$ did not have any influence on nonlinearity that was completely determined by the external field $H_e$. The dependence of the surface resistance on the magnetic field was linear, when $H_e << H_{c2}$. Lower critical field in this superconductor has magnitude of $H_{c1} < 1kG$, and therefore there are no any visible effects near this field in the graph [471].

In Fig. 11, in the low-temperature superconductor $In_xPb_{1-x}$, where these critical fields are positioned closer to each other, the $R_S(H_e)$ nonlinear effects were observed in two regions $H_e \geq H_{c1}$ and $H_e \leq H_{c2}$, utilising an *RF* resonator with a resonant frequency of about 170 *MHz* in Berezin, Il'ichev, Tulin, Sonin, Tagantsev, Traito [472]. The superconductor samples had a shape close to a disk with the thickness ~ 1 *mm* and diameter ~ 18 *mm*. For this $In_xPb_{1-x}$ alloy, the lower critical field $H_{c1}$ was equal to around 210 *Oe* and the upper critical field $H_{c2}$ was around 3150 *Oe*. It is clearly visible on the graph that the surface impedance *Rs* is proportional to ~ $H_e^{1/2}$ in the range $H_e > H_{c1}$. The nonlinear *S-type behaviour* of the surface resistance $R_S$ can be observed near the critical fields $H_{c1}$ and $H_{c2}$, this effect that has no complete theoretical explanation at present time [472].

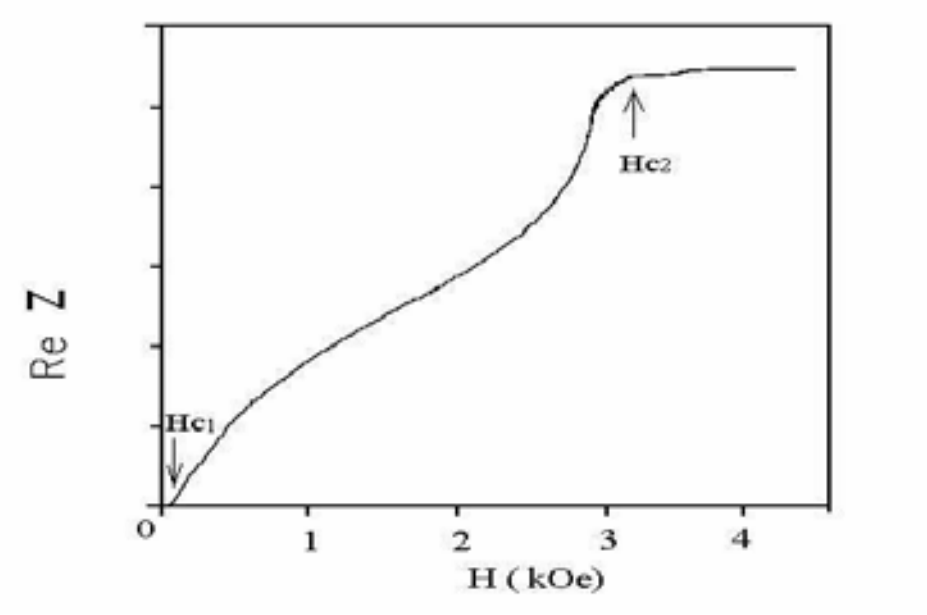

**Fig. 11.** Nonlinear dependence of surface resistance on external magnetic field in $In_xPb_{1-x}$ (x=0.19) (after [472]).

The superconductor samples had a shape close to a disk with the thickness ~ 1 *mm* and diameter ~ 18 *mm*. For this $In_xPb_{1-x}$ alloy, the lower critical field $H_{c1}$ was equal to around 210 *Oe* and the upper critical field $H_{c2}$ was around 3150 *Oe*. It is clearly visible on the graph that the surface impedance *Rs* is proportional to ~ $H_e^{1/2}$ in the range $H_e > H_{c1}$. The nonlinear *S-type behaviour* of the surface resistance $R_S$ can be observed near the critical fields $H_{c1}$ and $H_{c2}$, this effect that has no complete theoretical explanation at present time [472].

Other researchers focused on the understanding of change of amplitude of nonlinear oscillations at resonance frequency; asymmetric change of shape of resonance curve; harmonics and intermodulation generation, nonlinear dependence of surface resistance in *LTS* in Hein, Perpeet, Muller [231].

## 3.8. Nonlinear Surface Resistance $R_s$ in Oxypnictides Superconductors at Ultra High Frequencies.

The iron-based superconductors, including the oxypnictides LaOFeP with Tc=7 *K*, $La[O_{1-x}F_x]FeAs$ with *Tc=26 K*; $NdFeAsO_{1-y}$ with $T_C=54$ *K*; and iron-chalcogenide $FeSe_{0:5}Te_{0:5}$, which take the niche between the *LTS* and *HTS* in terms of critical temperature *Tc* values, were discovered by Japanese physicists in Japan in 2008 in Kamihara, Watanabe, Hirano, Hosono [462], Kito, Eisaki, Iyo [463].

Tab. 8 (a) shows a new class of the high temperature superconducting oxypnictides, including the crystal structure and physical characteristics of iron-based layered superconductors LaOFeP with Tc=7 *K* and $La[O_{1-x}F_x]FeAs$ with *Tc=26 K* in Kamihara, Watanabe, Hirano, Hosono [462], Kito, Eisaki, Iyo [463], Feng [464]. Tohyama [639] discussed the recent progress in iron pnictides.

**1)**

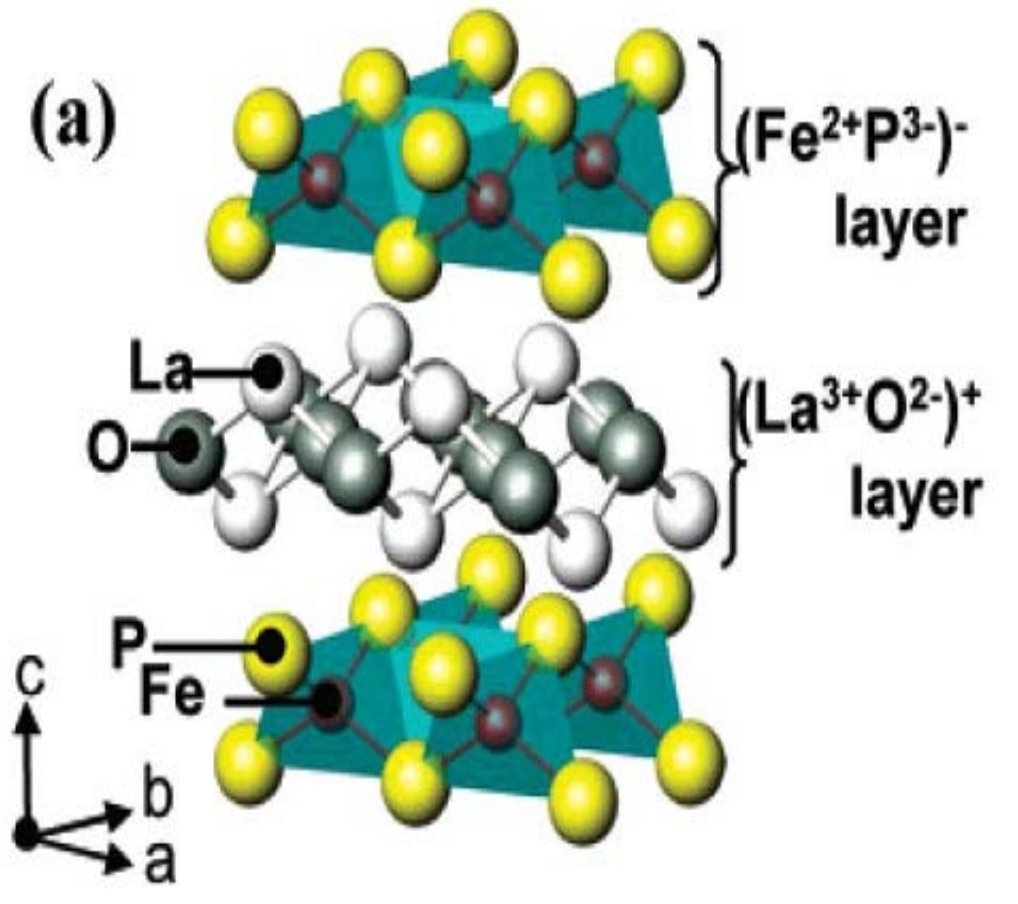

**Fig. 12.** Crystal structure of first discovered high-Tc iron-based layered pure superconductor LaOFeP with Tc = 4 *K*, and F doped LaOFeP with Tc = 7 K (after [462-464]).

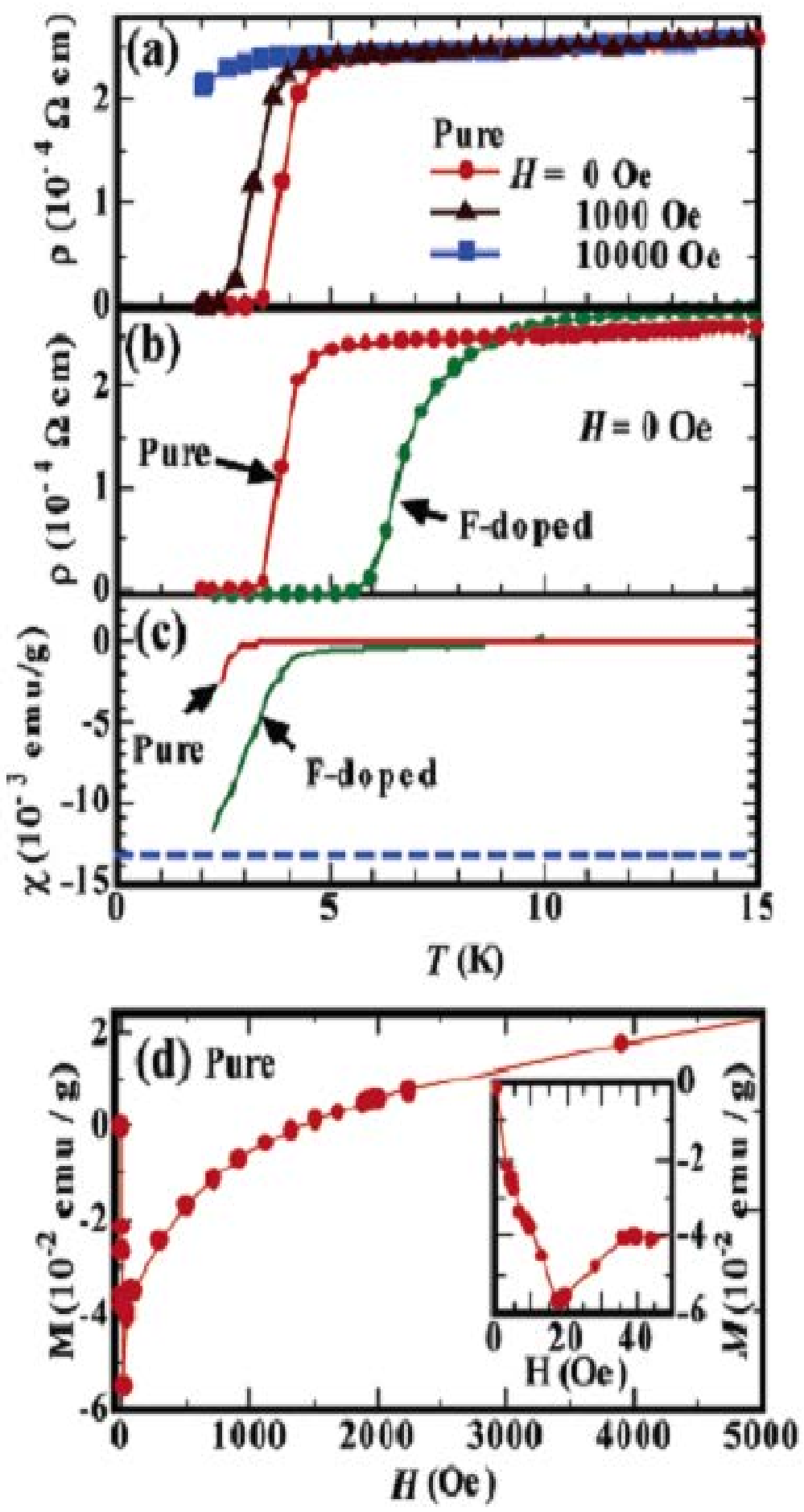

**Fig. 13.** Physical characteristics of first discovered high-Tc iron-based layered superconductor LaOFeP with Tc = 4 *K*, and F doped LaOFeP with Tc = 7 K (after [462-464]).

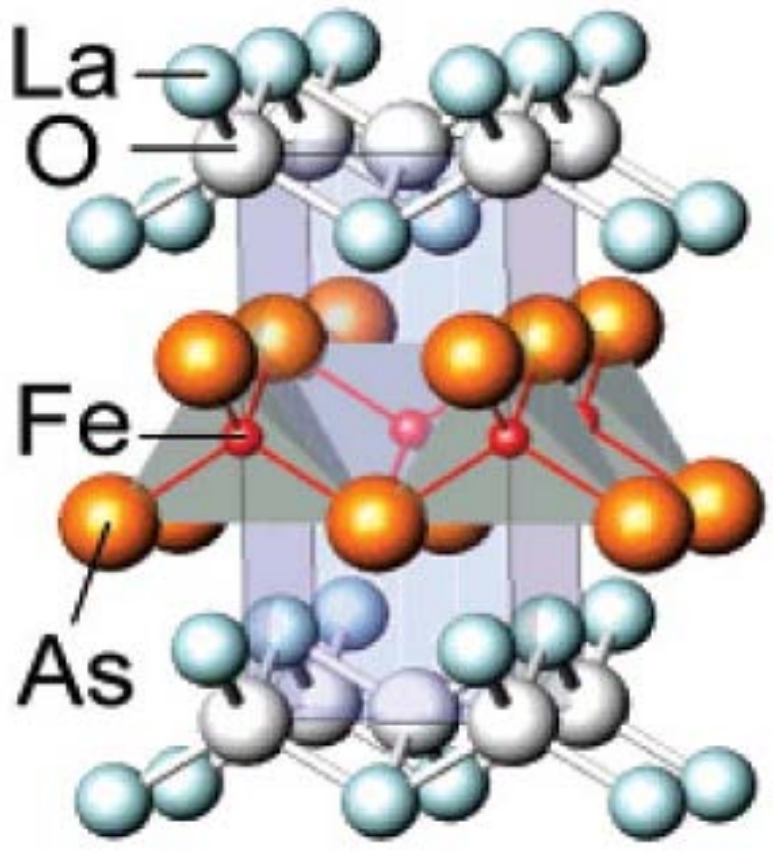

**Fig. 14.** Crystal structure of high-Tc iron-based layered superconductor La[$O_{1-x}F_x$]FeAs with Tc = 26 *K* (after[462-464]).

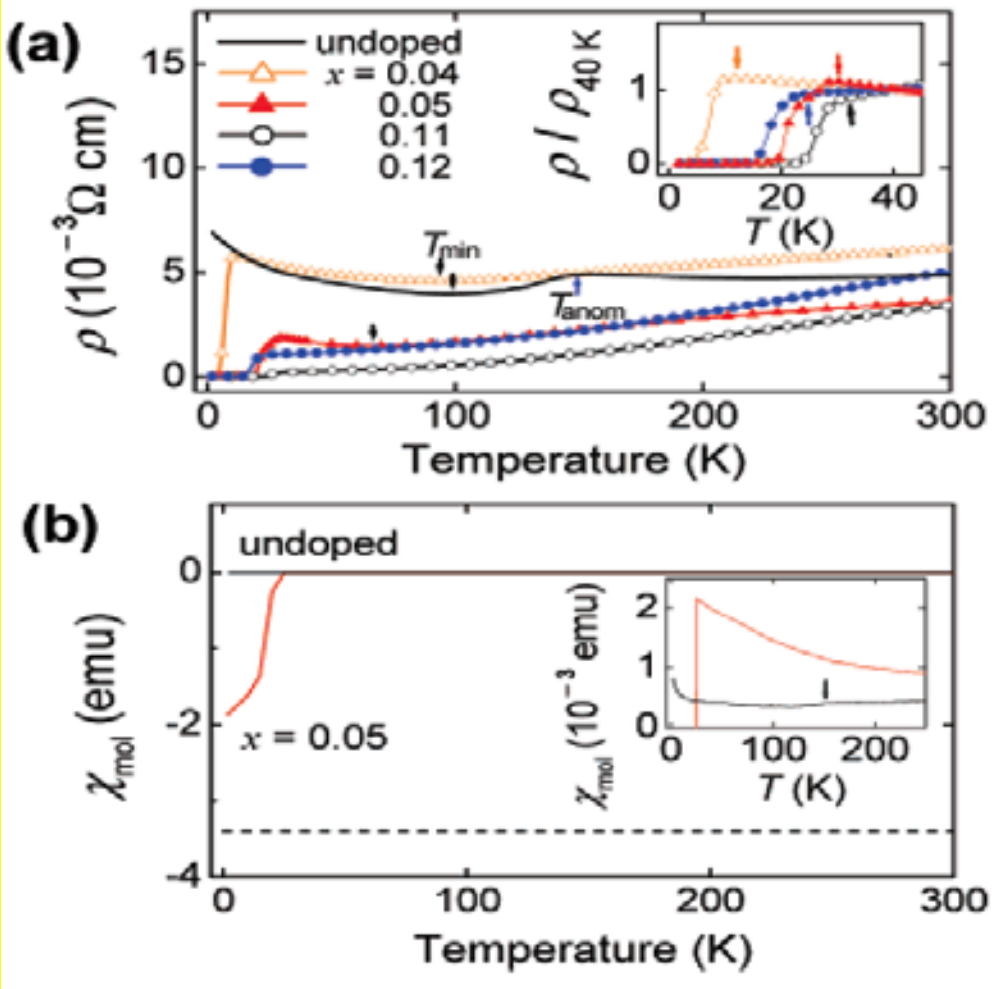

**Fig. 15.** Physical characteristics of high-Tc iron-based layered superconductor La[$O_{1-x}F_x$]FeAs with Tc = 26 *K* (after [462-464]).

| Crystal | $T_c$ (K) | $\lambda_{ab}$ (nm) | $\Delta_1$ (meV) | $\Delta_2$ (meV) | $\Delta_1/k_BT_c$ | $\Delta_2/k_BT_c$ | $x$ |
|---|---|---|---|---|---|---|---|
| #1 | 17.0 | 450 ± 90 | 2.93 | — | 2.00 | — | 1 |
| #2 | 15.6 | 510 ± 90 | 2.97 | 1.65 | 2.21 | 1.23 | 0.739 |
| #3 | 16.3 | 600 ± 90 | 2.98 | 1.10 | 2.12 | 0.785 | 0.896 |

**2)**

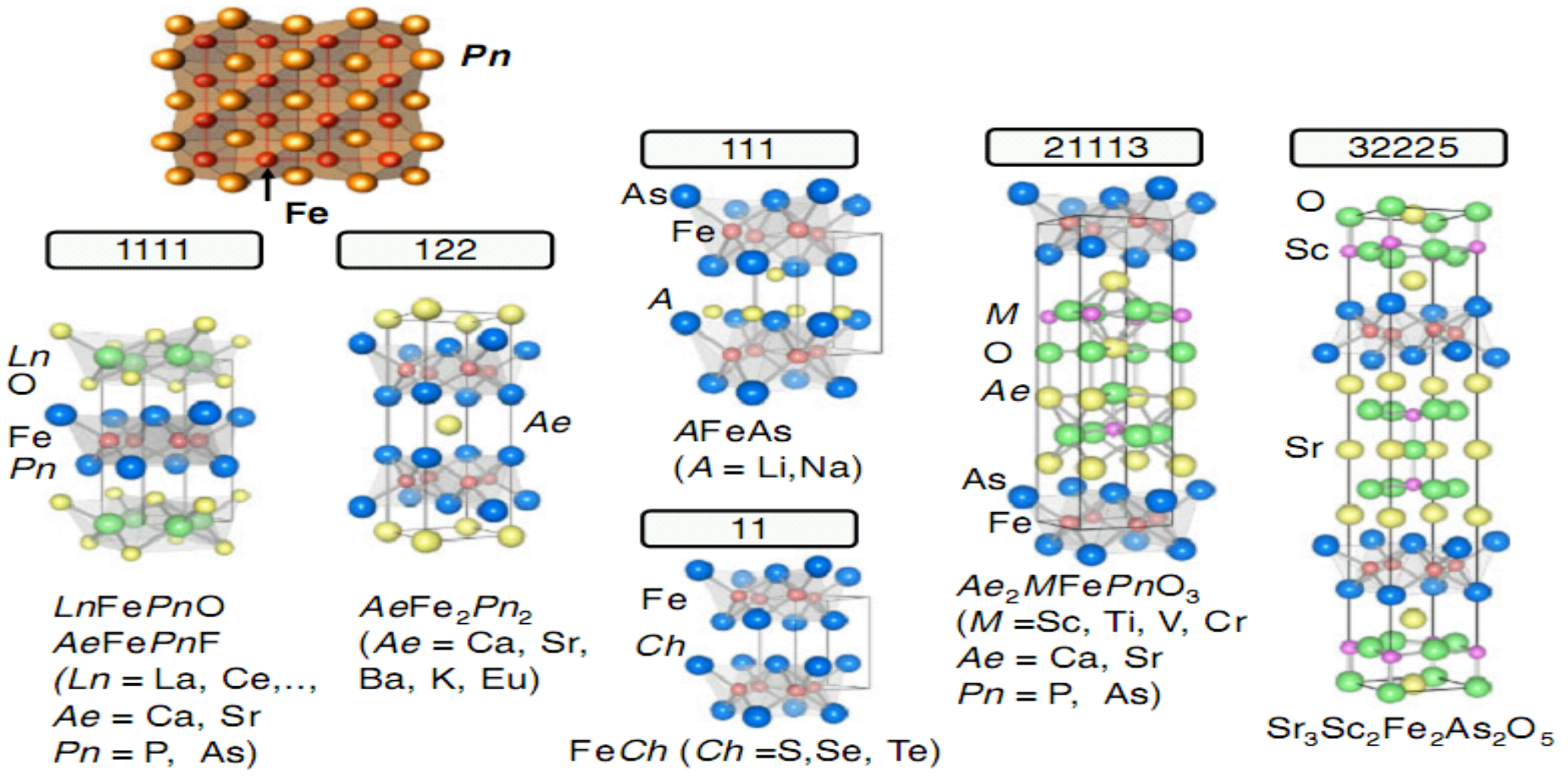

**Fig. 15⁺.** Crystal structures of representative iron-based superconductors, 1111, 122, 111, 11, 21113, 32225 type compounds (after [640]).

**Tab. 8. (a) 1)** A new class of the high temperature superconducting oxypnictides, inclu-ding the high-*Tc* iron-based layered superconductors La[$O_{1-x}F_x$]FeAs with *Tc=26K* (after [462-464]). **2)** Crystal structures of representative iron-based superconductors, 1111, 122, 111, 11, 21113, 32225 type compounds (after [640]).

Tanabe, Hosono [640] wrote an invited review paper and analyzed the frontiers of research on iron-based superconductors toward their application. The considerable efforts are presently directed on the research of high quality thin-film growth techniques for iron-based superconductors, including LiFeAs; oxypnictides La[$O_{1-x}F_x$]FeAs, NdFeAs$O_{1-y}$; and iron-chalcogenide $FeSe_{0:5}Te_{0:5}$ on different substrates, for example: $FeSe_{0:5}Te_{0:5}$ on non-oxide $CaF_2$ (100) in Japan in 2011 in Tsukada, Hanawa, Akiike, Nabeshima, Imai, Ichinose, Komiya, Hikage, Kawaguchi, Ikuta, Maeda [465]. Hanawa, Ichinose, Komiya, Tsukada, Imai, Maeda [641] researched the appropriate substrates for the *Iron Chalcogenide* superconducting thin films deposition. Some research programs are focused on the optimization of microwave properties of high quality iron-based superconductors thin-film NdFeAs(O,F) deposited on different substrates in Japan in 2011 Kawaguchi, Uemura, Ohno, Tabuchi, Ujihara, Takeda, Ikuta [466].

In Tab. 8 (b) shows the nonlinear microwave surface impedance measurements of LiFeAs single crystals made in Japan in 2011 in Imai, Takaheshi, Kitagawa, Matsubayashi, Nakai, Nagai, Uwatok, Machida, Maeda [467].

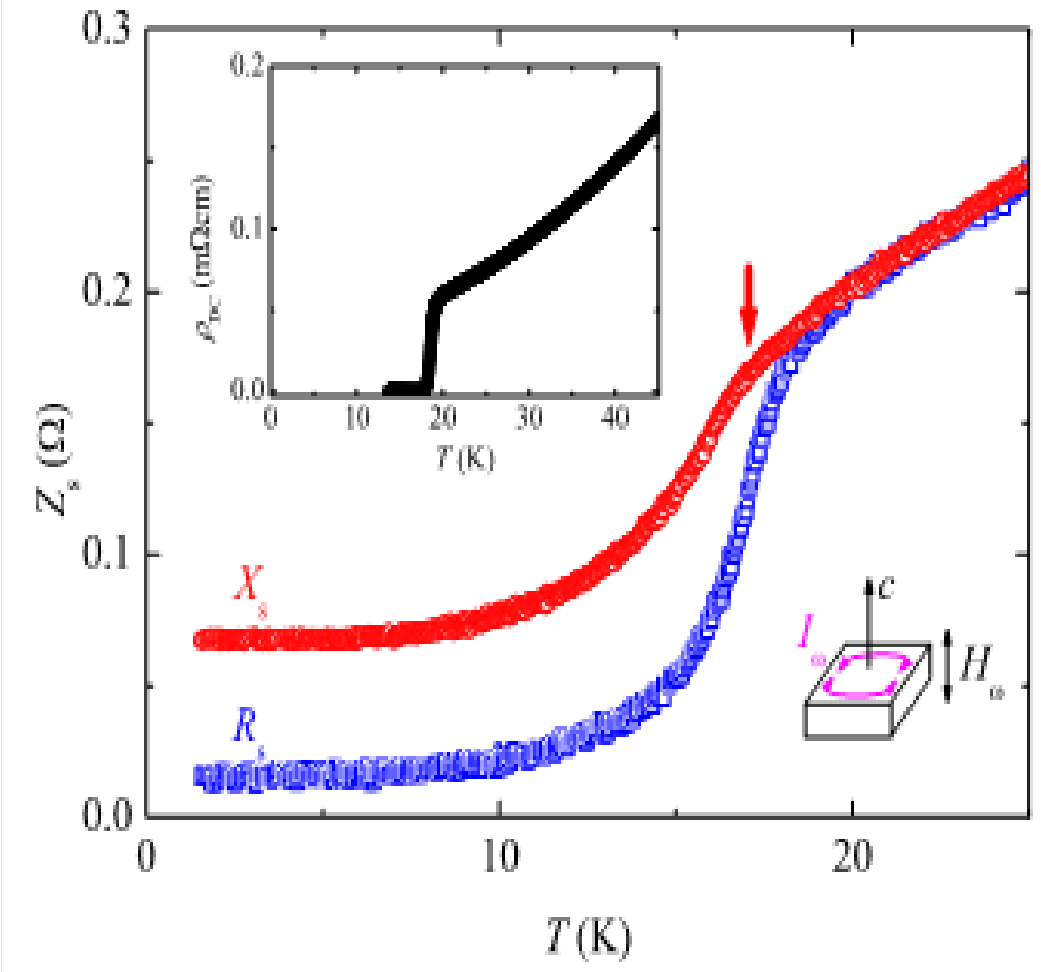

**Fig. 16.** Temperature dependence of *Rs* and *Xs* at 19*GHz* in crystal #1. The large red downward arrow exhibits *Tc* of crystal #1. The inset shows the temperature dependence of the dc resistivity (after [467).

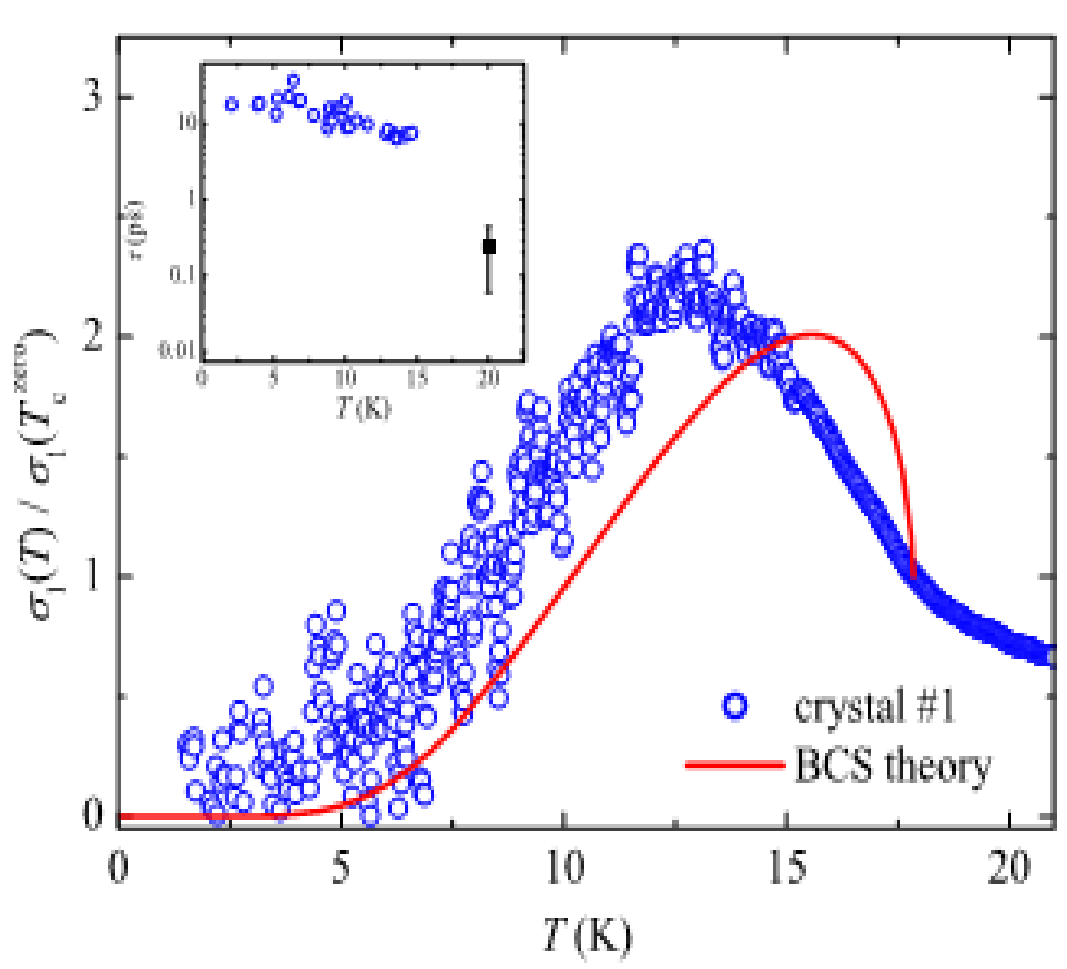

**Fig. 17.** Temperature dependence of normalized quasiparticle conductivity at 19*GHz* for crystal #1. The red solid line is a BCS calculation with $\tau(Tc)=2x10^{-13}s$. The inset shows the temperature dependence of $\tau$. $\tau$ in the superconducting state is calculated by eq. (2). $\tau$ in the normal state, on the other hand, is estimated at about $2x10^{-13}s$ from the data of $\rho_{DC}$ (after [467]).

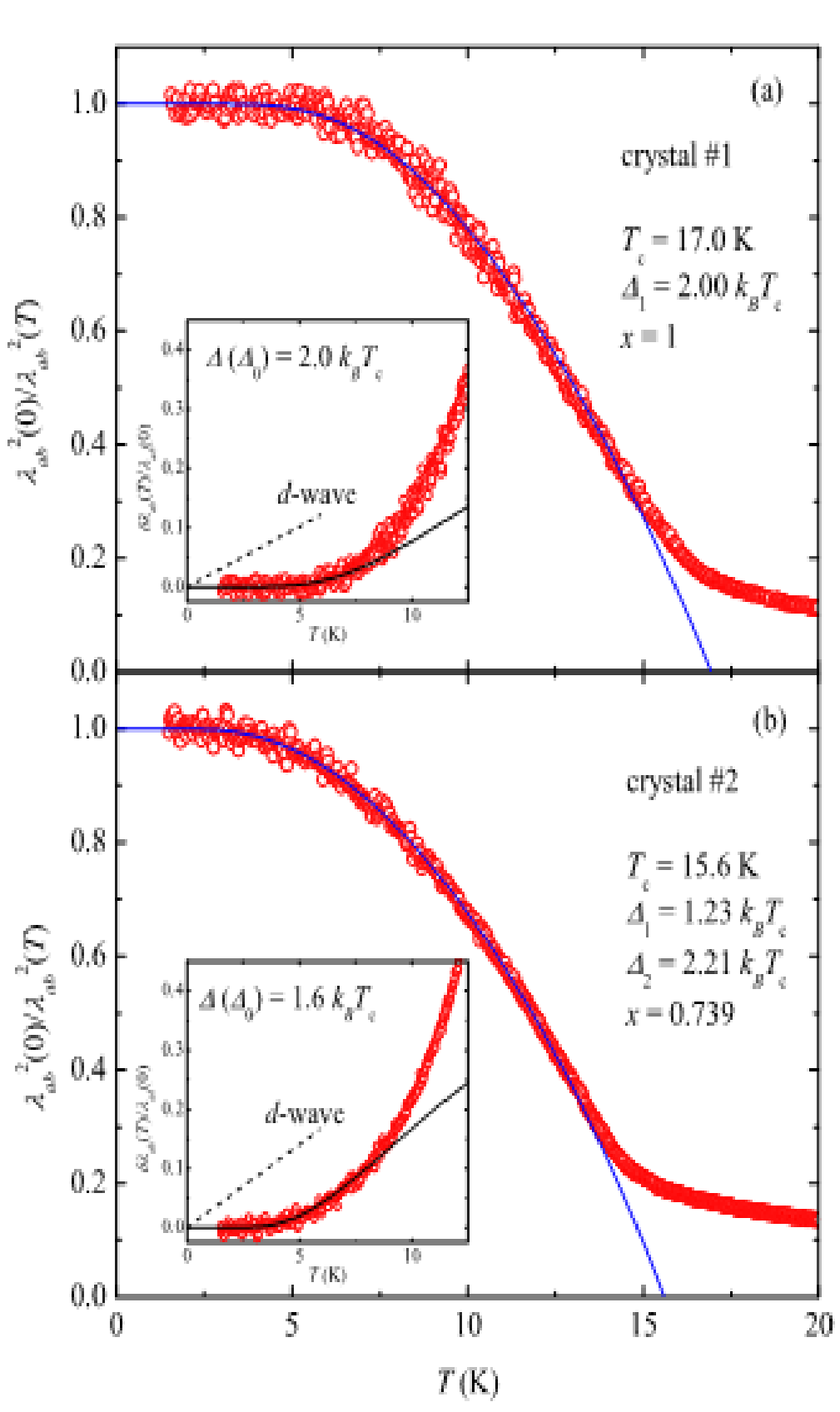

**Fig. 18.** Main panels: $\lambda_{ab}{}^2(0) / \lambda_{ab}{}^2(T)$ for (a) crystal #1 and (b) crystal #2 fitted to the simple two-gap model with (a) $\Delta_1=2.00k_BT_c$ or (b) $\Delta_1=1.23k_BT_c$, $\Delta_2=2.21k_BT_c$ (blue solid line). Above Tc, the normal-state skin-depth contribution gives a finite tail. Insets: $\delta\lambda_{ab}(T) / \lambda_{ab}(0)$ at low temperatures for (a) crystal #1 and (b) crystal #2 fitted to eq. (3) with $\Delta$= (a) $2.0k_BT_c$ or (b) $1.6k_BT_c$ (solid line). The dashed line represents the *T*-linear dependence expected in clean d-wave superconductors with maximum gap $\Delta_0$ = (a) $2.0k_BT_c$ or (b) $1.6k_BT_c$. (after [467]).

**Tab. 8 (b).** Microwave surface impedance measurements of LiFeAs single crystals by Imai, Takaheshi, Kitagawa, Matsubayashi, Nakai, Nagai, Uwatok, Machida, Maeda (after [467]).

## 3.9. Nonlinear Microwave Properties of High Temperature Superconductors.

The nonlinear microwave properties of *HTS* superconductors significantly depend on the temperature *T*, external magnetic field $H_e$, transport current magnitude $J_T$, and some other parameters. For instance, the selected method and technical conditions of *HTS* sample preparation may result in different number of impurities, varying number of domains with weak links, varying oxygen contents, which will have a considerable impact on the nonlinear microwave properties of *HTS* superconductors in Belk [468].

At ultra high frequency electromagnetic fields of high intensity, the surface resistance *Rs* and reactance *Xs* of superconductors become nonlinear, resulting in complex dependence of the surface resistance *Rs* and reactance *Xs* on microwave power of electromagnetic wave $R_S(P_{rf})$, which can change with the increase of temperature, changes of signal frequency, microwave power, *HTS* sample geometry. The $R_S(P_{rf})$ can also depend on the type of microwave resonator, which accommodates the *HTS* sample at microwaves. The intensity of external magnetic field $H_{ext}$ may have a significant influence on the nonlinear characteristics of *HTS* at microwaves. In addition, in the case of *HTS* thin films, the dielectric substrate, which accommodates the *HTS* thin film, must be characterized accurately, because it may also have a considerable influence on the physical characteristics of *HTS* thin films at microwaves in Takken [469]. Tab. 5 shows the characteristics of substrates for the use in *HTS* microwave filters in Zhao Xinjie, Li Lin, Lei Chong, Tian Yongjun [27].

| Material | Lattice constant/nm | Dielectric constant | $\tan\delta \times 10^{-5}$ | $f$/GHz | $T$/K |
|---|---|---|---|---|---|
| $\gamma$-$Al_2O_3$ | $a = 0.4758$ | $\varepsilon_{ab} = 11.5$ | 30 | | 300 |
| | $c = 1.299$ | $\varepsilon_c = 9.4$ | 0.15 | 9 | 77 |
| $LaAlO_3$ | $a = 0.3788$ | 24—25 | 30 | 10 | 300 |
| | | | 0.76 | 10 | 77 |
| | | | 0.30 | 10 | 10 |
| MgO | $a = 0.42$ | 9.5—10 | 0.6—910 | 10 | 4.2—300 |

**Tab. 9.** Substrates for the use in *HTS* microwave filters (after [27]).

Let us review the main research results on the nature of nonlinearities in *HTS* thin films at microwaves, obtained by different groups, in details.

Kong [470], Xia, Kong, Shin [471] distinguish the following types of nonlinearities in *HTS* superconductors:

1. ***Granular nonlinearity***: it occurs when the current *I* is above *Ic* (here *Ic* is the threshold current of the grain junctions). It is similar to a p-n junction's exponential *I-V* relation.
2. ***Vortex nonlinearity***: the vortex motion in the mixing state in type II superconductors (between the superconducting and the normal states) can cause a nonlinear *V - I* relation, when current density *J* is slightly greater than a critical current density *Jc*.
3. ***Intrinsic nonlinearity***: This is due to the dependence of the super-electron number density $n_s$ on the applied current density *J* resulting in the nonlinear voltage-current (*V - I*) relations and the nonlinear inductance-current (*L - I*) relations. The non-uniform distribution of current density, because of both the finite dimensions of superconductor and microstrip geometry as well as the non-uniform heating of thin film will certainly impact the character of $n_s(J)$ and *J(T)* dependences.

M. S. Dresselhaus, G. Dresselhaus [472] conducted the research program on the non-linear microwave and fast optics studies of $YBa_2Cu_3O_{7-\delta}$ superconducting films with accent on the following types of nonlinearities in *HTS* at microwaves:

1. ***Grain boundaries nonlinearity***: the effect of the angular dependence of the grain boundaries on the linear and non-linear microwave impedance of $YBa_2Cu_3O_{7-\delta}$ thin films.
2. ***Weak links nonlinearity***: the impact of coupled grain and distributed properties of long Josephson junctions (distributed resistively shunted junction model) on the non-linear microwave impedance of $YBa_2Cu_3O_{7-\delta}$ thin films.
3. ***Vortex nonlinearity***: the Josephson vortices influence on the non-linear microwave impedance of $YBa_2Cu_3O_{7-\delta}$ thin films. The impact of improved pinning of magnetic vortices in $YBa_2Cu_3O_{7-\delta}$ thin films on microwave properties of $YBa_2Cu_3O_{7-\delta}$ thin films.
4. ***Patterning nonlinearity***: the patterned edge effect on the non-linear microwave impedance of $YBa_2Cu_3O_{7-\delta}$ thin films (*Note*: the prevailing opinion is that the patterning has no significant influence on the nonlinearities.).

Dresselhaus, Oates, Sridhar [473] conducted the research program: "The studies of high-$T_c$ superconducting films for high power microwave applications," and concentrated on the following five types of sources of nonlinearities:

1.***Grain boundaries nonlinearity***:

a)Measurements and modelling of the nonlinear microwave properties of thin films of high-Tc materials in order to understand the loss mechanism and the nonlinear mechanisms.

b)Measurements of the nonlinear properties of grain boundaries as a function of angle in order to relate to films with defects. Authors of report [86] state that the measurements demonstrate that the low-angle grain boundaries ($\theta$<10°) have little effect on the RF power handling capabilities, while the high-angle grain boundaries ($\theta$=24°) cause large nonlinear losses due to Josephson vortices created by RF currents.

c)Measurements of linear and non-linear microwave impedance of the grain boundaries in an applied *DC* magnetic field in order to understand the vortex dynamics in the grain boundaries.

d)Measurements and modelling of the intermodulation distortion (*IMD*) in grain boundaries.

2.***Vortex nonlinearity***:

Measurements of linear and non-linear microwave impedance of the grain boundaries in an applied *DC* magnetic field in order to understand the vortex dynamics in the grain boundaries.

1) In the dielectric resonator, the current flows circularly, and the *RF* magnetic field lies in the plane of the film and points in the radial direction, presumably a direction that is unfavourable for flux penetration.

2) In microstrip resonator, in contrast, the current flows along the length direction of the stripline and the magnetic fields wrap around

the strip, parallel to the film plane at the center but perpendicular at the edges, a direction favouring flux penetration.

*Note:* Authors of report [473] write that *the nonlinear behaviour is the current driven rather than the magnetic field driven. It is still possible that at high values of the RF power and magnetic field, flux penetration still plays a significant role.*

3.***Patterning nonlinearity***:

Comparison between patterned and unpatterned films to understand the role of patterning in the measured nonlinearities. *Note*: Authors of report [473] write that the research on the effect of the patterning process on the nonlinearity of the microwave surface resistance $R_s$ of $YBa_2Cu_3O_{7-\delta}$ thin films has been completed. With the use of a sapphire dielectric resonator and a stripline resonator, the microwave $R_s$ was measured before and after the patterning process as a function of temperature and the RF peak magnetic field in the film. Authors of report [473] conclude that *the experimental and modelled results show that the patterning has no observable effect on the microwave residual $R_s$ or on the power dependence of $R_s$.*

4.***Impurities nonlinearity***:

Measurements of the nonlinear surface impedance of $YBa_2Cu_3O_{7-\delta}$ films doped with *Zn*, *Ni* and *Ca* impurities. Authors of report [473] express the following opinion that the impurity-doped films show slightly lower $R_s$ than the pure films. Overall, however, the reduction is relatively small and there is not a dramatic effect from the inclusion of the impurities in the films.

5.***Oxygen nonlinearities***:

Measurements of the nonlinear surface impedance of $YBa_2Cu_3O_{7-\delta}$ films with varying oxygen content. Authors of report [473] come to the conclusion that the underdoping increases the nonlinear surface resistance $R_s$, but affects the low power linear surface impedance very little, if at all. The optimally doped film and the overdoped film exhibit nearly the same nonlinearity, but the overdoped film shows a slight indication of better performance at high power.

D. E. Oates [474] analyzed the nonlinearities nature in microwave superconductivity stating that the nonlinear effects in superconducting devices are an important limitation to technical applications of these devices. Oates divides all the nonlinearities into the two main categories [474]:

1. ***Extrinsic nonlinearities***, caused by extrinsic effects in superconductors at microwaves: ***grain boundaries nonlinearity; twin boundaries nonlinearity; dislocation nonlinearity; normal phase inclusion nonlinearity; impurities nonlinearity; oxygen doping nonlinearity; possible patterning nonlinearity***.
2. ***Intrinsic nonlinearities***, caused by intrinsic effects in superconductors at microwaves: ***Cooper electron pair breaking nonlinearity***.

Discussing the measurements methods of $Z_s(H_{rf})$, Oates [474] writes that *the* ***nonlinear effects usually appear as changes in the $Z_s$ or as generation of harmonics and intermodulation distortion.*** The two distinct measurement approaches with application of different types of microwave resonators are normally used for accurate microwave characterization of *HTS* thin films at microwaves [474]:

1. ***Measurement methods for unpatterned HTS thin films***: dielectric resonators shielded with superconducting ground planes in Shen, Wilker, Pang, Holstein, Face, Kountz [475] (*TE* mode), parallel plate resonators in Cohen, Cowie, Gallop, Ghosh, Goncharov [476], disc resonators in Kaiser, Aminov, Baumfalk, Cassinese, Chalouoka, Hein, Kilesov, Medelius, Mitller, Perpeet, Piel, Wikborg [477] are used for the precise microwave characterization of *HTS* thin films at microwaves.
2. ***Measurement methods for patterned HTS thin films***: microstrip, coplanar waveguide, stripline resonators with planar transmission line geometries are employed for accurate microwave characterization of *HTS* thin films at microwaves in D. E. Oates, Anderson, Alfredo, Sheen, Ali [478], Sheen, Ali, D. E. Oates, Withers, Kong [479].

Tab. 10 provides the comparative analysis of main research results on the origin of nonlinearities in *HTS* thin films at microwaves [474, 480, 481].

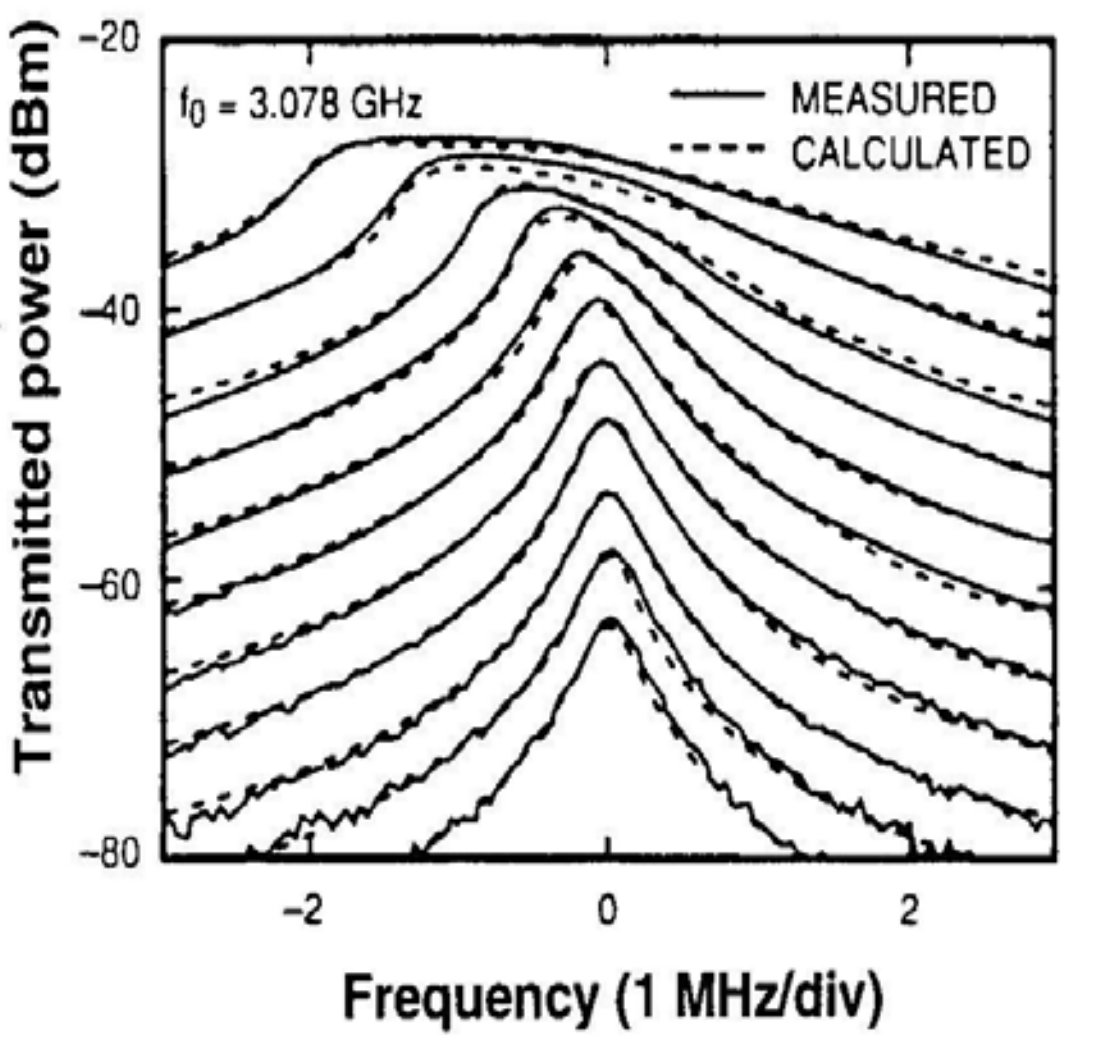

**Fig. 19.** Transmitted power vs. frequency for a $YBa_2Cu_3O_{7-\delta}$ resonator as a function of input power. The maximum power is +30 *dBm*, and the curves are in 5 *dBm* steps. The frequency is 3 *GHz*, and the temperature is 77 *K*. Solid lines are measured data; dashed lines are calculated (after [474, 482]).

The nonlinear effects are strong enough that the resonance curve is no longer Lorentzian, the *Q* can not be determined from the 3-*dB* points of the resonance curve, and another method must be used. The parameters can be extracted from the models [474, 482, 483], or time-domain measurement methods can be used, where the initial decay slope defines the high-power quality factor *Q*. The resonance frequency, the point, where the insertion loss is smallest, can be related to the changes in $\lambda$ [484].

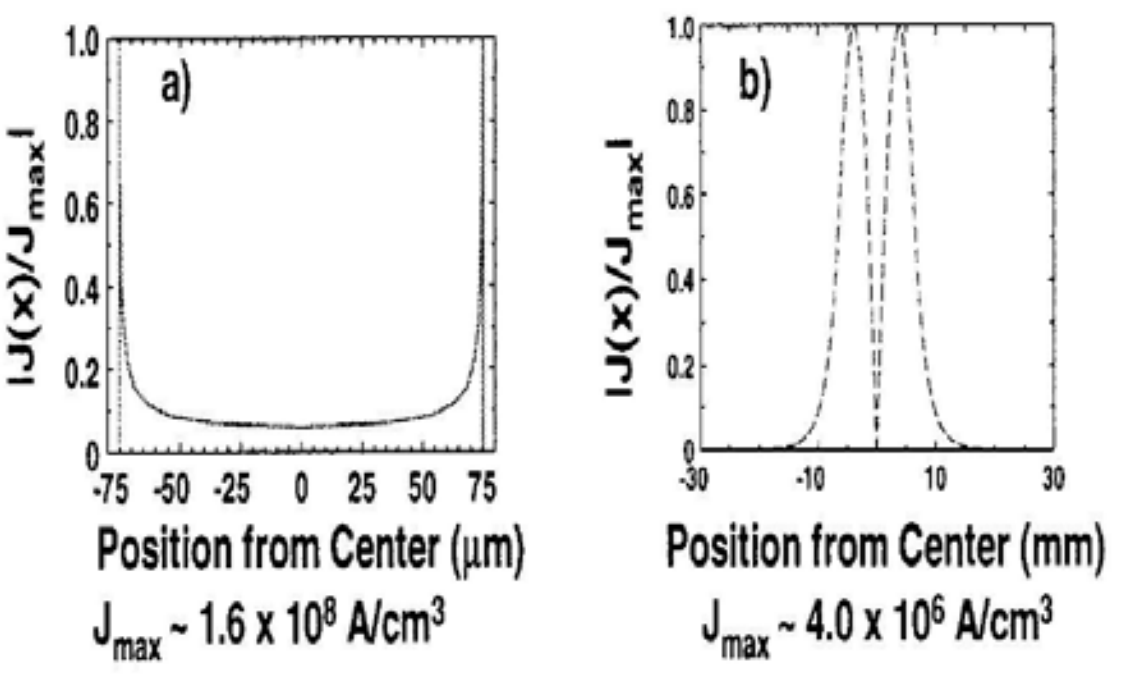

**Fig. 20.** Comparison of current distributions: (a) in a stripline resonator and (b) in a $TE_{010}$ dielectric resonator (after [474]).

The two measurements methods are used for accurate microwave characterization of *HTS* thin films at microwaves [474]:

***1. Measurement methods for unpatterned HTS thin films***: dielectric resonators shielded with superconducting ground planes [485] (*TE* mode), parallel plate resonators [486], disc resonators [487].

***2. Measurement methods for patterned HTS thin films***: microstrip, co-planar waveguide, stripline resonators [488, 489] with planar transmission line geometries

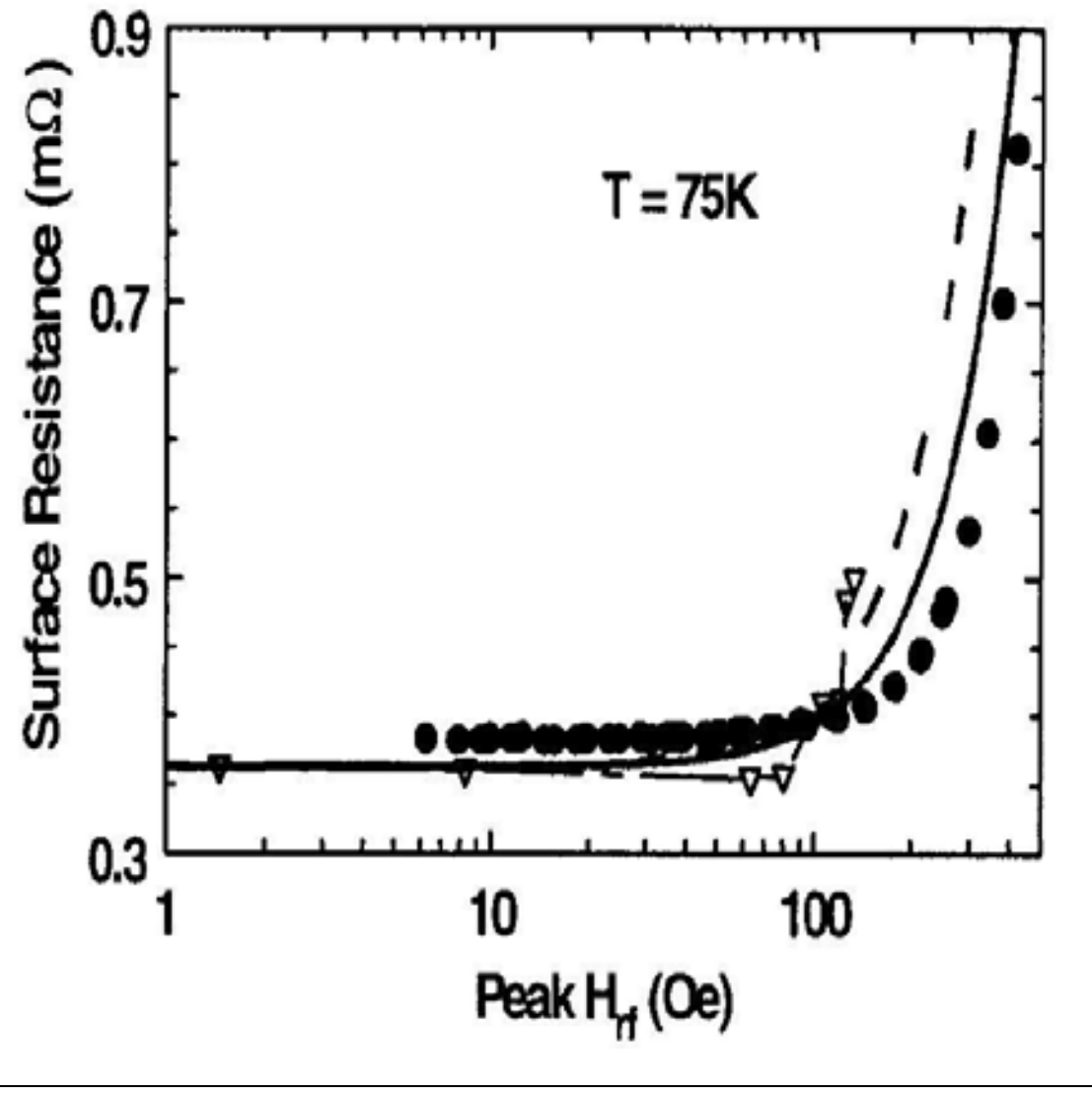

**Fig. 21.** Comparison of patterned and unpatterned films. Points are measured data. Solid and dashed lines are calculated from a model described in the text for the stripline and dielectric resonators respectively (after [474]).

The patterning process did not degrade the film properties [474].

***The observed nonlinearities are due to defects in the material and not due to intrinsic processes*** [490, 491, 492, 493]. This understanding is reconsidered and changed to some extend by D. E. Oates in [480].

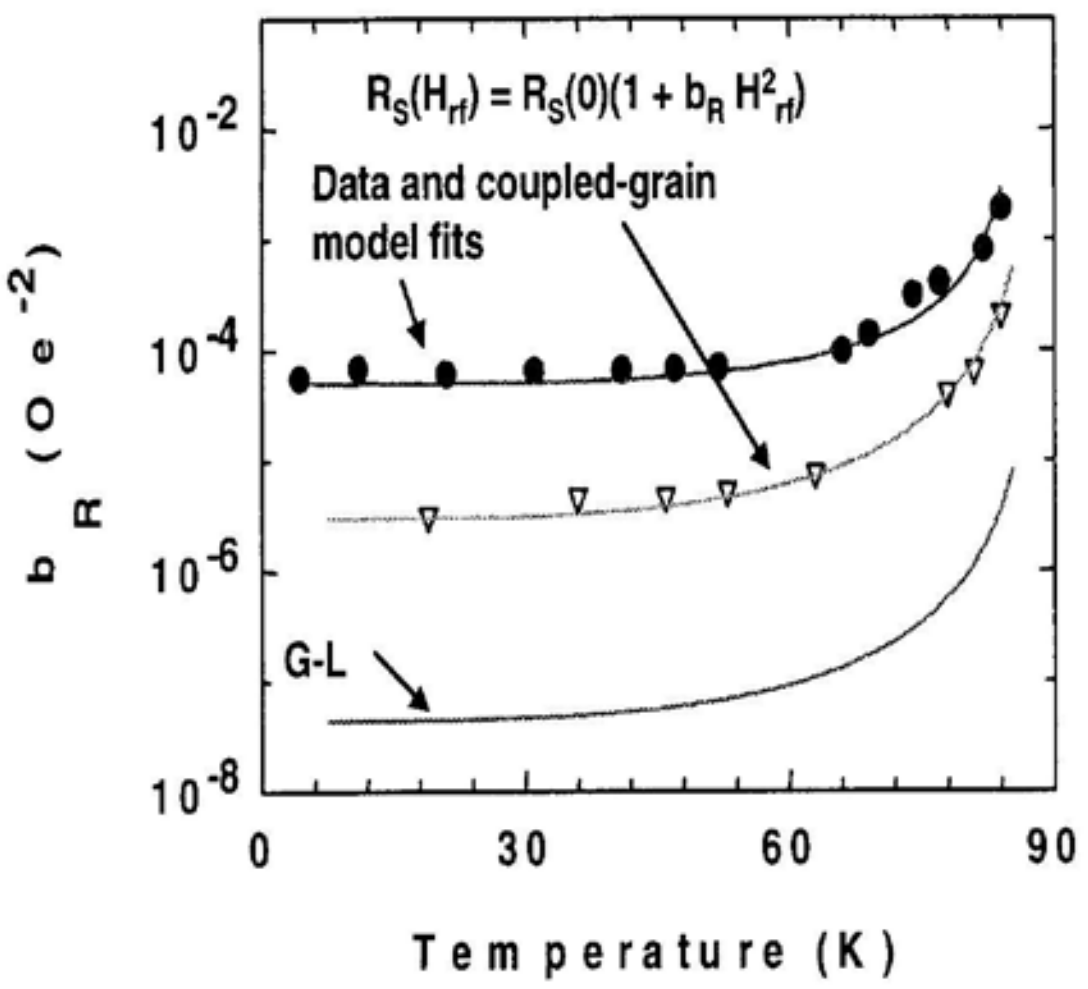

**Fig. 22.** Comparison of *G-L* prediction with simple quadratic fit using coupled-grain model (after [474]).

The *G-L* and nonlinear Meissner theories give predictions that are too small to explain the data. Since the intrinsic nonlinearities are not sufficient to explain the experimental results, most researchers attribute the causes to be various defects in the materials [474]. Examples are grain boundaries that can behave as Josephson junctions, twin boundaries, dislocations, inclusions of normal conducting material, or magnetic impurities [474]. This understanding is reconsidered and changed to some extend in recent research papers by D. E. Oates in [480].

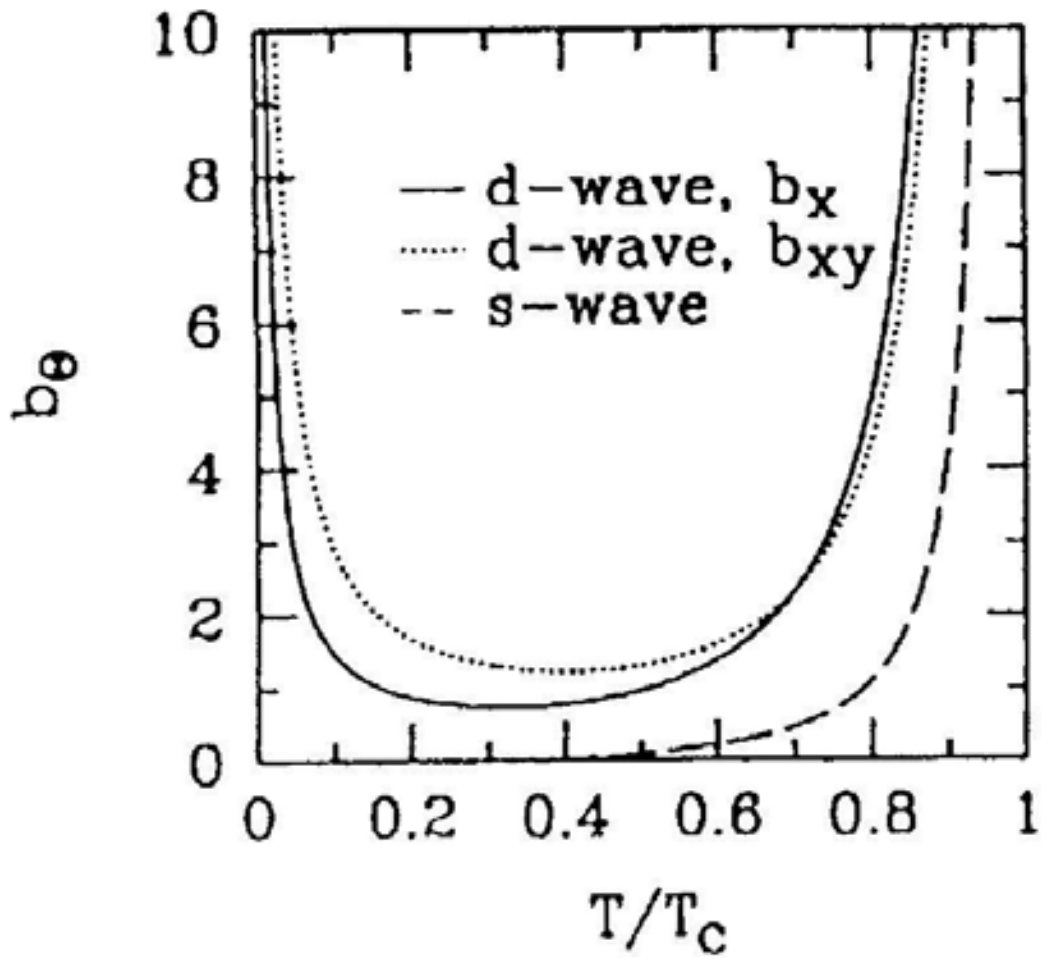

**Fig. 23.** Plot of the calculated [461] values of a temperature-dependent coefficient $b_\theta(T)$, which reflects the order parameter symmetry, for different magnitudes of the order parameter (after[474]).

There are the two possible causes of intrinsic nonlinearity [474]:

***1. Intrinsic nonlinearity*** due to Cooper pair breaking associated with the transport current in *s-wave symmetr*y superconductors described by *G-L* equations [494] in Fig. 22.

***2. Intrinsic nonlinearity*** as a result of the presence of nodes in the *d-wave symmetry* of the order parameter in superconductors, which lead to the enhanced Cooper pair breaking, resulting in the appearance of intrinsic nonlinearity, described by *Dahm-Scalapino theory* [495] in Fig. 23.

It was shown that because of the nodes in the energy gap, *d-wave symmetry* produces larger nonlinearity than *s-wave*, especially at low temperatures. But, the nonlinearities too small to explain the experiments, since experimentally nonlinearities occur at current densities two orders of magnitude smaller than the depairing current $J_c$.

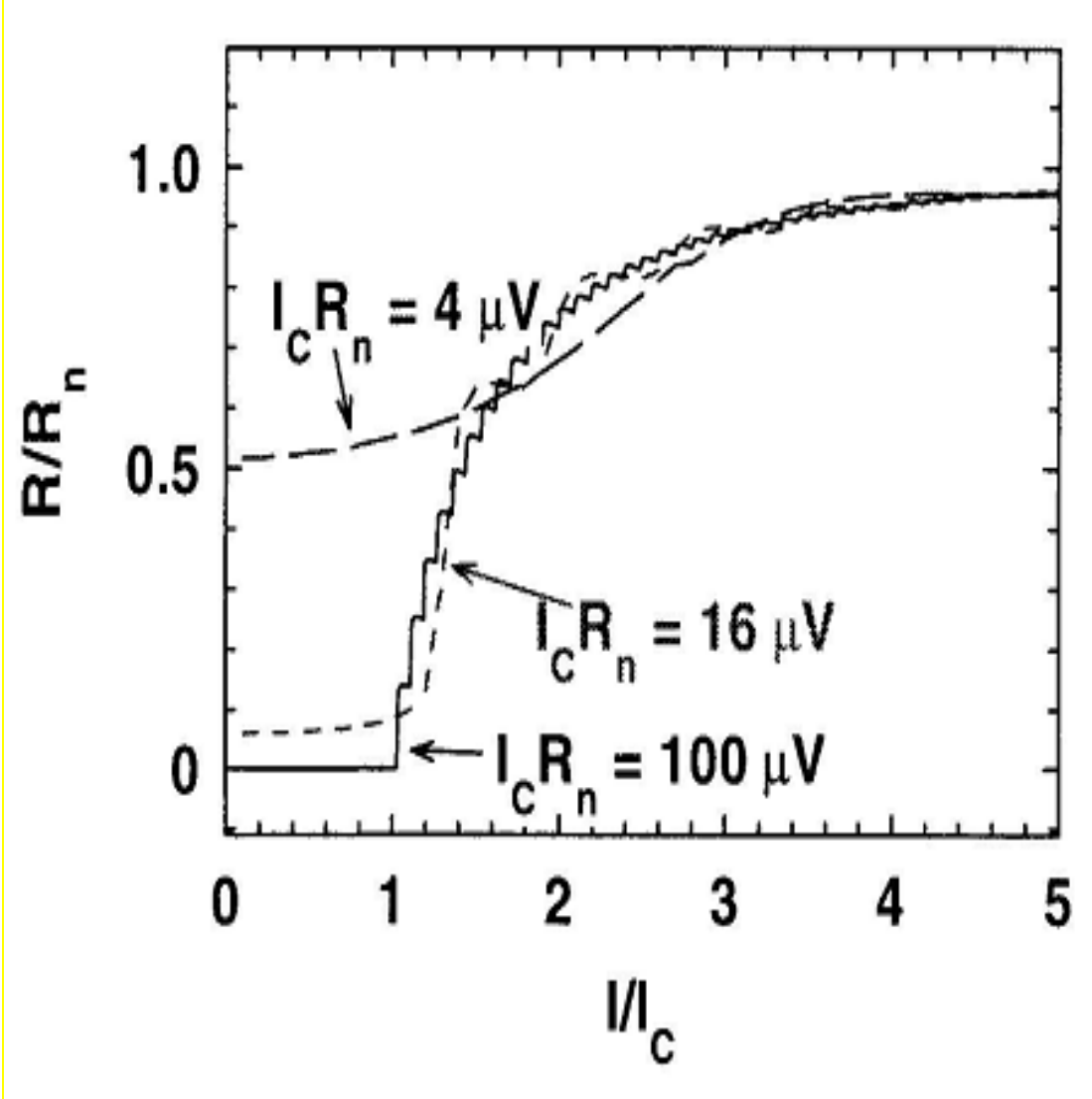

**Fig. 24.** Results of the calculation of the resistance of a resistively shunted Josephson Junction (after [474]).

Considering the ***extrinsic nonlinearities***, Oates [474] focuses on the ***grain boundaries nonlinearity*** modeled by the coupled-grain model mentioning that the coupled-grain model that treats a film as an effective Josephson medium, a network of superconducting grains separated by grain boundaries, which are Josephson-junction (*JJ*) weak links [496, 497] that are responsible for the power dependence.

When the *RF* current approaches $I_c$, the *RF* resistance rises from a low nonzero value at low current. When the *RF* current is many times $I_c$, the resistance asymptotically approaches the $R_n$ of the junction. It is easily seen than that a distribution of $I_c$ and $R_n$ can lead to a steadily increasing $R_s$ as is observed experimentally. In fact, a bimodal distribution of values could account for cases where the $R_s$ shows a plateau in the curve vs. $I_{rf}$.

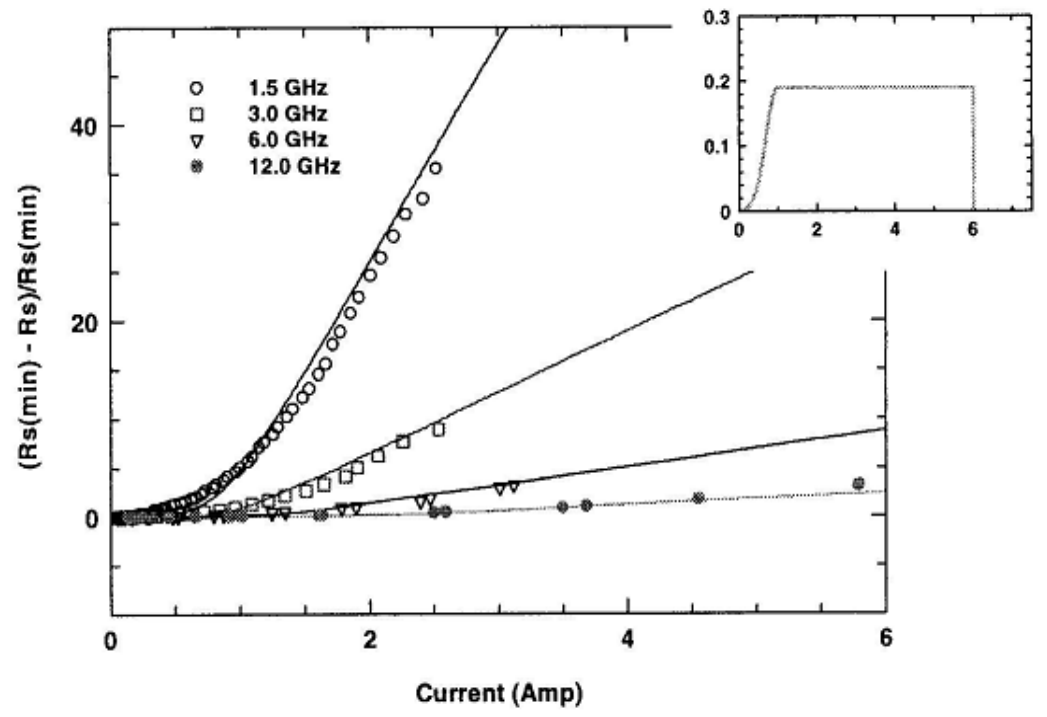

**Fig. 25.** Comparison with experiment of the surface resistance calculated by the extended coupled-grain model. The inset shows *P(Jc)* (after [474]).

Fig. 25 shows a comparison of the extended coupled-grain model with measurements of $R_s$ in a stripline resonator for various frequencies at 75 K. The model demonstrates very good agreement, including the variation in frequency [498].

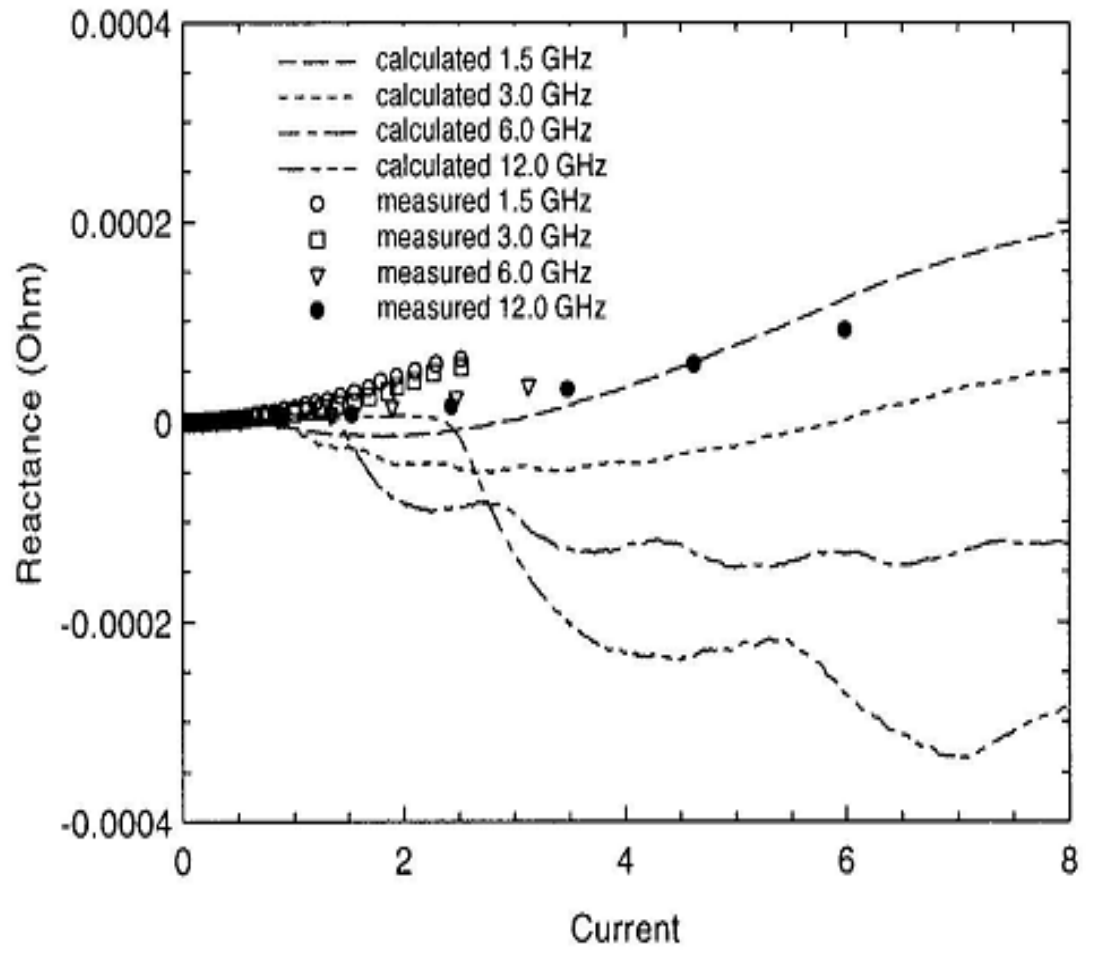

**Fig. 26.** Comparison of the surface reactance calculated by the extended coupled-grain model with experiment. The following parameters were used: $a$ = $10^{-5}$ m, $I_{cmin}$=0.1 A, $I_{c\ max}$ = 6.0 A, $I_{c\ mean}$ = 1.0 A, variance = 0.3 A, and $R_n$ = 2.5x10-5Ω. (after [474]).

The modeled reactance, compared with experiment in Fig. 26, did not show such good agreement, which the authors suggested may be due to a mechanism different from the coupled grains dominating the reactance [498]. This lack of correlation of the resistance and reactance has been observed by others [499].

The observed $Z_s(H_{rf})$ remains largely unexplained. These difficulties to model the $Z_s(H_{rf})$ lead to a possible conclusion that more than one process is responsible for the experimental observations: one process is responsible for the $R_s(H_{rf})$, and another entirely different one is responsible for the $X_s(H_{rf})$ [474].

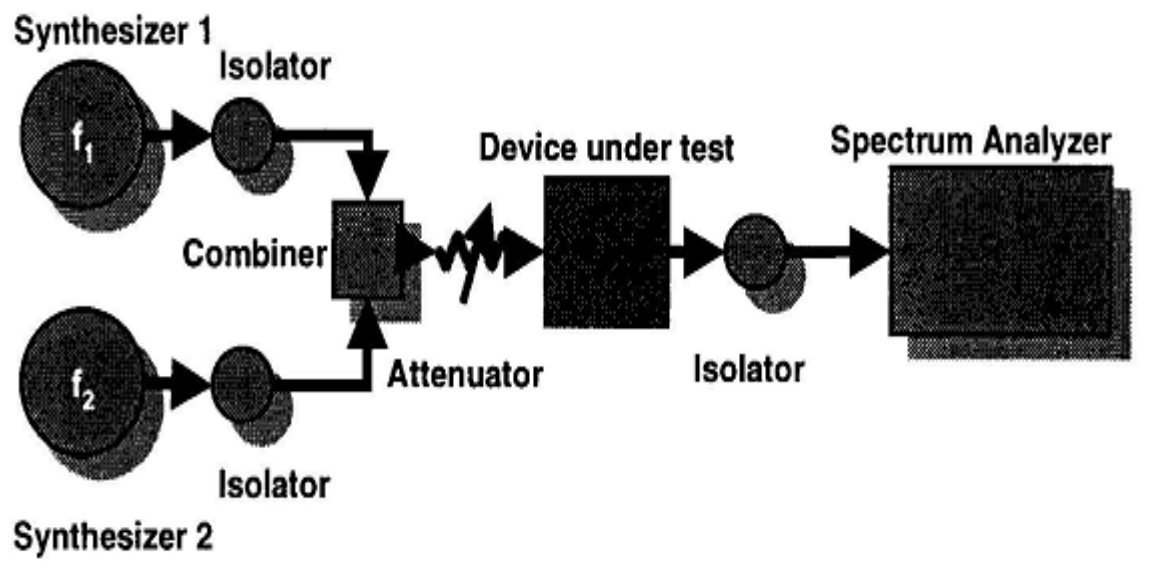

**Fig. 27.** Scheme of the *RF* set up for the method of measurement of intermodulation distortion in a microwave filter (after [474]). D. E. Oates [474] states that ***the harmonic and intermodulation generation are sensitive probes of nonlinear behavior.***

The term "intermodulation distortion" ("IMD") refers to the undesirable mixing of two signals whose mixing products lie within the bandpass. A case in point is mixing of signals lying outside the nominal bandpass with signals lying within the bandpass, thus producing added frequency components that contribute to distortion of the desired signals. IMD arises as a consequence of surface-impedance nonlinearity [500].

1)

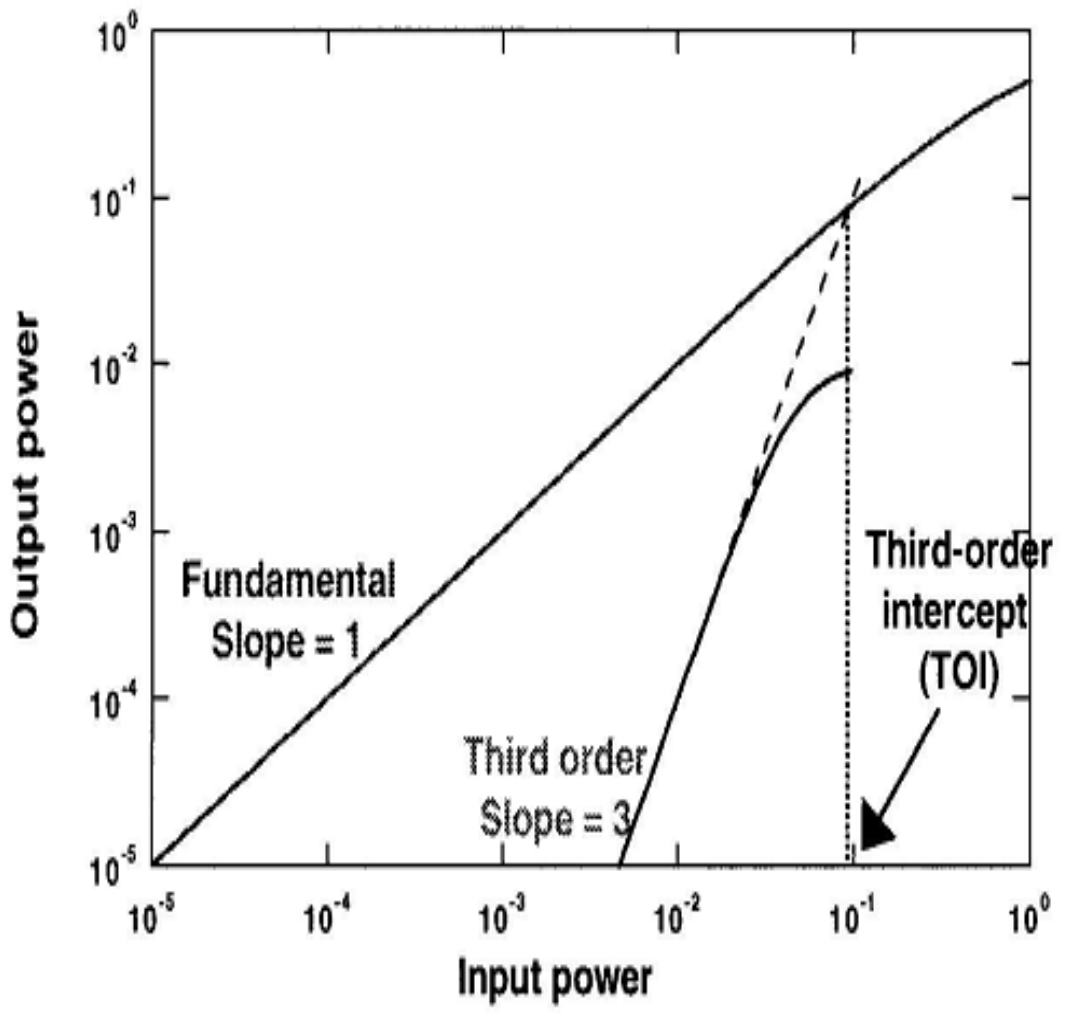

2)

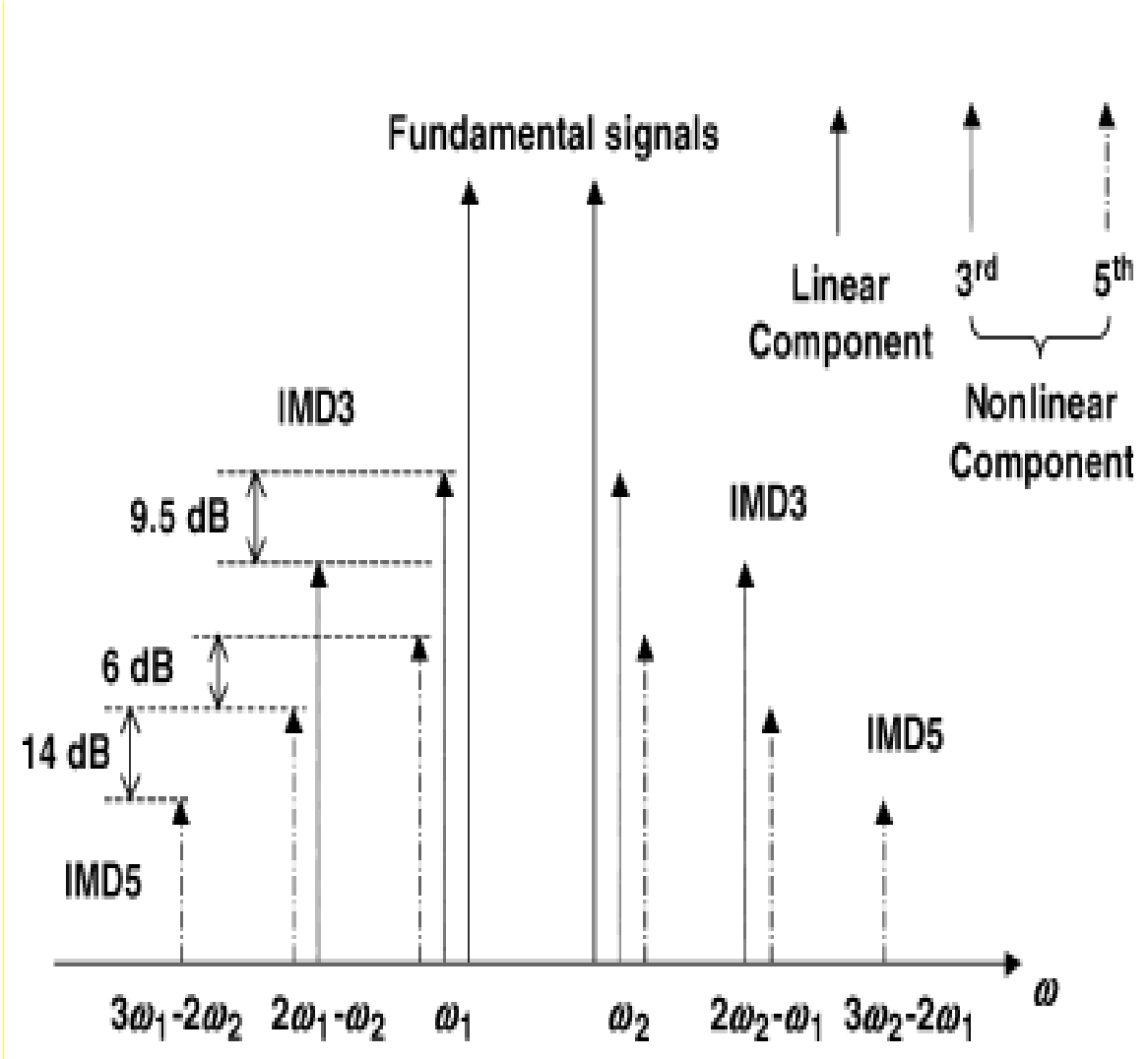

**Fig. 28.** 1) Schematic depiction of the results of the measurement of inter modulation products in a bandpass filter. Shown is the definition of the third-order intercept (after [474]).

2) Output spectrum of the nonlinear model in response to a two-tone fundamental signal represented by three- and fifth-order complex power series. The contributions of each order nonlinear component to fundamental signals, *IMD3*, and *IMD5* are described (after [501]).

***The 3rd order intercept point is derived by plotting the linear power gain and the 3rd order response of a two-tone intermodulation test and graphically interpreting the results. Intercept point is defined by extending the fundamental and third order response, output power vs. input power, in their linear region until they intersect.*** This is a two tone measurement. Third order products change at a rate of 3*dB* increase for every 1*dB* increase from input *RF* power (3:1), while the fundamental changes at a 1:1 rate. The system intercept point can be referred to the input or output of the system [502].

The measurement of *IM* in a high-$T_c$ filter: The output power is plotted against the input power, and it is assumed that the two input power levels are equal. Both the fundamental signal and the *IM* signals are plotted. On this log-log plot, the fundamental signal shows a slope of unity up to the maximum power that the filter can pass. The curve shows saturation effects, when the maximum power is reached. The third-order *IM* signals are expected to increase with slope three on the log-log plot. This is, however, not always the case with superconductors. The deviation from slope of three is discussed in more detail below. The *IM* signals also show saturation at sufficiently high power levels, in part due to energy conservation [474].

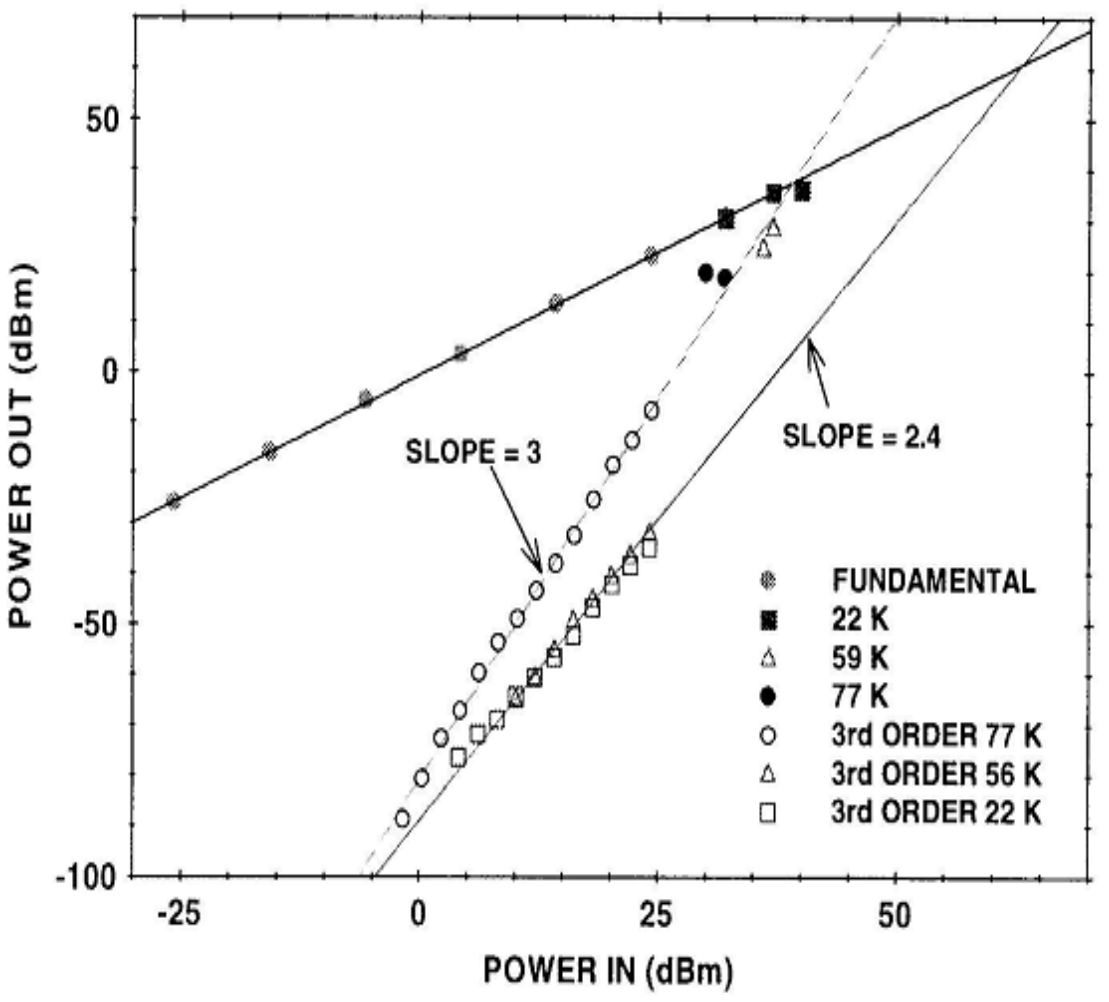

**Fig. 29.** Actual measurements of *IM* in a high-$T_c$ filter. This is a 5-pole filler, fabricated from *YBCO* deposited on $LaAlO_3$, designed with $10\Omega$ 1*mm* wide resonators for high-power-handling capability, and (after [474]).

This is a typical result for a filter and shows many of the features that are seen in general in high-$T_c$ filters and resonators. The *IM* signal shows a slope of three at high temperature, as expected, but at low temperatures the slope is closer to two. Slopes different from three are found often in *IM* measurements in high-$T_c$ devices. Usually the slope is less than three, but slopes greater than three also have been found. The measurable *IM* signals appear in the region of power where the filter behaves in a linear fashion at the fundamental frequency. That is, *IM* is visible and of sufficient magnitude that filter performance could be degraded in the region of power when the fundamental shows a slope of one. ***The IM is a more sensitive measure of nonlinearity than the measurements of loss, because in this region filter loss has not changed [474].***

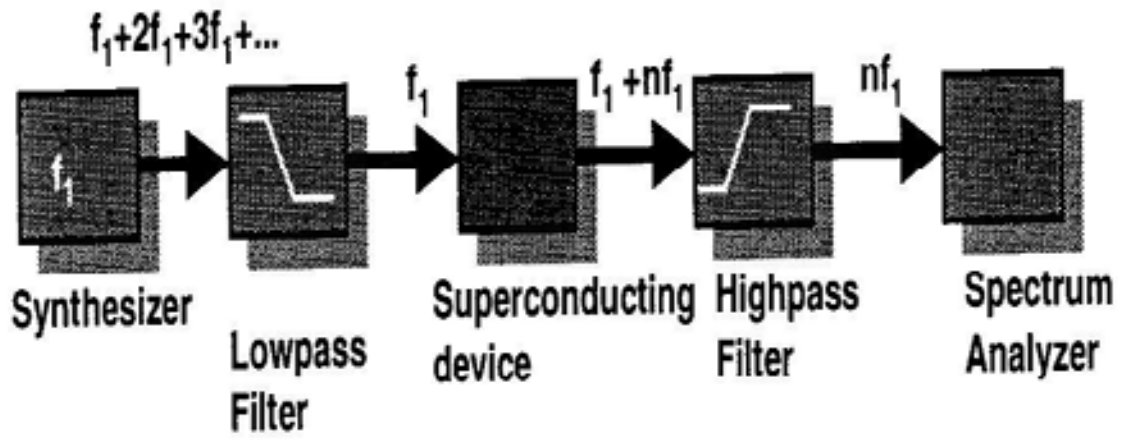

**Fig. 30.** Block diagram of third-harmonic measurement (after [474]).

The third-harmonic measurements are a useful measure of nonlinearity, and while not as important practically as third-order *IM*, third-harmonic signals presumably arise from the same mechanism, and thus are a useful characterization tool. Their measurement, moreover, is simpler than *IM* measurements, thus providing an attractive alternative [474]. Only one signal source is needed, but synthesizers are usually rich in harmonics, so a low-pass filter is used to remove them. However, with the high-pass filter at the output of the device, signal levels at the spectrum analyzer are low, so a larger dynamic range is possible for this measurement method compared to *IM* measurements [474].

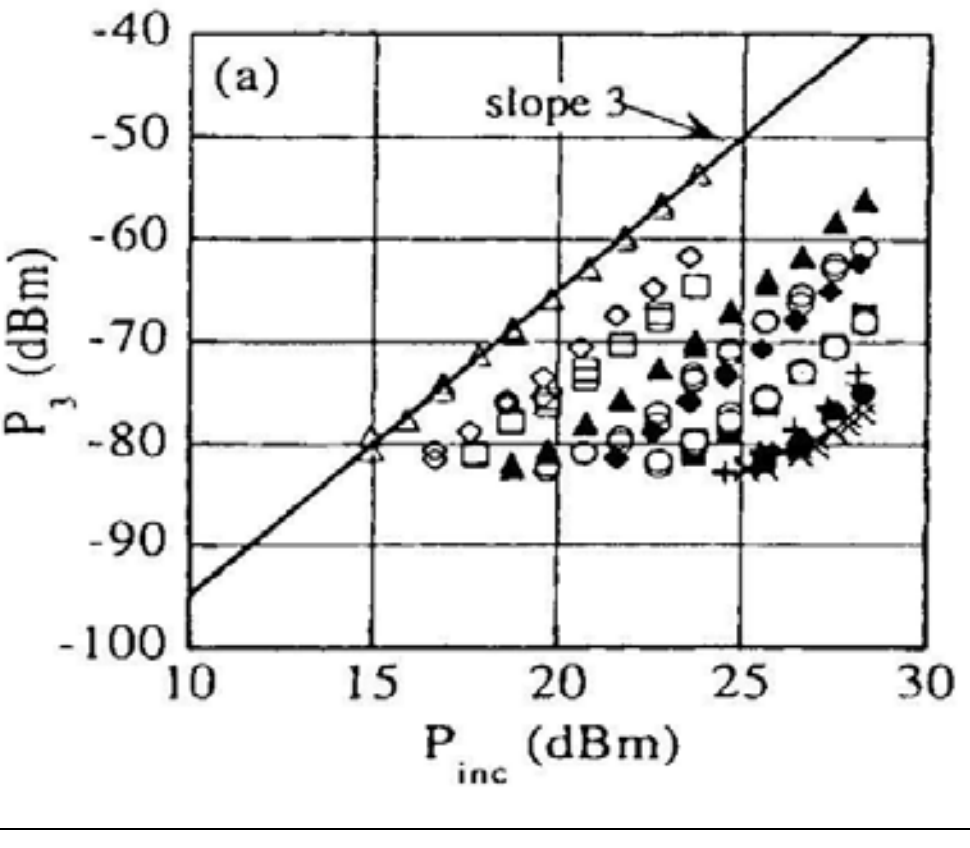

**Fig. 31.** Measurements of third harmonic in a coplanar delay line in Booth, Beall, Rudman, Vale, Ono [503] (after [474]).

The fundamental frequency in this case is 4 *GHz*, and the measurements are for *YBCO* on *LAO*. The figure shows results of measurements for several different delay lines of differing center conductor width and length in order to understand the effects of coplanar delay line geometry [503, 474].

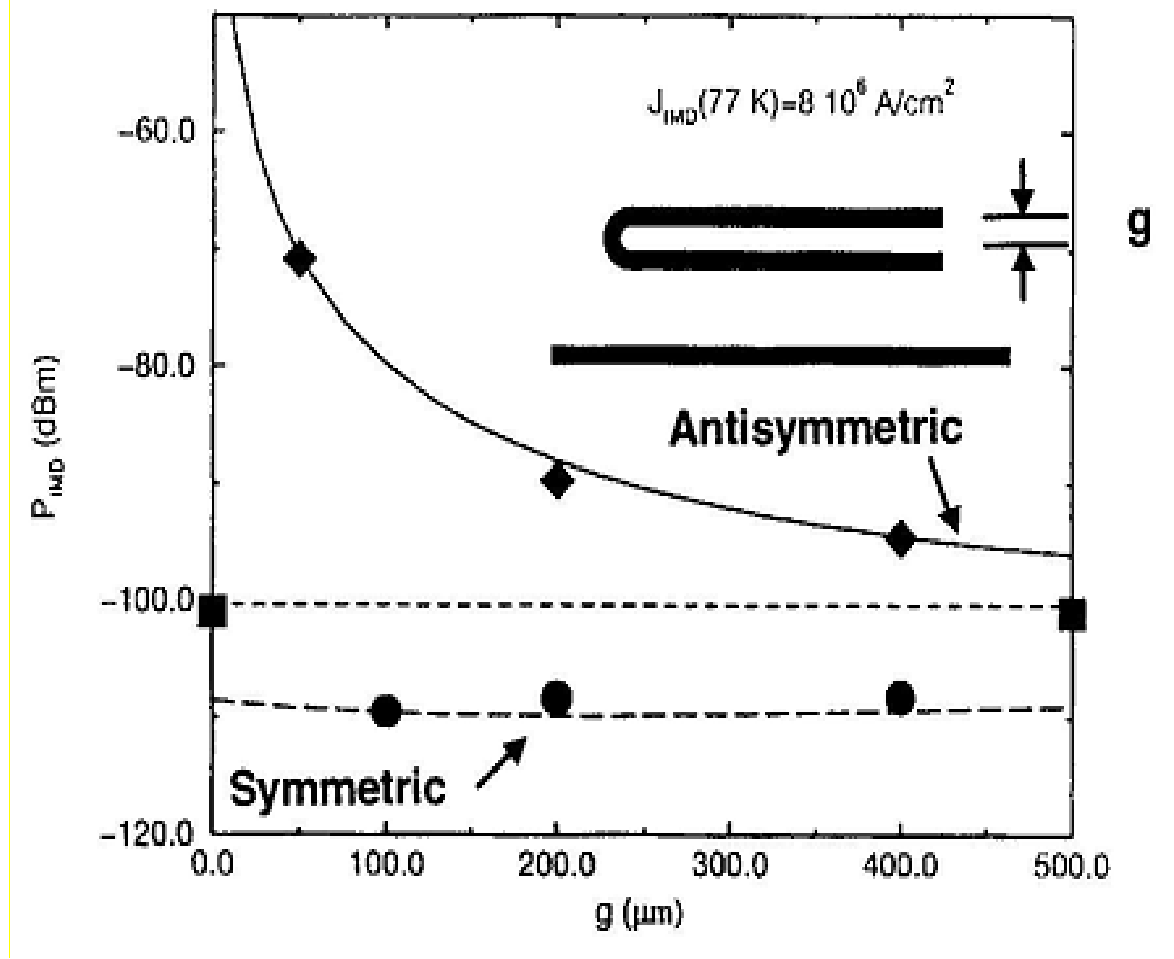

**Fig. 32.** *IM* power as a function of gap in the geometry shown in the inset (after [474]).

Discussing the approaches to the modeling of nonlinearities in *HTS* thin films at microwaves, Oates [474] notes that Dahm and Scalapino [495] have developed a model of *IM* derived from the intrinsic nonlinearity arising from the *d-wave* symmetry of the superconducting order parameter. Dahm and Scalapino model [495] takes into account the geometry, and thereby, the current distribution in the device [474]. Dahm, Scalapino, Willemsen [504] have tested this phenomenological theory using different geometries with substantially different current distributions and obtained good agreement using the same value of $J_{IMD}$. Measurements [504] were done as a function of the gap spacing between the microstrip line and the u-shaped resonator for frequencies at the fundamental of the resonator and the first overtone. In the former case the current distribution in the two legs of the resonator is anti-symmetric, and in the later, symmetric with significantly smaller current peaks. Agreement was good for the two extremes of current distribution using the same value of $J_{IMD}$ [474].

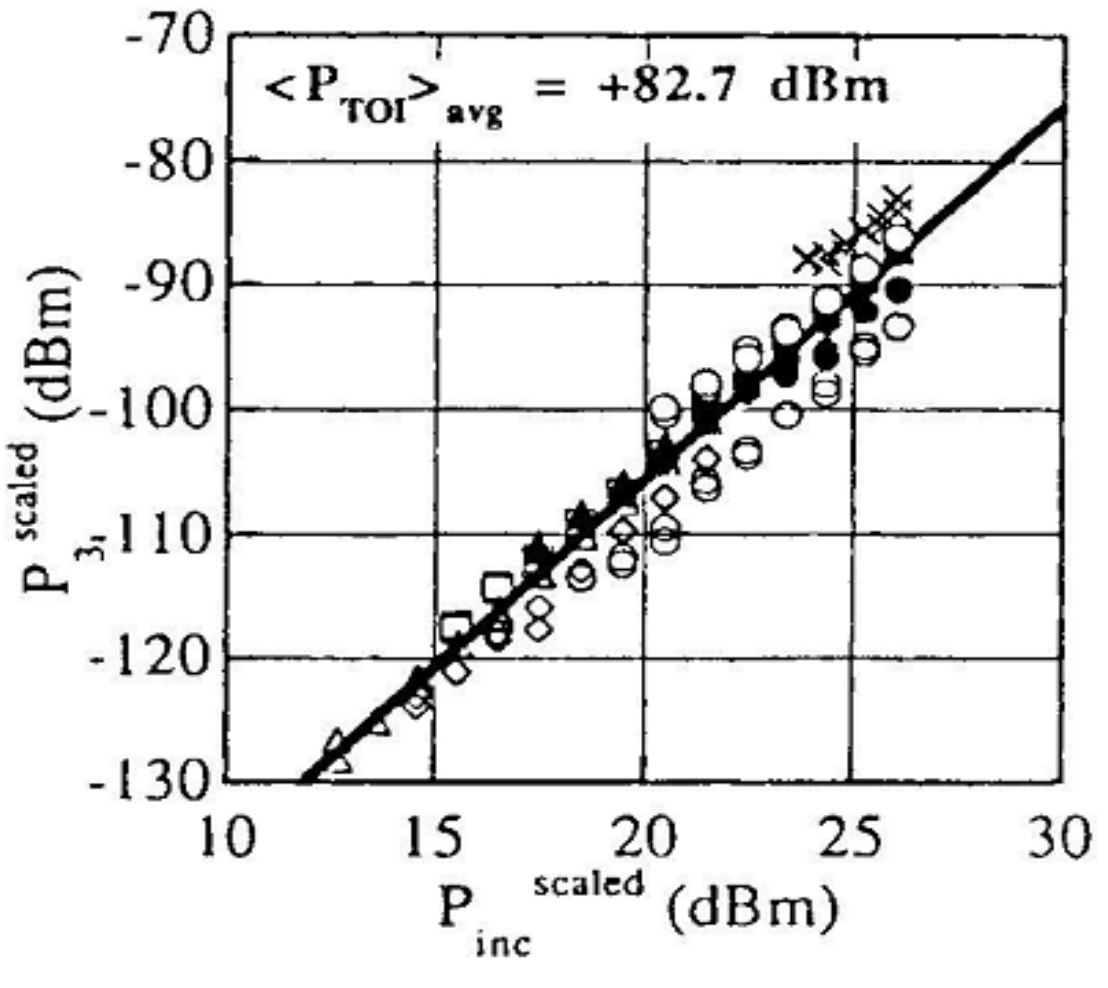

**Fig. 33.** Third harmonic measurements of Fig. 31, where the geometry effects have been scaled according to the equation using just one value of $J_c$ (after [474]).

D. E. Oates [474] writes that Booth, Beall, Rudman, Vale, Ono [503] also have used the phenomenological model of Dahm and Scalapino [495]to explain the dependence on geometry of the third-harmonic measurements shown in Fig. 31. The Booth, Beall, Rudman, Vale, Ono [503] result is shown in Fig. 33, where all of the geometries are scaled accordingly. From this it is seen that the model of Dahm and Scalapino [495] also is able to explain the third-harmonic measurements of Booth, Beall, Rudman, Vale, Ono [503]. This both verifies the model and verifies that ***the third-order IM and third-harmonic generation have a common origin*** [474]. D. E. Oates [474] emphasizes that the model of *IM* and third-harmonic generation proposed by Dahm and Scalapino [495] does not identify the origins of the nonlinearities. It only postulates that the nonlinearities cause a nonlinear inductance. The microscopic origins of $J_{IMD}$ used in this model remain unidentified [474]. D. E. Oates [474] concludes that ***while the proposed models describe some aspects of the observed nonlinearities, no single model is able to explain all the results***.

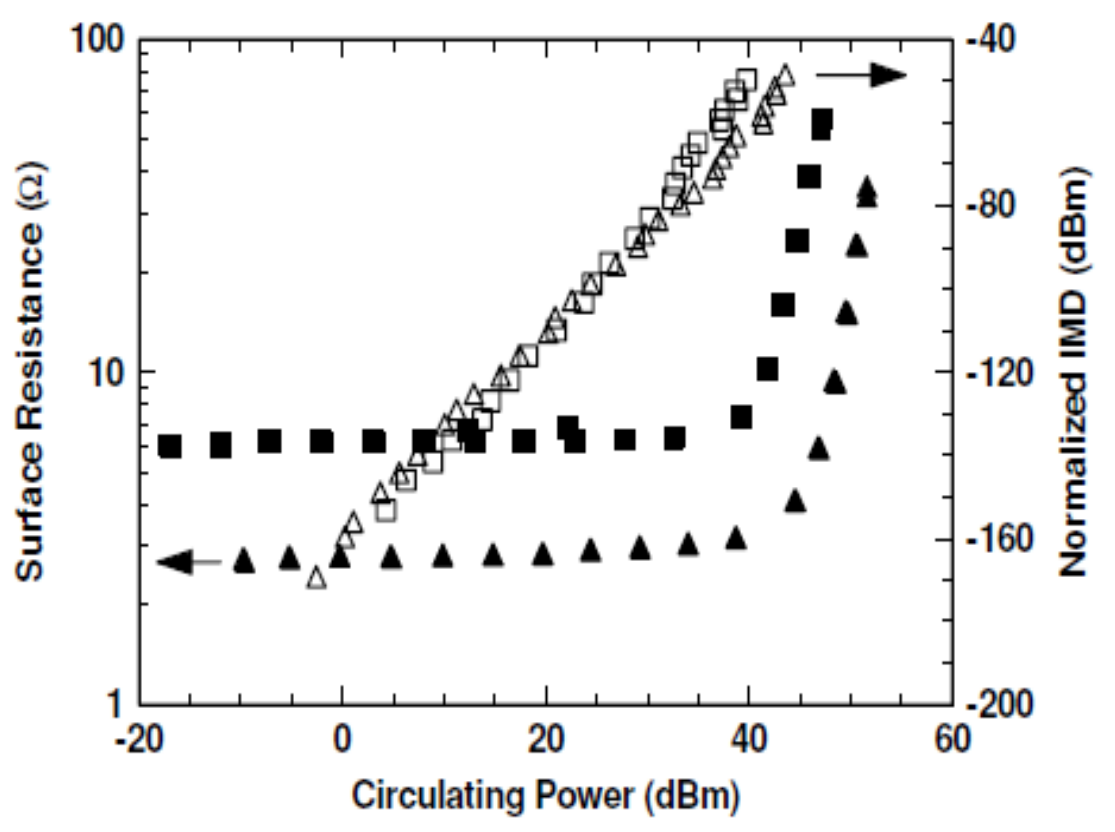

**Fig. 34.** Comparison of $R_S$ $(P_{circ})$ and *IMD* at temperature of 10*K* and 50*K*. Legend: (▲) $R_S$ at 10*K* and (■) $R_S$ at 50*K* both on left scale. (□) *IMD* at 10*K* and (△) IMD at 50*K* on the right scale. The film is *PLD YBCO* on *LAO* deposited by laser ablation in Koren, Gupta, Beserman, Lutwyche, Laibowitz [505] (after [480]).

***The IMD measurement is a most sensitive method to detect the nonlinearities in HTS thin films at microwaves*** in D. E. Oates [480].

For low circulating powers, below approximately 30 *dBm*, the surface impedance does not change appreciably, while the *IMD* is increasing visibly. At the circulating power of interest for applications, where the *IMD* is increasing rapidly, it is seen that the surface resistance measurements are not sensitive enough to resolve the nonlinearities. Since at these power levels $R_S$ is constant, a filter passband performance would be unchanged, but *IMD* would be present. The *IMD* is sensitive to changes in penetration depth $\lambda$ of a part in $10^5$ [480].

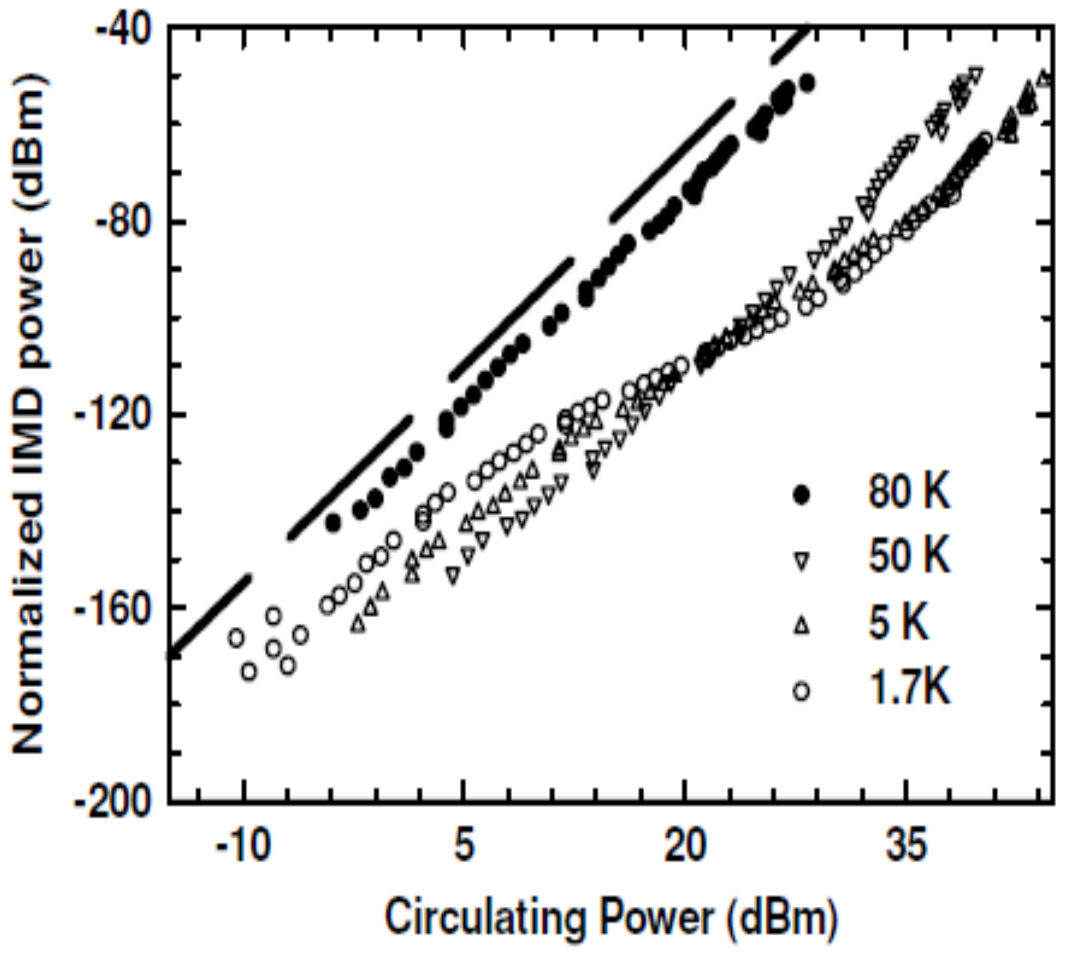

**Fig. 35.** Measured *IMD* vs. circulating power for various temperatures as indicated for *YBCO* on *LAO* in Koren, Gupta, Beserman, Lutwyche, Laibowitz [505]. The broken line is slope three and is included for reference (after [480]).

At low powers, the slope approaches three, indicating a quadratic dependence of the nonlinearity on current. For all powers at high temperatures, *T>70K* for this sample, the slope remains equal to three. For the temperatures below 70*K*, the slope changes over to approximately two at intermediate powers, and the *IMD* curves for the various temperatures converge and cross near 20 *dBm* circulating power. The slope two is indicative of a surface impedance that depends on $|I|$ rather than the $I^2$. It is also evident that at low power and low temperatures the *IMD* increases. Note that at 5-*dBm* circulating power the *IMD* at 1.7*K* is larger than that at 5*K* or 50*K* [480].

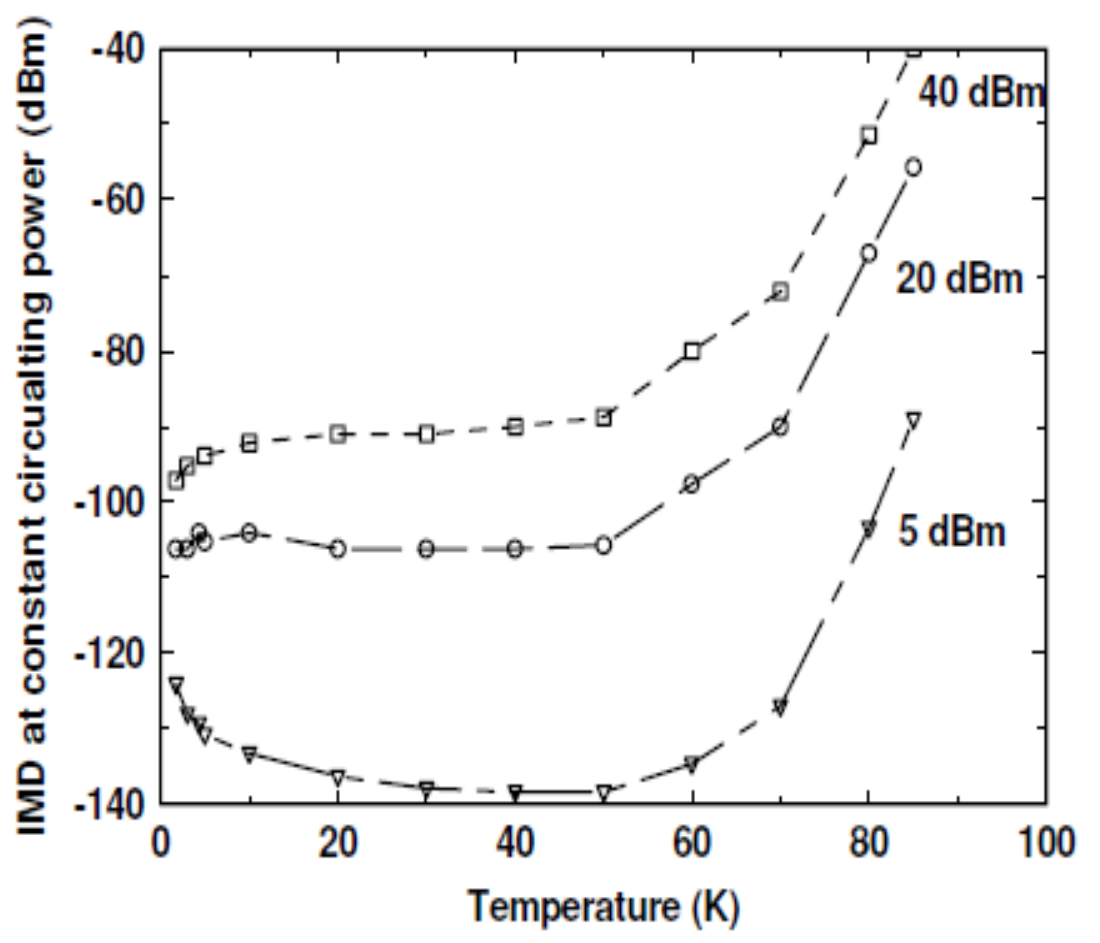

**Fig. 36.** *IMD* at constant circulating power vs. temperature for the powers indicated for laser-ablated *YBCO* on *LAO*, the film shown in Fig. 35 (after [480]). Shown here are sections through the temperature curves of Fig. 35 at constant circulating power for three different values of the circulating power 5, 20, and 40*dBm*. All measured temperatures were used. ***The temperature dependence at 5 dBm shows the increase in IMD at low temperature.*** This increase will be important in the discussion of the theory of nonlinearity in the high-*Tc* materials later in this paper. The other power levels show more conventional temperature dependence, either independent of temperature or increasing at higher temperatures [480].

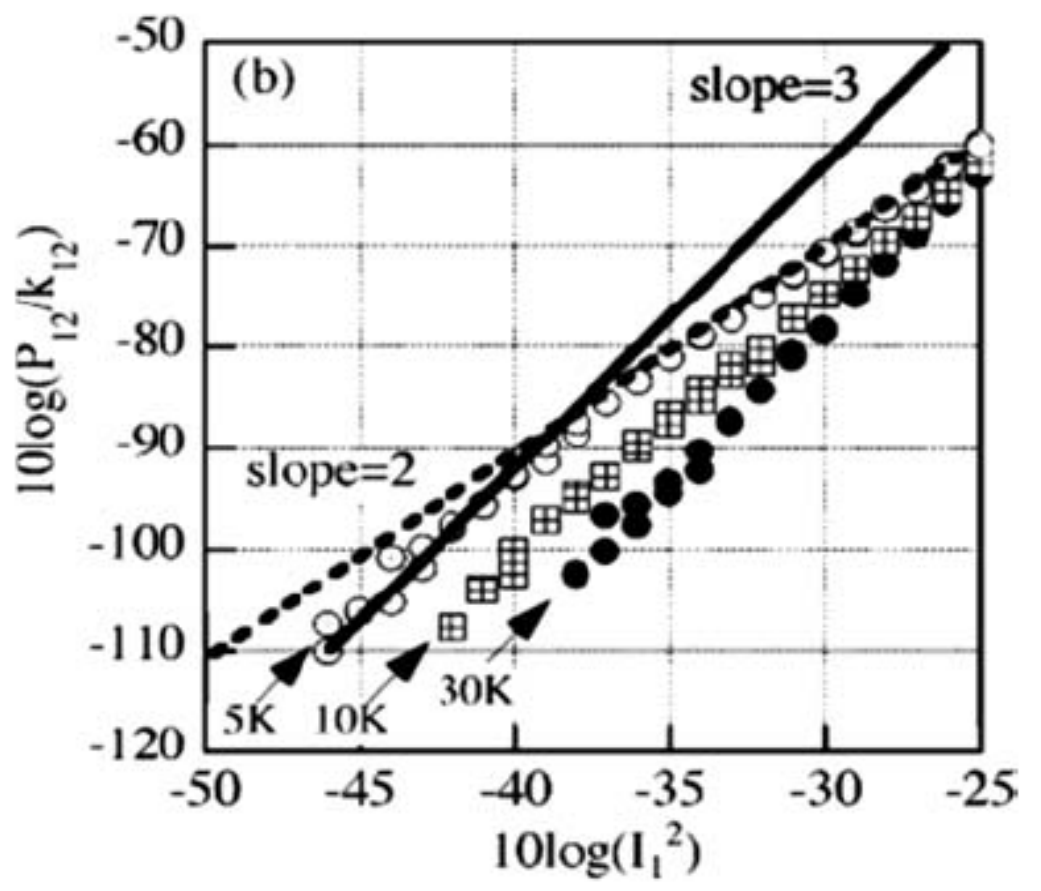

**Fig. 37.** Measurements of IMD vs. resonator current in *YBCO* film on *LAO* as reported by Leong, Booth, Schima [506], who used a coplanar resonator at 6*GHz*. (after [480]).

The *IMD* is a universal property of *HTS* thin films at microwaves in D. E. Oates [480].

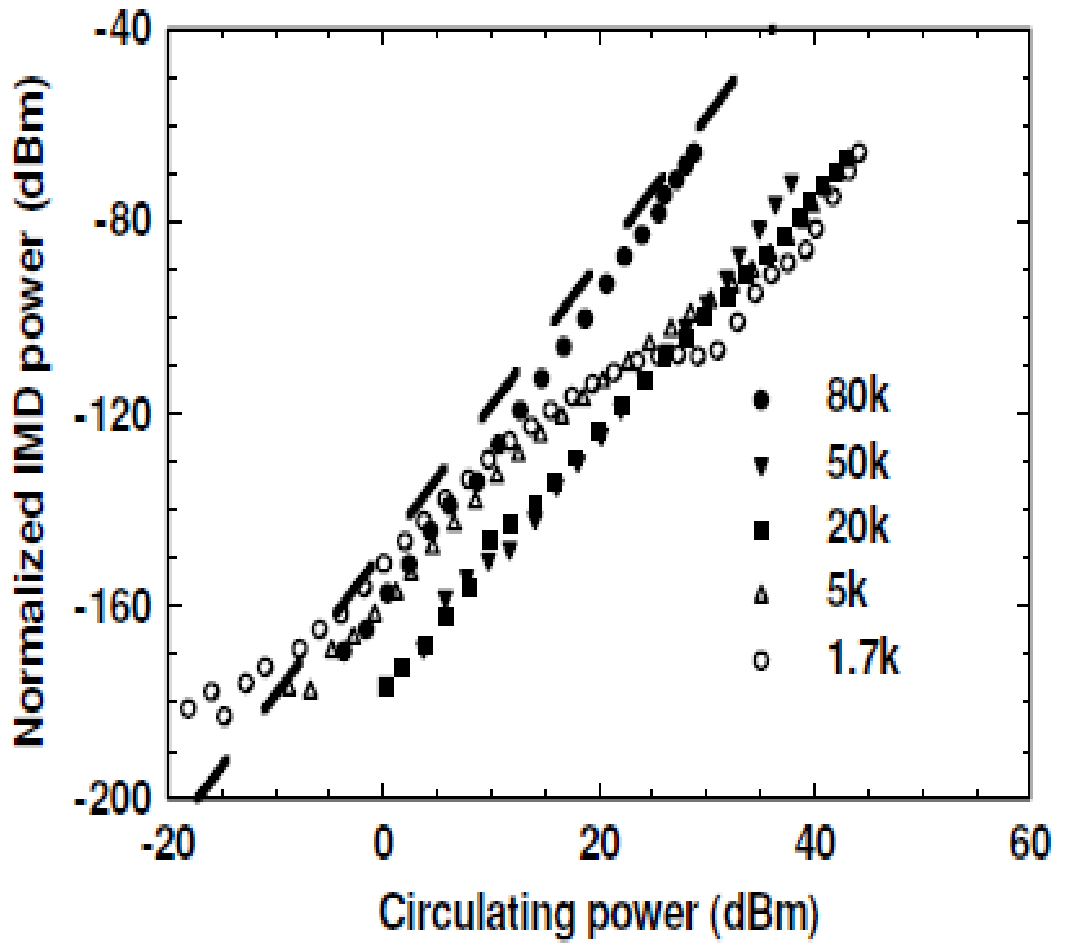

**Fig. 38.** Measurements of *IMD* vs. circulating power for temperatures as indicated for the *YBCO* on *LAO* film, grown by thermal evaporation, from Theva GmbH in Germany in Semerad, Knauf, Irgmaier, Prusseit [507]. The broken line is slope = 3 for reference (after [480]).

This film shows results similar to the film in Fig. 35, in that the *IMD* increases at low temperature and power, and the low-temperature *IMD* is close to slope three at low power and changes over to a slope of less than three at higher powers. At high temperatures for all powers and at high powers for all temperatures the slope becomes three again. ***This is further evidence of intrinsic behavior in general [480]***.

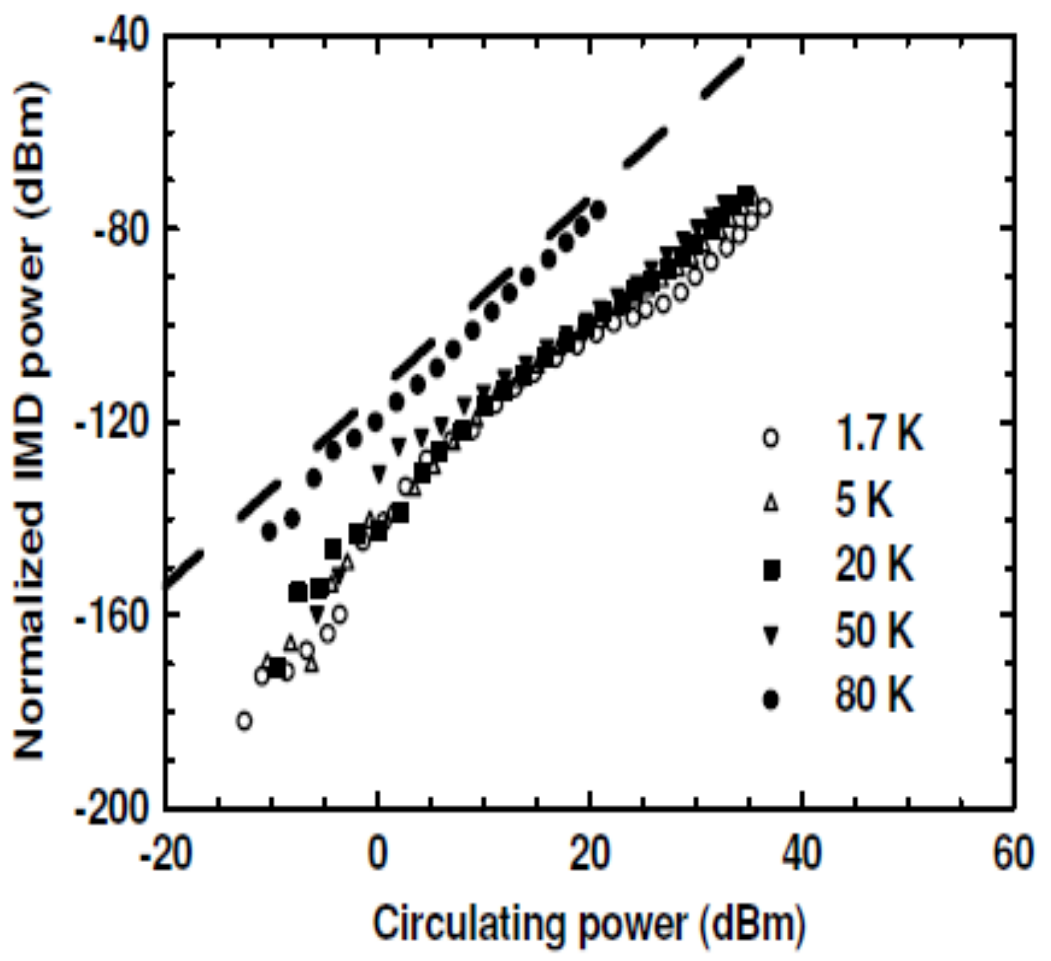

**Fig. 39.** Measurements of *IMD* vs. circulating power for temperatures as indicated for the *YBCO* on sapphire film from Theva GmbH in Germany in Semerad, Knauf, Irgmaier, Prusseit [507]. The broken line is slope = 2 for reference (after [480]).

This film does not show the increase in *IMD* at low temperatures and shows a slope of approximately two for all powers and temperatures. Also, note that ***the IMD is higher than that measured in the film on LAO*. *The film on sapphire is dominated by extrinsic sources of nonlinearity, which mask the intrinsic d-wave features shown on the films on LAO.*** The exact origins of the extrinsic *IMD* are not known, but one can speculate that weak links caused by the different thermal expansion coefficients of sapphire and *YBCO* are the cause [480].

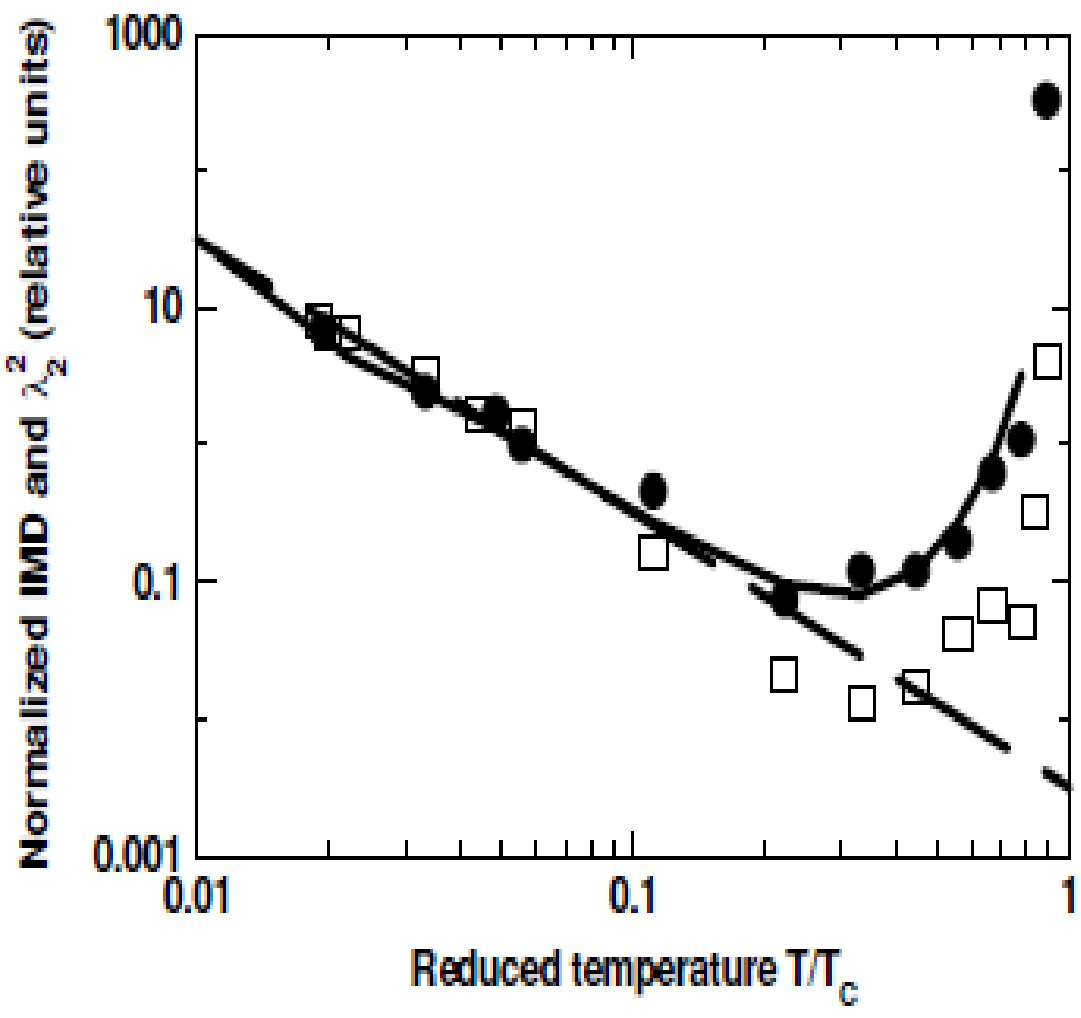

**Fig. 40**. Comparison of data and theory. The solid circles are the *IMD* data from the film of n Koren, Gupta, Beserman, Lutwyche, Laibowitz [505]. The open squares are the data from the Theva film in Semerad, Knauf, Irgmaier, Prusseit [507]. The solid line is the calculation of *d-wave theory* in Agassi and D. E. Oates [508, 509]. The dashed line shows a $1/T^2$ divergence for reference (after [480]).

The measured *IMD*-vs.-*T* data at 5*dBm* circulating power for the films in Figs. 35 and 38 (points) compared with the results of calculations from the *AO* theory (solid line) in relative units. The agreement with the temperature dependence is excellent. Especially to be noted is the $1/T^2$ dependence resulting from the $1/T$ dependence of $\lambda_2(T)$ at the lowest temperatures. The *IMD* $\sim\lambda^2$. The *DS* theory produces similar agreement. ***The*** $\boldsymbol{1/T^2}$ ***dependence at low power is a signature of the d-wave symmetry of the order parameter*** in D. E. Oates, Park, Koren [510].

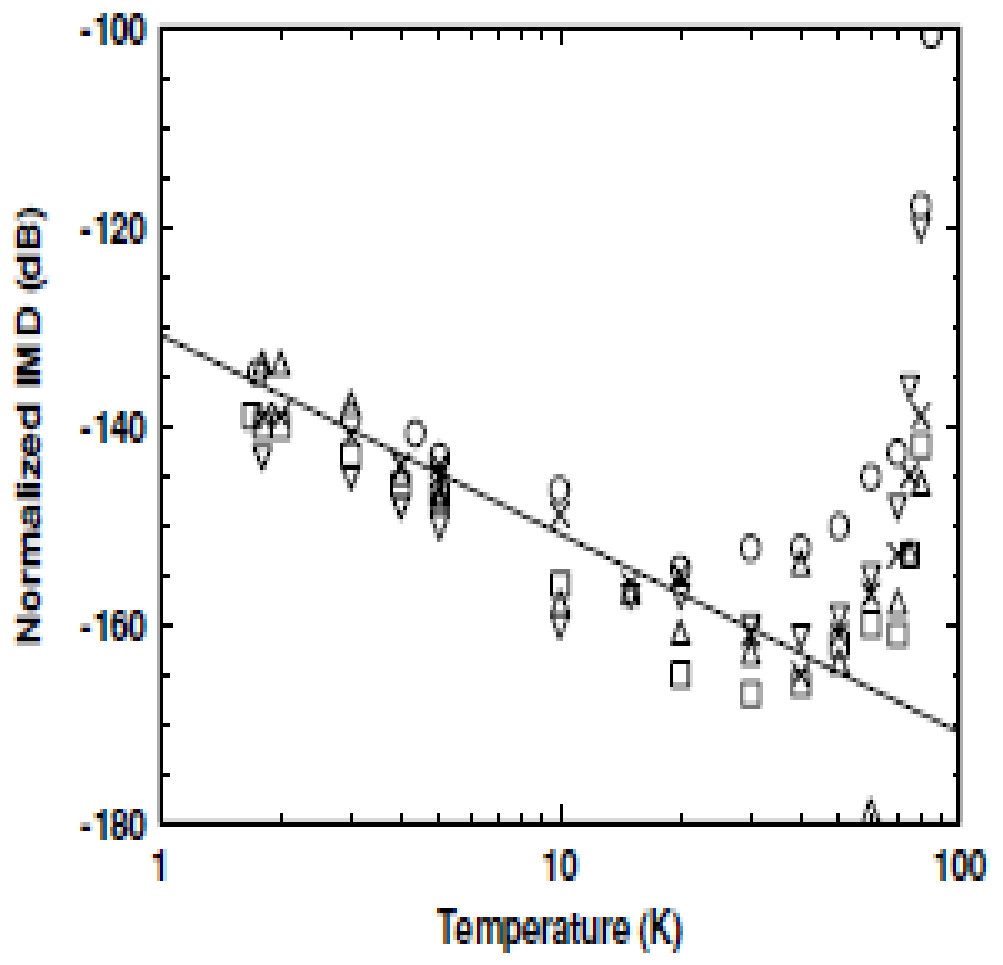

**Fig. 41.** *IMD* vs. temperature *T* for high quality *YBCO* films from various sources and various deposition methods. The resonance frequency is 1.5 *GHz*. (◦) Koren, Gupta, Beserman, Lutwyche, Laibowitz from Technion [505] *PLD*; (□) Semerad, Knauf, Irgmaier, Prusseit from Theva [507] *evaporated*; (∇) Chew, Goodyear, Edwards, Satchell, Blenkinsop, Humphreys from QinetiQ [511] *evaporated*; (×) Li, Suenaga, Ye, Foltyn, Wang [512] *PLD*. The solid line shows a $1/T^2$ behavior as expected from theory (after [480]).

All the films show the $1/T^2$ that is characteristic of the *d-wave* order parameter. ***This is a very significant new result indicating the general nature of the intrinsic nonlinearities and widespread appearance [480].*** There are however, significant variations of the level of *IMD* and the position of the minimum see below [480].

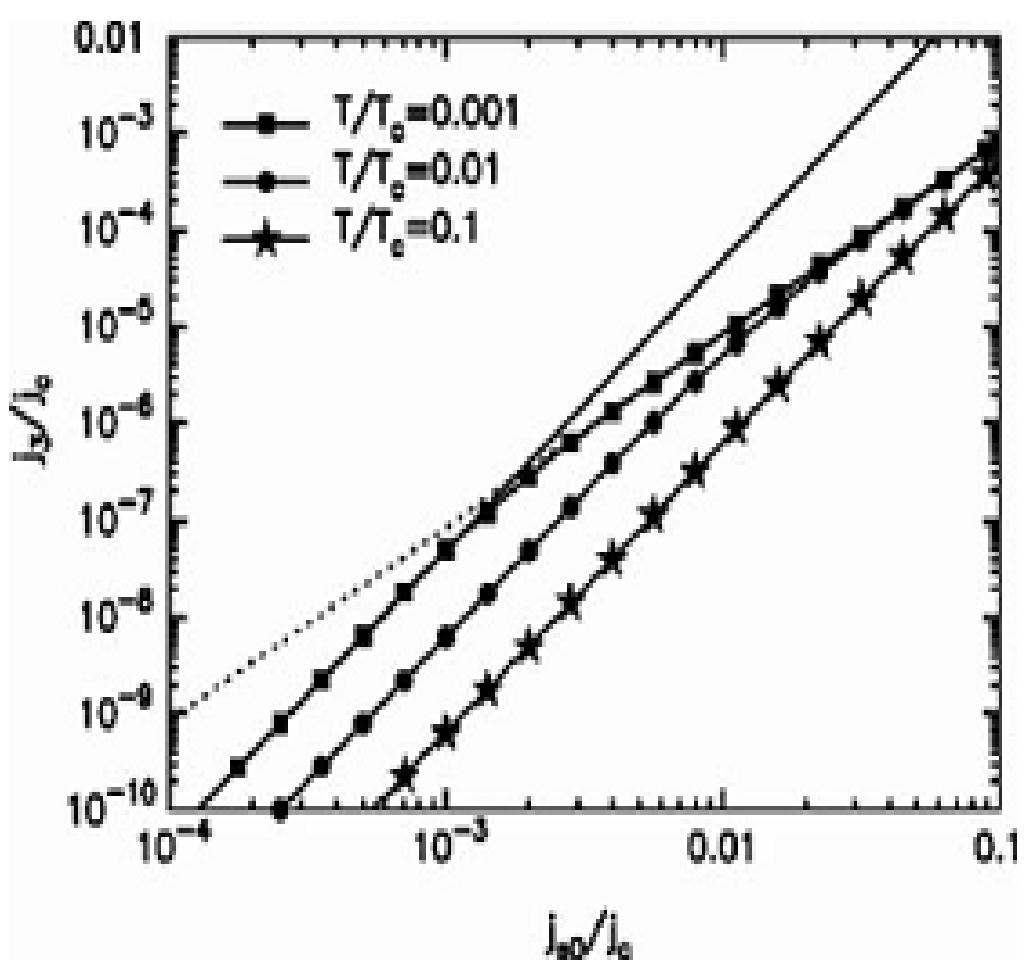

**Fig. 42.** *IMD* current vs. resonator current calculated by Dahm and Scalapino [495] for temperatures as indicated. The dotted line is slope two and the solid line is slope three (after [480]).

The features of the plot to be noted are the slope at low power is equal to three; as the power increases the slope changes over to slope two, and because of the change of slope, the curves for the various temperatures merge at high power. All of these features are shown in the data in Figs. 35 and 38, except at the highest powers in Fig. 35, where the *IMD* curves change to slope 3 and for temperatures above approximately 70*K* where the *IMD* shows slope 3 over the whole range of powers. This result suggests that a different mechanism governs the *IMD* at high power, perhaps flux penetration, and at high temperatures where the validity of the theories is not known [480].

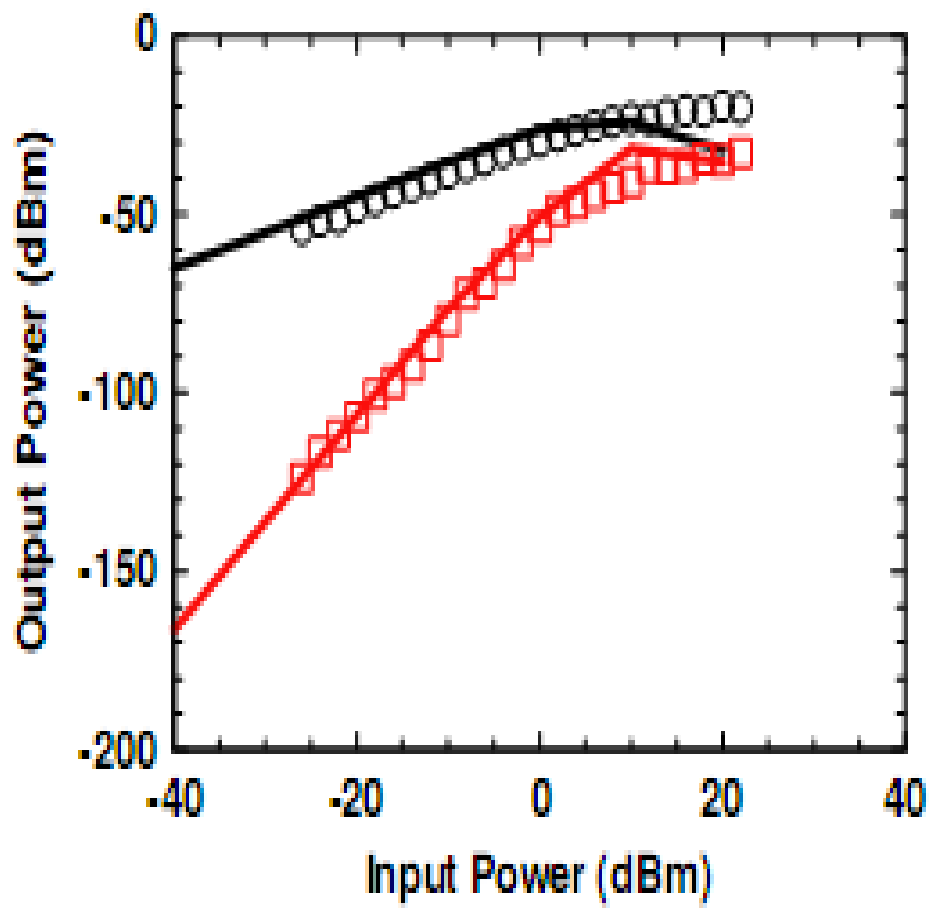

**Fig. 43.** Measured and calculated *IMD*. *YBCO* resonator film #1 laser ablated, at *T=50K*. Solid lines: calculations. Symbols: measured data. (◦) fundamental frequency, (□) third-order *IMD* (after [480]).

Fig. 43 plots the measured and calculated values of the output power at the fundamental frequency of the resonator and the measured and calculated values of the output power of one of the third-order *IMD* tones vs. input power $P_{in}$. There are no free parameters in the fitting. The agreement is excellent, verifying that the *IMD* can be derived from the nonlinear surface impedance and modeled accurately in a single resonator. In this measurement, the *IMD* depends on the third power of the input power as seen by the slope of three in the double logarithmic plot. This is not always the case. The calculation, because of the assumptions of the quadratic dependence of the *R* and *L* on current, can only yield a cubic dependence on power [480]. The *IMD* is dominated by the nonlinear inductance [480]. The level due to $L_2$ is approximately 10*dB* larger than that produced by $R_2$.This might not be a universal property of the *IMD* and might only hold for this film and this temperature. The subject has not been thoroughly investigated. The dominance of the nonlinear inductance has also been found by Booth, Beall, Rudman, Vale, Ono [503], Oates [480].

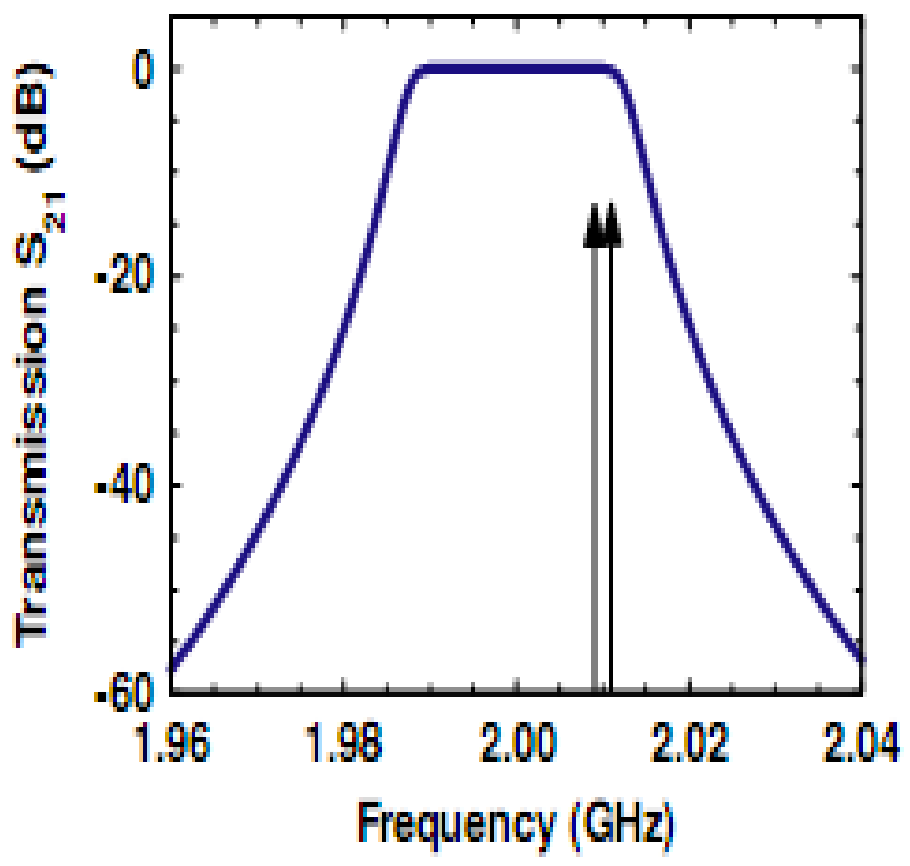

**Fig. 44.** Linear response of five-pole filter used for the calculation of filter *IMD*. The calculation used the measured surface resistance. The arrows show the position of the tones used in the *IMD* calculation (after [480]).

The calculation proceeds from the resonator results by the realization that the filters of interest for implementation in *HTS* are coupled resonator filters. Here, authors limit the simulation to a specific filter design. In order to calculate the *IMD* of the filter, one needs to select a particular set of filter parameters. Researchers have chosen a five-pole, *1%*-bandwidth, Chebychev filter design as a benchmark. Authors further assume that it is implemented with the same *λ/2* resonators as used in the resonator measurements so that we can use the same analytic functions for the resistance and inductance. The filter response, calculated using the measured surface impedance of the Koren, Gupta, Beserman, Lutwyche, Laibowitz film [505], is shown. The arrows in the figure show the frequencies of the tones used to calculate the *IMD*. They were chosen to be near the band edge because such frequencies generate the largest *IMD* [480].

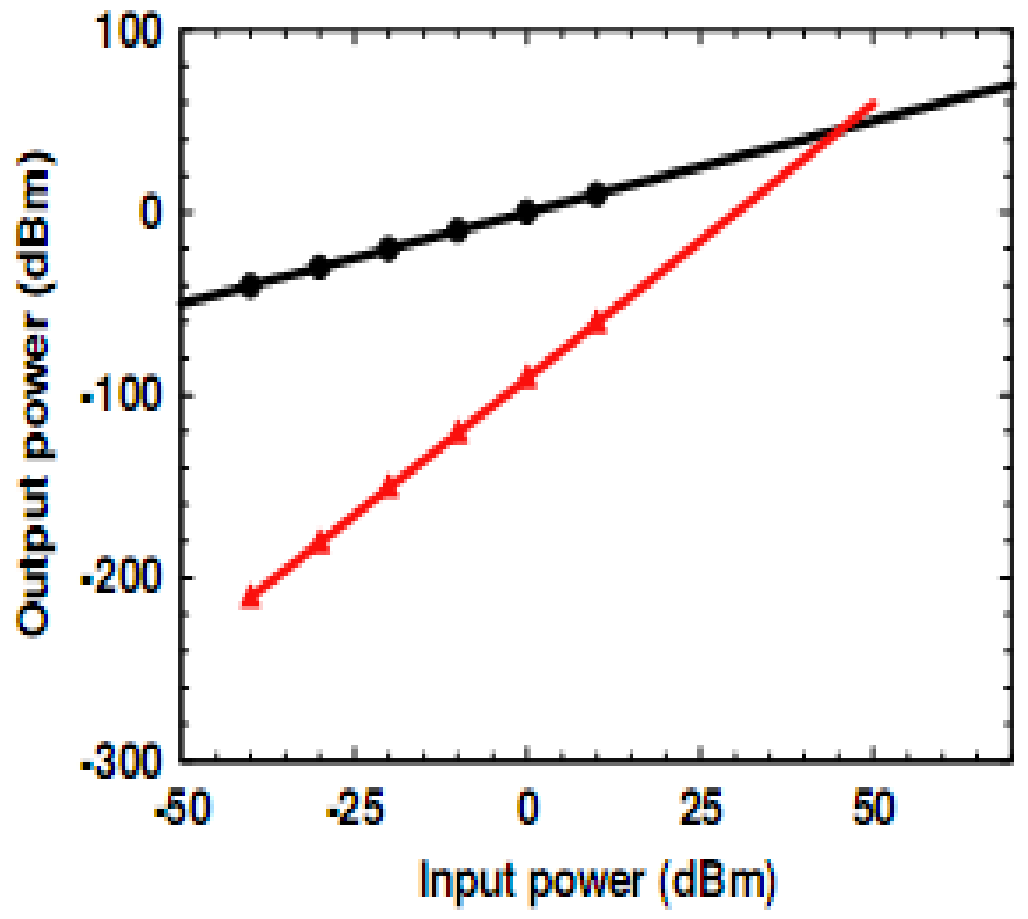

**Fig. 45.** Calculated *IMD* for five-pole filter shown in Fig. 44. Points are calculated values: (•) fundamental, (▲) third-order intermodulation. Solid lines are slope one and three, respectively (after [480]).

The *IMD* shows slope three in the double log plot because of the assumptions of the analytic form of the nonlinear resistance and inductance. The acceptability of this level of *IMD* is dependent on the exact system specifications and the nature of the interference the system is likely to encounter. For this case, the *IMD* is approximately 70*dB* below the fundamental at +10*dBm* input power for this set of input frequencies. The *IMD* level seems to be high enough that it cannot be ignored [480].

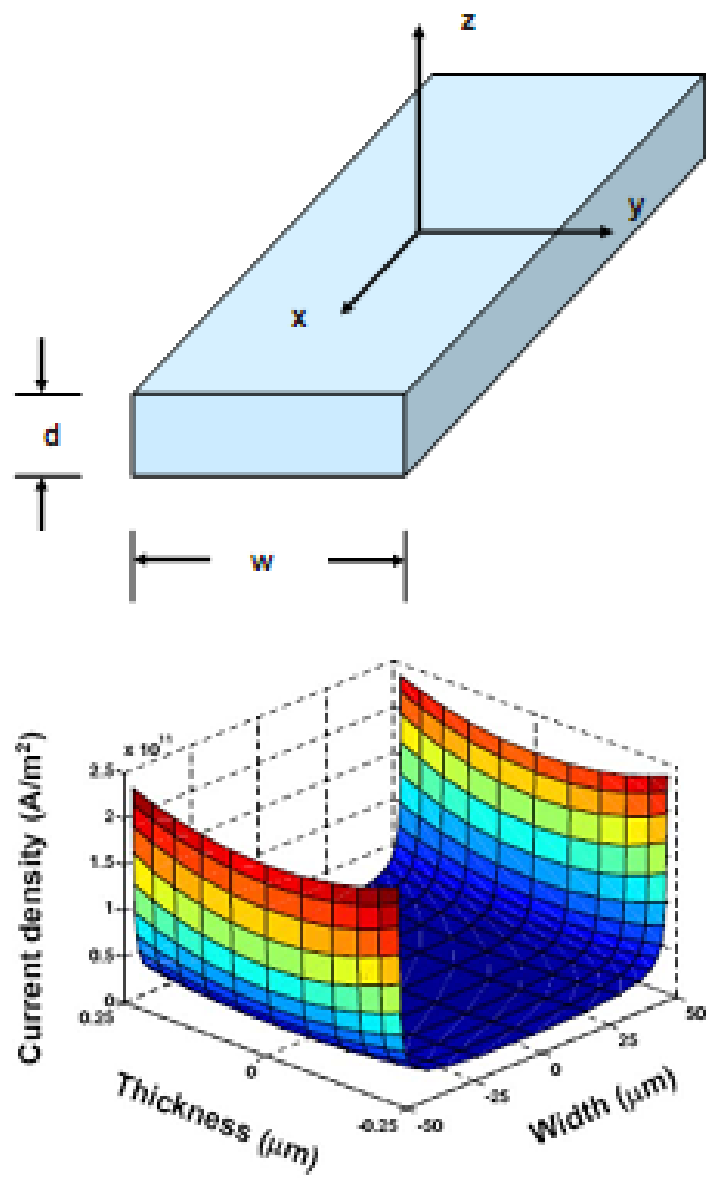

**Fig. 46. (a)** The choice of coordinate system and definition of the width *w* and thickness *d* parameter in a generic thin film strip w>>d. **(b)** An example of a numerically exact calculation of the pair-current density $J_s(y, z)$ for a typical strip, where w=100 *µm*, d=500 *nm*, and $\lambda_0$=250 *nm*. The circulating current is 0.026 *mA*. Note the negligible thickness dependence, which consequently is neglected throughout this work. Note also that the calculation is made with a linear (current-independent) penetration depth, and for the circulating currents considered in these experiments, we assume that the current distribution is independent of the circulating current, because the changes in $\lambda$ are much less than *0.1%*. Thus, the current density scales linearly with the total circulating current (after [513]).

The investigation of the nonlinear Meissner effect in a high-temperature superconductor is performed by measuring intermodulation distortion (*IMD*) power at 1.5 *GHz* in a series of $YBa_2Cu_3O_{7-\delta}$ stripline resonators of varying strip widths and by comparing the obtained results with the predictions of two qualitatively distinct theories of the nonlinear Meissner effect in [513].

***The main differences between the Dahm and Scalapino (DS) theory and Agassi and Oates (AO) theory, which describe the nonlinear Meissner effect: the DS theory implies local electrodynamics for the NLME and, consequently, for thin films [w>>d, Fig. 39(a)],*** $P_{IMD}$ ***is dominated by contributions from the current crowding at the strip edges. On the other hand, the AO theory implies a nonlocal electrodynamics for the NLME, and consequently*** $P_{IMD}$ ***is dominated by contributions from the strip midsection [513].***

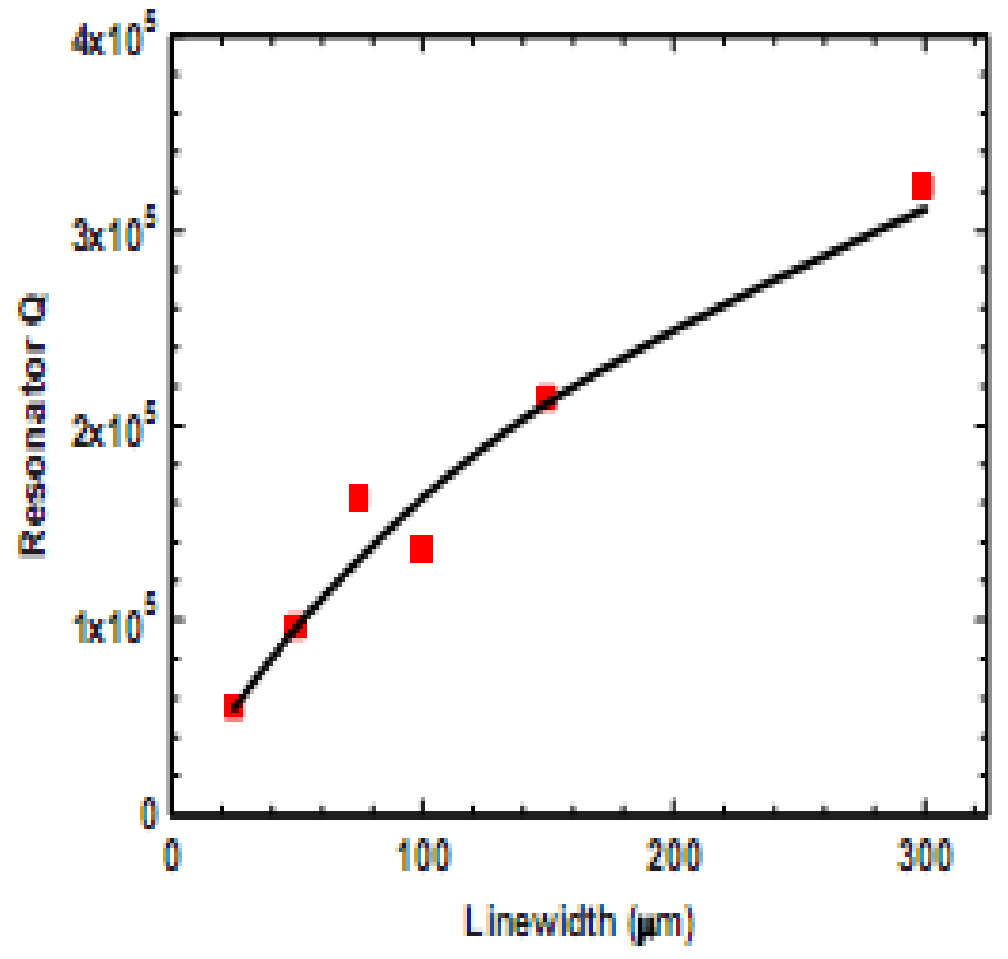

**Fig. 47.** Calculated and measured resonator $Q$ as a function of linewidth. Points are measured and line is calculated. The uncertainty of the measurement is estimated to be 10%. The calculation was normalized to the $Q$ of the 25-µm line. This is equivalent to determining the $R_S$ and $\lambda$ of the film from the measurement of the 25$\mu m$ linewidth (after [513]).

The measurements [513] were carried out on stripline resonators fabricated from a single $YBa_2Cu_3O_{7-\delta}$ film deposited on a 2-inch-diameter lanthanum aluminate $LaAlO_3$ (*LAO*) wafer. The $YBa_2Cu_3O_{7-\delta}$ thin films were deposited by a reactive-evaporation technique that has been described in Semerad, Knauf, Irgmaier, Prusseit [507].

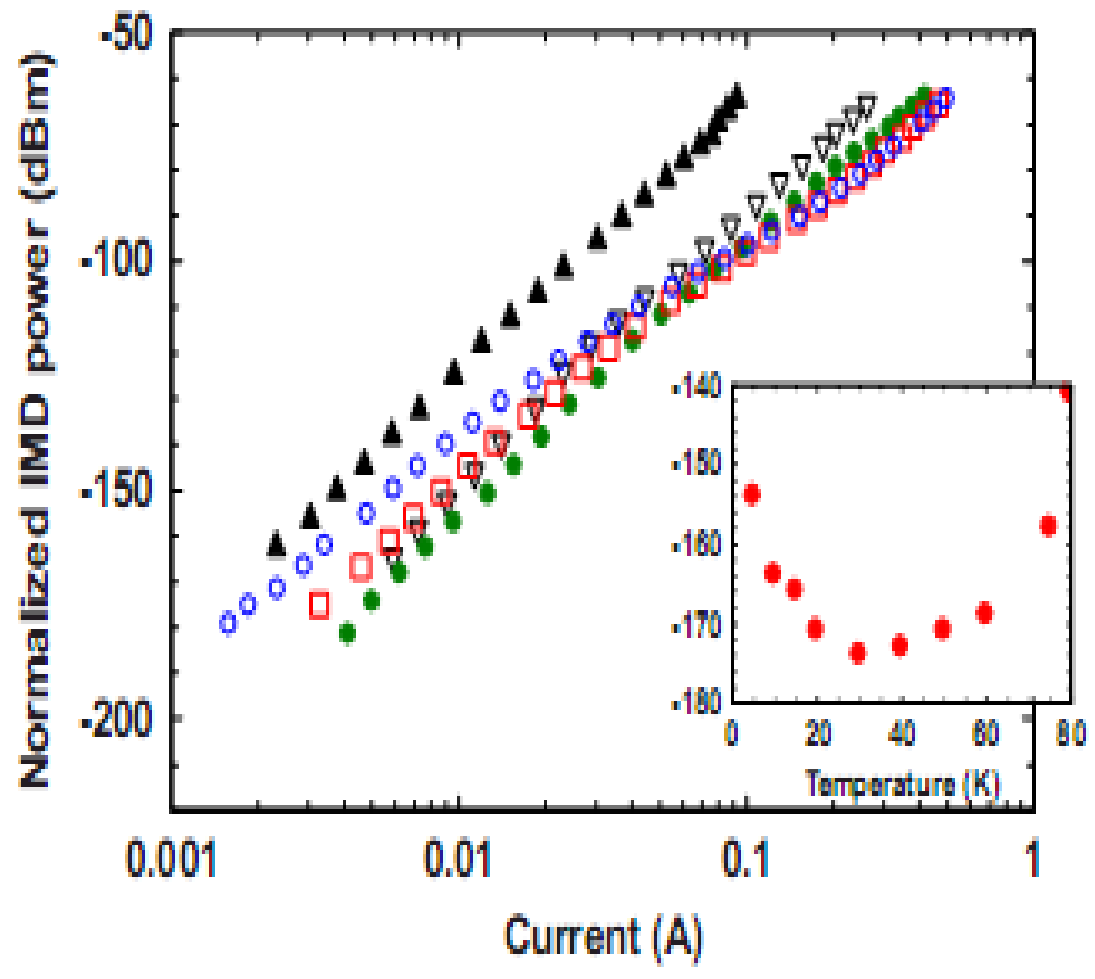

**Fig. 48.** An example of $P_{IMD}(dBm;\ I, T)$ data for a strip with $w=75\mu m$ plotted against the current *I*. The open circle, the open squares, the full circles, the open inverted triangles, and the full triangles correspond to the temperatures *T=5, 10, 30, 60, and 80 K*, respectively. The inset is a plot of $P_{IMD}(dBm;\ I=0.005\ A, T)$ against the temperature to demonstrate the low *T* divergence as a signature for intrinsic nonlinearity (after [513]).

The typical *IMD* power vs. *I* data in the low-power regime for the 75-µm-wide line for several selected temperatures is shown as in Fig. 48. [504]. ***To confirm that the measured IMD has an intrinsic origin, the inset shows the characteristic IMD power divergence at low temperatures*** $P_{IMD}(T\to 0\ K,\ w=75\mu m)\to T^{-2}$ ***[480].***

The points in the double logarithmic plot present the normalized *IMD* data vs. *I* at T=5 *K* for the series of the linewidth values quoted in the caption in Fig. 48. Note that for a given current, the normalized *IMD* increases as the line becomes narrower. This trend reflects the increased current density as the line narrows for fixed total current. In Fig. 48, the low-current data points are fitted with straight lines of slope 60 *dBm* per decade of current by a standard least-squares procedure. The slope of 60 *dBm* per decade of current corresponds to slope 3 of the $P_{NORM}$ vs. circulating power *P* on double logarithmic plots that has been discussed elsewhere [480] and the chosen current domain $0.0015 \le I \le 0.003\ A$ is consistent with the low-circulating-power considerations above. From these fitted lines, the *IMD* power vs. linewidth was extracted. This procedure has been repeated for each of the measured temperatures up to 60*K*. The data at T=5*K* are shown only. The other temperatures produce comparable results [513].

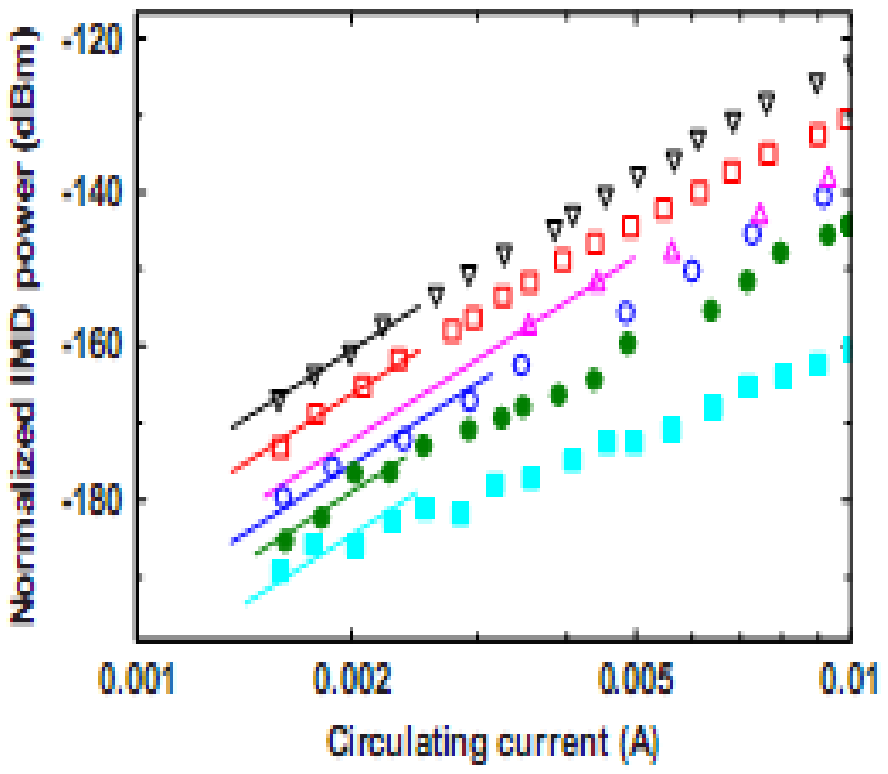

**Fig. 49.** Data points for $P_{IMD}(dBm;\ I\ ,T=5\ K,\ w)$ at the low-power regime and the corresponding fitted lines according to the methodology described in Sec. III. The linewidths are ∇25 µm, □50 µm, Δ75 µm, ○ 100 µm, • 150 µm, and ■300 µm (after [513]).

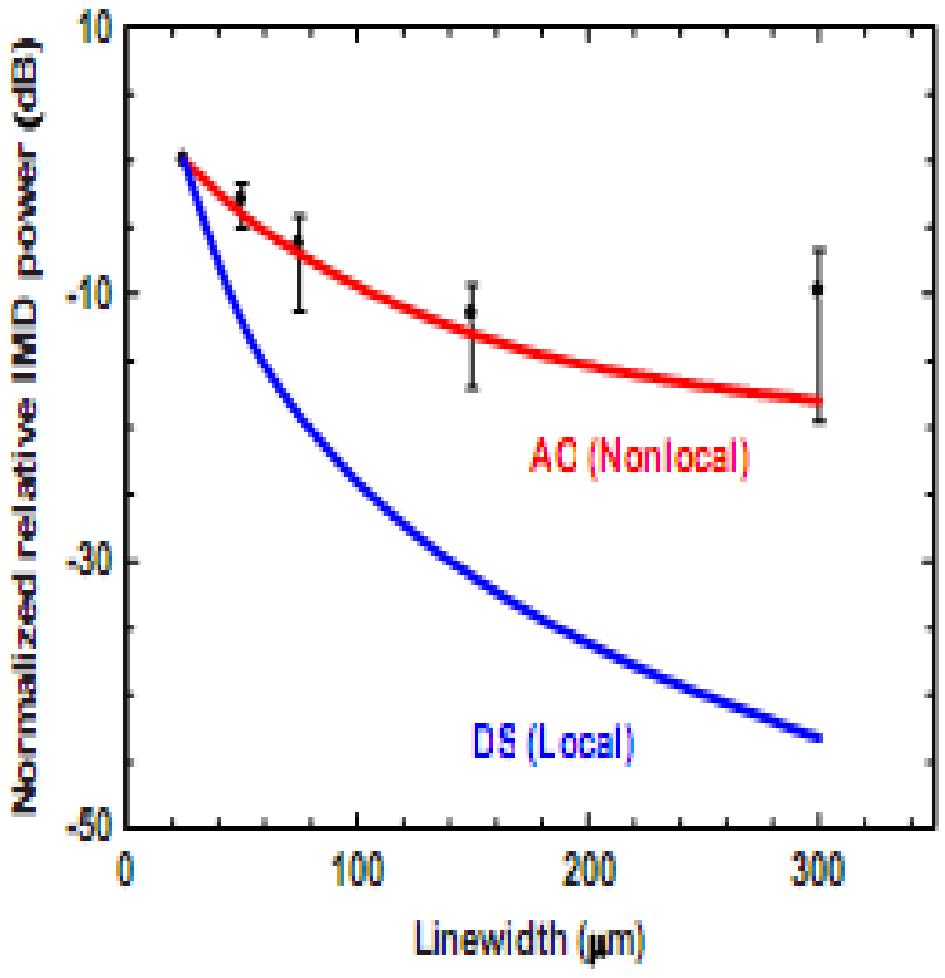

**Fig. 50.** Comparison of the $P_{IMD}(dBm;\ w)$ data with the theoretical predictions of the *DS* and *AO* theories. The 25-µm line is used as the reference. The data error bars represent the spread of the data points with T in the range $5 \le T \le 60\ K$. The calculations (solid lines) are with the exact numerical current density, such as in Fig. 46(b), using the method in [514] (after [513]).

The result of averaging over the temperatures of *5, 10, 15, 20, 30, 40, 50, and 60 K* [513].

The error bars on the data represent one standard deviation in Fig. 50. The chosen reference width is $w_0 = 25\ \mu m$. The only input parameters in the numerical calculations, involving the numerically calculated current density $J_S(y,\ z)$, are the penetration depth $\lambda_0(T)$ and the stripline thickness. The results are not very sensitive to either of the parameters. The *I* independence is automatically guaranteed by the linear fitting in Fig. 49, while the predicted *T* independence is used as a consistency check for the data analysis and is vindicated by the relatively small error bars in Fig. 50 and temperature variation in Fig. 51. For the researched samples, the corresponding values are $\lambda_0(T=0)=250$ *nm* and $d=500$ *nm*. As the data in Figs. 50 and 51 convincingly show, the difference between the local *DS* and the nonlocal *AO* theoretical predictions is considerably larger than the error bars, and the nonlocal *AO* theory predictions track the data. Furthermore, the *IMD* variation according to the *AO* theory is indeed smaller than those in the *DS* theory [513].

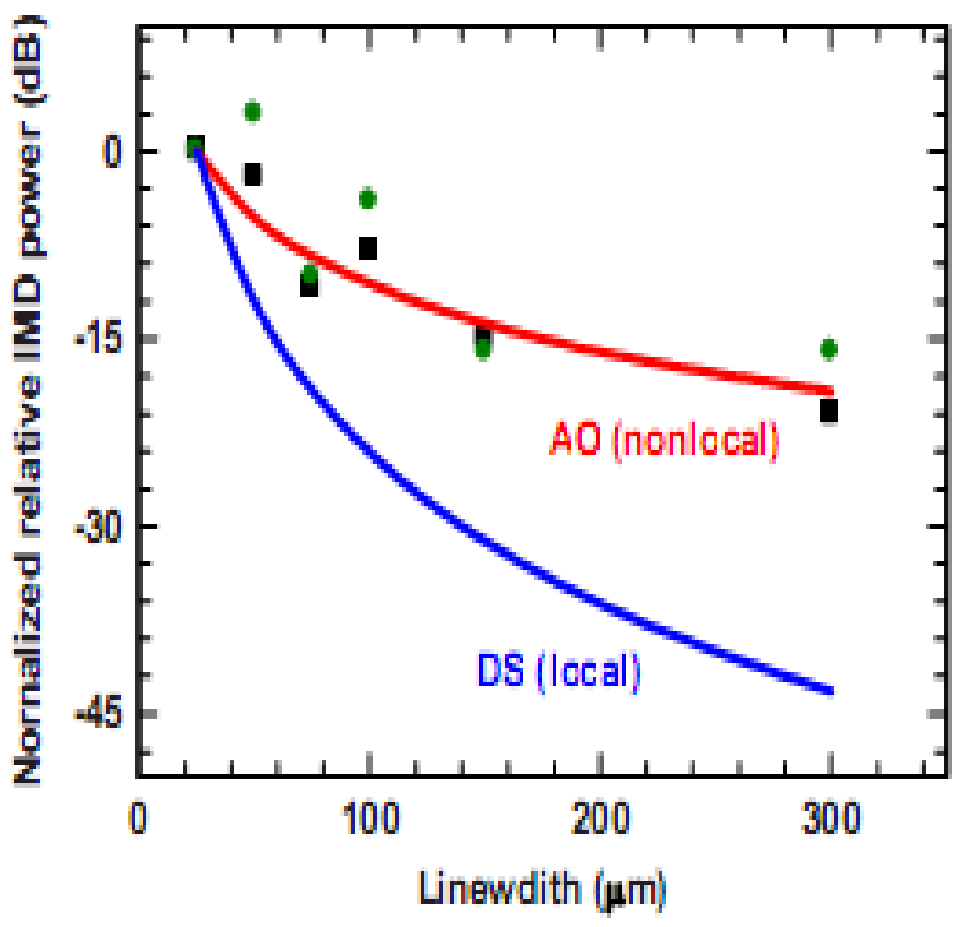

**Fig. 51.** Comparison of the $P_{IMD}(dBm;\ T,\ w)$ data at only 5 $K$ and 10 $K$ with the theoretical predictions of the *DS* and *AO* theories. The 25$\mu m$ line is used as the reference. The symbols • and ■ correspond to the temperatures $T$=5 $K$ *and* 10 $K$ (after [513]).

In Figs. 50 and 51, the $P_{IMD}(w)$ data agree with the nonlocal *AO* theory. This agreement adds to previous experimentally confirmed predictions of the *AO* theory regarding the temperature dependence of the $P_{IMD}(T)$ data, including the low-temperature divergence of $P_{IMD}(T \to 0K) \to T^{-2}$, [480] the non-monotonic slope of $P_{IMD}(P_{CIRC})$ [509], and indications that $P_{IMD}(d) \sim d^{-4}$ [505]. ***Since these IMD data sets represent distinct cuts of the*** $P_{IMD}(T, I,..; w, d,..)$ ***function in its multiple parameter space, the ability to account for them all in the low-power domain lends credence to validity of the nonlocal Agassi and Oates (AO) theory for intrinsic nonlinearity [513].***

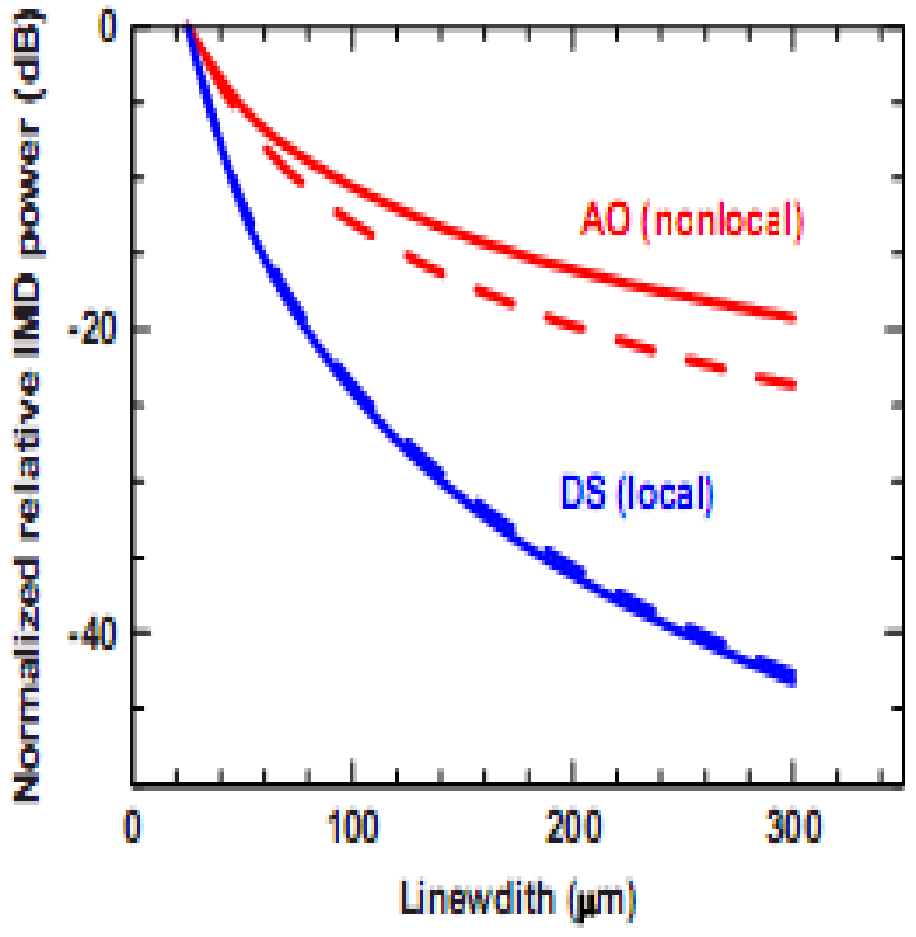

**Fig. 52.** Comparison of the exact numerical calculation (solid line) and the analytical approximation (broken line). The 25$\mu m$ line is used as the reference. Note the close resemblance of the exact numerical calculations and the analytical approximation (after [513]).

D. E. Oates *et al.* [513] explains that, beside the $Q_c$-values comparison in Fig. 40, a second check on the numerical calculations is an analytical approximation, displayed by the broken line in Fig. 52. As Fig. 52 shows, the expression in Eq. (4.2) [513] provides a very good approximation to full numerical calculations of the right-hand side of Eq. (2.7) [513]. Oates *et al.* [513] states that this provides another validation of the nontrivial exact numerical calculations.

**Tab. 10.** Comparative analysis of research results on the origin of nonlinearities in *HTS* thin films at microwaves by D. E. Oates (after [474, 480, 513]).

The authors of book would like to comment on all the important research findings by Oates in [480], by making the following conclusion:

1. High quality $YBa_2Cu_3O_{7-\delta}$ thin films on $LaAlO_3$ substrate (see Tabs. 1 [6]) exhibit *IMD* due to both the intrinsic nonlinearities at low temperatures and the extrinsic nonlinearities at high temperatures at high microwave powers [480, 513, 510].

2. High quality $YBa_2Cu_3O_{7-\delta}$ thin films on $Al_2O_3$ sapphire substrate (see Tabs. 1 [6]) exhibit *IMD* due to the extrinsic nonlinearities at any temperatures at high power microwaves [513], because of the induced defects in $YBa_2Cu_3O_{7-\delta}$ thin films on $Al_2O_3$ sapphire substrate [513, 515]. (Note: Author of dissertation tends to believe that the extrinsic and intrinsic nonlinearities are jointly responsible for the observed nonlinearities in high quality $YBa_2Cu_3O_{7-\delta}$ thin films on $Al_2O_3$ sapphire substrate. It is possible to assume that extrinsic nonlinearities resulting from induced defects may dominate in researched cases, but the nonlinearities of different origin may also contribute to the *IMD* to certain extend.)

3. High quality $YBa_2Cu_3O_{7-\delta}$ thin films on *MgO* substrate (see Tabs. 1 [6]) exhibit *IMD* due to the extrinsic nonlinearities at any temperatures at high power microwaves [119], because of substrate nonlinearities in *MgO* [516] (Note: Author of dissertation thinks that the appearance of nonlinearities in *MgO* depends on the temperature of *MgO* substrate. There are no nonlinearities in *MgO* substrate at normal operational temperature of $YBa_2Cu_3O_{7-\delta}$ thin film microwave resonators and filters (see Chapter 6). The nonlinearities in *MgO* substrate may only appear at low temperatures as in the researched case of *IMD* generation at low temperatures, which is considered by Hein, D. E. Oates, Hirst, Humphreys, Velichko [516]).

The authors of book would like to mention that the research results on nonlinearities nature in *HTS* thin films at microwaves, obtained by D. E. Oates [513], are in relatively good agreement with the early research findings presented by Andreone [517], however these research results have to be revised and updated significantly, because of recent research developments and findings in the research field, presented in the next Chapters in this dissertation.

In 2008, Oates, Agassi, Wong, Leese de Escobar, Irgmaier [513] focused their research interests on the investigation of the nonlinear Meissner effect in a high-temperature superconductor by measuring intermodulation distortion (*IMD*) power at 1.5 *GHz* in a series of $YBa_2Cu_3O_{7-\delta}$ stripline resonators of varying strip widths and by comparing the obtained results with the predictions of two qualitatively distinct theories of the nonlinear Meissner effect in [513].

D. E. Oates *et al.* [513] makes a historical overview on the *nonlinear Meissner effect* and explains the main differences between the *Dahm and Scalapino*

*(DS) theory* and *Agassi and Oates (AO) theory*, which describe the *nonlinear Meissner effect*, stating that the *nonlinear Meissner effect* (*NLME*), as manifested for example in a current (or magnetic field) dependence of the penetration depth $\lambda$, was first predicted for the cuprate high-temperature superconductors (*HTS*) by Yip, Sauls [518], Xu, Yip, Sauls [519], Stojkovic, Valls [520], and extended to the intermodulation distortion (*IMD*) power and harmonic generation by Dahm and Scalapino (*DS*) [495]. An alternative theoretical approach to the *NLME* was developed by Agassi and Oates (*AO*) [508], based on a perturbative expansion of the constitutive relation between the current and the vector potential. The *NLME* has been experimentally confirmed by *IMD* power measurements in high-quality $YBa_2Cu_3O_{7-\delta}$ (*YBCO*) films, specifically, by observation of the characteristic low-temperature divergence of the *IMD* power $P_{IMD}(T \rightarrow 0K) \sim T^{-2}$ [510]. This low-temperature *IMD* power divergence agrees with the predictions of both *DS* and *AO* theories. Notwithstanding this agreement, the *DS* and *AO* theories are qualitatively distinct. In particular, the *DS* theory implies local electrodynamics for the *NLME* and, consequently, for thin films [w>>d, Fig. 46(a)], $P_{IMD}$ is dominated by contributions from the current crowding at the strip edges. On the other hand, the *AO* theory implies a nonlocal electrodynamics for the *NLME*, and consequently $P_{IMD}$ is dominated by contributions from the strip midsection. The research, reported in [513], tests experimentally which of these two qualitatively distinct theories applies.

D. E. Oates *et al.* [513] states that since these *IMD* data sets represent distinct cuts of the $P_{IMD}$(T, I,..; w, d,..) function in its multiple parameter space, the ability to account for them all in the low-power domain lends credence to validity of the nonlocal Agassi and Oates (*AO*) theory for intrinsic nonlinearity [513].

D. E. Oates *et al.* [513] makes a remark that a practical implication of the experimental verification of the *AO* theory is that film edges are less important for $P_{IMD}$ than previously believed. For microwave receive-filter design, this result indicates that damage to the strip edges during the stripline patterning is not a major degrading factor in their performance.

D. E. Oates *et al.* [513] concludes that the presented experimental results for $P_{IMD}(w)$ clearly favor the *AO* theory over the alternative *DS* theory. On the other hand, both theories yield the same prediction for the low-temperature divergence $P_{IMD}(T \rightarrow 0K) \rightarrow T^{-2}$, which has been experimentally confirmed. As highlighted in the basic penetration-depth expansions, a key distinction between the DS and AO theories is that the former is local while the latter is nonlocal [508, 521].

D. E. Oates *et al.* [513] summarizes all the research findings in details saying that the measurements of the width dependence of the *IMD* power in a stripline resonator at a constant total current are analyzed and compared to predictions of two theories of the nonlinear Meissner effect. The measured strips were patterned from a single wafer to ensure uniformity of film quality. This data is compared to predictions of the local theory of Dahm and Scalapino [495], where the nonlinear penetration depth tracks the local current density, and of the theory of Agassi and Oates [508], where the nonlinear penetration depth is spatially constant. The parameter-free comparison between the data and theoretical predictions at several temperatures clearly favors the *AO* predictions. In Oates *et al.* [513] opinion, a key qualitative distinction between the *AO* and *DS* theories is the locality vs. non-locality.

The author of dissertation thinks that the most recently obtained research results on the origin of nonlinearities in superconductors at microwaves by Oates, Agassi, Wong, Leese de Escobar, Irgmaier [513], to some degree, complement the early vision on the main sources of nonlinearities in superconductors at microwaves expressed by Kong [470], Xia, Kong, Shin [471], namely that the nonlinearities, responsible for the *IMD* in high quality superconductors on the $LaAlO_3$ substrate at microwaves, are of ***intrinsic origin*** [470, 471]**,** resulting from the *d-wave* order parameter in Matsumoto, Shiba [522-524] and associated nodes in the energy gap of superconductor at low temperatures at microwaves [480, 495].

Velichko, Lancaster, Porch [46] wrote a review, which mostly concentrates on the nonlinear properties of $YBa_2Cu_3O_{7-\delta}$ (*YBCO*) films at microwaves. Authors mention that the thin films of other materials such as *TlBaCaCuO* and *HgBaCaCuO* also look promising for microwave device application and can offer rather good power handling capability, as substantial progress has been recently achieved in improving the microwave properties of those superconductors [46]. However, these materials are beyond the scope of review, mainly because the nonlinear studies performed on them are rather scarce and insufficient for reliable systematics of their nonlinear properties to be deduced. Therefore, Velichko, Lancaster, Porch [36] concentrated on physical properties of $YBa_2Cu_3O_{7-\delta}$ thin films at microwaves only.

Velichko, Lancaster, Porch [46] clarify that the microwave nonlinear properties of *HTS* films are usually studied by measuring:

1) ***Microwave power dependence of surface impedance*** $Z_S = Rs + i\,Xs$;
2) ***Harmonic generation***;
3) ***Intermodulation distortion (IMD)*** using both resonant and non-resonant transmission line techniques [495].

The author of dissertation would like to mention that Velichko, Lancaster, Porch [46] identify the following possible main sources of nonlinearities in $YBa_2Cu_3O_{7-\delta}$ thin films at microwaves:

1. ***Grain boundaries***
2. ***Patterning***
3. ***Impurity doping***
4. ***Oxygen content***
5. ***Heating***.

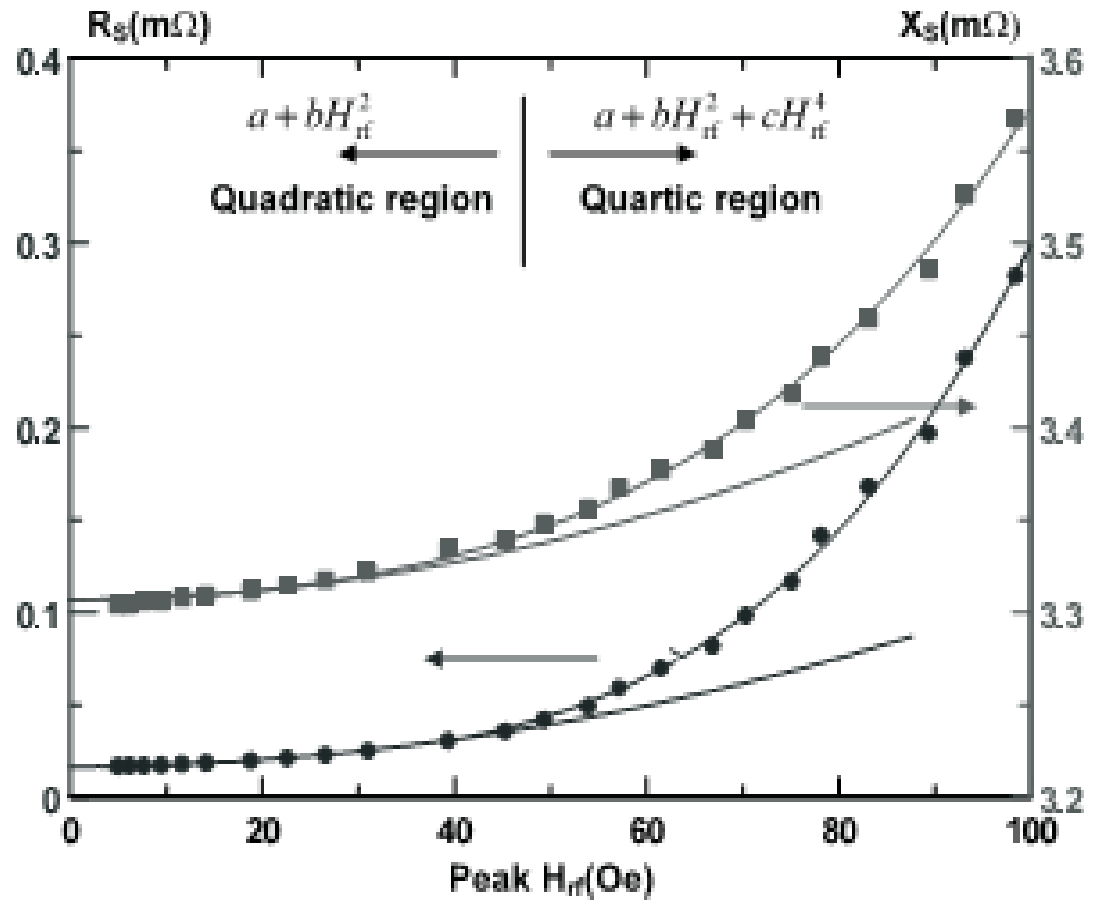

**Fig. 53.** Typical power dependence of an epitaxial $YBa_2Cu_3O_{7-\delta}$ film measured with the help of the stripline resonator at *f* = 1.5 *GHz* and T = 77.4 *K*. High quality films exhibit correlated $H_{2rf}$ -dependence of both *Rs* and *Xs* at low fields, which becomes steeper than $H_{2rf}$ with increased microwave power. Reprinted with permission from P. P. Nguyen *et al., Phys. Rev. B*, vol. **51**, 6686, 1995. © 1995 by the American Physical Society (after [525, 46]).

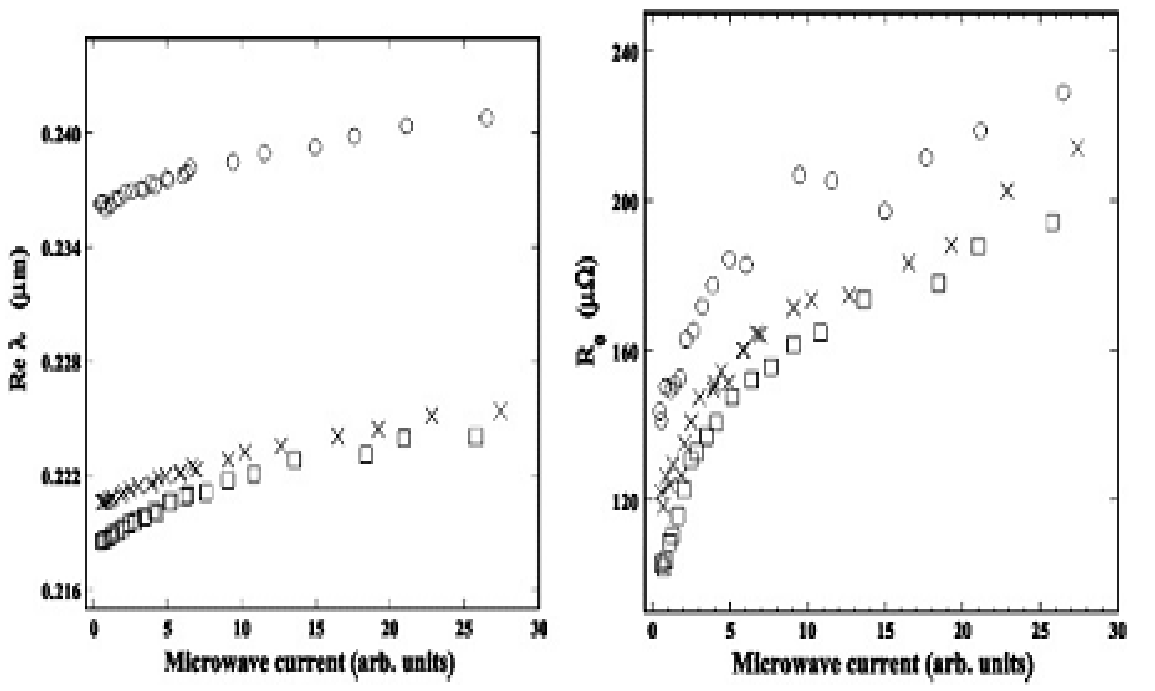

**Fig. 54.** $H_{rf}$ -field dependences of *Rs* and *Xs* for a high quality epitaxial thin film of $YBa_2Cu_3O_{7-\delta}$ in both zero and finite *DC* magnetic fields measured by using the parallel plate resonator at *f* = 5.7 *GHz* and T = 26 *K*: *squares*—zero field; *circles*—0.45*T* parallel to c-axis; *crosses*—0.45 T parallel to ab-plane. Reprinted with permission from M. Tsindlekht *et al., Phys. Rev. B*, vol. **61**, p. 1596, 2000. © 2000 by the American Physical Society (after [526, 46]).

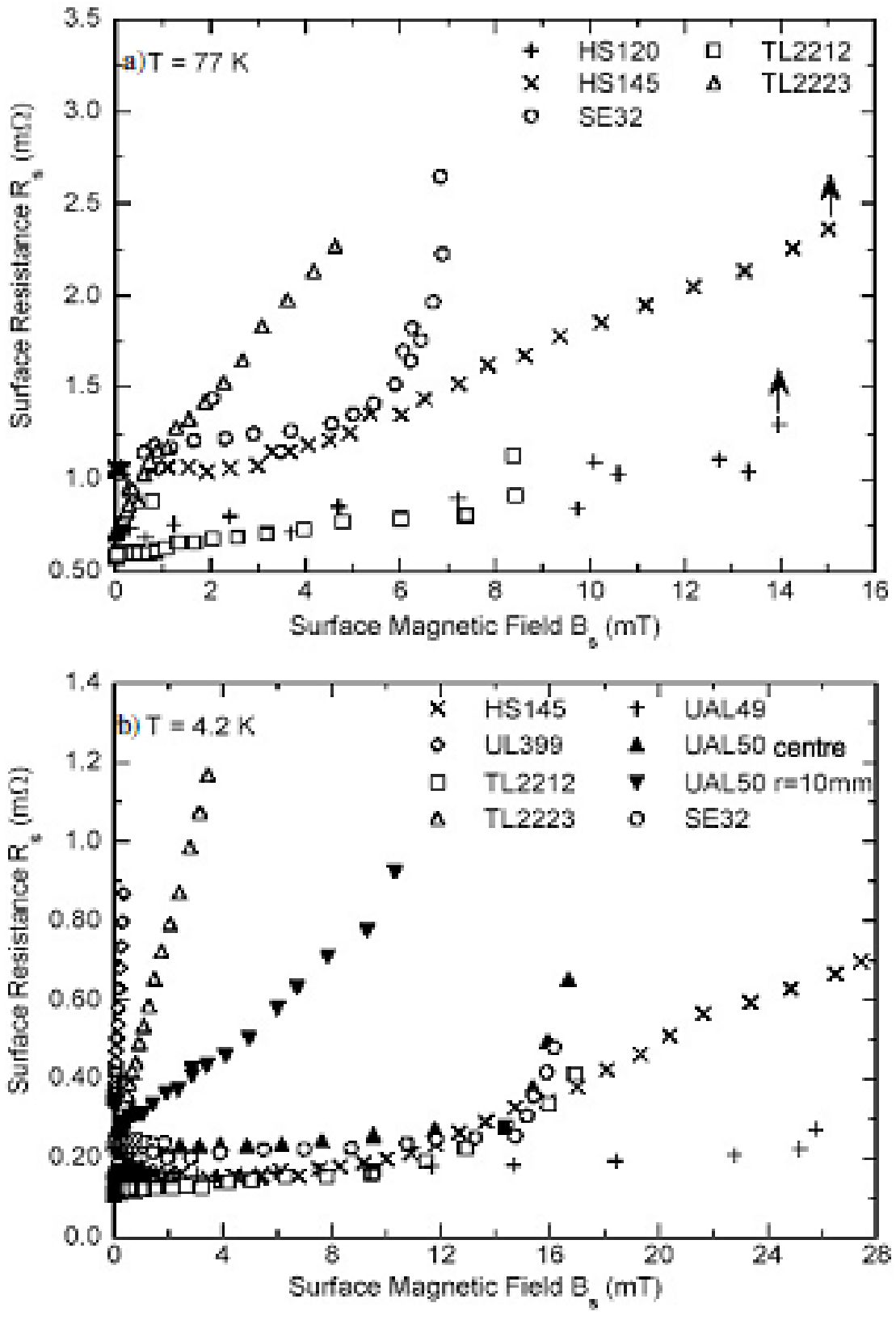

**Fig. 55.** A variety of different power dependences of *Rs* for different films on different substrates measured by using the dielectric resonator at 19 *GHz* and **(a)** 77*K* and **(b)** 4.2 *K* (see the legend in the Figure). Designations used in the Figure are as follows: *HS*— $YBa_2Cu_3O_{7-\delta}$ sputtered on $LaAlO_3$; *UAL*— $YBa_2Cu_3O_{7-\delta}$ laser deposited on $LaAlO_3$; *SE*— $YBa_2Cu_3O_{7-\delta}$ e-beam co-evaporated on MgO; *UM*— $YBa_2Cu_3O_{7-\delta}$ thermally evaporated on $CeO_2/Al_2O_3$; *Tl-2223*—$Tl_2Ba_2Ca_2Cu_3O_8$ two-step process on $LaAlO_3$;*Tl-2212*—$Tl_2Ba_2Ca_1Cu_2O_8$ two-step process on MgO; *UL*— $YBa_2Cu_3O_{7-\delta}$ laser deposited on $CeO_2/Al_2O_3$ [527]. Reprinted by permission from W. Diete, M. Getta, M. Hein, T. Kaiser, G. Muller, H. Piel, H. Schlick, Surface resistance and nonlinear dynamic microwave losses of epitaxial HTS films, *IEEE Trans. Appl. Superconductivity*, vol. **7**, pp.1236-39, 1997. (© 1997 IEEE.) (after [527, 46]).

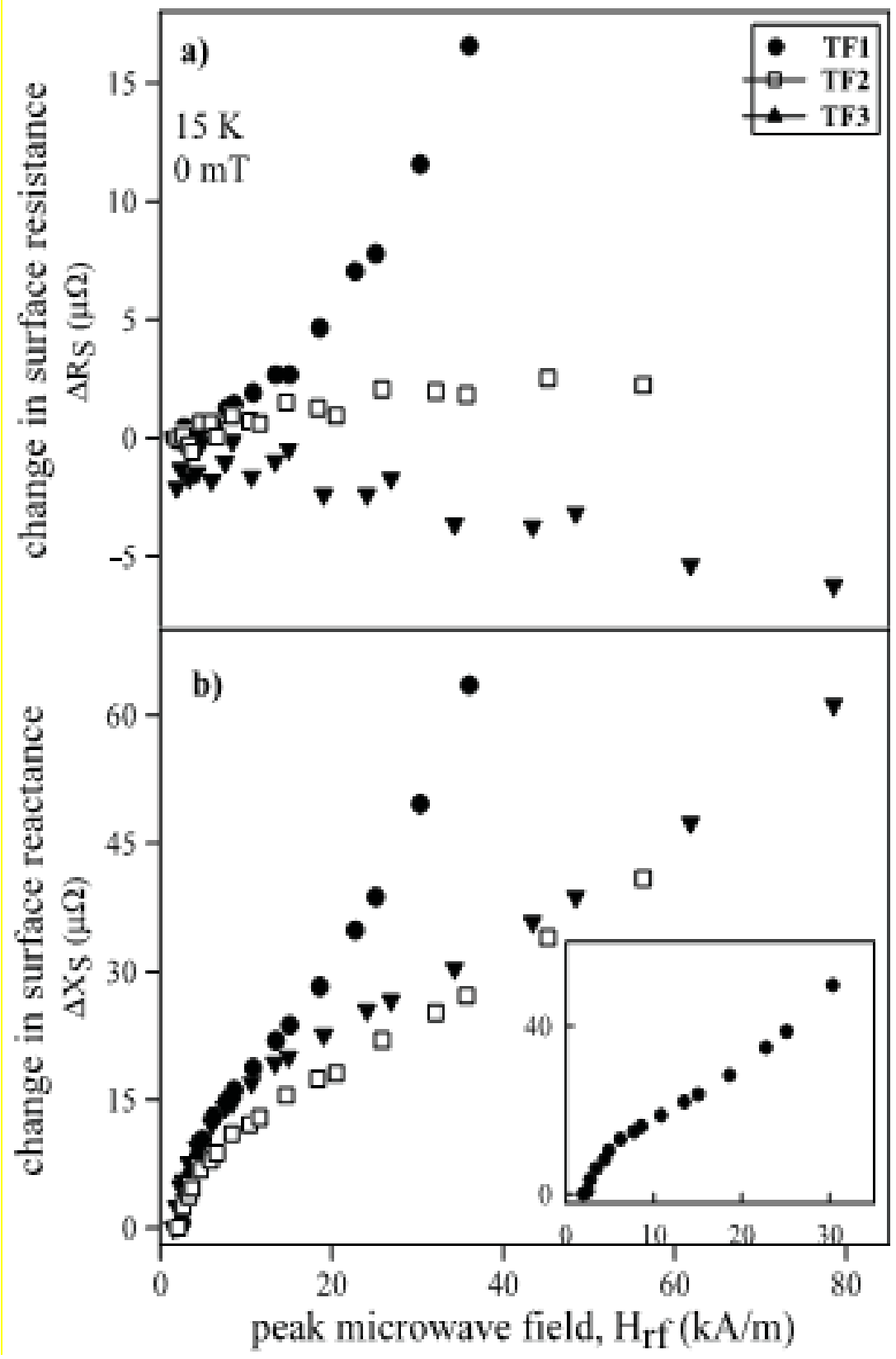

**Fig. 56.** $H_{rf}$ -dependences of the change in *Rs* and *Xs* for three epitaxial e-beam co-evaporated $YBa_2Cu_3O_{7-\delta}$ films on *MgO* substrates. The measurements are made by using the coplanar resonator technique at 8 *GHz* and 15 *K* [528]. Reprinted by permission from A. P. Kharel, B. Soon, J. R. Powell, A. Porch, M. J. Lancaster, A. V. Velichko, R. G. Humphreys, Non linear surface impedance of epitaxial HTS thin films in low dc magnetic fields, *IEEE Transactions on Applied Superconductivity*, vol. **9** (2), pp. 2121-2124, arXiv:cond-mat/9903309v1, 1999.
(© 1999 IEEE.) (after [528, 46]).

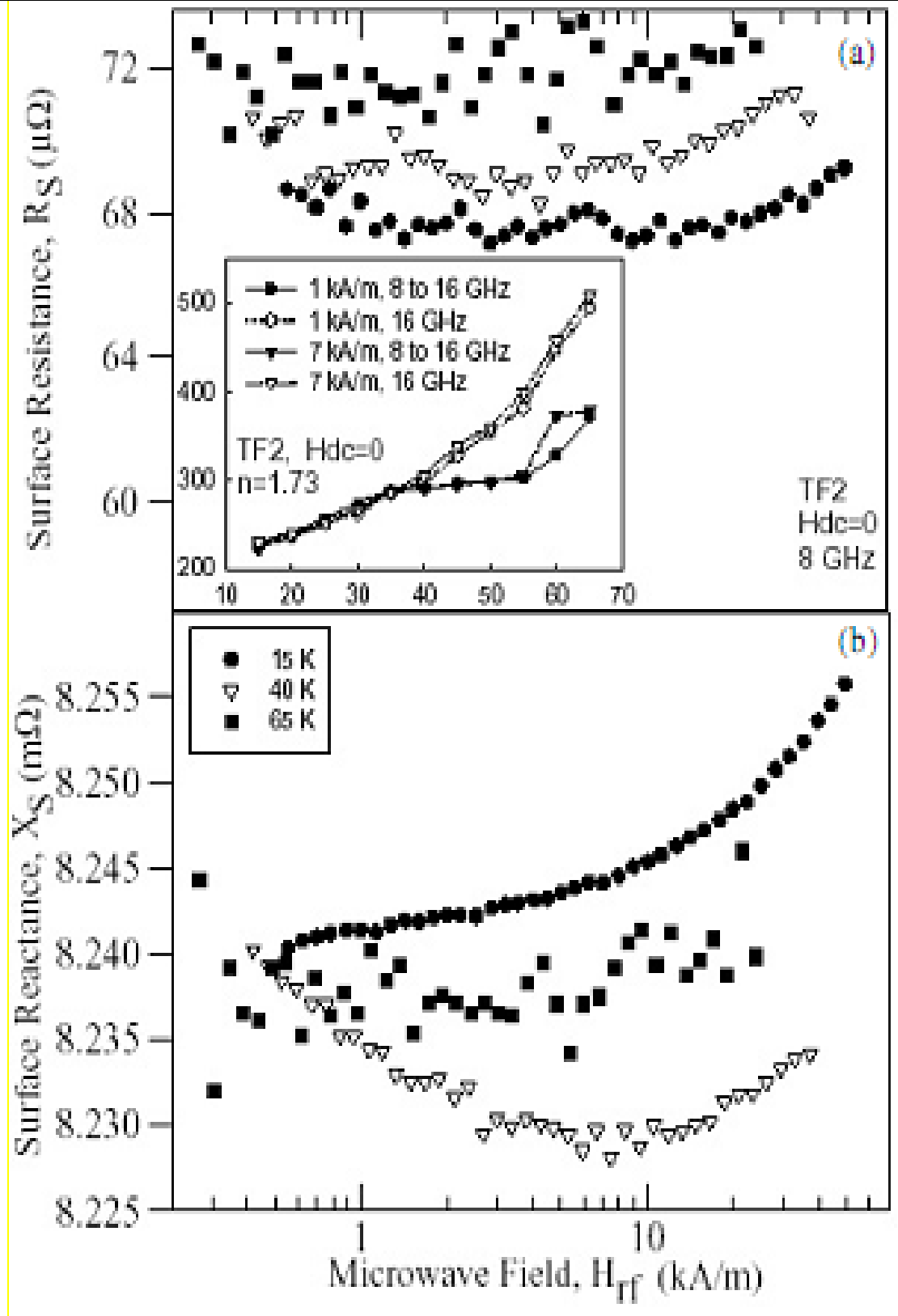

**Fig. 57.** *Rs($H_{rf}$ )* and *Xs($H_{rf}$ )* of a $YBa_2Cu_3O_{7-\delta}$ film on *MgO* similar in quality to that shown in Figure 48 at various temperatures (indicated in the Figure) in zero dc magnetic field. The inset in Figure 50(a) shows the *T* -dependence of *Rs* at low (1$kAm^{-1}$) and high (7 $kA\ m^{-1}$) values of $H_{rf}$ at two different frequencies of 8 *GHz* and 16 *GHz*. The data at 8 *GHz* are scaled to those at 16 *GHz* using the $\omega^n$-scaling law. Scaling exponents are given in the Figures [529]. Reprinted by permission from A. V. Velichko, A. Porch, M. J. Lancaster, R. G. Humphreys, Anomalous Features in Surface Impedance of YBaCuO Thin Films: Dependence on Frequency, RF and DC Fields, *IEEE Trans. Appl. Superconductivity*, vol. **11,** pp. 3497-3500, 2001. (© 2001 IEEE.) (after [529, 46]).

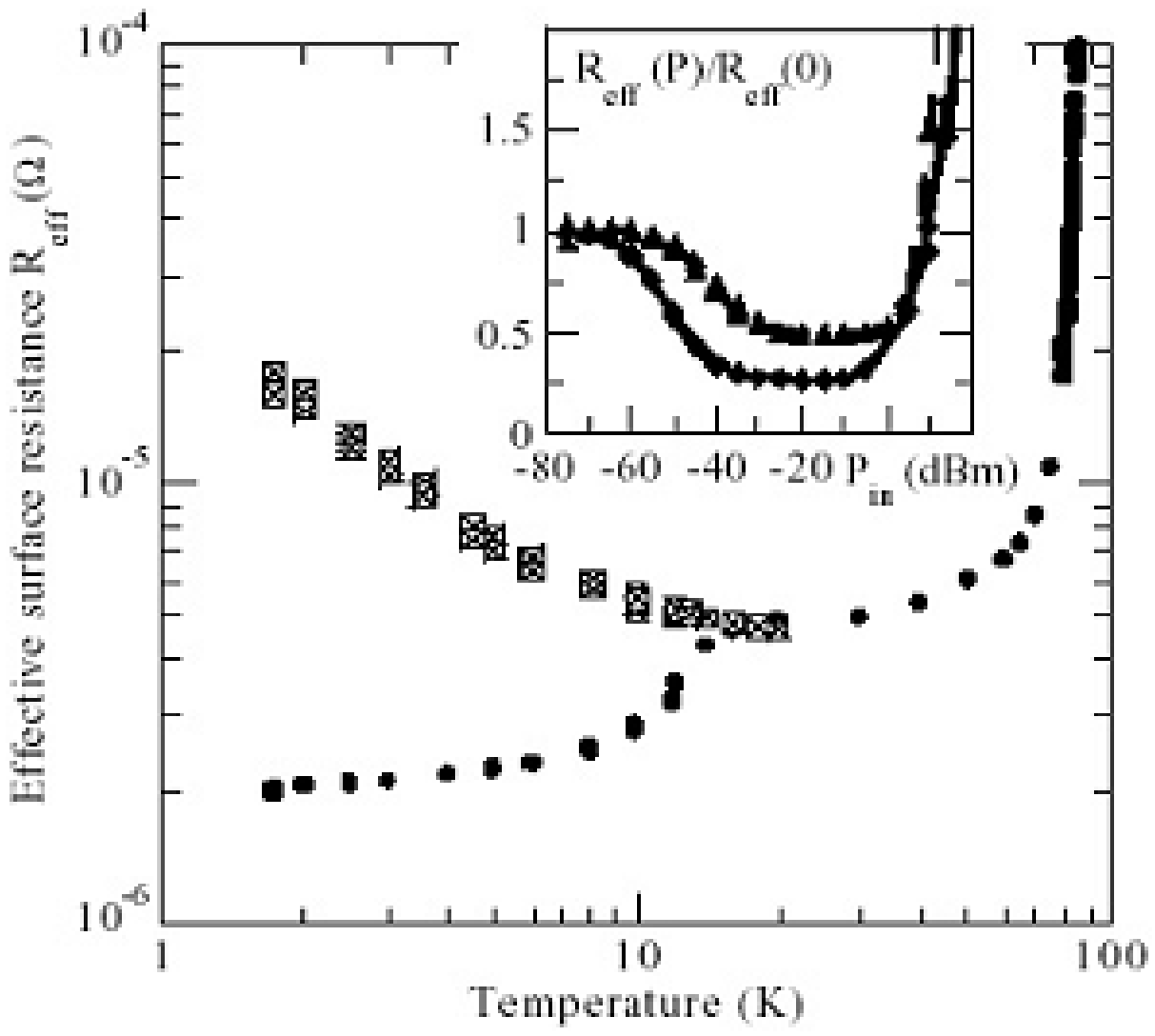

**Fig. 58.** Temperature dependences of the effective *Rs* of an e-beam co-evaporated $YBa_2Cu_3O_{7-\delta}$ film on *MgO* at 2.3 *GHz* and two input power levels: −60 *dBm—hatched squares*, and −20 *dBm—dots*. The inset shows the normalized effective *Rs* for the $YBa_2Cu_3O_{7-\delta}$ film (*diamonds*) and an *Nb* film on *MgO* (triangles) at 5 *K*. The absolute levels of effective *Rs* for the two films at 1.7 *K* were identical. Reprinted by permission from M. A. Hein, P. J. Hirst, R. G. Humphreys, D. E. Oates, A. V. Velichko, Nonlinear Dielectric Microwave Losses in MgO Substrate, *Appl. Phys. Lett.*, vol. **80**, no. 6, pp. 1007-09, 2002. (© 2002 AIP.) (after [536, 46]).

Hein et al. researched the abnormal nonlinear properties of $YBa_2Cu_3O_{7-\delta}$ films on *MgO* substrates in [530-536]. The power and temperature dependent surface impedance along with the two-tone intermodulation distortion were measured in stripline resonators at the several frequencies of 2.3 *GHz* - 11.2 *GHz* and at the temperatures of 1.7 *K* - *Tc*. All the measured films show a decrease of surface resistance *Rs* of up to one order of magnitude at increase of microwave current up to approximately 1*mA* for *T < 20 K*. The surface reactance *Xs* shows the weak increase in the same region. The usual nonlinear increase of *Rs* and *Xs* was observed at high currents in high microwave power range in Fig. 23 [530] In [531], authors proposed that the anomalous response results from dielectric losses in *MgO*, most probably due to the defect dipole relaxation. It was also observed that, above 20 *K*, the nonlinear response of microwave resonators was dominated by superconductor. Clear correlations were found between the nonlinear surface resistance, two-tone intermodulation (*IMD*) and oxygen content of films, which indicates that the superconducting order parameter has influence on the nonlinearities. Below 20 *K*, the dielectric loss tangent of *MgO* substrate dominates the nonlinear response of microwave resonators. At increase of *RF* power, the dissipation losses decrease sharply, and are accompanied by the enhanced *IMD*. Authors attribute these effects to the resonant absorption by the impurity states in *MgO* in [536].

Author of dissertation would like to note that Hein proposed that the well known ***two level systems*** [537] in dielectrics may play an important role in physics of nonlinearities of $YBa_2Cu_3O_{7-\delta}$ on *MgO* substrate system at low temperatures [530-536]. Recently, the two level systems were found in dielectrics of different compounds that lead to the appearance of maximum of absorption of ultra high frequency electromagnetic waves in low temperature superconductors at T<1 *K* in O'Connell *et al.*[538], Gao *et al.* [539]. In [539], it was shown that the thin layer appears in close proximity to the boundary between the superconductor and the dielectric in *Nb* film on sapphire substrate at low temperature, which exhibits the *two level system* properties.

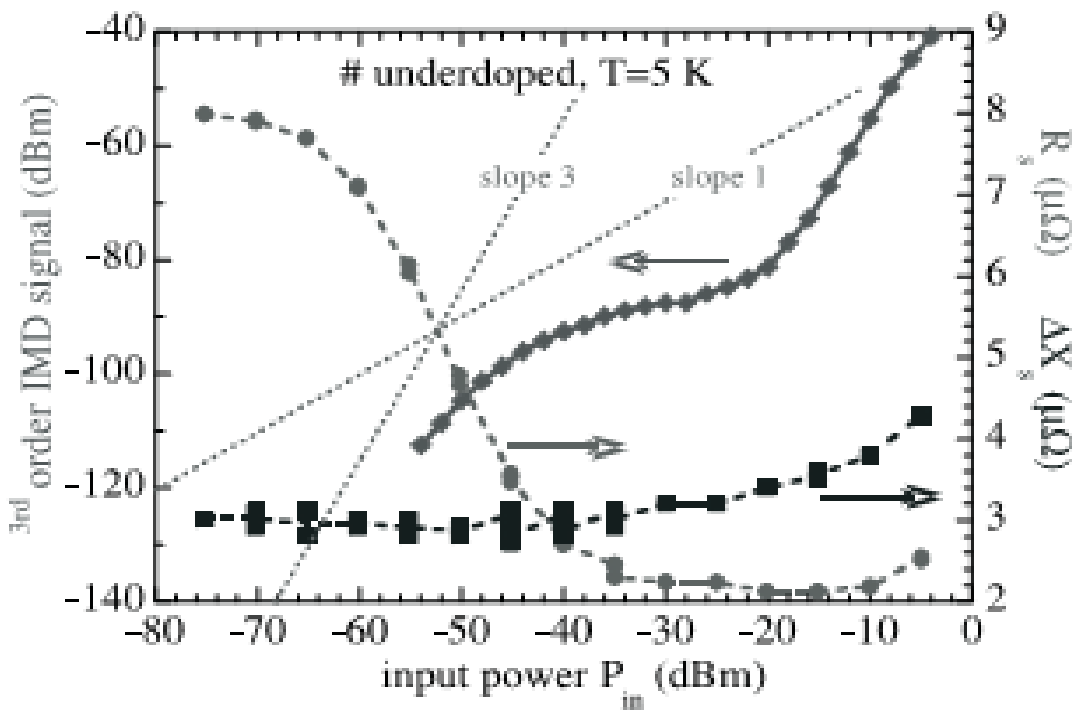

**Fig. 59.** Typical power dependence of the effective *Rs* (circles, dashed curve), *Xs* (squares, dashed curve), and intermodulation distortion (*IMD*) signal (diamonds, solid curve) for $YBa_2Cu_3O_{7-\delta}$ film on *MgO* at 2.3 *GHz* and 5 *K*. Reprinted by permission from M. A. Hein *et al*, *J. Superconductivity,* vol. **16**, p. 895, 2003.. Springer (© 2003 Science and Business Media.) (after [540, 46]).

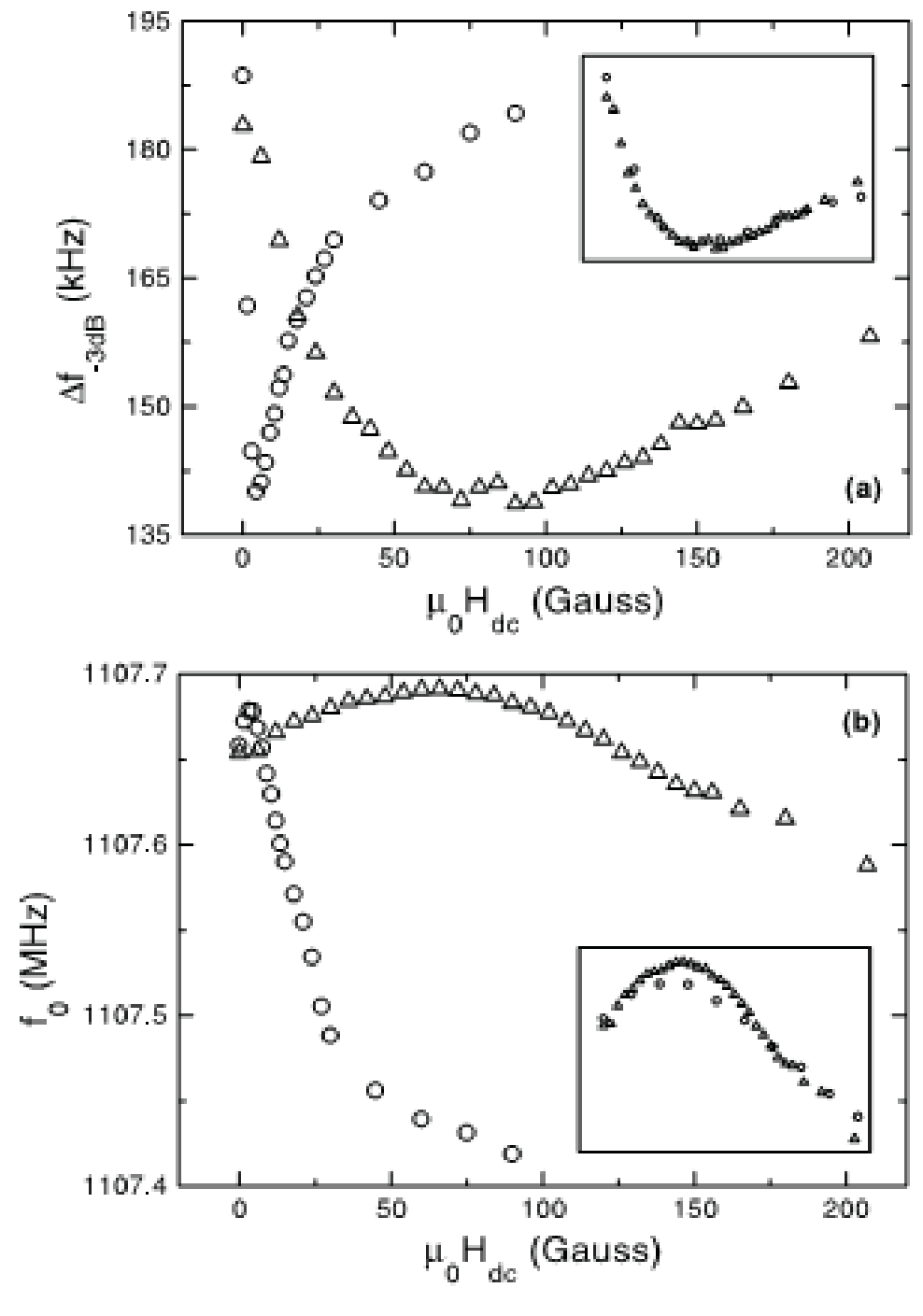

**Fig. 60.** *DC* magnetic field dependence of (a) $-3dB$ bandwidth and (b) the resonant frequency of $YBa_2Cu_3O_{7-\delta}$ microstrip resonator working at fundamental mode (~1.1*GHz*) with input power $P_{mw}=-10$ *dBm* at 77 *K*. The film is deposited by the *PLD* technique on $LaAlO_3$ substrate. Circles correspond to $H_{dc} // c$ and triangles correspond to $H_{dc} \perp c$. The insets show the curves normalized by $H_{dc}^{//}$ and $H_{dc}^{\perp}$, respectively [541]. Reprinted by permission from X. S. Rao, C. K. Ong, B. B. Jin, C. Y. Tan, S. Y. Xu, P. Chen, J. Lee and Y. P. Feng, *Physica C,* vol. **328**, p. 60, 1999. (© 1999 Elsevier) (after [541, 46]).

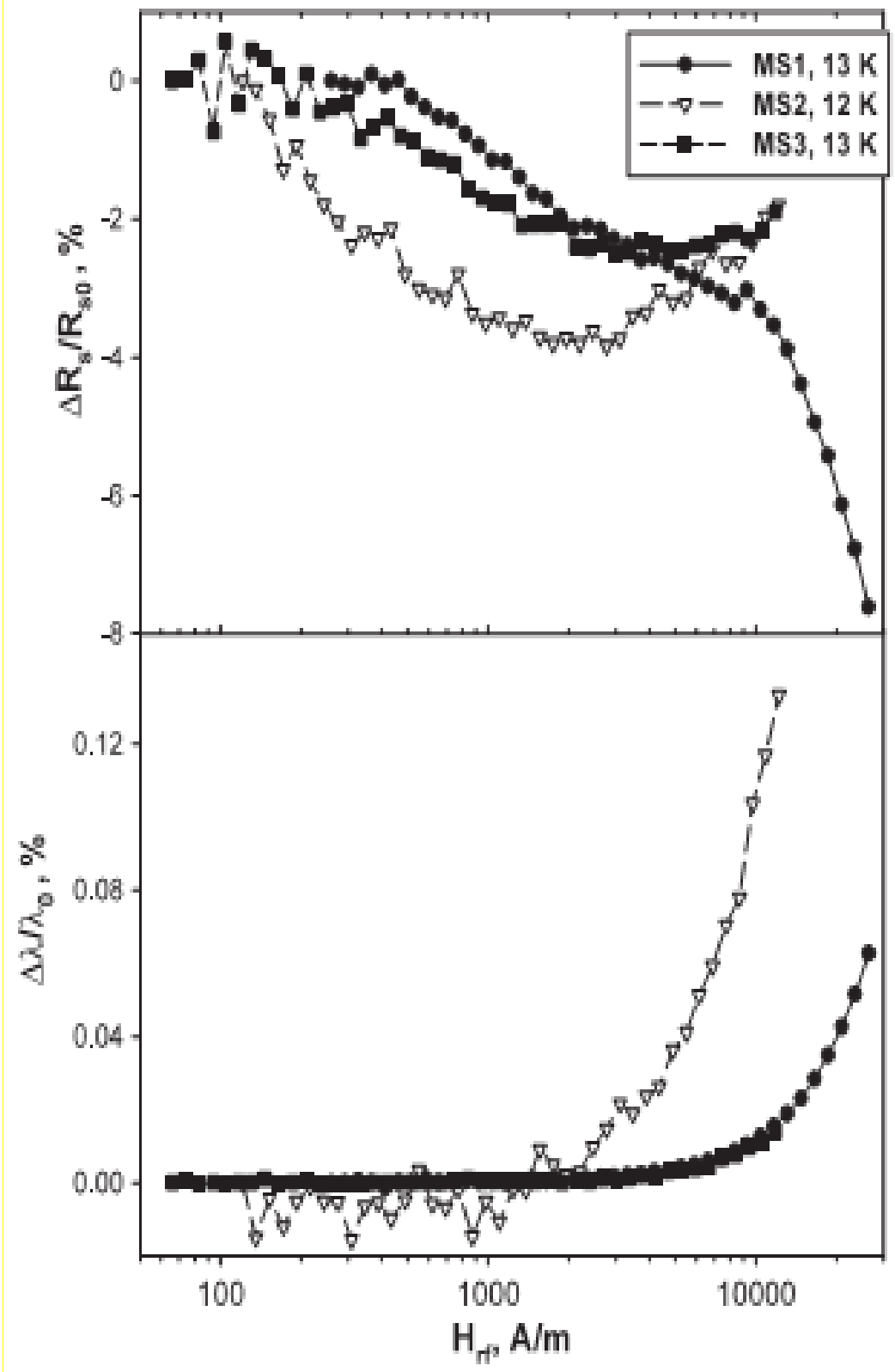

**Fig. 61.** Normalized changes in surface resistance $\Delta Rs = [Rs(H_{rf}) - Rs(0)]/Rs(0)$ and the penetration depth $\Delta\lambda = [\lambda(Hrf) - \lambda(0)]/\lambda(0)$ as a function of the microwave field $H_{rf}$ in zero dc field, 8 *GHz* and 13 *K*, for three $YBa_2Cu_3O_{7\text{-}\delta}$ films thermally evaporated on *MgO* substrates. Typical *Rs* values at the conditions stated above are ~40–50 μΩ [542]. Reprinted by permission from Velichko, Huish, Lancaster, Porch, *IEEE Trans. Appl. Supercond.* 13 3598. (© 2003 IEEE.) (after [542, 46]).

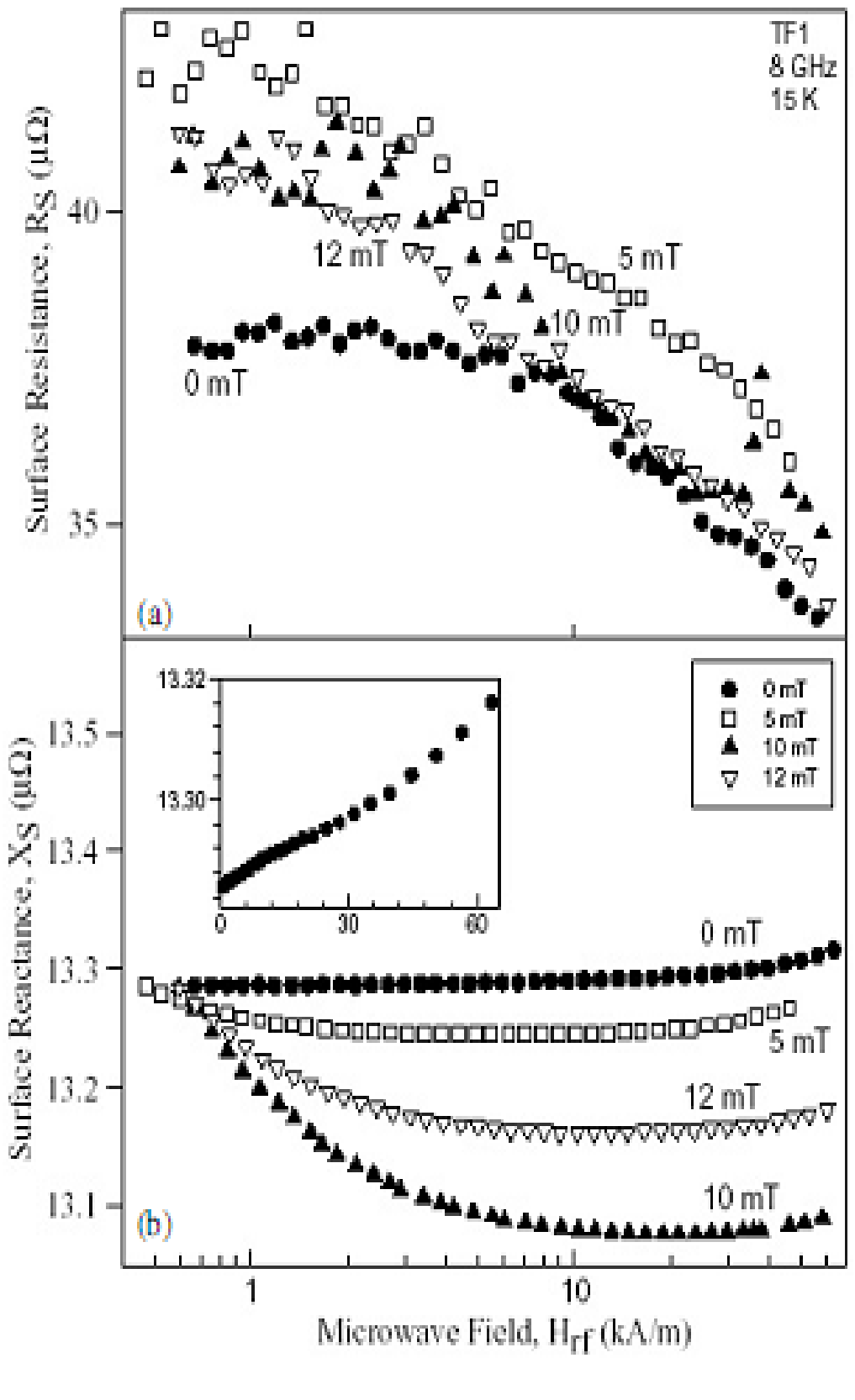

**Fig. 62.** $Rs(H_{rf})$ and $Xs(H_{rf})$ of a $YBa_2Cu_3O_{7\text{-}\delta}$ film on *MgO* at 8 *GHz* and 15 *K* in finite applied dc magnetic field (field cooled regime, dc fields are given in the Figures). The inset in **(b)** shows the 0mT curve on an expanded scale. Reprinted by permission from A. V. Velichko, D. W. Huish, M. J. Lancaster, A. Porch, Anomalies in Nonlinear Microwave Surface Impedance of YBCO Thin Films on MgO: Superconductor versus Substrate Effect, *IEEE Trans. Appl. Superconductivity*, vol. **13,** part 2, pp. 3598-3601, 2003.. (© 2001 IEEE.) (after [542, 46])..

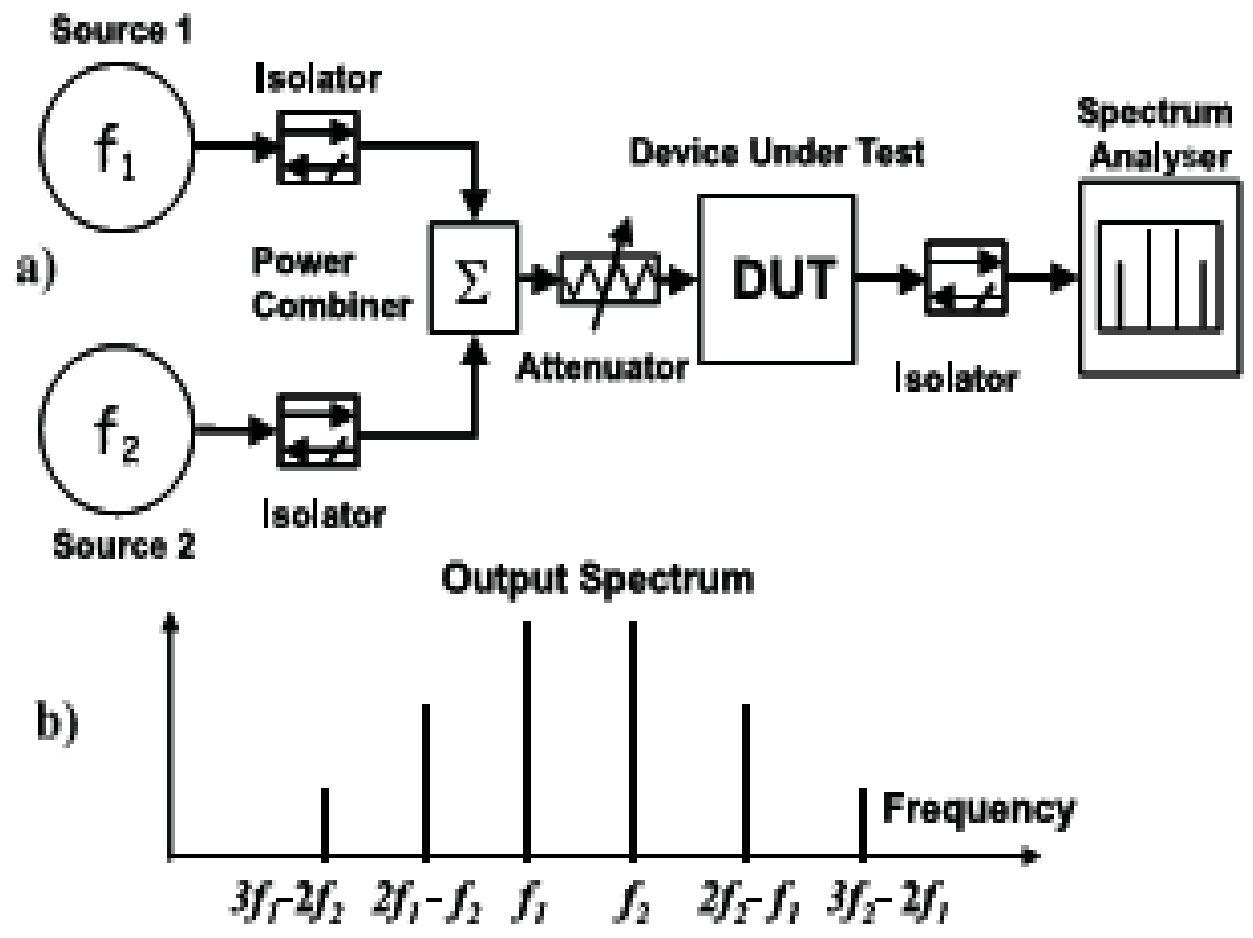

**Fig. 63. (a)** The experimental set-up and **(b)** schematic characteristic spectrum of the *IMD* products as observed in typical *IMD* measurements (after [46]).

The similar experimental set-up was already discussed in research review on the origin of nonlinearities in *HTS* thin films by D. E. Oates in Figs. 20-23 [46].

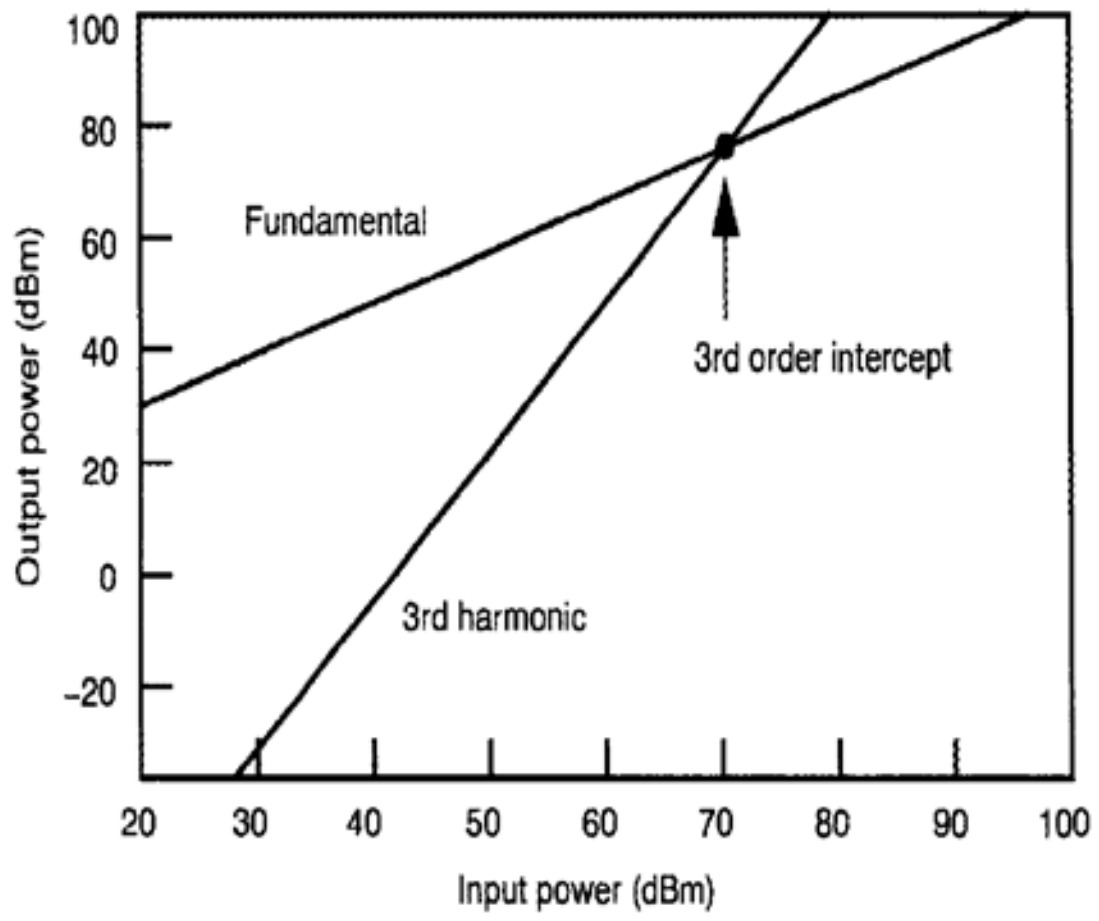

**Fig. 64. (a)** The 3[rd] order intercept plot at a certain fundamental *f* (after [474]).

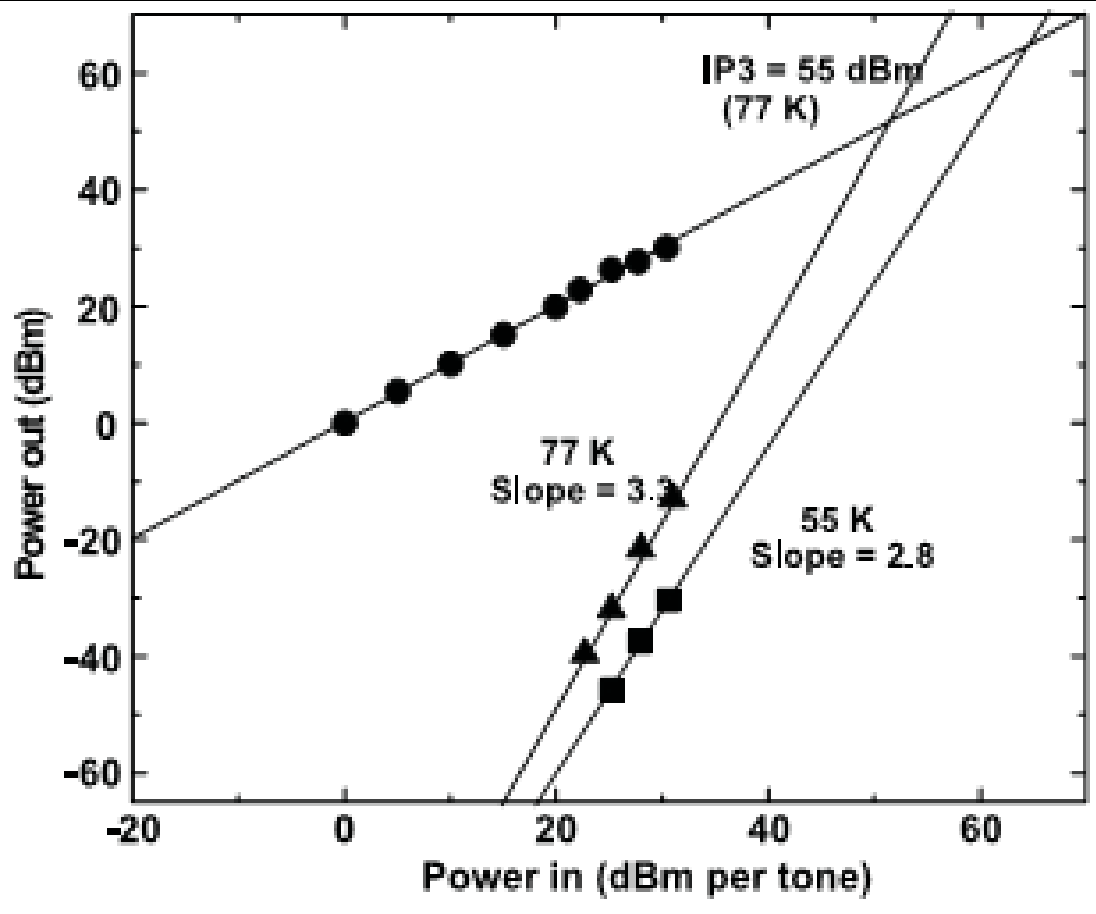

**Fig. 64. (b)** *IMD* measurement of three-pole $YBa_2Cu_3O_{7-\delta}$ band-pass filter with the mid-band frequency of 9.5 *GHz*. All measured powers are per tone. The upper line is the fundamental mode and the lower two lines are the third-order distortion products. Reprinted by permission from T. Yoshitake and S. Tahara, *Appl. Phys. Lett.*, vol. **67**, p. 3963, 1995. (© 1995 American Institute of Physics.) (after [543, 46]).

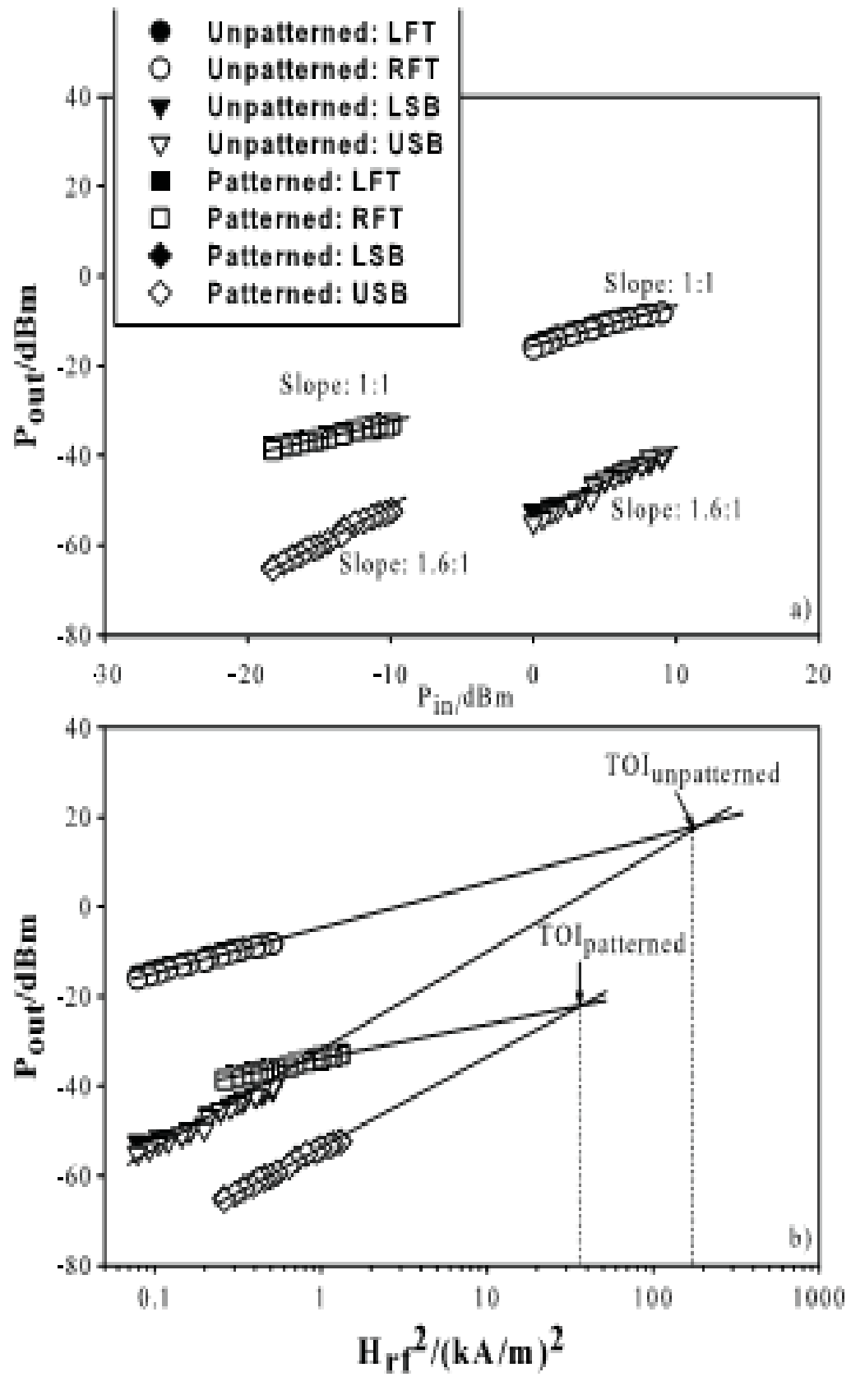

**Fig. 65.** Fundamental tones and third-order *IMD* products versus **(a)** the input power and **(b)** the square of the peak magnetic field in laser-ablated $YBa_2Cu_3O_{7-\delta}$ film on MgO at 12 K for the unpatterned (*f = 10 GHz*) and patterned (*f = 8 GHz*) states. The symbols used in both Figures are the same [544]. Reprinted with permission from M. Abu Bakar, A. V. Velichko, M. J. Lancaster, A. Porch, J. C. Gallop, L. Hao, L. F. Cohen and A. Zhukov, *Physica C*, vol. **372–376**, p. 692, 2002. (© 2002 Elsevier.) (after [544, 46]).

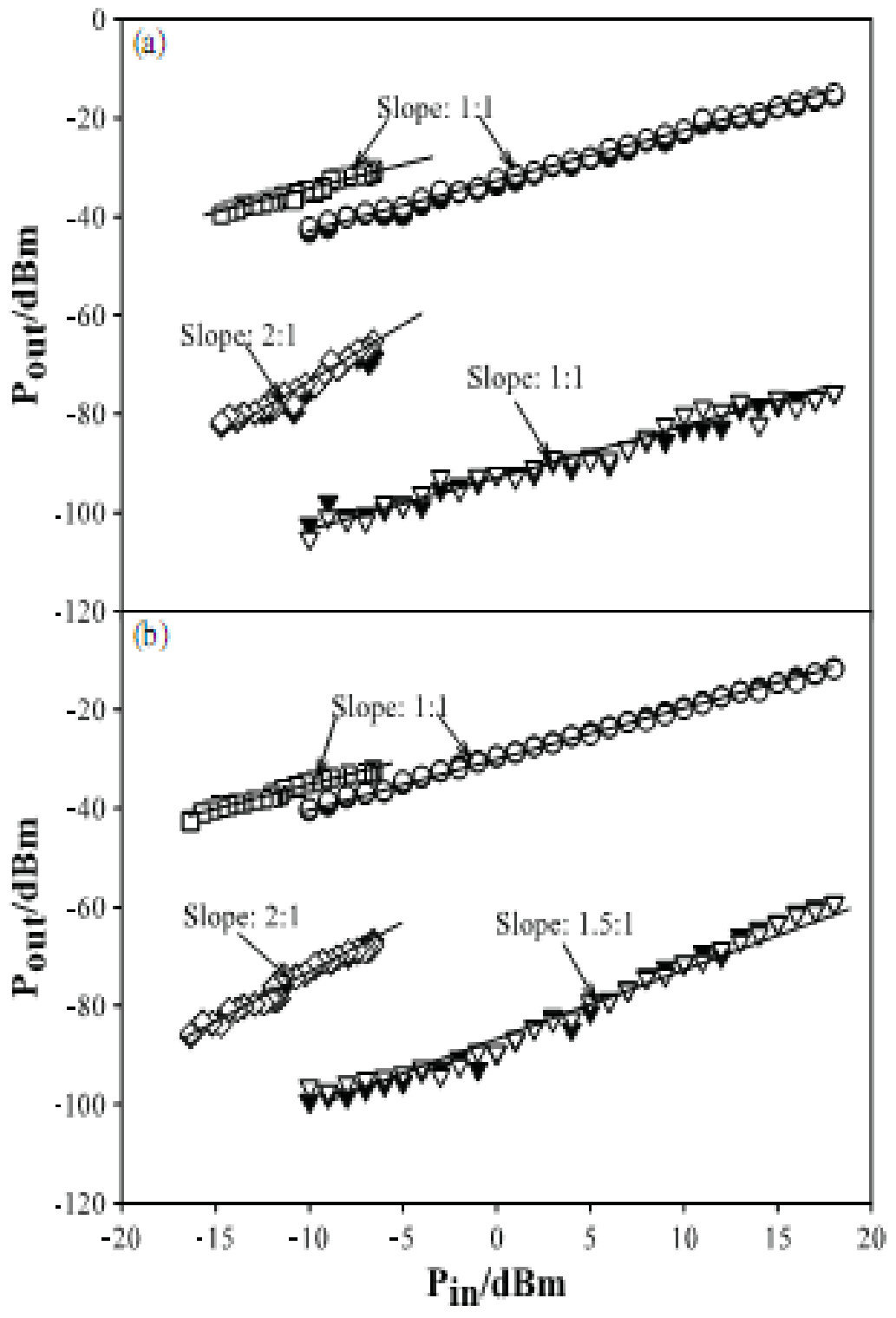

**Fig. 66.** Fundamental tones and third-order *IMD* products versus the input power for a magnetron sputtered $YBa_2Cu_3O_{7-\delta}$ film on $LaAlO_3$, at **(a)** 40 K and **(b)** 60 K for the unpatterned (*f* = 10 GHz) and patterned (*f* = 8 GHz) states. Reprinted with permission from M. Abu Bakar, A. V. Velichko, M. J. Lancaster, A. Porch, J. C. Gallop, L. Hao, L. F. Cohen and A. Zhukov, *Physica C*, vol. **372–376**, p. 692, 2002. (© 2002 Elsevier.) (after [544, 46]).

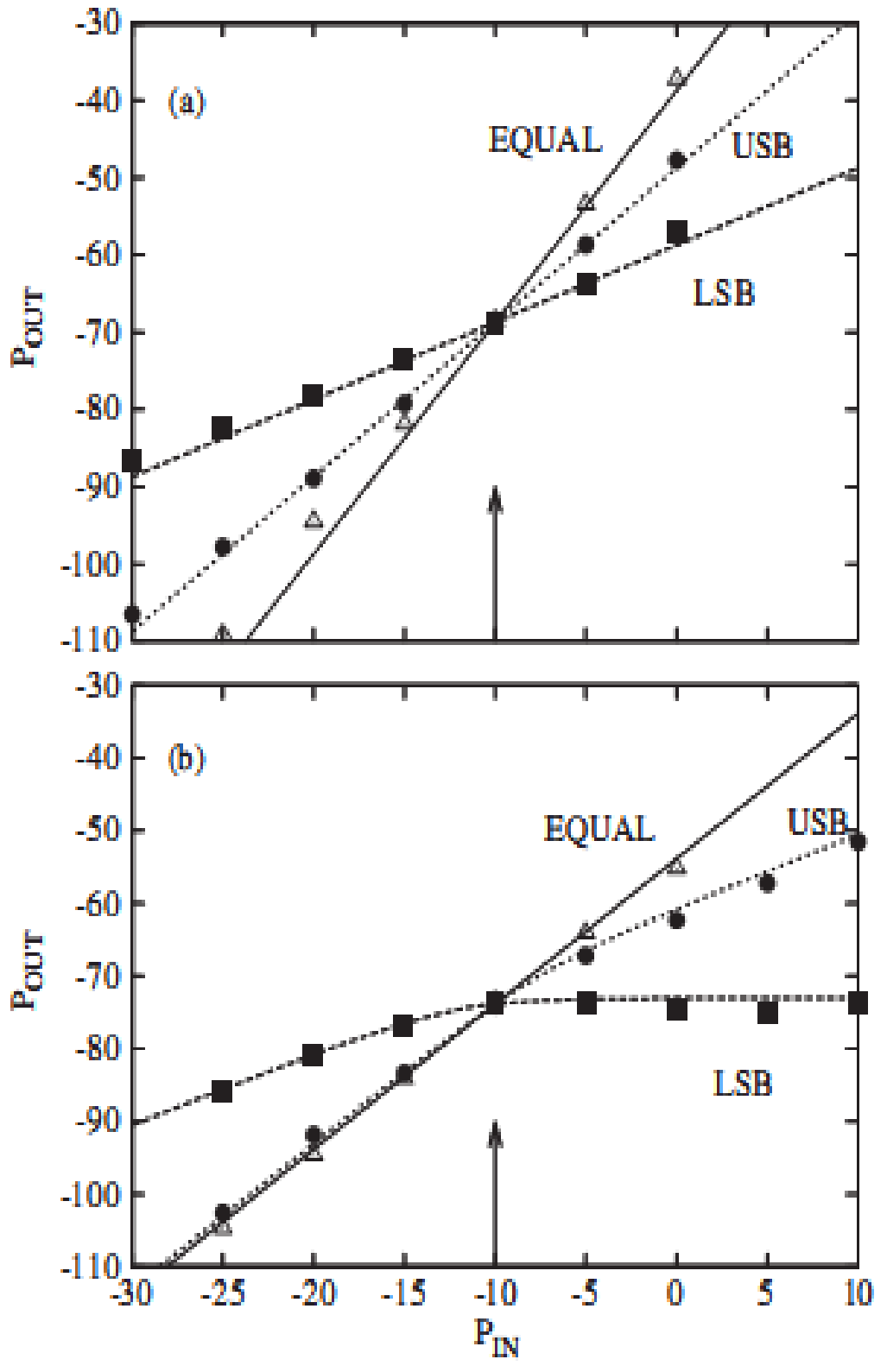

**Fig. 67.** Third-order *IMD* products $P_{out} = P_{2\omega1-\omega2}$ for (a)TBCCO film 1 at 851 *MHz*, 77 *K* and tone separation $\Delta f$ of 30 *KHz* and (b) TBCCO film 2 at 844 *MHz*, 27 K and $\Delta f$ = 30 *KHz*. The triangles represent the case of equal tones $P_{IN} = P_{\omega1}$. The circles and squares represent the *IMD* products for $P_{\omega1} = -10$ *dBm*, represented by the arrow, and varying $P_{IN} = P_{\omega2}$, showing the upper and the lower side band products (USB/LSB) at $2\omega_1 - \omega_2$ and $2\omega_2 - \omega_1$, respectively. The lines show the conventional behaviour in panel (a) and the theoretical results from equation (6) of [545] in panel (b). Note particularly the region in panel (b) where $P_{\omega1} \leq P_{\omega2}$ and $P_{2\omega1-\omega2}$ is independent of $P_{\omega2}$ [545] Reprinted by permission from B. A. Willemsen, K. E. Kihlstrom, T. Dahm, *Appl. Phys. Lett.*, vol. **74**, p. 753, 1999.
(© 1999 American Institute of Physics.) (after [545, 46]).

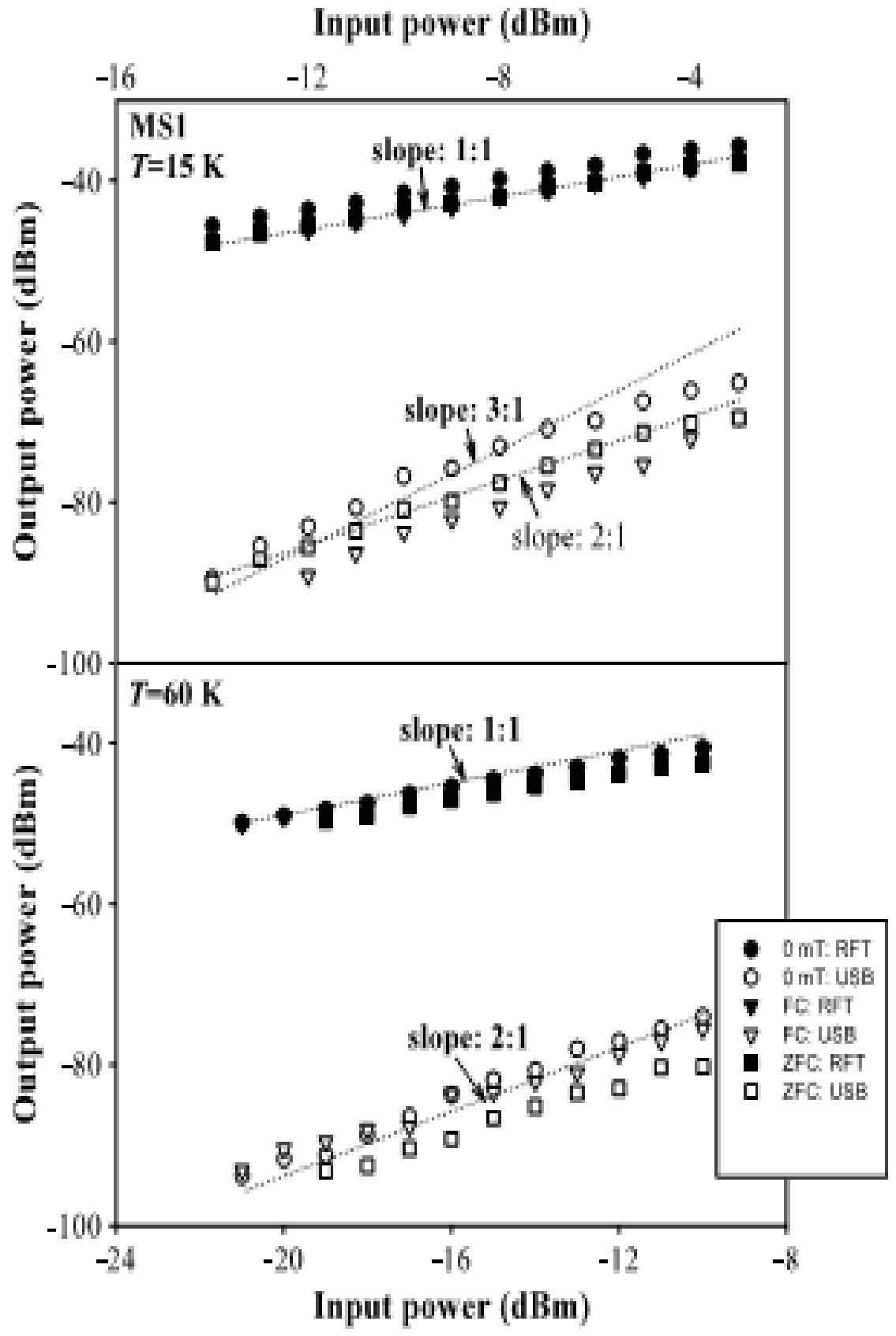

**Fig. 68.** Right fundamental tone (*RFT*) and upper side band (*USB*) third-order *IMD* products versus the input power for a magnetron sputtered $YBa_2Cu_3O_{7-\delta}$ film on $LaAlO_3$ measured by using the coplanar resonator technique at 8 *GHz* and 15 *K* and 60 *K* in zero and non-zero (10 *mT*) dc magnetic fields. Guidelines for 1:1, 2:1 and 3:1 *IMD* scalings are also shown. Reprinted by permission from M. Abu Bakar, A. V. Velichko, M. J. Lancaster, X. Xiong, A. Porch, R. J. Storey, Temperature and Magnetic Field Effects on Microwave Intermodulation in YBCO Films, *IEEE Trans. Appl. Superconductivity*, vol. **13,** part 2, pp. 3581-3584, 2003. (© 2003 IEEE.) (after [546, 46]).

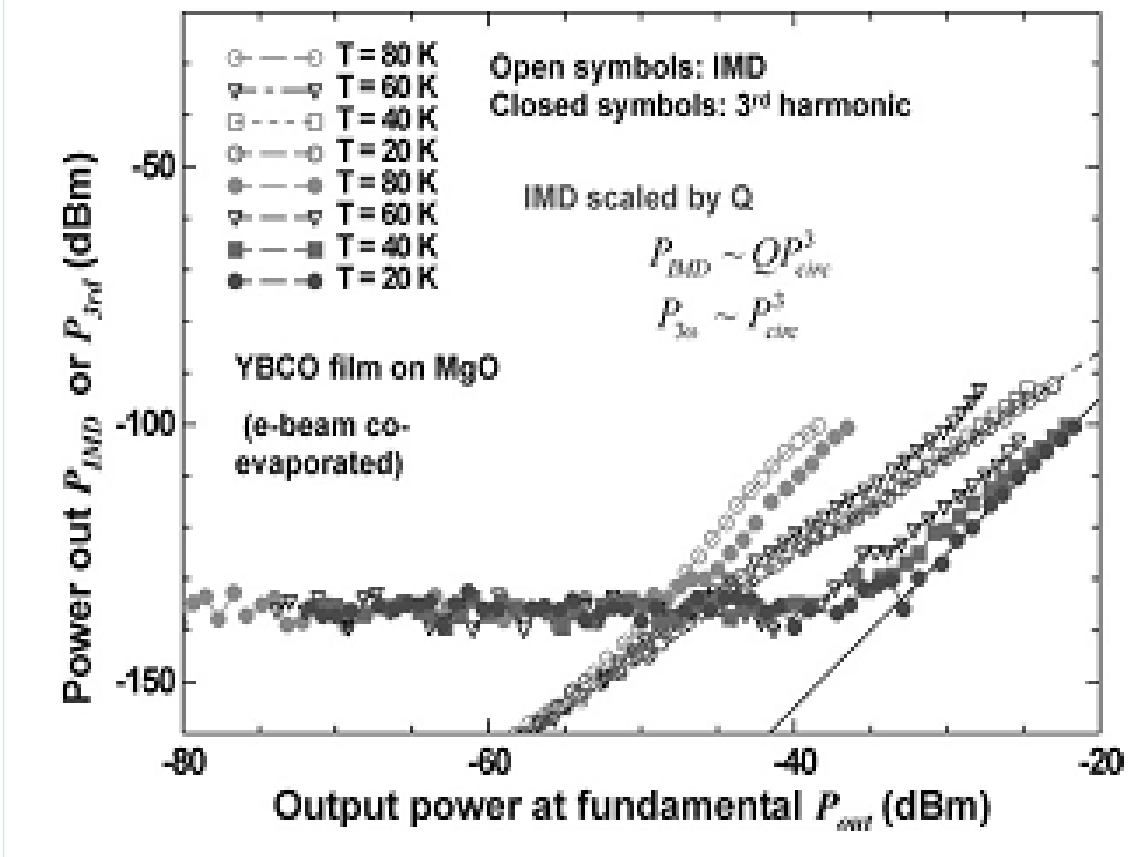

**Fig. 69.** Third-harmonic generation and third-order *IMD* products (*f* = 2.3 *GHz*) as functions of input power for $YBa_2Cu_3O_{7\text{-}\delta}$ e-beam co-evaporated on *MgO*. Measurements are made at various temperatures that are indicated in the Figure. Dashed and solid lines show the slopes of two and three respectively and are given solely for reference. Reprinted by permission from D. E. Oates, S. H. Park, M. A. Hein, P. J. Hirst, R. G. Humphreys, *IEEE Trans. Appl. Superconductivity*, vol. **13**, p. 311, 2003. (© 2003 IEEE.) (after [547, 46]).

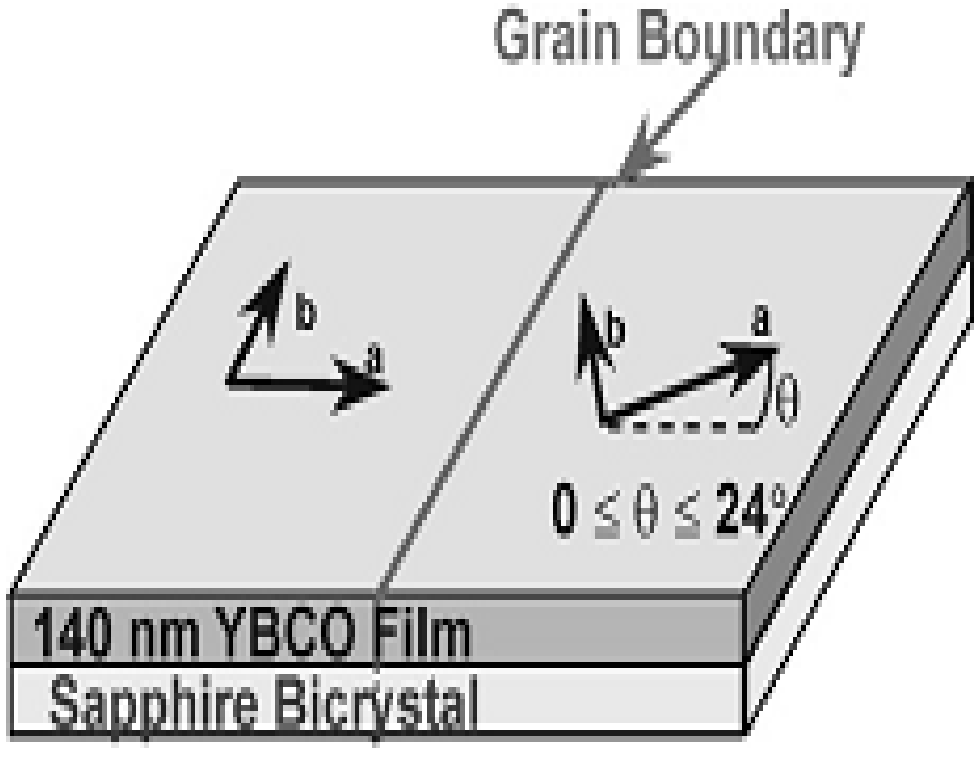

**Fig. 70.** Arrangement of the grain boundary in $YBa_2Cu_3O_{7\text{-}\delta}$ film on bi-crystal sapphire substrate (after [548, 46]).

Velichko, Lancaster, Porch [46] explain that, in order to investigate the influence of weak links (particularly low and high angle grain boundaries) on the nonlinear properties of *HTS* films, Xin, D. E. Oates, G. F. Dresselhaus, M. S. Dresselhaus [548] have measured the microwave response of an artificial $YBa_2Cu_3O_{7\text{-}\delta}$ junction prepared on a sapphire bi-crystal substrate with different misorientation angles ranging from 2° to 24°.The grain boundary (*GB*) was positioned at the centre of the resonant strip, so as only to affect the fundamental mode that has a peak of microwave current at this point. The rest of the films isolated from the effects of the *GB*. The *GB* arrangement is schematically shown in Fig. 63 [46].

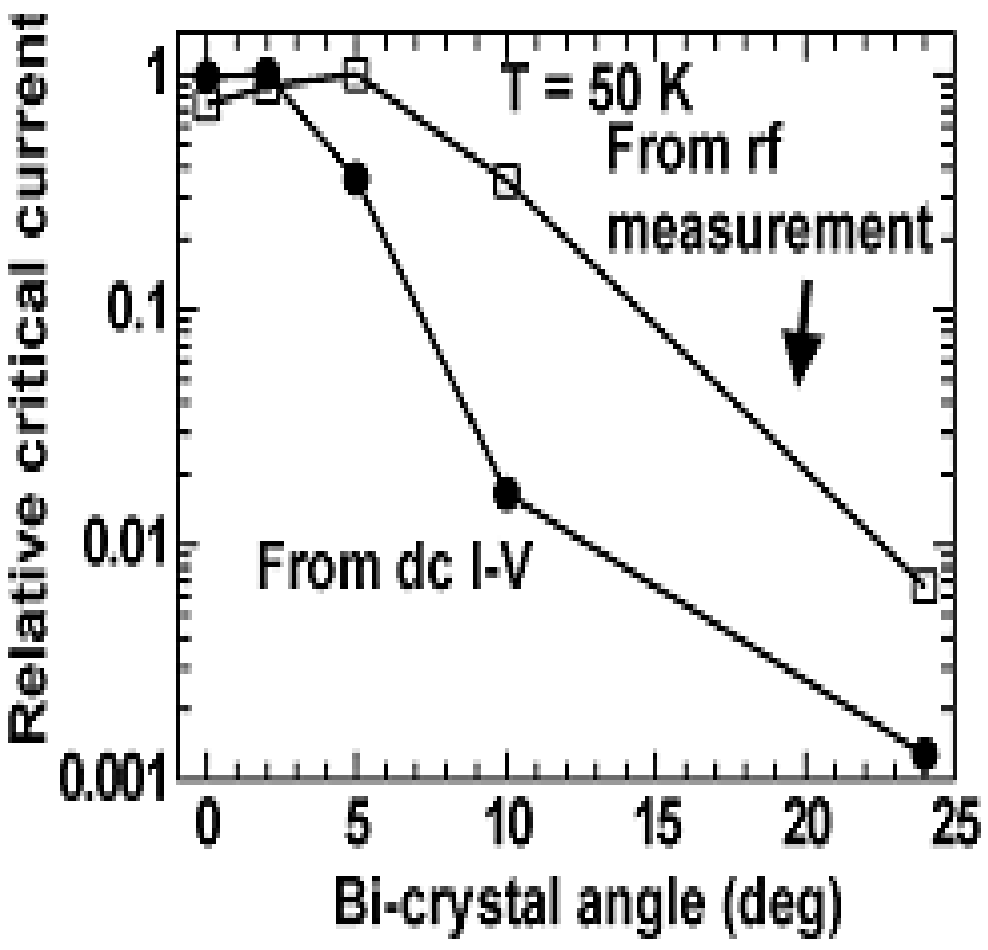

**Fig. 71.** The results for the critical currents as a function of the misorientation angle. The *RF* critical current shows weaker dependence on angle at small angles than the *DC* critical current. Reprinted with permission from D. E. Oates, M. A. Hein, P. J. Hirst, R. G. Humphreys, G. Koren, E. Polturak, Nonlinear microwave surface impedance of YBCO films: latest results and present understanding, *Physica C: Superconductivity*, vol. **372-376,** part 1, pp. 462-468, 2002. (© 2002 Elsevier) (after [549, 46])..

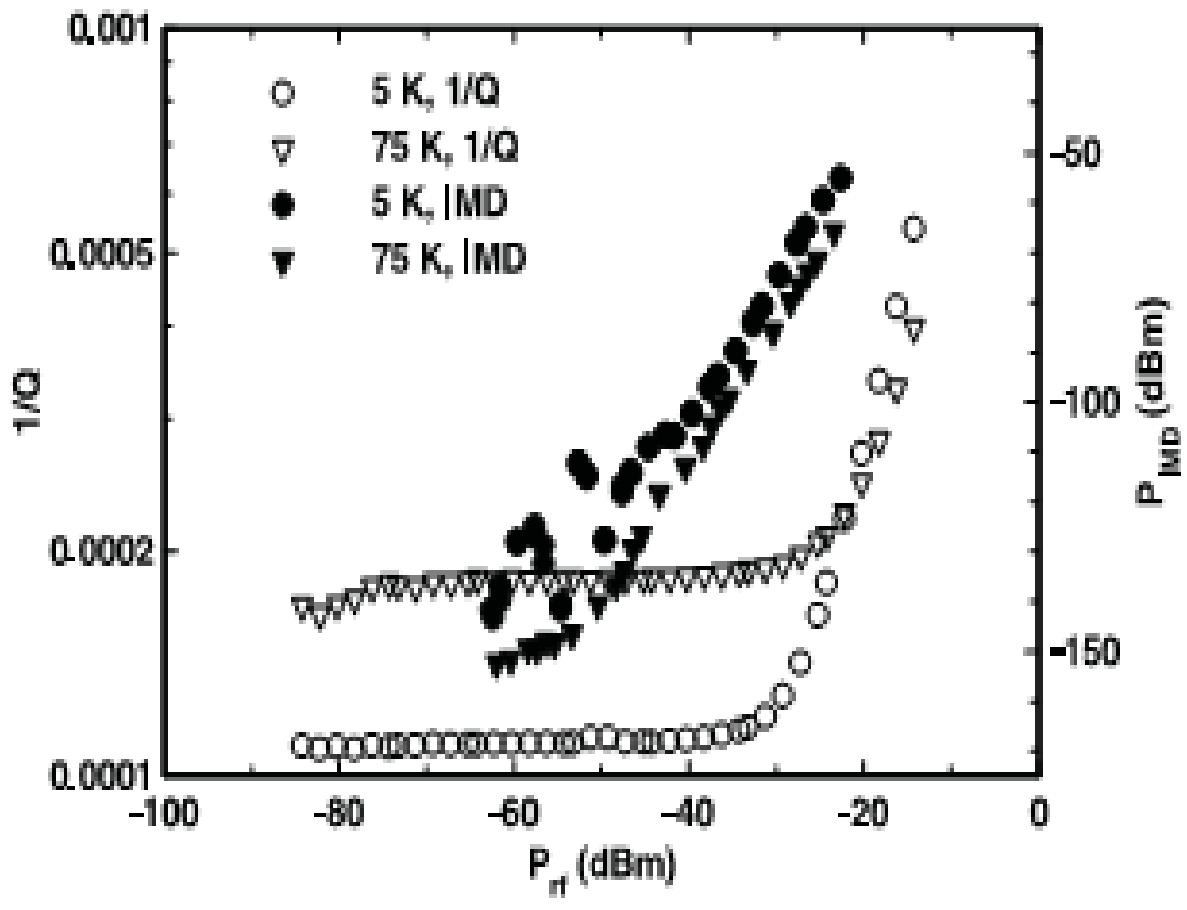

**Fig. 72.** Measured *1/Q* (left scale, open symbols) and $P_{IMD}$ (right scale, filled symbols) of a $YBa_2Cu_3O_{7-\delta}$ microstrip resonator (*f = 4.4 GHz*) as a function of the input microwave power $P_{rf}$ at 5K(circles) and 75 K (inverted triangles). Note that the marginally observable anomalous behaviour of *1/Q* at 5 *K* has a large impact on $P_{IMD}$, while the linear behaviour of *1/Q* at 75 *K* corresponds to the simple slope of three behaviour of $P_{IMD}$ [548]. Reprinted with permission from H. Xin, D. E. Oates, G. F. Dresselhaus, M. S. Dresselhaus, Microwave Intermodulation Distortion in Bicrystal YBCO Grain Boundary Junctions, *Phys. Rev. B,* vol. **65**, 214533, 2002. (© 2002 APS.) (after [548, 46])

The surface resistance and third-order *IMD* signal as a function of input power for a *YBCO* film on sapphire are shown in Fig. 72. It is clearly seen that the *IMD* signal arises at the input powers at which both *Rs* and *Xs* are still flat. This indicates that *IMD* measurements are a much more sensitive tool for studying nonlinear effects in HTS films [46].

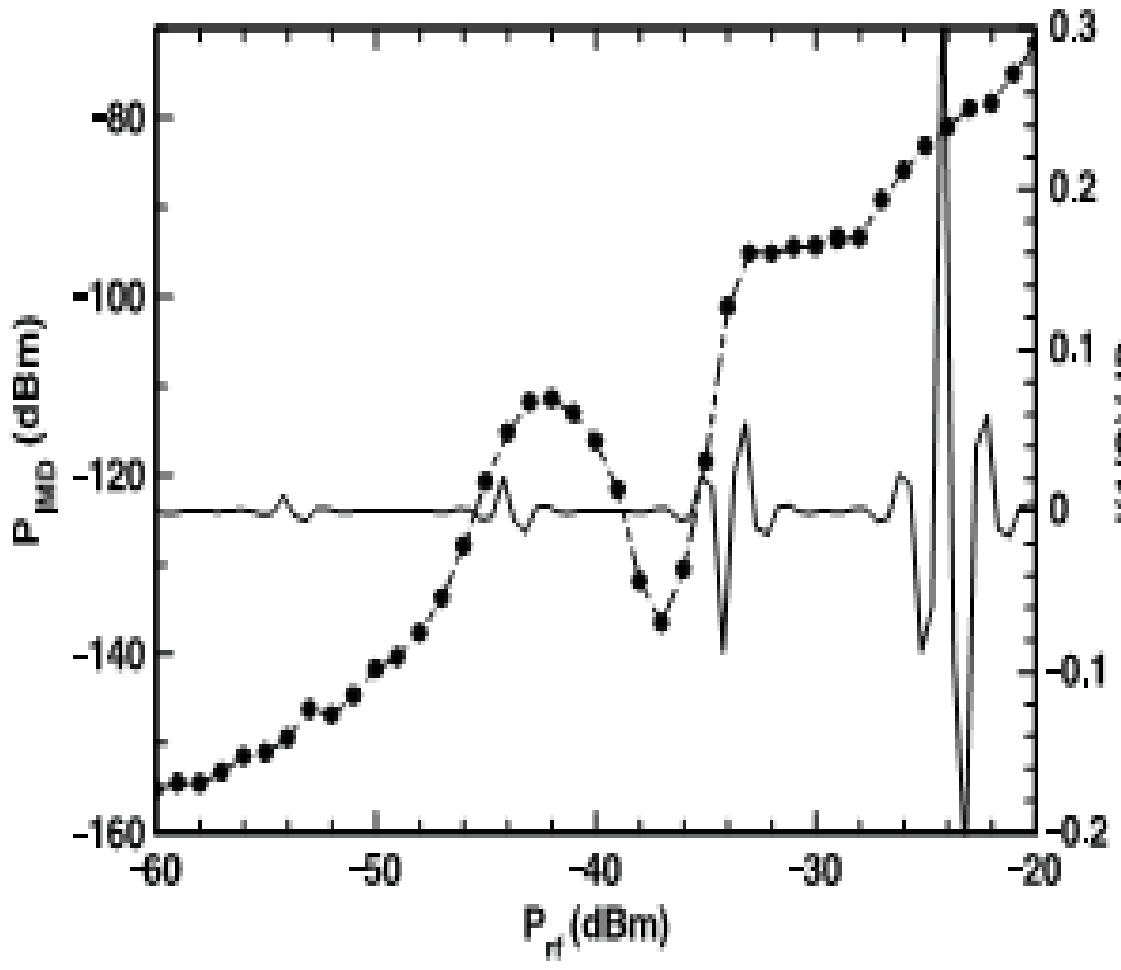

**Fig. 73.** Comparison of the measured $P_{IMD}$ (filled circles) and the first derivative of *1/Q* (solid curve) with respect to $P_{rf}$ as a function of the input power $P_{rf}$ for a 10° grain boundary at 4.4 *GHz* and 55 *K*. Notice that the structure seen in the *IMD* power corresponds to some of the features in the magnitude of $d(1/Q)/dP_{rf}$. Reprinted with permission from H. Xin, D. E. Oates, G. F. Dresselhaus, M. S. Dresselhaus, Microwave Intermodulation Distortion in Bicrystal YBCO Grain Boundary Junctions, *Phys. Rev. B*, vol. **65**, 214533, 2002. (© 2002 American Physical Society) (after [548, 46]).

Fig. 66 demonstrates the power dependence of the *IMD* signal together with $\partial Rs/\partial Pin$ for the same film as in Fig. 72. One can see that peculiarities in the *IMD* signal correlate well with similar features in $\partial Rs/\partial Pin$, signifying that the *IMD* signal is proportional to the derivative of *Rs* with respect to $P_{in}$. This explains why power-dependent *IMD* measurements are much more sensitive than nonlinear surface impedance measurements.

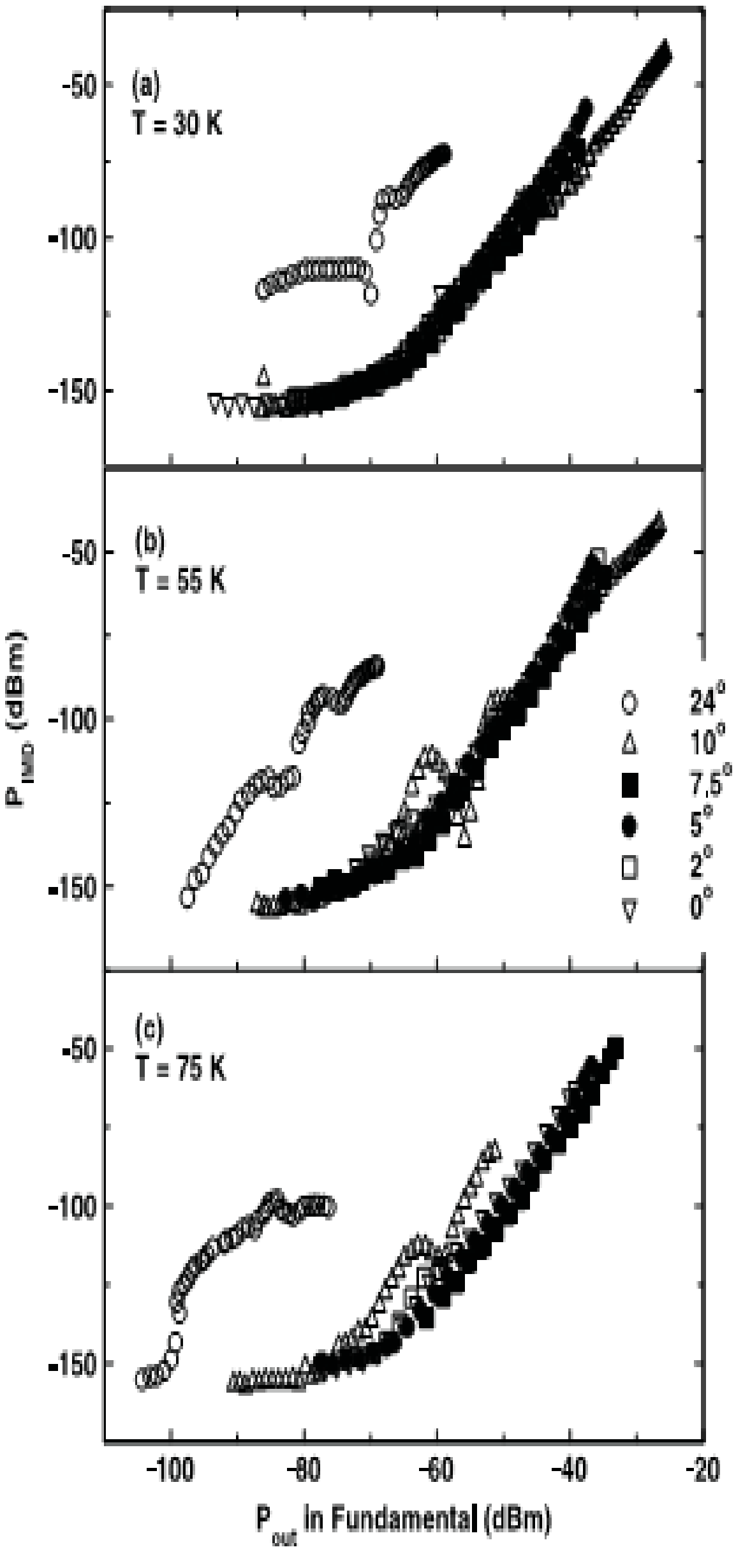

**Fig. 74.** Third-order *IMD* power as a function of the fundamental output power $P_{out}$ at 4.4 *GHz* and three different temperatures (a) 30 *K*, (b) 55 *K*, and (c) 75 *K* for the $YBa_2Cu_3O_{7-\delta}$ stripline resonators on sapphire bi-crystal substrate with different misorientation angles: open inverted triangles 0° (plain $YBa_2Cu_3O_{7-\delta}$ film on a single crystal sapphire substrate); *open squares*, 2°; *filled circles*, 5°; *filled squares*, 7.5°; *open triangles*, 10°; *open circles*, 24°. Reprinted with permission from H. Xin, D. E. Oates, G. F. Dresselhaus, M. S. Dresselhaus, Microwave Intermodulation Distortion in Bicrystal YBCO Grain Boundary Junctions, *Phys. Rev. B*, vol. **65**, 214533, 2002. (© 2002 American Physical Society) (after [548, 46]).

The microwave power dependence of the *IMD* products for a *YBCO* stripline resonator on a sapphire bi-crystal with misorientation angles varying from 0° to 24° at 4.4 *GHz* and 30, 55 and 75 *K* are shown in Fig. 74. At 30 *K* only, the 24° *GB* shows a difference from the rest of the film. At 55*K* and especially at 75*K* both 10° and 24° *GBs* deviate from the bulk of the film notably. This signifies a transition from a strongly coupled single crystal to weakly coupled *RSJ* (resistively shunted junction) behaviour for 10° *GBs* with *T* increased from 30 to 55 and 75 *K*. Oates concluded that low angle grain boundaries with the grain misorientation angle of less than 10° are not the source of microwave nonlinearity [46].

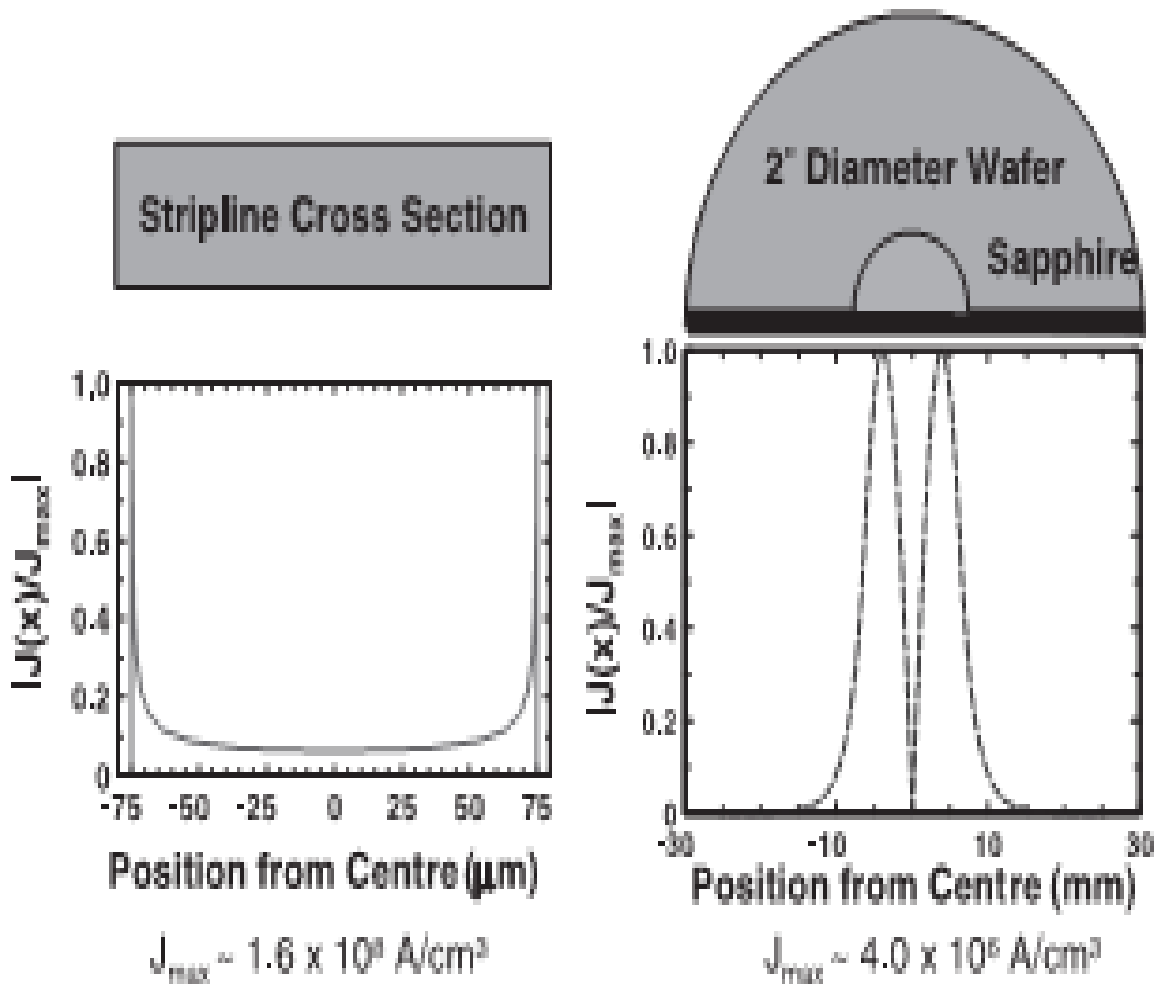

**Fig. 75.** Current distribution of both stripline resonator (left-hand side) and dielectric resonator (right-hand side). Reprinted by permission from H. Xin, D. E. Oates, A. C. Anderson, R. L. Slattery, G. Dresselhaus and M. S. Dresselhaus, *IEEE Trans. Microwave Theory Tech.*, vol. **48**, p. 1221, 2000. (© 2000 IEEE.) (after [550, 551, 46]).

The stripline resonator has peaks of microwave current at the edges of the strip with very low current in the middle of the strip widthwise. This means that if the edges are affected by the patterning then those changes would clearly transpire in the measurements. A dielectric resonator, on the other hand, has a radial peak of microwave current in close proximity to the centre of the disc, and virtually no current near the circumference of the disc.

The effect of patterning has been studied on a number of films and typical results of the power dependences of *Rs* measured with the two methods at 10.7 *GHz* and 75 *K* are shown in Fig. 75. It is seen that both methods give qualitatively the same result, with the onset of the nonlinearity taking place at about the same $H_{rf}$ for both resonators. Taking into account the drastically different field distribution for the two resonators, the authors concluded that ***patterning did not have any noticeable effect on the nonlinear properties of the film***. Besides, because in the case of the stripline $H_{rf}$ is mostly perpendicular to the film (particularly at the edges) and therefore promotes vortex penetration due to strong demagnetization effect, whereas for the dielectric resonator it is mostly parallel to the film surface with no or very little demagnetization effect, Xin *et al.* [550, 551] have also concluded that vortex penetration is not the mechanism responsible for the nonlinearity in *YBCO* films at low and intermediate (up to ~ 100 Oe) $H_{rf}$ –values [46].

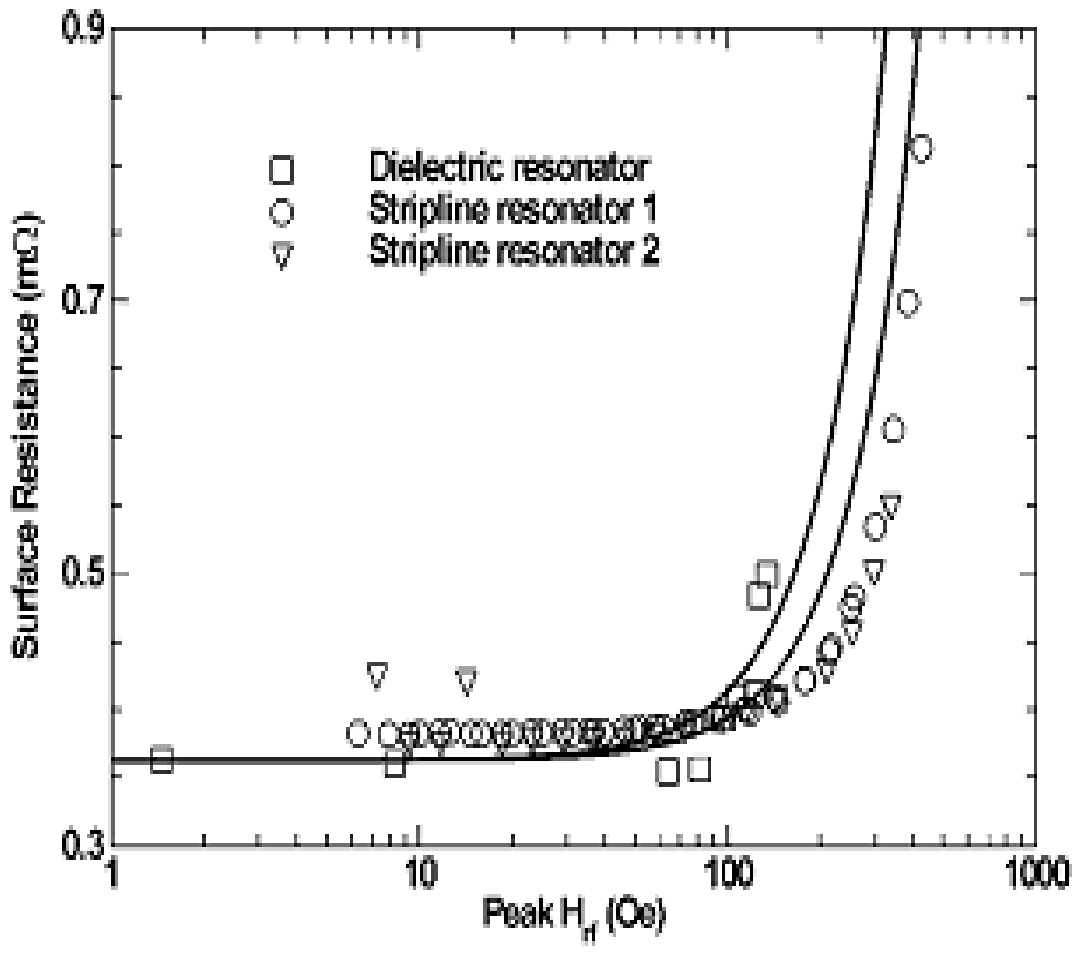

**Fig. 76.** *Rs* versus *RF* peak magnetic field $H_{rf}$ measurements in the dielectric resonator and the stripline resonator using the same $YBa_2Cu_3O_{7-\delta}$ film on sapphire for both resonators at 75 *K* and 10.7 *GHz*. The solid curves are computed curves using the model pre. Reprinted by permission from H. Xin, D. E. Oates, A. C. Anderson, R. L. Slattery, G. Dresselhaus and M. S. Dresselhaus, *IEEE Trans. Microwave Theory Tech.*, vol. **48**, p. 1221, 2000. (© 2000 IEEE.) (after [550, 551, 46]).

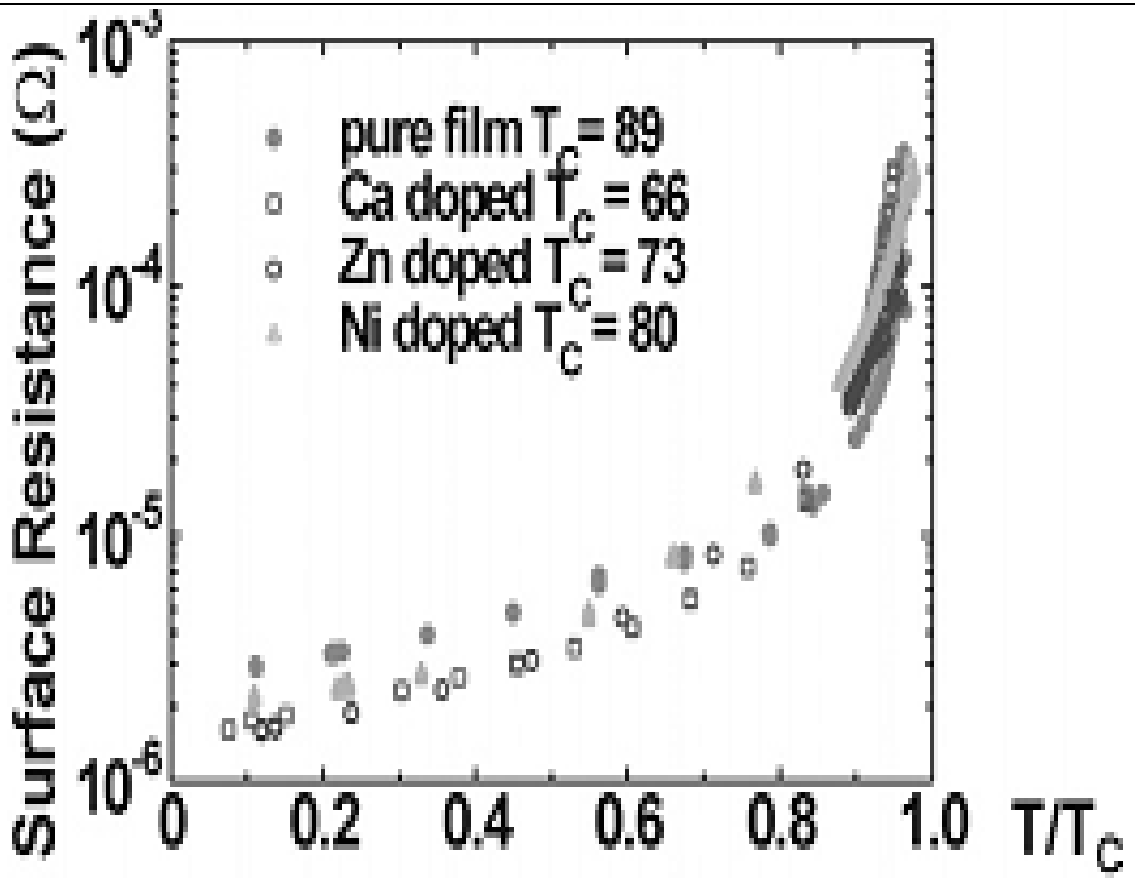

**Fig. 77.** Temperature dependence of *Rs* for $YBa_2Cu_3O_{7-\delta}$ films on *MgO* doped with *Ca*, *Ni* and *Zn*, obtained by using the stripline resonator at 1.5 *GHz* [549]. Reprinted from with permission from D. E. Oates, M. A. Hein, P. J. Hirst, R. G. Humphreys, G. Koren, E. Polturak, Nonlinear microwave surface impedance of YBCO films: latest results and present understanding, *Physica C: Superconductivity*, vol. **372-376,** part 1, pp. 462-468, 2002. (© 2002 Elsevier.) (after [549, 46]).

As is clear from the Figure, the ***doping*** effect is very weak in the films as opposed to that in single crystals. This is because the films are much more disordered compared to crystals and the scattering is already strong. Thus no effect of doping is seen for the low power surface resistance when plotted as a function of the reduced temperature [46].

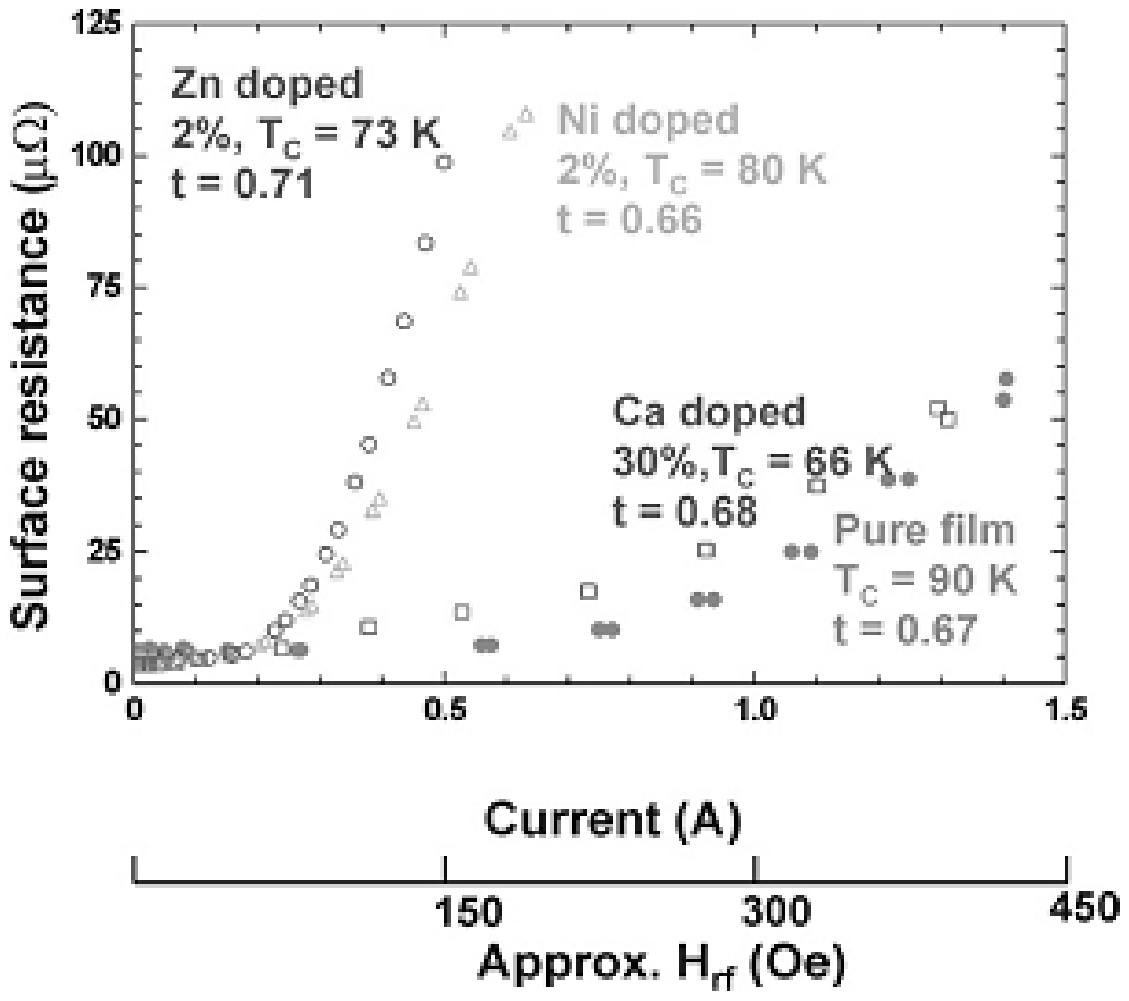

**Fig. 78.** *Rs* versus $H_{rf}$ at 1.5 *GHz* measured using the stripline resonators made of *YBCO* films on *MgO*. Both undoped (pure $YBa_2Cu_3O_{7-\delta}$) and doped (with *Ca*, *Ni* and *Zn*) films are measured. Reprinted with permission from D. E. Oates, M. A. Hein, P. J. Hirst, R. G. Humphreys, G. Koren, E. Polturak, Nonlinear microwave surface impedance of YBCO films: latest results and present understanding, *Physica C: Superconductivity,* vol. **372-376,** part 1, pp. 462-468, 2002. (© 2002 Elsevier.) (after [549, 46]).

It is seen that while doping with *Ca* slightly reduces the low power surface resistance *Rs (0)*, it has very little effect on the nonlinear performance. On the other hand, doping with *Ni* and *Zn*, although it has similarly negligible effect on *Rs (0)*, dramatically enhances the nonlinearity of the films. The effect of *Zn* doping is stronger than that of *Ni* doping. This is especially clearly seen in the *IMD* power dependences which are plotted in Fig. 72. The *Ni*-doped curve virtually coincides with that of the as-grown (pure) film at low powers and slightly deviates from it with increased $P_{in}$. The *Zn*-doped curve is substantially uplifted with regard to the pure film over the entire power range. ***This supports the conclusion that the nonlinearity is strongly enhanced by doping with Ni and Zn, and that Zn doping has a substantially stronger effect than that of doping with Ni.***

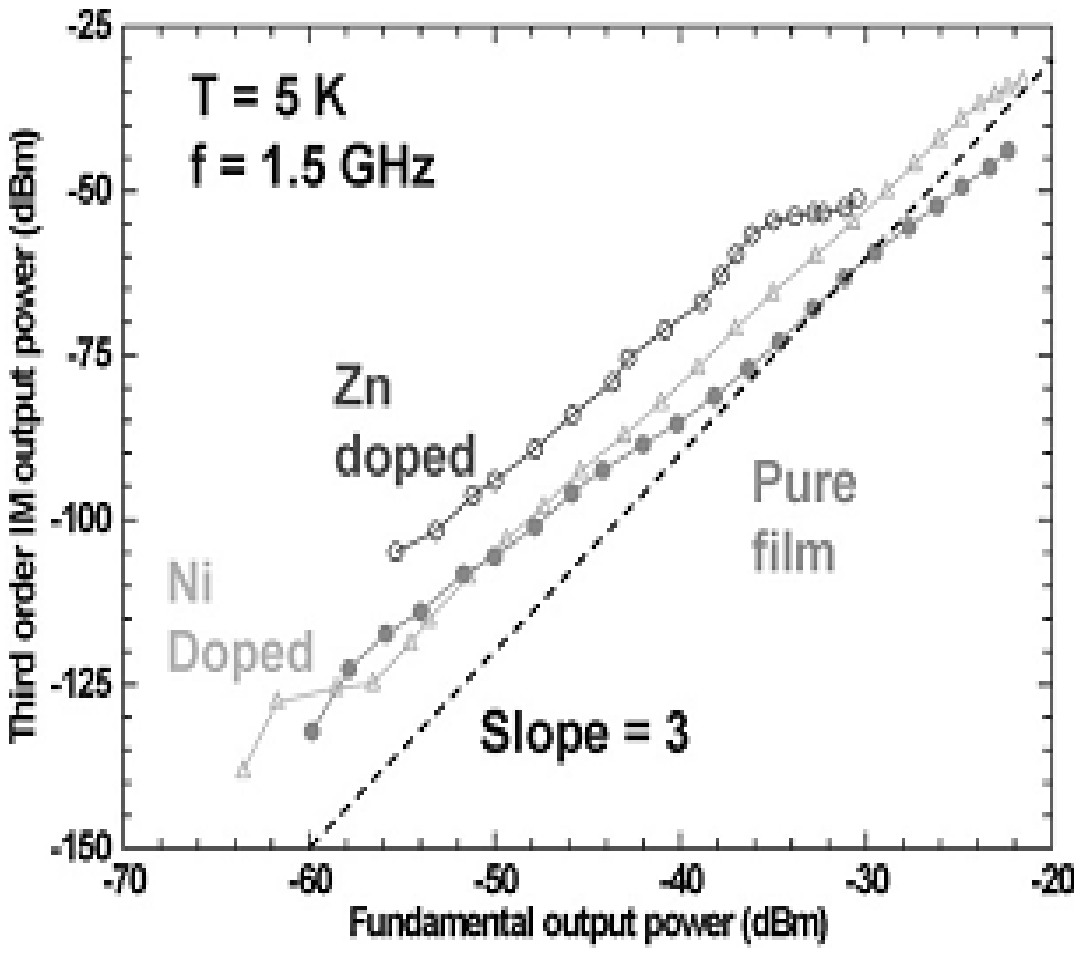

**Fig. 79.** Power dependence of the *IMD* products ( *f* = 1.5 *GHz*) for the same films as those shown in Figure 40. Doped films show a significant increase in *IMD*. The *Zn*-doped film is considerably worse than the *Ni*-doped film. Reproduced with permission from D. E. Oates and M.A. Hein private communication, 2002 [552]; D. E. Oates, M. A. Hein, P. J. Hirst, R. G. Humphreys, G. Koren, E. Polturak, Nonlinear microwave surface impedance of YBCO films: latest results and present understanding, *Physica C: Superconductivity*, vol. **372-376,** part 1, pp. 462-468, 2002. (© 2002 Elsevier.) (after [549,552, 46]).

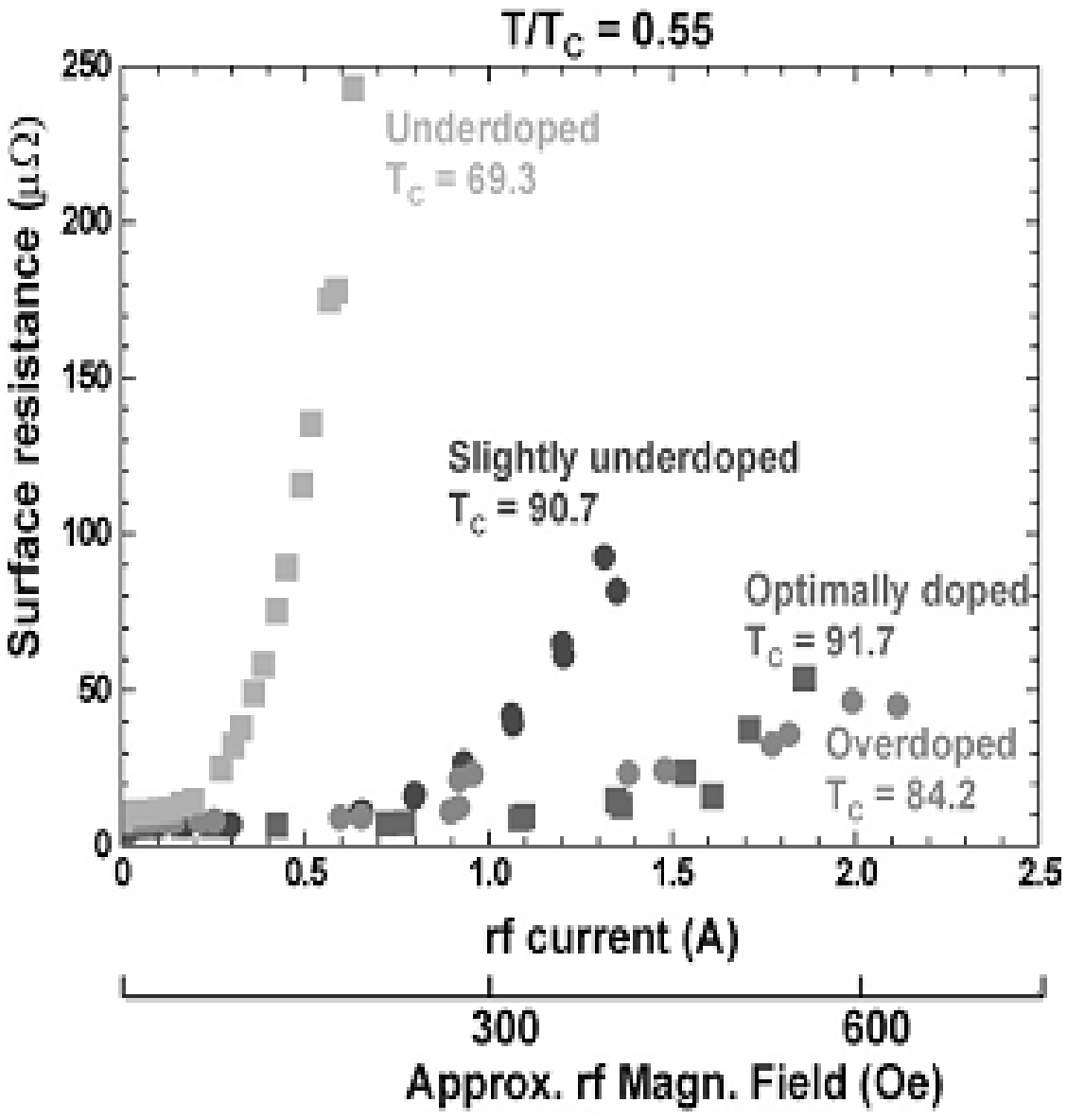

**Fig. 80.** The effects of ***oxygen doping*** on the nonlinear surface resistance. $T_{cs}$ for the films are given in the Fig. 91. The measurements are made by using the stripline resonator at 1.5 *GHz*. Reproduced with permission from D. E. Oates, M. A. Hein, P. J. Hirst, R. G. Humphreys, G. Koren, E. Polturak, Nonlinear microwave surface impedance of YBCO films: latest results and present understanding, *Physica C: Superconductivity*, vol. **372-376,** part 1, pp. 462-468, 2002. (© 2002 Elsevier.) (after [549, 15]).

The microwave current dependences for strongly ***oxygen*** underdoped, slightly underdoped, optimally doped and overdoped films at $t = T/Tc = 0.55$ and 1.5 *GHz* are shown in Fig. 73. It is seen that underdoping has a strong effect on the nonlinear performance of the film by weakening the power handling capability [549]. The stronger the underdoping, the less the threshold microwave current $I_{rf}^{*}$, above which the nonlinearity arises. On the other hand, overdoping improves nonlinearity in terms of $I_{rf}^{*}$, as can be seen from the graph [46].

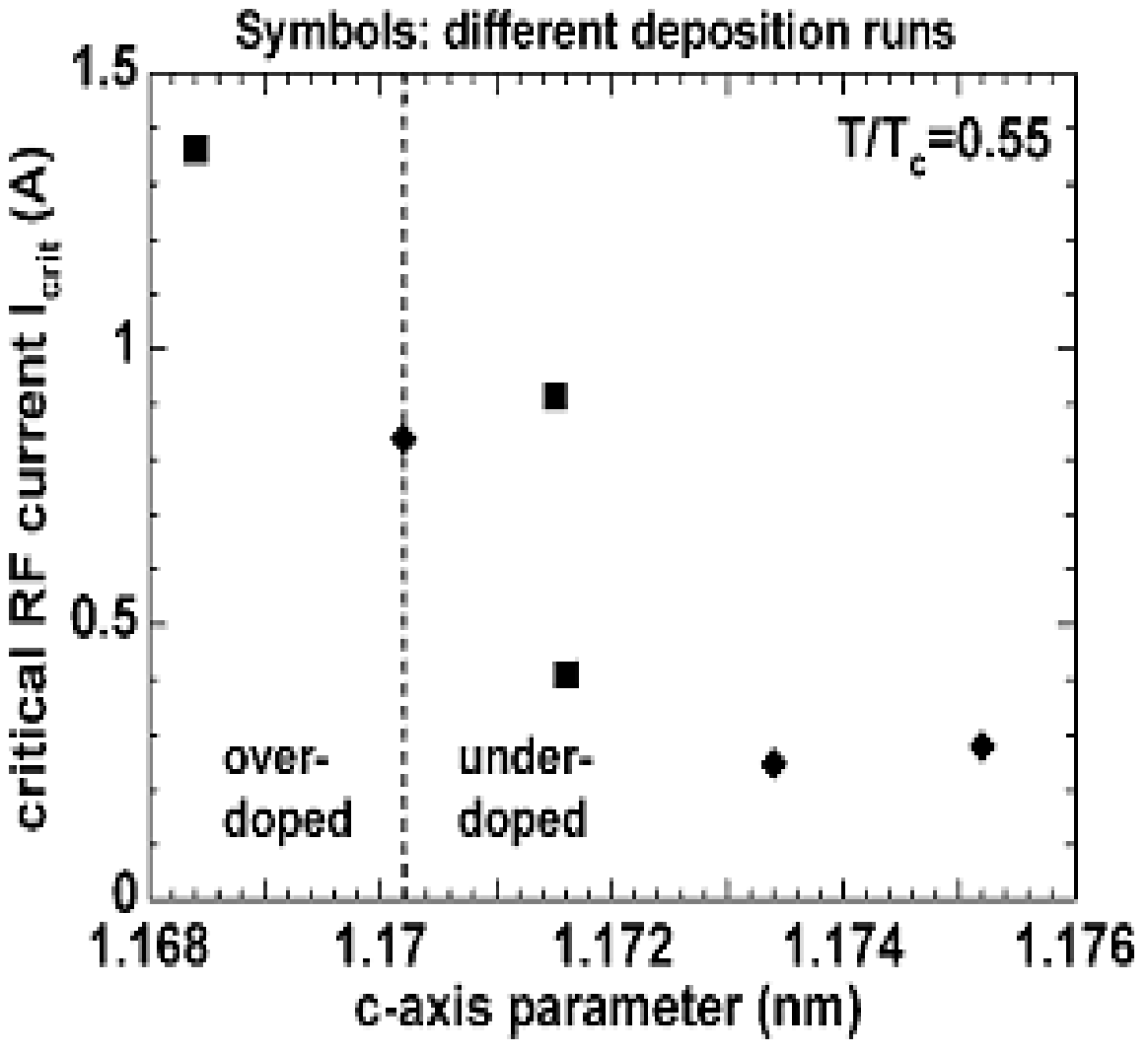

**Fig. 81.** *RF* critical current as a function of the c-axis length, which is directly related to the oxygen content of the films. The measurements are made by using the stripline resonator technique at 1.5 *GHz*. Reprinted with permission of *MIT* Lincoln Laboratory, Lexington, MA, as well as the authors, Oates and Hein (after [552, 46]).

Fig. 81 shows the dependence of the *RF* critical current, which is essentially the same as $I_{rf}^{*}$, on the oxygen content of the film. Although, there is a spread in the data, the tendency of decreasing $I_{rf}$ with increasing oxygen depletion is obvious [552, 46].

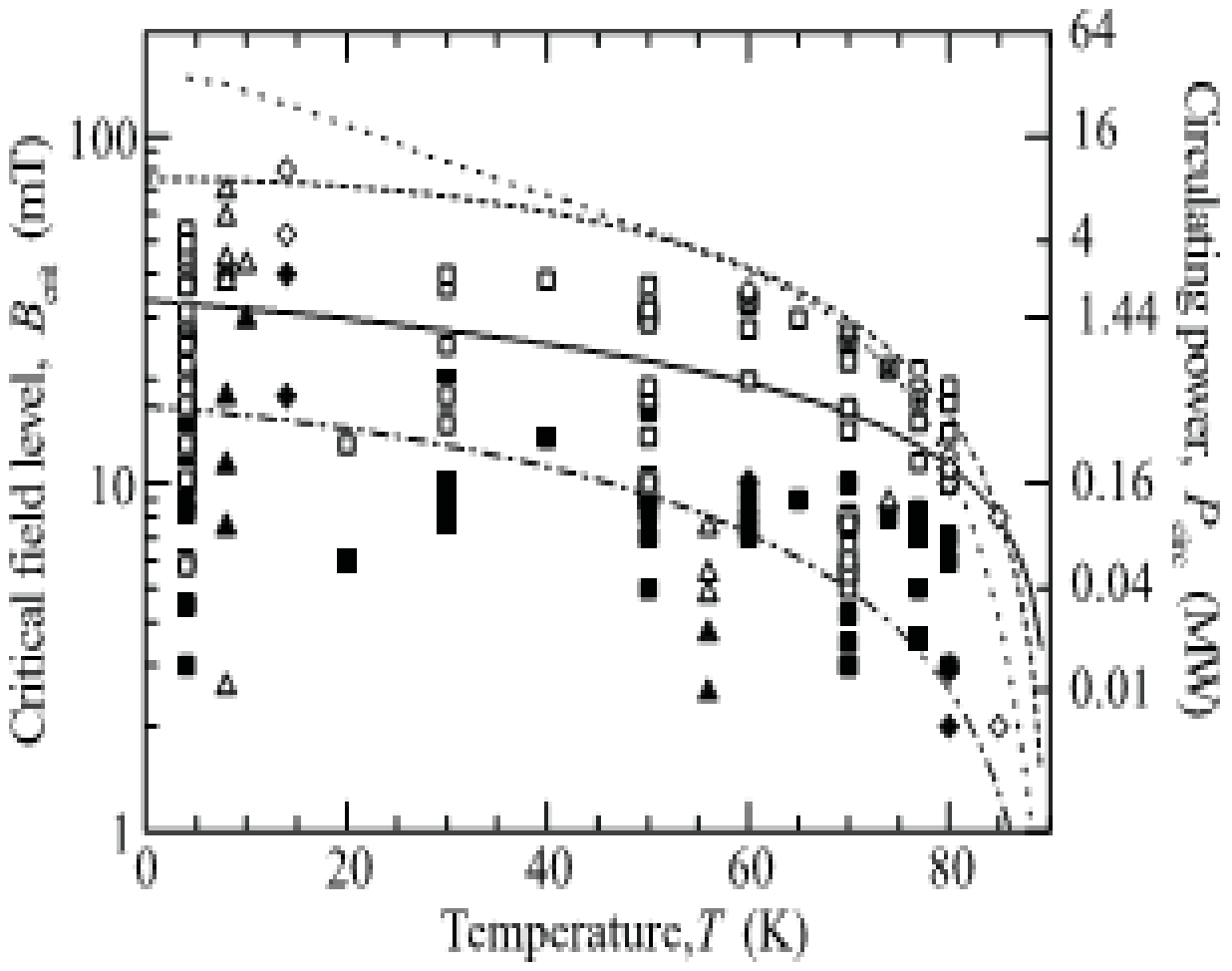

**Fig. 82.** Survey of the measured critical field amplitudes $B_{crit}(T)$ (symbols) in comparison with different estimated critical fields (curves) (after [553]). The right axis shows the circulating power corresponding to the *RF* field amplitude *Bs*. Full symbols represent those field levels at which *Rs* is increased by 20% above its low field value. Open symbols show the maximum achieved field levels. Data were measured with unpatterned films (squares), coplanar (triangles) and microstrip (diamonds) resonators. The curves denote the field limitations expected from weak links (dash–dotted), the lower critical field $B_{c1}$ (dashed), global heating (dotted) and local heating (solid). Reprinted with permission from Th. Kaiser, B. A. Aminov, A. Baumfalk, A. Cassinese, H. J. Chalouoka, M. A. Hein, S. Kilesov, H. Medelius, G. Mitller, M. Perpeet, H. Piel and E. Wikborg, Nonlinear power handling of $YBa_2Cu_3O_{7-\delta}$ films and microwave devices, *J. Superconductivity*, vol. **12,** pp. 343-351, 1999. (© 1999 Springer Science and Business Media) (after [553, 46]).

The ***microwave heating*** is another extrinsic mechanism for the nonlinear response in superconductors. High circulating power (up to tens of kilowatts) in microwave devices,

| | |
|---|---|
| | especially in resonators and filters, may lead to significant power dissipated in the material and, as a result, to heating. Heating may co-exist with other intrinsic and extrinsic nonlinear mechanisms, and will contribute to or dominate the power dependence of *Z* [46].<br>Heating can be global, when the whole interior or the bulk of the superconductor is heated, or local, when the enhanced dissipation takes place at poorly conducting regions such as weak links and defects. A global quench occurs when the power dissipated in the film is sufficient to induce a superconducting to normal state transition within the whole bulk of the film. A local quench occurs near defects where the local temperature due to heating exceeds the local critical temperature $T_c^{loc}$ of the surrounding superconducting matrix. This $T_c^{loc}$ can be noticeably less than the bulk $T_c$ of the film.<br>According to Hein [530], when heating is the dominant mechanism the shape of $R_s(H_{rf})$ is usually dictated by global heating. Local defects have negligible contribution to the total loss, and therefore do not affect the functional form of $R_s(H_{rf})$. On the other hand, normal conducting defects may significantly decrease the local quench fields, especially at intermediate and low *T* .Normally, at low *T* far away from the superconducting transition global quench fields remarkably exceed the local ones; however, in the vicinity of the superconducting transition the two fields become comparable (see Fig. 82). |

**Tab. 11.** Comparative analysis of research results on origin of nonlinearities in *HTS* thin films at microwaves by Velichko, Lancaster, Porch (after [46]).

Xin [554] noticed that the ***vortex penetration*** has no impact on the nonlinearities in $YBa_2Cu_3O_{7-\delta}$ films in a dielectric resonator at low and intermediate (up to ~ 100 Oe) $H_{rf}$ -values. However, in the case of a stripline resonator, taking to the account the demagnetization effect, the magnitude of effective magnetic field $H_{eff}$ may be significantly higher at low and intermediate (up to ~ 100 Oe) $H_{rf}$ – values, hence the Abricosov magnetic vortices may have dominant influence on the nonlinearities in $YBa_2Cu_3O_{7-\delta}$ films even at low and intermediate $H_{rf}$ – magnetic field values [530].

The author of dissertation would like to comment that the oxygen and its vacancies can diffuse into the Abricosov magnetic vortex core with normal metal phase from superconductor with superconducting phase in $YBa_2Cu_3O_{7-\delta}$ thin films at microwaves. This oxygen diffusion process was researched by the author of dissertation, and the conclusion is that the oxygen and its vacancies diffusion into the multiple Abricosov magnetic vortices cores can change the balance of superconducting and normal phases in superconductor, impacting the nonlinear physical properties of $YBa_2Cu_3O_{7-\delta}$ thin films at microwaves strongly in V. O. Ledenyov, D. O. Ledenyov, O. P. Ledenyov [555].

Porch, Huish, Velichko, Lancaster, Abell, Perry, Almond, Storey [556] conducted the research to understand the effects of residual surface resistance on the microwave properties of $YBa_2Cu_3O_{7-\delta}$ thin films. Authors used the two experimental techniques [556]:

1. ***Modulated Optical Reflectance (MOR) Technique***: The *MOR* technique measures the temperature dependence of the optical reflectivity of a material, which is related to the carrier concentration. For $YBa_2Cu_3O_{7-\delta}$, the technique is most sensitive at photon energies about 2 *eV* (i.e. wavelengths about 620 *nm*).
2. ***Coplanar Resonator (CPR) Technique***: The *CPR* technique has been used as a routine tool for characterizing patterned $YBa_2Cu_3O_{7-\delta}$ thin films at microwave frequencies. It requires only single-sided, small area films (1 or less) and has the attractive feature that the resonator response is dominated by the surface impedance of the film. For example, almost all the shift in resonant frequency as the temperature is varied is due to changes in $\lambda$, and no subtraction of extraneous losses is required when converting resonator factors into surface resistance; this is particularly important, when quantifying residual surface resistance.

Porch, Huish, Velichko, Lancaster, Abell, Perry, Almond, Storey [556] write that a combination of the modulated optical reflectance technique (for as-grown films) and the coplanar resonator technique (for patterned films) to assess three, high-quality, commercial $YBa_2Cu_3O_{7-\delta}$ thin films on *MgO* substrates. Whilst all three films exhibit similar, homogeneous *MOR* signals, with similar absolute values of the magnetic penetration depth $\lambda$, there is a large variability in $R_{s,0}$ and in the nonlinear $R_s$. However, by subtracting the residual microwave losses we extract quasiparticle conductivities $\sigma_1$ that vary little between the three films and, with the measured $\lambda$, are representative of their intrinsic microwave conductivity $\sigma=\sigma_1-j/\omega\mu_0\lambda^2$.

Porch, Huish, Velichko, Lancaster, Abell, Perry, Almond, Storey [556] state that ***the only significant variability in the 8 GHz surface impedance data of three high quality YBCO thin films on MgO substrates is in the residual at low and in the nonlinear at high microwave field levels***, which in general are not correlated. Subtracting the residual enables us to compute an intrinsic complex conductivity that is consistent for all three films, and also with the homogeneous *MOR* signals measured in the as-grown state. ***The common origin of the variability of the residual and nonlinear is likely to be due to granularity.*** It is not possible to probe this using *MOR* or measurements, which mainly yield information about the intrinsic properties of the intra-granular regions.

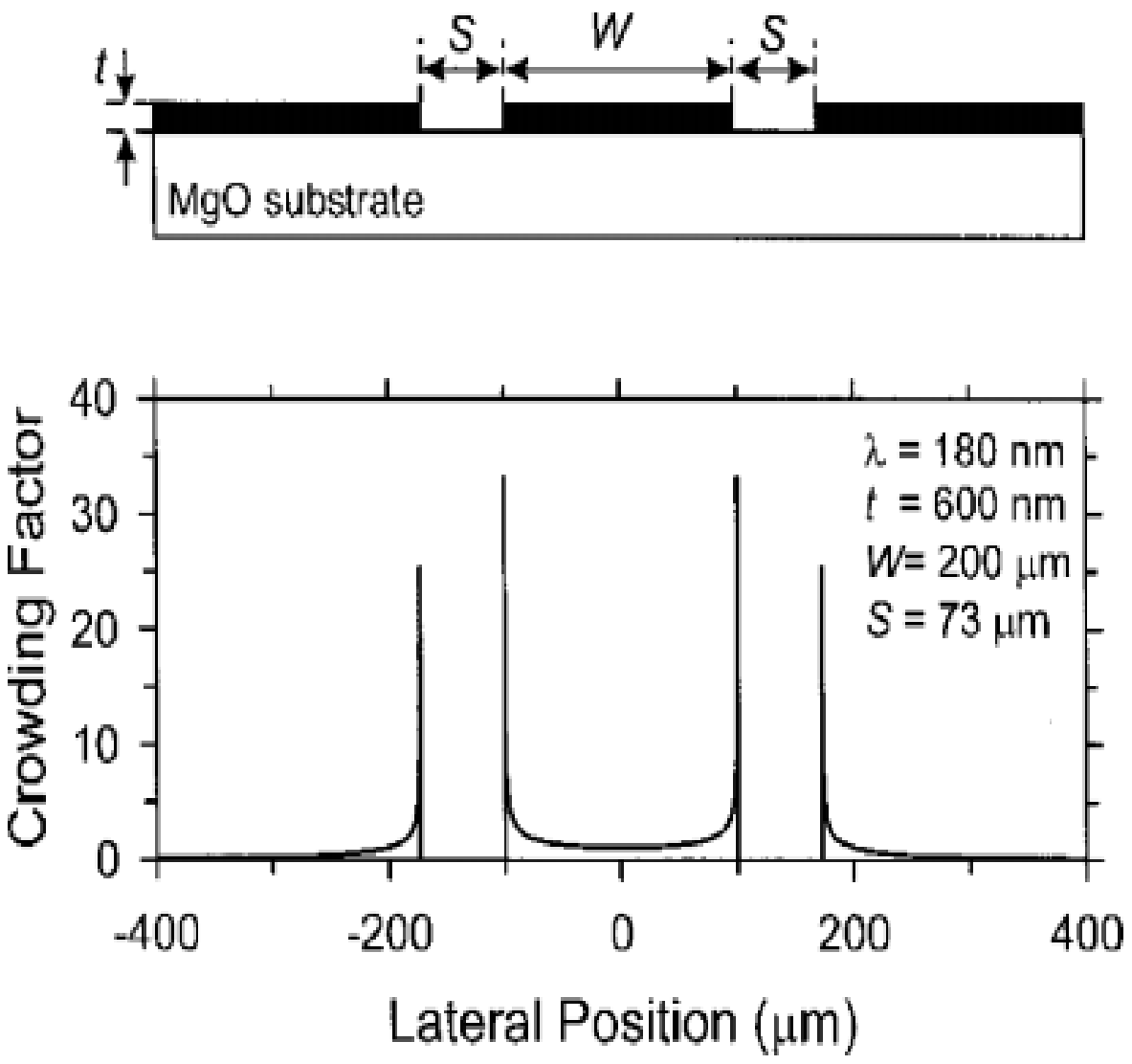

**Fig. 83.** Schematic diagram of the *CPR* cross section and the current distribution for *λ=180nm* and film thickness of *600 nm*; the crowding factor is the local current density divided by the average current density on the center strip (after [556]).

***The MOR signals of all three resonators are large and very homogenous, suggesting good electrical homogeneity of the researched HTS thin films*** [556].

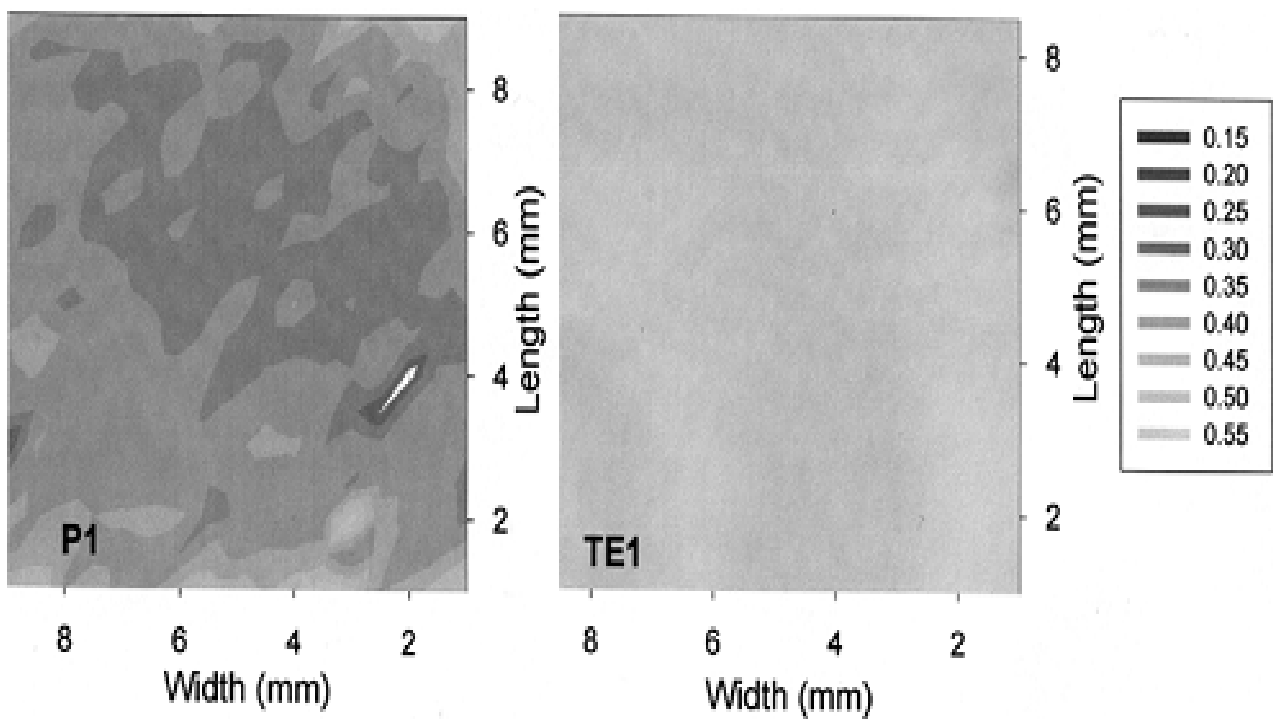

**Fig. 84.** The spatial dependence of the *MOR* signal (in units of V) for *TE1* compared with a highly granular film (P1) with poor microwave properties (after [556]).

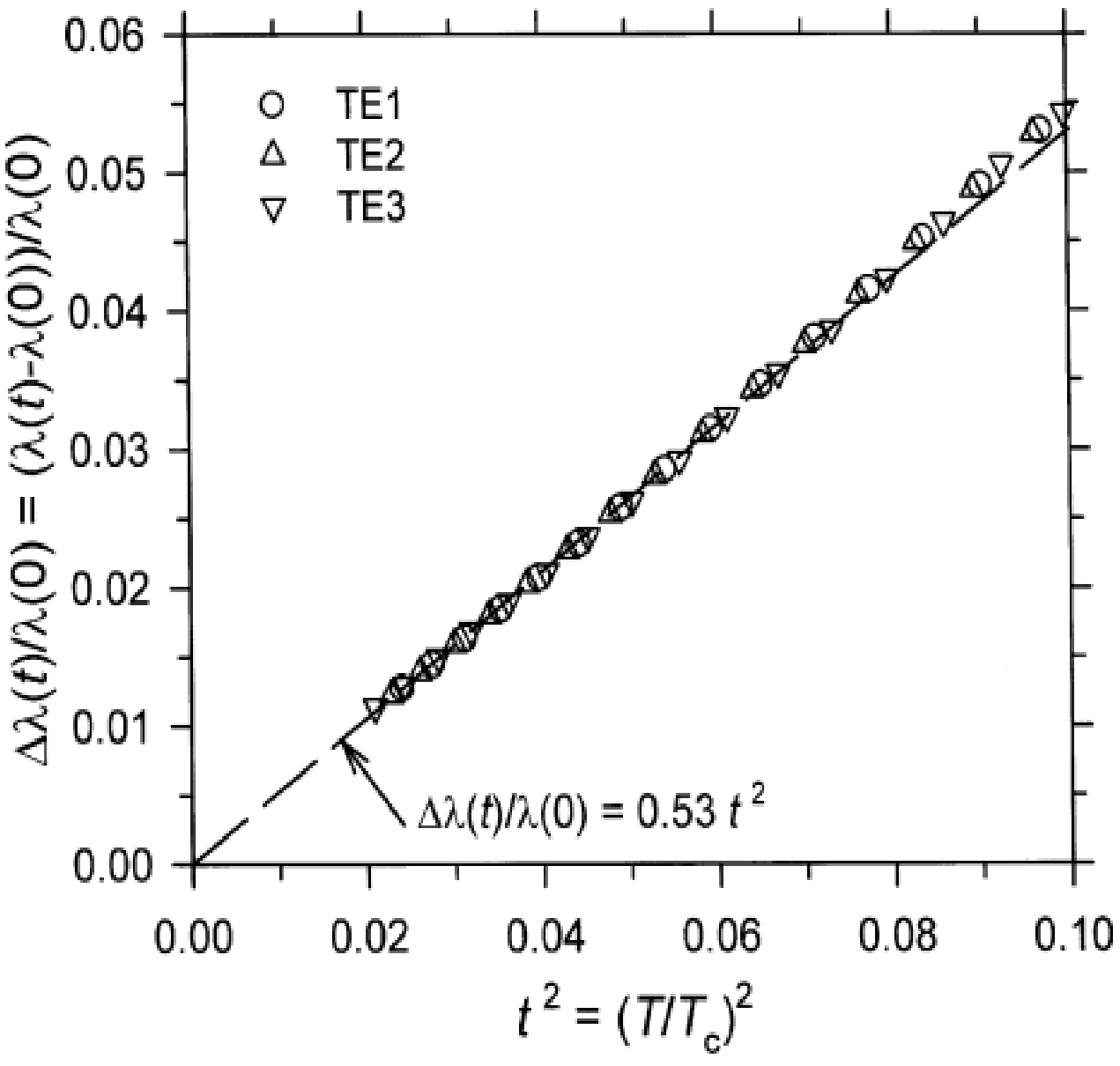

**Fig. 85.** Fits of *Δλ(t)=λ(0)* to *t*, from which *λ(0)* is calculated (after [556]).

Similar uniformity is found in *λ(T)* for all three films, where Fig. 85 shows the low-$T$ fits of *Δλ(T)/λ(0)* to $T^2$ used for the extraction of *λ(0)*, which is then used to obtain the full temperature dependence of *λ(T)* shown in Fig. 86. The obtained values are λ(0)=188, 187 and 178 nm for *TE1*, *TE2* and *TE3*, respectively. *λ(0)* for very high quality *YBCO* thin films extracted using the *CPR* technique are generally in the range 170–220 *nm*, so the values found here compare favorably [556].

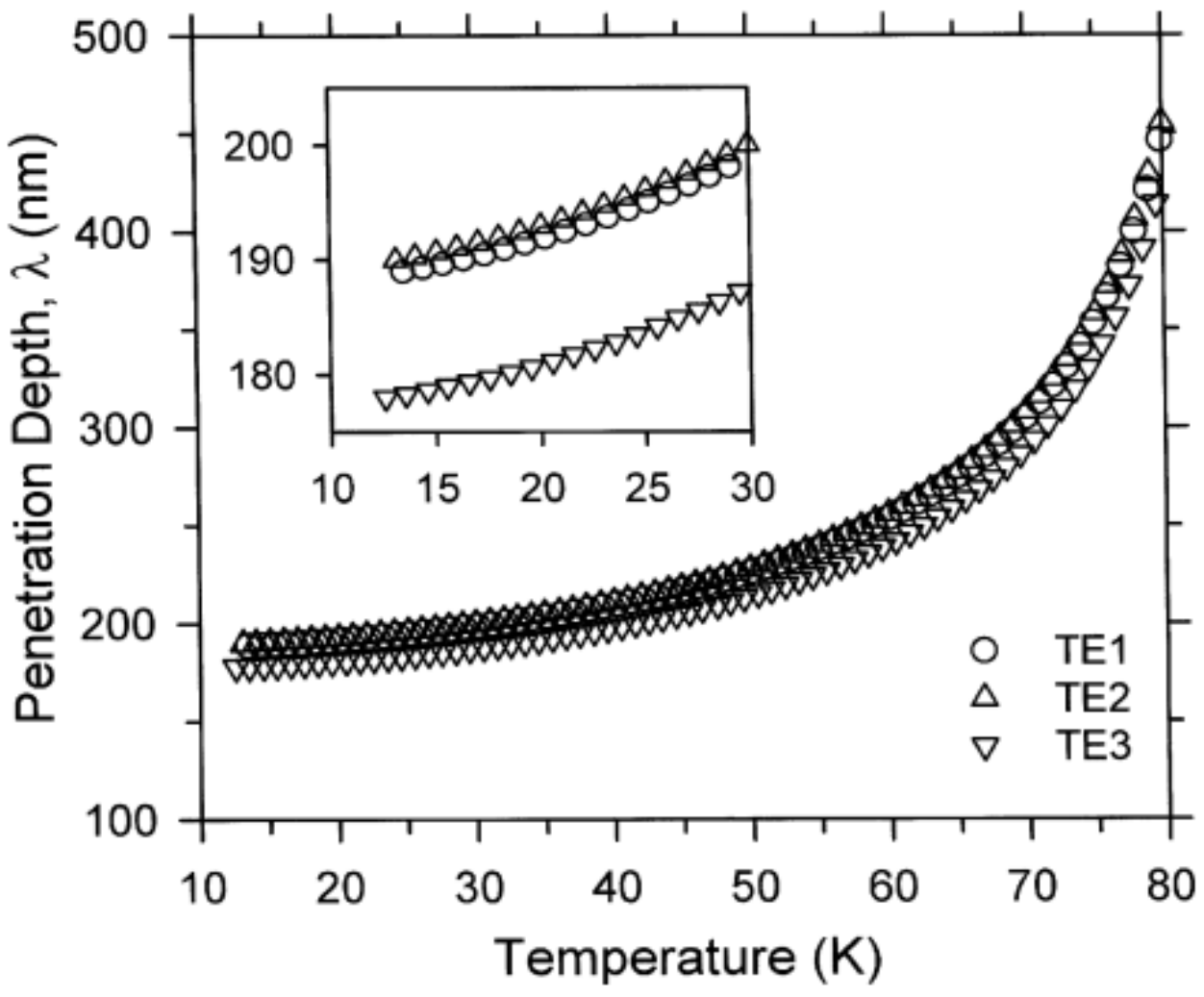

**Fig. 86.** $\lambda(T)$ for the three films, with slightly differing values of $\lambda(0)$ (after [556]).

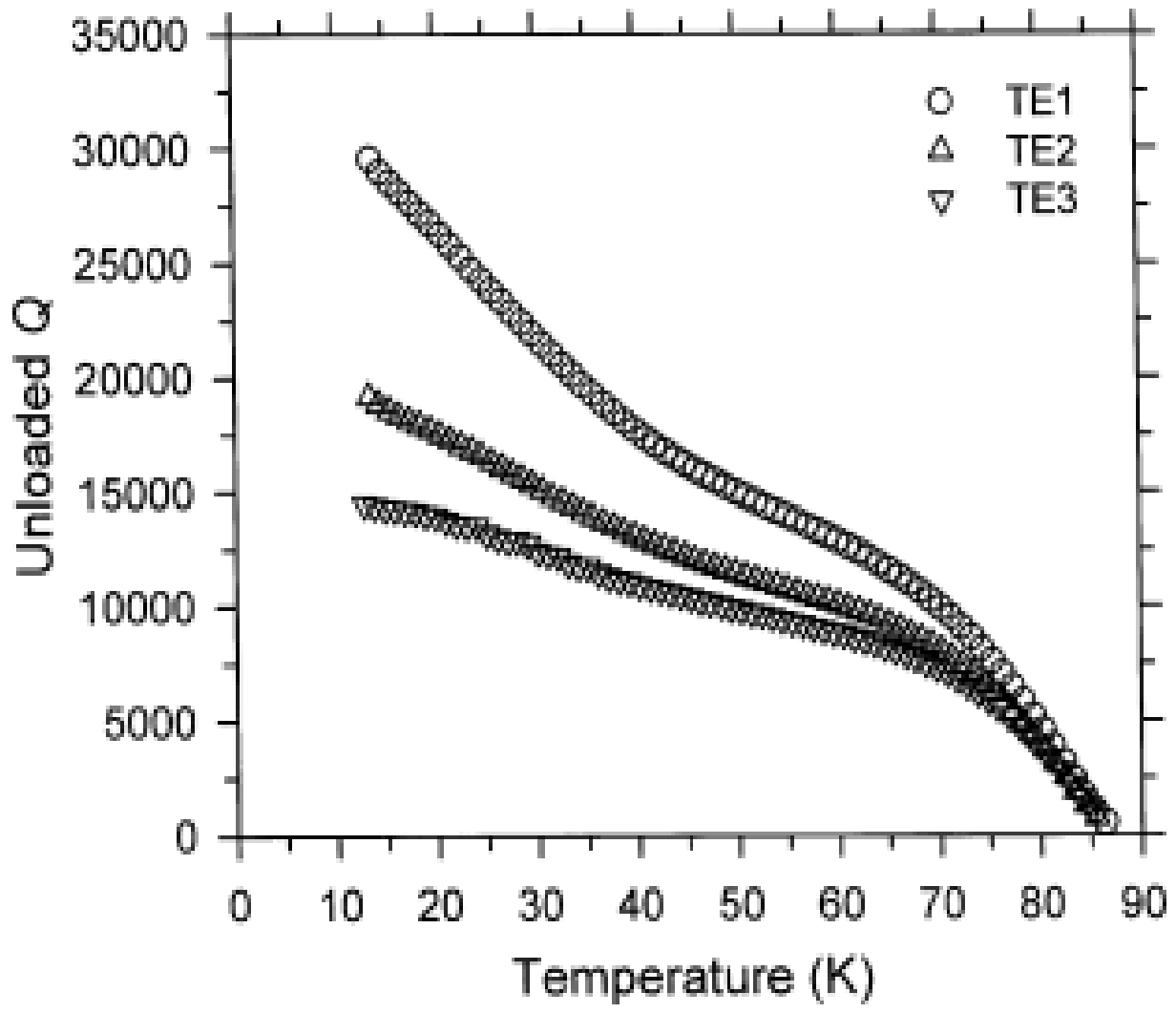

**Fig. 87.** $\lambda(T)$ for the three films, with slightly differing values of $\lambda(0)$ (after [556]).

The unloaded factors of the three *CPR*s at 8*GHz* as a function of *T* at low microwave input power varying at 15 *K* between 15 000 for *TE3*, to 30 000 for *TE1* [556]. Authors [532] note that this variability is unexpected and results in the $R_s(T)$ data of Fig. 88, with values at 15 *K* of 59, 83 and 101 for *TE1*, *TE2* and *TE3*, respectively; the variability is less at higher *T*, *e.g.* at 60 *K*, $R_s$ =109, 123 and 146 $\mu\Omega$, respectively [556].

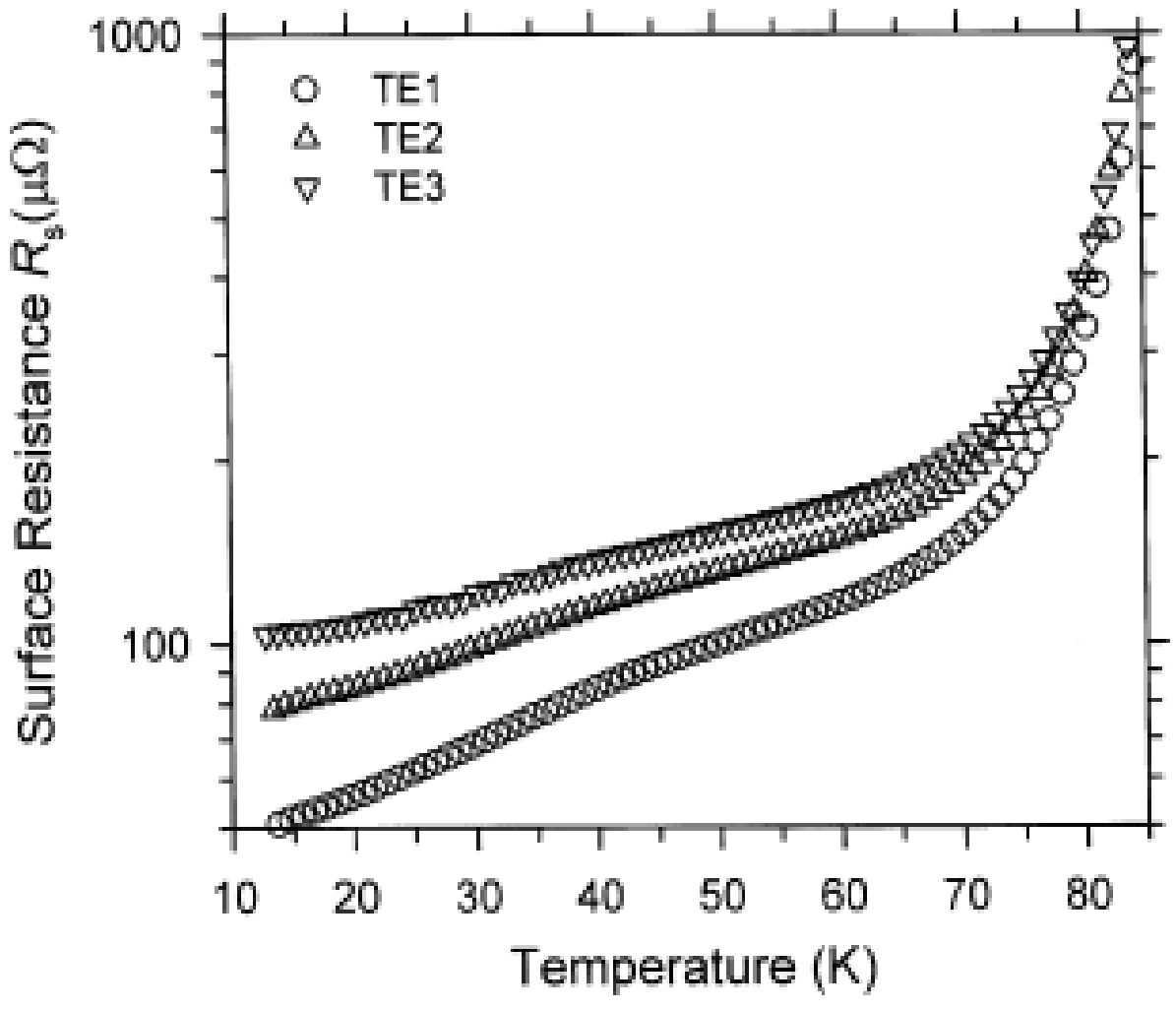

**Fig. 88.** Surface resistance at 8 *GHz* as a function of *T* calculated from *Q(T)* (after [556]).

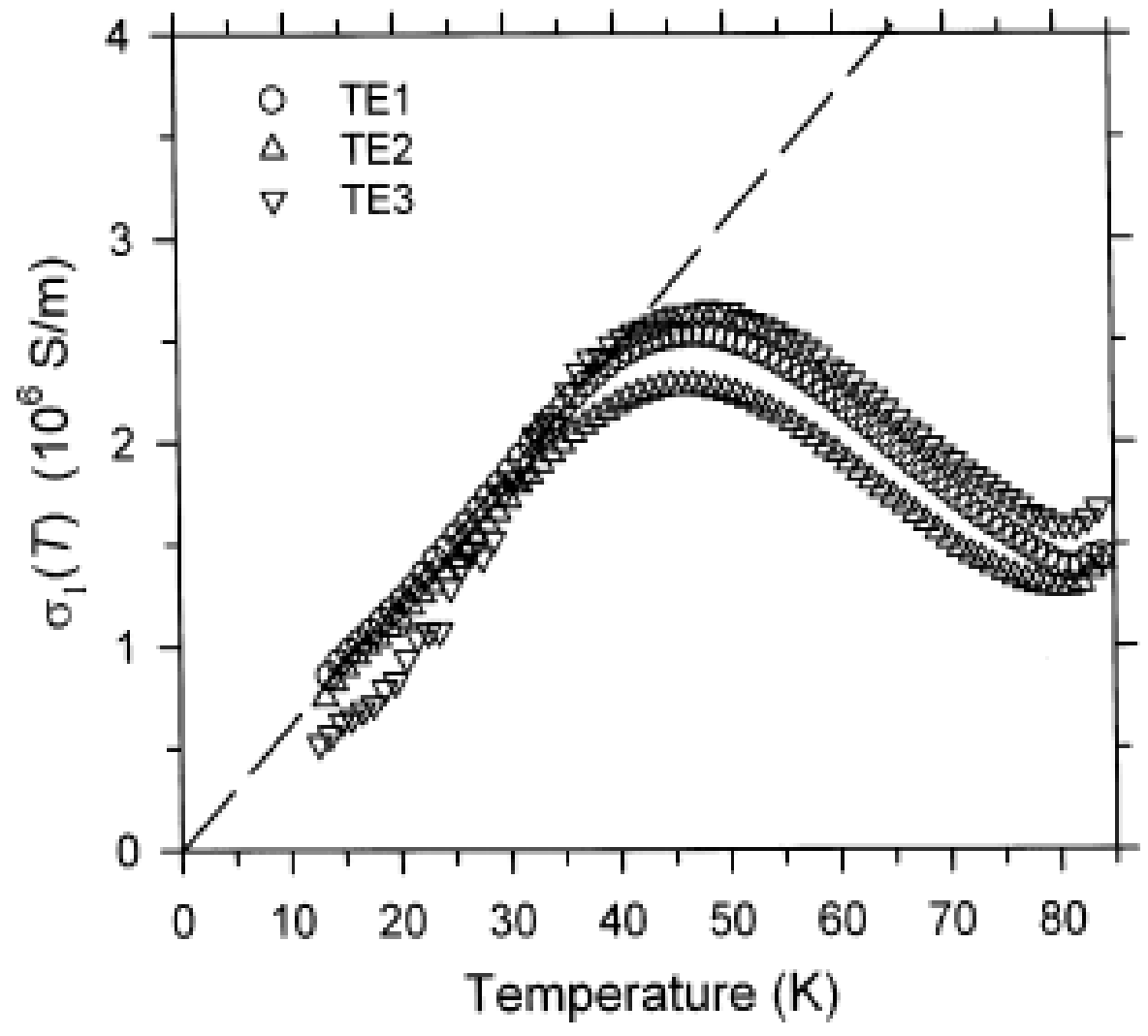

**Fig. 89.** Intrinsic quasiparticle conductivity $\sigma_1(T)$. The dotted line is a linear fit to the data for films $TE_1$ and $TE_2$ for $T< 35\ K$, where $\sigma_1 = 5.3\ x10^6 = T/T_c\ (S/m)$ (after [556]). The $R_s$ data exhibit a variable residual surface resistance, whose presence is uncorrelated with both the *MOR* signal and the inductive microwave response (i.e. λ). To attempt to extract the intrinsic $R_s$, we extrapolate the unloaded 3 *dB* bandwidth $f_B$ to $T=0K$; the residual bandwidth is then subtracted from $f_B(T)$, and $\sigma_1(T)$ is re-computed. The results are shown in Fig. 82, exhibiting little variability, similar to the *MOR* and $\lambda$ data. Authors [556] interpret this re-computed data as the intrinsic quasiparticle conductivity for these films, which is approximately below 35*K* for films $TE_1$ and $TE_2$. There is a broad peak around 40–45 *K* (also observed in single crystals), most likely due to competition between an increasing quasiparticle scattering time and a decreasing quasiparticle density as is reduced. The peak is suppressed in thin films compared to crystals due to a reduced mean free path.

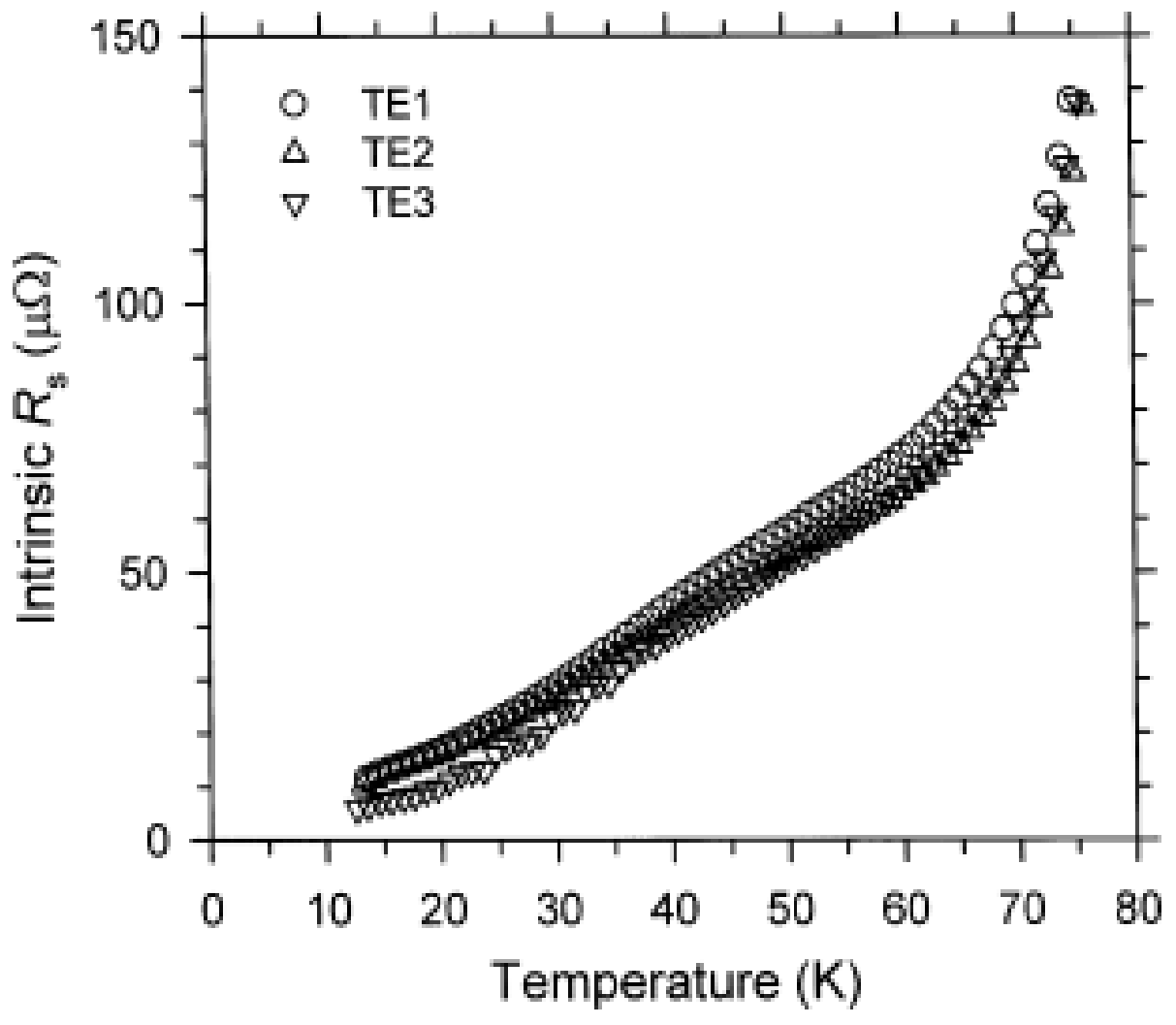

**Fig. 90.** The re-computed, intrinsic surface resistance $R_s(T)$ plotted on a linear scale up to 80 *K*. This reflects the *T* dependence of the product $\sigma_1\lambda^3$ (after [556]).

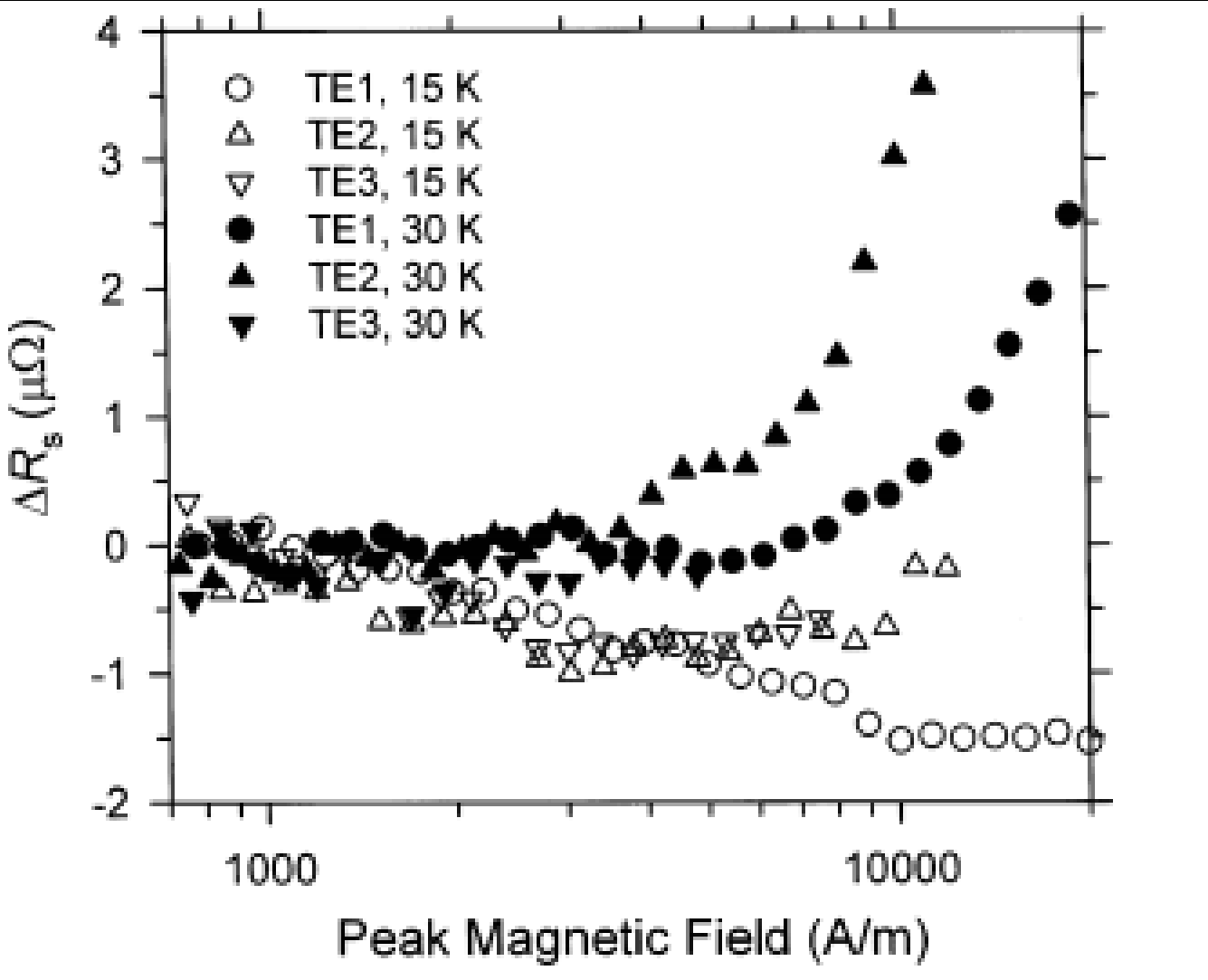

**Fig. 91.** Change of surface resistance as a function of peak magnetic field (after [556]). The results for the nonlinear $R_s$ at 8 *GHz* are shown in Fig. 91, where researchers plot $\Delta R_s=R_s(H)-R_s(0)$ at fixed temperatures of 15 and 30 *K* as a function of the peak microwave field within the *CPR*. ***There is no clear correlation between residual surface resistance and critical microwave field $H_c$ for the onset of nonlinearities. Both of these parameters are determined by extrinsic factors (e.g. a low density of large angle grain boundaries introduced during film growth), and there is no correlation of either with the homogeneity of the MOR signal or $\lambda(0)$.*** At 15 *K*, $\Delta R_s(H)$ initially decreases with increasing *H*. Whether this is due to *MgO* substrate nonlinearities, or is peculiar to *YBCO*, is impossible to tell from this data alone.

**Tab. 12.** Research results on sources of nonlinearities in *HTS* thin films at microwaves by Porch, Huish, Velichko, Lancaster, Abell, Perry, Almond, Storey (after [556]).

Zhuravel, Anlage, Ustinov [557], when researching the microwave current imaging in passive *HTS* components by Low-Temperature Laser Scanning Microscopy (*LTLSM*), have demonstrated that a typical superconducting microwave device has following different classes of defects and irregularities that can potentially cause enhanced nonlinear response [557]:

1. ***weak links,***
2. ***twin-domain blocks,***
3. ***in-plane rotated grains, and***
4. ***micro-cracks***.

The researchers mention that the ***peaked edge currents*** and ***RF field penetration into the weak links*** can also contribute to the nonlinear microwave properties of *HTS* thin films [557].

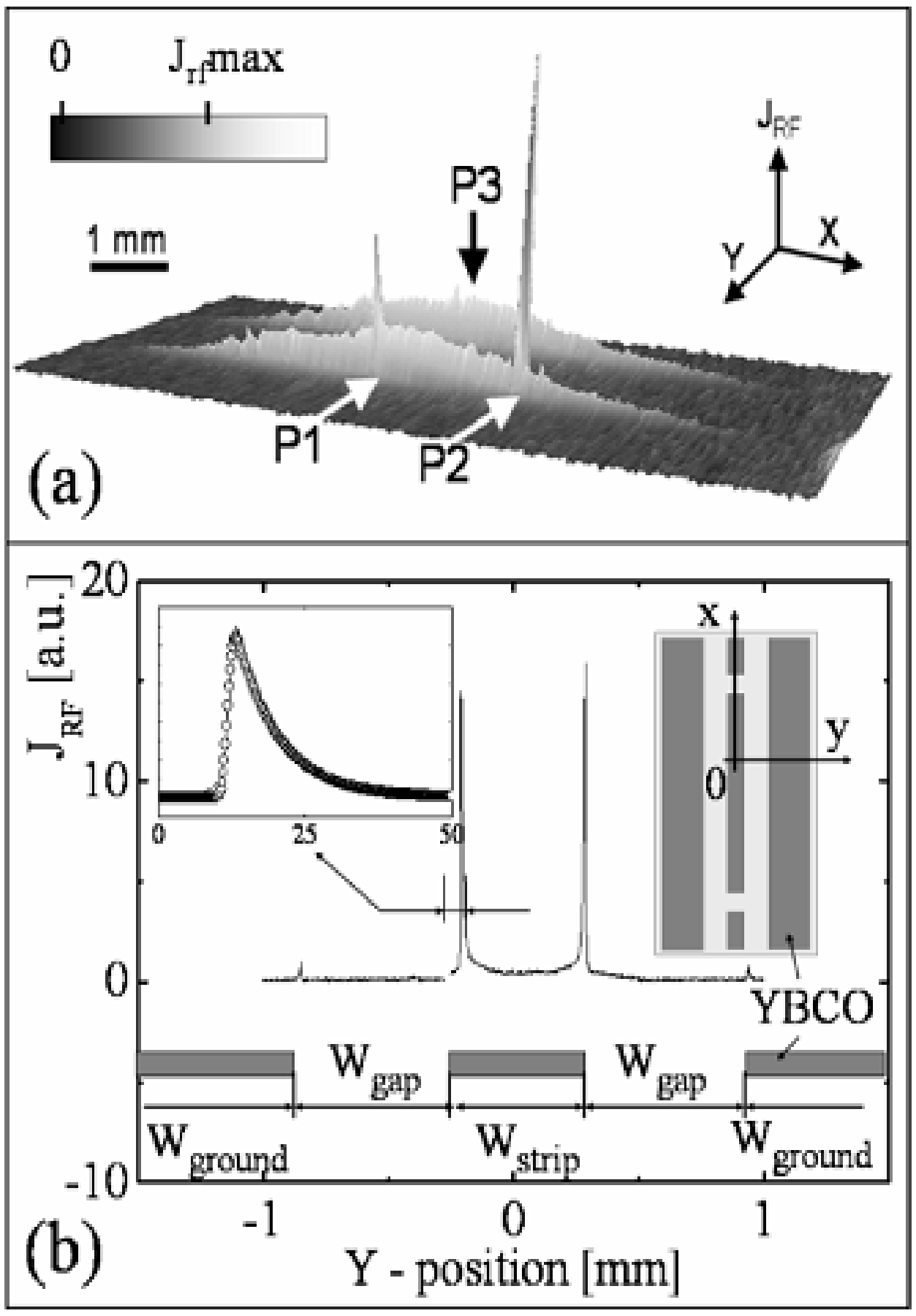

**Fig. 92.** *LTLSM* plots showing *RF* current distribution at *T = 79 K* in a $YBa_2Cu_3O_{7-\delta}$ film on *LAO* substrate **(a)** over a 1 x 8 $mm^2$ central area of the *CPW* resonator including the 0.5 x 7.75 $mm^2$ strip-line conductor, and **(b)** along a single y-scan across the *CPW* mid-section (after [557]).

Fig. 92 **(a)** represents a *3-D LTLSM* image showing the spatial variation of the *RF* current density along the *CPW* central strip-line at its resonance mode. The image was acquired at *T = 79 K*, $P_{RF}$ = -10 *dBm* and at a frequency that is 3 *dB* below the fundamental resonance peak $f_0$ = 5.2 *GHz* [557].

| | |
|---|---|
| 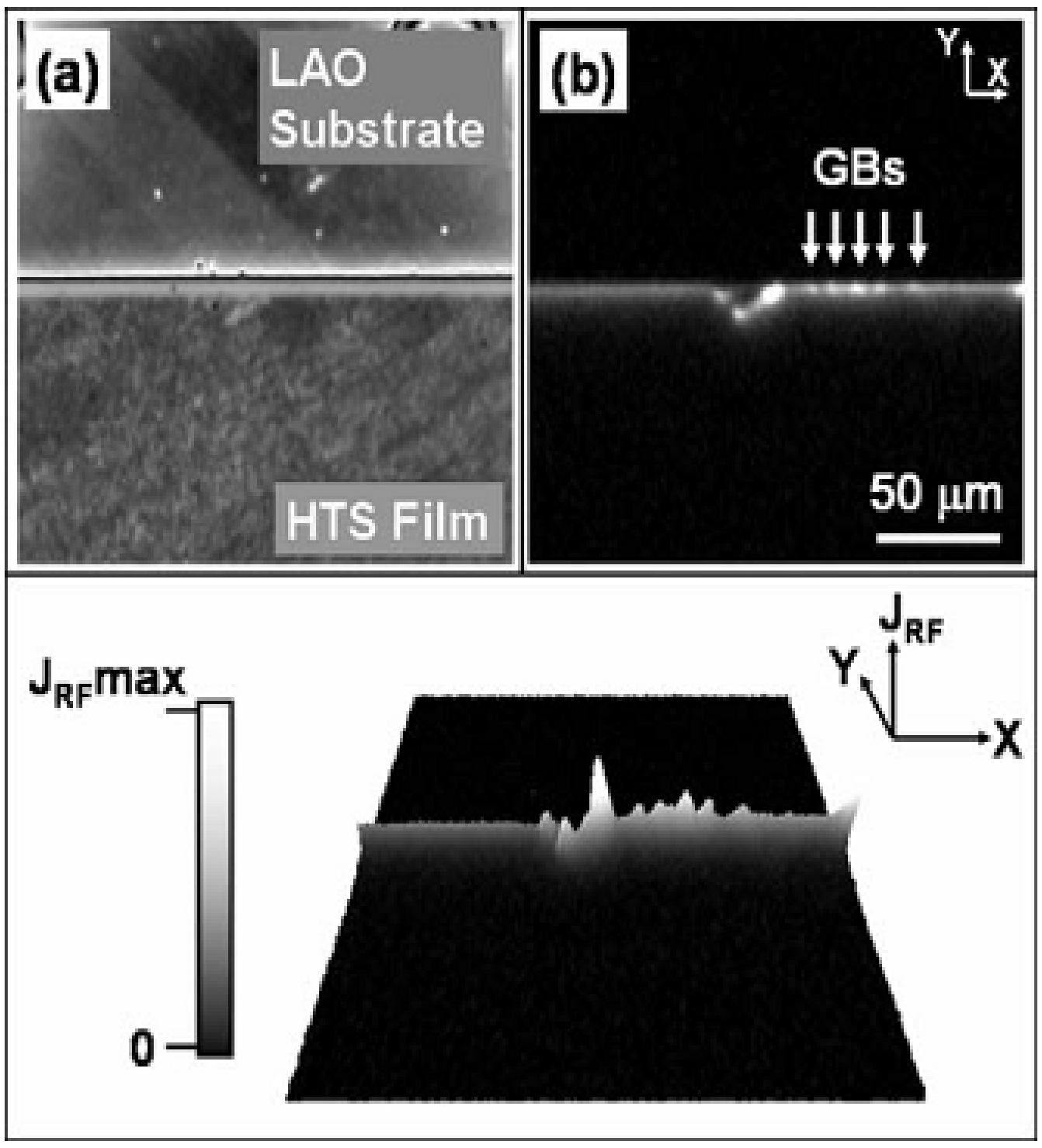 | **Fig. 93.** Detailed view of the $YBa_2Cu_3O_{7-\delta}$ film on *LAO* substrate showing (a) reflectivity, (b) *2D*- and (c) *3D*- photo-response of the $J_{RF}(x,y)$ in a 200 x 200 $mm^2$ area. The substrate clearly shows the linear patterns of twin domain blocks crossing the patterned edge of the $YBa_2Cu_3O_{7-\delta}$ thin film. The patterned edge is crenellated on the length scale of the twin domain blocks, as well as due to inhomogeneous associated with the grain boundaries (*GBs*) produced by substrate twinning (after [557]). |

**Tab. 13.** Research results on sources of nonlinearities in *HTS* thin films at microwaves by Zhuravel, Anlage, Ustinov (after [557]).

Tai, Xi, Zhuang, Mircea, Anlage [558] state that a measurement of the temperature dependent *3rd* order harmonic power is performed at the center of the *TBCCO* film by the magnetic write head probe with different exciting powers. The inset of Fig. 94 shows $P_{3f}(T)$ measured by the bare loop probe. A peak in $P_{3f}$ near $T_c$ shows up, as expected. ***This enhancement of $P_{3f}$ is due to the nonlinear Meissner effect near $T_c$.*** Due to $J_{NL}$ approaching zero and $\lambda(T)$ diverging at $T_c$, the *3rd* harmonic power will increase strongly. With the loop probe, the enhancement of $P_{3f}$ above background is only 15 *dB*, for 18 *dBm* fundamental input power. Such a small enhancement can be easily achieved by the magnetic head probe with only 9 *dBm* excited power, which means that the magnetic write head generates a more localized and intense field, inducing stronger surface currents on the sample. Despite these higher currents, there is no evidence of localized heating in the sample from the data in Fig. 94. In order to test the magnetic write head probe in a liquid Helium cooled environment, temperature dependent *3rd* order harmonic power is also measured in the center position of an $MgB_2$ thin film. In Fig. 95, a peak at 39.1 *K* shows up clearly near the $T_c$ of the film. This proves that the magnetic probe can function in the low temperature region. Comparison of this peak with that of *TBCCO*, one finds a much sharper transition, implying a narrow distribution of $T_c$ values in the $MgB_2$ thin film [558]. It is evident that ***the nature of $P_{3f}(T)$ dependence nonlinearities is due to the Cooper electron pair breaking resulting in the nonlinear Meissner effect near $T_c$ in researched HTS thin films at microwaves [558] (see Tab. 14).***

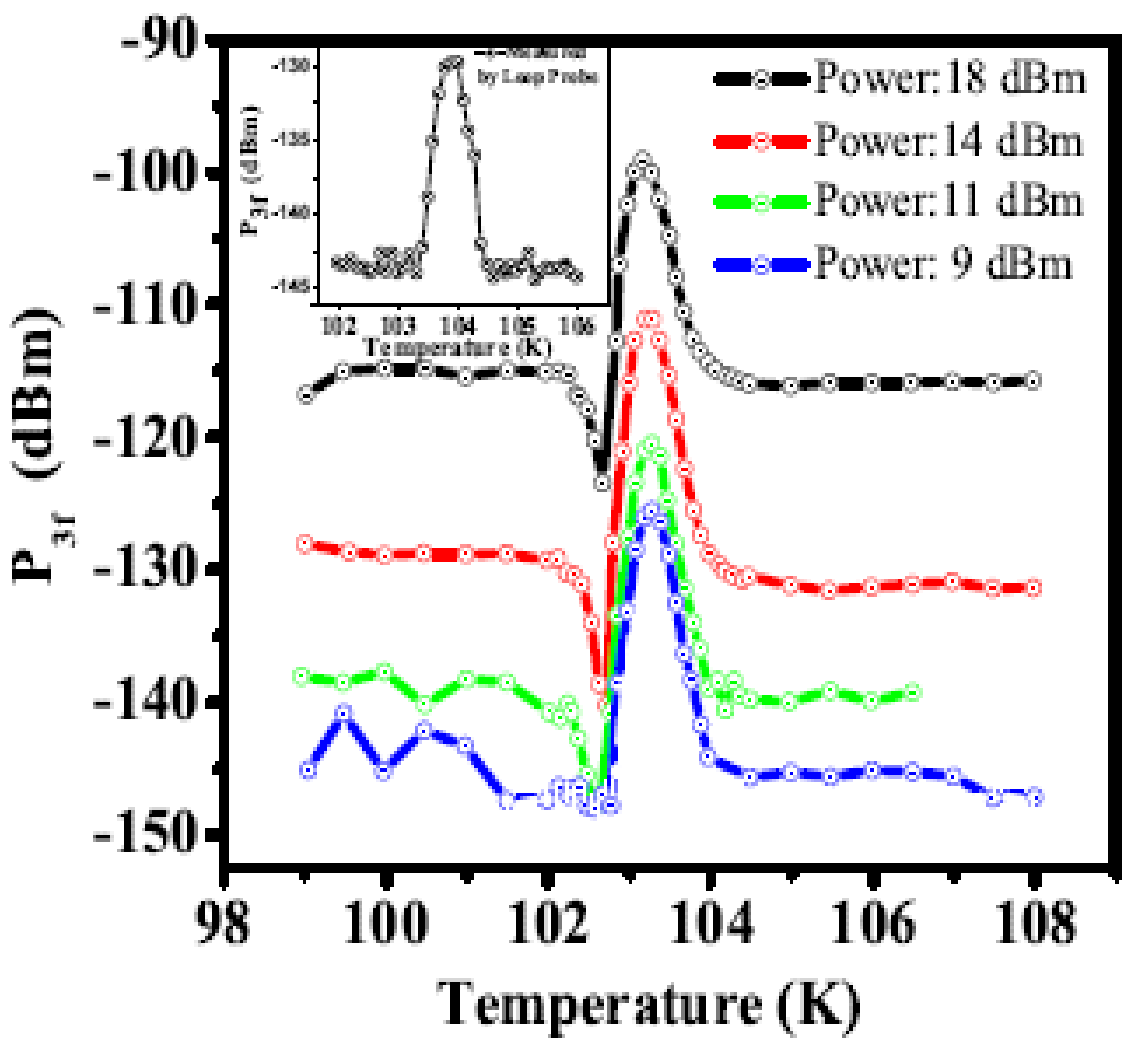

**Fig. 94.** Temperature dependence of third harmonic power $P_{3f}$ of a TBCCO film, measured by a magnetic write head probe with an excitation frequency of 3.5 *GHz*. Inset is measured by the bare loop probe with excitation frequency of 3.5 *GHz* and excited power 18 *dBm* (after [558]).

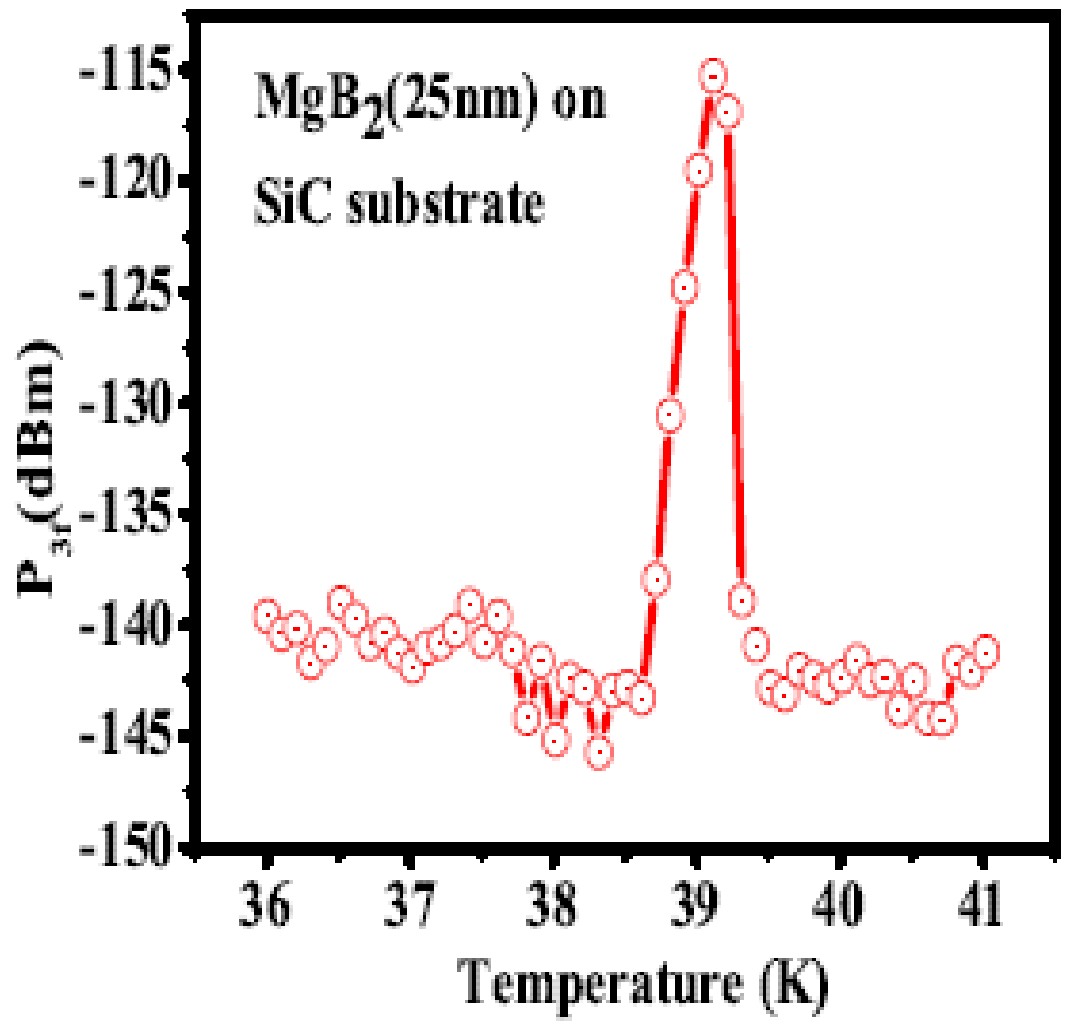

**Fig. 95.** Temperature dependence of third harmonic power $P_{3f}$ from an $MgB_2$ film, measured with the magnetic write head probe with an excitation frequency of approximately 3.5 *GHz* and a power of 12 *dBm* (after [558]).

**Tab. 14.** Research results on sources of nonlinearities in *HTS* thin films at microwaves by Tai, Xi, Zhuang, Mircea, Anlage (after [558]).

Golosovsky [559] at *Hebrew University of Jerusalem* researched the mechanisms of nonlinear high microwave power losses in high-Tc superconductors, mentioning that the two complementary models account for nonlinearities in 1997:

1. ***Extended coupled-grain model***, which assumes intrinsic granularity and grain boundaries acting as weak links having wide distribution of $I_cR_n$ products;

2. ***rf-critical state model***, based on *Abrikosov-Josephson vortices*.

***The author of dissertation would like to comment that, presently, the high quality $YBa_2Cu_3O_{7-\delta}$ thin film have almost no weak links, hence the Abricosov vortices, but not the Josephson vortices, have to be considered as a main possible source of nonlinearities in high quality HTS thin films at microwaves.***

Buks [560-566] with his research group at *Technion* focused on the investigation of nature of nonlinearities in microwave superconductivity in Abdo, Segev, Shtempluck, Buks [560], Zaitsev, Almog, Shtempluck, Buks [561], Abdo, Segev, Shtempluck, Buks [561, 562], Segev, Abdo, Shtempluck, Buks [563, 564], Bachar, Segev, Shtempluck, Shaw, Buks [565].

Abdo, Segev, Shtempluck, and Buks [560] mention that, in spite of the intensive study of nonlinearities in superconductors in the past decades, *there is a lack a coherent picture on origin of nonlinearities in superconductors at microwaves,* partly because the nonlinear mechanisms in superconductors are various, hence making the identification process of dominant factor mainly indirect, using eliminations [662].

Abdo, Segev, Shtempluck, and Buks [560] distinguish the following nonlinearities of ***intrinsic*** and ***extrinsic origins*** in HTS thin films at microwaves:

1. ***Meissner effect***,

2. ***Pair-breaking***,

3. ***Global and local heating effects***,

4. ***RF and DC vortex penetration and motion***,

5. ***Defect points, damaged edges***,

6. ***Substrate material***,

7. ***Weak links***.

Abdo, Segev, Shtempluck, and Buks [560] note that the weak links is a collective term representing various material defects located inside the superconductor such as *weak superconducting points* switching to normal state under low current density, *Josephson junctions* forming inside the superconductor structure, *grain boundaries*, *voids*, *insulating oxides*, *insulating planes*, which affect the conduction properties of *HTS*, causing extrinsic nonlinear effects [560].

Abdo, Segev, Shtempluck, and Buks [560] explain that the nonlinear effects in *HTS* were reported by several research groups. The authors [560] comment that the ***Duffing like nonlinearity*** was observed in different superconducting resonators, employing different geometries and materials, for example in a *HTS* parallel plate resonator in Cohen, Cowie, Purnell, Lindop, Thiess, Gallop [567], in a $YBa_2Cu_3O_{7-\delta}$ coplanar-waveguide resonator in Ma, Obaldia, Hampel, Polakos, Mankiewich, Batlogg, Prusseit, Kinder, Anderson, D. E.Oates, R. Ono, and J. Beall [568], in a $YBa_2Cu_3O_{7-\delta}$ thin film dielectric cavity in Wosik, Lei-Ming Xie, Miller, Long, Nesteruk [569], and also in a suspended *HTS* thin film resonator in Willemsen, Derov, Silva, Sridhar [671]. Other reported nonlinearities include: ***notches***, ***anomalies developing at the resonance lineshape*** and ***frequency hysteresis*** **[560]**.

Bachar, Shtempluck, Buks [570] used the chaos theory to research and understand the nature of nonlinearities in $YBa_2Cu_3O_{7-\delta}$ microwave resonator. Researchers [570] studied the superconducting $YBa_2Cu_3O_{7-\delta}$ microwave resonators, with an integrated microbridge, and their response to a monochromatic injected signal. In a bound region in the input power-input frequency plane, and close to a resonance frequency, ***a nonlinear phenomenon of self-sustained modulation of the reflected microwave power of the resonator*** is observed [570]. Authors [570] attribute this behavior to thermal instability in the microbridge, which alternately oscillates between normal- and super-conductive phases, and shifts a resonance frequency back and forth. Additional non-linear behavior of the reflected signal, such as ***period doubling bifurcation*** and ***jumping between different self-modulation frequencies***, is attributed to coupling of several closely lying resonance frequencies in the resonator [570]. Researchers [570] generalized a theoretical model, which can account for the various nonlinear phenomena in the experimental data, both in the case of single mode self-modulation, and where coupling between several modes is involved [570].

The author of dissertation would like to add a comment that the term: ***Duffing like nonlinearity,*** used by Abdo, Segev, Shtempluck, and Buks [560], means that this nonlinearity can be described by the ***Duffing equation*** in Duffing [571], Mosekilde [572], A. P. Kuznetsov, S. P. Kuznetsov, Ryskin [646], S. P. Kuznetsov [647, 648]

$$\frac{d^2x}{dt^2} + k\frac{dx}{dt} + ax + bx^3 = B\cos t\,.$$

*Duffing equation* describes the periodically forced nonlinear oscillator, and may have the steady state solutions as well as the persistent irregular chaotic solutions with arising sub-harmonics, ultra sub-harmonics and super-harmonics in Mosekilde [672]. *Duffing equation* [571] also applies to different electrical resonance circuits with magnetic flux saturation [672].

In connection with the above discussion on the *Duffing equation* [672], the authors of book made an assumption that, at certain conditions, the occurrence of new possible nonlinear dynamical phenomena due to:

1) *interior crisis*, which is associated with a sudden change in size of a chaotic attractor, and

2) *boundary crisis*, which is associated with a sudden disappearance of the attractor as it collides with its basin boundary,

may be observed in the *Ikeda map* in Ikeda [573], Ikeda, Daido, Akimoto [574] in addition to the *saddle node bifurcations* and *period-doubling bifurcations* in $YBa_2Cu_3O_{7-\delta}$ thin film microstrip resonators and in $YBa_2Cu_3O_{7-\delta}$ thin film dielectric cavity resonators at microwaves. In 2011, the innovative research on the new nonlinear dynamical phenomena has been conducted by the author of dissertation in V. O. Ledenyov, D. O. Ledenyov [575]. The modeling of output spectrum in response to two-tone fundamental input signals with equal microwave powers for different types of $YBa_2Cu_3O_{7-\delta}$ microstrip resonators at microwaves was performed, using the parallel computing techniques [575]. Modeling used the measured surface resistance *Rs* data for researched $YBa_2Cu_3O_{7-\delta}$ thin films at microwaves to study the intermodulation distortion (*IMD*), arising at appearance of nonlinear surface-impedance in *HTS* thin film at microwaves, which is observed at application of two-tone fundamental input signals with increased microwave power levels. Two conditions of *IMD* measurements in $YBa_2Cu_3O_{7-\delta}$ microstrip resonators were modelled: 1) in-band IMD modeling, 2) out-of-band *IMD* modeling. Duffing like nonlinearities were observed in $YBa_2Cu_3O_{7-\delta}$ microstrip resonators with different geometries in every researched case [575]. The modeling of the periodically forced nonlinear oscillator with application of Duffing equation results in the steady state solutions and in the persistent irregular chaotic solutions with arising sub-harmonics, ultra sub-harmonics and super-harmonics. The new nonlinear dynamical phenomena, including: 1) interior crisis, which is associated with a sudden change in size of a chaotic attractor, and 2) boundary crisis, which is associated with a sudden disappearance of the attractor as it collides with its basin boundary, were observed in the Ikeda map in addition to the saddle node bifurcations and period-doubling bifurcations in $YBa_2Cu_3O_{7-\delta}$ microstrip resonators at 1.985GHz in V. O. Ledenyov, D. O. Ledenyov [575], leading to the significant improvements in *HTS* microwave resonators design for applications in wireless communications and radars.

Kermorvant, van deer Beek, Mage, Marcilhac, Lemaitre, Briatico, Bernard, Villegas [576] identify the following possible mechanisms causing the nonlinearities of surface impedance in high $T_C$ superconductors in Tab. 15 [576]:

1. ***Intrinsic nonlinearity due to pair breaking***;
2. ***Weakly coupled grain***;
3. ***Josephson magnetic vortices in weak links***;
4. ***Abricosov magnetic vortices penetration into the grains***;
5. ***Non-Uniform heating***;
6. ***Heating of weak links***.

| Mechanism | Refs. | $R_s$ and $X_s$ microwave field dependence | $r$ value | Temperature dependence of $r$ | Frequency dependence of $r$ |
|---|---|---|---|---|---|
| Intrinsic nonlinearity | 14 and 15 | $\propto H_{rf}^2$ low power | $10^{-2}$ | Increase with $T$ | $\propto\omega$ |
| Pair breaking | | $\propto H_{rf}^4$ high power | | | |
| Weakly coupled grain | 16–18 | $\propto H_{rf}^2$ | $10^{-3}$ | Increase with $T$ | $\propto\omega$ |
| Vortices in weak link | 19–21 | $\propto H_{rf}$ | $\leqq 1$ | $T$ independent | $\omega$ independent |
| Vortex penetration to the grains | 19–21 | $\propto H_{rf}^n, n\sim 4$ | constant $\sim 1$ | $T$ independent | $\omega$ independent |
| Uniform heating | 22–25 | $\propto H_{rf}^2$ | $10^{-2}$ | Increase with $T$ | $\propto\omega$ |
| Heating of weak link | 22–25 | Unknown | $\simeq 1$ | $T$ independent | $\omega$ independent |

**Tab. 15.** Possible mechanisms causing nonlinearity of surface impedance of high $T_C$ superconductors (after [576]).

Kermorvant, van der Beek, Mage, Marcilhac, Lemaitre, Bernard, Briatico [577], using the dielectric resonator method, researched the nonlinearities in the surface impedance $Zs = Rs + jXs$ of $YBa_2Cu_3O_{7-\delta}$ thin films at 10 *GHz* as a function of the incident microwave power level and temperature [577]. Author of dissertation created the Tab. 16 with the summarized research results on sources of nonlinearities in *HTS* thin films at microwaves by Kermorvant, van der Beek, Mage, Marcilhac, Lemaitre, Bernard, Briatico [577].

| | |
|---|---|
| 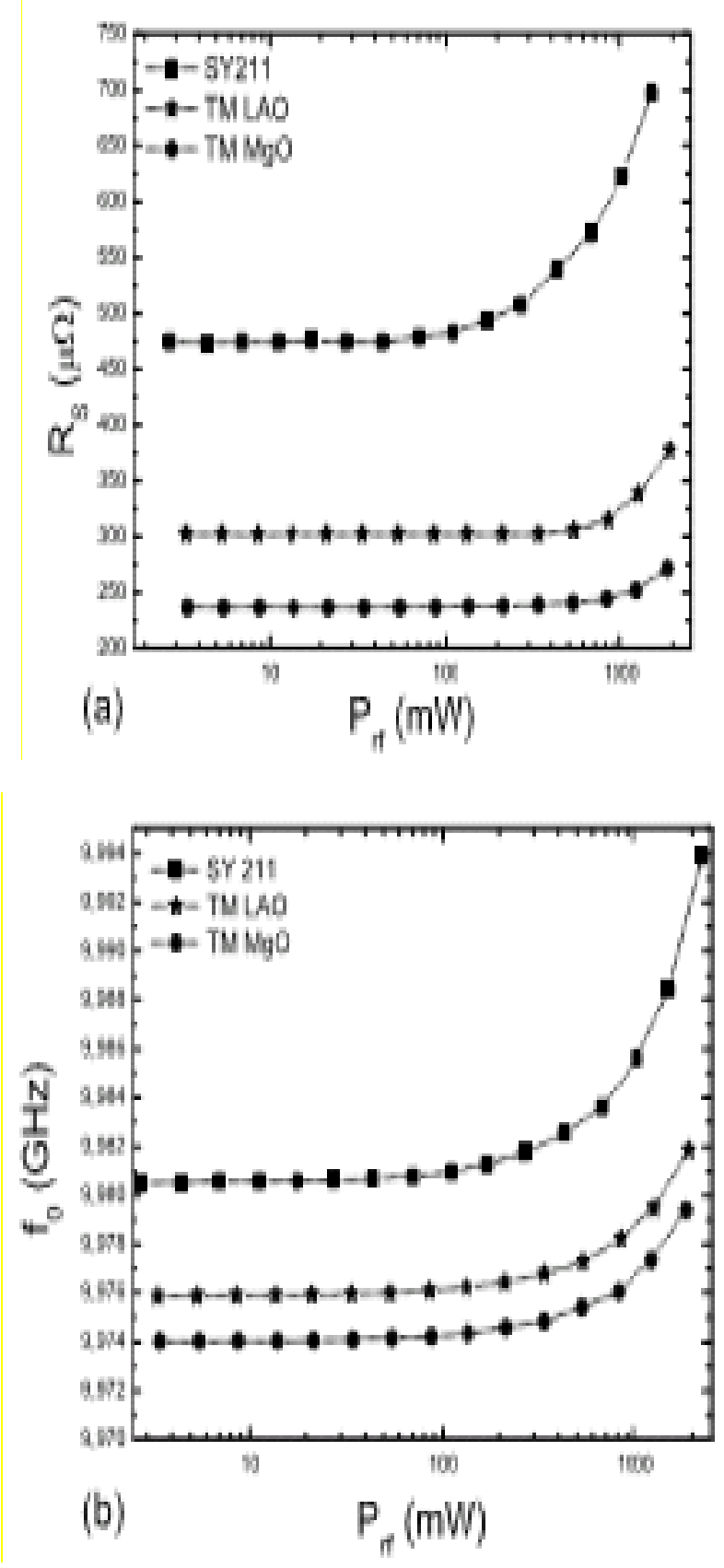  | **Fig. 96.** Dependence on microwave reactive power of *Rs* and $f_0$ for the investigated $YBa_2Cu_3O_{7-\delta}$ films on different substrates at a given temperature of 74 *K* (after [577]). Nonlinear dependence $R_s(P_{rf})$ is clearly observed at high levels of applied microwave power in Fig. 96. The microwave reactive power is defined by $P_{rf} = P_{incident, rf} \times Q_L$. The zero power limit is taken as those values of $P_{rf}$ below which $R_s(P_{rf})$ is essentially $P_{rf}$ -independent. Curves for different films present the same behavior, i.e. $R_s(P_{rf}\ )$ and $f_0(P_{rf})$ are independent of the microwave field in the zero field limit and become nonlinear (increase rapidly) above a threshold value of $P_{rf}$. Contrary to what is expected, the resonator frequency $f_0$ also increases with increasing microwave losses. Authors [577] ascertain that the increase of $f_0(P_{rf})$ is due to the heating of the rutile resonator by the $YBa_2Cu_3O_{7-\delta}$ film. |
| 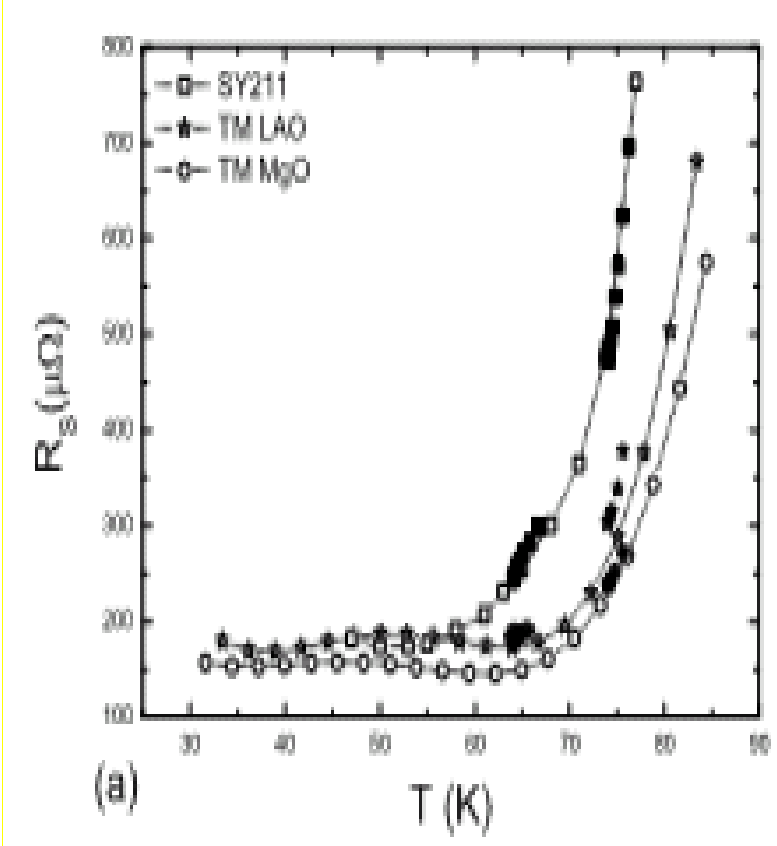  | **Fig. 97.** *Temperature* dependence of the *Rs* **(a)** and $f_0$ **(b)** in the regime of small microwave power (after [577]). The surface resistance of the $YBa_2Cu_3O_{7-\delta}$ films shows the usual monotonous increase with temperature for all samples. Concerning the temperature dependence of the resonant frequency, we observe an increase with temperature, as discussed previously. The only observed difference is the nearly constant frequency offset between the three curves. The estimation of the temperature of the resonator in the swept-power experiments is done using the curves measured as a function of temperature. Fig. 98 shows the temperature calculated from the variation of the resonant frequency [577]. The temperature obtained from |
| 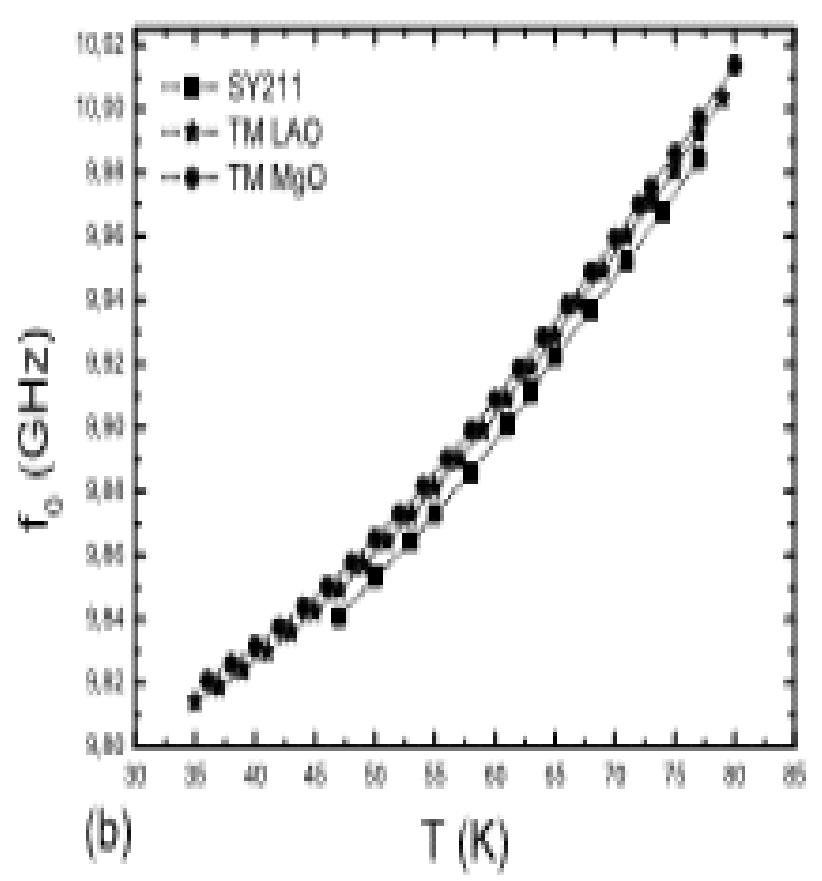 | the variation of the surface resistance gives exactly the same results [577]. This means that no intrinsic $P_{rf}$ dependence of $R_s$ is measured [577].<br>In Fig. 97(a), authors superpose surface resistance $R_s$ as a function of the measured temperature in the low power regime, and $R_s$ measured as a function of the estimated temperature in the swept power experiments at 74 *K* and 63 *K*. A very good agreement is observed showing that ***any observed nonlinearity is the consequence of Joule heating [577].*** |
| 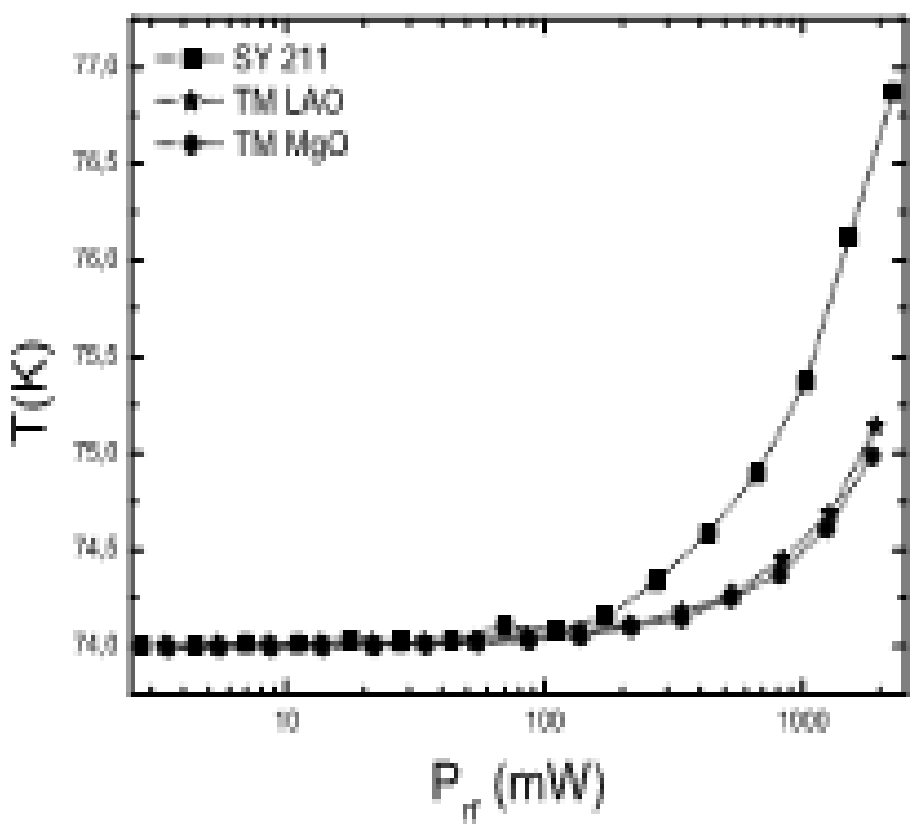 | **Fig. 98.** Estimated temperature increase as function of reactive microwave power *T($P_{rf}$)* (after [577]). |

**Tab. 16.** Research results on sources of nonlinearities in *HTS* thin films at microwaves by Kermorvant, van der Beek, Mage, Marcilhac, Lemaitre, Bernard, Briatico (after [577]).

The Laboratoire des Solides Irradiés at Saclay Institute of Matter and Radiation has a vision that the non-linear response to the high microwave power required for telecommunications is the major source of degradation of the signal, for which various causes have been proposed [578]:

1. ***The high frequency makes easier the excitation of the electrons in states where the electrical resistance is no longer zero.*** This effect is even more important for SHTC, because of the non-spherical symmetry of their ground state;

2. ***The screening current induced by a powerful electromagnetic field disrupts the vortex structure, allowing the penetration of the magnetic flux and hysteresis losses***;

3. ***The global or local warming of the material, because of very small periodic excursions of quasiparticles (unpaired electrons) or vortices induced by the microwave field***.

Kermorvant, van deer Beek, Mage, Marcilhac, Lemaitre, Briatico, Bernard, Villegas [579] explain that the Table 16 gives an overview of possible mechanisms leading to the nonlinear behavior of the surface impedance in Lahl, Wördenweber [580], D. E. Oates, Hein, Hirst, Humphreys, Koren, Polturak [581], Golosovsky [582]. First is the intrinsic nonlinearity due to pair breaking. The nonlinearity is then related to the increase in the quasiparticles density $n_{qp}$ in Dahm, Scalapino [583, 584], when the $H_{rf}$-induced current density is of the order of magnitude of the pair-breaking current density. A nonlinearity based on this effect has been predicted and investigated using a phenomenological expression for a nonlinear penetration depth. If the nonlinearity is dominated by this intrinsic mechanism, the r parameter should be small and strongly frequency dependent. This differs from the experimentally observed nonlinearities. Hysteretic losses are also proposed to be significant in Hylton, Kapitulnik, Beasley, Carini, Drabek, Gruner [585], Hylton, Beasley [586]. The weakly coupled-grain model holds that the large surface resistance of highly granular high-Tc superconductors as compared to single crystals can be explained by the different morphology. The polycrystalline samples can be modeled as a network of Josephson junctions. Nonlinear behavior is expected at *RF*-current densities that are limited by critical current density of the constituent Josephson junctions. This model yields a very small r parameter with a strong dependence on temperature and frequency. In the case of granular films the coupled-grain model describes the microwave nonlinearities fairly well. Nevertheless, it fails to describe strong nonlinearities in epitaxial films. Vortex penetration and creep into grain boundaries and/or weak links are also proposed as a possible source of microwave hysteretic losses. Vortex generation by the microwave magnetic field has been predicted in Dam, Huijbregtse, Klaasen, van der Geest, Doornbos, Rector, Testa, Freisem, Martinez, Stäuble-Pümpin, Griessen [587], Sridhar [588], McDonald, Clem, D. E.

Oates [589] and the *Bean model* has been extended to account for microwave nonlinearity in *HTS* films. The dependence of *Rs* on *Xs* is almost linear (with slope *r*). Experimental values are close to those predicted by the model. The last possible effect to explain nonlinearity is local or uniform heating in Hein [530], Wosik, Xie, Nesteruk, Li, Miller, Long [590], Pukhov [591, 592]. It has been proposed that heating can occur in superconducting thin films. This effect appears in the microwave frequency range, particularly in continuous mode, but also in pulsed mode, as a function of the pulse period. Heating is significant above a certain value of the incident microwave power and causes the transition to the normal state of weaker superconducting regions such as weak links or local defects. Heating and heat transfer to the substrate are shown to play an important role.

***The non-linear behaviour of $R_s(P_{rf})$ dependence is also due to the vortex penetration*** in *HTS* thin film, hence it is necessary to improve the material properties by optimization of the Abricosov magnetic vortices pinning in *HTS* thin films at microwaves as explained in Kermorvant [593, 595], Kermorvant, van der Beek, Mage, Marcilhac, Lemaitre Briatico, Bernard, Villegas [594].

Van der Beek, Konczykowski, Abal'oshev, Abal'osheva, Gierlowski, Lewandowski, Indenbom, Barbanera [596] write on the *Abricosov magnetic vortices* pinning optimization in *HTS* thin films that the high vortex pinning forces lead to high critical currents are an indispensable prerequisite for superconducting thin films if these are to be used in electronic and power applications. The ''volume'' pinning force exerted by material impurities on flux vortices opposes their motion, which is at the origin of flux noise and dissipation. ***Strong vortex pinning decreases noise in electronic and superconducting quantum interference devices (SQUID's) and leads to high quality factors necessary for the correct operation of radio-frequency and microwave filters and cavities*** *[596]*. The $YBa_2Cu_3O_{7-\delta}$ films combine a high critical current density $J_c$ with a high critical temperature $T_c$ and are thus ideal candidates for widespread application. The optimization of the critical current through identification and tailoring of defect microstructures which lead to high pinning and reduced vortex creep has therefore attracted a lot of interest in the high-temperature superconductor *HTC* community [596].

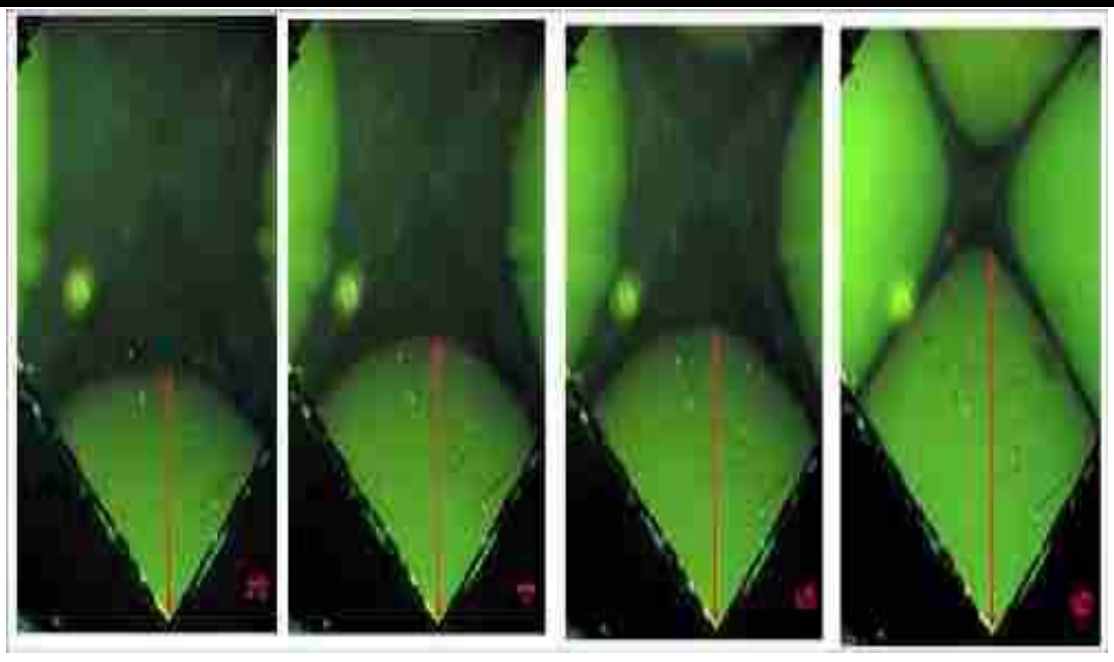

**Fig. 99.** Magneto optical imaging of the penetration of a static magnetic field, induced by the increasing power of the microwave field in a $YBa_2Cu_3O_{7-\delta}$ layer (*T=72 K*). The green light areas are penetrated by the magnetic flux, that is to say, by vortices. The green light areas are penetrated by the magnetic flux, that is to say, by vortices. The extension of the dark area in the center, free of vortices, shrinks with the amplitude of the microwave field (after [597]).

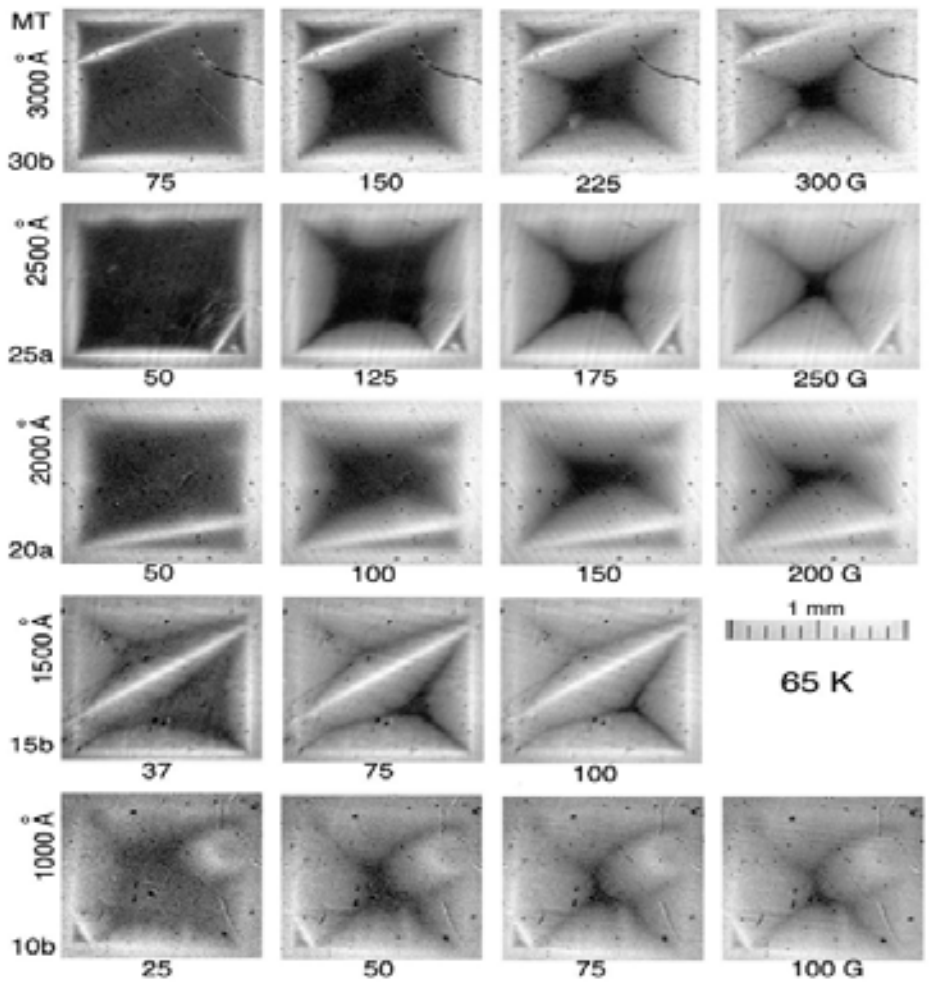

**Fig. 100.** Magneto-optical images of flux penetration into laser-ablated $YBa_2Cu_3O_{7-\delta}$ films of thickness *d5300 nm, 250 nm, 200 nm, 150 nm, and 100 nm*, respectively, at 65 *K*. Bright areas correspond to a high magnetic induction perpendicular to the film, while dark regions are areas of low flux density. Horizontal sequences correspond to constant film thickness and varying applied field $H_a/d$ the values are indicated below each frame, while the vertical sequences are deliberately chosen such that $H_a\ /d$ is approximately constant—a procedure which reveals the weaker pinning in the thinner films. This is most clearly seen for the lower fields. Weak links in the films are revealed as the bright lines of preferential flux penetration. (after [596]).

**Tab. 17.** Research results on sources of nonlinearities in *HTS* thin films at microwaves by Kermorvant [593, 595], Kermorvant, van der Beek, Mage, Marcilhac, Lemaitre Briatico, Bernard, Villegas [594], and by Van der Beek, Konczykowski, Abal'oshev, Abal'osheva, Gierlowski, Lewandowski, Indenbom, Barbanera [596] (after [555, 596]).

Most recently, Julien Kermorvant, Jean-Claude Mage, Bruno Marcilhac, Yves Lematre, Jean-Francois Bobo, and Cornelis Jacominus van der Beek reported the research results on the microwave heating-induced *dc* magnetic flux penetration in $YBa_2Cu_3O_{7-\delta}$ superconducting thin films [642]. Authors [642] developed an experimental set-up to simultaneously image the magnetic flux penetration into the superconductors, using the magneto-optical technique, and the measurement of their surface resistance in the range: 1 – 10 *GHz*. Authors [642] came to the conclusion that the application of a high-power microwave signal significantly enhances the *dc* flux penetration into $YBa_2Cu_3O_{7-\delta}$ due to the local heating of superconducting film in Tab. 18. Authors [642] decided that the local heating is indeed at the origin of the enhanced flux penetration and the nonlinear dependence $Rs(P_{rf})$.

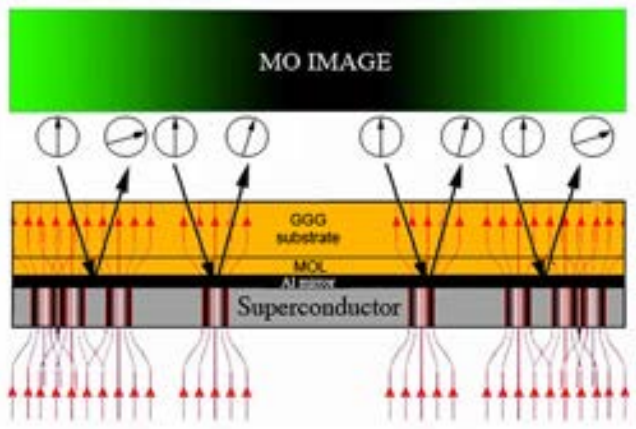

**Fig. 101**. Principle of MOI of superconductors. Thin drawn lines depict the magnetic flux as this traverses the superconductor and the MOL, thick black lines show the optical path of the impinging and reflected light, and the circled arrows illustrate the linear polarization direction of the light (after [642]).

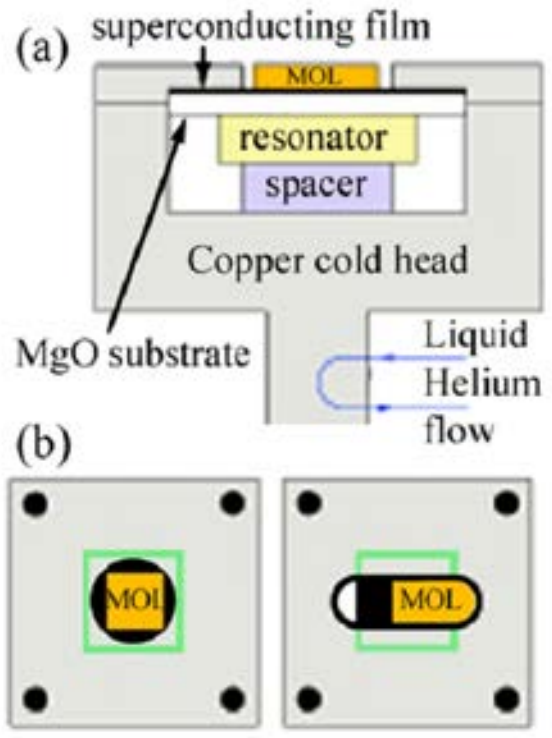

**Fig. 102**. Schematic view of the experimental assembly. Panel (a) shows a side view, while panel (b) depicts a top view for different lid apertures. The left-hand side cover allows the imaging of the central region of the sample, while the right-hand cover allows for the observation of the edge region (after [642]).

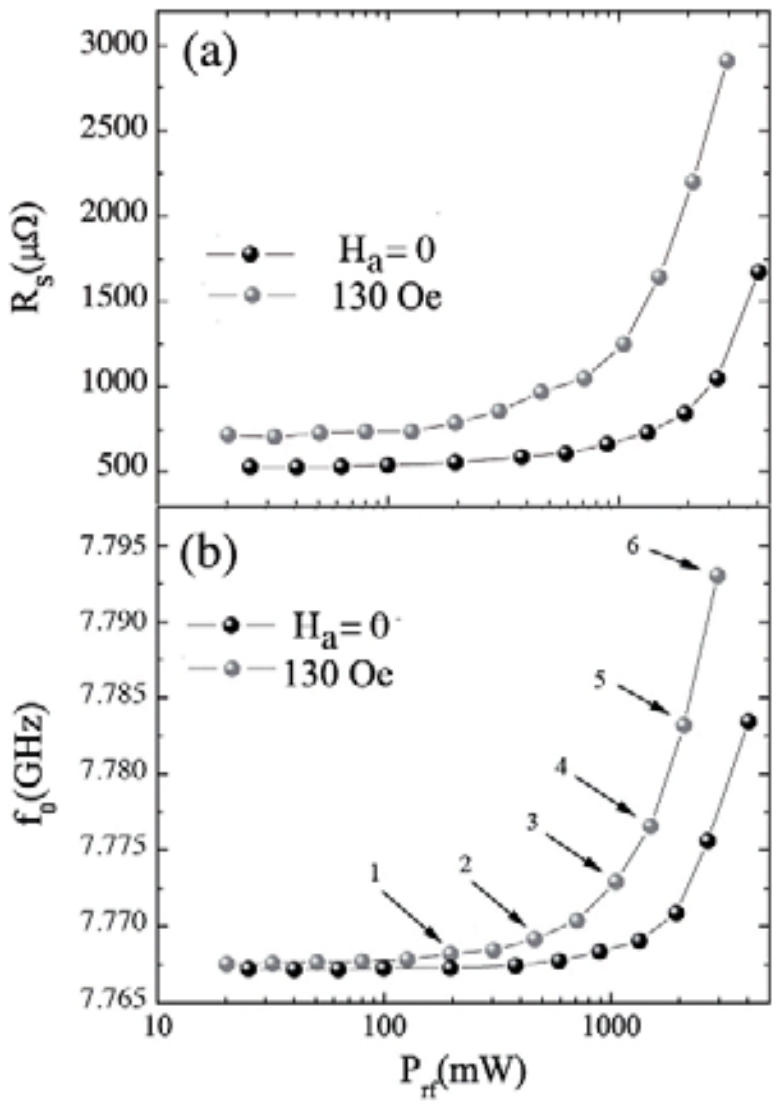

**Fig. 103**. (a) Surface resistance at 7 GHz as function of the input microwave power at 60 K, for zero applied field, and Ha = 130 Oe. (b) Resonant frequency as a function of the input microwave power at 60 K (after [642]).

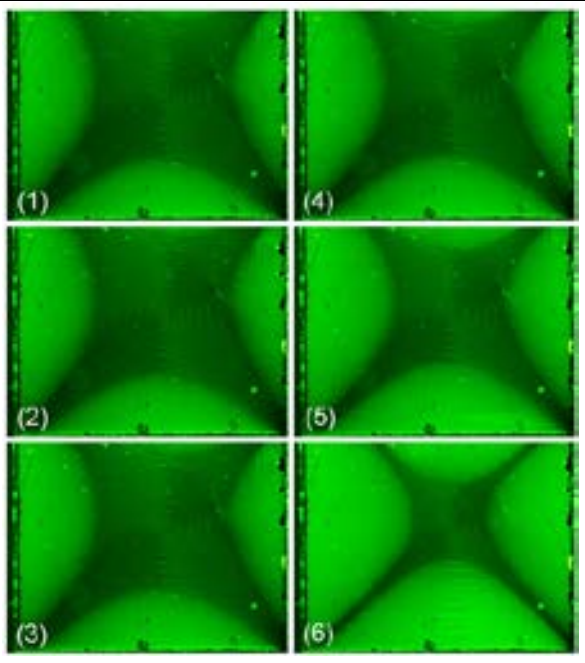

**Fig. 104**. MOI, showing the progression of the flux front as the microwave field power is increased. The images correspond to the microwave powers denoted by the arrows in Fig. 104(a): 1. No microwave field; 2. 10 dBm; 3. 14 dBm; 4. 17 dBm; 5. 20 dBm; 6. 24 dBm (after [642])..

**Tab. 18.** Research results on sources of nonlinearities during microwave heating-induced *DC* magnetic flux penetration in $YBa_2Cu_3O_{7-\delta}$ superconducting thin films by J. Kermorvant, J.-C. Mage, B. Marcilhac, Y. Lematre, J.-F. Bobo, C. J. van der Beek (after [642]).

Willemsen, Derov, Sridhar [598] investigated the critical-state flux penetration and linear and nonlinear microwave vortex response in $YBa_2Cu_3O_{7-d}$ films in Figs. 105-107 in Tab. 19. Researchers [598] report on microwave surface impedance measurements in patterned microstrip resonators to understand the vortex dynamics and flux penetration. Specially designed microstrip resonators patterned from thin films are used by authors as high *Q* structures to probe ***vortex response at microwaves*** in the presence of dc magnetic fields [598]. The high sensitivity of these structures enables measurement of the vortex response in fields from *0.1 mT* to several *T* [598]. The complex microwave surface impedance $Zs=Rs+iXs$ was measured as functions of magnetic field *H*, temperature *T*, and frequency $\omega$ (in the range 1 to 20 *GHz*) [598]. *Zs* is a measure of the total flux in the sample, and hence can lead to precise tests of flux penetration in thin films. In addition the results also yield information on vortex parameters such as *pinning force constants* and *viscosity* [598]. Authors [598] focus on the results for changes in surface impedance induced in $YBa_2Cu_3O_{7-\delta}$ thin films as the externally applied magnetic field is slowly increased from zero, as well as the hysteresis loops, which ensue, when the field is subsequently reduced. The response clearly shows evidence of two relevant field scales. As the field is increased an initial linear rise gives way to a superlinear behavior at *Hi* followed by a linear rise above *Hs* which persists to the highest fields available. While virgin response for $H>Hi$ and the magnitude of *Hs* appear to be consistent with the recent analytical expressions for thin strips in perpendicular fields, the hysteretic response is not so easily understood. The critical state model predicts a counter-clockwise hysteresis loop governed by a field scale of order *Hs*, while the experiments observe a clockwise hysteresis loop governed by the much smaller field scale *Hi*. The geometrical barriers are insufficient to explain the observed hysteresis loops. The results suggest that pinning at the film edges may be weaker than in the bulk [598]. Considering the linear electrodynamic response of vortices in thin films in the perpendicular geometry at microwaves, researchers proposed theoretical models to explain the observed nonlinearities [598].

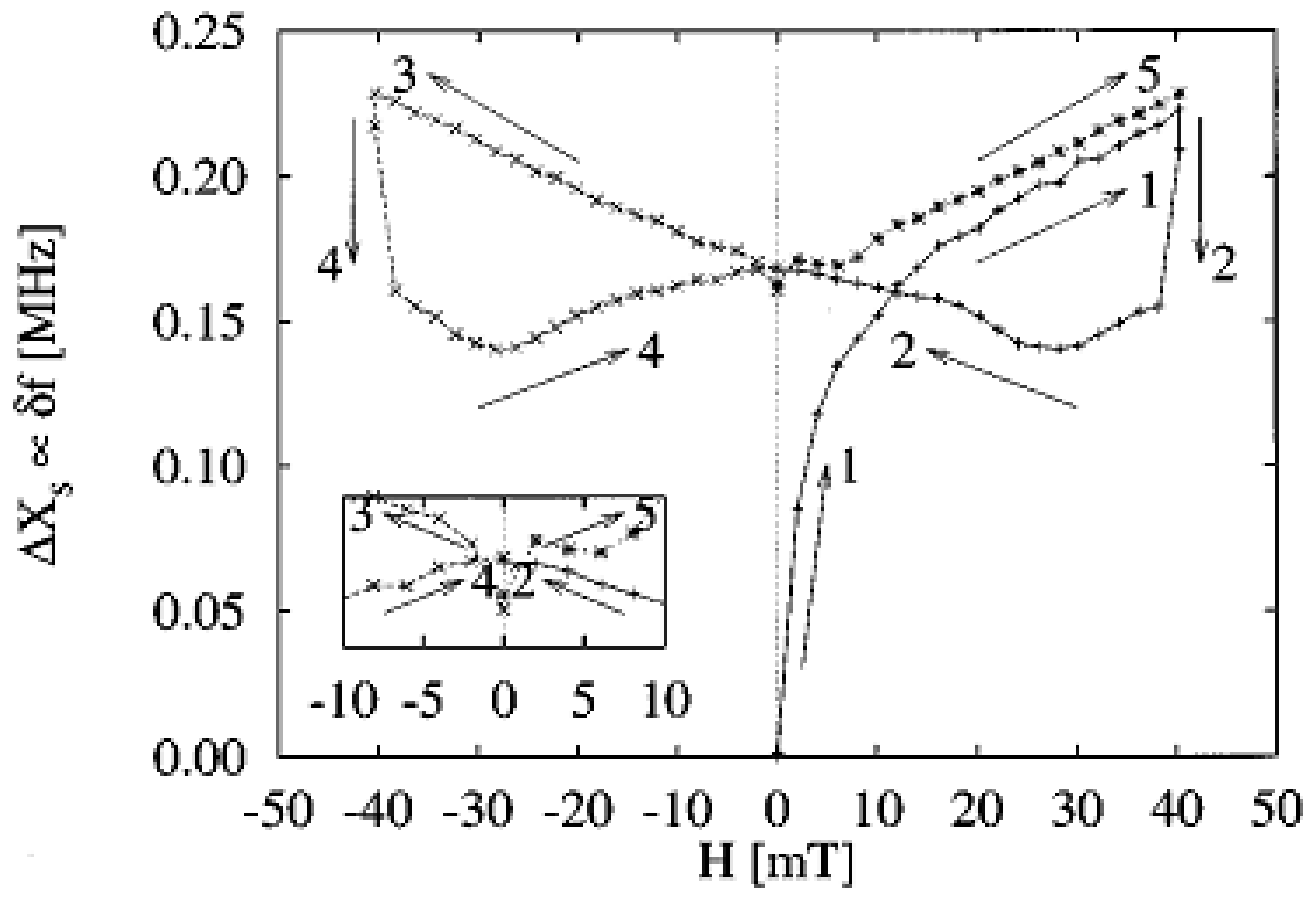

**Fig. 105.** Hysteretic response of *Xs(H)* for $H_{max}$~*Hi*. Results are presented for resonator N4 at 3.7 *GHz* and 35 *K*. (after [598]).

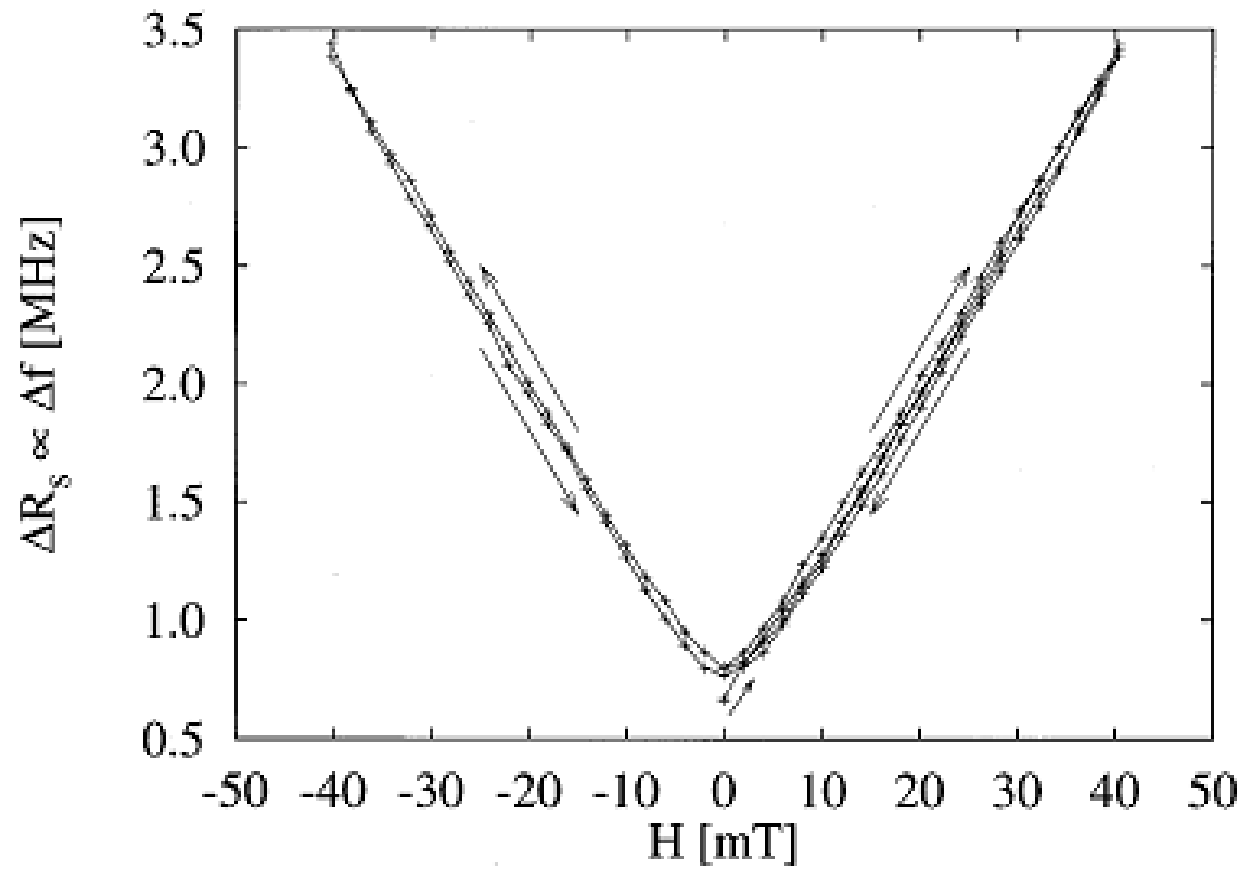

**Fig. 106.** Low-field hysteresis in *Rs(H)* at high temperatures. Results are presented for resonator N4 at 3.7 *GHz* and 84 *K* (after [598]).

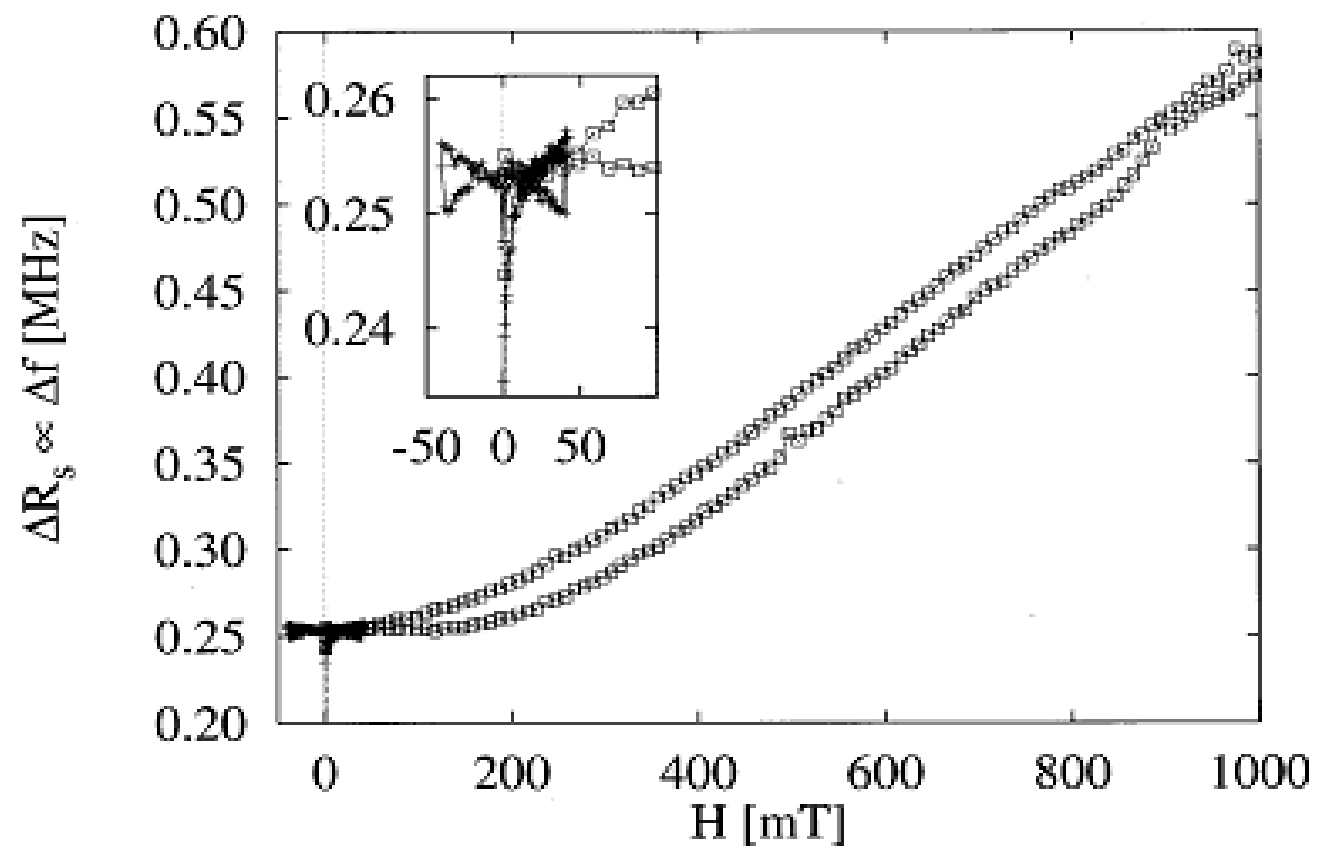

**Fig. 107.** Magnified view of the hysteresis data for *Rs(H)* dependence for moderate fields and low fields (inset) (after [598]).

**Tab. 19.** Research results on sources of nonlinearities, caused by magnetic vortex penetration, in *HTS* thin films at microwaves by Willemsen, Derov, Sridhar (after [598]).

Hua Zhao, Xiang Wang and Judy Z. Wu [599], when discussing the correlation of microwave nonlinearity and magnetic pinning in high-temperature superconductor thin film band-pass filters, mention that, unfortunately, *a nonlinear effect appears at moderate to high power levels, limiting the practical applications of the HTS passive microwave devices*. For example, in *rf* filters, the ***third-order intermodulation*** **(*IM3*)** is perceived to be the most serious problem to deal with due to the generation of spurious signals within the filter pass band, which leads to deterioration of the filter performance [599]. Hua Zhao, Xiang Wang and Judy Z Wu [599] conclude that the microwave third-order intermodulation nonlinearity has been studied on two-pole X-band filters fabricated on three different kinds of *HTS* films including Tl-2212, Hg-1212 and $YBa_2Cu_3O_{7-\delta}$. Despite dramatic differences in the electronic anisotropy and magnetic pinning of these materials, the *IP3* and *Jc* of each material follow the same trend at the reduced temperature scale at T >77 *K* [599]. ***This result suggests an intimate relationship between the microwave nonlinearity and magnetic pinning at elevated temperature near Tc.*** The surprisingly lower *IP3* values obtained on $YBa_2Cu_3O_{7-\delta}$/CeO2/$YBa_2Cu_3O_{7-\delta}$ tri-layer microstrip filters with improved magnetic pinning, as opposed to the single-layer $YBa_2Cu_3O_{7-\delta}$ filter of the same thickness and device geometry, might be attributed to extra microwave losses on the additional interfaces between $YBa_2Cu_3O_{7-\delta}$ and $CeO_2$ [558]. In Tab. 20, the author of dissertation summarized the research results on sources of nonlinearities in *HTS* thin films at microwaves by Hua Zhao, Xiang Wang and Judy Z. Wu [599].

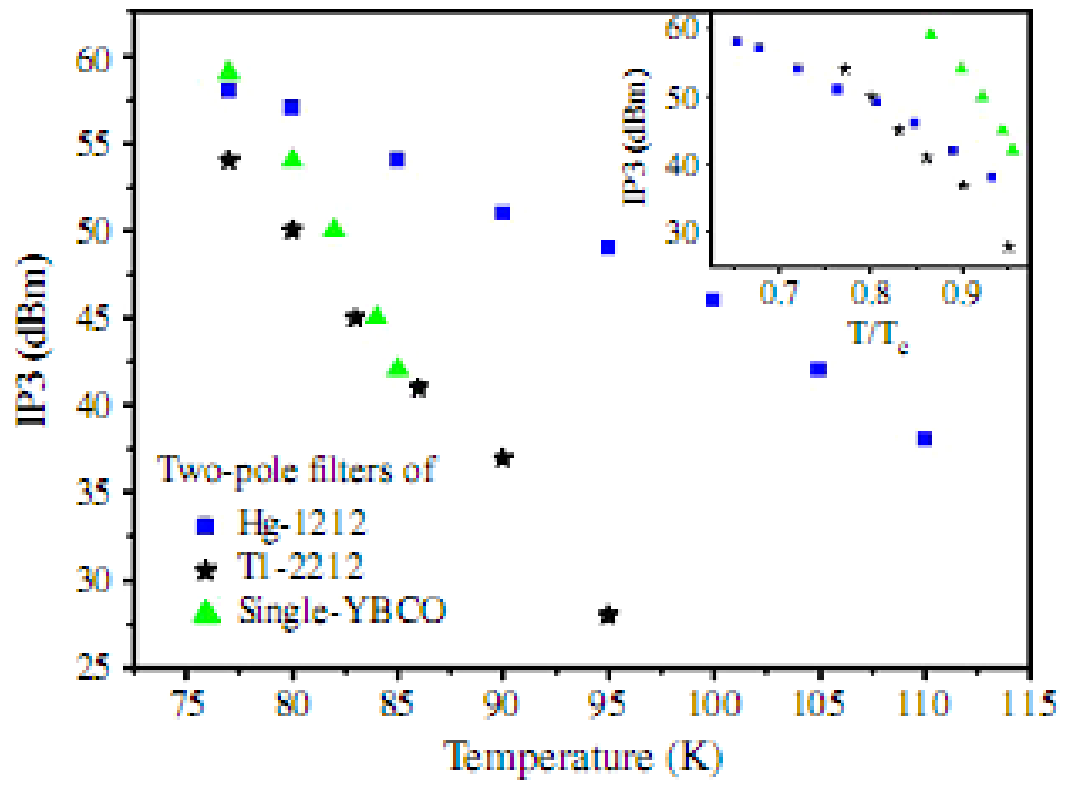

**Fig. 108.** Comparison of the *IP3* values for Hg-1212, Tl-2212 and $YBa_2Cu_3O_{7-\delta}$ filters as a function of temperature. Inset: plot of the *IP3* values for Hg-1212, Tl-2212 and $YBa_2Cu_3O_{7-\delta}$ filters at reduced temperature scale (after [599]).

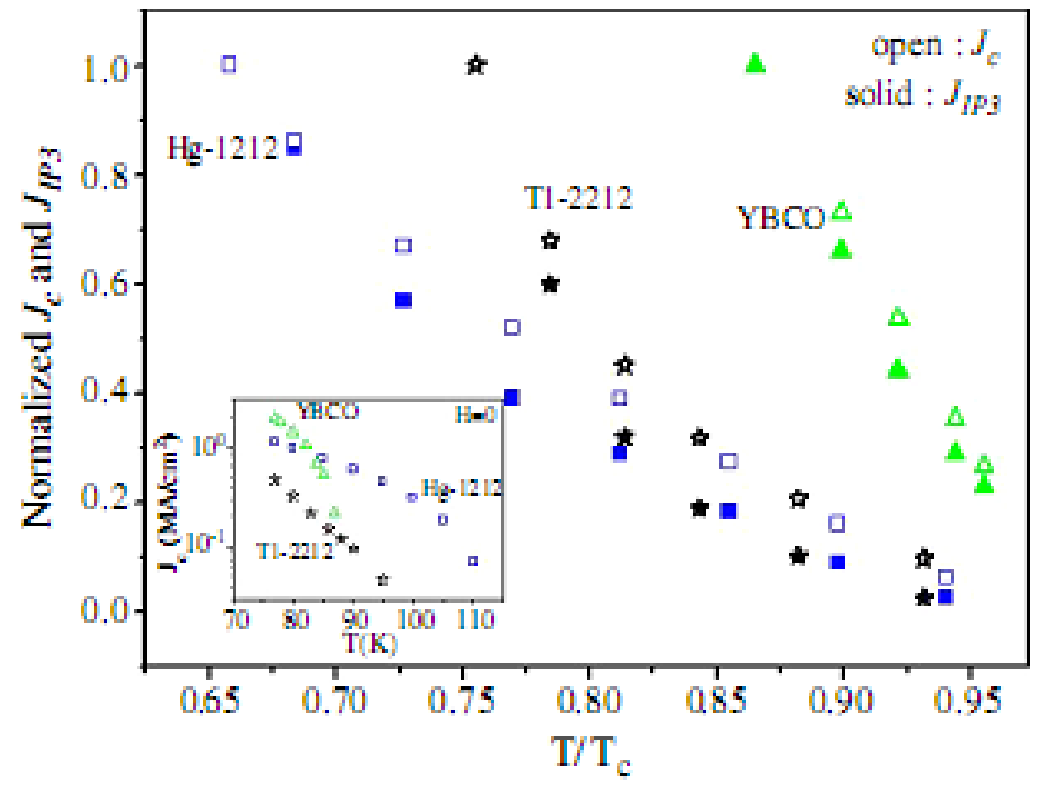

**Fig. 109.** Normalized $J_{IP3}/J_{IP3}$ (77 *K*) with *Jc/Jc* (77 *K*) against reduced temperature for Hg-1212, Tl-2212 and *YBCO* microstrip filters. Inset: the dc critical current density (*Jc*) values for Hg-1212, Tl-2212 and $YBa_2Cu_3O_{7-\delta}$ films before patterning at real temperatures (after [599]).

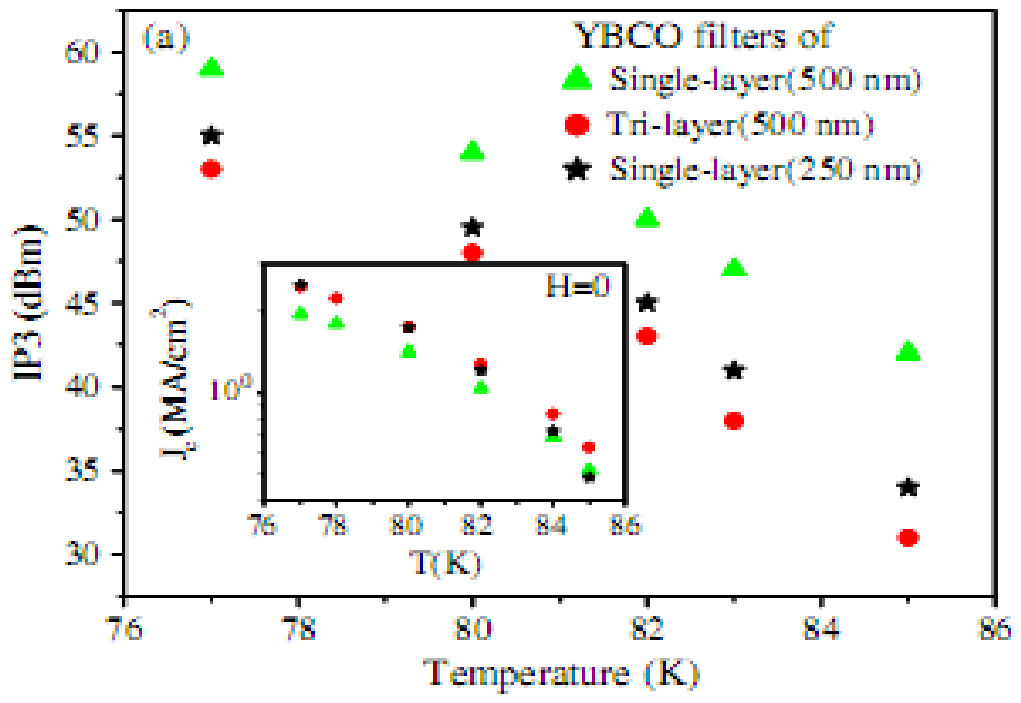

**Fig. 110.** The *IP3* values for single and tri-layer $YBa_2Cu_3O_{7-\delta}$ filters as a function of temperature. Inset: their Jc values versus temperature before patterning (after [599]).

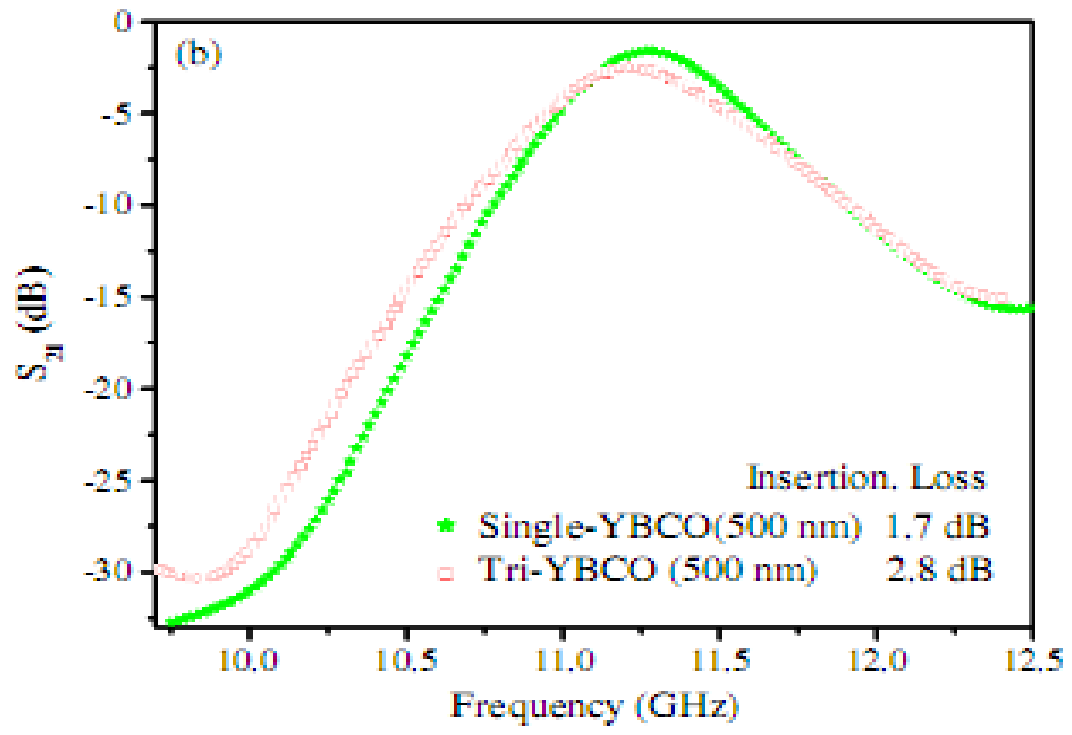

**Fig. 111.** Comparison of plots of the $S_{21}$ transmission coefficient versus frequency for two types of filters: single and tri-layer $YBa_2Cu_3O_{7-\delta}$ (after [599]).

**Tab. 20.** Research results on sources of nonlinearities in *HTS* thin films at microwaves by Hua Zhao, Xiang Wang and Judy Z. Wu (after [599]).

Futatsumori, Furuno, Hikage, Nojima, Akasegawa, Nakanishi, Yamanaka [600] mention that the planar *HTS* thin-film microwave filters are compact with the *steep skirt*, *low insertion loss* characteristics. However, the nonlinearities, observed in high-power transmitting filters or sometimes in low-power receiving filters, limit their practical applications. **The generation of intermodulation distortion (*IMD*) is one of the most important problems with regard to the nonlinear characteristics**, because technical specifications set strict limits of *IMD* generation to avoid *co-channel interference* or *adjacent-channel interference*. There are different origins of the nonlinearities, including:

1. ***Intrinsic effects*** such as *nonlinear Meissner effects* or *pair-breaking current* can play an important role in the generation of nonlinear characteristics; and

2. ***Extrinsic effects*** such as superconductor *weak links* or *Josephson junctions at grain boundaries* are analysed in [600].

In Tab. 21, the author of dissertation presents the main research findings on the *IMD* nonlinearities in microwave transmitting filters by Futatsumori, Furuno, Hikage, Nojima, Akasegawa, Nakanishi, Yamanaka [601], Futatsumori [602].

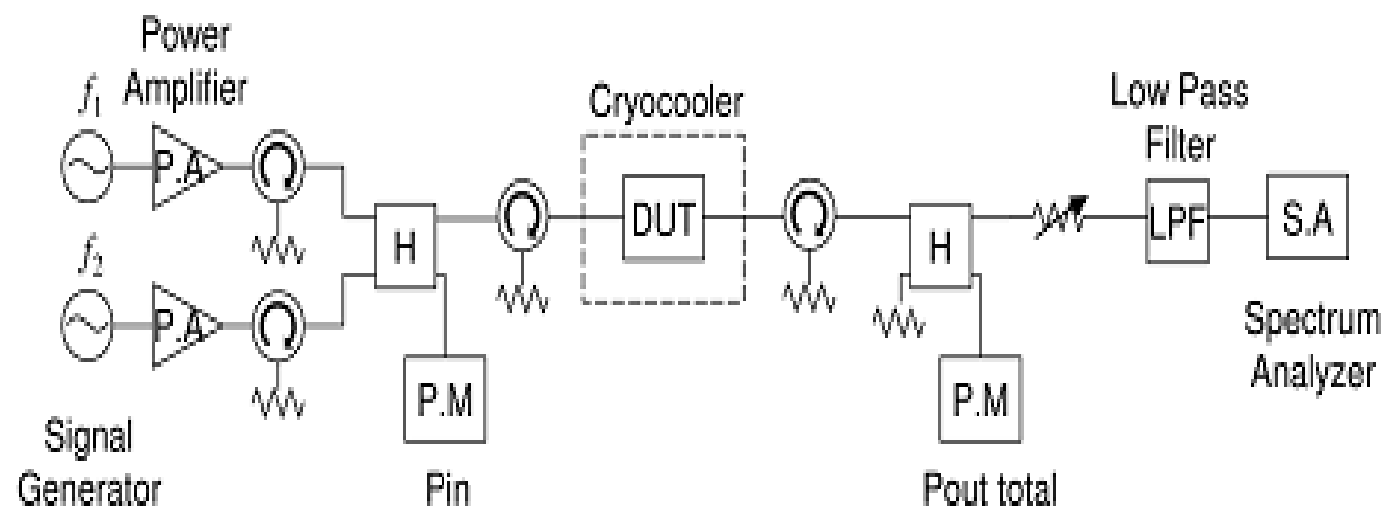

**Fig. 112.** Schematic diagram of an experimental setup for intermodulation distortion measurements using two-tone signal on 5-GHz three-pole HTS reaction-type filter, denoted by DUT (after [602]).

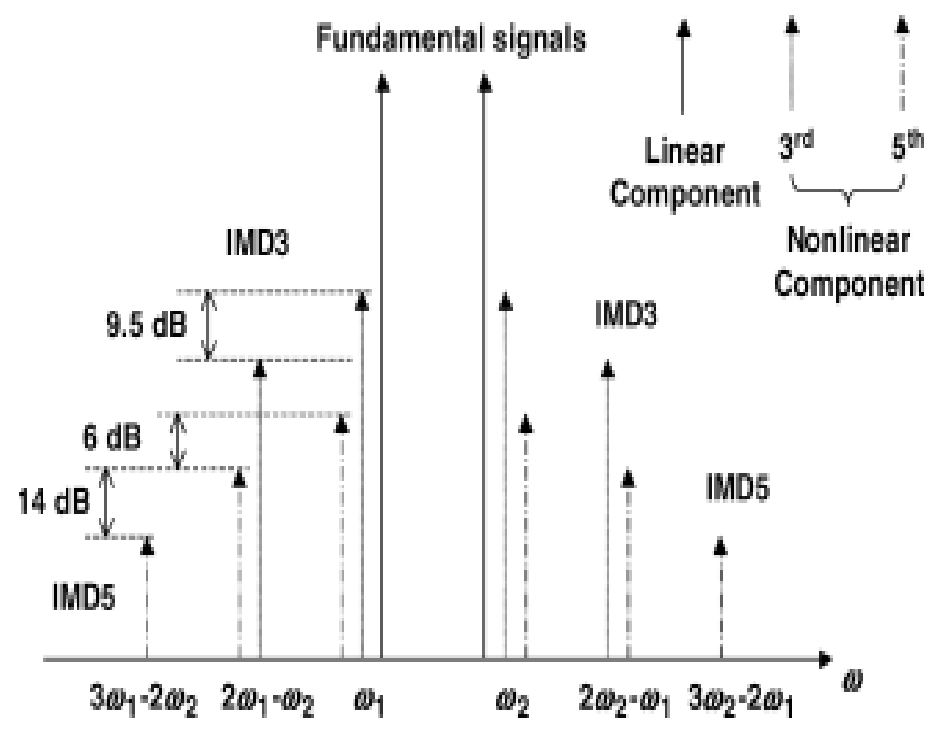

**Fig. 113.** Output spectrum of the nonlinear model in response to a two-tone fundamental signal represented by fifth-order complex power series. The contributions of each order nonlinear component to fundamental signals, IMD3, and IMD5 are described (after [601]).

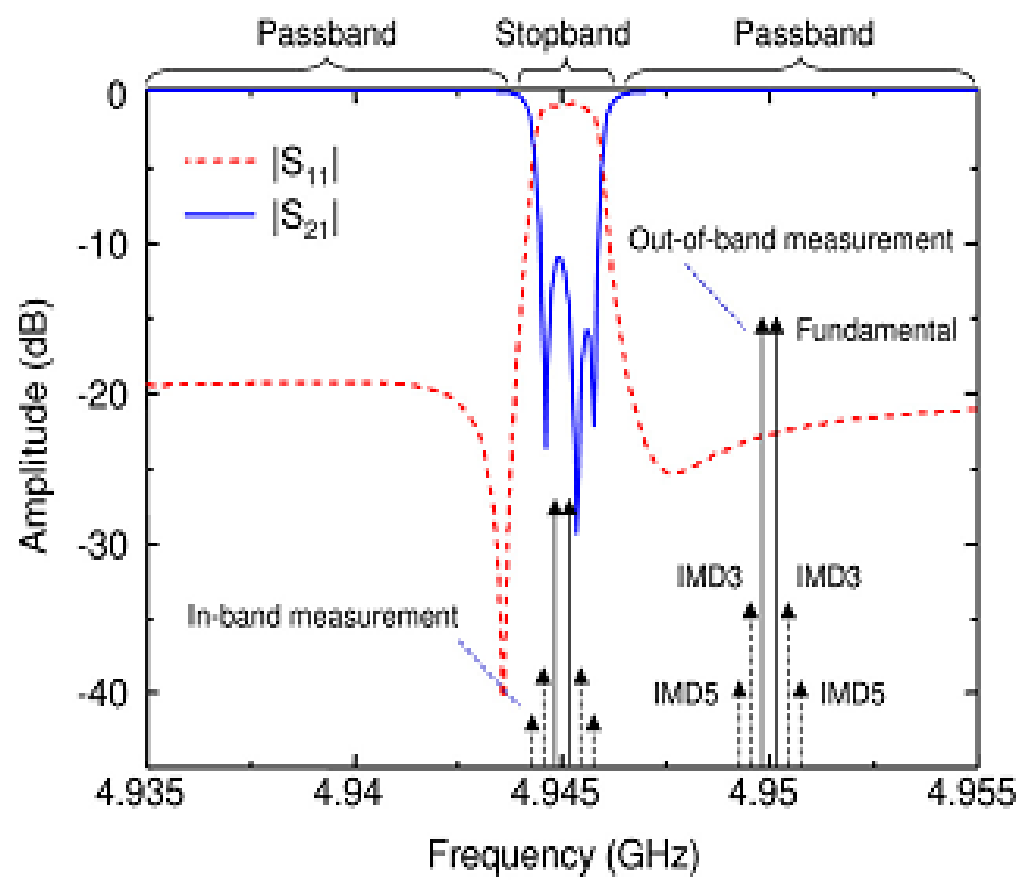

**Fig. 114.** Two conditions of intermodulation distortion measurements are shown on the measured S-parameter response. Two-tone signal is input at the centre of stopband region in the in-band measurements. In addition, the signal is input in the passband region (5MHz above offset from the 3dB band edge) in out-of-band measurements (after [602]).

<table>
<tr>
<td>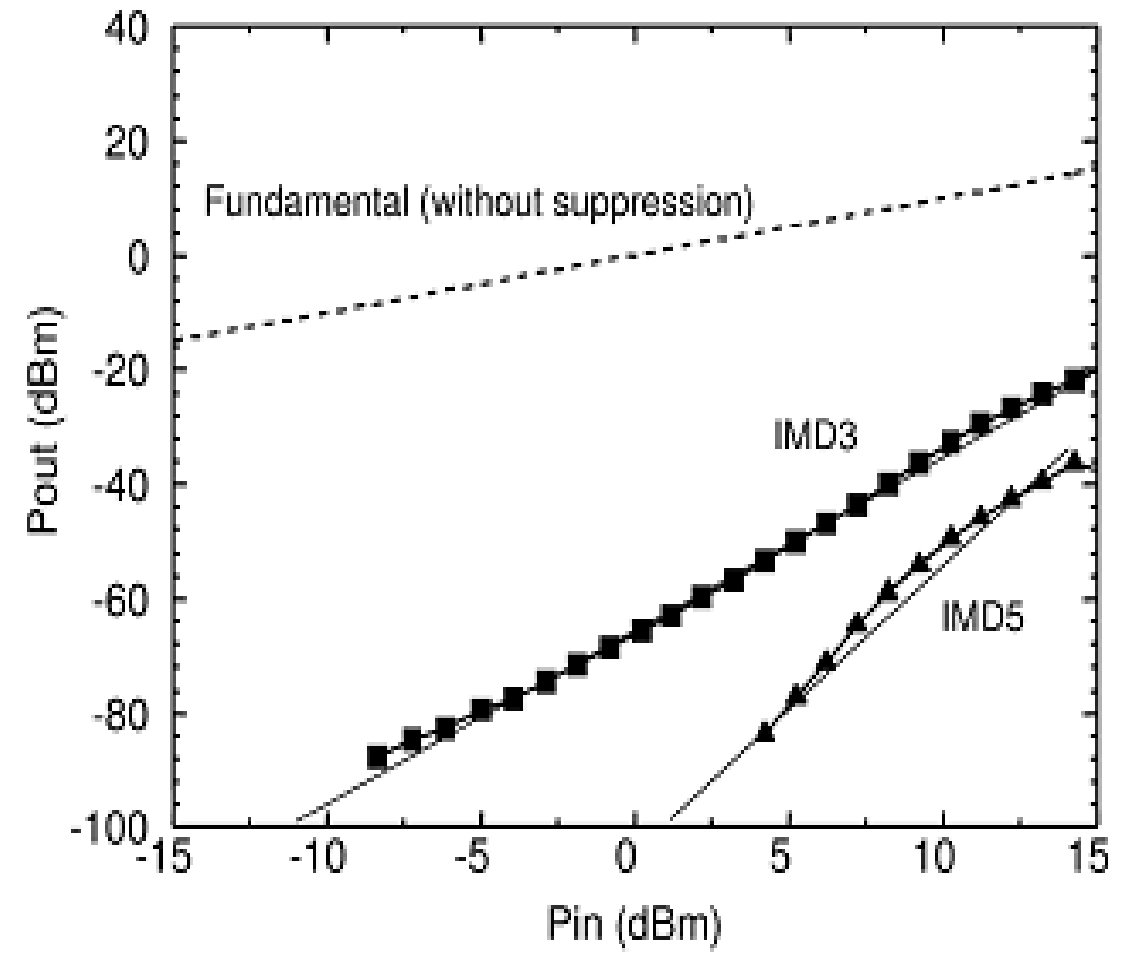
</td>
<td>Fig. 115. Measured intermodulation distortion values in the stopband region. The closed circles represent the amplitude of the fundamental signal. The closed boxes and closed triangles represent the amplitudes of IMD3 and IMD5, respectively. (The solid straight lines are the fitting results for the fundamental signal, IMD3, and IMD5 with slopes of 1, 3, and 5) (after [602]).</td>
</tr>
<tr>
<td>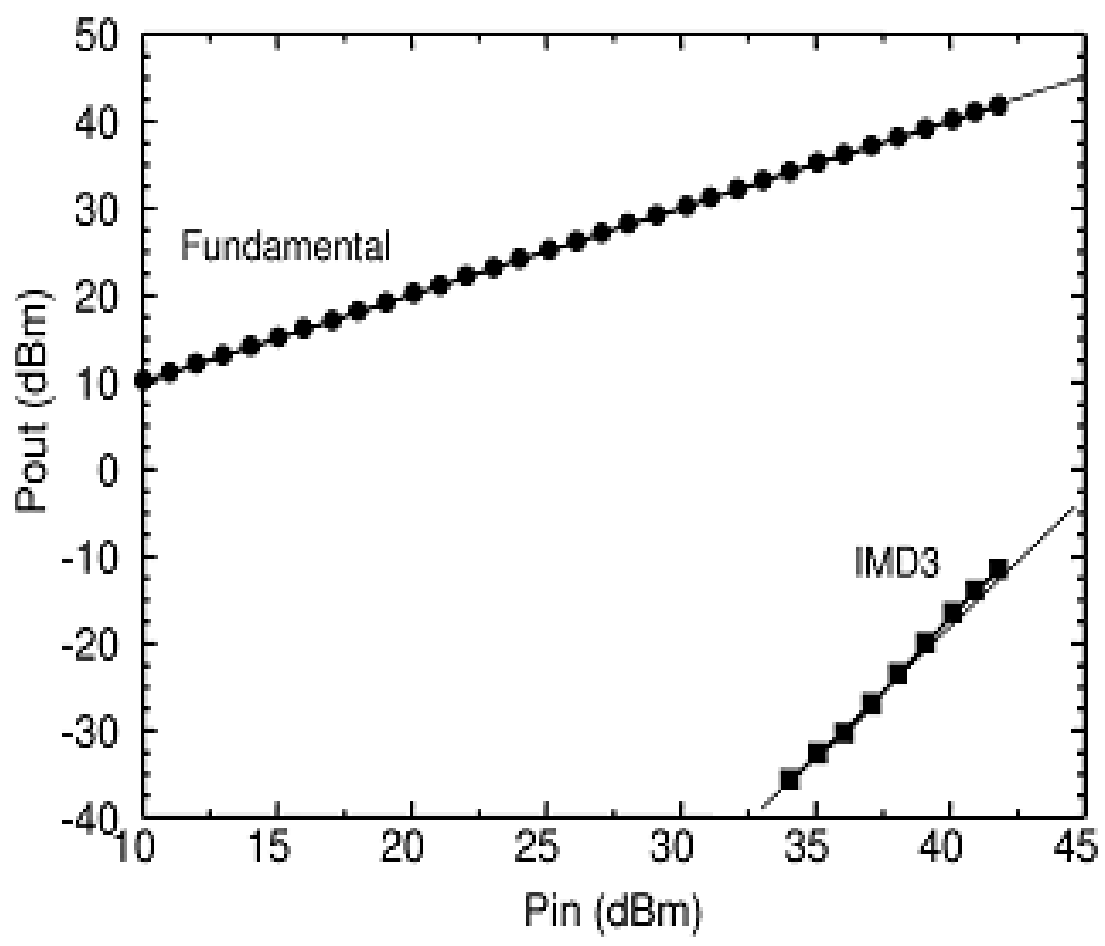
</td>
<td>Fig. 116. Measured intermodulation distortion values in the passband region. The closed circles represent the amplitude of the fundamental signal. The closed boxes and closed triangles represent the amplitudes of IMD3 and IMD5, respectively. (The solid straight lines are the fitting results for the fundamental signal, IMD3, and IMD5 with slopes of 1, 3, and 5) (after [602]).</td>
</tr>
</table>

**Tab. 21.** Research results on sources of nonlinearities in *HTS* thin films at microwaves by Futatsumori, Furuno, Hikage, Nojima, Akasegawa, Nakanishi, Yamanaka [601]; Futatsumori [602] (after [601, 602]).

In [643], C. Collado, J. Mateu, T. J. Shaw, J. O'Callaghan researched the HTS nonlinearities in microwave disk resonators. In [644], C. Collado, J.Mateu, R. Ferrús, and J. O'Callaghan completed their research on the prediction of nonlinear distortion in *HTS* filters for *CDMA* communication systems. In [645], C. Collado, J. Mateu, J. M. O'Callaghan made a comprehensive study on the analysis and simulation of the effects of distributed nonlinearities in microwave superconducting devices, using the phenomenological model of nonlinearities by which the penetration depth varies as a function of the current density in a superconductor.

Mateu, Booth, Collado, O’Callaghan researched the theoretical feasibility of suppressing the *IMD* generated by *HTS* materials with the use of a nonlinear dielectric (possibly a ferroelectric) [603]. The *HTS* front-end pre-select bandpass filter improves the sensitivity and selectivity of the receiver, increasing the coverage and improving the *quality of service* in Mateu, Booth, Collado, O’Callaghan [603], Willemsen [604]. The main limitation of ***the HTS microwave filter is its inherent nonlinear response, which produces intermodulation distortion (IMD), causing a serious problem in communication systems*** Willemsen, King, Dahm, Scalipino ***[605].*** *Quantifying and reducing this degradation is crucial for spreading the application of HTS filters.* The configurations explored avoid having to use a ferroelectric in the whole *HTS* planar layout, thus avoiding losses that the ferroelectric would produce in the tunable *HTS* filters in Mateu, Booth, Collado, O’Callaghan [603], Moeckly, Zhang [606], but this might not be the case in the use of ferroelectrics for *IMD* compensation Mateu, Booth, Collado, O’Callaghan [603], in Mateu, Booth, Moeckly [607].

In Tab. 22, the research results on sources of nonlinearities in *HTS* thin films at microwaves are presented by Mateu, Booth, Collado, O’Callaghan [603].

| | |
|---|---|
| 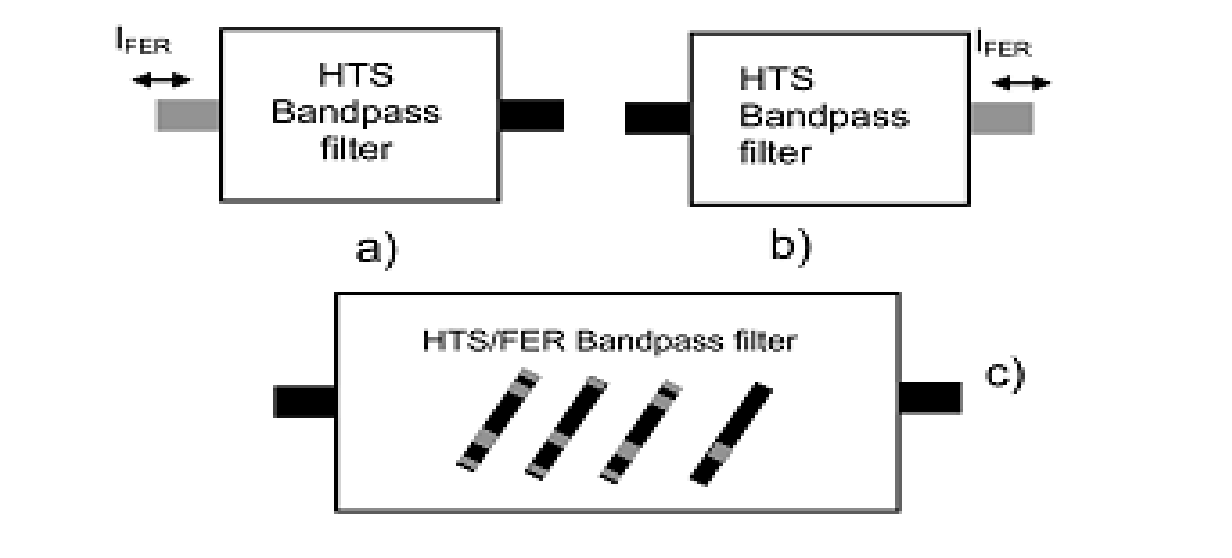  | **Fig. 117.** (a) Ferroelectric segment for nonlinear pre-distortion in an *HTS* band-pass filter. (b) Ferroelectric segment for nonlinear post-distortion in an *HTS* bandpass filter. (c) Intermediate ferroelectric compensation stages (after [603]). |
| 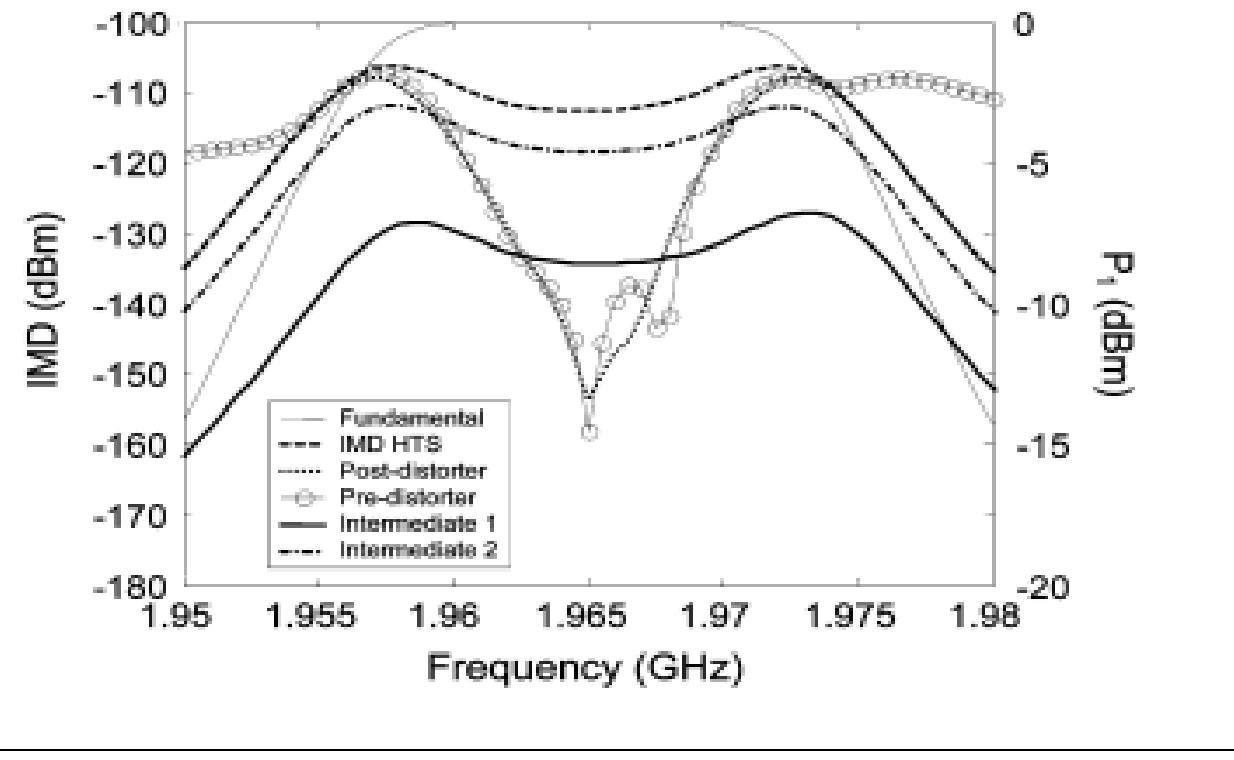  | **Fig 118.** Simulated response of the *HTS* microwave filter. *Thin solid line* represents the fundamental output power *P1*. *Dashed line*: *IMD* for an *HTS* microwave filter. *Dots and circles* represent the *IMD* for an *HTS* microwave filter with post- and pre-distortion ferroelectric stages, respectively. *Dotted–dashed and thick solid lines* represent the *IMD* for an *HTS* microwave filter with intermediated ferroelectric stages (after [603]). |

**Tab. 22.** Research results on sources of nonlinearities in *HTS* thin films at microwaves by Mateu, Booth, Collado, O’Callaghan (after [603]).

Mateu, Booth, Collado, O'Callaghan [603] conclude that *a significant reduction of the IMD in superconducting filters can be achieved with the use of ferroelectrics without degrading the quality factors in their resonators to the point, where the selectivity or the insertion loss is severely affected.* **The ferroelectrics may be used for maximum IMD reduction with a minimum effect on losses** [603].

Boot [608] evaluated the phase, using a large-signal vector network analyser and clarified that the nonlinear inductances, which are caused by *superconducting electron pair-breaking effects,* are dominant physical factors behind the nonlinear response of the measured $YBa_2Cu_3O_{7-\delta}$ coplanar waveguide transmission line at 76*K.* The possible routes for further reduction in nonlinear response of *HTS* thin films at microwaves were also proposed by Booth [608].

Agassi, D. E. Oates [609, 610] identify **the three critical design parameters for reducing the power level of *intermodulation distortion* (*IMD*)** in *HTS* filters at microwaves in Fig. 119:

(i) *thickness of the HTS film*,

(ii) *operation temperature of the HTS film*, and

(iii) *oxygen overdoping of the HTS film*.

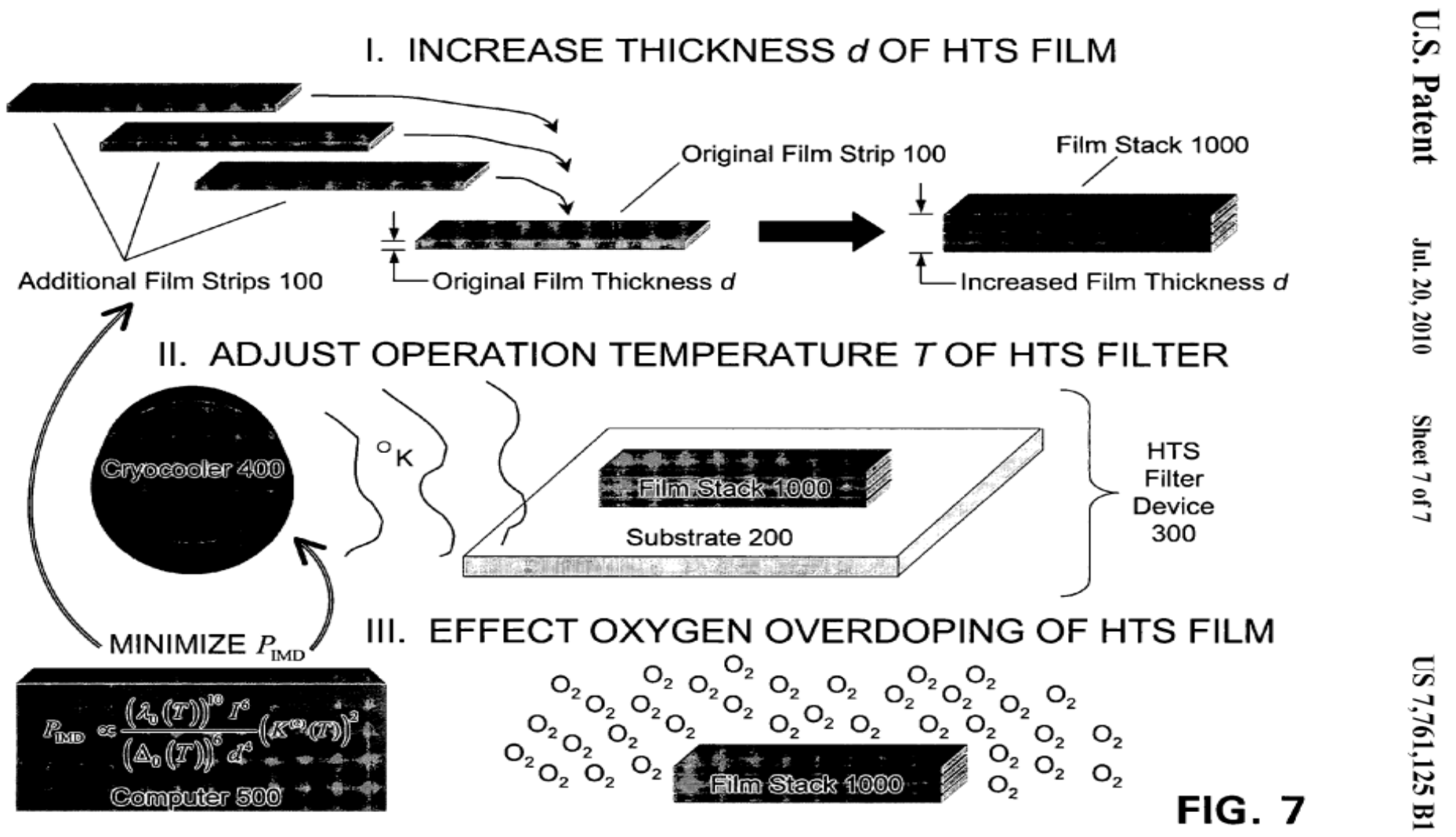

**Fig. 119.** Three critical design parameters for reducing the power level of *intermodulation distortion* (*IMD*) in *HTS* filters at microwaves (after [609]).

The author of dissertation would like to comment that, in some cases, the ***weak links*** between the *HTS* thin films in a stack with multiple layered *HTS* thin films may arise at applied increased microwave power levels, resulting in the new nonlinearities appearance at output port of the *HTS* microstrip resonator. Moreover, the increase of thickness of *HTS* thin film by creation of a stack with multiple layered *HTS* thin films may lead to the change of $Q_0$ quality factor of a microstrip resonator, shift of resonant center frequency, and change of resonance curve shape, which have to be taken to the detailed consideration during the *HTS* microstrip resonator designs.

The main contributions to the research toward the understanding of ***nonlinearities nature in microwave superconductivity*** are made by: Halbritter [70], Kong [470, 471], M. S. Dresselhaus, G. Dresselhaus [472]; Dresselhaus, D. E. Oates, Sridhar [473], D. E. Oates [474], Shen, Wilker, Pang, Holstein, Face, Kountz [475], Cohen, Cowie, Gallop, Ghosh, Goncharov [476], Kaiser, Aminov, Baumfalk, Cassinese, Chalouoka, Hein, Kilesov, Medelius, Mitller, Perpeet, Piel, Wikborg [477], D. E. Oates, Anderson, Alfredo, Sheen, Ali [478], Sheen, Ali, D. E. Oates, Withers, Kong [479], D. E. Oates [480], D. E. Oates, D. Agassi, E. Wong, A. Leese de Escobar, K. Irgmaier [481], Booth, Beall, Rudman, Vale, Ono [503], Dahm, Scalapino, Willemsen [504], Koren, Gupta, Beserman, Lutwyche, Laibowitz [505], Leong, Booth, Schima [506], Semerad, Knauf, Irgmaier, Prusseit [507], Agassi, D. E. Oates [508], Agassi, D. E. Oates [509], D. E. Oates, Park, Koren [510], Chew, Goodyear, Edwards, Satchell, Blenkinsop, Humphreys [511], Li, Suenaga, Ye, Foltyn, Wang [512], Oates, Agassi, Wong, Leese de Escobar, Irgmaier [513], Velichko, Lancaster, Porch [46], Nguyen, D. E. Oates, G. Dresselhaus, M. S. Dresselhaus, Anderson [525], Tsindlekht et al. [526], Diete, Getta, Hein, Kaiser, Muller, Piel, Schlick [527], Kharel, Soon, Powell, Porch, Lancaster, Velichko, Humphreys [528], Velichko, Porch, Lancaster, Humphreys [529], Hein [530], Hein, Hirst, Humphreys, D. E. Oates, Velichko [531], Hein, Hirst, Humphreys, D. E. Oates, Velichko [532], Hein, Getta, Kreiskott, Mönter, Piel, D. E. Oates, Hirst, Humphreys, Leed, Moond [533], Hein, Humphreys, Hirst, Park, D. E. Oates [534] Hein, Hirst, Humphreys, D. E. Oates, Velichko [535], Hein, Hirst, Humphreys, D. E. Oates, Velichko [536], O'Connell et al. [538], Gao et al. [539], Hein et al. [540], Rao, Ong, Jin, Tan, Xu, Chen, Lee, Feng [541], Velichko, D. W. Huish, M. J. Lancaster, A. Porch [542], Abu Bakar, Velichko, Lancaster, Porch, Gallop, Hao, Cohen, Zhukov [544], Willemsen, Kihlstrom, Dahm [545], Abu Bakar, Velichko,

Lancaster, Xiong, Porch, Storey [546], D. E. Oates, Park, Hein, Hirst, Humphreys [547], Xin, D. E. Oates, G. F. Dresselhaus, M. S. Dresselhaus [548], D. E. Oates, Hein, Hirst, Humphreys, Koren, Polturak [549], Xin, D. E. Oates, A. C. Anderson, R. L. Slattery, G. F. Dresselhaus, M. S. Dresselhaus [550], Xin, D. E. Oates, Anderson, Slattery, G. F. Dresselhaus, M. S. Dresselhaus [551], D. E. Oates, Hein [552], Kaiser, Aminov, Baumfalk, Cassinese, Chalouoka, Hein, Kilesov, Medelius, Mitller, Perpeet, Piel, Wikborg [553], Xin, D. E. Oates, Anderson, Slattery, G. F. Dresselhaus and M. S. Dresselhaus [554], V. O. Ledenyov, D. O. Ledenyov, O. P. Ledenyov [555], Porch, Huish, Velichko, Lancaster, Abell, Perry, Almond, Storey [556], Zhuravel, Anlage, Ustinov [557], Tai, Xi, Zhuang, Mircea, Anlage [558], Golosovsky [559], Abdo, Segev, Shtempluck, Buks [560], Zaitsev, Almog, Shtempluck, Buks [561], Abdo, Segev, Shtempluck, Buks [561, 562], Segev, Abdo, Shtempluck, Buks [563, 564], Bachar, Segev, Shtempluck, Shaw, Buks [565], Cohen, Cowie, Purnell, Lindop, Thiess, Gallop [567], Kermorvant, van deer Beek, Mage, Marcilhac, Lemaitre, Briatico, Bernard, Villegas [576], Kermorvant, van der Beek, Mage, Marcilhac, Lemaitre, Bernard, Briatico [577], Beek, Mage, Marcilhac, Lemaitre, Briatico, Bernard, Villegas [579], Dam, Huijbregtse, Klaasen, van der Geest, Doornbos, Rector, Testa, Freisem, Martinez, Stäuble-Pümpin, Griessen [587], Sridhar [588], McDonald, Clem, D. E. Oates [589], Wosik, Xie, Nesteruk, Li, Miller, Long [590], Pukhov [591, 592], Kermorvant [593, 595], Kermorvant, van der Beek, Mage, Marcilhac, Lemaitre Briatico, Bernard, Villegas [594], and by Van der Beek, Konczykowski, Abal'oshev, Abal'osheva, Gierlowski, Lewandowski, Indenbom, Barbanera [596], Willemsen, Derov, Sridhar [598], Hua Zhao, Xiang Wang and Judy Z. Wu [599], Futatsumori, Furuno, Hikage, Nojima, Akasegawa, Nakanishi, Yamanaka [600], Futatsumori, Furuno, Hikage, Nojima, Akasegawa, Nakanishi, Yamanaka [601], Futatsumori [602], Mateu, Booth, Collado, O'Callaghan [603], Willemsen [604], Boot [608], Agassi, D. E. Oates [609, 610], O. G. Vendik, Kozyrev, Popov [611], O. G. Vendik, Kozyrev, Popov [612], O. G. Vendik [613], O. G. Vendik, Likholetov, Karmanenko, Kolesov, Konson [614], O. G. Vendik, Popov [615], [O. G. Vendik, Kolesov [616], O. G. Vendik [617], O. G. Vendik, Kozyrev, Samoilova, Hollmann, Ockenfub, Wordenweber, Zaitsev [618], D.O. Ledenyov, Mazierska, Allen, Jacob [619, 620], and by some others.

## 4.1. Analysis of Main Research Results on Origin of Nonlinearities in HTS Thin Films at Microwaves: S-Shape Dependences of Surface Resistance on Magnetic Field $R_S(H)$ in HTS Thin Films at Microwaves.

The increase of the surface resistance $R_S$ and surface reactance $X_S$ with high microwave power levels in Cooke, Arendt, Gray, Bennett, Brown, Elliott, Reeves, Rollett, Hubbard, Portis [621], Portis, Cooke, Piel [622], sets some essential limitations to the application of *HTS* films in microwave device technologies. The understanding of the physical phenomena behind the nonlinear effects has significantly advanced, but the mechanisms of nonlinear surface impedance $Z_S$ in *HTS* films are not well known in Halbritter [623], Gaganidze, Heidinger, Halbritter, Shevchun, Trunin, Schneidewind [624]. The typical *quadratic nonlinear dependences* of the surface resistance on the magnetic field $R_S(H_{RF})$ for $YBa_2Cu_3O_{7-\delta}$ were measured by P. P. Nguyen, D. E. Oates, G. Dresselhaus, M. S. Dresselhaus in Fig. 120 [625]. The nonlinear dependences of quality factor on magnetic field $Q\ (H_{RF})$ for a stripline resonator made of $YBa_2Cu_3O_{7-\delta}$ at T=77.4*K* for different frequencies of 1.5*GHz* and 3 *GHz* are observed by P. P. Nguyen, D. E. Oates, G. Dresselhaus, M. S. Dresselhaus in Fig. 121 [625]. The dependence of the surface resistance of superconductor on magnetic field $R_S(H_{rf})$, which is similar to the *quadratic nonlinear dependence*, is observed by D. E. Oates, Anderson, C. Alfredo, D. M. Sheen, S. M. Ali in Fig. 122 [626]. Similar quadratic dependence of surface resistance $R_S$ on the peak magnetic field $H_{rf}$ for $YBa_2Cu_3O_{7-\delta}$ films is also presented by D. E. Oates in Fig. 123 [627]. The $R_S(H_{rf})$ dependences of unpatterned 2-inch-diameter 400-*nm*-thickness $YBa_2Cu_3O_{7-\delta}$ films were measured in the dielectric resonator at 10.7*GHz* as described in D. E. Oates [627]. The unpatterned films then were patterned using standard photolithography and wet chemical etching with 0.1% phosphoric acid to make stripline resonators with resonant frequency 1.5 *GHz*. The surface resistance $R_S$ was then measured as a function of *RF* power for the patterned striplines in D. E. Oates [627]. The comparison of the results before and after the pattering process directly clarified that the pattering process did not degrade the $YBa_2Cu_3O_{7-\delta}$ film properties. The modeling was performed for the two resonators: *dielectric resonator* and *stripline resonator*, assuming the same $R_S(H_{rf})$ for both the patterned and unpatterned $YBa_2Cu_3O_{7-\delta}$ films. The dependence of the surface resistance of $YBa_2Cu_3O_{7-\delta}$ thin films on the magnetic field $R_S\ (H_{rf})$ has the *quadratic behaviour* as explained by D. E. Oates [627]. However, the surface resistance $R_S$ of superconducting samples can exhibit the *S-shaped dependence*, if the power of the applied magnetic field $H_{rf}$ is strong enough in D. E. Oates [627]. The nonlinear increase of surface resistance of *HTS* films with the increase of microwave power levels is attributed to the nonlinear *S-shape dependence* of quality factor $Q$ of microwave resonator in D. E. Oates [627]. The following phenomena, may contribute to the nonlinearities: *thermal effects*, *weak links*, *non-homogeneities*, *non-equilibrium excitation*, *unconventional pairing* and others in D. E. Oates [627].

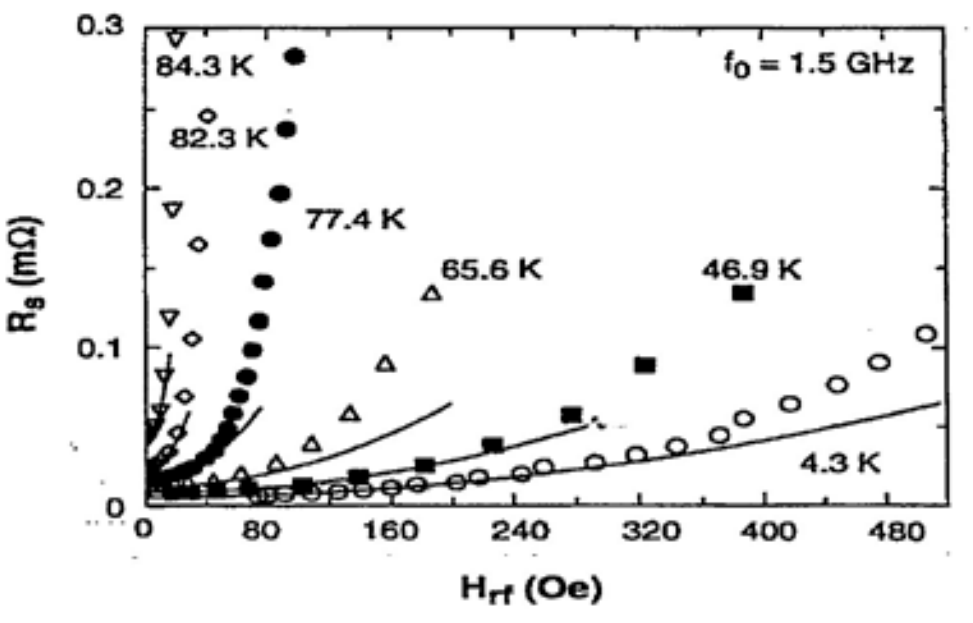

**Fig. 120.** Dependence of surface resistance on magnetic field $R_S(H_{RF})$ for a $YBa_2Cu_3O_{7-\delta}$ stripline resonator at different temperatures at frequency of 1.5 *GHz* (after [625]). Solid lines – quadratic fits.

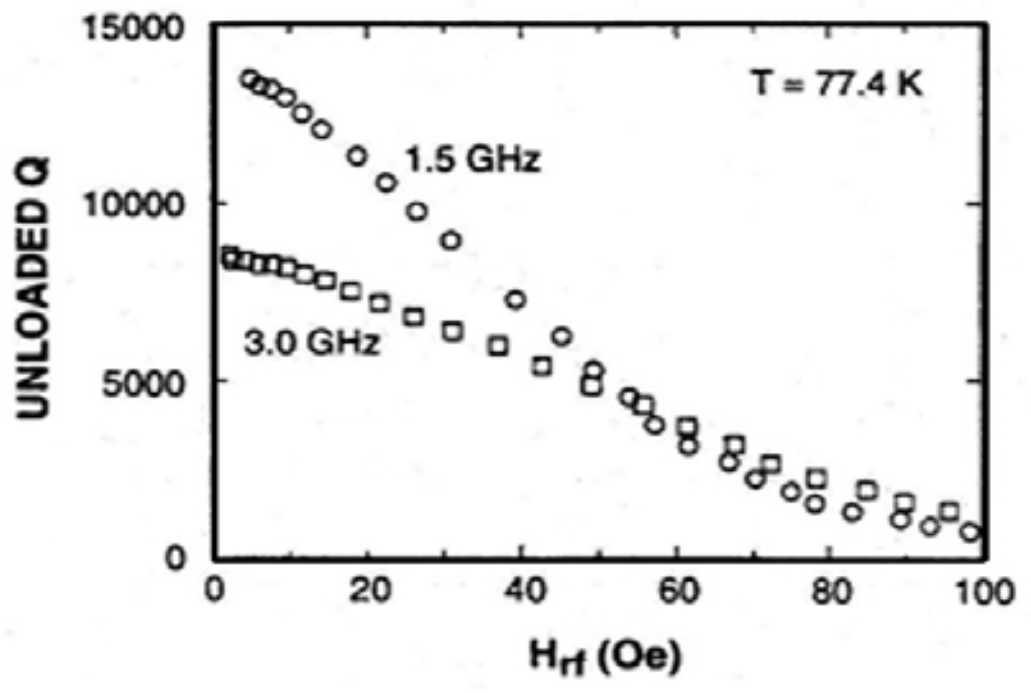

**Fig. 121.** Nonlinear dependences of quality factor $Q$ on magnetic field $H_{RF}$ for a stripline resonator made of $YBa_2Cu_3O_{7-\delta}$ at different frequencies 1.5*GHz* and 3*GHz* at T=77.4*K* (after [625]).

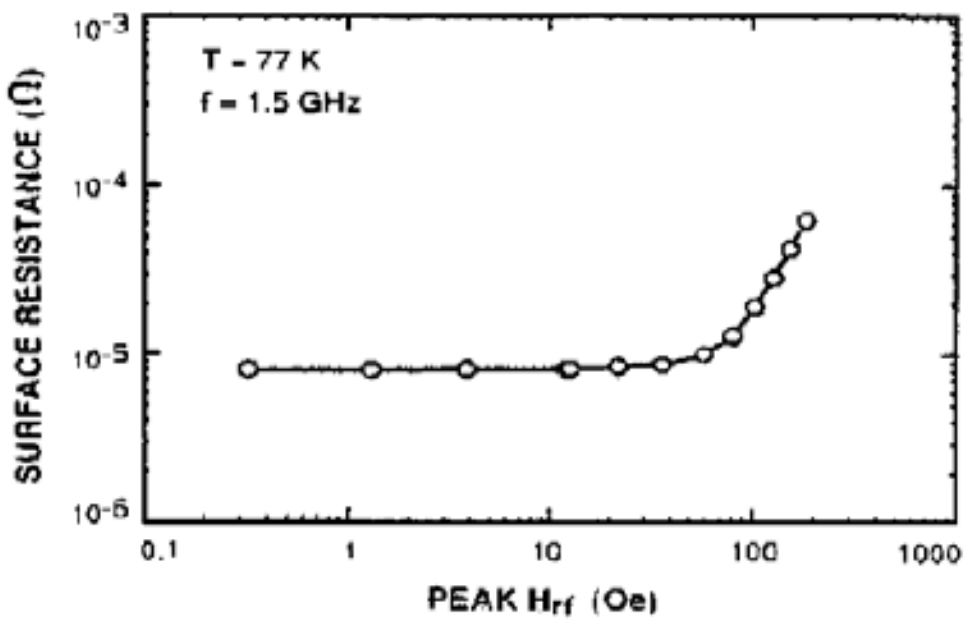

**Fig. 122.** Surface resistance $R_S$ of a stripline resonator as a function of peak $H_{RF}$ field at $YBa_2Cu_3O_{7-\delta}$ film edges at the frequency of 1.5 *GHz* at 77*K* (after [626]).

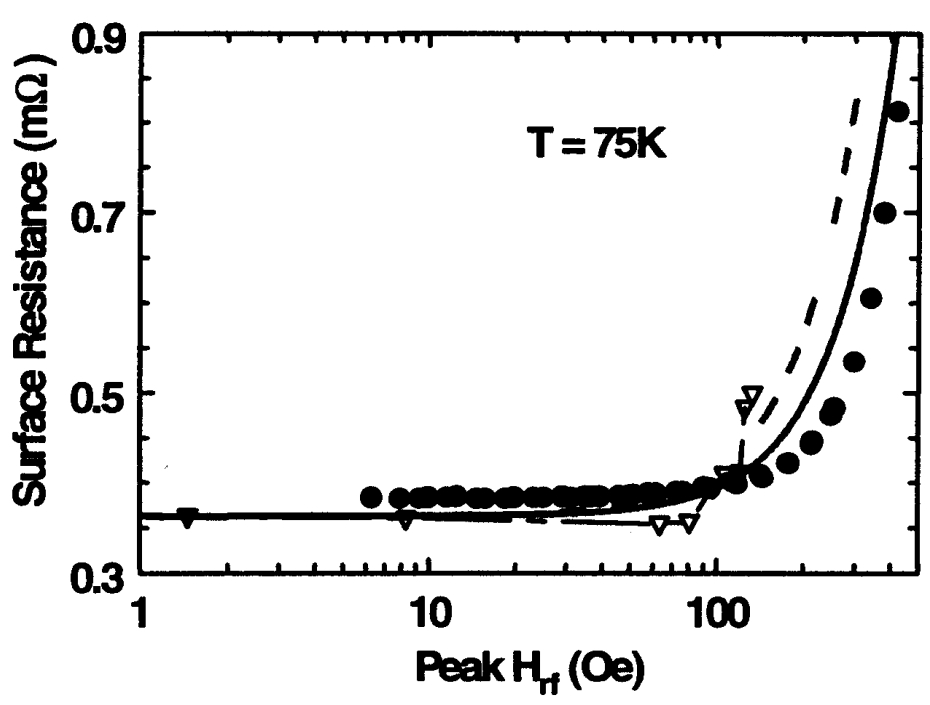

**Fig. 123.** Surface resistance $R_S$ vs. peak magnetic field $H_{rf}$ for $YBa_2Cu_3O_{7\text{-}\delta}$ films at 1.5*GHz* at 75*K*. Points are measured data. Solid and dashed lines are calculated from a model for the stripline and dielectric resonators respectively (after [627]).

**Tab. 23.** Analysis of main research results on origin of nonlinearities in *HTS* thin films at microwaves (after [625, 626, 627]).

The dependence of surface resistance of superconducting film on the electromagnetic field $R_S(H_{rf})$ at ultra high frequencies can be characterised by the four regimes:

1. ***Linear dependence in small field***;
2. ***Weak nonlinear dependence***;
3. ***Strong nonlinear dependence***;
4. ***Breakdown regime in highest saturated fields***.

It is believed that, in the linear regime, the surface resistance $R_S$ of *HTS* films is extrinsic and determined by the crystal defects such as the weak links at grain boundaries in crystal lattice of superconductor. ***The linear regime*** is well described by the coupled-grain model in Andreone, Cassinese, Di Chiara, Iavarone, Palomba, Ruosi [628]. The surface resistance $R_S$ gradually increases in ***the weakly nonlinear regime***, and its behaviour has the *quadratic dependence* as function of *RF* power usually. The weakly nonlinear regime supposedly originates because of the presence of defects such as the weak links at grain boundaries in crystal lattice of superconductor. This regime can be described by the extended coupled-grain model, which takes into the account the nonlinear inductance of the weak links in Delayen, Bohn [629]. ***The strong nonlinear regime*** expresses strongly nonlinear behaviour, where the surface resistance $R_S$ increases more rapidly. This mode is usually connected with the vortex generation phenomena by the strong microwave magnetic field. ***The breakdown regime in highest saturated fields***, the so-called breakdown region occurs at very high levels of microwave power, where the surface resistance $R_S$ increases very sharply. The breakdown regime is believed to have place because of the heating effect and the formation of normal-state domains in superconductor.

The ***nonlinear S-type behaviour*** of surface resistance $R_S$ can be observed near the critical magnetic fields $H_{c1}$ and $H_{c2}$. This effect has no complete theoretical explanation at the present time.

In Fig. 124, the microwave properties of $YBa_2Cu_3O_{7-\delta}$ thin films in the linear and nonlinear regimes in a *DC* magnetic field are shown in Tsindlekht, Sonin, Golosovsky, Davidov, Castel, Guilloux-Viry, Perrin [630]. It is shown that external *DC* magnetic field and applied electromagnetic field at ultra high frequencies have certain influences on the $YBa_2Cu_3O_{7-\delta}$ thin film properties in Fig. 124 [630].

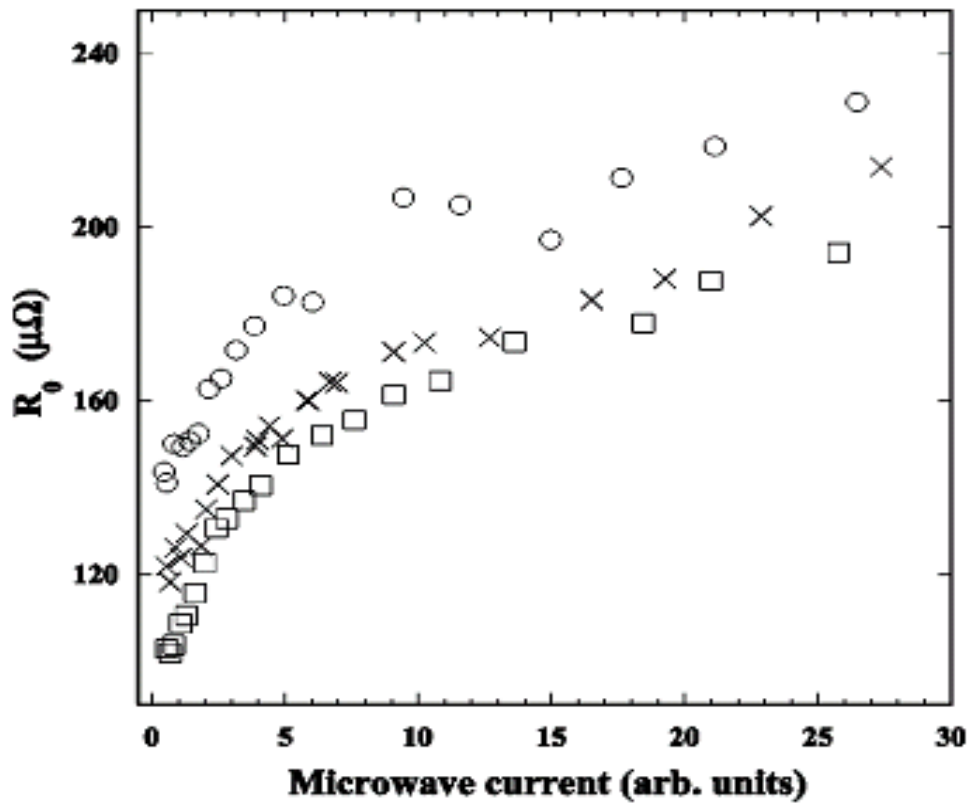

**Fig. 124.** $H_{rf}$-field dependences of surface resistance *R*s for high quality epitaxial thin film of $YBa_2Cu_3O_{7-\delta}$ in both zero and finite *DC* magnetic fields measured by using the parallel plate resonator at the frequency of 5.7 *GHz* and the temperature of 26 *K*: squares—zero field; circles—0.45*T* parallel to *c*-axis; crosses—0.45 *T* parallel to *ab*-plane (after [630]).

When the amplitude of an electromagnetic wave in a microwave resonator has sufficiently big magnitude, the electromagnetic wave has an influence on the surface resistance $R_S$, even, when the external magnetic field is zero ($H_e=0$). For example, in the conventional $Nb_3Sn$ superconductor, the first incurvation of the characteristic *S-type dependence* with the saturation effect can be observed in a range of small magnetic fields, and the second nonlinear behaviour can be registered at higher levels of magnetic field in Andreone, Cassinese, Di Chiara, Iavarone, Palomba, Ruosi [628]. The results are presented in Fig. 125 [628].

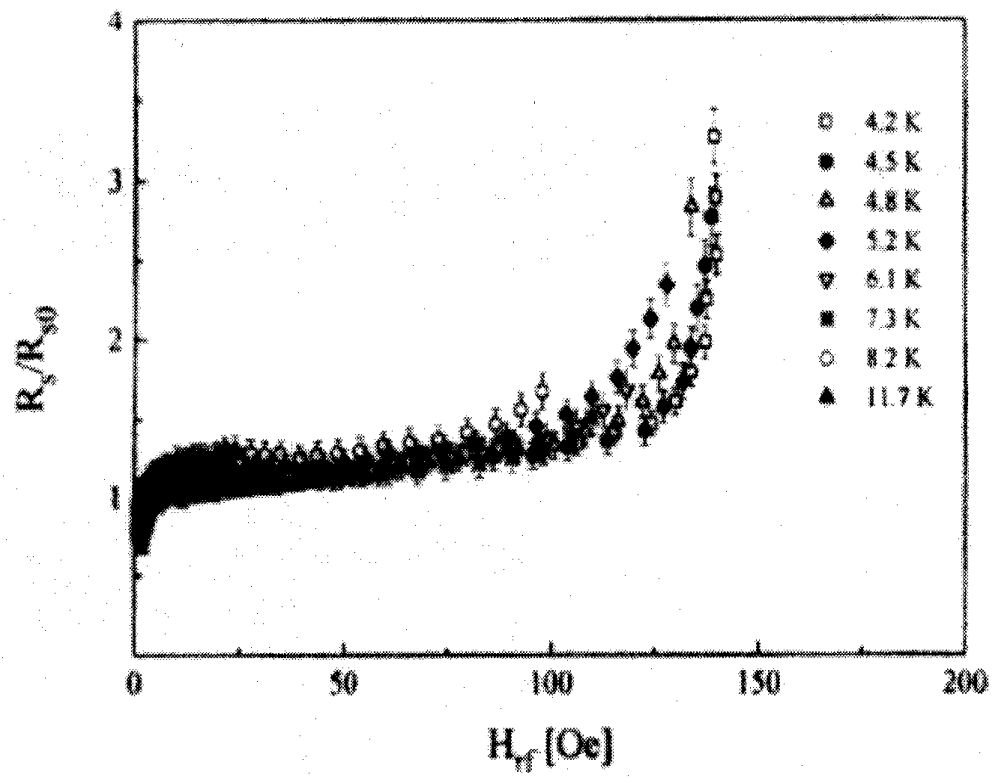

**Fig. 125.** Surface resistance $R_S$ of $Nb_3Sn$ film as a function of peak $H_{rf}$ measured by microstrip resonator at the frequencies of 1.38 *GHz* – 9.53 *GHz* and at the temperatures of 4.2 *K* – 11.7 *K* (after [628]).

In the range of small magnetic fields, the nonlinear behaviour of surface resistance $R_S$, which is visible in Fig. 125, begins at the critical field $H_{cJ1}$. The nonlinear effect occurs, when the quantum magnetic vortices begin to penetrate into the areas with the weak links at grain boundaries in crystal lattice of superconductor. These areas represent the so-called *Josephson junctions*, where the nonlinear behaviour of surface resistance $R_S$ can be explained by the influence of magnetic field $H_{rf}$ on the *Josephson junctions* in superconductor in Biondi, Garfunkel [631]. The nonlinear surface resistance $R_S$ of superconductor in the higher magnetic fields $H_{rf}$ can be observed, because of the phase transition effect in close proximity to the magnetic field $H_{c1}$, when the magnetic vortices begin to penetrate into the superconductor itself. In the superconductors, the following relation between the critical fields is valid: $H_{c1J}<H_{c1}$. It should be noted that the magnitude of the field $H_{c1J}$ depends on the thickness and resistance of the weak links area, and therefore the $H_{c1J}$ magnitude can change from one superconducting sample to another sample, depending on the quality of fabrication of superconducting thin films. While, the magnetic field $H_{c1}$ is the physical characteristic of superconductor.

In *HTS* films, the weak links at grain boundaries in crystal lattice of superconductor are easily created during the superconducting crystal growth process, and as a result, almost all the films display the nonlinear behaviour near the magnetic field $H_{c1J}$. Portis has summarized these dependences $R_S(H_{rf})$ in Fig. 126 in Portis [632].

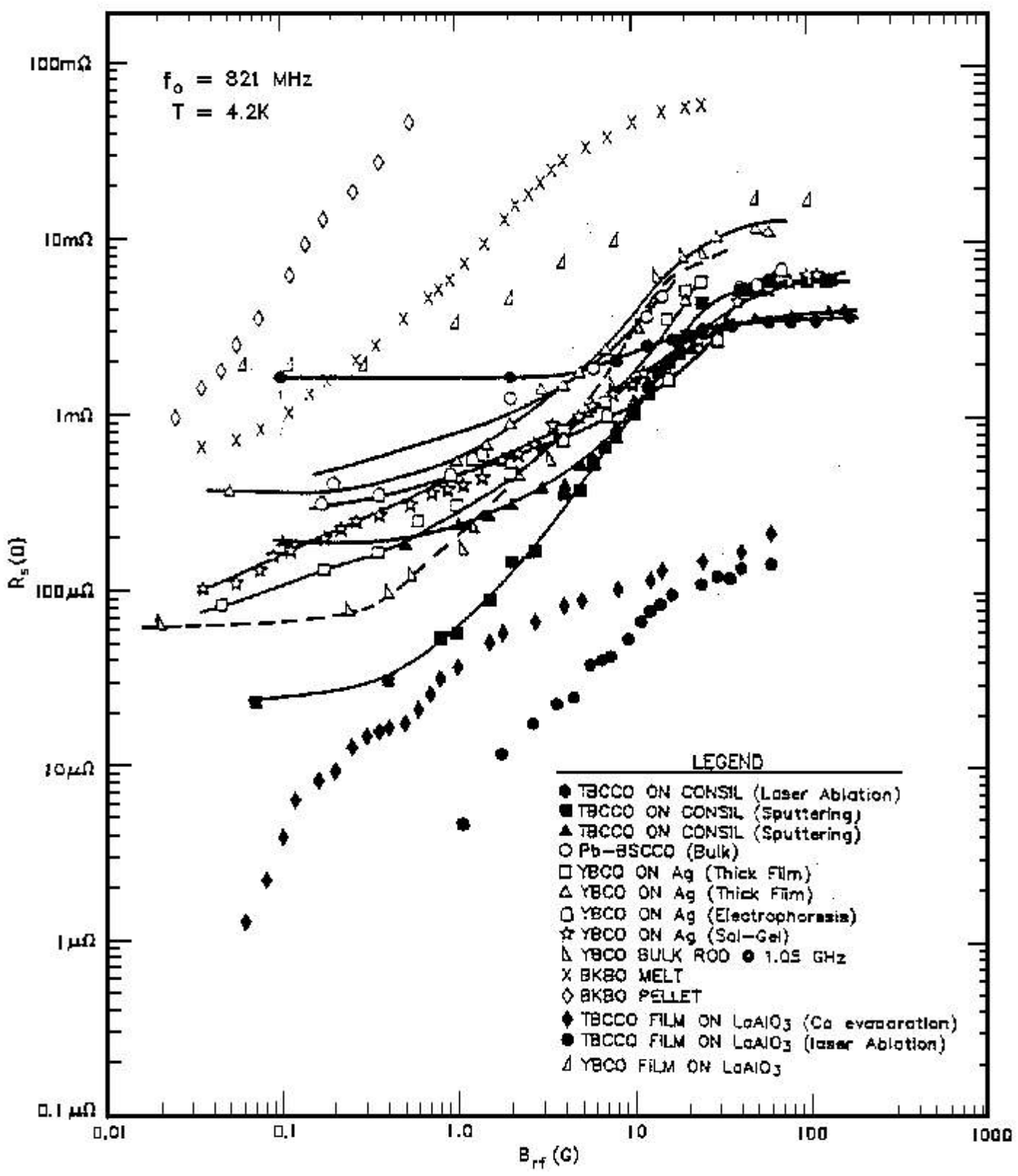

**Fig. 126.** Surface resistance $R_S$ vs. *RF* magnetic field $B_{rf}$ at the frequency of 821*MHz* and the temperature of 4.2*K* for a variety of *HTS* samples (after [632, 633]).

Portis demonstrated that, at low *RF* power level, the power-induced surface resistance $R_S$ has the quadratic dependence on the magnetic field $B_{rf}$ in granular or partially granular superconductors at ultra high frequencies [632]. At intermediate power level, the surface resistance $R_S$ increases linearly with the microwave field in Diete, Getta, Hein, Kaiser, Muller, Piel, Schlick [633]. Finally, at strong *RF* power level, the surface resistance $R_S$ saturates in magnetic field $B_{rf}$ between 10 *Oe* and 100 *Oe* at microwaves [633].

The similar *S-type dependence* of surface resistance $R_S$ of $YBa_2Cu_3O_{7-\delta}$ was observed in the quasi-optical resonator at 35 *GHz* by Velichko, Cherpak, Izhyk, Kirichenko, Chukanova, Zagoskin [634], where the $R_S(H_{rf})$ depends exponentially on squared magnetic field amplitude in eq. (3.4)

$$R_s(H_{rf}) = R_{s0}[1-\alpha\exp(-\frac{H_{rf}^2}{H_c^2})] \text{ , (for } H_{rf}>H_c\text{)} \quad (3.4)$$

where α and $H_c$ are the constants.

Halbritter [70] discussed the research results in Fig. 127 obtained by Delayen, Bohn [635], Delayen, Bohn, Roche [636]. It can be seen that the nonlinearity of surface resistance $R_S$ begins within the small fields $H < 1G$, and therefore the nonlinear effects in *HTS* can originate, because of the weak links related to the field $H_{c1J}$.

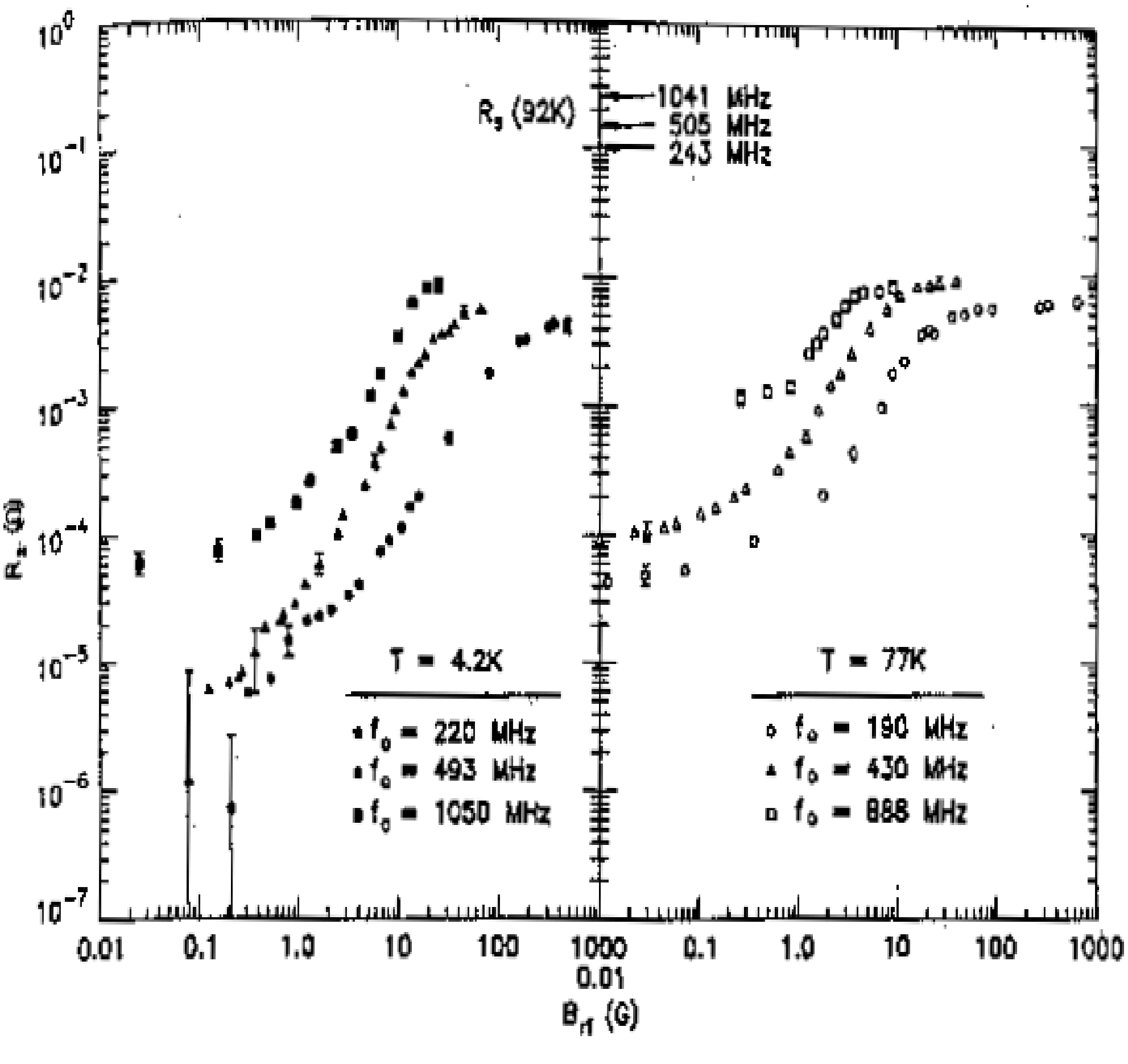

**Fig. 127.** Surface resistance $R_S$ of bulk $YBa_2Cu_3O_{7-\delta}$ vs. magnetic field $B_{rf}$ at the frequencies of *220MHz, 493MHz, 1050MHz* measured at 4.2*K*; and at the frequencies *190MHz, 430MHz, 888MHz* measured at 77*K* (after [635]).

Halbritter [637] has proposed that the observed behaviour of surface resistance $R_S$ can be classified into three regimes according to the $\Delta\lambda$ vs. $H$ dependence shown in Fig. 128:

I. $Rs \propto \omega^2 H^2$ and $Xs \propto \omega\mu_0 \lambda$;

II. $Rs \propto \omega^2 H$ and $Xs \propto \omega\mu_0 \lambda$;

III. $Rs \propto \omega^{1/2} |H_{sat} - H|$ and $Xs \propto \omega\mu_0 \lambda$.

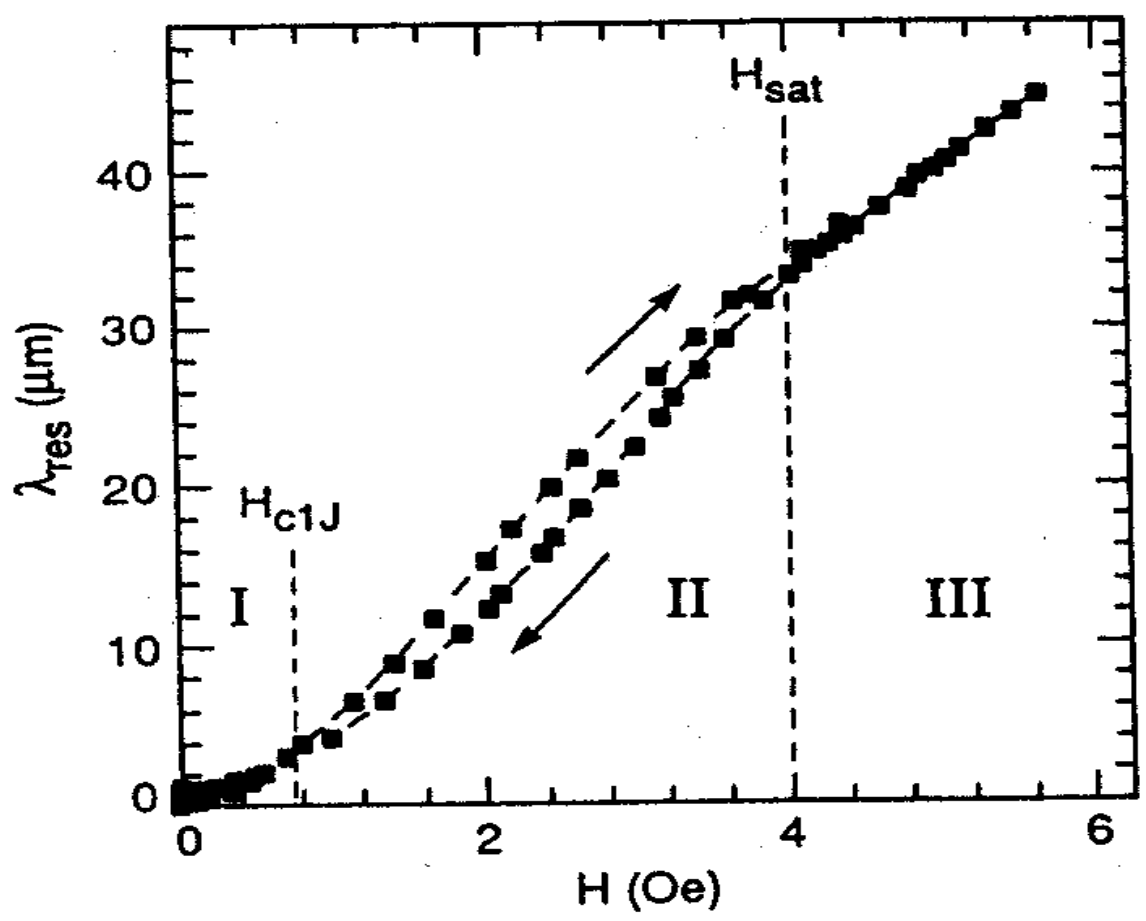

**Fig. 128.** Effect of magnetic field H on penetration depth λ displayed as Δλ vs. H (after [637]).

The $R_S(H_{rf})$ *S-shape nonlinear dependences*, displayed in Fig. 129, were firstly measured for the unpatterned films of different superconductors in the high sensitive *dielectric resonator* at frequency of 19 *GHz*; then the *stripline resonators* with the resonance frequency of 1.5 *GHz* were produced from these unpatterned *HTS* thin films by Diete, Getta, Hein, Kaiser, Muller, Piel, Schlick [633].

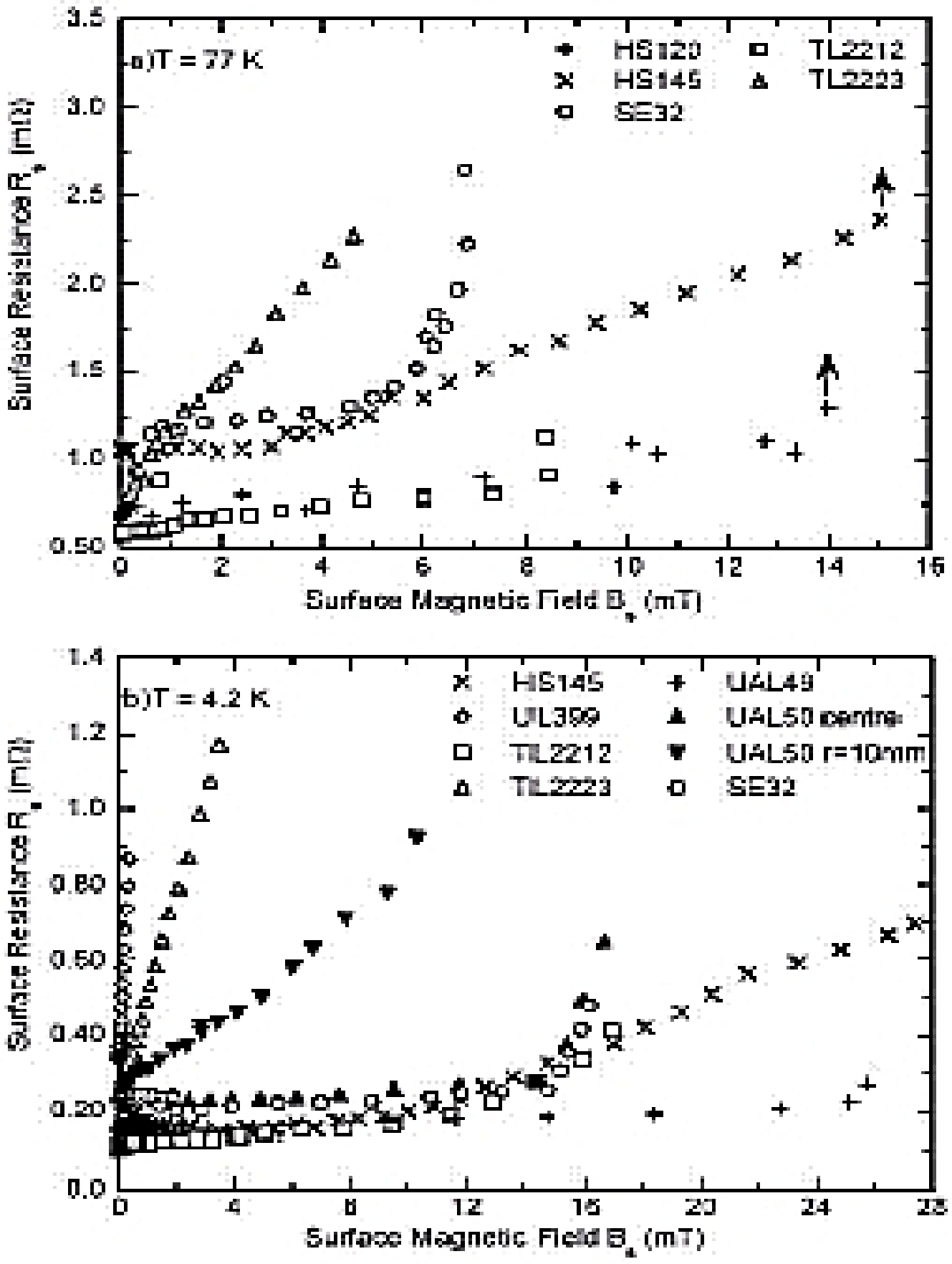

**Fig. 129.** Dependences $Rs(B_S)$ for different *HTS* thin films on different substrates measured with the use of dielectric resonator at 19 *GHz* and (a) 77 *K* and (b) 4.2 *K*. Designations used in the figure are as follows: *HS*— $YBa_2Cu_3O_{7-\delta}$ sputtered on $LaAlO_3$; *UAL*— $YBa_2Cu_3O_{7-\delta}$ laser deposited on $LaAlO_3$; *SE*—e-beam co-evaporated on *MgO*; *UM*— $YBa_2Cu_3O_{7-\delta}$ thermally evaporated on $CeO_2/Al_2O_3$; *Tl-2223*—$Tl_2Ba_2Ca_2Cu_3O_8$ two-step process on $LaAlO_3$; *Tl-2212*—$Tl_2Ba_2Ca_1Cu_2O_8$ two-step process on *MgO*; *UL*— $YBa_2Cu_3O_{7-\delta}$ laser deposited on $CeO_2/Al_2O_3$ (after [633]).

Hein [638] plotted the graph of dependences of the critical field on the temperature $B_c$ *(T)* for the various nonlinearity mechanisms in Fig. 130.

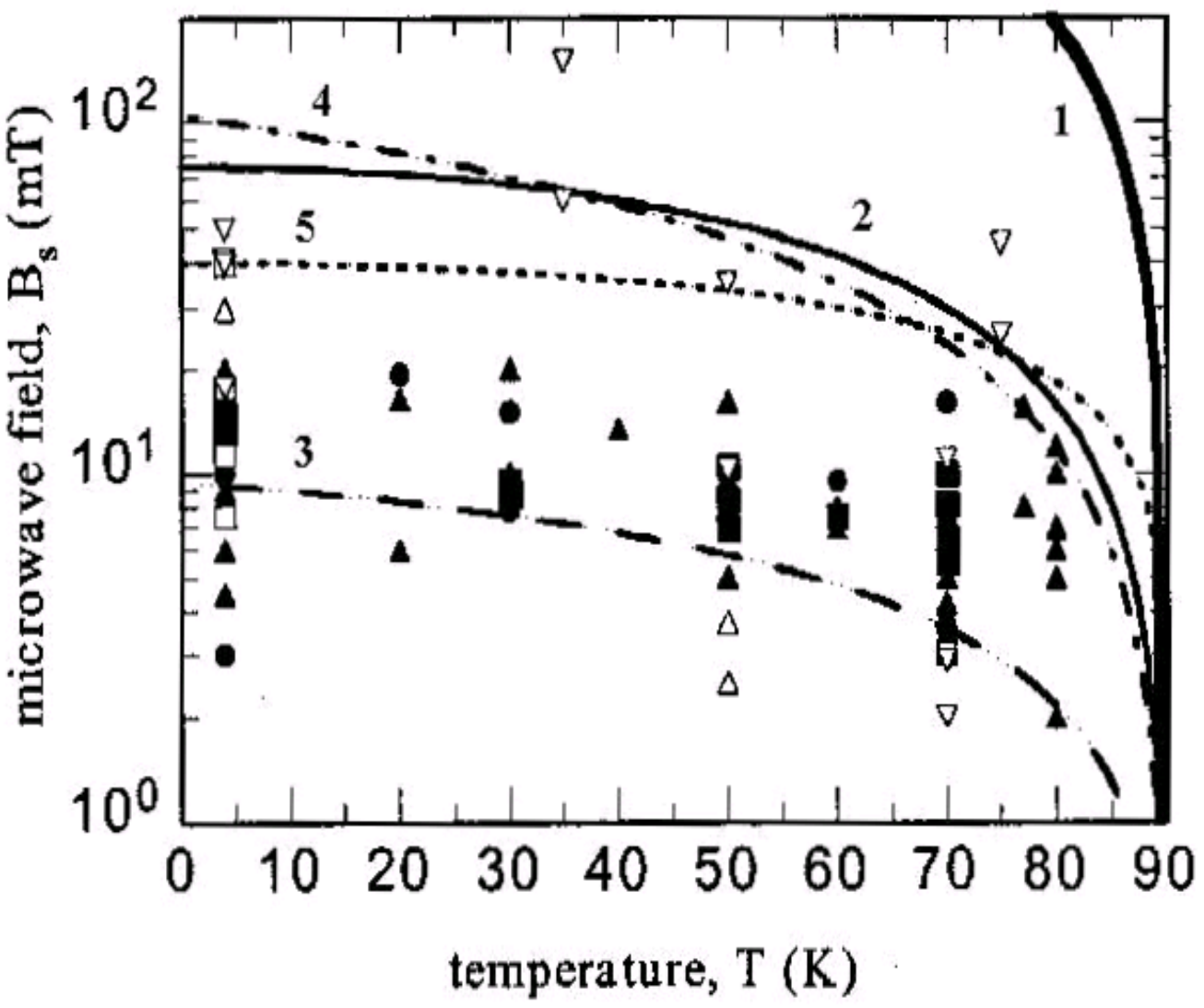

**Fig. 130.** Theoretical and experimental results for critical microwave fields: Superheating field $B_{sh}$ (1 - bold solid curve), bulk lower critical field $B_{c1}$ (2 - thin solid), Josephson critical field $B_{cJ}$ (3 - dashed), global quench field $B_q$~ (4 - dash-dotted), and local quench field $B_{qi}$ (5 - dotted). Symbols denote field amplitudes *B* measured with unpatterned (filled) and patterned (open) $YBa_2Cu_3O_{7-\delta}$ films on *LAO* (triangles), *CbS* (circles) and *MgO* (squares) at various laboratories (after [638]).

The results show that the field $H_{c1J}$ is reachable quite easily, therefore the nonlinearities near this area are observed more often. The $H_{c1J}$ area contribution to the nonlinear surface resistance depends on the relative volume of *m* sample occupied by weak links. In the polycrystalline samples, this value reaches m ~ 10 %, while in the high quality mono-crystal films, the value constitutes m < 0.1 %. ***Therefore, it is assumed that for the high quality HTS films, the nonlinearities, originated near the higher critical field $H_{c1}$, will mainly contribute to the nonlinear behaviour of superconductor in [662, 663].***

The researched nonlinearities in *HTS* thin films have to be taken to the consideration during the design of *HTS* microwave filters in Cryogenic Transceiver Front End (*CTFE*) in basestations in wireless communications as explained in Chapter 9.

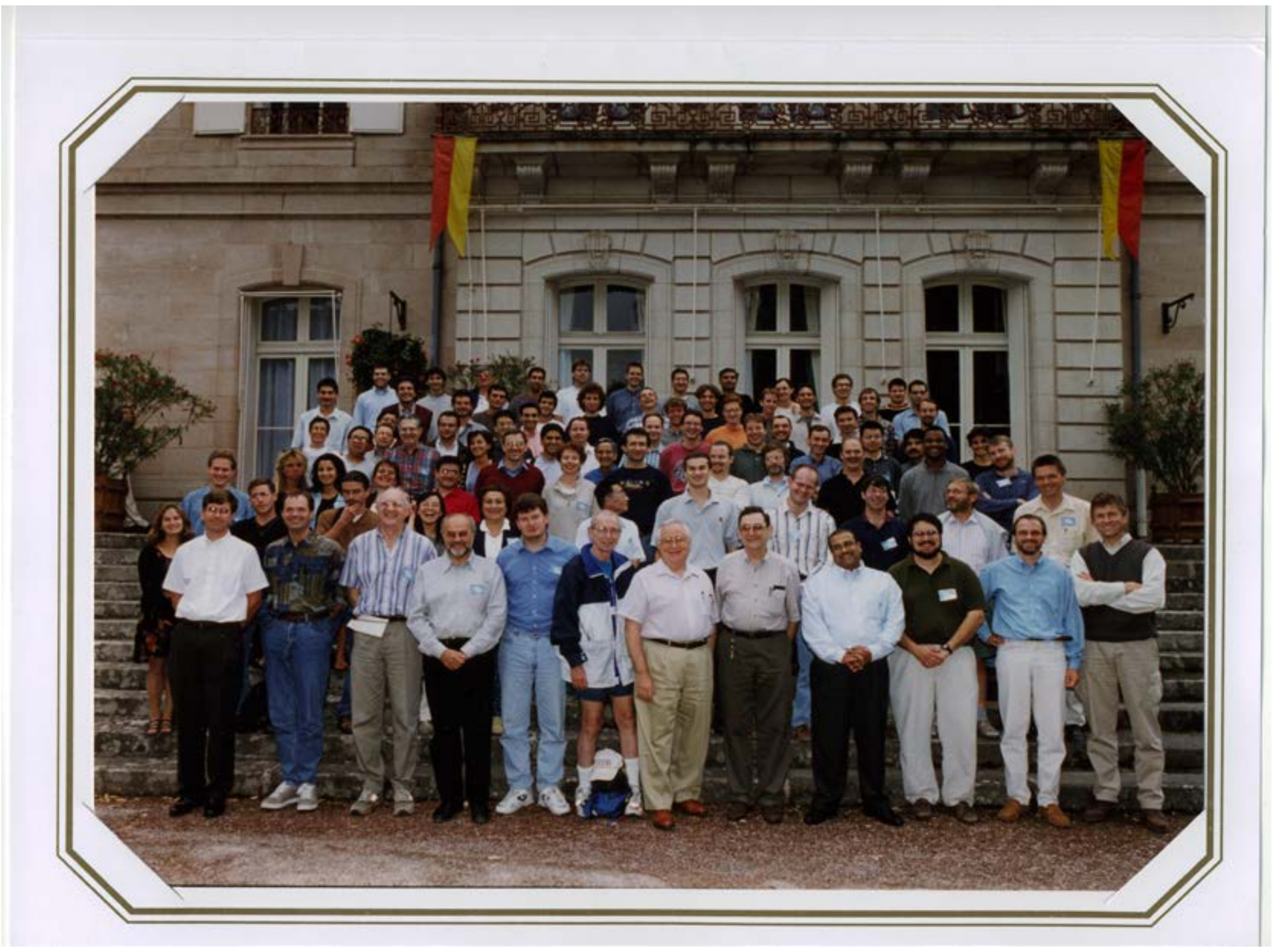

**Fig. 131.** Scientists from different countries, including Janina E. Mazierska, Harold Weinstock, Martin Nisenoff, John Clarke, Dimitri O. Ledenyov among other participants, who attended the NATO Advanced Study Institute (ASI) on Microwave Superconductivity in Millau, France.

**Summary.**

This Chapter 3 considers the microwave superconductivity, including the precise characterisation of physical properties of superconductors and technical applications of superconductors at microwaves. After a short historical overview on main research achievements in microwave superconductivity during last century, the central attention is focused on the synthesis of *HTS* thin films for microwave applications as well as the accurate characterisation of *HTS* thin films at microwaves, including the dependencies of surface resistance *Rs* on the frequency *f*, temperature *T* and microwave power *P*. Main parameters for accurate characterization of *HTS* thin films at microwaves are also described in details. The special consideration is given to the research on the nature of nonlinearities in *HTS* thin films at microwaves. Several physical mechanisms, that are believed to be responsible for the origin of nonlinear effects in *HTS* materials at microwaves, were reviewed and discussed comprehensively. It is explained that, going from the underlying physical mechanism responsible for the nonlinearities origination, the nonlinearities in *HTS* thin films at microwaves can be classified in two groups: 1) ***extrinsic nonlinearities***, 2) ***intrinsic nonlinearities.***

The ***extrinsic nonlinearities*** are mainly observed in the ***HTS thin films of relatively poor quality of fabrication***, which represent a network of superconducting grains connected by the weak links with the *Josephson junctions*, hence these relatively poor quality high temperature superconductors can be characterized by the *coupled-grain model*. In such relatively poor quality high temperature superconductors at microwaves, the nonlinearities origination mainly depends on the following ***extrinsic effects****: the non-uniform microwave heating, weakly coupled grains, local heating of weak links; and an* ***intrinsic effect****: Josephson magnetic vortices origination in weak links*.

The ***intrinsic nonlinearities*** can play an important dominant role in the generation of nonlinear characteristics in the *HTS* thin films of high quality of synthesis, which have almost no weak links. In high quality *HTS* thin films at microwaves, the nonlinearities may be induced by the following ***intrinsic effects:*** *nonlinear Meissner effects* or *superconducting electron pair-breaking current*, caused by the change of density of superconducting electrons due to the applied magnetic field $H_{rf}$, however this effect is considerably small, and nonlinear behaviour of high quality superconductors can not be fully attributed to this phenomena, even though it can explain the quadratic dependence $R_S(H_{rf})$. It is assumed by the author of dissertation that the ***intrinsic nonlinearities***, which are dependent on the magnetic field, applied in high quality *HTS* thin films at microwaves, appear due to the magnetic field $H_{rf}$ penetration into the interior of *Type II* superconductor, leading to the appearance of *Abricosov magnetic vortices* in close proximity to the magnitude of critical magnetic field $H_{C1}$. These ***intrinsic nonlinearities*** are believed to be connected with the nonlinear change of magnetic momentum of *Type II* superconductor.

Discussing the Duffing like nonlinearity, the authors of book considered the possibility of origination of new chaos induced phenomena in the nonlinear dynamic system such as the superconducting microwave resonator. In 2008, we theoretically predicted that, at the certain conditions, the new possible nonlinear dynamical phenomena can be observed on the Ikeda map: 1) interior crisis, which is associated with a sudden change in size of a chaotic attractor, and 2) boundary crisis, which is associated with a sudden disappearance of the attractor as it collides with its basin boundary. The recently reported experimental data confirmed our important theoretical predictions. In our opinion, the possible physical nature of all the nonlinear effects in *HTS* thin films at microwaves is still not clarified yet, that is why the research on nonlinearities in microwave superconductivity is so important for *HTS* microwave filters design in Cryogenic Transceiver Front End (*CTFE*) in basestations in wireless communications. Some discussions on the modern designs of *HTS* microwave filters are conducted in following chapters and the *HTS* microwave filters design layouts with frequency responses are demonstrated together with a number of encountered design problems and limitations, because of nonlinearities appearance in *HTS* thin films at microwaves.

In the next Chapters, the research will be focused on the accurate characterisation of nonlinear properties of *HTS* thin films at microwaves, using the dielectric and microstrip resonators. The experimental results and theoretical modelling of the nonlinear *S-shape dependences* of surface resistance $R_s(H_{rf})$ for the case of $H_{rf} \sim H_{c1}$ will also be presented.

# CHAPTER 4

# LUMPED ELEMENT MODELING OF NONLINEAR PROPERTIES OF HIGH TEMPERATURE SUPERCONDUCTORS IN MICROWAVE RESONANT CIRCUITS

## 4.1. Introduction.

As described in Chapter 3, the characterisation of superconducting materials in high frequency electromagnetic fields is usually conducted with use of different types of microwave resonators: *dielectric resonator* and *microstrip resonator*.

Resonance systems can have different geometrical shapes such as rectangular, round, and some others with different technical characteristics. Resonance systems are also differentiated on either *transmission* or *reflective* types in relation to the electromagnetic wave propagation characteristics.

Electrical scheme of the two port transmission resonator is presented in Fig.1 [1].

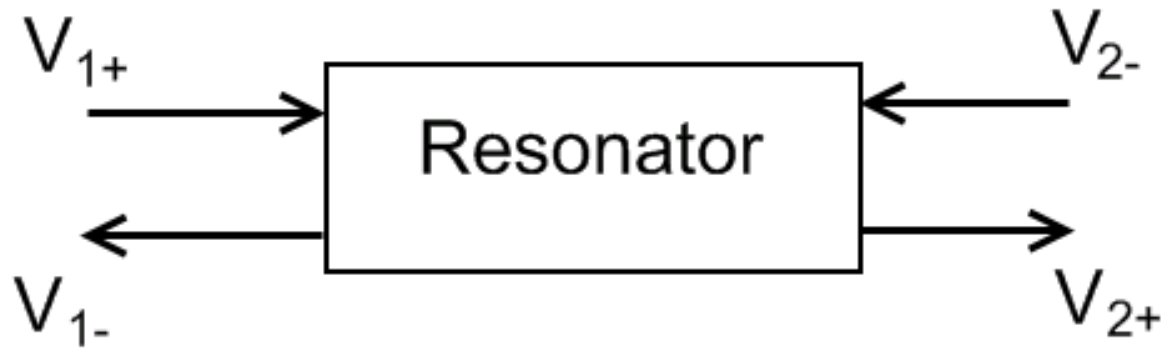

**Fig. 1.** Electrical scheme of two ports network for *S*-parameters definition (after [1]). The amplitudes $V_1$ and $V_2$ of direct and reflected signals are connected by the linear expression

$$\begin{bmatrix} V_{1-} \\ V_{2-} \end{bmatrix} = \begin{bmatrix} S_{11} & S_{12} \\ S_{21} & S_{22} \end{bmatrix} \begin{bmatrix} V_{1+} \\ V_{2+} \end{bmatrix},$$

where reflections coefficients equal to $S_{11}=(V_{1-}/V_{1+})$, when $V_{2+}=0$, and $S_{22}=(V_{2-}/V_{2+})$, when $V_{1+}=0$. Transmission coefficients are $S_{21}=(V_{2-}/V_{1+})$, when $V_{2+}=0$ and $S_{12}=(V_{1-}/V_{2+})$, when $V_{1+}=0$.

All resonance systems are characterised by certain parameters such as the *resonance frequency*

$$f_0= \omega_0/2\pi,$$

and the *quality factor*

$$Q = f_0/bf,$$

where $bf$ is the bandwidth of the resonance curve at - 3$dB$ level as shown by the author of dissertation in Fig. 2

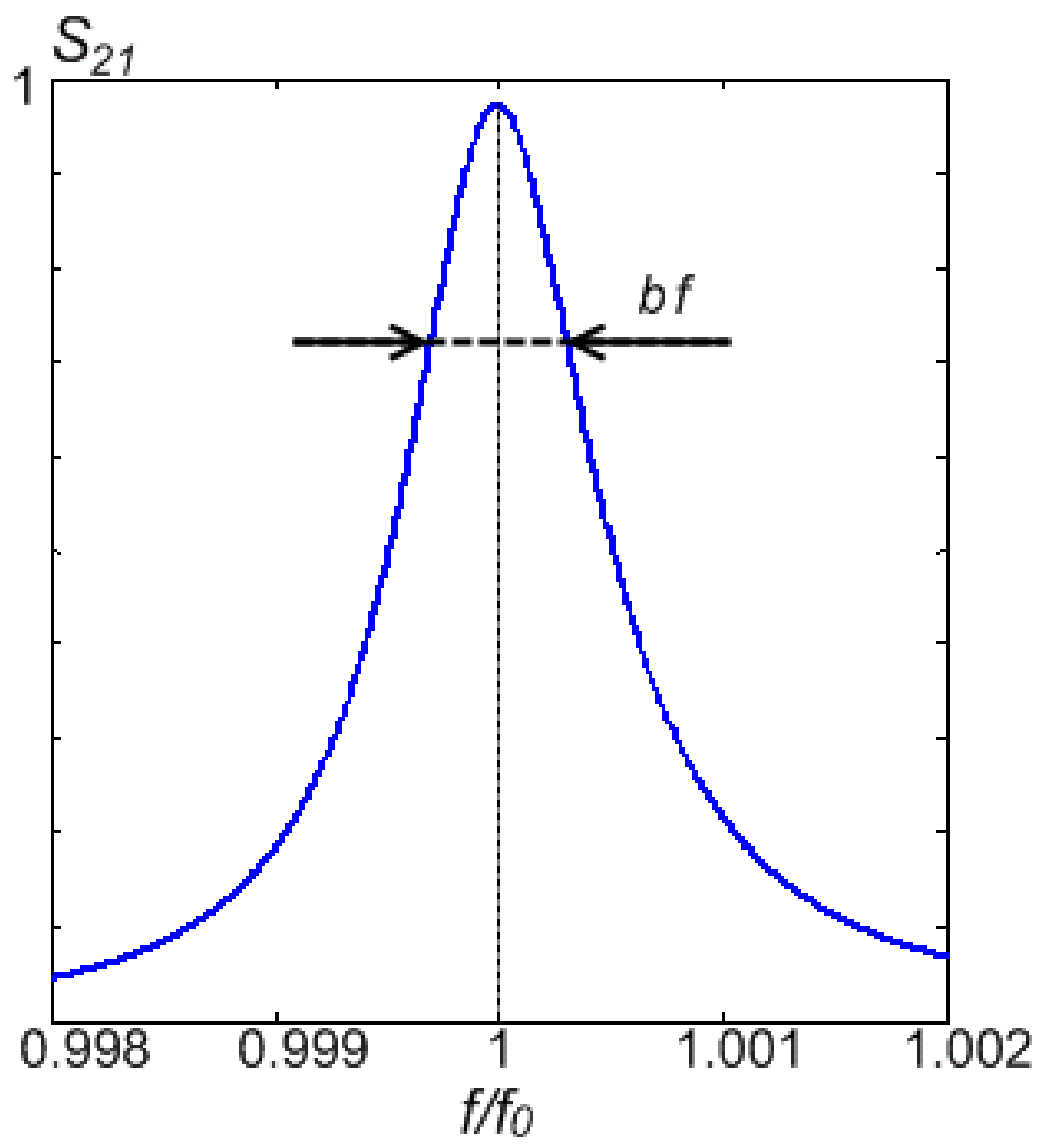

**Fig. 2.** Linear *Lorenztian resonance curve* with resonance frequency $f_0$ and band width $bf$.

Resonance circuit has the characteristic impedance

$$Z_r = R_r + jX_r,$$

where $R_r$ is the resonator resistance and $X_r$ is the resonator reactance, which are defined for the system of the resonator with sample. At the different magnitudes of the temperature $T$, microwave power $P$ and other parameters, the experimentally measured resonance frequency shifts on

$$\Delta f_0(P) = f_0(P_{min}) - f_0(P)$$

and the bandwidth *bf* increases as the quality factor *Q* of resonance system decreases.

In the nonlinear case, at action of external factors, the shape of resonance curve changes so that it becomes different from the *Lorentzian shape*, but it can still be characterized by the *RF* parameters of resonance circuit.

Resistance of system of the microwave resonator with a sample is

$$R_r = A / 2Q$$

and the reactance is

$$X_r = A \cdot \Delta f_0 / f_0,$$

where *A* is the geometrical factor. After taking to the account all the corrections, connected with energy absorption of electromagnetic wave in resonator's surface, dielectrics and other losses in resonator, it is possible to find the surface resistance $R_S$ of superconductor sample and its reactance $X_S$ as a function of microwave power *P* of ultra high frequency electromagnetic wave, and then to research the nonlinear effects in *HTS* thin film samples.

The ***cavity resonator*** can be filled with a high quality dielectric puck, usually placed at the centre, to increase the accumulated electromagnetic wave energy and concentrate the field within the dielectric area. *Quality factor* of resonance system with a sample is determined as a ratio of the accumulated wave energy within free resonator volume to the dissipated energy in the resonator walls and researched sample for a period of oscillation. Therefore, systems with the high quality factors are more sensitive to the change of physical properties of researched superconductor samples, and as a result, these systems are more widely used for the accurate microwave characterisation of physical properties of superconductors.

The ***microstrip resonators*** have smaller quality factors *Q* than the cavity ones, but they produce the higher fields at the same signal power levels. To some extend, this effect is an advantage in research on the nonlinearities, which arise at elevated electromagnetic field strengths and big magnitudes of currents, generated in superconducting sample. The *microstrip resonator* filters are used in the receiver in *Cryogenic Transceiver Front End* (*CTFE*). The *Sliced Microstrip Line Resonator* (*MSLR*) filters are designed and used as tunable high-power transmit filters in *4G* mobile communication systems [35].

***One of the most important tasks in research of superconducting materials is a transition from the physical characteristics, obtained for a resonant system, to the physical characteristics, which correspond to the researched superconductor.*** Main measurement parameter of superconductors is the ***surface resistance*** $\boldsymbol{R_S}$. As the surface resistance $R_S$ of superconductors has a small value, the resonance systems with the high quality factors $Q$ are employed for its research. The obtaining of the precise quantitative data on the surface resistance $R_S$ dependence on the temperature $T$, magnetic field density $H$, magnitude of currents $J$ on the surface of superconductor, and other parameters represents a formidable task, in which the value of surface resistance $R_S$ has to be retrieved from the measured characteristics of microwave resonator.

Generally, knowing the structure and distribution of fields in microwave resonator along with the geometrical shape of sample and its exact position in resonator, it is possible to get the parameters of superconductor by direct calculation of magnitudes of fields. However, this task is complicated and, in mathematical sense, does not possess appropriate convergence. The distribution of fields in resonator is well known and can be easily calculated in the case, when their frequency is either close or matches the resonance one. When these fields are shifted from resonance frequency, their distribution becomes essentially more complicated, the integer of half wave is not within the resonator's length, and the calculations become approximate. In this case, the energy stored in resonator is represented as a complex function of frequency $f$. In order to find the *quality factor* $Q$ of microwave resonator with superconductor, it is necessary to know the width of resonance curve $S_{21}(\omega)$, which directly depends on function $P(\omega)$, where $S_{21}$ is the direct transmission coefficient. Taking into the account that the characteristic values of surface resistance $R_S$ for superconductors are rather small, such calculations must be conducted with high accuracy in order to detect changes in magnitude of surface resistance $R_S$.

In these conditions another computational method of resonance system and its characteristics is more convenient. It is based on the possible representation of a resonator near any of its resonance frequencies as the lumped *RLC* element circuit. This method is less complex for calculations and enables to extract the superconductor parameters based on the experimental data, obtained for a resonator with the load. This approach has been employed in this dissertation.

## 4.2. Lumped Element Equivalent Circuits for Representation of Superconductors at Microwaves.

The equivalent circuit for a superconductor can be determined going from the physical properties of superconductor at microwaves. In the *two fluid model* [2], the conductivity of superconductor

$$\sigma_S = \sigma_1 - j\sigma_2,$$

and its impedance

$$Z_S = R_S + jX_S,$$

are complex variables at external electromagnetic field. Therefore, it is convenient to represent the superconductor as an equivalent electric network, consisting of the active resistance and inductance, which are connected in parallel, for the calculations of the conductivity $\sigma_S$ of a superconductor at external electric field. The equivalent circuit for the calculations of conductivity $\sigma_S$ is shown in Fig. 3 in [3]

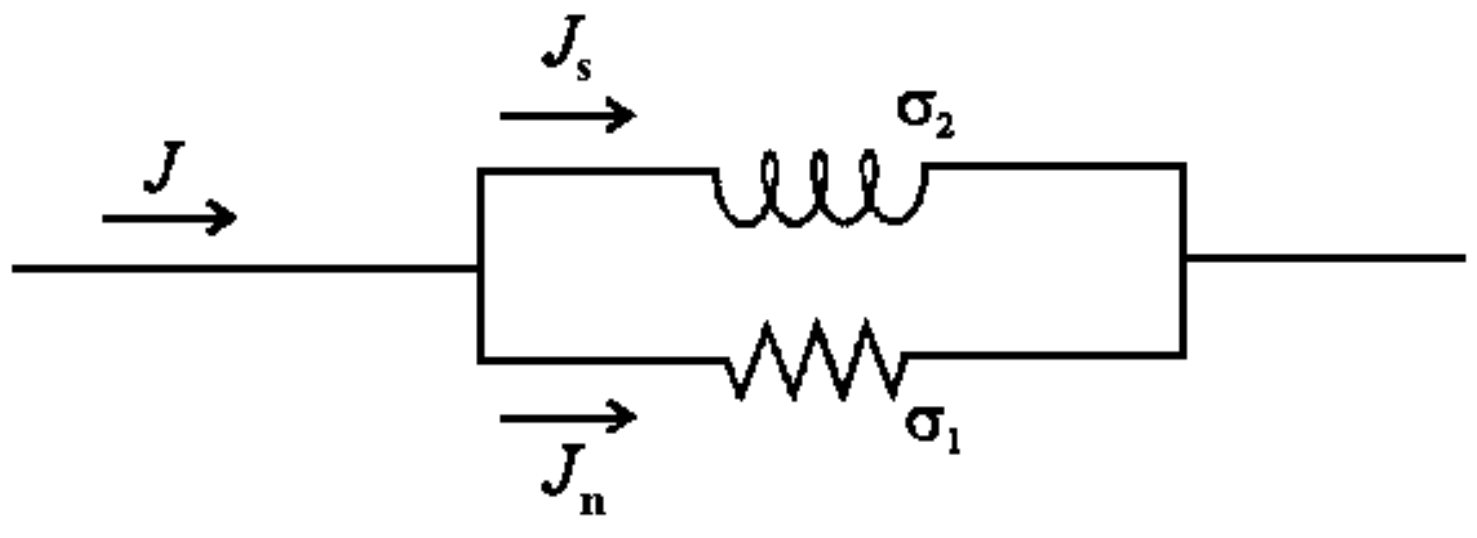

**Fig. 3.** Equivalent circuit representation of superconductor conductivity $\sigma_S$ in *two fluid model* (after [3]).

The surface impedance $R_S$ of a superconductor is

$$Z_s = R_s + jX_s$$

It can be represented as the resistance and inductance connected in series as illustrated in Fig. 4.

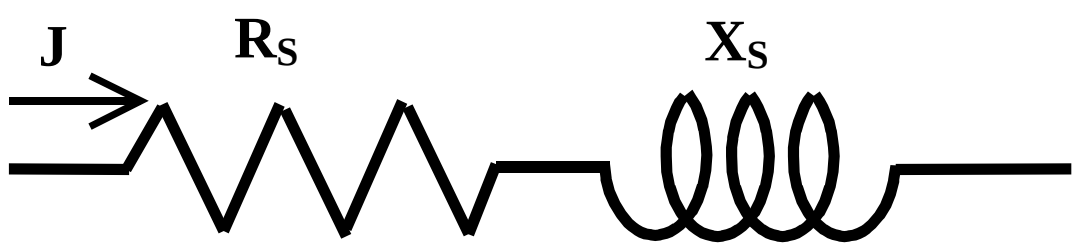

**Fig. 4.** Equivalent circuit representation of superconductor impedance (after [4]).

The real and imaginary components of the conductivity and impedance can be expressed as in [4]:

$$\sigma_1 = \frac{2\omega\mu_0 R_s X_s}{(R_s^2 + X_s^2)^2}$$
$$\sigma_2 = \frac{\omega\mu_0 (X_s^2 - R_s^2)}{(R_s^2 + X_s^2)^2} \quad (4.1)$$

and

$$R_s = \frac{1}{2\lambda_L}\frac{\sigma_1}{\sigma_2^2}$$
$$X_s = \left(\frac{\omega\mu_0}{\sigma_2}\right)^{1/2} = \omega\mu_0\lambda_L = \omega L_S \quad (4.2)$$

where, $\lambda_L = (\mu_0\omega\sigma_2)^{-1/2}$ is the magnetic field penetration depth.

The *two fluid model* approach was used in [5], where the transition from the parallel equivalent circuit representation in Fig. 5 to the series one in Fig. 6 was made. It was used for the analysis of properties of the *NbN* superconducting stripline resonators at the frequency of 1.146*GHz* and the $YBa_2Cu_3O_{7-\delta}$ superconductor at the frequency of 3.0778*GHz*.

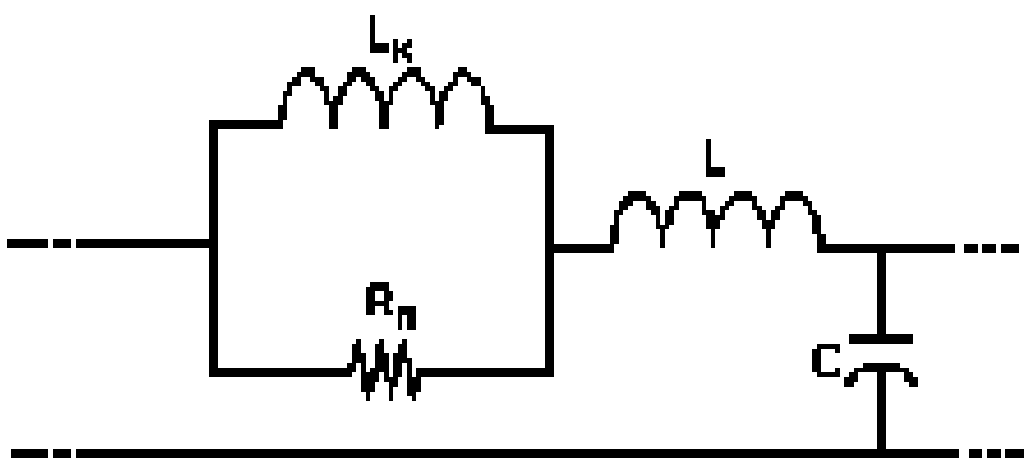

**Fig. 5.** Standard parallel equivalent circuit representation of superconductor connected to a series resonant *LC* circuit (after [5]).

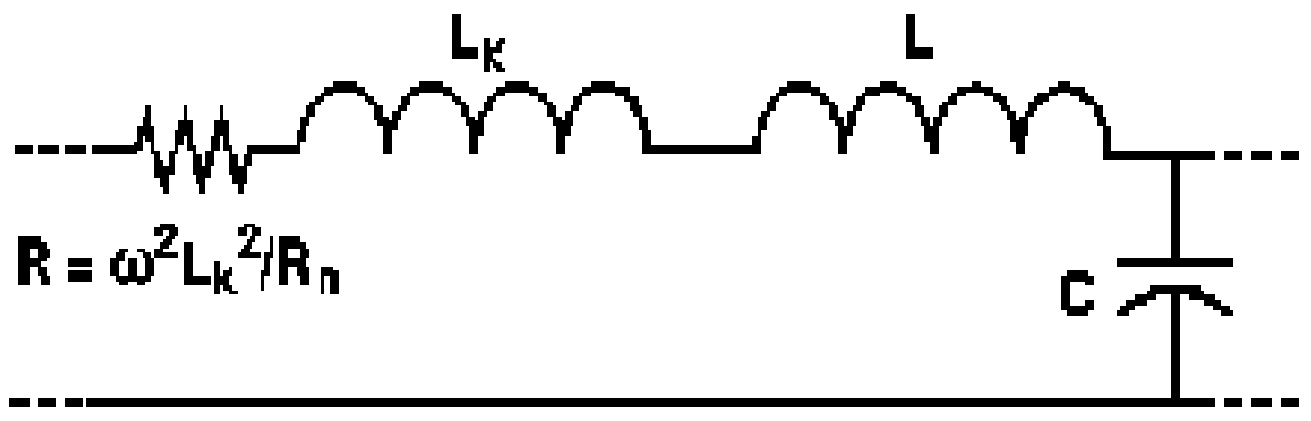

**Fig. 6.** Series impedance equivalent circuit representation of superconductor (after [5]).

Van Duzer and Turner [2] proposed the more complex equivalent circuit for representation of a superconductor shown in Fig. 7.

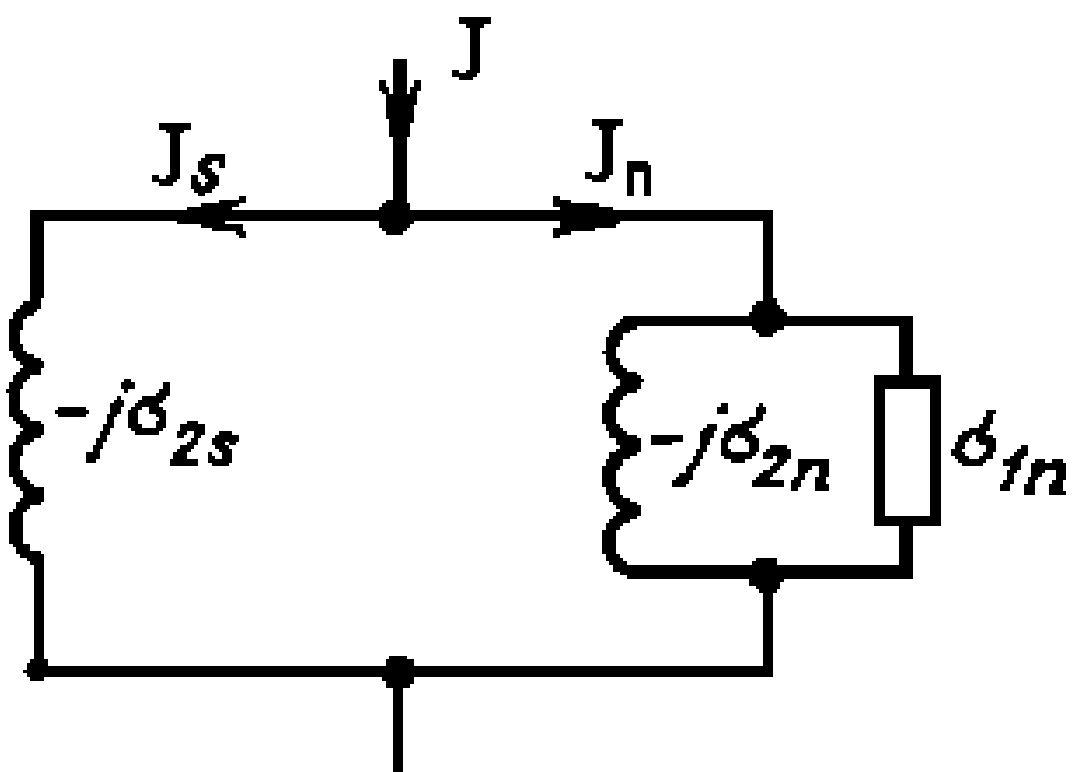

**Fig. 7.** Equivalent circuit for the current distribution between normal and superconducting electron subsystems in a superconductor (after [2]).

In this representation, the additional inductance with the conductivity $\sigma_{2n}$ is introduced to the part of circuit that corresponds to the normal electrons. The introduction of the additional element enables to take into account the effect of influence of the magnetic field on normal electrons.

Nguyen [6] employed a representation of *HTS* superconductor shown in Fig. 8. This approach is suited for calculations of conductivity of *HTS* thin films, when the Josephson inter-granular contacts play an important role in creation of nonlinearities in *HTS* thin film. In [6], the surface impedance of $YBa_2Cu_3O_{7-\delta}$ thin films was researched at the frequencies of 1 - 17*GHz* and temperature range of 4.2 – 91*K*.

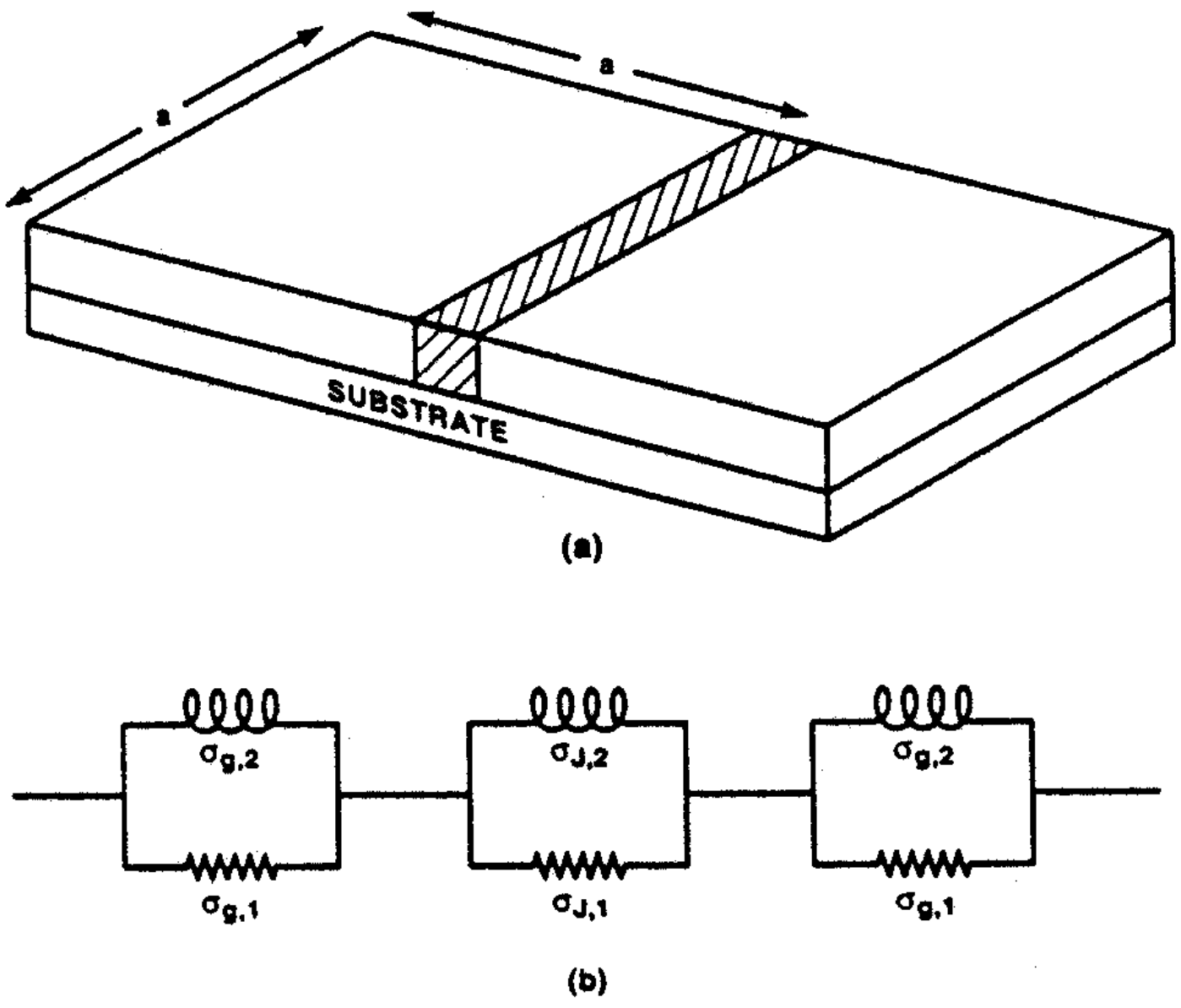

**Fig. 8.** (a) Conditional representation of $YBa_2Cu_3O_{7-\delta}$ superconducting thin film and (b) its equivalent circuit (after [6]).

A similar circuit was used in order to investigate the Josephson contacts between the grains in crystal lattice of a superconductor, which exhibit the weak superconducting properties in [7]. Authors [7] proposed a model of *Josephson coupling* between the grains to analyse the surface impedance of polycrystalline $YBa_2Cu_3O_{7-\delta}$ thin films as shown in Fig. 9.

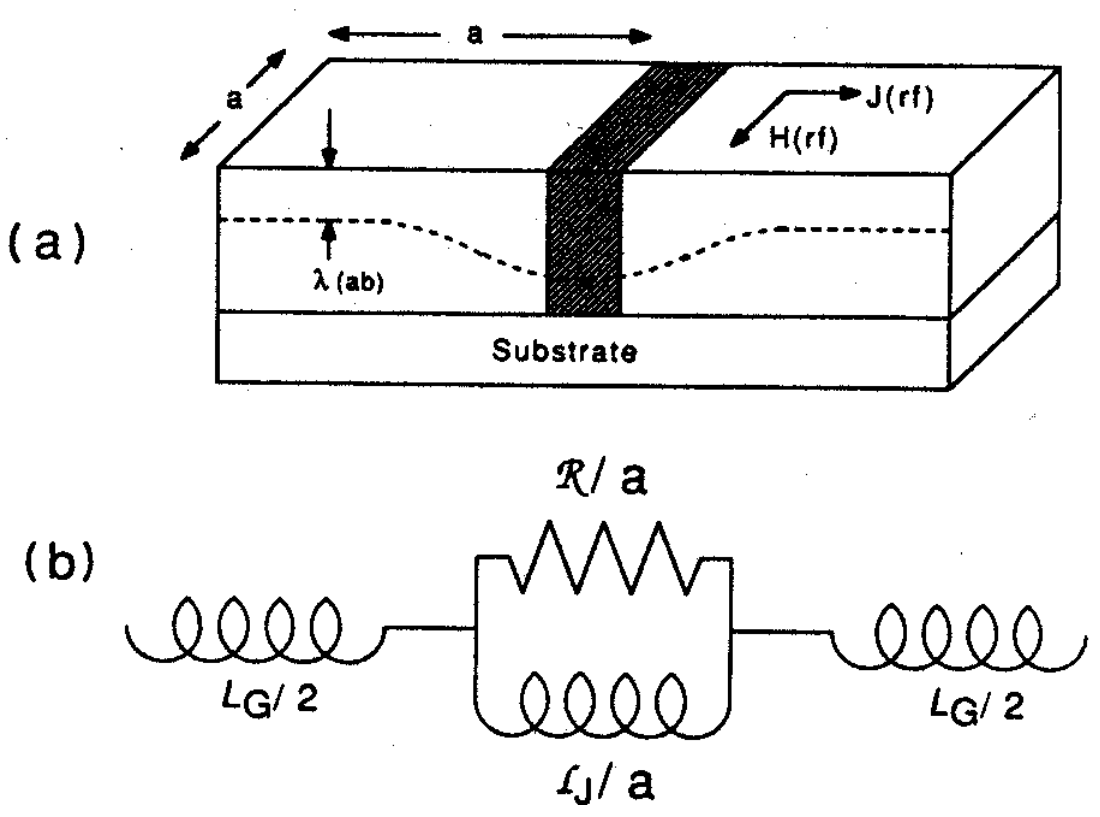

**Fig. 9.** (a) Superconducting film with grain boundary; (b) equivalent circuit of superconducting film with grain boundary structure (after [7]).

The equivalent circuit for the residual resistance analysis of $YBa_2Cu_3O_{7-\delta}$ polycrystalline superconductor is shown in Fig. 10. in [8]

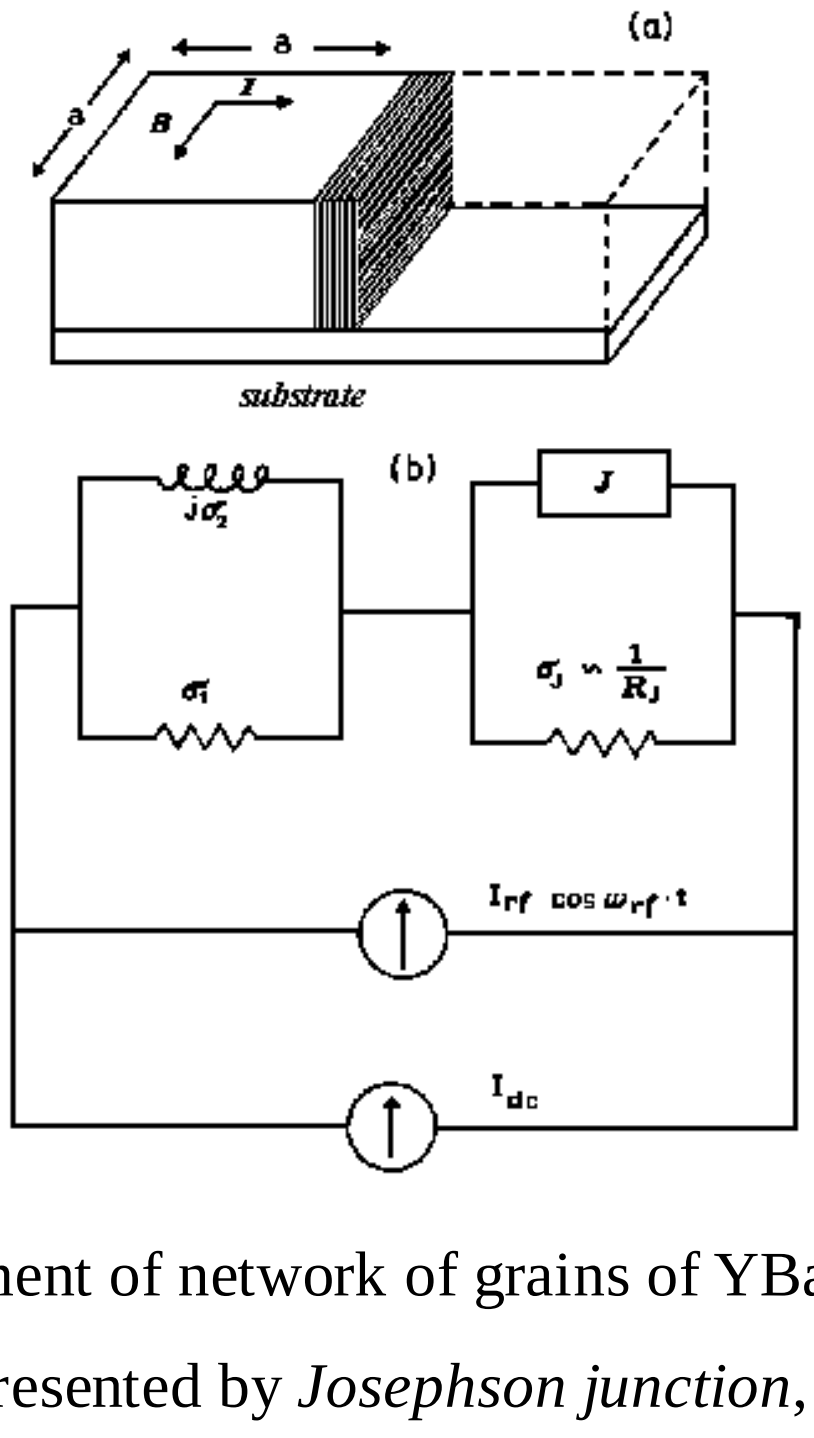

**Fig. 10.** (a) Basic element of network of grains of $YBa_2Cu_3O_{7-\delta}$ polycrystalline superconductor is represented by *Josephson junction*, (b) Equivalent circuit for $YBa_2Cu_3O_{7-\delta}$ polycrystalline superconductor (after [8]).

The equivalent circuit in Fig. 11 was employed in order to research the nonlinear microwave response of $YBa_2Cu_3O_{7-\delta}$ thin films at the frequencies of 1.5–20 *GHz* in [9]. In the *RSJ* model equivalent circuit representation: $Z_g$ is the surface impedance of each one of the neighbouring grains; $L_j$ is the series inductance per square of junction region; $l_j$ is the Josephson kinetic inductivity of junction. Permittivity $\varepsilon_j$ and resistivity $\rho_j$ of *Josephson junction* are in shunt. The *Telegraphers equations* for the shown transmission line are analogous to the *Maxwell equations* of *Josephson junction* with appropriate boundary conditions.

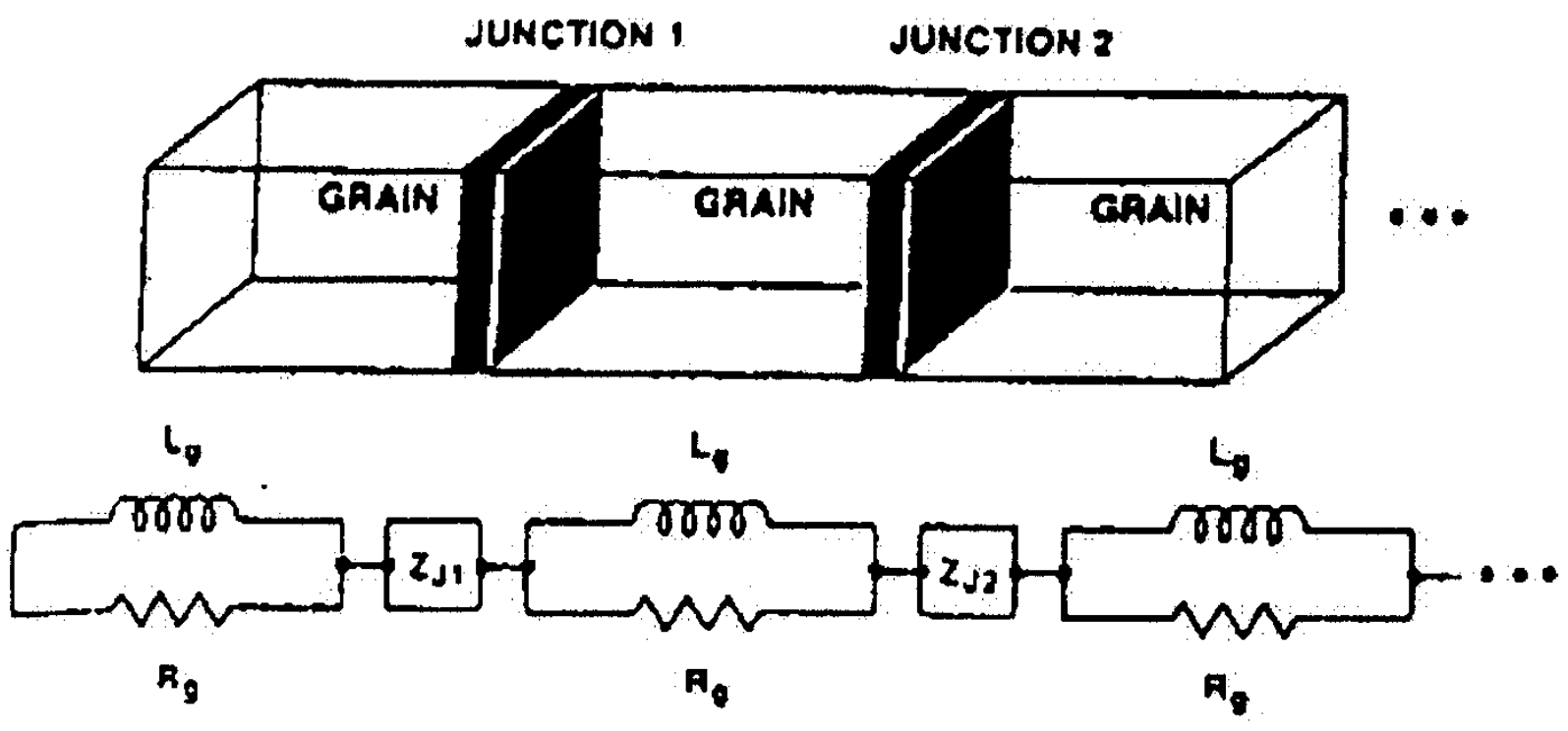

**Fig. 11.** (a) Network representation of coupled grain model with distributed resistively shunted Josephson junctions: *RSJ* model, and (b) *RSJ* model equivalent circuit (after [9]).

The circuit, called "$\alpha$, $\beta$ - model," for the characterisation of properties of non-ideal superconductors for application in particles accelerators is shown Fig. 12 in [10]. This is one of the variations of a network, which was firstly proposed in [7].

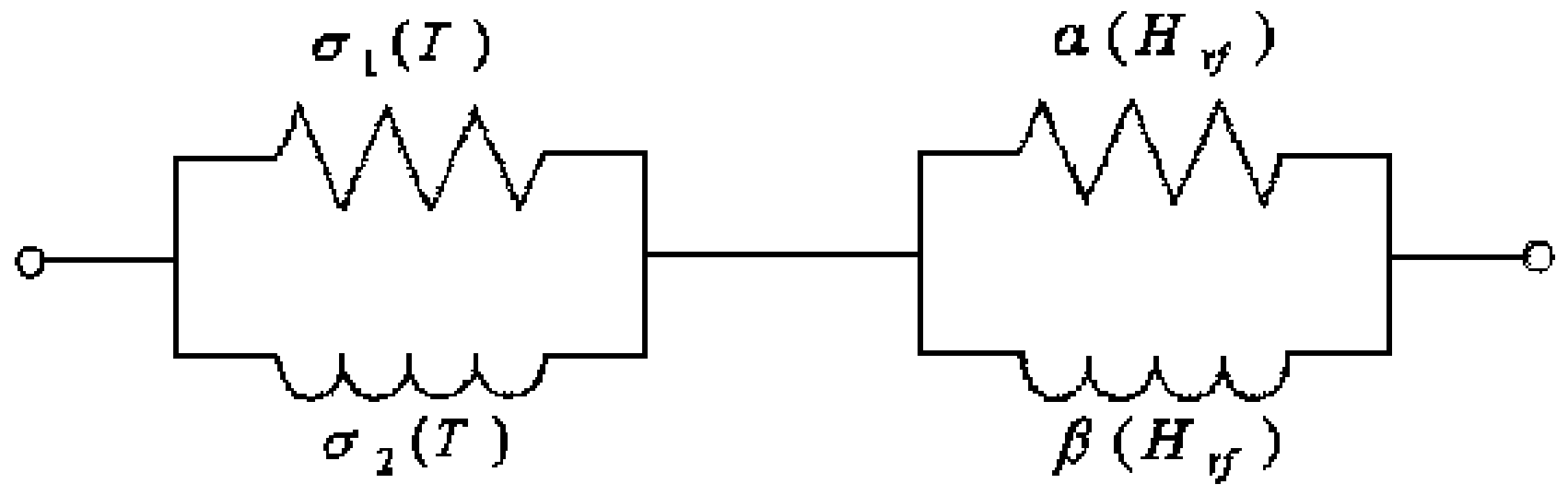

**Fig. 12.** Equivalent circuit of *α, β* - model to describe the electrodynamics of non-ideal superconductor (after [10]).

Dissipation processes that can occur in the superconducting part of equivalent circuit, shown in Fig. 11, were taken into account by Ma and Wolf in [11]. Therefore, the updated equivalent representation of superconductor was utilised in the form of parallel impedance network in Fig. 13 in [11]. The presented approach is based on the theoretical assumption that the superconducting electrons, as well as normal ones, have small resistance and their impedance can be expressed as $Z_s = R_s + j\omega L_s$ and $Z_n = R_n + j\omega L_n$ respectively. Main difference from the two fluid model is that the parallel scheme for two distinct impedances of the normal electrons $Z_n$ and superconducting electrons $Z_S$ is proposed in Fig. 13 in [11] instead of the series scheme $R_S$ and $X_S$ is shown in Fig. 4. in [4]

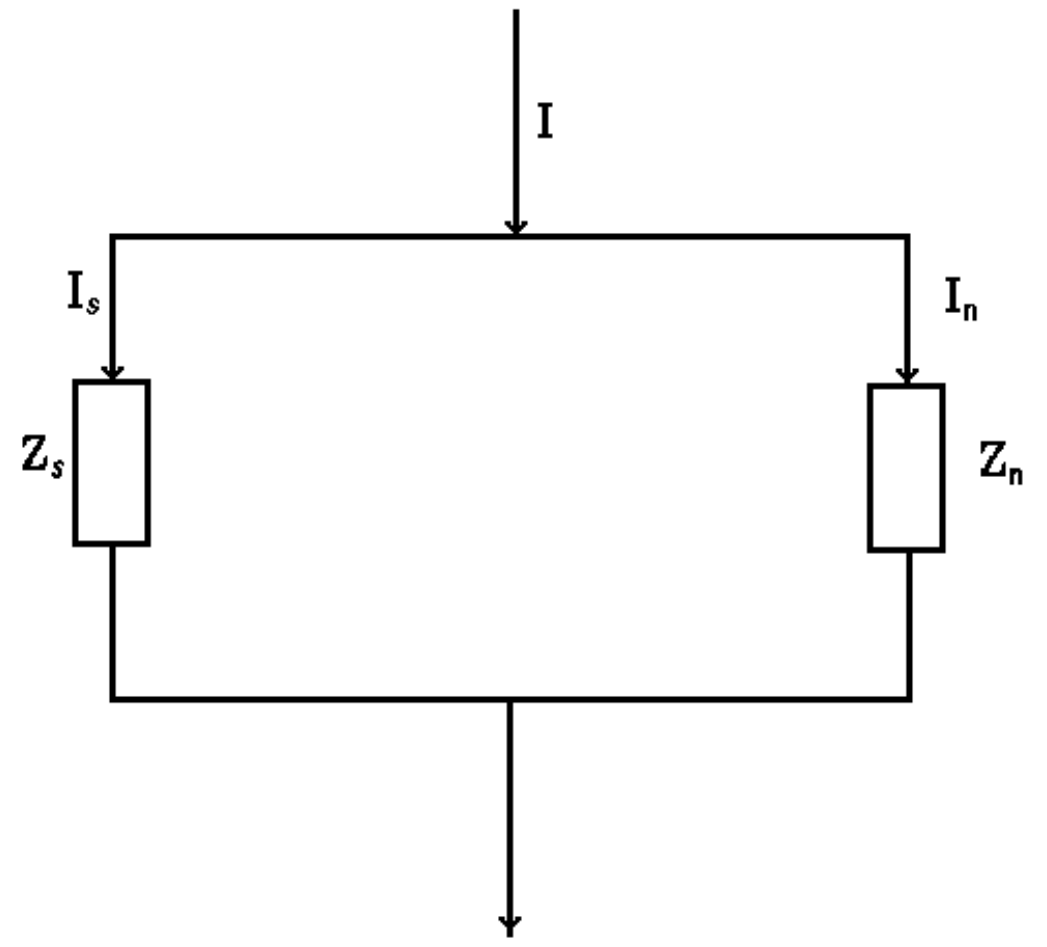

**Fig. 13.** Equivalent circuit in form of parallel impedance network for representation of superconductor at ultra high frequencies (after [11]).

All the networks listed above can be used for the analysis of properties of superconductors and play an important role in characterisations of superconductors.

As it will be shown in next Chapters 4.3 and 4.4, the microwave resonator itself, without superconducting sample, can be represented as either a series or a parallel network depending on the purpose of research task. It will also be explained that the series network is more appropriate for the representation of microwave resonator in this thesis. The review on possible network variations would also be provided.

## 4.3. Lumped Element Equivalent RLC Circuits for Modeling of Microwave Resonators.

Microwave resonators are electromagnetic components, which exhibit the resonant behaviours with frequency, and are used in filters, oscillators, and tuned amplifies [34]. Microwave resonators are often made from sections of waveguides or transmission lines [34]. At microwaves, the resonators are usually designed using the discrete lumped elements that behave like lumped circuit elements such as a series or a parallel combination of inductors and capacitors [34].

As known, the ideal resonance circuit has an infinite number of resonant frequencies, which correspond to the different wave modes excited in the network. In the systems with the high quality factors *Q*, these resonant frequencies are located far enough from each other in relation to the width of resonance curve. In this case, according to the known electromagnetic theory (see, for example, [12]), any microwave resonator, near each of its own resonance frequencies, can be represented as either a parallel or a series equivalent *RLC* lumped element circuit. The frequency dependence of the normalised imaginary part $\overline{\mathbf{X}}$ (reactance) of impedance $Z = R+jX$ for an ideal resonance system without loses ($R=0$) can be presented as a graph shown in Fig. 14 in [12]

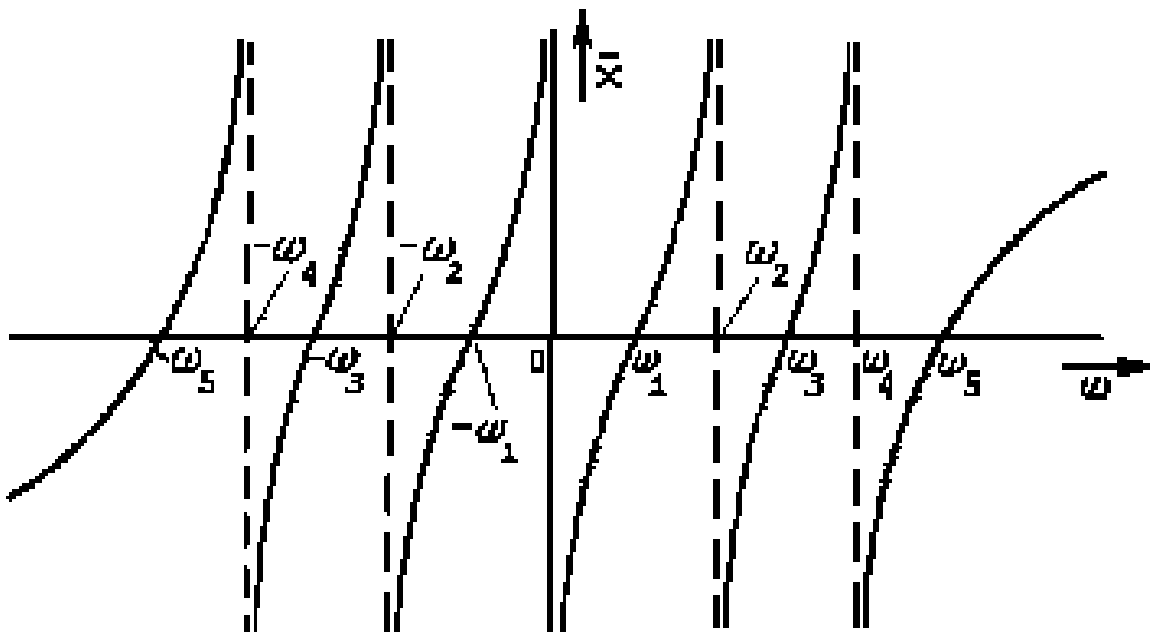

**Fig. 14.** Frequency dependence of normalised imaginary part $\overline{X}$ (reactance) of impedance $Z = R+jX$ for an ideal resonance system without loses ($R=0$) (after [12]).

As can be seen,

$$X(-\omega) = -X(\omega).$$

The equation to define the reactance $\overline{X}$ in a general format can be expressed as (4.3)

$$\overline{X} = A\omega\frac{(\omega^2-\omega_1^2)(\omega^2-\omega_3^2)...(\omega^2-\omega_{2n-1}^2)}{\omega^2(\omega^2-\omega_2^2)(\omega^2-\omega_4^2)...(\omega^2-\omega_{2n-2}^2)} \qquad (4.3)$$

where $\omega_2$, $\omega_4$, ... $\omega_{2n-2}$ are the positive poles, $-\omega_2$, $-\omega_4$, ... $-\omega_{2n-2}$ are the negative poles, $\omega_1$, $\omega_2$, ... $\omega_{2n-1}$ are the positive zeros, $-\omega_1$, $-\omega_2$, ... $-\omega_{2n-1}$ are the negative zeros, and $A$ is a constant.

On the basis of the theory of residues of complex functions, the equation (3.1) can be represented near the reactance poles $X \to \infty$ as (3.4)

$$\overline{X} = L\omega + \frac{a_0}{\omega} + \frac{2a_2\omega}{\omega^2-\omega_2^2} + ... + \frac{2a_{2n-2}\omega}{\omega^2-\omega_{2n-2}^2}, \qquad (4.4)$$

where $a_i$ is the residues of given function near its corresponding poles.

An equivalent network of the equation (4.2) corresponds to a series representation of the parallel $L_iC_i$ circuits shown in Fig. 15.

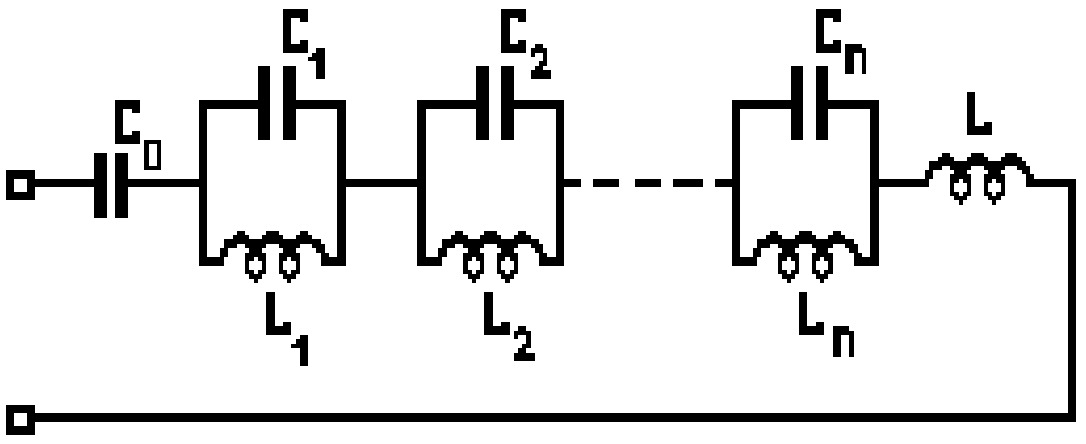

**Fig. 15.** Equivalent circuit representation for expression (4.4) (after [12]).

The corresponding elements of the circuit in Fig. 14 can be expressed as

$$C_0 = -1/a_0,\ C_i = -1/2a_i,\ \omega_{2i}^{\ 2} = 1/L_iC_i,\ L = A.$$

The values of $L_i$ and $C_i$ are determined as a particular set for each resonant frequency.

The normalized reactive admittance, $\bar{\mathbf{b}} = -1/\bar{\mathbf{X}}$, has poles near zeros of the function X(ω). The equation for **b** can be written as (4.5).

$$\bar{\mathbf{b}} = -\frac{1}{A\omega}\frac{\omega^2(\omega^2-\omega_2^2)(\omega^2-\omega_4^2)...(\omega^2-\omega_{2n-2}^2)}{(\omega^2-\omega_1^2)(\omega^2-\omega_3^2)...(\omega^2-\omega_{2n-1}^2)} \qquad (4.5)$$

Following the residue theory, the equation (3.5) can be represented as:

$$\bar{\mathbf{b}} = C\omega + \frac{b_0}{\omega} + \frac{2b_1\omega}{\omega^2-\omega_1^2} + ... + \frac{2b_{2n-1}\omega}{\omega^2-\omega_{2n-1}^2} \qquad (4.6)$$

where $\boldsymbol{b_i}$ is the residues of the given function near its corresponding poles.

An equivalent network for the equation (4.6) corresponds to a parallel representation of series $L_iC_i$ circuits shown in Fig. 16 [12]

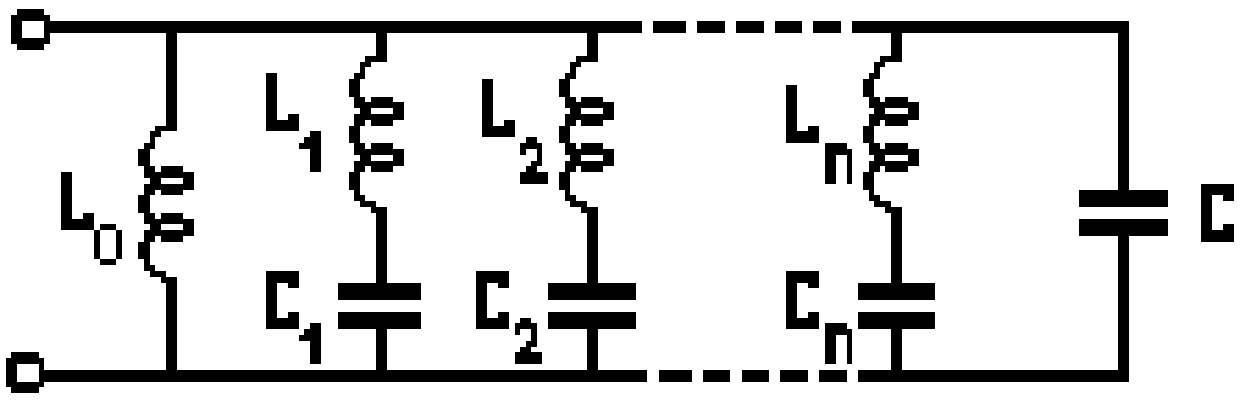

**Fig. 16.** Equivalent circuit representation for expression (3.6) (after [12]).

The corresponding elements of the equivalent circuit in Fig. 16 can be written as:

$$L_0 = -1/b_0,\ L_i = -1/2b_i,\ \omega_{2i-1}^2 = 1/L_iC_i,\ C = -1/A.$$

This approach indicates that the equivalent circuits with the lumped elements can be successfully used for simulations of resonant microwave systems.

In practical case, it is also necessary to take into account the energy dissipation effects and that $R \neq 0$. Therefore, the equivalent networks for both a parallel and a series circuits should be transformed to the forms shown in Fig. 17. in [12]

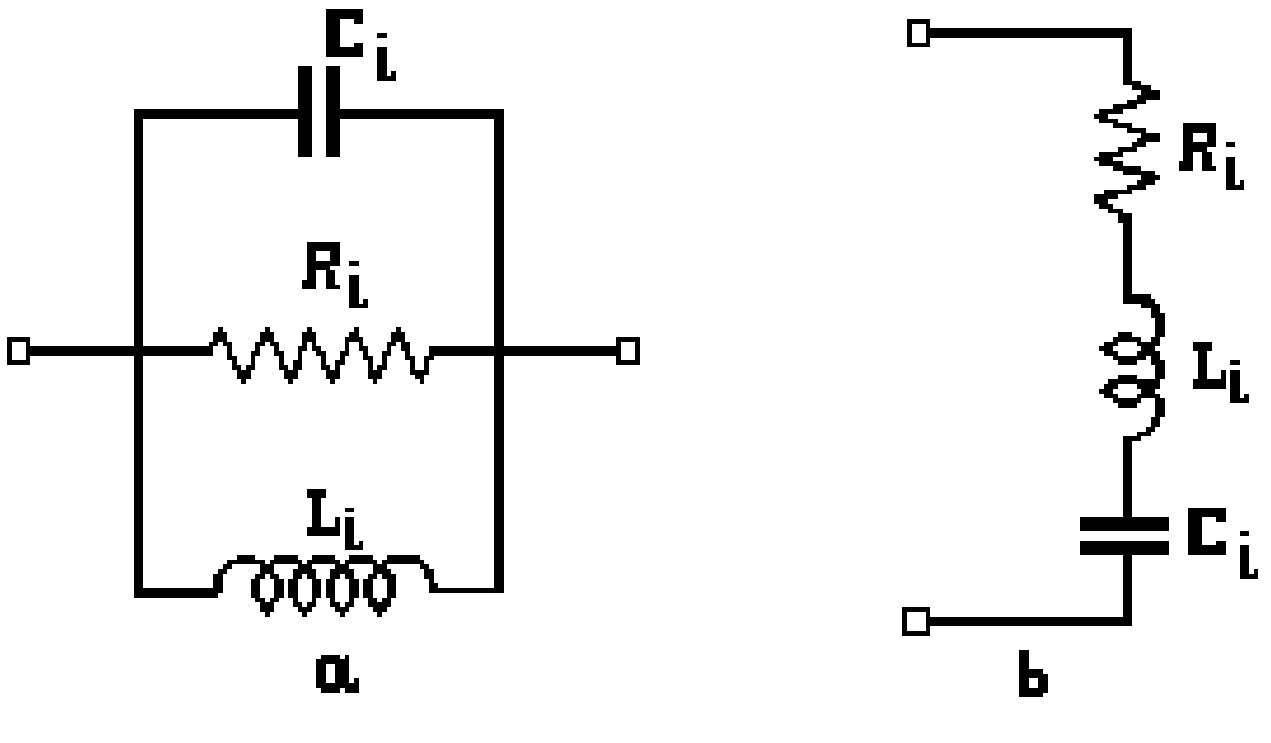

**Fig. 17.** Parallel (a) and series (b) equivalent lumped element networks for representation of microwave resonators taking into account dissipation effects (after [12]).

In this case, the quality factor $Q$ of the parallel resonance equivalent circuit in Fig. 17 (a) is:

$$Q_i = \omega_{2i} C_i R_i, \tag{4.7}$$

and the quality factor $Q$ of the series resonance equivalent circuit in Fig.3.17 (b) is:

$$Q_i = \omega_{2i-1} \frac{L_i}{R_i} \tag{4.8}$$

From these equations, it can be seen that the quality factor $Q$ of the parallel resonant network increases, when the resistance $R$ goes up, while for the series resonant circuit, the quality factor $Q$ becomes higher, when the resistance $R$ decreases. This property of the resonant circuits indicates that one of the networks might be more suitable than other for particular given case.

In practical measurements, the microwave resonator in the transmission mode of operation needs to be coupled to both a source and a load. The result of incorporation of the microwave resonator into the system is that the measured loaded quality factor $Q_L$ differs from the unloaded $Q_0$-factor

$$Q_0 = Q_L(1+\beta_1+\beta_2),$$

where $\beta_1$ and $\beta_2$ are the coupling coefficients of the microwave resonator in the system. The coupling coefficients depend on the energy leakage through the input and output ports of microwave resonator respectively:

$$\beta = P_{ext}/P_0 = Q_0/Q_{ext},$$

where $P_{ext}$ is the power dissipated in external (to the resonator) circuitry and $Q_{ext}$ is the $Q$-factor of this circuitry in [15].

**In order to research the properties of the dielectric materials, they have to be placed in a microwave resonator in the position, where the magnitude of strength of electric field is highest. The dielectrics possess high values of the resistance *R*, and therefore, the parallel resonant equivalent circuit is more suitable for an adequate representation of the microwave resonator-sample system.** In this case, the quality factor $Q$ will be higher, if the resistance $R$ is higher.

On the other hand, **the superconducting materials have very small values of surface resistance $R_S$, and as a result, the superconductors need to be placed in the position, where the strength of magnetic field reaches its maximum in microwave resonator. In the case of superconductors, it is more convenient to represent a microwave resonator-sample system in form of the series resonant equivalent circuit with the small total resistance value.** The quality factor $Q$ of the microwave resonator-superconductor system will be higher, if the surface resistance $R_S$ of superconducting sample is lower [32, 33].

It should be noted that the both equivalent circuits shown in Fig. 17 (a, b) can be transformed one into another without the missing data, and therefore, either network can be utilised, based on the subscribed tasks of a particular research.

## 4.4. Parallel Lumped Element Equivalent Circuits for Modeling of Microwave Resonators.

In [12], J. L. Altman adopted the parallel resonant equivalent circuit representation of transmission resonator with the input and output networks connected in the transformer mode. The parallel resonant equivalent circuit representation is shown in Fig. 18. [12]

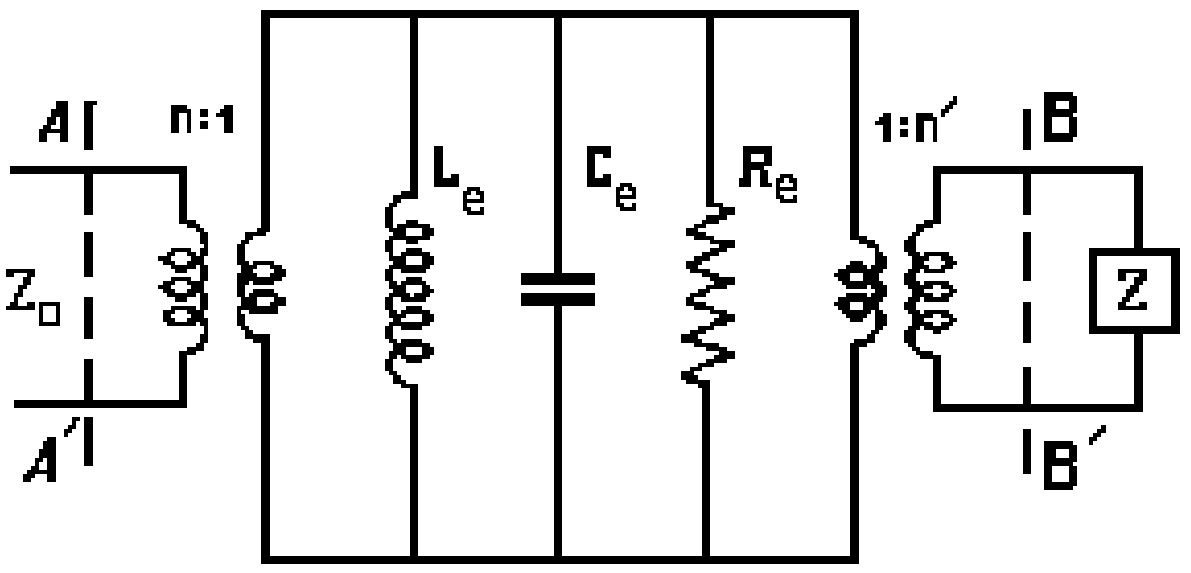

**Fig. 18.** Parallel resonant equivalent circuit representation of transmission resonator with input and output networks connected in transformer mode (after [12]).

The description of resonator cavities by Portis [13] in his research on the microwave properties of the high-temperature superconductors is followed by the lumped elements circuit analysis for the two-port microwave resonator coupled to the input and output transmission lines shown in Fig. 19.

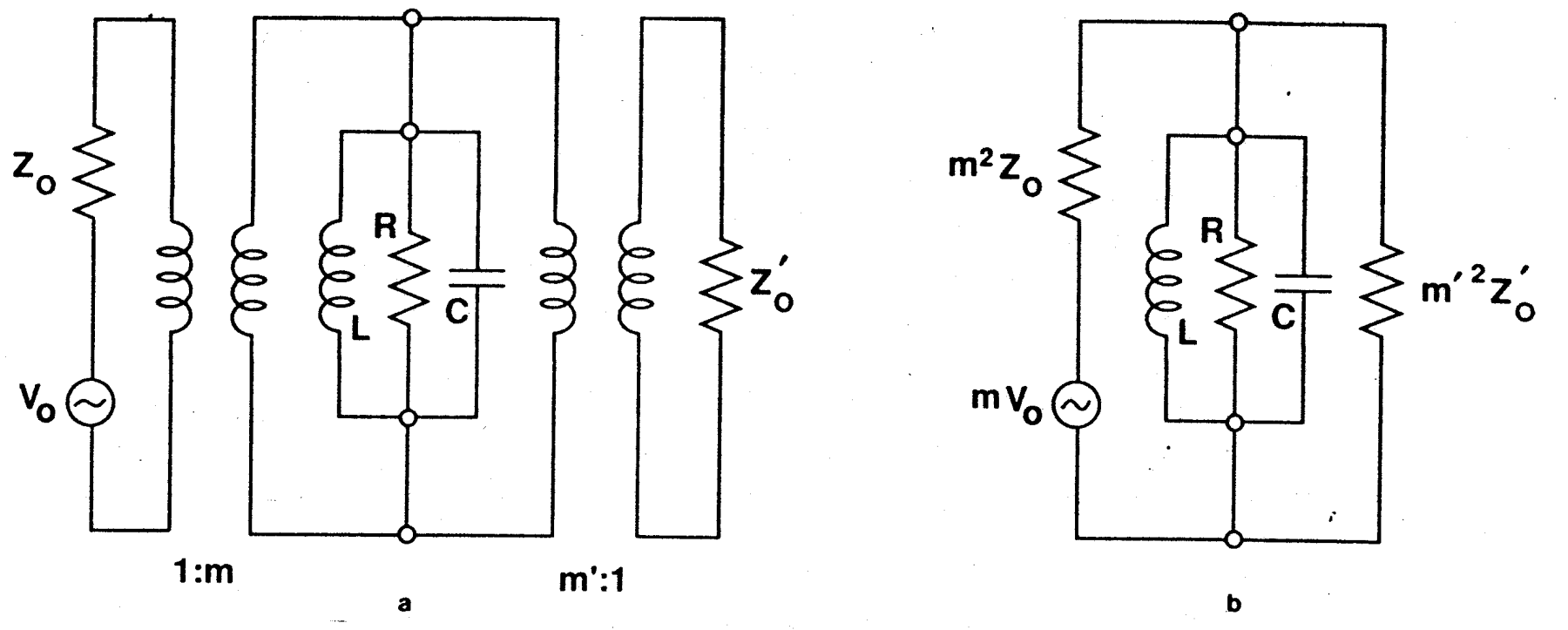

**Fig. 19**. (a) Equivalent circuit of two-port resonator coupled to input and output transmission lines. (b) Transformation of driving voltage and loads into shunt with microwave resonator (after [13]).

The magnetic energy, stored in microwave resonator, is associated with the inductance *L*. The electrical energy is associated with the capacitance *C*, and the dissipation of energy takes place on the resistor *R*. Microwave resonator is driven from an input line of impedance $Z_0$ through an ideal transformer of turns *1:m*, and loaded by the output line of impedance $Z'_0$ through an ideal transformer of turns m':1. The microwave source is matched to the input line and represented by the voltage generator $V_0$. The detector is matched to the output line. In parallel with the resistance $R$ are the load resistances $m^2 Z_0$ and $m'^2 Z'_0$.

Portis also used the transmission line representation, in which the effect of the elevated levels of the microwave power $P$ on the surface impedance of granular and thick-film high temperature superconductors was researched as shown in Fig. 20 [14].

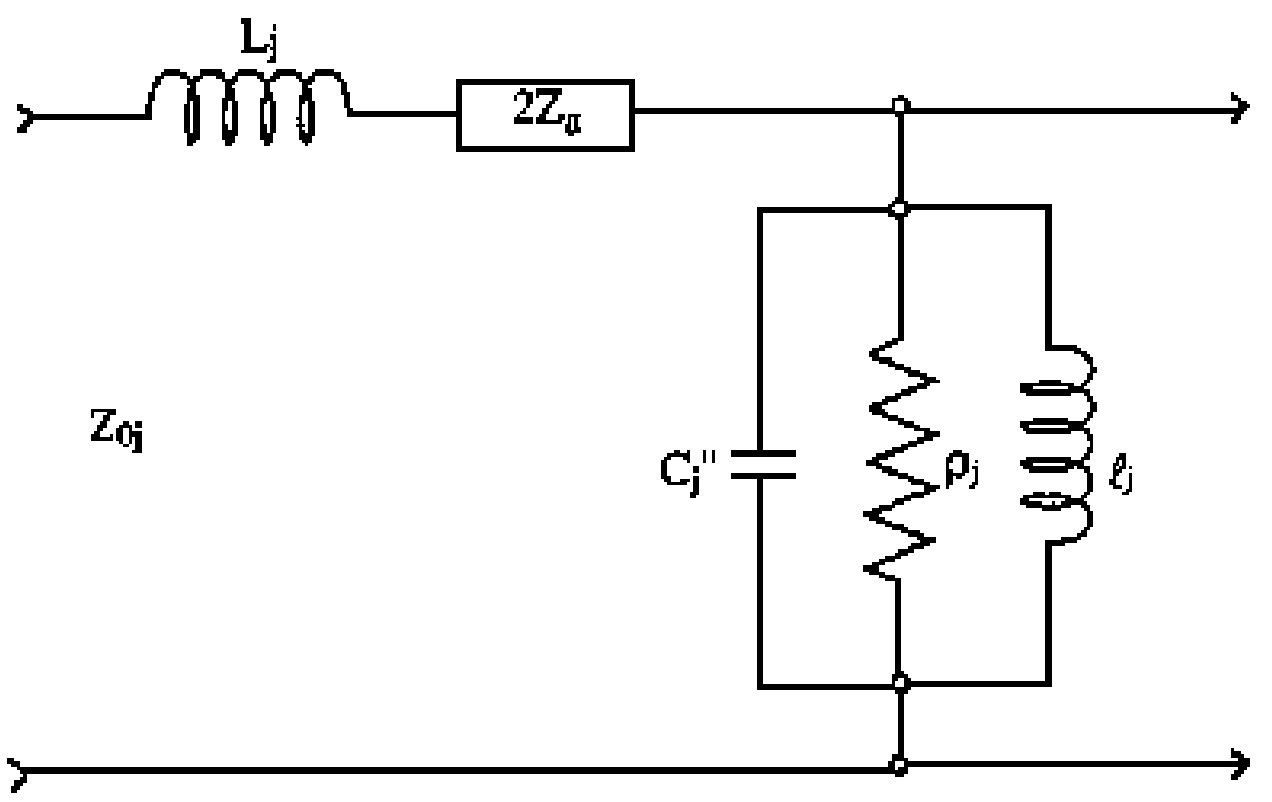

**Fig. 20.** Transmission line representation of inter-granular junction with input impedance $Z_{0j}$ (after [14]).

The following lumped elements circuit model of dielectric resonator system was presented by Leong and Mazierska in Fig. 21 [15]

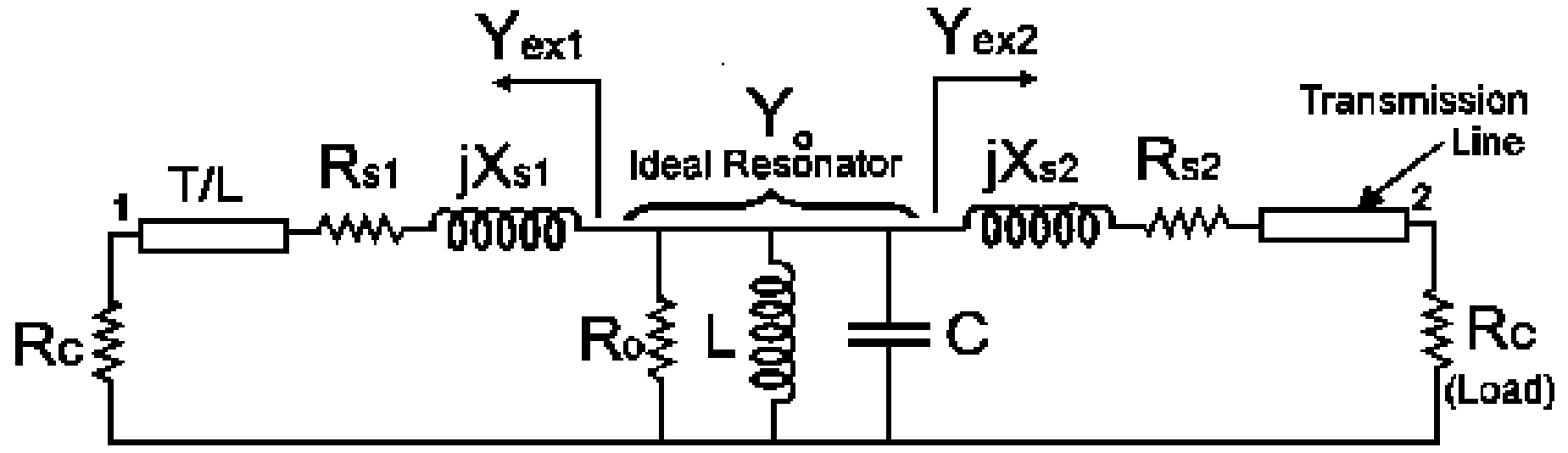

**Fig. 21.** Equivalent lumped elements model of transmission dielectric resonator system (after [15]).

In Fig. 21, the microwave dielectric resonator coupled to the transmission line is presented by the parallel *RLC* network. The coupling losses of each input and output ports are represented by the resistances $R_{S1}$ and $R_{S2}$, and the coupling reactances are modelled by the $X_{S1}$ and $X_{S2}$. The transmission lines of each input and output ports have the characteristic impedances equal to the load $R_C$.

A simple parallel equivalent circuit with coupling capacitors in Fig. 22 was used by D. E. Oates to describe the microwave resonators near the resonance frequency for both the linear and nonlinear operation modes in review [16].

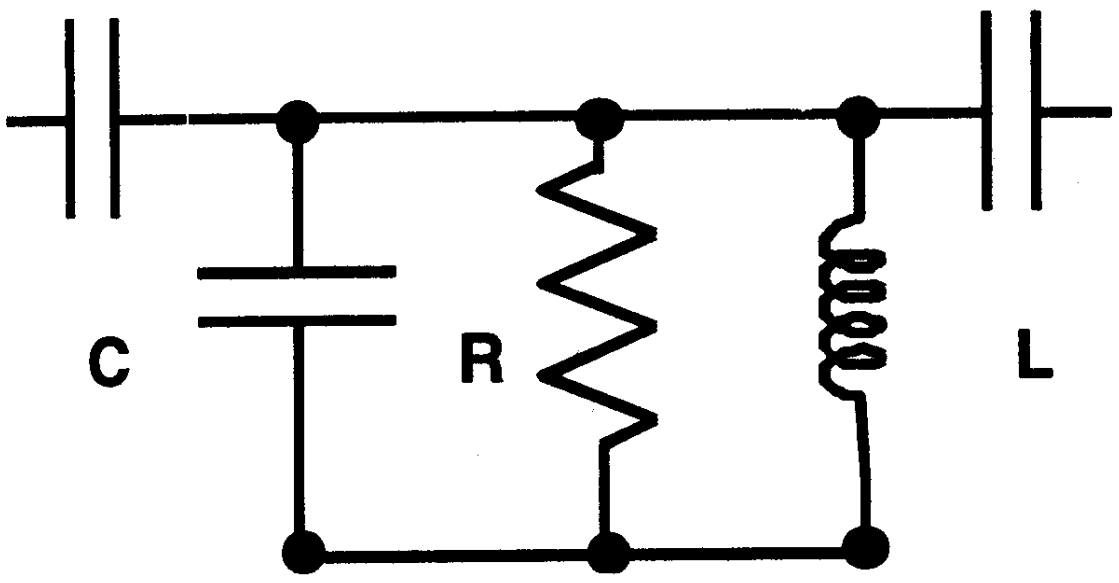

**Fig. 22.** Parallel equivalent circuit of microwave resonator operating near resonance frequency for both linear and nonlinear operation modes (after [16]).

## 4.5. Series Parallel Lumped Element Equivalent Circuits for Modeling of Microwave Resonators.

C. P. Poole demonstrates the calculations of the equivalent series network for the microwave resonator utilised in research on the electron spin resonance in his book [17]. The two variations of the equivalent resonant circuit are presented below in Fig. 23 and Fig. 24. The first case involves the representation of the series equivalent resonant circuit, which is connected to the input and output coupling networks through the transformer, while in other case, the coupling parts are integrated into the series circuit itself.

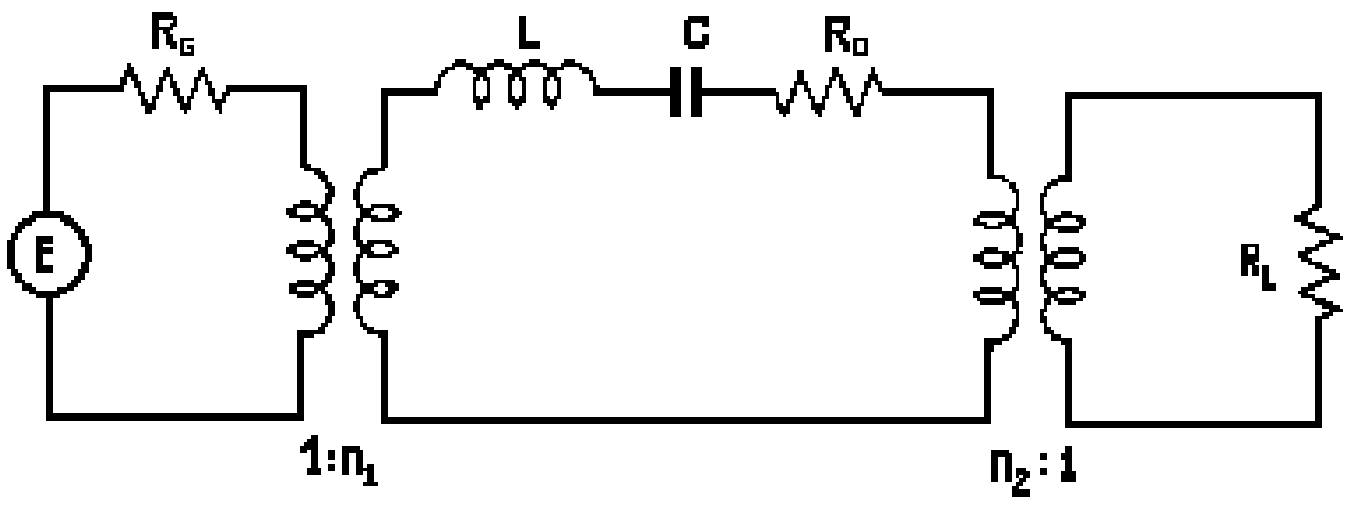

**Fig. 23.** Series equivalent circuit of microwave resonator, connected to the input and output coupling networks, for research on electron spin resonance (after [17]).

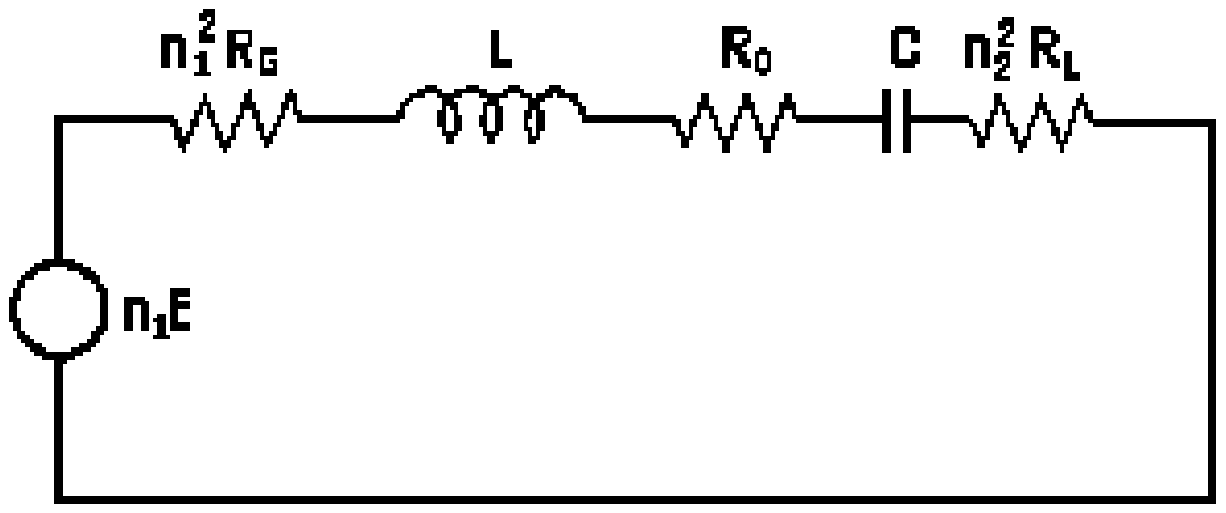

**Fig. 24.** Series equivalent circuit of microwave resonator with integrated coupling parts (after [17]).

O. G. Vendik [18] used a simplified series equivalent circuit for analysis of transmission microstrip resonators. In this work, authors discussed the nonlinear phenomena in a half-wavelength resonator based on a superconducting microstrip line shown in Fig. 25 and 26.

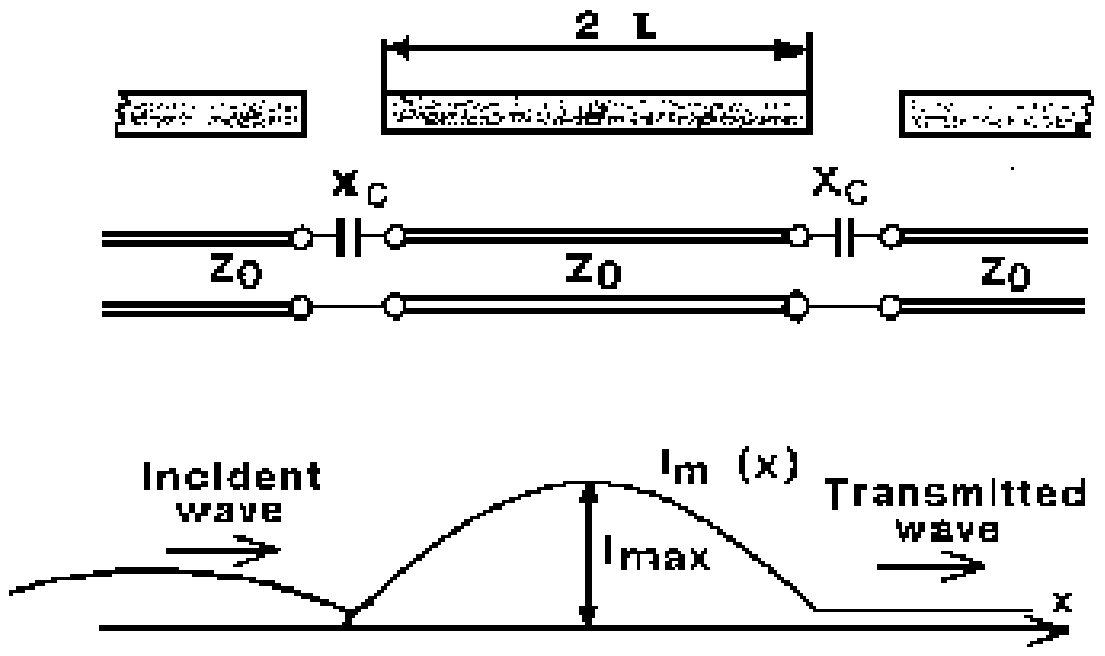

**Fig. 25.** Transmission microstrip resonator with scheme of current distribution (after [18]).

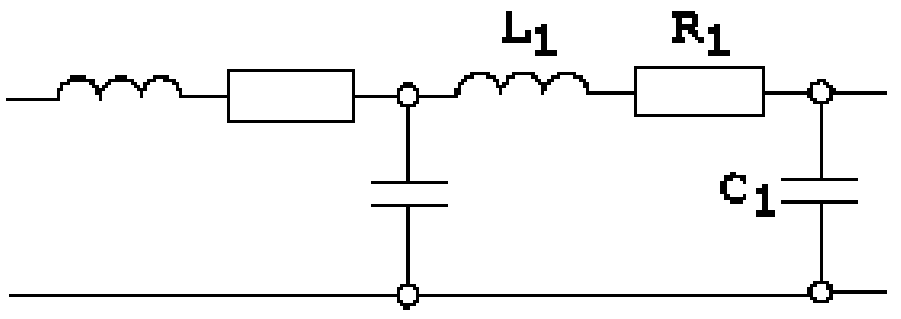

**Fig. 26.** Series equivalent circuit of transmission microstrip resonator (after [18]).

A series equivalent circuit was employed by C. Collado [19] in the application to the harmonic balance algorithms for the simulation of nonlinearities in the *HTS* devices. Fig. 27 shows a series equivalent circuit of transmission line section with incorporated nonlinear elements that account for the nonlinearities in *HTS* [19].

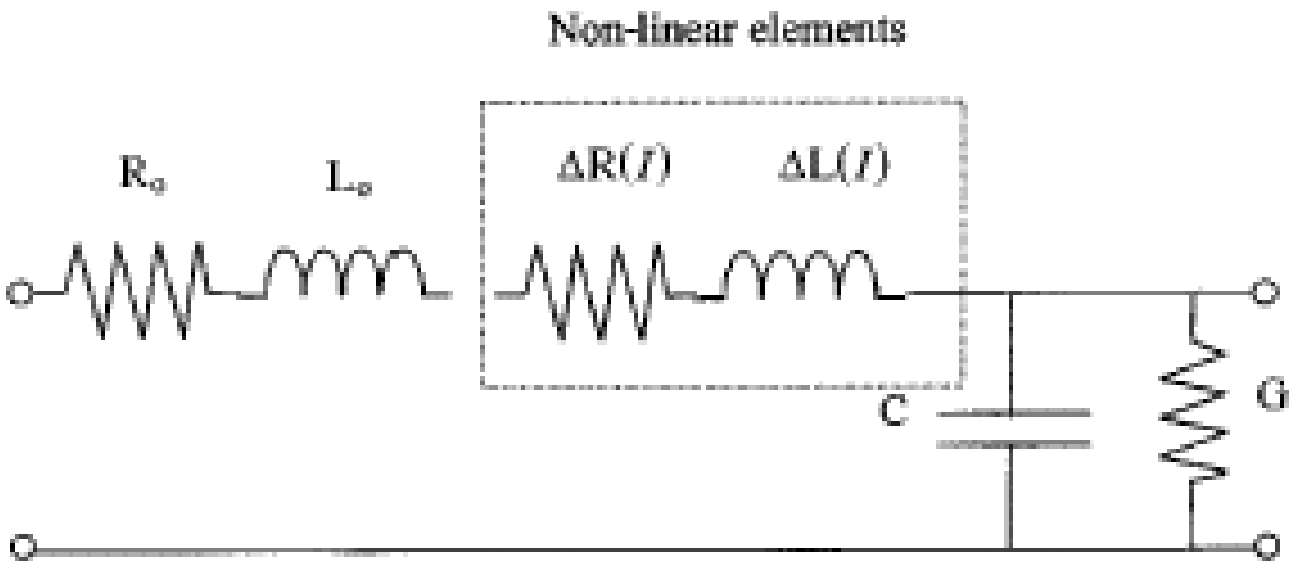

**Fig. 27.** Series equivalent circuit of infinitesimal section of transmission line. Elements $\Delta R(I)$ and $\Delta L(I)$ account for nonlinearities in *HTS* (after [19]).

A complete network, suggested by Collado [19], to analyse the response of *HTS* nonlinear transmission line to a large-signal source is shown in Fig. 3.28.

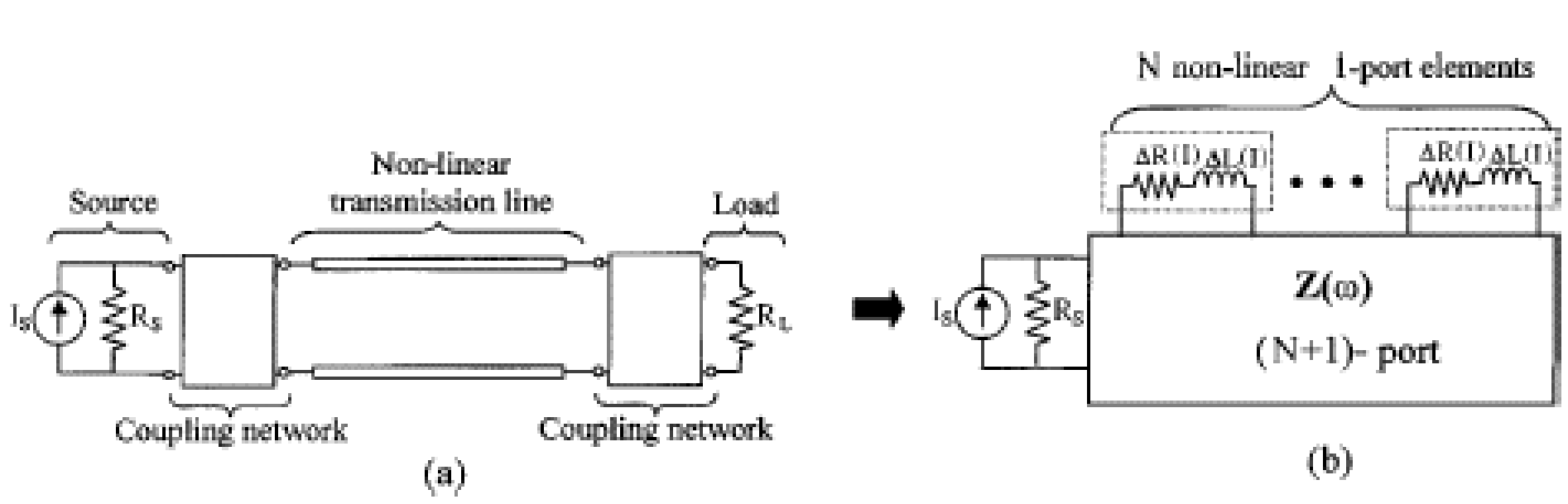

**Fig. 28.** Equivalent circuit of *HTS* nonlinear transmission line with input and output coupling networks (after [19]).

J. Mateu [20] analysed the nonlinear effects in *HTS* using the equivalent circuit of *HTS* line resonators shown in Fig. 29. The equivalent network was simulated to investigate the inter modulation distortion (*IMD*) effects in microwave resonators.

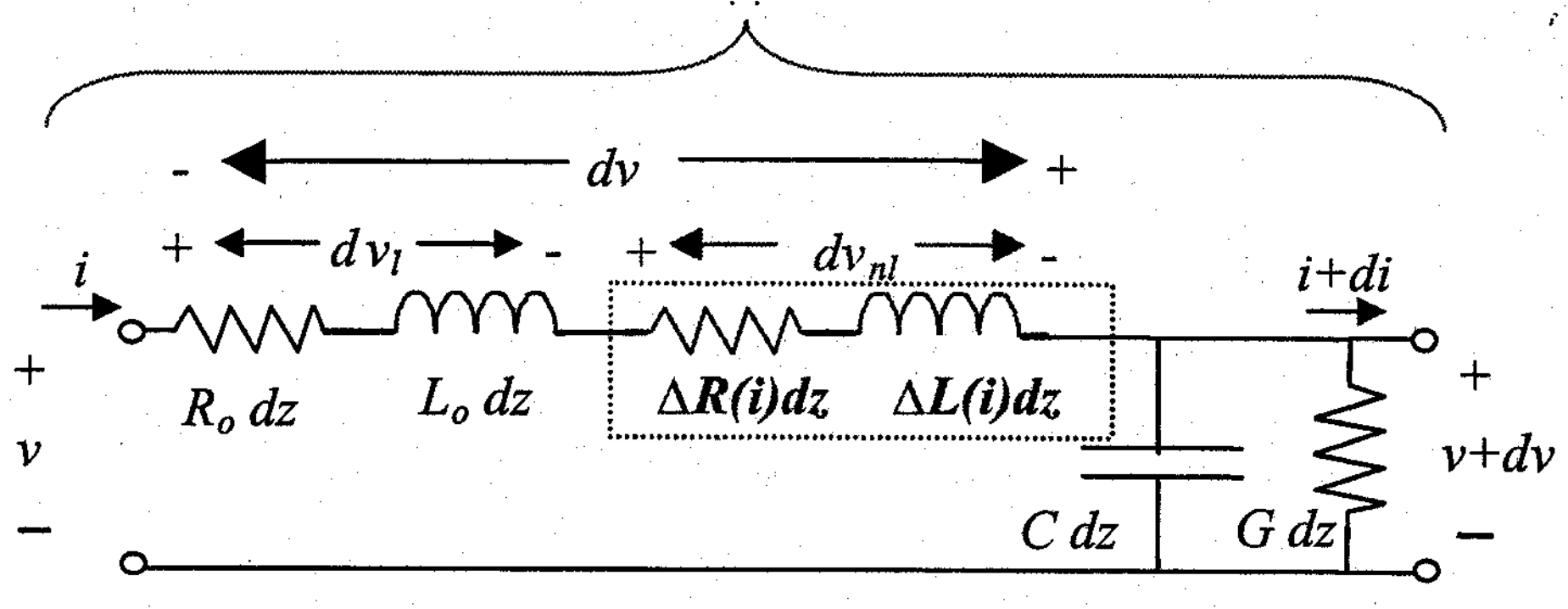

**Fig. 29.** Equivalent circuit to research Inter Modulation Distortion (IMD) in microwave resonators, where *ΔR(i)* and *ΔL(i)* are nonlinear functions of current ***i*** (after [20]).

J. C. Booth also utilised the series quasi-linear transmission line model in research on the nonlinear effects in *HTS* microwave devices in [21]. The equivalent network is depicted in Fig. 30. The quadratic dependence of nonlinear *R(i), L(i)* elements was adopted to calculate the magnitude and phase of nonlinear signal.

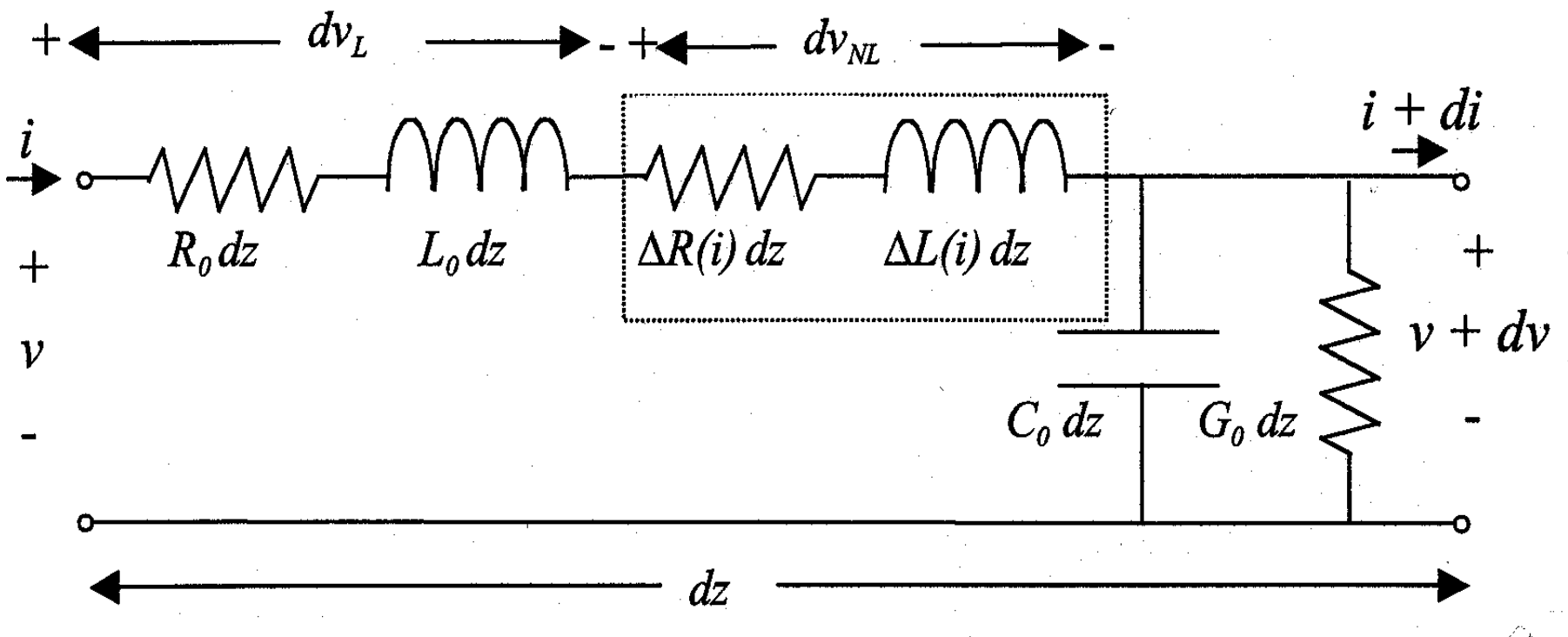

**Fig. 30.** Series quasi-linear transmission line model of *HTS* microstrip resonator (after [21]).

Mateu [30] described and compared two alternative methods of analyzing the dielectric-loaded resonator cavities for measurement of nonlinear intermodulation distortion in *HTS* films. Mateu [30] proposed that the electric field, connected with the superconductor impedance $Z_S$, will increase in microwave resonator, when the transport current $j_s$ is in superconductor surface. The nonlinear part of electric field is defined by superconducting current, which passes through the nonlinear resistive $R_{NL}$ and reactive $L_{NL}$ elements

$$E_{NL}(j_S) = R_{NL}(j_S)\cdot j_S + \frac{\partial}{\partial t}\left(L_{NL}(j_S)\cdot j_S\right),$$

where $R_{NL}=\Delta R_S\cdot|j_S|^n$, $L_{NL}=\Delta L_S\cdot|j_S|^n$

$n$, $\Delta R_S$, $\Delta L_S$ are the parameters that characterize the nonlinearity in the *HTS*. In this research, the consideration and selection of equivalent circuit model for the resonance cavity was made. The equivalent circuit is shown in Fig. 31.

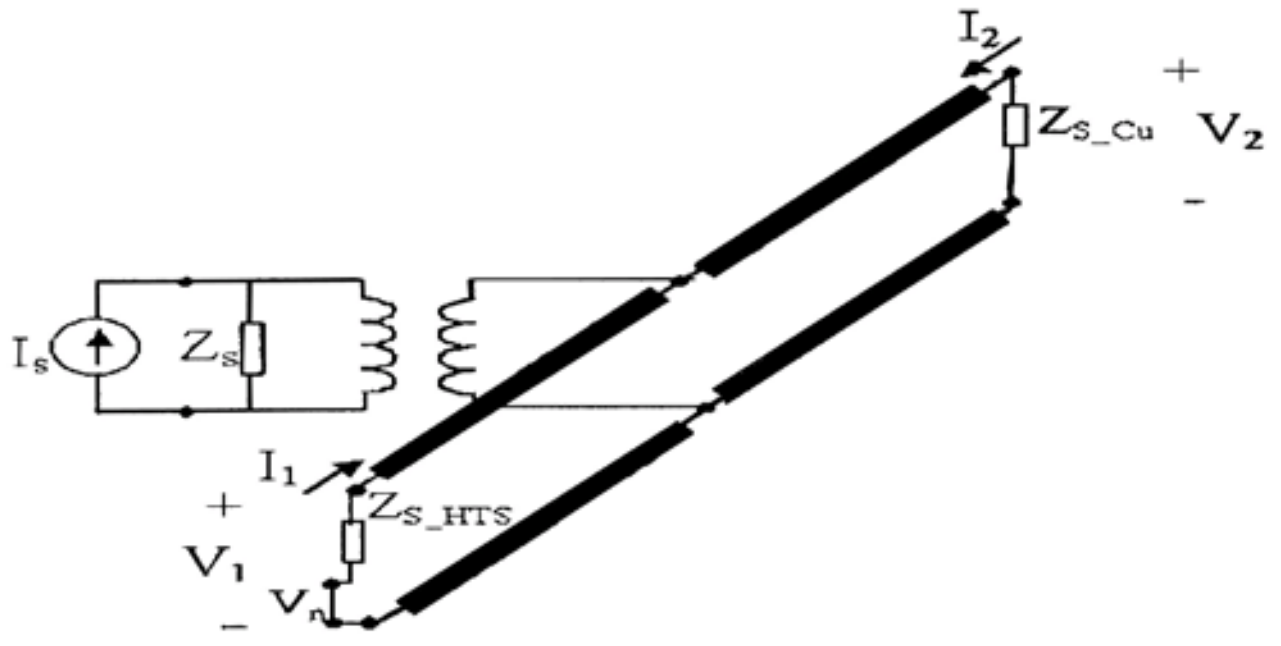

**Fig. 31.** Equivalent circuit of microwave resonator cavity with superconductor (after [30]).

The transmission line model for axial propagation of electromagnetic wave in microwave resonator cavity is used in [30]. In Fig 31, the transmission line terminates the surface impedance of endplates. The nonlinearities of *HTS* are modeled as a linear one-port series network with its surface resistance $R_S$. The results of research are compared with the data, obtained in the research on the $YBa_2Cu_3O_{7-\delta}$ thin film of 700 *nm* on *MgO* substrate, using the parameters: $n=1$, $\Delta L_{NS} = 3.5\cdot 10^{-16}$ *Hm/A* and $\Delta R_{NL} = 0$ in [30]. Discrepancies between the analytical and numerical values of power of intermodulation products were on the order of 1% magnitude at low power $P$ of electromagnetic wave. These techniques were applied to a rutile resonator for nonlinear characterisation of unpatterned *10mm×10mm HTS* thin films.

Ohshima, Kaneko, Lee, Osaka, Ono, Saito [36] used a conventional π-type equivalent circuit to represent a three-pole forward-coupled *HTS* filter [36].

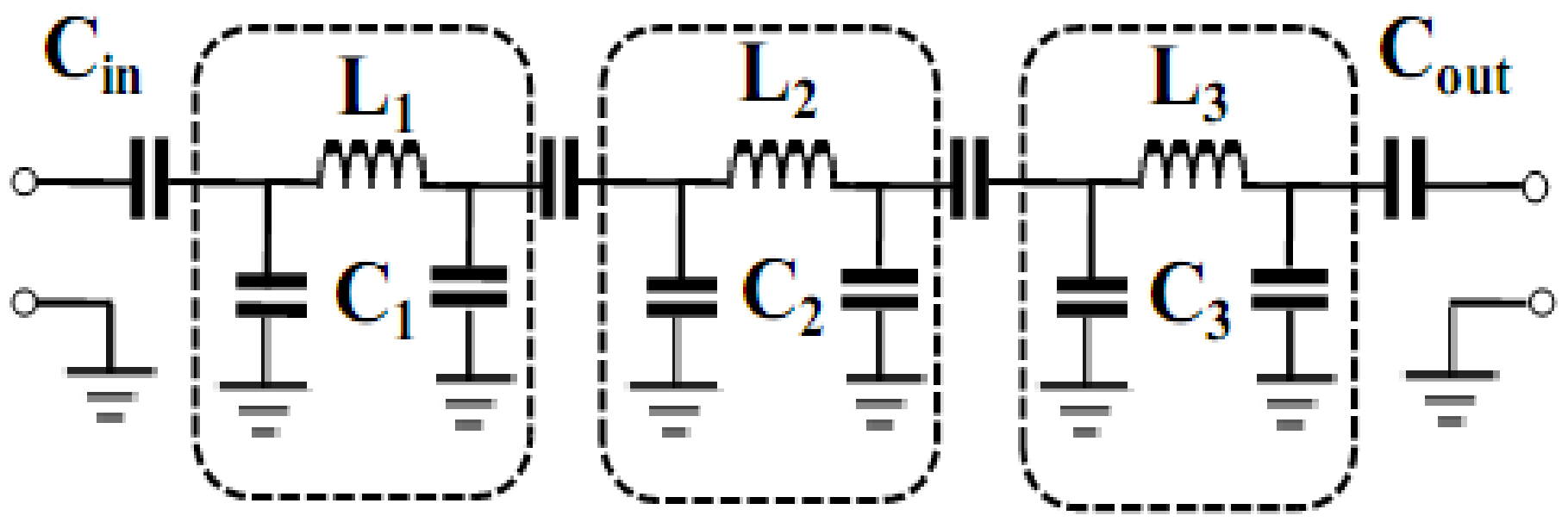

**Fig. 32.** Equivalent circuit of microstrip filter with three *HTS* resonators (after [36]).

## 4.6. r- Parameter for Nonlinear Lumped Element Models Analysis.

It is possible to introduce an analogy between the microwave characteristics of superconductor surface and the conditional series resonator, which is described by the magnitudes of characteristic parameters surface resistance $R_S$ and reactance $X_S$. In this case, the quality factor $Q$ of series resonator can be presented as

$$Q = X_S / R_S.$$

Quality factor for $YBa_2Cu_3O_{7-\delta}$ single crystal will be equal to $Q \approx 10^2$ in the *GHz* frequency range at low temperature $T << T_C$ approximately (see Fig. 1 in Ch. 3). Taking to the account that fact that, from the energy point of view, the quality factor is

$$Q = W_d / W_s,$$

where $W_d$ is the dissipated wave energy and $W_s$ is the stored wave energy in surface layer of superconductor. Therefore, it becomes clear that the main part of the energy is concentrated in the surface inductance, and a small part of energy is dissipated on the surface resistance $R_S$ in superconductor.

The dimensionless parameter $\boldsymbol{r_H}$ is defined as the ratio in [22, 23]

$$r_H = \Delta X_S(P)/\Delta R_S(P).$$

In [23], the inverted parameter $\boldsymbol{r_G}$ was introduced as the ratio

$$r_G = \Delta R_S(P)/\Delta X_S(P),$$

with the purpose to differentiate the mechanisms responsible for the nonlinear effect in superconductor at the different power levels *P*.

In further discussion, the parameter $\boldsymbol{r_H}$ is equal to the differential quality factor $Q_{Sdif}$ on superconductor surface

$$r_H = Q_{Sdif}.$$

It is necessary to note that, in normal metal, the magnitude of temperature dependent differential quality factor $Q_{Sdif}$ is 1

$$Q_{Sdiff}(T) = \Delta X_S(T)/\Delta R_S(T) = 1.$$

In single crystal $YBa_2Cu_3O_{7-\delta}$, the dependence of differential quality factor as a function of temperature $Q_{Sdiff}(T)$ is approximately equal to $Q_{Sdiff}(T) \sim 30$ at low temperature $T < 40$ *K*, in agreement with data provided in Fig. 1 in Ch. 3.

It has to be noted that magnitude of differential quality factor, which depends on temperature, $Q_{Sdif}(T)$ will be negative in temperatures range $40\ K < T < 70\ K$, because $\Delta R_S(T) < 0$ in this temperatures range. There is nothing unexpected in the characteristic behaviour of parameter $r = Q_{Sdiff}$, because the magnitude of parameter $\boldsymbol{r}$ tells us information on the direction of change of superconductor impedance at the action of the temperature *T* or microwave power *P* of electromagnetic wave in the *R-X* space. This is demonstrated by author of dissertation in Fig. 33.

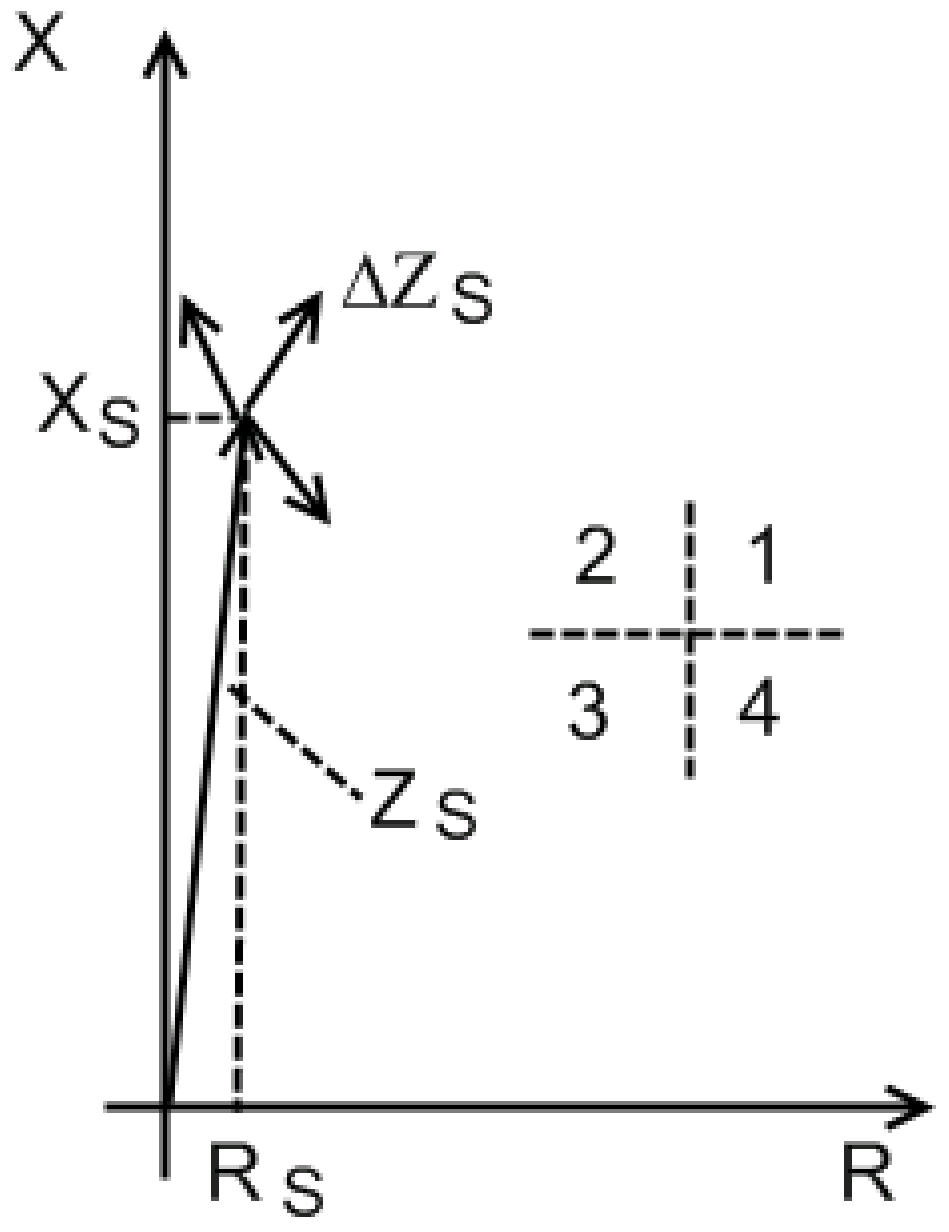

**Fig. 33.** Change of vector of impedance $Z_S$ on magnitude $\Delta Z_S$ at action of external parameters (temperature *T*, microwave power *P*) in superconductor [31].

As it follows from Fig. 33, the magnitude of differential quality factor $Q_{Sdiff}=\Delta X_S/\Delta R_S$ will be negative in the case, when the increments for values of surface resistance $\Delta R_S$ or reactance $\Delta X_S$ are to be negative for one value, but positive for another. In other words, the magnitude of differential quality factor $Q_{Sdiff}$ is positive, when the vector $\Delta Z_S$ is located in first and third quadrants in *R*-*X* space, but it is negative, when the vector $\Delta Z_S$ is in second and forth quadrants in *R*-*X* space. Absolute magnitude of differential quality factor $Q_{Sdiff}$ does not depend on length of the vector $\Delta Z_S$, but is defined by its direction, because the absolute magnitude of differential quality factor $Q_{Sdiff}$ is equal to the tangent of rotation angle $\varphi$ of vector $\Delta Z_S$ with respect to the axe *R*. Magnitude of differential quality factor $Q_{Sdiff}$ is small in the case, when vector $\Delta Z_S$ is almost parallel to the axe *R*, if value $\Delta R_S$ is big, but value $\Delta X_S$ is small as it shown by the author of dissertation in Fig. 34. Magnitude of differential quality factor $Q_{Sdiff} = 1$, when vector $\Delta Z_S$ on rotated on 45 degrees. Magnitude of differential quality factor $Q_{Sdiff} >> 1$, when main influence by external factor (temperature *T*, microwave power *P*) is focused on $\Delta X_S$, it means that vector $\Delta Z_S$ is directed almost in parallel to the axis $X$. This situation is shown by author of dissertation in Fig. 33 [31].

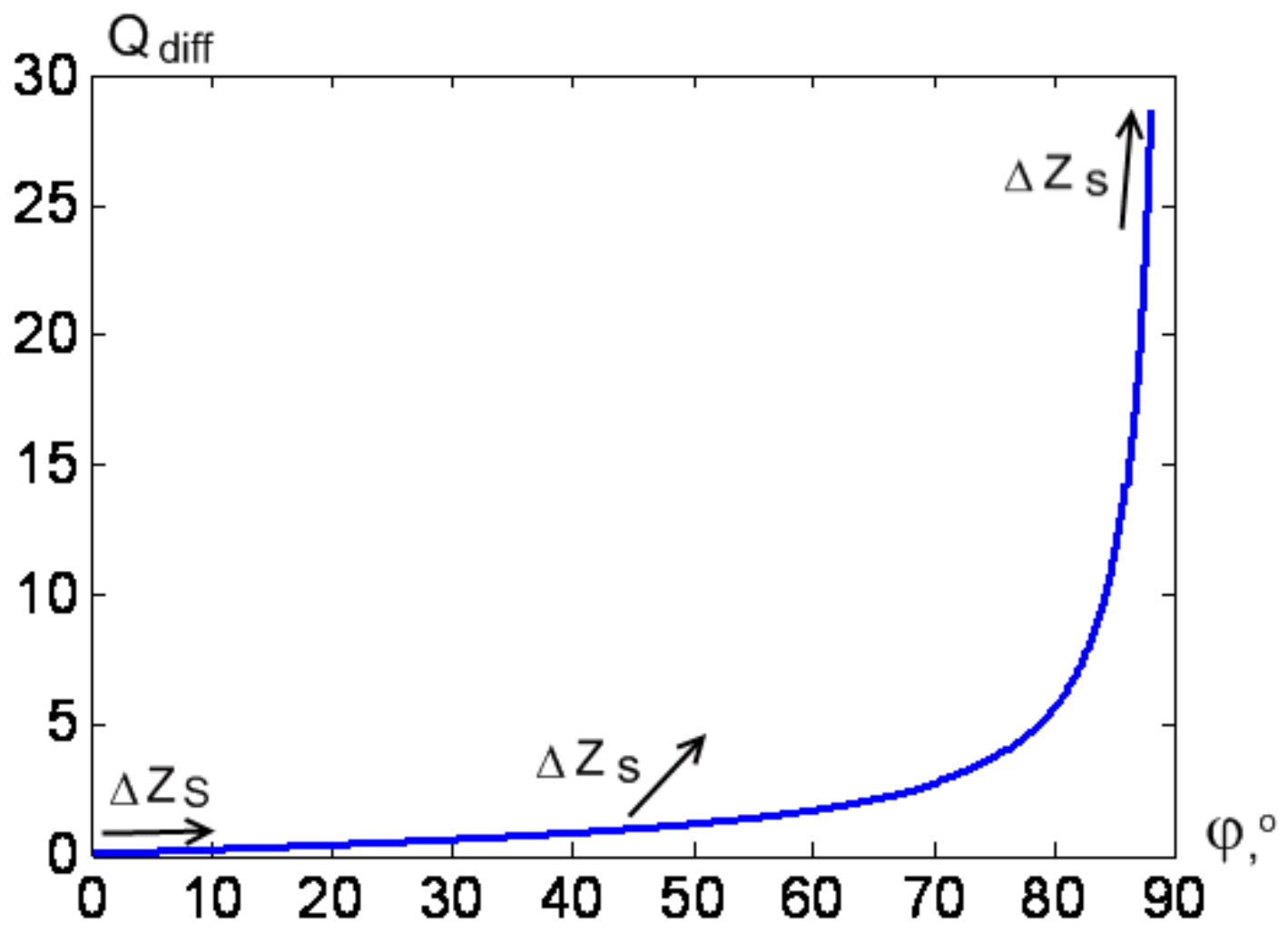

**Fig. 34.** Dependence of magnitude of differential quality factor $Q_{Sdiff}$ on orientation of vector $\Delta Z_S$ in $R$-$X$ space. ($\varphi$ is the angle between vector $\Delta Z_S$ and axe $R$) [31].

Among various mechanisms, proposed to describe the microwave power dependence in high-$T_c$ superconductors, each one has the characteristic parameter **r**=$Q_{Sdiff}$. Not only the magnitudes of the parameter ***r*** are different, but also the frequency and temperature dependences have characteristic signatures. In this context, the magnitudes of the parameter ***r*** can basically be used to differentiate among these mechanisms.

The intrinsic nonlinearity mechanism [24], based on the *BSC* theory, states that the nonlinear effect arises at the large transport currents, because of the breaking of Cooper pairs into the quasiparticles, thus affecting the complex conductivity, with the values of parameter ***r*** being around $10^2$.

In [25], Golosovsky considered: What are the magnitudes of parameter ***r***, which correspond to the different mechanisms defining the nonlinear phenomena in superconductors?

In the uniform microwave heating mechanism, the superconducting films have the finite surface resistance $R_S$, and therefore absorb the microwave power, resulting in the increase of the film's temperature. In this case, the values of parameter ***r*** correspond to ***r*** *>> 1*, particularly *50-100*.

Another mechanism to consider is a model with the vortices in weak links, based on the fact that strong microwave magnetic field leads to the creation of Josephson vortices, because of the weak links (grain boundaries). The nonlinearity arises from the dependence of vortex concentration on microwave magnetic field, where the dissipation of microwave energy is due to the viscous drag of vortices. This model, which was developed by J. Halbritter, predicts the almost temperature and frequency independent parameter $r \geq 1$ in [22].

As for the *bulk vortex penetration model* that assumes that the large microwave magnetic field causes the magnetic vortices to penetrate the bulk, not just the grain boundaries lead to the periodical change of magnetic flux through the sample. Vortices leave the grains at decrease of the field, but their motion lags after the field due to the viscous forces, and the hysteresis occurs. In [25], the value of the parameter ***r*** for purely hysteretic losses is in the range of 1, and depends on the geometric form of hysteresis curve only. If the effect of viscosity is dominant, then the situation is similar to the normal-state skin effect for which parameter $r = 1$.

As far as the local heating of weak links is concerned, where, at high microwave current, a certain weak link may be switched to the normal state, it concludes that, if this weak link blocks the path of current, there will be large local heating, which may also switch the surrounding material to the normal state. The creation of normal state domain, the size of which depends on the microwave current, can lead to the nonlinear effects. The order of the **r** parameter was found to be around *1* for a thick film and about $6 \cdot 10^{-3}$ for thin films.

An example of r-parameters dependence was analysed by Wosik in [26]. Microwave power-handling capabilities of $YBa_2Cu_3O_{7-\delta}$ thin superconducting films, up to 150 *W* of the input power, using the 14 *GHz* shielded dielectric cavity was researched. In [26], the simple nonlinear *R(P)-L(P)-C* equivalent circuit model to simulate a nonlinear cavity response was used. The *RF* magnetic field reached ~ 500*Oe*. Measurements were conducted on the $YBa_2Cu_3O_{7-\delta}$ superconducting films on $LaALO_3$, $NdGaO_3$ and $Al_2O_3$ substrates in the $TE_{011}$ mode. In linear case,

Lorenztian resonance curves were observed. At high power *P*, the form of curves changed and the nonlinear effects in $f_0(P)$ and within the bandwidth were detected. In [27], the nonlinear electromagnetic response was mathematically modelled using a simple equivalent circuit and the iterate procedure for calculation process. The simulation results are shown in Fig. 35.

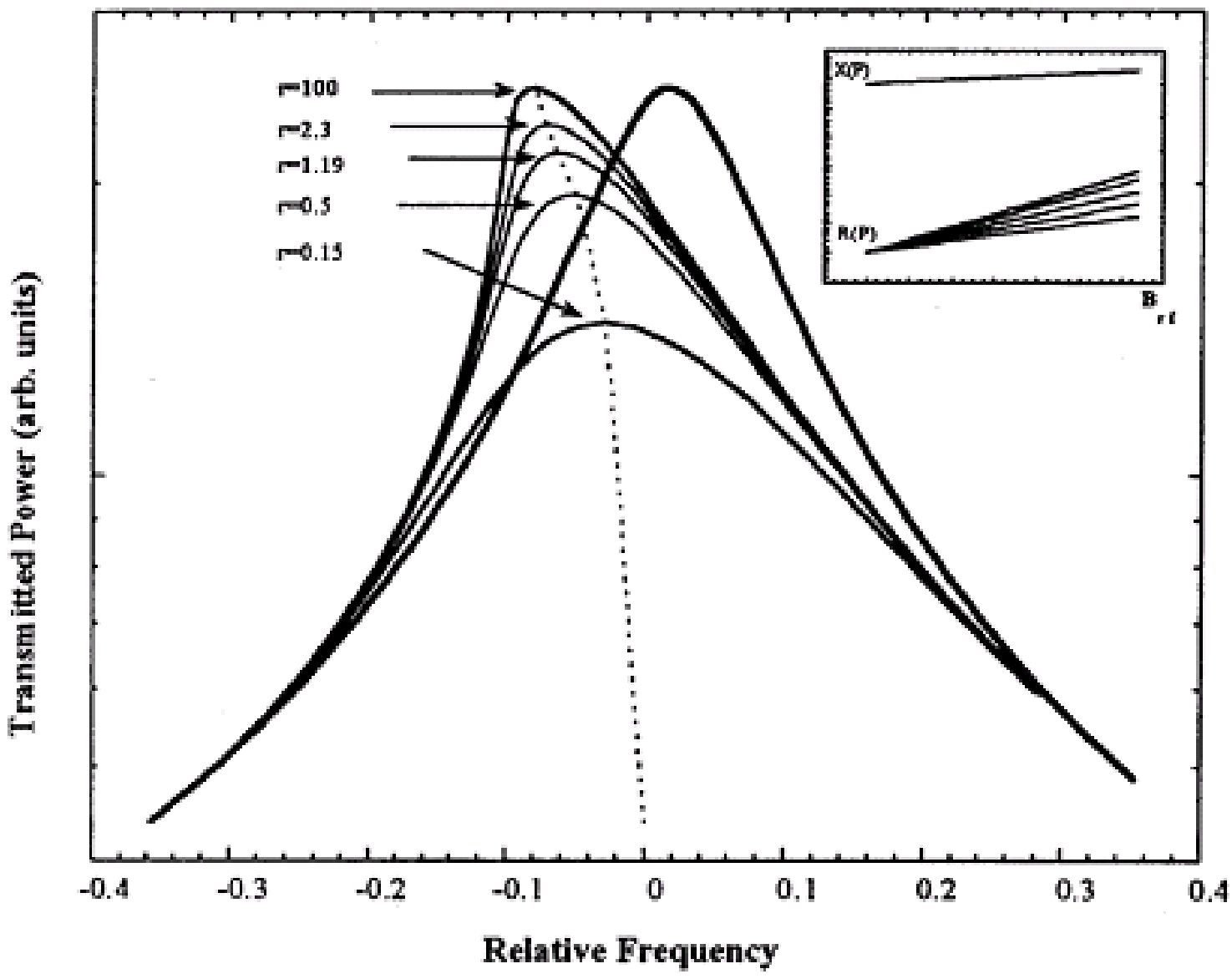

**Fig. 35.** Simulation of transmitted power *P* vs. frequency *f* for different r-parameters in microwave resonator (after [27]).

As shown in Fig. 34, **the different values of parameter *r* are believed to correspond to the different mechanisms of nonlinearity. Therefore, there is a possibility to detect, research and describe the nonlinear properties of superconducting films in terms of model parameters involved.** In Fig. 35, the initial resonance curve has *Lorenztian shape*. In this case, the parameter ***r*** as a function of power *P* is equal to 0

$$r = Q_{Sdiff} \equiv 0,$$

because, in linear case, increase of signal power doesn't lead to the change of resonance frequency *f* and quality factor *Q* of resonator, that is it doesn't result in change of shape of resonance curve. At the same time, in considered case, parameter ***r*** can be determined as a function of temperature *T*, and because characteristics of superconductor sample depend on the temperature, then even in linear case, it will lead to the appearance of increment of parameter ***r***

$$r(T) = \Delta X_S(T) / \Delta R_S(T).$$

This influence of temperature may change the parameter ***r*** in nonlinear case, when the applied microwave power *P* of electromagnetic wave results in the temperature change of superconductor sample and leads to the appearance of the parameter ***r*** with magnitude, which is different from 0

$$r \neq 0.$$

The resonance curves have more complex dependencies in nonlinear case, because their resonance frequencies decrease, and the shapes of resonance curves become asymmetric. As it is shown in Fig 34, in case of ***r*** *= 100*, the total amplitude of resonance curve and its width ***bw*** change a little, hence the increment of parameter ***r***, *ΔR*, is small, while the shift of resonance frequency is big enough, that results in the considerable change of reactance *ΔX* and magnitude of the parameter ***r***

$$r = \Delta X / \Delta R.$$

In the case ***r***=*0.15*, the resonance frequency changes less, and the magnitude of reactance *ΔX* decreases in comparison with the magnitude of reactance *ΔX* in the first case, when r = 100. From other side, the amplitude of resonance curve becomes significantly smaller, that is why width of resonance curve becomes bigger at 3 *dB* level, this results in increase of surface resistance $R_S$, and parameter ***r*** becomes considerably smaller. **It is necessary to note that the parameter *r* is not universal parameter for description of nonlinear phenomena, because parameter *r* is only differential quality factor, which formally describes the direction of movement of vector of impedance $Z_S$ in *R-X* space, and is not directly connected with the physical mechanisms, which are the root causes of nonlinearities in superconductors.**

In [28, 29], the researchers observed that the surface resistance $R_S$ and reactance $X_S$ are not correlated, and expressed some doubts on the application of the *r-parameter* for the accurate characterization of superconductors.

Some other research findings on the non - linear surface impedance of the *YBCO* superconductors by the research group at *James Cook University* in *Townsville* in *Australia* have been recently described in Jacob [37].

## Summary

As it can be seen from the review presented in this chapter, the systems of microwave resonators with superconducting samples can be adequately represented by the resonant ***lumped element circuits***. It is shown that the ***two-fluid model*** approach is adopted for the equivalent representation of superconducting materials at microwaves usually. The review on the equivalent circuit models, proposed by a number of researchers, including research proposals by Wosik, Booth, Mateu, O'Callaghan to represent the superconducting resonators, is completed. ***The main conclusion of chapter is that the series equivalent networks with the small internal resistance are more suitable for the analysis and accurate characterization of physical properties of superconductors at microwaves. The parallel equivalent circuit representations are more appropriate for the precise characterisation of dielectrics or semiconductors at microwaves.*** Modeling of the nonlinear properties of the microwave resonators and superconducting samples at microwaves can be accurately completed, if the equivalent circuit parameters *R, L* and *C* are represented as nonlinearly dependent on either the superconducting current $I_S$ or the magnetic field $H_{RF}$. The dependence of these *R, L* and *C* parameters on the current $I_S$ or magnetic field $H_{RF}$ will be investigated in details in Chapter 5.

Considering the problems, related to the accurate characterization of nonlinear properties of *HTS* thin films at microwaves, it is explained that the practical use of *r-parameter* for the numerical characterization of nonlinearities in superconductors is limited, because the physical behavior of surface resistance *Rs* is not in good agreement with the approximation results, obtained by the *r-parameter* computing, due to several discussed reasons, namely, because the *r*-parameter is the *differential quality factor* only, which formally describes the direction of movement of vector of impedance $Z_S$ in the *R-X* space, and it is not directly connected with the physical mechanisms, which are the root causes of nonlinearities origination in superconductors.

# CHAPTER 5

# DEVELOPMENT OF A LUMPED ELEMENT MODEL FOR ACCURATE MICROWAVE CHARACTERISATION OF SUPERCONDUCTORS

## 5.1. Introduction.

The main subject considered in Chapter 5 is the development of a lumped element model, which describes the nonlinear properties of *HTS* materials in a microwave resonator accurately. This model has to be relatively simple for use by electronics engineers for development of superconducting devices with advanced characteristics. There are two separate issues to consider: one is relatively simple – the modeling of a microwave resonator; second one is more complex – the modeling of high temperature superconductor with nonlinear properties, which depend on microwave power. The basic model of *HTS* is the two fluid model, introduced by Gorter and Casimir, is described in Chapter 2, but it does not consider the nonlinear properties of *HTS*. The proposed approach to characterise the nonlinear properties of *HTS* has to be incorporated in a new model with the appropriate lumped element circuits with distinct functional dependences (linear, quadratic and exponential) of surface resistance of *HTS* on microwave power: $R_s(P)$. The new research approach to develop the advanced lumped element model for accurate microwave characterization of superconductors in resonance circuits at ultra high frequencies is described in Chapter 5.

## 5.2. General Assumptions and Discussions about Superconducting Microwave Resonators.

The designed equivalent lumped element model possesses the two sets of parameters that have to be arranged separately for detailed investigation. The first set includes the parameters to represent a dielectric resonator with its own component characteristics in operation at ultra high frequencies. The second set defines the parameters to characterise the superconductor with its unique microwave properties. Both sets of parameters are integrated into a unified model to describe behaviour of the system at ultra high frequencies accurately.

Practical realisation of system for microwave characterisation of *HTS*, and development of techniques to extract the necessary system parameters have to be undertaken with a great care to satisfy a number of criteria. This is due to the fact that the measurement technique used for accurate microwave characterization of superconductors has to enable the low uncertainty, high accuracy, and high reproducibility. A measurement set up should not modify the physical properties of the superconductor under test, and all the effects of measurement set up have to be adequately reflected in applied computation process.

There is a number of measurement techniques used for accurate microwave characterisation of superconductors, which can be divided into two categories: the *resonant methods* and the *transmission technique*. In the resonant approach various types of resonators can be used such as the copper cavity [1], micro- stripline and slotlines [2, 3], parallel-plate [4], confocal [5], dielectric rod resonators [6] and other [7]. Each resonator type has its advantages and disadvantages that result in different accuracy, sensitivity and complexity of measurement set up. In the transmission technique, introduced by Glover and Tinkham [8] for quasi-optical frequencies and mostly used as a waveguide technique for *HTS* films [9-11], the complex conductivity $\sigma$ of thin film is determined. The accuracy of this technique is considered lower than that of the resonant method [12].

Presently, the dielectric resonator technique is the most popular measurement method for accurate microwave characterisation of *HTS* films due to its high sensitivity, accuracy, easy mode identification, and wide range of surface resistances $R_S$ that can be precisely measured [6]. The dielectric resonator technique was proposed as a *possible standard for precise characterisation of HTS thin films for microwave applications* by J. E. Mazierska [6]. There are two classical designs of dielectric resonators used for microwave characterisation of *HTS*, namely the *Hakki-Coleman* dielectric resonator [12] (*HCDR*) shielded version, and the one-plated open-ended dielectric resonator [13, 14]. *Hakki-Coleman* type dielectric resonator, which is shown in Fig. 1, is used for accurate microwave characterisation of *HTS* thin films in this thesis

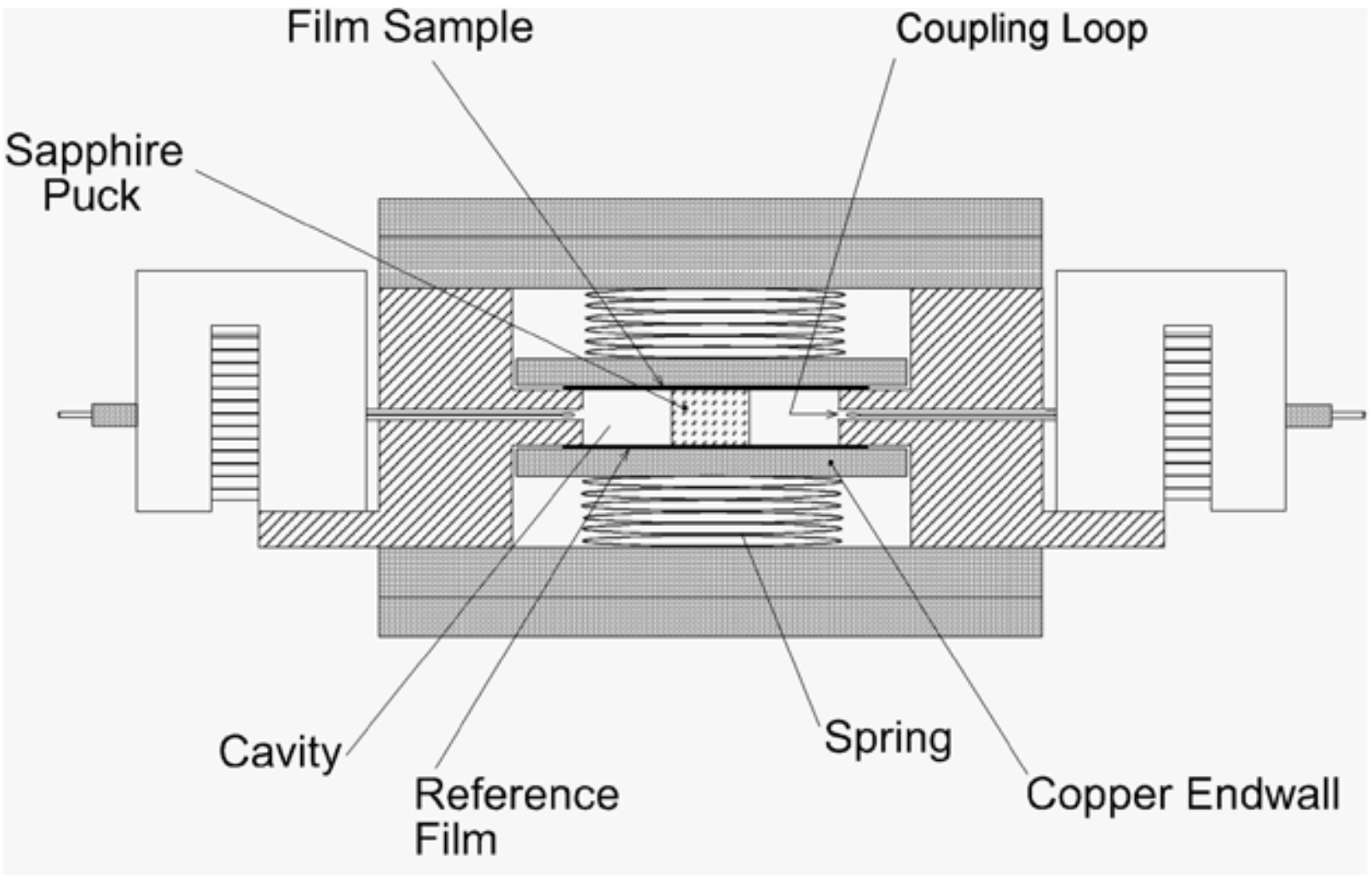

**Fig. 1.** *Hakki-Coleman dielectric resonator* with *HTS* thin films (after [6]), where reference and sample films are superconductors.

Dielectric resonator systems consist of a very low loss and temperature-stable dielectric puck located centrally to a shielded cavity with two superconducting films under test of similar quality placed at the top and bottom of cavity plates. Extremely high *Q*-factor values are achievable by dielectric resonators, because the permittivity of the dielectric is typically much higher than the permittivity of free space, resulting in a situation, when electromagnetic microwave energy is stored within the dielectric. It leads to the case, when the big part of total microwave energy losses occur in *HTS* samples, enabling to conduct very sensitive measurements of surface resistance $R_S$ of *HTS*.

In measurements with application of resonance techniques, the surface resistance $R_S$ of *HTS* thin films is typically determined from the unloaded *quality factor Q* of a resonator, geometrical factor and physical constants, which characterize the materials of a dielectric resonator. Equations for the calculations are derived from the analysis of losses in microwave resonators. Total loss of the resonator can be expressed as in eq. (5.1) [15]:

$$\frac{1}{Q_0} = \frac{1}{Q_C} + \frac{1}{Q_d} + \frac{1}{Q_r}, \tag{5.1}$$

where $Q_c$, $Q_d$, and $Q_r$ are the *quality factors*, associated with the conductor, dielectric and radiation losses respectively. The total conductor loss ($1/Q_c$) is equal to the sum of losses in the superconductor sample and normal conductor parts of resonator cavity in eq. (5.2)

$$\frac{1}{Q_C} = \frac{1}{Q_S} + \frac{1}{Q_m} \tag{5.2}$$

Using definition (4.1), the loss in *HTS* thin films ($1/Q_s$) can be expressed as in eq. (5.3):

$$\frac{1}{Q_S} = \frac{1}{Q_0} - \left[\frac{1}{Q_m} + \frac{1}{Q_d} + \frac{1}{Q_r}\right] \tag{5.3}$$

The loss in a superconductor sample is related to the surface resistance $R_S$ of *HTS* sample and geometrical factor $A_S$, which depends on the geometry of sample in eq. (5.4) [16]:

$$\frac{1}{Q_S} = \frac{R_S}{A_S} \tag{5.4}$$

Integration of equation (4.4) into (4.3) provides the expression to calculate the surface resistance $R_s$ of superconductor from the unloaded $Q$-factor and predetermined components loss in eq. (5.5)

$$R_S = A_S\left\{\frac{1}{Q_0} - \left[\frac{1}{Q_m} + \frac{1}{Q_d} + \frac{1}{Q_r}\right]\right\}, \tag{5.5}$$

where the metal wall loss ($1/Q_m$) is related to the surface resistance $R_m$ of the metal, and the geometrical factor $A_m$ is associated with the geometry of the walls in eq. (5.6) [16]:

$$\frac{1}{Q_m} = \frac{R_m}{A_m} \tag{5.6}$$

Dielectric losses ($1/Q_d$) can be described by eq. (5.7) [16]:

$$\frac{1}{Q_d} = p_e \tan\delta , \tag{5.7}$$

where the constant $p_e$ is the electric energy filling factor of the dielectric part, calculated from geometry, the loss tangent ($tan\delta$) is obtained from the measurements. A low loss dielectric material is usually used such as sapphire with the loss tangent less than order of $10^{-7}$ at $77K$, so that the dielectric loss becomes negligibly small compared to other losses in the microwave resonator.

The radiation losses, if they are present, can be difficult to determine accurately, but they can be minimised or even eliminated by appropriate design and choice of the resonator.

Integrating the expressions for the metal and dielectric losses (5.6) and (5.7) into the formula (5.5), we can obtain the expression for the surface resistance $R_s$ of *HTS* thin films as it is shown in (5.8). The geometrical factors $A_s$ and $A_m$ depend on the design and dimensions of the resonator. The constant $p_e$ for dielectric resonators in trapped state is approximately equal to one, and is weakly dependent on the resonator dimensions only.

The surface resistance $R_s$ can be expressed as in eq. (5.8)

$$R_S = A_S \left\{ \frac{1}{Q_0} - \frac{R_m}{A_m} - p_e \tan\delta \right\}, \tag{5.8}$$

where $Q_0$ is the unloaded quality factor, which can be determined from the measured loaded $Q_L$-factor and two port coupling coefficients $\beta_1$ and $\beta_2$ in eq. (5.9):

$$Q_0 = Q_L(1 + \beta_1 + \beta_2) \quad (5.9).$$

The use of a weak coupling usually means the measurements with low signal-to-noise ratio, which can significantly contribute to uncertainties in the experimental results. Therefore, the use of coupling at sufficiently elevated levels to achieve a appropriate signal-to-noise ratio is desirable during measurement procedures.

Some parameters, such as the *geometric factors*, utilised for calculations of surface resistance $R_s$ of *HTS* thin films, based on dielectric resonator techniques, require knowledge of the electromagnetic (*EM*) field distribution in the microwave resonator.

## 5.3. Electromagnetic Waves in a Dielectric Resonator.

A correlation between parameters of a dielectric resonator, functioning at a certain mode, and the component values of equivalent networks can be determined using characteristic expressions for electromagnetic fields at the given mode. Dimension distributions of electric and magnetic fields in a microwave resonator can correspond to the oscillations, defined as the *transverse electric mode* (*TE*), *transverse magnetic mode* (*TM*) and more complex hybrid and other modes. Measurements for microwave characterisation of *HTS* materials, using dielectric resonators, are typically done in the $TE_{011}$ or $TE_{01\delta}$ modes. The $TE_{011}$ mode is the dominant one of the *TE* family and provides more convenient way of identification than higher modes. These *TE* modes are insensitive to electrical contacts on metallic enclosure such as contacts of resonator walls with cavity plates.

The *EM* fields can be described by the *Helmholtz equations* in eq. (5.10)

$$\nabla^2 E = -k_0^2 E \qquad \nabla^2 H = -k_0^2 H, \tag{5.10}$$

where *E* and *H* represent electric and magnetic fields intensities respectively, *k* is the wave number in eq. (5.11)

$$k_0 = \omega(\mu_0 \varepsilon_0)^{1/2} = \frac{\omega}{c}, \tag{5.11}$$

where $\omega$ is the wave circular frequency, *c* is the light velocity in free space, $\mu_0$ and $\varepsilon_0$ are the magnetic permeability and dielectric constant of free space respectively.

A dielectric rod, cantered inside the cavity with the radius $\rho_c$ and height *l*, has the radius $\rho_d < \rho_c$ as shown in Fig. 2

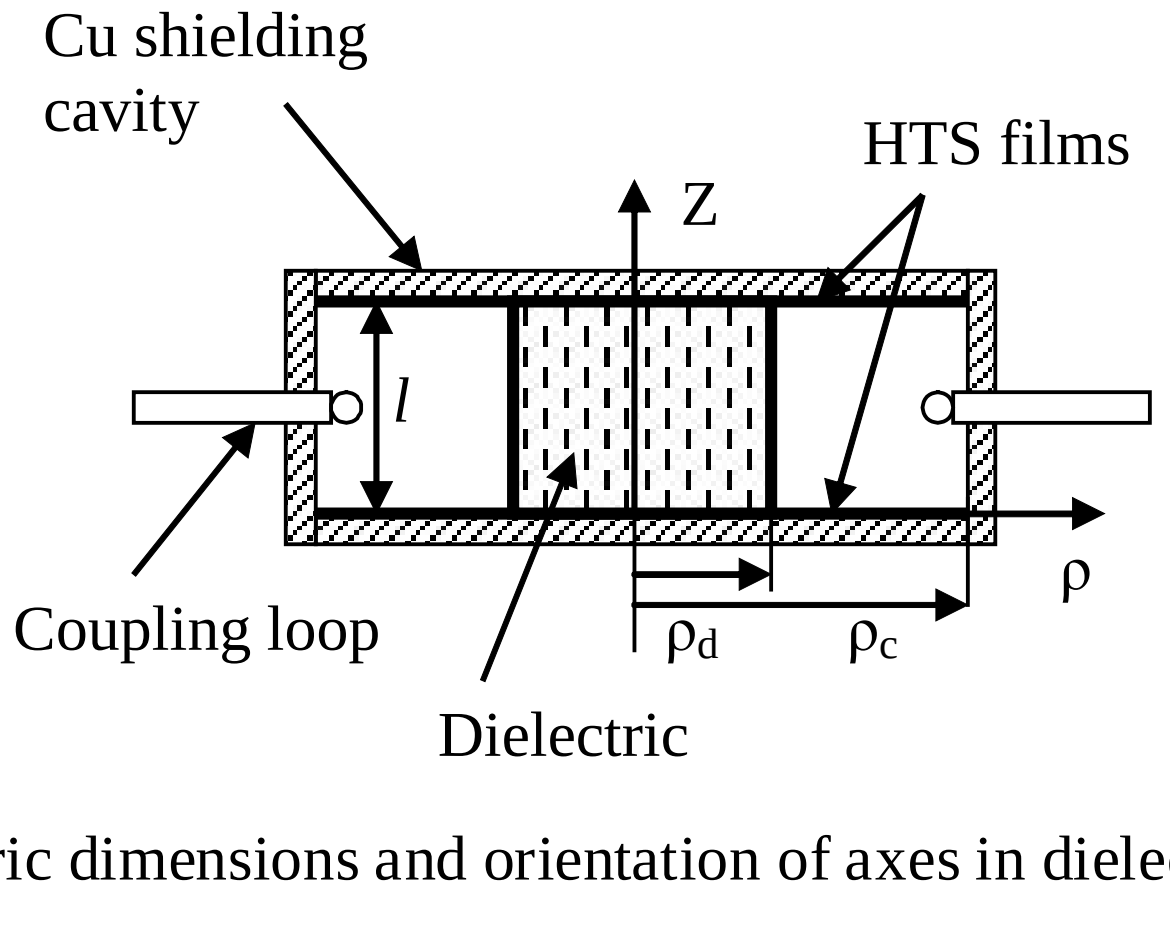

**Fig. 2.** Dielectric dimensions and orientation of axes in dielectric resonators.

Using the cylindrical coordinates (ρ, ϕ, z), the components of electric and magnetic fields can be defined as follows:

For $\rho < \rho_d$ in eq. (5.12)

$$
\begin{aligned}
&E_{\rho 1} = 0 \\
&E_{\phi 1}(\rho, z) = -jA(\frac{\omega\mu_0}{k_1}) J_1(k_1\rho)\sin(\beta z) \\
&E_{z1} = 0 \\
&H_{\rho 1}(\rho, z) = -A(\frac{\beta}{k_1}) J_1(k_1\rho)\cos(\beta z) \\
&H_{\phi 1} = 0 \\
&H_{z1}(\rho, z) = AJ_0(k_1\rho)\sin(\beta z)
\end{aligned}
\tag{5.12}
$$

For $\rho > \rho_d$ in eq. (5.13)

$$
\begin{aligned}
&E_{\rho 2} = 0 \\
&E_{\phi 2}(\rho, z) = jB(\frac{\omega\mu_0}{k_2}) F_1(k_2\rho)\sin(\beta z) \\
&E_{z2} = 0 \\
&H_{\rho 2}(\rho, z) = B(\frac{\beta}{k_2}) F_1(k_2\rho)\cos(\beta z) \\
&H_{\phi 2} = 0 \\
&H_{z2}(\rho, z) = BF_0(k_2\rho)\sin(\beta z)
\end{aligned}
, \tag{5.13}
$$

where in eq. 5.14)

$$
\begin{aligned}
&F_0(k_2\rho) = I_0(k_2\rho) + K_0(k_2\rho)\frac{I_1(k_2\rho_C)}{K_1(k_2\rho_C)} \\
&F_1(k_2\rho) = -I_1(k_2\rho) + K_1(k_2\rho)\frac{I_1(k_2\rho_C)}{K_1(k_2\rho_C)}
\end{aligned}
, \tag{5.14}
$$

where, $J_m$ is the $m$-th order *Bessel function* of the first kind, $I_m$ and $K_m$ are the $m$-th order modified *Bessel and Hankel functions* respectively [17], $\beta = \pi/l$, $l$ is the height of the cavity.

The wave number in the dielectric can be expressed as

$$k_1^2 = k_0^2 - \beta^2$$

and outside of the cavity as

$$k_2^2 = \beta^2 - k_0^2,$$

where $k_0 = \omega / c$, $c$ is the velocity of electromagnetic wave in free space.

Components of the electric and magnetic fields depend on time as $e^{j\omega t}$ and are shifted in phase by $\pi/2$. The fields in the cylindrical resonator are related by the *Maxwell equations* in eq. (5.15). Taking into the account eqs. (5.12) and (5.13)

$$rotE_\phi = -\frac{\partial H_{(\rho,z)}}{\partial t} \tag{5.15}$$

The density of shifting currents can be expressed as in eq. (5.16)

$$J_{\phi,disp} = -\frac{\partial D_\phi}{\partial t} \tag{5.16}$$

where, $D_\phi = \varepsilon_d E_\phi$.

***These formulas represent a complete set of equations for accurate characterisation of electromagnetic properties of HTS thin films in a dielectric resonator at microwaves.***

Microwave resonators are usually employed for the microwave characterisations of superconducting materials, because the extremely high quality factor $Q$ of superconducting resonators results in high accuracy of measurements.

## 5.4. Equivalent Lumped Element Models for Representation of Superconductor in Hakki-Coleman Dielectric Resonator at Microwaves.

As discussed in Chapter 4, two basic lumped element *RLC* circuits can be used for modelling of microwave electromagnetic responses of any resonator with *HTS* films. Therefore, microwave resonators, including the *Hakki-Coleman dielectric resonator*, can be represented as either a parallel or a series *RLC* network in Fig. 3

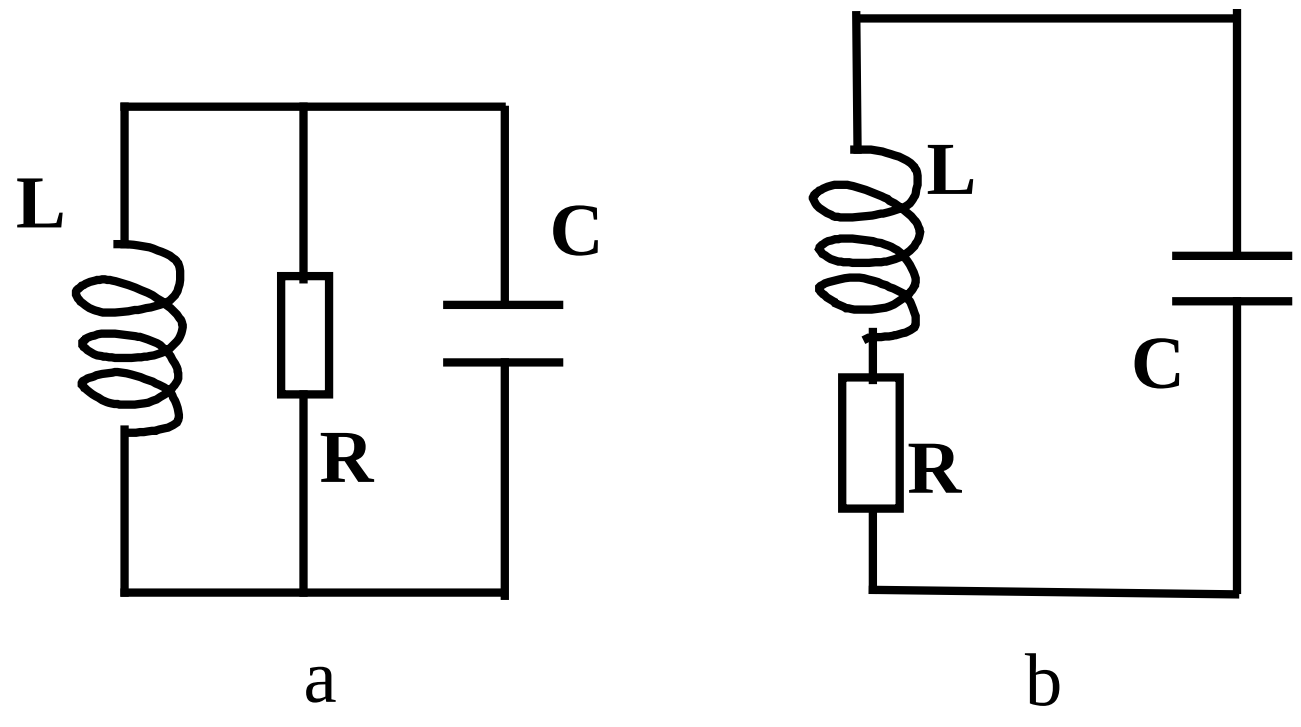

**Fig. 3.** Parallel (a) and series (b) circuits representing a dielectric resonator.

The series *RLC* circuit has been selected for the research work, described in this thesis. As it is known from the circuit theory [18], the quality factor $Q$ is higher, if the resistance $R$ has a big magnitude in the parallel network in eq. (5.17)

$$Q_{par} = \frac{R}{\omega L}. \tag{5.17}$$

It means that the resistive parts of dielectric resonator, the cavity walls and top / bottom plates, need to have low losses in order to achieve corresponding high quality factor $Q$ values of a microwave resonator. Therefore, the representation of resonator as a series circuit is more appropriate for considered case, because the quality factor $Q$ is higher, if the resistance $R$ is lower in series network in eq. (5.18)

$$Q_{ser} = \frac{\omega L}{R} \tag{5.18}$$

Based on the above assumptions, the series circuit, depicted on Fig 3 (b), was chosen as a standard equivalent model of the dielectric resonator with superconductor for accurate microwave characterization of *HTS* thin films in this thesis. In the case of *HCDR*, the equation for quality factor $Q$ is (see eq. (3.4) in Ch. 3)

$$\frac{1}{Q} = \frac{R_S}{A_S} + \frac{1}{Q_{par}} + \frac{\tan\delta}{A_{diel}},$$

where $\tan\delta$ is very small for high quality dielectrics, usually $\tan\delta < 10^{-5} - 10^{-6}$ for temperature $T < 77K$.

The basic model of *HTS* is the *two fluid model*, consisting of a parallel combination of an inductor and a resistor [19-22], as shown in Fig. 4, where the inductive part represents superconducting electrons and the resistive part is responsible for normal electrons in *HTS* materials.

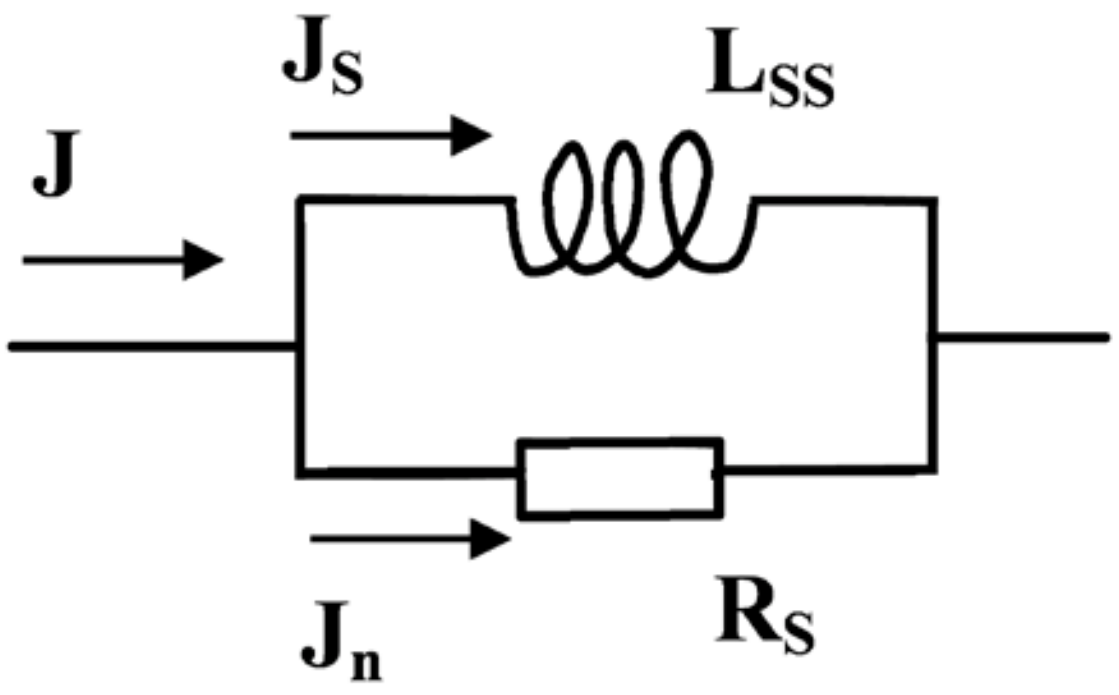

**Fig. 4.** Typical equivalent *LR* representation of a superconductor as described in [19-22].

The above equivalent representation of superconductors is in a good agreement with the theoretical two fluid model especially in the case of *HTS* [20]. However, at temperatures close to absolute zero, the losses in a superconductor are very small and the resistance of normal electrons $R_S$ is close to zero, meaning that the reactance $\omega L_{SS}$ can be short-circuited by surface resistance $R_S$, and in this case superconducting properties will not be adequately represented. In order to avoid this scenario, the superconductor needs to be modelled by an inductance, related to the magnetic flux, which arises in magnetic field that penetrates into the *HTS* thin film even at the temperature of zero *Kelvin*. Therefore, the author of dissertation proposes that an additional element $L_{Sn}$ has to be added in series to the parallel combination of $L_{SS}$ and $R_S$ as shown in Fig. 5.

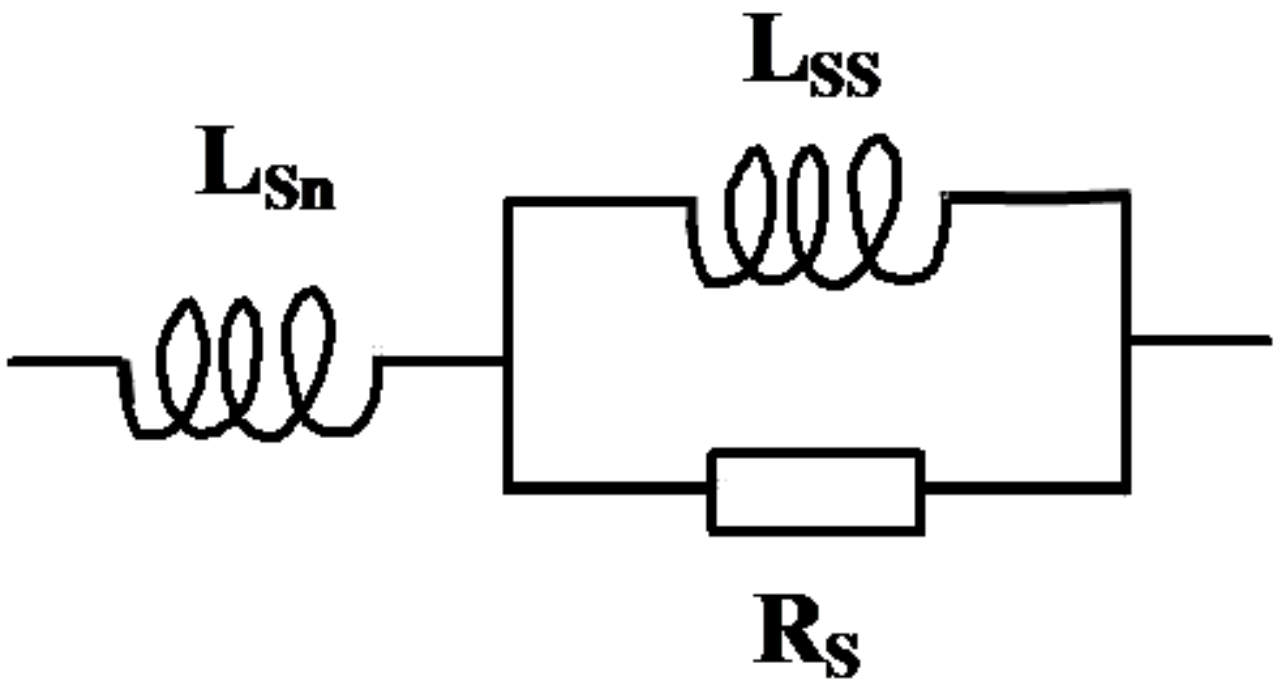

**Fig. 5.** Equivalent lumped element model: series *L* - parallel *LR* of a superconductor.

In addition to the model depicted on Fig. 5, author of dissertation developed a slightly modified network, where the inductance $L_{sn}$ is embedded into a parallel combination of $L_{ss}$ and $R_s$ as shown in Fig. 6.

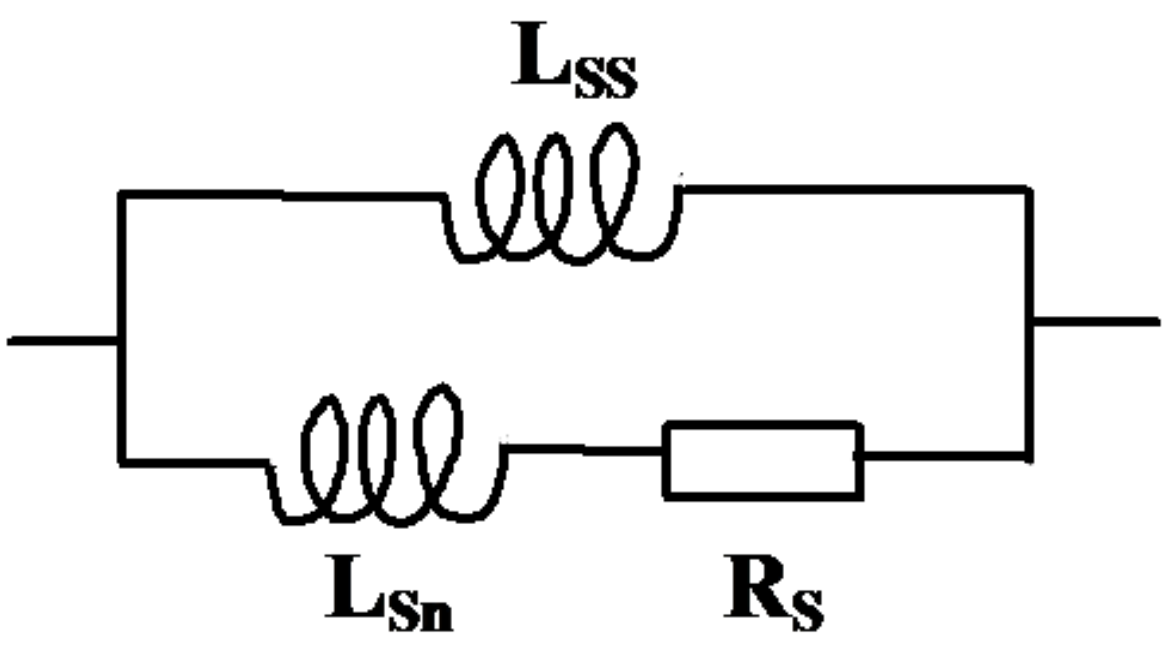

**Fig. 6** Equivalent lumped element model: parallel *L*- parallel *LR* of a superconductor.

In this model, the superconducting current path with no resistance is represented by the inductance $L_{ss}$, and the normal current path is modelled by impedance with the inductance $L_{sn}$ and resistance $R_s$.

All three lumped element models, shown in Figs. 4, 5 and 6, have been researched in the microwave resonator-superconductor system with the results presented in this thesis [32, 33].

## 5.5. Proposed Equivalent Lumped Element Models of Superconductor in Microwave Resonator.

Some possible equivalent lumped element models of superconductor resonators, published in literature by other researchers, were reviewed in Chapter 3.

In Chapter 4, the three lumped element models to represent the dielectric resonator with superconductors were developed by author of dissertation, going from the above discussed assumptions. Three configurations are presented below.

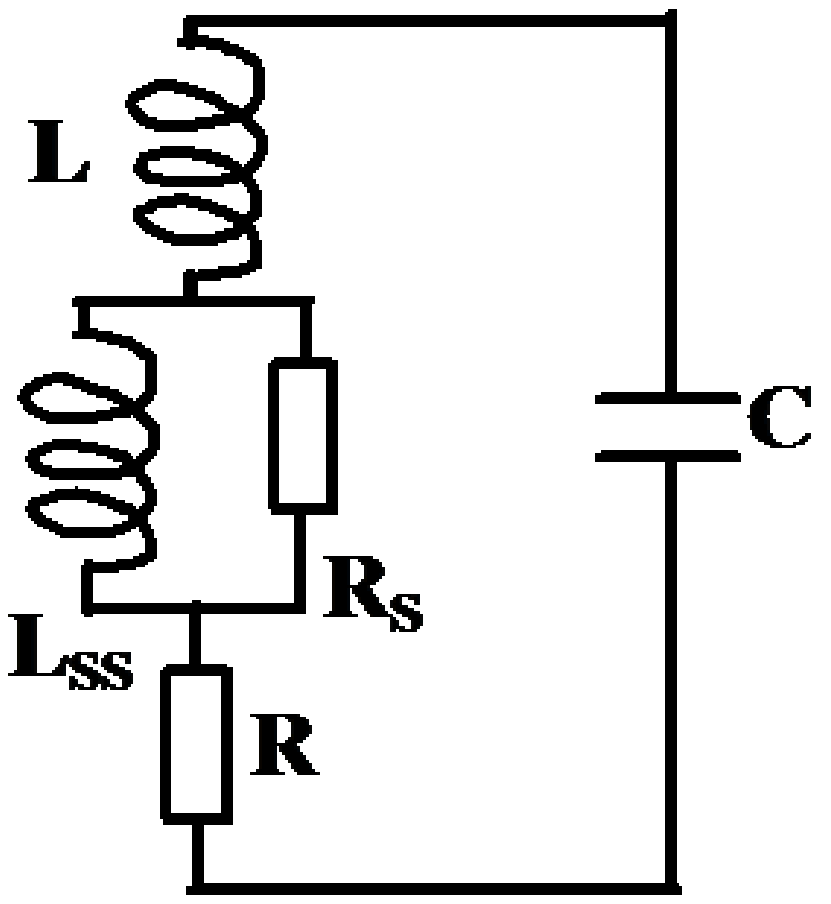

**Fig. 7.** Equivalent lumped element model of resonator circuit, which represents a superconductor in a dielectric resonator.

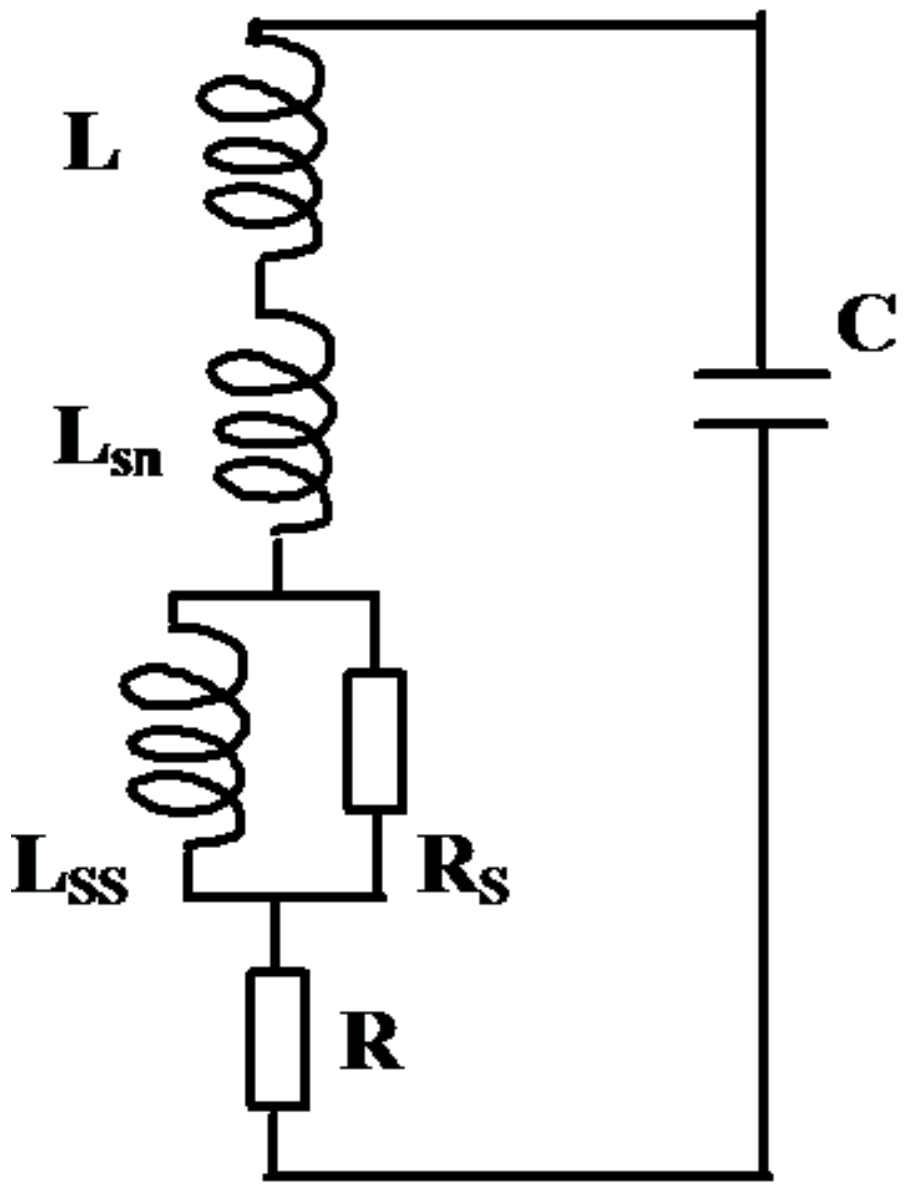

**Fig. 8.** Equivalent lumped element model of resonator circuit, which represents a superconductor in series with its inductance *L* in a dielectric resonator (*L*-$\sigma$ model).

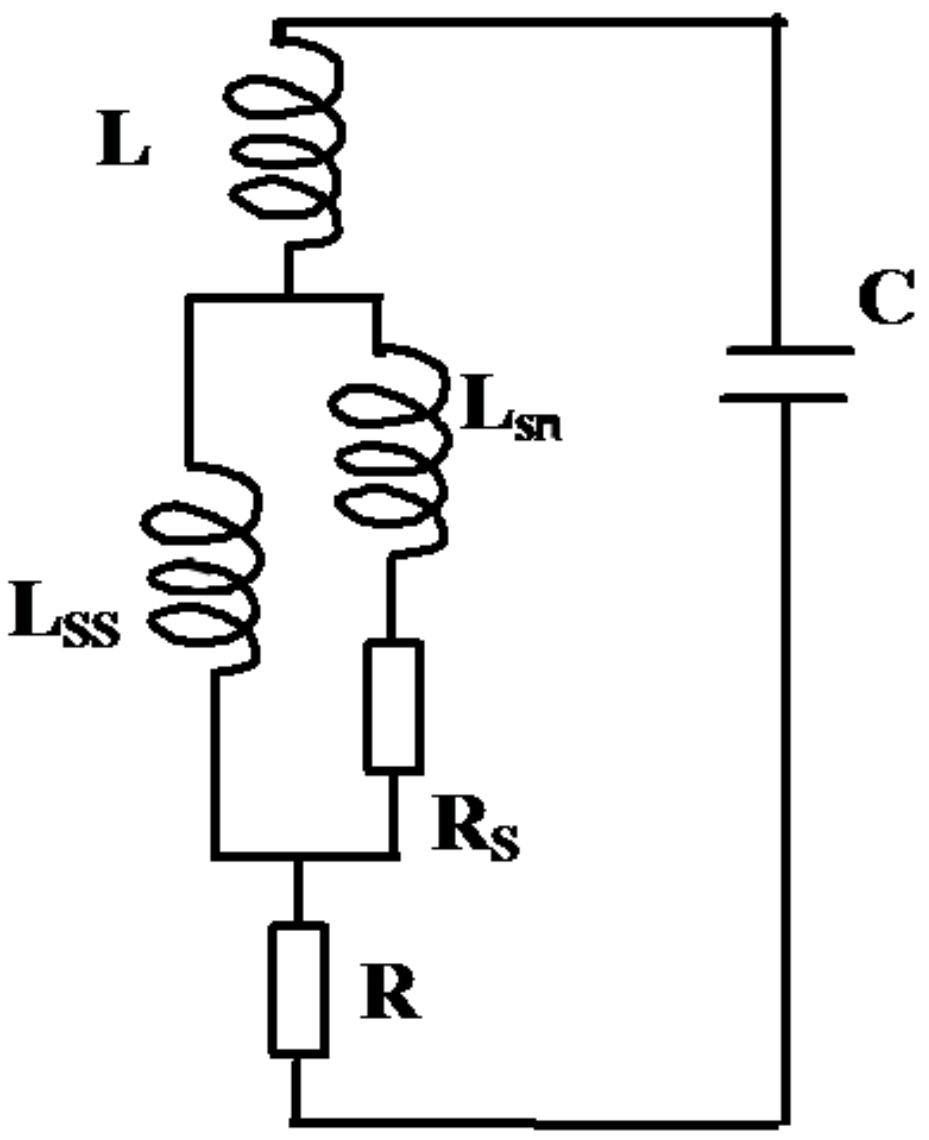

**Fig. 9.** Equivalent lumped element model of resonator circuit, which represents a superconductor in a normal part of its impedance *Z* in series with its inductance *L* in a dielectric resonator (*L*-*Z* model).

The circuit, shown in Fig. 7, is simplest usually researched case, when a resonator itself, is represented by a series *RLC network* with a parallel combination of inductance $L_s$ and surface resistance $R_s$ to model a superconductor.

Fig. 8 shows a resonator with an incorporated superconductor, modelled by the inductance $L_{sn}$ in series to parallel combination of inductance $L_{ss}$ and surface resistance $R_s$ ($L$-$\sigma$ model).

The modified model with the inductance $L_{sn}$, integrated into the parallel network is given in Fig. 9 (*L-Z* model). Usually, both the inductances $L_{sn}$ and $L_{ss}$ have relatively small values as compared to the inductance $L$ of the resonator, and they model the magnetic flux

$$\Phi = (L_{sn}+L_{ss})\times I_{rf}$$

that penetrates into the superconductor ($I_{rf}$ is the *RF* current). The resonance frequency of a resonator is determined mostly by the inductance $L$ with a very small contribution from the inductances $L_{ss}$ and $L_{sn}$. The element $R$ reflects the surface resistance $R_s$ of normal metal walls of the dielectric resonator.

Formula, to express transmitted *RF* power for the circuit depicted in Fig. 8 for the $L$-$\sigma$ model, is given below in eq. (5.19). It was derived, using the circuit theory (see Chapter 3), and mathematical algorithms created in the *Maple* software [27, 28] with application of appropriate calculation approach in form of algebraic expression:

$$P = \frac{1}{2}\frac{V^2\omega^2C^2(R\omega^2L_{SS}^2 + RR_S^2 + L_{SS}^2\omega^2R_S)}{R_S^2 - 2\omega^2(L+L_{Sn})CR_S^2 - 2R_S^2\omega^2L_{SS}C + R^2\omega^4C^2L_{SS}^2 + \omega^4(L+L_{Sn})^2C^2R_S^2 + R_S^2\omega^4L_{SS}^2C^2 - 2\omega^4L_{SS}^2LC + \omega^2L_{SS}^2 + 2R\omega^4C^2L_{SS}^2R_S + R^2\omega^2C^2R_S^2 + \omega^6(L+L_{Sn})^2C^2L_{SS}^2 + 2\omega^4(L+L_{Sn})C^2R_S^2L_{SS}} \quad (5.19)$$

Author of dissertation came up with the conclusion that this solution is stable in wide range limits of varying parameters, and it provides accurate microwave characterisation of superconductors.

Derived equation, describing the transmitted microwave power for the *L-Z* model, shown in Fig. 9, is expressed in eq.(5.20):

$$P = \frac{1}{2}\frac{\omega^2 C^2 V^2 (L_{SS}^2 \omega^2 R_S + RR_S^2 + R\omega^2 L_{SS}^2 + 2R\omega^2 L_{SS} L_{Sn} + R\omega^2 L_{Sn}^2)}{\begin{gathered} 2\omega^2 L_{SS} L_{Sn} + R^2\omega^4 C^2 L_{SS}^2 + R_S^2 + \omega^2 L_{SS} + \omega^2 L_{Sn}^2 + 2\omega^6 LC^2 L_{Sn}^2 L_{SS} + \omega^4 L_{SS}^2 C^2 R_S^2 + \\ + 2\omega^4 LC^2 R_S^2 L_{SS} + 2\omega^6 L^2 C^2 L_{SS} L_{Sn} + 2\omega^6 LC^2 L_{SS}^2 L_{Sn} + 2R\omega^4 C^2 L_{SS}^2 R_S + R^2 \omega^4 C^2 L_{Sn}^2 \\ + \omega^4 L^2 C^2 R_S^2 - 4\omega^4 L_{SS} L_{Sn} C L_{Sn} + 2R^2\omega^4 C^2 L_{SS} L_{Sn} - 2\omega^4 L_{SS}^2 LC - 2\omega^4 L_{SS}^2 C L_{Sn} - 2\omega^4 L_{Sn}^2 LC \\ - 2\omega^4 L_{Sn}^2 L_{SS} C + R^2\omega^2 C^2 R_S^2 + \omega^6 L^2 C^2 L_{SS}^2 + \omega^6 L^2 C^2 L_{Sn}^2 + \omega^6 L_{SS}^2 C^2 L_{Sn}^2 - 2R_S^2 \omega^2 LC - 2R_S^2 \omega^2 L_{SS} C \end{gathered}} \tag{5.20}$$

As it was found by the author of dissertation, this solution is stable in wide range limits of parameters deviation, and it describes the changes in resonance frequency shift and shape of resonance curve during modeling of nonlinear properties of dielectric resonator with superconductors more precisely. The presented surface resistance $R_s$ on microwave power $P$ dependencies $R_s(P)$ are used for the accurate microwave characterisation of superconductors [32, 33].

## 5.6. Microwave Power vs. Electromagnetic Field Dependence of Lumped Element Model Parameters.

In order to model the nonlinear behaviour of *HTS* materials, the following elements $R_{ss}$, $L_{ss}$, and $L_{sn}$ of equivalent lumped element model circuits need to be expressed as a function of *RF* magnetic field $H_{rf}$, or microwave power $P$, where $P \propto H_{rf}^2$. There have been various mathematical dependencies proposed to describe the magnetic field dependence in terms of *linear* [23, 24], *quadratic* [23, 24] and *exponential* representations [25, 26]. In this thesis, all three types of *RF* field dependences are used and analysed to define $R_s$, $L_{ss}$, and $L_{sn}$. The expressions for each of the dependences are presented below in eqs. (5.21 – 5.23).

***Linear dependence*** in eq. (5.21)

$$\begin{aligned} R_S &= R_{S0}(1+\rho H(i-1)); \\ L_{SS} &= L_{SS0}(1+lH(i-1)); \\ L_{Sn} &= L_{Sn0}(1+lH(i-1)); \end{aligned} \qquad (5.21)$$

***Quadratic dependence*** in eq. (5.22)

$$\begin{aligned} R_S &= R_{S0}(1+\rho_1 H(i-1)+\rho_2 H^2(i-1)); \\ L_{SS} &= L_{SS0}(1+l_1 H(i-1)+l_2 H^2(i-1)); \\ L_{Sn} &= L_{Sn0}(1+l_1 H(i-1)+l_2 H^2(i-1)); \end{aligned} \qquad (5.22)$$

***Exponential dependence*** in eq. (5.23)

$$\begin{aligned} R_S &= R_{S0}(1+a\exp(bH(i-1))); \\ L_{SS} &= L_{SS0}(1+c\exp(dH(i-1))); \\ L_{Sn} &= L_{Sn0}(1+c\exp(dH(i-1))); \end{aligned} \qquad (5.23)$$

where $\rho$, $\rho 1$, $\rho 2$, $l$, $l1$, $l2$, $a$, $b$, $c$, $d$ are fitting constants, $R_{S0}$ is the initial value of surface resistance, $L_{SS0}$ is the initial values of superconducting surface inductance and $L_{Sn0}$ is the initial values of normal metal surface inductance.

The expressions (5.21 – 5.23) were integrated into the transmitted microwave power representations eqs. 5.19, 5.20 to model the nonlinear effects in superconductors embedded into a dielectric resonator.

## 5.7. Modeling and Identification of Lumped Element Model Parameters for Superconducting Hakki-Coleman Dielectric Resonator.

Simulations of the transmitted microwave power in a dielectric resonator, containing the *HTS* materials, using the models shown in Figs. 8 and 9, along with the derived equations (5.19) and (5.20), have been performed using special software developed in *Matlab* [29, 30, 31]. An iterative computational method has been applied to obtain the transmitted microwave power $P$ on frequency $f$ dependence. At each frequency point $\omega_i$, the values of the complex inductance $L$ and resistance $R$ were computed, using the magnitude of microwave power $P$ taken at the previous frequency point, $P(\omega_{i-1})$. Fig. 10 shows the results of simulated microwave power responses for the circuits in Fig. 8 and 9 respectively, assuming an ideal and three complex function dependences eq. (5.21) – (5.23) used to manifest the nonlinear phenomena.

The following system parameters have been applied:

1) Dielectric resonator (based on the *Hakki-Coleman sapphire dielectric design* of 10*GHz*): $R=3\times10^{-4}\Omega$, $C=10^{-12}F$, $L=2.5\times10^{-10}H$.
2) *HTS* film – $YBa_2Cu_3O_7$: $L_{ss0}=L_{sn0}=1.8\times10^{-13}H$ ($\lambda=1.4\times10^{-7}m$), $R_{so}=5\times10^{-4}\Omega$, $\rho=1.6\times10^{-2}$, $l=1.1\times10^{-3}$, $\rho_1=4.1\times10^{-2}$, $l_1=1.8\times10^{-3}$, $\rho_2=3.2\times10^{-7}$, $l_2=2\times10^{-8}$, $a=9.5\times10^{-1}$, $b=2\times10^{-2}$, $c=3.2\times10^{-2}$, $d=2.5\times10^{-2}$, $f_0=10GHz$.

The ideal curve reflects the case, where components of $R_s$, $L_{ss}$ and $L_{sn}$ do not depend on *RF* magnetic field.

Results of simulations for $L$-$\sigma$ and $L$-$Z$ models for all types of nonlinear dependencies and provided parameters are shown in Figs. 10 and 11.

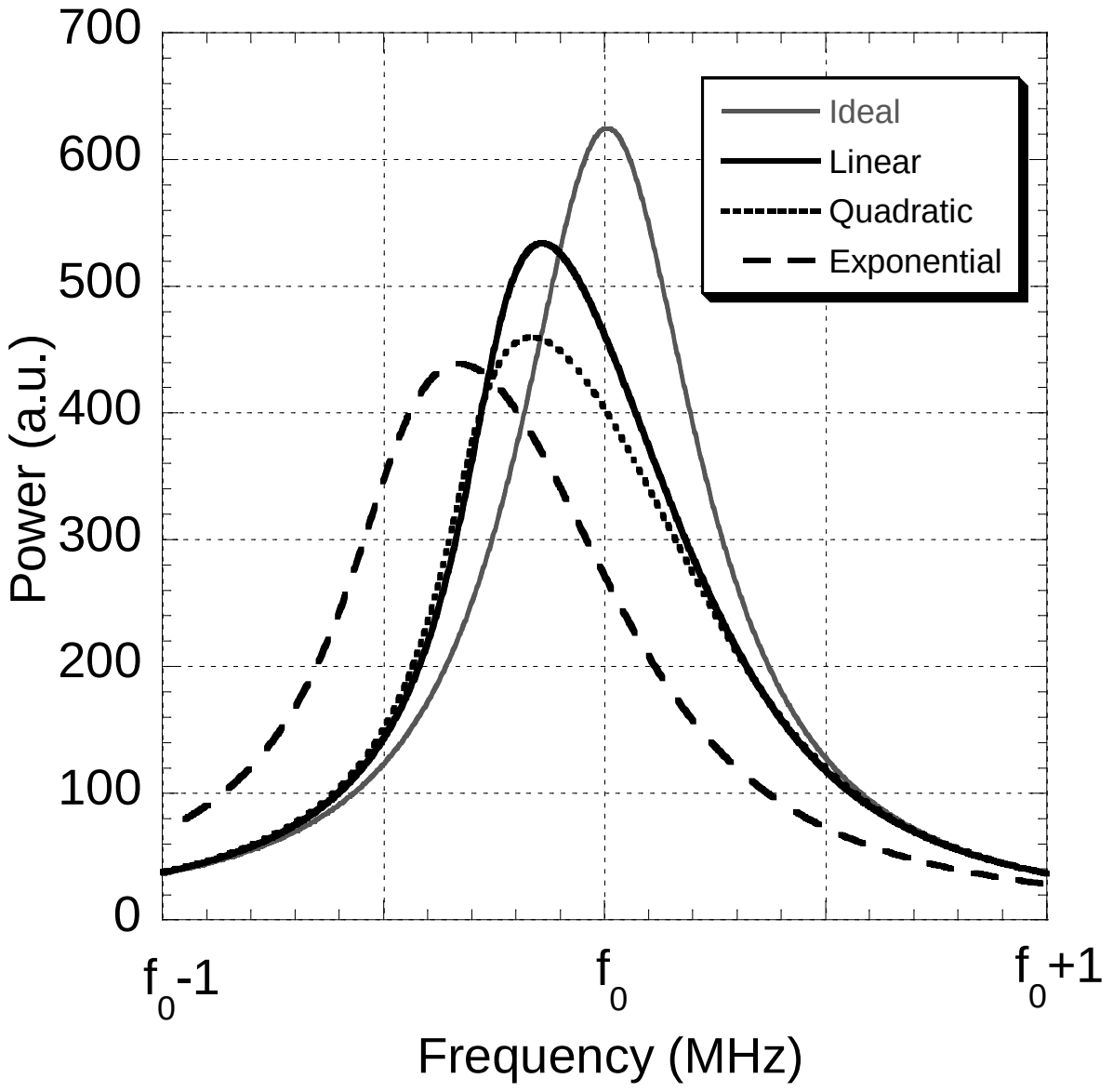

**Fig. 10.** Simulated dependence of microwave power on frequency *P(f)* for *L-σ* model of network in Fig. 8 ($f_0$ = 10 *GHz*).

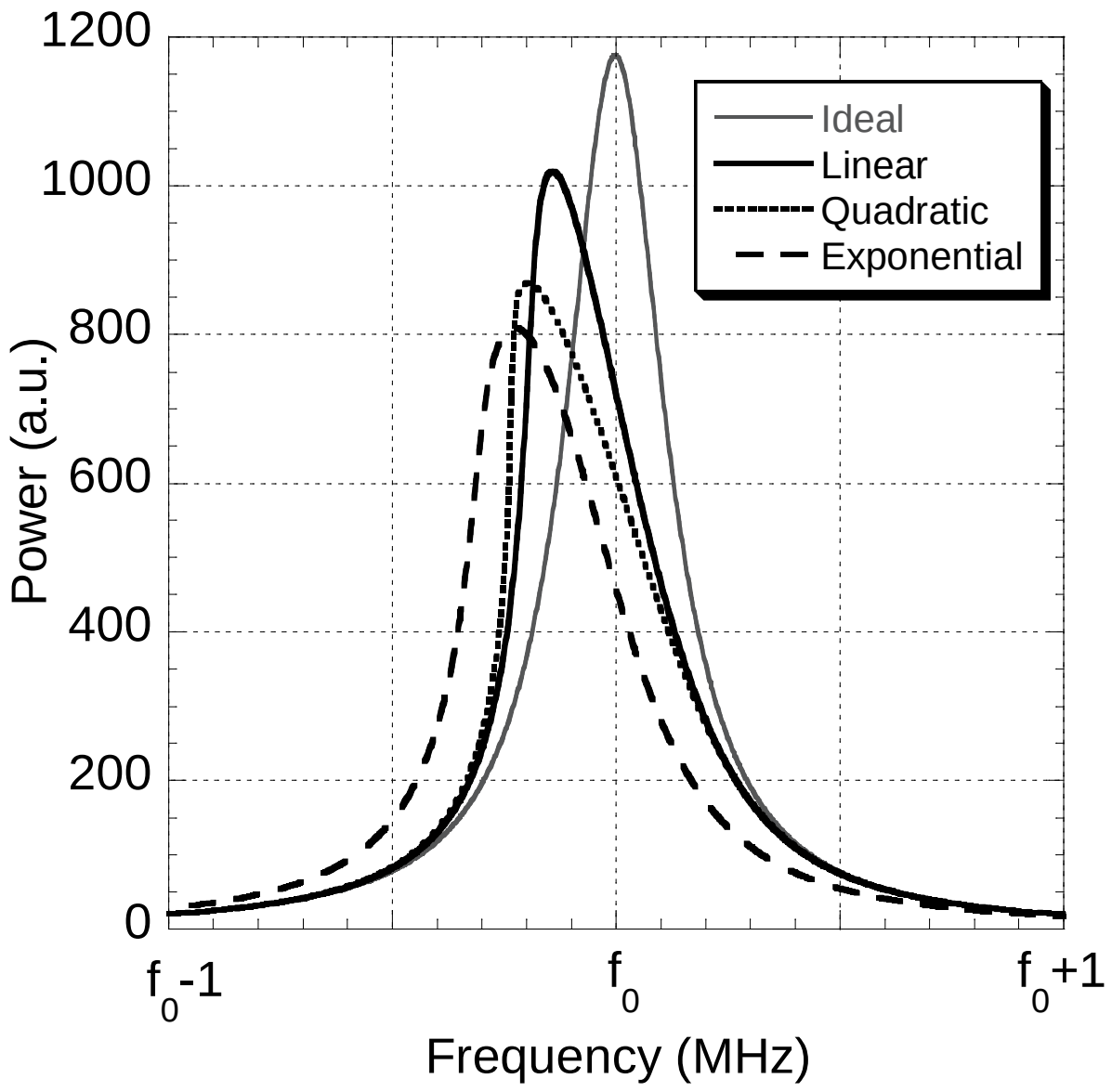

**Fig. 11.** Simulated dependence of microwave power on frequency *P(f)* for *L-Z* model of network in Fig. 9 ($f_0$ = 10 *GHz*).

As can be seen in Figs. 10 and 11, the linear function results in the smallest influence on nonlinear properties in both models, while the exponential dependence illustrates the strongest nonlinear effects in both models. The high quality factor $Q$ is expressed in the *L-Z* model, shown in Fig. 11, because the additional inductance is in parallel circuit and has a less influence on the total inductance $L$ of resonance circuit.

In order to verify the fitting parameters of the models in terms of their physical correlations with the simulated responses and to investigate their influence on the obtained results, the simulations have been performed for the two models with varying values of the following components: $\rho$, $\rho 1$, $\rho 2$, $l$, $l1$, $l2$, $a$ and $c$ in Fig. 12 and 13. Each parameter was changed at a time for the linear, quadratic and exponential complex dependences.

The following Figs. 5.12 – 5.14 illustrate the modeled responses for the linear (5.21), quadratic (5.22), and exponential (5.23) dependences of microwave power on frequency $P(f)$ for $L$-$\sigma$ model of network in Fig. 8.

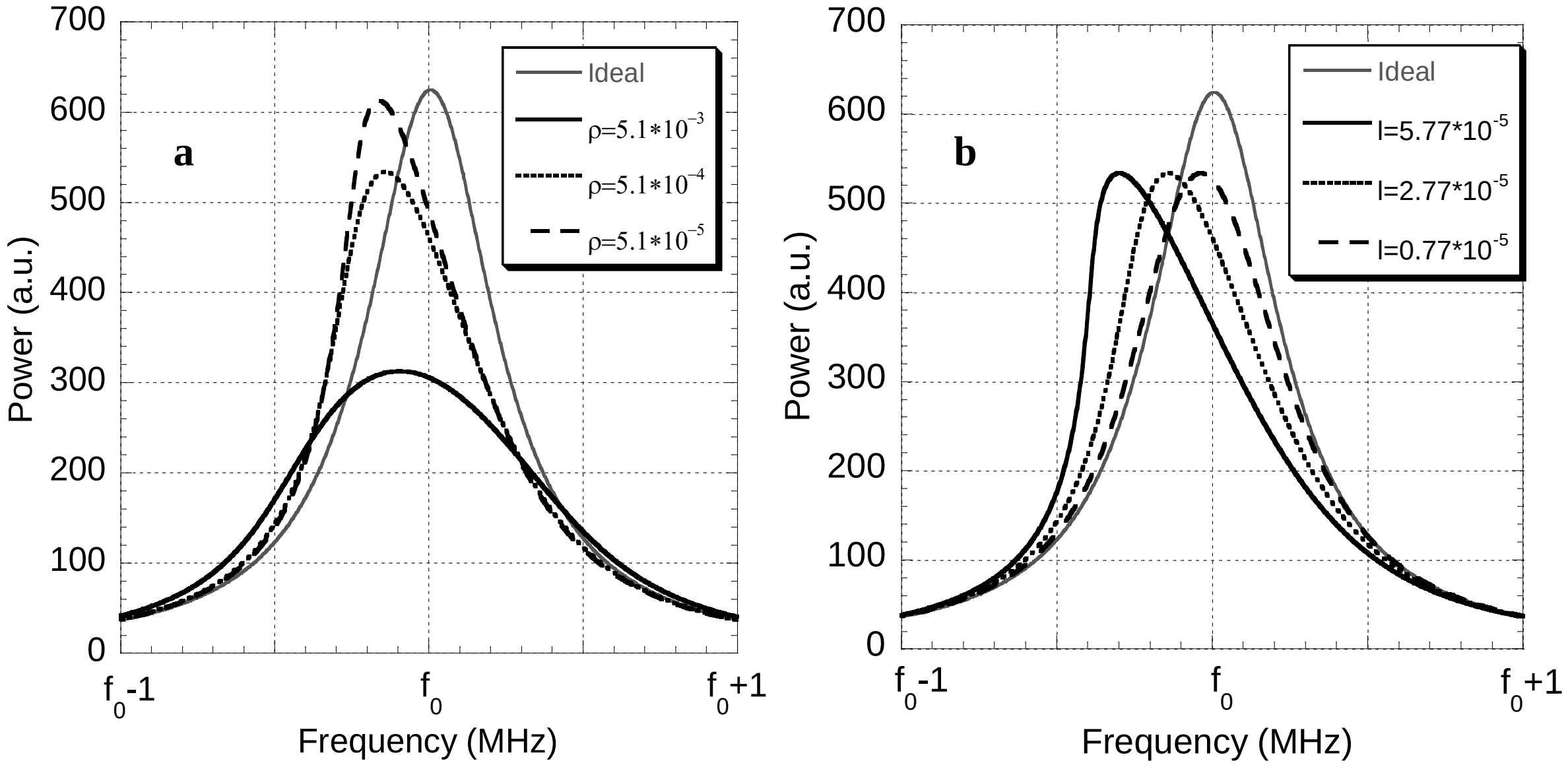

**Fig. 12.** Simulated dependences of microwave power on frequency $P(f)$ for varying fitting parameters for linear dependence for $L$-$\sigma$ model of network in Fig. 8 ($f_0$=10*GHz*).

The ideal case reflects the unperturbed function, where the nonlinear dependences were not applied. Fig. 12(a) illustrates the parameter $\rho$ related to $R_s$ change, while the alteration of the parameter $l$ related to both $L_{ss}$ and $L_{sn}$ is shown in Fig. 12(b).

In Fig 13, the nonlinear dependences *P(f)* for varying fitting parameters for quadratic dependence for $L$-$\sigma$ model of network in Fig. 8 ($f_0$=10*GHz*) are shown.

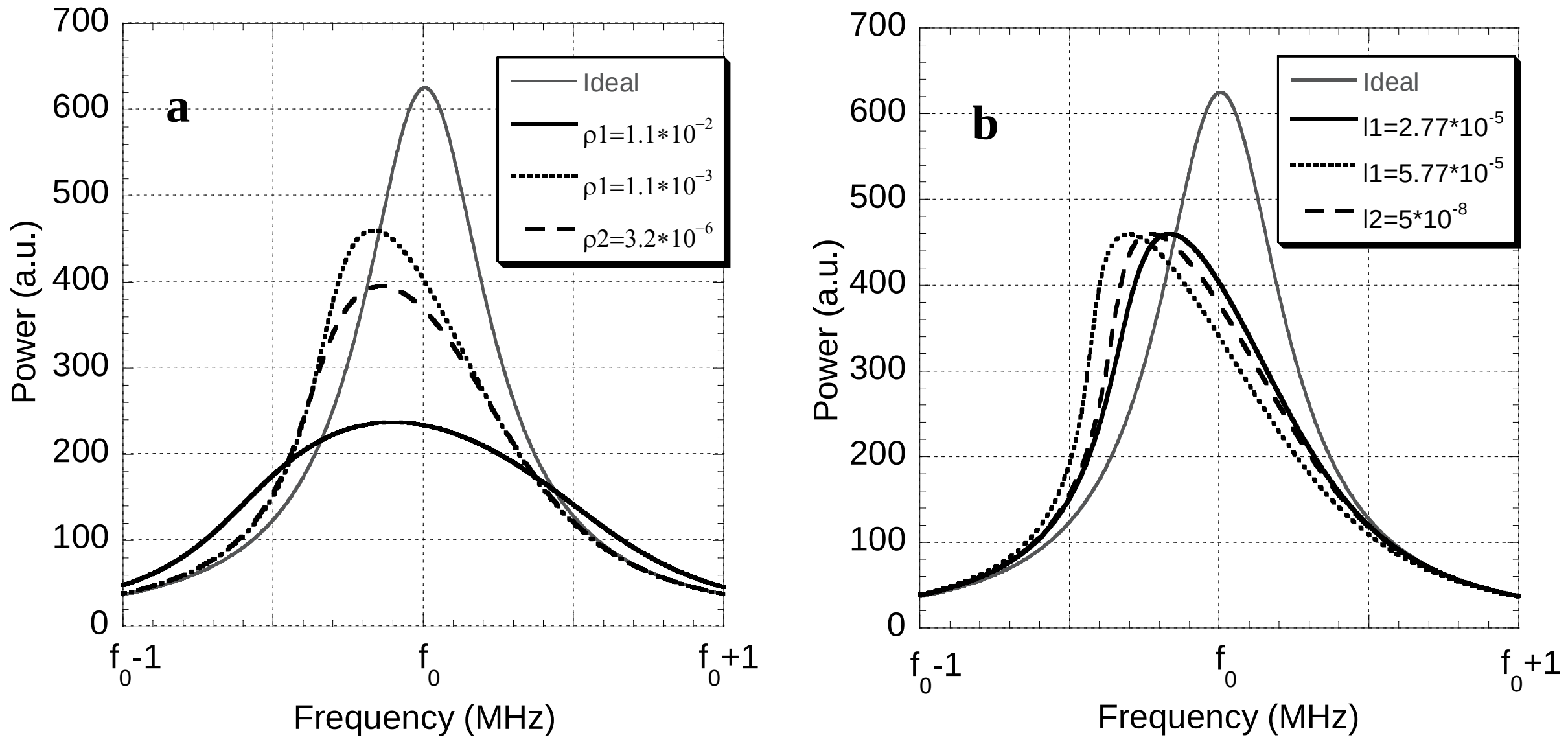

**Fig. 13.** Simulated dependences of *RF* power on frequency *P(f)* for varying fitting parameters for quadratic dependence for $L$-$\sigma$ model of network in Fig. 8 ($f_0$=10*GHz*).

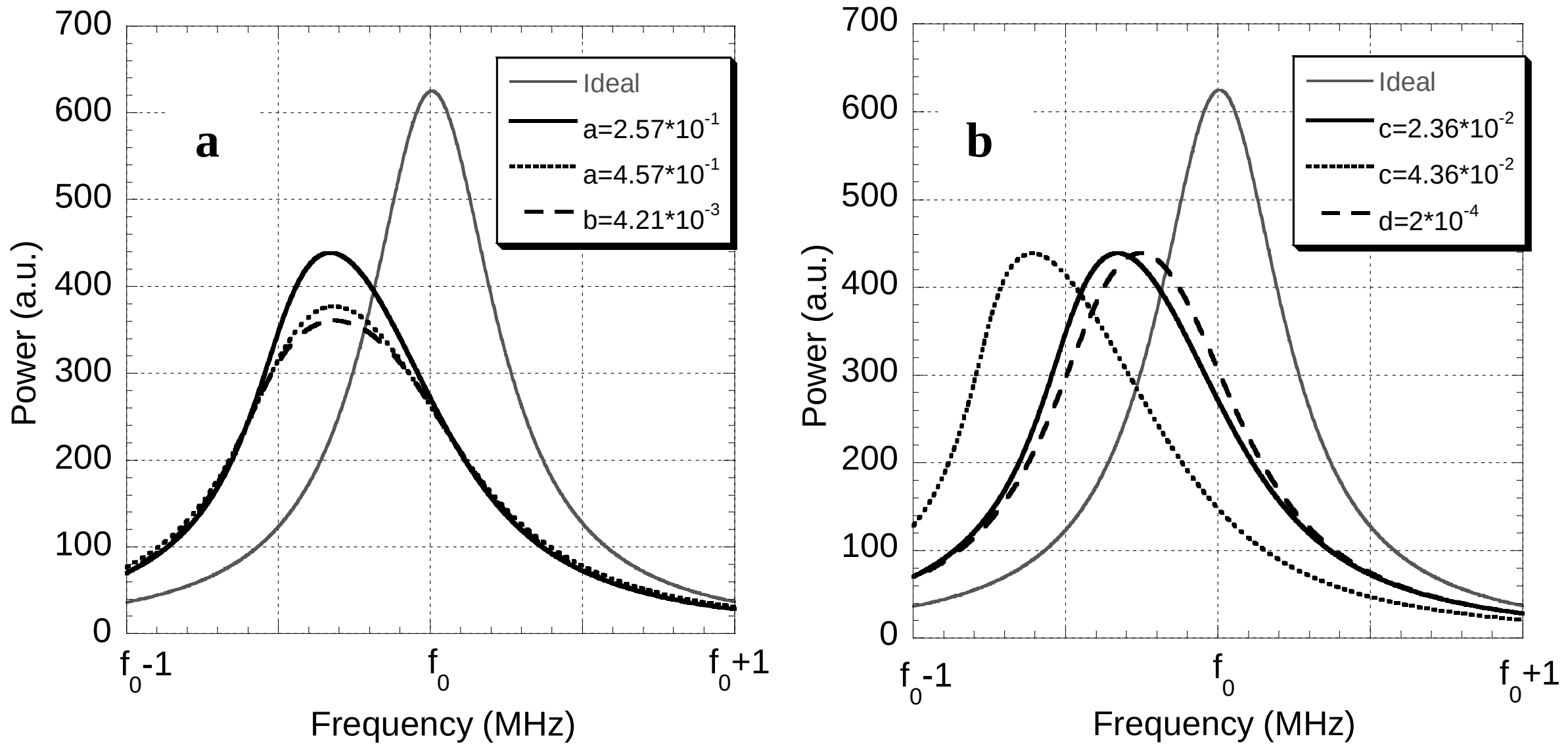

**Fig.14.** Simulated dependences of *RF* power on frequency *P(f)* for varying fitting parameters for exponential dependence for $L$-$\sigma$ model of network in Fig. 8 ($f_0$=10*GHz*).

Fig. 13(a) reflects the case, where the parameters $\rho_1$ and $\rho_2$, related to the complex $R_s$ of the quadratic case, were changed. Components of the nonlinear inductances ($L_{ss}/L_{sn}$) such as $l_1$ and $l_2$ are investigated in Fig. 13(b). Fig. 14 illustrates the two cases, where the fitting parameters of the exponential case *a*,*b*, and *c*,*d* were varied in relation to the nonlinear network elements *R* and $L_{ss}/L_{sn}$ .

In Fig 15, the nonlinear dependences *P(f)* for varying fitting parameters for linear dependence for *L-Z* model of network in Fig. 9 ($f_0$=10*GHz*) are shown [31].

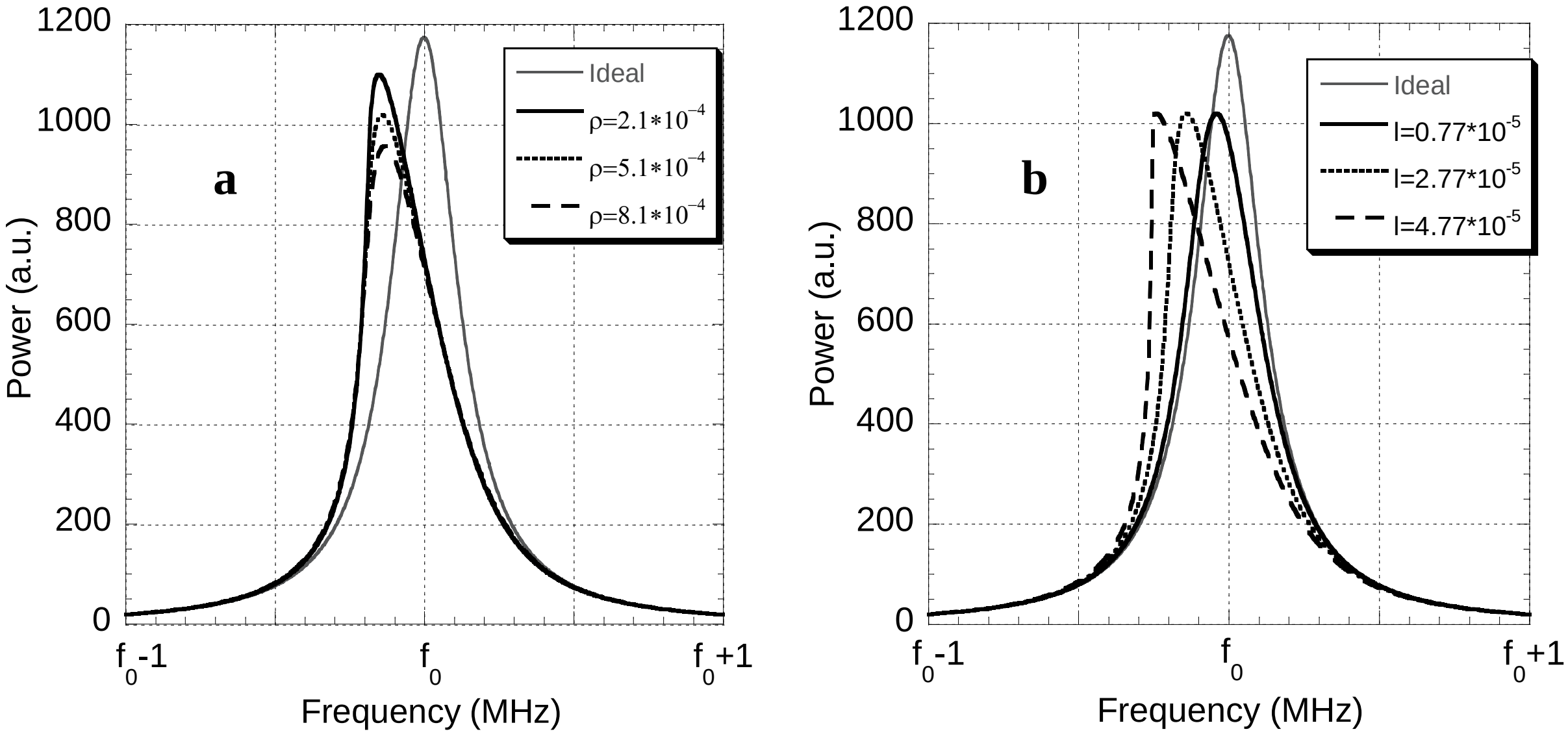

**Fig. 15.** Simulated dependences of *RF* power on frequency *P(f)* for varying fitting parameters for linear dependence for *L-Z* model of network in Fig. 9 ($f_0$=10*GHz*)[31].

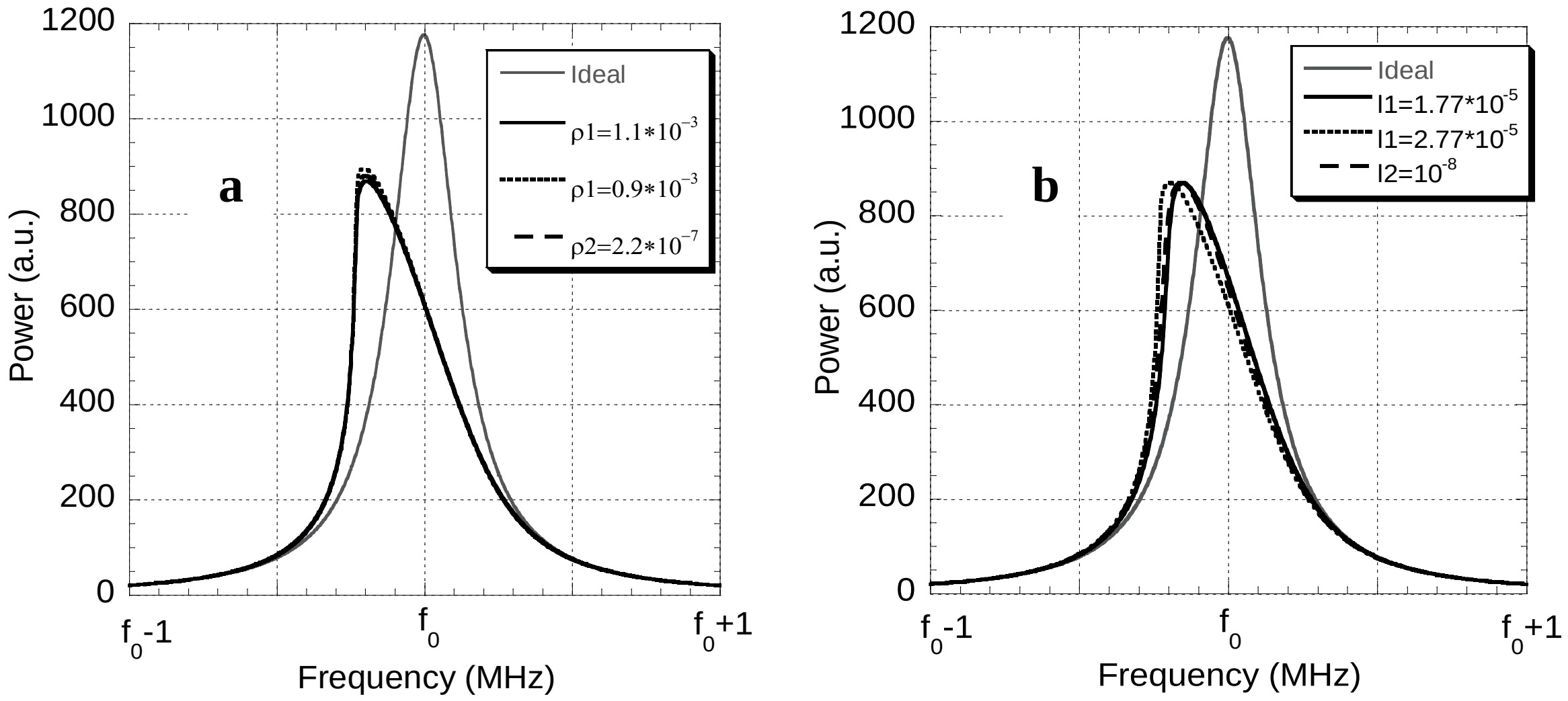

**Fig. 16.** Simulated dependences of microwave power on frequency *P(f)* for varying fitting parameters for quadratic dependence for *L-Z* model of network in Fig. 9 ($f_0$=10*GHz*) [31].

Modelled responses depicted in Fig. 15 and Fig. 16 show the influence of variations of fitting parameters in terms of the nonlinear network components $R_s$ and $L_{ss}/L_{sn}$ respectively. The parameter $\rho$ is related to the linear case of $R_s$; $\rho_1$ and $\rho_2$ are linked to the quadratic representation. The parameters $l$ is represented in the linear dependence ($L_{ss}/L_{sn}$), while the quadratic dependence includes two fitting parameters $l_1$ and $l_2$.

In Fig. 17, the simulated dependences of microwave power on frequency *P(f)* for varying fitting coefficients for exponential nonlinear dependence for *L-Z* model of network in Fig. 9 ($f_0$=10*GHz*), which include the two pairs of fitting parameters such as *a, b* and *d, c* related to the $R_s$ and $L_{ss}/L_{sn}$ components respectively, are shown [31].

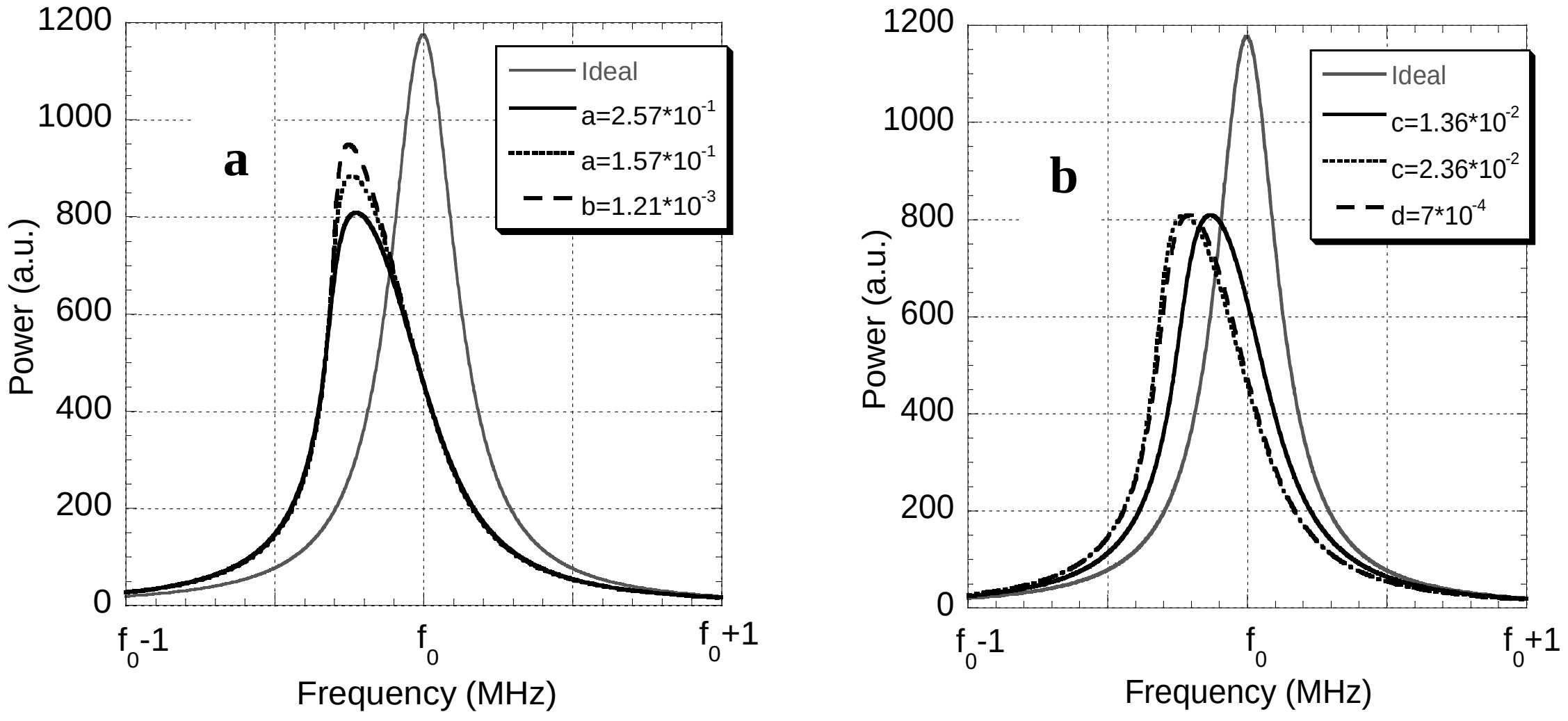

**Fig. 17.** Simulated dependences of microwave power on frequency *P(f)* for varying fitting coefficients for exponential nonlinear dependence for *L-Z* model of network in Fig. 9 ($f_0$=10*GHz*) [31].

Obtained results show that the varying of the parameter, related to surface resistance $R_s$, affects the amplitude of the resonance responses only, as expected, while changing the parameters, referred to $L_s$ tilts the resonance curves towards either higher or lower frequencies, due to $X(P)=\omega L(P)$. The outlined behaviour, based on the parameters variations, confirms the application of proper physical approach is engineered for the proposed models.

Author of dissertation made the calculation of parameter ***r*** = *ΔX/ΔR* for exponential nonlinear dependences. Results are shown in Fig. 18 [31].

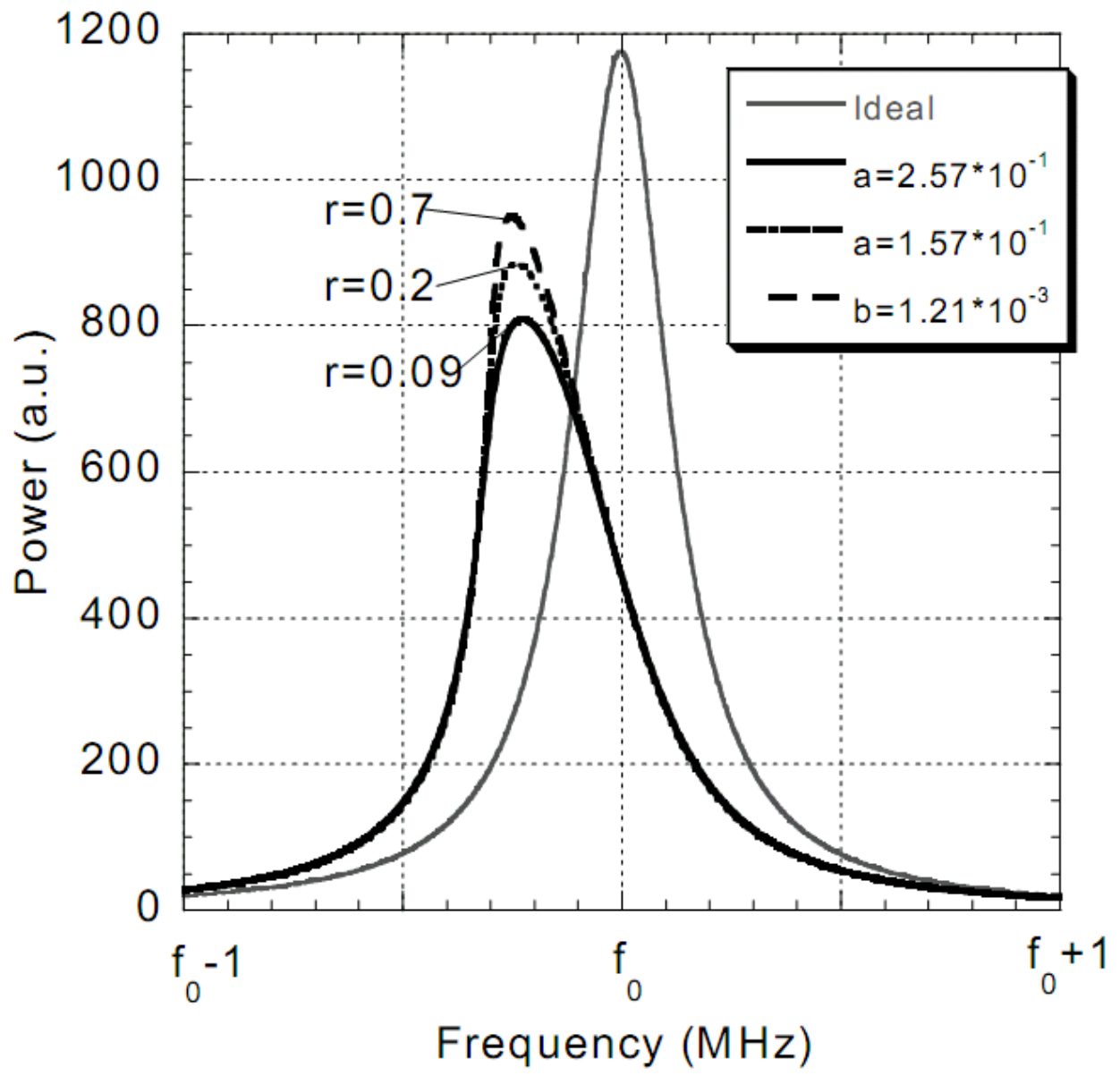

**Fig. 18.** Simulated change of value of ***r***-parameters on frequency for exponential nonlinear dependences of microwave power vs. frequency *P(f)* for *L-Z* model of network in Fig. 9 ($f_0$=10*GHz*) [31].

All the numerical values of fitting coefficients are picked up in a way that allows to observe the changes in shapes of resonance curves in simulation at deviation frequency limits Δ*f*=± 1 *MHz* at resonance frequency $f_0$ = 10 *GHz*. The increase of the magnitudes of fitting coefficients in nonlinear dependencies leads to the type of nonlinear dependencies, which was not observed in *HTS* thin films in author's researches. Therefore, the selected fitting coefficients, and hence the shapes of resonance curves in simulations in *Matlab*, closely approximate the experimental results, obtained by author of dissertation during the measurements toward the accurate microwave characterization of *HTS* thin films at James Cook University in Australia.

## Summary

The development of an advanced lumped element model, which accurately describes the nonlinear properties of *HTS* materials in a microwave resonator is completed in this chapter.

The basic concept of modelling is based on a ***Hakki-Coleman type dielectric resonator*** incorporating the superconductors. Mathematical modelling of *Hakki-Coleman* resonator is comprehensively described and analysed, and the most suitable equivalent network for representation of a microwave resonator is determined.

A number of researches by other authors towards the development of a lumped element model for simulations of nonlinear microwave characteristics of superconducting materials are examined firstly. Among a variety of proposed equivalent networks, derived power equations, and presented computer simulations, which are discussed in the chapter, the primary task is to select an appropriate model for more accurate microwave characterization of superconductors with nonlinear phenomena. A number of criteria have been applied on the basis of engineering, physical and mathematical considerations to accomplish this task. Three possible lumped element models representations of a superconductor, proposed by J. Wosik, J. C. Booth, J. Mateu, are investigated to create the accurate model of resonator. The three networks, which are depicted in Fig. 7, 8, 9, have been designed to model nonlinear properties of high temperature superconducting thin films by the author of dissertation. Transmitted power equations (5.19, 5.20) have been derived for the computer modelling and simulation for each model. Nonlinear components, which closely describe superconducting properties, have been integrated into the models, and the simulated responses have been researched, namely three characteristic dependences of the superconductor lumped elements on the *UHF* magnetic field are introduced to describe nonlinearities imposed by superconductors. Microwave power responses for all the researched samples were obtained using a special software program, which was developed by author of dissertation in the *Matlab* [31].

Discussing and analysing the computer modelling and simulation results, it is necessary to note that, going from the obtained results, the network illustrated in Fig. 4.9 has been identified as the most suitable to describe electromagnetic behaviour of the system consisting of the microwave resonator with *HTS* thin films. First circuit, which is depicted in Fig. 7, examines the basic properties of the structure, but it lacks comprehensiveness in expression of some physical effects at low temperatures. The updated design was proposed in Fig. 8, where a superconductor is represented with an additional component, $L_{sn}$ inductance in series to the parallel combination of $L_{ss}$ and $R_s$. In this approach the mathematical representation of the model in terms of the transmitted power equation is similar to the previous circuit without $L_{sn}$. For further investigation, a modified lumped element model was developed as illustrated in Fig. 9, where the component $L_{sn}$ was integrated into the parallel superconducting network. This design produces a more precise formulation of the transmitted power analysis. The performed computer simulations of the networks have also shown that the model, depicted in Fig. 9, expresses a higher quality factor. It has also been found that the exponential nonlinear dependence of superconducting circuit components, $R_s$, $L_{ss}$ and $L_{sn}$ gives the strongest simulated nonlinear effect. Further researches in this thesis are based on the chosen lumped element model network illustrated in Fig. 9.

Simulation results indicate that the nonlinear resonance curves to characterize properties of microwave resonators can be obtained with the help of synthesis of equivalent lumped element models of resonant circuits and modeling of their nonlinear properties using different types of nonlinear dependencies with fitting coefficients. The main problem is that physical phenomena in superconductors can not always be closely approximated by linear, quadratic and exponential dependencies in process of simulation, because the nature of nonlinearities is not clearly understood and may require its future research. However, it is possible to simulate the nonlinear properties of *HTS* thin films with good approximation and find the appropriate fitting coefficients for simulated nonlinear dependences, if the superconductor thin film is well researched and technology of its fabrication is advanced to the high level. The accurate characterization of *HTS* thin films is important for creation of new passive and active electronic devices with the *HTS* thin films, which will function in linear and nonlinear regimes of operation and have advanced technical characteristics.

# CHAPTER 6

# EXPERIMENTAL AND THEORETICAL RESEARCHES ON MICROWAVE PROPERTIES OF MgO SUBSTRATES IN A SPLIT POST DIELECTRIC RESONATOR AND NONLINEAR SURFACE RESISTANCE OF $YBa_2Cu_3O_{7-\delta}$ THIN FILMS ON MgO SUBSTRATES IN A DIELECTRIC RESONATOR AT ULTRA HIGH FREQUENCIES

## 6.1. Introduction.

Microwave characterization of superconducting thin films as well as superconducting single- and poly-crystals in microwave resonators attracts a considerable research interest, because of possibility to employ this advanced measurement technique during an optimization of design of microwave devices with the superconductors in microwave circuits in the micro- and nano-electronics. In this research approach, the selection of certain type of lumped element model of equivalent circuit, which can accurately characterize the superconductor in a dielectric resonator, is required. It is necessary to find the fitting coefficients to describe the nonlinear properties of superconductor during the simulation. This method is acceptable in the case, when the calculations are done for *HTS* thin films with repeatable physical properties. It can be used for one class of microwave resonators only. This simulation technique requires both the new calculations of magnitudes of magnetic fields and distribution of magnetic fields in a microwave resonator as well as the finding of new fitting parameters, if a different class of resonator devices is to be used as described in Chapter 4.

General approach for the accurate microwave characterisation of superconductors can be realized using the experimental research results on nonlinear properties of *HTS* thin films at ultra high frequency electromagnetic fields in combination with the better understanding of nature of physical phenomena behind the nonlinearities in superconductors.

Thus, it is necessary to research the dependence of surface impedance of superconductors *Zs = Rs + jXs* on the microwave power of *UHF* magnetic field, temperature and other parameters. The research is aimed to understand the physical phenomena, which originate the nonlinearities and define the nonlinear properties of $YBa_2Cu_3O_{7-\delta}$ superconductors. Therefore, it would be possible to model the nonlinearities more accurately and apply the simulation results during the design of new advanced electronic devices with superconductors. The accuracy of a model can only be as good as the accuracy of identification of its parameters. In Chapter 4, the numerical characterization of formulated lumped model parameters was described, and the way to determine necessary system values was shown. The determination of the resolved component and fitting coefficients, using the adequate identification physical models, is an important issue as it enables more effective microwave characterizations of the model. As there is a big number of different fitting coefficients, components and parameters involved in the detailed system description in the physical model, however the close approximation to reduce the total number of parameters to a small number of most important parameters: the critical magnetic field, energy gap and other can be used. The nonlinear phenomena in *HTS* thin films at microwaves can then be modeled accurately.

It has to be mentioned that many physical mechanisms, which can be connected with the appearance of nonlinearities in superconductors at microwaves, were researched, including the overheating effects, *Josephson phenomena* in weak links of grain boundaries, transport of magnetic vortices, de-pairing of Cooper pairs at action of current, and some others. However, the universal understanding of nature of nonlinearities in superconductors at microwaves is not achieved yet [1-3].

The experimental and theoretical research results on nonlinear resonance response of $YBa_2Cu_3O_{7-\delta}$ thin films on *MgO* substrates in a dielectric resonator at microwaves are presented in Chapter 6. The nonlinear properties of *HTS* thin films were also researched in [4-8]. The *HTS* thin films are generally characterized by advanced superconducting properties, but the nonlinear phenomena were discovered in these superconducting thin films at low temperatures $T < 15$ *K*. In the case of nonlinear phenomena, the energy absorption in a microwave resonator at small magnitudes of microwave signal power was bigger than at medium magnitudes of microwave power. The unusual nonlinear energy absorption in a microwave resonator, which is characterized by sharp increase of surface resistance $R_s$ was observed in a range of big magnitudes of microwave power. In [4-8], it was proposed that the *two levels system of excitations*, which can make an influence on the absorption of ultra high frequency electromagnetic wave, arises in surface layers of dielectric in the case of $YBa_2Cu_3O_{7-\delta}$ thin films on *MgO* substrate in a microwave resonator resulting in these effects. Usually, the two levels systems of excitations appear in dielectrics at very low temperatures T < 1 *K* [9, 10]. Author of dissertation decided to research the nonlinear phenomena in $YBa_2Cu_3O_{7-\delta}$ thin films on *MgO* substrate at high microwave powers up to 30 *dBm* at temperatures T = 25*K* and 50*K*, which are higher than the temperature used in [4-8]. The ultra high frequency measurement system and procedure as well as the results of experimental researches on microwave properties of pure *MgO* substrates in the split post dielectric resonator (*SPDR*) at ultra high frequency *f* =10.48 *GHz* are presented in the beginning of research paper. Then, the results of experimental researches on nonlinear resonance response of $YBa_2Cu_3O_{7-\delta}$ thin films on *MgO* substrates in the *Hakki-Coleman dielectric resonator (HCDR)* at *f* = 25*GHz* are provided. The similar mechanism of nonlinearity is observed in the researched $YBa_2Cu_3O_{7-\delta}$ thin films on *MgO* substrates despite of the use of the high temperatures T = 25*K* and 50*K*, as in the research results presented in [4-8]. The measurement results and accuracy issues are discussed comprehensively, including the evaluation of uncertainties.

## 6.2. Ultra High Frequency Measurement System and Procedure.

The ultra high frequency experimental measurement system for research on nonlinear resonance response of $YBa_2Cu_3O_{7-\delta}$ thin films on *MgO* substrates in a dielectric resonator at microwave power consisted of the following equipment:

- Vector Network Analyser (HP 8722C)
- Temperature Controller (Conductus LTC-10) fitted with a resistive heating element and two silicon temperature diode sensors

- Vacuum Dewar
- Close cycle cryogenic laboratory system (*APC-HC4*) suitable for measurements in a wide range of temperatures (10 K– 300 *K*)
- Computer system (*IBM-PC*) fitted with a *GPIB* card utilised for the computer control of temperature controller and network analyser, and S-parameter measurement data transfer from network analyser to computer.

The measurement set up was precisely calibrated by measuring the voltage standing wave ratio $VSWR=\frac{1+|\Gamma|}{1-|\Gamma|}$, ($\Gamma$ is the reflection coefficient), before accurate experimental measurements of *S*-parameters as schematically shown in Fig. 1.

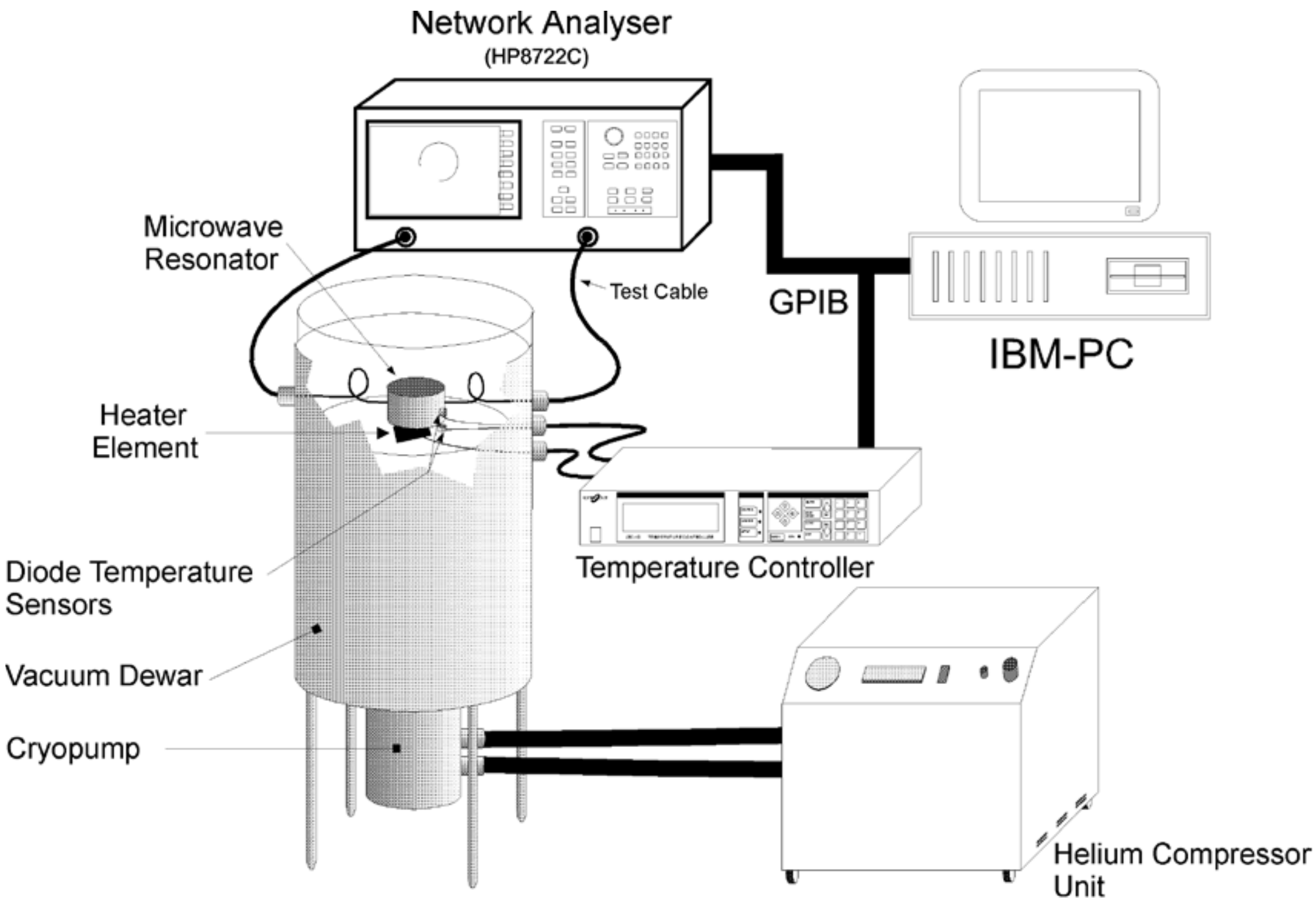

**Fig. 1.** Cryogenic measurement system for research on nonlinear resonance response of $YBa_2Cu_3O_{7-\delta}$ thin films on *MgO* substrates in a dielectric resonator at microwaves.

*Hakki-Coleman dielectric resonator* (*HCDR*) and *microstrip resonator* for measurements of *HTS* thin films and *split post dielectric resonator* (*SPDR*) for measurements of dielectric *MgO* substrates are placed inside the vacuum dewar and firmly fixed to the brass platform using a rigid strip and fastening screw bolts. To achieve a good thermal contact between the resonator and platform, a piece of indium foil placed between the base of resonator and platform as shown in Fig 2.

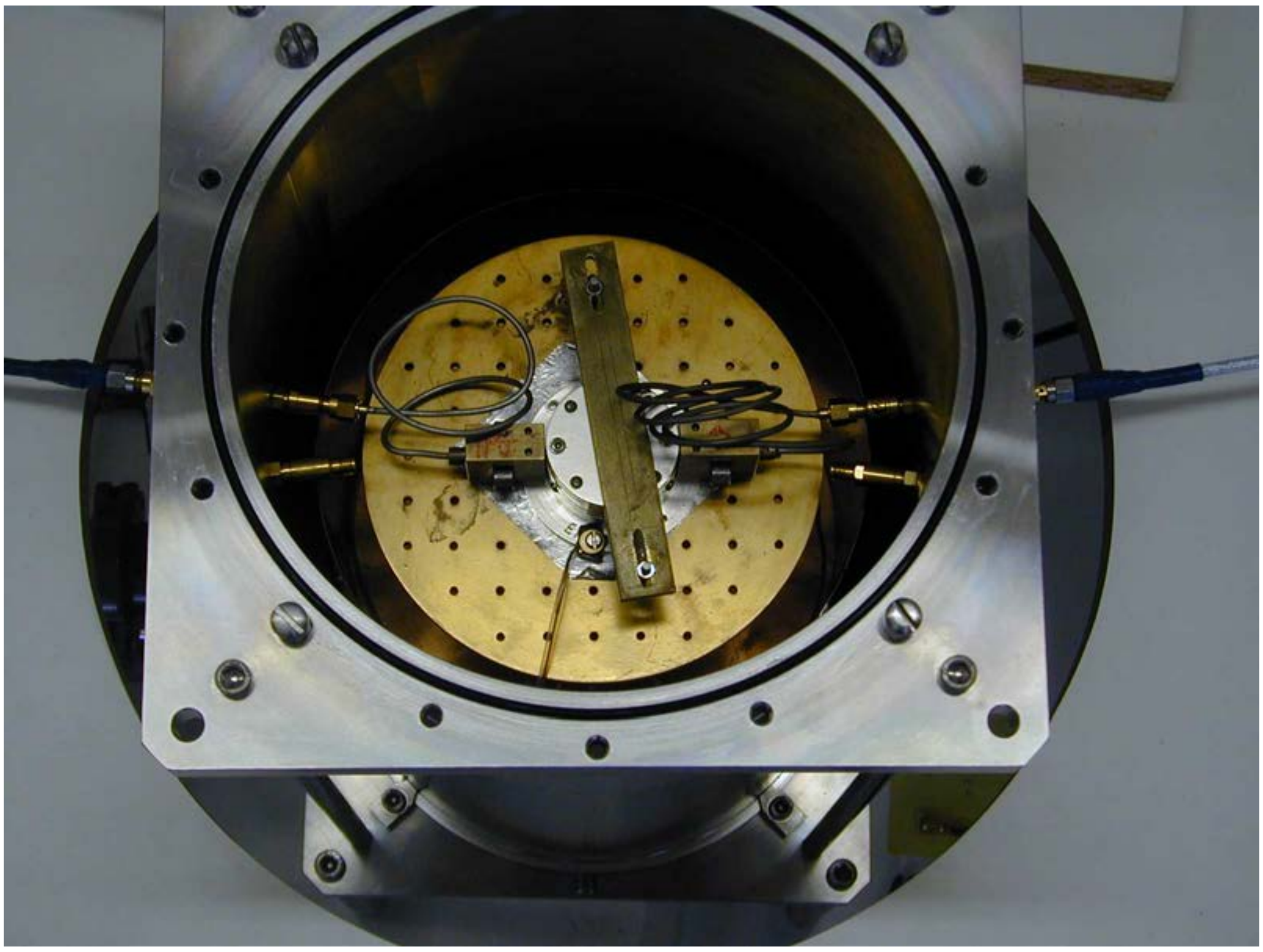

**Fig. 2.** *Hakki-Coleman dielectric resonator* mounted inside vacuum dewar with *RF* cable interconnections.

The loaded $Q_L$-factor and coupling coefficients $\beta_1$ and $\beta_2$ of resonator were obtained from multi-frequency measurements of $S_{21}$, $S_{11}$ and $S_{22}$ parameters measured around the resonance using the *Transmission Mode Q-Factor* (*TMQF*) Technique [11, 12]. The *TMQF* method enables to obtain accurate values of surface resistance accounting for factors such as noise, delays due to uncompensated transmission lines, and crosstalks occurring in measured data. The unloaded $Q_0$-factor was calculated from the exact equation

$$Q_0 = Q_L(1+\beta_1+\beta_2), \tag{5.1}$$

using the *TMQF* method for all temperatures. Dielectric resonator with embedded superconducting samples, mounted inside the vacuum dewar, was cooled down to temperature $T$ around 12 $K$, and the S parameters were measured at resonance frequency up to temperature $T = 85\ K$. The *RF* power of source signal was -5 *dBm* and the number of points was 1601.

Surface resistance $R_s$ of $YBa_2Cu_3O_{7-\delta}$ thin films on *MgO* substrates has been computed using the software *SUPER* [13], based on the equation (6.2)

$$R_S = A_S \left\{ \frac{1}{Q_0} - \frac{R_m}{A_m} - p_e \tan\delta \right\}. \tag{6.2}$$

The geometric factors $A_s$, $A_m$, and $p_e$ were computed using the incremental frequency rules as follows in eq. (6.3) [14]

$$\begin{aligned} A_S &= \frac{\omega^2 \mu_0}{4} \bigg/ \frac{\partial \omega}{\partial L} \\ A_m &= \frac{\omega^2 \mu_0}{2} \bigg/ \frac{\partial \omega}{\partial a} \\ p_e &= 2\left|\frac{\partial \omega}{\partial \varepsilon}\right| \frac{\varepsilon_r}{\omega} \end{aligned} \tag{6.3}$$

The data allows to get magnitudes of surface resistance $R_s$ of superconductor thin films at microwaves, and investigate the dependences of surface resistance on temperature $R_S(T)$, and nonlinear dependences of surface resistance on microwave power $R_S(P_{rf})$. In the case of dielectrics, it is possible to research the characteristics of dielectrics, including the influence of microwave power on dielectrics.

## 6.3. Precise Microwave Characterization of MgO Substrates for HTS Circuits with Superconducting Post Dielectric Resonator.

Accurate data on complex permittivity of dielectric substrates are needed for efficient design of *HTS* microwave planar circuits. Author of dissertation has researched *MgO* substrates from three different manufacturing batches, using a dielectric resonator with superconducting parts recently developed for precise microwave characterization of laminar dielectrics at cryogenic temperatures.

The measurement fixture has been fabricated using a $SrLaAlO_3$ post dielectric resonator with $DyBa_2Cu_3O_7$ endplates and silver-plated copper sidewalls to achieve the resolution of loss tangent measurements of $2{\cdot}10^{-6}$. The *MgO* substrates are essentially free of twinning, strain defects and air bubbles; they do not require buffer layers for *HTS* films; their orientation is typically along the (100) planes.

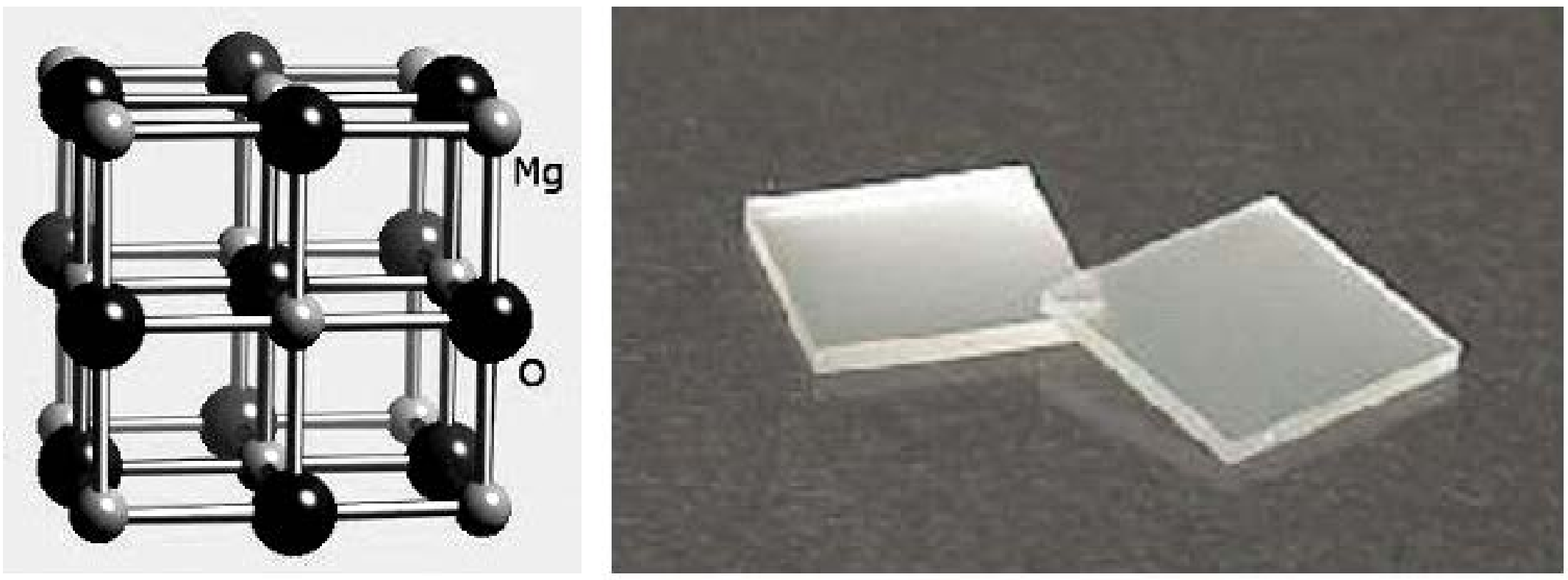

**Fig. 3.** Crystal structure of magnesium oxide and view of its substrates.

Recently, the split cavity technique [15] and the split post dielectric resonator (*SPDR*) [16, 17] have been successfully introduced for microwave characterisation of low loss single-crystal substrates. The *SPDR* technique can work at lower frequencies than the split cavity method for smaller complex permittivity dielectrics. However, it requires a very precise centering of the posts and the best loss tangent resolution reported is only $10^{-5}$. A single post fixture (open ended resonator) avoids the need of precise centering and aligning of the posts but was constructed in the past for testing of bulk dielectrics only [18].

Based on the concepts of the single-post dielectric resonator for dielectric rods and the split post dielectric resonator (*SPDR*) for planar dielectrics, author of dissertation developed a post resonator, as described in [19], for precise measurements of *HTS* substrates. The resonator was optimized for measurements of low loss planar dielectrics at cryogenic temperatures and was intended for testing substrates for *HTS* thin films before the deposition of *HTS* thin films. A schematic diagram of the designed resonator is shown in Fig. 4.

In author's case, the split post dielectric resonator (*SPDR*) contains one $SrLaAlO_3$ dielectric rod resonating at frequency of 10.48*GHz* on which the dielectric substrate under test is placed. The split post dielectric resonator was designed to measure dielectric properties of substrates of thickness of 0.7 *mm* or smaller, and diameters in the range from 35 *mm* to 55 *mm* in Fig. 4

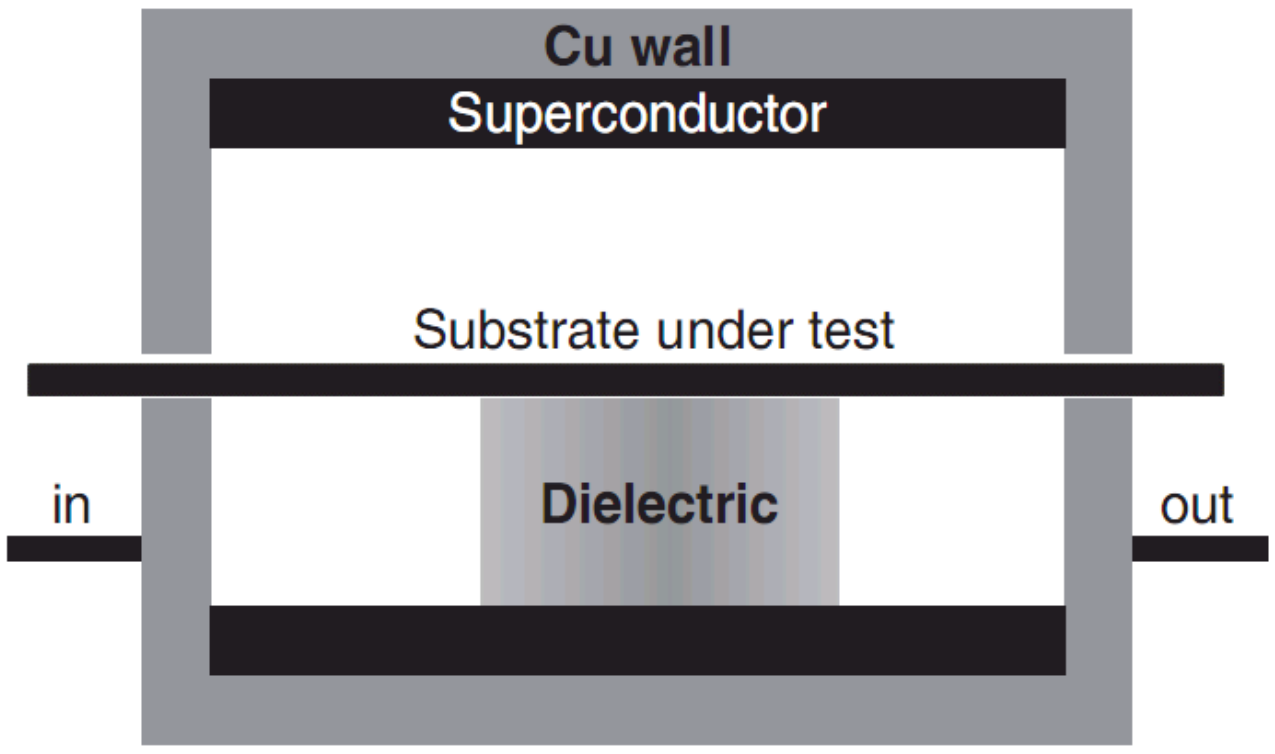

**Fig. 4.** Diagram of cryogenic split post dielectric resonator (after [19]).

A pair of $DyBa_2Cu_3O_7$ films was used as the end plates to increase sensitivity of dielectric loss tangent measurements. The constructed resonator was used for precise microwave measurements of relative permittivity $\varepsilon'_r$ and tan $\delta$ of *MgO* substrates described below. The *TMQF* technique [20] to eliminate noise, un-calibrated cables and adaptors and impedance mismatch from the measured data, to ensure high precision in unloaded $Q_0$-factor computations, was used for data processing. A simplified data logging of *S* parameters was performed as described in [21].

Author of dissertation measured the nine *MgO* substrates from three different batches manufactured by *Tateho Chemical Industries Co. Ltd.* in Japan, as a function of temperature from $T = 14$ *K* to 84 *K*. The substrates had the average thickness of 0.508 *mm* for each batch. Spatial variations in thickness of ±1 $\mu m$ were observed; measurements of thickness were made with ±1 $\mu m$ uncertainty. The measurement procedure was the same as described in [22] and for details on the transmission mode Q-factor technique readers are referred to [20]. The *RF* input power level used for measurements dielectric characteristics was −5 *dBm* and for its power dependence was -20 *dBm* ÷ +15 *dBm*.

The real relative permittivity of the substrates under test has been computed numerically based on the rigorous electro-magnetic modeling of the cryogenic post dielectric resonance structure using the *Rayleigh–Ritz technique* [23] as explained in [22]. The maximum absolute uncertainty in $\varepsilon'_r$ resulting from the analysis for our resonator is ±0.2% [24]. Taking into consideration the relative uncertainty of the substrate's thickness we assess the total uncertainty in computations of permittivity to be±0.5%. Computed values of permittivity $\varepsilon'_r$ of the tested *MgO* substrates are shown in Fig. 5.

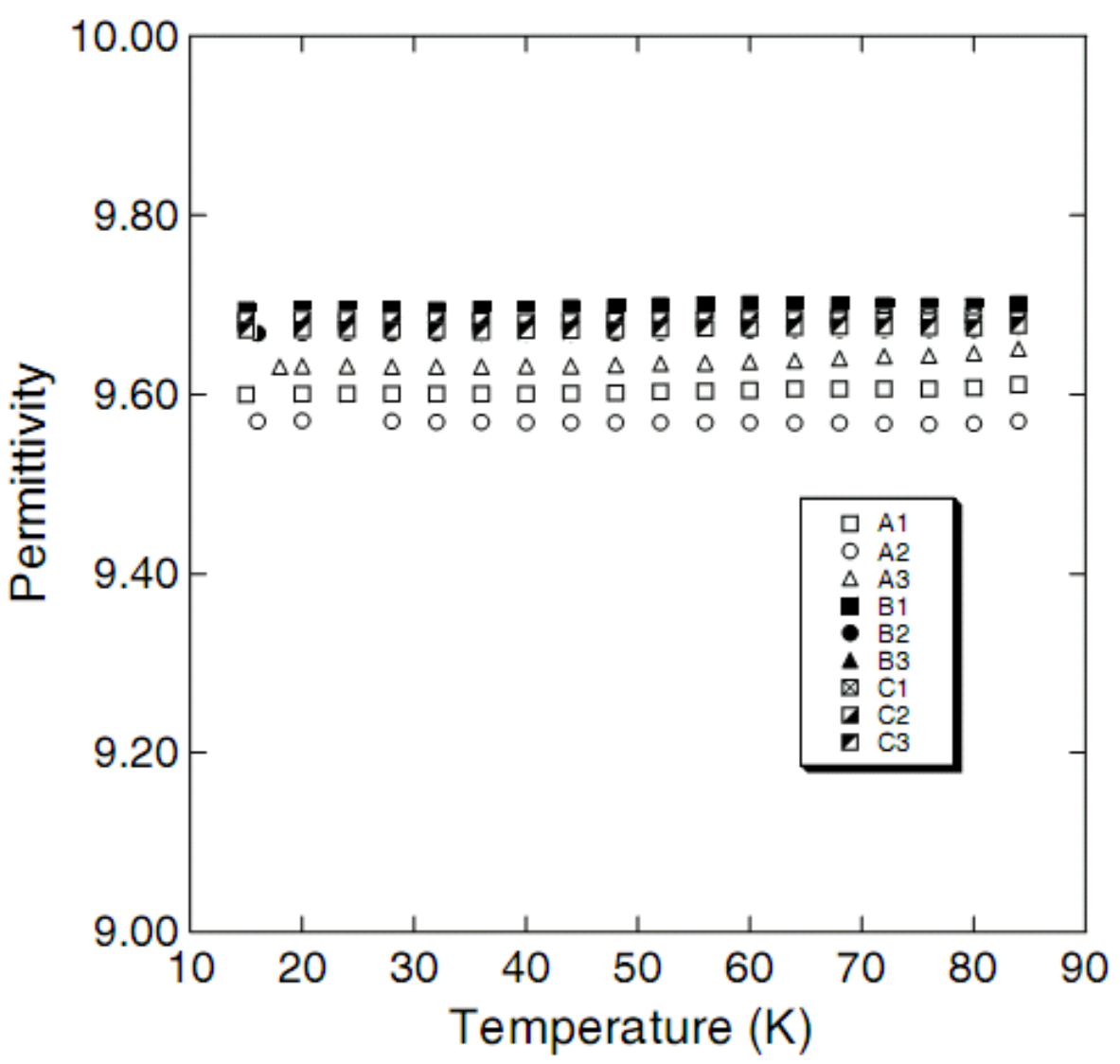

**Fig. 5.** Dependence of measured permittivity $\varepsilon'_r$ on temperature of MgO substrates.

The samples exhibited constant values of real relative permittivity, when temperature was varied from 15 *K* to 84 *K* with values of $\varepsilon'_r$ in the range from 9.57 to 9.68 with the maximum difference of 1% (or ±0.5%), hence within the measurement uncertainty. The average $\varepsilon'_r$ value of 9.63 differs from the average of the values reported for the substrates (9.71) by 0.85% and is smaller than highest literature data by 2.8%, confirming existing variations in *MgO* substrates.

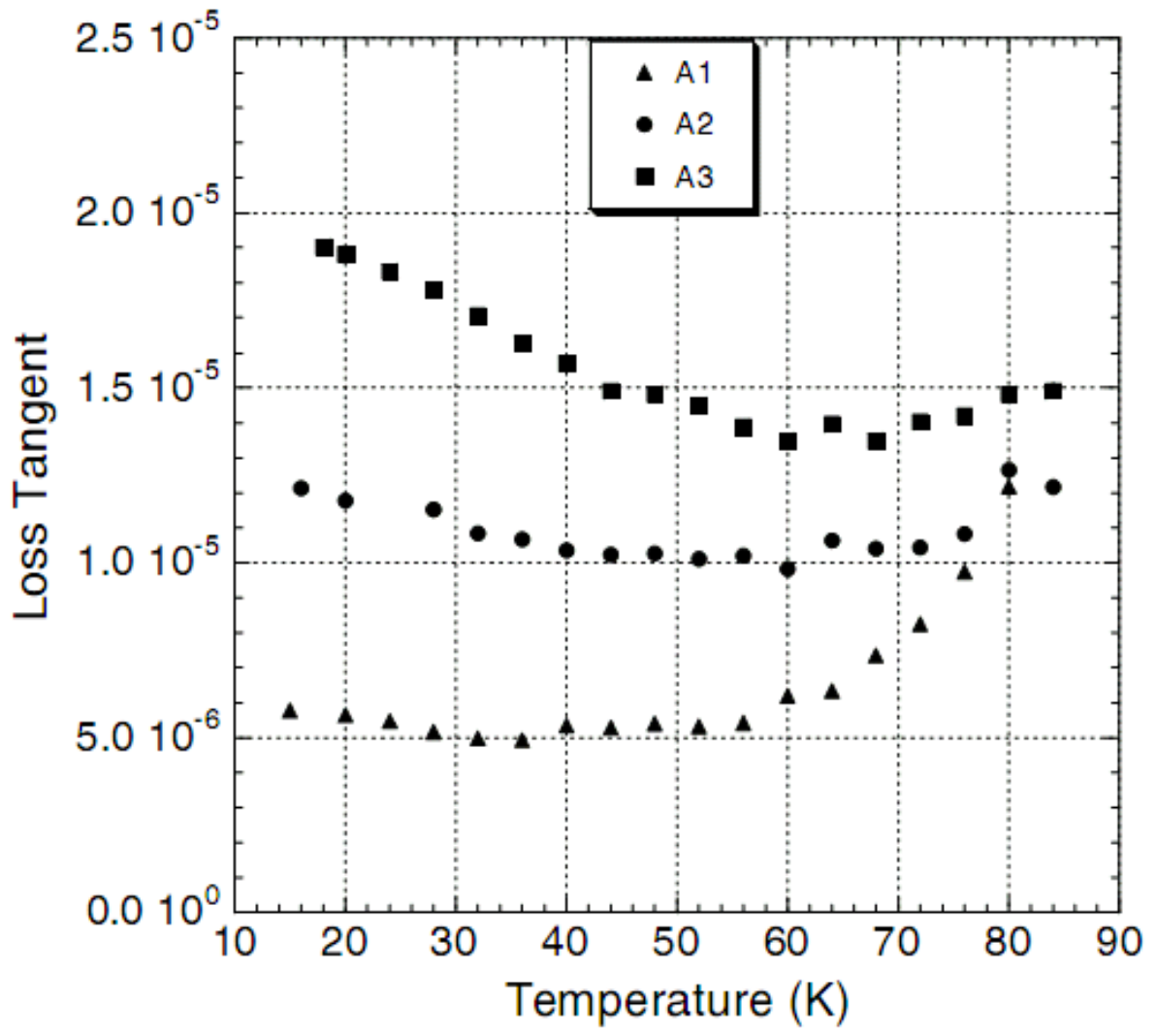

**Fig. 6.** Dependence of measured loss tan δ on temperature of MgO substrates of batch A.

The loss tangents of the *MgO* substrates for batches *A*, *B* and *C* at a frequency of 10.48 *GHz*, computed from the measured unloaded $Q_0$-factors, are given in Figs. 6 – 8. The details of the computation procedure are discussed in [22].

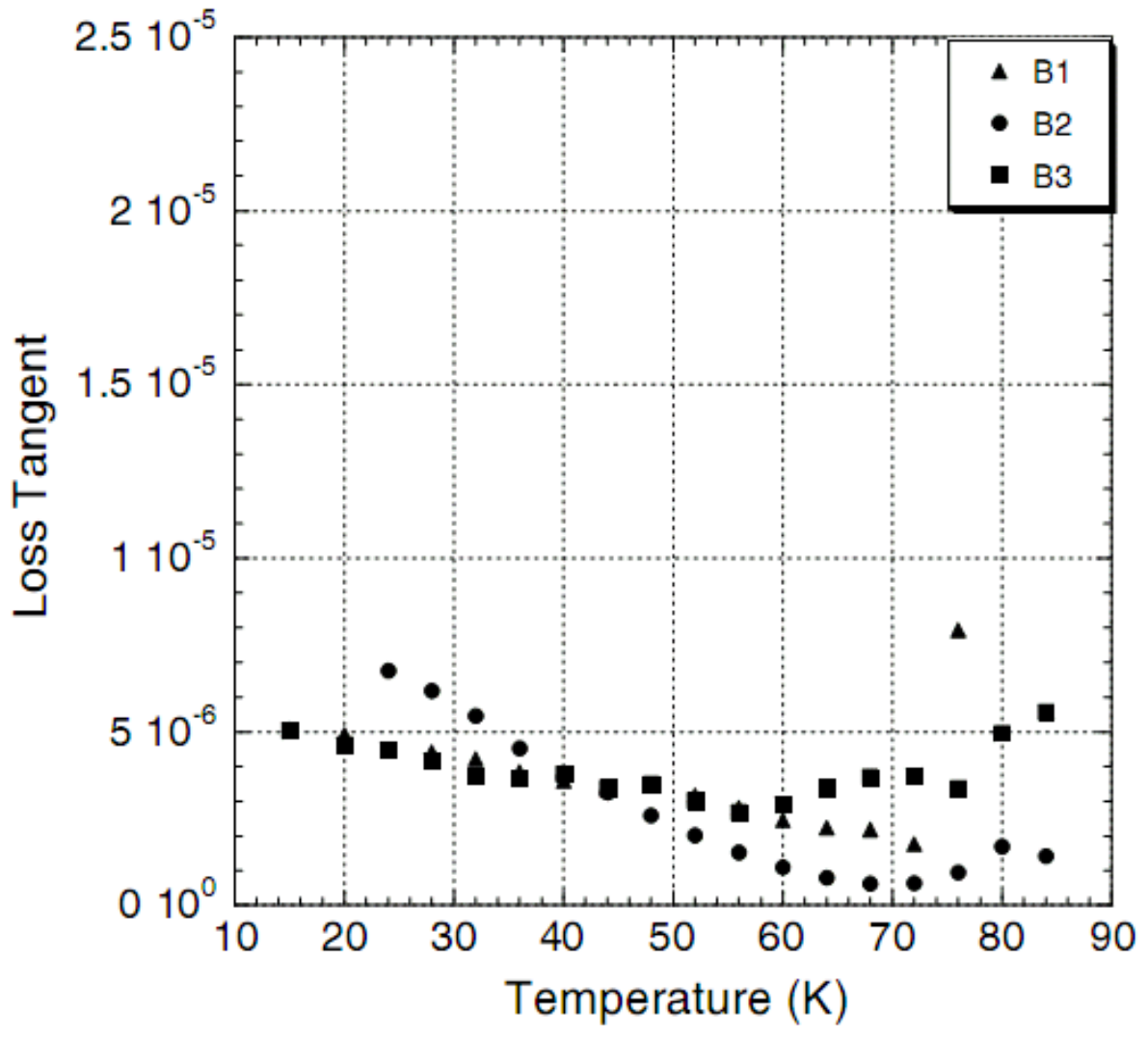

**Fig. 7.** Dependence of measured loss tan $\delta$ on temperature of *MgO* substrates of batch *B*.

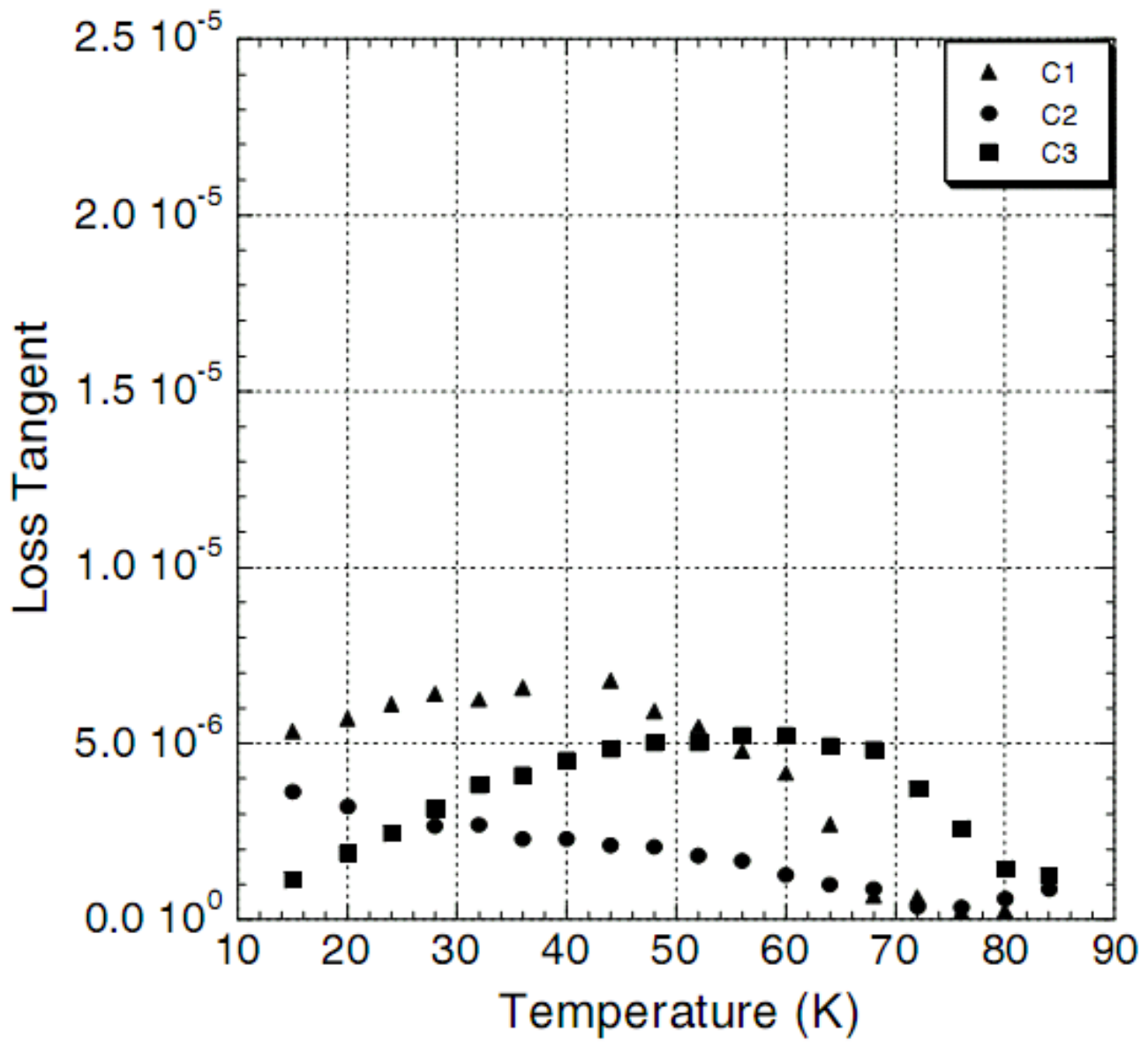

**Fig. 8.** Dependence of measured loss tan *δ* on temperature of *MgO* substrates of batch *C*.

The tested substrates exhibited low values of the loss tangent. The losses of batch A were of the order of $1 \times 10^{-5}$, while for batches *B* and *C* losses were below $5 \times 10^{-6}$ .The minimum values of tan δ for batches *A, B* and *C* were 4.93 × 10−6 ,$6.33 \times 10^{-7}$ and $3.8 \times 10^{-7}$ respectively. The temperature dependence of tan δ exhibited no evident pattern. However, it could be said that the loss tangent of batch A exhibited a minimum in the temperature dependence. As an example, tan*δ* of sample *A1* decreased from $5.79{\times}10^{-6}$ to $4.93{\times}10^{-6}$ at temperature 36 *K*, and then increased to $1.21{\times}10^{-5}$ at temperature 80 *K*. For sample *A1*, the minimum occurred at a temperature of 36 *K*, and for samples *A2* and *A3* between temperatures 60 *K* and 70*K*. The samples of batch *B* showed a more or less similar pattern to *A*; batch *C* however exhibited rather maxima versus temperature. The samples from batch *A* exhibited large variation in losses at lower temperatures (up to four times), much bigger than batches *B* and *C*. This is also illustrated in Fig. 9, where a ‘standard deviation’ in loss tangent values, calculated for the three batches, is given. (Author of dissertation uses ‘SDT’ in inverted commas as a number of samples was too small to perform a proper statistical analysis.)

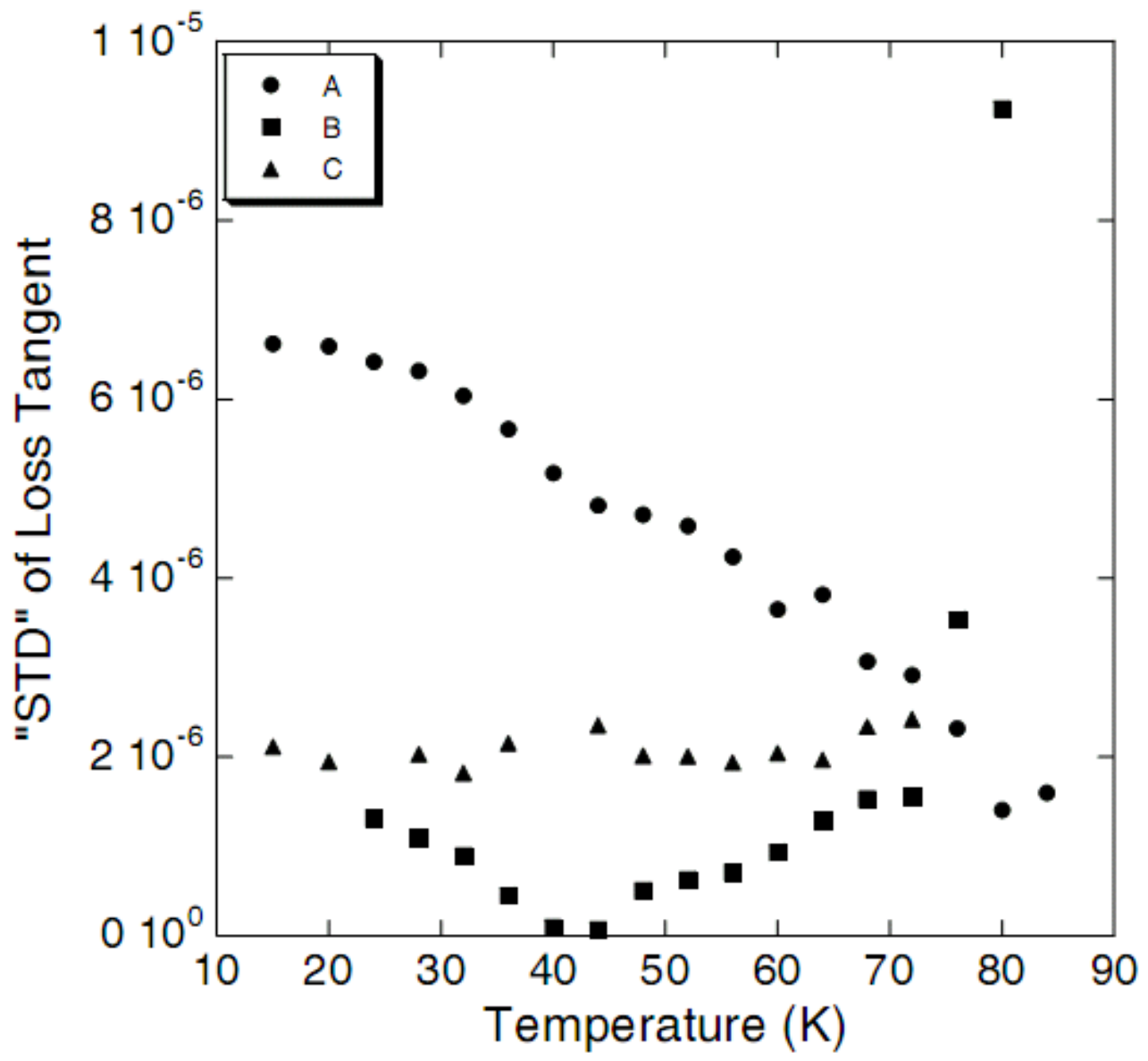

**Fig. 9.** Dependence of 'Standard Deviation' in tan $\delta$ values on temperature for three *MgO* batches.

The uncertainty in measured tan δ, ($\Delta_r$ tan $\delta$) values was between 9% and 90% for tan$\delta$ above 2 × $10^{-6}$ (assuming 1% error in unloaded $Q_0$-factor measurements and 1% uncertainty in the computed energy filling factor for the substrates). The large value of $\Delta_r$ tan δ is due to the principle of the loss tangent computation based on a difference between the $Q_0$ of the empty resonator and $Q_{0s}$ of the resonator with a substrate under test. For measured values below the resolution of the resonator, 2 × $10^{-6}$, the uncertainty dramatically increases, even exceeding 500%. Hence, in practice, for the substrates of batches *B* and *C*, we assess the upper bound for the losses. A summary comparison of the real relative permittivity, loss tangent, and relative uncertainties of the *MgO* substrates under test is given in Tab. 1 for temperatures of 15 *K*, 52 *K* and 72 *K*.

| | Substrate | 15 K | 52 K | 72 K |
|---|---|---|---|---|
| Real relative permittivity | A1 | $9.60 \pm 0.5\%$ | $9.60 \pm 0.5\%$ | $9.61 \pm 0.5\%$ |
| | A2 | $9.57 \pm 0.5\%$ | $9.57 \pm 0.5\%$ | $9.57 \pm 0.5\%$ |
| | A3 | $9.63 \pm 0.5\%$ | $9.63 \pm 0.5\%$ | $9.64 \pm 0.5\%$ |
| | B1 | $9.69 \pm 0.5\%$ | $9.70 \pm 0.5\%$ | $9.70 \pm 0.5\%$ |
| | B2 | $9.67 \pm 0.5\%$ | $9.67 \pm 0.5\%$ | $9.67 \pm 0.5\%$ |
| | B3 | $9.68 \pm 0.5\%$ | $9.68 \pm 0.5\%$ | $9.69 \pm 0.5\%$ |
| | C1 | $9.68 \pm 0.5\%$ | $9.68 \pm 0.5\%$ | $9.69 \pm 0.5\%$ |
| | C2 | $9.68 \pm 0.5\%$ | $9.68 \pm 0.5\%$ | $9.68 \pm 0.5\%$ |
| | C3 | $9.67 \pm 0.5\%$ | $9.67 \pm 0.5\%$ | $9.68 \pm 0.5\%$ |
| $\tan\delta$ | A1 | $5.79 \times 10^{-6} \pm 27\%$ | $5.32 \times 10^{-6} \pm 35\%$ | $8.24 \times 10^{-6} \pm 30\%$ |
| | A2 | $1.20 \times 10^{-5} \pm 14\%$ | $1.01 \times 10^{-5} \pm 19\%$ | $1.04 \times 10^{-5} \pm 24\%$ |
| | A3 | $1.90 \times 10^{-5} \pm 9\%$ | $1.45 \times 10^{-5} \pm 14\%$ | $1.40 \times 10^{-5} \pm 18\%$ |
| | B1 | $5.07 \times 10^{-6} \pm 30\%$ | $3.17 \times 10^{-6} \pm 58\%$ | $1.77 \times 10^{-6} \pm 134\%$ |
| | B2 | | $2.02 \times 10^{-6} \pm 91\%$ | $6.5 \times 10^{-7} \pm 360\%$ |
| | B3 | $5.03 \times 10^{-6} \pm 30\%$ | $3.02 \times 10^{-6} \pm 61\%$ | $3.74 \times 10^{-6} \pm 64\%$ |
| | C1 | $5.33 \times 10^{-6} \pm 29\%$ | $5.48 \times 10^{-6} \pm 34\%$ | $6.5 \times 10^{-7} \pm 331\%$ |
| | C2 | $3.63 \times 10^{-6} \pm 42\%$ | $1.8 \times 10^{-6} \pm 101\%$ | $3.7 \times 10^{-7} \pm 564\%$ |
| | C3 | $1.13 \times 10^{-6} \pm 131\%$ | $5.0 \times 10^{-6} \pm 37\%$ | $3.6 \times 10^{-6} \pm 60\%$ |

**Tab. 1.** Measured parameters of *MgO* substrates at 15, 52, and 72 *K*.

Author of dissertation precisely measured the complex permittivity of *MgO* substrates from differing batches using the recently developed superconducting post resonator. The resonator was optimised to achieve the high resolution and high sensitivity ($Q_0$ of the empty resonator was $2.4 \times 10^5$ at 14 *K*) to allow for accurate microwave characterization of low loss planar dielectrics. The constructed resonator and developed *EM* analysis enabled computations of the real part of relative permittivity with uncertainty of ±0.5% (allowing for ±0.3% uncertainty in thickness of substrates). The tested *MgO* substrates exhibited constant values of $\varepsilon'_r$ with temperature in the range from 14 to 84*K*. The average value of permittivity $\varepsilon'_r$ was 9.63 with variation of ±0.5% between the samples, equal to the measurement uncertainty. The average permittivity $\varepsilon'_r$ differs by 2.8% as compared to the highest literature data. The differences of 0.5% and 2.8% in permittivity $\varepsilon'_r$ would result in a change of resonance frequency $f_{res}$ of an *HTS* resonator deposited on this substrate from a predicted value of 1.850 to 1.845 *GHz* and 1.826 *GHz* respectively.

Author of dissertation examined the losses in tested *MgO* substrates as a function of input microwave power from −20 to +15 *dBm* in Fig 10.

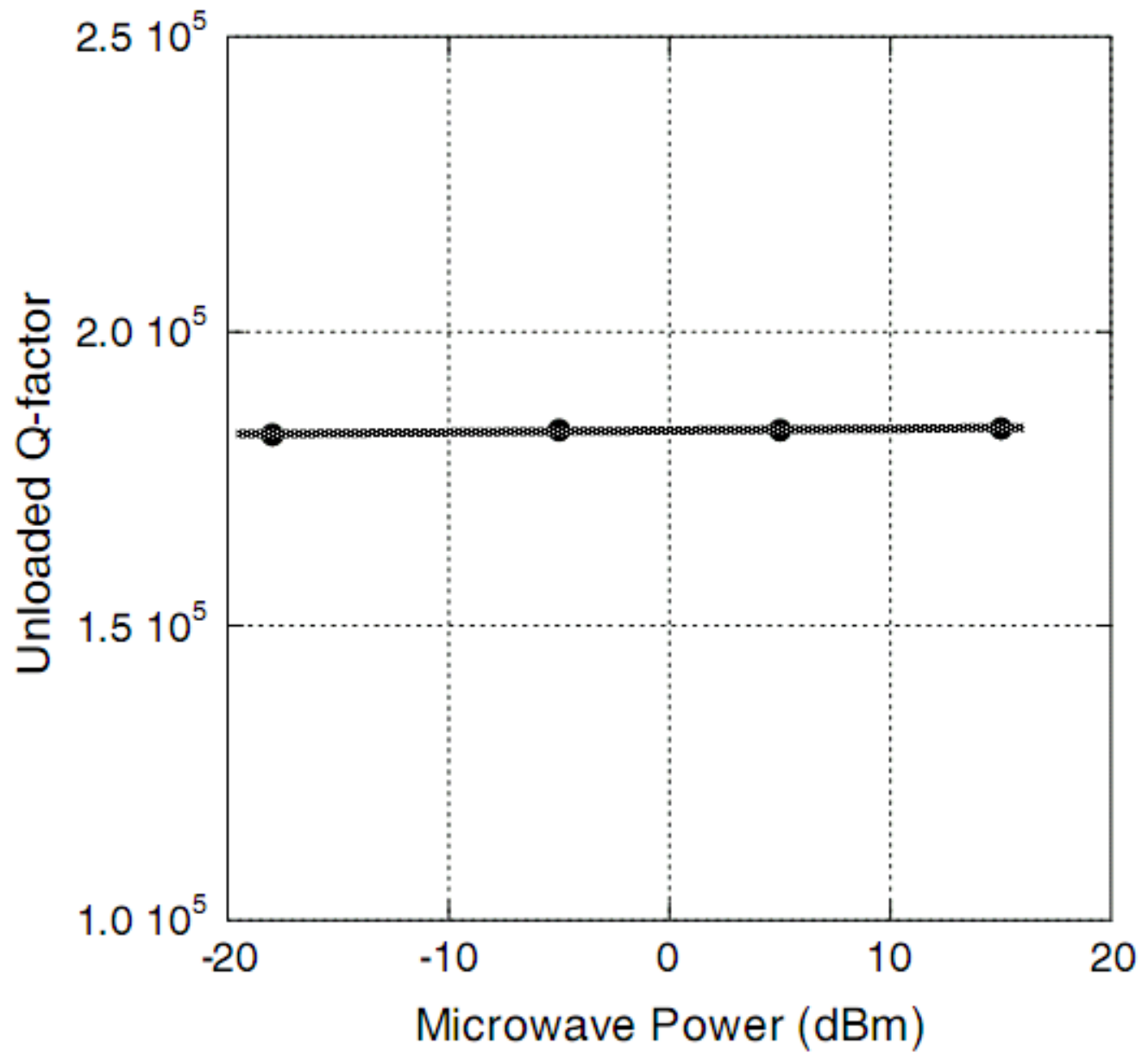

**Fig 10.** Dependence of $Q_0$-factor versus microwave input power of split post dielectric resonator at frequency $f$ = 10.4 *GHz* and temperature T = 24 *K*.

The obtained dependence of the unloaded $Q_0$-factor of split post dielectric resonator with sample *A2* at a temperature of 24 *K* does not show any nonlinear behaviour with the increased power up to 15 *dBm*. It is possible to make a conclusion, based on the obtained experimental results, that the developed cryogenic post dielectric resonator has proved to be a useful tool for precise characterization of *MgO* substrates prior to deposition of *HTS* thin films for use in design of *HTS* passive / active electronic devices for applications in wireless communication systems and high performance computing systems.

## 6.4. Experimental Research on Nonlinear Surface Resistance $R_S$ of $YBa_2Cu_3O_{7-\delta}$ Superconducting Thin Films on MgO Substrates in a Dielectric Resonator at Ultra High Frequencies.

The high temperature superconducting thin films of $YBa_2Cu_3O_{7-\delta}$ type on 10x10 $mm^2$ *MgO* substrates were used for investigation of their nonlinear properties in the dielectric resonator with sapphire rod at high microwave power up to 30 *dBm* at ultra high frequencies up to 25 *GHz* in Fig. 11. The high quality superconducting thin films were manufactured by *THEVA GmbH* in Germany with $YBa_2Cu_3O_{7-\delta}$ superconductor layer thickness of 700*nm* and *MgO* substrate thickness of 0.5 mm. The critical temperature of the *HTS* thin films stated by the manufacturer was around $T_c$ = 87*K* with critical current $J_c$ ~ 2.3 *MA/cm*$^2$.

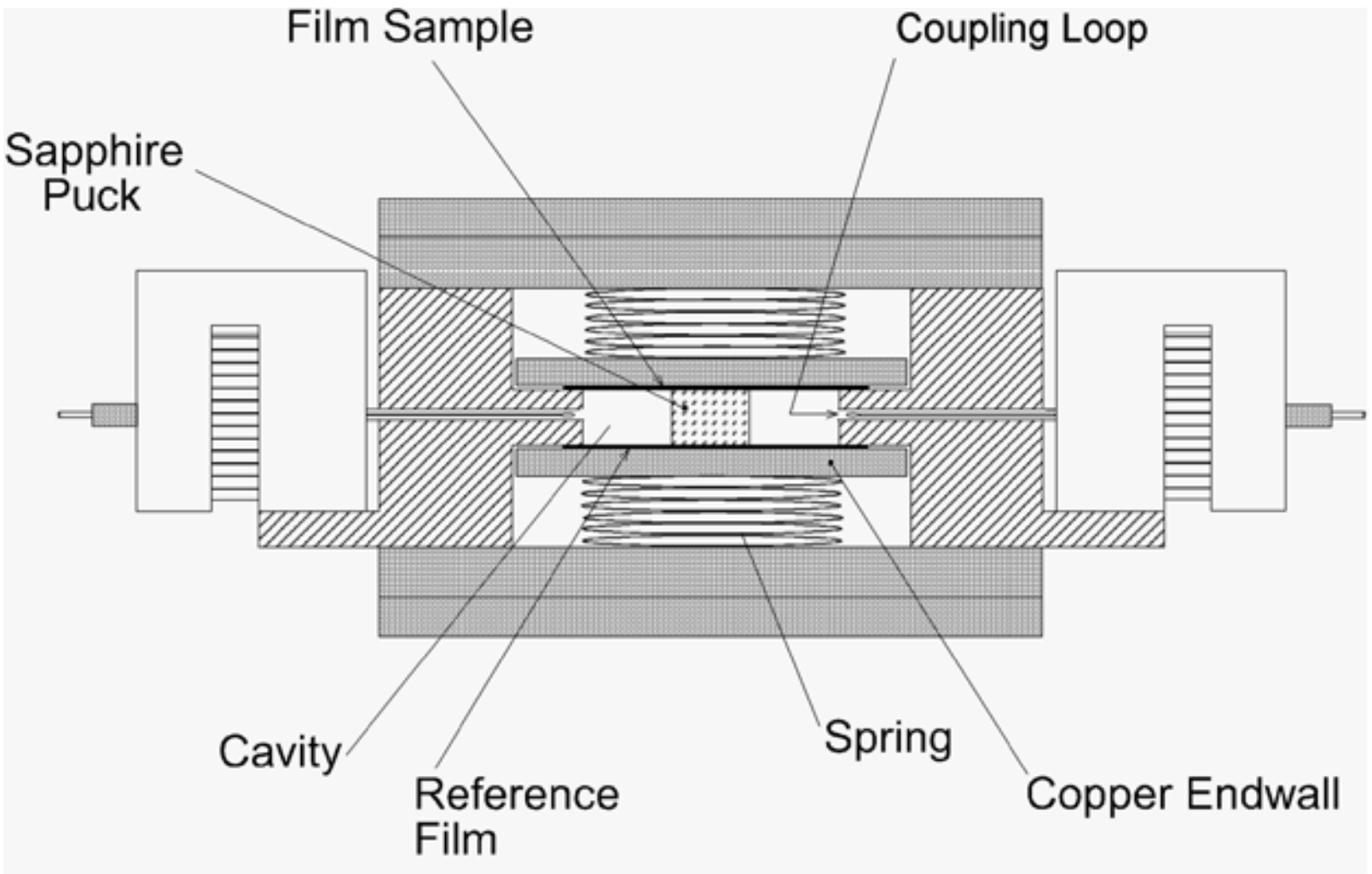

**Fig. 11.** Hakki-Coleman dielectric resonator with *HTS* films (after [6]), where reference and sample films are the superconductors.

## 6.5. Temperature Dependence of Surface Resistance $R_S(T)$ of $YBa_2Cu_3O_{7-\delta}$ Superconducting Thin Films on MgO Substrates in a Dielectric Resonator at Ultra High Frequencies.

The six $YBa_2Cu_3O_{7-\delta}$ on *MgO* thin films have been utilised for the microwave measurements. All thin films were divided into two groups, and each group was measured on the "*round robin*" rotation basis by pairs to enable determination of microwave parameters for each sample. For the first group of three samples, the temperature dependences of sums of surface resistances $R_{s1}+R_{s2}$, $R_{s1}+R_{s3}$, $R_{s2}+R_{s3}$ were obtained in the process of measurements in temperature range from 12*K* to 85*K*. Dependences of surface resistances on temperature: $R_{s1}(T)$, $R_{s2}(T)$ and $R_{s3}(T)$ were derived from the above data. The same measurements were conducted with the second group of samples. The obtained data for dependences of surface resistances on temperature for each tested sample are shown in Figs. 12 and 13. From the measurement results, it can be seen that the values of the surface resistance $R_s$ are slightly different for each sample. The residual surface resistance at temperature 12*K* can be related to the quality of microwave properties of samples [25]. The sample *1* has the lowest surface resistance $R_s$ in the entire temperature range with surface resistance $R_s$ values in the range of 0.3-0.5 $m\Omega$, while the second and third samples have visibly higher values within the range of up to 1.8-2 $m\Omega$ at 75 *K*. The sample 6 of second bunch of films have values of surface resistance $R_s$ presented in Fig. 13 with the lowest surface resistance $R_s$ for this group, but it is still higher of that sample 1 of Fig. 12.

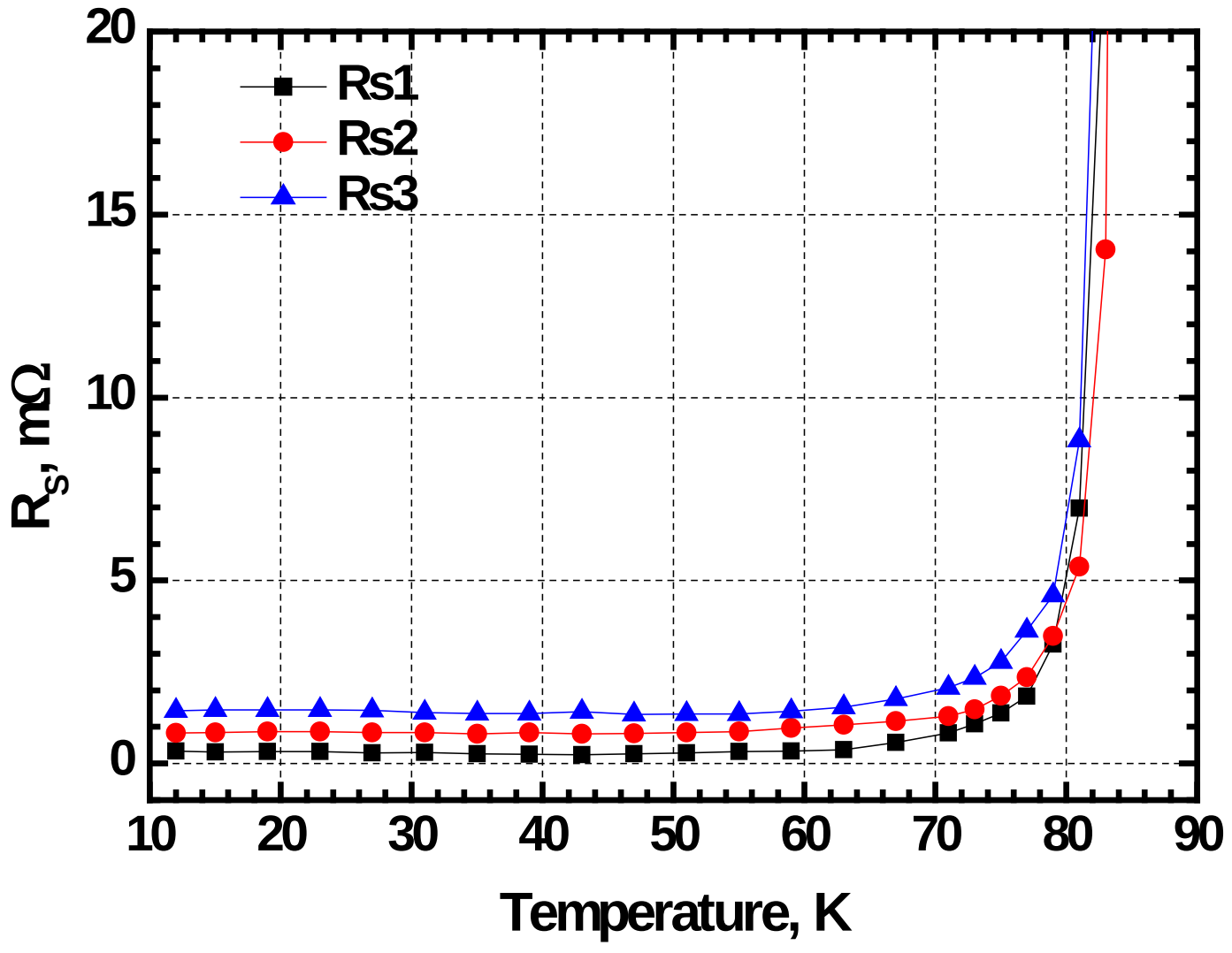

**Fig. 12.** Temperature dependence of surface resistance $R_s(T)$ of $YBa_2Cu_3O_7$ thin films on *MgO* substrate (1-3) at 25*GHz* at -5*dBm* [36].

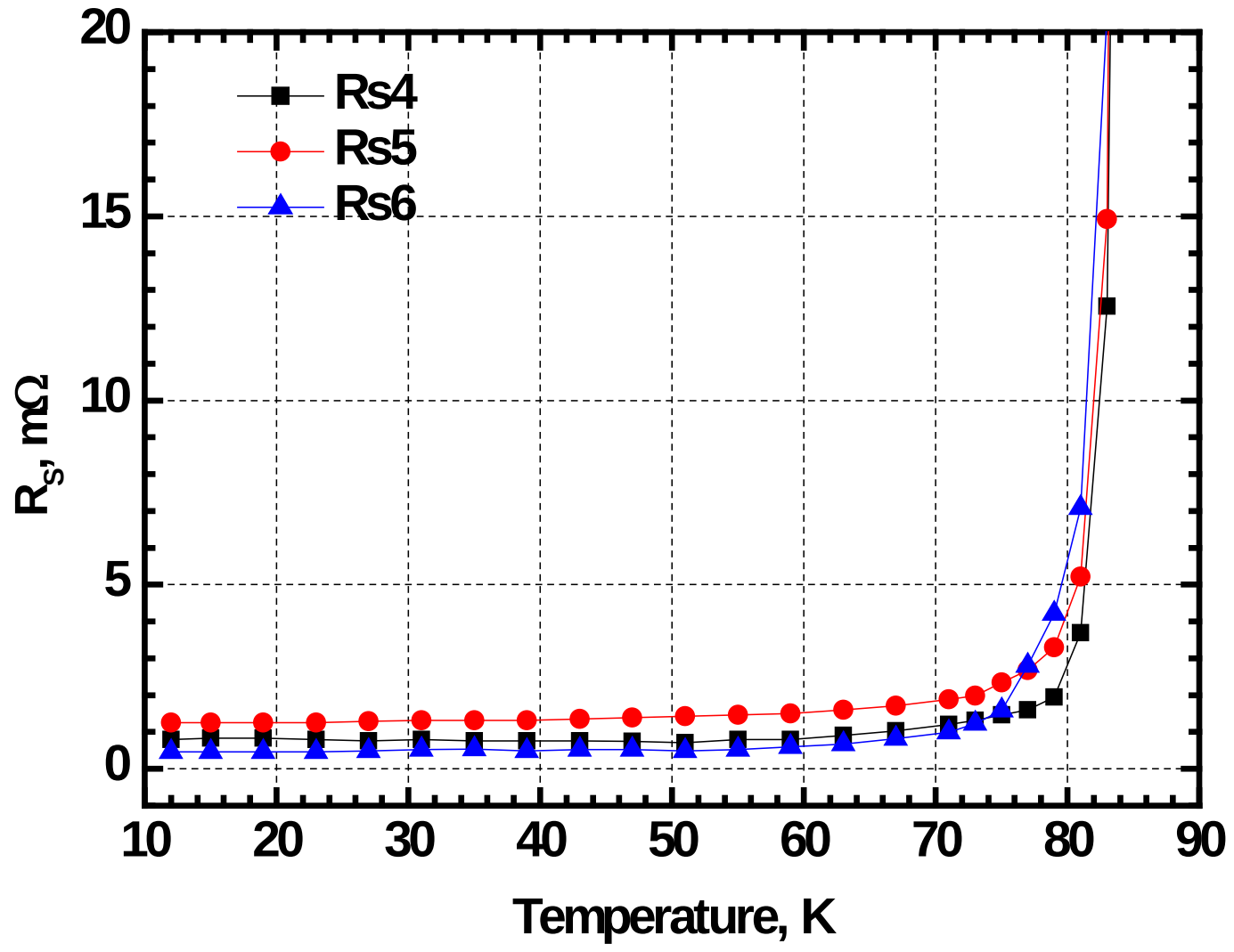

**Fig. 13.** Temperature dependence of surface resistance *Rs(T)* of $YBa_2Cu_3O_7$ thin films on *MgO* substrate (4-6) at 25*GHz* at -5*dBm* [36].

It can also be observed from the obtained results that the rate of change of surface resistance $R_s$ on approaching to critical temperature $T_c$ is different for various samples. The beginning of surface resistance curvature is likely to indicate the temperature up to which the sample will exhibit stable performance at microwaves. Based on this assumption, the samples 2, 3 and 5 are expected to show good microwave properties. It should also be mentioned that the temperatures beyond 80 *K* are not of main interest as they do not reflect the adequate values of surface resistance $R_s$ due to limited superconducting properties of *HTS* thin films near the critical temperature $T_{c.}$

## 6.6. Microwave Power Dependence of Surface Resistance $R_S$(P) of $YBa_2Cu_3O_{7-\delta}$ Superconducting Thin Films on MgO Substrates in a Dielectric Resonator at Ultra High Frequencies.

The accurate measurements of surface resistance *Rs* of $YBa_2Cu_3O_{7-\delta}$ superconductor thin films on *MgO* substrate as a function of microwave power *Rs(P)* were conducted utilizing the measurement system depicted in Fig. 1. The additional microwave power amplifier, connected to the cable, leading to the input port of microwave resonator under test inside the vacuum dewar, enabled measurements from - 18*dBm* to + 30*dBm* microwave signal amplitude level. It should be mentioned that at higher microwave power levels, the attenuator was placed at output port of microwave resonator to prevent the high microwave power signals entering the input port of vector network analyser system. The attenuator had 20*dBm* attenuation characteristics. The amplifier had a maximum of +20*dBm* to +25*dBm* microwave signal amplification range, and enabled measurements from 0dBm to the maximum range in our experiments +30 *dBm*. Appropriate scale modifications were made to the output signal in order to numerically compensate the attenuation effect.

The experimental results of surface resistance *Rs* vs. microwave power *P* measurements are presented below in Figs. 14, 15.

Fig. 14 displays that the microwave power *P* is dependent on the surface resistance *Rs* of $YBa_2Cu_3O_7$ superconductor thin films (1-2) at different temperatures *T = 25K* and 50*K*. The (1-2) pair of *HTS* thin film samples shows almost nonlinear dependence of surface resistance $R_s(P)$ at temperature T of 25*K* and 50*K* with the minimal dependence $R_S(P)$ at ~ 10dBm.

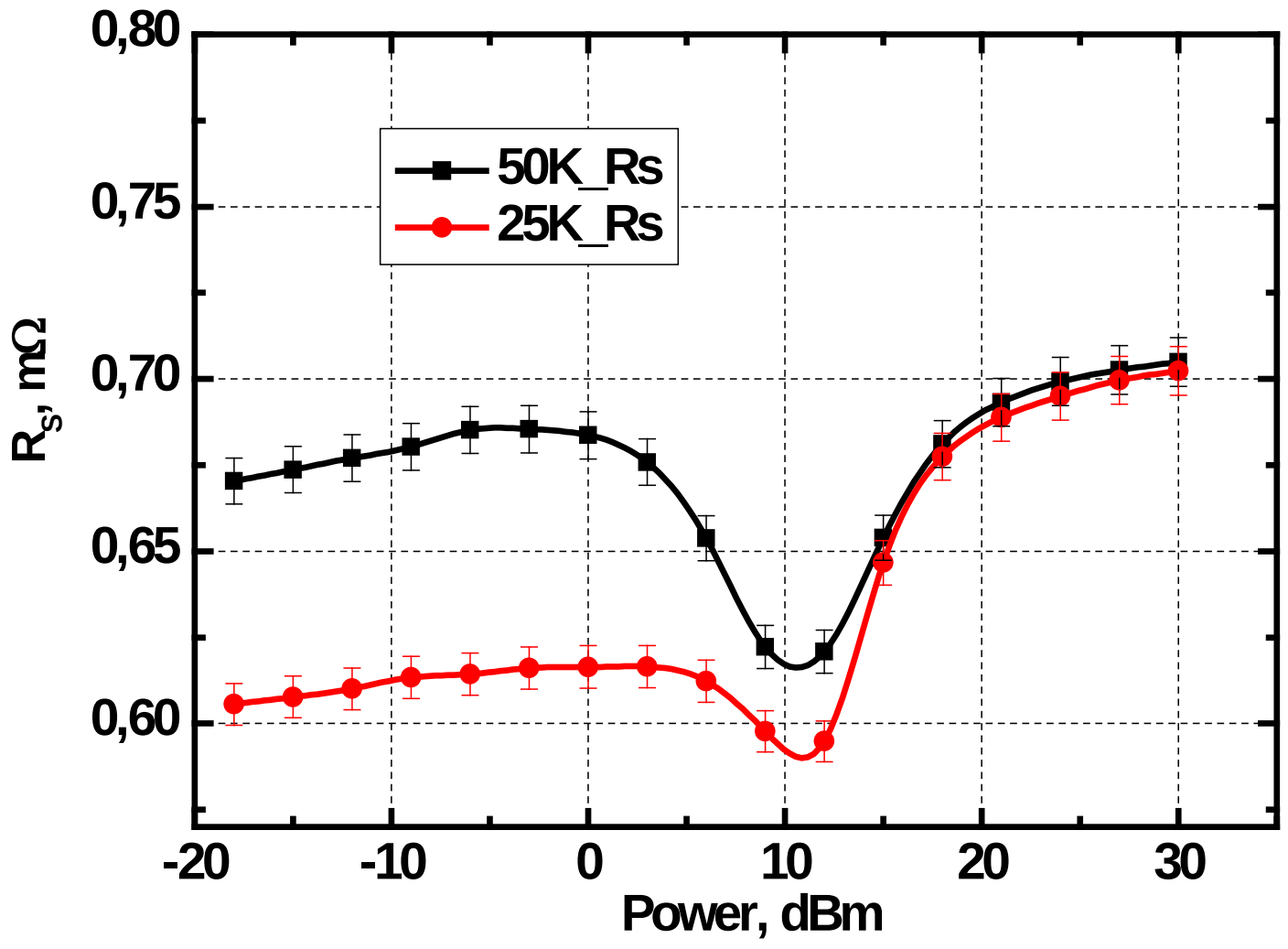

**Fig. 14.** Microwave power dependence of surface resistance *Rs(P)* of $YBa_2Cu_3O_{7-\delta}$ thin films on *MgO* substrate at 25*GHz* at *T* = 25 *K*, 50 *K* (1-2) [37].

Fig. 15 demonstrates that the samples (1-3) reveal nearly identical surface resistance $R_s(P)$ values at both temperatures *T* of 25*K* and 50*K*.

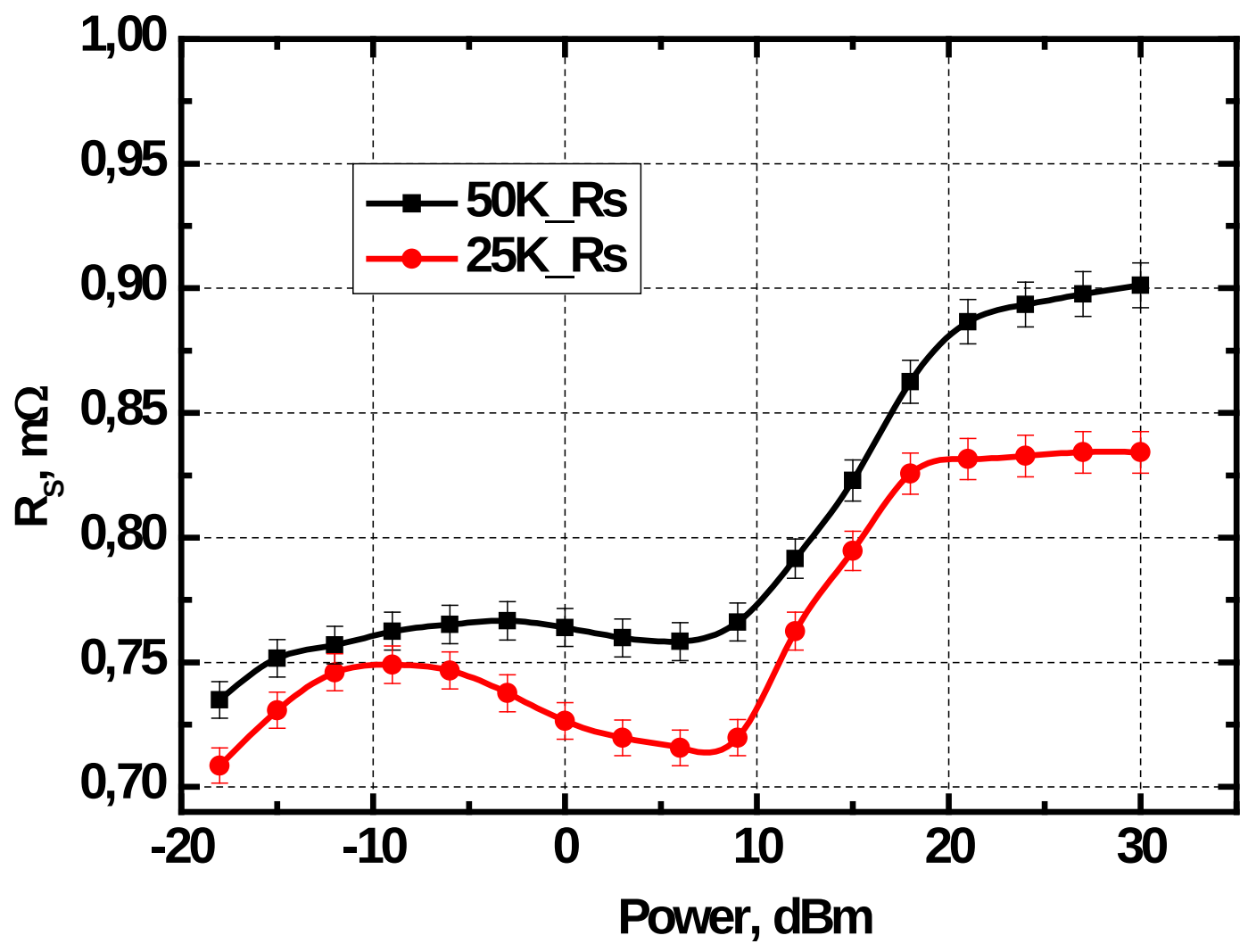

**Fig. 15.** Microwave power dependence of surface resistance *Rs(P)* of $YBa_2Cu_3O_{7-\delta}$ thin films on *MgO* substrate at 25*GHz* at *T* = 25 *K*, 50 *K* (1-3) [37].

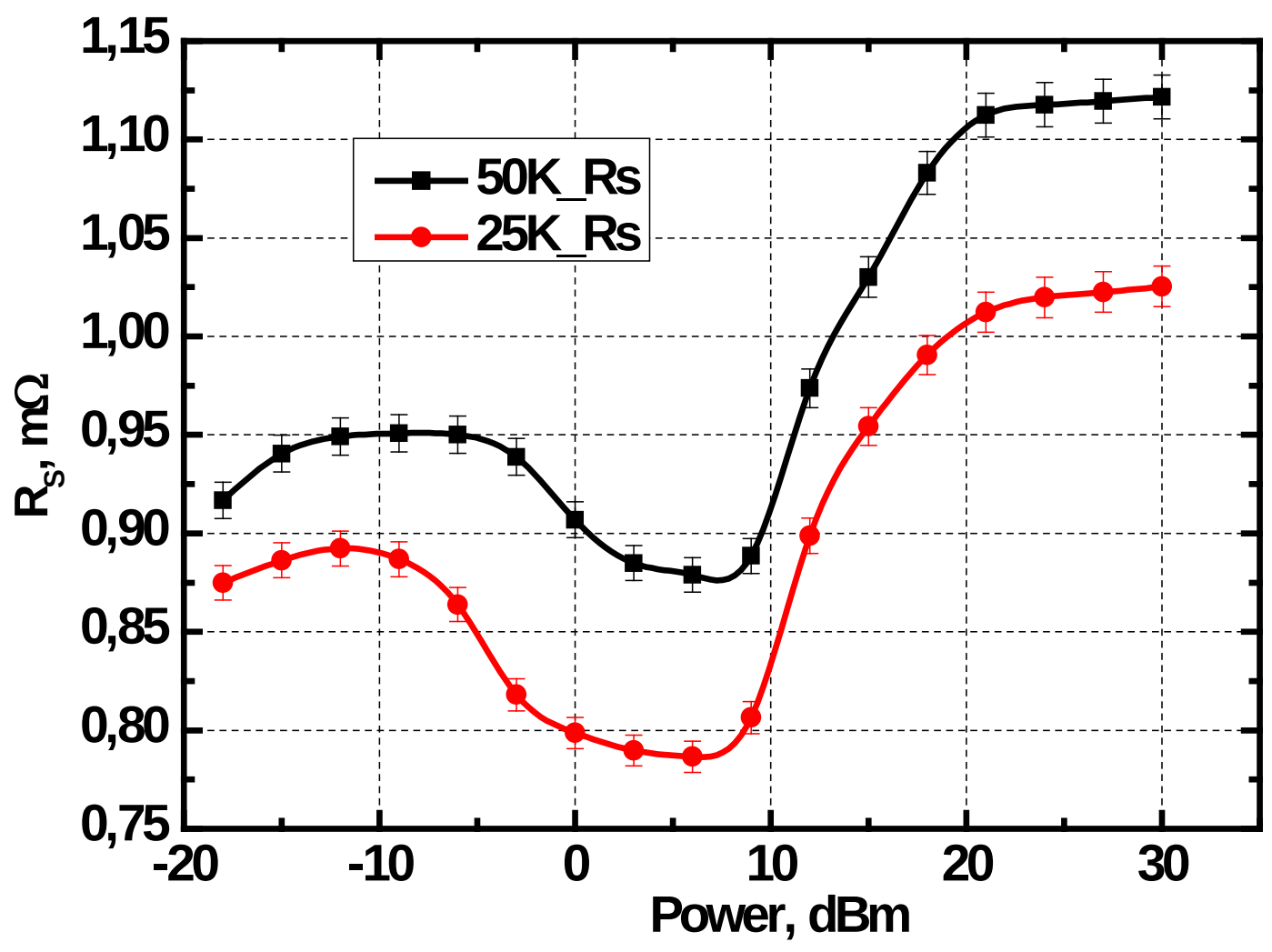

**Fig. 16.** Microwave power dependence of surface resistance $R_s(P)$ of $YBa_2Cu_3O_{7-\delta}$ thin films on *MgO* substrate at 25*GHz* at T = 25 *K*, 50 *K* (2-3) [37].

Fig. 16 shows the nonlinear dependence of surface resistance on microwave power $R_s(P)$ of $YBa_2Cu_3O_{7-\delta}$ superconductor thin film samples (2-3). In the range of microwave signal level 0-10 *dBm*, the surface resistance $R_s$ has a minimum, which is the same as for other samples, that is in agreement with data obtained in [4-8].

Fig. 17 demonstrates that the microwave power dependent surface resistance $R_s(P)$ of $YBa_2Cu_3O_{7-\delta}$ (4-5) is slightly changing at around 1mΩ at temperatures of 25*K* and 50*K*. At higher temperature range of 75*K*, the samples (4-5) have bigger values of surface resistance $R_s$ at around 1.5 mΩ with relatively steady behaviour.

Next graph, shown in Fig. 18, illustrates that the superconducting $YBa_2Cu_3O_{7-\delta}$ samples (4-6) reveal almost identical values of surface resistance $R_s$ at wide range of microwave power with a small temperature difference. Although, it should be pointed out that the slight variations of surface resistance $R_s$ were detected in microwave power handling at the range of 1-1.3 *mΩ* for all referred temperatures. The obtained experimental results are in good agreement with data measured by other research groups.

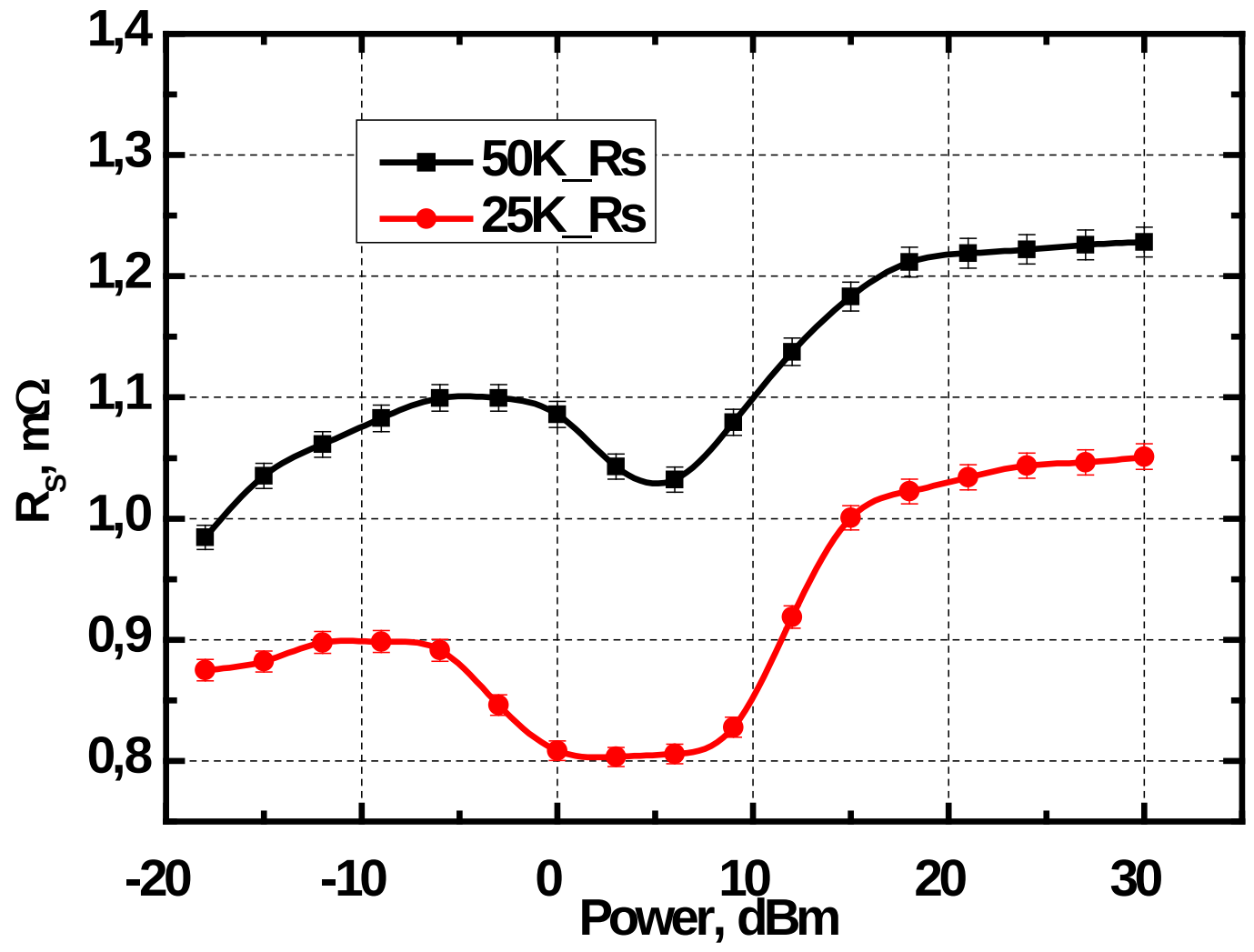

**Fig. 17.** Microwave power dependence of surface resistance $R_s(P)$ of $YBa_2Cu_3O_{7-\delta}$ thin films on *MgO* substrate at 25*GHz* at T = 25 *K*, 50 *K* (4-5) [37].

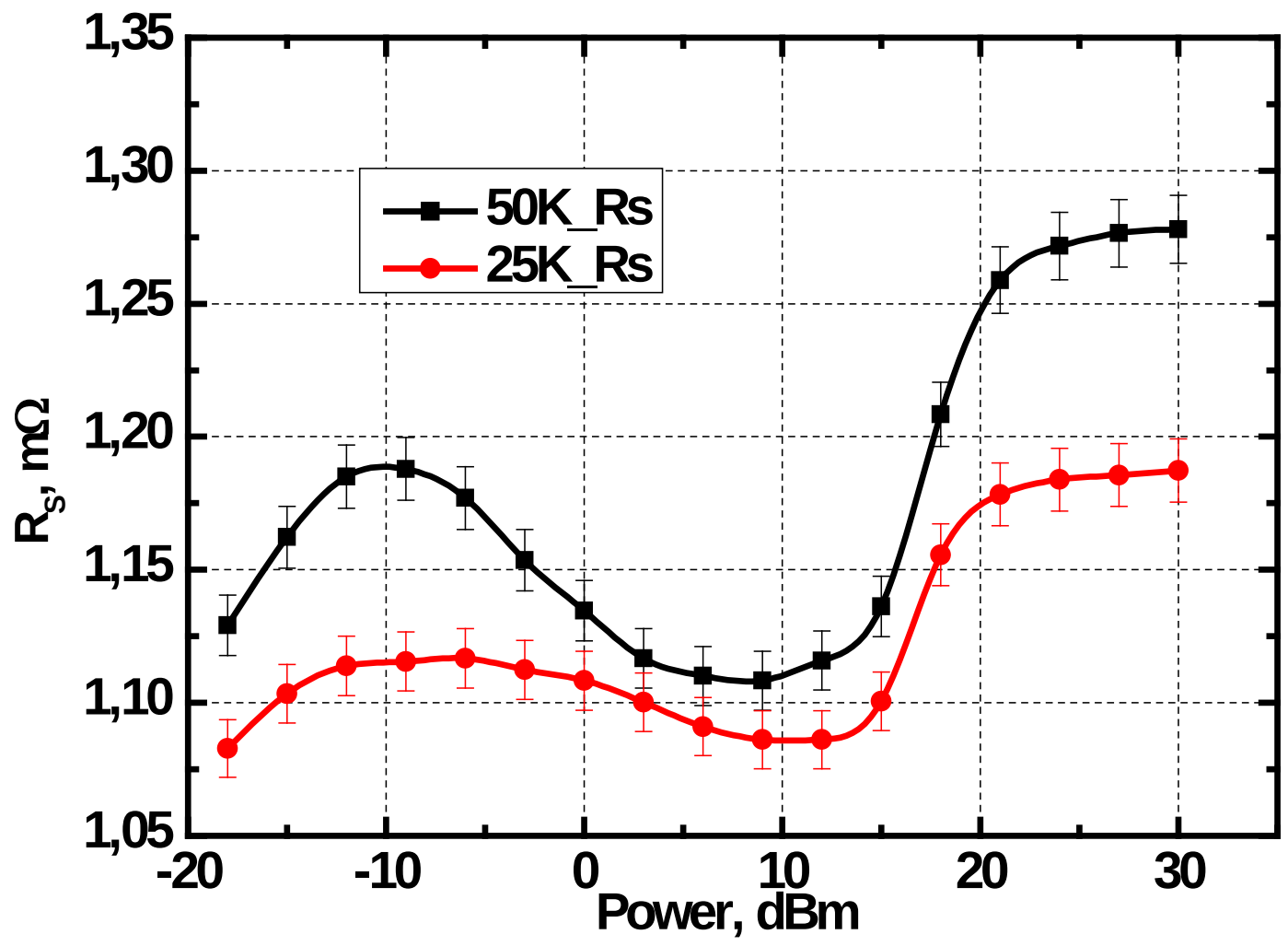

**Fig. 18.** Microwave power dependence of surface resistance $R_s(P)$ of $YBa_2Cu_3O_{7-\delta}$ thin films on *MgO* substrate at 25*GHz* at T = 25 *K*, 50 *K* (4-6) [37].

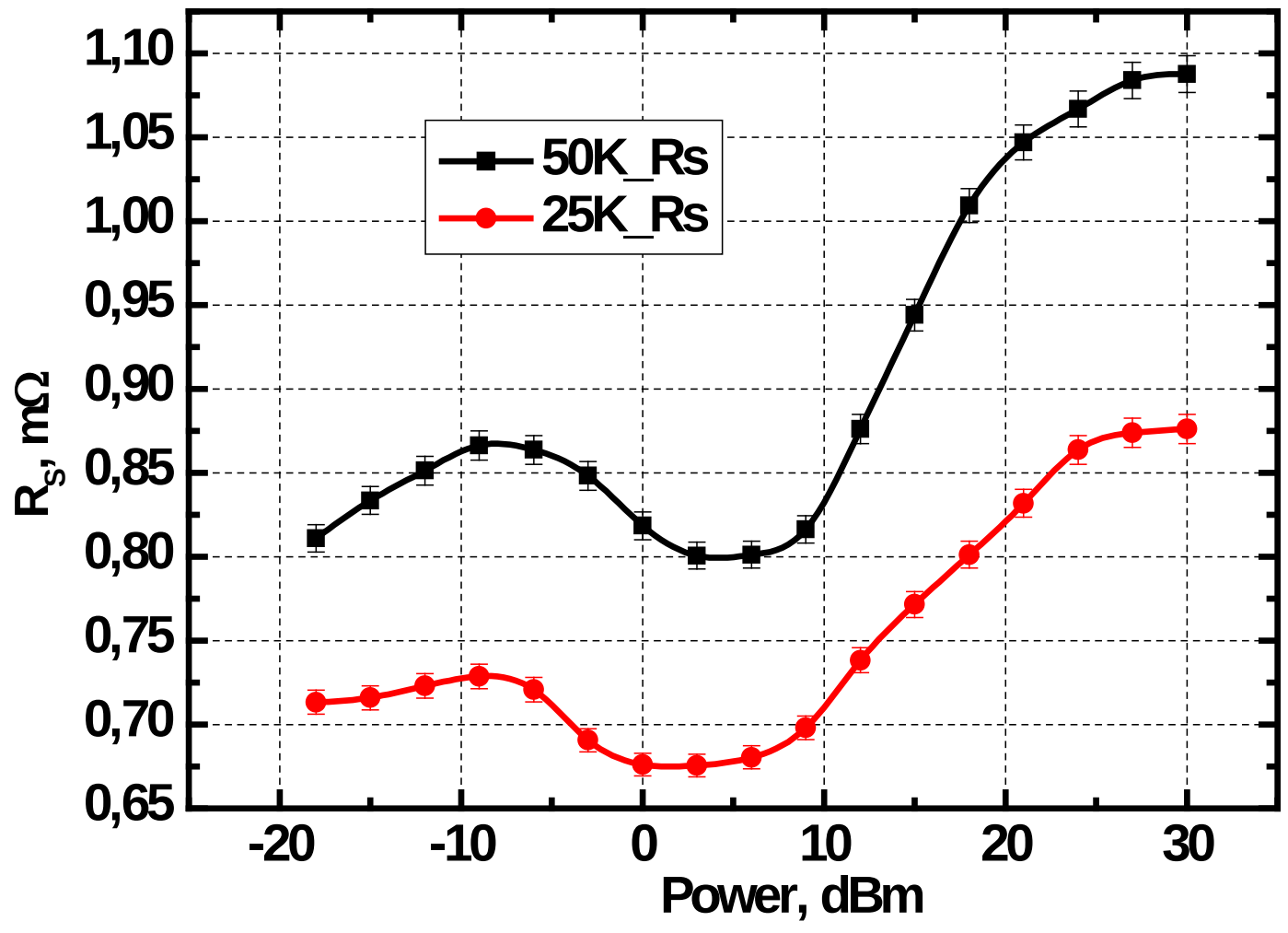

**Fig. 19.** Microwave power dependence of surface resistance $R_s(P)$ of $YBa_2Cu_3O_{7-\delta}$ thin films on *MgO* substrate at 25*GHz* at T = 25 *K*, 50 *K* (5-6) [37].

Fig. 19 displays the *RF* power dependent surface resistance $R_s$ of $YBa_2Cu_3O_{7-\delta}$ thin films (5-6). Some uncharacteristic behaviour is detected at increased temperature range: the surface resistance $R_s$ abruptly decreases for nearly 0.3 *m*$\Omega$ drop at *RF* power level around 8-10 *dBm* from the highest to lowest levels of surface resistance $R_s$ at wide temperature range.

## 6.7. Measurement Accuracy.

The accuracy of measured results is an important issue to consider, when conducting novel research projects. In order to identify the range of acceptable variations in the obtained data, the formulating analysis is made in reference to the equations and parameters related to the obtained results.

The surface resistance has been described earlier in this chapter (eq. 6.1):

$$R_S = A_S \left\{ \frac{1}{Q_0} - \frac{R_m}{A_m} - \frac{1}{Q_d} \right\} \tag{6.4}$$

where $Q_d = 1/p_e \tan\delta$.

In order to determine the magnitude of the correction to the $R_S$ value, i.e. $\Delta R_S$ or relative $\Delta R_S/R_S$, it is necessary to represent all the values in the formula (6.4) with their corresponding independent $\Delta$ corrections, as shown in eq. (6.5).

$$R_S + \Delta R_S = (A_S + \Delta A_S)\left( \frac{1}{Q_0 + \Delta Q_0} - \frac{R_m + \Delta R_m}{A_m + \Delta A_m} - (p_e + \Delta p_e)(\tan\delta + \Delta\tan\delta) \right) \tag{6.5}$$

To eliminate fractions in the denominators, the expression $(A_S+\Delta A_S)$ can be modified as $A_S(1+|\Delta A_S/A_S|)$ in eq. (6.6)

$$\frac{1}{Q_0 + \Delta Q_0} = \frac{1}{Q_0\left(1 + \frac{\Delta Q_0}{Q_0}\right)} = \frac{1}{Q_0}\left(1 - \frac{\Delta Q_0}{Q_0}\right) = \frac{1}{Q_0}\left(1 + \left|\frac{\Delta Q_0}{Q_0}\right|\right) \tag{6.6}$$

It follows from the rule that $1/(1+\Delta x)=1-\Delta x$, which is valid, if $x<<1$. Because the $\Delta x$ may be $\pm|\Delta x|$, we use only $+|\Delta x|$ and the error has a maximum value.

The second part of the equation (6.5) will look like in eq. (6.7)

$$\frac{R_m + \Delta R_m}{A_m + \Delta A_m} = (R_m + \Delta R_m)\frac{1}{A_m}\frac{1}{1+\frac{\Delta A_m}{A_m}} = R_m(1+\left|\frac{\Delta R_m}{R_m}\right|)\frac{1}{A_m}(1+\left|\frac{\Delta A_m}{A_m}\right|) \tag{6.7}$$

and the last term can simply be written as in eq. (6.8)

$$\left(p_e + \Delta p_e\right)\left(\tan\delta + \Delta\tan\delta\right) = \frac{1}{Q_d}\left(1+\left|\frac{\Delta p_e}{p_e}\right|\right)\left(1+\left|\frac{\Delta\tan\delta}{\tan\delta}\right|\right) \tag{6.8}$$

After substitutions and multiplications, all related terms higher than first degree of Δ can be discarded as in eq. (6.9)

$$1+\left|\frac{\Delta R_S}{R_S}\right| = \frac{A_S}{R_S}\left(1+\left|\frac{\Delta A_S}{A_S}\right|\right)\left[\begin{array}{l}\frac{1}{Q_0}\left(1+\left|\frac{\Delta Q_0}{Q_0}\right|\right) - \frac{R_m}{A_m}\left(1+\left|\frac{\Delta R_m}{R_m}\right|\right)\left(1+\left|\frac{\Delta A_m}{A_m}\right|\right) \\ -\frac{1}{Q_d}\left(1+\left|\frac{\Delta p_e}{p_e}\right|\right)\left(1+\left|\frac{\Delta\tan\delta}{\tan\delta}\right|\right)\end{array}\right] \tag{6.9}$$

It can be observed that $|\Delta R_S/R_S|$ is a function of the six independent variables $\Delta A_S$, $\Delta Q_0$, $\Delta R_S$, $\Delta A_m$, $\Delta p_e$ and $\Delta\tan\delta$.

Now, the coefficients for these variables need to be found as well as the expression for $|\Delta R_S/R_S|$ as a multidimensional vector, where the mentioned independent variables are vectors, needs to be formulated. To calculate the coefficients, the term $|\Delta R_S/R_S|$ will be present in the left side of the formula (6.7). In the right side though, only a variable in interest will be left, and all the others will supposed to be zeros. Thus, a sum of six terms will be obtained in the right part, where each term depends on its own variable only, as shown in eq. (6.10).

$$\left|\frac{\Delta R_S}{R_S}\right| = C_1\left|\frac{\Delta A_S}{A_S}\right| + C_2\left|\frac{\Delta Q_0}{Q_0}\right| + C_3\left|\frac{\Delta R_m}{R_m}\right| + C_4\left|\frac{\Delta A_m}{A_m}\right| + C_5\left|\frac{\Delta p_e}{p_e}\right| + C_6\left|\frac{\Delta\tan\delta}{\tan\delta}\right| \tag{6.10}$$

It follows from eq. (6.9) that the $C_1$, $C_2$, $C_3$, $C_4$, $C_5$ can be written as in eq. (6.11):

$$
\begin{aligned}
C_1 &= \left(\frac{A_S}{R_S}\right)\left(\frac{1}{Q_0} - \frac{R_m}{A_m} - \frac{1}{Q_d}\right),\\
C_2 &= \left(\frac{A_S}{R_S Q_0}\right),\\
C_3 = C_4 &= \left(\frac{R_m A_S}{R_S A_m}\right),\\
C_5 = C_6 &= \left(\frac{A_S}{R_S Q_d}\right).
\end{aligned}
\tag{6.11}
$$

In multidimensional space, the vector length square is equal to the sum of squares of the orthogonal vectors in eq. (6.12)

$$
\left|\frac{\Delta R_S}{\Delta R}\right|^2 = C_1^2\left|\frac{\Delta A_S}{A_S}\right|^2 + C_2^2\left|\frac{\Delta Q_0}{Q_0}\right|^2 + C_3^2\left|\frac{\Delta R_m}{R_m}\right|^2 + C_4^2\left|\frac{\Delta A_m}{A_m}\right|^2 + C_5^2\left|\frac{\Delta p_e}{p_e}\right|^2 + C_6^2\left|\frac{\Delta \tan\delta}{\tan\delta}\right|^2 \tag{6.12}
$$

or in eq. (6.13)

$$
\begin{aligned}
\left|\frac{\Delta R_S}{R_S}\right|^2 &= \left(\frac{A_S}{R_S}\right)^2\left(\frac{1}{Q_0} - \frac{R_m}{A_m} - \frac{1}{Q_d}\right)^2\left|\frac{\Delta A_S}{A_S}\right|^2 + \left(\frac{A_S}{R_S Q_0}\right)^2\left|\frac{\Delta Q_0}{Q_0}\right|^2\\
&+ \left(\frac{R_m A_S}{R_S A_m}\right)^2\left(\left|\frac{\Delta R_m}{R_m}\right|^2 + \left|\frac{\Delta A_m}{A_m}\right|^2\right) + \left(\frac{A_S}{R_S Q_d}\right)^2\left(\left|\frac{\Delta p_e}{p_e}\right|^2 + \left|\frac{\Delta \tan\delta}{\tan\delta}\right|^2\right)
\end{aligned}
\tag{6.13}
$$

The magnitude of $|\Delta R_S/R_S|$ is equal to the square root of the right part of eq. (6.12).

The next step requires making certain assumptions about the magnitudes of the involved terms. $\Delta As$, $\Delta Am$ and $\Delta p_e$ are measured with the accuracy *0.01*, that is to say *1%*; $\Delta tan\ \delta$ has a very small value, because tan $\delta$ is very small for the dielectrics $(10^{-6} – 10^{-8})$ at low temperatures and $Q_d$ is very large ~ $10^7$.

Therefore, the $C_5$ and $C_6$ are very small, and these members make a very little contribution to the error of experiment and they can be disregarded. It is also possible to discount both the $\Delta tan\ \delta$ value and all the elements with the $\Delta tan\ \delta$. In this case, the first term is determined by the fact that the uncertainty (error) $\Delta As$, which is equal to *0.01*, is divided by $Q_0$ equalled to around $10^4$, that might be even bigger for dielectric resonators. It is clear that this error can be neglected. The second term can be disregarded as well, as it includes $Q_0$ in power of two in the denominator. The following element of the equation will be proportional to $\Delta As$, and consequently, equal to around *0.01*. The fourth term is proportional to $\Delta Am$ and is equal to approximately *0.01*. Therefore, the following two members are present in the $\Delta Rs$ expression in eq. (6.14)

$$\left|\frac{\Delta R_S}{R_S}\right| \approx \left[\left(\frac{A_S}{R_S}\right)^2\left(\frac{R_m}{A_m}\right)^2\left|\frac{\Delta A_S}{A_S}\right|^2 + \left(\frac{R_m A_S}{R_S A_m}\right)^2\left(\left|\frac{\Delta R_m}{R_m}\right|^2 + \left|\frac{\Delta A_S}{A_S}\right|^2\right)\right]^{1/2}$$
$$= \left[\left(\frac{R_m A_S}{R_S A_m}\right)^2\left(\left|\frac{\Delta A_S}{A_S}\right|^2 + \left|\frac{\Delta R_m}{R_m}\right|^2 + \left|\frac{\Delta A_S}{A_S}\right|^2\right)\right]^{1/2} \quad (6.14)$$

The biggest error will be in the case, when all the elements values are added, having taken them by modulo. It is clear that the main error and uncertainty in the *Rs* magnitude depends on the $\Delta Am$ и $\Delta As$ (their values are expressed in *ohms* –same as for the *R*), i.e. on the geometrical factors of normal metal and superconductor, which are determined by the geometry of both the sample and the resonator, and does not depend on the signal level at first approximation.

It is assumed that the geometrical factors do not depend on microwave power or other parameters of the formula at certain temperatures. Then the absolute value of the correction does not depend on microwave power either and is equal to the total value of the absolute corrections.

From the analysis, ***it is evident that the error values will not have practical grounds to exceed a certain range that is approximately equal to around 1% in our case.*** The error band limits at *1%* are shown in Figs. 13 - 18. The additional information can be found in [38].

## 6.8. Discussions on Experimental Results on Microwave Power Dependence of Surface Resistance Rs(P) of $YBa_2Cu_3O_{7-\delta}$ Thin Films on MgO Substrate at Ultra High Frequencies.

The characterisation of microwave behaviour of surface resistance $R_s$ in relation to magnitudes of microwave power is an important issue in the field of microwave superconductivity. The overall obtained experimental results display the relatively steady flat behaviour of surface resistance $R_s$ indicating the high microwave power handling capabilities of $YBa_2Cu_3O_{7-\delta}$ superconducting thin films. However, the nonlinear effects were detectable for some researched *HTS* samples at elevated microwave power levels. It should also be mentioned that the anomalous microwave power / magnetic field effect was observed in researched samples in some cases, where the surface resistance $R_s$ decreases, while the microwave power $P$ increases. It is assumed that the extreme behaviour of $R_s(P)$ is caused by the competition of counteracting effects. The increase of surface resistance $R_s$ is usually expected, because of transition of weak superconducting domains into the normal state in *HTS* thin films at microwaves. The positive field slopes of surface resistance $R_s$ could be evoked by the penetration and motion of *Abricosov magnetic vortices* in high quality *HTS* thin films and /or by the microwave heating in poor quality *HTS* thin films [26, 27]. In contrast to the usual dependence of surface resistance $R_s$, which increases with the increase of microwave power $P$, the decrease of magnitude of surface resistance $R_s(P)$ in microwave power range from -5*dBm* to 10*dBm* is a known unconventional feature.

The different mechanisms were proposed to explain the nonlinearities in microwave superconductivity [28]. In [29], the negative change of the penetration depth and of the surface resistance $R_s$ were researched using the properly chosen networks of grain- boundary *Josephson junctions*, leading to the frequency dependent current redistribution around the dissipative circuit elements. This mechanism could be applied to poor quality *HTS* samples, and be attributed to the weak coupling at boundaries between the grains in *HTS* thin films. Nevertheless, the negative changes of surface resistance on microwave power $R_s(P)$ also occurred in *HTS* thin films, where the granularity played only a minor role. Thus, it is unclear, whether this approach can explain the co-existence of both the nonlinear $R_s(P)$ dependence and the good microwave power handling capabilities of *HTS* thin films.

The nonequilibrium quasi-particle relaxation represents another possible mechanism for explanation of field enhanced superconductivity. The explanation of $R_s(P)$ dependence in *HTS* thin films, using this approach, was proposed in [30].

The observed negative slopes $dR_s/dH < 0$ are attributed in [24] to the current or magnetic field induced ordering of free spins, which may be present in $YBa_2Cu_3O_{7-\delta}$ thin films. Ordering of the impurity spins is found to disturb the spin-flip transitions, and thus to suppress the probability of magnetic pair breaking [31]. The suppression of pair breaking due to the magnetic alignment of impurity spins could lead to the recovery of superconductivity by reducing the number density of normal carriers and increasing the number density of *Cooper pairs* at a correlated rate. Overall, the results on the anomalous dependence of surface resistance on magnetic field $R_s(H)$ reveal the complicated behaviour in [32], leaving the general understanding still open to the discussions.

Typical dependence of the surface resistance $R_s$ of a superconducting film on microwave field $H$ has three regions [33, 34]. Two regions can be roughly distinguished for $R_s(H)$ dependence. The first one expresses the linear behaviour at field strengths below a critical value $H/H_c<1$, where $R_s$ and $X_s$ are the constants, and the nonlinear behaviour at $H/H_c>1$, where $Z_s=R_s + jX_s$ can often be approximated by a power law with temperature dependent coefficients. At field levels above $H_c$, the $R_s(H)$ often increases steeply with the degree bigger than two and the saturation, this increase likely indicates a transition from the superconducting to the normal state. The linear region $R_s$ is of significant importance for the construction of microwave devices, which need to sustain the high microwave field levels at optimum performance.

It is visible that there are two peaks of $R_s(P)$ dependencies at low and high levels of microwave powers. Therefore, we can suppose that the peak or step in surface resistance on microwave power dependence $R_s(P)$ is connected with the absorption of electromagnetic waves on "weak links" (*Josephson junctions*) at low powers. The different mechanism was presented in [4-8], where the increasing $R_S$ in a low microwave power range was connected with the dissipation of electromagnetic wave in the *MgO* substrate. The results are presented in Fig. 20 and Fig. 21.

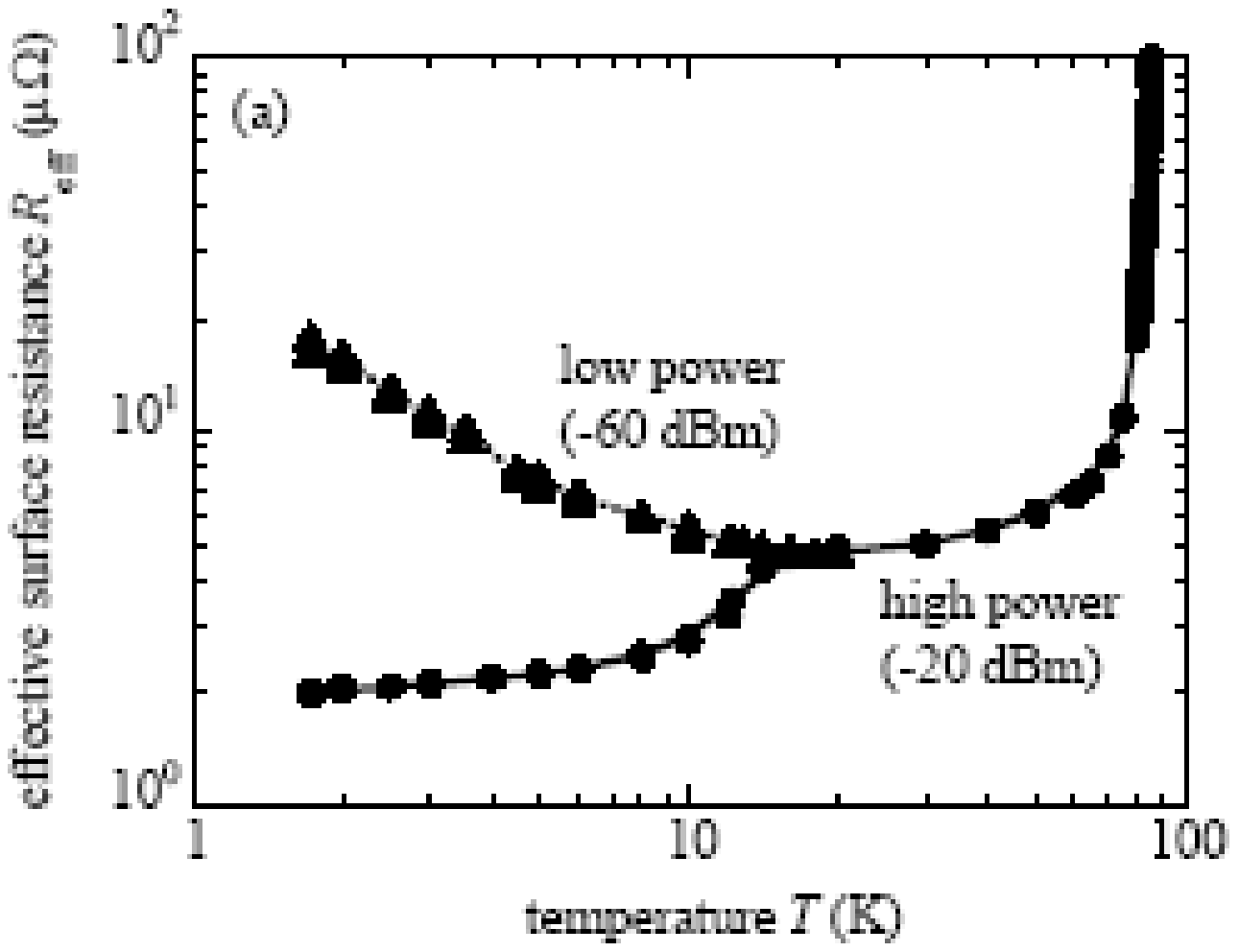

**Fig. 20.** Typical data for effective surface resistance $R_{eff}$ for $YBa_2Cu_3O_{7-\delta}$ on *MgO* substrate at frequency of 2.3 *GHz* at temperatures *T* below 20 *K* (after [4-8]).

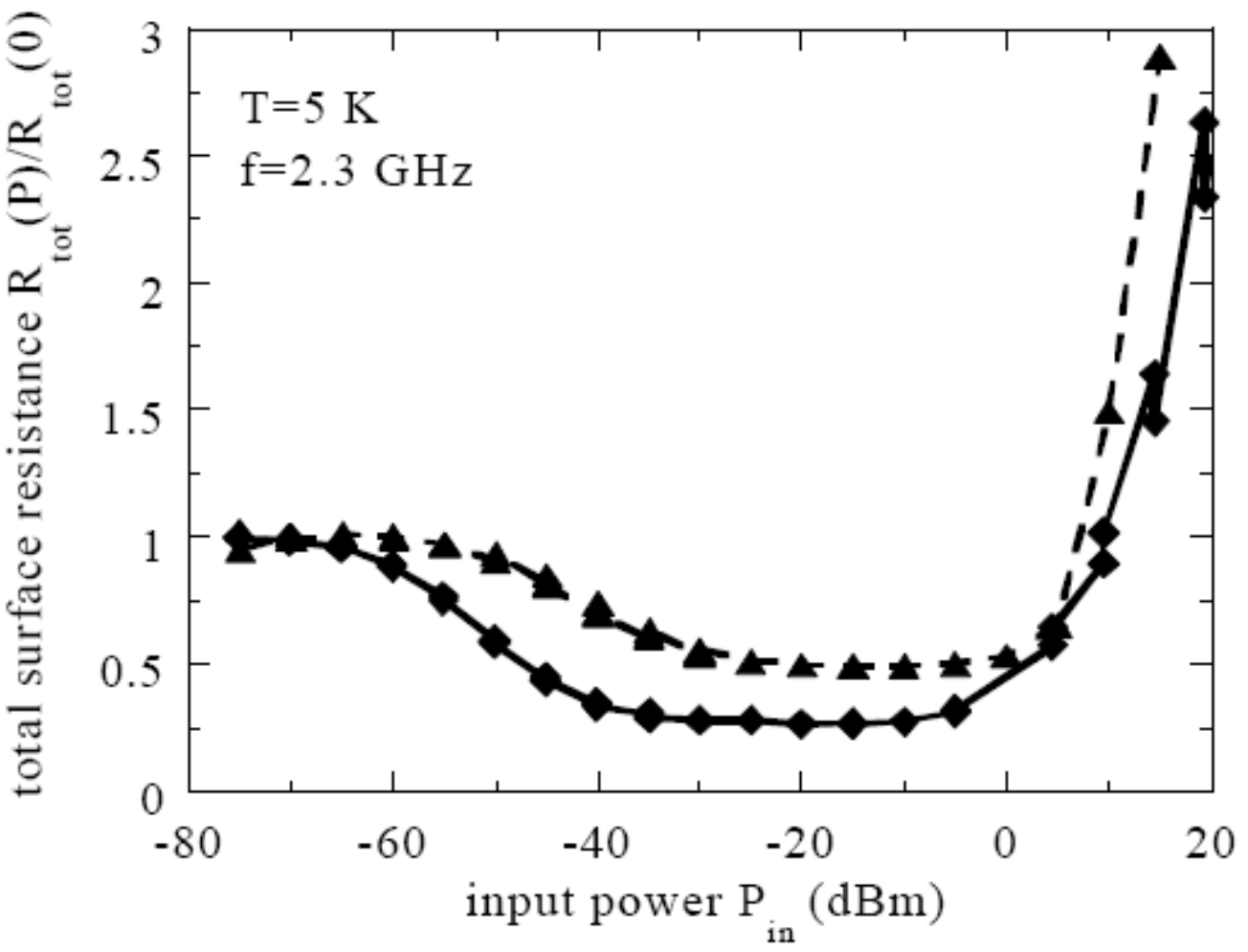

**Fig. 21.** Dependence of total surface resistance $R_{tot}$ for $YBa_2Cu_3O_{7-\delta}$ (diamonds) and *Nb* film (triangles) on MgO substrate on microwave input power $P_{in}$ at $f$=2.3$GH_Z$ at $T$=5$K$ (after [4-8]).

Authors of book believe that the nonlinearities in surface resistance on microwave power dependence $R_s(P)$ in [38], which were observed in $YBa_2Cu_3O_{7-\delta}$ thin films on *MgO* substrate at microwaves during the experimental researches on microwave superconductivity at James Cook University in Australia, are similar to the nonlinear phenomena researched in [4-8]. The nonlinear effects were observed at *slightly* increased microwave power level $P = 10dBm$ at frequency of 25*GHz* at higher temperatures up to T = 50 *K* in course of experimental work at Electrical and Computer Engineering Department at James Cook University in Australia. It is evident that the researched $YBa_2Cu_3O_{7-\delta}$ thin films on *MgO* substrate have ***high quality***, because the nonlinearities were only observed at high magnitudes of microwave power and high temperatures.

In addition, the authors would like to attract the readers' attention to the most recent research paper by Krupka, Wosik, Jastrzebski, Ciuk, Mazierska and Zdrojek, who made an interesting research on the complex conductivity of *YBCO* films in the normal and superconducting states, probed by the microwave measurements, in [39].

Finally, the authors would like to highlight the new experimental results by Mazierska, Leong, Ledenyov, Rains, Zuchowski and Krupka, who reported on the microwave measurements of the surface resistance and complex conductivity of *NdBaCuO* superconducting thin films in [40].

## Summary

The presented research on the nonlinear properties of the $YBa_2Cu_3O_{7-\delta}$ thin films on the *MgO* substrates at the ultra high frequency $f = 25$ *GHz* focuses on the development of practical measurement set up, including the selection of measurement methodology, creation of lumped element model for accurate characterization of $YBa_2Cu_3O_{7-\delta}$ thin films on *MgO* substrates, design of reproducible accurate *Hakki-Coleman dielectric resonator,* calculation of geometrical factors of microwave resonator, calibration of measurement system, verification of microwave resonator performance, design of data processing network and actual experimental work to acquire data on the nonlinear resonance response of $YBa_2Cu_3O_{7-\delta}$ thin films on *MgO* substrates in a dielectric resonator at microwaves [38].

The compromise between accuracy and simplicity was restricted by the size of *HTS* thin films and frequency limitations of the measurement instrumentation. The experimental results are analyzed and discussed comprehensively. As far as the measurement procedure is concerned, it is necessary to note that the surface resistance $R_s$ derived from a single measurement in a *Hakki-Coleman dielectric resonator* is the average value of the two *HTS* thin films used in a microwave resonator. If the three different *HTS* thin films are used in the three possible pair combinations then the values of surface resistance $R_s$ of each individual *HTS* thin film under test can be derived accurately.

Authors of book would like to state that the *S*-type dependence of surface resistance on microwave power $R_s(P)$, which is connected with $H_{C1}$, can be observed in *HTS* thin films in a dielectric resonator at higher microwave power levels [38]. It was found that, on dependence $R_s(P)$, the minimum of surface resistance value $R_s$ is observed in $YBa_2Cu_3O_{7-\delta}$ thin films on *MgO* substrate in the range of microwave signal powers $P = 0 \div 10$ *dBm* at the temperatures $T = 25$ *K*, *50 K* [38]. This minimum of $R_S$ is similar to one, which was observed in range of microwave signal powers $P = -40 \div 0$ *dBm* at temperature $T = 5$ *K* in [4-8]. In the researched case, the minimum of surface resistance $R_s$ appears at the high microwave signal power levels *P=0÷10 dBm* and at the high temperatures *T = 25 K, 50 K* [38]. ***Authors of book propose that the minimum of surface resistance $R_s$ is because of interaction of the system of magnetic dipoles, including the Josephson and Abricosov vortices, with the ultra high frequency signal, but it is not connected with the electrical dipole two level system.*** The same type dependences are also visible on other included graphics.

In relation to the accuracy of surface resistance $R_s$ measurements, the authors of book have to comment that the increase of accuracy of surface resistance $R_s$ measurements requires a trade-off between the aspect ratio of the dielectric rod and the cavity to dielectric rod ratio, (*CDDR),* in [38]. The error band limits on the graphics for the derived surface resistance were calculated and set to $\Delta R_S/R_S=1\%$, and for measured quality factor to $\Delta Q/Q=1\%$ in [38]. The detailed analysis of physical mechanisms responsible for the *S*-type dependences will be presented in forthcoming research papers.

The authors would like to highlight the fact that the surface resistance *Rs* can nonlinearly increase due to the transition by the sub-surface layer of the *HTS* thin film in a mixed state with the Abricosov and Josephson magnetic vortices generation at an increase of the microwave signal power *P* above the magnitude of 8 *dBm*. Authors assume that some additional energy losses have place, because of the microwave power dissipation on the normal metal cores of the magnetic vortices.

Finally, let us to note that the future research on the advanced materials by the researchers at James Cook University is mainly focused on the microwave measurements of the surface resistance and complex conductivity of *NdBaCuO* superconducting thin films in [40].

# CHAPTER 7

# EXPERIMENTAL AND THEORETICAL RESEARCHES ON NONLINEAR SURFACE RESISTANCE OF $YBa_2Cu_3O_{7-\delta}$ THIN FILMS ON MgO SUBSTRATES IN SUPERCONDUCTING MICROSTRIP RESONATORS AT ULTRA HIGH FREQUENCIES

## 7.1. Introduction.

Microwave nonlinear resonance response of $YBa_2Cu_3O_{7-\delta}$ thin films on *MgO* substrates in a *Hakki-Coleman dielectric resonator* (*HCDR*) at ultra high frequencies was researched in Chapters 5, 6. Firstly, the research on the microwave properties of *MgO* substrates in *split post dielectric resonator* (*SPDR*) at $f$ = 10.48 *GHz* was completed. It was found that the *MgO* substrates did not contribute to nonlinear properties of $YBa_2Cu_3O_{7-\delta}$ thin films on *MgO* substrates at ultra high frequencies. Then, the research on the nonlinear surface resistance of $YBa_2Cu_3O_{7-\delta}$ thin films on *MgO* substrates in *Hakki-Coleman dielectric resonator* (*HCDR*) at $f$ = 25*GHz* was conducted. The main result of research is that the $YBa_2Cu_3O_{7-\delta}$ thin films have nonlinear characteristics in form of *S*-type dependence of surface resistance on microwave power *Rs(P)* at elevated microwave power levels, when the magnetic field $H_{rf}$ is higher than $H_{c1}$. The accuracy issues of surface resistance $R_s$ measurement in a dielectric resonator at ultra high frequencies were comprehensively discussed. The concise theoretical explanation of observed minimum of surface resistance $R_s$ in *HTS* thin films at microwaves was proposed.

Chapter 7 focuses on the experimental and theoretical researches on the nonlinear surface resistance $R_s$ of $YBa_2Cu_3O_{7-\delta}$ superconducting thin films on *MgO* substrates in microstrip resonators at ultra high frequencies. As it is noted by Zhou [454], the very compact high-*Q* resonators and filters using planar *HTS* stripline and patch have been recently deployed in microwave subsystems for high-performance radio communication systems, where the filters with very sharp skirts, low insertion loss, small size, reduced weight are of great importance, that is why the nonlinear material properties of *HTS* at high power levels for applications in microstrip filters in transceivers are currently being extensively investigated [454]. Chapter 7 starts with the consideration of influence by ultra high-frequency magnetic field $H_{rf}$ of an electromagnetic wave on the nonlinear surface resistance $R_s(H_{rf})$ in $YBa_2Cu_3O_{7-\delta}$ superconducting thin films on *MgO* substrate in close proximity to the critical magnetic field $H_{c1}$ in the microstrip resonators at microwaves.

## 7.2. Experimental and Theoretical Researches on Nonlinear Surface Resistance of $YBa_2Cu_3O_{7-\delta}$ Thin Films on MgO Substrates in Superconducting Microstrip Resonators at Ultra High Frequencies.

Microstrip resonators made of $YBa_2Cu_3O_{7-\delta}$ superconducting thin films on *MgO* substrates, represent a third type of microwave resonators, which was intensively used to conduct the experimental and theoretical researches on the nonlinear properties of high temperature superconducting thin films at microwaves.

The same measurement set up as in Chapter 6 was employed for the experimental researches on the superconducting microstrip resonators in Chapter 7.

The microstrip resonators have smaller geometrical volumetric dimensions, in which the electromagnetic wave field is concentrated, in comparison with the *Hakki-Coleman dielectric resonator* (*HCDR*), that is why ***the geometric factor of microstrip resonators is significantly smaller. The magnitude of quality factor in the microstrip resonators is smaller than in the dielectric resonators, and is in order of magnitude of a few thousands approximately***, while it is $Q \sim 10^5$ and above in the dielectric resonators. During the research on nonlinear phenomena in superconductors, the above described difference had no influence on the physical characteristics of high temperature superconducting thin films, while the nonlinear phenomena significantly depends on the magnitude of electromagnetic fields, which act on the *HTS* thin films at applied microwave power. In the dielectric resonators with a big magnitude of quality factor, the intensities of magnetic and electric fields are also strong, hence a number of quantums $N_f$ of electromagnetic field – the photons with the frequency *f*, which is close to the resonance frequency $f_0$, inside the dielectric resonator, is big. The *HTS* thin films researched in a dielectric resonator were not patterned, and the electromagnetic fields were changing fluently along the *HTS* thin films, resulting in the disappearance of patches against the electromagnetic fields or currents distributions, which can lead to an appearance of nonlinear phenomena. In contrary, in the microstrip resonators, ***the HTS thin films were patterned, and the increased density of electrical current was normally observed near to the edges of HTS thin films, resulting in a strong appearance of nonlinear phenomena.*** However, the total energy, reserved by the electromagnetic field in a microstrip resonator, is smaller than in a dielectric resonator. Also, the quality factor and a number of photons $N_f$ in a microstrip resonator is smaller than in a dielectric resonator.

In Figs. 1 (a) and 1 (b), the geometrical form of microstrip resonators with the resonance frequency $f_0$ =1.985*GHz* is shown. The microstrip resonators were manufactured from $YBa_2Cu_3O_{7-\delta}$ high-temperature superconductor thin films by *Theva Gmbh* in Germany, and researched by the author of dissertation at Department of Electrical and Computer Engineering, James Cook University in Australia. The thickness of $YBa_2Cu_3O_{7-\delta}$ thin film was equal to 700 *nm* and its width was 0.49 *mm*. The film was deposited on a substrate with the width 0.5 *mm* made from *MgO*. The size of a microstrip resonator was approximately 10x8.5 $mm^2$.

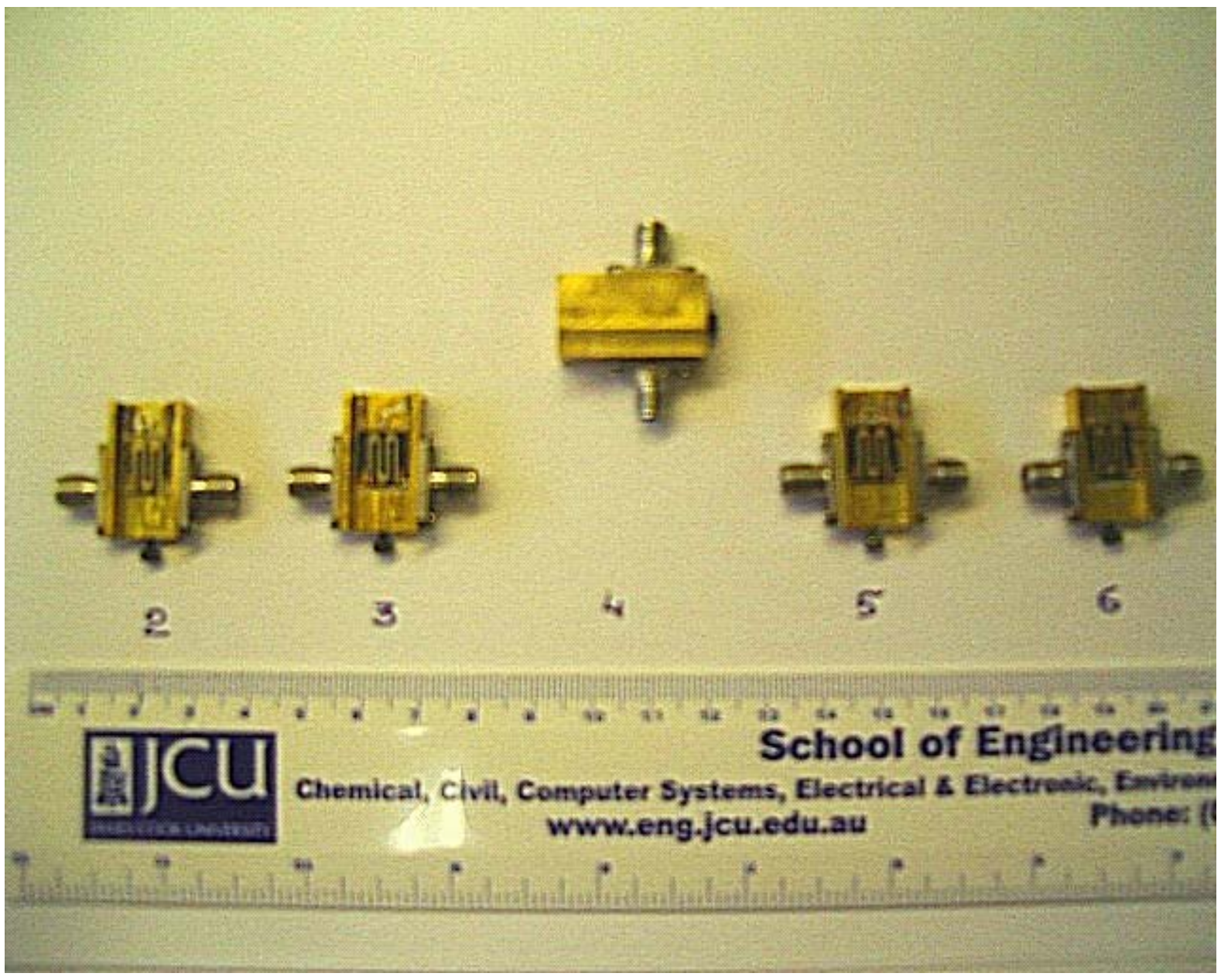

**Fig. 1.a.** Researched microstrip resonators with resonance frequency $f_0$ = 1.985*GHz* made from $YBa_2Cu_3O_{7-\delta}$ thin films on *MgO* substrate.

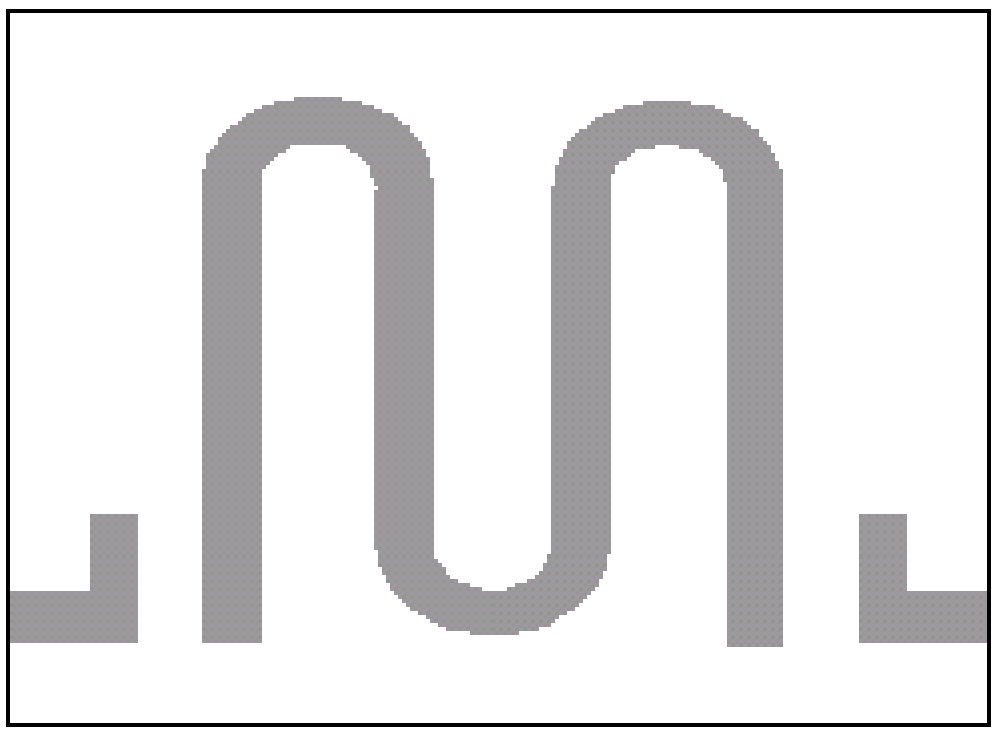

**Fig. 1.b.** Geometric form of microstrip resonators with resonance frequency $f_0$ = 1.985*GHz* made of $YBa_2Cu_3O_{7-\delta}$ thin films on *MgO* substrate researched at Department of Electrical and Computer Engineering at James Cook University in Australia.

In the computer simulation and modeling of nonlinear effects in superconducting microstrip resonators, the microstrip resonators are usually considered as the elementary standing half-wave microwave resonators, in which the superconducting sample looks like a linear plane unit. The microstrip resonator made of *HTS* thin film with composite form can easily be transformed into this relatively simple half-wave straight lines microwave resonator and modeled by an equivalent circuit during the mathematical modeling [1]. It is shown in Fig. 2.

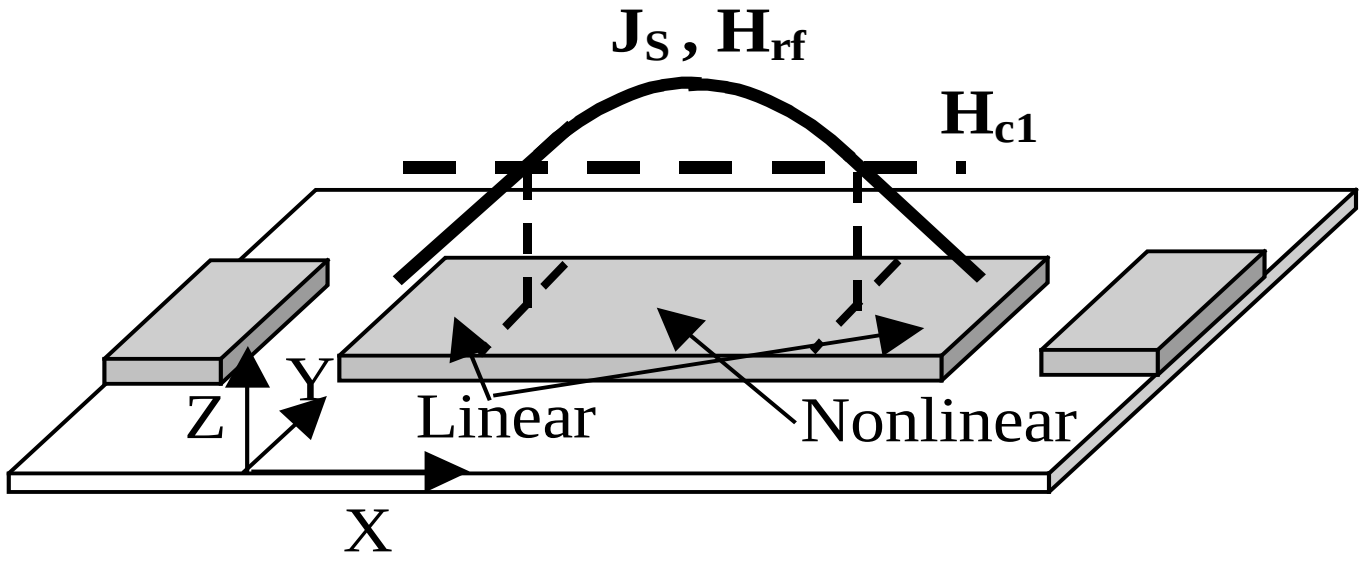

**Fig. 2.** Standing half-wave microstrip resonator and distribution of current density along resonating element.

The part of microstrip resonator, where the magnitude of current density is biggest, and where the magnetic field $H_{rf} \geq H_{c1}$, gives a most important contribution to the nonlinear phenomena. The current density distribution along microstrip resonator is described in eq. (7.1):

$$J_X = J_{max} \sin(2\pi x / \lambda) \sin(\omega t), \qquad (7.1)$$

where $\lambda$ is the wave length and $\omega$ is the cyclic wave frequency.

The nonlinear dependence of surface resistance $R_S$ on amplitude of magnetic field $H_{rf} \propto \sin(\omega t)$ of electromagnetic wave $R_S(H_{rf})$ is a result of an appearance of nonlinear effects in *HTS* thin film at microwaves [30-33]. In the case of a microstrip resonator, it is necessary to take into an account the change of electromagnetic wave amplitude along the *X* axis, which is proportional to $H_{rf} \propto \sin(2\pi x / \lambda)$. In this case, the space discontinuity of an electromagnetic wave will lead to the situation, when a part of sample, where $H_{rf} < H_{c1}$, will give a linear contribution to the surface resistance $R_S$ only. In the central part of a sample, where $H_{rf} \geq H_{c1}$, the surface resistance $R_S$ will be nonlinear. It is possible to write the common expression for the average surface resistance $R_S$ in eq. (7.2):

$$< R_S(H_{rf}) >= \frac{\int_0^{\pi/2}\int_0^{\lambda/4} R_S(t,x)dtd\ell}{\pi/2} = \frac{\int_0^{\pi/2}\int_0^{\lambda/4} \left(R_{SLin}(x,t) + R_{SNlin}(x,t)\right)d\ell dt}{\pi/2}, \quad (7.2)$$

where $R_{SNlin}$ is the nonlinear surface resistance.

The allocation of superconducting current and distribution of magnetic field, which is bound with the current, may vary in *Y-Z* plane of *HTS* thin film at microwaves as shown in Fig 3.

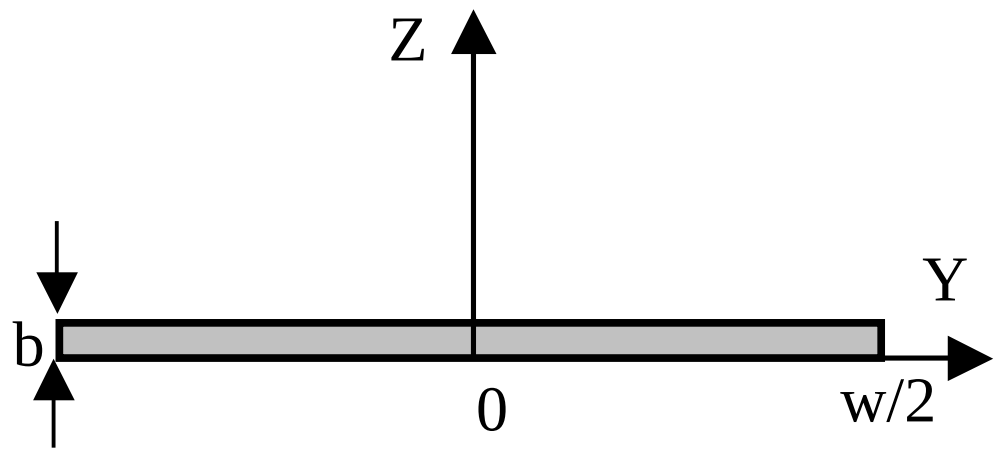

**Fig. 3.** Cross section of *HTS* thin film used in microstrip resonator. Superconducting current transports in direction of *X* axis.

Near to the centre of *HTS* thin film, the current depends on ***y*** as in eq. (7.3) in [1]

$$J_S(y) = \frac{J_S(0)}{\sqrt{1-(2y/w)^2}}, \quad (7.3)$$

where ***y*** is counted from the centre of the film, *w* is the width of *HTS* thin film, the thickness of a film is equal to b in Fig. 3. It is visible, that the current magnitude is incremented as it approaches the edge of the *HTS* thin film. In the area of *HTS* thin film edges, this dependence can not be applied, since it tends to the infinity. In this area, it can be transformed into the expression in eq. (7.4)

$$\mathbf{J_S(y) = J_S(0)\left(\frac{1.165}{\lambda}\right)\left(\frac{wb}{a}\right)^{1/2} \exp\left(-\frac{\left(\frac{w}{2}-|y|\right)b}{a\lambda^2}\right)}, \qquad (7.4)$$

where "a" is the constant ~ 1, $\lambda$ is the penetration depth of magnetic field.

The formula (7.4) features the distribution of current density in close proximity to the edges ~ $\lambda$ of *HTS* thin film. Since $\lambda << w$, this area is rather small in comparison with total square of *HTS* thin film. These formulas (7.3) and (7.4) are valid in the case, when $\lambda \sim b$. In our case, the thickness of *HTS* thin film in a microstrip resonator was equal to 700 *nm*, and the penetration depth $\lambda \sim$ *200 nm*. Thus, it is possible to suppose, that the current density is uniform along the *Z*-axis and featured by the indicated formulas. The simulation result of dependence of current density distribution on cross section of *HTS* thin film in a microstrip resonator with resonance frequency $f_0$ = 1.985*GHz* made of $YBa_2Cu_3O_{7-\delta}$ thin films on *MgO* substrate is shown in Fig. 4 [19, 20].

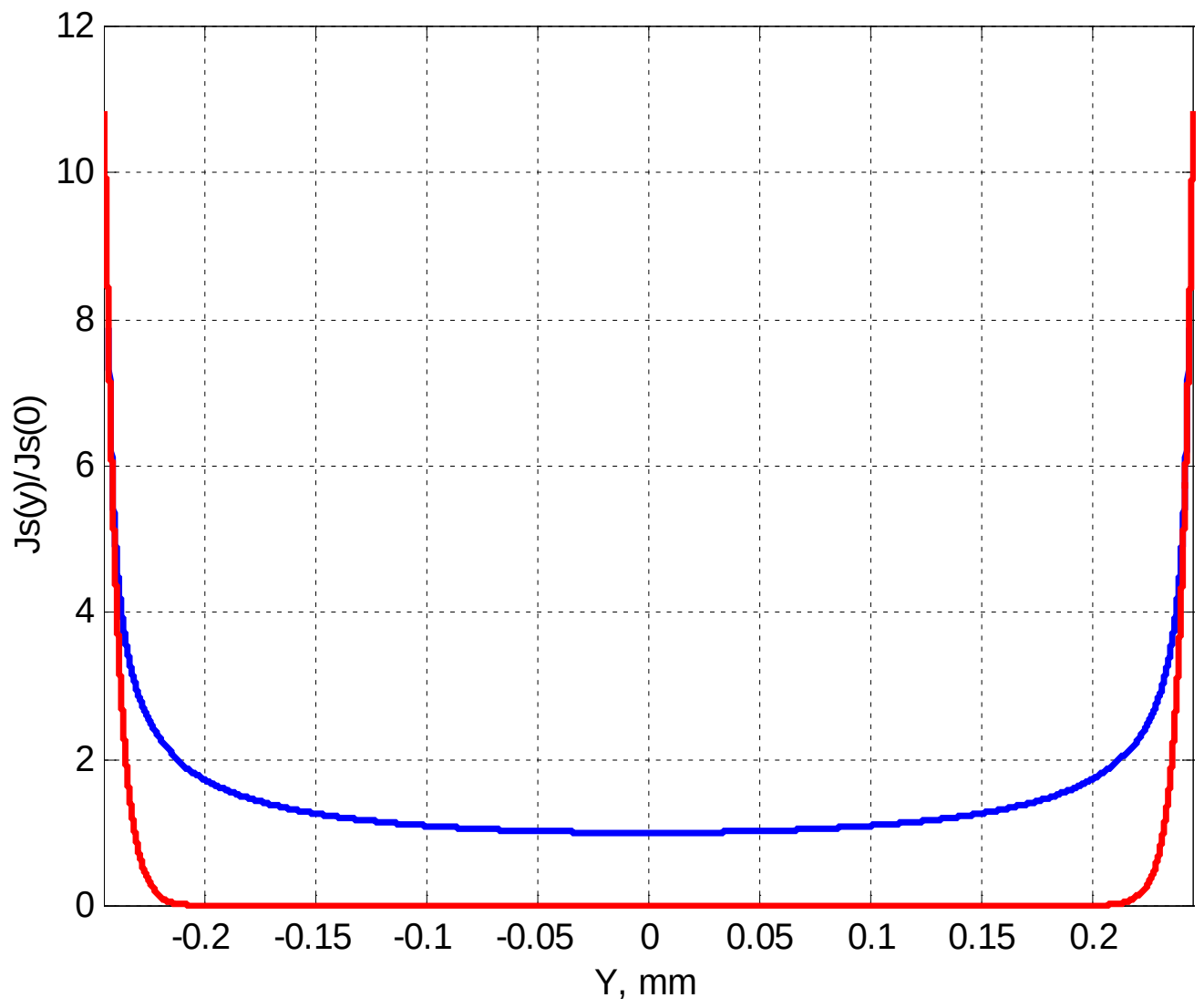

**Fig. 4.** Modeling of current density distribution on cross section of superconducting microstrip resonator with resonance frequency $f_0$ = 1.985*GHz* made of $YBa_2Cu_3O_{7-\delta}$ thin films on *MgO* substrate [19].

The upper curve corresponds to the expression (7.3). The lower curve concerns the formula (7.4) and features the allocation of superconducting current near to the edges of *HTS* thin film, when $y \approx \pm w/2$. The first curve passes to the second curve in the points

$$\mathbf{y = \pm(\frac{w}{2} - \frac{a\lambda^2}{2b})}\,.$$

It is visible, that the density of current varies slightly along the *Y*-axis, and on majority of *HTS* thin film, the ratio $J_S(y)\,/J_S(0) \sim 1$. In close proximity to the edges of *HTS* thin film, the density of current increases sharply. It is clear, that such behaviour of current density is stipulated by the boundary conditions, according to which the magnetic field $H_{rf}$ will not penetrate into the superconductor and only, when the $H_{rf}$ becomes more than $H_{c1}$, this penetration starts in the form of magnetic vortices. Let's note that as soon as it happens, the current density on the edges of *HTS* thin film is reallocated and will penetrate together with the magnetic vortices into the superconducting sample on the bigger depth than $\lambda$, and the current density peak on the boundary will be flattened. Therefore, it is possible to suppose, that in the case of our research interest, the current density is distributed almost uniformly (almost homogeneously) on the cut of *HTS* thin film along *Y*-axis, and the average magnitude of current density exceeds its value at the centre of *HTS* thin film on a very small magnitude.

## 7.3. Ultra High Frequency and Microwave Power Dependences of Transmission Coefficient $S_{21}(f, P)$ in $YBa_2Cu_3O_{7-\delta}$ Superconducting Microstrip Resonator at Microwaves at Different Temperatures.

The *S*-parameters are widely used for the characterization of microwave devices in microwave circuits in micro- and nano-electronics. There are the following *S-parameters* [29]:

$S_{11}$ = input reflection coefficient with the output matched.

$S_{21}$ = forward transmission coefficient with the output matched.

$S_{22}$ = output reflection coefficient with the input matched.

$S_{12}$ = reverse transmission coefficient with the input matched.

The *S*-parameters are important, because of the several reasons [29]:

1. *S*-parameters are determined with resistive terminations. This obviates the difficulties involved in obtaining the broadband open and short circuit conditions required for the *H, Y,* and *Z*-parameters.

2. Parasitic oscillations in active devices are minimized, when these devices are terminated in resistive loads.

3. Equipment is available for determining *S*-parameters since only incident and reflected voltages need to be measured.

Among the *S*-parameters, the forward transmission coefficient $S_{21}$ is an important characteristic of a microwave resonator. It is possible to obtain exact information about the microwave signal transmission properties, exhibited by any microwave resonator, going from the calculation and measurement of forward transmission coefficient $S_{21}$.

The two-port network model was selected to analyze the *transmission line* with S-parameters application as shown in Fig. 5

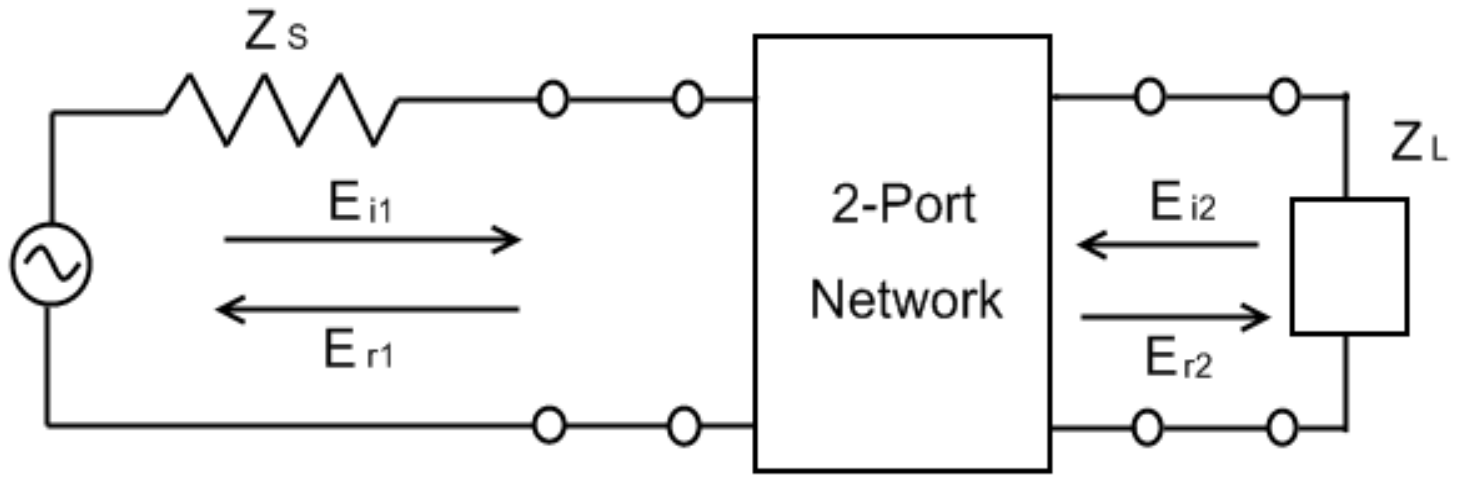

**Fig. 5.** Model of two-port network inserted into transmission line with incident and reflected electromagnetic waves schematically shown at microwaves.

In Fig. 5, the source with impedance $Z_s$ generates the incident wave, which travels along the transmission line to the load with impedance $Z_L$ in the network. The voltage amplitudes of incident waves are $E_{i1}$ and $E_{i2}$ and of reflecting waves are $E_{r1}$ and $E_{r2}$. The variables $a$ and $b$ and $S$-parameters are used to analyze the electromagnetic wave propagation through the network

$$\begin{bmatrix} b_1 \\ b_2 \end{bmatrix} = \begin{bmatrix} S_{11} & S_{12} \\ S_{21} & S_{22} \end{bmatrix} \begin{bmatrix} a_1 \\ a_2 \end{bmatrix},$$

where $a_{(1,2)} = E_{i(1,2)} / Z_0^{1/2}$ and $b_{(1,2)} = E_{r(1,2)} / Z_0^{1/2}$, $Z_0$ is the characteristic impedance of the transmission line.

The forward transmission coefficient $S_{21}$ is

$$S_{21} = \frac{b_2}{a_1} \Big|_{a_2=0} = \frac{E_{r2}}{E_{i1}}.$$

The measurement set up with cryogenic system for accurate microwave characterisation of microstrip resonators at microwaves included:

- Vector Network Analyser (*HP 8722C*),
- Microwave Amplifier,
- Temperature Controller (Conductus *LTC-10*) fitted with a resistive heating element and two silicon temperature diode sensors,

- Vacuum Dewar,
- Close cycle cryogenic laboratory system (*APC-HC4*) suitable for measurements in a wide range of temperatures (10 *K*– 300 *K*),
- Computer system (*IBM-PC*) fitted with a *GPIB* card utilized for the control of temperature controller and network analyser, and measured *S*-parameters data transfer from network analyser to computer.

The measurement set up was precisely calibrated by measuring the voltage standing wave ratio $VSWR=\frac{1+|\Gamma|}{1-|\Gamma|}$, ($\Gamma$ is the reflection coefficient), before accurate experimental measurements of *S*-parameters as schematically presented in Fig. 6.

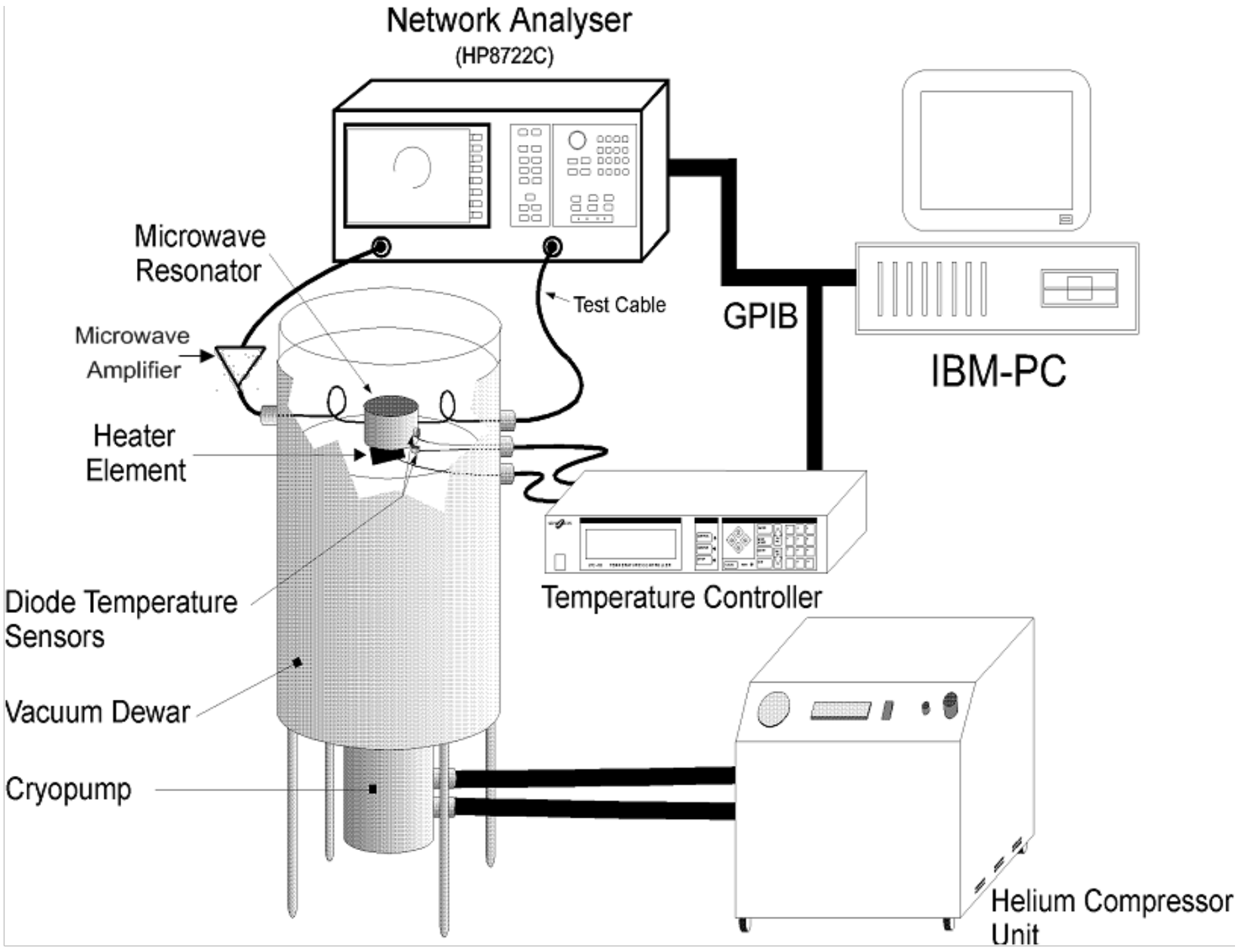

**Fig. 6.** Cryogenic measurement system for research on nonlinear resonance response of $YBa_2Cu_3O_{7-\delta}$ thin films on *MgO* substrates in microstrip resonators at microwaves.

The geometric form of researched microstrip resonators with resonance frequency $f_0$ = 1.985*GHz* made of $YBa_2Cu_3O_{7-\delta}$ thin films on *MgO* substrate is pictured in Fig. 1.b. The thickness of $YBa_2Cu_3O_{7-\delta}$ thin film was equal to 700 *nm* and its width was 0.49 *mm*. The *HTS* thin film was deposited on a substrate with the width 0.5 mm made from *MgO*. The size of a microstrip resonator was approximately 10x8.5 $mm^2$.

In Fig. 7, the characteristic experimental resonance dependences of transmission coefficient on ultra high frequency $S_{21}(f)$ in $YBa_2Cu_3O_{7-\delta}$ superconducting microstrip resonator at different microwave power levels at temperature of 25 *K* are shown. These dependences $S_{21}(f)$ were measured at different microwave power levels from -18 *dBm* up to +30 *dBm* in ultra high-frequencies range of 2*GHz* at temperature of 25*K*. It is visible that the microwave signal strength has a relatively little influence on the value of transmission coefficient $S_{21}$ at low temperature of 25*K*. All the microwave measurements $S_{21}(f)$ were carried out at the temperatures of 25*K*, 35*K*, 45*K*, 55*K*, 65*K*, 75*K* and 82*K*. In some cases, the microwave measurements of transmission coefficient on frequency $S_{21}(f)$ in researched microstrip resonators were conducted at the temperature of 13*K*.

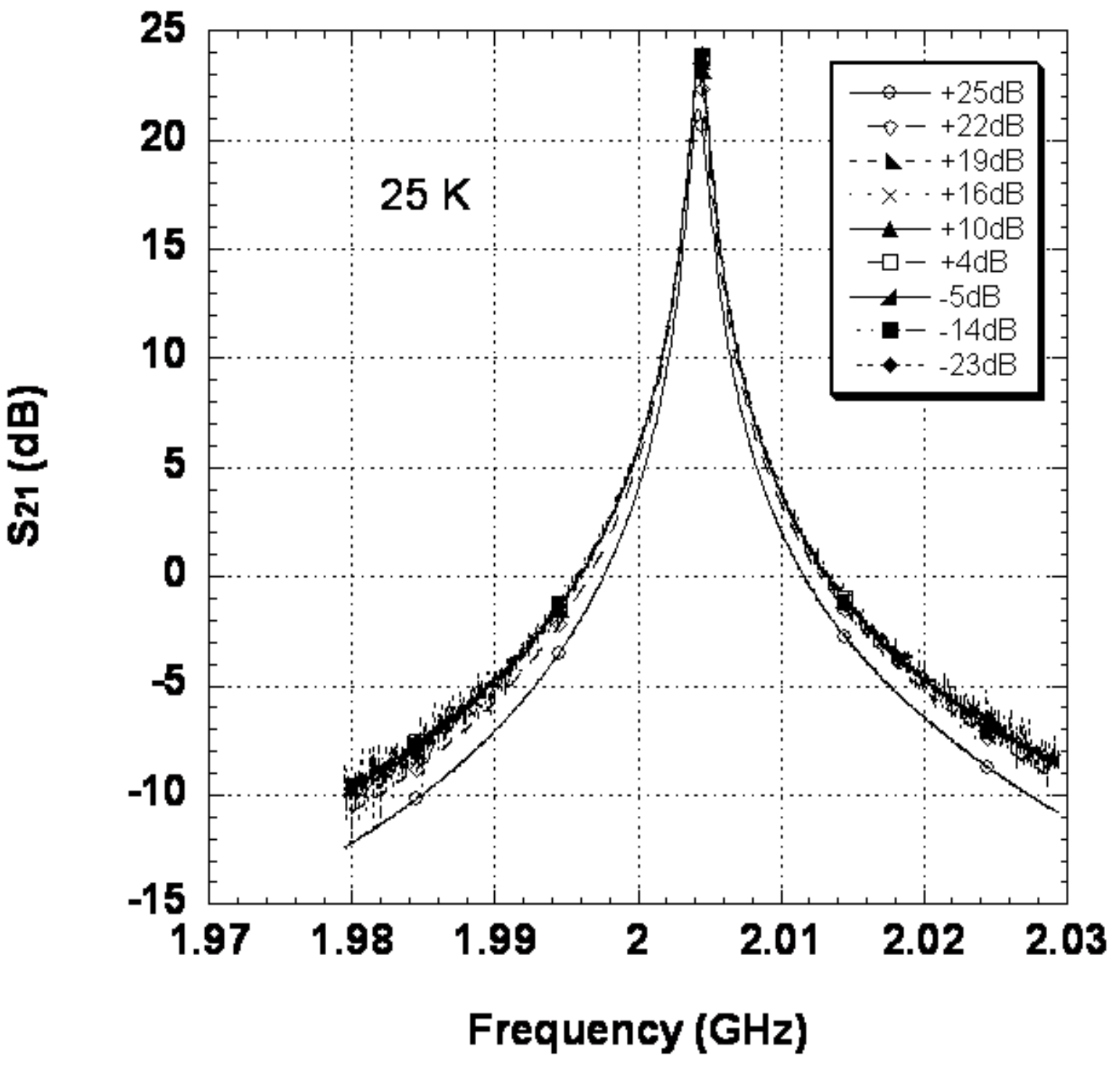

**Fig. 7.** Experimental dependence of forward transmission coefficient on ultra high frequency $S_{21}(f)$ in $YBa_2Cu_3O_{7-\delta}$ superconducting microstrip resonator at different microwave power levels at temperature of 25 *K*.

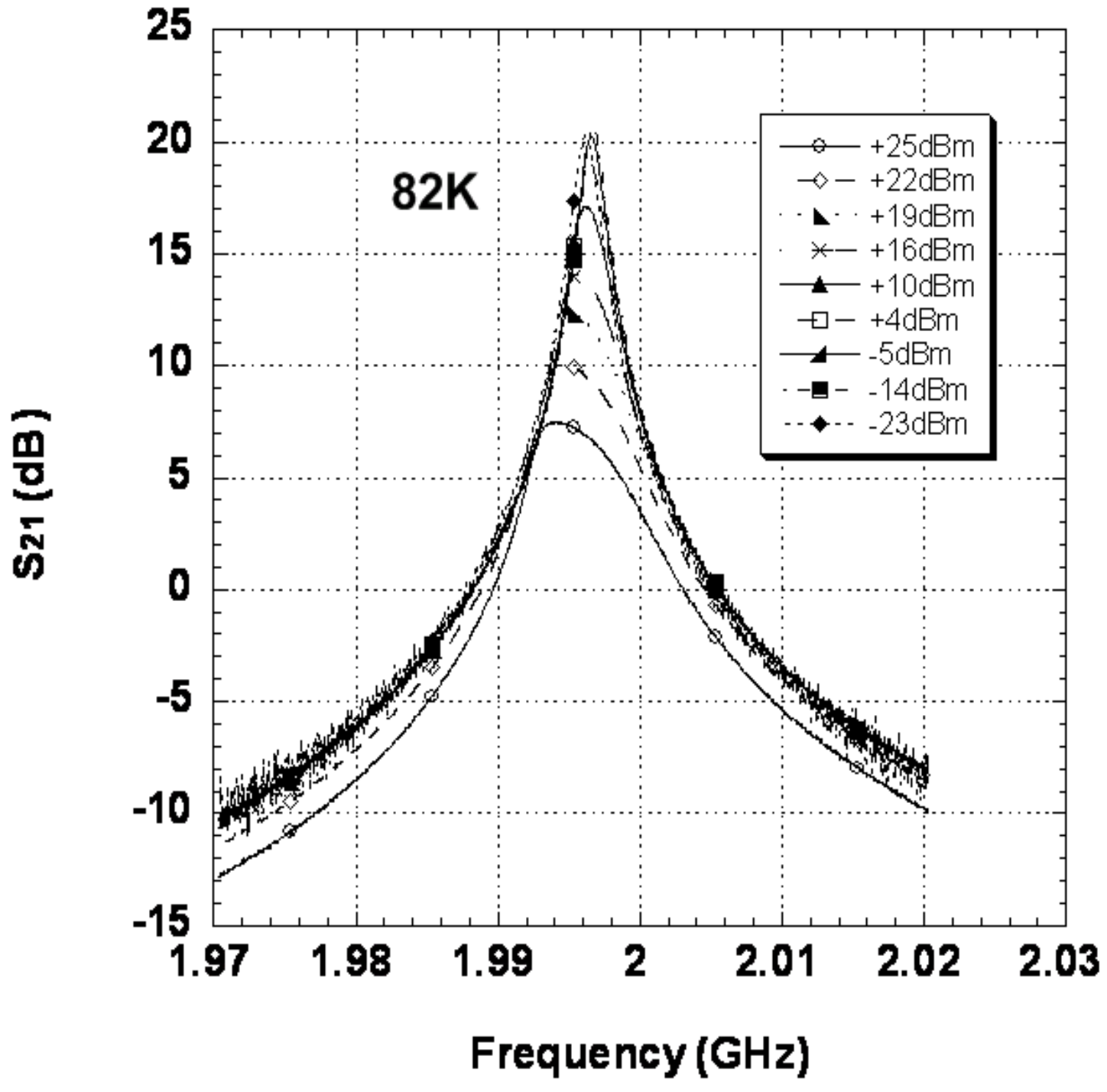

**Fig. 8.** Experimental dependence of forward transmission coefficient on ultra high frequency $S_{21}(f)$ in $YBa_2Cu_3O_{7-\delta}$ superconducting microstrip resonator at different microwave power levels at temperature of 82 $K$.

In Fig. 8, the experimental dependence of transmission coefficient on ultra high frequency $S_{21}(f)$ in $YBa_2Cu_3O_{7-\delta}$ superconducting microstrip resonator at different microwave power levels at temperature of 82$K$ is presented. The microwave signal strength has a strong influence on value of transmission coefficient $S_{21}$ at the temperature of 82$K$ close to the critical temperature $T_c$ of a *HTS* thin film at microwaves.

Author of dissertation performed the experimental measurements on transmission coefficient dependences as functions of both the frequency and the microwave signal strength $S_{21}(f)$, $S_{21}(P)$ at microwaves at intermediate temperatures, which were extensively used for the analysis of the resonant frequency shift, surface resistance $R_S$, and noise performances of superconducting microstrip resonators. The obtained data on these dependences $S_{21}(f)$, $S_{21}(P)$ measured at microwaves are comprehensively studied and discussed below.

### 7.4. Shift of Resonant Frequency $f_0$, Change of Quality Factor *Q*, and Change of Surface Resistance $R_s$ Depending on Applied Microwave Power $H_{rf}$ in $YBa_2Cu_3O_{7-\delta}$ Superconducting Microstrip Resonators at Ultra High Frequencies.

As it is visible from Fig. 8, the increase of microwave signal power of electromagnetic wave in a microstrip resonator reduces in an increase of width of resonance curve and exhibits in a shift of resonance frequency $f_0$. The decrease of amplitude and the broadening of width of curve $S_{21}(f)$ are because of the increase of surface resistance $R_S$ of superconductor; but the decrease of resonant frequency $f_0$ is stipulated by an increase of effective penetration depth of magnetic field $H_{rf}$ during the infiltration of *Abricosov magnetic curls* into the superconductor. It reduces in both the change of effective inductance of a microstrip resonator and the shift of resonant frequency of a microstrip resonator.

The characteristic experimental dependences of resonance frequency on microwave signal power $f_0(P)$ measured in $YBa_2Cu_3O_{7-\delta}$ superconducting microstrip resonator at microwaves at different temperatures are shown in Fig. 9. The resonance frequency shift in $f_0(P)$ dependence is proportional to the change of effective penetration depth of magnetic field $H_{rf}$ into a superconductor in Fig. 9.

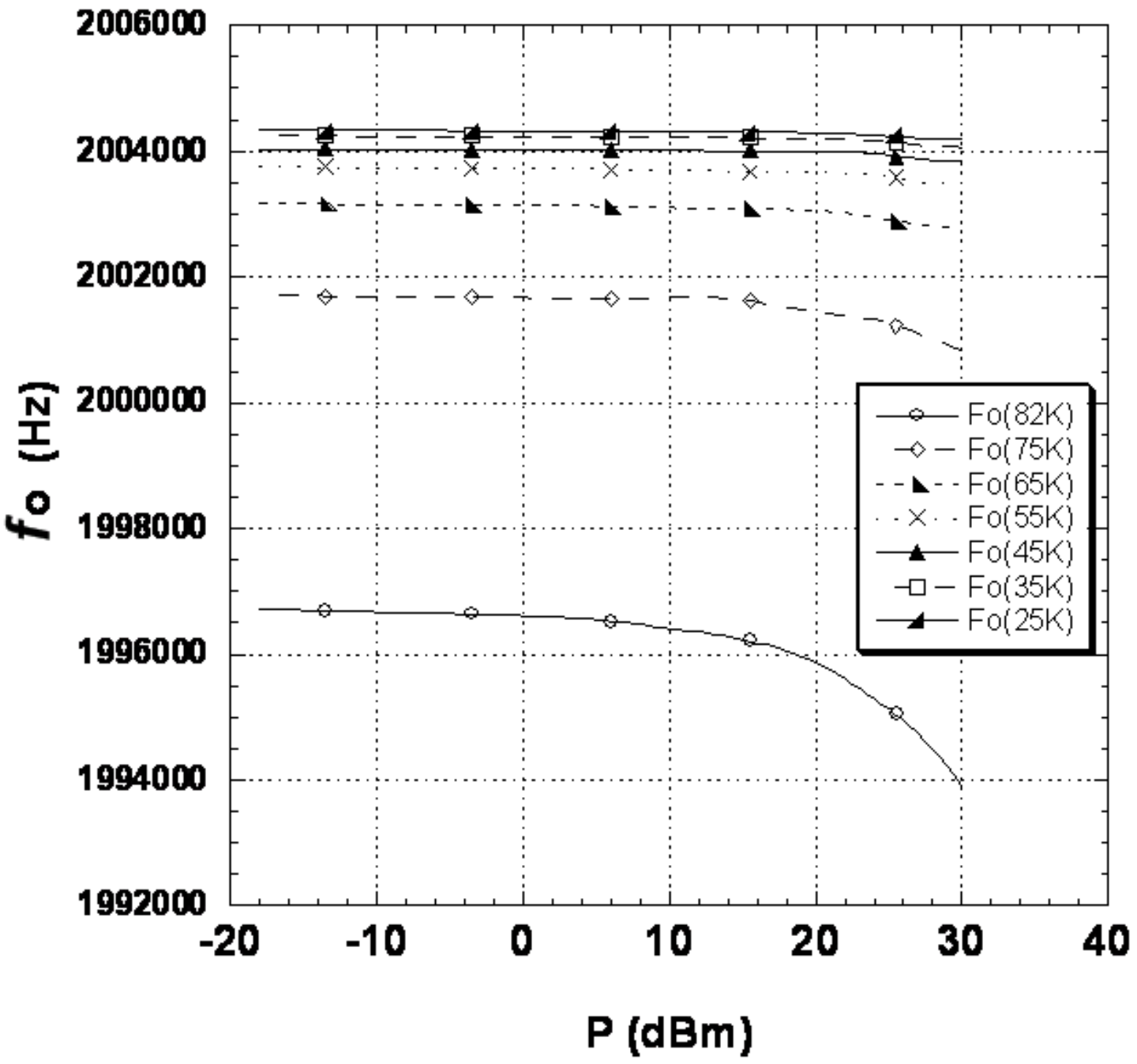

**Fig. 9.** Experimental dependence of resonance frequency $f_0$ on microwave signal power $P$ in $YBa_2Cu_3O_{7-\delta}$ superconducting microstrip resonator at microwaves at different temperatures. Shift of resonance frequency is observed.

In Fig. 9, it is visible that the strongest influence by microwave signal power $P$ of electromagnetic wave on resonance frequency $f_0$ happens at the high temperatures. The influence is small enough at low temperatures in Fig. 9. The nature of phenomenon is well understood since the superconductor physical properties such as the energy gap, critical magnetic field appear weak in the range of high temperatures.

The researched dependence of quality factor on microwave signal power *Q(P)* in $YBa_2Cu_3O_{7-\delta}$ superconducting microstrip resonator at frequency $f \approx 2GHz$ for a wide range of temperatures is shown in Fig. 10.

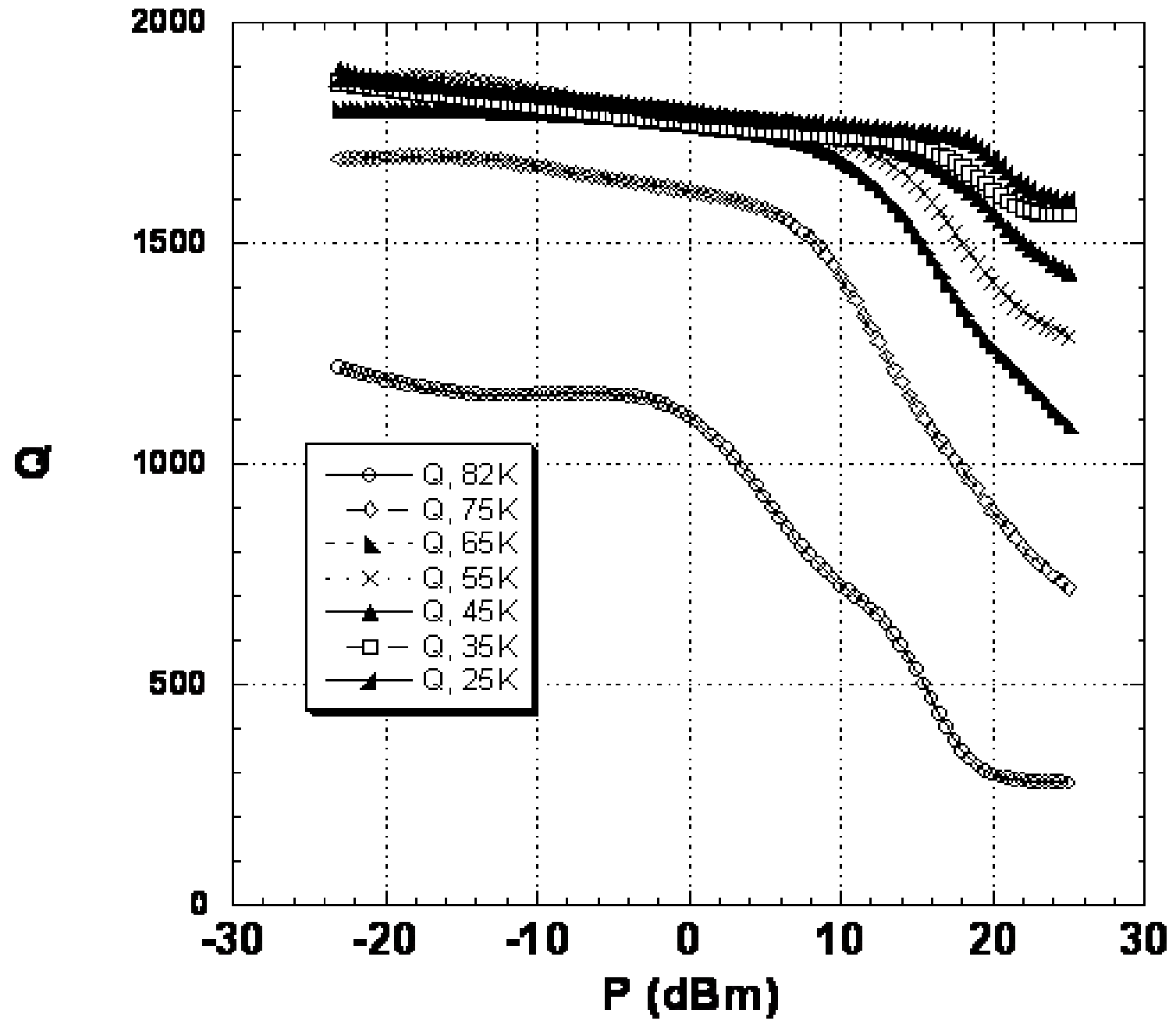

**Fig. 10.** Experimental dependence of quality factor on microwave signal power *Q(P)* in $YBa_2Cu_3O_{7-\delta}$ superconducting microstrip resonator at frequency $f \approx 2GHz$ at different temperatures.

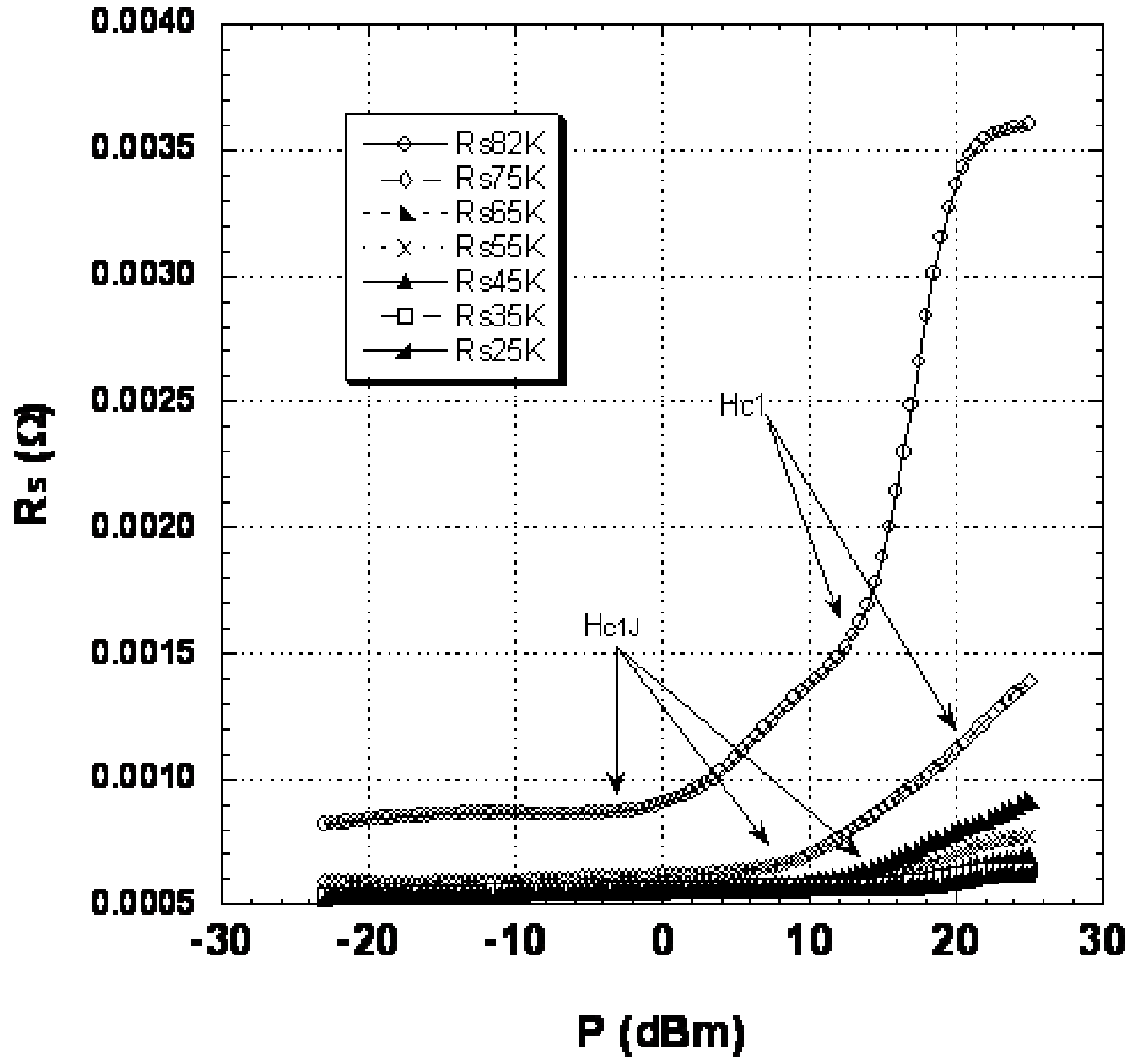

**Fig. 11.** Experimental dependence of surface resistance on microwave signal power $R_S(P)$ in $YBa_2Cu_3O_{7-\delta}$ superconducting microstrip resonator at frequency $f \approx 2GHz$ at different temperatures.

The data on the dependence of surface resistance of *HTS* thin film on microwave signal power $R_s(P)$ in $YBa_2Cu_3O_{7-\delta}$ superconducting microstrip resonator at frequency $f \approx 2GHz$ at different temperatures are reduced in Fig. 11. It follows from the theory of microwave resonators that the surface resistance is $R_s \sim 1/Q$. Therefore, the surface resistance $R_s$ has a small value at low temperatures. In the range of researched temperatures, the surface resistance $R_s$ of *HTS* thin films appears to be responsive to the applied microwave signal power of an electromagnetic wave, and it becomes strongly nonlinear at high microwave power levels in Fig. 11. In Fig. 11, the magnitudes of microwave signal power *P* at which the nonlinear properties were observed in $YBa_2Cu_3O_{7-\delta}$ microstrip resonators, are depicted by the arrows. In these points, the magnetic field $H_{rf}$ of an electromagnetic wave appears to be equal to the critical magnetic fields $H_{c1J}$ and $H_{c1}$ of *HTS* thin film in view of its demagnetization factor.

In Fig. 12, the experimental dependences of surface resistance on magnetic field $R_s(H_{rf})$ in $YBa_2Cu_3O_{7-\delta}$ superconducting microstrip resonator at frequency $f \approx 2GHz$ at different temperatures are presented. It is notable that the magnitude of magnetic field $H_{rf}$, at which the nonlinearities of surface resistance $R_s$ in *HTS* thin films are observed, decreases at an increase of temperature of a microstrip resonator. The magnitude of magnetic field $H_{rf}$, at which the nonlinearities of surface resistance $R_s$ in *HTS* thin films are registered, correlates well with the critical magnetic fields $H_{c1}$ and $H_{c2}$ as it follows from the theory of superconductivity.

In Fig. 13, the dependences $dR_s/dH_{rf}\,(H_{rf})$, representing the ratio of $dR_s/dH_{rf}$ derivatives as a function of magnetic field $H_{rf}$ , in $YBa_2Cu_3O_{7-\delta}$ superconducting microstrip resonator at frequency $f \approx 2GHz$ at different temperatures were constructed, using the results presented in Fig. 12

The obtained graphics of experimental dependencies $dR_s/dH_{rf}\,(H_{rf})$ in $YBa_2Cu_3O_{7-\delta}$ superconducting microstrip resonator at frequency $f \approx 2GHz$ at different temperatures confirm a conclusion that ***the nonlinearities of surface resistance $R_s$ in $YBa_2Cu_3O_{7-\delta}$ superconducting microstrip resonator at microwaves are somehow interlinked with some unusual features of physical properties of superconductors appearing near to the critical magnetic fields $H_{c1}$ and $H_{c2}$ in HTS thin films at microwaves.***

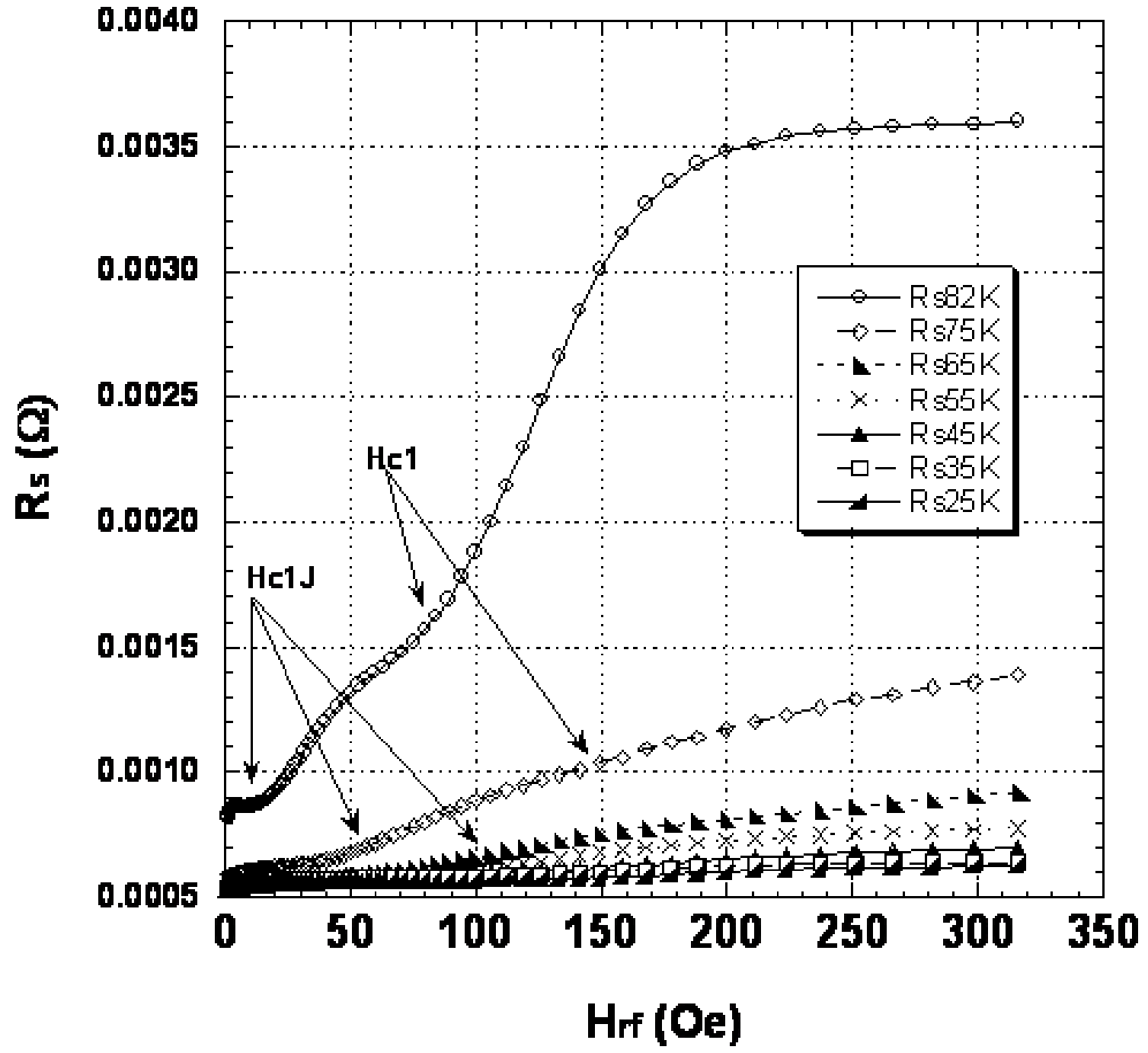

**Fig. 12.** Experimental dependence of surface resistance on magnetic field $R_s(H_{rf})$ in $YBa_2Cu_3O_{7-\delta}$ superconducting microstrip resonator at frequency $f \approx 2GHz$ at different temperatures.

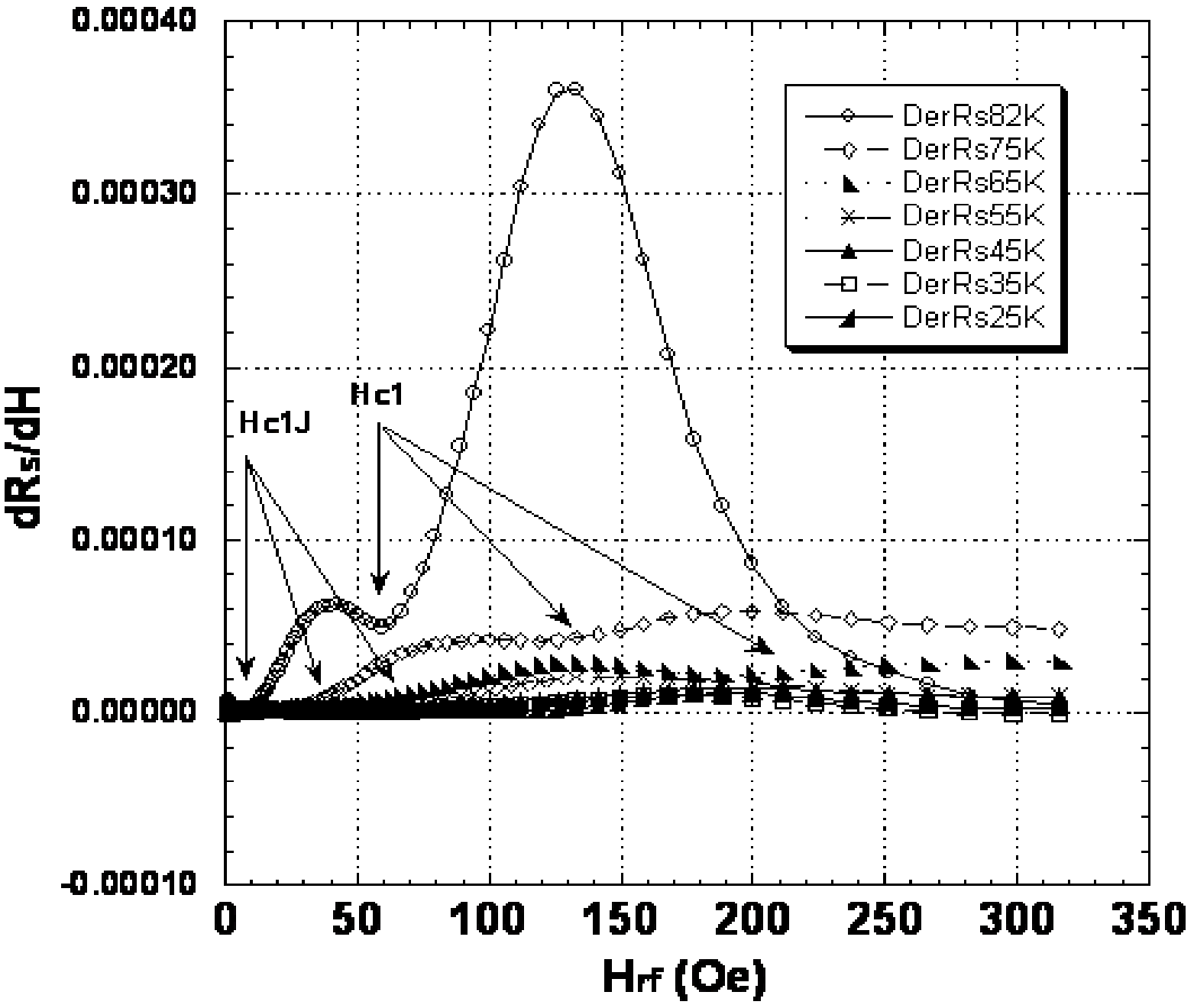

**Fig. 13.** Experimental dependences $dR_s/dH_{rf}$ $(H_{rf})$, representing the ratio of $dR_s/dH_{rf}$ derivatives as a function of magnetic field $H_{rf}$ , in $YBa_2Cu_3O_{7-\delta}$ superconducting microstrip resonator at frequency $f \approx 2GHz$ at different temperatures.

## 7.5. Modeling of Nonlinear Dependence of Surface Resistance on External Magnetic Field $R_S(H_e)$ in Proximity to Critical Magnetic Fields $H_{c1}$ and $H_{c2}$ in HTS Microstrip Resonators at Ultra High Frequencies.

In the course of research, it was observed that there is a nonlinear dependence of surface resistance $R_s$ on the external constant magnetic field $H_e$ or on the magnetic field of an applied electromagnetic field $H_{rf}$ near to the critical magnetic field of superconductor $H_{c1}$. In the region close to the critical magnetic field of superconductor $H_{c2}$, the magnetic field $H_{rf}$ is almost always much less than the critical magnetic field of superconductor $H_{c2}$, because of the fact that the magnetic field $H_{rf}$ can be comparable with the critical magnetic field of superconductor $H_{c2}$ at the temperatures close to the critical temperature of superconductor $T \sim T_c$ only. It is possible to disregard the influence by the $H_{c2}$ on the surface resistance $R_s$ in *HTS* thin films at low temperatures. The nonlinear change of surface resistance $R_s$ under the operation of external magnetic field $H_e$, in both these cases, in close proximity to the magnetic fields $H_{c1}$ and $H_{c2}$, is stipulated by the fact that there are phase transitions of second type at critical magnetic fields $H_{c1}$ and $H_{c2}$ in a superconductor, and the superconductor is in a mixed state in the range between the critical magnetic fields from $H_{c1}$ up to $H_{c2}$. At an increase of the external magnetic field $H_e$, when it becomes higher than the critical magnetic field $H_{c1}$, the quantum magnetic lines (*Abricosov magnetic vortices*) start to appear and penetrate into the superconductor in Fig. 14. The area of *Abricosov magnetic vortex core*, which has a radius $r \sim \xi$, is similar to the normal metal by its properties ($\xi$ – the correlation length of a superconductor). The total number of *Abricosov magnetic vortices* and the amount of normal phase, connected with the presence of the normal cores of the *Abricosov magnetic vortices*, have a nonlinear nature of change depending on the applied external magnetic field $H_e$ or magnetic field of an electromagnetic wave $H_{rf}$, leading to the nonlinear increase of a surface resistance $R_s$ in a superconductor, because the normal cores of *Abricosov magnetic vortices* have the essentially bigger surface resistance $R_{sn}$ than the surface resistance $R_s$ of superconductor Fig. 15. At an increase of the external magnetic field $H_e$ up to the critical magnetic field $\sim H_{c2}$, i.e. at transition from the mixed state to the normal state in a superconductor, the superconductivity phenomena disappears and the density of *Abricosov magnetic lines* increases. Also, the *Abricosov magnetic lines* start to overlap each other in such a way that the surface resistance $R_s$ increases non-linearly in this case, and when the external magnetic field becomes comparable to critical magnetic field $H_e \sim H_{c2}$, the magnitude of surface resistance $R_s$ comes close to the resistance of normal metal. Thus, the total value of surface resistance $R_s$ in a superconductor, in both cases, is not equal to the simple sum of the resistances of the superconducting and normal phases, which are present in a superconducting sample. To find the total value of surface resistance $R_s$ in a superconductor, it is necessary to take into the account the fact that the distance on which there is a local decrease of superconducting properties is essentially smaller than both the remaining space parameters of this task as well as the length of a microwave penetration into the superconducting sample. Indeed, the penetration depth $\lambda_L$ of magnetic field $H_{rf}$ and the length of the electromagnetic wave $\lambda_{rf}$ in a superconductor have the values of about $10^{-5}$ *cm*, which essentially exceed the value $2{\cdot}10^{-7}$ *cm* in the high-temperature superconductors at low temperatures. Therefore, the influence by the resistance of normal cores of Abricosov magnetic vortices should be averaged on the area of a superconducting sample ~ $10^{-5}$ *cm*.

**To estimate the effect of these phase transitions on the surface resistance $R_s$, it is necessary to consider the average model of a superconductor, in which as the author of dissertation supposes that the increasing external magnetic field will penetrate into a superconductor and suppress its superconducting energy gap <Δ> close to the low critical magnetic field $H_{c1}$.** It is assumed that the similar suppression of superconducting properties of a sample by external magnetic field has place at the phase transition from mixed state to normal state in close proximity to the upper critical magnetic field $H_{c2}$. Considering the interaction between the magnetic fields in a superconductor, the change of magnitude of external magnetic field $H_e$ leads to the proportional change of a number of *Abricosov magnetic vortices*, which has smoothly varying character. **In both cases, the suppression of superconductivity phenomena happens as a measure of change of average free energy of a superconductor in the external magnetic field. Therefore, it is more convenient to solve all the research problems, taking** **into the account the influence by the external magnetic field on the free energy of electronic system in a superconductor.** As it is known in the research of electrodynamics of superconductors at ultra high frequencies by Abrikosov, Gor'kov, Khalatnikov [2] and by Mattis, Bardeen [35], based on the microscopic theory of superconductivity by Bardeen, Cooper, Schrieffer (*BCS)* [21], the surface resistance $R_s$ of a superconductor depends on the energy gap as in eq. (7.5)

$$\Delta R_S(T) = R_{s0} + A \cdot \exp\{-\Delta(T)/kT\}, \quad (7.5)$$

where A is the constant.

This expression is correct for the range of the temperatures $T \le 0.7\ T_c$. At these temperatures, almost all the normal electrons, which participate in transport current, are in the superconducting state. Therefore, in the two-fluid Gorter-Casimir theory [23, 24], the density of superconducting electrons is

$$n_S(T)=n\,(1-(T/Tc)^4),$$

where *n* is the density of electrons in the normal state.

A little bit different, but the similar dependence exists in the *BCS* theory. To precede to the expression (5) for the surface resistance *Rs* in terms of the thermodynamic description, it is necessary to multiply the multi-pair numerator and denominator of expression in the exponent by $n_s$. Then, the formula for the change of surface resistance $\Delta R_S$ as a function of temperature *T* and magnetic field *H* can be written as in eq. (7.6)

$$\Delta R_S(T, H) = R_{S0} + A \exp\{-n_S(T, H)\cdot\Delta(T)/ n_S(T,H)\cdot kT\}. \quad (7.6)$$

In the indicated temperature limit, the density of superconducting electrons is approximately equal to the density of electrons in the normal state $n_S \approx n$, hence in a denominator, there is the energy equal, on an order of magnitude, to the thermal energy. In the multi-pair numerator, there is the energy equal to the energy of condensation of electrons during the transition to a superconducting state. It is equal to the difference of the free energies of the normal and superconducting states for the single volume of a sample

$$n_S(T,H)\cdot\Delta(T) = F_n(T,H) - F_S(T,H).$$

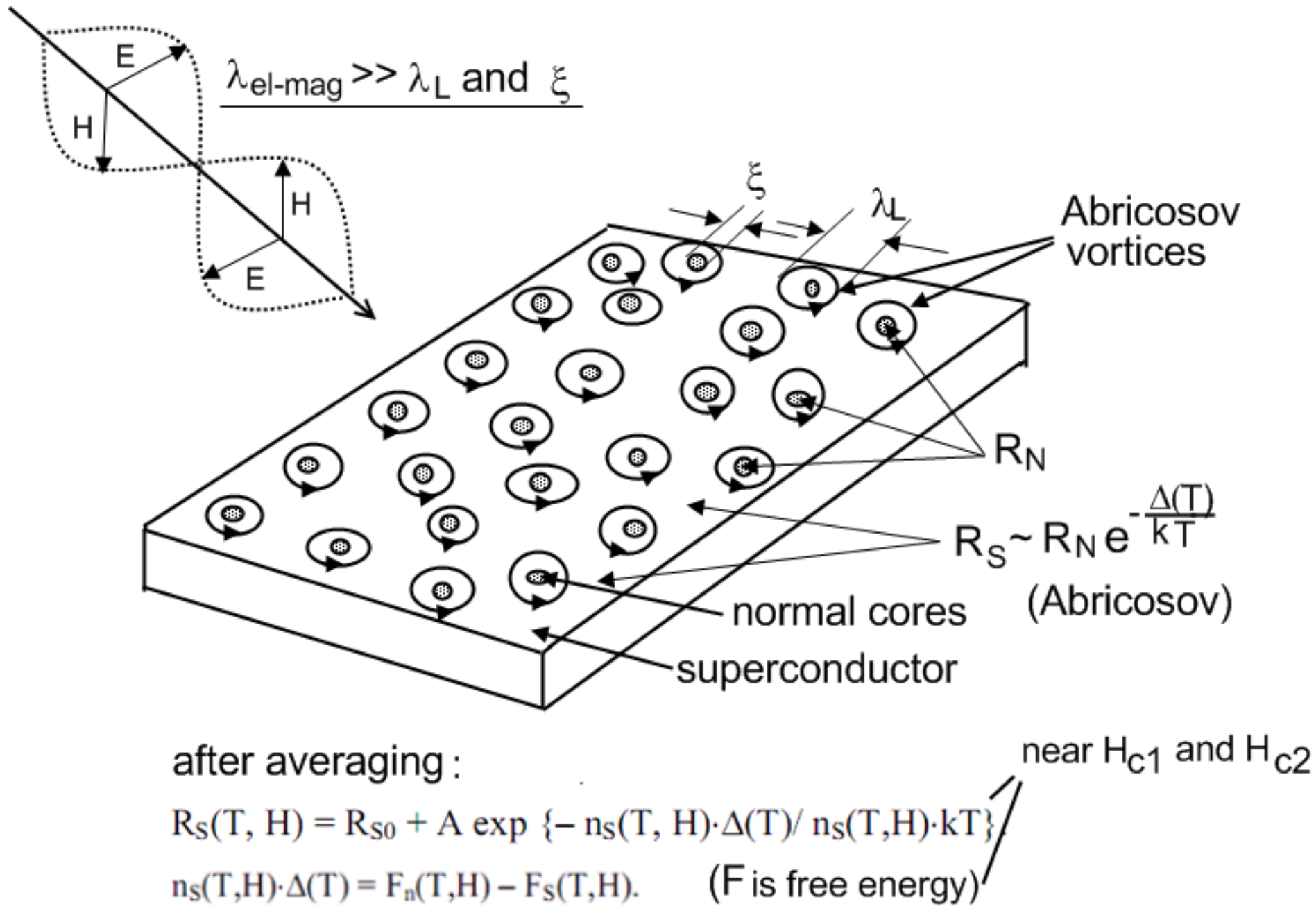

**Fig. 14**. Effect of Abricosov magnetic vortices generation on nonlinear surface resistance *Rs(He)* in close proximity to critical magnetic fields $H_{c1}$ and $H_{c2}$.

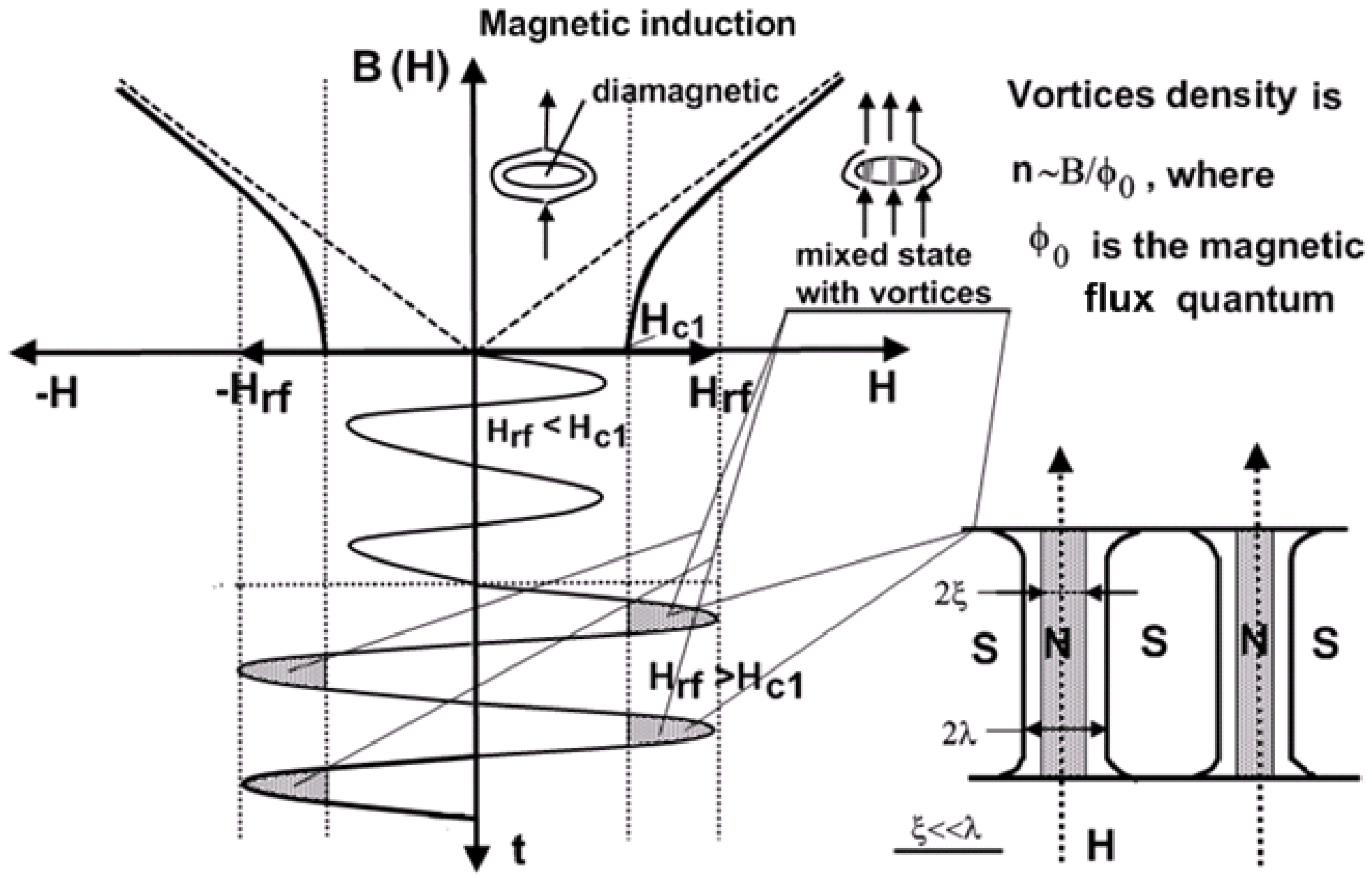

**Fig. 15**. Nature of nonlinear surface resistance *Rs(He)* appearance due to Abricosov magnetic vortex generation in close proximity to critical magnetic fields $H_{c1}$ and $H_{c2}$. In the case: $H_{rf} > H_{C1}$ , the $H_{rf}$ penetrates into superconductor, generating Abricosov magnetic vortices and resulting in nonlinear surface resistance appearance, because of normal metal cores of Abricosov magnetic vortices.

### 7.6. Modeling of Nonlinear Dependence of Surface Resistance on External Magnetic Field $R_S(H_e)$ in Close Proximity to Low Critical Magnetic Field $H_{c1}$ in HTS Microstrip Resonators at Ultra High Frequencies.

Near to the low critical field $H_{c1}$, it is convenient to represent the free energy in the magnetic field as the Gibbs free energy $G_s$, and as it was scored above, the Abricosov magnetic lines contribute to this energy. This contribution will reduce the value the density of superconducting electrons $n_s$ by the value $G_{M1}$, which was counted up, for example, in [4]. It can be written as in eq. (7.7)

$$\mathbf{G_{M1}} = \frac{2\phi_0}{\sqrt{3}\lambda^2} \frac{(\mathbf{H_e} - \mathbf{H_{C1}})}{4\pi\left\{\ln\left[\frac{3\phi_0}{4\pi\lambda^2(\mathbf{H_e} - \mathbf{H_{C1}})}\right]\right\}^2}, \quad (7.7)$$

where $\phi_0$ is the quantum of flux of a magnetic field, $\lambda$ is the penetration depth of a magnetic field.

Then, the surface resistance $R_S$, which is close to the low critical magnetic field $H_{C1}$, will change as in eq. (7.8)

$$R_S(T,H) \propto A\cdot\exp[-(n_S(T,H)\cdot\Delta(T) - G_{M1}) / n\cdot kT] \propto B\exp(G_{M1} / n\cdot kT), \quad (7.8)$$

where $B$ is some value equal to $B = A\exp[-(n_s\cdot\Delta)/n\cdot kT]$.

Let's emphasize that each mechanism of dissipation gives a contribution to the total resistance, and since the surface impedance of a superconductor is a sum of the resistances, hence the additional term of sort $R_s(H) \sim B\cdot\exp(G_{M1}/nkT)$ will correspond to this mechanism. Let's write

$$G_{M1} / n = g_{M1},$$

where $g_{M1}$ is the energy, which is bound with the penetration of *Abricosov magnetic lines* [28] into a superconducting sample and came on by one electronic state. Then, the following expression may be written in eq. (7.9)

$$R_s\,(H) \sim B \exp\,(g_{M1} / kT) \qquad (7.9)$$

At $H_e = H_{c1}$, we have $g_{M1} = 0$ and the exponential curve in expression (9) is equal to 1, and at the increase of a magnetic field it also increases together with the increase of the $g_{M1}$. Such a functional connection is not convenient for the representation of common resistance as the sum of the resistances from the different mechanisms. For this purpose, it is necessary to reduce this function in a sort, when at $g_{M1}=0$, it also is equal to *0*. Doing so, it is possible to shift this function on *–1* on an axis of ordinates. In the outcome, it will look like in eq. (7.10)

$$R_s\,(H) \sim B\,\{\exp\,(g_{M1} / kT) - 1\} \qquad (7.10)$$

Thus, its functional connection has not changed. Let's conduct normalization of this function on *1*, we will divide it on $exp(g_{M1} / kT)$. Then, it is possible to obtain the expression for the transition close to $H_{C1}$ at $H \geq H_{C1}$ in eq. (7.11)

$$\mathbf{R_{S1}(H) \approx B\frac{\exp(g_{M1}/kT)-1}{\exp(g_{M1}/kT)} = B\left\{1-\exp(-\frac{a_1 \cdot g_{M1}}{kT})\right\}} \qquad (7.11)$$

where $a_1$ is the fitting parameter.

The expression (7.11) features a change of the surface resistance $R_s$ in in $YBa_2Cu_3O_{7-\delta}$ superconducting microstrip resonator at frequency $f \approx 2GHz$ at different temperatures in close proximity to the low critical magnetic field $H_{c1}$.

In Fig. 16, the simulation of dependence of surface resistance on magnetic field *Rs(H)* in $YBa_2Cu_3O_{7-\delta}$ superconducting microstrip resonator at frequency $f \approx 2GHz$ at different temperatures in close proximity to the critical magnetic field $H_{C1}=50a.u.$ with the residual resistance $R_{s0}=0.1a.u.$, according to the equation (7.6) [19].

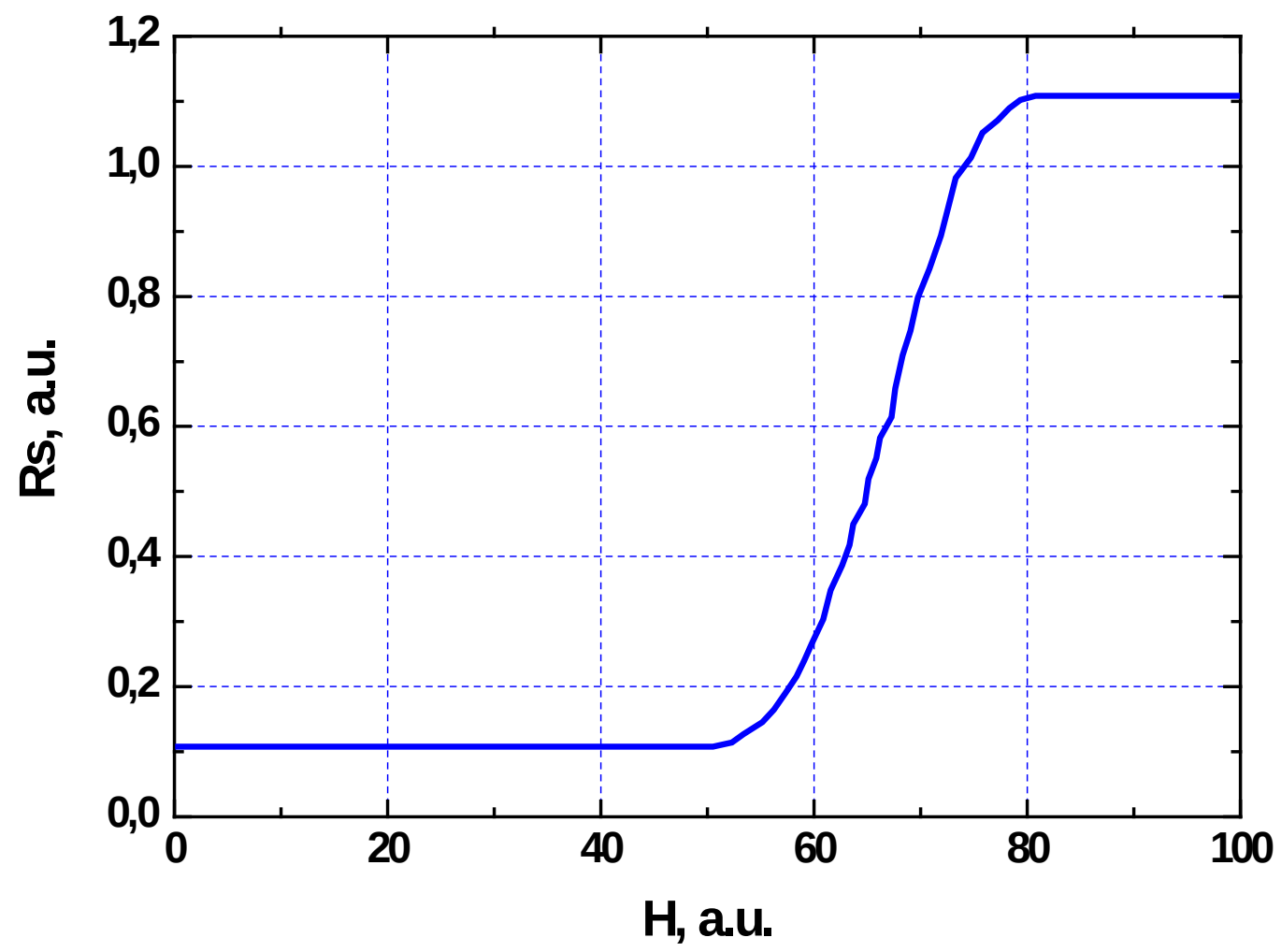

**Fig. 16.** Modeling of dependence of surface resistance on magnetic field *Rs(H)* in $YBa_2Cu_3O_{7-\delta}$ superconducting microstrip resonator at frequency $f \approx 2GHz$ at different temperatures in close proximity to critical magnetic field $H_{C1}=50a.u.$ with residual resistance $R_{s0}=0.1a.u.$ according to equation (7.6) [19].

## 7.7. Modeling of Nonlinear Dependence of Surface Resistance on External Magnetic Field $R_S(H_e)$ in Close Proximity to High Critical Magnetic Field $H_{c2}$ in HTS Microstrip Resonators at Ultra High Frequencies.

Let's consider, now, the variation of the surface resistance $R_s$ near to the high critical magnetic field $H_{c2}$. In this case, the change of free energy of a superconductor is featured by the expression in eq. (7.12) (see, for example [4, 5])

$$\mathbf{G_{M2} = \frac{(H_{C2} - H_e)^2}{8\pi(2\kappa^2 - 1)\beta_A}}, \qquad (7.12)$$

where $H_e \leq H_{c2}$ , $\beta_A \geq 1$ and κ is the *Ginzburg-Landau parameter,* which is equal to 70 for the high-temperature superconductors.

Let's pay attention to the fact that in the hard superconductors, the high critical magnetic field $H_{c2}$ much more exceeds the thermodynamic critical magnetic field $H_c$. Therefore, going from the common thermodynamic theory reasoning, the superconductivity phenomena should not exist under the external magnetic field, which is much more higher than the thermodynamic critical magnetic field: $H_e >> H_c$, because the energy gap $\Delta$ in superconductor should be equal to zero at the thermodynamic critical magnetic field, which is much less than the high critical magnetic field: $H_c << H_{c2}$. Indeed, the expression

$$n_s(T)\,\Delta(T) - H_c^2(T)/8\pi = 0$$

is derived from the thermodynamics theory of superconductors. However, in the case of *Type II* superconductors, the external magnetic field will penetrate into the superconducting sample in the form of *Abricosov magnetic lines* and it will be reduced in the fields $H_e >> Hc$ lowering the free energy of the *Type II* superconductors in this field in comparison with the normal metal. Thus, from the thermodynamics point of view, there is an apparent increase of energy gap in a superconductor, and the superconductivity phenomenon exists up to the high critical magnetic field $H_{c2}$. Therefore, in the expression for the surface resistance $R_s$, we must consider the magnetic field energy with the positive sign, and then we will obtain the expression in eq. (7.13)

$$\mathbf{R_s(H) \propto A_1 \cdot exp[-(n_s \cdot \Delta + G_{M2}) / n \cdot kT] \propto B_1\ exp(-G_{M2} / n \cdot kT).} \quad (7.13)$$

Let's move from the total change of free energy of a superconductor $G_{M2}$ to the change of free energy of a superconductor by the one electronic state $g_{M2}$ in eq. (7.14)

$$\mathbf{R_{S2}(H) \approx B_1\, exp(-\frac{a_2 \cdot g_{M2}}{kT})}, \quad (7.14)$$

where $B_1$ is the amplitude of change for surface resistance $R_s$, $a_2$ is the fitting coefficient defining the speed of change $R_s(H)$ and approximately equal to

$$a_2 \approx b / (50^2),$$

where b is the coefficient, which matches the dimensions of the multi-pair numerator and denominator.

According to this dependence, the surface resistance $R_s$ increases exponentially, when the external magnetic field $H_e$ increases and comes nearer to the high critical magnetic field $H_{c2}$. In this case, the energy gap $\Delta = 0$ at $H_e = H_{c2}$, and the energy $g_{M2} = 0$, and the surface resistance $R_s$ reaches a maxima. The total value of the surface resistance $R_s$ of a superconductor in this point becomes equal to a surface resistance of normal metal $R_n$. The sort of dependence of the surface resistance on the magnetic field $R_s(H)$ near to high critical magnetic field $H_{c2}=5000a.u.$ at different coefficients $a_1 = 0.001$ (red curve), $0.01$ (green curve), $0.05$ (blue curve); $B_1 = 0.5, R_{s0} =0.1$ is shown in Fig. 17 [19]

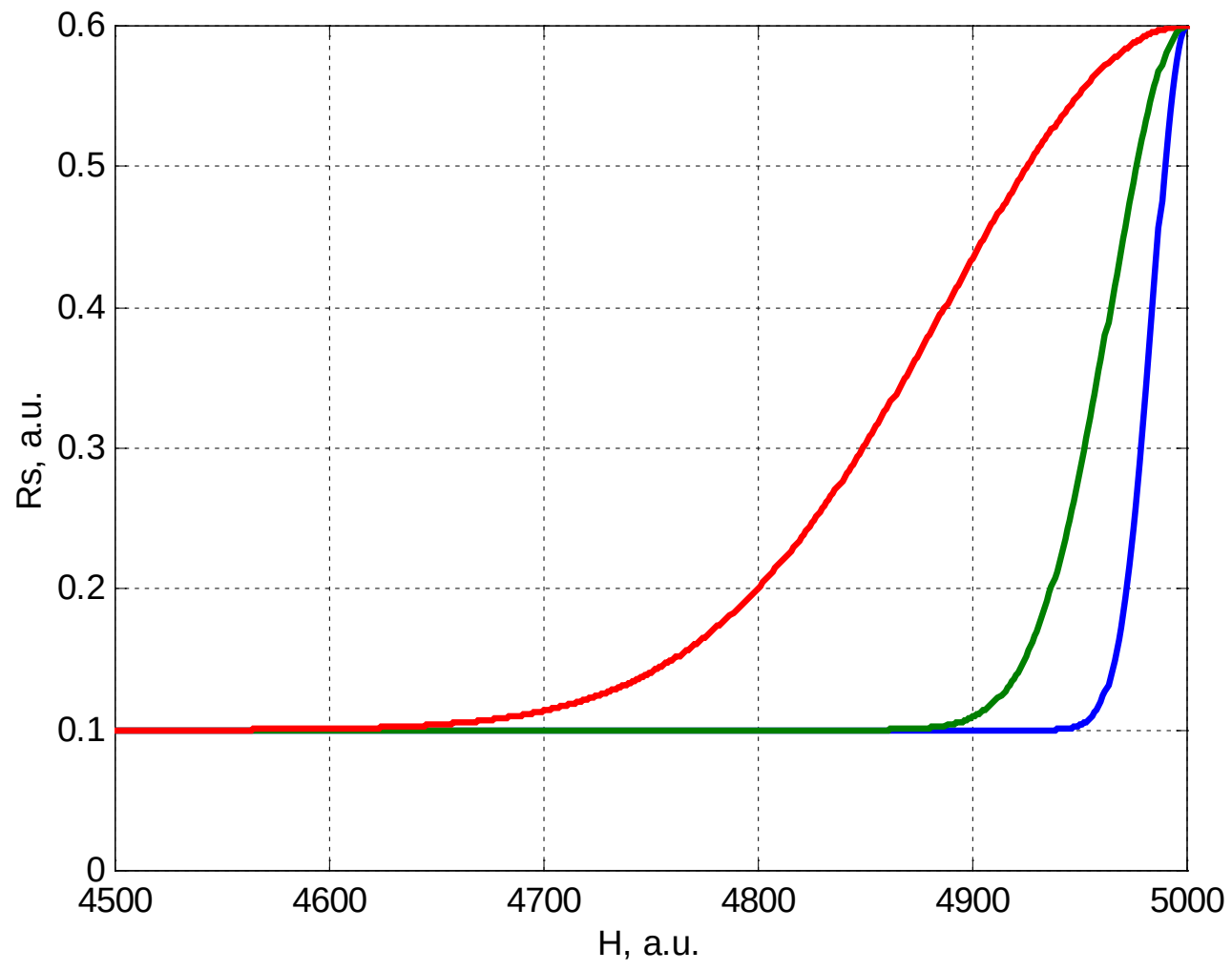

**Fig. 17.** Dependence of surface resistance on magnetic field *Rs(H)* near to high critical magnetic field $H_{c2}$ =*5000 a.u.* at different coefficients $a_1$ = 0.001 (red curve), 0.01 (green curve), 0.05 (blue curve); $B_1 = 0.5$, $R_{s0} = 0.1$ [19].

In Fig. 17, the curves concern the superconductors with the different stationary constants $\kappa$. The most steep blue conditional dependence *Rs(H)* corresponds to $\kappa \geq 3$, and the broad red conditional dependence *Rs(H)* corresponds to $\kappa \sim 15$.

In technical low-temperature superconductors, the high critical magnetic field $H_{c2} \approx 10^5 Oe$, and the slope of curves essentially depends on the temperature as it follows from eq. (7.14). At low temperatures, the *Rs(H)* curves become more abrupt in eq. (14).

The simulations of experimental results, obtained for the conventional superconducting alloys: titanium-vanadium $Ti_{0.6}V_{0.4}$ [6] and indium-lead $In_{0.19}Pb_{0.8}$ [7], which are the Type II superconductors - the same as the high temperature superconductors, were performed with the aim to check the authors original approach to simulate the nonlinear dependence of surface resistance of *HTS* thin films on external magnetic field $Rs(H_e)$ in simple case, when the amplitude of magnetic field $H_{rf}$ is smaller than the amplitude of external magnetic field $H_e$. As it is shown below, the developed software works well for $Ti_{0.6}V_{0.4}$ and $In_{0.19}Pb_{0.8}$ [19].

The common dependence $R_s/R_n(H)$ for a conditional superconductor, including the cases of the critical magnetic fields $H_{c1}$ and $H_{c2}$ is represented in Fig.

18 [19]. The critical magnetic fields are supposed to be equal to $H_{c1}$ = *50 a.u.* and $H_{c2}$ = *5000 a.u.* Then, the value of stationary constant κ can be easily found $\kappa \approx 7$ from the ratio $H_{c2}/H_{c1} \approx \kappa^2$[4].

The conditional dependence of surface resistances ratio on magnetic field $R_s/R_n(H)$ in a range between the critical magnetic fields $H_{c1} \div H_{c2}$ has a linear sort of dependence $R_s/R_n(H) \propto H$. The nonlinear change of surface resistance $R_s/R_n(H)$ occurs close to the low critical magnetic field $H_{c1}$=*50a.u.* and high critical magnetic field $H_{c2}$=*5000a.u.* in Fig. 18. In this case, it is possible to suppose that the surface resistance $R_s$ is proportional to a number of Abricosov magnetic vortices in a superconductor at microwaves.

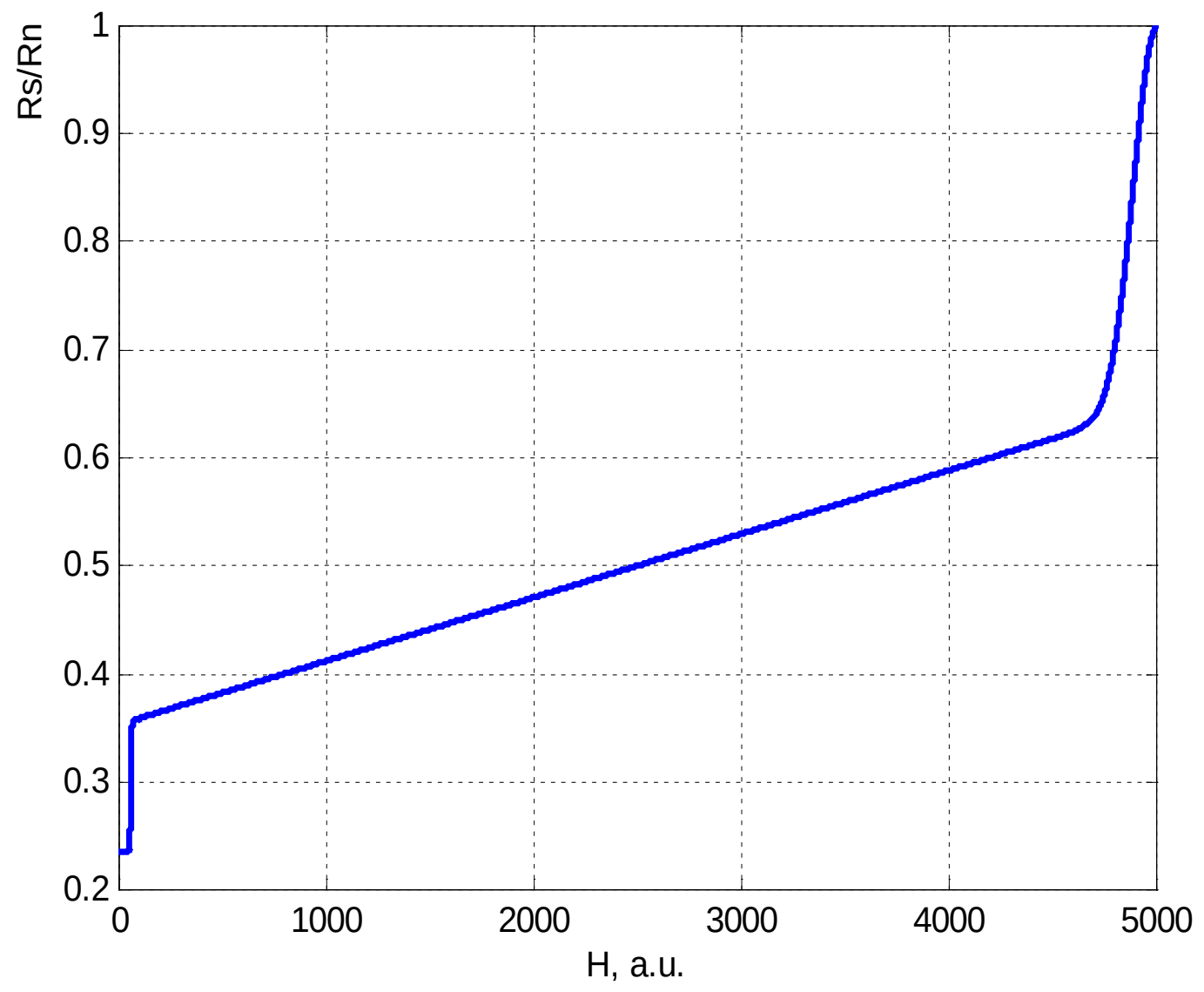

**Fig. 18.** Modeling of conditional dependence of superconductor surface resistance to normal metal resistance ratio on magnetic field $R_s/R_n(H)$ with linear sort of dependence $R_s/R_n(H)$ between critical magnetic fields $H_{c1} \div H_c$ and nonlinear dependence $R_s/R_n(H)$ close to low critical magnetic field $H_{c1}$=*50* and high critical magnetic field $H_{c2}$=*5000*. Dependence $R_s/R_n(H)$ represents symmetric function between critical magnetic fields $H_{c1}$ and $H_{c2}$ [19].

The matching of simulation of dependences $R_s(H)$ [19] with the experimental measurements of $R_s(H)$ for the *Type II* superconducting alloy $Ti_{0.6}V_{0.4}$ [6] is presented in Figs. 19 and 20. The purpose is to check that the developed software can approximate the $R_s(H)$ dependences for the *Type II* superconductors, including $YBa_2Cu_3O_{7-\delta}$ and $Ti_{0.6}V_{0.4}$, in close proximity to critical magnetic fields $H_{c1}$ and $H_{c2}$ at ultra high frequencies at any temperature properly. It is necessary to note that authors didn't perform the experimental measurements of dependence $R_s(H)$ in $YBa_2Cu_3O_{7-\delta}$ superconducting microstrip resonators in proximity to critical magnetic field $H_{c2}$, because the magnitude of critical magnetic field $H_{c2} \sim 10^6$ *Oe* is too high for $YBa_2Cu_3O_{7-\delta}$ and it cannot be reached with the used measurement setup.

Such high magnetic fields can be reached in experiments with pulse generators of external magnetic field $H_e$, but in this case, the additional screening currents with big amplitude will appear in the superconducting sample. As a result the superconducting properties of *HTS* thin films will degrade, and the accompanying thermal effects will have a main influence on the superconducting nonlinear properties of *HTS* thin films, whereas the influence by *Abricosov magnetic vortices* on the superconducting nonlinear properties of *HTS* thin films will be negligible.

Author conducted the modeling of $R_s(H_{rf})$ dependence behaviour in *Type II* superconductors using the available experimental data for the *Type II* conventional superconducting alloys: titanium-vanadium $Ti_{0.6}V_{0.4}$ [6] and indium-lead $In_{0.19}Pb_{0.8}$ [7] with similar magnetic properties, were used for the comparative analysis of experimental and simulated dependences $R_s(H)$ in *Type II* superconductors in proximity to critical magnetic field $H_{c2}$ at microwaves.

In Fig. 19, the experimental dependence of surface resistance on magnetic field $R_s(H)$ in $Ti_{0.6}V_{0.4}$ at frequency of 14.4*GHz* at temperature $T = 4.2K$ is shown by blue curve [6] and simulation of experimental dependence of surface resistance on external magnetic field $R_s(H)$ in $Ti_{0.6}V_{0.4}$ at frequency of 14.4*GHz* at temperature $T=4.2K$ is approximated by red curve. The linear relation at $H_e < H_{c2}$ and formula (6.14) at $H_e \leq H_{c2}$ are utilized to simulate $R_s(H)$ dependence in $Ti_{0.6}V_{0.4}$ at frequency of 14.4*GHz* at temperature $T = 4.2K$ [19].

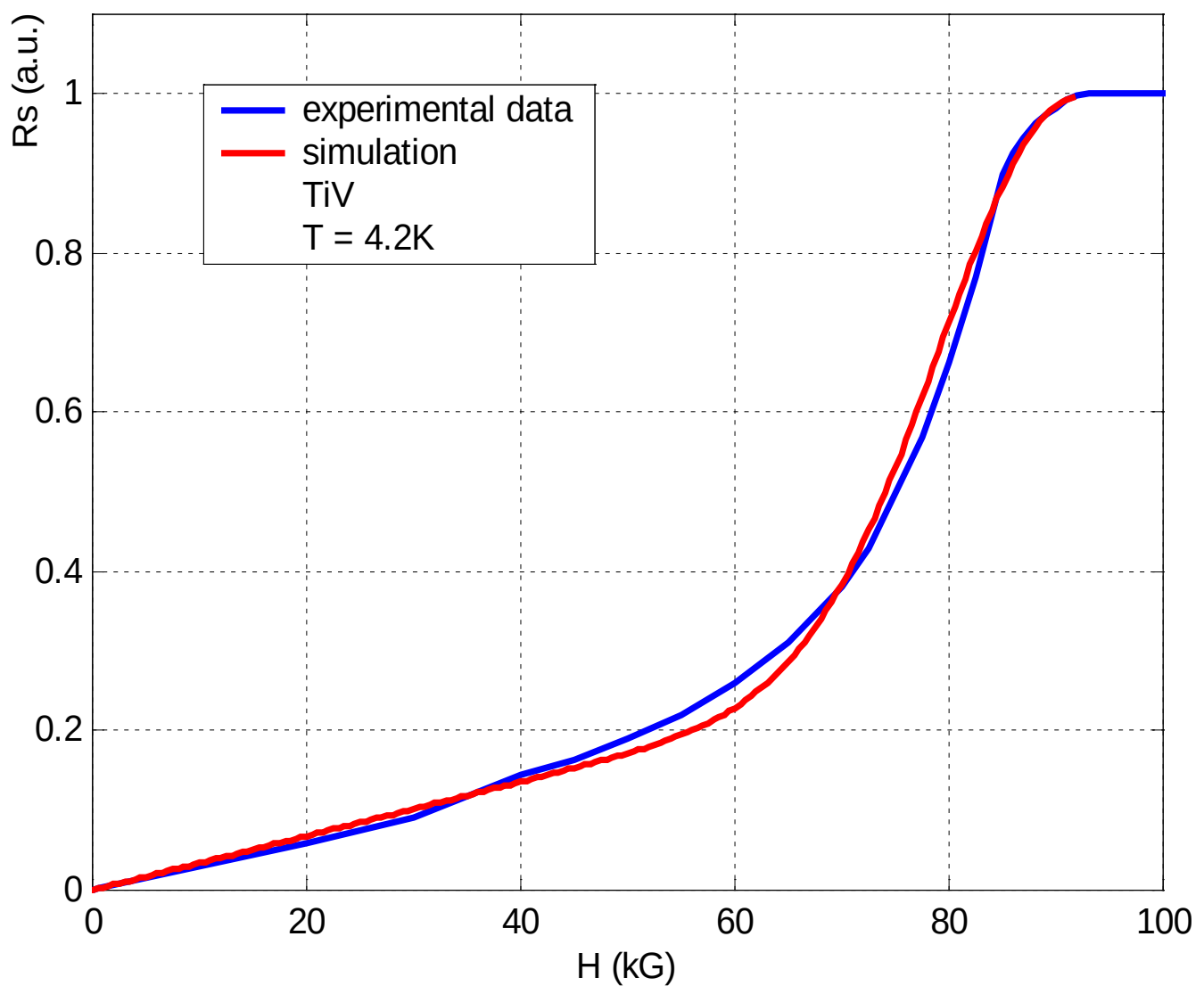

**Fig. 19.** Experimental dependence of surface resistance on magnetic field $R_s(H)$ in $Ti_{0.6}V_{0.4}$ at frequency of 14.4*GHz* at temperature $T = 4.2K$ is shown by blue curve [6] and simulation of experimental dependence of surface resistance on magnetic field $R_s(H)$ in $Ti_{0.6}V_{0.4}$ at frequency of 14.4*GHz* at temperature $T = 4.2K$ is approximated by red curve. Linear relation at $H_e < H_{c2}$ and formula (6.14) at $H_e \leq H_{c2}$ are utilized to simulate $R_s(H)$ dependence in $Ti_{0.6}V_{0.4}$ at frequency of 14.4*GHz* at temperature $T = 4.2K$ [19].

In Fig. 20, the experimental dependence of surface resistance on magnetic field $R_s(H)$ in $Ti_{0.6}V_{0.4}$ at frequency of 14.4*GHz* at temperature $T = 1.18K$ is shown by blue curve [6], and simulation of experimental dependence of surface resistance on magnetic field $R_s(H)$ in $Ti_{0.6}V_{0.4}$ at frequency of 14.4*GHz* at temperature $T=1.18K$ is approximated by red curve [19, 20]. The linear relation at $H_e < H_{c2}$ and formula (14) at $H_e \leq H_{c2}$ are utilized to simulate $R_s(H)$ dependence in $Ti_{0.6}V_{0.4}$ at frequency of 14.4*GHz* at temperature $T = 1.18K$ [19]. In this case, the dependence of surface resistance on the external magnetic field $R_s(H)$ close to high critical magnetic field $H_{c2}$ at microwaves becomes more steep as it follows from the formula (7.14).

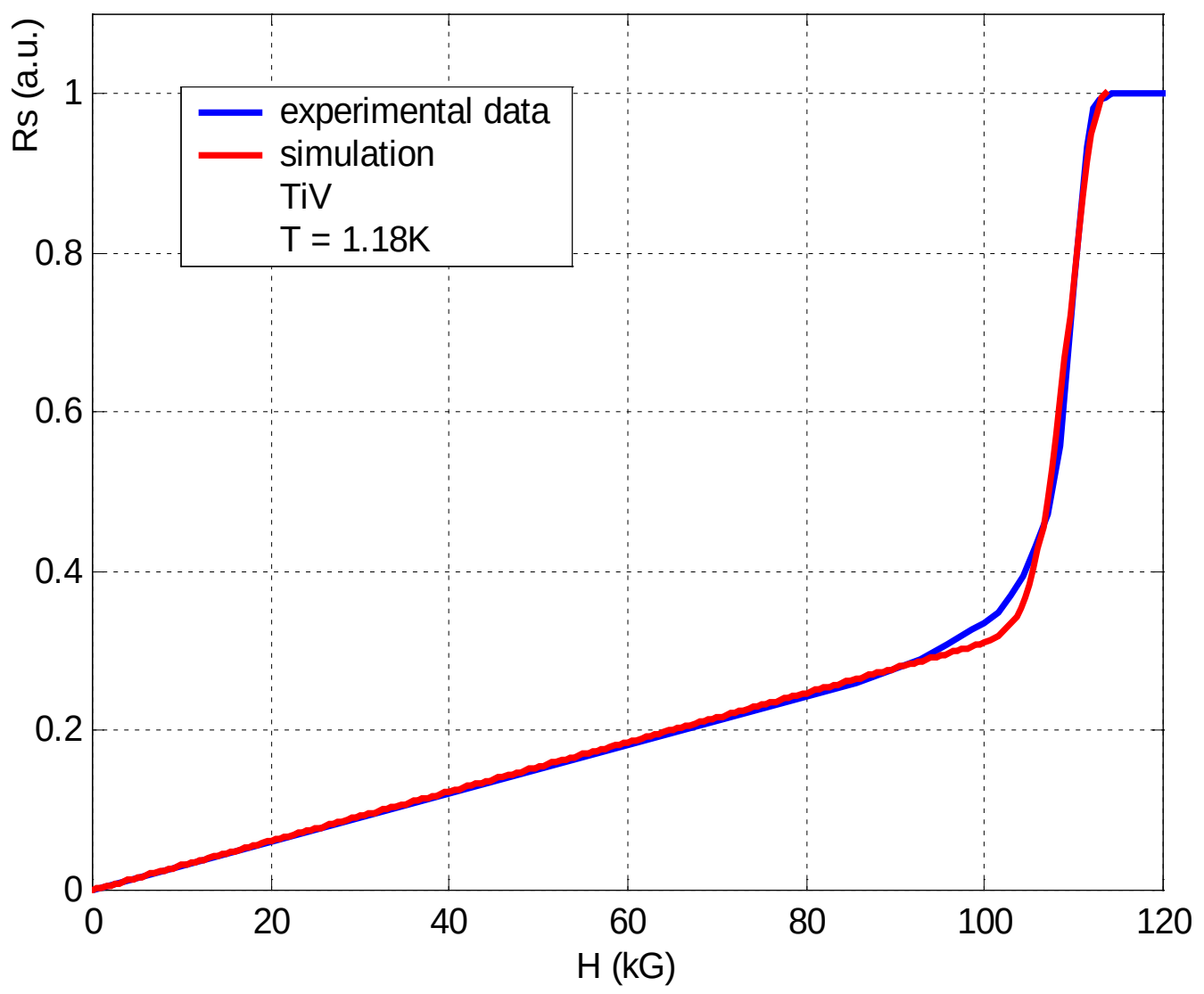

**Fig. 20.** Experimental dependence of surface resistance on magnetic field $R_s(H)$ in $Ti_{0.6}V_{0.4}$ at frequency of 14.4*GHz* at temperature $T = 1.18K$ is shown by blue curve [6], and simulation of experimental dependence of surface resistance on magnetic field $R_s(H)$ in $Ti_{0.6}V_{0.4}$ at frequency of 14.4*GHz* at temperature $T = 1.18K$ is approximated by red curve. Linear relation at $H_e < H_{c2}$ and formula (7.14) at $H_e \leq H_{c2}$ are utilized to simulate $R_s(H)$ dependence in $Ti_{0.6}V_{0.4}$ at frequency of 14.4*GHz* at temperature $T = 1.18K$ [19].

In some cases, the dependence of surface resistance on magnetic field $R_s(H)$ in *HTS* thin films between the critical magnetic fields $H_{c1}$ and $H_{c2}$ can not be represented as a linear dependence, but can be approximated as $R_s\ (H) \propto (H)^{1/2}$, when the dissipation of energy of an electromagnetic wave in a mixed state of superconductor is interlinked to the driving of *Abricosov magnetic vortices* by transport current in a superconductor at microwaves [7]. This case was also modeled by author of dissertation.

In Fig. 21, the conditional dependence of superconductor surface resistance to normal metal resistance ratio on magnetic field $R_s/R_n(H)$ in *HTS* thin film at microwaves with the dependence $R_s/R_n(H) \sim (H)^{1/2}$ between the critical magnetic fields $H_{c1} \div H_c$ and the nonlinear dependence $R_s/R_n(H)$ close to the low critical magnetic field $H_{c1}=50$ and high critical magnetic field $H_{c2}=3500$ is shown. The dependence $R_s/R_n(H)$ in *HTS* thin film at microwaves represents an almost symmetric function between the low critical magnetic field $H_{c1}$ and the high critical magnetic field $H_{c2}$ [19].

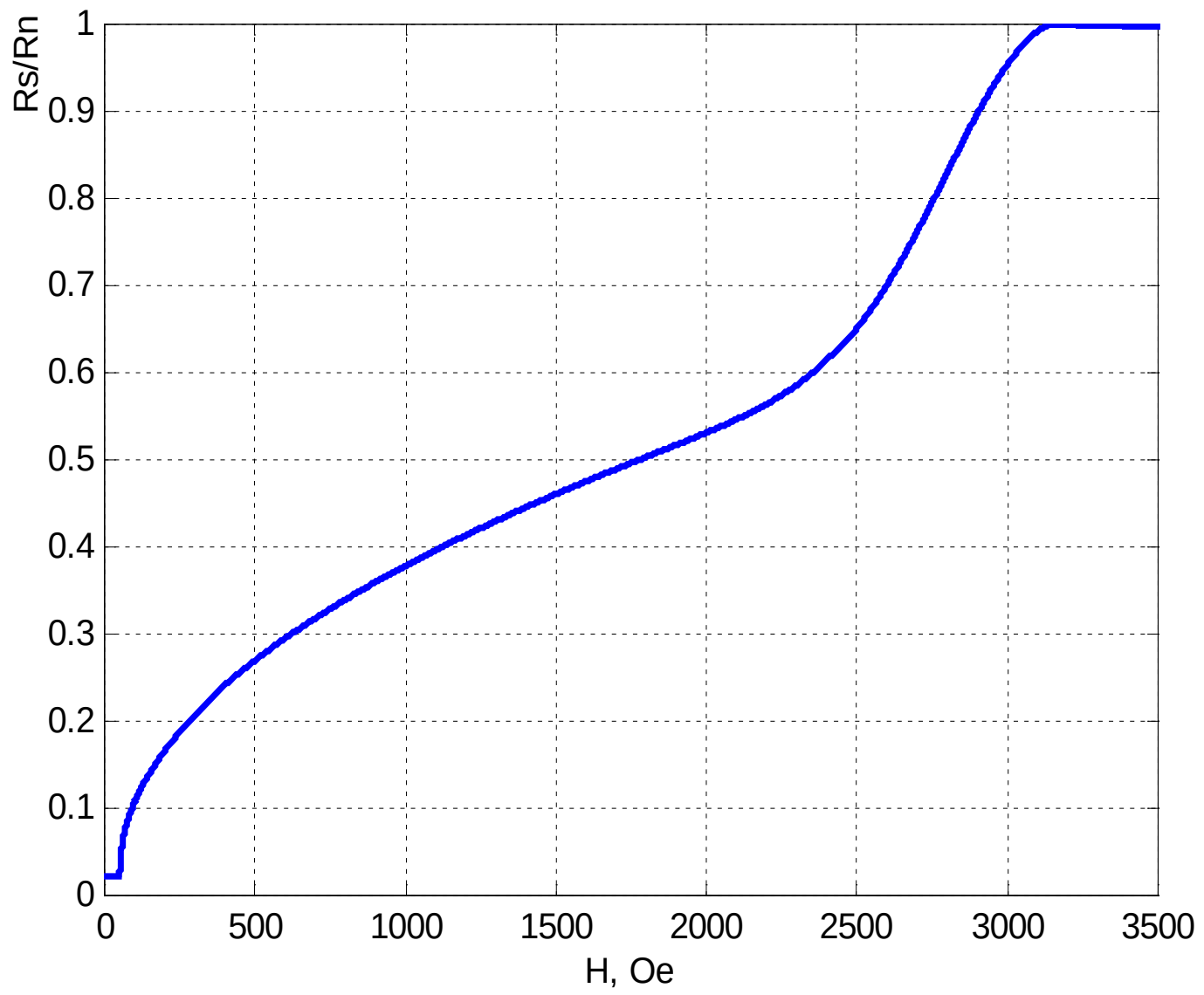

**Fig. 21.** Modeling of conditional dependence of superconductor surface resistance to normal metal resistance ratio on magnetic field $R_s/R_n(H)$ in *HTS* thin film at microwaves with dependence $R_s/R_n(H) \sim (H)^{1/2}$ between critical magnetic fields $H_{c1} \div H_c$ and nonlinear dependence $R_s/R_n(H)$ close to low critical magnetic field $H_{c1}$=50*Oe* and high critical magnetic field $H_{c2}$=3500 *Oe* in *HTS* thin film at microwaves. ***Dependence $R_s/R_n(H)$ in HTS thin film at microwaves represents almost symmetric function between critical magnetic fields $H_{c1}$ and $H_{c2}$.*** [19].

In Fig. 22, the experimental dependence of surface resistance on magnetic field $R_s/R_n(H)$ $In_{0.19}Pb_{0.8}$ at frequency 170 *MHz* at temperature of 4.2*K* is shown by black curve [7] and simulation of dependence of superconductor surface resistance to normal metal resistance ratio on magnetic field $R_s/R_n(H)$ with dependence $R_s/R_n(H) \sim (H)^{1/2}$ between critical magnetic fields $H_{c1} \div H_c$ and the nonlinear dependence $R_s/R_n(H)$ close to the low critical magnetic field $H_{c1}=50$ *Oe* and high critical magnetic field $H_{c2}$=3000 *Oe* in case of $In_{0.19}Pb_{0.8}$ at frequency of 170 *MHz* at temperature of 4.2*K* is shown by blue curve [19].

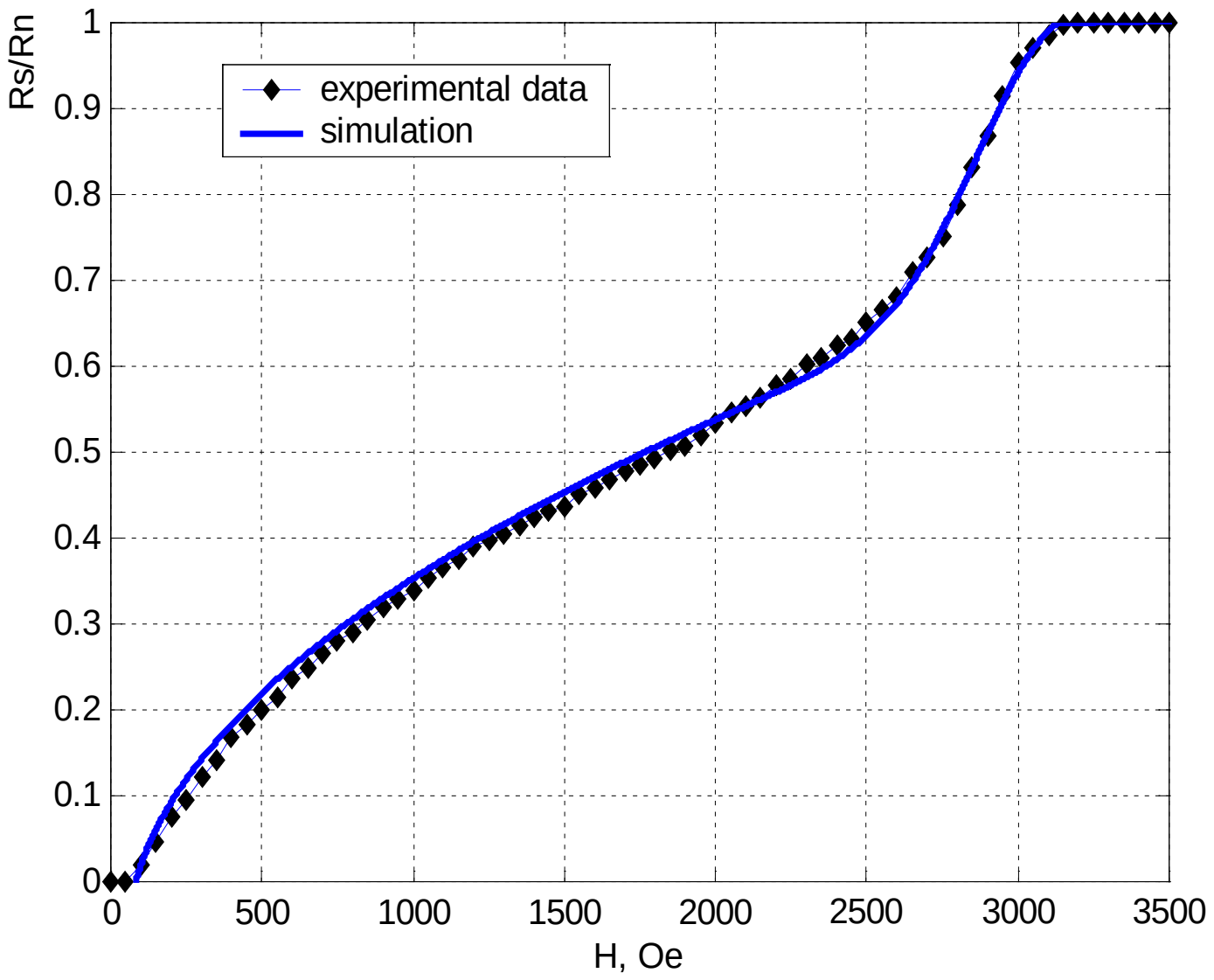

**Fig. 22.** Experimental dependence of surface resistance on magnetic field $R_s/R_n(H)$ $In_{0.19}Pb_{0.8}$ at frequency 170 *MHz* at temperature of 4.2*K* is shown by black curve [7] and simulation of dependence of superconductor surface resistance to normal metal resistance ratio on magnetic field $R_s/R_n(H)$ with dependence $R_s/R_n(H) \sim (H)^{1/2}$ between critical magnetic fields $H_{c1} \div H_c$ and the nonlinear dependence $R_s/R_n(H)$ close to the low critical magnetic field $H_{c1}=50\ Oe$ and high critical magnetic field $H_{c2}=3000\ Oe$ in case of $In_{0.19}Pb_{0.8}$ at frequency 170 *MHz* at temperature of 4.2*K* is shown by blue curve [19].

The presented simulation results on the dependence of surface resistance on external magnetic field $R_s(H_e)$ in close proximity to critical magnetic field $H_{c2}$ are in good agreement with the experimental data for *Type II* superconductors in the case, when the amplitude of magnetic field of *UHF* electromagnetic wave $H_{rf}$ is much less than the magnitude of external magnetic field $H_e$ created by solenoid.

The next researched case is when the magnetic field of electromagnetic wave has values comparable or even high than the low critical magnetic field $H_{rf} \sim H_{c1}$ or even $H_{rf} > H_{c1}$, and the transition to the nonlinear region happens under the influence of the magnetic field $H_{rf}$ of an electromagnetic wave at microwaves.

## 7.8. On Influence of Magnetic Field $H_{rf}$ on Nonlinear Surface Resistance $R_S$ of $YBa_2Cu_3O_{7-\delta}$ Thin Films on MgO Substrates in Close Proximity to Low Critical Magnetic Field $H_{c1}$ in Superconducting Microstrip Resonators at Ultra High Frequencies.

The *HTS* thin film in a microstrip resonator, in which the variable magnetic field $H_{rf}$ of an electromagnetic wave has the amplitude $H_{max}$, which is close or exceeds the magnitude of the lower critical magnetic field $H_{c1}$ in a superconductor at microwaves is researched. It is assumed that the *Abricosov magnetic vortices* have a considerable influence on the nonlinear surface resistance of $YBa_2Cu_3O_{7-\delta}$ thin films on *MgO* substrate in proximity to low critical magnetic field $H_{c1}$ in superconducting microstrip resonators at ultra high frequencies. The processes of *Abricosov magnetic vortices* appearance in a superconductor at microwaves takes time duration equal to the period of an electromagnetic wave. The characteristic time of *Abricosov magnetic curl* origination (the nucleation time) in high-temperature superconductors is equal to time $10^{-11}$ *sec*, and it is much less than the period of an electromagnetic wave even in the *GHz* frequency range. By reviewing this researched case, it is necessary to take into the account that, if the electromagnetic wave amplitude is higher than the amplitude of critical magnetic field $H_{max} > H_{c1}$, then, because of the sine change of amplitude of magnetic field $H_{rf}$ in time, the magnetic field $H_{rf}$ will be less than the critical magnetic field $H_{c1}$ $(H_{rf} < H_{c1})$ in the initial part of the period of oscillation, after that, it will be equal to critical magnetic field $H_{c1}$ $(H_{rf} = H_{c1})$, and in the field of a maxima, it will be more than critical magnetic field $H_{c1}$ $(H_{rf} > H_{c1})$. Thus, the area of maxima of an electromagnetic wave gives a significant contribution to both the nonlinear absorption of electromagnetic wave as well as the nonlinear surface resistance $R_s$, where the expression $H_{rf} > H_{c1}$ is valid as it is shown in Fig. 23.

In Fig. 23, the modeling of time dependence of the electromagnetic waves with amplitudes 45, 65, 100 a.u. during their propagation in the region with the nonlinear surface resistance $R_s$ in *HTS* thin films in microstrip resonator in proximity to low critical magnetic field $H_{c1}=50$ *a.u.* at microwaves is shown. In the case of negative magnetic field $H_{rf}$, there are the linear and nonlinear areas of electromagnetic wave amplitudes [19].

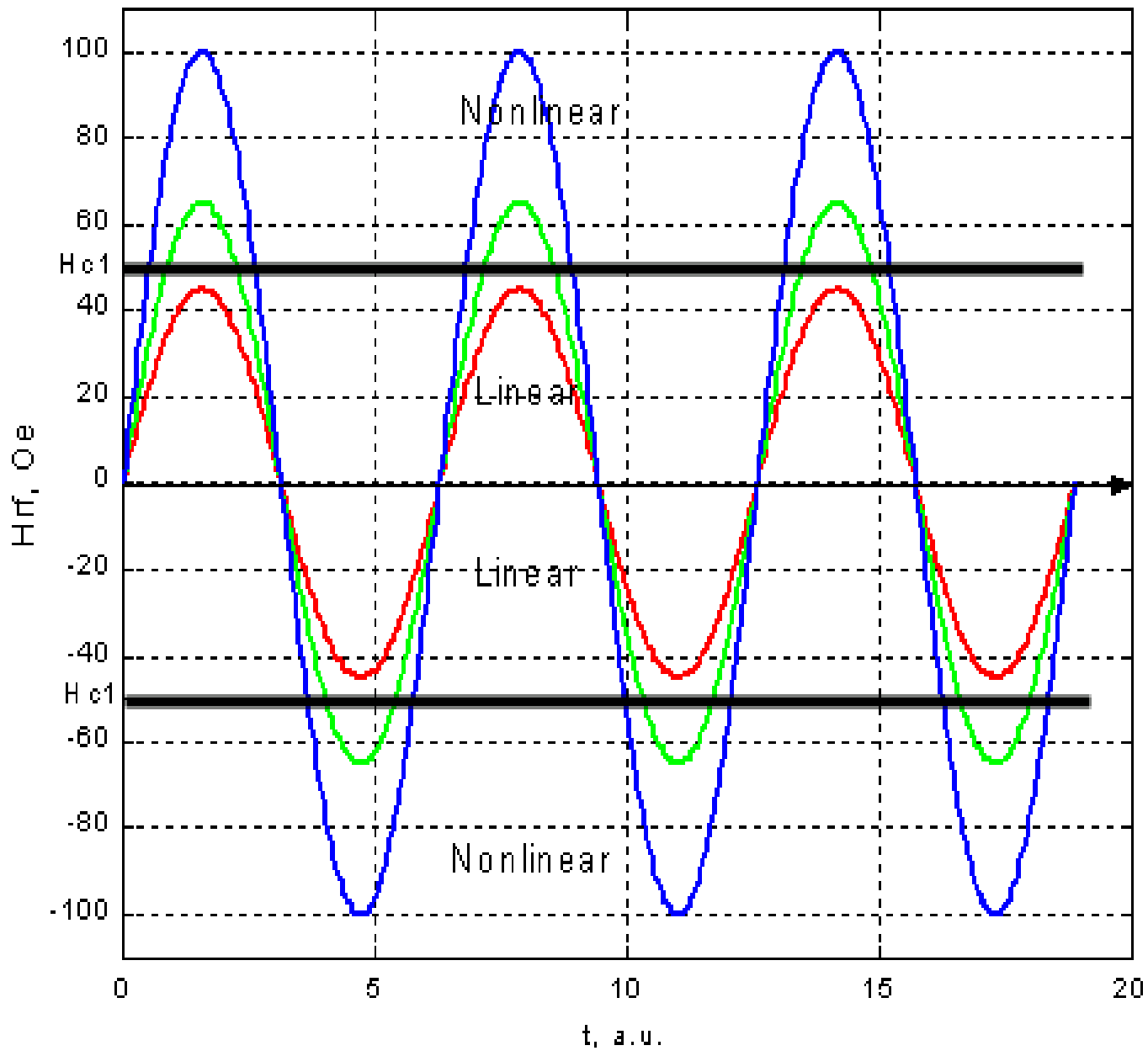

**Fig. 23.** Modeling of time dependence of magnetic fields of electromagnetic waves $H_{rf}$ with amplitudes *45a.u.* (red curve), *65a.u.* (green curve), *100a.u.* (blue curve) during their propagation in region with nonlinear surface resistance $R_s$ in *HTS* thin films in microstrip resonator in proximity to low critical magnetic field $H_{c1}$=*50a.u.* at microwaves [19].

In the three-dimensional graphs in Fig. 24, the modeling of surface resistance dependence on the time and magnetic field $R_s(t, H_{rf})$ for magnetic fields $H_{rf}$ with amplitudes 45a.u. (red curve), 65a.u. (green curve), 100a.u. (blue curve) during their propagation in the region with nonlinear surface resistance $R_s$ in *HTS* thin films in microstrip resonator in proximity to low critical magnetic field $H_{c1}$=*50a.u.* at microwaves is performed. The simulation was conducted for the dependence $R_s(H)$ following from the eq. (7.11) and supposing that $H_{c1}$=*50Oe*. The magnitude of surface resistance $R_s$ is shown along the axis *Z*. The magnitude of surface resistance $R_s$ changes non-linearly (blue and green curves), and depends on the amplitude of magnetic field $H_{rf}$ of ultra high frequency electromagnetic wave. It is visible, that the dependence $R_s(H)$ changes sharply, when the amplitude of ultra high frequency electromagnetic wave reaches lower critical magnetic field $H_{c1}$ (blue and green curves) [19].

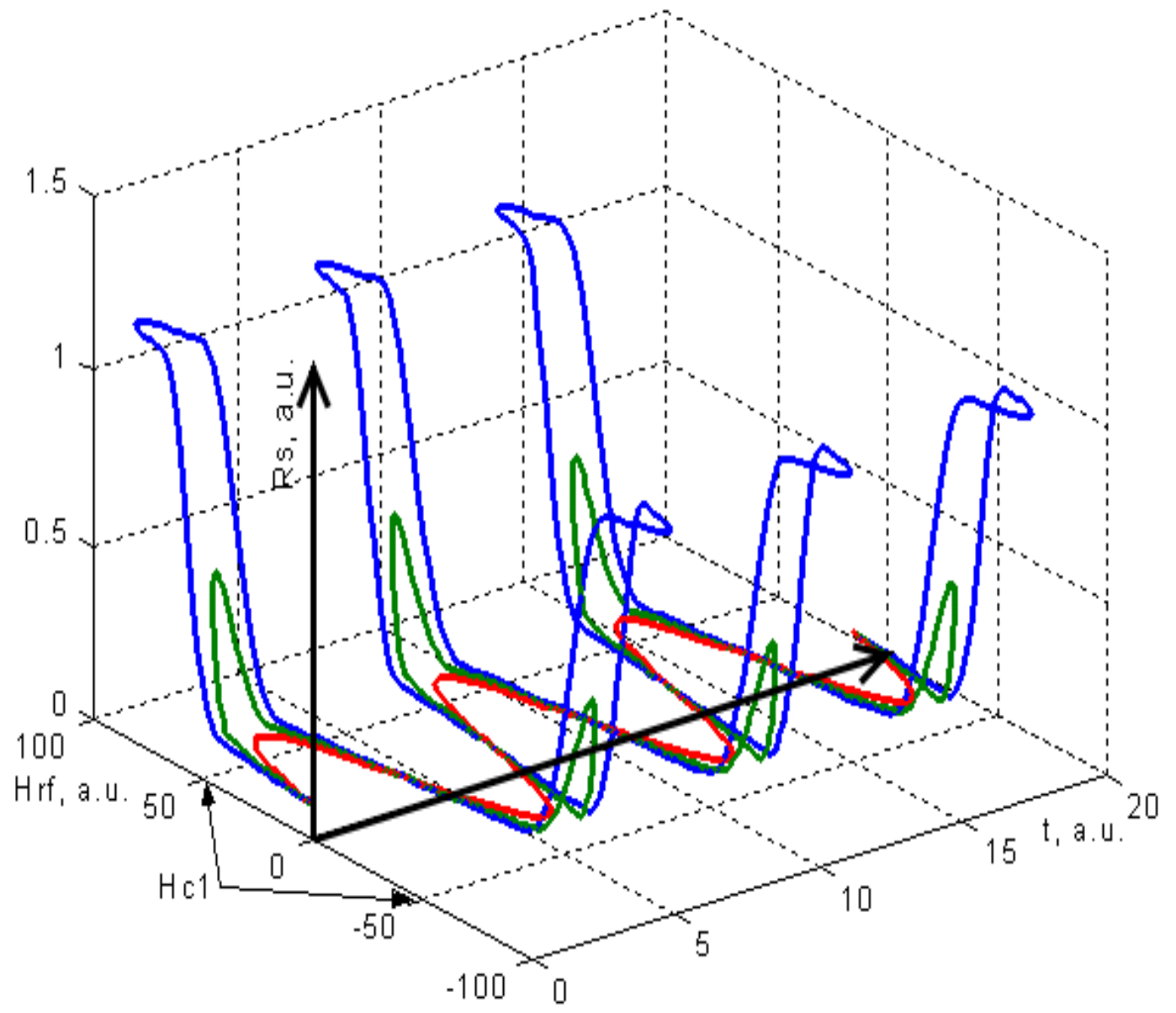

**Fig. 24.** Modeling of surface resistance dependence on time and on magnetic field $R_s(t,H_{rf})$ for magnetic fields $H_{rf}$ with amplitudes 45a.u. (red curve), 65a.u. (green curve), 100a.u. (blue curve) during their propagation in region with nonlinear surface resistance $R_s$ in *HTS* thin films in *microstrip resonator* in proximity to low critical magnetic field $H_{c1}$ *=50a.u.* at microwaves.

In Fig. 25, the modeling of dependence of surface resistance on time $R_s(t)$ for magnetic fields $H_{rf}$ with amplitudes *45a.u.* (red curve), *65a.u.* (green curve), *100a.u.* (blue curve) during their propagation in the region with nonlinear surface resistance $R_s$ in *HTS* thin films in *microstrip resonator* in proximity to low critical magnetic field $H_{c1}$ *=50a.u.* at microwaves is provided. In the experiment, the average value of surface resistance $<R_s>$ is measured, which is equal to the integral, taken on any selected curve, i.e. to the square of amplitude of surface resistance in Fig. 25 [19].

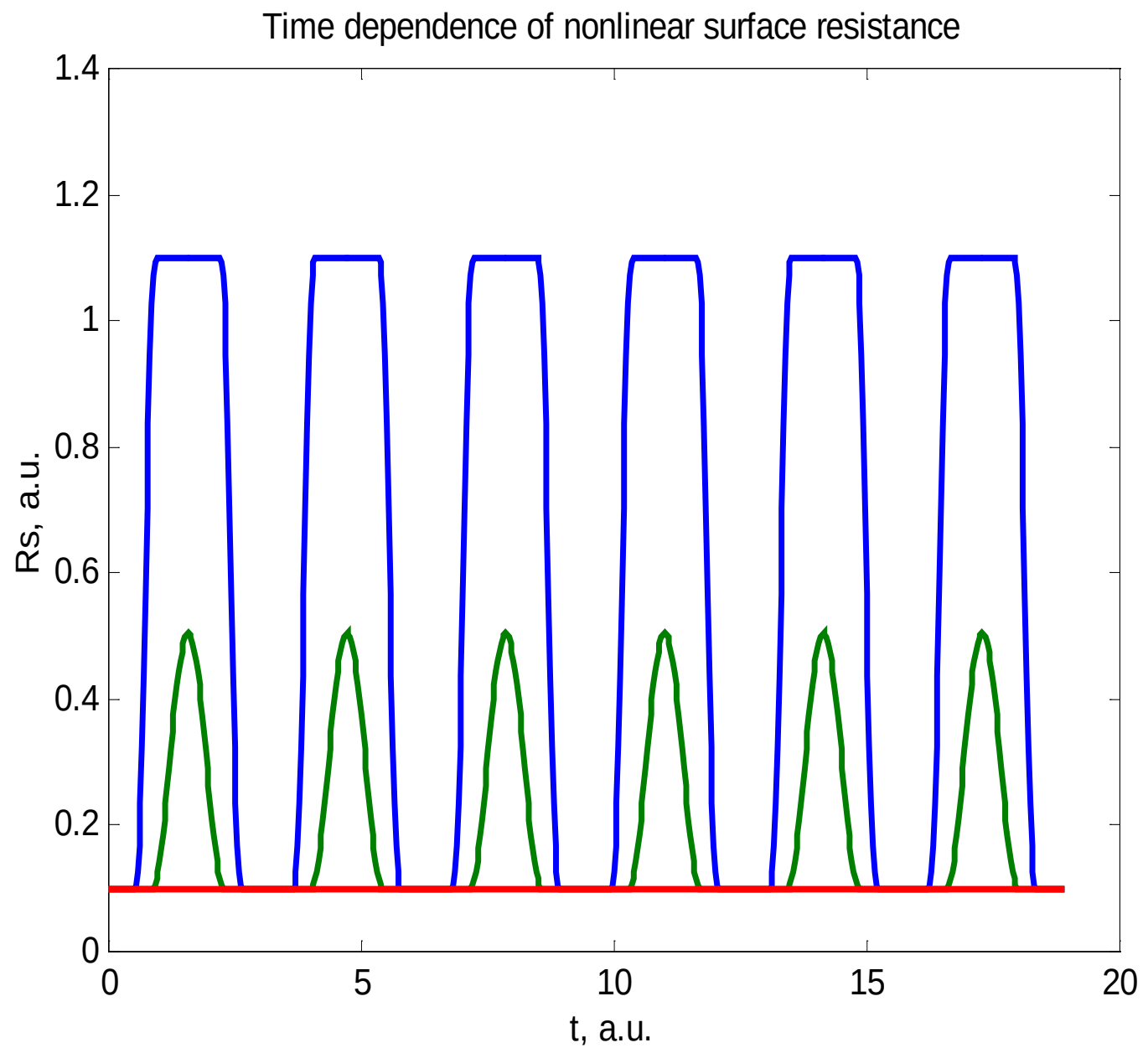

**Fig. 25.** Modeling of time dependence of surface resistance $R_s(t)$ in linear (red curve) and nonlinear (green and blue curves) cases for magnetic fields $H_{rf}$ with amplitudes *45a.u.* (red curve), *65a.u.* (green curve), *100a.u.* (blue curve) during their propagation in region with nonlinear surface resistance $R_s$ in *HTS* thin films in *microstrip resonator* in proximity to low critical magnetic field $H_{c1}$=50a.u. at microwaves. Residual resistance $R_0 = 0.1$ [19].

The calculation of change of square of surface resistance curves depending on the amplitude of magnetic field $R_s(H_{rf})$, was conducted for several selected cases. It was found that the average value of surface resistance $R_s$ can be written as in eq. (7.15)

$$<R_S(H_{rf})>=\frac{\int_0^{\pi/2} R_S(t)dt}{\pi/2}=\frac{R_0T_n+\int_{\varphi_n}^{\pi/2} R_S(t)dt}{\pi/2}, \qquad (7.15)$$

where $\varphi_n = arcsin(H_{c1}/H_{rf})$ is the phase angle of electromagnetic wave counted from the beginning of period of oscillation at which the magnetic field $H_{rf}(t)$ becomes equal to the critical magnetic $H_{c1}$. The calculation of square of amplitude of surface resistance is completed for a quarter of period from 0 up to $\pi/2$.

In Fig. 26, the modeling of time dependence of surface resistance $R_s(t)$ in linear (red curve) and nonlinear (green and blue curves) cases for magnetic fields $H_{rf}$ with amplitudes *45a.u.* (red curve), *65a.u.* (green curve), *100a.u.* (blue curve) during their propagation in the region with nonlinear surface resistance $R_s$ in *HTS* thin films in microstrip resonator in proximity to low critical magnetic field $H_{c1}$=50a.u. at microwaves is shown in the oscillation period limits from 0 up to π [19]

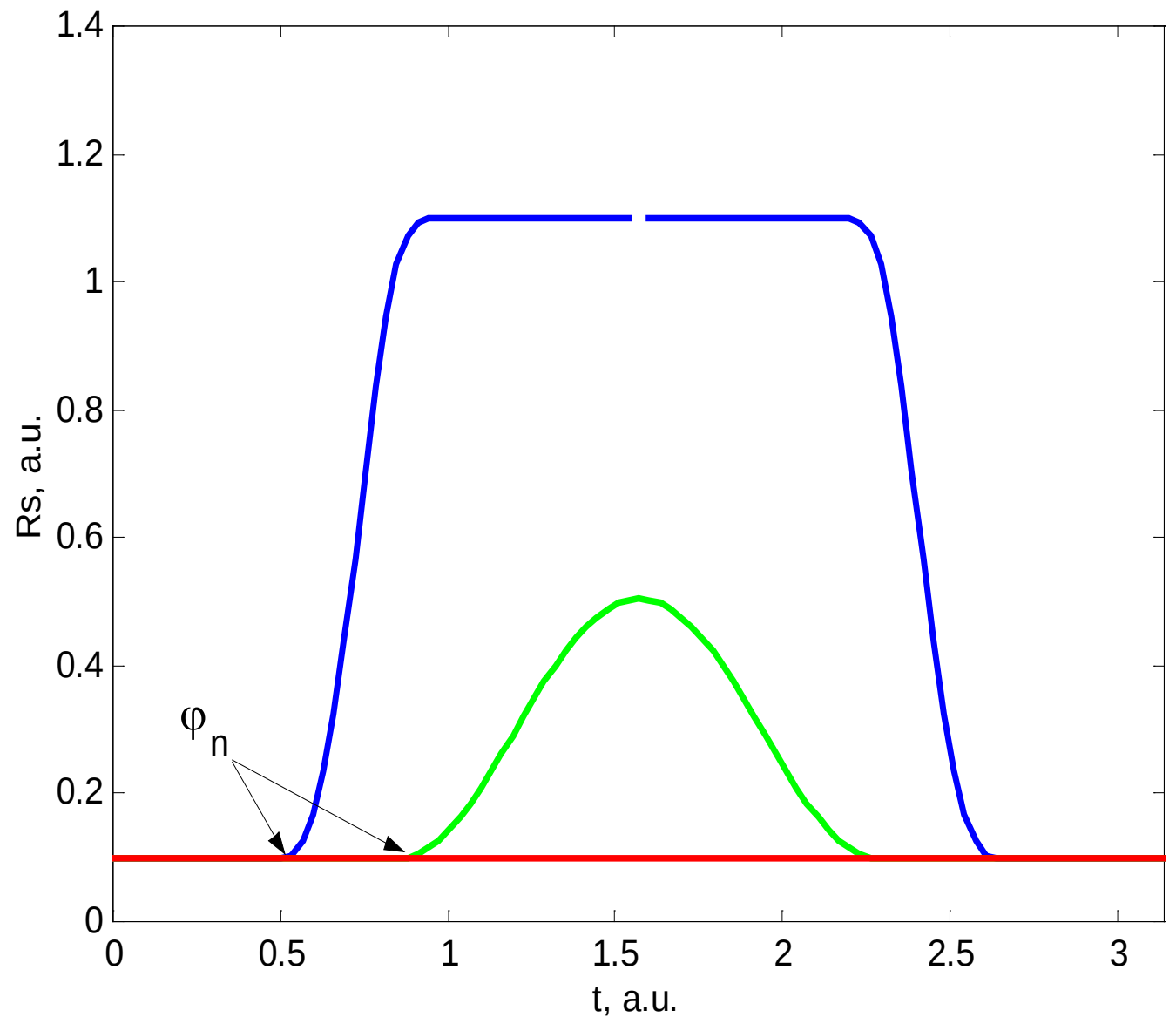

**Fig. 26.** Modeling of time dependence of surface resistance $R_s(t)$ in linear (red curve) and nonlinear (green and blue curves) cases for magnetic fields $H_{rf}$ with amplitudes *45a.u.* (red curve), *65a.u.* (green curve), *100a.u.* (blue curve) during their propagation in region with nonlinear surface resistance $R_s$ in $YBa_2Cu_3O_{7-\delta}$ superconducting microstrip resonator in proximity to low critical magnetic field $H_{c1}$*=50a.u.* at microwaves is shown in oscillation period phase limits from 0 up to π. Residual resistance $R_0 = 0.1$ [19].

In Fig. 26, the simulated curves are symmetric in relation to the point $\varphi=\pi/2$, where they reach a maxima. As it was mentioned above, the $\varphi_n=arcsin(H_{c1}/H_{rf})$ is the phase angle of electromagnetic wave counted from the beginning of period of oscillation at which the magnetic field $H_{rf}(t)$ becomes equal to the critical magnetic $H_{c1}$. The phase angle $\varphi_n$ is a function of magnetic field of electromagnetic wave $H_{rf}$, and when the magnitude of $H_{rf}$ increases, the value of phase angle $\varphi_n$ decreases, and therefore the total area of nonlinear mechanism is incremented, and the integral in a right member of the equation (7.15) is incremented. When the amplitude of magnetic field of an electromagnetic wave $H_{rf}$ is small, and the magnetic field of an electromagnetic wave is equal or high than critical magnetic field $H_{c1} \leq H_{rf}$, $\varphi_n = \pi/2$, and this integral in a right member of the equation (7.15) is equal to $0$, hence the lower and upper limits of integration are equal. In this case, there is no influence on surface resistance $R_s$ by nonlinearities in $YBa_2Cu_3O_{7-\delta}$ superconducting microstrip resonator at microwaves.

In Fig. 27, the dependence of phase angle of electromagnetic wave on magnetic field of electromagnetic wave $\varphi_n(H_{rf})$ in $YBa_2Cu_3O_{7-\delta}$ superconducting microstrip resonator for case, when magnetic field is above or equal to low critical magnetic field $H_{rf} \geq H_{c1} = 50Oe$ at microwaves is shown [19].

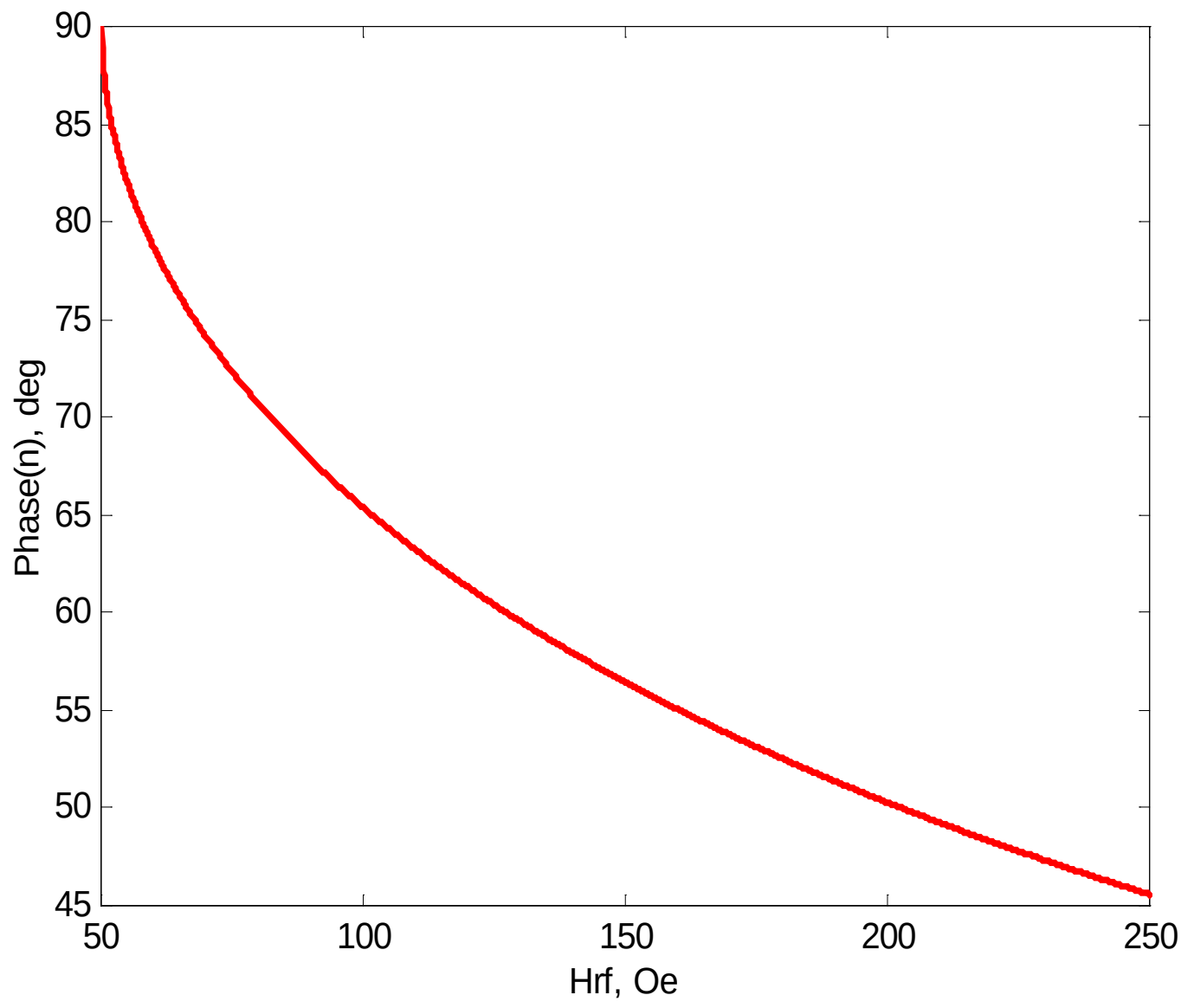

**Fig. 27.** Modeling of dependence of phase angle of electromagnetic wave on magnetic field of electromagnetic wave $\varphi_n(H_{rf})$ in $YBa_2Cu_3O_{7-\delta}$ superconducting microstrip resonator for case, when magnetic field is above or equal to low critical magnetic field $H_{rf} \geq H_{c1} = 50Oe$ at microwaves [19].

In Fig. 28, the modeling of dependence of an average nonlinear resistance $<R_s(H_{rf})>$, which was modeled on the base of equation (7.15), is shown [19].

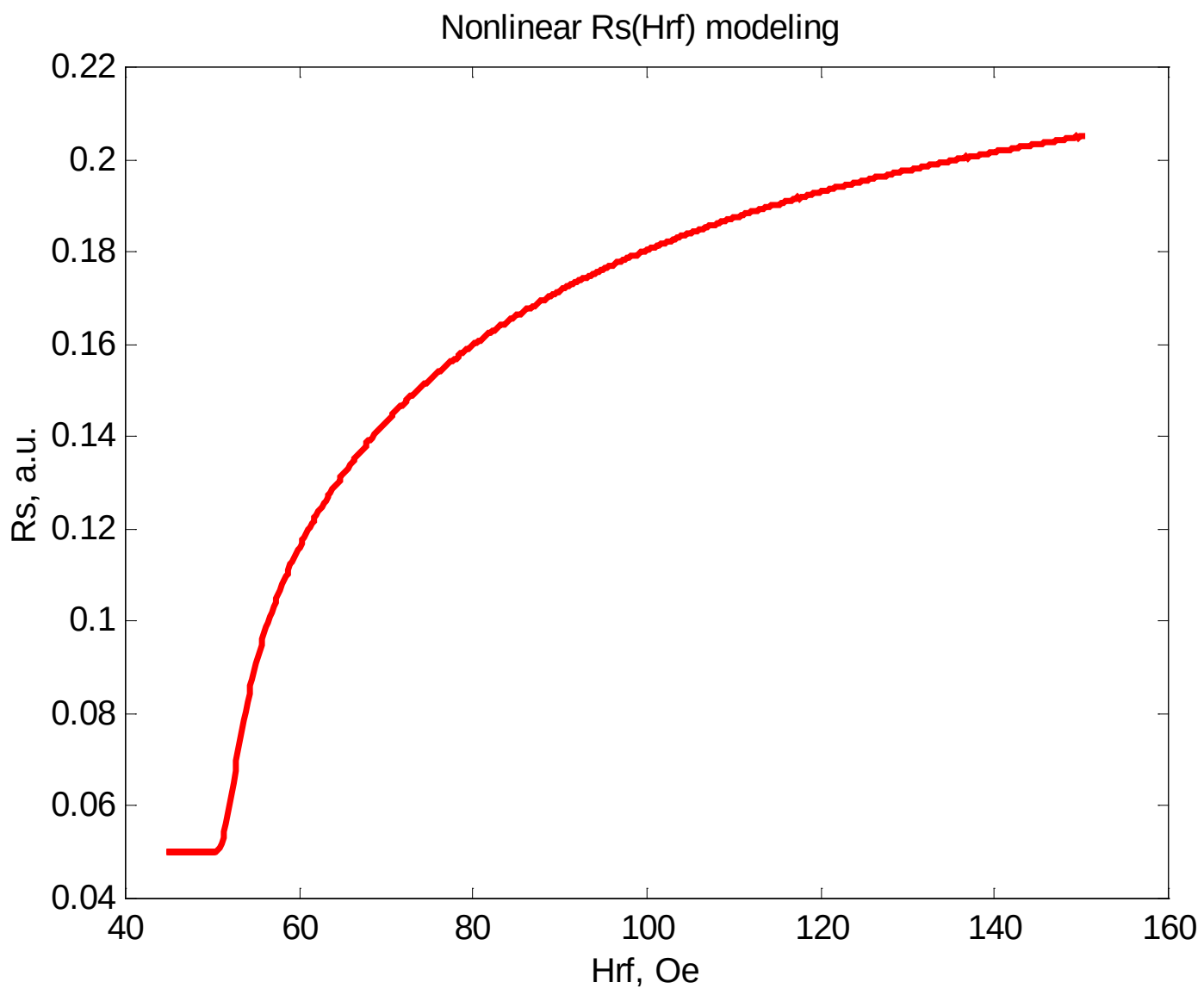

**Fig. 28.** Modeling of dependence of average surface resistance on magnetic field $R_s(H_{rf})$ in $YBa_2Cu_3O_{7-\delta}$ superconducting microstrip resonator with low critical magnetic field $H_{c1} = 50Oe$ at microwaves [19].

In Fig. 28, it is visible, that the average surface resistance $R_s$ increases at an increase of magnitude of magnetic field $H_{rf}$, and the average surface resistance $R_s$ has the S-type dependence, which is frequently observed in experiments.

In Fig. 29, the sharp change of the simulated dependence $dR_s/dH_{rf}$ $(H_{rf})$, representing the ratio of $dR_s/dH_{rf}$ derivatives as a function of magnetic field $H_{rf}$, in $YBa_2Cu_3O_{7-\delta}$ superconducting microstrip resonator at frequency $f \approx 2GHz$ at selected temperature of 82$K$ is clearly visible. The modeling of dependence $dR_s/dH_{rf}$ $(H_{rf})$ is performed for the experimental case described in Fig. 13 [19].

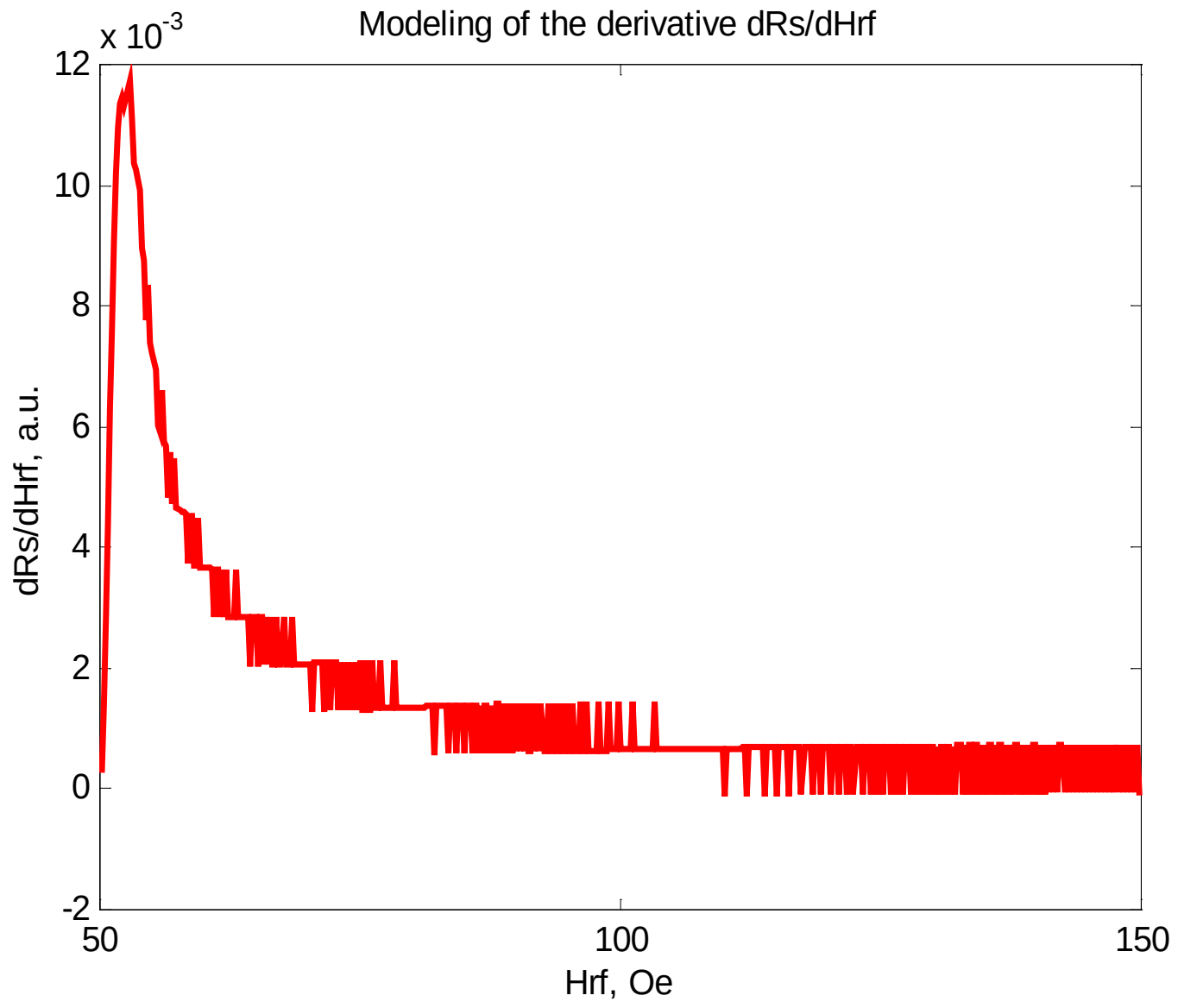

**Fig. 29.** Modeling of dependences $dR_s/dH_{rf}$ $(H_{rf})$, representing ratio of $dR_s/dH_{rf}$ derivatives as function of magnetic field $H_{rf}$ , in $YBa_2Cu_3O_{7-\delta}$ superconducting microstrip resonator at frequency $f \approx 2GHz$ at selected temperature of 82*K*. Modeling of dependence $dR_s/dH_{rf}$ $(H_{rf})$ is performed to simulate experimental dependence $dR_s/dH_{rf}$ $(H_{rf})$ in Fig.13 [19].

In Fig. 29, it is visible, that the maximum of dependence $dR_s/dH_{rf}$ $(H_{rf})$ is close to the critical magnetic field $H_{c1}$. This circumstance allows us to use these derivatives as the convenient tool in analysis of experimental data in [7, 8].

In Fig. 30, the graph of dependence of $log$ $(R_s)$ on $log$ $(10*H_{rf}/H_{c1})$ in $YBa_2Cu_3O_{7-\delta}$ superconducting microstrip resonator with low critical magnetic field $H_{c1}$ = $50Oe$ at microwaves is constructed on the logarithmic scale. The modeling of dependence of $log$ $(R_s)$ on $log$ $(10*H_{rf}/H_{c1})$ is performed for the $R_s(H_{rf})$ curve shown in Fig. 28 [19].

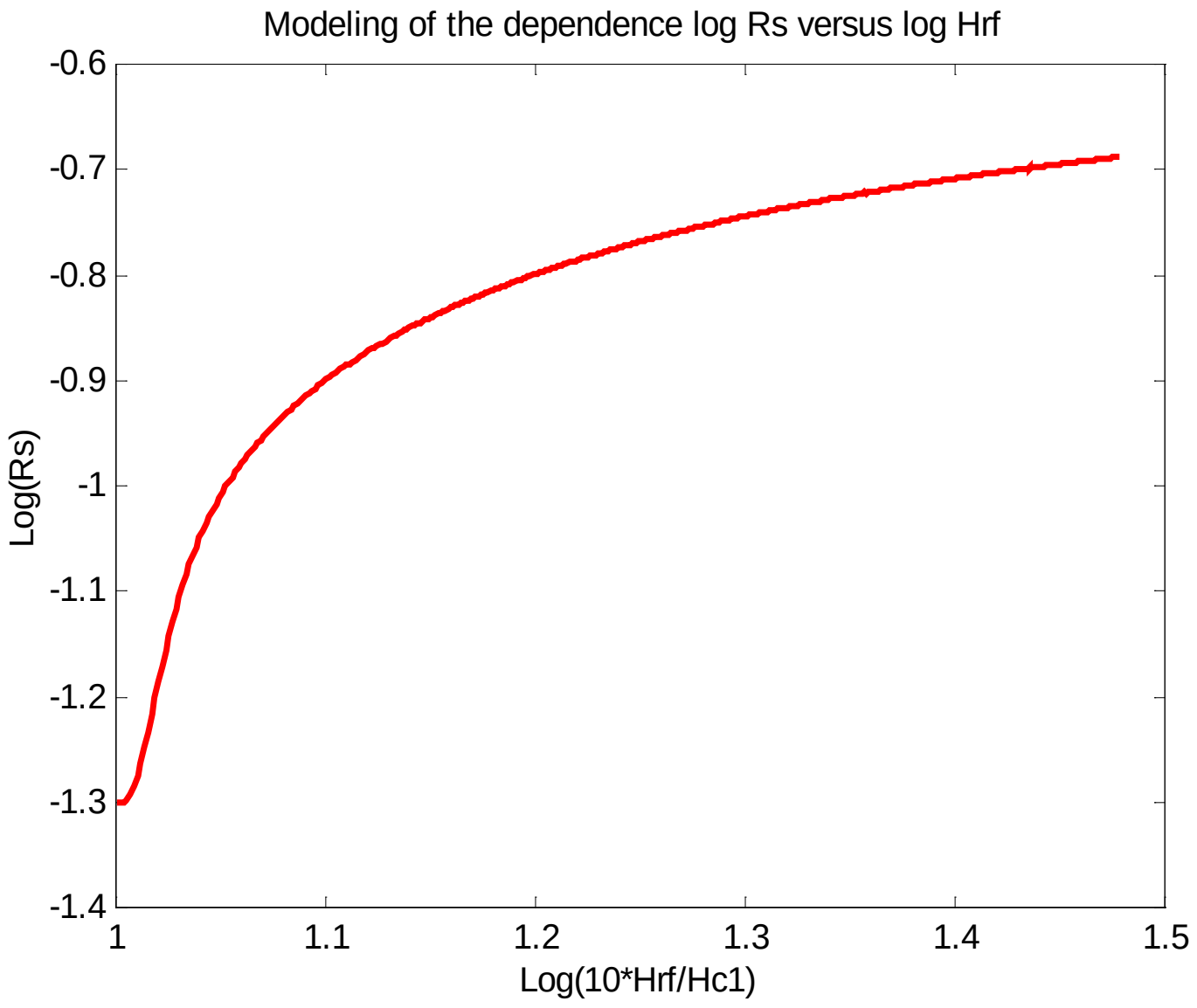

**Fig. 30.** Modeling of dependence of *log* $(R_s)$ *vs. log* $(10{*}H_{rf}/H_{c1})$ in $YBa_2Cu_3O_{7-\delta}$ superconducting microstrip resonator with low critical magnetic field $H_{c1} = 50Oe$ at microwaves. Modeling of dependence of *log* $(R_s)$ *vs. log* $(10{*}H_{rf}/H_{c1})$ is performed for $R_s(H_{rf})$ curve in Fig. 28 [19].

The presented simulations were created using the formula (6.11) in the *Matlab*. The results of simulations demonstrated that the nonlinear surface resistance in $YBa_2Cu_3O_{7-\delta}$ thin films on *MgO* substrates may have a little bit different dependence in the dielectric resonators than in the microstrip resonators. The more sluggish transition of surface resistance $R_s$ from the linear area to the nonlinear area with smaller derivative $dR_s(H_{rf})/dH_{rf}$ is observed in the case of a microstrip resonator.

In Fig. 31, the characteristic nonlinear dependences of relative surface resistance on magnetic field $R_s(H_{rf})/H_{rf}(300\ Oe)$ *vs.* $H_{rf}$ in $YBa_2Cu_3O_{7-\delta}$ superconducting dielectric and microstrip resonators with the low critical magnetic field $H_{c1}=50Oe$ at microwaves: the red curve corresponds to the dependence $R_s(H_{rf})/H_{rf}(300\ Oe)$ *vs.* $H_{rf}$ in the case of a dielectric resonator; the blue curve corresponds to the dependence $R_s(H_{rf})/H_{rf}(300\ Oe)$ *vs.* $H_{rf}$ in the case of a microstrip resonator are shown [19].

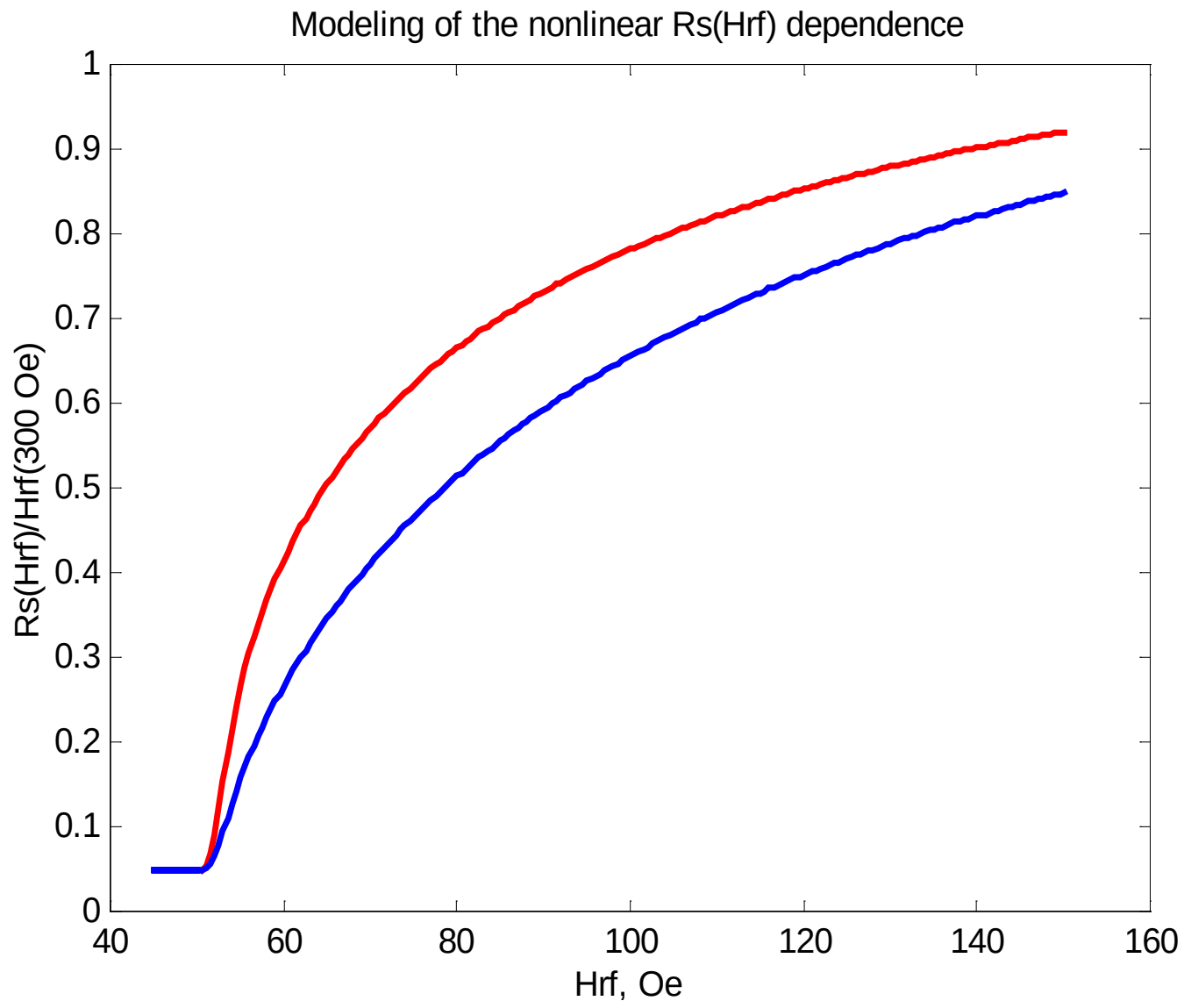

**Fig. 31.** Modeling of nonlinear dependence of relative surface resistance on magnetic field $R_s(H_{rf})/H_{rf}(300\ Oe)$ *vs.* $H_{rf}$ in $YBa_2Cu_3O_{7-\delta}$ superconducting dielectric and microstrip resonators with low critical magnetic field $H_{c1} = 50Oe$ at microwaves: red curve corresponds to dependence $R_s(H_{rf})/H_{rf}(300\ Oe)$ *vs.* $H_{rf}$ in case of dielectric resonator; blue curve corresponds to dependence $R_s(H_{rf})/H_{rf}(300Oe)$*vs.*$H_{rf}$ in case of *microstrip resonator* [19].

In Fig. 32, the nonlinear dependences of relative surface resistance on magnetic field $log(R_s(H_{rf})/H_{rf}(300\ Oe))$ *vs. log* $(10*H_{rf}/H_{c1})H_{rf}$ in $YBa_2Cu_3O_{7-\delta}$ superconducting *dielectric* and *microstrip* resonators with low critical magnetic field $H_{c1}=50Oe$ at microwaves are shown at a logarithmic scale: the red curve corresponds to the dependence $log(R_s(H_{rf})/H_{rf}(300\ Oe))$ *vs. log* $(10*H_{rf}/H_{c1})$ in the case of a *dielectric resonator*; the blue curve corresponds to the dependence $log(R_s(H_{rf})/H_{rf}(300\ Oe))$ *vs. log* $(10*H_{rf}/H_{c1})$ in the case of a *microstrip resonator*. Modeling of nonlinear dependences $log(R_s(H_{rf})/H_{rf}(300\ Oe))$ *vs. log* $(10*H_{rf}/H_{c1})$ is performed for $R_s(H_{rf})/H_{rf}(300\ Oe)$ *vs.* $H_{rf}$ curves shown in Fig. 31 [19].

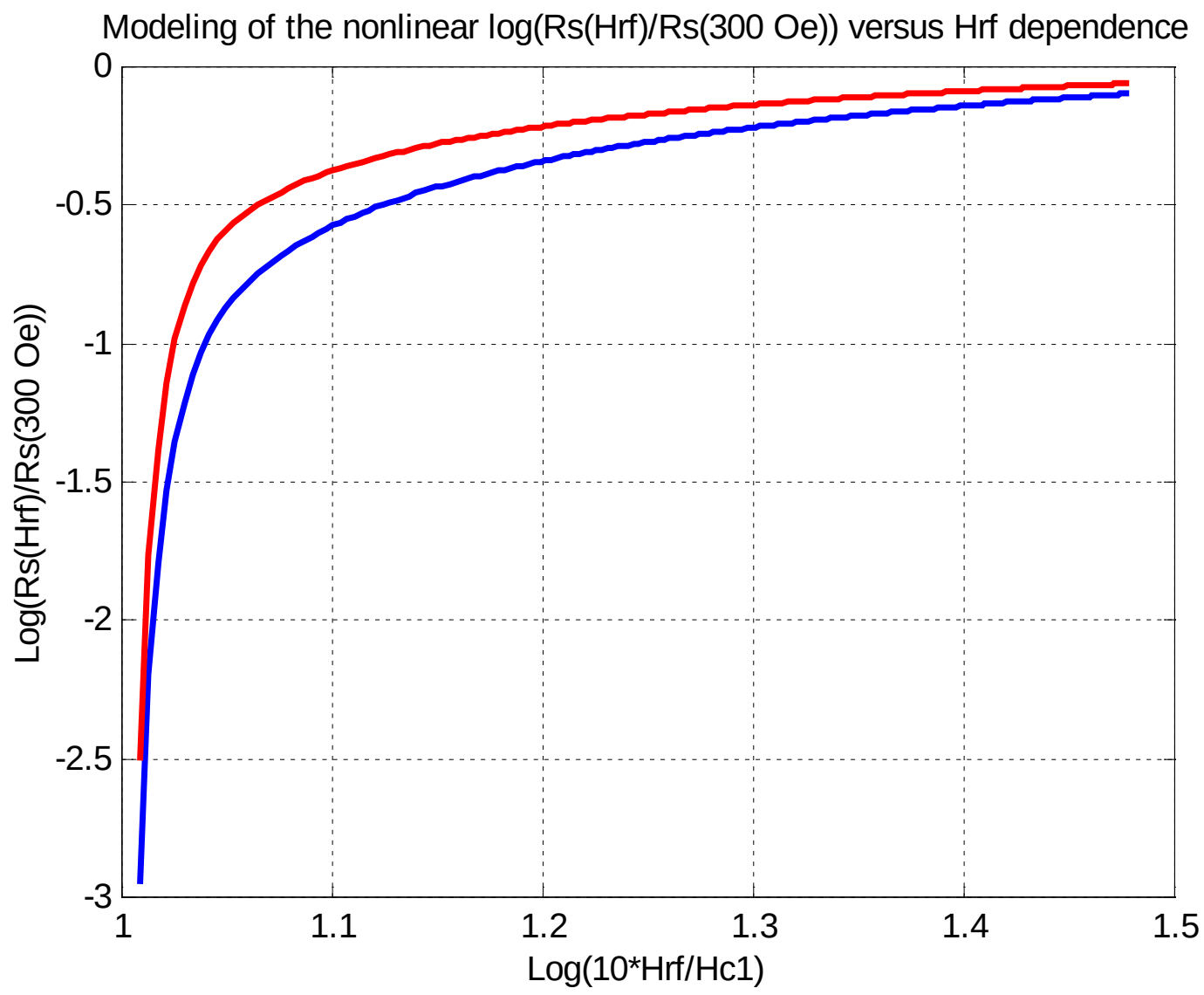

**Fig. 32.** Modeling of nonlinear dependences of relative surface resistance on magnetic field *log(R*$_s$*(H*$_{rf}$*)/H*$_{rf}$*(300 Oe)) vs. log (10*H*$_{rf}$*/H*$_{c1}$*)* in $YBa_2Cu_3O_{7-\delta}$ superconducting dielectric and microstrip resonators with low critical magnetic field $H_{c1}=50Oe$ at microwaves shown at a logarithmic scale: red curve corresponds to dependence *log(R*$_s$*(H*$_{rf}$*)/H*$_{rf}$*(300 Oe)) vs. log (10*H*$_{rf}$*/H*$_{c1}$*)* in case of dielectric resonator; blue curve corresponds to dependence *log(R*$_s$*(H*$_{rf}$*)/H*$_{rf}$*(300 Oe)) vs. log (10*H*$_{rf}$*/H*$_{c1}$*)* in case of microstrip resonator. Modeling of nonlinear dependences *log(R*$_s$*(H*$_{rf}$*)/H*$_{rf}$*(300 Oe)) vs. log (10*H*$_{rf}$*/H*$_{c1}$*)* is performed for $R_s(H_{rf})/H_{rf}(300\ Oe)$ *vs.* $H_{rf}$ curves in Fig. 31 [19].

In Fig. 33, the modeling of dependences $dR_s/dH_{rf}(H_{rf})$, representing the ratio of $dR_s/dH_{rf}$ derivatives as a function of magnetic field $H_{rf}$ , in $YBa_2Cu_3O_{7-\delta}$ superconducting dielectric and microstrip resonators with low critical magnetic field $H_{c1} = 50Oe$ at microwaves: the red curve corresponds to the dependence $dR_s/dH_{rf}(H_{rf})$ in the case of a *dielectric resonator*; the blue curve corresponds to the dependence $dR_s/dH_{rf}(H_{rf})$ in the case of a *microstrip resonator* are displayed. The modeling of dependences $dR_s/dH_{rf}(H_{rf})$ is performed for $R_s(H_{rf})/H_{rf}(300\ Oe)$ *vs.* $H_{rf}$ curves shown in Fig. 31. The dependence of derivatives on magnetic field $dR_s/dH_{rf}(H_{rf})$ is smaller in the case of a microstrip resonator than in the case of a dielectric resonator, because there is a more smoothly varying transition to the nonlinear area at increase of magnetic field $H_{rf}$ of an electromagnetic wave in a microstrip resonator. The maximums of curves $dR_s/dH_{rf}(H_{rf})$ are located in the points corresponding to the magnitudes of magnetic field equal to the low critical magnetic field $H_{C1}$ in $YBa_2Cu_3O_{7-\delta}$ thin films [19].

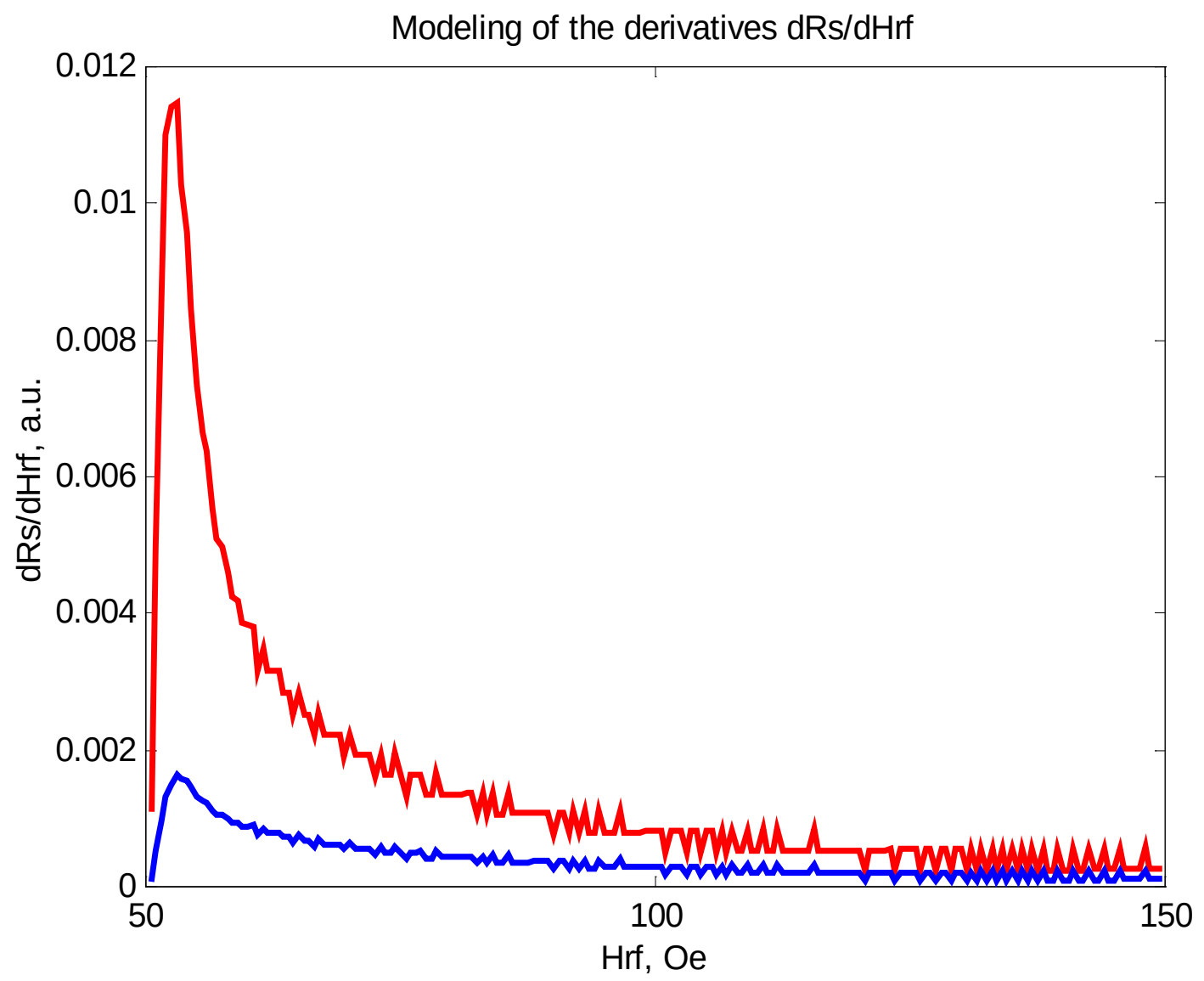

**Fig. 33.** Modeling of dependences $dR_s/dH_{rf}(H_{rf})$, representing ratio of $dR_s/dH_{rf}$ derivatives as function of magnetic field $H_{rf}$ , in $YBa_2Cu_3O_{7-\delta}$ superconducting dielectric and microstrip resonators with low critical magnetic field $H_{c1} = 50Oe$ at microwaves: red curve corresponds to dependence $dR_s/dH_{rf}(H_{rf})$ in case of dielectric resonator; blue curve corresponds to dependence $dR_s/dH_{rf}(H_{rf})$ in case of *microstrip resonator*. Modeling of dependences $dR_s/dH_{rf}(H_{rf})$ is performed for $R_s(H_{rf})/H_{rf}(300\ Oe)$ *vs.* $H_{rf}$ curves shown in Fig. 31 [19].

In Fig. 34, the modeling of dependences $dR_s/dH_{rf}(H_{rf})$, representing the ratio of $dR_s/dH_{rf}$ derivatives as a function of magnetic field $H_{rf}$, in $YBa_2Cu_3O_{7-\delta}$ superconducting *microstrip resonator* with low critical magnetic field $H_{c1} = 50Oe$ at microwaves is constructed on a different scale with better resolution. The modeling of dependences $dR_s/dH_{rf}(H_{rf})$ is performed for $R_s(H_{rf})/H_{rf}(300\ Oe)$ *vs.* $H_{rf}$ curve for a microstrip resonator in Fig. 33 [19].

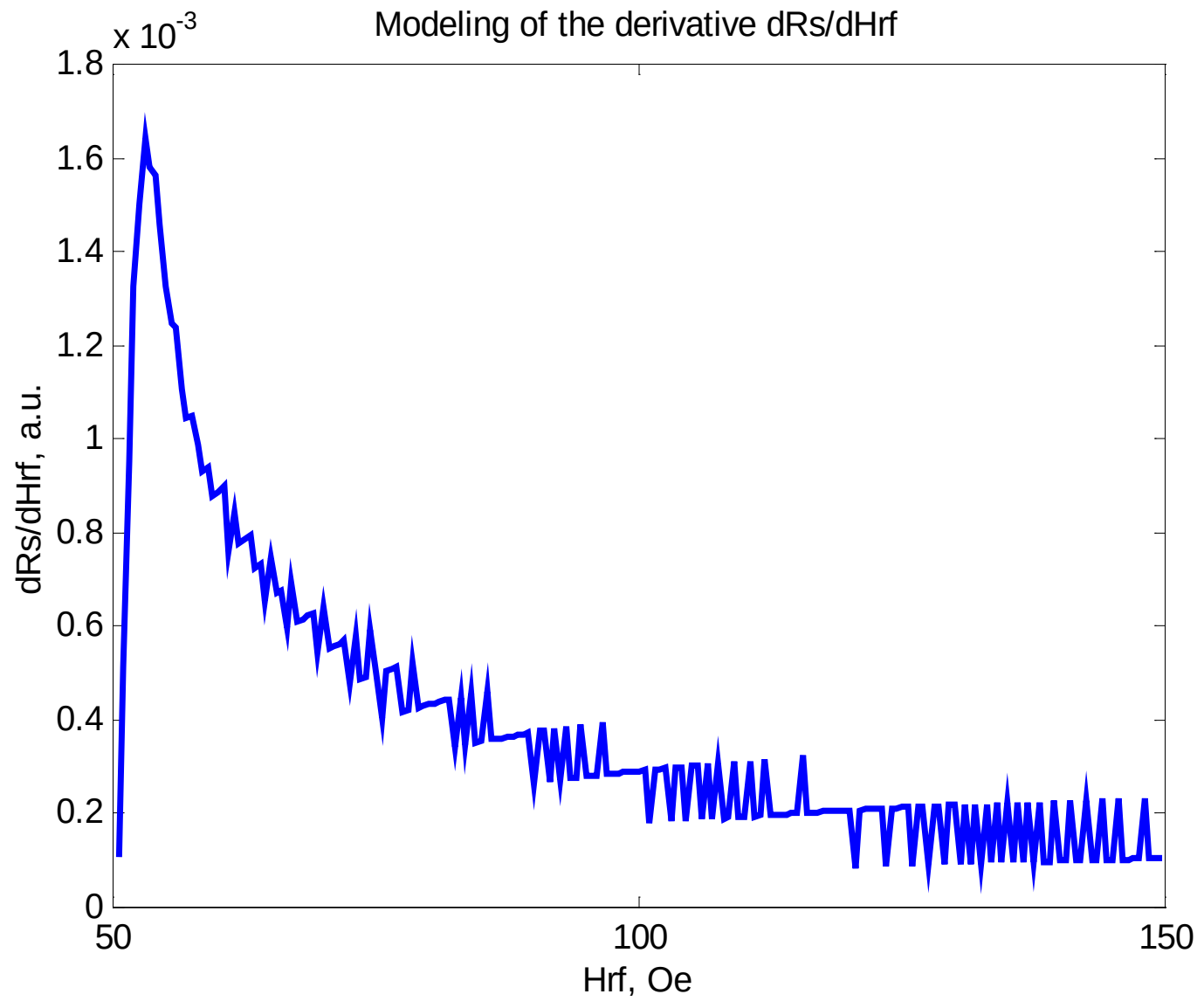

**Fig. 34.** Modeling of dependences *dR*$_s$*/dH*$_{rf}$*(H*$_{rf}$*)*, representing ratio of *dR*$_s$*/dH*$_{rf}$ derivatives as function of magnetic field $H_{rf}$ , in $YBa_2Cu_3O_{7-\delta}$ superconducting microstrip resonator with low critical magnetic field $H_{c1}$ = *50Oe* at microwaves constructed on a different scale. Modeling of dependences *dR*$_s$*/dH*$_{rf}$*(H*$_{rf}$*)* is performed for *R*$_s$*(H*$_{rf}$*)/H*$_{rf}$*(300 Oe) vs. H*$_{rf}$ curve for *microstrip resonator* in Fig. 33 [19].

Let's analyze the difference between the dielectric resonator and the microstrip resonator in the case, when the applied magnetic field is high or equal to the critical magnetic field $H_{rf} \geq H_{c1}$ in $YBa_2Cu_3O_{7-\delta}$ superconducting dielectric and microstrip resonators with low critical magnetic field $H_{c1}$ = *50Oe* at microwaves.

In Fig. 35, the modeling of nonlinear dependence of relative surface resistance on magnetic field *R*$_s$*(H*$_{rf}$*)/H*$_{rf}$*(300 Oe) vs. H*$_{rf}$ in $YBa_2Cu_3O_{7-\delta}$ superconducting dielectric and microstrip resonators in close proximity to the low critical magnetic field $H_{c1}$ = *50Oe* at microwaves: the red curve corresponds to the dependence *R*$_s$*(H*$_{rf}$*)/H*$_{rf}$*(300 Oe) vs. H*$_{rf}$ in the case of a dielectric resonator; the blue curve corresponds to the dependence *R*$_s$*(H*$_{rf}$*)/H*$_{rf}$*(300 Oe) vs. H*$_{rf}$ in the case of a microstrip resonator [19].

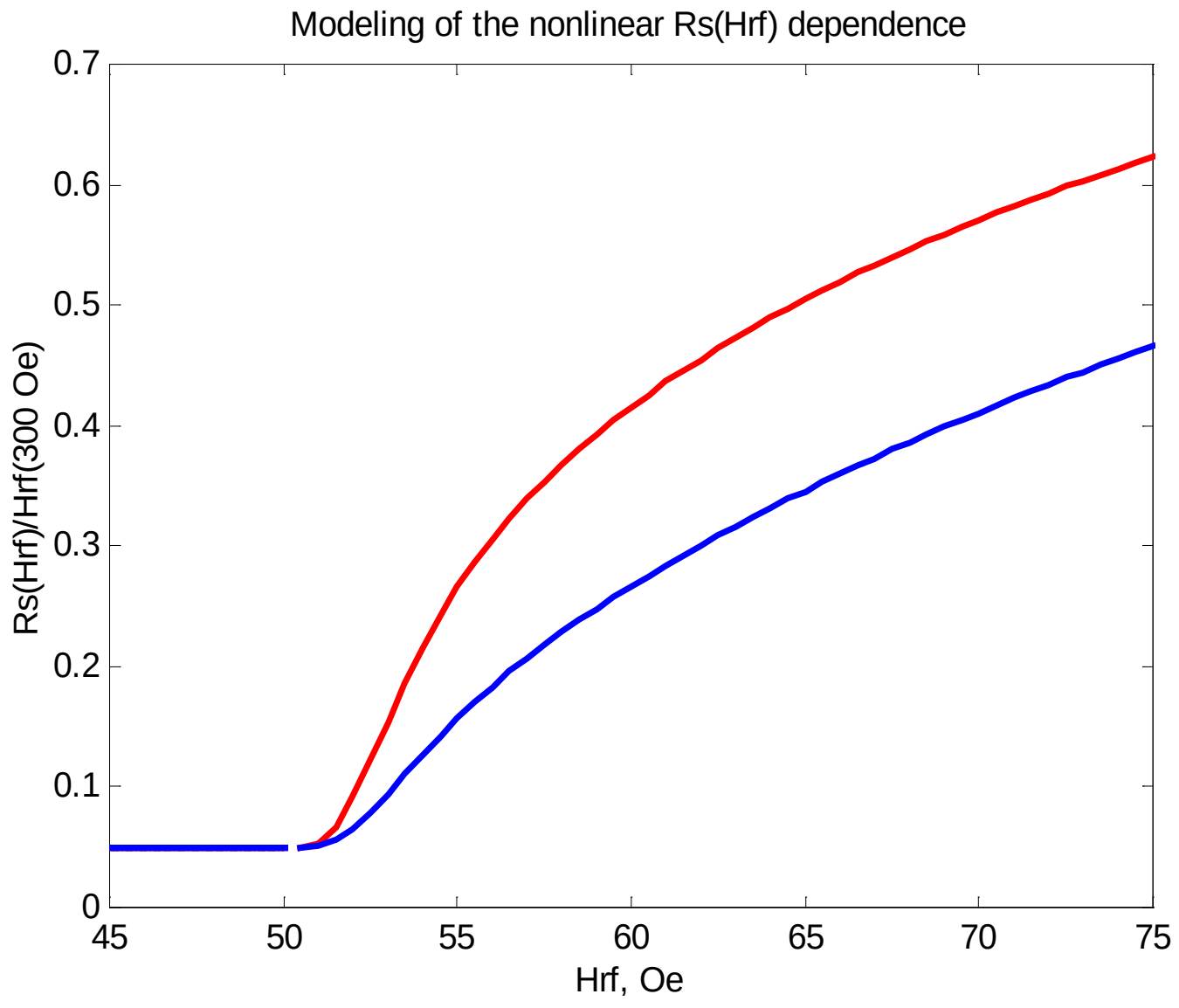

**Fig. 35.** Modeling of nonlinear dependence of relative surface resistance on magnetic field $R_s(H_{rf})/H_{rf}(300\ Oe)$ *vs.* $H_{rf}$ in $YBa_2Cu_3O_{7\text{-}\delta}$ superconducting dielectric and microstrip resonators in close proximity to low critical magnetic field $H_{c1}=50Oe$ at microwaves: red curve corresponds to dependence $R_s(H_{rf})/H_{rf}(300\ Oe)$ *vs.* $H_{rf}$ in case of dielectric resonator; blue curve corresponds to dependence $R_s(H_{rf})/H_{rf}(300\ Oe)$ *vs.* $H_{rf}$ in case of *microstrip resonator* [19].

In Fig. 36, the modeling of nonlinear dependences of relative surface resistance on magnetic field $log(R_s(H_{rf})/H_{rf}(300\ Oe))$ *vs.* $log\ (10{*}H_{rf}/H_{c1})$ in $YBa_2Cu_3O_{7\text{-}\delta}$ superconducting *dielectric* and *microstrip resonators* in close proximity to the low critical magnetic field $H_{c1}=50Oe$ at microwaves shown at a logarithmic scale: the red curve corresponds to the dependence $log(R_s(H_{rf})/H_{rf}(300Oe))$ *vs.* $log\ (10{*}H_{rf}/H_{c1})$ in the case of a dielectric resonator; the blue curve corresponds to the dependence $log(R_s(H_{rf})/H_{rf}(300\ Oe))$ *vs.* $log(10{*}H_{rf}/H_{c1})$ in the case of a microstrip resonator. The modeling of nonlinear dependences $log(R_s(H_{rf})/H_{rf}(300\ Oe))$ *vs.* $log\ (10{*}H_{rf}/H_{c1})$ is performed for $R_s(H_{rf})/H_{rf}(300\ Oe)$ *vs.* $H_{rf}$ curves in Fig. 35 [19].

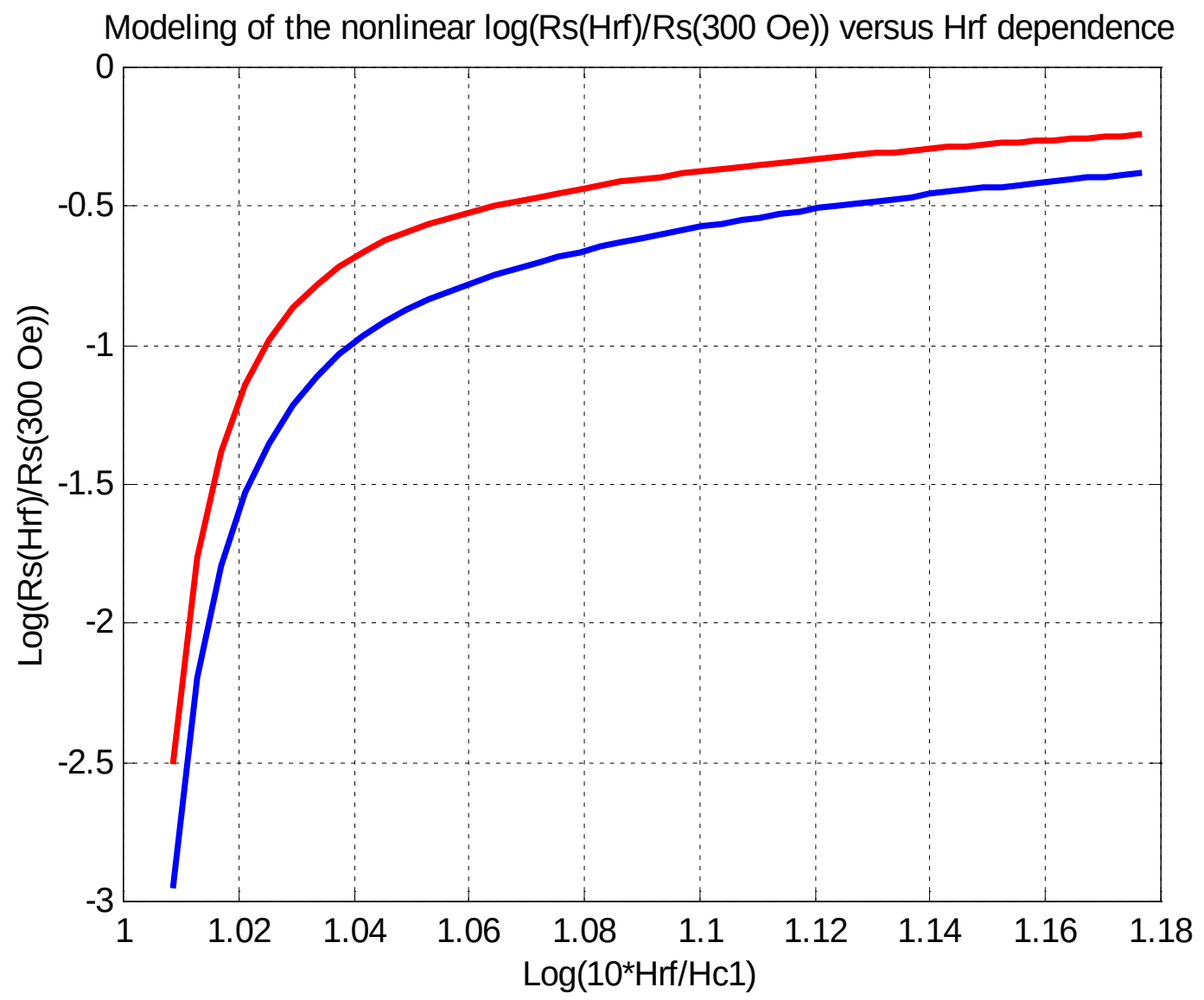

**Fig. 36.** Modeling of nonlinear dependences of relative surface resistance on magnetic field $log(R_s(H_{rf})/H_{rf}(300\ Oe))$ *vs.* $log\ (10{*}H_{rf}/H_{c1})$ in $YBa_2Cu_3O_{7-\delta}$ superconducting dielectric and microstrip resonators in close proximity to low critical magnetic field $H_{c1}=50Oe$ at microwaves shown at a logarithmic scale: red curve corresponds to dependence $log(R_s(H_{rf})/H_{rf}(300\ Oe))$ *vs.* $log\ (10{*}H_{rf}/H_{c1})$ in case of dielectric resonator; blue curve corresponds to dependence $log(R_s(H_{rf})/H_{rf}(300\ Oe))$ *vs.* $log\ (10{*}H_{rf}/H_{c1})$ in case of microstrip resonator. Modeling of nonlinear dependences $log(R_s(H_{rf})/H_{rf}(300\ Oe))$ *vs.* $log\ (10{*}H_{rf}/H_{c1})$ is performed for $R_s(H_{rf})/H_{rf}(300\ Oe)$ *vs.* $H_{rf}$ curves in Fig. 35 [19].

In Fig. 37, the modeling of dependences $dR_s/dH_{rf}(H_{rf})$, representing the ratio of $dR_s/dH_{rf}$ derivatives as a function of magnetic field $H_{rf}$, in $YBa_2Cu_3O_{7-\delta}$ superconducting dielectric and microstrip resonators in close proximity to the low critical magnetic field $H_{c1} = 50Oe$ at microwaves: the red curve corresponds to the dependence $dR_s/dH_{rf}(H_{rf})$ in the case of a dielectric resonator; the blue curve corresponds to the dependence $dR_s/dH_{rf}(H_{rf})$ in the case of a microstrip resonator are demonstrated. The modeling of dependences $dR_s/dH_{rf}(H_{rf})$ is performed for $R_s(H_{rf})/H_{rf}(300\ Oe)$ *vs.* $H_{rf}$ curves shown in Fig. 35 [19].

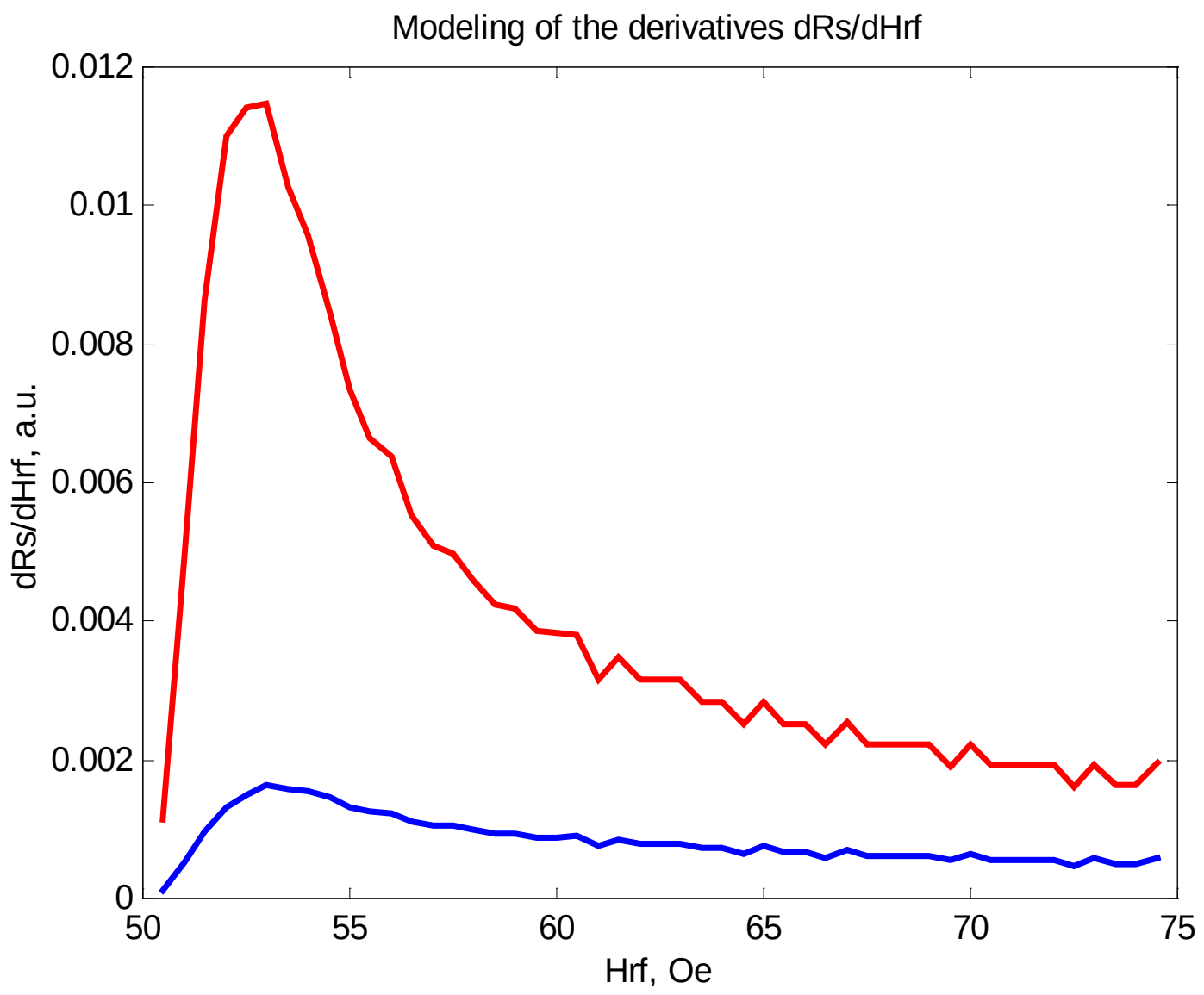

**Fig. 37.** Modeling of dependences $dR_s/dH_{rf}(H_{rf})$, representing the ratio of $dR_s/dH_{rf}$ derivatives as a function of magnetic field $H_{rf}$ , in $YBa_2Cu_3O_{7-\delta}$ superconducting dielectric and microstrip resonators in close proximity to low critical magnetic field $H_{c1} = 50Oe$ at microwaves: red curve corresponds to dependence $dR_s/dH_{rf}(H_{rf})$ in case of dielectric resonator; blue curve corresponds to dependence $dR_s/dH_{rf}(H_{rf})$ in case of microstrip resonator. Modeling of dependences $dR_s/dH_{rf}(H_{rf})$ is performed for $R_s(H_{rf})/H_{rf}(300\ Oe)$ *vs.* $H_{rf}$ curves in Fig. 35 [19].

In Fig. 38, the modeling of dependences $dR_s/dH_{rf}(H_{rf})$, representing the ratio of $dR_s/dH_{rf}$ derivatives as a function of magnetic field $H_{rf}$, in $YBa_2Cu_3O_{7-\delta}$ superconducting microstrip resonator in close proximity to low critical magnetic field $H_{c1} = 50Oe$ at microwaves is constructed on a different scale with better resolution. The modeling of dependences $dR_s/dH_{rf}(H_{rf})$ is performed for $R_s(H_{rf})/H_{rf}(300\ Oe)$ *vs.* $H_{rf}$ curve for a microstrip resonator in Fig. 37 [19]

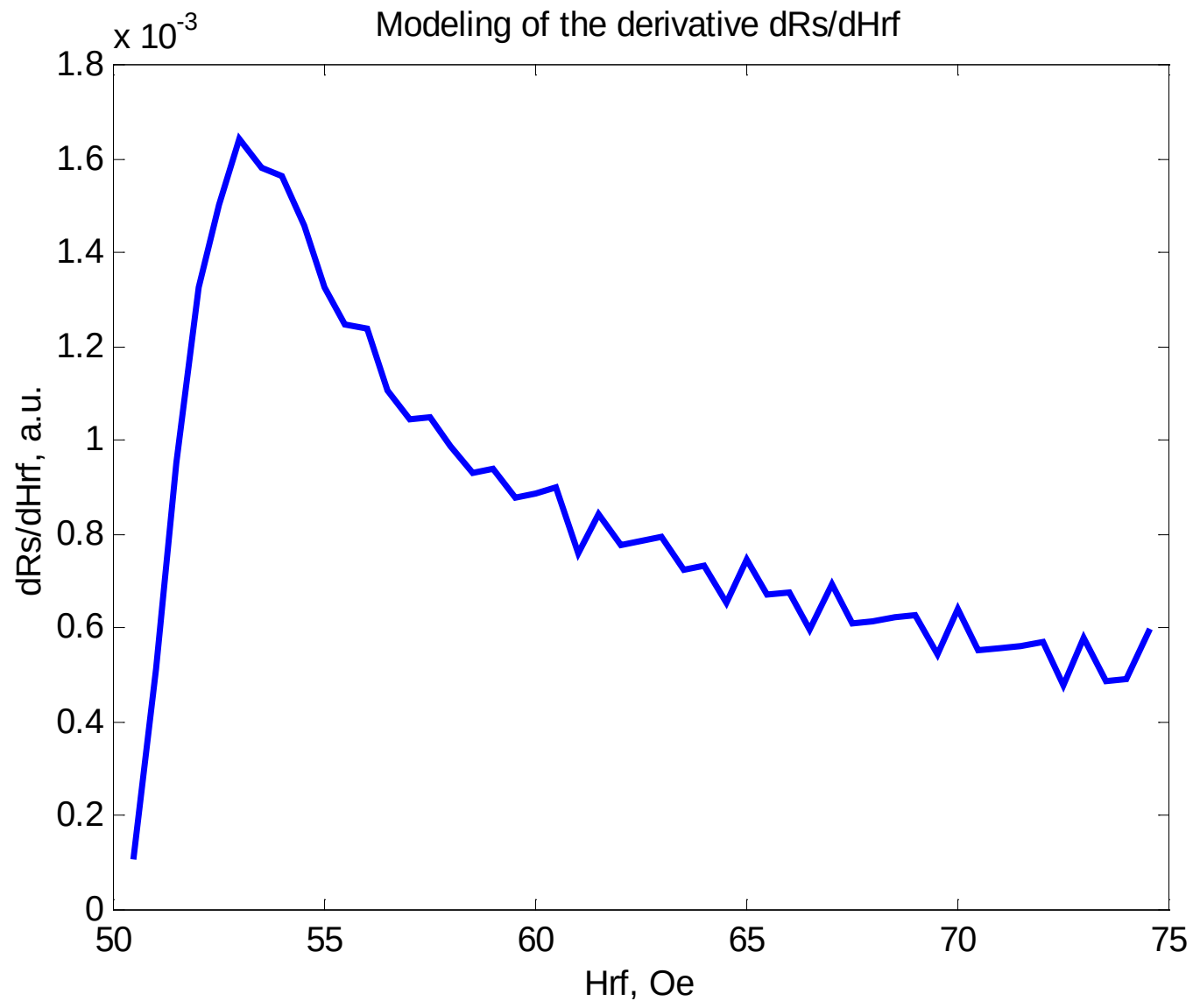

**Fig. 38.** Modeling of dependences $dR_s/dH_{rf}(H_{rf})$, representing ratio of $dR_s/dH_{rf}$ derivatives as function of magnetic field $H_{rf}$ , in $YBa_2Cu_3O_{7-\delta}$ superconducting microstrip resonator in close proximity to low critical magnetic field $H_{c1} = 50Oe$ at microwaves constructed on a different scale. Modeling of dependences $dR_s/dH_{rf}(H_{rf})$ is performed for $R_s(H_{rf})/H_{rf}(300\ Oe)$ *vs.* $H_{rf}$ curve for microstrip resonator in Fig. 37 [19].

In these research, author of dissertation conducted a number of computer modelings [19, 20], and conducted a comparative analysis of the simulated results with experimental results with the purpose to analyze the obtained research results and understand the nature of nonlinear surface resistance $R_s$ in $YBa_2Cu_3O_{7-\delta}$ thin films on *MgO* substrate in a microstrip resonator.

The characteristic experimental *S*-shape dependences of surface resistance on applied magnetic field $R_s(H_{rf})$ in $YBa_2Cu_3O_{7-\delta}$ thin films on *MgO* substrate in a microstrip resonator were observed in [9, 10]. The similar $R_s(H_{rf})$ curves were measured by author of dissertation in Figs. 11 and 12. The typical *S*-dependences of the surface resistance on the magnetic field $R_s(H_{rf})$ in $YBa_2Cu_3O_{7-\delta}$ thin films on *MgO* substrates in the microstrip and dielectric resonators were modeled by author of dissertation in [19]. The circumstance that, the typical *S*-dependences $R_s(H_{rf})$ are observed in rather small fields, can be interlinked to the fact that the plane *HTS* thin film samples could have a major demagnetization factor and consequently, the effective magnetic fields $H_{rf}*$ on their surface have higher magnitude than the magnetic field $H_{rf}$ of an electromagnetic wave. Therefore, these magnetic fields can reach the magnitudes comparable to the magnitude of low critical magnetic field $H_{rf}{\sim}H_{c1}$, at which the transitions to the nonlinear mode in rather small applied magnetic fields are registered in $YBa_2Cu_3O_{7-\delta}$ thin films on *MgO* substrate in a microstrip resonators at microwaves.

As it is visible from the presented research, the computer modeling results are in good agreement with the experimental data and allow author of dissertation to attempt to clarify the nature of *S*-type dependences of nonlinear surface resistance on magnetic field $R_s(H_{rf})$ in $YBa_2Cu_3O_{7-\delta}$ thin films on *MgO* substrates in superconducting microstrip resonators at microwaves.

## 7.9. Discussion on Physical Nature of Nonlinear Surface Resistance $R_S$ of $YBa_2Cu_3O_{7-\delta}$ Thin Films on MgO Substrates in Superconducting Microstrip Resonators at Ultra High Frequencies.

One of the main purposes of research is the investigation on nonlinear properties of $YBa_2Cu_3O_{7-\delta}$ superconductor thin films at microwaves, which can be used to develop the advanced passive and active electronic devices operating at ultra high frequencies. The nonlinear surface resistance $R_s$ in $YBa_2Cu_3O_{7-\delta}$ superconductor thin films at microwaves creates a number of difficulties in development of electronic devices with application of *HTS* thin films, because of the generation of harmonic oscillations and intermodulation products leading to an appearance of different kinds of microwave signal distortions as a result of changing nonlinear surface resistance $R_s$ in *HTS* thin films at microwaves.

The physical nature of nonlinear phenomena in *HTS* thin films is not clear and may have different origins. For example, the thermal over-heating of *HTS* thin films during the increase of microwave power of an ultra high frequency electromagnetic wave, which can appear in low quality *HTS* thin films with not optimized compounds, can be considered as a main source of nonlinearities. However, the overheating effects were not observed in high quality $YBa_2Cu_3O_{7-\delta}$ thin films, made by *Theva Gmbh* in Germany, researched by author of dissertation at *JCU* in Australia. Also, the transport of superconducting currents with big magnitudes, which are close to the values of critical currents of de-pairing in a superconductor, can originate the nonlinearities in *HTS* thin films at microwaves. However, the probability of presence of this mechanism behind the nonlinearities origination in researched $YBa_2Cu_3O_{7-\delta}$ thin films at microwaves is low.

The research on physical properties of both $YBa_2Cu_3O_{7-\delta}$ thin films as well as *MgO* substrates used for deposition of *HTS* thin films was conducted with the purpose to obtain veracious data on true causes of nonlinearities in *HTS* thin films at microwaves. Some authors proposed that the *MgO* substrates may contribute to the appearance of nonlinearities in *HTS* thin films at low temperature $T<10\ K$ as it was mentioned by D. E. Oates. In this case, the interesting effect, when the measured surface resistance $R_s$ decreases in a certain range of microwave powers at an increase of applied microwave power $P$ of an electromagnetic wave is observed. This effect may be connected with the existence of system of dielectric two levels excitations in the transition layer between the superconductor and the *MgO* substrate, probably, that is why the described physical behaviour of surface resistance $R_s$ at microwaves was observed by different authors [11-15].

The author of dissertation experimental and theoretical researches on the nonlinear surface resistance $R_s$ of $YBa_2Cu_3O_{7-\delta}$ thin films in a *Hakki-Coleman dielectric resonator* (*HCDR*), made it possible to find the minimum of surface resistance $R_s$ in $R_s(P)$ dependence in $YBa_2Cu_3O_{7-\delta}$ thin films on *MgO* substrate in range of microwave signal powers $P=0\div10\ dBm$ at frequency $f = 25\ GHz$ at temperatures $T = 25\ K$, 50 $K$. This minimum of surface resistance $R_S$ is similar to one, which was reported in a dielectric resonator in a range of microwave signal powers $P = -40 \div 0\ dBm$ at temperature $T = 5\ K$ in [12-13] and its physical nature may be connected with the interaction of an electromagnetic wave with electrical dipoles in *MgO* substrate at microwaves.

In this research, the preference in computer modeling was given to the physical mechanism, which is connected with the penetration of *Abricosov magnetic vortices* into a superconductor, and their nonlinear influence on surface resistance $R_s$ of $YBa_2Cu_3O_{7-\delta}$ thin films on *MgO* substrates in dielectric and microstrip resonators at microwaves, because the nonlinear phenomena were experimentally observed in close proximity to the critical magnetic fields $H_{CJ1}$, $H_{C1}$ and $H_{C2}$ in $YBa_2Cu_3O_{7-\delta}$ thin films on *MgO* substrates at microwaves only. ***Author proposes that the minimum of surface resistance $R_s$ is because of interaction of the system of magnetic dipoles, including the Josephson and Abricosov magnetic vortices, with ultra high frequency signal, but it is not connected with the electrical dipoles of two level system.*** In other words, there are two types of excitations in *HTS* thin films at microwaves:

1) *Josephson magnetic vortices* start to appear at relatively small critical magnetic field $H_{CJ1}$ [24-27];

2) *Abricosov magnetic vortices* originate at stronger critical magnetic field $H_{C1}$ [28].

The two types of excitations interact with the magnetic field of an electromagnetic wave, and can absorb or irradiate the energy of electromagnetic field in *HTS* thin films at microwaves, because of the interaction of the system of magnetic dipoles, which was observed in $YBa_2Cu_3O_{7-\delta}$ thin films on *MgO* substrate in a range of microwave signal powers $P = 0 \div 10\ dBm$ at frequency $f = 25\ GHz$ at temperatures $T = 25\ K,\ 50\ K$. The author of dissertation proposes that there is a minimum of surface resistance $R_S$, when the microwave power of an ultra high frequency electromagnetic wave is big enough and the *Josephson magnetic vortices* are present, but the Abricosov magnetic vortices, which will increase the surface resistance $R_S$, are not present in *HTS* thin films at microwaves. In this case, the minimum of surface resistance $R_S$ is connected with the decrease of stochastic scattering of electromagnetic field, absorbed by magnetic excitations, because the forced coherent counter-radiation of electromagnetic wave from system of magnetic vortical excitations to ultra high frequency field of an electromagnetic wave has place at increase of microwave power of an electromagnetic wave - the same way as in the case of interaction of ultra high frequency electromagnetic wave with electrical dipole two level system at microwaves. In this situation, during the registered absorption of an electromagnetic wave, the surface resistance $R_S$ decreases in *HTS* thin films at microwaves.

In quantum physics, this irradiation is described as the forced coherent radiation in the field of an electromagnetic wave from the quantum system of excited dipoles. The probability of this irradiation is proportional to a number of photons $N_f$ with the frequency *f*, which are reserved in the volume of a microwave resonator.

In the case of microstrip resonator, a number of photons is significantly smaller, because the volume of microstrip resonator is small, therefore the effect of decrease of surface resistance $R_S$, because the process of increase of a number of photons at an increase of microwave power of an electromagnetic wave has no time to appear in *HTS* thin film at microwaves, when the critical magnetic field $H_{C1}$ is reached. As a result the strong nonlinear increase of surface resistance $R_S$ in *HTS* thin film at microwaves is to begin, because of the origination of Abricosov magnetic vortices. That is why, in a microstrip resonator, the decrease of surface resistance $R_S$ of *HTS* thin film at an increase of microwave power *P* of an electromagnetic wave is not observed. At origination of *Abricosov magnetic vortices*, it is possible to conduct the calculation of surface resistance $R_S$ of *HTS* thin film and evaluate the change of value of surface resistance $R_S$ at a change of magnitude of magnetic field $H_{rf}$ using the well known expression for calculation of surface resistance $R_S$ in the *BCS* theory [2, 21]. This theoretical analysis is presented below.

Author of dissertation assumes that, in a complex case with the present *Abricosov magnetic vortices* in *HTS* thin film at microwaves, the simple evaluation of nonlinear properties of *HTS* thin film with the use of ***r**-parameter* theory is not quite correct, because the *r-parameter theory* is based on the supposition that the geometric factor $A_S$ of a microwave resonator does not change at nonlinear response

$$A_S \approx L/\lambda_L,$$

where *L* is the size of resonator, $\lambda_L$ is the penetration depth of magnetic field of electromagnetic wave in superconductor. The geometric factor $A_S$ of a microwave resonator depends on the penetration depth $\lambda_L$ of magnetic field of an electromagnetic wave in a superconductor at microwaves.

In close proximity to the critical magnetic field $H_{C1}$, the *Abricosov magnetic vortices* begin to penetrate to the whole thickness of *HTS* thin film, and the magnetic field is no longer concentrated in the thin surface layer $\lambda_L$ of *HTS* thin film. The time of nucleation of an *Abricosov magnetic vortex* is $\tau_N \le 10^{-11}$ *sec* for *HTS* thin film with its thickness ~$10^{-5}$ *cm*. Therefore, there is a process of origination and disappearance of *Abricosov magnetic vortices* at an applied microwave power of an electromagnetic wave with frequency $f \le 10^{11}$ *Hz*. The origination and disappearance of *Abricosov magnetic vortices* or the change of orientation of magnetic moments **m** of *Abricosov magnetic vortices* are accompanied by the dissipation of energy of an electromagnetic field inside *HTS* thin film by means of different physical mechanisms, which include the heating of *HTS* thin film, because of the irradiation of non-equilibrium phonons, relaxation of currents in normal cores of *Abricosov magnetic vortices* in a superconductor, etc. The mean free path *l* of normal electron excitations in high temperature superconductors is not more than $l < 10$ *nm* at low temperatures $T<10K$, and it becomes shorter at high temperatures. This mean free path *l* of normal electron excitations is comparable with the coherence length $\xi$~2*nm* in high temperature superconductors. In this case, all the physical processes in superconductors can be considered in terms of the local theory, and the research on nonlinearity of surface resistance $R_S$ in *HTS* thin films at microwaves can be done with the use of equations from the *BCS* theory of a superconductor, assuming that these equations are valid in every local point of a superconductor. The absorption of ultra high frequency currents, generated by an electromagnetic wave, takes place in the normal cores of *Abricosov magnetic vortices* in *HTS* thin films at microwaves. The origination of a new *Abricosov magnetic vortex* has an influence on the thermodynamic potential of a superconductor in the external magnetic field. A total number of *Abricosov magnetic vortices*, and hence a total number of normal cores, which may absorb an electromagnetic wave, is proportional to the change of thermodynamics potential of a superconductor at action of magnetic field of electromagnetic wave at microwaves. **Therefore, the superconductor's energy gap, averaged by volume, depends on the external magnetic field $H_{rf}$. The superconductor's energy gap, is changing strongly, when the magnetic field of an electromagnetic wave reaches value, which is comparable with the** **magnitude of critical magnetic field $H_{C1}$ at microwaves. Then, the effective surface resistance $R_s$, can be expressed by the formula, used in the *BCS* theory [2, 21], as it was done in this research.** The author's theoretical model allows to describe the experimentally researched nonlinear dependence of surface resistance on applied magnetic field $R_S(H_{rf})$ in *HTS* thin films at microwaves accurately.

It is possible to suppose that some other theoretical models on the origination of nonlinearities in *HTS* thin films at microwaves, for example, because of the existence of nodes $\Delta=0$ in the *d-wave* superconductors can also be proposed. The surface resistance $R_s$ must be much bigger in close proximity to these nodes in the momentum space than on the rest of the *Fermi surface*. At consideration of influence by the magnetic field $H_{rf}$ of an electromagnetic wave on the dynamics of electrons in these nodes, it can be noted that **the nonlinear transformation of superconducting currents into normal currents may have place in this region in *HTS* thin films at microwaves.** This physical process has to have an impact on nonlinear properties of *HTS* thin films at microwaves. These phenomena were discussed during the consideration of *intermodulation distortion (IMD) effects* in a superconducting microstrip resonator in [16].

In the author's opinion, the interaction of system of magnetic dipoles of *Josephson vortices* with electromagnetic wave may represent another possible theoretical mechanism, which can be responsible for the origination of nonlinear phenomena in close proximity to the *Josephson critical magnetic field* $H_{CJ1}$. This theoretical mechanism may be proposed to describe the effect of a decrease of surface resistance $R_s$ at an increase of magnitude of microwave signal power $P$ in range of 0–10*dBm*. This effect was observed experimentally as described in Chapter 5. It is possible to suppose that the *Josephson magnetic vortices,* oriented along the direction of applied magnetic field $H_{rf}$ of an electromagnetic wave, begin to penetrate in HTS thin film by grain boundaries at microwaves. The direction of magnetic field vector $\boldsymbol{H}_{rf}$ is changing to - $\boldsymbol{H}_{rf}$ during the change of phase $\varphi$ of an electromagnetic wave on $\pi$ radians. In this case, the intrinsic magnetic momentum of each *Josephson magnetic vortex* $\boldsymbol{m}$ changes to $\boldsymbol{-m}$. This process has a shape of hysteresis loop as it is shown by authors in Fig. 39. The *Josephson magnetic vortices*, directed along the vector of magnetic field $\boldsymbol{H}_{rf}$, are in state 1 during the first semi-period of oscillation of an electromagnetic wave. Theses *Josephson magnetic vortices* don't disappear at once, but transit to the non-equilibrium state *1′*, preserving the direction. This state has increased energy in the external magnetic field $\boldsymbol{H}_{rf}$, because the energy *E* of each *Josephson magnetic vortex* in external magnetic field $\boldsymbol{H}_{rf}$ is equal to the scalar multiplication

$$E = -\mu_0\, \mathbf{m} \cdot \mathbf{H}_{rf},$$

where $\boldsymbol{m}$ is the intrinsic magnetic momentum of Josephson magnetic vortex, $\boldsymbol{H}_{rf}$ is the vector of magnetic field of an electromagnetic wave. At transition to state *2* from state *1′*, and then, during the next semi-period, at transition to state *1* from non-equilibrium state *2′*, there is the irradiation of electromagnetic energy. Authors denote that the dipole magnetic moments of *Josephson magnetic vortices* **m**·change their direction with frequency *f*, which is equal to the oscillation frequency of an electromagnetic wave at microwaves. In agreement with the theory of magnetic-dipole irradiation, the intensity of magnetic-dipole irradiation is proportional to ~ $f^4$, as in the case of electric-dipole irradiation [17]. Therefore, this theoretical mechanism has to be observed with a big probability in *HTS* thin films at ultra high frequencies rather than at low frequencies. This supposition was confirmed during the first author's experimental research, the minimum of surface resistance $R_s$ was observed in $YBa_2Cu_3O_{7-\delta}$ thin films on *MgO* substrates in a *Hakki-Coleman dielectric resonator* (*HCDR*) at frequency $f_{HCDR}$=25*GHz* at temperatures *T = 25 K*, 50*K* as described in Chapter 6, but it was not registered in a microstrip resonator (*MSR*) at frequency $f_{MSR}$=2*GHz* in Chapter 7.

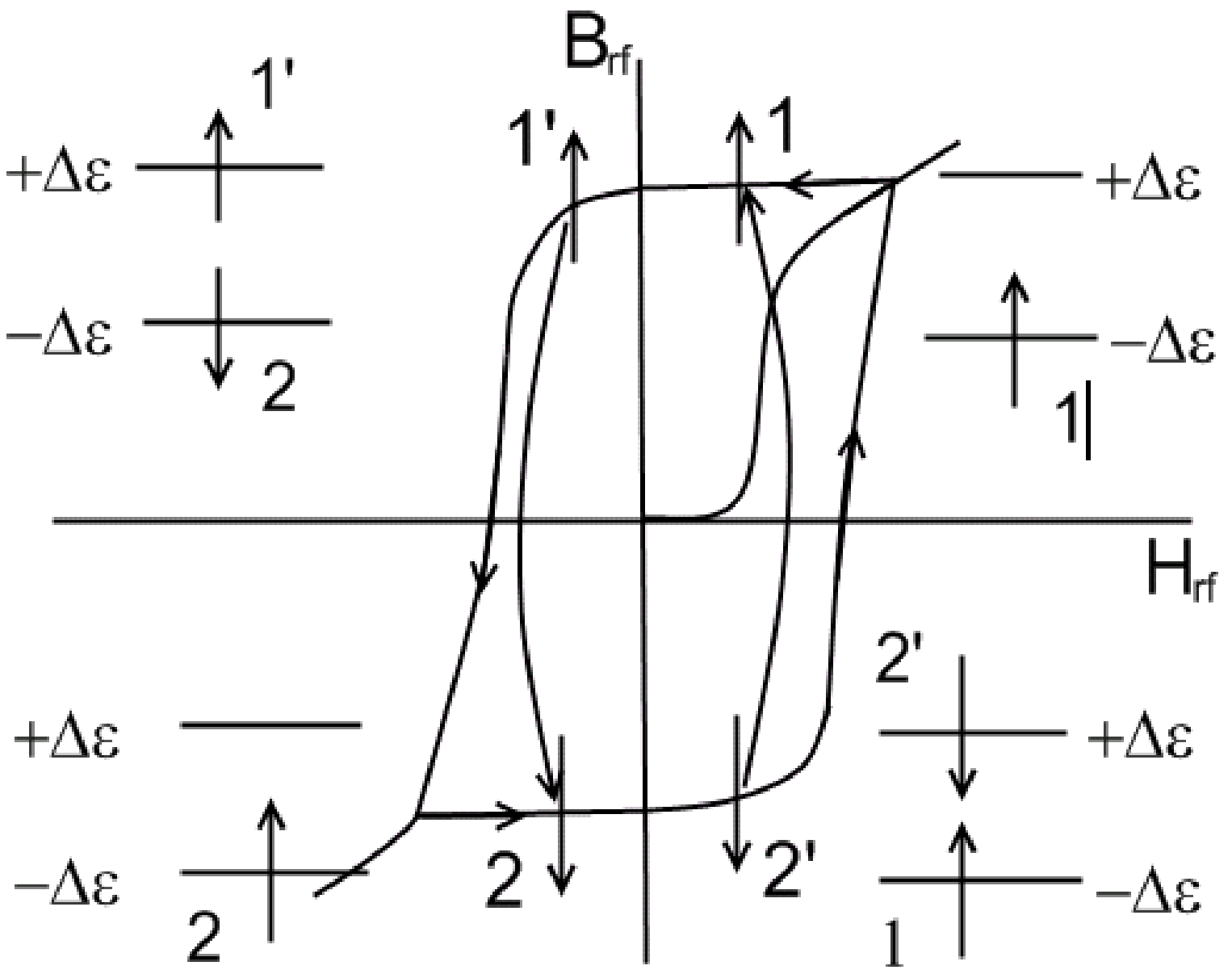

**Fig. 39.** Hysteresis loop of magnetic induction $B_{rf}$ , which depends on magnetic field $H_{rf}$ for *Josephson magnetic vortices* in high temperature superconductor thin films at action of ultra high frequency electromagnetic wave. Transition of *Josephson magnetic vortices* to equilibrium states from non-equilibrium states at change of magnetic field $H_{rf}$ are shown by arrows. Energy levels for all the orientations of intrinsic magnetic moment of *Josephson magnetic vortex* $\boldsymbol{m}$ and vector of magnetic field of electromagnetic wave $\boldsymbol{H_{rf}}$ are also shown.

The probability $W$ of this theoretical mechanism realization in *HCDR* vs. microstrip resonator is equal to the ratio

$$(f_{HCDR} \,/\, f_{MSR})^4 \approx 2.45\ 10^4.$$

Therefore, the exchange by energies between the electromagnetic field and the *Josephson magnetic vortex* system without loss of energy is more probable in *HCDR,* and it leads to the decrease of both the energy dissipation and the surface resistance $R_s$ in a range of microwave powers from 0 to 10*dBm* as described in Chapter 6.

The statistics of photons plays an important role in the case of interaction of the electromagnetic wave with a superconductor at microwaves. Besides, the probability *W* grows up, if a number of present photons in dielectric resonator increases

$$W \propto (N+1),$$

where *N* is the number of photons in a microwave resonator. The nature of process is connected with the *Einstein effect* on the incident radiation of electromagnetic wave quantums [18]. This effect can be observed in a dielectric resonator with a big volume of cavity and high quality factor, hence with a big reserved energy and a big number of photons *N* at microwaves. Author of dissertation would like to note that the reserved energy and number of photons *N* are small in a microstrip resonator, hence the *Einstein effect* was not observed in a microstrip resonator during the experiments at *Electrical and Computer Engineering Department* at *James Cook University* in Townsville, Australia.

A number of photons *N* in a microstrip resonator is proportional to the magnitude of microwave signal power of an electromagnetic wave, therefore, at sufficiently big magnitude of microwave signal power, when $N>>1$, the main part of photons will return to the microstrip resonator from superconductor sample. This process will lead to both the decrease of dissipation of photons in a superconducting sample, and the decrease of effective value of surface resistance $R_S$ of *HTS* thin film at microwaves. However, at further increase of magnitude of microwave signal power of an electromagnetic wave, the magnetic field of an electromagnetic wave reaches the value of low critical field $H_{C1}$, that results in an origination of a big number of *Abricosov magnetic vortices* in a superconductor at microwaves, and hence in an increase of surface resistance $R_S$ of *HTS* thin film at microwaves, which is caused by the *Abricosov magnetic vortices* action. Modeling of described process was completed by the author of dissertation, and it was shown that the dependence of surface resistance on magnetic field $R_s(H_{rf})$ in *HTS* thin film at microwaves is the *S-shape dependence*. The research results, obtained during the computer modeling, are in good agreement with the experimental results measured by the author of dissertation at *Electrical and Computer Engineering Department* at *James Cook University* in Townsville, Australia.

## Summary.

In Chapter 7, the experimental and theoretical researches on the nonlinear physical properties of high quality $YBa_2Cu_3O_{7-\delta}$ thin films on *MgO* substrates in a microstrip resonator at microwaves are completed.

The experimental dependences of transmission coefficient on ultra high frequency and microwave power $S_{21}(f, P)$ in a microstrip resonator, made of $YBa_2Cu_3O_{7-\delta}$ thin films on *MgO* substrates, are measured at different temperatures as shown in Figs. 7 and 8. The shift of resonant frequency in $f_0(P)$ dependence in Fig. 9; the change of value of quality factor in $Q(P)$ dependence in Fig 10; and the change of magnitude of surface resistance in $R_s(P)$ dependence in Fig. 11 at sufficiently intense applied microwave power levels *P* in a microstrip resonator, made of $YBa_2Cu_3O_{7-\delta}$ thin films on *MgO* substrate, are observed. In Fig. 12, the experimental *S-shape dependencies* of surface resistance on applied magnetic field $R_s(H_{rf})$ in a $YBa_2Cu_3O_{7-\delta}$ microstrip resonator at frequency $f_0$=1.985*GHz* at different temperatures are shown. In Fig. 13, the $dR_s/dH_{rf}$ $(H_{rf})$ dependencies in $YBa_2Cu_3O_{7-\delta}$ microstrip resonator at frequency $f \approx 2GHz$ at different temperatures confirm a supposition that ***the nonlinearities of surface resistance at increased levels of applied microwave power $R_s(H_{rf})$ in $YBa_2Cu_3O_{7-\delta}$ microstrip resonator are somehow interlinked with some unusual features of physical properties of superconductors, appearing near to the critical magnetic fields $H_{c1}$ and $H_{c2}$ in HTS thin films at microwaves.*** The influence by different geometrical forms of *HTS* thin film on the dependence of surface resistance on applied microwave power $R_s(P)$ in a $YBa_2Cu_3O_{7-\delta}$ microstrip resonator is evaluated.

The modeling of dependences of surface resistance on magnetic field $R_s(H_{rf})$ in close proximity to the critical magnetic fields $H_{CJ1}$, $H_{C1}$ and $H_{C2}$ in $YBa_2Cu_3O_{7-\delta}$ thin films on *MgO* substrates in Figs. 16, 17, 23-38; as well as in *titanium-vanadium* $Ti_{0.6}V_{0.4}$ and *indium-lead* $In_{0.19}Pb_{0.8}$ *Type II* superconductors in Figs. 18, 19, 20 in a microstrip resonator at microwaves, is performed in the *Matlab*. The comparative analysis of modeling data with experimental results is done in every considered case.

The superconductors, based on complex oxide ceramic, are the *Type-II* superconductors [37]. Therefore, the applied magnetic field can penetrate in the thickness of superconducting film in the form of *Abrikosov magnetic vortex lattice* [36]. The presence of *Abricosov magnetic vortex lattice* in the film leads to:

1) the additional dissipation of energy, and

2) the nonlinear increase of energy losses by an electromagnetic wave at surface layer of an *HTS* thin films at microwaves.

The first proposed theoretical model by the author of dissertation is based on the interaction of the applied ultra high frequency (*UHF)* high intensity electromagnetic waves with the generated *Abricosov magnetic vortices*, which may explain the physical nature of nonlinearities in microwave superconductivity, and help to accurately characterize the nonlinear surface resistance on magnetic field dependence $R_s(H_{rf})$ in close proximity to critical magnetic fields $H_{CJ1}$, $H_{C1}$ and $H_{C2}$ in high quality $YBa_2Cu_3O_{7-\delta}$ thin films on *MgO* substrates at microwaves. The inhomogeneous distribution of magnetic flux, induced by generated *Abricosov magnetic vortices* in *HTS* thin film at microwaves, may increase the nonlinearity in $R_s(H_{rf})$ dependence. This theoretical model is proposed and discussed by the author of dissertation comprehensively.

The second considered theoretical model is based on the original proposition by the author of dissertation about the possible origination of nonlinearities in the surface resistance on magnetic field dependence $R_s(H_{rf})$ in *HTS* thin films at microwaves, because of the nodes $\Delta=0$ existence in the *d-wave* superconductors.

In the authors of book opinion, the interaction of system of magnetic dipoles of *Josephson magnetic vortices* with applied ultra high frequency (*UHF*) electromagnetic wave may represent the third possible theoretical mechanism, which can be responsible for the origination of nonlinear phenomena in close proximity to the *Josephson critical magnetic field* $H_{CJ1}$. This theoretical mechanism is proposed to describe the effect of a decrease of surface resistance $R_s$ at an increase of magnitude of microwave power *P* in range of *0–10dBm* in high temperature superconducting (*HTS)* thin films at microwaves.

Apart from the design of microwave filters, the low loss *HTS* thin films with nonlinear properties can be used for metamaterials creation to design the microwave sensors based on the *split-ring resonator* (*SSR*) resonators with evanescent wave amplification, near-perfect lens, hyperlensing, transformation optics [38].

# CHAPTER 8

# NOISE IN SUPERCONDUCTING MICROWAVE RESONATORS

## 8.1. Introduction.

Microwave superconductivity has developed into a fascinating fundamental research subject with numerous technical applications in the active and passive microwave devices in the microwave circuits in the micro- and nano-electronics over the recent decades. The Chapter 8 considers the *nature of noise* in the superconducting microwave resonators. The main known noise sources in the *solid state devices* as well as in the *microwave resonators* are re-viewed in this *Introduction*, before the detailed consideration on the research results on both the *new possible noise sources* and the *differential noise in HTS microwave resonators*.

The *stochastic processes* like the actual *noise fluctuations* in the physical system may be mathematically treated using the notion on the *ensemble* of statistically similar processes, observed simultaneously over the same period of time [10, 83, 84, 85, 86, 87, 88, 89]. There are two different averaging procedures that can be applied to the member functions of an *ensemble* [10]:

1) *Time averaging*, performed on a single member function of the ensemble;

2) *Ensemble averaging*, where the averages are expected values computed at fixed times throughout the observation interval from the probability measures (Gaussian function).

***The knowledge of the mean value and covariance function is sufficient to specify all the statistical properties of the stochastic process such as the actual noise fluctuations in physical system [10].***

The first- and second time averages of the $i^{th}$ member of the *ensemble* are the mean value

$$\left\langle x^{(i)}(t)\right\rangle = \lim_{T\to\infty}\frac{1}{T}\int_{-T/2}^{T/2} x^{(i)}(t)dt$$

and the ***autocorrelation function*** is a measure of the "memory" of process can be written as

$$\phi_n^{(i)}(\tau) \equiv \langle x^{(i)}(t)x^{(i)}(t+\tau)\rangle = \lim_{T\to\infty}\frac{1}{T}\int_{-T/2}^{T/2} x^{(i)}(t)x^{(i)}(t+\tau)dt$$

where *T* is the duration of the observation interval in Buckingham [10].

Thus, the *noise* in the physical systems can be described in terms of a fluctuation $\mathcal{L}(t)$ with an average value $<\mathcal{L}>=0$, and a white spectrum determined by the autocorrelation function

$$<\mathcal{L}(t)\mathcal{L}(t+\tilde{t})> = 2\Upsilon\delta(\tilde{t})$$

where $\Upsilon = k_B T/R_n$, in application to the *Josephson junction* in the case of ultra high frequencies in Van der Ziel [53]; Kubo, Toda, Hashitsume [60]; Fischer [61].

The noise can originate in two- and three- terminal solid state devices at low temperature, for example, in the *Josephson Junctions* and *Single Electron Transistors*.

Considering the inherent noise in two terminal device like the *Josephson Junction* (*JJ*), it is necessary to note that many different sources of noise in the *dc* and *ac Josephson effects* are observed. The noise in the *dc Josephson effect* was theoretically treated by Anderson, Goldman [42]; Ivanchenko, Zilberman [43]; Ambegaokar, Halperin [44, 45]; Lee [46]; and experimentally researched by Galkin, Borodai, Svistunov, Tarasenko [41]; Anderson, Goldman [42]. The thermal fluctuations, appearing as a result of equilibrium energy exchange with the environment, affect the phase relationship between the two sides of *JJ* [43, 44, 45]. Thermal noise contributions in *JJ* lead to the *thermally activated phase slippage* (*TAPS*) driving the *JJ* to the quasi-static state [47]. The spectral density of the voltage fluctuations across a current biased *resistively shunted Josephson Junction* (*RSJ*) was theoretically treated by Likharev, Semenov [48]; Vystavkin, Gubankov, Kuzmin, Likharev, Migulin, Semenov [49, 50]; Koch, Van Harlingen, Clarke [51]. The nonlinear Langevin equation for the phase, in which the thermal fluctuations (main source of noise in *RSJ*) are represented as a part of the source term, was derived by Koch, Van Harlingen, Clarke in [51]. The noise rounding of the current-voltage characteristic of the *RSJ* was researched by Koch, Van Harlingen, Clarke in [52]. The noise in the *ac Josephson effect* was theoretically investigated by Scalapino [39]. A *Langevin treatment of noise* in the *ac Josephson effect* was provided by Stephen [40].

The voltage biased *Single Electron Transistors (SET)* transistor is a three-terminal device, where the bias voltage *V* and the gate voltage *U* are input variables and the output variable is the current [30]. The current noise in the *SET* appears strongly to depend on *dI/dU* [62]. The current noise is large, if the *dI/dU* is large. The current noise is small, if the *dI/dU* is small. Most of the *SET*s have no the telegraph noise, exhibiting the *1/f* noise spectrum researched by Visscher, Verbrugh, Lindeman, Hadley, Mooij [31]; Zimmerli, Eiles, Kautz, Martinis [32]. However, the telegraph noise may occur in some *SETs*. The main contribution to the noise is from the *Two Level Fluctuations*, which induce a time dependent charge on the island [33, 34, 35].

Considering the noise sources in *SQUIDs* made of *HTS* thin films, Clarke focuses on the *thermal noise* and *1/f noise* [76]. Clarke summarizes the accumulated knowledge [69-76] on *1/f* noise sources saying that there are at least two separate sources of *1/f* noise [76]:

1) *1/f* fluctuations in critical current of the *Josephson junctions* [70]. An electron becomes trapped on a defect and subsequently released during tunneling throughout the potential barrier in *Josephson junction*. While the trap is occupied, there is a local change in the height of the tunnel barrier and hence in the critical current density of that region [76]. As a result, the presence of a single trap causes the critical current of the junction to switch randomly back and forth between two values, producing a random telegraph signal [76]. In other words, the traps in the potential barrier enable electrons to tunnel in certain voltage range, a process producing both leakage of current and *1/f* noise [76].

2) Motion of magnetic flux lines trapped in the body of the *SQUID* [76]. This mechanism manifests itself as a magnetic flux noise [76]. The level of *1/f* flux noise appears to depend strongly on the microstructure of the thin film [76]. Ferrari *et al* [72-75] researched the magnetic flux noise in *SQUIDs* made of $YBa_2Cu_3O_{7-\delta}$ thin films. Ferrari *et al* [74] described the origin of magnetic flux noise in terms of the motion of flux quanta in the film. Each vortex hops independently between two pinning sites under thermal activation, a process that produces a random telegraph signal with a Lorentzian power spectrum [74].

Clarke [76] notes that there is important practical difference between the two sources of *1/f* noise: *critical current noise* can be reduced by a suitable modulation scheme, whereas *magnetic flux noise* can not.

The original research on the noise sources in *HTS* microwave resonators by author of dissertation starts with a detailed consideration on the noise theory in *RF* devices and systems [10] with particular interest in the nature of noise in *LTS* and *HTS* microwave resonators [37]. The comprehensive theoretical description of main noise sources in microwave resonators is provided below. ***In a superconducting microstrip resonator, the resonance frequency is impacted by the superposition of permittivity of dielectric substrate and complex conductivity of LTS/HTS thin films.*** Many different sources of noise generation in *HTS* microwave resonators were experimentally researched and theoretically modeled [10]. The frequency noise in modern high performance *LTS* superconducting resonators is two to three orders of magnitude above the fundamental limit of *generation-recombination noise* [37]. The noise has been conjectured to arise from *electric dipole two-level systems* (*ETLS*) in surface dielectrics in *LTS* microstrip resonators [54]. The surface has been shown to be a dominant source of noise in Gao *et al* [55].

Author of dissertation makes an innovative theoretical proposition that ***the magnetic dipole two-level systems (MTLS)*** in *HTS* thin films, based on the *Abricosov magnetic vortex* or *Josephson magnetic vortex*, may be a dominant noise source in superconducting microwave resonators at certain conditions [59]. This assumption may add more clarification to the understanding of shape of resonance curve in superconducting microwave resonators and the nature of nonlinearities in microwave superconductivity.

Also, the author of dissertation presents some original data on the ***differential noise*** modeling and suppression in *HTS* microwave filters. The comprehensive research on physical characteristics of differential noise in superconducting microwave resonators is also provided, using the comparative study of experimental and computer modeling results obtained by author of dissertation at *ECE* at *JCU* in Australia.

## 8.2. Noise in Passive and Active Electronic Devices and Systems.

Electronic passive and active multiport microwave devices can exhibit random fluctuations of voltage and current at their terminals. These fluctuations are usually referred to as the ***noise***. All electronic devices, including the microwave filters, exhibit the noise. The noise originates in a device, because of the existing nature of random microscopic behaviour of the charge carriers, when interacting with the electronic components constituting the system. It is usually called the ***charge noise***.

***Noise spectrum*** is commonly observed as a randomly varying function of total charge on time, in other words as random fluctuations in either the voltage across the terminals of a device, or the current flowing through the terminals of a device. It is referred to as the ***telegraph noise*** [1-10].

The three most commonly encountered types of noise are:

1. ***Thermal noise*** arises from random velocity fluctuations of the charge carriers (i.e. electrons and/or holes) in a resistive material. The effect can be explained by the Brownian motion of the charge carriers due to the thermal energy within the material. In other words, it caused by random fluctuation of charge carriers due to thermal agitation. Thermal noise is present, when the resistive element is in the thermal equilibrium with its surroundings, and is often referred to as ***Johnson noise*** [3-5] It is expressed as $kT$, where $k$ is the *Boltzmann's constant*, T is the temperature.
2. ***Shot noise*** is always associated with direct current flow, and in this sense is a non-equilibrium form of noise, as electrical currents do not flow uniformly and do not vary smoothly in time varying at microscopic levels in unpredictable ways, which are considered as the noise. In other words, the shot noise is caused by random variation of *dc* current carriers. Shot noise was firstly discussed by Schottky [2].
3. ***Flicker noise*** appears due to the generation of intermodulation products of electromagnetic signals in a device. It is expressed as $1/f$ or $1/f^2$.

Let us discuss the thermal, shot, flicker noises power spectral densities, in other words, the thermal, shot, flicker noises powers per unit bandwidth in details.

The physical origins of ***thermal noise*** and ***shot noise*** are different, but the structures of the two types of noise waveforms are similar in a way that they both can be represented as a random pulse train consisting of a sequence of similarly shaped pulses randomly distributed in time.

If the noise waveform is represented by the function $x(t)$ and the pulse shape function is $f(t)$ (note that $f(t) = 0$ for $t < 0$, assuming that the event, giving rise to the pulse, occurs at $t = 0$ and that system is in a non-equilibrium state), then the random pulse train can be represented as the linear superposition in eq. (1)

$$x(t) = \sum_k a_k f(t - t_k) \qquad (1)$$

where $\alpha_k$ is the amplitude of the $k^{th}$ pulse in the sequence and $t_k$ is the time at which the $k^{th}$ event occurs.

The distribution of the $t_k$ is governed by the *Poisson law.* A waveform, which structure can be represented by equation (1), shows a number of interesting properties, in particular, its power spectral density can be expressed as in eq. (2)

$$\overline{S_x(\omega)} = 2v\overline{\alpha^2}\left|F(j\omega)\right|^2, \quad (2)$$

where $\omega$ is the angular frequency, $F(j\omega)$ is the *Fourier transform* of the pulse shape function $f(t)$, $v$ is the mean rate of events, $\overline{\alpha^2}$ is the mean-square value of the pulse amplitudes, and the over-bar on the left hand side indicates an average taken over a large number of trials (i.e. ensemble average). This expression is known as *Carson's theorem* [8, 9].

The pulses arising from the discrete events responsible for thermal and shot noises show the flat power spectral densities up to very high frequencies. The level of the power spectral densities in the two cases, that is the value of the term on the right hand side of the above equation (2), is determined from the physics of the noise mechanisms.

In the case of ***thermal noise***, the discussion concentrates on statistical mechanisms and the law of equipartition of energy, which states that every system at an absolute temperature $\theta$, which is in thermal equilibrium with its surroundings, contains thermal energy amounting, on average, to $k\theta$ per degree of freedom, where $k$ is the *Boltzmann's constant.* The result for the power spectral density of the open circuit noise voltage across the terminals of a resistor $R$ is in eq. (3)

$$\overline{S_v(\omega)} = 4k\theta \, \mathrm{R} \qquad (3)$$

Thus, the resistor can be represented as in Fig. 1(a), where a serial noise voltage generator has a spectral density given by eq. (3). By a simple circuit transformation, the noisy resistor can be represented as in Fig. 1(b), where the parallel noise current generator has a spectral density in eq. (4)

$$\overline{S_i(\omega)} = 4k\theta \, \mathrm{G} \qquad (4)$$

In this expression the conductance $G=1/R$.

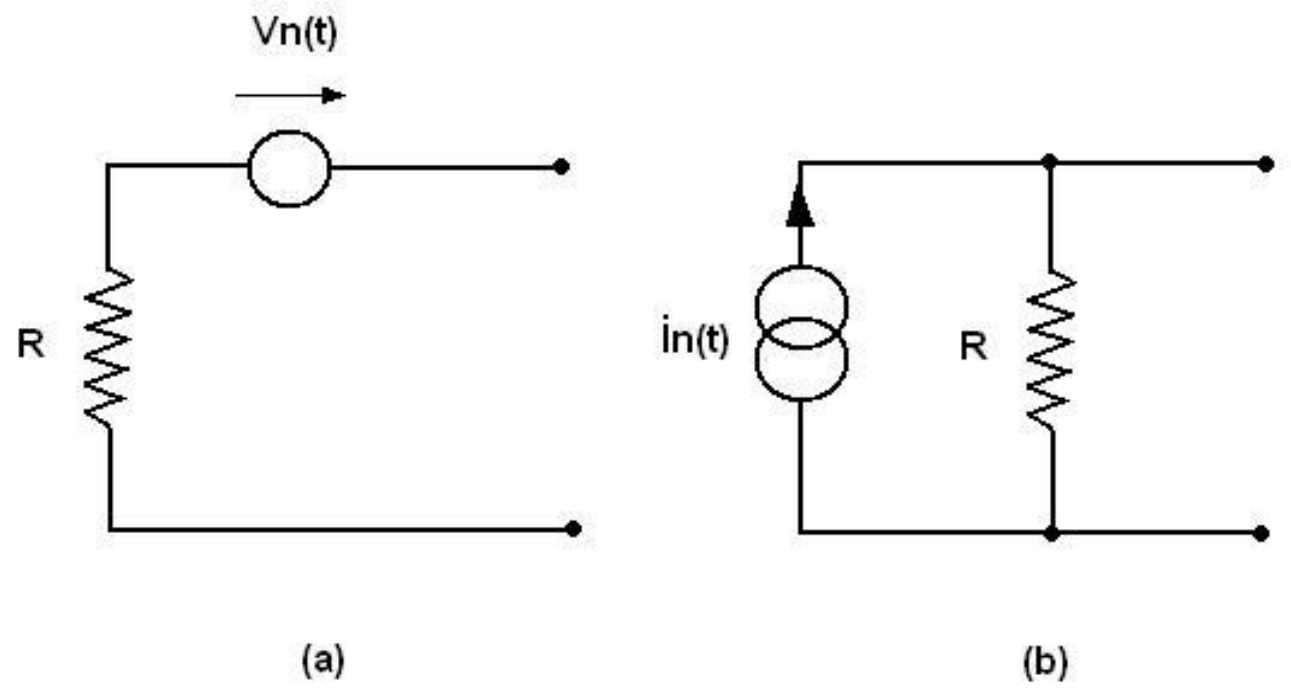

**Fig. 1.** Thermal noise in a resistor *R*, represented by (a) serial noise voltage generator and (b) parallel noise current generator (after [10]).

Equations (3) and (4) were first derived by *Nyquist* based on the thermodynamics theory taking into account the exchange of energy between the resistive elements in equilibrium [6, 7]. *Nyquist's* macroscopic approach to the problem of *thermal noise* is different from the microscopic view. *Nyquist's theorem* involves the resistance *R*, while the maximum available noise power from the resistance in a frequency interval *df* is independent of *R*. This effect can be seen in Fig. 2, which shows the resistance *R* in parallel with a noise current generator feeding into a matched noiseless load.

From the simple circuit analysis the power dissipated in the load in a frequency range *df* is in eq. (5)

$$dP = \frac{\overline{S_i(\omega)}R}{4}df = k\theta df \qquad (5)$$

where the expression on the right hand side has been substituted for $S_i(\omega)$ expressed in eq. (4).

The value of *Boltzmann's constant* is $k=1.38x10^{-23} joule/degK$, which, when multiplied by $\theta=293K$ - corresponding to room temperature, gives the maximum available noise power from the resistance in a frequency band of $1Hz$ as $4x10^{-21}W$.

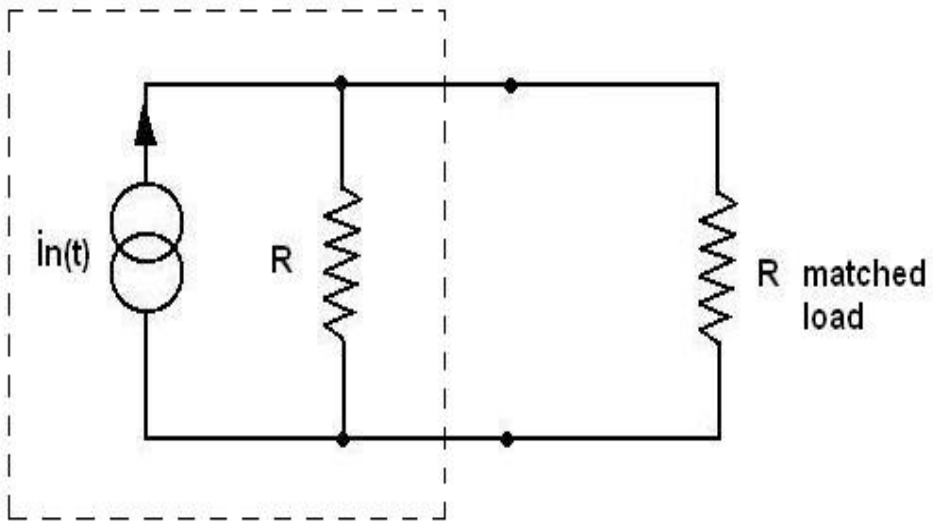

**Fig. 2.** Noisy resistance, enclosed in broken lines, connected to a noiseless matched load (after [10]).

The power spectral density of a ***shot noise***, when the mean current is *I*, is in eq. (6)

$$\overline{S_i(\omega)} = 2qI \qquad (6)$$

where *q* is the magnitude of the electronic charge. Note that the power spectral density of an impulse process is equal to in eq. (7)

$$\overline{S_x(\omega)} = 2v\overline{\alpha^2} \qquad (7)$$

as the Fourier transform of an impulse is unity. The spectrum of an impulse process is flat up to an indefinitely high frequency. Such spectrum is sometimes said to be "white". Therefore, the equation (7) follows almost immediately from the equations (2) and (6) since the mean rate of pulses is *I/q* and the pulse amplitudes are all equal to *q*, giving $\alpha^2 = q^2$ [10].

A type of noise that occurs in a wide variety of systems *1/f* is sometimes known as the *current noise, flicker noise, intermodulation noise, contact noise or excess noise* [1-10]. The ***1/f noise*** is called so because it shows a power spectral density, which varies with frequency as $/f/^{-\alpha}$, where α usually falls between 0.8 and 1.2. This dependence has been observed in frequencies as low as $10^{-6}$ *Hz*. The upper limit of its existence is difficult to establish due to the fact that it is usually masked by thermal noise or some other types of noise. There are various theoretical difficulties associated with the treatment of *1/f* noise, mostly concerning convergence of integrals [10]. The *1/f* noise may originate, because of change of mobility or density of charge carriers in *HTS* thin films. The quantum theory of *1/f* noise was formalized by Handel [56, 57, 58, 10], who proposed that the charge carriers may interact with quantized electromagnetic field, resulting in both the charge carriers scattering by arbitrary potential barriers and the low frequency photon irradiation by charge carriers. The *1/f* noise may be produced because of modulation of transport current in the element due to the energy emitted by photons [56, 57, 58, 10].

Author of dissertation proposes the ***new quantum theory of 1/f noise*** in *HTS* thin films at microwaves stating that the charge carriers (*Cooper electron pairs*) may interact with quantized electromagnetic field in *HTS* thin films at microwaves, resulting in both the charge carriers scattering by arbitrary potential barriers on the *N-S* boundaries, created by *Abricosov or Josephson magnetic vortices* with normal cores in *HTS* thin films at microwaves. This process may be accompanied by the low frequency photon irradiation by charge carriers. **The *1/f* noise may be produced as a result of modulation of transport current by the emitted energy of photons** [59].

In the case, when the quantum energy of the electromagnetic field is $\hbar\omega >> kT$, the ***quantum noise*** has to be taken into consideration. The spectral density of *quantum noise* may be represented in eq. (8) [11]

$$\overline{S(\omega)} = \frac{R\hbar\omega}{2\pi}\coth\left(\frac{\hbar\omega}{2kT}\right) = \frac{R}{\pi}\left[\frac{1}{2}\hbar\omega + \frac{\hbar\omega}{\exp(\hbar\omega / kT) - 1}\right] \quad (8)$$

In addition to *thermal noise*, *shot noise*, *1/f noise*, *quantum noise* introduced above, several other types of noise exist such as the *generation-recombination noise* due to the random trapping of charge carriers in semiconductor materials, *burst* *noise*, *avalanche noise* due to the impact ionization, and *non-equilibrium Johnson noise* exhibited by hot electrons in high electric fields [10], phase noise [82].

Noise in electronic circuits is usually regarded as detrimental, and indeed it often imposes practical limitations to circuit performance. The noise effect may be caused by one or more sources within the system. A *noisy electronic circuit* with a pair of input and a pair of output terminals, i.e. a two-port network, is shown in Fig. 3(a)

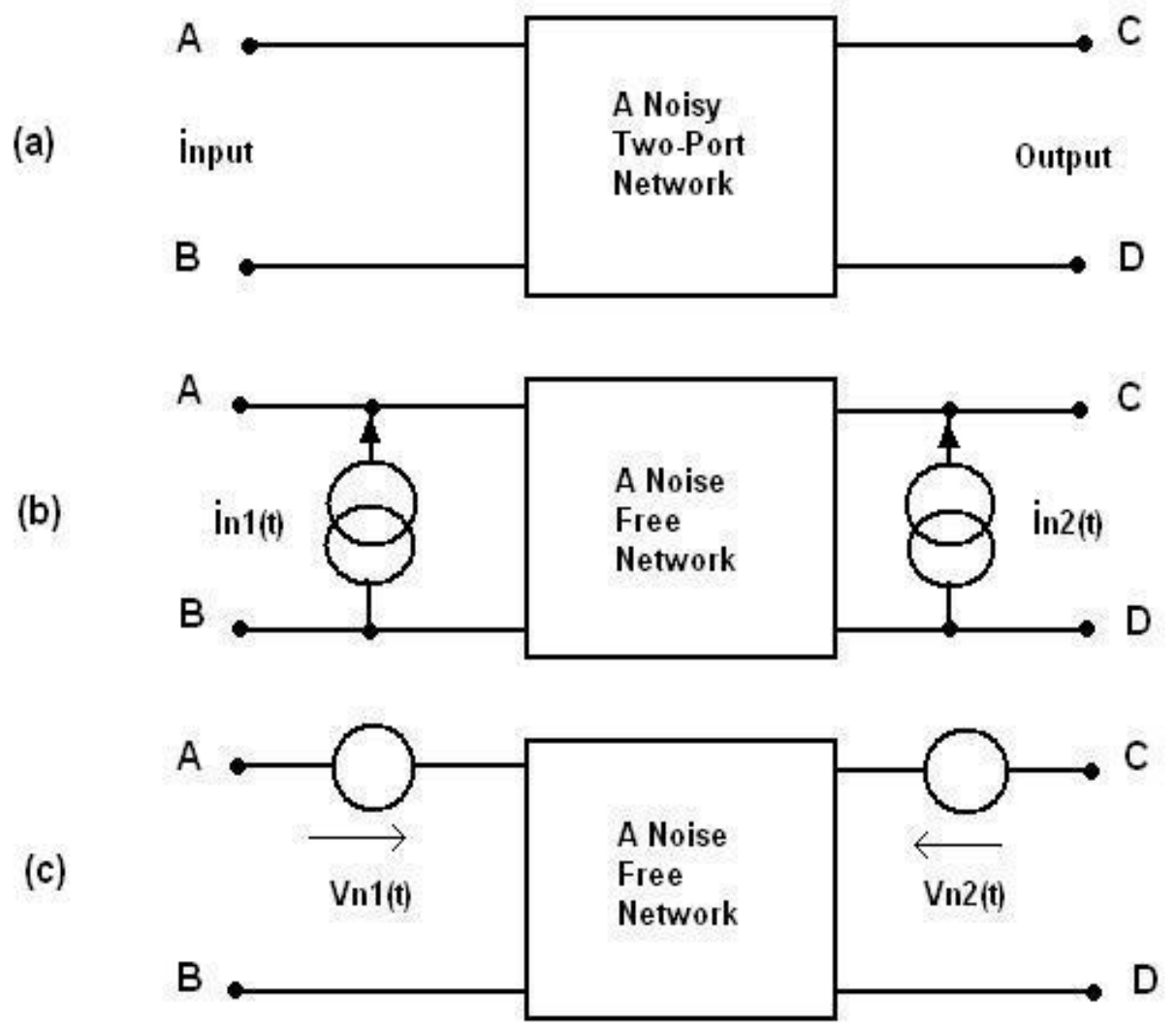

**Fig. 3.** (a) A noisy two-port; and two equivalent circuits with the noise represented by (b) current generators at the ports and (c) voltage generators at the ports (after [10]).

A convenient way of representing the noisy system is illustrated in Fig. 3(b), where the network is shown as noise-free and the noise is represented by the noise current generators $i_{n1(t)}$ and $i_{n2}(t)$ at the input and output ports, respectively. These two current generators may show some degree of correlation, because the mechanisms causing rise to the noise at the two ports could be to some extent common. An alternative representation of the noisy system is presented in Fig. 3(c), where the network is again shown as a noise-free, but now the noise is represented by the noise voltage generators $v_{n1}(t)$ and $v_{n2}(t)$, which may also be correlated, at the input and output ports, respectively.

In order to specify the noise generators at the ports, the details of the network and the characteristics of the internal noise sources must be known. The internal sources are associated with the electronic passive and active devices within the system, and in general these sources are strongly device-dependent, even though the physical mechanisms responsible for the noise may be common to a range of microwave devices [10].

Summarizing the above discussed research findings, it has to be noted that, in a simple case, the sources of noise generated in a two port device, can be modeled by the two types of input noise generators in Figs. 4 and 5 [22]:

1. A series voltage source:

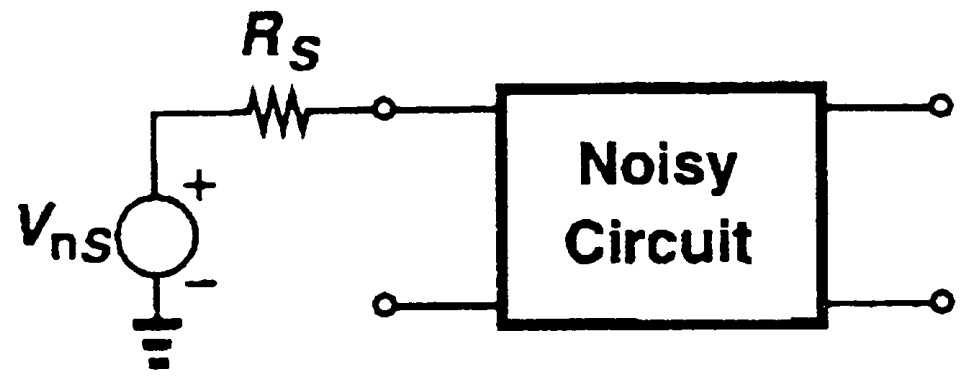

**Fig. 4.** Modeling of noise by a series voltage source (after [22]).

2. A parallel current source:

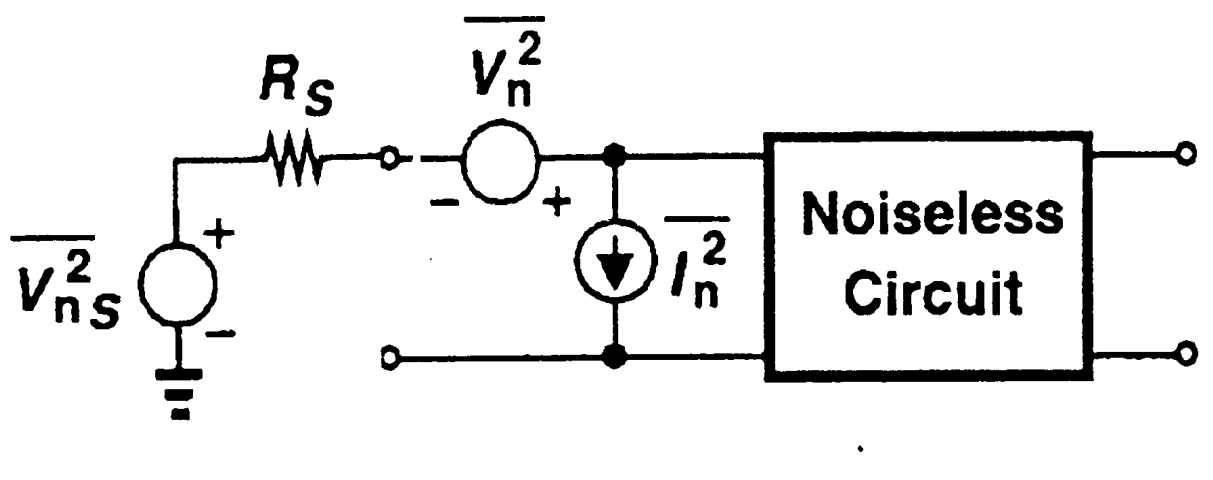

**Fig. 5.** Modeling of noise by a parallel current source (after [22]).

## 8.3. Constant Noise Figure Circuits of a Linear Two Port Network at Smith Chart.

The *Cryogenic Transceiver Front End* (*CTFE*) block diagram with *HTS* microwave transmitting / receiving filters and low noise / *RF* power amplifiers is shown in Fig. 1 in Chapter 9. Presently, the *High Temperature Superconducting* (*HTS*) microwave filter with cooled *Low Noise Amplifiers* (*LNA*) configuration has been used to improve the spatial coverage, mobile user's capacity and energy consumption by the *CDMA/UMTS/LTE* base stations in *4G/5G* wireless networks [77]. Chaloupka [80, 81] comments that, in the receiver in the *Cryogenic Transceiver Front End* (*CTFE*), the information carrier signal must be amplified to exceed a required minimum of microwave signal level, but the amplification is always associated with a decrease in *signal-to-noise (SNR)* ratio, hence the receiver must be optimized with respect to both a *sufficient output microwave power level* and a *low noise figure.* In this sub-chapter, the concept of ***noise figure*** is discussed in application to the networks, formed by the microwave filters and amplifiers in the *Cryogenic Transceiver Front End* (*CTFE*). The concept of *noise figure* was introduced to characterize the radio receivers in [23]. The concept was further developed in [24, 25, 26, 27, 28], where *the noise figure* was defined as *a figure of merit quantifying the noise properties of four terminal networks* [10]. In practice, *the noise figure* of two port network is defined for a specified frequency as *the ratio of total output noise power per unit bandwidth to output noise power per unit bandwidth due to the input termination* [10].

The ***noise factor*** is an important characteristic in microwave devices design. The *noise factor* is defined as *the ratio of the S/N (Signal-to-Noise) ratio at the input to the S/N ratio at the output of two port network* in eq. (9)

$$F = \frac{S/N\ in}{S/N\ out} \quad (9)$$

The ***noise figure*** (*NF*) is defined as *the noise factor expressed in dB* in eq. (10) [22]

$$NF = 10\log(F)$$
$$F = 10^{\frac{NF}{10}} \quad (10)$$

The ***equivalent noise temperature*** is another form of expressing *circuit noise* in eq. (11)

$$T = T_0(F - 1) \quad (11)$$

where $T_0 = 290K$.

Let consider the constant *noise figure* circuits at the ***Smith Chart*** [20, 21]. In Fig. 6, the linear two port network is shown

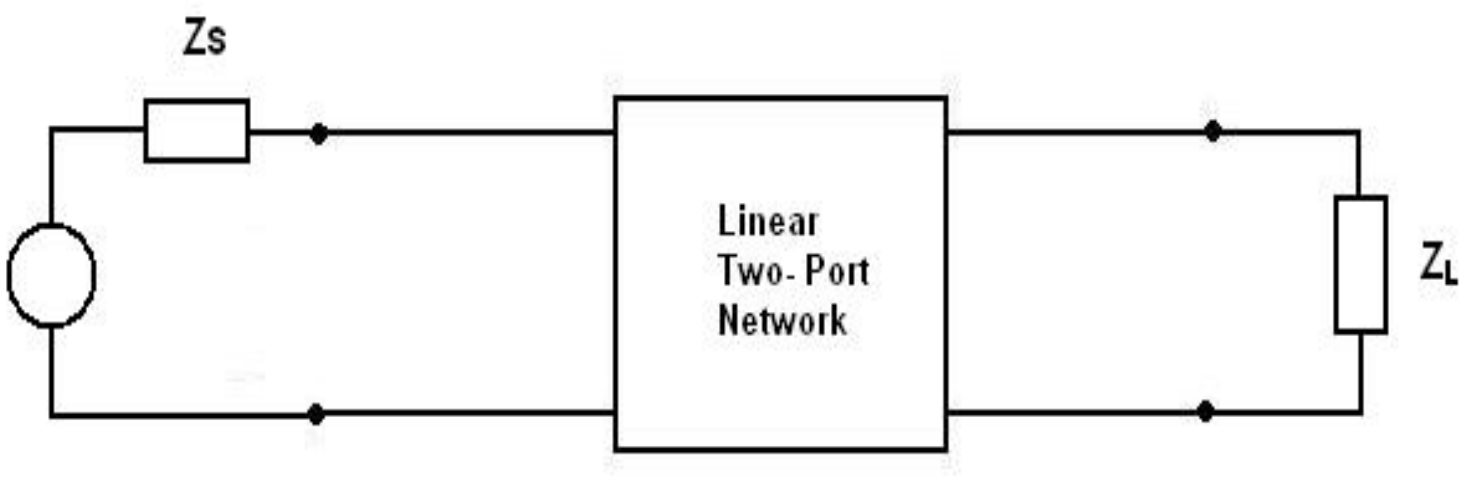

**Fig. 6.** Linear two-port network (after [12]).

The *noise figure* for this network can be formulated as in eq. (12):

$$NF = NF_{\min} + \frac{r_n}{g_s}\left[\left(g_s - g_0\right)^2 + \left(b_s - b_0\right)^2\right] \quad (12)$$

where $r_n$ is the equivalent input noise resistance of the two port, $G_s$ and $b_s$ represent the real and imaginary parts of the source admittance, and $g_0$ and $b_0$ represent the real and imaginary parts of that source admittance, which results in the minimum *noise figure*, $NF_{min}$ in eq. (13)

$$NF - NF_{\min} = \frac{r_n}{g_s}\left[\left(g_s - g_0\right)^2 + \left(b_s - b_0\right)^2\right] \quad (13)$$

Let us substitute:

$$Y_S = \frac{1-\Gamma_S}{1+\Gamma_S},\ Y_0 = \frac{1-\Gamma_0}{1+\Gamma_0} \quad (14)$$

$$NF - NF_{\min} = 4r_n \frac{|\Gamma_S - \Gamma_0|^2}{\left(1-|\Gamma_S|^2\right)|1+\Gamma_0|^2} \quad (15)$$

The equivalent noise resistance, $r_n$, can be found by making one additional noise figure reading with a known source reflection coefficient. If a 50-ohm source were used for example, $\Gamma_S$ = 0 and this expression could be used to calculate $r_n$. For $\Gamma_S$ = 0 in eq. (16)

$$r_n = \left[NF_{\Gamma_S=0} - NF_{\min}\right]\frac{|1+\Gamma_0|^2}{4|\Gamma_0|^2} \quad (16)$$

To determine a family of noise figure circles, a noise figure parameter $N_i$ needs to be defined in eq. (17).

$$N_i = \frac{NF_i - NF_{\min}}{4r_n}|1+\Gamma_0|^2 \quad (17)$$

where $NF_i$ is the value of the desired noise figure circle and $\Gamma_0$, $F_{min}$, and $r_n$ are as previously defined.

With a value for *Ni* determined, the center and radius of the circle can be found using the following expressions in eqs. (18, 19).

$$C_{F_i} = \frac{\Gamma_0}{1+N_i} \quad (18)$$

$$R_{F_i} = \frac{1}{1+N_i}\sqrt{N_i^{\,2} + N_i\left(1-|\Gamma_0|^2\right)} \quad (19)$$

It can be seen that $N_i=0$, where $NF_i=NF_{min}$, and the center of the $NF_{min}$ circle with zero radius is located at $\Gamma_0$ on the *Smith Chart*. The centers of the other noise figure circles lie along the $\Gamma_0$ vector.

The plot shown in Fig. 7 gives the noise figure for a particular device for any arbitrary source impedance at a particular frequency. For example, given the source impedance of 40+j50 *Ohms*, the noise figure would be 5 *dB*. Likewise, a source of 50 ohms would result in a noise figure of approximately 3.5 *dB*.

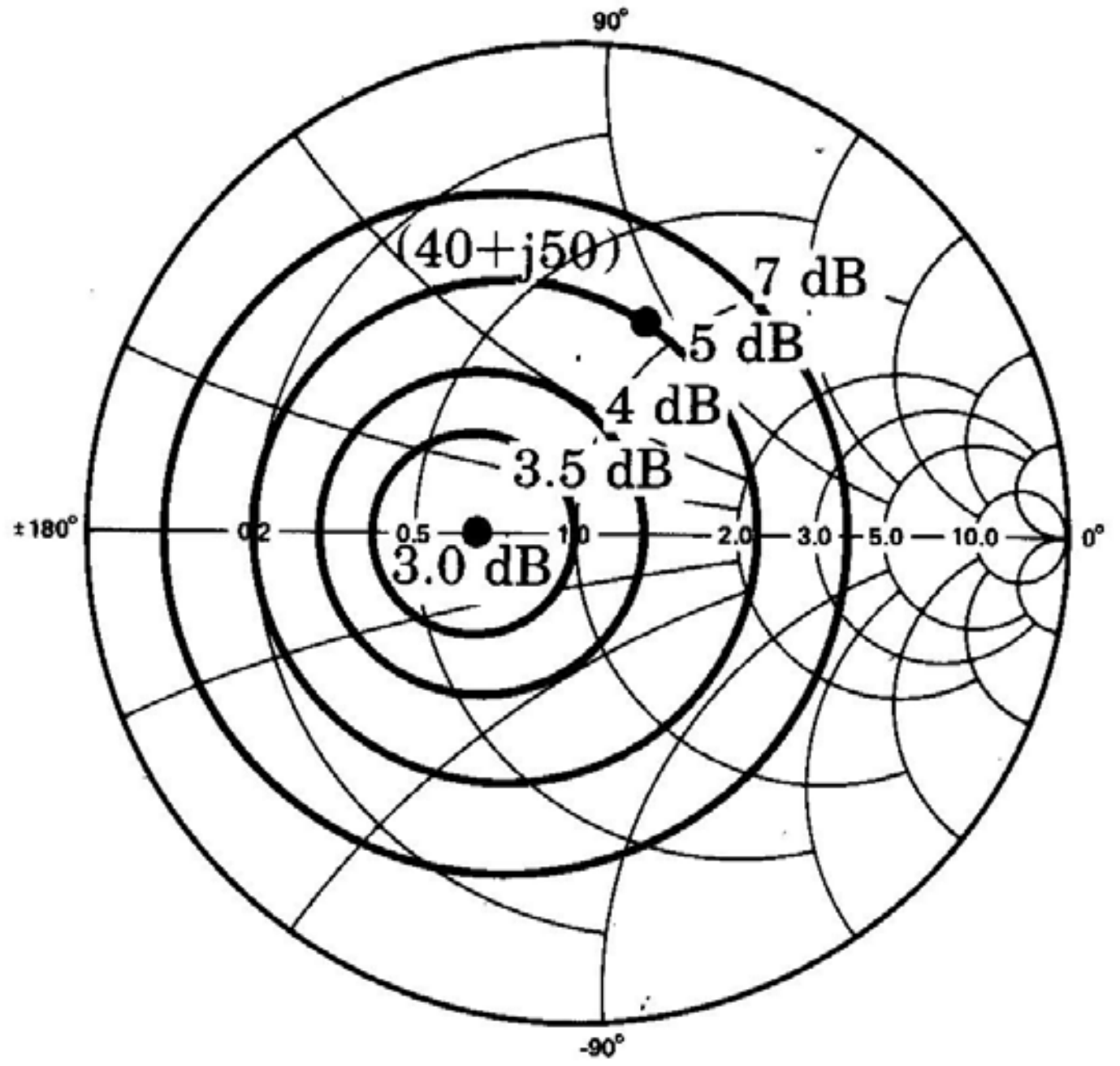

**Fig. 7.** Noise figure for a particular device for any arbitrary source impedance at a particular frequency at Smith chart (after [12]).

In the case of an active microwave device design such as an amplifier, the constant gain circles can now be overlaid on these noise figure circles as it is illustrated in Fig. 8. The resulting plot clearly indicates the tradeoffs between gain and noise figure that have to be made in the design of low noise stages of an amplifier. In general, the maximum gain and minimum noise figure can not be obtained simultaneously. In this given example, designing for maximum gain results in a noise figure of about 6 *dB*, while designing for minimum noise figure results in approximately 2 *dB* less than maximum gain.

The relative importance of the two design objectives, gain and noise figure, will dictate the compromise that must be made in the design process.

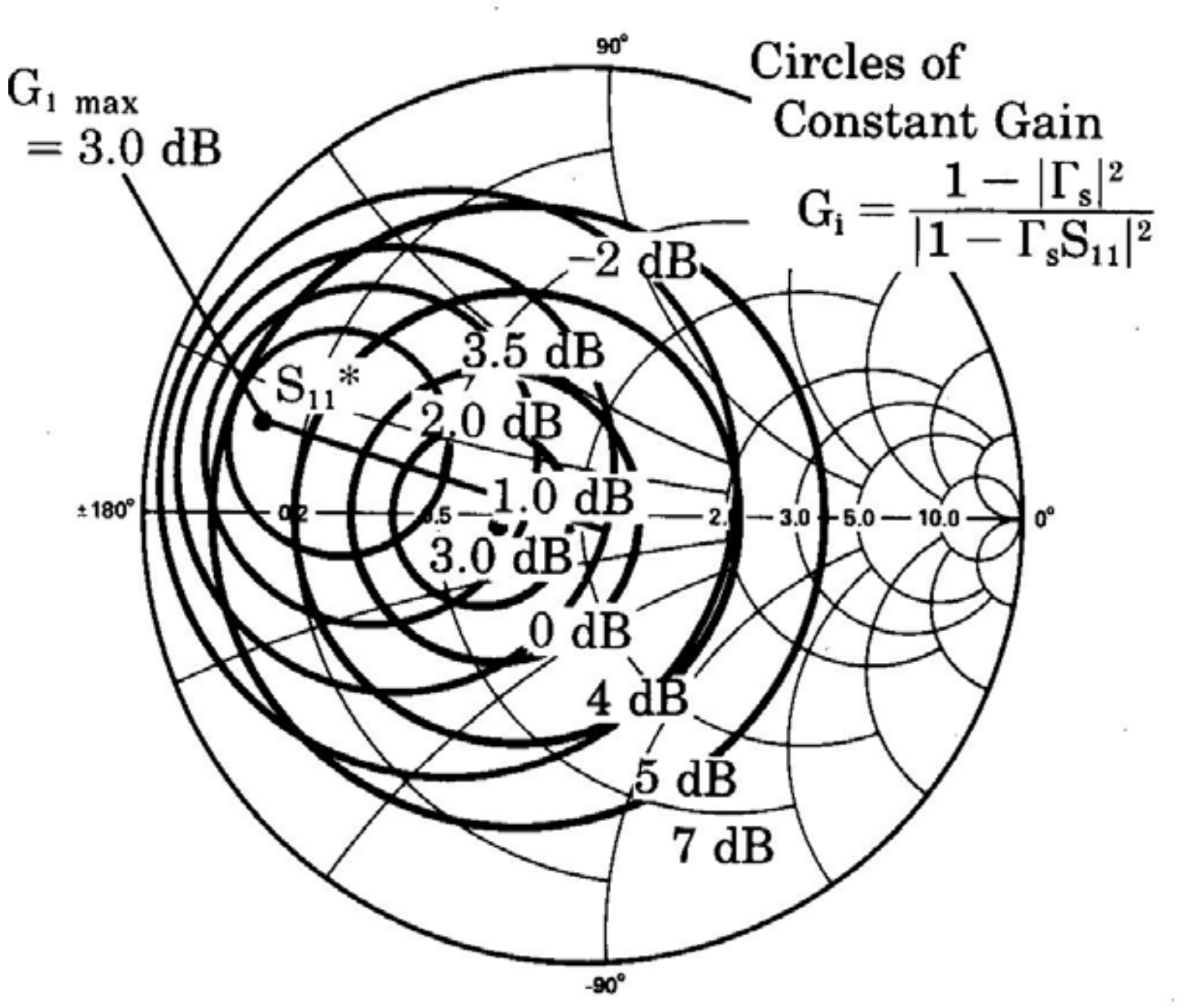

**Fig. 8.** Overlay of constant gain circles on noise figure circles at *Smith chart* (after [12]).

It is also important to remember that the contributions of the second stage to the overall amplifier noise figure can be significant, especially if the first stage gain is low in eq. (20).

$$NF_{overall} = NF_1 + \frac{NF_2 - 1}{G_1} \quad (20)$$

It is, therefore, not always advisable to minimize the first stage noise figure, if the cost in gain is too great. Very often, there is a better compromise between the first stage gain and the *noise figure,* which results in a lower overall *noise figure* for an amplifier [12].

## 8.4. Theory of Response Function. Real and Imaginary Parts of Response Function.

The *theory of response function* is comprehensively discussed in this sub-chapter before the consideration of topics related to the modeling of *differential noise* in *HTS* microstrip resonators in next sub-chapter. The electromagnetic wave in the vacuum coming in contact with a physical body such as a dielectric, metal, or superconductor will penetrate into it. The propagation conditions of the wave depend on certain average physical characteristics of the element's body such as the dielectric and magnetic permeability, conductivity and others. Any material environment can be characterized in electrodynamics by the *Maxwell equations*, which can be written differentially as in eqs. (21-24) by Landau, Lifshits [14]:

$$\mathbf{rot\,E} = -\,\partial\mathbf{B}/\partial\mathbf{t}, \qquad (21)$$

$$\mathbf{rot\,H} = \partial\mathbf{D}/\partial\mathrm{t} + \mathbf{j}, \qquad (22)$$

$$\mathbf{div\,D} = \rho, \qquad (23)$$

$$\mathbf{div\,B} = 0, \qquad (24)$$

where $\boldsymbol{E}$ and $\boldsymbol{B}$ are the vectors of strengths of the electric and magnetic fields accordingly, the induction of an electric field $\boldsymbol{D}$ is $\varepsilon\varepsilon_0\boldsymbol{E}$**,** and the induction of a magnetic field $\boldsymbol{B}$ is $\mu\mu_0\boldsymbol{H}$, where ε and μ are the electric and magnetic relative permeabilities of environment, **j** is the density of an electric current, $\boldsymbol{j}=\sigma\boldsymbol{E}$, $\rho$ is the density of electric charge, $\sigma$ is the conductivity.

In a dielectric, in absence of the external charges ($\rho=0$) and currents ($j=0$), the wave equations for $\boldsymbol{E}$ and $\boldsymbol{H}$ are written in the following way in eqs. (25, 26):

$$\Delta\mathbf{E} - (\varepsilon\mu/c^2)\,\partial^2\mathbf{E}/\partial t^2 = 0, \qquad (25)$$

$$\Delta\mathbf{H} - (\varepsilon\mu/c^2)\,\partial^2\mathbf{H}/\partial t^2 = 0, \qquad (26)$$

where the operator $\Delta=\partial^2/\partial x^2+\partial^2/\partial^2 y+\partial^2/\partial z^2$, $c$ is the speed of light in vacuum, $t$ is time.

Thus, it is visible that, in the environment, the wave spreads with a phase velocity $v_{ph} = c/(\varepsilon\mu)^{1/2}$ or $v_{ph} = c/N$, where $N$ is the index of refraction of the environment, $N=(\varepsilon(\omega,\mathbf{k})\ \mu(\omega,\mathbf{k}))^{1/2}$. $N$ is the complex value $N = N' + iN''$.

The plane simple harmonic waves can be written as in eqs. (27, 28):

$$\mathbf{E} = \mathbf{E_0} \exp\{i(\mathbf{kr} - \omega t)\}, \qquad (27)$$

$$\mathbf{H} = \mathbf{H_0} \exp\{i(\mathbf{kr} - \omega t)\}, \qquad (28)$$

where $E_0$, $H_0$ are the amplitudes of electric and magnetic fields of the electromagnetic waves, $\omega$ is the frequency and $\boldsymbol{k}$ is the wave vector.

Frequency and wave number are bound among themselves by the relation $k=\omega/v_{ph} = 2\pi/\lambda$, where $\lambda$ is the wavelength. The amplitudes of waves can be complex at distribution in the medium with the absorption as the complex values can be represented by both the wave number and the wavelength. The dielectric constant $\varepsilon$ and permeability $\mu$ are complex values in the environment with a dissipation of energy, i.e. $\varepsilon=\varepsilon'+i\varepsilon''$ and $\mu=\mu'+i\mu''$ in the medium with the absorption. If the environment has a complex index of refraction and the wave is spread in direction $z$, the equations (27, 28) can be written as in eqs. (27 ', 28 ')

$$\mathbf{E} = \mathbf{E_0} \exp\{i\omega (z/v_{ph} - t)\} \exp\{-z/\delta\}, \qquad (27')$$

$$\mathbf{H} = \mathbf{H_0} \exp\{i\omega (z/v_{ph} - t)\} \exp\{-z/\delta\}, \qquad (28')$$

where $v_{ph} = c/N'$ and $\delta = c/(\omega N'')$ is the electromagnetic field penetration depth.

In the composite medium, the permeability can have a dispersion, i.e. the frequency dependence $\varepsilon=\varepsilon(\omega)$ and $\mu=\mu(\omega)$ (temporary dispersion) and the dependence on a wave vector $\varepsilon=\varepsilon(\boldsymbol{k})$ *u* $\mu=\mu(\boldsymbol{k})$ (spatial dispersion). In this case, in eqs. (29, 30)

$$\mathbf{D}_{\mathbf{k}\,\omega}= \varepsilon_0\ \varepsilon(\omega, \mathbf{k})\ \mathbf{E}_{k\,\omega}, \qquad (29)$$

$$\mathbf{B}_{\mathbf{k}\,\omega}= \mu_0\ \mu(\omega, \mathbf{k})\ \mathbf{H}_{\mathbf{k}\,\omega}. \qquad (30)$$

If the dispersions are not present, the inductions in an instant *t* in the given point of space *r* depend on the fields in the same instant in the same point of space, i.e. the local conditions of correlation of these values are fulfilled. At presence of the space and temporary dispersions, they are linked by the *Fourier-conversions* as in eqs. (41, 42):

$$\varepsilon(\mathbf{r}, t) = \int \varepsilon(\omega, \mathbf{k}) \exp(i\mathbf{kr} - i\omega t)\, d^3k\, d\omega / (2\pi)^4 , \quad (41)$$

$$\mu(\mathbf{r}, t) = \int \mu(\omega, \mathbf{k}) \exp(i\mathbf{kr} - i\omega t)\, d^3k\, d\omega / (2\pi)^4 \quad (42)$$

and then in eqs. (43, 44)

$$\mathbf{D}(\mathbf{r}, t) = \varepsilon_0 \int \varepsilon(\mathbf{r} - \mathbf{r'}, t - t')\, \mathbf{E}(\mathbf{r'}, t')\, dV'\, dt', \quad (43)$$

$$\mathbf{B}(\mathbf{r}, t) = \mu_0 \int \mu(\mathbf{r} - \mathbf{r'}, t - t')\, \mathbf{B}(\mathbf{r'}, t')\, dV'\, dt', \quad (44)$$

where $dV=dxdydz$.

It can be seen that the inductions are determined by the response of the environment to the effect of a wave within *V*. In this case nonlocal dependence of values is implemented. Thus, for the execution of the principle of causality, the following requirements should be fulfilled in eqs. (45, 46):

$$\varepsilon(\mathbf{r} - \mathbf{r'}, t - t') \text{ at } t>t' \text{ and } \varepsilon(\mathbf{r} - \mathbf{r'}, t - t')=0 \text{ at } t<t' \quad (45)$$

$$\mu(\mathbf{r} - \mathbf{r'}, t - t') \text{ at } t>t' \text{ and } \mu(\mathbf{r} - \mathbf{r'}, t - t')=0 \text{ at } t<t'. \quad (46)$$

Permeability can depend on an electromagnetic wave amplitude $\varepsilon=\varepsilon(E)$ and $\mu=\mu(H)$, that reduces in nonlinear effects.

$$\mathbf{D} = \varepsilon_0\varepsilon\mathbf{E} = \varepsilon_0(1+\alpha)\mathbf{E},$$

where $\alpha\boldsymbol{E}=\boldsymbol{P}$ - polarization of the environment, therefore $\varepsilon= 1+\alpha$, where $\alpha$- electrical polarizability.

The magnetic strength is equal to

$$\mathbf{B}=\mu_0\mu\mathbf{H} = \mu_0(1+\chi)\mathbf{H},$$

where $\chi\boldsymbol{H}=\boldsymbol{J}$ is the magnetic moment of the environment and $\mu=1+\chi$, where a $\chi$-magnetizability of the environment.

In the field of nonlinear effects, the amplitude of the electric field $E_0$ and the amplitude of the magnetic field $H_0$ can influence the values $\alpha$ and $\chi$, which characterize response of the system to the external field. Generally, both the electrical polarizability, and the magnetizability are complex values (($\alpha=\alpha'+i\alpha''$, $\chi=\chi'+i\chi''$), both their real and imaginary parts are linked by the *Kramers-Kronig integral relations* in eqs. (47, 48):

$$\alpha'(\omega) = \frac{1}{\pi} P \int_{-\infty}^{+\infty} \frac{\alpha''(\xi)}{\xi - \omega} d\xi, \qquad (47)$$

$$\alpha''(\omega) = -\frac{1}{\pi} P \int_{-\infty}^{+\infty} \frac{\alpha'(\xi)}{\xi - \omega} d\xi, \qquad (48)$$

where $P$ denotes the principal value of the integral, $\omega$ and $\xi$ are the appropriate frequencies. Here, under $\alpha$ - we understand a generalized susceptibility of the system, which is entered as an aspect ratio between the gauged physical quantity $X$ and the external time-dependent force **f**(t), X=a**f**. The perturbing operator is added to the *Hamiltonian* of the system as a result of the operation of this force in eq. 49

$$\hat{\mathbf{V}} = -\hat{\mathbf{x}}\mathbf{f}(\mathbf{t}), \qquad (49)$$

where $\hat{\mathbf{x}}$ is the operator of the given physical quantity, **f**(t) is the generalized force, which is a function of time. Then, the average value <x> can be represented as in eq. (50)

$$\langle \mathbf{x} \rangle = \hat{\alpha}\mathbf{f} = \int_{0}^{\infty} \alpha(\tau)\mathbf{f}(\mathbf{t}-\tau)\mathbf{d}\tau, \quad (50)$$

where $\tau$ is the time.

Any time-dependent perturbation as $exp(-i\omega t)$ can be restricted to the total of monochromatic components, utilizing the *Fourier-decomposition*. Having substituted $f=f_{\omega}exp(-i\omega t)$ and $<x>_{\omega}exp(-i\omega t)$, we will obtain in eq. (51)

$$\langle \mathbf{x} \rangle_{\omega} = \alpha(\omega)\mathbf{f}_{\omega}. \qquad (51)$$

The function $\alpha(\omega)$ is determined as in eq. (52)

$$\alpha(\omega) = \int_{0}^{\infty} \alpha(\mathbf{t})\exp(\mathbf{i}\omega\mathbf{t})\mathbf{dt}. \qquad (52)$$

According to the most common physical representations this function $\alpha(\omega)$ is named as a generalized susceptibility and is the complex value. It determines the behaviour of the environment under the operation of perturbation. For the given time *t*, the susceptibility is in eq. (53):

$$\alpha(\omega)_t=\alpha_0(t)\exp(i\omega t)= \alpha_0(t)(\cos(\omega t)+i\sin(\omega t))= \alpha'_t +i\alpha''_t. \quad (53)$$

Denote that in eq. (54)

$$\alpha(-\omega) = \alpha^*(\omega), \qquad (54)$$

where * is a sign of complex conjugate.

The real part $\alpha'$ is symmetric in eq. (55)

$$\alpha'(-\omega) = \alpha'(\omega), \qquad (55)$$

and the imaginary part is anti-symmetric in eq. (56)

$$\alpha''(-\omega) = -\alpha''(\omega). \qquad (56)$$

It is possible to obtain the following differential relations between α' and α" at a given response time in eqs. (57, 58)

$$\alpha'_t = (1/t)\, d\alpha''_t/d\omega, \qquad (57)$$

$$\alpha''_t = -(1/t)\, d\alpha'_t/d\omega. \qquad (58)$$

These relations link among themselves the substantial α' and imaginary α" parts of a generalized susceptibility in a given instant *t*, which is defined from the moment of the beginning of process of the given force influence on the system, or, for example, in case of a resonator with a wave, passing through it. This time is set in distance up to the measuring point of a response function of the system. At fixed distances up to the measuring point this time is a constant and does not render effect on a sort of functions defined by means of the measurement. The examples of calculation and analysis of derivatives $d\alpha''_t/d\omega$ for the analysis of a sort of resonance curves in the case of an electronic spin resonance can be found in the monograph [15].

It is necessary to note that during the experimental research of the values α' and α", the important fact is that a subject of interest is not a complete type of dependence of these functions on the frequency in the whole frequency band, but their behaviour near certain characteristic frequency $\omega_0$, at which the resonant response of the system to an external action is observed. There is a resonant exchange of the energy between the wave and the body at this frequency. The average power of a wave, which is dispersed in the environment, is bound to the imaginary part α" by the expression in eq. (59)

$$P=(\omega/2)\, \alpha''(\omega)\, |f_0|^2. \qquad (59)$$

In reality, any physical process is accompanied by the dissipation of energy, and consequently $P > 0$, hence the important output follows that the value α" is distinct from zero and is positive, i.e. α">0 for all the positive frequency ranges ω. Therefore, the vital conclusion is that for the negative frequencies, which are physically bound to the processes happening with the change of the sign of time, α<0 and consequently the environment returns energy to the wave, that for example, is characteristic for the environment with the inversion energy distribution of states and appears in laser systems.

It can be assumed that at a small dissipation, the value $P(\omega-\omega_0)$ is symmetric in relation to $\omega_0$, but remains anti-symmetric in relation to $\omega=0$. The value α' behaves anti-symmetrically in relation to $\omega_0$ and symmetric in relation to $\omega=0$. Usually, in the resonant systems, the value $P(\omega-\omega_0)$ is featured by the *Lorentz dependence* $P(\omega-\omega_0)\propto 1/(\omega-\omega_0)^2$ or by the *Gauss dependence* $P(\omega-\omega_0) \propto exp\{-(\omega-\omega_0)^2\}$. In this case, the *Fourier factorization* happens at $sin\{(\omega-\omega_0)t + \pi/2\}=cos(\omega-\omega_0)$ for α", and at $cos\{(\omega-\omega_0)t + \pi/2\}= - sin(\omega-\omega_0)$ *for* $\alpha'$. Thus, by the virtue of the fast decrease $\alpha''(\omega)$ at a scrap from $\omega_0$, the integration is spread to the area of frequencies $\Delta\omega$, which is considerably smaller in comparison with $\omega_0$.

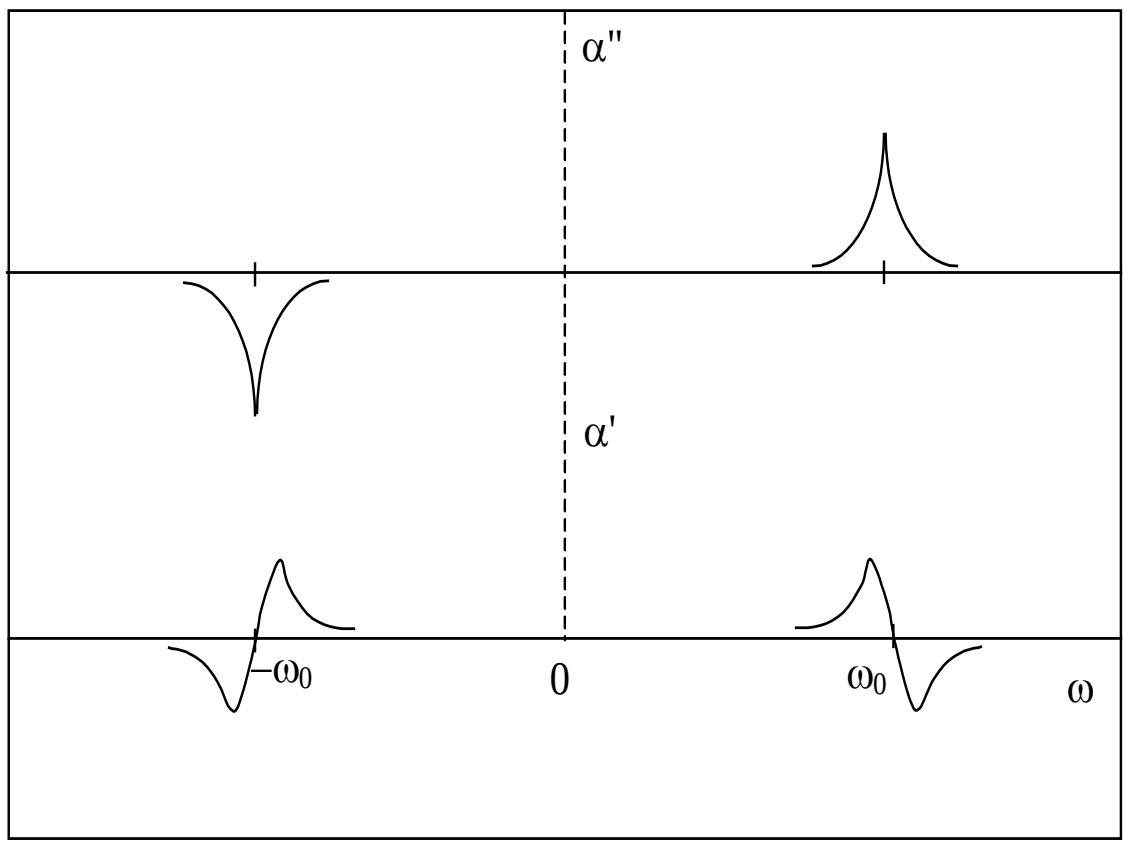

**Fig. 9.** Conditional graph of dependence of response functions on ultra high frequency α'(ω) and α"(ω).

The nonlinear effects in the interaction of a wave with the environment reduce in change of the form of a resonance curve $P(\omega-\omega_0)$ that is stipulated by the influence of nonlinear interaction on $\alpha''(\omega-\omega_0)$, and simultaneously reduces in the change of dependence $\alpha'(\omega-\omega_0)$. Thus, the differential relations (57-58) between $\alpha''$ and $\alpha'$ are valid. Also, it is possible to utilize them in the experiment.

## 8.5. Differential Noise in $YBa_2Cu_3O_{7-d}$ Superconducting Microstrip Resonator and its Nonlinear Dependence on Microwave Power: Experimental Results, Computer Modeling and Discussions.

***Resonances*** are regions in the phase space of a dynamical system in which the frequencies of some angular variables become nearly commensurate in Haller [78]. These regions have a profound effect on the time-scale dynamics of the system, since they are rich sources of highly complex oscillation processes accompanied by energy transfer [78].

In this sub-chapter, the experimental and theoretical researches have been conducted to investigate the real and imaginary parts of a ***response function*** of a microstrip resonator at microwaves

$$\alpha = \alpha' + i\alpha''.$$

According to the equations (27 - 28), the real and imaginary parts $\alpha'$ and $\alpha''$ can be obtained one from another one by the derivation. Therefore, the function $\alpha'=dS_{21}/df$ is a ratio of transmission coefficient derivative to ultra high frequency derivative, where $S_{21} \sim \alpha''$, frequency $f = \omega/2\pi$.

Such experimental dependence of transmission coefficient on ultra high frequency $S_{21}(f)$ for $YBa_2Cu_3O_{7-\delta}$ superconducting microstrip resonator at microwave power $P = -23dBm$ at temperature $T=82K$ is shown in Fig. 10.

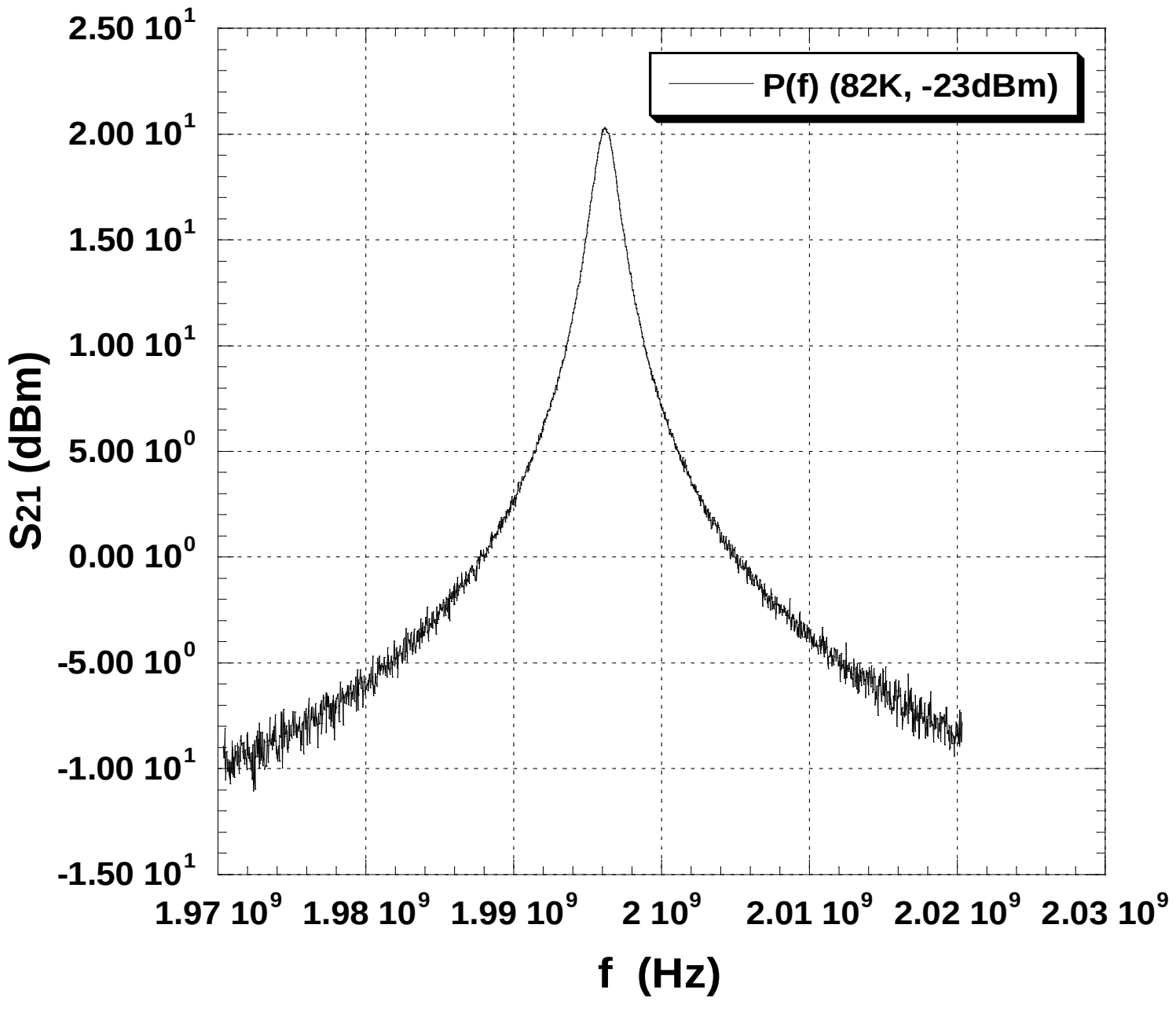

**Fig.10.** Experimental dependence of transmission coefficient on ultra high frequency $S_{21}(f)$ in $YBa_2Cu_3O_{7-\delta}$ superconducting microstrip resonator at microwave power of -23 *dBm* at temperature *T=82K.*

Modeling of dependence of ratio of transmission coefficient derivative to ultra high frequency derivative on ultra high frequency $(dS_{21}/d(f))(f)$ for $YBa_2Cu_3O_{7-\delta}$ superconducting microstrip resonator at microwave power $P=-23dBm$ at temperature $T=82K$ is presented in the *Origin* in Fig. 11 [16].

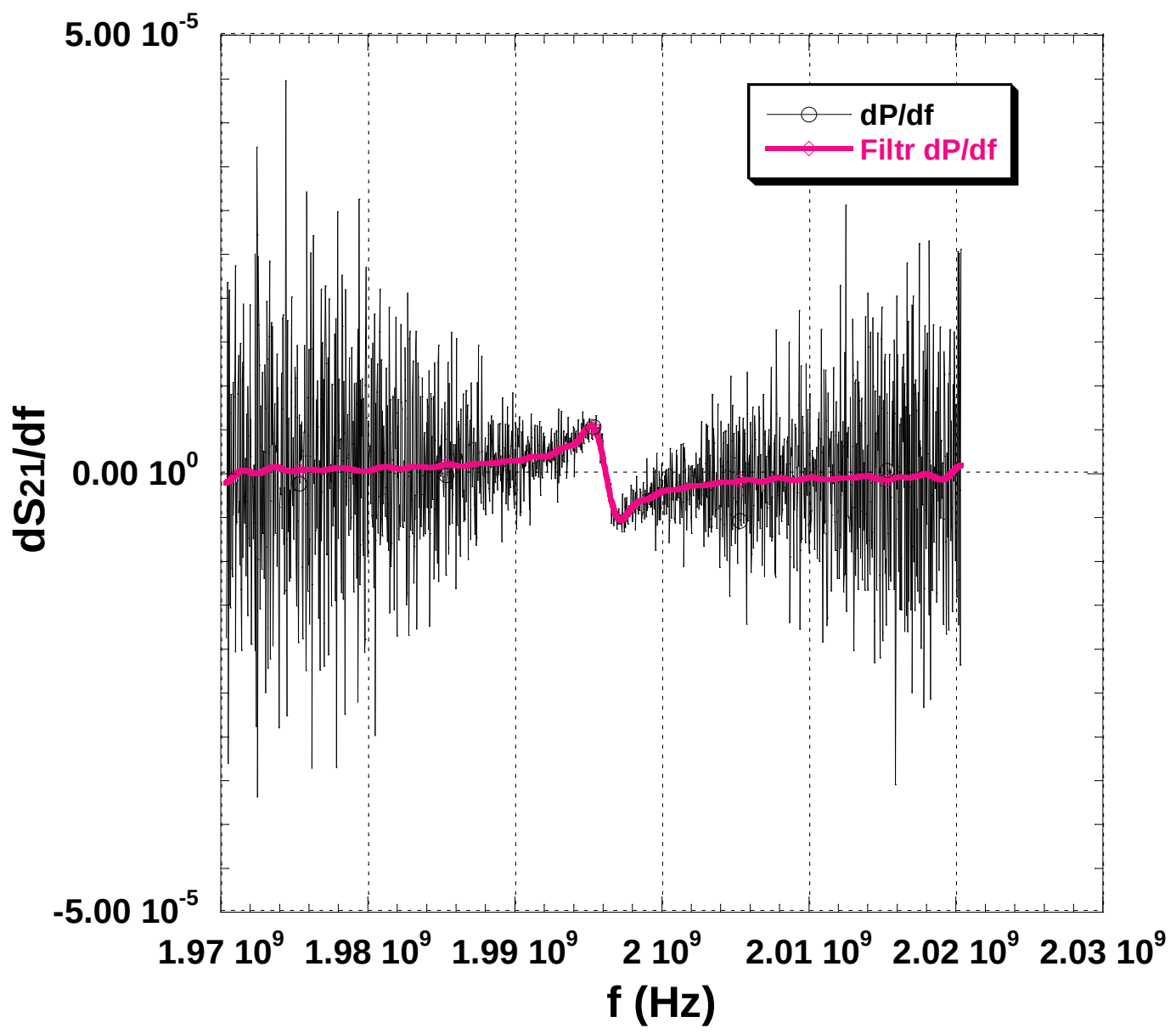

**Fig. 11.** Modeling of dependence of ratio of transmission coefficient derivative to ultra high frequency derivative on ultra high frequency *(dS₂₁/df)(f)* in $YBa_2Cu_3O_{7-\delta}$ superconducting microstrip resonator at microwave power of -23*dBm* at temperature *T=82K* [19].

In Fig. 11, the red curve illustrates the average dependence of microwave power on ultra high frequency *dP/df*, which is obtained after the filtering of the main dependence show in black.

In Fig. 12, the experimental dependence of transmission coefficient on ultra high frequency $S_{21}(f)$ in $YBa_2Cu_3O_{7-\delta}$ superconducting microstrip resonator at microwave power of *+25 dBm* at temperature *T=82K* is shown.

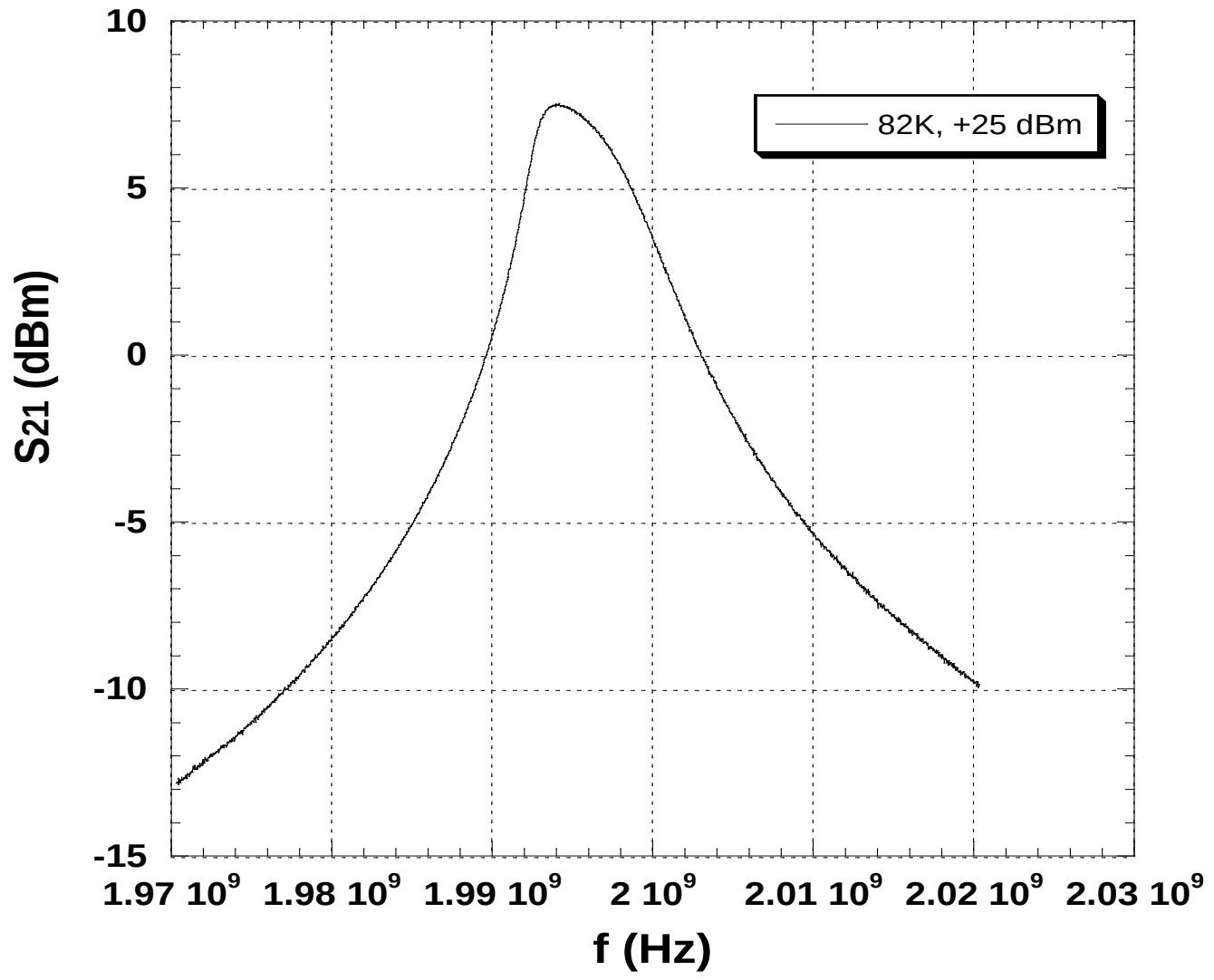

**Fig. 12.** Experimental dependence of transmission coefficient on ultra high frequency $S_{21}(f)$ in $YBa_2Cu_3O_{7-\delta}$ superconducting microstrip resonator at microwave power of +25 *dBm* at temperature *T=82K.*

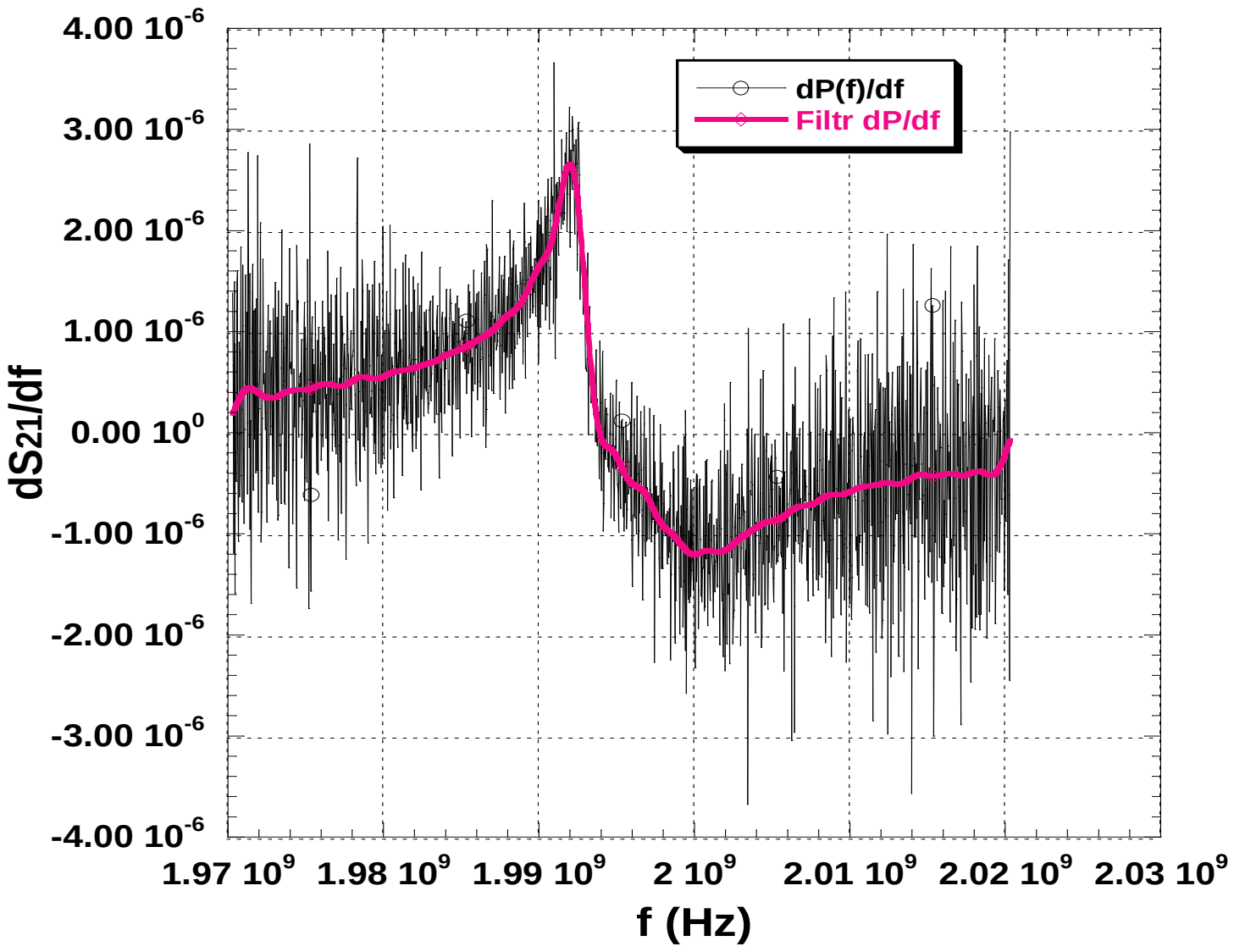

**Fig. 13.** Modeling of dependence of ratio of transmission coefficient derivative to ultra high frequency derivative on ultra high frequency $(dS_{21}/df)(f)$ in $YBa_2Cu_3O_{7-\delta}$ superconducting microstrip resonator at microwave power of +25*dBm* at temperature *T=82K* [19].

In Fig. 13, the average curve after the filtering of the outcomes of the main black curve is also shown.

In Fig. 11 and Fig. 13, the bigger magnitudes of differential noises are clearly visible. The corresponding amplitude response of transmission coefficient $S_{21}$ appears at a change of ultra high frequency of microwave signal in $YBa_2Cu_3O_{7-\delta}$ superconducting microstrip resonator [19].

In Fig. 14, the modeling of dependence of absolute value of difference between the ratio of transmission coefficient derivative to ultra high frequency derivative and filtered ratio of transmission coefficient derivative to ultra high frequency derivative on ultra high frequency $abs(dS_{21}/df - (dS_{21}/df)_{filtr})(f)$ at microwave power of -23 *dBm* at temperature *T=82K* is shown for more clear visual representation of differential noise in $YBa_2Cu_3O_{7-\delta}$ superconducting microstrip resonator [19].

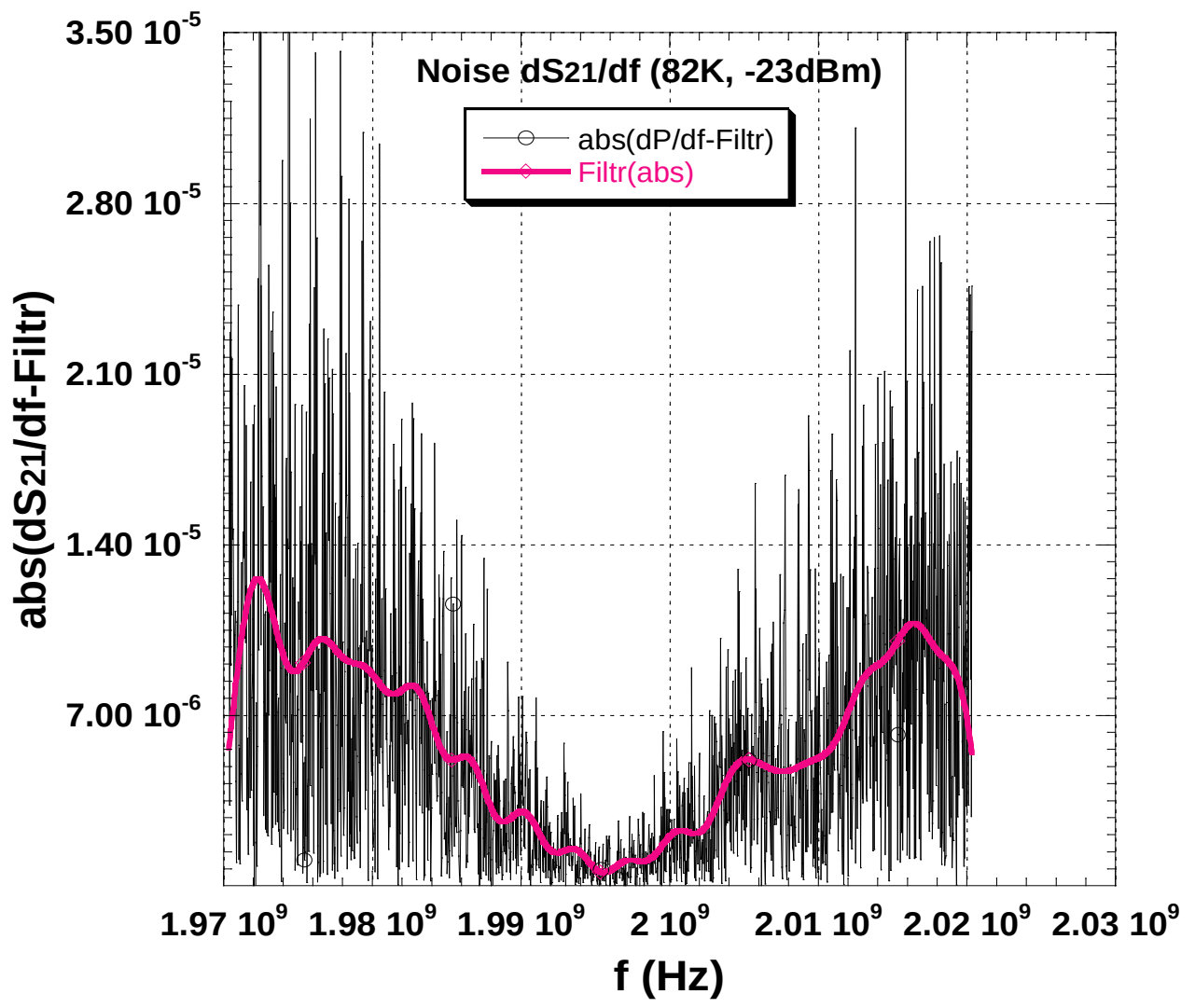

**Fig. 14.** Modeling of dependence of absolute value of difference between ratio of transmission coefficient derivative to ultra high frequency derivative and filtered ratio of transmission coefficient derivative to ultra high frequency derivative on ultra high frequency $abs(dS_{21}/df - (dS_{21}/df)_{filtr})(f)$ in $YBa_2Cu_3O_{7-\delta}$ superconducting microstrip resonator at microwave power of -23 *dBm* at temperature *T=82K* [19].

In Fig. 14, the average differential noise, which has smoother varying dependence, is presented by red curve [19].

In Fig. 15, the modeling of dependence of absolute value of difference between the ratio of transmission coefficient derivative to ultra high frequency derivative and filtered ratio of transmission coefficient derivative to ultra high frequency derivative on ultra high frequency $abs(dS_{21}/df - (dS_{21}/df)_{filtr})(f)$ in $YBa_2Cu_3O_{7-\delta}$ superconducting microstrip resonator at microwave power of +25 *dBm* at temperature *T=82K* is shown [19]. In this case, the $YBa_2Cu_3O_{7-\delta}$ superconducting microstrip resonator is in the nonlinear mode of operation, however in the contrary to the expected results, the absolute value of differential noise varies in order of magnitude approximately.

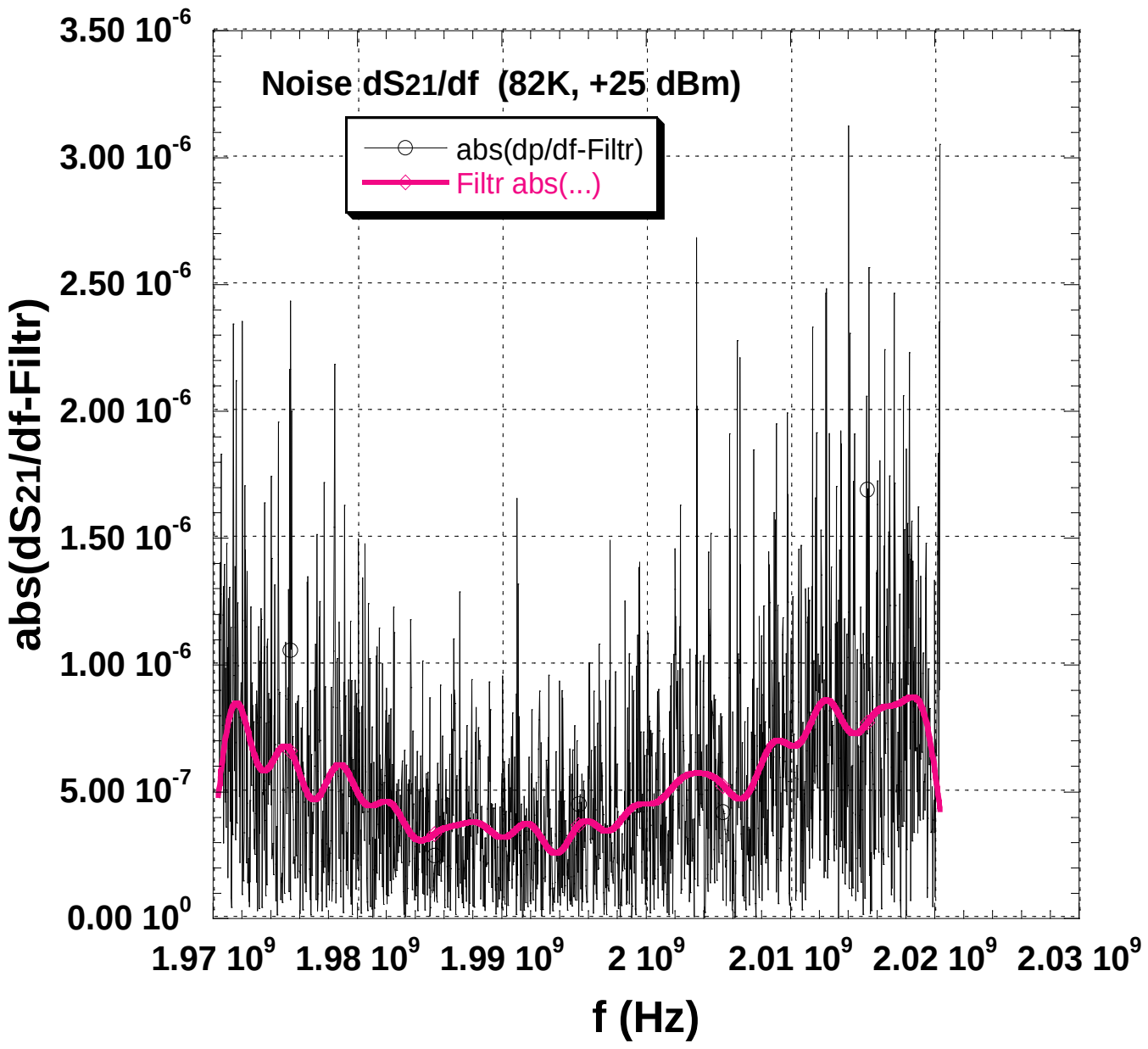

**Fig. 15.** Modeling of dependence of absolute value of difference between ratio of transmission coefficient derivative to ultra high frequency derivative and filtered ratio of transmission coefficient derivative to ultra high frequency derivative on ultra high frequency $abs(dS_{21}/df - (dS_{21}/df)_{filtr})(f)$ in $YBa_2Cu_3O_{7-\delta}$ superconducting microstrip resonator at microwave power of +25 *dBm* at temperature *T=82K* [19].

The computer modeling results are very innovative and interesting. They ***show that the differential noise in the nonlinear superconducting microwave resonators is suppressed.*** Also, it can be observed that the average curves $abs(dS_{21}/dP - (dS_{21}/dP)_{filtr})$, obtained at filtering process, have the *oscillating shapes* in Fig. 14 and Fig. 15. The average curves $abs(dS_{21}/dP - (dS_{21}/dP)_{filtr})$ oscillations can be connected with the phenomena of electromagnetic wave multi-photon absorption by superconducting electrons under the circumstances, when the superconducting energy gap $\Delta(T, P)$ decreases in the electron spectrum as a result of action of the temperature $T$ and microwave power $P$ of electromagnetic wave, and the condition of absorption for a number of photons $N$ and a number of superconducting pairs of electrons $K$ is satisfied in eq. (30)

$$N\hbar\omega \approx K\, 2\Delta(T, P) \qquad (30)$$

where $N$ and $K$ are the integers $N, K = 1,2,3...$

According to eq. (8), in the case, when $\hbar\omega > kT$ the spectral density of *quantum noise* may be represented as in eq. (31)

$$\overline{S\ (\omega)} = \frac{R\hbar\omega}{2\pi}\coth\left(\frac{\hbar\omega}{2kT}\right) = \frac{R}{\pi}\left[\frac{1}{2}\hbar\omega + \frac{\hbar\omega}{\exp(\hbar\omega/kT)-1}\right] \ (31)$$

In this case, it can be supposed that $N\hbar\omega > kT$, where $N >> 1$. In eq. 31, the first term in right hand side is $R\,N\,\hbar\omega/2\pi$, and it is not observed due to the fact that it arises from the zero point fluctuations. The second term is $R\ N\hbar\omega/\pi(\exp(N\,\hbar\omega/kT - 1))$. This function is flat for $N\hbar\omega << kT$, but for $N\hbar\omega > kT$, it rapidly approaches zero. Therefore, the thermal noise in this quantum case becomes small in comparison with the classical case.

In the case of the electromagnetic wave with a high level of microwave power, the number of photons $N$ is big, and the oscillations of the average curve $abs(dS_{21}/dP-(dS_{21}/dP)_{filtr})$ are more frequent. It is possible to suppose that this phenomenon may have some other nature, which has to be explained and clarified in further research studies.

The *differential noise* in two terminal devices such as the $YBa_2Cu_3O_{7-\delta}$ superconducting microstrip resonators is researched by author of dissertation firstly. Some other experimental results on origin of noise in superconducting two and three terminals devices were also presented in research works by other authors [17, 18].

It should be mentioned that the research efforts toward the creation of new $YBa_2Cu_3O_{7-\delta}$ superconducting microwave resonators generated considerable interest towards the understanding of nature and origin of *quantum noise* [11].

## Summary.

In course of research, the primary objective to investigate the nature of physical characteristics of noise in superconducting microwave resonators and filters is completed successfully, including the three main research results:

1. The new source of noise: ***the magnetic dipole two-level systems (MTLS) in HTS thin films,*** based on the *Abricosov or Josephson magnetic vortices*, in superconducting microwave resonators is proposed first [59].

2. The new ***quantum theory of 1/f noise in HTS thin films at microwaves***, based on the original proposition by the author of dissertation, is formalized: The charge carriers (Cooper electron pairs) may interact with quantized electromagnetic field, generated by *Abricosov and Josephson magnetic vortices* in *HTS* thin films at microwaves [63, 64, 65], resulting in both the charge carriers scattering by arbitrary random discrete potential barriers on the *N-S boundaries*, created by *Abricosov or Josephson magnetic vortices* with normal cores in *HTS* thin films at microwaves. This process may be accompanied by the low frequency photon irradiation by charge carriers and *Cooper electron pair* breakup. The *1/f* noise may be produced as a result of modulation of transport current by the emitted energy of photons [59, 68].

3. The novel original research results on the nature of ***differential noise*** in $YBa_2Cu_3O_{7-\delta}$ thin films at frequency of 25 *GHz* are also presented. It is shown that the magnitude of differential noises for the standard signal, which is transmitted through the superconducting microstrip resonator, is significantly suppressed in the case, when the resonator with a high level of reserved microwave energy is in the nonlinear regime. This conclusion is not trivial due to the fact that in the nonlinear case in a microwave resonator certain intermodulation and high harmonics generation processes may be present, which, at the first view, have to result in the high level of noise [79].

Presently, the intensive research and development efforts (*R&D*) are focused on the creation of nonlinear devices with the use of *HTS* thin films [66]. In the case of microwave resonators made of *HTS* thin films, the nonlinearity of $S_{21}(f)$ in *HTS* microstrip resonator opens a wide range of technical applications in communication systems and radars [63, 64, 65, 67]. The discovered effect of suppression of *differential noise* in the nonlinear regime of microstrip resonator operation may be used for the development of microwave passive and active devices with low levels of differential noise based on the superconducting thin film technology. Author of dissertation notes that some properties of the investigated superconducting microstrip resonator system may be interconnected with the quantum phenomena occurring in the microwave resonator influenced by multi-photon processes, when $N\ \hbar\omega > kT$, as the thermal noise approaches zero and the total noise becomes relatively small.

Finally, it makes sense to note that the noise may also appear in the superconducting digital circuits such as the superconducting digital filters, where the character of noise depends on many factors like the type of nonlinearity, filter structure, type of arithmetic, representation of negative numbers, and properties of input signal, however this topic is beyond the scope of consideration of present dissertation [13].

# CHAPTER 9

# DESIGN OF HIGH TEMPERATURE SUPER-CONDUCTING (HTS) MICROWAVE FILTERS IN CRYOGENIC TRANSCEIVER FRONT END IN WIRELESS COMUNICATION SYSTEMS

## 9.1. Introduction.

The High Temperature Superconducting (*HTS*) microwave filters designs for application in wireless communication systems have been researched in [1-15]. In this chapter, the *HTS* microwave filters in *CTFE* are described:

1. Block diagram for the Cryogenic Transceiver Front End (*CTFE*) in Fig 1 [17].
2. Microwave filtering requirements for digitally modulated signals [28, 29].
3. General engineering requirements to the *HTS* microwave filters design [17].
4. Technical advantages of *HTS* microwave vs. metallic filters [1-17]
5. Miniature *HTS/LTS* microwave filters characteristics and types [36, 100].
6. Modern *HTS* microwave filters in Cryogenic Receiver Front Ends (*CRFE*) at basestation in wireless networks [1-25, 30, 37, 40, 49-51, 54, 55, 57, 90-96], and space applications [11, 80, 82].
7. *HTS* microwave filters tuning and trimming techniques [4, 10, 25, 37, 45, 71, 76].
8. Thermal stability of the *HTS* microwave filters [11].
9. List of researchers, who made significant research contributions to the advanced innovative designs of high performance *HTS* microwave filters [1-100].

## 9.2. Cryogenic Transceiver Front End (CTFE) with HTS Transmitting and Receiving Microwave Filters in Wireless Communication Systems.

The *HTS* microwave electromagnetic signal filter [1-15] is one of the essential microwave components in modern wireless communication systems in which the complete and independent measurement of the entire signal space to identify and decode the information in the spectral transmission sequences over the wireless channel is made [130]. The main functions of microwave filter are to select the information signal carrier in the frequency domain and amplify it by the resonance. In the increasingly dense electromagnetic spectrum, the microwave filtering allows the transceiver to tune into the only frequency band, containing the desired signals, while rejecting all other signals, when receiving or transmitting information over the wireless channel. The High Temperature Superconducting *(HTS)* microwave filters in the Cryogenic Transceiver Front End (*CTFE*) have the significantly improved microwave characteristics due to the low energy losses in the *HTS* thin films, comparing to the metallic microwave filters in Figs. 1, 3 in [17]:

1. *Better signal-to-noise ratio,*
2. *Greater input signal selectivity,*
3. *Better adjacent channel leakage power ratio* (in transmitter),
4. *Small insertion losses,*
5. *Compact tunable design,*
6. *Greater capacity throughput due to lack of in/out of channel interference.*

Nisenoff, Pond [120] write that the superconducting filters in combination with cryogenically cooled *LNA*s have the following technical advantages:

1) Decreased noise figure for the system within the bandpass of *HTS* filter, and

2) Much steeper frequency roll-off on the edges of bandpass of *HTS* filter due to the higher order filter that can be used and the higher *Q* values of *HTS* resonators.

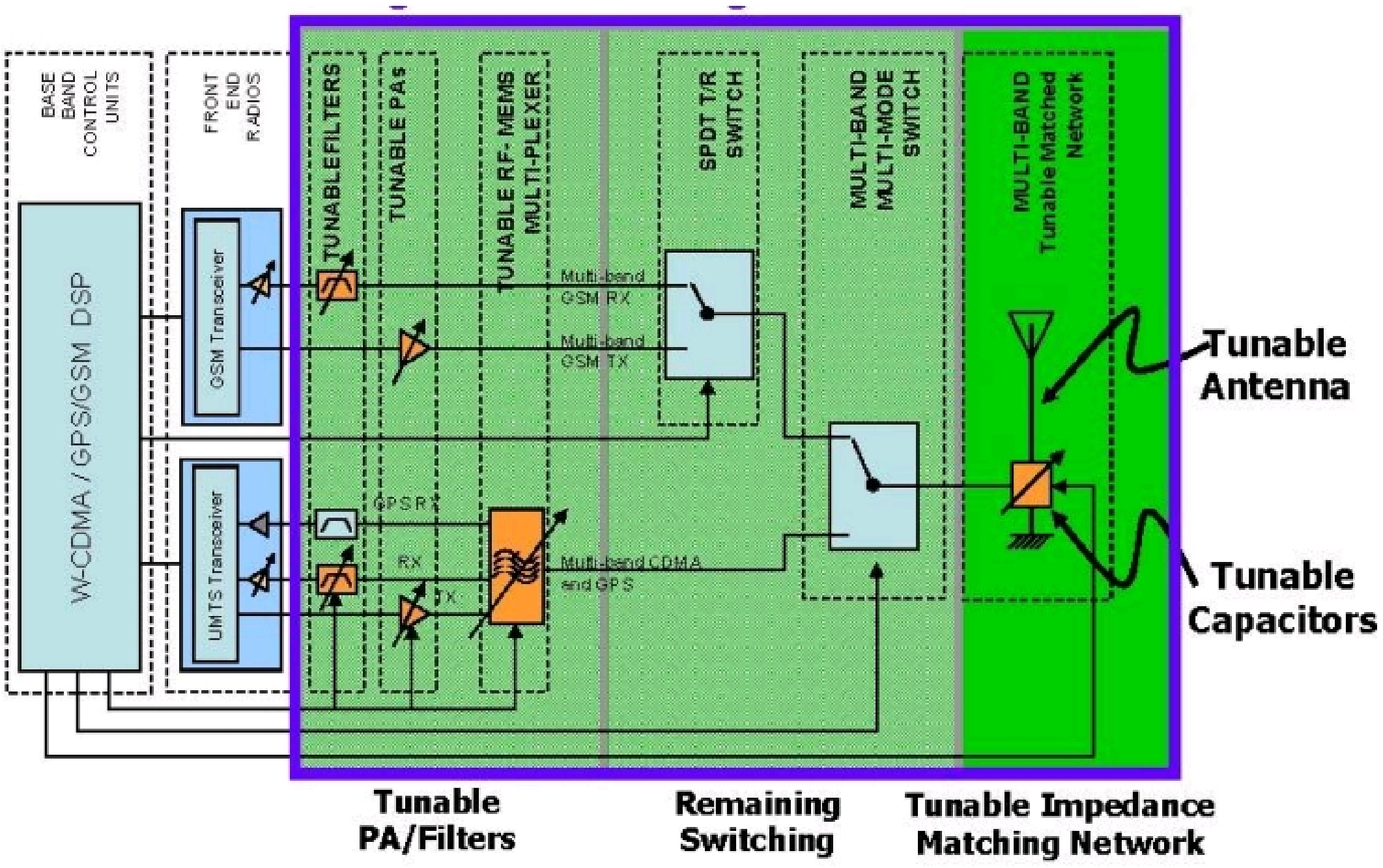

**a)**

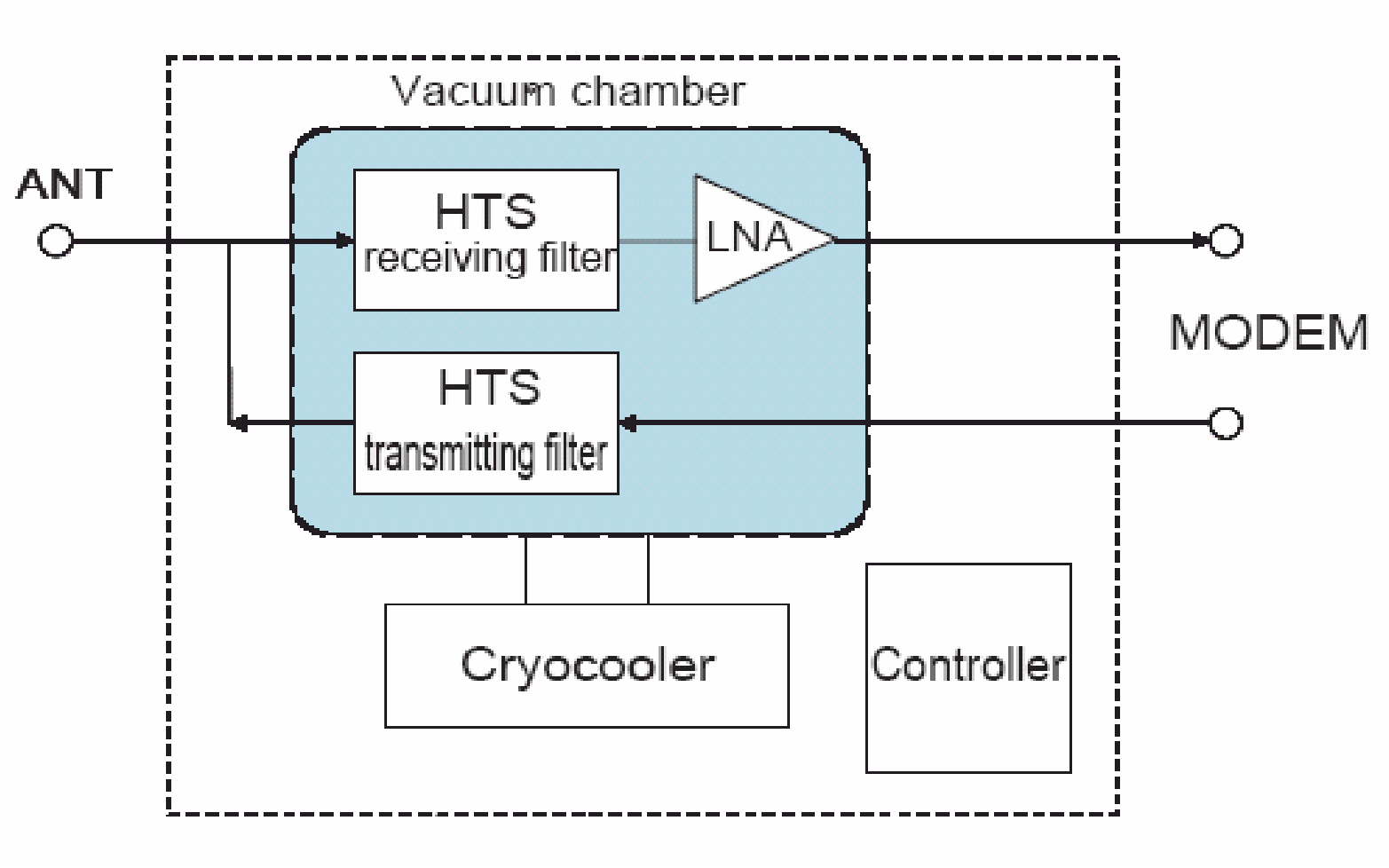

**b)**

**Fig. 1. a)** Reconfigurable transceiver design (after Jim Hwang [131, 132-134]); **b)** Cryogenic Transceiver Front End (*CTFE*) block diagram that implements *HTS* microwave transmitting and receiving filters (after Yamanaka, Kurihara [17]).

The general categories of microwave filters include: *low-pass* filters, *high-pass* filters, *band-pass* filters, *band-reject* filters, indicative of the type of frequencies that are selectively passed or rejected by the microwave filter [115].

There are many different types of microwave filters, including:

1. *Butterworth*,
2. *Chebyshev I and II*,
3. *Elliptic*,
4. *Bessel*,
5. *Gaussian*.

Some types of microwave filters, according to their frequency response functions, are shown in Fig. 2 [29].

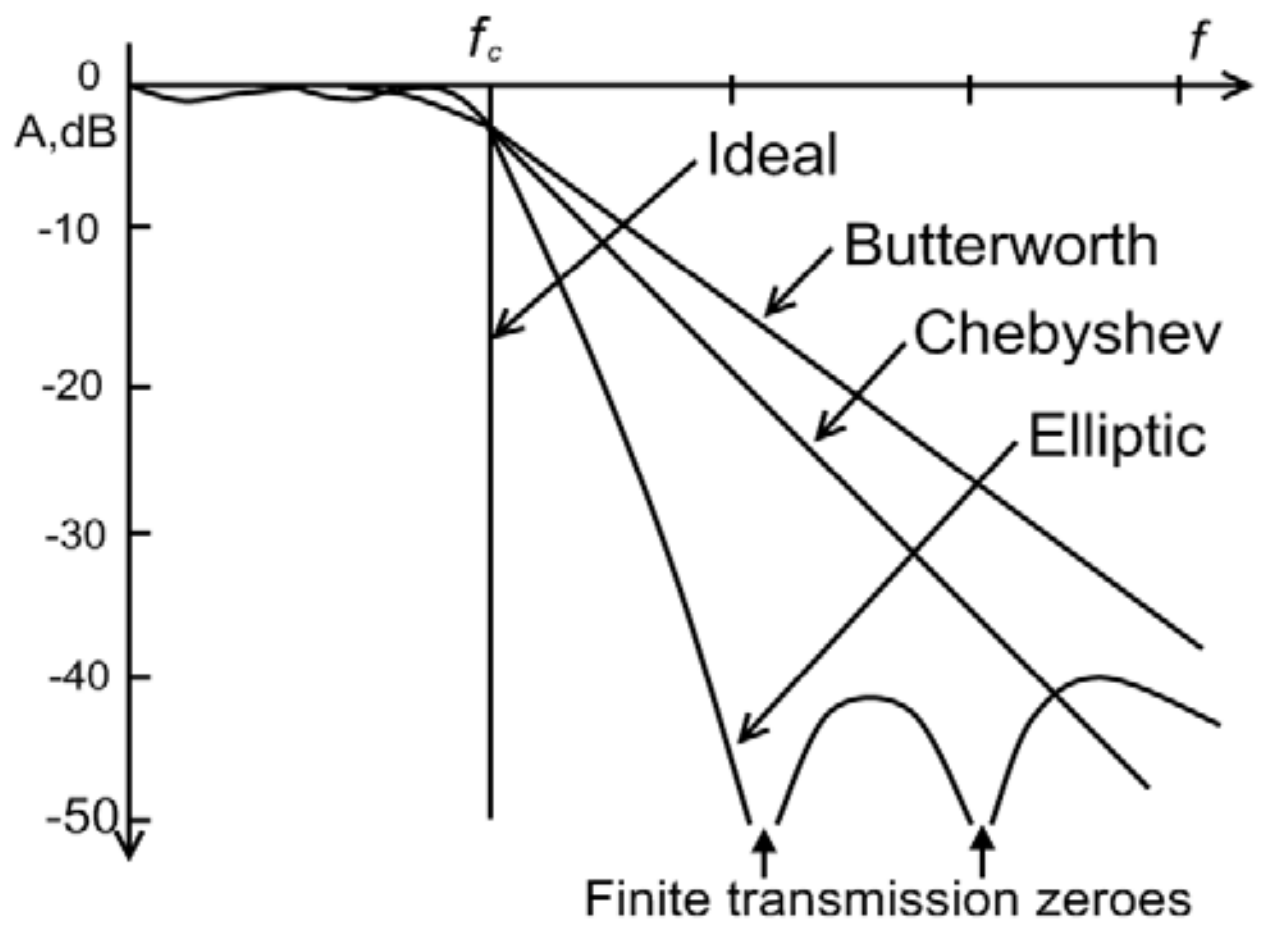

**Fig. 2.** Ideal, Butterworth, Chebyshev, Elliptic Types of Microwave Filters According to their Response Functions (after [29]).

The microwave filters can also be distinguished by different categories based on the resonator type and functional principles, including: the planar *HTS* filters such as microstrip filters and stripline filters; cavity filters; coaxial filters; combline filters; airline filters; dielectric puck filters; Micro Electro Mechanical (*MEMS*) filters; Surface Acoustic Wave (*SAW*) filters; Film Bulk Acoustic Resonator (*FBAR*) filters; bulk acoustic wave filters; quasi-lumped elements filters; and many others in Tsuzuki, Willemsen [115].

In Tab. 1, the detailed information about the most important technical characteristics of a microwave filter is provided [29].

| Terminology | Symbol | Unit | Explanation of Terms |
|---|---|---|---|
| Center Frequency | $f_0$ | Hz | The frequency in the center of the band, or the point where the loss is at its lowest point. |
| Pass-Bandwidth | $f_c$ | (3dB) BW | Signifies a difference between the two frequencies, where the attenuation becomes 3dB below minimum loss. |
| Insertion Loss | IL | dB | Passband loss Insertion Loss = 20 LOG($V_2/V_1$). |
| Ripple | - | dB | The difference between the maximum peaks and minimum valleys. |
| Attenuation Bandwidth | (20dB) BW | Hz | The bandwidth at a specified level of attenuation. |
| Stopband Attenuation | - | dB | The bandwidth at a specified level of attenuation. |
| Spurious Response | SR | dB | The difference in decibels between the insertion loss and the spurious signal in the stopband. |
| Selectivity | - | dB | The shape factor of the filter. The attenuation bandwidth divided by the pass bandwidth |

**Tab. 1.** Microwave filter technical characteristics (after [29]).

The *RF* receiver in the Cryogenic Transceiver Front End (see Fig. 1) has to satisfy a number of *RF* technical requirements such as [29, 99]:

1. ***Good sensitivity for a given BER***. Sensitivity of a receiver can be defined as the minimum detectable signal required to maintain a reasonable signal to noise ratio or Bit Error Ratio (*BER*). Estimated receiver sensitivity = $KT_{dBm/Hz}$ + $NF_{dB}$ + $10log(bandwidth)$ + $E_S/N_{0\,dB}$ $+10log(symbols\ per\ second) – 10log(bandwidth\ Hz)$.

2. ***Good adjacent channel selectivity***. Selectivity is the property of receiver that allows to separate the information signal carrier on one frequency from those on all other frequencies in frequency domain. Selectivity of any receiver is determined by the narrowest selectivity of microwave filter. Microwave filter shape factor must be sharp enough to pass the desired signal and harmonics with minimum phase and amplitude distortions.

3. ***Good dynamic range***. Dynamic Range of a receiver is the range of signals amplitudes from the smallest to the largest to which the receiver can respond. Dynamic range of receiver depends on the dynamic range of linear *Low Noise Amplifier* (*LNA*), which depends on the width of energy gap and mobility of charge carriers in the used solid state *GaAs, GaN, SiGe, SiC, InN, AlN* semiconductors in low noise *High Electron Mobility Transistors* (*HEMT*) [61, 99, 131]. Dynamic range of *LNA* is limited at the low end by the noise, and at the upper end by the phenomenon known as gain compression. It has to be bigger than the pass band of microwave filter at front end system of receiver. Microwave filter's pass-band loss adds directly to the noise figure of the *LNA*, hence the combining of microwave filters in cascade is not advisable.

4. ***Free of near and far intermodulation spurious responses.*** The ideal filter for processing of rectangular digital pulses is raised cosine filter, because it reduces bandwidth of digitally modulated spectrum for close channel spacing as well as it reduces the inter-symbol interference since there is zero response, when the next pulse arrives.

6. ***Low noise figure.*** Noise figure (*NF*) is a measure of degradation of the signal-to-noise ratio (*SNR*), caused by components in microwave signal chain. The noise figure is defined as the ratio of the output noise power of a device to the portion thereof attributable to thermal noise in the input termination at standard noise temperature $T_0$ (usually 290 *K*).

7. ***High IP3.*** The higher the third-order intercept point $IP_3$, the better the intermodulation suppression in amplifier, where the intercept point is defined by extending the fundamental and third order responses on the chart of output power vs. input power in their linear region until they intersect. This is a characteristic of amplifier, which depends on physical characteristics of energy gap and mobility of carriers in semiconductor transistors in amplifier. In practice, the $IP_3$ can vary because of changing temperature of transistor in amplifier as a result of high microwave power irradiation with big digital data streams over wireless channel.

8.***Proper stage gains, local oscillator (LO), microwave filters rejection parameters.***

*The HTS* microwave planar filters with novel hairpin, spiral and patch resonators with *Q*-factors up to 90,000 and insertion loss of 0.2 *dB* together with low noise amplifiers (*LNA*) constitute Cryogenic Receiver Front End (*CRFE*) in Fig. 1.

***Finding a desirable combination between the chosen HTS microwave filter and low noise amplifier (LNA) in Cryogenic Receiver Front End (CRFE) is a challenging engineering task, because most of the above listed technical requirements have to be satisfied in real designs [99]***. Most of *HTS* microwave filters in *Cryogenic Receiver Front End* (*CRFE*) are *the* ***raised cosine microwave filters***, because the raised cosine microwave filter is a realizable microwave filter with an access bandwidth over the *Nyquist bandwidth*, which can be best used for processing of digitally modulated information carrier with the high order amplitude-phase modulation techniques: *BPSK, QPSK, π/4 DQPSK, 16QAM, 32QAM, 64QAM, 128 QAM, 256QAM, 512QAM, 1024QAM* in Fig. 3 [29, 62-65].

The use of higher-order modulations increases the design requirements on the radio transmitter and receiver in terms of ***linearity*** [97]. The application of the carefully designed *HTS* microwave filters in combination with linear *High power Bipolar Transistor* (***HBT***) *amplifier* (*HPA*) at transmitter as well as the *HTS* microwave filters with multi-stage *low noise High Electron Mobility Transistor* ***HEMT*** *amplifier* (*LNA*) at receiver in *Cryogenic Transceiver Front End* (*CTFE*) can decrease the ***phase noise***, caused by the nonlinearities, and lower the ***adjacent channel leakage power ratio*** (*ACLR*) significantly. The ***Passive Intermodulation*** (*PIM*) products power level at receiver, generated by transmitted signal, has to be at least 20*dB* below the received information carrier signal level in receiving band at receiver in *Cryogenic Transceiver Front End* (*CTFE*). Hence, the *PIM* signal level at receiver has to be 100-150*dB* below the information carrier signal level at transmitter in *Cryogenic Transceiver Front End* (*CTFE*) [99].

Yamanaka, Kurihara from *Fujitsu Ltd*., Japan [17] formulated the following technical requirements for *HTS* microwave filters in Cryogenic Receiver Front End (*CRFE*) in basestations in wireless communications networks [17]: 1. *To satisfy the specified power handling capability;* 2. *Miniaturization of dimensions to a minimum;* 3. *Small intermodulation distortion;* 4. *Circuit and packaging structure such as multi-pole filter, to provide sharp and required frequency characteristics, while utilizing its high-Q parameter;* 5. *Efficient and high reliability least compact cryocooler;* 6. *Cryogenic packaging structure with a minimum thermal load*.

Yamanaka, Kurihara [17] commented on the *HTS* microwave filters in Cryogenic Transmitter Front End (*CTFE*), mentioning that the dual resonant mode generation and provisioning of attenuation poles outside of the passband were researched for planar circuit resonators that had disc patterns as candidates for the resonator used in passband filters to achieve miniaturization and power operation by Akasegawa *et al.* [21] and Ishii *et al.* [22]. Prototyping of transmitting *HTS* microwave filter resulted in characteristic of less than -70 *dBc* (at 10 *W*) in third intermodulation distortion (*IMD3*) in the 5 *GHz* band. Side lobe power of transmitted microwave signal is relatively small; a study on a band rejection type filter has been conducted using split open ring planar-circuit resonators of the reaction type by Futatsumori *et al.* [23]. Additionally, *the research and developments (R&Ds)* have been undertaken on the superconducting tunable transmitting filters by Yamanaka *et al.* [24], and the cryocooler of the pulse-tube type by Matsumoto *et al.* [26]. The tunable transmitting *HTS* microwave filter is aimed at varying the center frequency of passband frequency by several hundreds *MHz* in a low microwave band, and is expected to be applicable to cognitive software defined radio and by Ohsaka *et al.* [25]. The cryocooler is aimed at achieving both a high efficiency and miniaturization by Matsumoto *et al.* [26]. ***Maximum frequency selectivity and maximum receiver sensitivity can be reached*** ***in HTS microwave filters at the same time [17, 90].*** Conventional filter technologies sacrifice the sensitivity, when the selectivity is increased.

Ohshima [20] discussed the present advances in *HTS* microwave technologies in research projects by different universities and companies:

1. ***Superconductor Technologies Inc. (STI)*** of the United States has installed superconducting filters in more than 7000 of its base stations (http://www.suptech.com/home.htm).

2. In Korea, a venture company, ***RFtron***, is developing a superconducting filter system. Equipment for commercial use has not been produced yet, but the level of its technology is high. (http://www.rftron.com/).

3. In China, ***Tsinghua University*** and ***Zongyi Superconductivity Science and Technology Co.*** have developed a high temperature superconducting (*HTS*) filter system and conducted a field test at five base stations in Beijing. This system has been awarded a prize by the *Chinese Information Industry Ministry* as "2007 Top-ten Chinese *IT* Technology Inventions."

4. In Germany, ***Cryoelectra GmbH*** is continuing its efforts. Recently, *Cryoelectra* gave a presentation at an academic meeting on a superconducting front end for *CDMA* basestations. (http://www.cryoelectra.de).

In North America, the ***Northrop Grumman*** corporation conducts a number of large scale *R&D* programs toward the *HTS* microwave filters development for military communications and radars in the USA [12, 13]. The ***Northrop Grumman, NASA Lewis Center*** and ***Superconductor Technologies Inc.*** have been engaged in joint cooperation military *R&D* programs, aiming to effectively share the technical, labour and financial resources during the *HTS* microwave filters design cycle. The ***Lincoln Laboratory*** at ***Massachusetts Institute of Technology***, and ***Naval Research Laboratory*** in Washington, DC in the USA continue their active research programs on *HTS* microwave filters, and have accumulated considerable knowledge base and practical expertise in the field. In Canada, the ***Com Dev Ltd.*** produces *HTS* microwave filters and multiplexers, using the *HTS* thin films made by ***DuPont*** (U.S.A.) and the cryo-coolers by ***Lockheed Martin*** (U.S.A.), and with research support from ***Waterloo University***, for numerous space applications. In Asia, the ***Fujitsu Ltd., Tohoku Seiki Industries Ltd., Toshiba, NTT, NHK, JRC*** corporations are involved in a considerable number of *R&D* programs in microwave superconductivity in collaboration with ***Hokkaido University***, ***Yamagata University***, ***Yamanashi University***, ***Tokyo University*** in Japan. In P. R. China, the ***Microwave Center, University of Electronic Science and Technology of China,*** Chengdu***;*** ***National Laboratory of Superconductivity, Institute of Physics, Chinese Academy of Sciences***, Beijing; ***Department of Material Science and Chemical Engineering, Yanshan Univeristy***, Qinhuangdao participate in the Chinese space program research on *HTS* microwave filters. In Europe, the German ***Cryoelectra GmbH, Theva GmbH, Rohde and Schwartz GmbH*** are considered as prime contractors on *HTS* microwave filters design for a wide range of radar and wireless communications. The *R&D* programs on *HTS* microwave filters design with strong emphasis on the airborne radars and *LTE* wireless communication networks have been completed at ***Ericsson Components AB*** in Sweden. The ***Thales*** corporation has similar *R&D* programs on *HTS* microwave filters design for space, airborne and terrestrial applications in France [101]. The **Boeing** partners with the ***Northrop Grumman*** to develop *HTS* radars & missile guidance systems in frames of *MDI* initiative in the U.S.A.

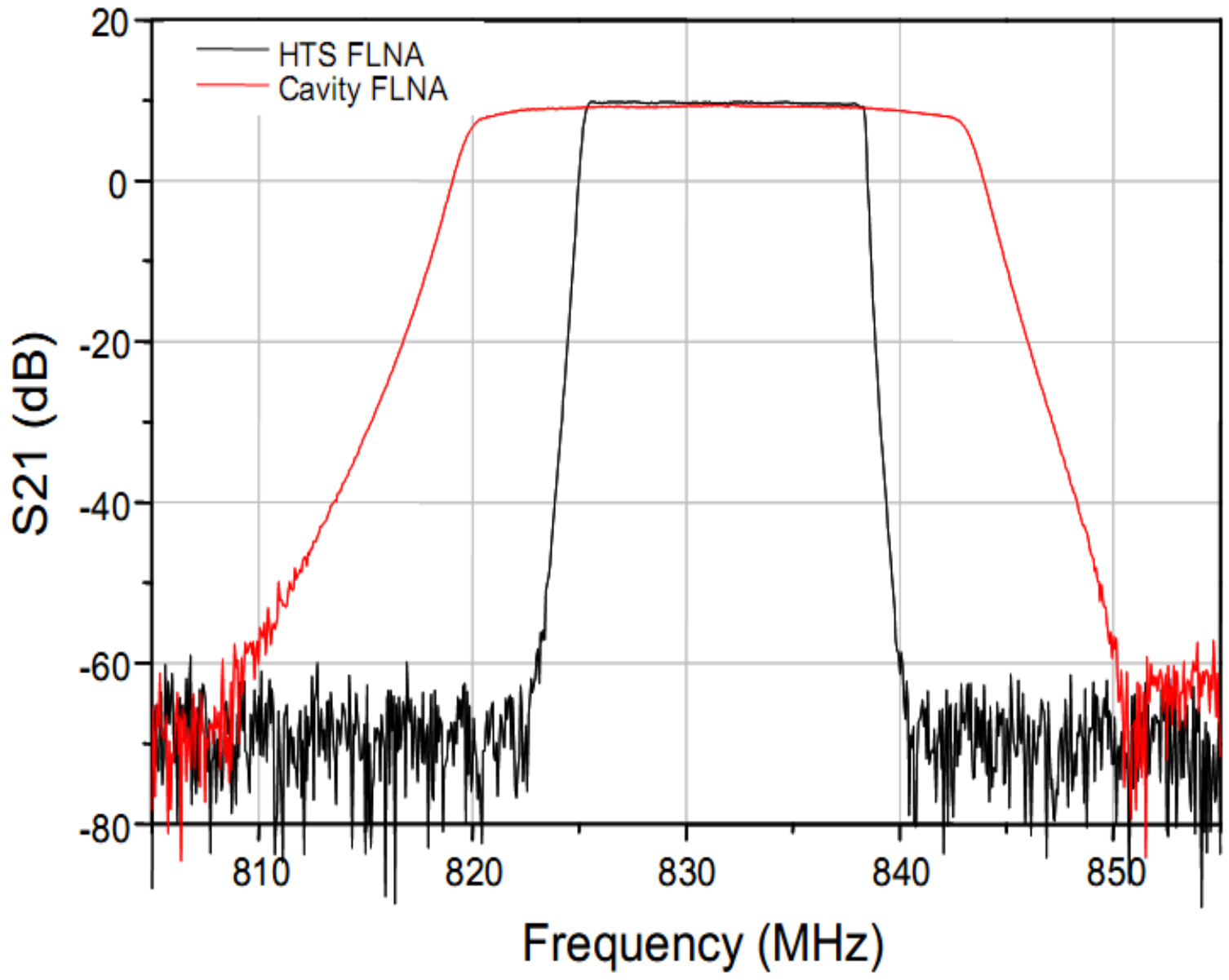

**Fig. 3.** Illustration of advantages by *HTS* microwave filters over metallic microwave filters in *CRFE*, leading to effective use of energy and better spectrum efficiency in wireless communications. Measured $S_{21}$ coefficients of both a 16-pole *HTS CDMA RF* filter, and a metallic *CDMA RF* filter of same operational bandwidth (after [79]).

## 9.3. Miniaturization of Dimensions of High Temperature Superconducting (HTS) Microwave Filters.

Miniaturization of dimensions *HTS* microwave filters to a minimum is an important *R&D* task in Chaloupka [119], Chaloupka, Kaesser [15]. Chaloupka, Kolesov [9] write that the trade-off between miniaturization and a high quality factor *Q* for passive *RF* components can be overcome by means of planar *HTS* circuits. In the frequency range between about 0.1 to 2 *GHz* an even higher degree of miniaturization is obtained, if *HTS* lumped elements (*LE*) instead of distributed *HTS* elements are employed [9]. Chaloupka, Kolesov [9] express an opinion that ***the design of LE circuits is governed by upper limits for the achievable capacitance and inductance values.*** These upper limits can be characterized by means of a characteristic figure of merit $\Lambda$. A significant extension to higher capacitances and higher inductances, and therefore to lower $\Lambda$ values, could be obtained, if an *HTS* multilayer technology for the fabrication of low-loss conductor-insulator-conductor capacitors and spiral inductors were available Chaloupka [119].

Lancaster, Huang, Porch, Avenhaus, Hong, Hung [36] distinguish the three main types of *HTS* microwave filters, discussing the miniature superconducting microwave filters designs:

1. ***Delay line filters***: a component, which uses the coiling or meandering principle for miniaturization is the delay line filter. The impedance of the microstrip line varies along the line length, it is this variation, which causes the filtering action.
2. ***Lumped element filters:*** Lumped elements are by definition much smaller than the wavelength at which they operate. Hence, at high frequencies, where the wavelength is short, filters based on lumped elements will be physically small. It turns out that, where the line widths are limited by the patterning process, the centre frequencies of filters are in the several tens of gigahertz range. At these narrow line widths superconductors are able to help overcome the loss associated with the finite resistance of the conductors.
3. ***Filters based on slow wave structures:*** a transmission line formed of discrete inductors and capacitors.

Tab. 2 shows the designs of three main types of *HTS* microwave filters [36].

| | | |
|---|---|---|
| ***Delay line microwave filter:*** a component, which uses the coiling or meandering principle for miniaturization. | 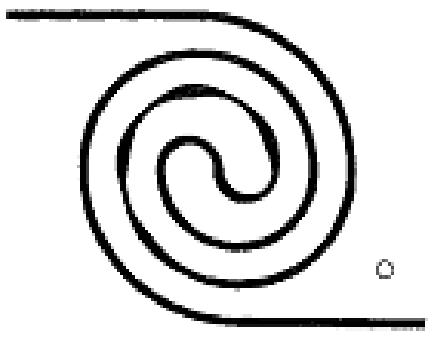<br>**Fig. 4.** *HTS* single transmission line microstrip delay line filter (after [35, 36]).<br>Lancaster, Huang, Porch, Avenhaus, Hong, Hung [36] explain that the filter is designed to give a linear phase response over a 4 *GHz* bandwidth centred on 10 *GHz* in [35]. The filter is made of 0.35 *pm* thick *YBCO* on a 1 *in* square, 300*pm* thick *MgO* substrate with double-sided deposition of the superconductor [35]. Time delays of only several nanoseconds have been demonstrated using this design [35]. | 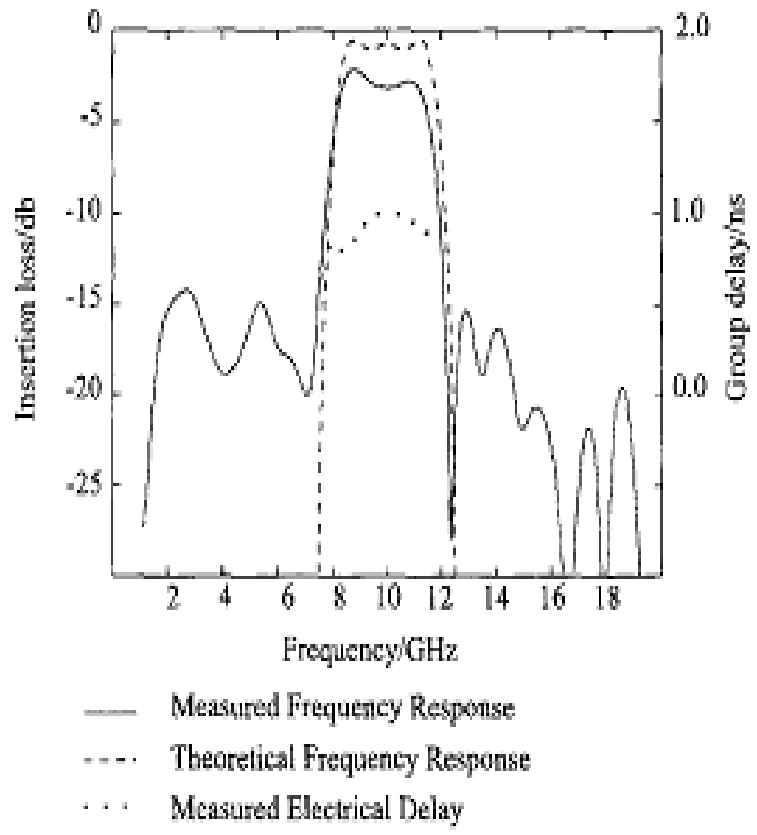<br><br>**Fig. 5.** Frequency response of the *HTS* microstrip linear phase delay line filter shown in Fig. 4 (after [35, 36]). |
| ***Delay line microwave filter:*** a component, which uses the coiling or meandering principle for miniaturization. | 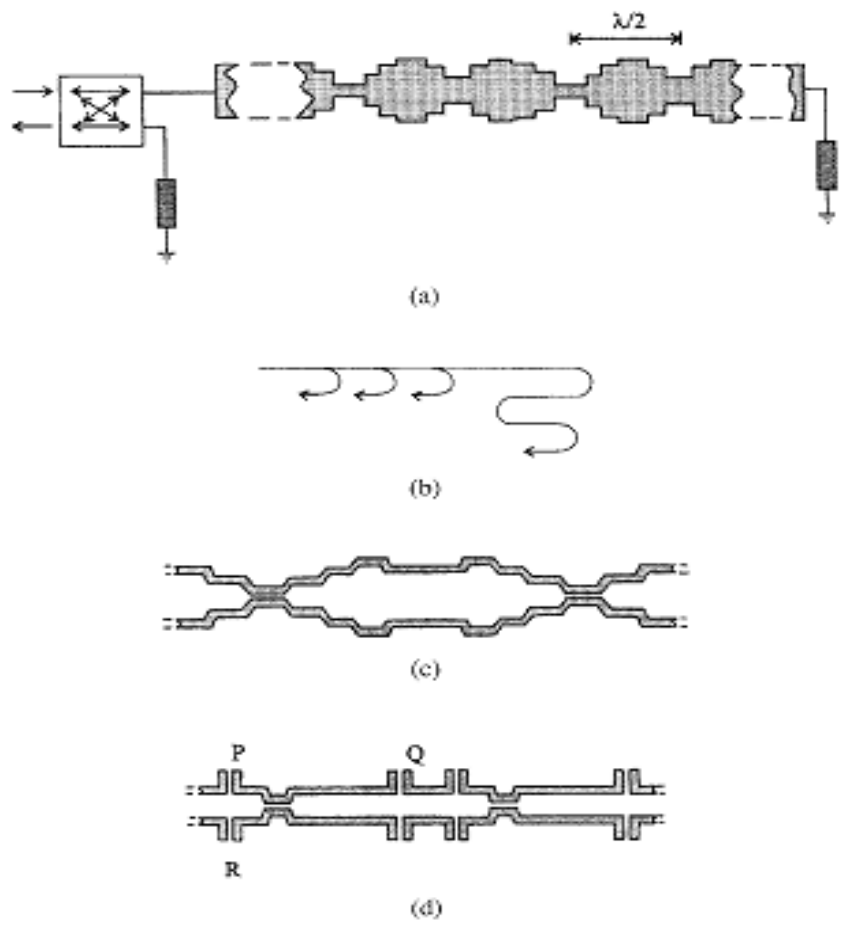<br><br>**Fig. 6.** *HTS* linear phase delay line filter:<br>**a)** The input and output of the filter occur at the same port as shown, and a directional coupler is required in order to separate them. The other end of the line is matched with a 50 $\Omega$ load;<br>**b)** Each impedance step causes a reflection of the forward propagating wave propagation paths, and passbands occur, when these reflections interfere constructively;<br>**c)** Dual delay line filters; d) For narrow band filters resonant sections can be incorporated within the delay line (after [2, 36]). | 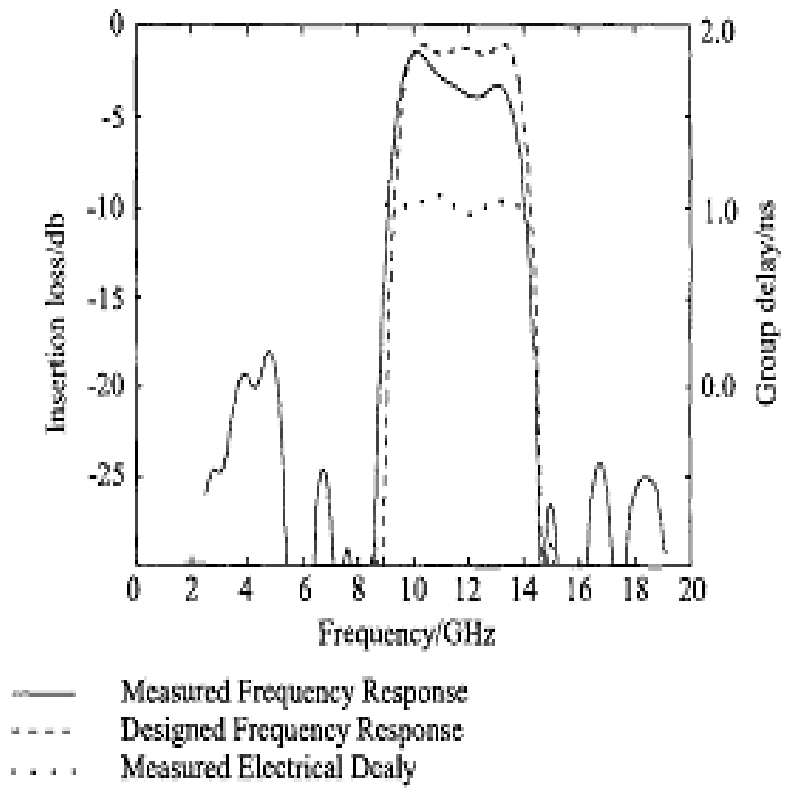<br><br>**Fig. 7.** Frequency response for a *HTS* linear phase delay line filter using a coplanar line delay line (after [2, 36]). |
| ***Lumped element microwave filter:*** a component, which consists of much smaller lumped elements than the operational wavelength. | 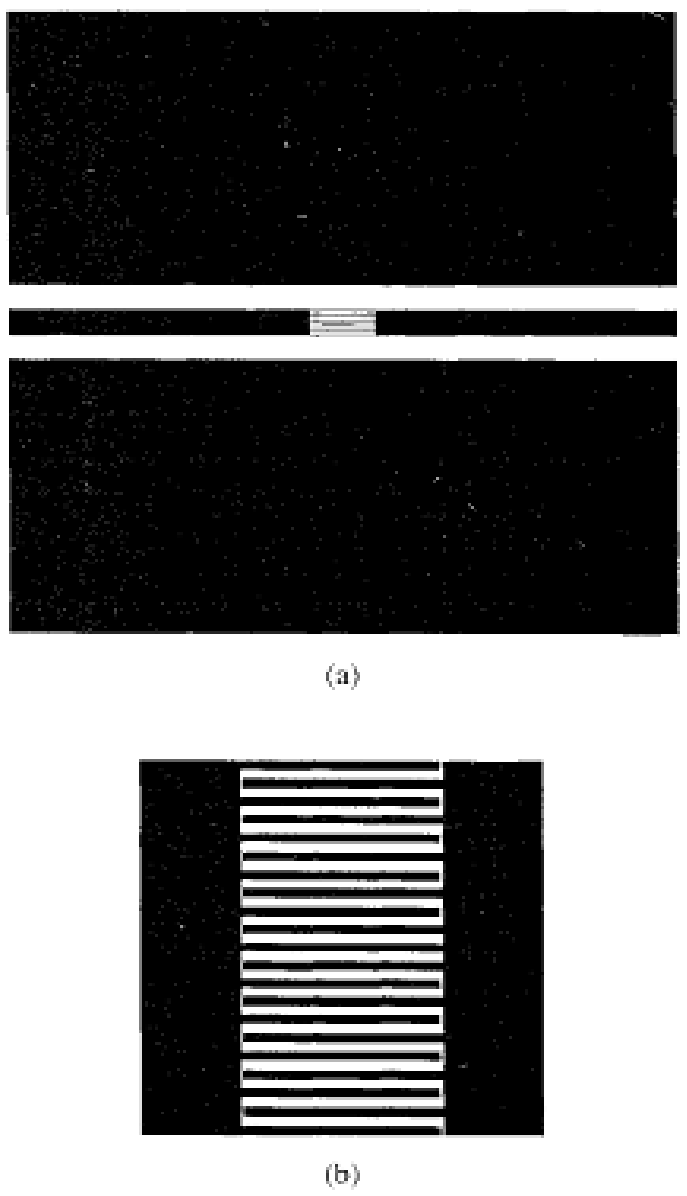<br><br>**Fig. 8. a)** Lumped element bandstop filter with **b)** showing an enlargement of the central resonant section (after [31, 36]).<br>Lancaster, Huang, Porch, Avenhaus, Hong, Hung [36] explain that, in Fig. 8 (b), lumped element resonator consists of a number of interdigital fingers forming a capacitor. The central finger connects both sides of the capacitor acting as an inductor and hence forming a parallel resonant circuit. This circuit can be used to estimate the quality factors available for superconducting lumped element circuits; although different geometry will obviously produce different quality factors. The losses in this circuit come mainly from the inductor because of the high current density on this element. This structure can in fact be very simply made into a band-stop filter [31] as shown in Fig 8 (a), here the element shown in Fig. 8 (b) is placed centrally in a coplanar transmission line. The whole $YBa_2Cu_3O_{7-\delta}$ structure fits on a 1*cm* square *MgO* substrate. In this particular example there are 20 fingers in the interdigital capacitor each with a length of 1 *mm* and width 10 $\mu m$. The coplanar line is 0.41 *mm* wide with a 0.16 *mm* gap between the ground plane and central strip. | 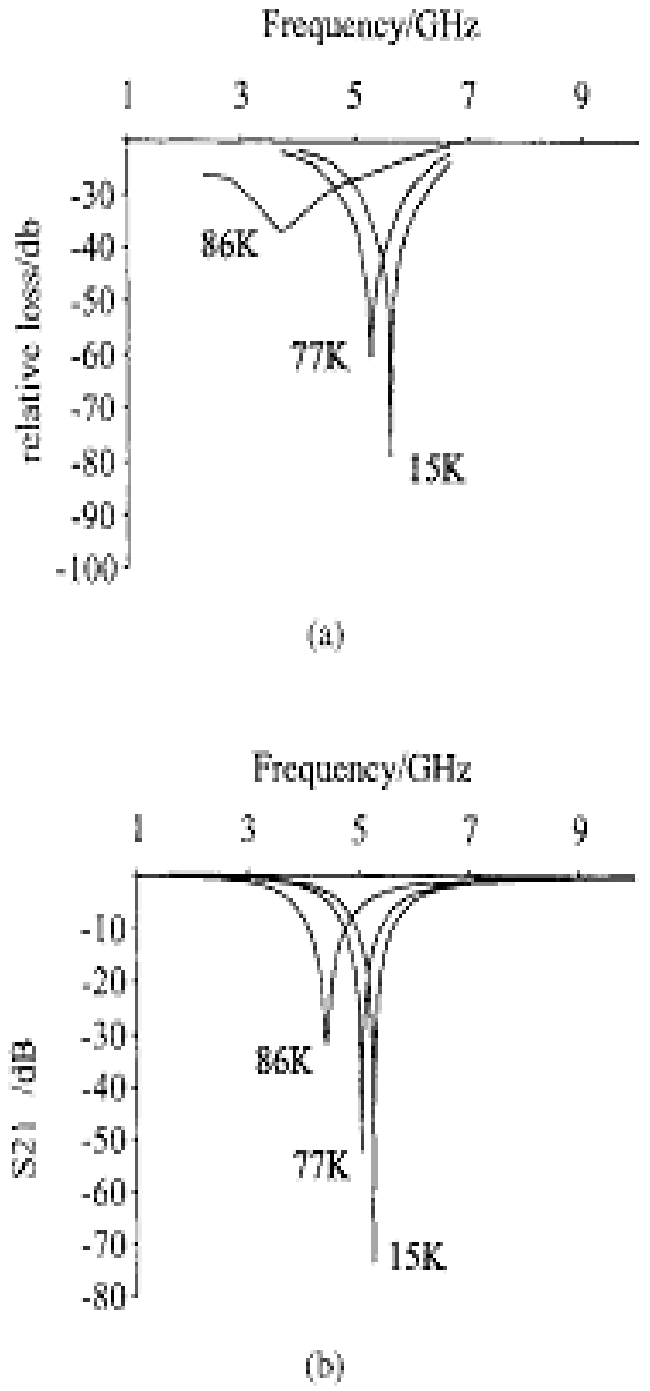<br><br>**Fig. 9.** Bandstop filter performance. **a)** The insertion loss at a number of different temperatures and **b)** the modelled response (after [31, 36]).<br>Lancaster, Huang, Porch, Avenhaus, Hong, Hung [36] explain that the frequency response of the filter is shown in Fig. 9 (a). The bandstop response is centred at about 5 *GHz* and varies substantially with temperature. This temperature variation is due to the field penetration into the strip inductor as the superconducting penetration depth alters. Because the film thickness (0.35 $\mu m$) is of the order of the penetration depth, these changes can not be taken into account by conventional methods; as the volume current distribution in the inductor needs to be calculated. To model the frequency shift the numerical calculation based on the coupling of multiple transmission lines described in references [34], [38, [39] is used. The results of the calculations are shown in Fig. 9 (b). There is a 32% change in resonant frequency between 15 *K* and 86*K*. The stop band performance varies in temperature range, but the maximum stopband rejection is more than 50*dB*. The power dependence of the filter is good with only a 0.03 % change in center frequency for input power varying from -45 *dBm* to -10*dBm*. The corresponding change in insertion loss is 16% at 15*K* and 2% at 77*K*. |

***Microwave filter based on slow wave structures:*** a transmission line formed of discrete inductors and capacitors.

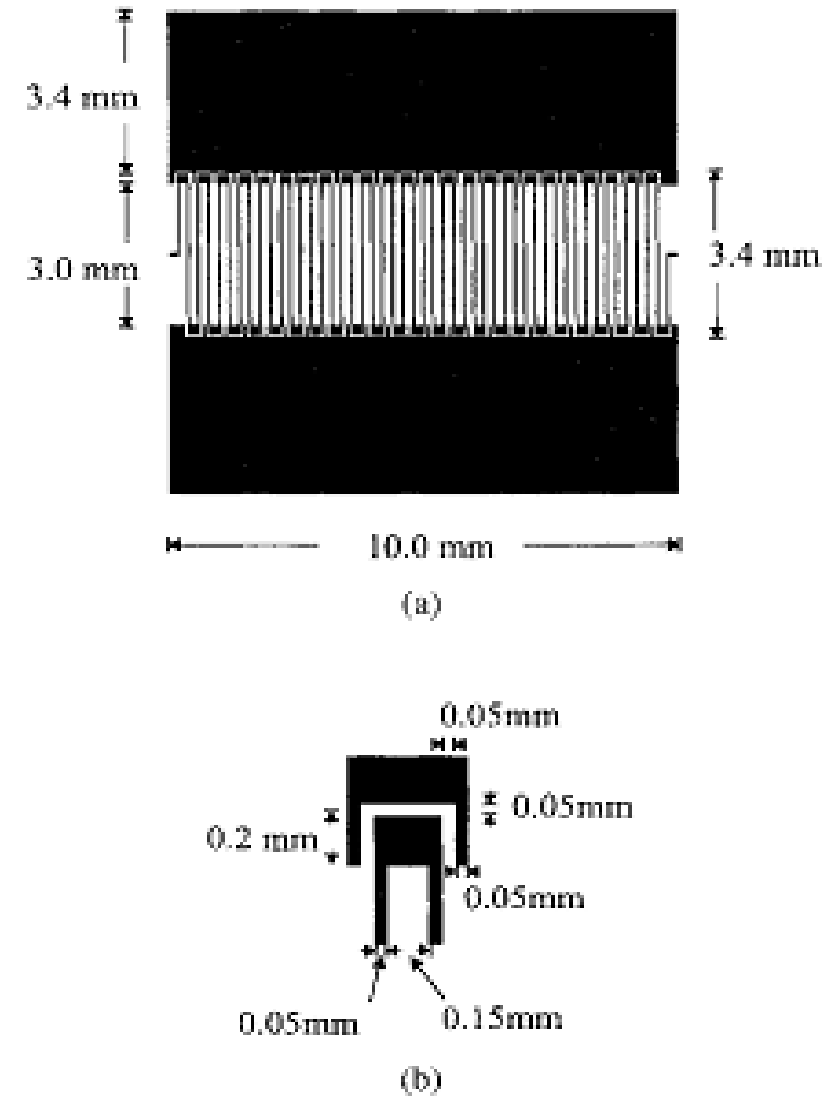

**Fig. 10. a)** A coplanar slow wave resonator.
**b)** the capacitive portion of the structure (after [36]).

Lancaster, Huang, Porch, Avenhaus, Hong, Hung [36] explain that a coplanar slow wave resonator is simply a transmission line formed of discrete inductors and capacitors. The inductors are the narrow vertical tracks and the capacitance is gained from the narrow gap between the coplanar ground plane and the central conductor. Making a transmission line in this way allows independent reduction in the velocity by increasing both the capacitance and inductance per unit length. Fig. 10 (b) is a device in the form of a resonator, convenient for measuring the slowing of the wave by looking at the resonant frequencies. It should be noted that as the wavelength decreases to around and smaller than the unit cell in the slow wave line, it no longer behaves like a transmission line [36].

The *HTS* device shown in Fig. 10 is made of thin film $YBa_2Cu_3O_{7-\delta}$ on a 1 *cm* square *MgO* substrate and deposited by laser ablation. The length of the resonator is about 10 *mm*. The fundamental resonant frequency was 1 *GHz* at 77 *K*, providing a velocity reduction factor of 15 over the free space velocity.

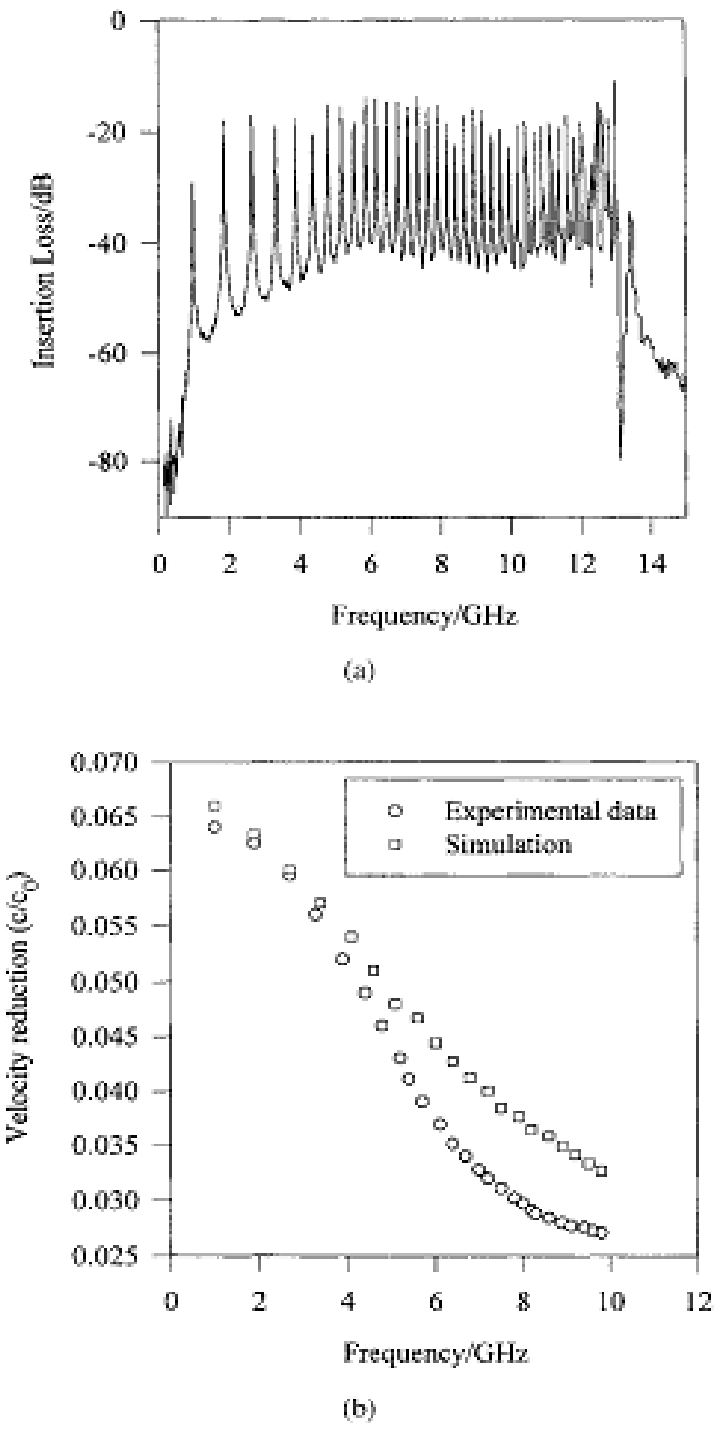

**Fig. 11.** The performance of the slow wave resonator in Fig. 8: **a)** Insertion loss as a function of frequency. **b)** Velocity reduction as a function of frequency (after [36]).

Lancaster, Huang, Porch, Avenhaus, Hong, Hung [36] write that the Fig. 11(a) shows the frequency response of the resonator, a large number of harmonic resonances can be seen, they are dispersive as seen by the non-equal frequency difference between each resonance. A cut-off frequency occurs when the wavelength of operation is equal to the length of one period in the structure. The dispersion is complex and is governed not only by the basic response of the inductive and capacitive sections but also by these elements becoming larger in terms of wave length as the frequency increases. In addition coupling between these elements plays a role. Fig. 11(b) shows this dispersion in the form of the velocity reduction over free space velocity as a function of frequency. In Fig. 11(b) the values of reduced velocity with this coplanar structure are quite large enabling a large reduction in filter size, if it is used in a filter structure. Fig. 11(b) also shows a numerical computation of this velocity, in this case no account has been taken of internal inductance effects [36].

<table>
<tr>
<td>Miniature slow wave microwave resonators</td>
<td>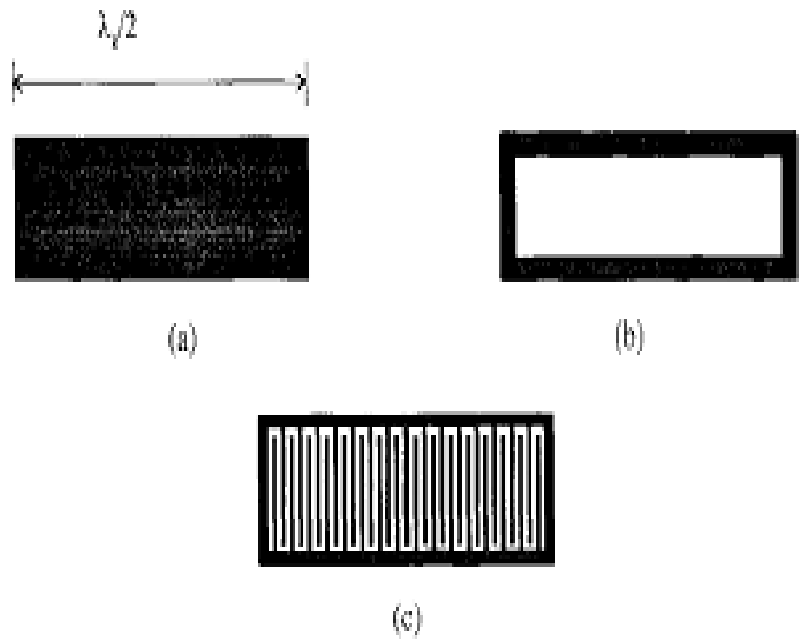<br><br>Fig. 12. (a) Conventional half wavelength microstrip resonator. (b) Modified resonator forming a microstrip square loop resonator. (c) Capacitively loaded loop resonator forming a slow wave structure (after [36]).<br>Lancaster, Huang, Porch, Avenhaus, Hong, Hung [36] explain that the Fig. 12 shows how a similar effect can be achieved using microstrip. Consider the standard microstrip resonator in Fig. 12(a). The effect of removing the central portion to produce the loop of Fig. 12(b) is only small. It effectively turns the standard patch into a loop resonator. The frequency reduction is small as the width of the patch is small compared with its length. To reduce the frequency of the resonator, the loop can be loaded with capacitive fingers as shown in Fig. 12(c). The velocity reduction on this type of transmission line is controlled by the number of fingers within the loop. Copper resonators of this type have shown a 25% reduction in frequency around 4GHz with 31 fingers in the loop [32]. Coupled resonators have also been demonstrated [33] showing that conventional design techniques can be used to design coupled slow wave lines. A superconducting resonator of this type with outside dimensions 4 by 1 mm and 195 fingers each of 10 pm width and 890 km long resonates at 10.53 GHz with a Q in excess of 1200 at 77 K. This represents about a 25% reduction in size over the conventional microstrip resonator [36].</td>
<td></td>
</tr>
</table>

| | | |
|---|---|---|
| ***Compact slow wave microwave filters*** | 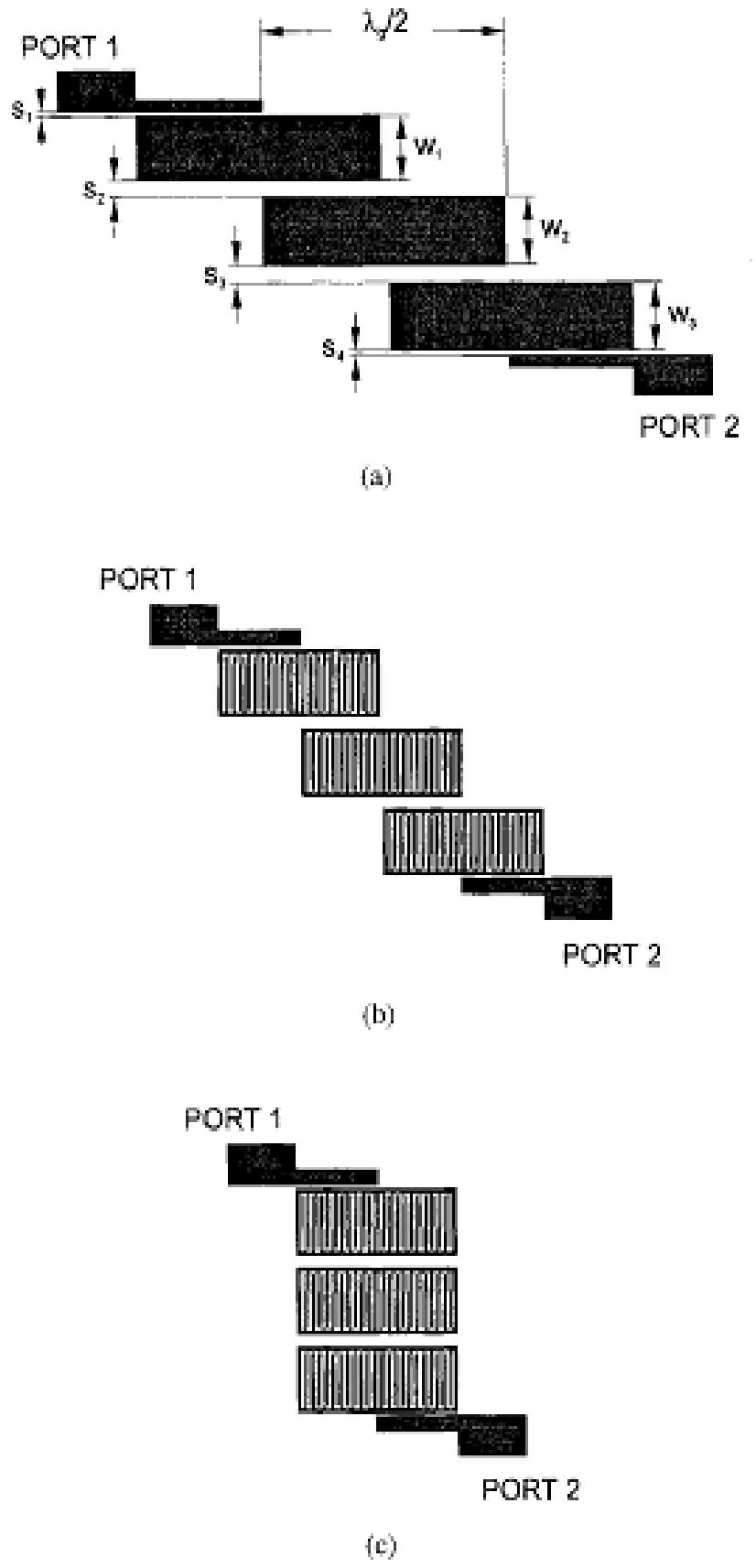<br>**Fig. 13.** (a) Edge coupled microstrip filter with conventional resonators. (b) Compact filter with slow wave line resonators. (c) Alternative filter (after [36]).<br>Lancaster, Huang, Porch, Avenhaus, Hong, Hung [36] elaborate that the Fig. 13 shows two filters based on the microstrip capacitively loaded loop. One with the standard edge coupled design in Fig. 13(b), and the other one in line version of the same filter in Fig. 13(c). A conventional microstrip filter is shown to the same scale in Fig. 13(a) for comparison purposes. These filters are made of copper to demonstrate the principle and are all centered on a frequency of 3.4 *GHz* [36]. | 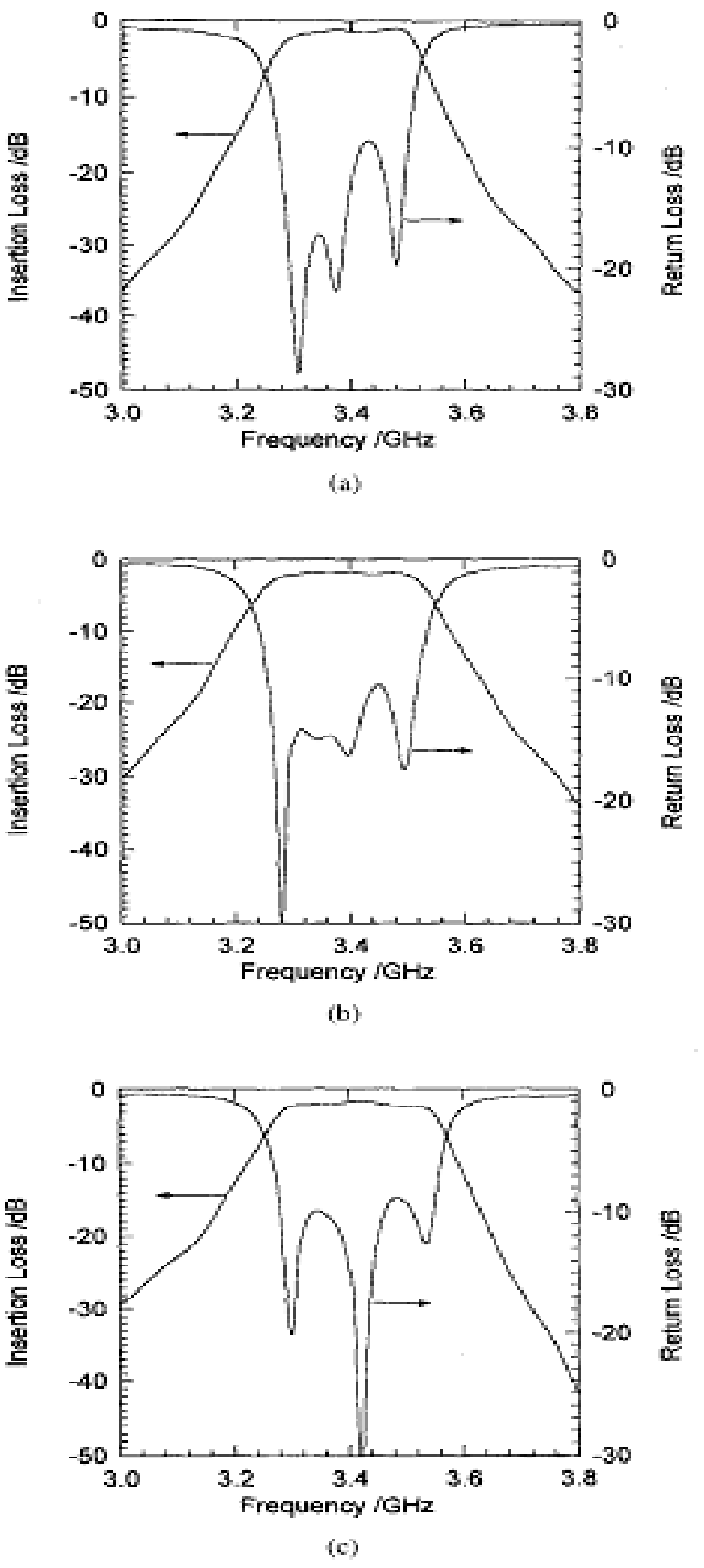<br>**Fig. 14.** Measured frequency responses for the microstrip filters shown in Fig. 13. (a) Conventional filter. (b) Compact filter. (c) In-line compact filter (after [36])<br>Lancaster, Huang, Porch, Avenhaus, Hong, Hung [36] write that the frequency response of all three filters is shown in Fig. 14. As can be seen all demonstrate good low loss performance with excellent return loss [36]. |

**Tab. 2.** Designs of three main types of *HTS* microwave filters:

1. Delay line microwave filters;

2. Lumped element microwave filters;

3. Microwave filters based on slow wave structures (after [36]).

## 9.4. Review on the State-of-the-Art High Temperature Superconducting (HTS) Microstrip Filters Designs.

In Australia, the *CSIRO* has strong telecommunications and radio astronomy research programs, which led to the development of *HTS* microwave filters for use in radio telescopes and mobile communications in Fig. 15 [67].

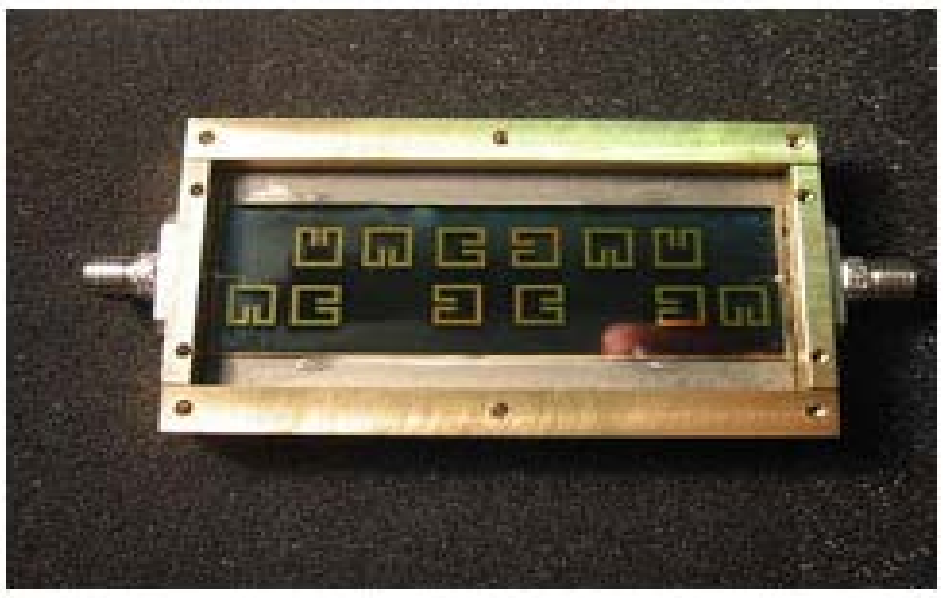

**Fig. 15**. A photograph of a microwave filter made at *CSIRO* in Australia [67].

At the *Microwave and Electronic Material Research Group*, *Electrical and Computer Engineering Department*, *James Cook University*, Townsville, Australia, the innovative research, led by Prof. Janina E. Mazierska (Personal Chair), is mainly focused on the accurate characterisation of physical properties of *HTS* thin films in *HTS* microwave filters in wireless communications in Fig. 16 [14, 72, 78, 79].

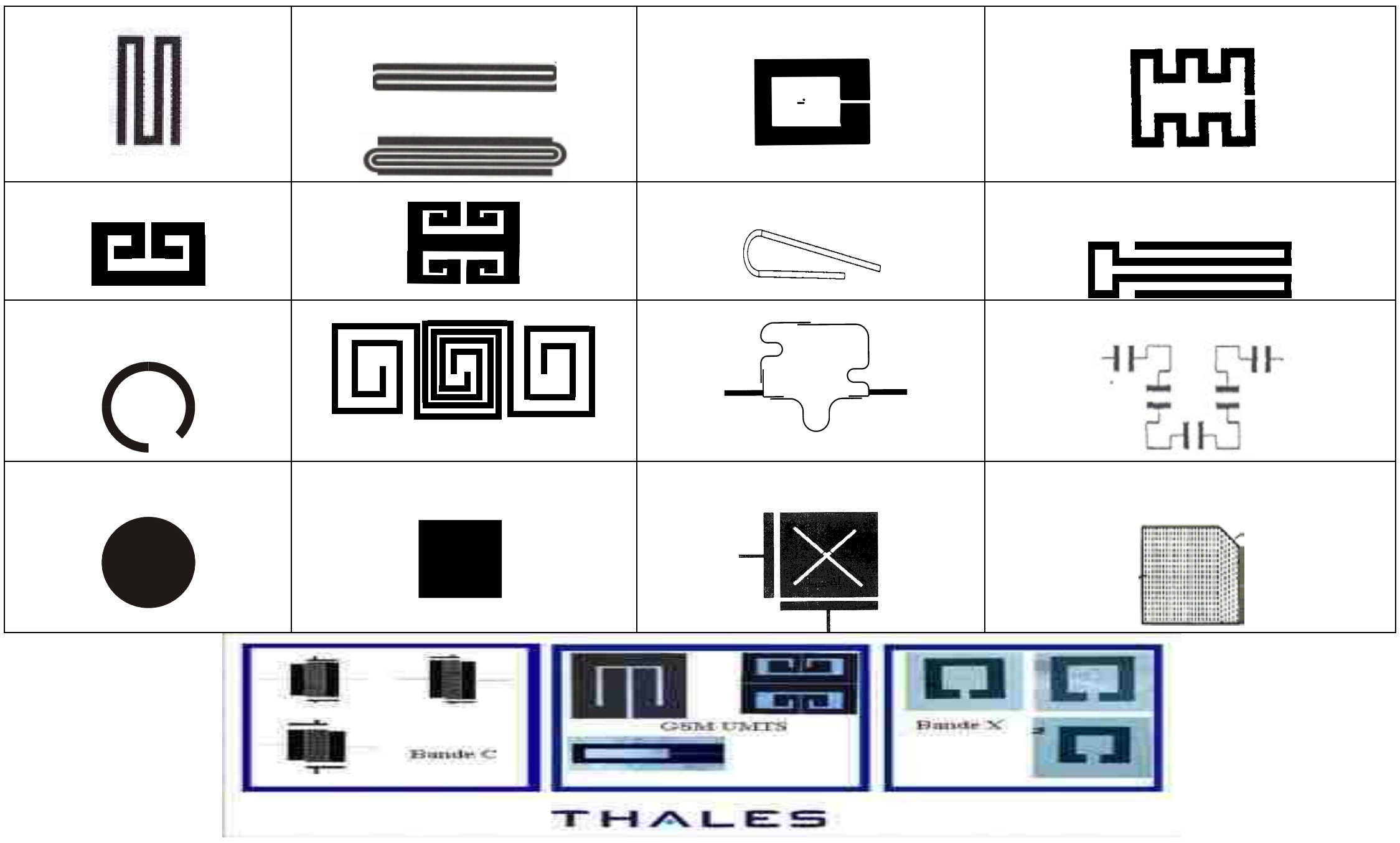

**Fig. 16.** Examples of *HTS* microwave resonators layouts for use in *HTS* microwave filters (after [14, 101]).

In Tab. 3, the author of dissertation shows the schematic of layouts of *HTS* microwave resonators and multi pole filters with their frequency responses.

| Publication Reference | HTS Microwave Filters Layout Geometries | HTS Microwave Filters Frequency Responses |
|---|---|---|
| W G Lyons, R S Withers, in Fossheim K ed, Superconducting Technology. 10 Case Studies, *World Scientific Publishing Pte Ltd*, Singapore, 1991. | 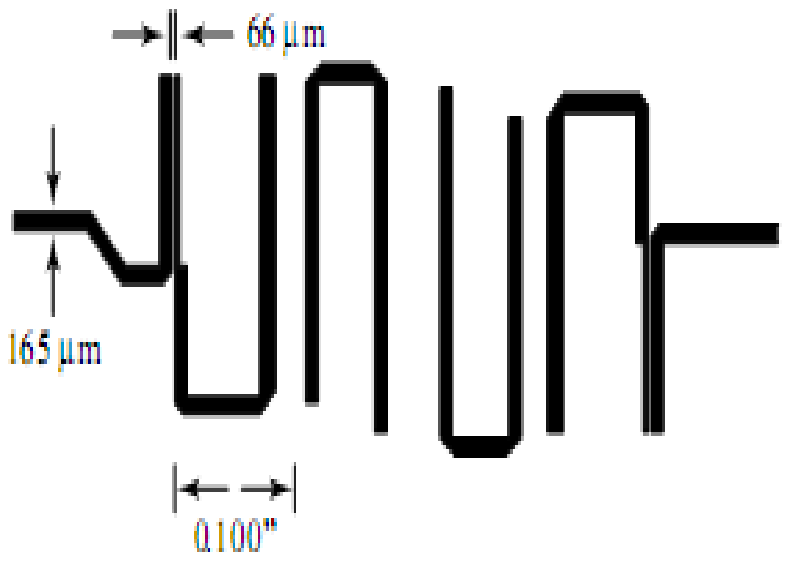 **Fig.17.** Layout of HTS microstrip filter (after[1]). | 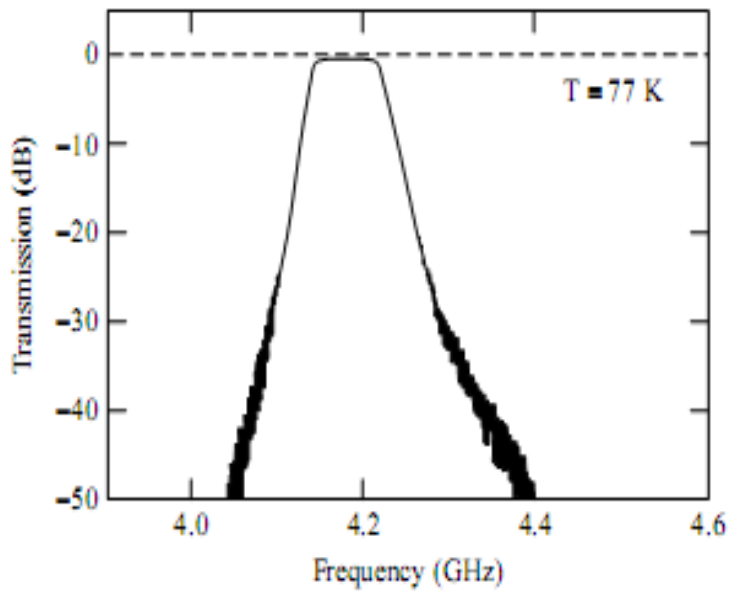 **Fig.18.** Frequency response (after [1]). |
| Sh Ohshima, S Takeuchi, M Osaka, H Kinouchi, S Ono, J F Lee, A Saito, Examination of the Resonator Structure for a Superconducting Transmitting Filter, *EUCAS2007*, pp. 1-6, 2007. | 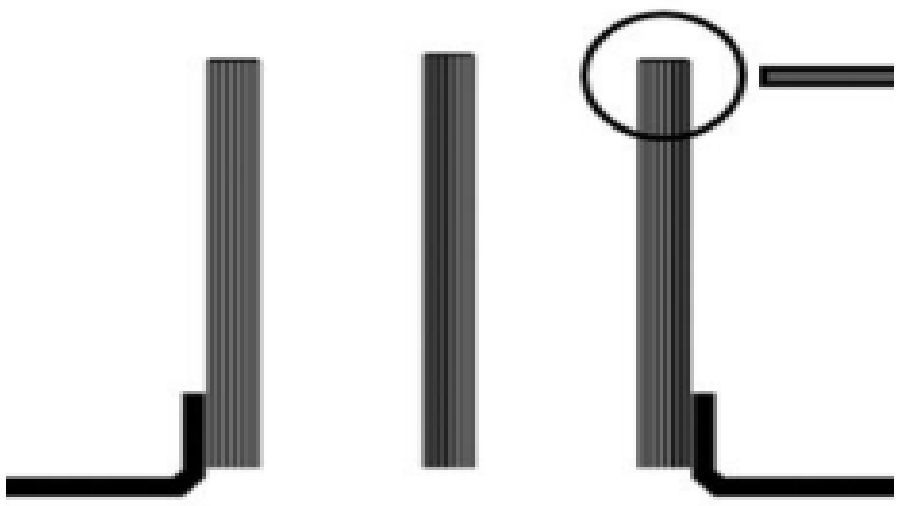 **Fig. 19.** Layout of HTS microstrip filter (after[110]). | 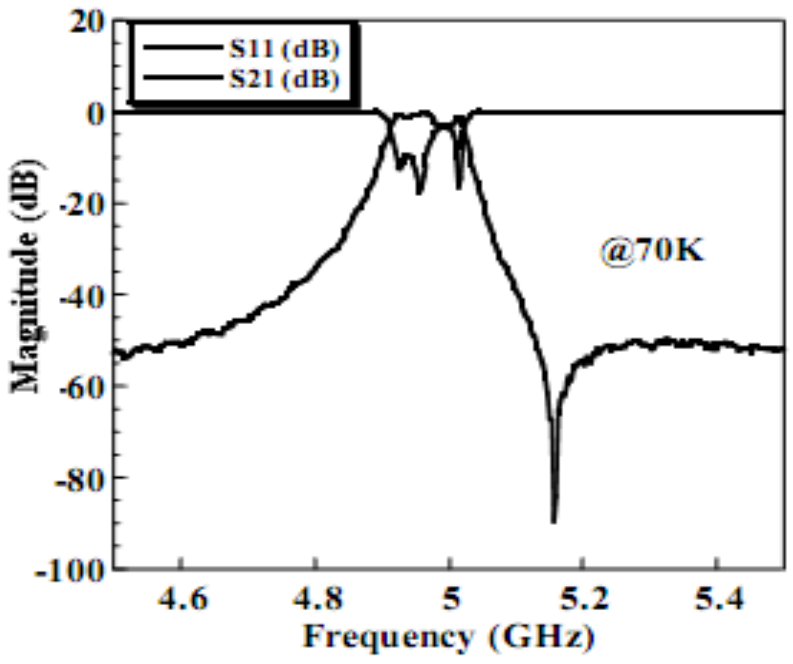 **Fig. 20.** Frequency response (after[110]). |
| N Sekiya, K Yamamoto, S Kakio, A Saito, S Ohshima, Miniaturized sharp-cutoff transmit HTS filter with a stripline structure, *Physica C*, doi:10.1016/j. physc.2011.05. 162, 2011. | 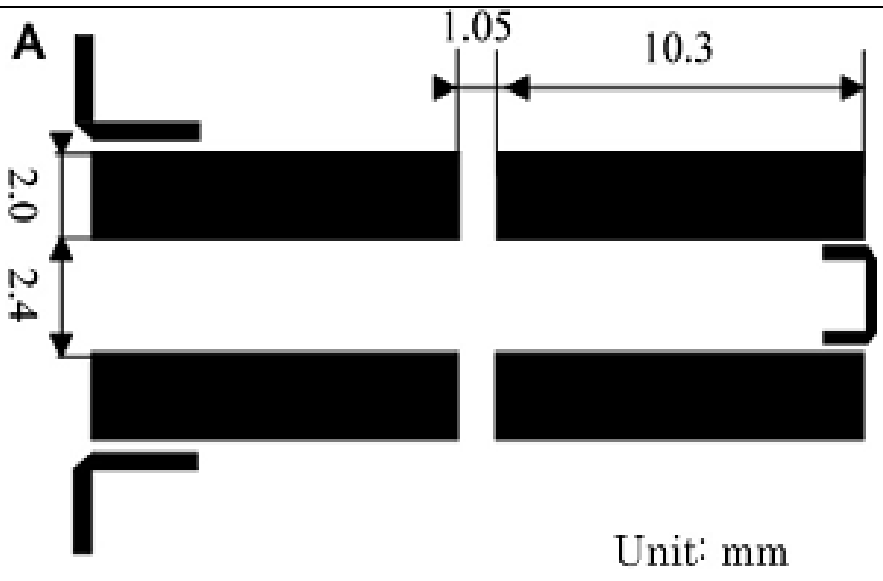 **Fig. 21.** Layout of HTS stripline filter (after[111]). | 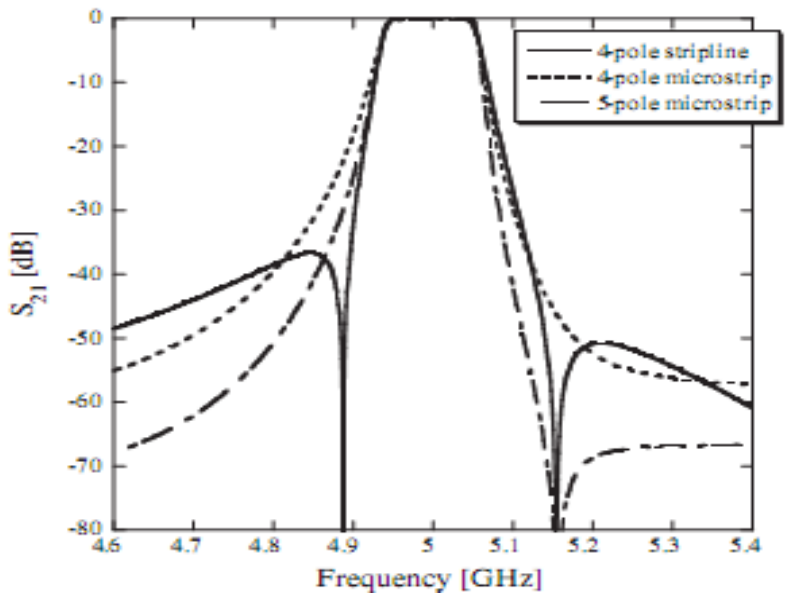 **Fig. 22.** Frequency response (after[111]). |
| N Sekiya, N Imai, S Kakio, A Saito, S Ohshima, Miniaturized transmit dual-mode HTS patch filter with stripline structure, *Physica C*, doi:10.1016/j. physc.2011.05. 164, 2011. | 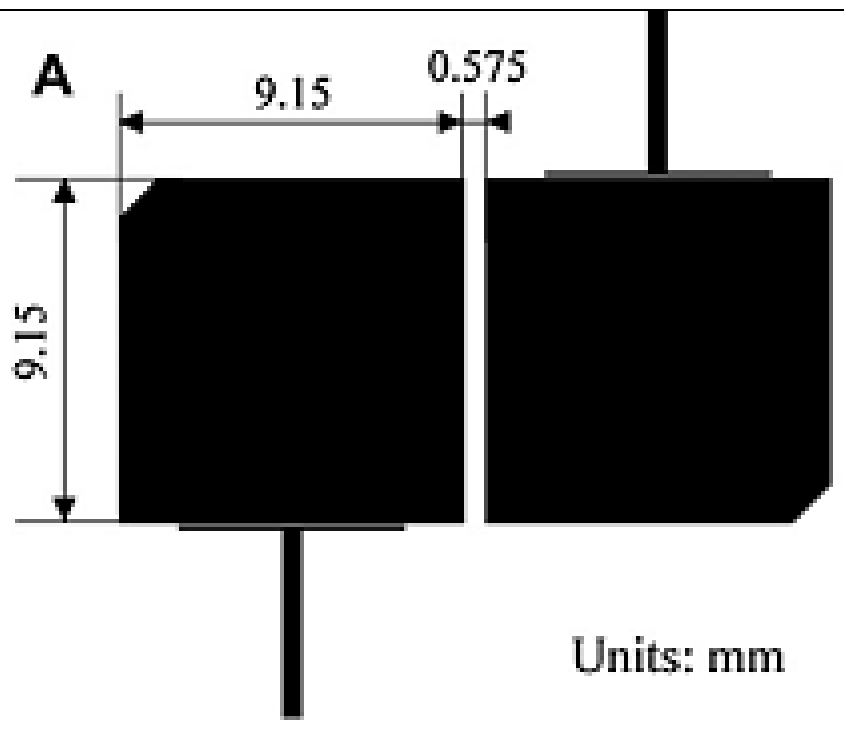 **Fig. 23.** Layout of HTS stripline filter (after[112]). | 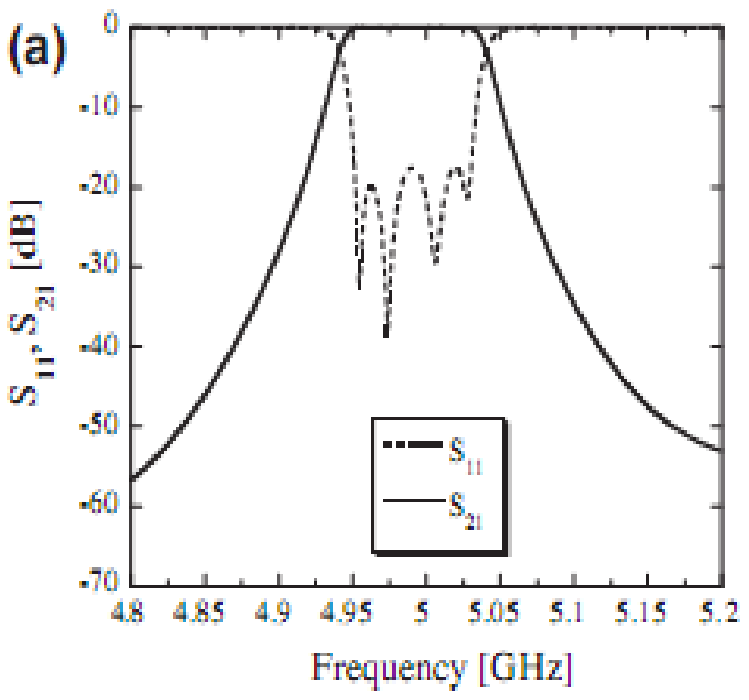 **Fig. 24.** Frequency response (after[112]). |

<table>
<tr>
<td>H Harada, N Sekiya, S Kakio, S Ohshima, Center frequency and bandwidth tunable HTS filter, Physica C, doi:10.1016/j. physc.2011.05. 165, 2011.</td>
<td>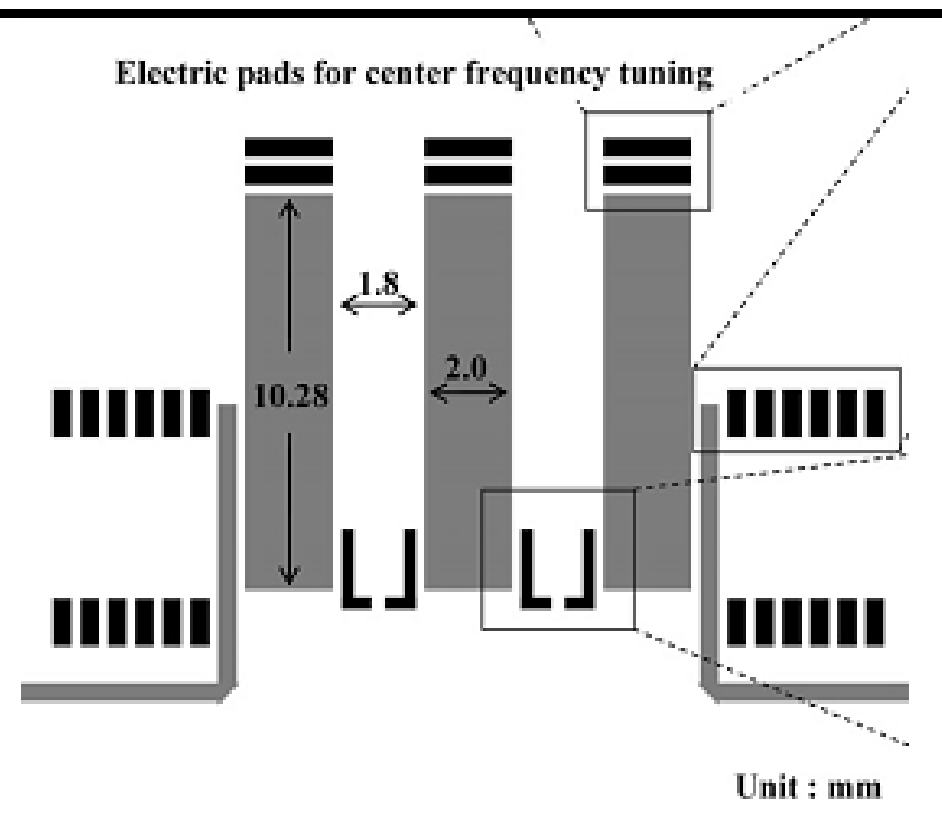<br>Fig. 25. Layout of HTS microstrip filter (after[113]).</td>
<td>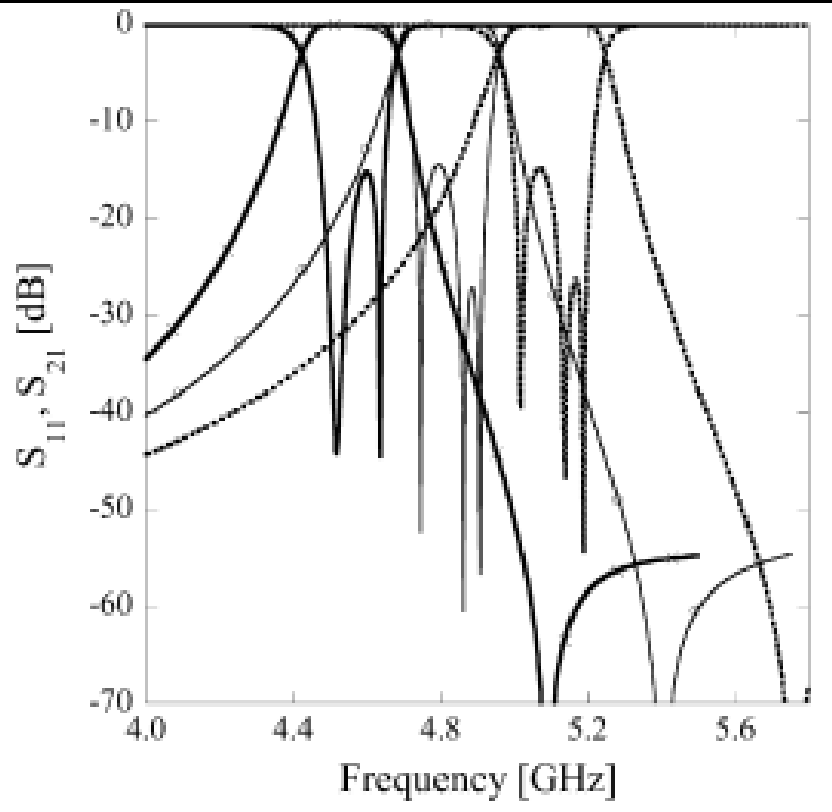<br>Fig. 26. Frequency response (after[113]).</td>
</tr>
<tr>
<td>A P Knack, Design and Implementation of HTS Technology for Cellular Base Stations: An Investigation into Improving Cellular Communication, Ph.D. degree thesis, Massey University, New Zealand and James Cook University, Australia, 2006.</td>
<td>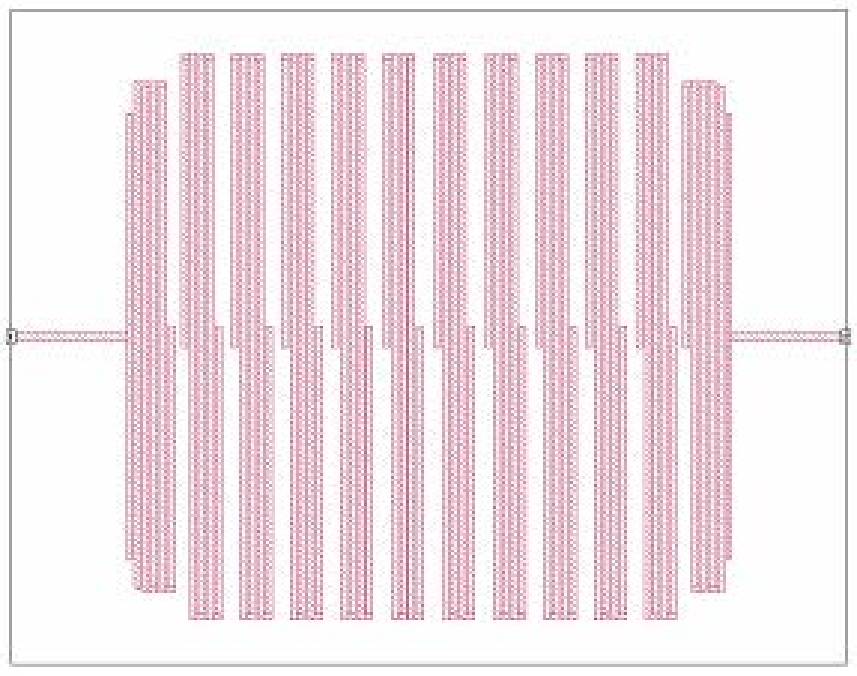<br>Fig.27. Layout of HTS microstrip filter (after[79]).</td>
<td>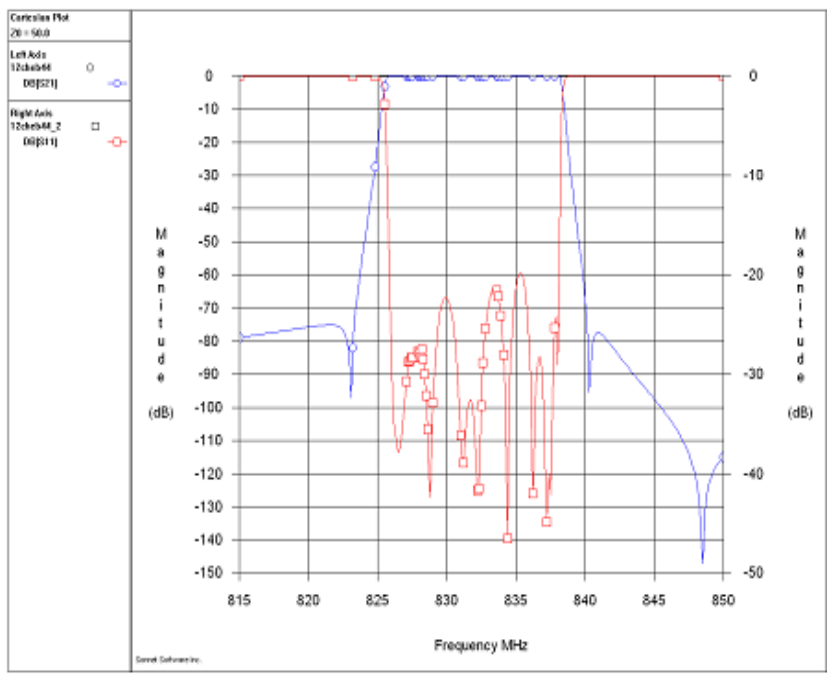<br>Fig.28. Frequency response (after [79]).</td>
</tr>
<tr>
<td>A P Knack, Design and Implementation of HTS Technology for Cellular Base Stations: An Investigation into Improving Cellular Communication, Ph.D. degree thesis, Massey University, New Zealand and James Cook University, Australia, 2006.</td>
<td>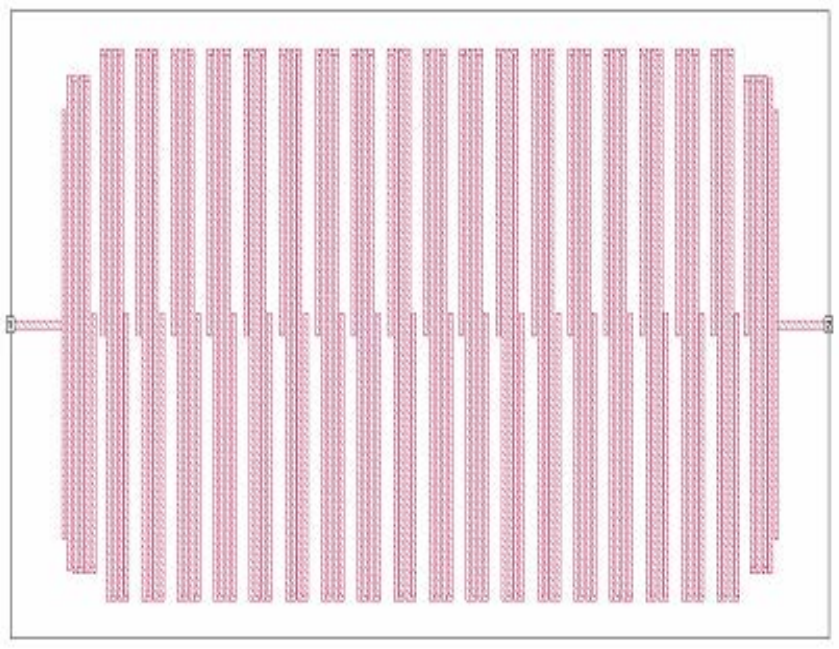<br>Fig.29. Layout of HTS microstrip filter (after[79]).</td>
<td>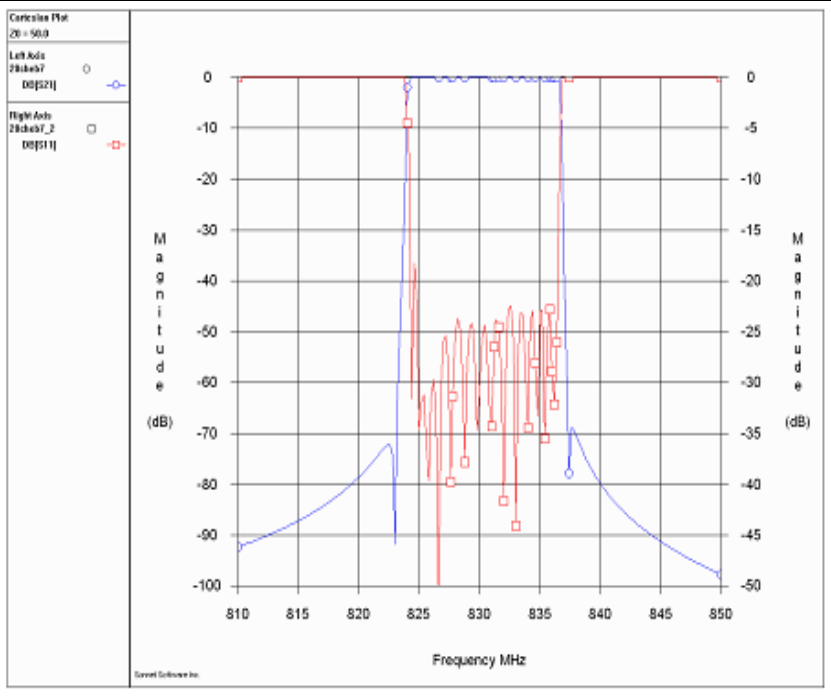<br>Fig.30. Frequency response (after [79]).</td>
</tr>
<tr>
<td>A P Knack, Design and Implementation of HTS Technology for Cellular Base Stations: An Investigation into Improving Cellular Communication, Ph.D. degree thesis, Massey University, New Zealand and James Cook University, Australia, 2006.</td>
<td>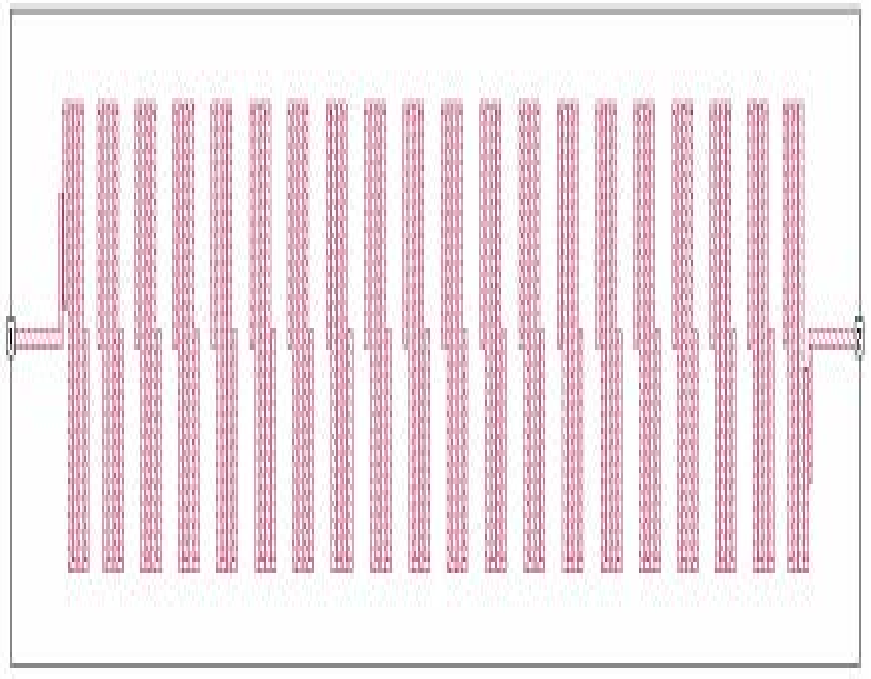<br>Fig.31. Layout of HTS microstrip filter (after[79]).</td>
<td>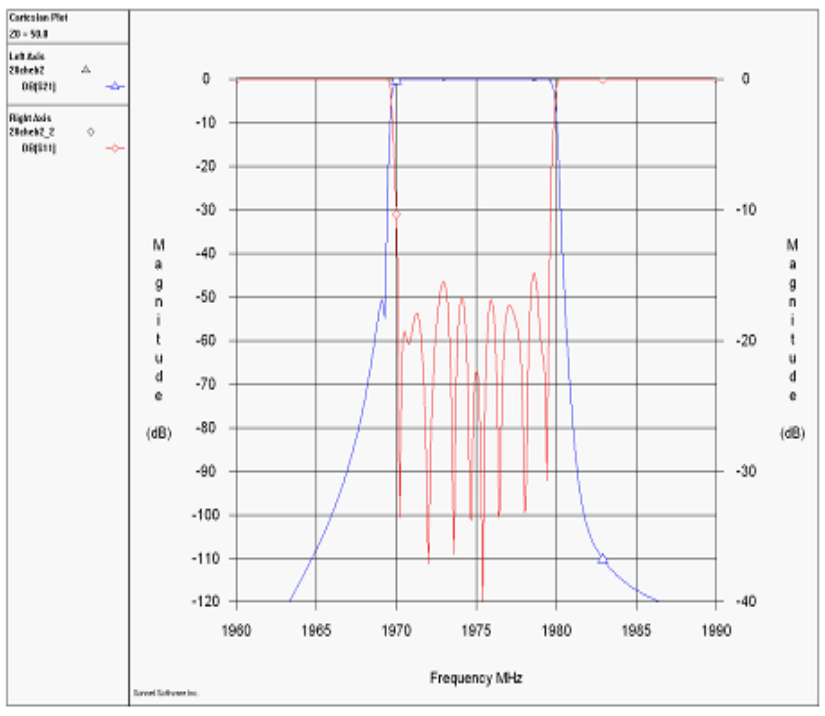<br>Fig.32. Frequency response (after [79]).</td>
</tr>
</table>

A P Knack, Design and Implementation of HTS Technology for Cellular Base Stations: An Investigation into Improving Cellular Communication, Ph.D. degree *thesis, Massey University, New Zealand and James Cook University, Australia*, 2006.

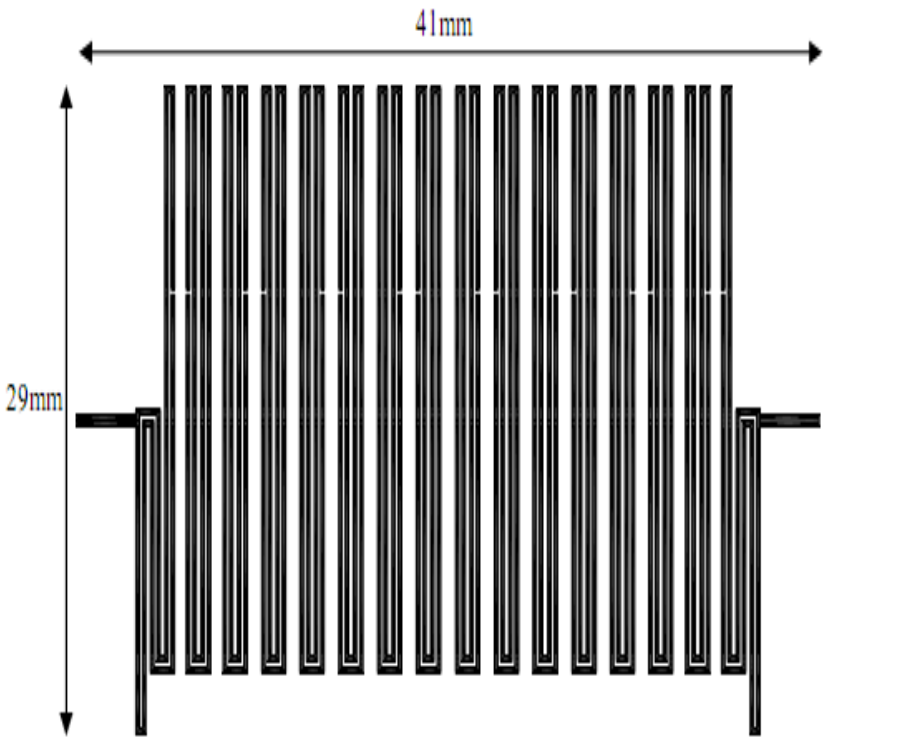

**Fig.33.** Layout of HTS microstrip filter (after[79]).

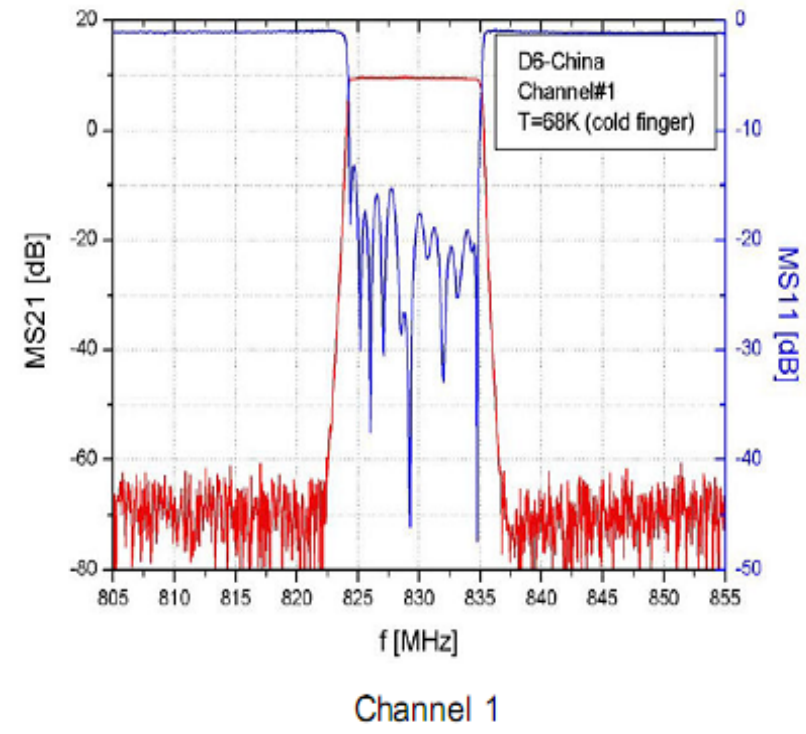

**Fig.34.** Frequency response (after[79]).

M V Jacob, J Mazierska, A P Knack, S Takeuchi, Miniaturized 10Pole Superconducting Filter on MgO Substrate for Mobile Communication, *Proceedings o IEEE Region 10 TENCON 2004*, Chiang Mai, Thailand, *IEEE Press*, ISBN: 0-7803-8561-6, pp. 554 – 557, 2004.

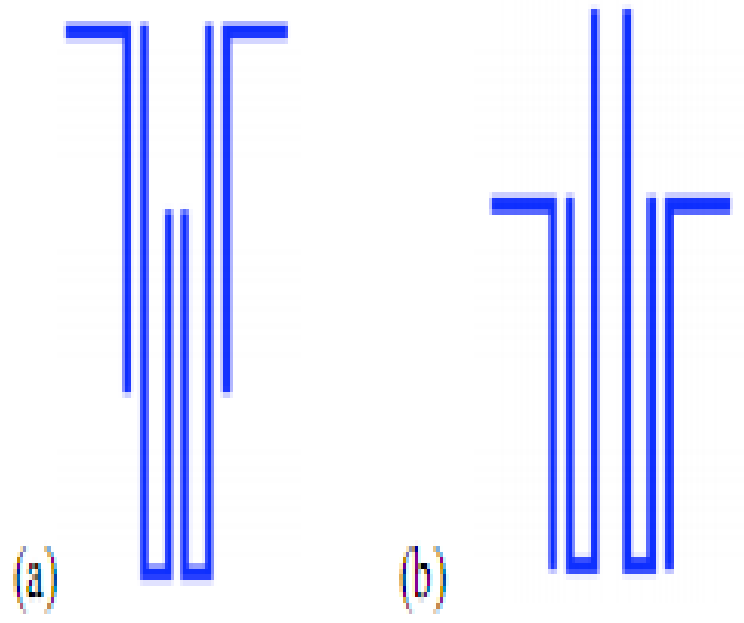

**Fig.35.** Layout of HTS microstrip filter (after[72]).

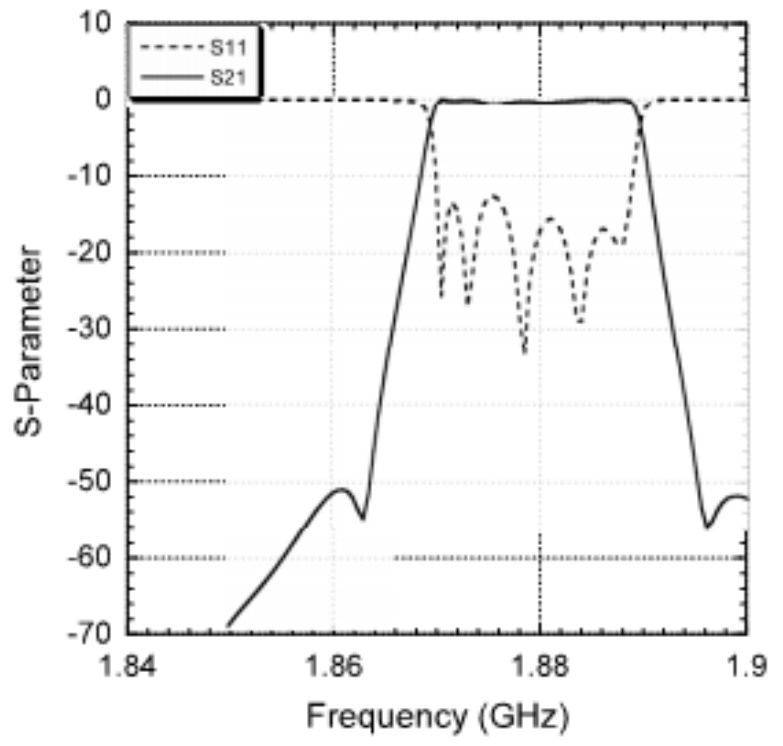

**Fig.36.** Frequency response (after [72]).

S Pal, Novel designs of high-temperature superconducting bandpass filters for future digital communication services, J. Indian Inst. Science, vol. **86**, pp. 257–264, 2006.

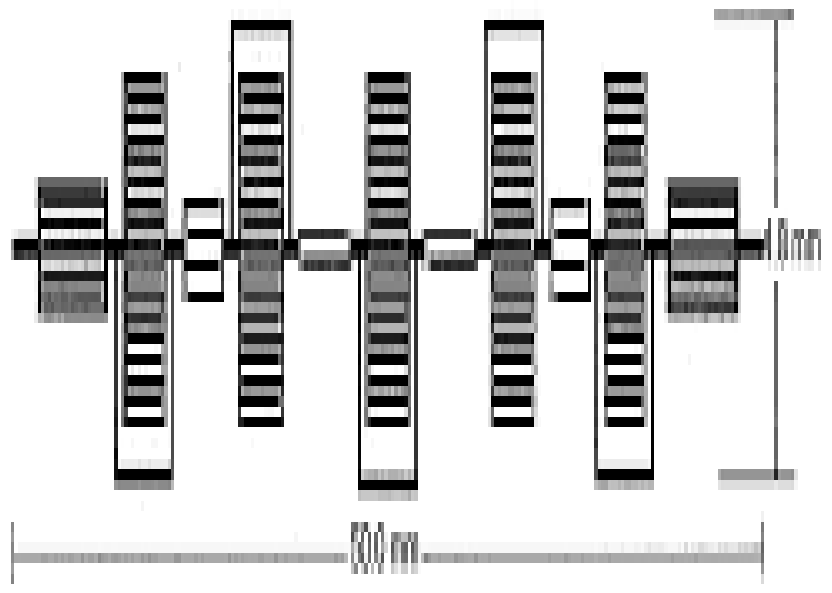

**Fig.37.** Layout of HTS microstrip filter (after[58]).

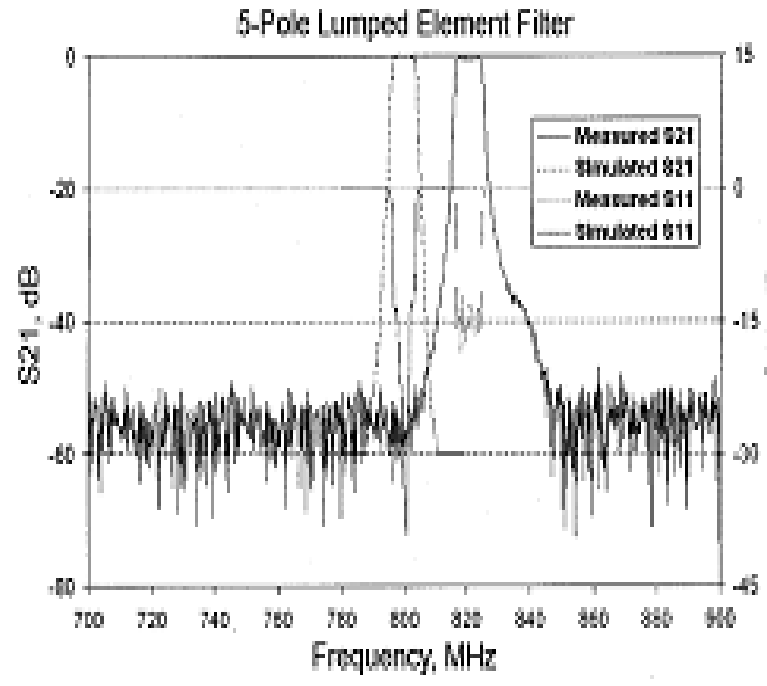

**Fig.38.** Frequency response (after [58]).

S Pal, Novel designs of high-temperature superconducting bandpass filters for future digital communication services, J. Indian Inst. Science, vol. 86, pp. 257–264, 2006.

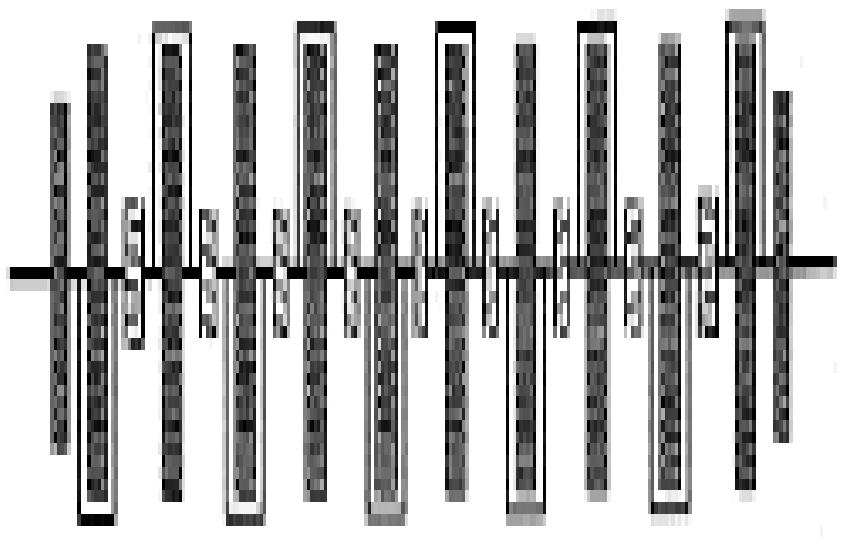

**Fig.39.** Layout of HTS microstrip filter (after[58]).

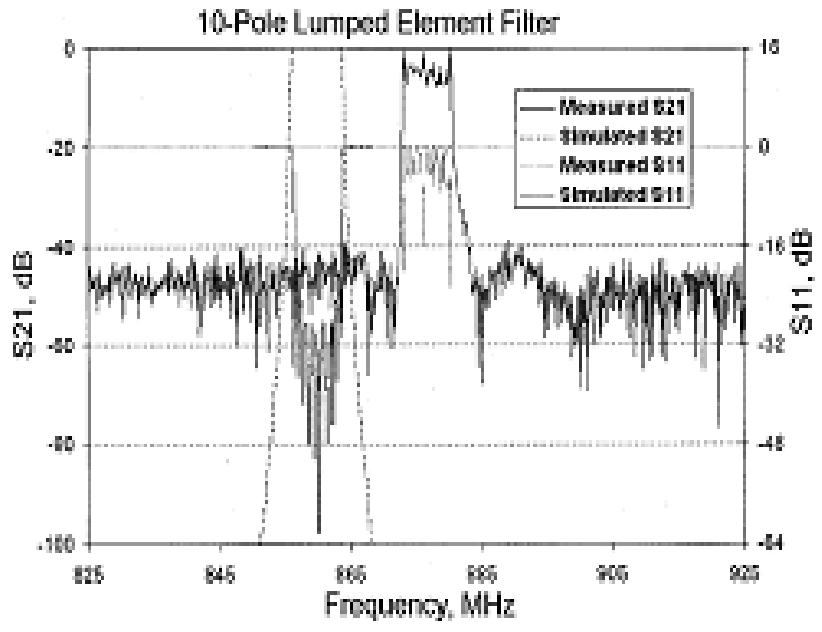

**Fig.40.** Frequency response (after [58]).

<table>
<tr>
<td>S Pal, Novel designs of high-temperature superconducting bandpass filters for future digital communication services, J. Indian Inst. Science, vol. 86, pp. 257–264, 2006.</td>
<td>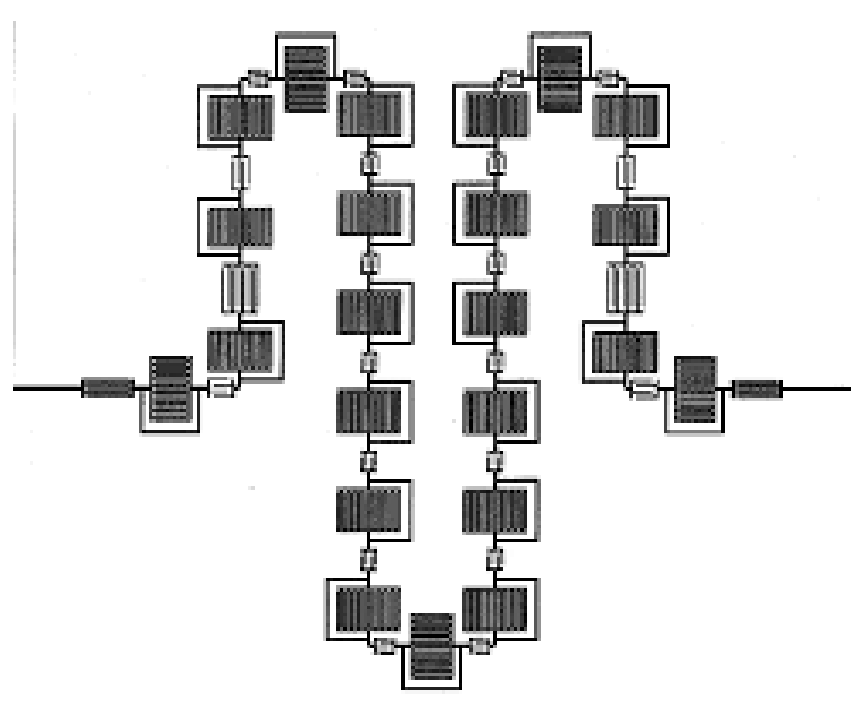<br>Fig.41. Layout of HTS microstrip filter (after[58]).</td>
<td>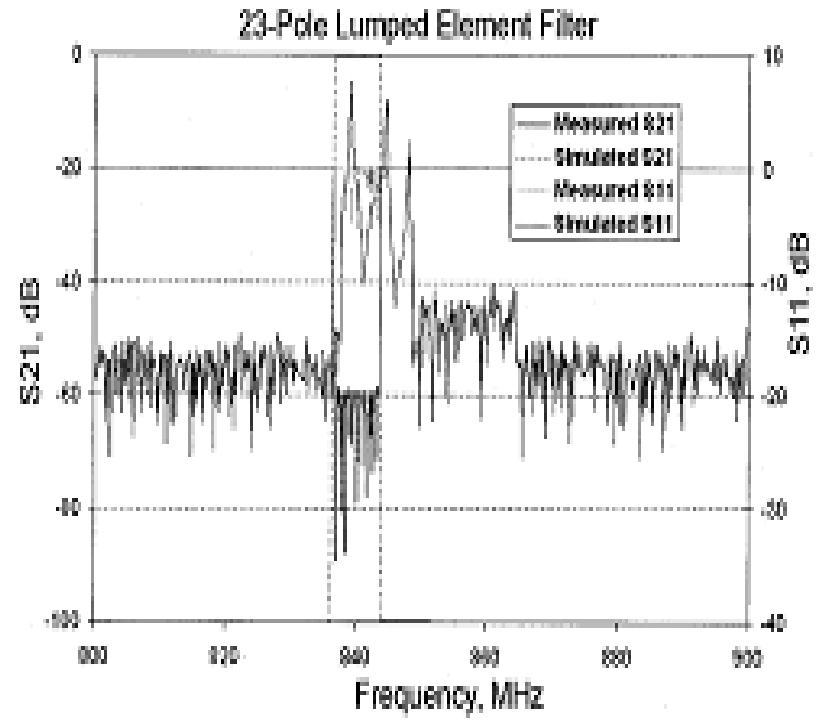<br><br>Fig.42. Frequency response (after [58]).</td>
</tr>
<tr>
<td>O G Vendik, I B Vendik, D V Kholodniak, Applications Of High-Temperature Superconductors In Microwave Integrated Circuits, Advanced Study Center Co Ltd, 2000.</td>
<td>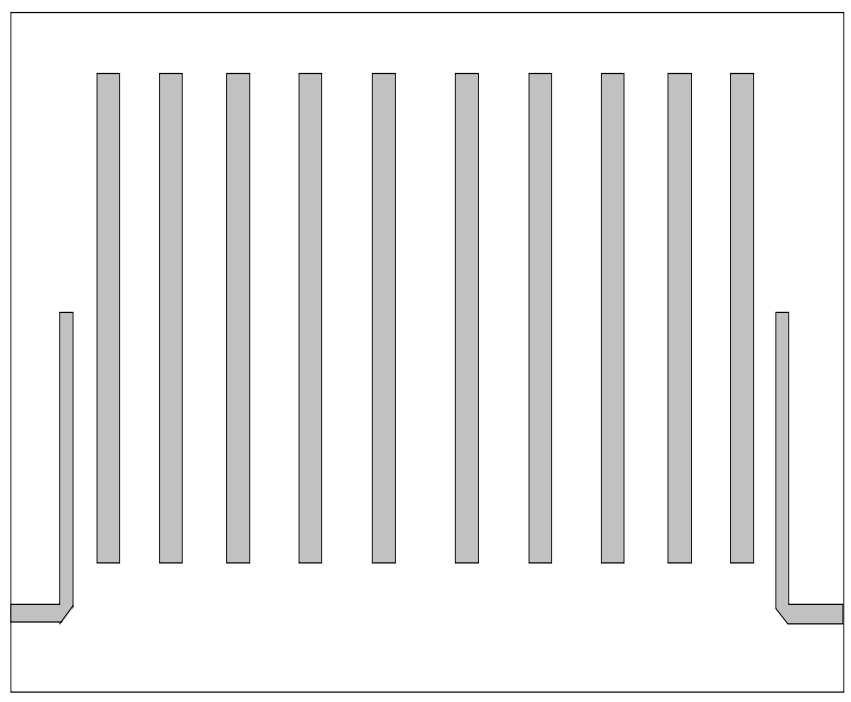<br>Fig.43. Layout of HTS microstrip filter (after[45]).</td>
<td>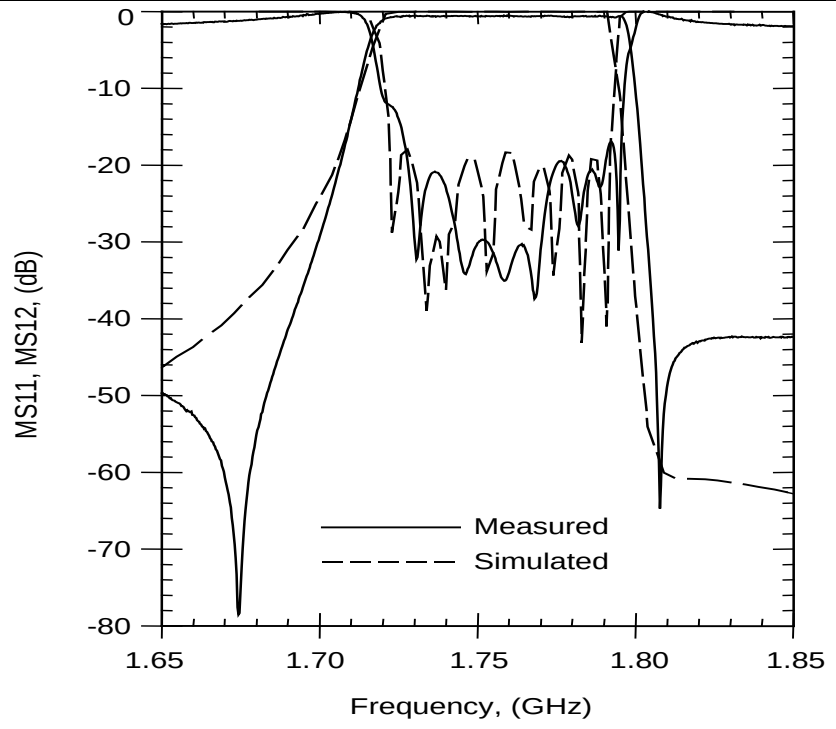<br><br>Fig.44. Frequency response (after [45]).</td>
</tr>
<tr>
<td>O G Vendik, I B Vendik, D V Kholodniak, Applications Of High-Temperature Superconductors In Microwave Integrated Circuits, Advanced Study Center Co Ltd, 2000.</td>
<td>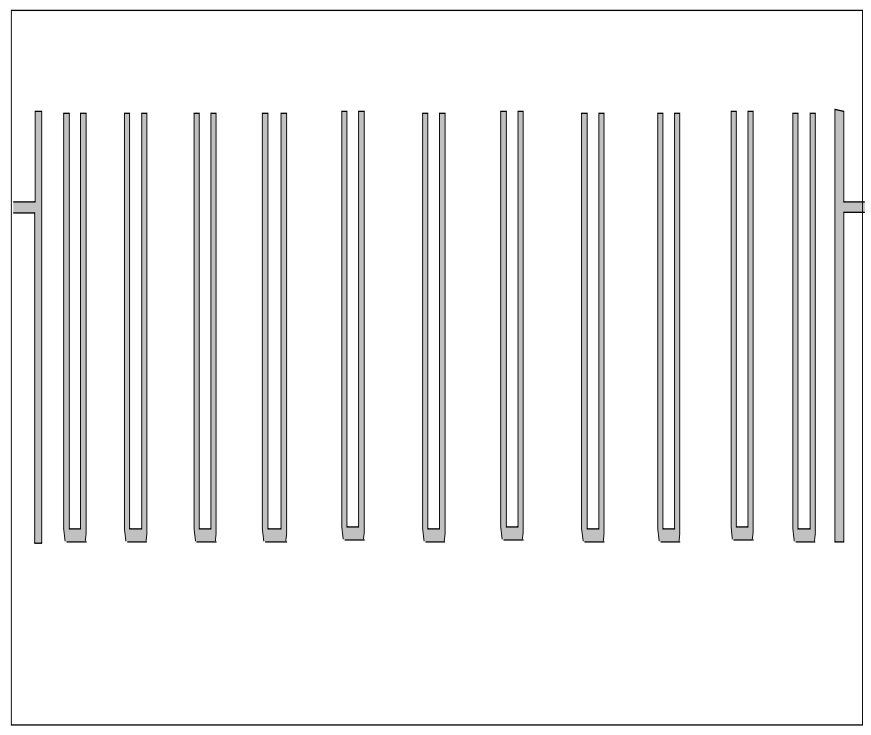<br>Fig.45. Layout of HTS microstrip filter (after[45]).</td>
<td>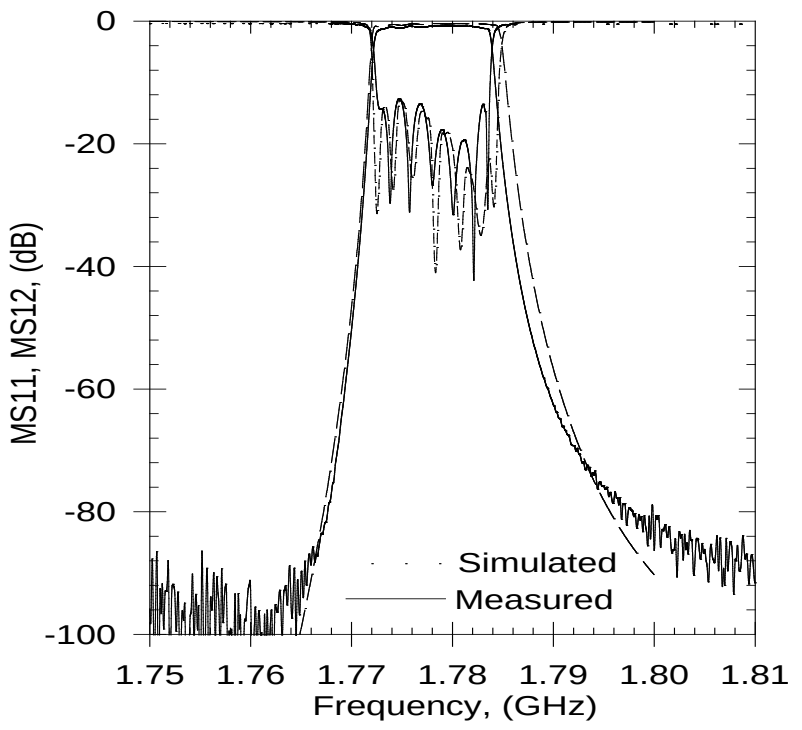<br><br>Fig.46. Frequency response (after [45]).</td>
</tr>
<tr>
<td>I B Vendik, M Goubina, A Deleniv, D Kaparkov, D Kholodniak,V Kondratiev, S Gevorgian, E Kollberg, A Zaitsev, R Wordenweber, Modelling and investigation of HTS planar reso-nators and filters on sapphire sub-strate, Supercond Science Tech., vol. 12, pp. 394–99, 1999.</td>
<td>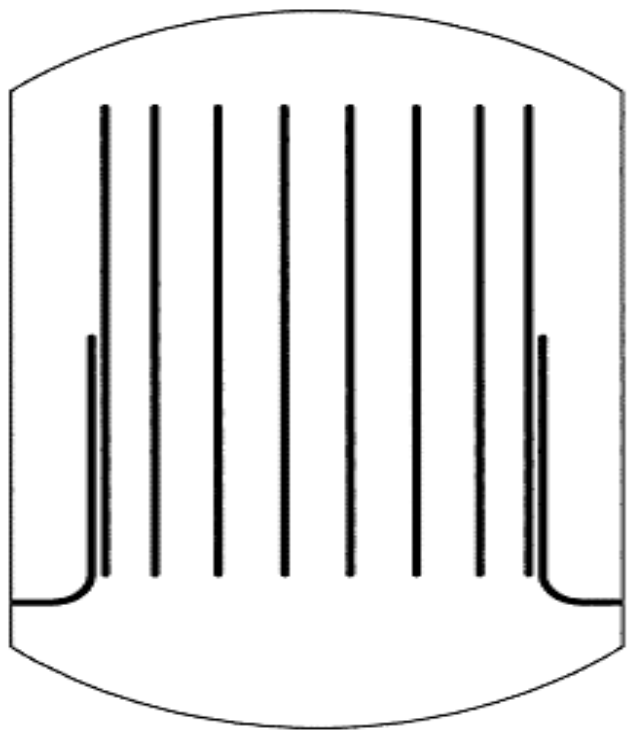<br>Fig.47. Layout of HTS microstrip filter (after[44]).</td>
<td>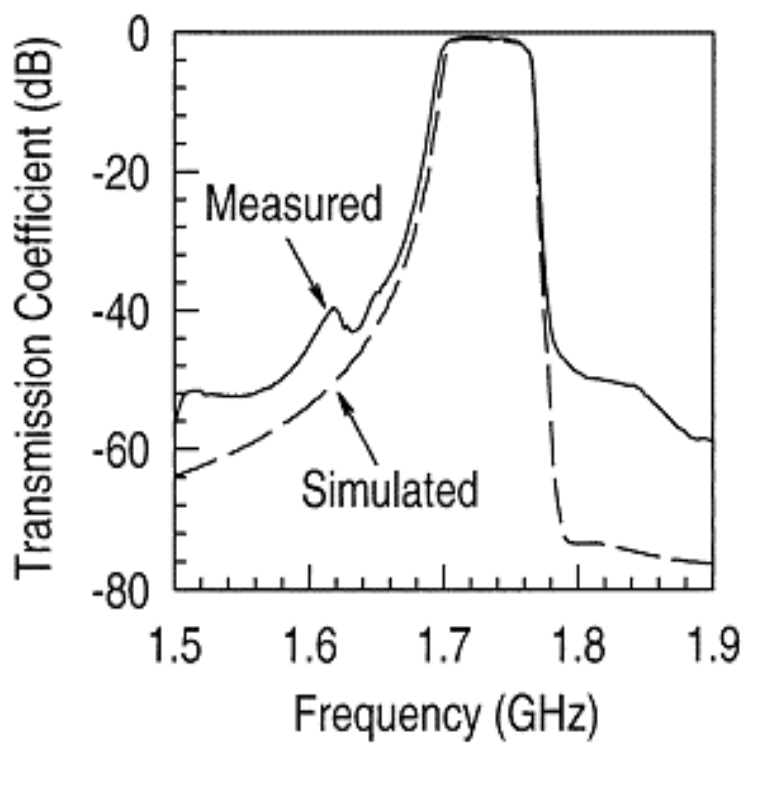<br><br>Fig.48. Frequency response (after [44]).</td>
</tr>
</table>

| | | |
|---|---|---|
| I B Vendik, M Goubina, A Deleniv, D Kaparkov, D Kholodniak,V Kondratiev, S Gevorgian, E Kollberg, A Zaitsev, R Wordenweber, Modelling and investigation of HTS planar resonators and filters on sapphire substrate, *Supercond Science Tech.*,, vol. **12,** pp. 394–99, 1999. | 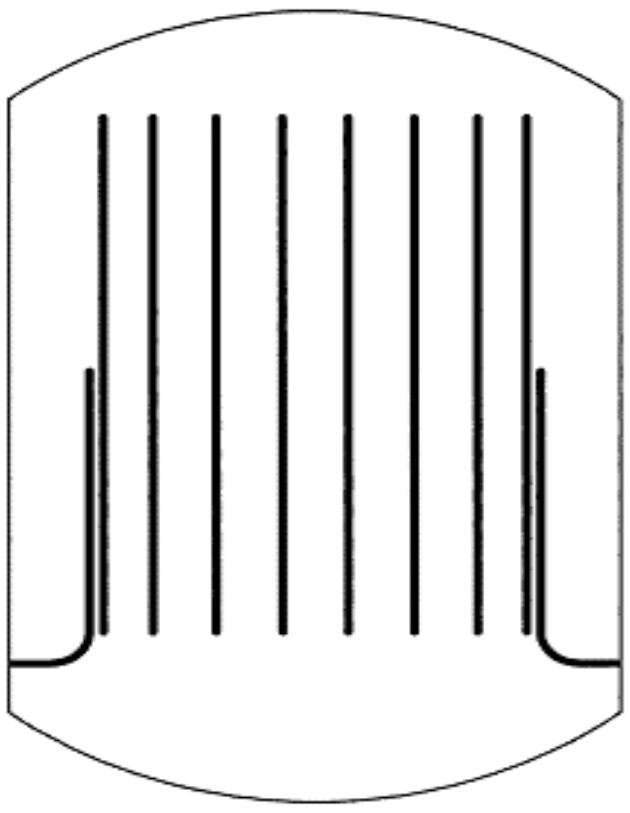<br>**Fig.49.** Layout of HTS microstrip filter (after[44]). | 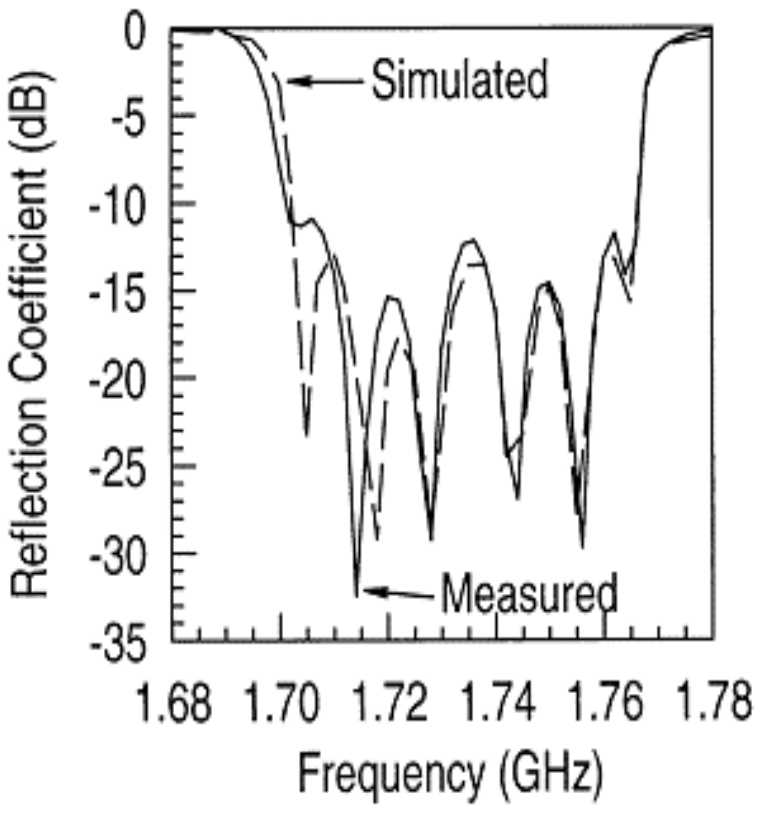<br><br>**Fig.50.** Frequency response (after[44]). |
| M F Sitnikova, I B Vendik, O G Vendik, D V Kholodnyak, P A Tural'chuk, I B Kolmakova, P Yu Belyavsky, A A Semenov, Modeling and Experimental Research of Microstrip Resonators and Filter Based on HTS, *ZhETP Pisma Redaktsiyu*, vol.**36**, no 18, pp. 67-74, 2010. | 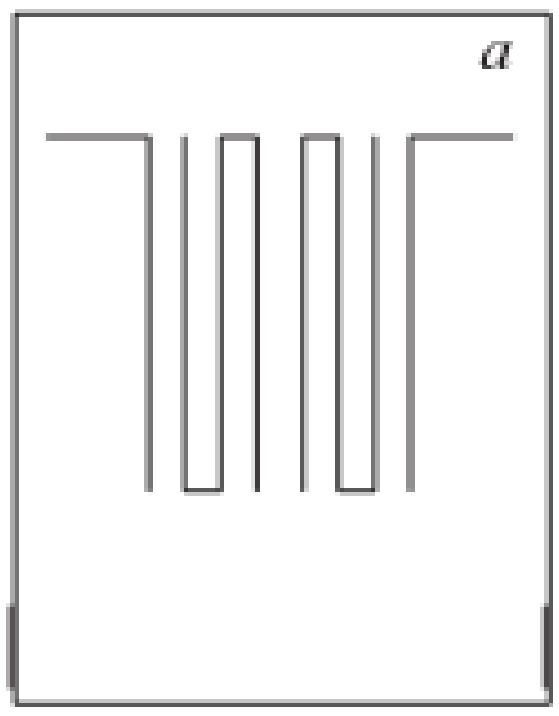<br><br>**Fig.51.** Layout of HTS microstrip filter (after[47]). | 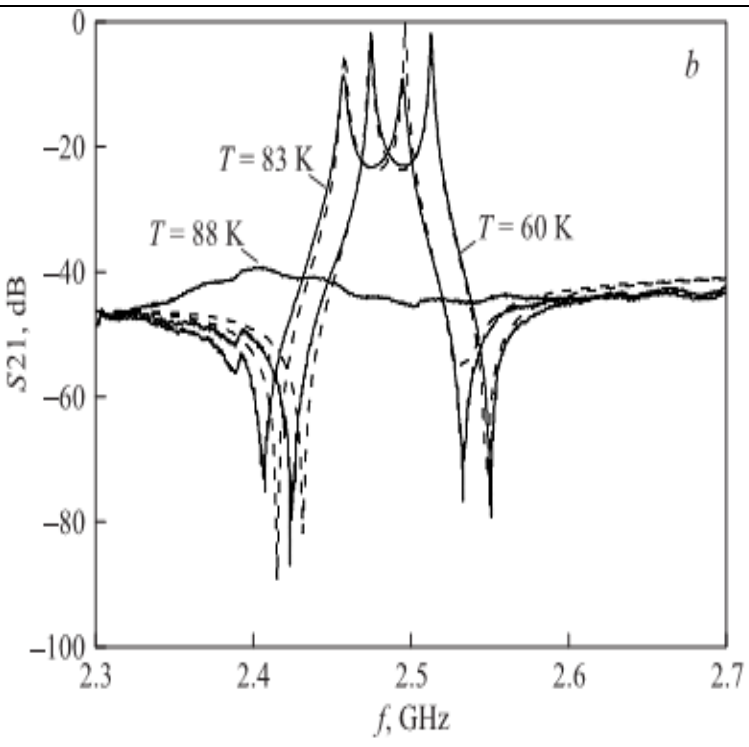<br><br>**Fig.52.** Frequency response (after [47]). |
| R W Simon, R B Hammond, S J Berkowitz , B A Willemsen, Superconducting Microwave Filter Systems for Cellular Telephone Base Stations, *Proceedings of the IEEE*, vol. **92**, pp. 1585-1596, 2004. | 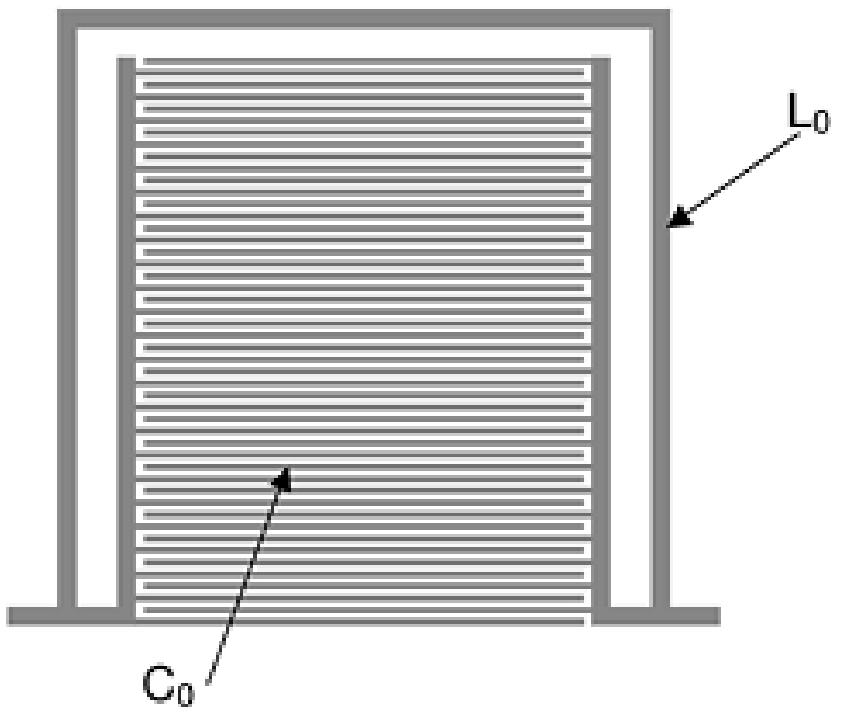<br><br>**Fig.53.** Layout of HTS microstrip filter (after[49]). | 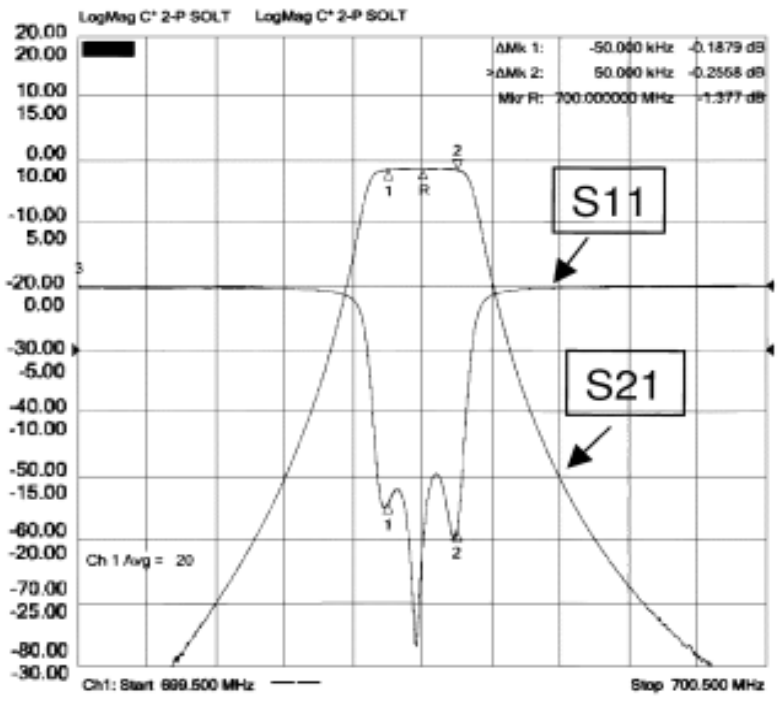<br><br>**Fig.54.** Frequency response (after [49]). |
| R W Simon, R B Hammond, S J Berkowitz , B A Willemsen, Superconducting Microwave Filter Systems for Cellular Telephone Base Stations, *Proceedings of the IEEE*, vol. **92**, pp. 1585-1596, 2004. | 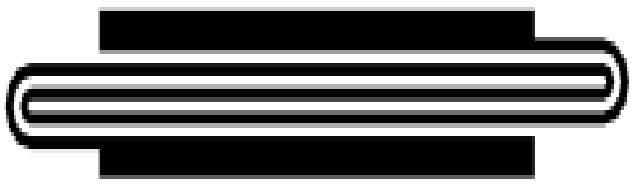<br>**Fig.55.** Layout of HTS microstrip filter (after[49]). | 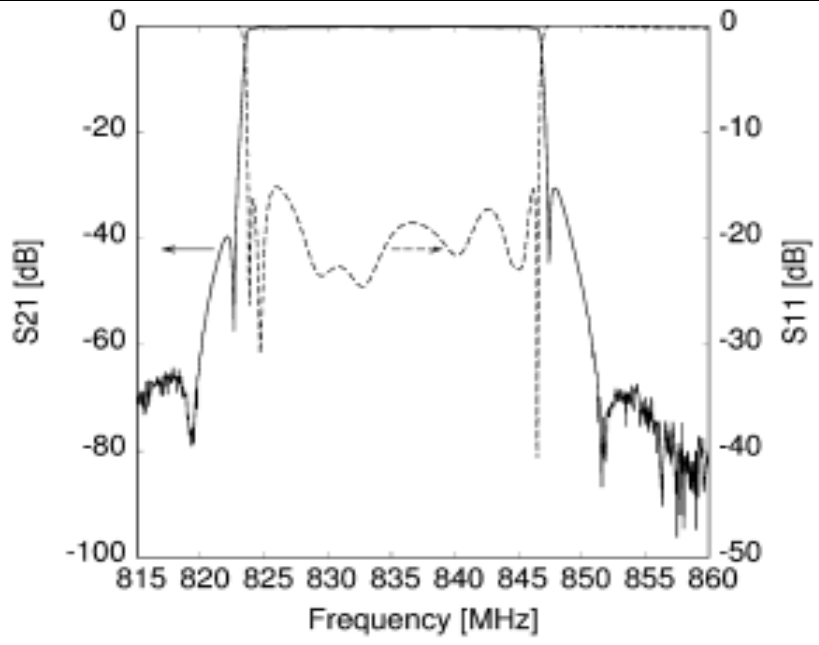<br><br>**Fig. 56.** Frequency response (after [49]). |
| R W Simon, R B Hammond, S J Berkowitz , B A Willemsen, Superconducting Microwave Filter Systems for Cellular Telephone Base Stations, *Proceedings of the IEEE*, vol. **92**, pp. 1585-1596, 2004. | 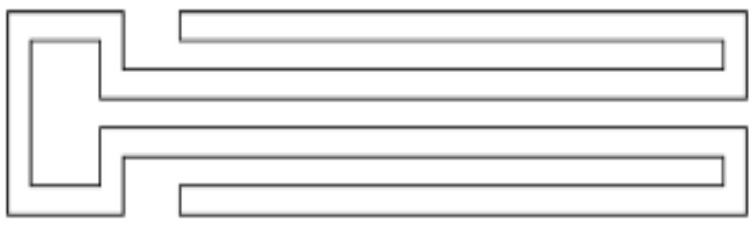 **Fig.57.** Layout of HTS microstrip filter (after[49]). | 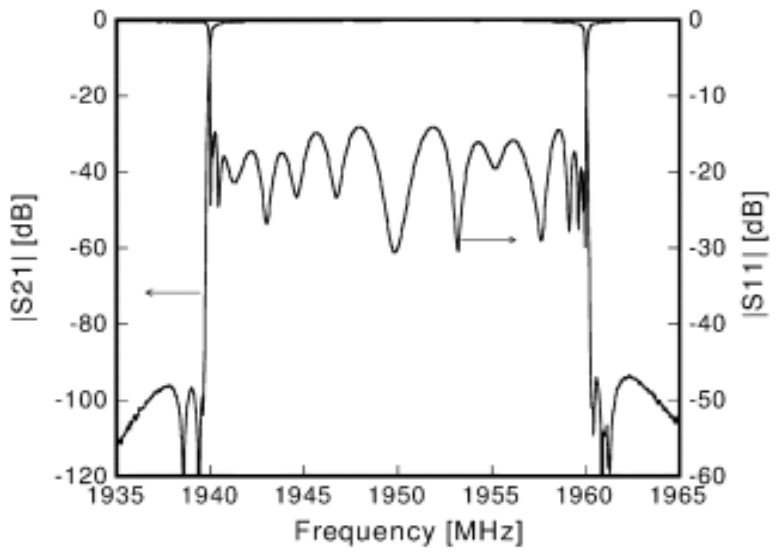  **Fig.58.** Frequency response (after [49]). |
| J P Shivhare, P P Vaidya, S Srinivasulu, C N Lai, D Balasubramanyam, High-temperature superconducting microwave filters for communication payloads of satellites, *Supercond Science Technol*, **15**, pp 983–985, 2002. | 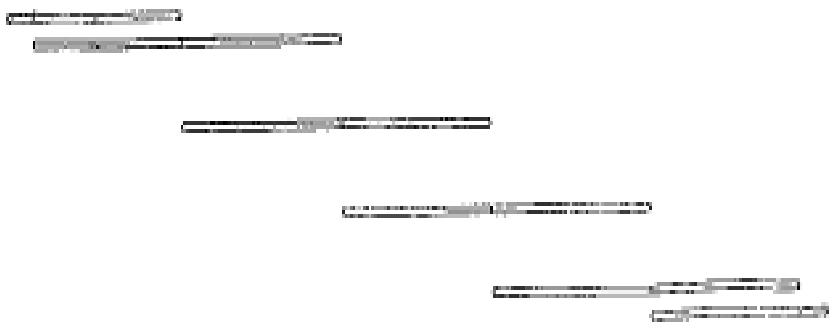 **Fig.59.** Layout of HTS microstrip filter (after[53]). | 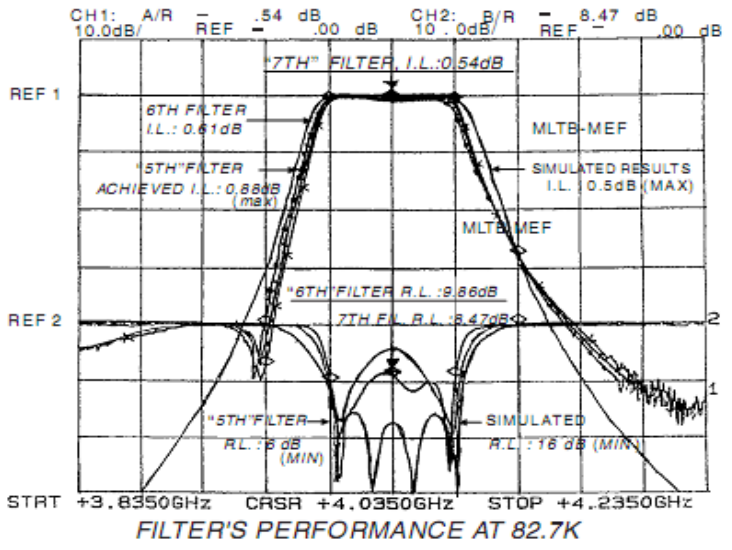  **Fig.60.** Frequency response (after [53]). |
| C X Zhou, H H Xia, T Zuo, X J Zhao, L Fang, S L Yan, Development of X-band high temperature superconducting filters, *Chinese Science Bulletin*, vol. **55**, no. 2, pp. 168-171, doi: 10.1007/s11434-009-0568-6, 2010. | 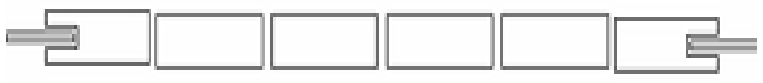 **Fig.61.** Layout of HTS microstrip filter (after[52]). | 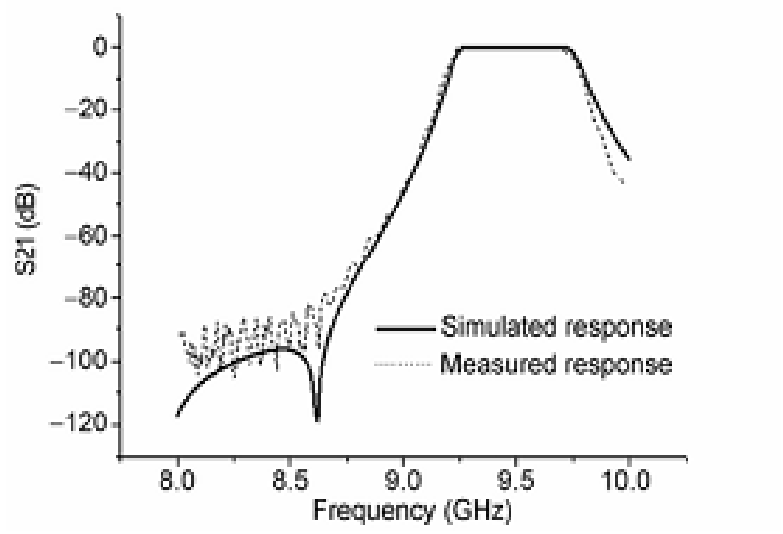  **Fig.62.** Frequency response (after [52]). |
| C X Zhou, H H Xia, T Zuo, X J Zhao, L Fang, S L Yan, Development of X-band high temperature superconducting filters, *Chinese Science Bulletin*, vol. **55**, no. 2, pp. 168-171, doi: 10.1007/s11434-009-0568-6, 2010. | 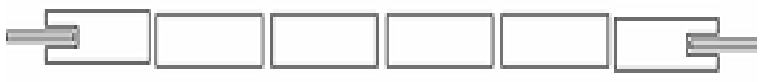 **Fig.63.** Layout of HTS microstrip filter (after[52]). | 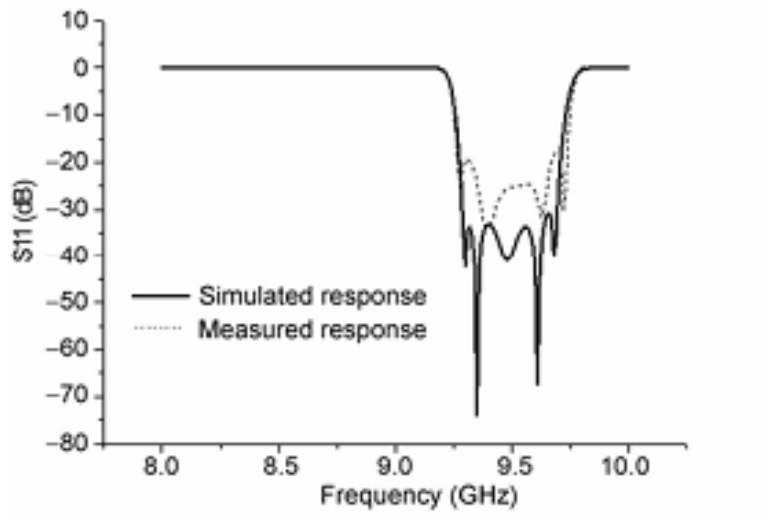  **Fig.64.** Frequency response (after [52]). |
| E M Prophet, J Musolf, B F Zuck, S Jimenez, K E Kihlstrom, B A Willemsen, Highly -Selective Electro-nically-Tunable Cryogenic Filters Using Monolithic, Discretely-Switchable MEMS Capacitor Arrays *IEEE Trans Appl Supercond*, vol **15**, pp. 956–59, 2005. | 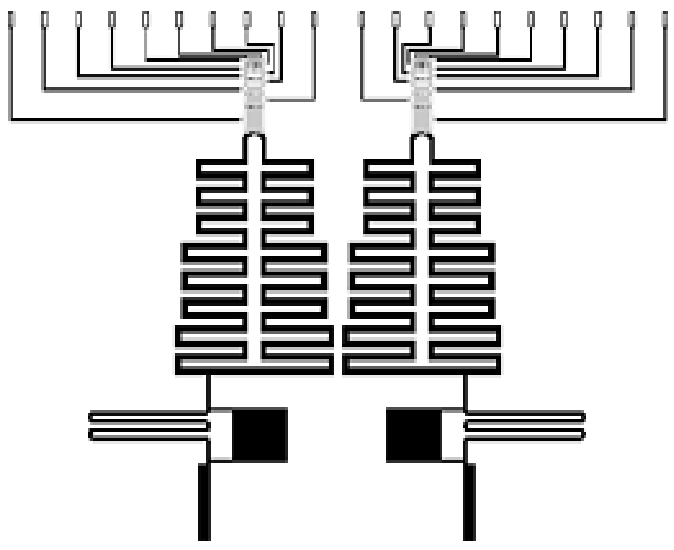 **Fig.65.** Layout of HTS microstrip filter (after[51]). | 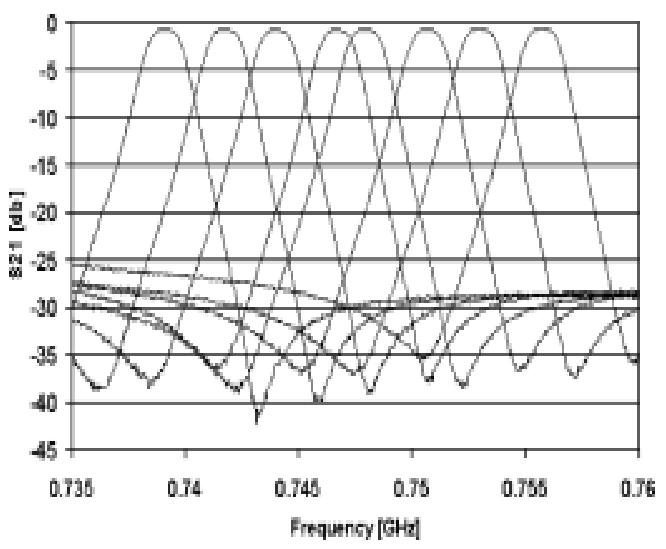  **Fig.66.** Frequency response (after [51]). |
| G Tsuzuki, M P Hernandez, B A Willemsen, Tuning Fork Filter Design for Hand Scribe Tuning, 2005 *IEEE MTT-S International Microwave Symposium Digest*, vol. **3,** pp 1471-1474, 2005. | 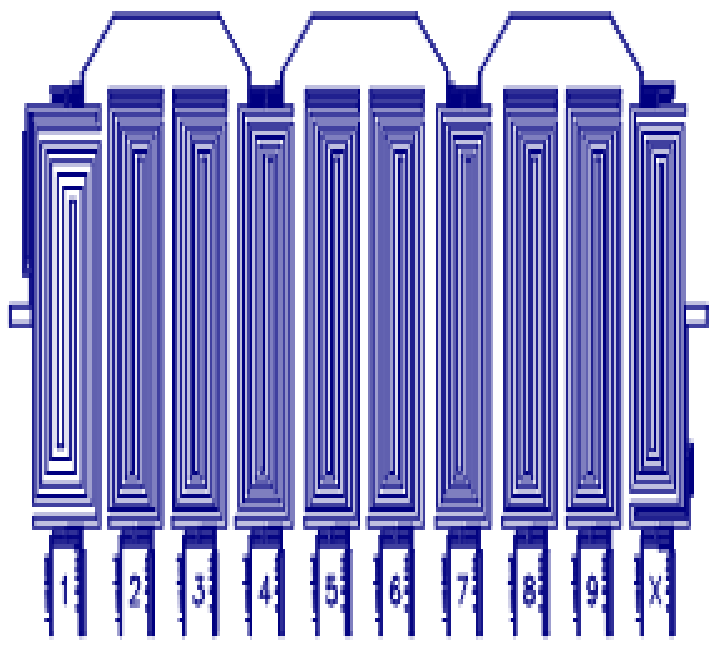<br>**Fig.67.** Layout of HTS microstrip filter (after[76]). | 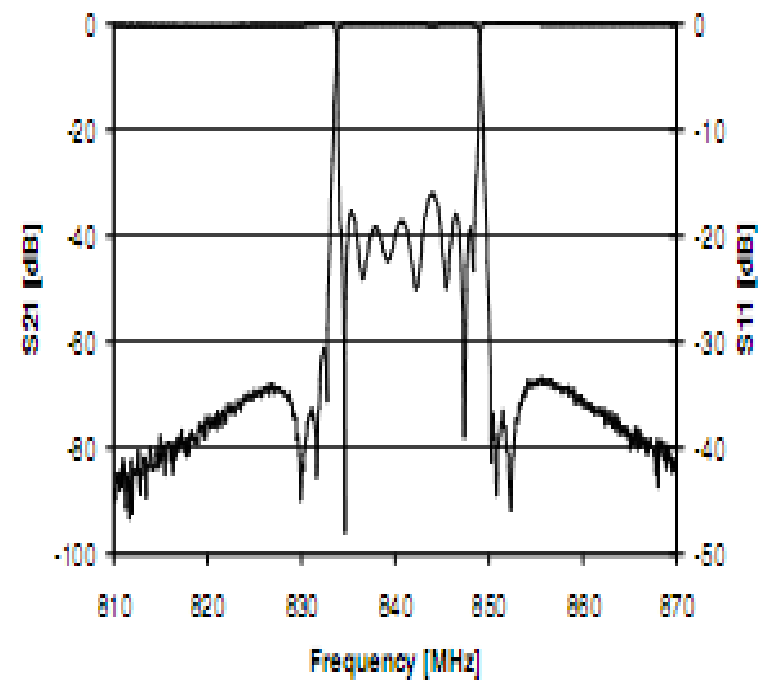<br><br>**Fig.68.** Frequency response (after [76]). |
| Sh. Futatsumori, M. Furuno, T. Hikage, T Nojima, A. Akasegawa, T. Nakanishi, K. Yamanaka, Precise Measurement of IMD Behavior in 5-GHz HTS Resonators and Evaluation of Nonlinear Microwave Characteristics, *IEEE Transactions on Applied Superconductivity*, vol. **19**, no. 3, pp. 3595-3599, 2009. | 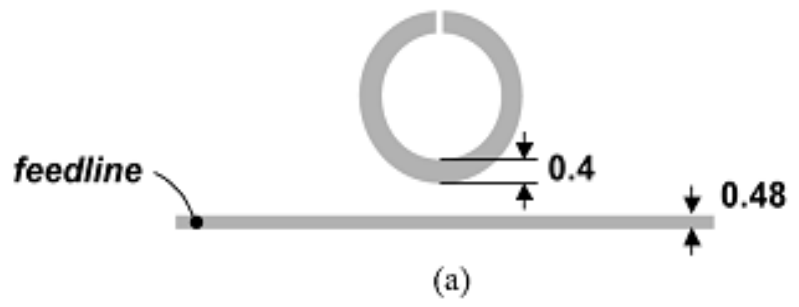<br><br>**Fig.69.** Layout of HTS microstrip filter (after[90]). | 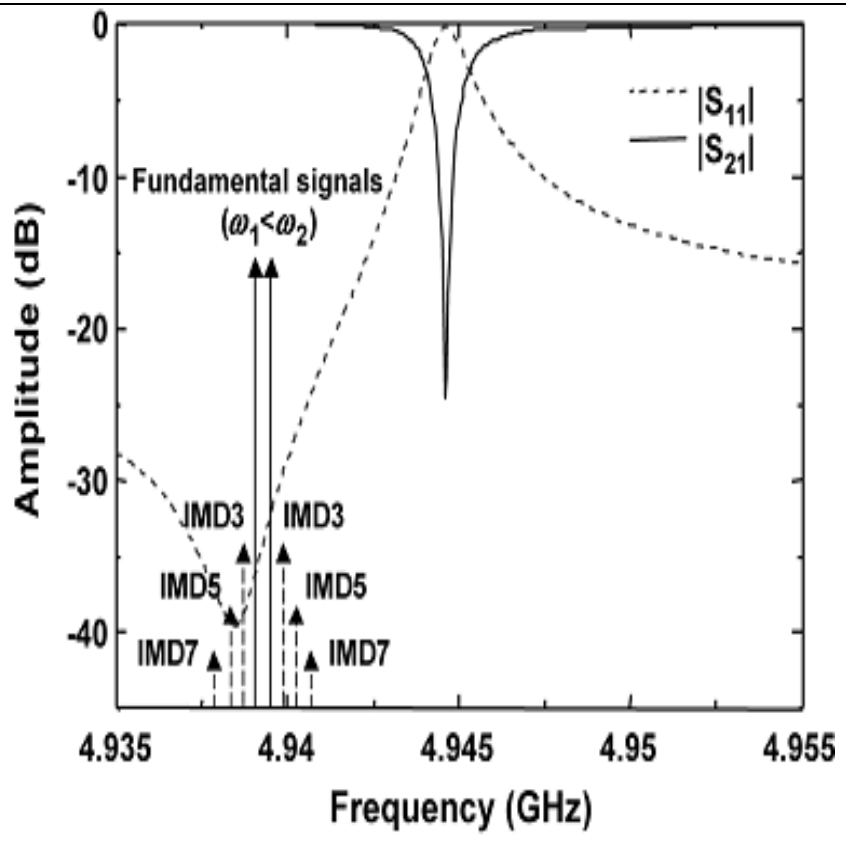<br><br>**Fig.70.** Frequency response (after [90]). |
| Sh. Futatsumori, T. Hikage, T. Nojima, A. Akasegawa, T. Nakanishi, K. Yamanaka, ACLR Improvement of a 5-GHz Power Amplifier Using HTS Reaction-Type Transmitting Filters, *Proceedings of the 38th EuMA*, Amsterdam, The Netherlands, pp. 1145-1148, 2008. | 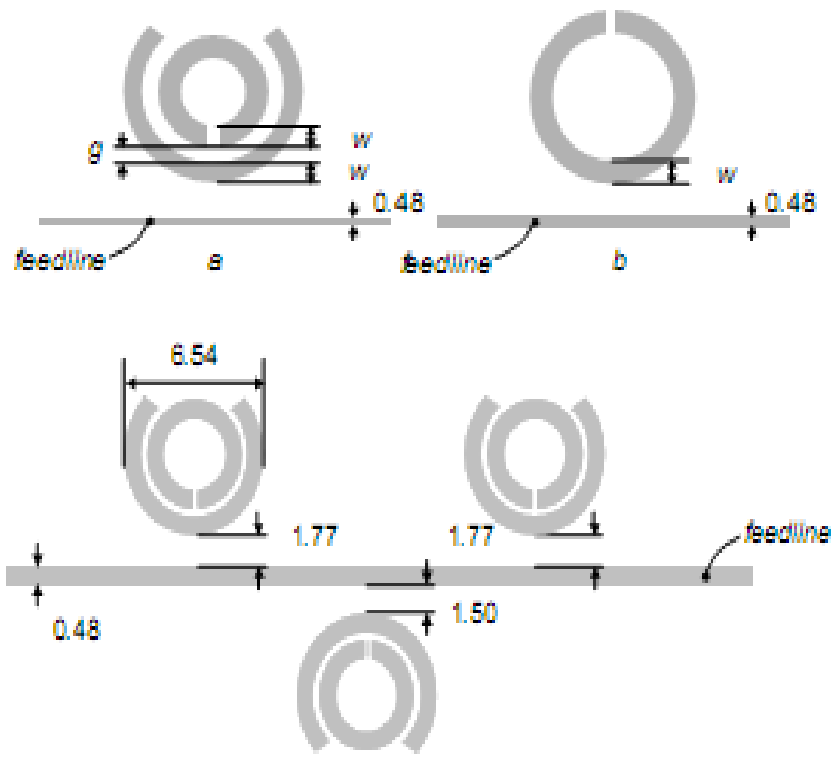<br><br>**Fig.71.**Layout of HTS microstrip filter (after[92]). | 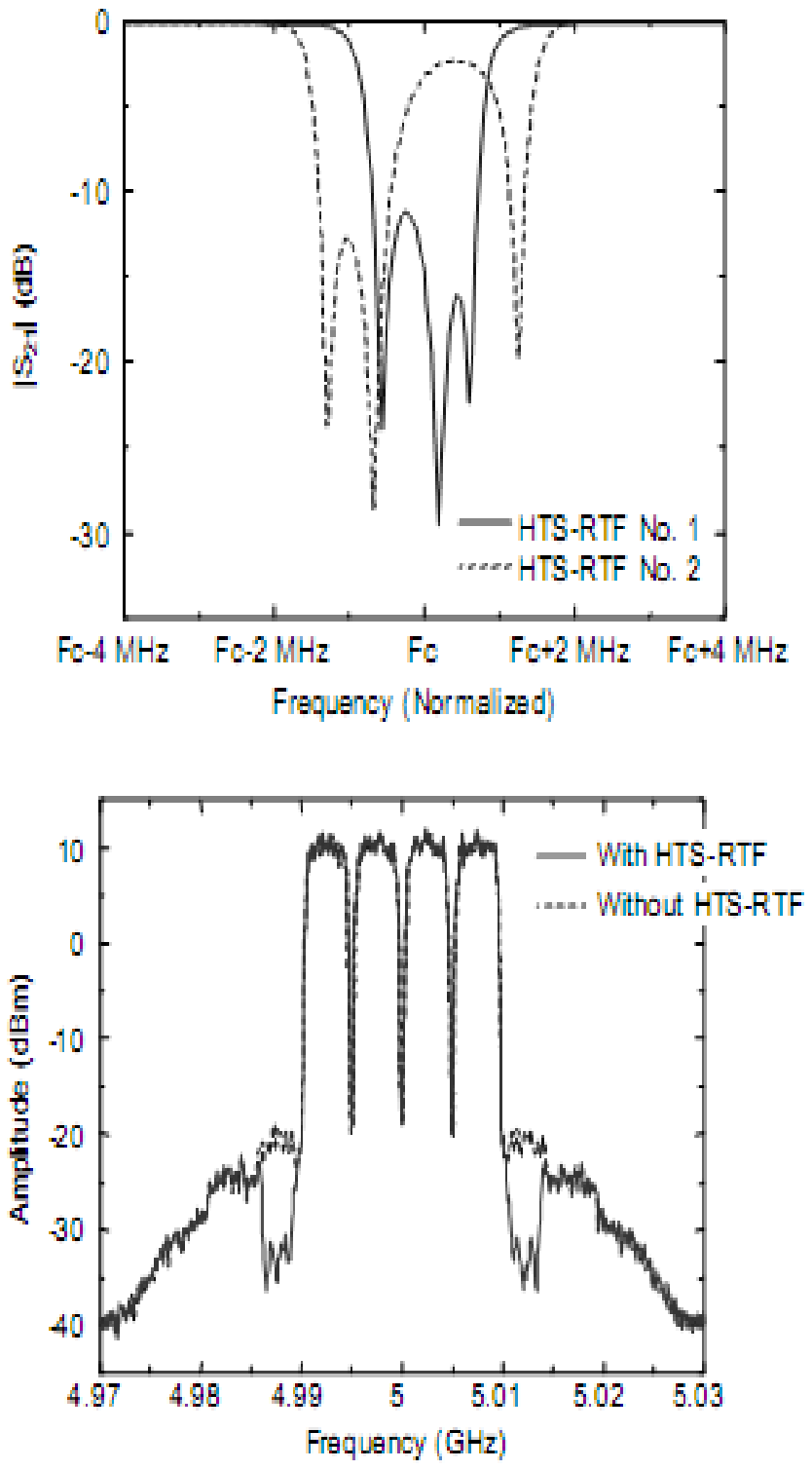<br><br>**Fig.72.** Frequency response (after [92]). |
| Sh. Futatsumori, A 5 GHz high-temperature superconducting reaction type transmitting filter based upon split open-ring resonators, *Superconductors Science Technology*, vol. **21**, no. 4, pp. 1–8, 2008. | 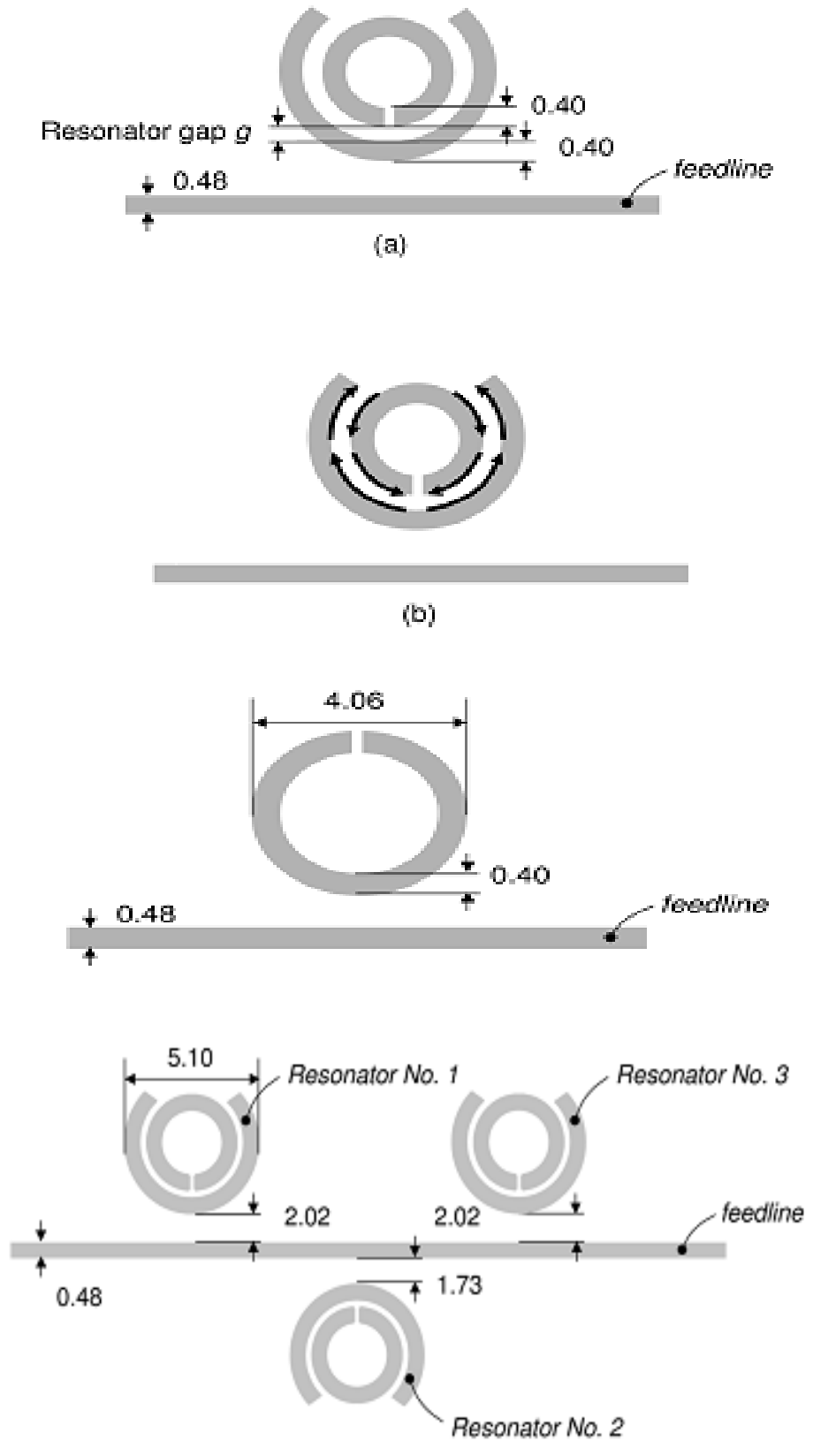<br><br>**Fig.73.** Layout of HTS microstrip filter (after[91]). | 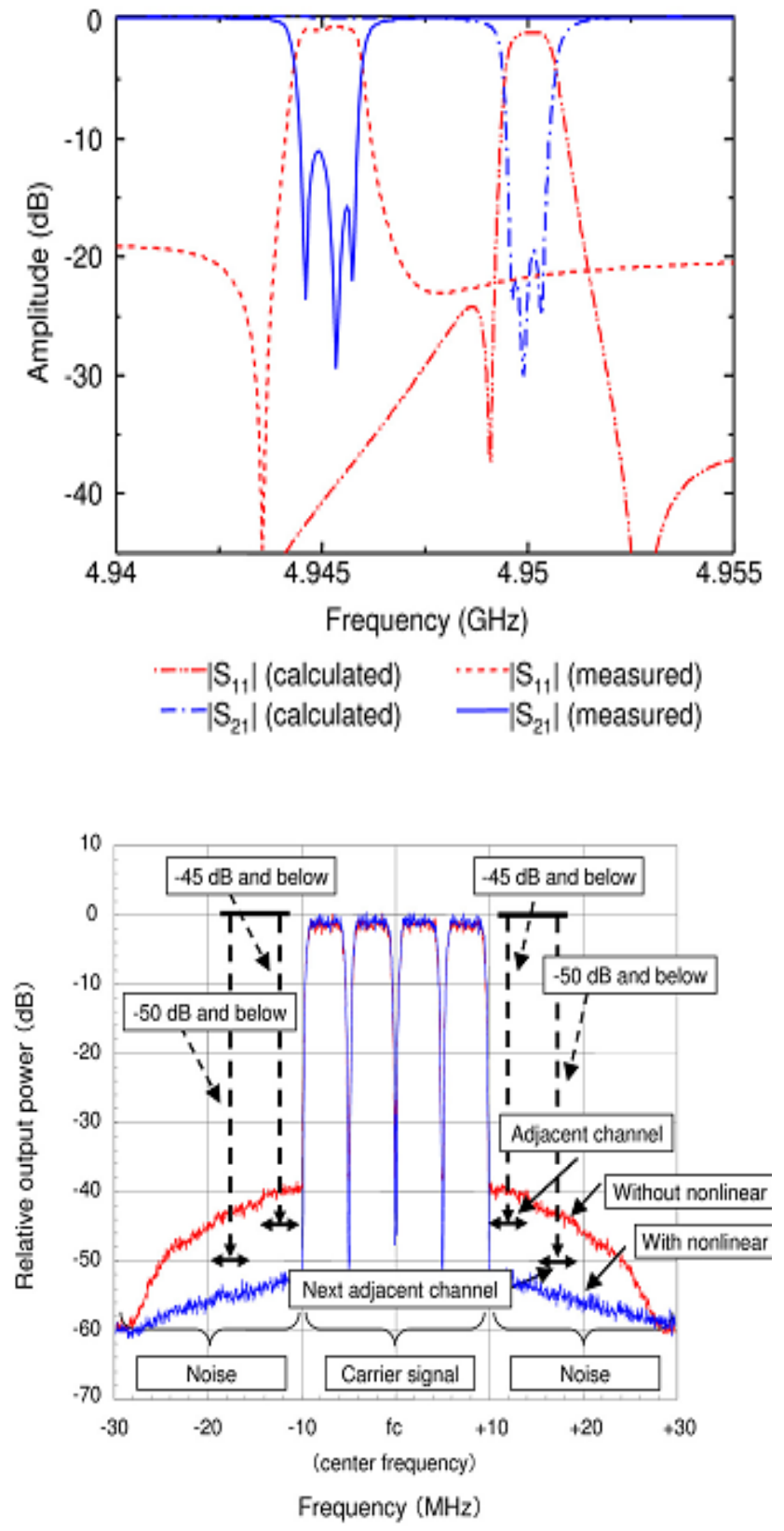<br><br>**Fig.74.** Frequency response (after [91]). |
| Sh. Futatsumori, T. Hikage, T. Nojima, High-Temperature Superconducting Reaction-type Transmitting Filter Consisting of Novel Split Open-ring Resonators, *Proceedings of Asia-Pacific Microwave Conference APMC2006*, Pacifico Yokohama, Yokohama, Japan, pp. 1-4, 2006. | 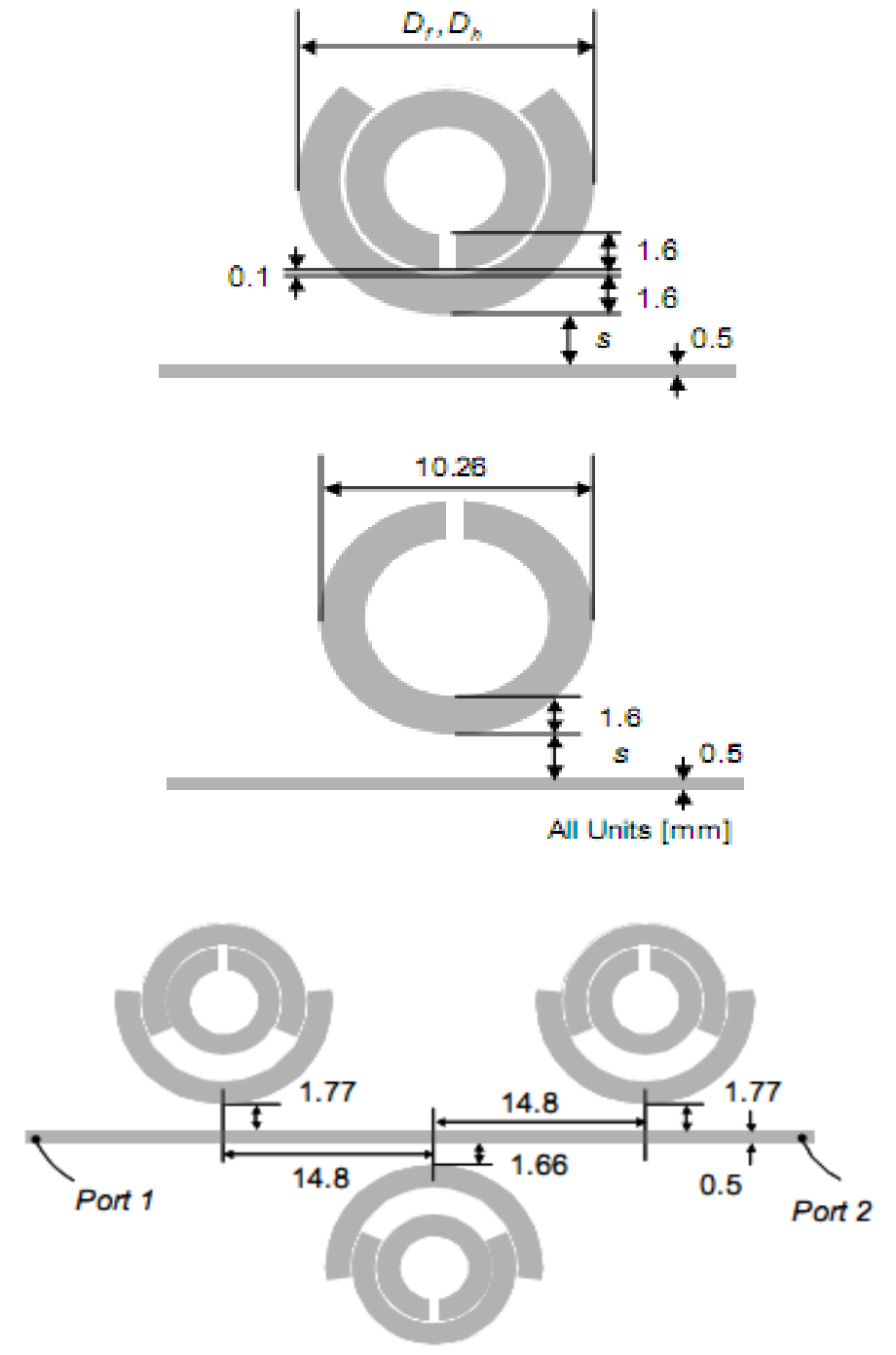<br><br>**Fig.75.** Layout of HTS microstrip filter (after[93]). | 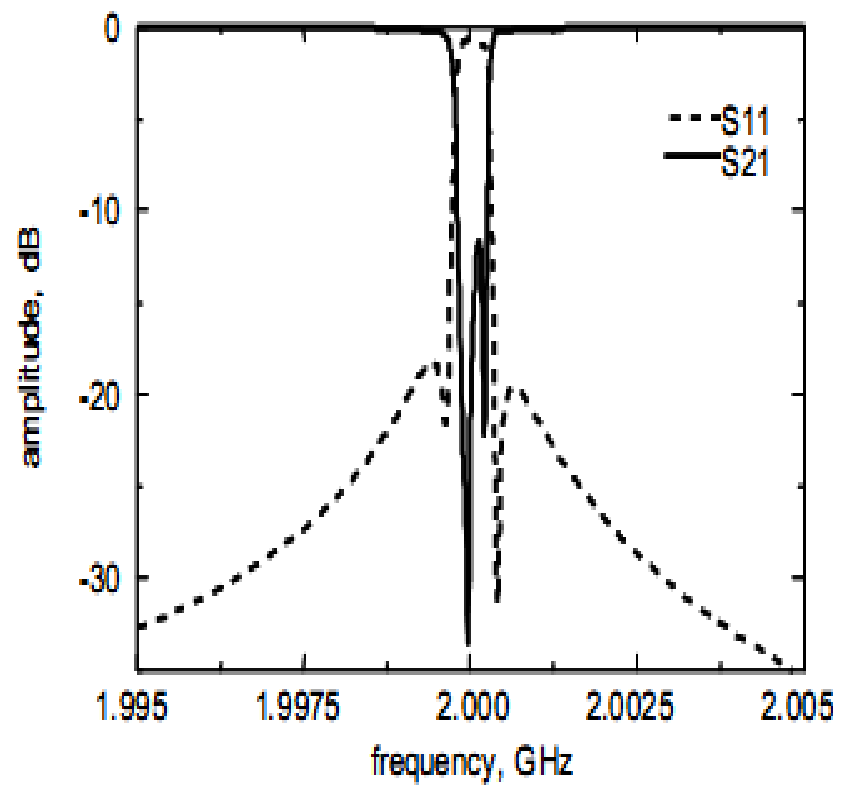<br><br>**Fig.76.** Frequency response (after [93]). |

**Tab. 3.** Schematic of layouts of *HTS* microwave resonators and multi pole filters together with their frequency responses for technology overview purposes only.

Tanaka, Tatsunokuchi, Akiya, Saito, Ohshima [122] analyzed the present achievements in the design of transmitting *HTS* microwave filters layouts with high power handling capabilities:

1. Liang, Zhang, Shin, Johansson, Withers, Oates, Anderson, Polakos, Mankiewich, de Obaldia, Miller [123] reported a BPF microstrip filter with a wide microstrip line capable of handling of microwave power of 10W;
2. Setsune and Enokihara [124] as well as Yamanaka, Ishii, Akasegawa, Nakanishi, Baniecki, Kurihara [125] reported a dual-mode disk-resonator filter with a power-handling capability over *10-100W* by decreasing the current concentration at the edge of the microwave resonator;
3. Saito, Lee, Teshima, Ohshima [126] reported a microwave filter with high microwave power, over 20W, obtained by using the disk ring microwave resonator;
4. Guo, Wei, Zhang, Cao, Jin, Peng, LM. Gao, BX. Gao [127] reported a 5-pole *HTS BPF,* using optimized linewidths of resonator 1-5 by calculating current density distribution. They obtained an HTS filter that could handle power of 3 W.
5. Takeuchi, Osaka, Kinouchi, Ono, Saito, Akasegawa, Nakanishi, Kawakami, Yamanaka, Kurihara, Ohshima [128], and Ohshima, Uno, Endo, Takeuchi, Ono, Saito, Sekiya [129] reported a high power microwave filter with a sliced microstrip line to decrease the current concentration. The power-handling capability of a sliced microstrip line filter was more than *7W*.

Jacob, Mazierska, Knack, Takeuchi [72] express some interesting comments on the *HTS* microwave filters design strategy saying that ***it is important to minimise the physical size of a filter without compromising the out of band rejection***. High *Out of Band Rejection* (*OBR)* can be achieved by increasing the number of poles, but this leads to higher insertion loss and asymmetrical frequency response [72]. However, the increase in the number of poles enlarges the physical size of the filter, resulting in increased cost of the filter. Therefore, it is not an easy task to achieve high out-of-band rejection without compromising the insertion loss and miniaturization of the filter. This can be achieved by novel resonator designs, and in some cases by elliptic or quasi-elliptic filter magnitude approximations [72].

Sekiya, Yamamoto, Kakio, Saito, Ohshima [111] developed the miniaturized sharp-cutoff transmit *HTS* filter with a stripline structure. Sekiya, Imai, Kakio, Saito,

Ohshima [112] developed the miniaturized transmit dual-mode *HTS* patch filter with stripline structure. The miniaturization of high pole *HTS* microwave filters with increased microwave power handling capability was achieved by using the design approach with the stripline (*SL*) resonators with shorter distance between resonators, resulting from a weak coupling. Researchers [111, 112] made some modifications to the *HTS SL* microwave filter by adding the increased spacing between the *SL* resonators to reduce maximum surface current. It is reported that the developed *HTS* microwave filters with stripline (*SL*) resonators have smaller size in comparison with the *HTS* microstrip line (*MSL*), dual-mode or bulk resonators [111, 112].

In some *HTS* microwave filters designs, there are very sharp passband-to-stopband transitions, but the passband profile is not good. The reasons for this kind of response can be manifold as mentioned by Pal from Oxford University [57, 58]:

1. Firstly, as the layout covers almost entire surface of the film, the film quality needs to be accurately uniform over the entire surface, otherwise one single non-superconducting inclusions or any form of defects will either break the current path or may cause severe impedance mismatch to kill the response altogether. The etching process needed to be checked thoroughly, as non-uniform etching rate was observed across the filter surface from different etchants. Non-uniform etching may cause over etching or undercutting of the dimensions of the filter. Sometimes, it is observed that because of over etching some dimensions of the filter are corrupted with various forms, which severely affect its performance. Dry etching could be a better proposition for these kind of delicate designs.

2. Secondly, the design of a 23-pole lumped element filter demands uniform dielectric constant, substrate height and film thickness across the entire surface. Any form of deviations of these film and substrate parameters would have non-uniform impacts in the elements; as a result, the filter response may be severely affected. Another interesting point to be noted is the substrate used here, that is, lanthanum aluminate, which has twinned regions with dielectric constant different from its non-twinned regions across the substrate surface. This non-uniformity of the dielectric profile of the lanthanum aluminate substrate may be large enough for the thin and small transmission line sections of this compact design to perform differently.

3. Thirdly, the filter is very compact; unwanted inter-element parasitic coupling from the non-adjacent elements could be sufficiently large to cause far enough impedance mismatch among the constituent elements. This problem could not be solved even running *EM* simulation on individual parts, unless the entire design layout is run on simulation.

Pal [58] makes the following final comment on the comparative analysis of 5-, 10- and 23-pole *HTS* band-pass *RF* filters saying that, from the measured responses, it was observed that the 5-pole filter performed best in terms filter response shape, followed by 10 and 23 poles. This is because the constituent elements of a lower-order filter are less closely spaced and also occupy less surface area than that of a higher-order filter; therefore, are less demanding for uniform substrate and film parameters to perform efficiently. With increase in the number of poles in a filter, the selectivity of the filter increases as observed from the measured responses of 5-, 10- and 23-pole filters. In order to avoid the sensitivity with the substrate and film parameters in lumped element filter designs, resonators with relatively bigger dimensions will be helpful. Author of dissertation would like to state that the above mentioned findings by Pal [58] are in good agreement with the theoretical research findings on microstrip filters by Hong and Lancaster [8].

Nakakita, Kanamori, Hashimoto, Fuke, Kayano, Aiga, Terashima, Yamazaki [116] reported about the *HTS* microstrip filters application in digital terrestrial television transposer for digital-terrestrial-television broadcasting (*DTTB*) in Japan. The filter consists of 12 microstrip resonators fabricated from $YBa_2Cu_3O_y$ thin films deposited on an $Al_2O_3$ (sapphire) substrate.

The author of dissertation added the detailed information on the references to the Tab. 4 in Mazierska, Jacob [14], which presents the layout design specifications and technical characteristics of *HTS* microstrip filters.

Tab. 4 [14] in intended for the *state-of-the-art HTS* microstrip filters technology overview purposes only.

| Publication Reference | Pole No. | Filter Type | Sub-strate Type | Size, mm | $f_0$, GHz | BW, MHz | IL, dB | OBR, db/ MHz |
|---|---|---|---|---|---|---|---|---|
| M. Klauda *et al.* "Superconductors and Cryogenics for Future Communication Systems", *IEEE Transactions on Microwave Theory and Techniques*", vol. **48**, No. 7, pp. 1227- 1239,2000. | 9 | Elliptic | Sapp-hire | | 1.93 | 20 | 0.1 | 15dB/ MHz |
| M. Reppel and J. C. Mage, "Superconducting Microstrip Bandpass Filter on LAO with High out of Band Rejection", *IEEE Microwave and Guided Letters*, vol. **10**, No. 5, pp.180-182, 2000. | 8 | Quasi-elliptic | LAO | 39x12 | 1.8 | 15 | 0.3 | 60dB/ 5MHz |
| I. B. Vendik, A. N. Deleniv, V. O. Sherman, A. A. Svishchev, V. V. Kondratiev, D. V. Kholodniak, A. V. Lapshin, P. N. Yudin, B. C. Min, Y. H. Choi and B. Oh, "Narrownband YBCO Filter with Quasi Elliptic Characteristic", *IEEE Trans-actions on Applied Super-conductivity*, vol. **11**, No. 1, pp. 477-480, 2001. | 12 | Quasi-elliptic | LAO | 35x20 | 1.775 | 10 | 0.5 | 75dB/ 2MHz |
| M. Repel and H. Chaloupka, "Novel Approach for Narrowband Superconducting Filters", *IEEE MTT-S Digest* 1999, TH2D-6, 1999. | 4 | Quasi-elliptic | LAO | 20x20 | 1.972 | 5 | 0.4 | - |
| M. Repel and H. Chaloupka, "Novel Approach for Narrowband Superconducting Filters", *IEEE MTT-S Digest* 1999, TH2D-6, 1999. | 9 | Quasi-elliptic | LAO | 39x22.5 | 1.715 | 10 | 0.25 | - |
| M. Repel and H. Chaloupka, "Novel Approach for Narrowband Superconducting Filters", *IEEE MTT-S Digest* 1999, TH2D-6, 1999. | 9 | Quasi-elliptic | LAO | 39x22.5 | 1.777 | 15 | 0.25 | - |
| G. Tsuzuki, S. Ye, S. Berkowitz: "Ultra Selective 22-Pole, 10-Transmission Zero Superconducting Bandpass Filter Surpasses 50-Pole Chebyshev Filter" t *IEEE Transactions on Microwave Theory and Techniques*, vol. **50** pp 2924 – 2929, 2002. | 22 | Cheby-shev | LAO | Dia 2" | 1.95 | 20 | 0.2 | 30dB/ 100kHz |
| G. Tsuzuki, M. Suzuki and N. Sakakibara, "Superconducting Filter for IMT-2000 Band", *IEEE Transactions on Microwave Theory and Techniques*, vol. **48**, no. 12, pp. 2519-2525, 2000. | 16 | Cheby-shev | MgO | Half of 3" | 1.93 | 20 | 0.14 | 30dB/ 1.3MHz |
| G. Tsuzuki, M. Suzuki and N. Sakakibara, "Superconducting Filter for IMT-2000 Band", *IEEE Trans-actions on Microwave Theory and Techniques*, vol. **48**, No. 12, pp. 2519-2525, 2000. | 32 | Cheby-shev | MgO | Dia 3" | 1.93 | 20 | 0.32 | 30dB/ 0.5MHz |
| H. Kayano, H. Fuke, F. Aiga, Y. Terashima, R. Kato and Y. Suzuki, "2 GHz Superconducting Bandpass Filter with Parallel Resonators Structure", *Proceedings of Asia Pacific Microwave Conference,* APMC2000, Sydney, pp. 592-595, 2000. | 2 | Patch | LAO | Dia 3" | 1.90 9 | 9.38 | 0.77 | 30dB/ 25MHz |
| H. Kayano, H. Fuke, F. Aiga, Y. Terashima, R. Kato and Y. Suzuki, "2 GHz Superconducting Bandpass Filter with Parallel Resonators Structure", *Proceedings of Asia Pacific Microwave Conference,* APMC2000, Sydney, pp. 592-595, 2000. | 9 | Cheby-shev | LAO | 46x18 | 1.78 4 | 11 | 0.8 | 34dB/ 1MHz |
| A. Andreone, A. Cassinese, M. Iavarone, P. Orgiani, F. Palomba, G. Pica, M. Salluzzo and R. Valio, "Development of L-Band and C-Band Superconducting Planar Filters for Wireless Systems", *Proceedings of Asia Pacific Microwave Conference* APMC2000, Sydney, pp. 581-586, 2000. | 2 | Cheby-shev | LAO | 10x10 | 3.7 | 1% and 10% | | 30dB/ 45MHz |
| A. Andreone, A. Cassinese, M. Iavarone, P. Orgiani, F. Palomba, G. Pica, M. Salluzzo and R. Valio, "Development of L-Band and C-Band Supercon-ducting Planar Filters for Wire-less Systems", *Proc. of Asia Pacific Micro-wave Conference* APMC2000, Sydney, pp. 581-586, 2000. | 7 | Cheby-shev | LAO | Dia 50" | 1.76 | 75 | | 30dB/ 5MHz |
| K. Setsune and A. Enokihara, "Elliptic Disc Filters of HTS Films for Power Handling Capa-city over 100 W", *IEEE Transa-ctions on Micro-wave Theory and Techniques*, vol. **48**, No. 7, pp. 1256-1264, 2000. | 4 | Cheby-shev | LAO | Dia 14 | 2.38 | 24 | 0.21 | 30dB/ 40MHz |
| T. Kasser, IMS 2002, Presentation (private commu-nication) 2002. | 13 | Quasi-elliptic | | | 1.95 | 5 | | 80dB/ 1MHz |
| T. Kasser, IMS 2002, Presentation (private commu-nication) 2002. | 17 | Quasi-elliptic | | 22x70 | 1.93 | 20 | | 30dB/ 1MHz |
| J. S. Hong, M. J. Lancaster, D. Jedamzik, R. B. Greed and J. C. Mage, "On the Performance of HTS Microstrip Quasi-elliptic Function Filters for Mobile Communications Applications", *IEEE Transactions on Microwave Theory and Techniques*", vol. | 8 | Quasi-Elliptic | LAO | 39x23.5 | 1.77 8 | 15 | | 30dB/ 17.5MHz |
| **48**, No. 7, pp. 1240-1246, 2000. | | | | | | | | |
| J. S. Hong, M. J. Lancaster, D. Jedamzik, R. B. Greed and J. C. Mage, "On the Performance of HTS Microstrip Quasi-elliptic Function Filters for Mobile Communications Applications", *IEEE Transactions on Microwave Theory and Techniques*", vol. **48**, No. 7, pp. 1240-1246, 2000. | 8 | Quasi-Elliptic | MgO (0.3) | 39x23.5 | 1.778 | 15 | 15.5 | 36.5dB/ 2MHz |
| G. H. Shipton, "Superconducting Filters for Base Stations in the USA", *IEEE IMS MTT-S Workshop Notes*, Workshop on High Performance and Emerging Filter Technologies for Wireless, Phoenix, 2001. | 10 | Quasi-Elliptic | MgO | 34x18 | 842 | 15 | 0.4 | 60dB/ 2MHz |
| M. Barra, A. Cassinese, M. Cirillo, G. Panariello, R. Russo and R. Vaglio, "Super-conducting Dual Mode Dual Stage Cross Slotted Filters", Microwave and Optical Tech-nology letters, vol. 33 (6), pp. 389-392, 2002. | 4 | Cheby-shev | LAO | 22.3x10 | 3.32 | 1% | 0.9 | 50dB/ 20MHz |
| R. S. Kwok, D. Zhang, Q. Huang, T. S. Kaplan, J. Lu and G. C. Liang, "Superconducting Quasi-Lumped Element Filter on R-Plane Sapphire", *IEEE Transactions on Microwave Theory and Techniques*, vol. 47, No. 5, pp. 586-591, 1999. | 6 | Cheby-shev | Sapp-hire | 43x10 | 1.857 | 20 | 0.2 | 60dB/ 15MHz |
| H. T. Kim, B. C. Min, Y. H. Choi, S. H. Moon, S. M. Lee, B. Oh, J. T. Lee, I. Park and C. C. Shin, "A Compact Narrowband HTS Microstrip Filter for PCS Applications", *IEEE Trans-actions on Applied Super-conductivity*, vol. 9, No. 2, pp. 3909-3912, 1999. | 11 | Cheby-shev | LAO | Dia 50 | 1.778 | 11.5 | 0.6 | 80dB/ 20MHz |
| B. Lippmeier and Mal Sinclair "HTS Microwave Filter for ANTF", CSIRO, Australia, private communication, 2001. | 8 | Quasi-elliptic | LAO | Dia 50 | 2.260 | 68 | 0.7 | 55dB/ 10MHz |
| H. Kanaya, T. Shinto, K. Yoshida, T. Uchiyama and Z. Wang, "Miniaturised HTS Coplanar Waveguide Bandpass Filters with Highly Packed Meander-Lines", *IEEE Transactions on Applied Superconductivity*, vol. 11, pp. 481-484, 2001. | 11 | Cheby-shev | MgO | 10x10 | 1.92 | 15 | 0.1 | 60dB/ 20MHz |

**Tab. 4.** Layout design specifications and technical characteristics of state-of-the-art *HTS* microstrip filters developed by different research groups (after [14]).

Mazierska, Jacob [14] conclude that the High Temperature Superconducting (*HTS*) materials exhibit exceptionally low surface resistance $R_s$, allowing for manufacturing of small size microstrip planar filters with very sharp skirts of *Q*-factors even 90,000 and insertion loss of 0.2*dB*. This results in considerable out-of-band signals rejection, and very small in-band receiving/transmitting signals degradation correspondingly. Therefore, the *HTS* planar filters based on novel hairpin, spiral and patch resonators, enclosed in air evacuated cryo-coolers together with low noise amplifiers make superior cryogenic front end receivers for wireless communications as compared to conventional hardware [14]. Mazierska, Jacob [14] note that even though the market penetration of *HTS CRFE* receivers is currently relatively small, this may change rapidly, when this new technology receives wide acceptance by the *4G* wireless service provides [14].

## 9.5. Tuning and Trimming Techniques for High Temperature Superconducting (HTS) Microstrip Filters Characteristics Optimization.

Summarizing the discussion on the *HTS* microwave filters for wireless communications, it is necessary to discuss the *HTS* microwave filters ***tuning*** and ***trimming*** used with the aim of *HTS* microwave filters characteristics optimization.

O. G. Vendik, I. B. Vendik, Kholodniak [45] provided some comments on the design of *HTS RF* filters by stating that the main advantage of the *HTS* filters consists in their extremely low in-band insertion loss level. Planar-structure filter configurations are characterized by a small size and weight. Development of such filters requires precise modeling, which should take into consideration the *HTS* film contribution to the propagation parameters of the transmission line sections used as a basis for microwave filter resonators. The phenomenological description of the microwave surface impedance [102-108] permits high-accuracy prediction of the *HTS* filter characteristics for any desired temperature [45].

Vendik [45] explains the *HTS* microwave filters *tuning* and *trimming* techniques, stating that the design and testing of planar microwave filters makes often use of specially introduced trimmers. Usually, the trimmers are small dielectric screws [45]. By properly rotating the screw, one can change the distance between the screw and a filter component [45]. In this way, the couplings between the filter components and their resonant frequencies can be varied so as to correct the filter characteristics [45]. It is clear that the trimming procedure is poorly suited for mass production of microwave components [45].

Willemsen [71] provides a comprehensive analysis on the techniques used for the *HTS* microwave filters *tuning* and *trimming*, making a comment that it is very common for high performance filters to make use of mechanical tuning elements to correct for manufacturing tolerances, which can affect the center frequency of the filter, as well as its return loss. Willemsen comments that, at *STI*, the dielectric tuners, using sapphire cylinders above the *HTS* microwave filters were initially incorporated [74]. Willemsen [71] notes that these work very well, providing sufficient tuning range to adjust center frequency and return loss. However, they still require precise, fine-detail machining to incorporate the fine threads, required for fine positioning of dielectric rod, and since these actuators form an integral part of the *RF* enclosure, their cost is shipped with final product [71].

Willemsen [71] comments that another approach was developed at Conductus, where the tuning elements are small pieces of *HTS* attached to a spring clip [75]. The actuator for the spring clips was designed as a separate piece which could be removed from the part once the filter was tuned. In this way, some of the cost is shifted from the product to the tool required to make the part, which can then be amortized over a large number of products [71].

In [76], Willemsen describes an approach for *computer aided filter trimming*, which allows the filter to be trimmed for center-frequency and return-loss without any additional mechanical tuners. The additional features required for trimming are fabricated along with the filter and thus do not add any cost. This approach leads to well tuned filters in their test housing, but by the time the filters are integrated with the rest of the *RF* chain required for the system that includes bond-wires, *LNAs*, cryocables, switches and connectors the return loss of the chain is degraded [71]. In this respect, the mechanical tuner approach can lead to better system performance, as it allows the filter to be tuned so as to compensate for any mismatches in the other components in the *RF* chain [71]. As a first approach, the pre-tuned filters were integrated with the rest of the components in the standard mechanically tuned filter housing, removing 8 or 9 of the 10 tuners required for each filter for a substantial reduction in labour and materials cost for trimming each filter [71].

In parallel, Willemsen [71] developed some approaches for computer-aided trimming of the whole *RF* chain. The resonator frequencies alone are sufficient to optimize the *RF* chain from our mechanical tuning experience, so researchers should not have to add any additional trimming features to the filters. The complexity to the objective function, used in the curve fitting, is added, and the $S_{11}$ and $S_{21}$ are only considered, when fitting, it happens, because the used simplified model of the *RF* chain does not model $S_{12}$ or $S_{22}$ well [71]. Willemsen [71] optimizes only the resonator frequencies to trim the *RF* chain, leaving the extracted coupling values, gain and transmission-line parameters constant during the optimization with very high yields (>95%) [71].

Ohshima, Kaneko, Lee, Osaka, Ono, Saito [117] examined the effectiveness of a trimming library technique for automatically trimming a forward-coupled *HTS* filter with three resonators and obtained the following results in Fig. 77 (1):

1) The sapphire rod trimmers were able to correct the ripple of the filter. The correction sapphire rods were set on the resonator edge and between the resonators.

2) The effectiveness of the trimming library method was confirmed.

3) The trimming rod should be set on the resonator edge and moved up and down.

4) Automatic tuning of such a filter is difficult with rod trimming.

Harada, Sekiya, Kakio, Ohshima [113] developed a bandwidth and a center frequency tuning method for use during the *HTS* microstrip filter in Fig. 77 (2). It is evident that the *HTS* microwave filters *tuning* and *trimming* are very complex and time consuming design stages during the *HTS* microwave filters design [113, 114].

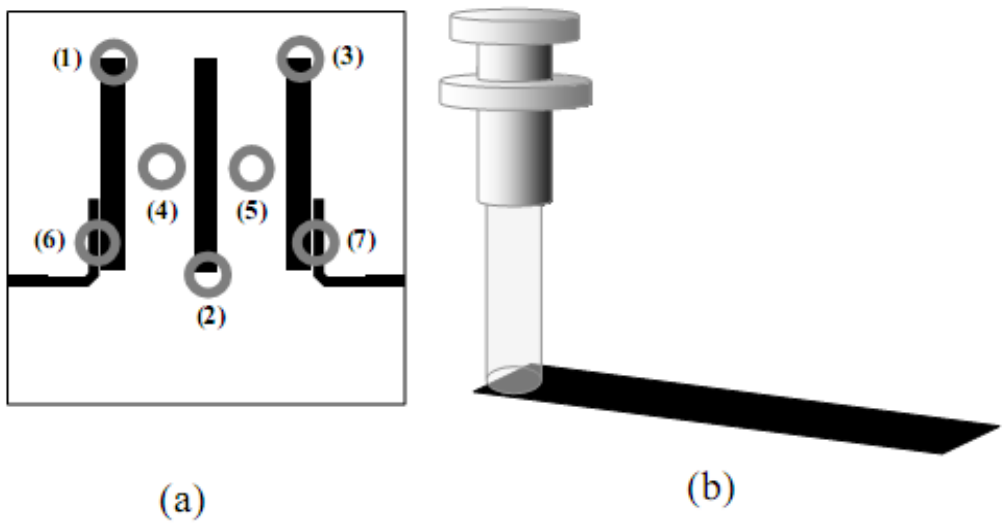

1)

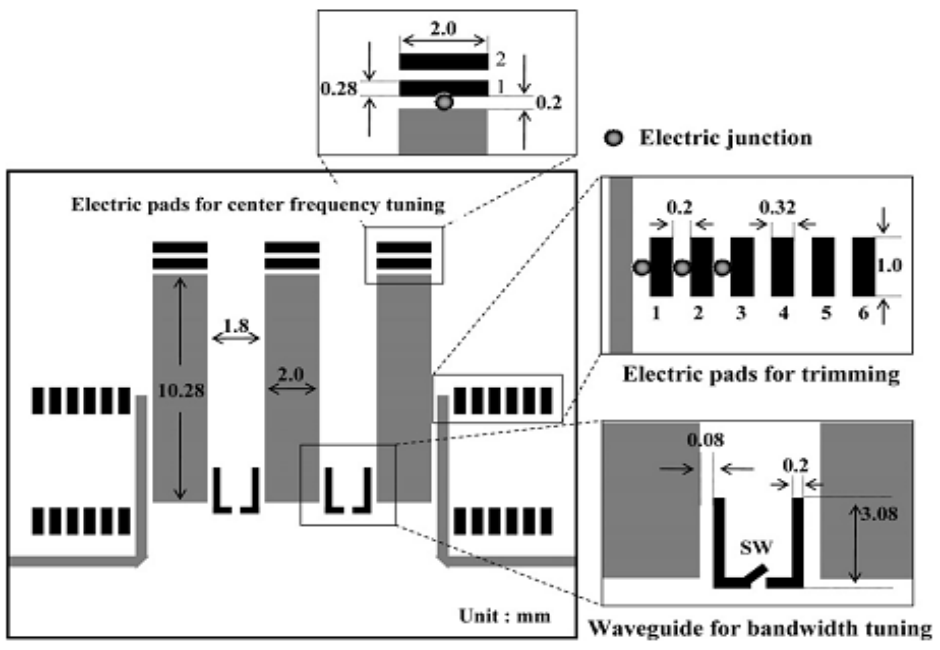

2)

**Fig. 77. 1)** Schematic drawing of filter configuration of a forward-coupled filter **(a)** and sapphire trimming rod **(b)**: The circles in **(a)** indicate the positions of the rod trimmer (after [117]). **2)** Layout of center frequency and bandwidth tunable filter with trimming and tuning elements (after [113]).

Tanaka, Tatsunokuchi, Akiya, Saito, Ohshima [122] optimized the 4-pole *HTS BPF* microstrip filter layout to improve its microwave power handling capability by expanding the width of microwave strip line of second resonator in Fig. 78.

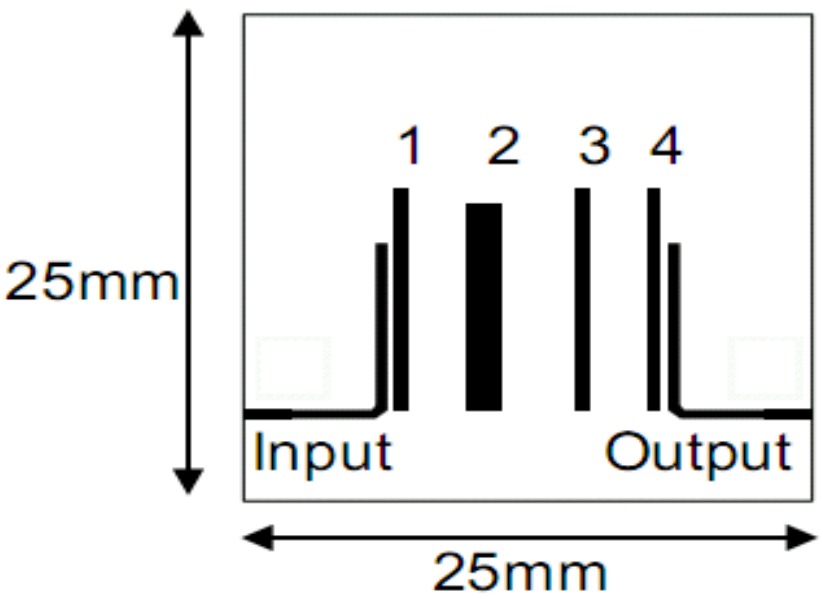

1)

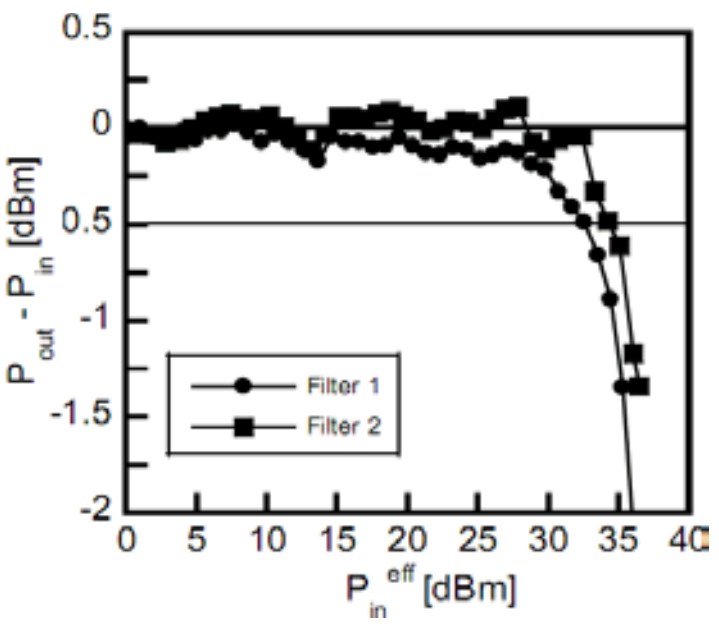

2)

**Fig. 78.** 4-pole *HTS BPF* microstrip filter layout (1) with improved power handling capability (2) by expanding width of strip line of second resonator (after [122]).

D. E. Oates, Anderson, Dionne [118] used the lightly coupled *ferroelectric films* to trim the individual resonators in magnetically tuned *HTS* microstrip filters.

Zhao, Li, Lei, Tian [30] noticed that the operation temperature of *HTS* microwave components should be below $0.8\ T_C$ as the surface resistance $R_s$ increases rapidly above this temperature as explained by Piel, Müller [66]. Mansour [11], discussing the *thermal stability* of the *HTS* microwave filters, makes his comments that the inductance of a superconductor line has two parts: a *geometric* *inductance* and a *kinetic inductance*. Magnitudes of kinetic inductance depend on the superconductivity penetration depth, which is a function of temperature *T.* At the *T* close to the *Tc*, the penetration depth increases rapidly, resulting in a strong increase in the kinetic inductance, hence *the thermal stability of HTS filters is essential.*

The designs of high performance *HTS* microwave filters were performed by Bonetti, Williams [77], Lyons, Withers [1], Chaloupka, Kolesov [9]; O. G. Vendik, I. B. Vendik, Kholodnyak [45]; I. B. Vendik, Kondratiev, Svishchev, Leppyavuori, Y(J)akku [41]; I. B. Vendik, Kondrat'ev, Kholodnyak, Gal'chenko, Svishchev, Deleniv, Goubina, Leppävuori, Hagberg, Yakku [42]; Sitnikova, I. B. Vendik, Vendik O G, Kholodnyak, Tural'chuk, Kolmakova, Belyavsky, Semenov [47]; Kermorvant, Van der Beek, Mage, Marcilhac, Lemaitre, Briatico, Bernard, Villegas [109]; Pal [57, 58]; Tsuzuki, Hernandez, Willemsen [50]; Simon, Hammond, Berkowitz, Willemsen [68]; Prophet, Musolf, Zuck, Jimenez, Kihlstrom, Willemsen [69]; Knack [78, 79]; Zhou, Xia, Zuo, Zhao, Fang, Yan [80]; Mateu, Booth, Collado, O'Callaghan [81]; Shivhare, Vaidya, Srinivasulu, Lai, Balasubramanyam [82], Xubo Guo, Xiaoping Zhang, Bicsong Cao, Zhenghe Feng, Bin Wei, Huili Peng, Shichao Jin [83], Wang, Wu, Lin, Chang [84], Sekiya, Yamamoto, Kakio, Saito, Ohshima [85], Jacob, Mazierska, Knack, Takeuchi [72], Futatsumori [91], Futatsumori, Furuno, Hikage, Nojima, Akasegawa, Nakanishi, Yamanaka [92], Futatsumori, Hikage, Nojima [93], Yamanaka, Kurihara, Akasegawa [96], Mansour [10, 11, 98, 99, 100], Talisa [12, 13], B. Oh, H.T. Kim, Y.H. Choi, S.H. Moon, P.H. Hur, M. Kim, S.Y. Lee, A.G. Denisov [121], and by many others.

The modern trends and achievements in *R&D* towards the *HTS* microwave filters design and fabrication were reviewed by Withers [1, 2]; Shen [3]; Lancaster [4]; I. B. Vendik, O. G. Vendik, Kaparkov [5], Jha [6], Van Duzer, Turner [7]; Hong, Lancaster [8]; Hong, Lancaster, Jedamzik, Greed [37]; Weinstock, Ralston [2]; Weinstock, Nisenoff [9, 10, 11, 12, 13]; Chaloupka, Kolesov [9]; Chaloupka, Kaesser [15]; Mazierska, Jacob [14]; Yamanaka, Kurihara [16, 17]; Xia, Zhou, Zuo [86], Talisa [12, 13], Mansour [10, 11, 98, 99, 100], Scharen [55], Knack [78, 79], Ohshima, Kaneko, Lee, Osaka, Ono, Saito [117], D. E. Oates, Anderson, Dionne [118], Nisenoff, Pond [120], and by some others.

**Summary.**

Chapter 9 presents a short review with some information on the design, tunning and trimming techniques to develop the High Temperature Superconducting (*HTS*) microwave filters for application in Reconfigurable Cryogenic Transceiver Front End (*RCTFE*) in wireless communication systems. The *state-of-the-art HTS* microwave filters designs issues are reviewed with the aim to show up the frequently used *HTS* microwave filters design layouts with their frequency responses, and describe a number of possible design problems and limitations, which can arise, because of the nonlinearities appearance in *HTS* thin films at increased microwave power levels.

Considering the Reconfigurable Cryogenic Transceiver Front End (*RCTFE*) in Fig 1, it is explained that the *HTS* microwave filters are developed for the use in the *receiver* as well as *transmitter*. The most effective filtering of digitally modulated microwave signals can be achieved with the application of the *raised cosine HTS microwave filters.* The application of carefully designed *HTS* microstrip filters in combination with Low Noise Amplifier (*LNA*) in the receiver in Reconfigurable Cryogenic Transceiver Front End (*RCTFE*) can make it possible to reach the *better signal-to-noise ratio*; *greater input signal selectivity*; and s*mall insertion losses,* leading to the much more better quality of digital communications with low Bit Error Rate (*BER*) over wireless channel. The use of carefully designed *HTS* microstrip filters in combination with high power linear amplifier in the transmitter in Reconfigurable Cryogenic Transceiver Front End (*RCTFE*) can decrease the *phase noise*, caused by the nonlinearities; lower the *adjacent channel leakage power ratio* (*ACLR*) significantly, introduce the *small insertion losses,* resulting in the high quality digital communications with low Bit Error Rate (*BER*) over wireless channel, small co-channel interference generation, and effective energy consumption at basestation. All the main advantages of the *HTS* microwave filters applications in Reconfigurable Cryogenic Transceiver Front End (*RCTFE*) are summarized below:

1. *Better signal-to-noise ratio,*
2. *Greater input signal selectivity,*

3. *Small insertion losses*,

4. *Better adjacent channel leakage power ratio* (in transmitter), and

5. *Compact design*.

Discussing the miniaturization of dimensions of High Temperature Superconducting (*HTS*) microwave filters, the author of dissertation analysed the modern *HTS* microstrip filters designs made by other researchers, and focused on the three main types of *HTS* miniature microwave filters in Tab. 2:

1. *Delay line filters,*

2. *Lumped element filters,*

3. *Filters based on slow wave structures.*

Presenting the information on the advanced layouts of *HTS* microstrip resonators and filters in Reconfigurable Cryogenic Transceiver Front Ends (*RCRFE*), the author of dissertation provided some selected examples of the *HTS* microwave resonators layouts for the use in *HTS* microwave filters in Fig. 16, and the schematic of layouts of *HTS* microwave multi-pole filters with their frequency responses in Tab. 3. Next Tab. 4 contains the layout design specifications and technical characteristics of *HTS* microstrip filters, and it is intended for the *state-of-the-art HTS* microstrip filters technology overview purposes only.

Summarizing the research review on the *HTS* microwave filters in Reconfigurable Cryogenic Transceiver Front Ends (*RCRFE*) in wireless communications, the author of dissertation discussed the important problems of *HTS* microwave filters ***tuning*** and ***trimming*** with the purpose of *HTS* microwave filters characteristics optimization. It is explained that the *HTS* microwave filters *tuning* and *trimming* are very complex and time consuming stages during the *HTS* microwave filters research and development (*R&D*) process. The *trimmers* are small dielectric screws, for example the sapphire cylinders above the *HTS* microwave filters, which can be rotated with the aim to change the distance between the screw and a filter component. Using this approach, the couplings between the filter components can be varied, resulting in a change of resonant frequencies of *HTS* microwave resonators, thus the *HTS* microwave filter's central frequency, return loss and some other characteristics can be adjusted as required. The *tuning elements* can also be made of small pieces of *HTS* attached to a spring clip.

Discussing the issues on the *thermal stability* of the *HTS* microwave filters, it is noted that, at temperatures close to the critical temperature *Tc*, the penetration depth $\lambda$ in superconductor increases rapidly, resulting in a strong increase in the kinetic inductance, which may lead to the change of *HTS* microwave filter's main parameters, hence *the thermal stability of HTS microwave filters is essential.* The normal operational temperature of *HTS* microwave filters is equal to *0,8Tc*.

A complete list of leading researchers, who made significant research contributions to the advanced innovative designs of High Temperature Superconducting *(HTS)* microwave filters is compiled at the end of Chapter 9.

Considering the numerous reviewed *HTS* microstrip filters designs, completed by different research groups, the author of dissertation would like to make the following important comments on possible innovative strategies to design novel highly linear *HTS* microstrip filters with enhanced microwave power handling capabilities, reduced microwave power level of intermodulation distortion (*IMD*), high interception point three (*IP3*), low insertion loss characteristics and compact design at high magnitudes of applied microwave power.

1. Improvement of material properties of High Temperature Superconductors (*HTS*) through the synthesis of *high quality HTS* thin films without crystal defects or imperfections and with big magnitude of critical magnetic field $H_{c1}$, leading to small non-linear response of microwave resonator to high level of microwave power.

2. Improvement of material properties of High Temperature Superconductors (*HTS*) through the use of composite *HTS* thin films with *green phase nano-clusters, oxide nano-clusters*, *nano-impurities* to improve the *Abricosov magnetic vortices* pinning, and hence increase the magnitude of critical current *Ic*, leading to small non-linear response of microwave resonator to high microwave power at $H_{rf}>H_{c1}$.

3. Optimization of geometrical parameters of *HTS* thin films by *increasing the thickness of HTS thin films*, using of *stacks with many HTS thin films layers*, or by *increasing the width of microstrip line in HTS resonator* in *HTS* microstrip filter.

4. Optimization of geometric structure of *HTS* thin films, deposited on substrate, by creation of *sliced HTS microstrip lines.* For example, the use of *sliced HTS microstrip lines* for the design of microwave resonators in a transmitting filter.

5. Optimization of geometric structure of *HTS* thin films in *HTS* microstrip filter by use of *HTS microstrip line with slots* instead of solid microstrip line to create the *split resonators*, aiming to reduce the peak current density at outer edges of *HTS* microstrip and avoid high current density *HTS* microstrip discontinuities.

6. Optimization of geometry of *HTS* microstrip filter by *rounding all the edges and corners* of *HTS* thin films in microstrip filters in receivers and in transmitters. For instance, the use of *HTS split open-ring resonators* (*SRR*) in microstrip filters in a transmitter.

7. Miniaturization of high pole *HTS* microwave filters with increased microwave power handling capability, by using the stripline (*SL*) resonators with shorter distance between resonators, resulting from a weak coupling. The modified *SL* filter with increased spacing between resonators to reduce maximum surface current has smaller size in comparison with the *HTS* microstrip line (*MSL*), dual-mode or bulk resonators.

8. Optimization of *HTS* microstrip filter design, using the different design techniques toward the intermodulation distortion reduction in *HTS* microstrip filters. The *ferroelectrics* may be used for maximum *IMD* reduction with a minimum effect on losses in *HTS/FER* microstrip filters. For example, the suppression of *IMD,* generated by *HTS* materials, with the use of a *nonlinear ferroelectric segment* for nonlinear pre-distortion in *HTS* band-pass filter.

9. Introduction of *quality control* during microwave filters fabrication to avoid the edge defects in *HTS* thin films, which can lead to the decrease of critical magnetic field $H_{c1}$. The absence of edge defects in *HTS* thin films will significantly improve the characteristics of HTS microstrip filter.

10. Maintenance of *thermal stability* of *HTS* microwave filters by constantly adjusting the real operational temperature to normal operational temperature, which is equal to *0,8Tc*, where *Tc* is the critical temperature of *HTS*.

11. Improvement of the *packaging* of *HTS* microwave filters to avoid the vibration related problems with their tunning in space and airborne applications.

Let us add that the *RCRFE* can play important role in the *5G* wireless communications networks, which will use the adaptive transceiver technology, reaching the *1,056 Gbps* wireless data throughput at *28 GHz* band in [135].

## Appendix I.

At what magnitude of external magnetic field $H_e$ will the *HTS* samples of different geometric form have a transition to the mixed state of superconductor in a microwave resonator? This problem is important for the understanding of next problem: Why does the electromagnetic wave with the same amplitude of magnetic field $H_{rf}$ and same microwave power P in both a cavity resonator and a microstrip resonator make different influences on the nonlinear effects? The solution of problem is stipulated by the fact that in a cavity resonator the magnetic field is parallel to *HTS* thin film, whereas in the microstrip resonator it is directed perpendicularly to *HTS* thin film.

It is well known, that any magnetized sample as an ellipsoid in an external magnetic field has the demagnetization factors, directed along the principal axis, which are in the total equal to the unity in [1]

$$N_x + N_y + N_z = 1.$$

Thus, the magnetic field, which is directed, for example, toward the *Z*-axis, will have the value on the surface of equator of a superconductor in the diamagnetic state equal to

$$H_{iz} = H_e / (1 - N_z). \quad \text{(A.1)}$$

For a plane ellipsoid of revolution $(x^2 + y^2)/a^2 + z^2/c^2 = 1$, for which semi-axis a > c, the demagnetization factors are equal to $N_x = N_y = (1–N_z)/2$. The value $N_z$ is

$$\mathbf{N_z = \frac{1+k^2}{k^3}(k - arctg(k))}, \quad \text{(A.2)}$$

where $k = [(a/c)^2 - 1]^{1/2}$. The plane spherical ellipsoid with the ratio $a/c=10$, situated in the external magnetic field $H_e < H_{c1}$, which is parallel to the *Z*-axis, is shown by author of dissertation in Fig. 1:

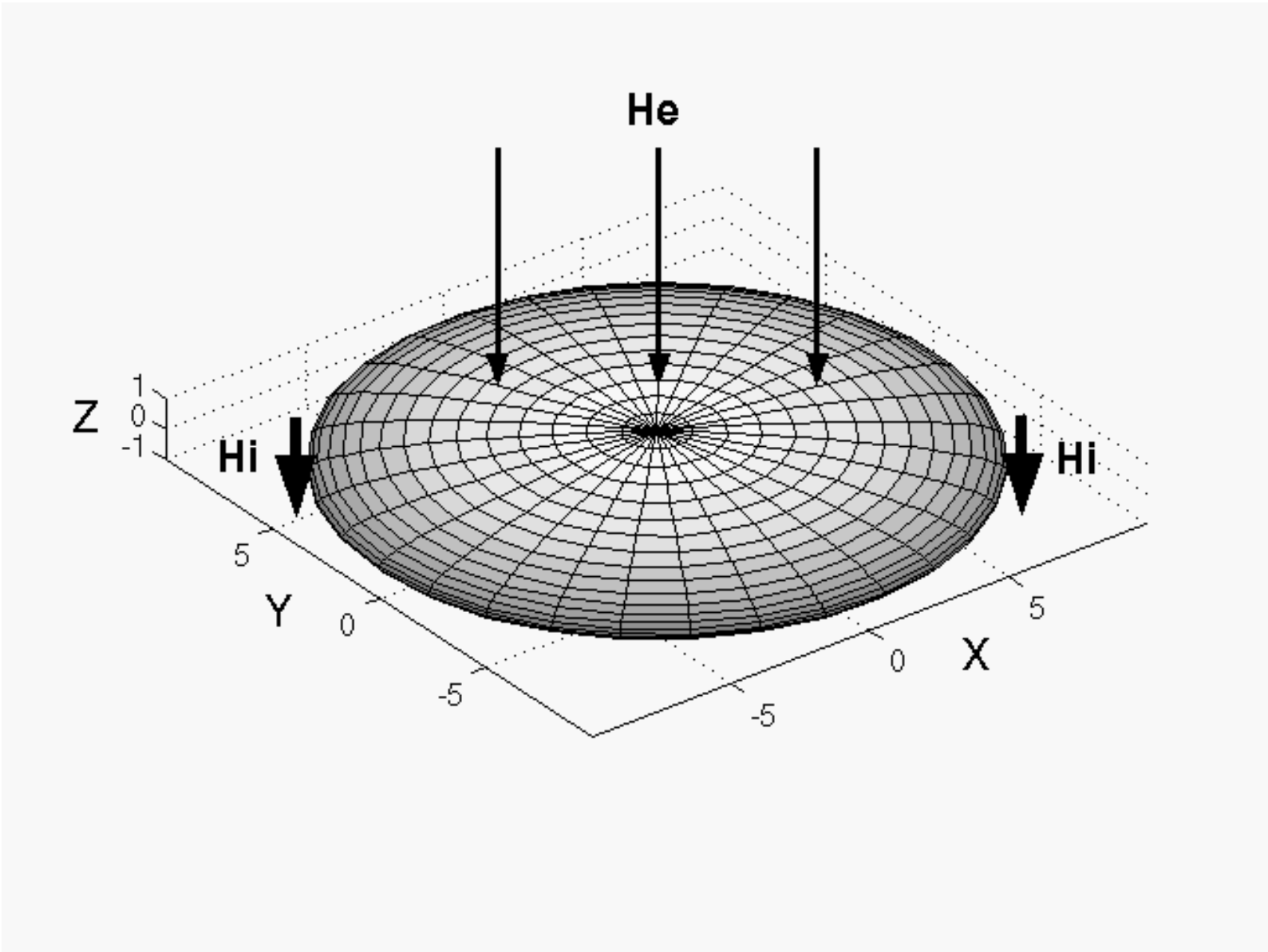

**Fig. 1.** Superconducting plane spherical ellipsoid in external magnetic field $H_e$ || *Z* and $H_e < H_{c1}$, *a/c =10.*

The external magnetic field $H_e$ will penetrate into the sample, when the magnetic field $H_i$ on the equator of an ellipsoid becomes equal to magnitude of critical magnetic field $H_{c1}$. In this situation, the external magnetic field $H_e$ is less than critical magnetic field $H_{c1}$ and equal to

$$H_e = (1 - N_z)\, H_{c1}. \qquad (A.3)$$

Utilizing the equation (A.2), we shall discover the value of the demagnetization factors for the plane ellipsoids with the ratio of semi-axis from 2 up to 100. The graph of this dependence is shown in Fig. 2.

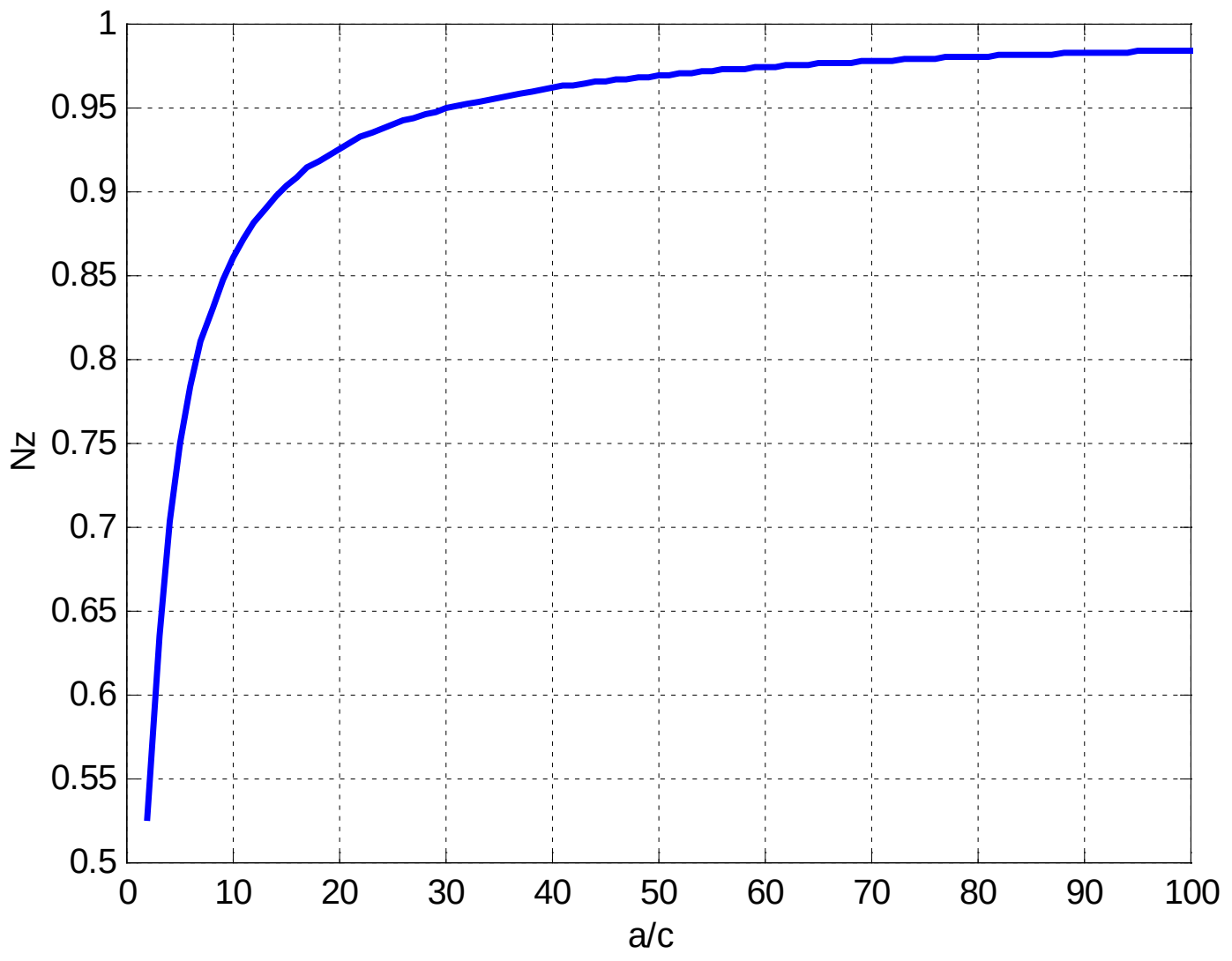

**Fig. 2.** Demagnetization factor $N_z$ of the plane ellipsoid versus the ratio *a/c* for external magnetic field $H_e$ || $Z$.

It is visible, that for $a/c = 10$, the demagnetization factors $N_z = 0.8608$. If the critical magnetic field of a superconductor $H_{c1} = 50$ *Oe*, the field $H_i = H_{c1}$ on the equator of an ellipsoid, when the external field $H_{ez} = H_{c1}\ (1–N_z) = 6.96$ *Oe*. Therefore, at such geometry the appropriate nonlinear behaviour $R_s(H_{rf})$ will begin to be observed in a small magnetic field in comparison with $H_{c1}$. The graph of dependence of decrease of relative external field $H_i/H_{c1}$, at which there is a mixed state in a *HTS* sample, on the value a/c for the orientation $H_e$ || $H_z$ is shown in Fig.3.

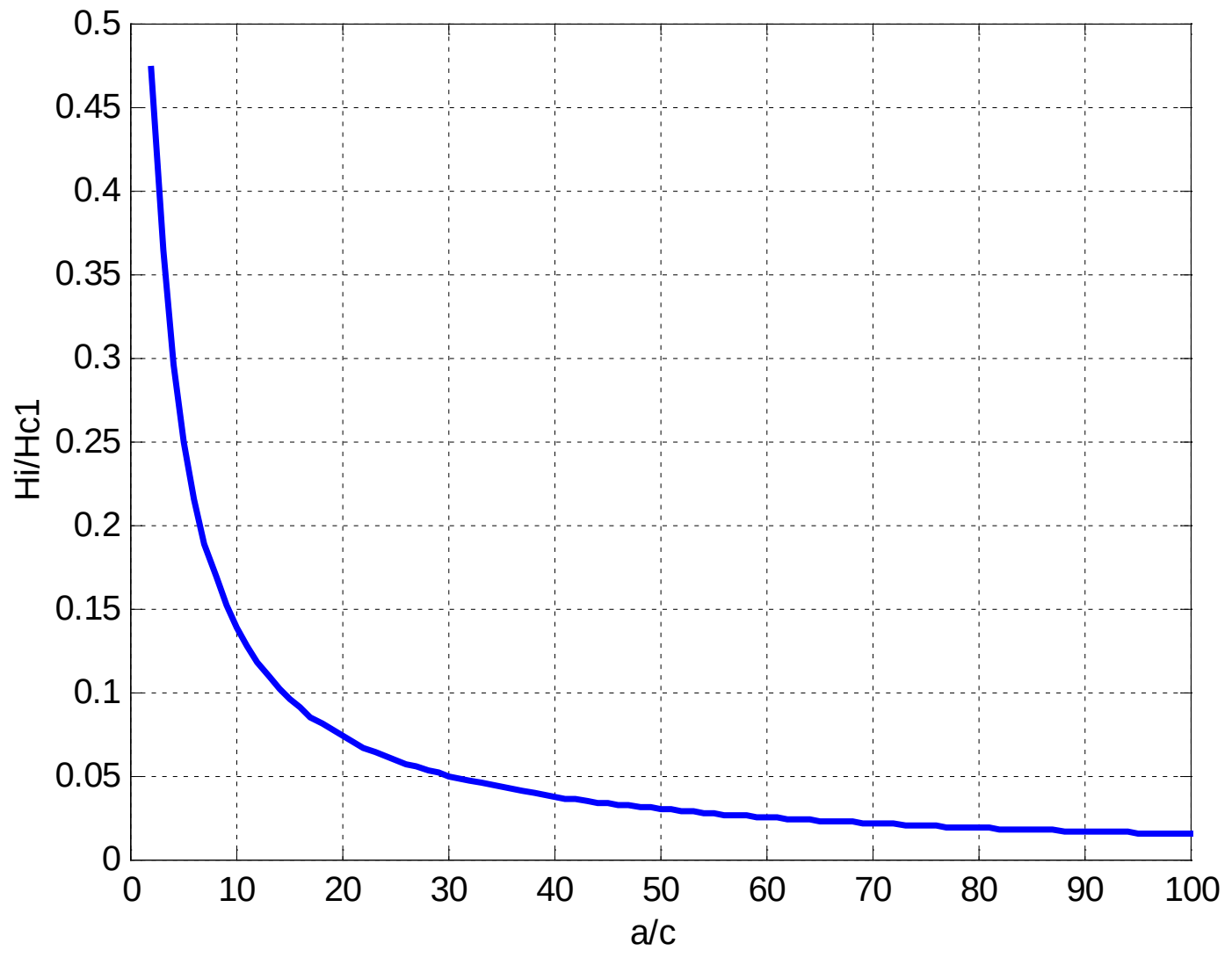

**Fig. 3.** Dependence of the relative critical magnetic field $H_i/H_{c1}$ on the value of ratio $a/c$ for the superconducting plane spherical ellipsoid.

Thus, the critical magnetic field of *HTS* sample depends on its geometrical form and the direction of external magnetic field $H_e$. In the dielectric resonator used in the measurements setup at Department of Electrical and Computer engineering at James Cook University in Australia, the magnetic field $H_{rf}$ of an electromagnetic wave was parallel to the plane of an *HTS* thin film, and consequently, such effects did not appear to be observed. Therefore, for the given above parameters of a *HTS* sample, the demagnetization factor for the magnetic field $H_{rf}$, directed along the *X*-axis, will be $N_x = 0.0696$ and $H_i = 46.52\ Oe$, if $H_{c1} = 50\ Oe$, i.e. the critical magnetic field of a superconductor and the magnetic field of a superconducting sample are equal approximately. However, in the works by other authors, there could be situations, when the external magnetic field $H_e$ was oriented without the registration of demagnetization factor, which could have an effect on the estimation of the value of the low critical magnetic field $H_{c1}$. For example, it could lead to the supposition that the low critical magnetic field $H_{c1}$ has a small value, whereas actually it happened, because of the influence of a demagnetization factor of *HTS* sample. Therefore, the understanding of this problem is important for the experimental researches. The results of this discussion will be utilized by author of dissertation in further work on the estimation of low critical magnetic field $H_{c1}$ in the experiments with the superconducting microstrip resonators.

The low critical magnetic field $H_{c1}$ can also have a small value, because of the presence of areas with the weak superconducting properties (weak-links) in *HTS* thin films. These areas are located on the grain boundaries of crystals, creating the *HTS* thin film, and can take about 10-15 % of superconductor sample's size. They have the small critical magnetic fields, small critical currents, and frequently represent the systems, which can be circumscribed as the Josephson junctions in [2].

Authors of book would like to denote that, in the important case of the microstrip resonators, the magnetic field $H_{rf}$ is generated in the plane thin film of a superconductor by the ultra high frequency current, which is transporting on its surface. In this case, it is necessary to take into the account both the surveyed earlier time dependence of an electromagnetic wave's amplitude and the space discontinuity of a magnetic field $H_e$ with the purpose of finding the precise value of nonlinear surface resistance $R_s$ and evaluation of influence by magnetic field $H_{rf}$ of electromagnetic wave on $YBa_2Cu_3O_{7-\delta}$ thin films on *MgO* substrates in microstrip resonator at microwaves.

## Appendix II

This appendix contains some brief historical and biographical facts about the scientists, who made the research contributions in the field of superconductivity.

| | | |
|---|---|---|
| 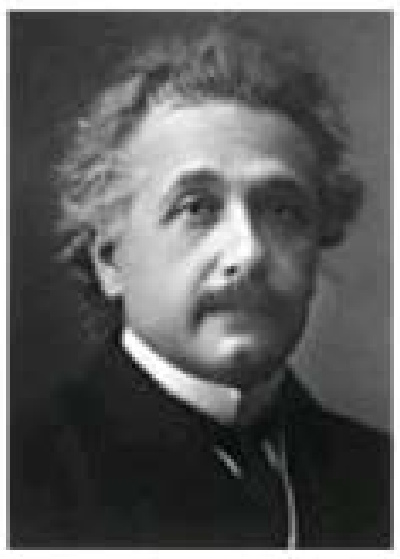 Albert Einstein (1879-1955) | 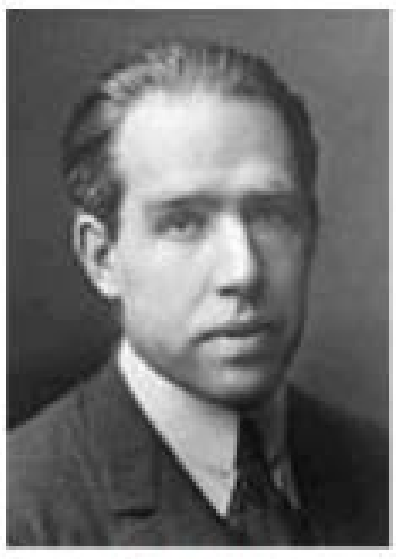 Niels Bohr (1885-1962) | 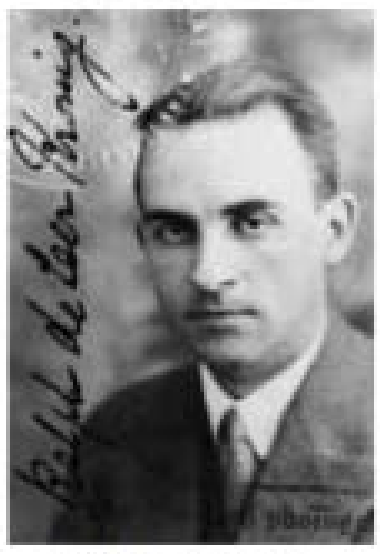 Ralph Kronig (1905-1995) |
| 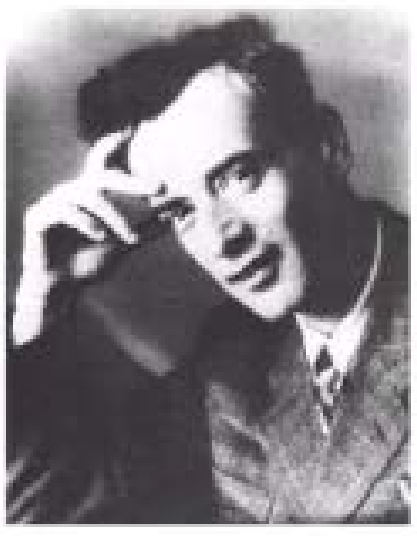 Lev D. Landau (1908-1968) | 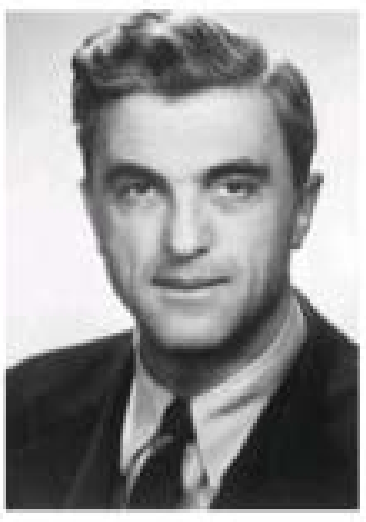 Felix Bloch (1905-1983) | 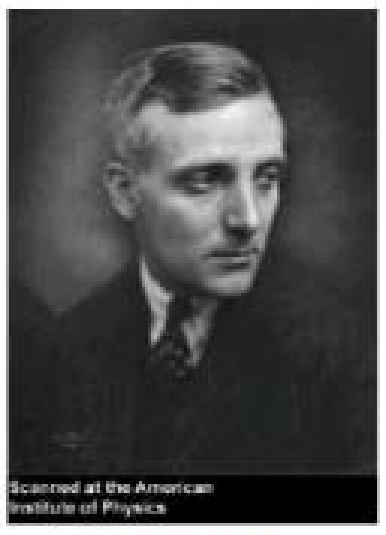 Léon Brillouin (1889 -1969) |
| 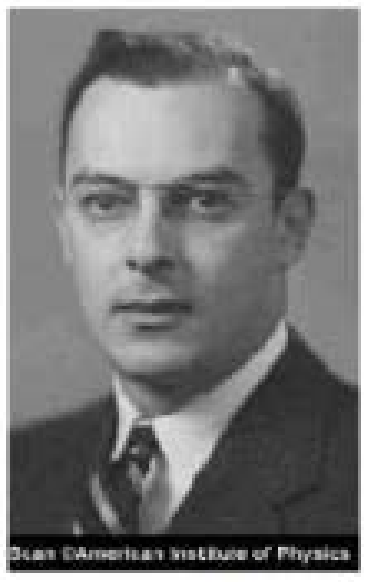 John Bardeen (1908-1991) | 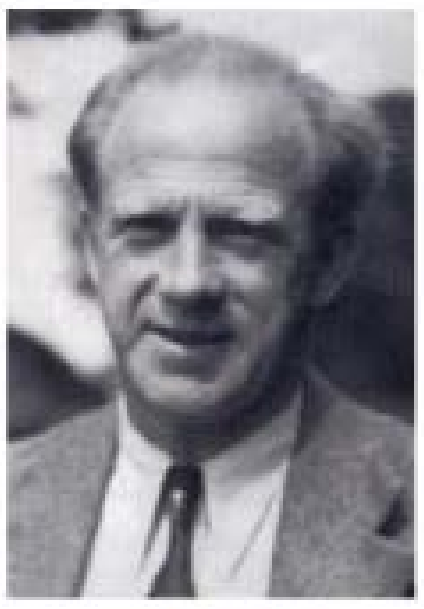 Werner Heisenberg (1901-1976) | 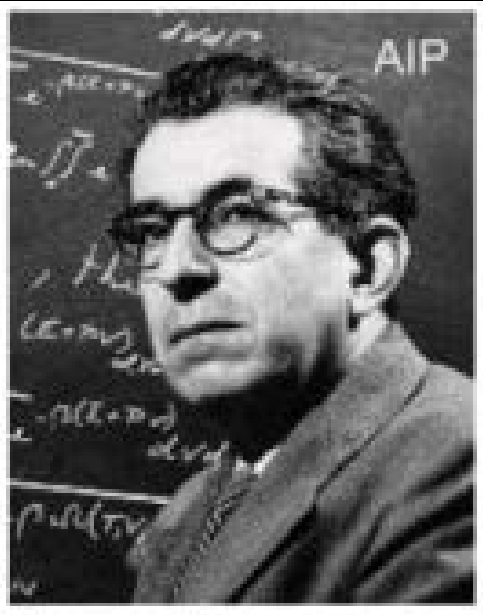 Fritz London (1900-1954) |
| 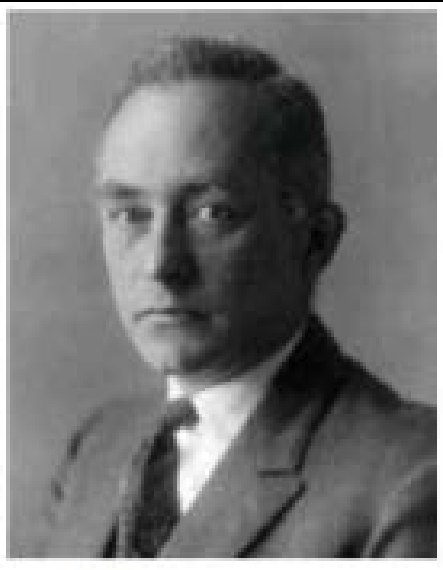 Max Born (1882-1970) | 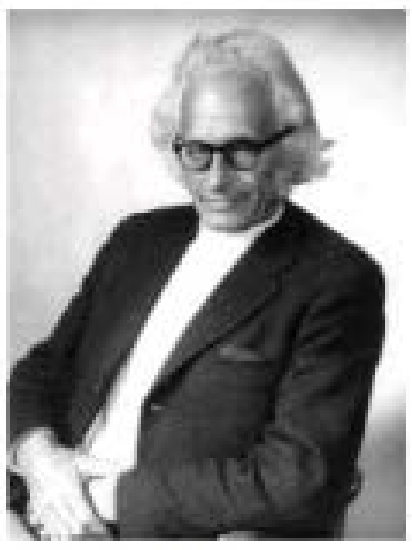 Herbert Fröhlich (1905-1991) | 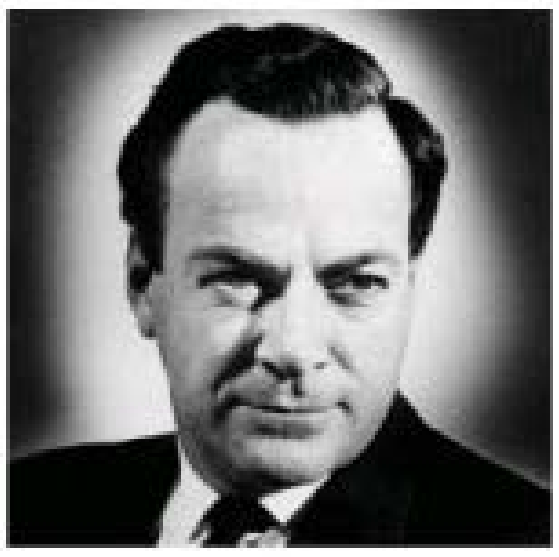 Richard Feynman (1918-1988) |

**Tab. 1.** Well known scientists, who attempted to create the microscopic theories of superconductivity (after [8]).

| | | |
|---|---|---|
| 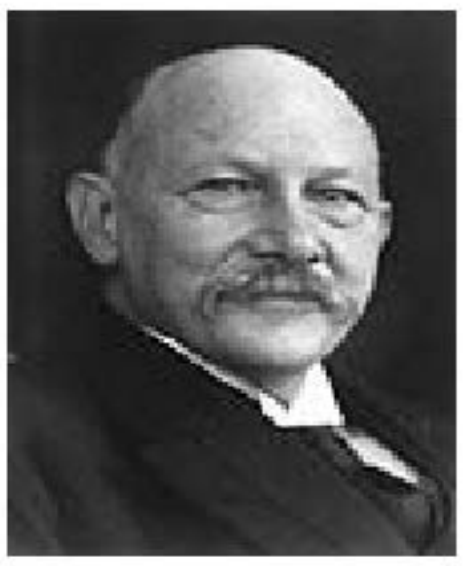 H. Kamerlingh Onnes | 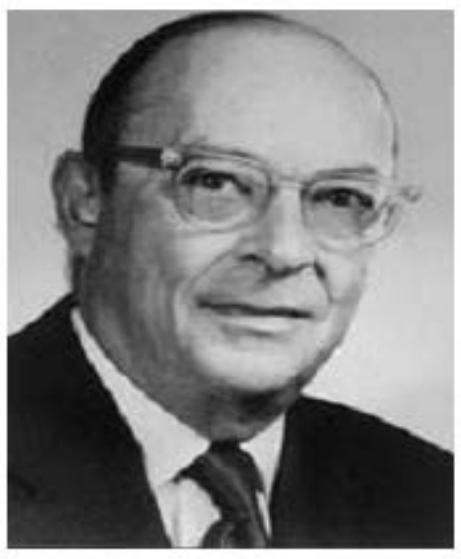 John Bardeen | 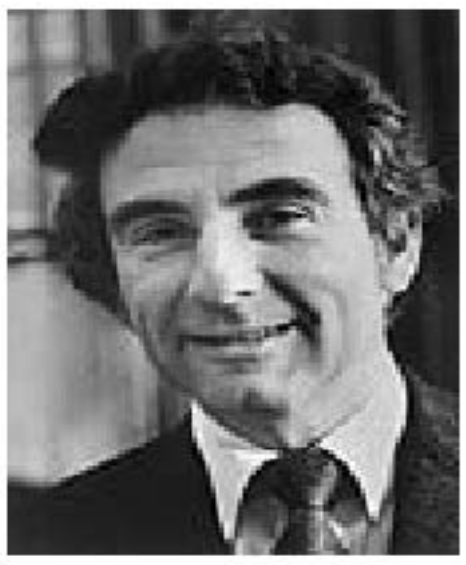 Leon N. Cooper |
| 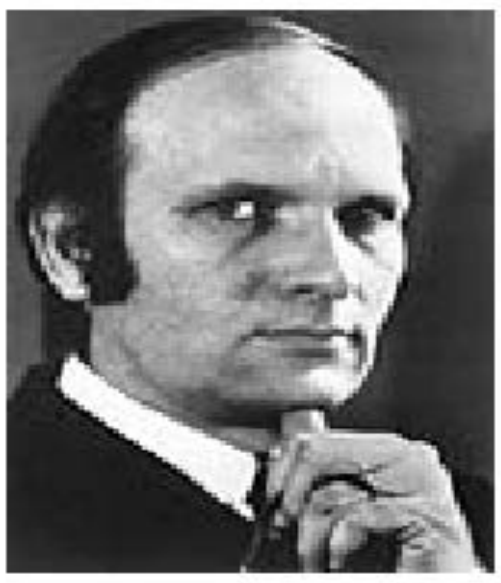 J. Robert Schrieffer | 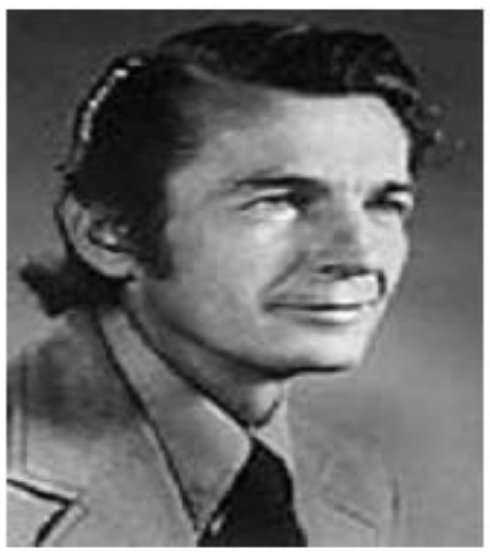 Ivar Giaever | 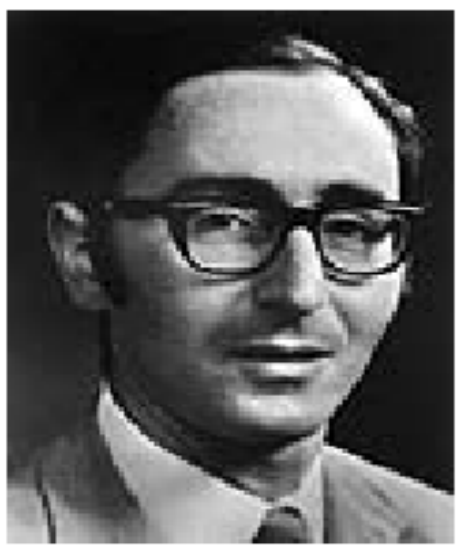 Brian D. Josephson |
| 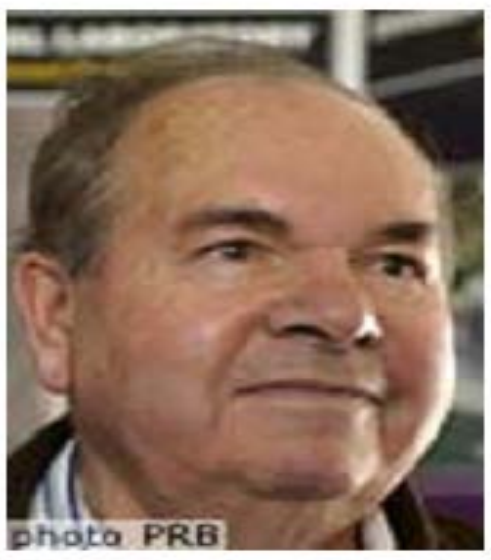  Alexei Abrikosov | 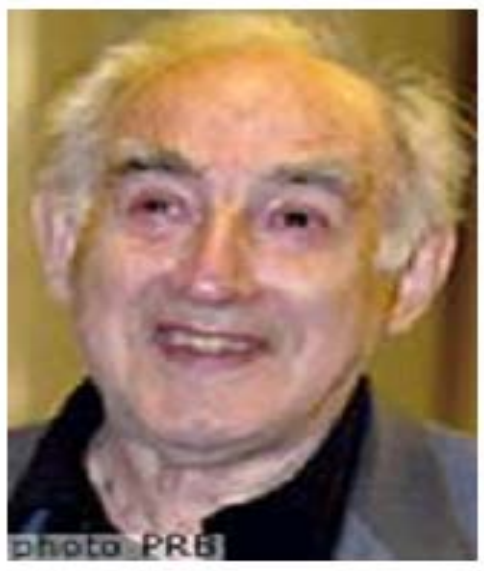  Vitaly Ginsburg | 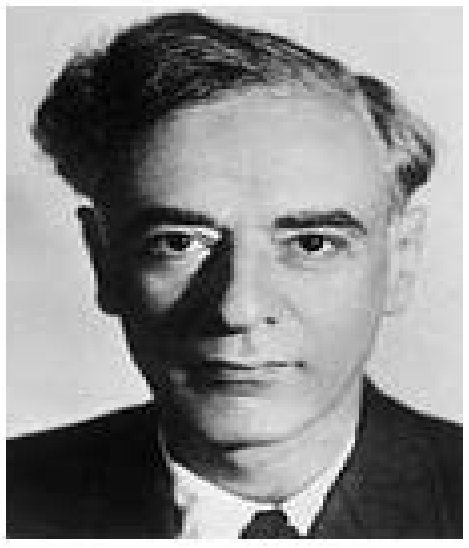 Lev D. Landau |
| 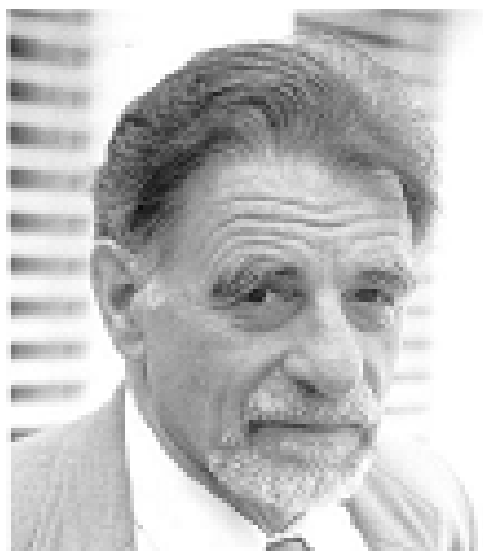 K. A. Muller | 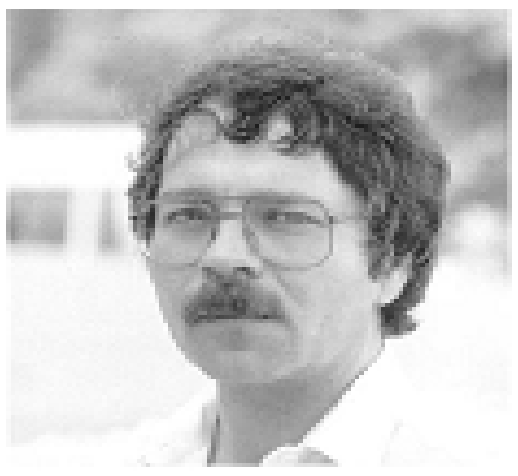 G. Bednorz | 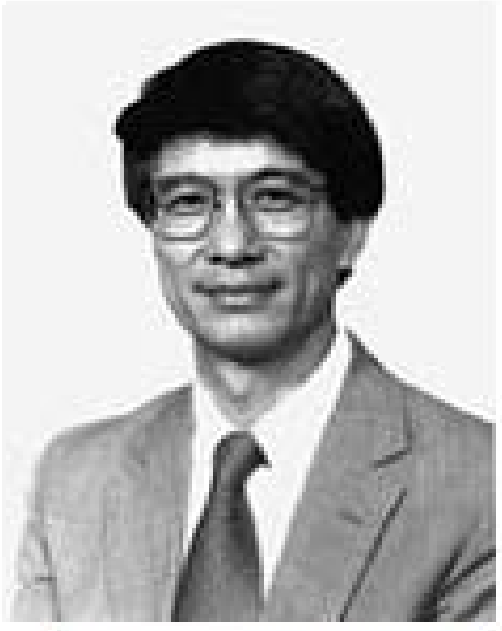 C. W. Chu |

**Tab. 2.** Prominent scientists, who made the significant contributions to the research on the understanding on the nature of the superconductivity (after [4, 5]).

| | | |
|---|---|---|
| 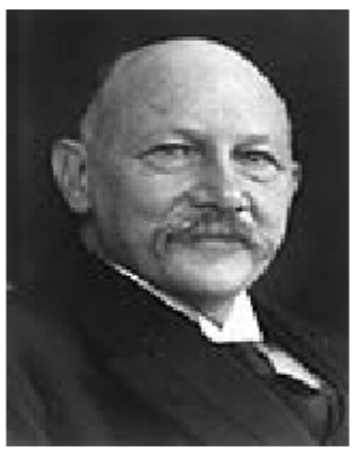<br>Heike Kamerlingh Onnes | 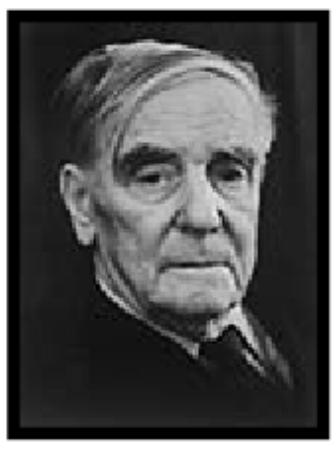<br>Pyotr L. Kapitsa | 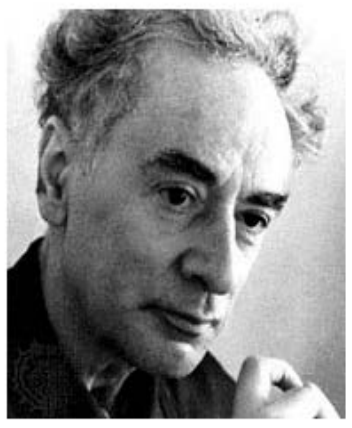<br>Lev D. Landau |
| 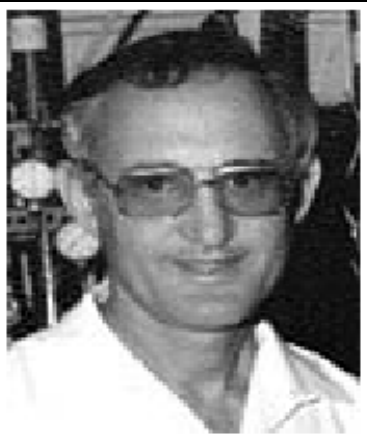<br>David M. Lee | 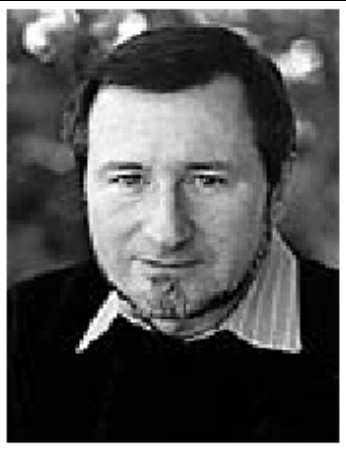<br>Douglas D. Osheroff | 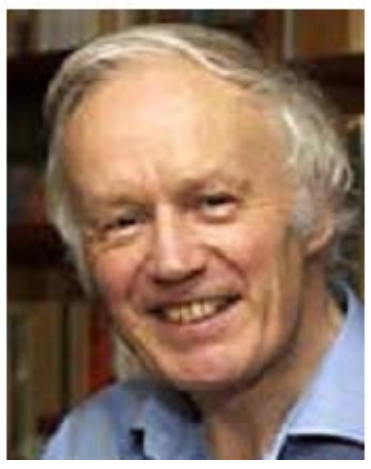<br>Anthony Leggett |
| 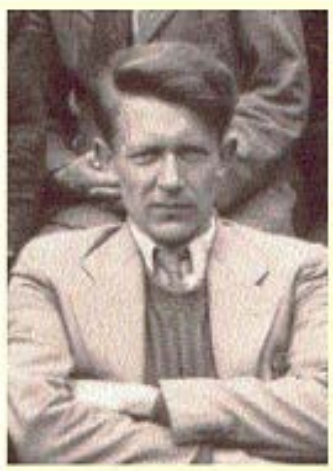<br>J. F. Allen | 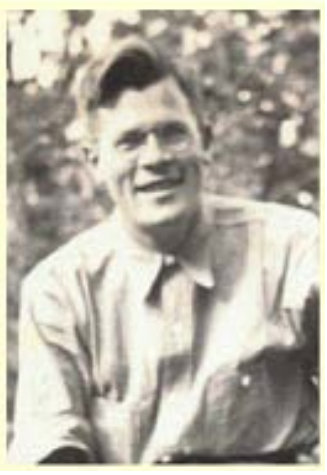<br>A. D. Misener | 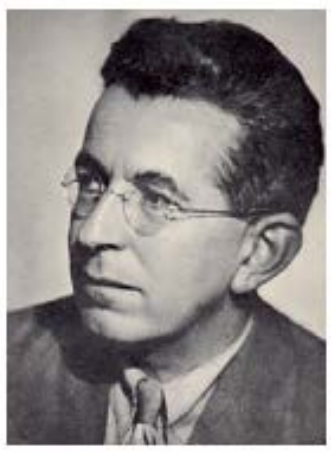<br>Fritz London |
| 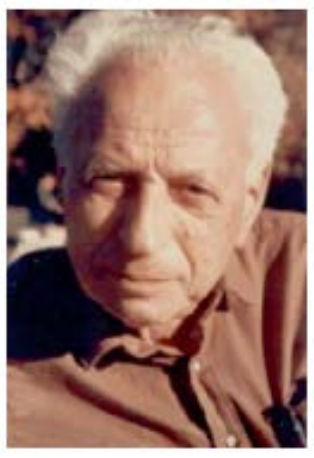<br>L. Tisza | 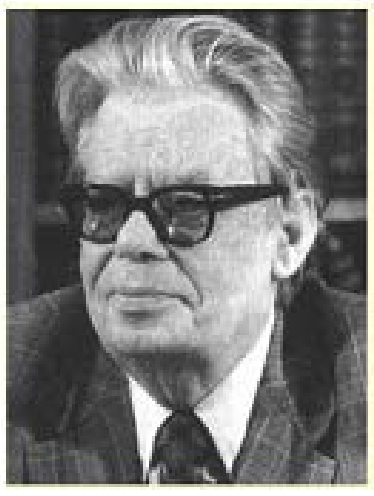<br>N. N. Bogoliubov | 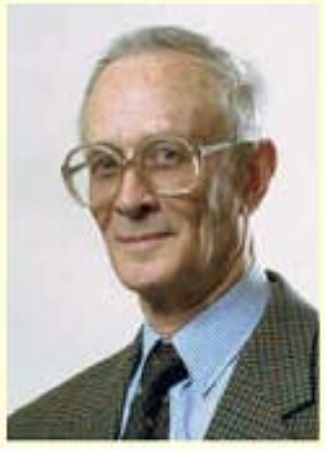<br>S. T. Beliaev |
| 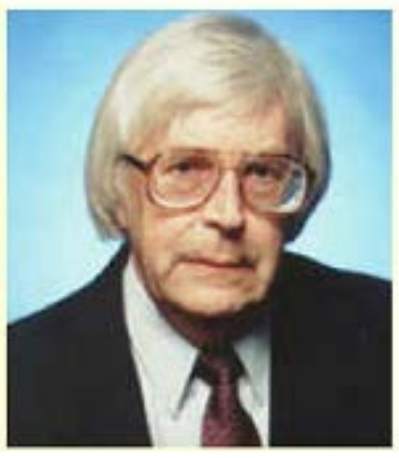<br>L. P. Gor'kov | 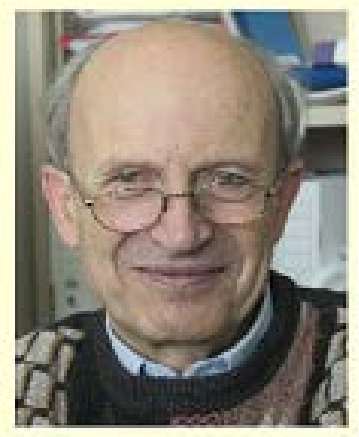<br>Philippe Nozieres | 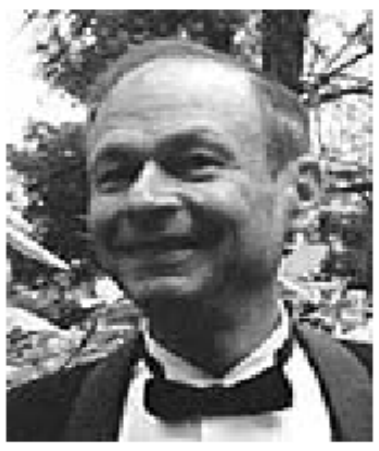<br>Robert Richardson |

**Tab. 3.** World renowned scientist, who contributed to the research on the *He*, $^{3}He$, $^{4}He$ and *superfluidity* (after [4, 5]).

| | | |
|---|---|---|
| 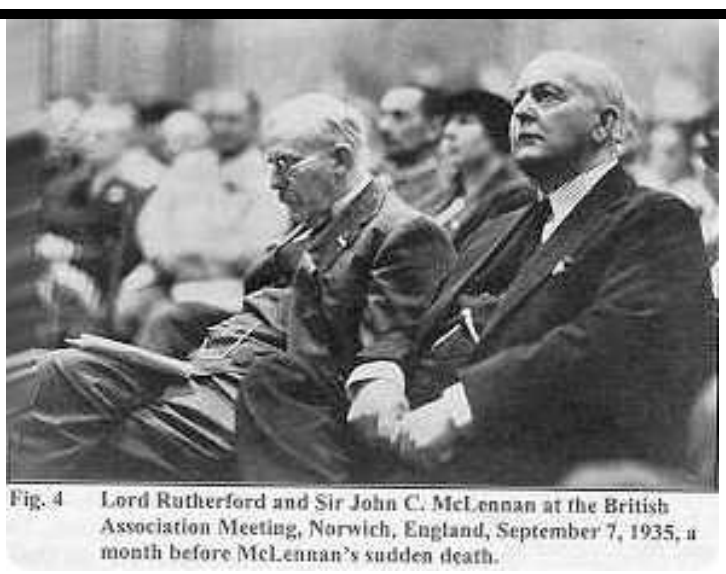  J.C. McLennan (right) | 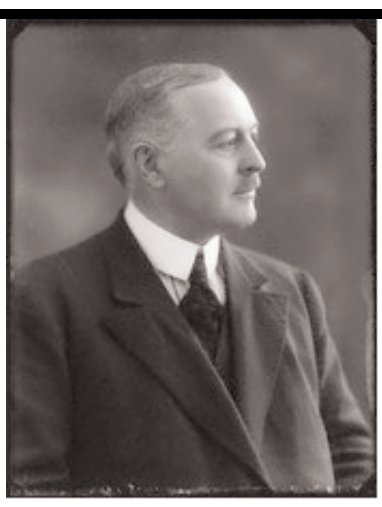 Sir John C. McLennan | 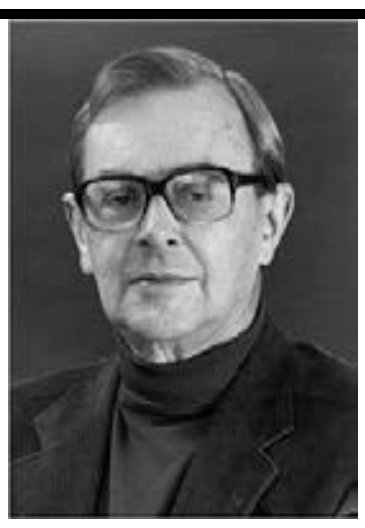 A.B. Pippard |
| 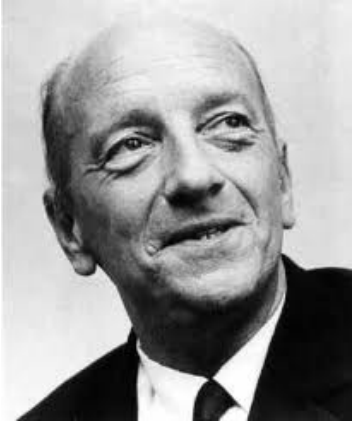 Cor Gorter | 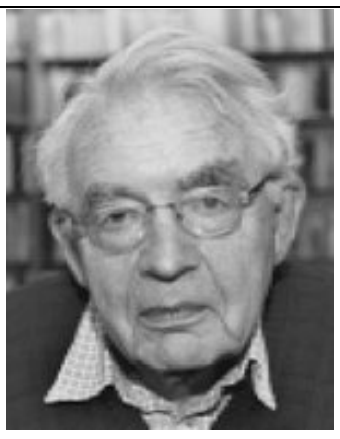 H.B.G. Casimir | 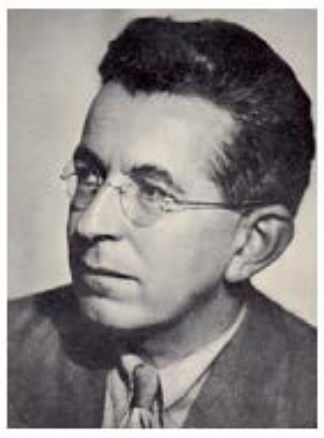 Fritz London |
| 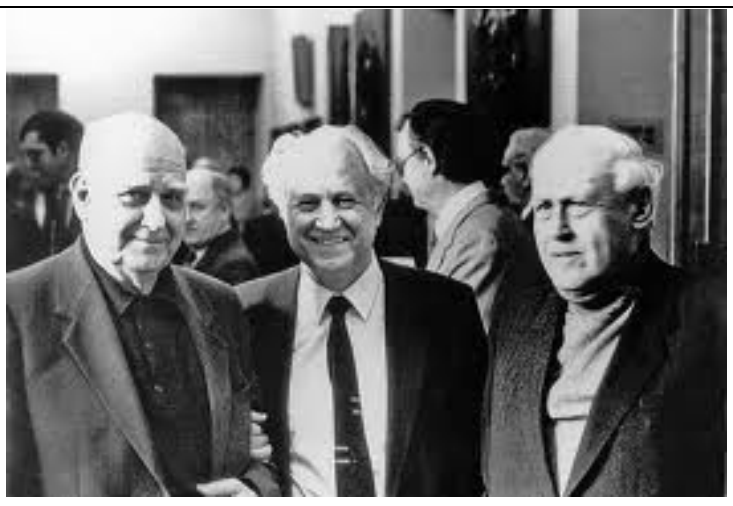 B. G. Lazarev (left) | 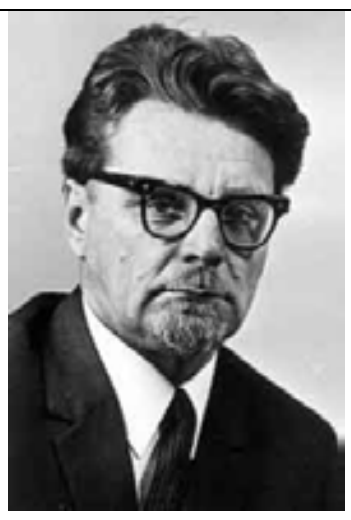 A. A. Galkin | 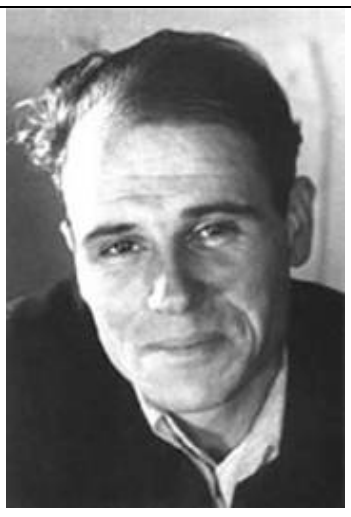 V.I. Khotkevich |
| 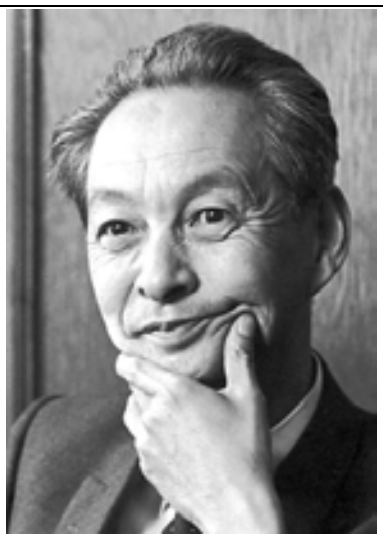 Sin-Itiro Tomonaga | 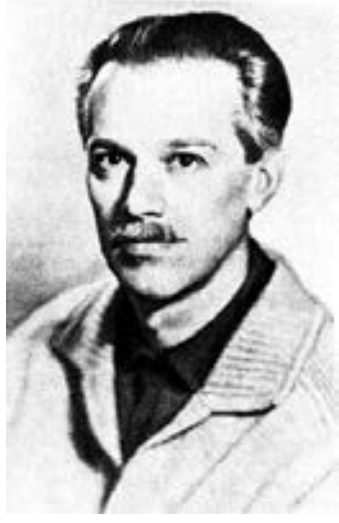 M. S. Khaikin | 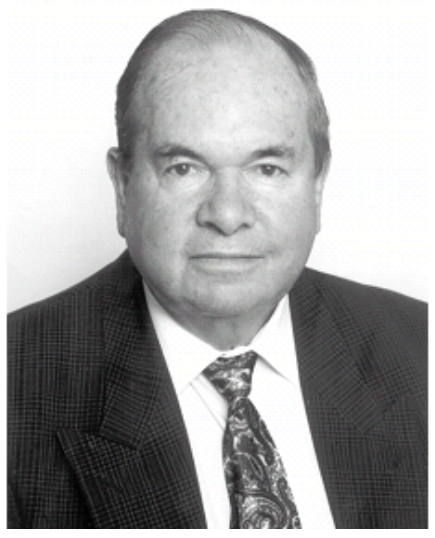 A.A. Abricosov |
| 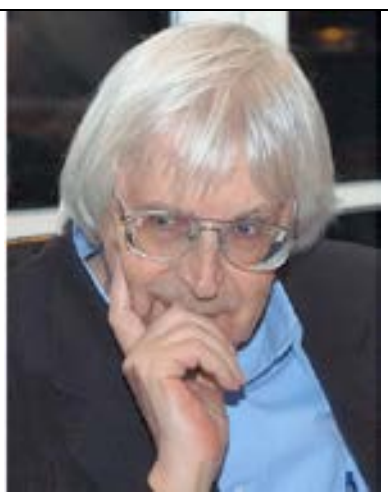 L.P. Gor'kov | 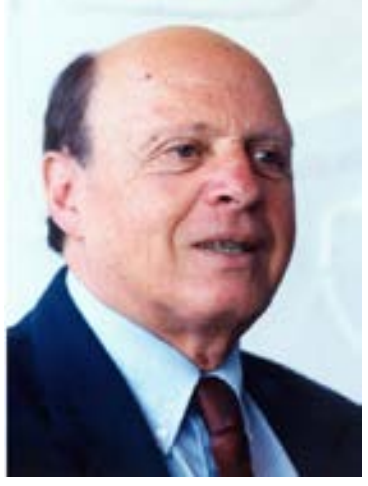 I. M. Khalatnikov | 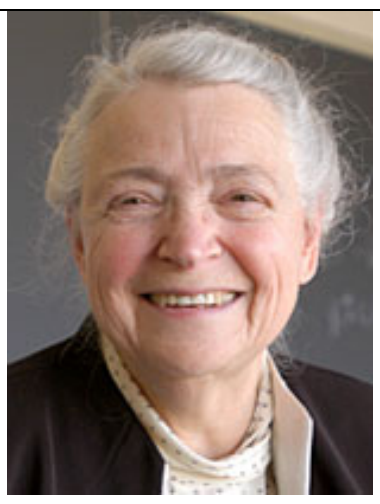 M. Dresselhaus |

**Tab. 4:** Brilliant scientists, who made the groundbreaking researches in the field of microwave superconductivity (after [9-23]).

The detailed information on the biographies of the Nobel laureates in the field of superconductivity, including the portraits of scientists provided by the *Nobel foundation* [1, 2, 3, 6]:

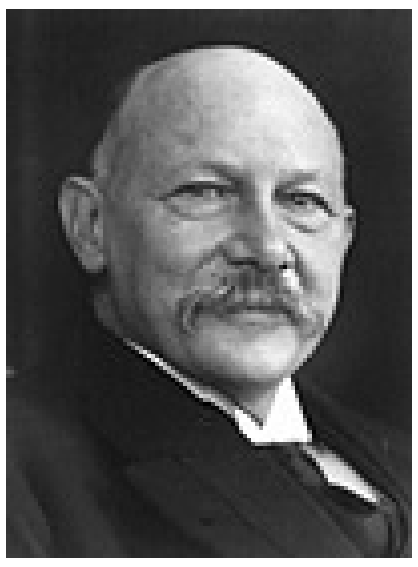

***Heike Kamerlingh Onnes*** was born in Groningen, Holland in 1853, and received his bachelor's degree from University of Groningen in 1871. Already at the age of 18 he received a gold medal in a science competition sponsored by the University of Utrecht, in which he investigated '...methods for determining vapour density... '. The subject was to have importance in his later research. In 1908 he became the first to liquify the inert gas helium with a boiling point of 4.2K, thereby opening the door to investigations of all kinds of matter at low temperatures, and in particular laying the foundation for the discovery of superconductivity. The interest in low-temperature physics was by no means a new. Michael Faraday had taken deep interest in condensation of gases. James Dewar in Edinburgh, and later at the Royal Institution, was a prominent expert and had, among other things, invented his famous thermos flask, commonly called a 'dewar' even today. Onnes was influenced early by the theoretical work of van der Waal, which pointed in the direction of low temperature physics in gas-liquid systems. He was appointed professor at Leiden in 1882, a post he would hold for 42 years. His slogan *Door meten tot weten* (through measurement to knowledge) defined the style and spirit of his laboratory. Having gained access to temperatures down to just above $1K$ by reducing the vapour pressure above the helium bath, a natural task was to continue the investigations of the low temperature electrical resistance in metals. A whole new domain of temperatures was at hand. A series of investigations by Dewar and co-workers had given resistance curves for a number of metals down to about $-200^{\circ}$C. Their properties in the new temperature domain needed to be measured. Mercury was chosen by Onnes for its high purity which could be obtained through evaporation.

Contained in long capillary glass tubes it would freeze to a solid filament. Its electrical resistance could be measured with standard experimental methods. Superconductivity was discovered in 1911. Reports given at the prestigious Solvay conference the same year did not cause the stir we might have expected. The fact that one was dealing with a new state of metals was not quite appreciated. The available theoretical apparatus had not been forged. But this did not distract Onnes and co-workers from pursuing the subject which became known as superconductivity. Gradually it was realized that the transition to the superconducting state was a fairly normal occurrence among metals. Onnes had high hopes of using superconductors to build high-field magnets, up to 100 000G. Unfortunately, since he worked with type I superconductors this was impossible due to limitations in critical current. Heike Kamerlingh Onnes received the Nobel prize in physics in 1913 for ‘his investigations on the properties of matter at low temperatures which led, inter alia, to the production of liquid helium’. The whole history of the discovery of superconductivity by Kamerlingh Onnes and co-workers, as well as many other related historically interesting events, have been vividly described by Dahl [3].

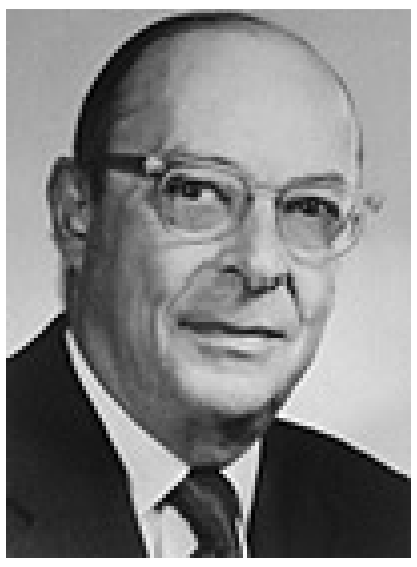

***John Bardeen*** was born in Madison, Wisconsin in 1908. He studied electrical engineering at the University of Wisconsin, receiving his BS degree in 1928, and his MS degree in 1929. After a few years in geophysics he did his graduate studies in mathematical physics at Princeton University, and received his PhD in 1936. After research periods at Harvard University, at University of Minnesota, and at the Naval Ordonnance Lab in Washington DC, he came to Bell Labs in New Jersey in 1945. Here he joined the solid state physics group, and became interested in semiconductor research. In a collaboration with Brattain and Shockley he discovered the transistor effect in semiconductors in 1947, and together they laid the foundation for the modern age of electronics and computers. In 1951 he left Bell Labs to become Professor of electrical engineering and physics at University of Illinois, Urbana. Here he set up the team with Cooper and Schrieffer which was to develop the first successful microscopic theory of superconductivity, later referred to as the BCS theory. John Bardeen's influence on solid state physics, electrical engineering and technology was monumental. He received a number of prestigious awards and prizes. In 1956 he was awarded the Nobel prize in physics with Brattain and Shockley for research leading to the invention of the transistor, and in 1972 he shared the Nobel prize in physics with Cooper and Schrieffer for the theory of superconductivity. John Bardeen is the only person to have received the Nobel prize twice in the same prize domain. The transistor is often called the most important invention of the 20th century. John Bardeen was named by Life Magazine among the 100 most influential people of the 20th century. [*Sources*: The web pages of the Nobel e-museum is a rich source for further information about the scientific career of John Bardeen, and the impact of his work.]

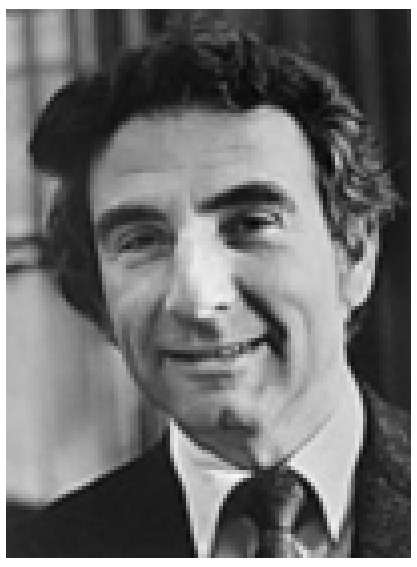

***Leon N. Cooper*** was born in New York in 1930. He attended Columbia University where he received his AB in 1951, AM in 1953, and PhD in 1954. During 1954–55 he was a member of the Institute of Advanced Study at Princeton. He held a post doctoral position as a Research Associate with John Bardeen at Urbana, University of Illinois during 1955–57, and served as an Assistant Professor at Ohio State University 1957–58. Since 1958 he has been Professor at Brown University. In his own account, his interest in superconductivity began with meeting John Bardeen at Princeton in 1955. Until then he had no previous knowledge of the field. His background was in field theory, exactly what Bardeen was looking for. His first task upon arriving at Urbana at the age of 26, was to learn the basics of superconductivity. He became convinced, as was Bardeen, that the essence of the problem was an energy gap in the single particle spectrum as evidenced by the exponentially decreasing heat capacity towards T = 0K. From a lecture by Pippard he learned that the facts of superconductivity appeared to be simple. In this respect it was an advantage that the isotope effect had been established, while all the exceptions found later were not yet known. Therefore, a phonon mechanism, as had been discussed by Bardeen and Fröhlich, would seem like the right idea. But first he made the important proof, later referred to as the 'Cooper problem': The zero degree instability of the fermion system against formation of a bound electron pair in the presence of the slightest attractive interaction between two electrons placed outside an already full Fermi distribution. An intense collaboration with Bardeen and his young student Robert Schrieffer started, with the aim to develop a theory for the electron-phonon interaction, and for superconductivity. In 1957 their famous 'BCS'-paper was published. The pairing due to the previously envisaged hypothetical attraction between two electrons was identified as an electron–phonon scattering event by which electrons with opposite momenta and spin in a thin shell near the Fermi-surface formed a short lived binding. The effect of this, happening all over the Fermi-surface, was to create a new ground state, the superconducting state. The new theory had all the right properties, the energy gap, the Meissner effect, the penetration depth, the coherence length, the isotope effect, the prediction for ultrasonic attenuation, the coherence factors in NMR, etc. Cooper was appointed Professor at Brown University in Providence, Rhode Island in 1958 and has remained there since. He has later changed field entirely, becoming the Director of Brown University's Center for Neural Science, founded in 1973 to study animal nervous systems. The center created an interdisciplinary environment with students and faculty interested in neural and cognitive sciences towards an understanding of memory and other brain functions. Professor Cooper holds a number of honorary doctorates. He shared the Nobel prize in physics with Bardeen and Schrieffer in 1972. [*Sources:* A personal interview for this book conducted by one of the authors (K.F.) in 2001. In addition: The Nobel e-museum, and the scientific literature].

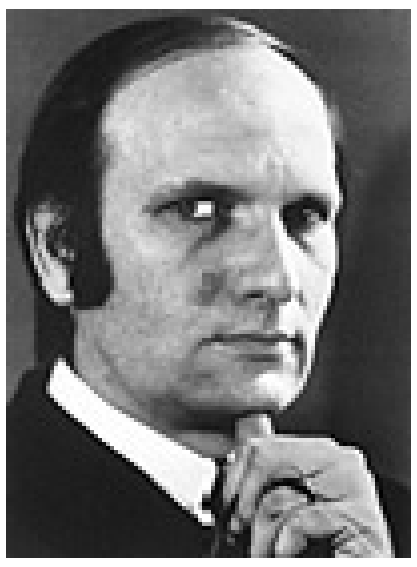

***Robert J. Schrieffer*** was born in Illinois in 1931. In 1940 the family moved to New York, and in 1947 to Eustis, Florida. Schrieffer's original plan was to make a career in electrical engineering, and started on an engineering education at MIT in 1949. His interest in this field came from personal experience as a radio amateur on a homemade 'ham' radio in his young teenage years. In those years he had also made somewhat daring experiments in rocket science. But at MIT he discovered, through his own reading, the challenges and fascinations of physics, and made the switch to physics after two years. Under John C. Slater he did his bachelor's thesis on the structure of heavy atoms. He became interested in solid state physics, and began graduate studies with John Bardeen at University of Illinois. He did research, both theoretical and experimental, on semiconductors the first two years. In his third year on the advice of Bardeen he started collaboration with Leon Cooper, and the three together were committed to solving the superconductivity problem. While the three men struggled with the superconductivity problem, the young PhD student Robert Schrieffer felt uneasy about progress, and without telling his adviser, Professor Bardeen, he conducted a separate research project in ferromagnetism as a safeguard against a possible total failure to solve the superconductivity problem. When Bardeen was about to go away for a meeting in December 1956 he suggested Schrieffer should go on working on the superconductivity problem for yet another month before changing subject, because he felt they might be able to solve the problem. While Bardeen was away, Schrieffer happened to be on a visit to New York. Sitting on the subway he realized that the Tomonaga approach for the integration between pions and nucleons might be the way to go in a consistent way. He wrote down the wave function, now known as the BCS wave function, and calculated the energy of the system. It had the same form as the Cooper solution, but was exponentially stronger. He felt this was an interesting development. On his return he told Cooper, and then Bardeen. Bardeen said: ‘I think that’s the answer. That solved it!’ During the next 11 days they worked out the thermodynamics and other properties. First they calculated the condensation energy in terms of the gap. Using the results of Tinkham and Glover who had recently measured the energy gap by infrared absorption, numerical values could be determined. Their analysis agreed with experiment. The paper was published in Physical Review in 1957, the famous ‘BCS’ paper. Schrieffer emphasizes that the BCS theory has a much wider validity than just the phonon mechanism, referring to the applicability of the BCS theory in totally different systems like in nuclear matter and in neutron stars. Schrieffer has had a distinguished career. He spent the first couple of years after his thesis work on the BCS theory at the University of Birmingham and at the Niels Bohr Institute in Copenhagen, and then at the University of Chicago and the University of Illinois. In 1962 he joined the faculty of the University of Pennsylvania and became a professor there. In 1980 he was appointed Professor at University of California in Santa Barbara where he served as Director of the Institute of Theoretical Physics from 1984 to 1989. He was later called on to become University Professor at Florida State University in Tallahassee, Florida, and is Chief Scientist at the National High Magnetic Field Laboratory (NHMFL) since 1992. Professor Schrieffer holds several honorary doctorates, and a number of prestigious awards. He received the Nobel prize in physics for 1972, shared with Bardeen and Cooper for the theory of superconductivity. [*Sources*: A personal interview for this book conducted by one of the authors (K.F.) in his office at NHMFL in 2001. In addition: The Nobel e-museum, and the scientific literature.]

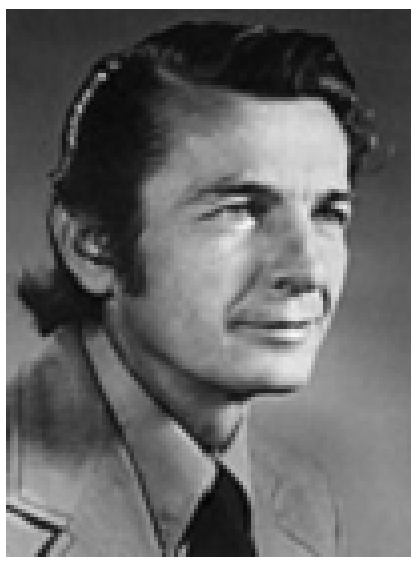

***Ivar Giaever***’s career in physics is a highly unusual one. He was born in Bergen, Norway in 1929, but the family moved to Toten north of Oslo within a year.

After high school, his priority was to study electrical engineering at Norway's leading engineering school, the Norwegian Institute of Technology, in Trondheim (since 1995 incorporated into the Norwegian University of Science and Technology). However, the competition to get in was very strong, and due to the equivalent of a C in a Norwegian language course, he was not admitted to study in the Department of Electrical Engineering, but had to accept Mechanical Engineering, a subject which he simply was not interested in. Consequently, by his own account, he spent his student years in Trondheim from 1948 to 1952 mostly doing other things than study, like playing bridge, billiard, chess, and poker. He became a local champion in all but the last one. Still, when he showed up for exams, he passed, and made it to a degree in mechanical engineering, after which he married. Now, the post-war housing situation in the cities of Norway after the Nazi occupation was extremely difficult. Having found a job in Oslo turned out to be of no help. The young family decided to emigrate to Canada in 1954, where Giaever, after a brief period in an architect's office, joined the Canadian General Electric's Advanced Engineering Program. Soon Giaever discovered that salaries were better south of the border, and moved to General Electric Company in Schenectady, where he as an employee, and now also as a serious and hard working student, took the company's demanding engineering courses, level A, B and C. Next, he worked as an applied mathematician on various assignments. He was greatly attracted by the opportunity to do research within the company with its impressive staff of skilled scientists at the GE Research and Development Center. Having joined the center in 1958, and concurrently started to study physics at Rensselaer Polytechnical Institute, he earned his PhD degree in 1964. It was during class in a physics course several years earlier, where superconductivity was being taught, that the idea struck him how to measure the superconducting energy gap which the BCS theory had recently predicted. His mentor at the research centre had told him that electrons could tunnel through thin barriers between semiconductors, a notion he could hardly believe since he was not yet familiar with quantum mechanics. Giaever now saw the possibility to try it out in a superconductor–insulator–metal contact, and in the process measure this important quantity, the gap. To his relief he also calculated that the predicted gap size, being in the millivolt range, was perfectly suited for the experiment he planned. All experimental facilities he needed were around, along with the support of highly skilled scientists. Giaever made his thin film structure as planned, and could soon, in 1960, 'measure the energy gap in a superconductor with a voltmeter', as he put it. Against the background sketched above, this was quite an achievement, and became next to a scientific sensation in the physics community. The mechanical engineer, now physics student, had done an experiment the experts could only have wished to do, but did not conceive. Adding to this the great ability Giaever has to communicate his work orally, has made him an attraction at meetings, at universities and research institutes. No doubt, his story can be taken as yet another example of the 'American dream' come true. The fact that he is always open and candid about his unusual background as a physicist has added a special flavour to his story and his work. His follow-up research on the density of states near the superconducting gap demonstrated even further that his discovery was no accident. But he is the first to insist that some element of luck was involved, and comments that this is needed to succeed. One should not be tempted to think his success came easy, however. Many years of hard work at General Electric was behind it all. Ivar Giaever continued his tunneling experiments for several years and contributed immensely to the progress in superconductivity. Giaever also was the first to published measurement showing a finite current between superconductors in a zero voltage condition, what later became known as the DC Josephson effect. However, due to the fact that the whole idea of a zero voltage supercurrent across a barrier between two superconductors had not yet been formulated, Giaever never laid claims on having discovered the Josephson effect. Ivar Giaever shared the Nobel prize in physics in 1973 with Leo Esaki and with Brian D. Josephson. Giaever had by then already left superconductivity, and had started work in biophysics during a stay at Cambridge, UK in 1969. His special area has been the behavior of protein molecules at solid interfaces. He left General Electric in 1988 to become an Institute Professor at Rennselaer Polytechnique Institute in Troy, and has for a number of years, concurrently, been a Professor at the University of Oslo, Norway. He is the recipient of numerous honorary degrees and prestigious prizes. [*Sources*: A personal interview for this book conducted by one of the authors (KF) in Schenectady, 2001. In addition: The Nobel e-museum, and the scientific literature.]

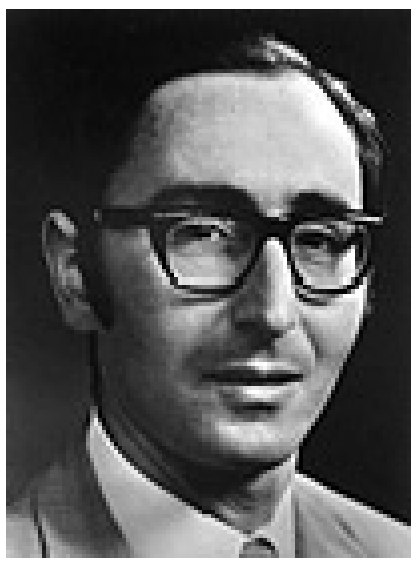

***Brian D. Josephson*** was born in Cardiff, UK in 1940. He received his BA degree at the University of Cambridge in 1960, and his PhD, also in Cambridge, in 1964. Josephson had already shown exceptional talent as a teenage scientist, but in an area far removed from where he was to gain international fame. He had the good fortune to do his PhD research with Pippard at the Royal Society Mond Laboratory. Furthermore, already during his second year as a research student the laboratory had Professor Phil Anderson as an inspiring visitor, in 1961–62. Josephson gives much credit to Anderson for his own interest in superconductivity, in particular for introducing him to the concept of 'broken symmetry' in superconductors. This concept was already inherent in Anderson's pseudo-spin formulation of superconductivity theory from 1958. In particular, Josephson wondered if the broken symmetry could be observed experimentally. He concluded that while the absolute phase angle of pseudo-spins would be unobservable, the consequences of a phase difference might not. At this point he learned about Giaever's tunnelling experiment from 1960. However, Pippard had considered the tunnelling of Cooper pairs through a thin barrier and found the probability to be very small. When Anderson showed him a paper by Cohen, Falicov and Philips where they had confirmed Giaever's formula for the current in his superconductor–insulator–metal contact, Josephson understood how he could calculate the current through a barrier between two superconductors. The expression he arrived at contained three terms, two of which were already known from previous work, but a third one was new. This term was unexpected: a current which was proportional to the sine of the phase difference across the barrier. The coefficient of this term was an even function of the voltage, and could not be expected to vanish at zero voltage. The obvious interpretation was that this was a supercurrent, and it appeared with the same order of magnitude as the quasiparticle current seen by Giaever. This was surprising, considering earlier suggestions by Pippard. At the age of 22, Josephson made the famous prediction of the supercurrent through an insulation barrier between two superconductors, known today as the Josephson effect. He made the prediction of both a DC effect and an AC effect. While the former would appear at zero applied voltage the latter should be present under the application of a small DC voltage. His predictions were confirmed, and became the basis for whole new fields of superconductivity research and technology. In later years there have been discussions among scientist whether the Nobel prize to Josephson should rightly have been shared with Anderson. To this question Anderson answers a clear ‘no’. He explained us that such opinions might stem from the fact that he had re-derived some of the results that Josephson had already found. The reason for doing so, he explained us, was that Josephson had not published all his findings, some of which were only reported in his thesis. Brian Josephson has held academic positions at the University of Illinois, the University of Cambridge and various visiting professorships. He is Professor of Physics at the University of Cambridge since 1974. [*Source*: The Nobel e-museum, and interview by one of the authors (K.F.) with P.W. Anderson in 2001.]

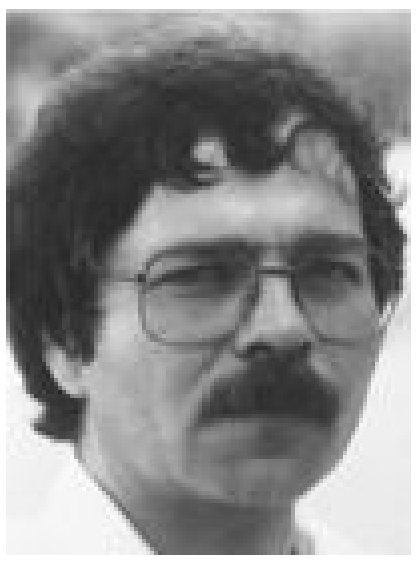

***J. George Bednorz*** was born in Neuenkirchen, Germany in 1950 as the fourth child. His parents had involuntarily been separated during the turbulences after World War II, but were happily reunited in 1949. In his youth he was influenced by his mother’s music interest, and came to play both piano, violin and trumpet. His fascination with science was awakened not by physics, but by chemistry. He felt that doing experiments in chemistry stimulated his practical interests, and could have unexpected results. He started to study chemistry at the University of Münster in 1968, but ended up with majoring in crystallography. During two periods as a summer student at the IBM Zurich Research Laboratory in Rüschlikon, and later as a diploma student in 1974, he worked under the guidance of Hans Jörg Sheel in the Physics Department headed by K. Alex Müller, a scientist he deeply respected. His diploma work was on SrTiO3, a great specialty of Müller's who became so pleased with young Bednorz' work that he encouraged him to continue his research on perovskite materials towards a PhD, supported by IBM, at the Swiss Federal Institute of Technology (ETH) under the combined supervision of Professor Heini Gränicher and Müller. His thesis work was on the crystal growth and solid solutions of perovskite type compounds, investigating structural, dielectric and ferroelectric properties. Upon completion, he joined the IBM lab in Rüschlikon in 1982. This would not seem like a good background for superconductivity research. But already while Bednorz was a student at ETH in 1980, Heini Rohrer at the IBM laboratory had asked him if he could prepare crystals of SrTiO3 doped with Nb for the purpose of studying the superconducting properties of this material under varying doping conditions, with Gerd Binnig. Bednorz responded that 'if Nature allows, you will get it'. After a couple of days the material was ready, and the superconducting transition temperature had increased by a factor 4! This also had the interesting implication that the gradient of Tc versus doping was very steep. But when Bednorz joined the IBM laboratory in 1982, this line of research had been stopped, since now Rohrer and Binnig were working on the scanning tunnelling microscope, also a work to be awarded the Nobel prize. However, in 1983 Alex Müller, having spent two sabbatical years at the IBM laboratory in Yorktown Heights, New York where he had done work on granular superconducting Al, approached Bednorz again, and asked if he would join him in an attempt to go new ways in superconductivity. The idea was to exploit a polaronic interaction using Jahn–Teller ions, a field championed by Harry Thomas. Müller thought the mechanism might work in perovskites. Bednorz immediately agreed to collaborate. From then on a systematic effort was being made. This was a low cost project carried out as a side effort along with other ongoing work by both. Naturally, the first attempt was to go for classical Jahn–Teller systems like the lanthanum nickelates. Here, La was replaced by Y. Later also the B-position was modified. The idea was to modify the bandwidth. After one year the project was in danger of being stopped since the results were discouraging: All compounds were insulators. Bednorz now suggested to use copper instead of nickel to achieve the Jahn–Teller effect. Electrical conduction was obtained, but no superconductivity. Bednorz needed a break and went to the library. Here, he discovered the work by a French group, Raveau and co-workers, on the Ba-La-Cu-O compounds, and realized they should modify the A-position of the ABX3 instead of the B-position. Already in the first measurement, in January 1986, a dip in the resistivity was found at 11K. Since they did not have a magnetometer at the time, the test for diamagnetism could not be performed until a SQUID magnetometer had been acquired in September. However, the results were stable and reproducible. They felt confident that superconductivity had been discovered. Still, when each of them gave talks in different places in Germany in the fall of 1986 there was almost no response. This changed dramatically after the Japanese group headed by Tanaka at University of Tokyo in the fall of 1986 announced results that confirmed superconductivity in lanthanate. Their own work also showed the Meissner effect. From now on superconductivity was a matter of public interest. A new era had started. George Bednorz has continued as a scientist at the IBM laboratory in Rüschlikon near Zurich. He is the recipient of numerous awards and prizes, and shared the Nobel prize in physics with K. Alex Müller in 1987. [*Sources*: Interview for this book by one of the authors (K.F.) in 2001, and scientific collaboration. In addition, the Nobel e-museum and the scientific literature.]

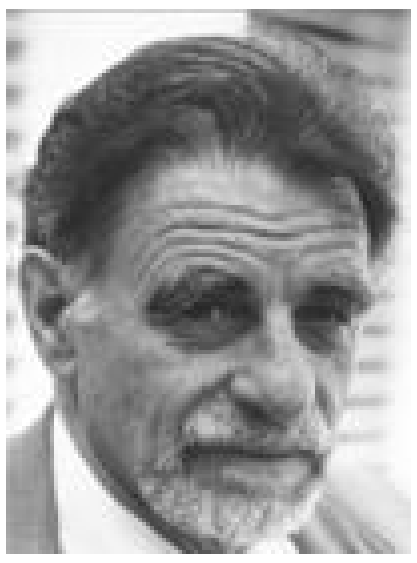

***K. Alex Müller*** was born in Basel, Switzerland in 1927, and lived first in Saltzburg where his father studied music, later in Lugano where he became fluent in the Italian language. His mother died when he was eleven, after which he attended Evangelical College in Schiers, in the Swiss mountains. He remained there until the end of the war. He was fascinated by the radio, and wanted to become an electrical engineer, but his chemistry tutor, Dr Saurer, convinced him to study physics. After military service he enrolled in the Physics and Mathematics Department of the Swiss Federal Institute of Technology (ETH). The freshman class was three times too big, and the process of elimination was correspondingly tough. They were called the ‘atom bomb semester’ for obvious reasons. M¨ uller had excellent teachers, like Scherrer, Kanzig and Pauli, and did his diploma work with Professor G. Busch on the Hall effect in grey tin, followed later by PhD work on paramagnetic resonance (EPR) in Busch’s group. Here he identified the impurity present in the perovskite SrTiO3 a fact he took much advantage of later. Upon completion of his PhD and after graduation in 1958 he worked at Battelle Memorial Institute in Geneva, where after he came to the IBM laboratory in Rüschlikon in 1963. He remained there until his official retirement from IBM, after which he is a professor at University of Zurich. He was a key person in the research which took place in the late 1960s and in the 1970s and early 1980s on understanding the critical properties of phase transitions. Again SrTiO3 was the vehicle, and it became the best studied of all, specially its properties related to the structural phase transition near 105K. With Thomas he identified the order parameter and worked out the Landau theory for this system. He was and is a world leading scientist as far as structural transitions in perovskites is concerned. This competence was not wasted, as it turned out, when he undertook the challenge with Bednorz to discover high-Tc compounds. From the time superconductivity was discovered in oxygen deficient $SrTiO_3$ at Bell Labs in 1964, he had an eye on this subject, but did not get directly involved in superconductivity until his 2-year long sabbatical at the IBM lab in Yorktown Heights at the end of the 1970s, at which time he studied Tinkham’s textbook from A to Z, as he said to us: ‘... like a graduate student after the age of 50’. Now he started research on superconductivity for the first time, in granular Al. His interest in the subject did not diminish after this. Some time after his return to the IBM lab in R¨ uschlikon, having heard a talk by Harry Thomas at a meeting in Erice, he was inspired to invite George Bednorz to collaborate in a search for superconductors among Jahn-Teller perovskites. We refer to his own account in Chapter 2, and to the account given by Bednorz above about the ensuing progress. The work that Binnig and Bednorz had done on his ‘old friend’ among perovskites, $SrTiO_3$ –a work he had followed closely as a manager–was also on his mind when he suggested the collaboration which would turn out such spectacular results, ending with the sensational developments in late 1986 and in early 1987: the discovery of record breaking high-Tc perovskite superconductivity in $La_{2-x}Ba_xCuO_4$.AlexM¨ uller has achieved the rare position to be a world-leading scientist in two totally different fields of physics. Those who have the privilege to know him, have experienced his profound ability to combine knowledge from different areas of physics into a penetrating understanding of complicated subjects. The award of the Nobel prize in physics to Müller and Bednorz in 1987, attests to the fact that the spectacularly important and unexpected is often to be found in such combination of knowledge. Alex Müller holds on to his original ideas about the (bi)polaronic mechanism for superconductivity in the cuprate superconductors, a view that undeniably led to their great success. In his view, the observed isotope effect as well as the so-called stripe domains attest to the correctness of this basis for superconductivity in cuprate superconductors. M¨ uller is the recipient of numerous awards in addition to the Nobel prize. He holds an honorary doctor degree at 17 universities. [*Sources*: Interview for this book by one of the authors (K.F.) in 2001, scientific collaboration and personal correspondence; the Nobel e-museum, and the scientific literature.]

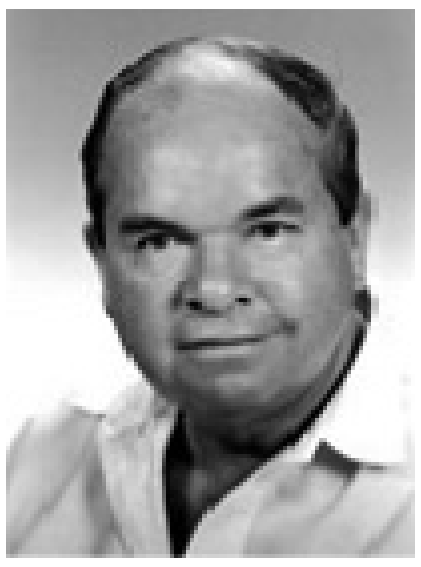

***Alexei A Abrikoso***v grew up in a well known family in Moscow, where he was born in 1928. He lived in Moscow all his life until he emigrated to the US in 1991. Both his parents were medical doctors. His father was quite famous, and received the Golden statue of socialist labor, and the Stalin prize, but was not politically active. Upon Lenin’s death, he performed the autopsy. Young Alexei’s mother told him that under no circumstance should he become a medical doctor, for reasons he still does not know. Consequently, he excluded a medical career from the start, but already at the age of ten he was convinced he would become a scientist. He was already very interested in the life of great scientists and inventors, like Faraday and Edison, and already at that age he dreamed of winning the Nobel prize, and becoming a member of The Royal Society, completely unrealistic, he thought. With time he was to achieve both. He graduated from high school at the age of 15. He had great talents in mathematics, but entered, at this young age, the Institute of Power Engineering, partly to avoid the looming danger of being drafted in the future when he would reach such age; and then transferred to Moscow University after the war ended, still only 17 years old. Already as a very young man, still looking like a small boy, according to his own description, he was accepted by the great Lev Landau who understood what talents were at hand. At an unusually early age he passed Landau's famous "theoretical minimum", and stayed close to him. However, the KGB did not allow him to work on the hydrogen bomb with the Landau group due to suspicions against an uncle of his, a diplomat, of whose existence young Alexei had himself, at that time, no idea! Eventually, he did his PhD with Landau after all, and later was a postdoc in his group. He received his first degree in 1951 from the Institute for Physical Problems (Moscow, Russia) for the theory of thermal diffusion in plasmas and then the next degree, Doctor of Physical and Mathematical Sciences, in 1955, from the same Institute for a thesis on quantum electrodynamics at high energies. In 1975 he was awarded the Honorable Doctorate from the University of Lausanne, Switzerland. During his long scientific life he has explored successfully many fields but mainly the theory of solids: superconductors, metals, semimetals and semiconductors. He is very famous for the discovery of the theoretical foundation for Type II superconductors and their magnetic properties (the Abrikosov vortex lattice), for which he was awarded the Noble Prize in Physics 2003. This work was published in 1957, but the results were achieved already in 1953, without Abrikosov being allowed to publish them. His boss, Landau, did not initially believe the results, and was not convinced about them until he learned that Richard Feynman in the US had the idea that quantized vortices in superfluids could be responsible for driving the lambda transition from a superfluid phase in Helium II to a normal fluid. In this case, contrary to normal practice, Landau read the Feynman paper himself, and believed the results. Landau never apologized. In his view Abrikosov had not come up with the simple physical arguments which he required, and which would make it obvious why Abrikosov's solution was correct. Even after Landau's acceptance, there were many more obstacles ahead before recognition was fully achieved. In Abrikosov's own mind it was only reached upon the publication of decoration experiments by Essmann and Träuble in 1967, in which a regular lattice of vortices was clearly demonstrated. In 1991 Abrikosov moved to the US and joined the Materials Science Division as an Argonne Distinguished Scientist at the Condensed Matter Theory Group of the Materials Science Division where he is still active. In Argonne he has worked on the theory of high-Tc superconductors, properties of colossal magnetoresistance in manganates and, together with experimentalists there, discovered the so called "quantum magnetoresistance" in silver chalcogenides. Abrikosov has been elected a Member of the National Academy of Sciences (USA) and the Russian Academy of Sciences, Foreign Member of the Royal Society of London and the American Academy of Arts and Sciences. He has been awarded numerous Russian and International Awards and the Honorable Citizenship of Saint Emilion (France). [*Sources*: A personal interview with Abrikosov by one of the authors of this book (KF) in 2003, the Nobel e-Museum, and the scientific literature].

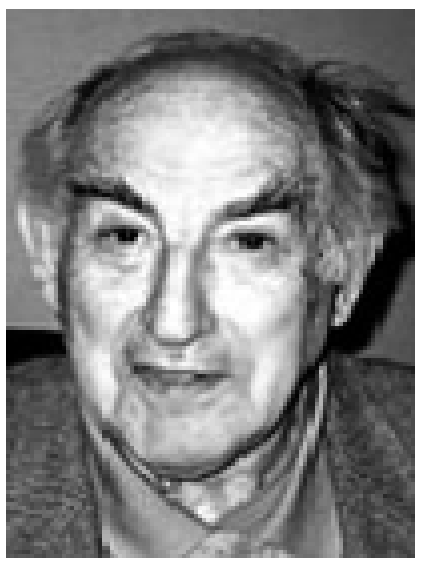

***Vitaly L. Ginzburg*** was born in 1916, and grew up in Moscow during revolutionary times, under the establishment of the Soviet Union. His father was an engineer, and his mother a medical doctor. Very unfortunately she died when he was still only four years old. Except for two years of evacuation during the war, he has lived all his life in Moscow. Times were difficult after the revolution. Before the revolution their family had a four room apartment, after it they had to share it with two more families. They did not starve, but the food they had to eat was far below traditional Russian standards. In 1931 the government decided that those who had finished seven years of elementary school should go to a special school to be trained to be workers, instead of receiving higher education. But Ginzburg went to work as a technician in a laboratory instead, and educated himself enough to enter Moscow University in 1933 at the age of 17. He finished there in 1938. He originally doubted his abilities to be a theoretical physicist, but after some encouraging work on quantum electrodynamics he was accepted by the famous physicist I.E. Tamm, head of the P. N. Lebedev Physical Institute, belonging to the Academy of Sciences. From 1938 Ginzburg studied to be a theorist, and defended his Candidate of Science thesis in 1940, and his Doctor of Science in 1942. He became a deputy under Tamm, and remained in the Lebedev institute for the rest of his career and life, still active there at the age of 87. After Tamm died in 1971, Ginzburg became the director of the institute until 1988, when he retired. Andrei Sakharov was at the same institute, but could not be the head since he was a dissident. In 1943 Ginzburg started work in superconductivity, trying to follow up Landau's work in superfluids which in its turn had been inspired by Kapitza's discovery of superfluidity in helium. First he worked on the thermoelectric effect. Eventually his interest focused on the application of Landau's general theory of second order phase transitions. His first application of this theory was in ferroelectrics where he used polarization as the order parameter and established the famous Ginzburg criterion for the validity of the Landau expansion. Superconductivity was a far less obvious case. He wanted to expand the energy in the superfluid density. But in quantum mechanics the density is the square of the wave function. So he had to use the square of the still unknown $\psi$-function for the density. Hence the energy was expanded in a series in even powers of $\psi$. Landau agreed with this development. But according to his recollection they disagreed on the matter of the charge to put into the quantum mechanical momentum in the kinetic energy term. Ginzburg thought of the charge as an effective charge which could be different from unity. Landau insisted there was no reason why it would not be unity. Hence that is stated in the paper. Out of modesty Ginzburg prefers to call their theory "ψ-theory" instead of Ginzburg-Landau theory. This theory has become monumentally important in superconductivity. It is usually applied as a mean field theory, but computationally it can be generalized to include fluctuations, and to also treat dynamical problems in superconductivity. Its wide applicability in high-Tc superconductivity has come as both a surprise and a blessing to this field where the coherence length is so short that initially there were serious doubt as to its validity in such cases. Theoretical progress in the field of high-temperature superconductivity, particularly on the microscopic origins of the phenomenon, has been very slow indeed. It has been one of the major outstanding issues in physics for nearly two decades, since its discovery in 1986. However, the Ginzburg-Landau model has been enormously fruitful in uncovering and understandning the plethora of novel vortex phases that can appear in extreme type-II superconductors such as the high-Tc cuprates, where disorder and thermal fluctuation effects are pronounced. This is extremely important for intelligently engineering of superconductors for large-scale applications. Vitaly L. Ginzburg shared the Nobel prize in physics with Abrikosov, and with Anthony Leggett in 2003, for inventing the Ginzburg-Landau model. It is fair to say that the Nobel prize for this work was extraordinarily well deserved, and much overdue. Ginzburg has in additon, received a number of awards and honors. [*Sources*: Interview of Ginzburg by one of the authors of this book (KF) in 2003, the Nobel e-Museum, and the scientific literature.]

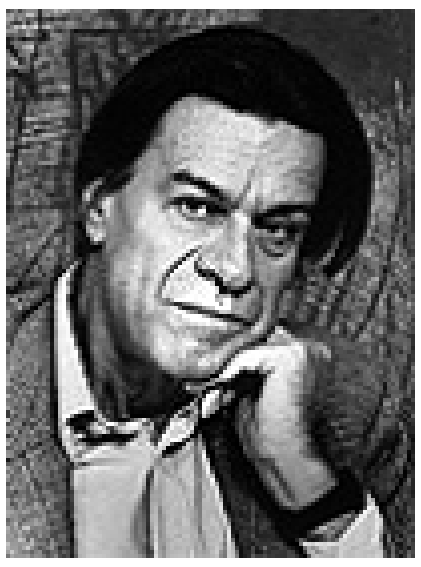

***Pierre-Gilles de Gennes*** was born in Paris in 1932. In the 1960s he was one of the leading scientists in the field of superconductivity, culminating his research in that field by publishing his famous textbook, Superconductivity of Metals and Alloys in 1966, still a classic in the field. He did not receive the Nobel prize in superconductivity, but rather for his contributions to the understanding of ordering in soft matter, in 1991. However, his impact on the field of superconductivity could well be characterized as being at the Nobel prize level. de Gennes' background had some unusual elements: During the war his family moved from Paris to a small village, Barcelonette in the French mountains, partly because of the German occupation, but more importantly because of a health problem. This had the consequence that the young de Gennes did not go to school until the age of 11 to 12. Instead, his mother taught him literature and history which she was very interested in, but no science. He was admitted to high school at an unusually early age. He liked science, but felt no particular push. However, as he explained us: 'The attraction of science was perhaps that it allows a precision test. In our field, when you say something you may advance bold assumptions. Later you can check it out.' Before studying at the university he attended a school which gave untraditional science schooling with a direct observational approach to nature. He did his PhD in magnetism, and was influenced by several prominent scientists, among them Abragam and Friedel. He mentions also Edmund Bauer as a specially influential figure in his career. During his military service he studied the BCS theory, and was ready to enter the field upon completion of the service. He set up a very powerful group at Orsay where he created an unusually effective collaboration between experimentalists and theorists. Later, in 1968 he undertook research in liquid crystals, followed by studies of polymers. He moved on to fields like the dynamics of wetting, the physical chemistry of adhesion, and granular materials. Much of his research has been in what we now call complex systems. Pierre-Gilles de Gennes has written 10 textbooks on different subjects in physics. Few scientists have mastered such a broad palette. de Gennes is a towering figure in French and international science. He is a Professor at the Coll` ege de France since 1971, and Director of Ecole Superieure et de Chimie Industrielle de la Ville de Paris. [*Sources*: Interview for this book by one of the authors (K.F.) in 2001. The Nobel e-museum, and the scientific literature.]

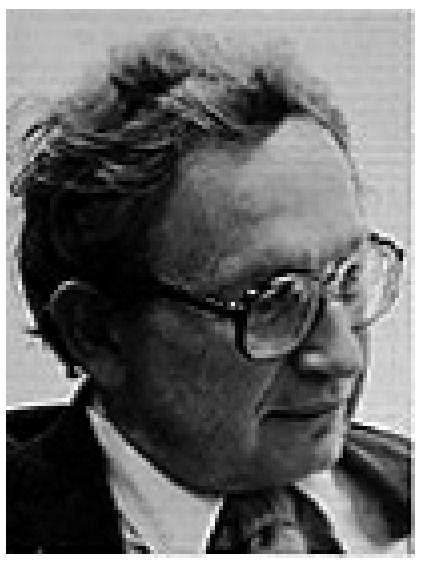

***Philip W. Anderson,*** born in 1923, grew up in an intellectually stimulating and outdoors loving college environment, with college teachers in the near family on both sides. His father was a professor of plant pathology at University of Illinois in Urbana. His mother came from a similar background. Among the family friends were several physicists. After high school he had an intention of majoring in mathematics, but at Harvard it turned out differently. This was during the wartime, 1940-43, and electrical engineering and nuclear physics were important subjects. Anderson chose electronics and went to the Naval Research Laboratory in Washington DC to build antennas during 1940-43. Back at Harvard from 1945 to 1949 he enjoyed both the courses, and the friendship of people like Tom Lehrer, the mathematician turned popular singer with a knack for political humor. He chose van Vleck as his thesis adviser due to greater accessibility than Schwinger, got married and settled down to learn modern quantum field theory which turned out to be useful even in experimental problems. This was at the birth of many-body physics, an area where he was later to be a major participant and leading scientist. Having completed his thesis he went to Bell Labs to work with a number of outstanding scientists like William Shockley, John Bardeen, Charles Kittel, Conyers Herring, Bernd Matthias, and Gregory Wannier. Here he also became acquainted with the work of Neville Mott and Lev Landau. At about the same time both he and his wife became quite active politically in the Democratic party. They worked enthusiastically for the candidacy of Adlai Stephenson towards the presidential election in 1952, and were active in several other connections. Anderson's initial interest in superconductivity came from association with the experimentalist Bernd Matthias at Bell Labs with whom he first worked on ferroelectricity. After the BCS-paper came out he made a study of gauge invariance which they had not considered, and which was a concern among theorists. Also, he was a key figure in the development of a pseudo-spin formalism for superconductivity towards the end of the 50's. This line of thinking has later been successful in completely different fields of physics. His paper on super-exchange from 1959 is a landmark piece of work. He contributed to the development of a theory for d-wave and p-wave superfluid phases of helium-3. Anderson's name is also associated with the Higgs phenomenon. With Kim he did highly original studies of the dynamics of quantized magnetic flux in superconductors in the early 60's. He coined names like "dirty superconductor", "spin glass" and probably also the name "condensed matter", and of course was the inventor of the theory for "Anderson localisation", producing the famous paper on Scaling Theory of Localization together with the "Gang of Four": Abrahams, Anderson, Licciardello and Ramakrishnan. His stay in Cambridge around 1962 was instrumental in inspiring Brian D. Josephson to develop his theory for Cooper pair tunneling between superconductors, the DC and the AC Josephson effects. He has worked extensively on the Kondo problem, solving it by a "poor man's scaling" approach, as well as inventing the co-called Anderson impurity and Anderson lattice model for heavy fermions. From more recent years his efforts to create a theory for high-Tc cuprate superconductivity, the so-called RVB-theory, stands out as a major effort in his career. Anderson's influence on condensed matter physics has been of profound importance. He is often characterized as one of the most influential minds in all of theoretical physics in the second half of the 20th century. In short, there is hardly an area in condensed matter physics worth mentioning which this truly outstanding scientist has not contributed significantly to. Anderson shared the Nobel prize in physics with John van Vleck and Sir Neville Mott in 1977. [*Sources*: Interview with Anderson by one of the authors of this book (*KF*) in 2001, the Nobel e-Museum, and the scientific literature].

# Appendix III

Prof. Lev D. Landau, the Nobel Prize laureate, made his contributions to both *the* ***theory of superfluidity of liquid Helium*** $^4He$ as well as *the* ***theory of superconductivity***.

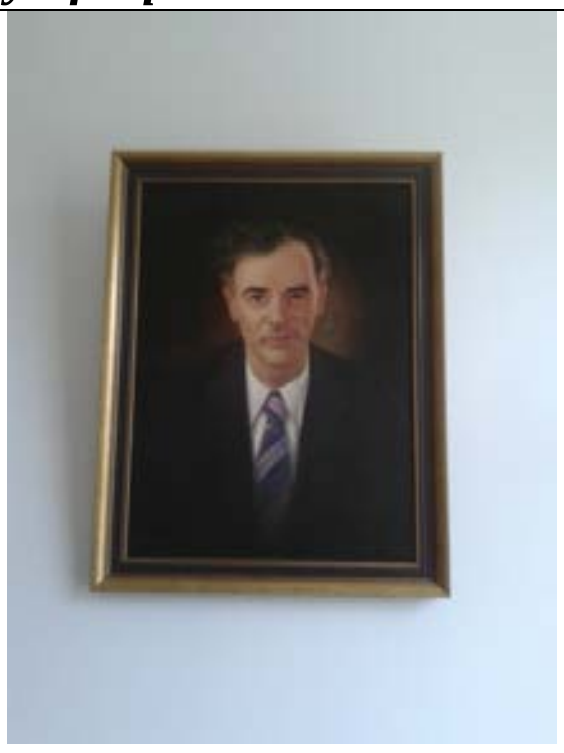

***Fig. 1.*** *The picture of the oil painting of Prof. Lev D. Landau.*

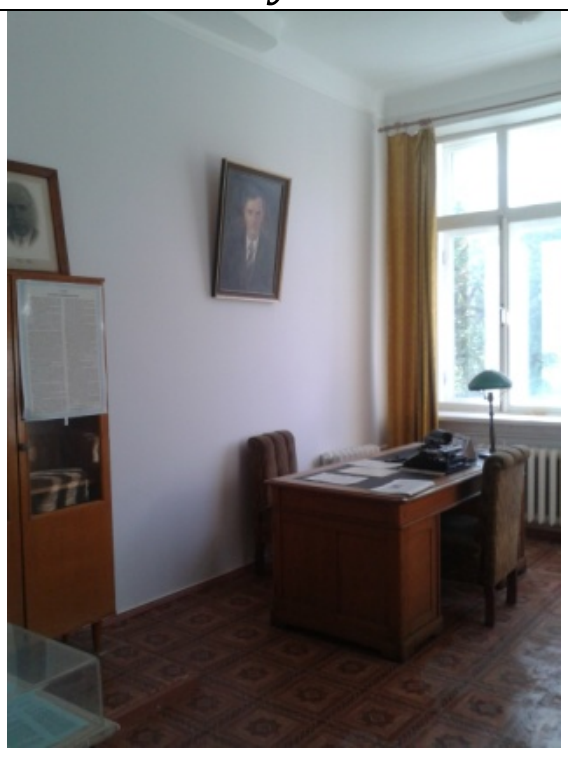

***Fig. 2.*** *The head office of Prof. Lev D. Landau.*

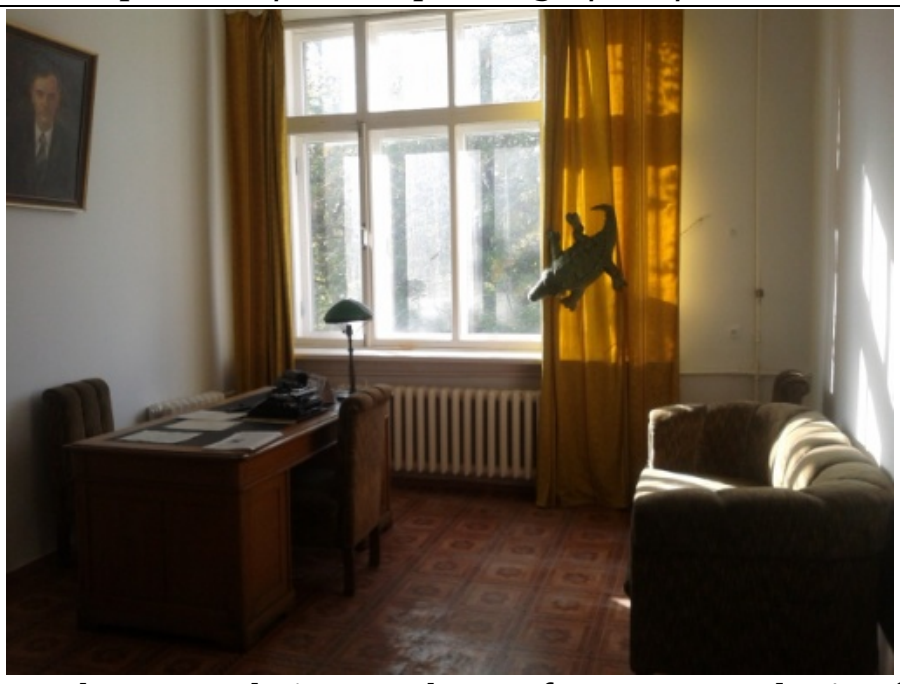

***Fig. 3.*** *The general view on the Prof. Lev D. Landau's office.*

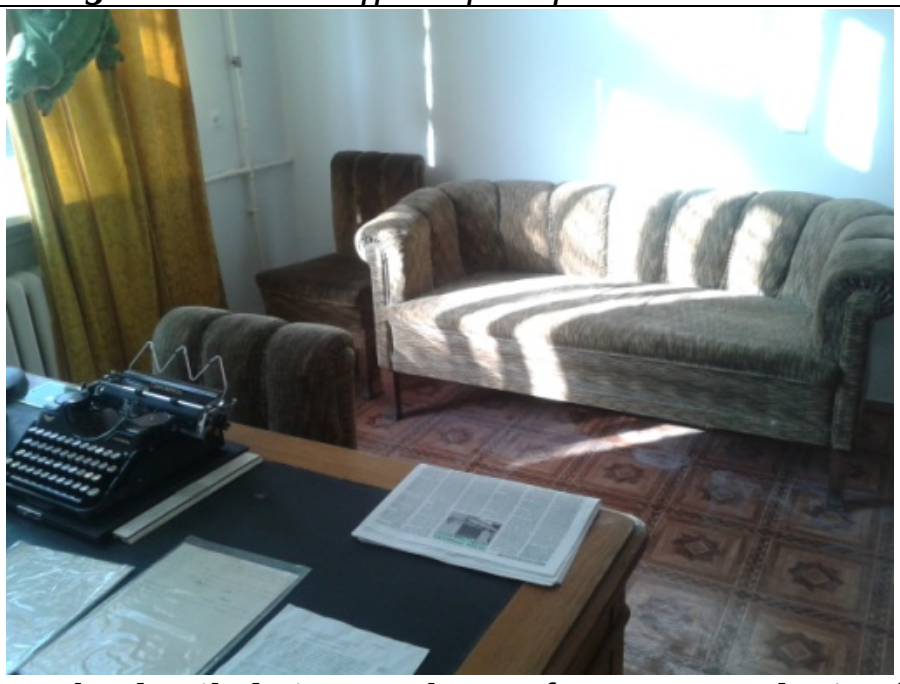

***Fig. 4.*** *The detailed view on the Prof. Lev D. Landau's office.*

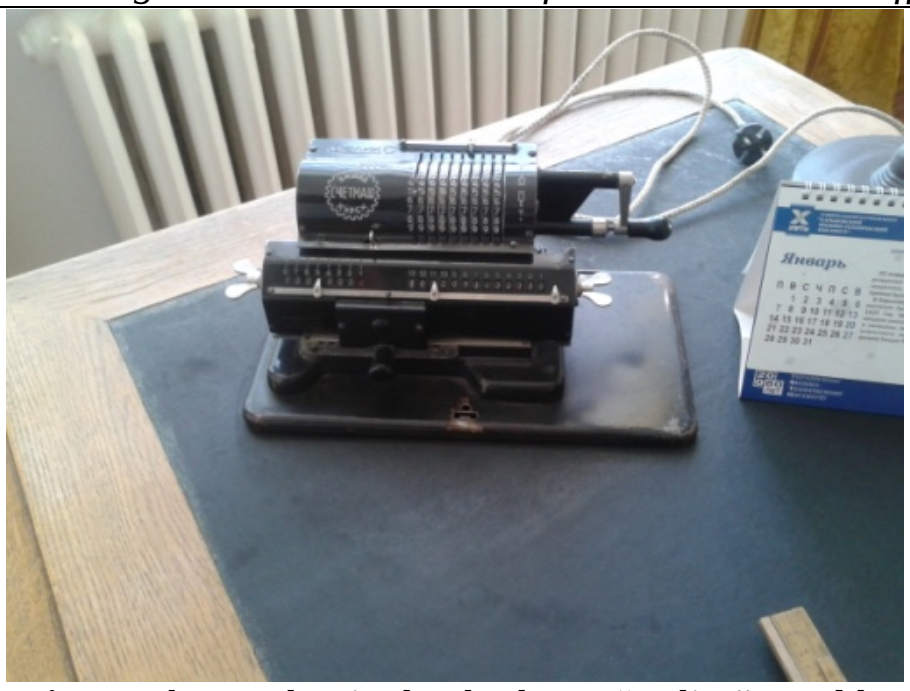

***Fig. 5.*** *The mechanical calculator "Felix," used by Prof. Lev D. Landau to make the theoretical calculations.*

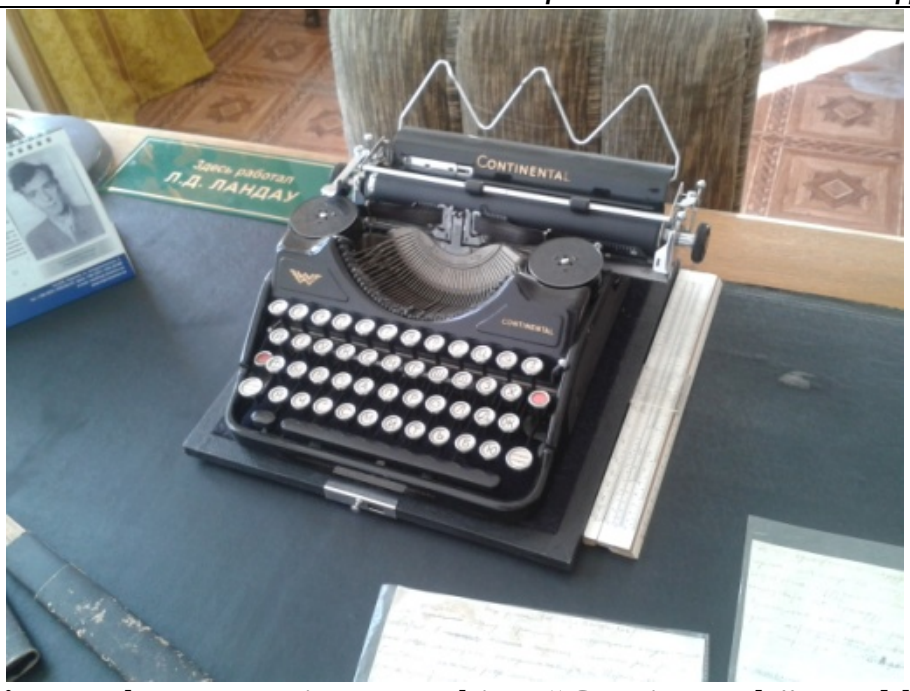

***Fig. 6.*** *The type written machine "Continental," used by Prof. Lev D. Landau to write his research papers and books.*

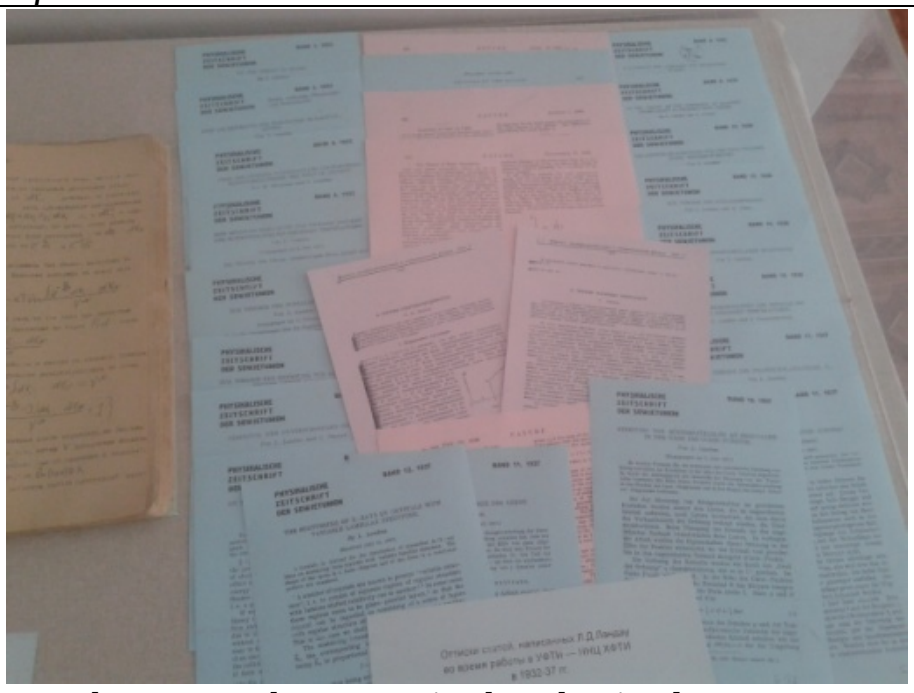

***Fig. 7.*** *The research papers in the physics by Lev D. Landau*

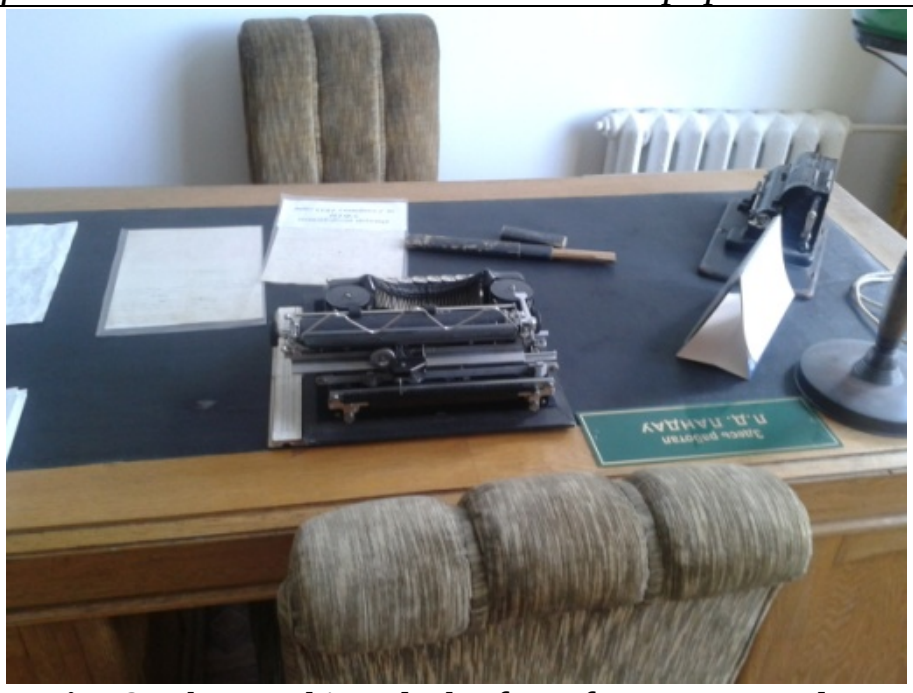

***Fig. 8.*** *The working desk of Prof. Lev D. Landau.*

***Tab. 1.*** *The photographs of Prof. Leo D. Landau office at the National Scientific Center Kharkov Institute of Physics and Technology in the City of Kharkov in Ukraine in 1932 – 1937 (Copyright © Dimitri O. Ledenyov and Viktor O. Ledenyov).*

## Appendix IV.

### The measurement and calculation of magnitude of surface resistance of superconducting thin film made of *NdBaCuO*.

Let us discuss the measurement and calculation of magnitude of the surface resistance of superconducting thin film made of ***NdBaCuO*** by making the following notes:

***1.*** The *dielectric resonator* with the known parameters, including the *resonant frequency* $f_0$ and the *geometric factor* ***A****,* has been used for the measurements.

***2.*** The accurate characterization of the four individual superconducting $YBa_2Cu_3O_{7-\delta}$ thin films $Y_1$, $Y_2$, $Y_3$ and the ***NdBaCuO*** thin film has been performed during the experimental measurements.

***3. We can conduct the measurements, using the microwave resonators with the two embedded superconducting thin films only.*** The accurate characterization of a *single thin film* can be done by the *round-robin scheduling method* in the case of a dielectric resonator with the two thin films under the measurement.

***4.*** We research the physical properties of a *dielectric resonator with the two superconducting thin films*, measuring the dependences of its resonant frequency $f_0$, $S_{11}$, $S_{21}$ and $S_{22}$ on the temperature ***T*** in a certain temperatures range under the specified *RF* signal power ***P*** from ***-5dBm*** to ***+25dBm***.

***5.*** Using the *standard calculations method*, we can find the dependence of the *quality factor* $Q_0(T)$ *of an unloaded dielectric resonator* on the *temperature* for every selected temperature ***T****,* and compute the *surface resistances of the investigated superconducting thin films* $Rs(Y_1, Y_2)$, $Rs(Y_1, Y_3)$, $Rs(Y_2, Y_3)$ and $Rs(Y_1, Nd)$, $Rs(Y_2, Nd)$. In view of the fact that the s*eries equivalent scheme of a dielectric resonator* is being used to analyze the superconducting thin films, then $Rs = As/Q_0$. In the *series equivalent scheme*, the *surface resistances of both embedded superconducting thin films* contribute to the *total magnitude of measured surface resistance*. It is necessary to sum up the *surface resistances of the two superconducting thin films* to obtain the *total surface resistance of a dielectric resonator* as it is shown in the equations

$$\boldsymbol{Rs(Y_1,Y_2) = [Rs(Y_1, Y_1) + Rs(Y_2, Y_2)]/2,}$$

$$\boldsymbol{Rs(Y_1, Y_3) = [Rs(Y_1, Y_1) + Rs(Y_3, Y_3)]/2,}$$

$$\boldsymbol{Rs(Y_2, Y_3) = [Rs(Y_2, Y_2) + Rs(Y_3, Y_3)]/2.}$$

*where* ***$Rs(Y_1, Y_1)$, $Rs(Y_2, Y_2)$ and $Rs(Y_3, Y_3)$*** are the resistances of virtual resonators with the identical films ***$(Y_1, Y_1)$***, ***$(Y_2, Y_2)$*** and ***$(Y_3, Y_3)$.***

Going from the *physics*, the *surface resistance* ***Rs*** *describes the energy dissipation in a dielectric resonator*, hence *all the dissipative processes have to be summed up and they have a positive magnitude only*.

The same way, we can write the expressions in the cases of the ***(Y, Nd)*** superconducting thin films

$$\boldsymbol{Rs(Y_1, Nd) = [Rs(Y_1, Y_1) + Rs(Nd, Nd)]/2,}$$

$$\boldsymbol{Rs(Y_2, Nd) = [Rs(Y_2, Y_2) + Rs(Nd, Nd)]/2.}$$

where ***$Rs(Y_1, Y_1)$*** is the *resistance of the virtual resonator with identical films* ***$(Y_1, Y_1)$***.

We obtain the following equations for surface resistances Rs of films

$$\boldsymbol{Rs(Y_1,Y_1) = Rs(Y_1,Y_2) + Rs(Y_1, Y_3) – Rs(Y_2, Y_3),}$$

$$\boldsymbol{Rs(Y_2,Y_2) = Rs(Y_1, Y_2) + Rs(Y_2, Y_3) –Rs(Y_1, Y_3),}$$

$$\boldsymbol{Rs(Y_3,Y_3) = Rs(Y_1, Y_3) + Rs(Y_2, Y_3) – Rs(Y_1, Y_2).}$$

In right part of equation, we have all measured data. In other words, in left part of equations, we calculate the *surface resistance of a dielectric resonator with the two virtual embedded superconducting thin films* ***$(Y_1, Y_1)$***, ***$(Y_2, Y_2)$*** and ***$(Y_3, Y_3)$***. Therefore, the *surface resistance* ***Rs*** of a *single superconducting thin film* is in the two times smaller, because the ***surface resistance Rs is the additive value and it is necessary to take into account the sizes of two films at the calculation of the geometric factor As for a microwave resonator, then, for example***

$$\boldsymbol{Rs(Y_1, Y_1) = Rs(Y_1) + Rs(Y_1),}$$

***where $Rs(Y_1)$ is the calculated (not measured) resistance of an individual film.***

We selected the *superconducting thin films*, ***$Y_1$*** and ***$Y_2$***, among the measured superconducting thin films, because these *superconducting thin films* have the almost identical dependences of the *surface resistance* on the *temperature* ***Rs(T)*** at the specified *RF* signal power level.

In addition, we conducted the experimental measurements of the ***Rs(T)*** for the ***($Y_1$, Nd)*** and ***($Y_2$, Nd)*** at the specified *RF* signal power levels.

Going from the measured data, we get the following equations

$$\boldsymbol{Rs(Y_1, Nd) = [Rs(Y_1, Y_1) + Rs(Nd, Nd)]/2,}$$

$$\boldsymbol{Rs(Y_2, Nd) = [Rs(Y_2, Y_2) + Rs(Nd, Nd)]/2.}$$

After the summation, we obtain the expression

$$\boldsymbol{Rs(Y_1, Nd) + Rs(Y_2, Nd) = [Rs(Y_1, Y_1) + Rs(Y_2, Y_2)]/2 + Rs(Nd, Nd).}$$

Going from the above expression, we derive the equation to describe a *dielectric resonator with the two embedded superconducting thin films*

$$\boldsymbol{Rs(Nd, Nd) = Rs(Y_1, Nd) + Rs(Y_2, Nd) - [Rs(Y_1, Y_1) + Rs(Y_2, Y_2)]/2\ .}$$

It is necessary to note that we have all measured data in the right part of obtained equation.

# CONCLUSION

The main goal of thesis: to research the nonlinear properties of $YBa_2Cu_3O_{7-\delta}$ high temperature superconducting (*HTS*) thin films on *MgO* substrates at ultra high frequencies, facilitating the creation of electronic devices with advanced characteristics for the use in information communication technologies, is fully accomplished.

**Theoretical research in the dissertation consisted of:**

1. Introduction to the superconductivity theory. Analysis of present state of research on nonlinear microwave properties of *HTS* materials in Chapters 1, 2, 3.

2. Study of equivalent circuit models, proposed by a number of researchers, Wosik, Booth, Mateu, O'Callaghan, Ohshima to represent the superconducting resonators at microwaves is completed. The main research conclusion is that the series equivalent networks with small internal resistance are more suitable for analysis and accurate characterization of physical properties of superconductors at microwaves, but the parallel equivalent circuit representations are more appropriate for precise characterisation of dielectrics or semiconductors at microwaves. The consideration of limitations of *r*-parameter application for nonlinear models analysis in *HTS* microwave resonators is also described in Chapter 4.

3. Review on lumped element modelling of nonlinear properties of *HTS* thin films in microwave resonant circuits. Creation of both a lumped element model of superconductor and a lumped element model of microwave resonators for accurate characterisation of nonlinear properties of *HTS* thin films in microwave resonators. Derivation of equations for linear, quadratic, and exponential dependences of microwave power on frequency *P(f)* for different models of network was made in *Maple*. Modeling of linear, quadratic, and exponential dependences of microwave power on frequency *P(f)* for different models of network was made in *Matlab* in Chapter 5.

4. Modeling and identification of lumped element model parameters for superconducting *Hakki-Coleman dielectric resonator (HCDR)*. Simulation of nonlinear microwave responses of system with three types of *R* and *L* element dependences (linear, quadratic & exponential) in *Matlab*. Simulations showed that the lumped element approach adequately describes nonlinear properties of concrete *HTS* thin films in Chapter 6.

5. The detailed consideration on measurement accuracy issues in Chapter 6.

6. In Chapter 7, the simplified *RF* model, based on the *BCS* theory, to determine nonlinear behaviour of *Rs(P)* of *HTS* thin films near $H_{c1}$ and $H_{c2}$ was created. Simulation was conducted in *Matlab*. Simulation results demonstrated that the curves have *S*-type dependences for *Rs(P)* close to experimental results in microstrip resonators at microwaves.

7. The new source of noise: *the magnetic dipole two-level systems (MTLS) in HTS thin films,* based on the *Abricosov or Josephson magnetic vortices,* in superconducting microwave resonators is proposed firstly in Chapter 8.

8. The new *1/f noise quantum theory in HTS* at microwaves in Chapter 8.

9. The new computer modeling research results on the nature of *differential noise* in $YBa_2Cu_3O_{7-\delta}$ thin films at frequency of 25*GHz* are presented in Chapter 8.

10. Some theoretical aspects on the accurate characterization of nonlinearities in microwave resonators were considered in the *Appendix 1*.

**Experimental research in the dissertation included:**

1. Research on microwave properties of *MgO* substrates in *split post dielectric resonator* (*SPDR*) at $f$ = 10.48 *GHz*. The *MgO* substrates did not contribute to nonlinear properties in Chapter 6.

2. Research on nonlinear surface resistance of $YBa_2Cu_3O_{7-\delta}$ thin films on *MgO* substrates in *Hakki-Coleman dielectric resonator* (*HCDR*) at $f$ = 25*GHz*. $YBa_2Cu_3O_{7-\delta}$ thin films have nonlinear characteristics in form of *S*-type dependence *Rs(P)* at elevated microwave power levels, when $H_{rf}$ is higher than $H_{c1}$. The *Rs(P)* and $R_s(T)$ charts were created in *OriginPro* and *KaleidaGraph* in Chapter 6.

3. Research on nonlinear surface resistance of $YBa_2Cu_3O_{7-\delta}$ superconducting thin films on *MgO* substrates in microstrip resonators at $f_0$ = 1.985*GHz*. Microstrip resonators expressed *S*-type nonlinearity in *Rs(P)* dependence at the same power levels, when $H_{rf}$ is higher than $H_{c1}$ in Chapter 7.

4. Complex experimental setup with the application of the *Hewlett Packard Spectrum Analyser* with the use of the *GPIB* card and *LabView* software from *National Instruments* was constructed to perform the actual measurements. The calibration of measurements set up was performed. The proprietary dielectric and microstrip resonators were developed in the *GHz* frequency range in Chapters 6, 7.

Thesis's main conclusion is that the magnitude of critical magnetic field $H_{c1}$ of a superconductor need to be firstly measured to predict the nonlinear behaviour of any *HTS* thin films at microwaves as this is the level of magnitude of magnetic field at which the nonlinear effects arise. Also, it was discovered that the nonlinear processes arise in the certain range of microwave powers of an electromagnetic wave at which the variable magnetic field in a microwave resonator reaches the value of a critical magnetic field of a superconductor. Since, in a superconductor, there is a set of three critical magnetic fields: $H_{C1J}$ – the critical field of penetration of the Josephson magnetic vortices in the Josephson junctions on boundaries of crystal grains of a superconductor, $H_{C1}$ – the lower critical magnetic field of penetration of the *Abricosov magnetic vortices* in a superconductor, $H_{C2}$ - the upper critical magnetic field, at which the magnetic vortices disappear and superconductor becomes the normal metal. Thus, there are three regions, in close proximity to these critical magnetic fields, in which the nonlinear phenomena can be observed. The critical magnetic field $H_{C1J}$ in high-quality superconductors is not so essential, because of the fact that there are no enough the grain boundaries with the weak superconductivity in the high-quality superconductors, and accordingly, the influence of this magnetic field on the nonlinear phenomena is small. The influence of nonlinearity, connected with the *Josephson magnetic vortices,* is reduced to zero in the high quality superconducting thin films. Therefore, author did not give a separate detailed reviewing of this mechanism. The critical magnetic field $H_{C1}$ has an essential influence on the nonlinear properties as it is bound with the full volume of a superconductor. The same statement is true in relation to the critical magnetic field $H_{C2}$. Therefore, the author has researched the features of nonlinear properties of superconductors at the magnetic fields $H_{C1}$ and $H_{C2}$. The experimental researches by the author were carried out without the application of external magnet systems, and consequently concern to the nonlinearity originating near to the critical magnetic field $H_{C1}$, which plays an essential role in the microwave resonators, where the increase of a signal amplitude reduces in the reach of this critical magnetic field $H_{C1}$ and subsequent manifestation of nonlinear properties of superconductors. The field $H_{C2}$ has a very high value in high-temperature superconducting devices, and it cannot be reached by simple increase of a signal strength even in the high-*Q* resonator cavities, and the application of an external magnetic field $He \sim H_{C2}$, which can be generated by some external magnet systems, is necessary. The author has utilised the experimental outcomes by other researchers, which had no theoretical explanations till now, for the matching of his calculations with the nonlinear behaviour of a surface resistance $R_S$ close to the critical magnetic field $H_{C2}$. The author has used a well known formula from the *BCS* theory for the expression of the dependence of the surface resistance $R_S$ on the value of a superconducting energy gap $\Delta(r)$ to describe the behaviour of the surface resistance $R_S$ near to the critical magnetic fields. Since the energy gap $\Delta(r)$ is interlinked to the thermodynamic characteristics of a superconductor, hence under the conditions, when the penetration depth $\lambda$ of an electromagnetic wave is much greater than the coherence length $\xi$ of a superconductor, that is a characteristic case for the high-temperature superconductors, it is possible to change the dependence of an energy gap $\Delta(r)$ from the distance by the average thermodynamic value of an energy gap $\Delta(r)$, which essentially varies near to the critical magnetic field. It allows to disregard the process of calculation: How does the penetration of each magnetic vortex into the superconductor change its surface resistance $R_S$ in the local area $\sim \xi$, where the value of coherence length $\xi$ is approximately equal to the radius of the normal core of a magnetic vortex. The obtained dependences for both the critical magnetic fields $H_{C1}$ and $H_{C2}$, well feature a sort of dependence of the surface resistance $R_S$ near to these critical magnetic fields, as it is shown by the author in the given thesis. It is clarified that there is a nonlinearity of the surface resistance $R_S$ in the superconductor resonators at the microwave power range at which the magnitude of magnetic field $H_{rf}$ is close to the value of critical magnetic field $H_{C1}$. Under this condition, the superconductor is in a nonlinear state during a fraction of an oscillation period of an electromagnetic wave, when $H_{rf} > H_{C1}$. The same kind of the nonlinearity, which has some other dependence on the magnetic field, arises and near to the critical magnetic field $H_{C2}$, when the external magnetic field $H_e$ reaches the value $H_{C2}$. In this case, the magnetic field of an electromagnetic wave is much less in comparison with the external magnetic field $H_{rf} << H_e$, and the superconductor is in the nonlinear state during all the period of oscillation of an electromagnetic wave. The two characteristic dependences are researched comprehensively: 1) the surface resistance on external magnetic field $R_S(H_e)$ in close proximity to the critical magnetic field $H_{C1}$, and 2) the surface resistance on external magnetic field $R_S(H_e)$ in close proximity to the critical magnetic field $H_{C2}$.

It is assumed that optimization of Abricosov magnetic vortices pinning, including the *Abricosov magnetic vortices* - nanoscopic objects with lateral dimensions of a few *nm* and spiral like shape – pinning on green phase nano clusters with 5-10*nm* diameter can increase critical current $I_C$ in *HTS* thin films and enhance microwave power handling capabilities of microwave resonators at high magnitudes of applied microwave power in mobile and space wireless communication systems.

The theoretical computer simulation results, derived in the given research work, well feature the experimental data registered by the author at the measurements of superconductor parameters in the dielectric and microstrip resonators in close proximity to the critical magnetic field $H_{C1}$; and the experimental data, published by other researchers, on the measurements of superconductor parameters near to the critical magnetic field $H_{C2}$. Thus, it is assumed that it is enough to find the values of critical magnetic fields $H_{C1}$ and $H_{C2}$ for the used superconductors, and then to take an advantage of derived mathematical expressions with the purpose to obtain the dependences of the nonlinear surface resistance on the external magnetic fields $R_S(H_e)$ close to these critical magnetic fields to predict the operational characteristics of *HTS* in electronic devices. The author has checked the derived mathematical expressions for $YBa_2Cu_3O_{7-\delta}$ thin films surface resistance $R_S$ in the temperature range from 15*K* up to 82*K*. The magnitudes of magnetic fields, retrieved by the author, at which the nonlinearities start to appear in the superconductor thin films at microwaves, well coincide with the critical magnetic fields of the $YBa_2Cu_3O_{7-\delta}$ superconductor reported by other researchers.

The successful application of *HTS* thin films in emerging *RF* devices makes it possible to achieve the better technical characteristics in linear regime of their operation, when the magnitude and phase changes are only imposed on complex input signals without the new signals generation. The problem is that *RF* devices in nonlinear mode of operation can shift the complex input signals in frequency domain and/or generate the harmonics and intermodulation products, causing the severe signal distortion. Therefore, the determination of magnitude of critical magnetic field $H_{c1}$ of a superconductor can provide the exact data on the threshold magnitudes of microwave power at which the *RF* device with *HTS* will begin to exhibit the nonlinear behaviour, if driven with elevated microwave input signal power. The better linear specifications of *RF* devices can be achieved by improving the quality of *HTS* thin film synthesis as well as by precisely controlling and maintaining the applied level of microwave input signal power and operational temperature of *RF* component to gain stable functional performance of *RF* devices with *HTS* at microwaves. The research on the nonlinearities allows predicting the physical behavior of *HTS* thin films in electronic devices, having only obtained the data on the values of the critical magnetic fields of a superconductor at microwaves.

New innovative strategies to design novel highly linear *HTS* microstrip filters with enhanced microwave power handling capabilities, reduced microwave power level of intermodulation distortion (*IMD*), high interception point three (*IP3*), low insertion loss characteristics and compact design are:

1. Improvement of material properties of High Temperature Superconductors (*HTS*) through the synthesis of *high quality HTS* thin films without crystal defects or imperfections and with big magnitude of critical magnetic field $H_{c1}$, leading to small non-linear response of microwave resonator to high level of microwave power.

2. Improvement of material properties of High Temperature Superconductors (*HTS*) through the use of composite *HTS* thin films with *green phase nano-clusters, oxide nano-clusters, nano-impurities* to improve the *Abricosov magnetic vortices* pinning, and hence increase the magnitude of critical current *Ic,* leading to small non-linear response of microwave resonator to high microwave power at $H_{rf}>H_{c1}$.

3. Optimization of geometrical parameters of *HTS* thin films by *increasing the thickness of HTS thin films,* using of *stacks with many HTS thin films layers,* or by *increasing the width of microstrip line in HTS resonator* in *HTS* microstrip filter.

4. Optimization of geometric structure of *HTS* thin films, deposited on substrate, by creation of *sliced HTS microstrip lines.* For example, the use of *sliced HTS microstrip lines* for the design of microwave resonators in a transmitting filter.

5. Optimization of geometric structure of *HTS* thin films in *HTS* microstrip filter by the use of *HTS microstrip line with slots* instead of solid microstrip line to create the *split resonators*, aiming to reduce the peak current density at outer edges of *HTS* microstrip and avoid high current density *HTS* microstrip discontinuities.

6. Optimization of geometry of *HTS* microstrip filter by *rounding all the edges and corners* of *HTS* thin films in microstrip filters in receivers and in transmitters. For instance, the use of *HTS split open-ring resonators* (*SRR*) in microstrip filters in a transmitter.

7. Miniaturization of high pole *HTS* microwave filters with increased microwave power handling capability, by using the stripline (*SL*) resonators with shorter distance between resonators, resulting from a weak coupling. The modified *SL* filter with increased spacing between resonators to reduce maximum surface current has smaller size in comparison with the *HTS* microstrip line (*MSL*), dual-mode or bulk resonators.

8. Optimization of *HTS* microstrip filter design, using the different design techniques toward the intermodulation distortion reduction in *HTS* microstrip filters. The *ferroelectrics* may be used for maximum *IMD* reduction with a minimum effect on losses in *HTS/FER* microstrip filters. For example, the suppression of *IMD*, generated by *HTS* materials, with the use of a *nonlinear ferroelectric segment* for nonlinear pre-distortion in *HTS* band-pass filter.

9. Introduction of *quality control* during microwave filters fabrication to avoid the edge defects in *HTS* thin films, which can lead to the decrease of critical magnetic field $H_{c1}$. The absence of edge defects in *HTS* thin films will significantly improve the characteristics of HTS microstrip filter.

10. Maintenance of *thermal stability* of *HTS* microwave filters by constantly adjusting the real operational temperature to normal operational temperature, which is equal to $0,8T_c$, where $T_c$ is the critical temperature of *HTS*.

11. Improvement of the *packaging* of *HTS* microwave filters to avoid the vibration related problems with their tunning in space and airborne applications.

In the author's opinion, the progress toward the application of *HTS* thin films in *HTS* microwave resonators and novel electronic device technologies will continue to redefine the strategic boundaries of hi-tech *ICT* industry for many decades to come.

## LIST OF FIGURES

### CHAPTER 1

**CHAPTER 2**

**CHAPTER 3**

A. V. Velichko, A. Porch, M. J. Lancaster, R. G. Humphreys, Anomalous Features in Surface Impedance of YBaCuO Thin Films: Dependence on Frequency, RF and DC Fields, *IEEE Trans. Appl. Superconductivity,* vol. **11,** pp. 3497-3500, 2001. (© 2001 IEEE.) (after [529, 46]).

**Fig. 58.** Temperature dependences of the effective *Rs* of an e-beam co-evaporated $YBa_2Cu_3O_{7-\delta}$ film on *MgO* at 2.3 *GHz* and two input power levels: −60 *dBm—hatched squares*, and −20 *dBm—dots*. The inset shows the normalized effective *Rs* for the $YBa_2Cu_3O_{7-\delta}$ film (*diamonds*) and an *Nb* film on *MgO* (triangles) at 5 *K*. The absolute levels of effective *Rs* for the two films at 1.7 *K* were identical. Reprinted by permission from M. A. Hein, P. J. Hirst, R. G. Humphreys, D. E. Oates, A. V. Velichko, Nonlinear Dielectric Microwave Losses in MgO Substrate, *Appl. Phys. Lett.*, vol. **80**, no. 6, pp. 1007-09, 2002. (© 2002 AIP.) (after [536, 46]).

**Fig. 59.** Typical power dependence of the effective *Rs* (circles, dashed curve), *Xs* (squares, dashed curve), and intermodulation distortion (*IMD*) signal (diamonds, solid curve) for $YBa_2Cu_3O_{7-\delta}$ film on *MgO* at 2.3 *GHz* and 5 *K*. Reprinted by permission from M. A. Hein *et al*, *J. Superconductivity,* vol. **16**, p. 895, 2003.. Springer (© 2003 Science and Business Media.) (after [540, 46]).

**Fig. 60.** *DC* magnetic field dependence of (a) −3*dB* bandwidth and (b) the resonant frequency of $YBa_2Cu_3O_{7-\delta}$ microstrip resonator working at fundamental mode (~1.1*GHz*) with input power $P_{mw}$ =−10 *dBm* at 77 *K*. The film is deposited by the *PLD* technique on $LaAlO_3$ substrate. Circles correspond to $H_{dc} // c$ and triangles correspond to $H_{dc} \perp c$. The insets show the curves normalized by $H_{dc}^{//}$ and $H_{dc}^{\perp}$, respectively [541]. Reprinted by permission from X. S. Rao, C. K. Ong, B. B. Jin, C. Y. Tan, S. Y. Xu, P. Chen, J. Lee and Y. P. Feng, *Physica C*, vol. **328**, p. 60, 1999. (© 1999 Elsevier) (after [541, 46]).

**Fig. 61.** Normalized changes in surface resistance $\Delta Rs = [Rs(H_{rf}) - Rs\ (0)]/Rs\ (0)$ and the penetration depth $\Delta\lambda = [\lambda(Hrf) - \lambda(0)]/\lambda(0)$ as a function of the microwave field $H_{rf}$ in zero dc field, 8 *GHz* and 13 *K*, for three $YBa_2Cu_3O_{7-\delta}$ films thermally evaporated on *MgO* substrates. Typical *Rs* values at the conditions stated above are ~40–50 μΩ [542]. Reprinted by permission from Velichko, Huish, Lancaster, Porch, *IEEE Trans. Appl. Supercond.* 13 3598. (© 2003 IEEE.) (after [542, 46]).

**CHAPTER 4**

**CHAPTER 5**

**CHAPTER 6**

**CHAPTER 7**

**Fig. 2.** Standing half-wave microstrip resonator and distribution of current density along resonating element.

**Fig. 3.** Cross section of HTS thin film used in microstrip resonator. Superconducting current transports in direction of X axis.

**Fig. 4.** Modeling of current density distribution on cross section of superconducting microstrip resonator with resonance frequency $f_0$ = 1.985GHz made of $YBa_2Cu_3O_{7-\delta}$ thin films on MgO substrate [19, 20].

**Fig. 5.** Model of two-port network inserted into transmission line with incident and reflected electromagnetic waves schematically shown at microwaves.

**Fig. 6.** Cryogenic measurement system for research on nonlinear resonance response of YBa2Cu3O7-d thin films on MgO substrates in microstrip resonators at microwaves.

**Fig. 7.** Experimental dependence of transmission coefficient on ultra high frequency $S_{21}(f)$ in $YBa_2Cu_3O_{7-\delta}$ superconducting microstrip resonator at different microwave power levels at temperature of 25 K.

**Fig. 8.** Experimental dependence of transmission coefficient on ultra high frequency $S_{21}(f)$ in $YBa_2Cu_3O_{7-\delta}$ superconducting microstrip resonator at different microwave power levels at temperature of 82 K.

**Fig. 9.** Experimental dependence of resonance frequency $f_0$ on microwave signal power P in $YBa_2Cu_3O_{7-\delta}$ superconducting microstrip resonator at microwaves at different temperatures. Shift of resonance frequency is observed.

**Fig. 10.** Experimental dependence of quality factor on microwave signal power $Q$(P) in $YBa_2Cu_3O_{7-\delta}$ superconducting microstrip resonator at frequency $f \approx$ 2GHz at different temperatures.

**Fig. 11.** Experimental dependence of surface resistance on microwave signal power $R_S$(P) in $YBa_2Cu_3O_{7-\delta}$ superconducting microstrip resonator at frequency $f \approx$ 2GHz at different temperatures.

**Fig. 12.** Experimental dependence of surface resistance on magnetic field $R_s$ ($H_{rf}$) in $YBa_2Cu_3O_{7-\delta}$ superconducting microstrip resonator at frequency $f \approx$ 2GHz at different temperatures.

**CHAPTER 8**

## CHAPTER 9

**APPENDIX I**

**APPENDIX II**

N/A

**APPENDIX III**

# LIST OF TABLES

**CHAPTER 3**

**CHAPTER 4, 5**

N/A

**CHAPTER 6**

**CHAPTER 7, 8**

N/A

**CHAPTER 9**

**APPENDIX I**

N/A

## APPENDIX II

## APPENDIX III

# SELECTED RESEARCH PUBLICATIONS

The selected research publications, which have been published by Dimitri O. Ledenyov in the process of international collaboration with other researchers in the field of microwave superconductivity. In addition, a number of complex software programs in *Matlab* have been developed by the authors of book in 2000-2015.

# SUBJECT INDEX

# AUTHOR INDEX

# EXTENDED INDEX

# INDEX TO SYMBOLS

$a$ radius of circular cross section of long superconducting wire

$a_2$ fitting coefficient defining the speed of change $R_s(H)$

$\alpha$ electrical polarizability

$\alpha \mathbf{E} = \mathbf{P}$ polarization of environment

$\alpha_k$ amplitude of the $k^{th}$ pulse in the sequence

a constant ~ 1

$\alpha'(\omega)$ and $\alpha''(\omega)$ response functions, which depend on ultra high frequency

$\overline{\alpha^2}$ mean-square value of the pulse amplitudes

**A**(**r**, t) vector potential

A constant in formula for $R_s$

$A_s$, $A_m$, and $p_e$ geometric factors of microwave resonator

$\Delta A_S$, $\Delta Q_0$, $\Delta R_S$, $\Delta A_m$, $\Delta p_e$ and $\Delta \tan\delta$ six independent variables used in derivation of $|\Delta R_S/R_S|$ function

$b^*$and $b$ creation and destruction operators for phonons of wave vector $q$

b coefficient, which matches the dimensions of the multi-pair numerator and denominator

β $\beta = \pi/l$, $l$ is the height of the cavity in dielectric microwave resonator

$\beta_A$ constant $\beta_A \geq 1$

$\beta_1$ and $\beta_2$ coupling coefficients of microwave resonator

**B** magnetic induction

$B$ value $B = A\exp[-(n_s\Delta)/n \cdot kT]$

$B_1$ amplitude of change for surface resistance

$c$ velocity of light

$c^*_{k\sigma}$ and $c_{k\sigma}$ creation and annihilation Fermi operators

$\delta$ delta-function

δ electromagnetic field penetration depth $\delta = c/(\omega N'')$

$\Delta(r)$ energy gap in BCS theory

$\Delta_r \tan\delta$ uncertainty in measured $\tan\delta$ of *MgO* substrates

$\Delta R_S$ magnitude of correction to the $R_S$ value

$\Delta R_s/R_s$ error band limits on graphs for derived surface resistance $\Delta R_s/R_s = 1\%$

$\Delta Q/Q$ error band limits on graphs for measured quality factor $\Delta Q/Q = 1\%$

$d$ effective dimension of inhomogeneity

**D** electric field

$\varepsilon_0$ dielectric constant of free space

$\varepsilon'_r$ real relative permittivity of MgO substrates

$\varepsilon$ energy of electron of wave vector $q$ in normal state

$\varepsilon\varepsilon_0\mathbf{E}$ induction of an electric field

$e$ charge of an electron

**E** electric field

$E$ energy

$E_F$ Fermi level energy

**f**(t) generalized force, which is a function of time

$f$ frequency

$f_c$ pass frequency bandwidth

$f_{res}$ resonance frequency

$f_0$ center frequency

$1/f$, $1/f^2$ intermodulation noise

$\phi_0$ magnetic flux quantum

Φ magnetic flux

$F_N$ density of free energy of normal metal

$F_S$ density of free energy of superconductor
$F(j\omega)$ *Fourier transform* of the pulse shape function $f(t)$
$F$ noise factor
$\boldsymbol{F_L}$ Lorentz's force
$\boldsymbol{F_P}$ pinning force of Abricosov magnetic vortices
σ spin of an electron
$\sigma_1$ and $\sigma_2$ real and imaginary parts of complex conductivity
$g$ electron phonon coupling constant
$g_0$ and $b_0$ real and imaginary parts of that source admittance, which results in minimum noise figure, $F_{min}$
$g_{M1}$ energy, which is bound with the penetration of *Abricosov magnetic lines* into a superconducting sample
$G_s$ Gibbs free energy
$G_s$ and $b_s$ real and imaginary parts of source admittance
$\Gamma_S$ reflection coefficient
h Planck's constant
$\hbar$ Planck's constant divided by 2π
$\hbar\omega$ average phonon energy in BCS
**H** magnetic field
$H_{c1}$ lower critical magnetic field
$H_{c1J}$ Josephson lower critical magnetic field
$H_{c2}$ upper critical magnetic field
$H_e$ external magnetic field
$H_{max}$ maximum magnetic field
$H_{rf}$ applied magnetic field
$H_{rf}$* effective applied magnetic field
$H$ Hamiltonian
$i_{n1(t)}$ and $i_{n2}(t)$ noise current generators
$I_C$ critical current
$I_m$ *m*-th order modified *Bessel function*
$J_c$ critical current density
$J_m$ *m*-th order *Bessel function* of the first kind
***K*** vector of reciprocal lattice
$K$ number of superconducting pairs of electrons
$K_m$ *m*-th order modified *Hankel function*
***k*** wave vector
k Ginzburg-Landau parameter
$k_B$ Boltzmann constant $k_B=1.38x10^{-23}$*joule/degK*
$l$ orbital momentum number of an electron
$l$ mean free path of normal electron excitations in high temperature superconductors
$L$ size of resonator
λ penetration depth
$\lambda_L$ penetration depth in London limit
$\mathcal{L}(t)$ fluctuation, which describes the noise
$\mu_0$ magnetic permeability
$\mu\mu_0\boldsymbol{H}$ induction of magnetic field **B**
***m*** and ***–m*** intrinsic magnetic momentums of each Josephson magnetic vortex
$n_{k\sigma} = c^*_{k\sigma}c_{k\sigma}$ particle number operator
$n_S$ density of superconducting electrons in the two-fluid Gorter-Casimir theory
$N$ electron density
$N$ number of photons in a microwave resonator
$N$ index of refraction of environment
$N_i$ noise figure parameter
$N(0)$ density of states at the Fermi level for electrons with one spin orientation
$NF$ noise figure
$P$ principal value of integral
$P(f)$ microwave power
$\varphi_n$ phase angle of electromagnetic wave counted from the beginning of period of

| | |
|---|---|
| | oscillation at which magnetic field $H_{rf}(t)$ becomes equal to critical magnetic $H_{c1}$ |
| $Q_0$ | unloaded quality factor of microwave resonator |
| $Q_L$ | loaded quality factor of microwave resonator |
| $Q_d$ | quality factor $Q_d = 1/p_e tan\delta$ |
| $Q(P)$ | quality factor of microwave resonator |
| $Q_c$, $Q_d$, and $Q_r$ | quality factors of microwave resonator, associated with the conductor, dielectric and radiation losses respectively |
| $r$ | given point of space |
| $r_n$ | equivalent input noise resistance of the two port |
| $\mathbf{r}$ | parameter $\mathbf{r} = \Delta X/\Delta R$ |
| $\rho$ | density of electric charge |
| $R_n$ | surface resistance of normal metal |
| $R_S$ | surface resistance of superconductor |
| $R_{eff}$ | effective surface resistance of superconductor |
| $R_{tot}$ | total surface resistance of superconductor |
| $<R_s>$ | average value of surface resistance of superconductor |
| $R_{SNlin}$ | nonlinear surface resistance |
| $S_{11}$ | input reflection coefficient with the output matched |
| $S_{21}$ | forward transmission coefficient with the output matched |
| $S_{22}$ | output reflection coefficient with the input matched |
| $S_{12}$ | reverse transmission coefficient with the input matched |
| $S(\omega)$ | spectral density of microwave power |
| $\tau$ | time |
| $\tau_N$ | time of nucleation of an *Abricosov magnetic vortex* |
| $t$ | time |
| $t_k$ | time at which the $k^{th}$ event occurs |
| tan $\delta$ | loss tangent of *MgO* substrates |
| $T$ | temperature |
| $T_c$ | critical temperature |
| $TE_{011}$ or $TE_{01\delta}$ | modes of electro-magnetic wave oscillation in microwave resonator |
| $v$ | mean rate of events |
| $v_{ph}$ | phase velocity $v_{ph} = c/(\varepsilon\mu)^{1/2}$ or $v_{ph} = c/N$ |
| $V$ | volume |
| $V_{\boldsymbol{kl}}$ | matrix element for scattering from a state $(\boldsymbol{l}\uparrow,-\boldsymbol{l}\downarrow)$ to one with $(\boldsymbol{k}\uparrow,-\boldsymbol{k}\downarrow)$ |
| $<V>$ | average energy of Cooper electron pair |
| $V$ | energy potential in *BCS theory* |
| *VSWR* | voltage standing wave ratio |
| $\omega_q$ | radian frequency of a phonon of wave vector $q$ |
| $\omega_C$ | critical frequency |
| $W$ | probability |
| $\xi_0$ | superconducting electrons coherence length |
| $\boldsymbol{\xi_{GL}}$ | Ginzburg-Landau coherence length |
| $\chi$ | magnetizability of environment |
| $\chi\boldsymbol{H}=\boldsymbol{J}$ | magnetic moment of environment |
| $\hat{\mathbf{x}}$ | operator of given physical quantity |
| $X$ | gauged physical quantity X=a**f** |
| $X_s$ | surface reactance |
| $\Psi(\mathbf{r})$ | wave operator for electrons |
| $\Upsilon$ | coefficient $\Upsilon=k_BT/R_n$ |
| $Z_S$ | surface impedance of superconductor $Zs = Rs + jXs$ |
| $Z_s$ | impedance of source in network |
| $Z_L$ | impedance of load in network |
| $Z_r$ | characteristic impedance of resonance circuit $Z_r=R_r+jX_r$ |
| $Z_0$ | characteristic impedance of microstrip transmission line |

# GLOSSARY

**1 nm** One Nanometre ($10^{-9}$ metre)
**10GEA** 10 Gigabit Ethernet Alliance
**1R** Re-amplification
**2D** Two (2) Dimension
**2R** Re-amplification, Re-shaping
**3D** Three (3) Dimension
**3G** Third Generation
**3GPP** Third Generation Partnership Project
**3R** Re-amplification, Re-shaping and Re-timing
**4G** Fourth Generation
**4GMF** 4G Mobile Forum
**AAA** Authentication, Authorization, and Accounting
**AAAC** Authentication, Authorization, Accounting and Charging
**ACLR** Adjacent channel leakage power ratio
**ADC** Analog to Digital Converter
**ADSL** Asymmetric Digital Subscriber Line
**AFM** Atomic Force Microscope
**AM** Amplitude Modulation
**AMP** Asymmetric Multiprocessing
**Ångstrom** $10^{-10}$ of a metre
**ANSINSP** American National Standards Institute Nano-technology Standards Panel
**API** Application Programming Interface
**APON** Asynchronous Transfer Mode Passive Optical Network
**APP** A Posteriori Probability decoder
**ASIC** Application-Specific Integrated Circuits
**ATM** Asynchronous Transfer Mode
**ASSP** Application-Specific Standard Product
**ATM** Asynchronous Transfer Mode
**BASP** Broadband Access Service Provider
**BCI** Brain Computer Interface
**BER** Bit Error Rate
**BPSK** Binary Phase Shift Keying modulation
**BPON** Broadband Passive Optical Network
**BWA** Broadband Wireless Access
**C60** Carbon 60 buckyball or Buckminster fullerene
**CAD** Computer Aided Design
**CD** Compact Disc
**CDG** Code Division Multiple Access Development Group
**CDM** Code Division Multiplex
**CDMA** Code Division Multiple Access
**CDN** Content Distribution Network
**CIDR** Classless Inter-Domain Routing
**CIGS** Copper-Indium-Gallium-Diselenide
**CMA** Constant Modulus Algorithm
**CMOS** Complementary Metal Oxide Semiconductor
**CNH** Carbon Nanohorn
**CNT** Carbon Nanotube
**CPFSK** Continuous Phase Frequency Shift Keying
**CPU** Central Processing Unit
**CR** Cognitive Radio
**CRC** Cyclic Redundancy Check Coding
**CRLH** Compact Right Left Handed metamaterial structures
**CRT** Cathode Ray Tube
**CTFE** Cryogenic Transceiver Front End
**CRFE** Cryogenic Receiver Front End
**CW** Continuous Wave
**CWT** Continuous Wavelet Transform
**CZT** Chirp-Z Transform
**DAB** Digital Audio Broadcasting
**DARPA** Defense Advanced Research Projects Agency
**DCD** Direct Conversion Device
**DCT** Discrete Cosine Transform

**DDoS** Distributed Denial of Service
**DFB** Distributed Feedback QW
**DiffServ** Differentiated Services
**DMFC** Direct Methanol Fuel Cell
**DMR** Digital Modular Radio
**DNA** Deoxyribonucleic Acid
**DOA** Direction of Arrival
**DOCSIS** Data Over Cable Service Interface Specification
**DoS** Denial of Service
**DPSK** Differential Phase Shift Keying
**DRAM** Dynamic Random Access Memory
**DSC** Dye Solar Cell
**DSI** Data Storage Institute
**DSL** Digital Subscriber Line
**DSP** Digital Signal Processing
**DSSS** Direct Sequence Spread Spectr.
**DTN** Delay Tolerant Networking
**DTTB** Digital Terrestrial Television Broadcasting
**DVB-H** Digital Video Broadcasting-Handheld
**DVD** Digital Versatile Disk
**DWDM** Dense Wavelength Division Multiplexing
**DWT** Discrete Wavelet Transform
**EDGE** Enhanced Data Rates for GSM Evolution
**eDRAM** Embedded Dynamic Random Access Memory
**EDT** Embedded Data Technology
**EEPROM** Electrically Erasable Programmable Read Only Memory
**EFM** Ethernet First Mile
**ENUM** Electronic Numbering
**EPC** Electronic Product Code
**EPON** Ethernet Passive Optical Net.
**EPROM** Erasable Programmable Read Only Memory
**ERC** Economic Review Committee
**ESA** European Space Agency
**ESES** Electronically Steered Electronically Scanned radar
**eSRAM** Embedded Static Random Access Memory
**EDCT** Even Discrete Cosine Transf.
**EUDCH** Enhanced Uplink Data Channel
**FARADS** Forwarding Directives, Associations, Rendezvous, and Directory Service
**FBO** Facilities-Based Operators
**FCC** Federal Communications Commission
**FDD** Frequency Division Duplex
**FDDI** Fibre Distributed Data Interface
**FDMA** Frequency Division Multiple Access
**FDTD** Finite Difference Time Domain image reconstructiontechnique
**FEC** Forward Error Correction
**FED** Field Emission Display
**FeRAM** Ferroelectric Random Access Memory
**FET** Field Effect Transistor
**FF** Form Factor
**FFT** Fast Fourier Transform
**FHSS** Frequency Hopping Spread Spectrum
**FIR** Finite Impulse Response filter
**FOLED** Flexible OLED
**FOMA** Freedom of Mobile Multimedia Access
**FPGA** Field Programmable Gate Array
**FSK** Frequency Shift Keying modulation
**FSO** Free Space Optics
**FTTH** Fibre-To-The-Home
**GaAr** Gallium Arsenide
**GaN** Gallium Nitride
**GB** Gigabyte
**Gb/in$^2$** Gigabit per square inch
**Gbit** Gigabit or Gb
**GDP** Gross Domestic Product
**GHz** Gigahertz
**GMSK** Gaussian Minimum Shift Keying modulation
**GN** Gaussian Noise
**GNI** Gross National Income
**GPON** Gigabit Passive Optical Net.
**GPRS** General Packet Radio Service
**GPS** Global Positioning System

**GSM** Global System for Mobile Communications
**GSMA** Global Mobile Suppliers Association
**GUI** Graphical User Interface
**HAMR** Heat-Assisted Magnetic Recording
**HBT** Hetero-junction Bipolar Transistor
**HCI** Human Computer Interface
**HDD** Hard Disk Drive
**HDTV** High Definition Television
**HEMT** High Electron Mobility Trans.
**HEP** High Energy Physics
**HFC** Hybrid Fibre Coaxial
**HSDPA** High Speed Downlink Packet Access
**HTMT** Hybrid Technology Multi-treaded Architecture
**HTS** High Temperature Superconductor
**HVD** Holographic Versatile Disk
**I/O** Input / Output
**IC** Integrated Circuit
**ICANN** Internet Corporation for Assigned Names and Numbers
**ICT** Information and Communications Technology
**IDA** Institute for Defense Analysis
**IEC** International Electro-technical Commission
**IEEE** Institute of Electrical and Electronics Engineers
**IETF**Internet Engineering Task Force
**IFFT** Inverse Fast Fourier Transform
**IHPC** Institute of High Performance Computing
**IIR** Infinite Impulse Response Filter
**IME** Institute of Microelectronics
**IMRE** Institute of Materials Research & Engineering
**IMT-2000** International Mobile Telecommunications-2000
**Info-MICA** Information-Multilayered Imprinted CArd
**IntServ** Integrated Services
**IP** Internet Protocol
**IP3** Interception Point 3
**IPSec** IP Security
**IPv4** Internet Protocol version 4
**IPv6** Internet Protocol version 6
**I/Q** In-phase and Quadrature phase imbalance constellation diagrams
**IRAM** Intelligent Random Access Memory
**ISDB-T** Integrated Services Digital Broadcasting Terrestrial
**ISDN** Integrated Services Digital Net
**ISI** Inter Symbol Interference
**ISM** Industrial, Scientific, Medical
**ISO** International Standards Org.
**ISP** Internet Service Provider
**IST** Information Society Technol.
**IT** Information Technology
**ITRS** International Technology Roadmap for Semiconductor
**ITU** International Telecom. Union
**ITU-R** International Telecommunication Union: Radio communication Standardization Sector
**ITU-T** International Telecommunication Union: Telecommunication Standardization Sector
**JCU** James Cook University
**JTRS** Joint Tactical Radio System
**Kbps** Kilobits per second
**KHz** Kilohertz
**Lambda** Light wavelength variable
**LAN** Local Area Network
**LBS** Location Based Services
**LCD** Liquid Crystal Display
**LCOS** Liquid Crystal On Silicon
**LCP** Liquid Crystal Polymer
**LDPC** Low Density Parity Check codes in error correction
**Li-ion** Lithium Ion
**Li-polymer** Lithium Polymer
**LLR** Long Likelihood Ratio
**LMS** Least Mean Square adaptive algorithm object
**LNA** Low Noise Amplifier
**LSI** Large Scale Integration
**LTCC** Low Temperature Co-fired Ceramics

**LTE** Latest Technology Evolution
**LTS** Low Temperature Supercond.
**mA** milliampere
**MAMMOS** Magnetically Amplifying MO System
**MAN** Metropolitan Area Network
**Mbps** Megabits per second
**MBWA** Mobile Broadband Wireless Access
**MDI** Missile Defense Initiative
**MEMS** Micro Electro Mechanical System
**MFC** Micro Fuel Cell
**MIMD** Multiple Instructions Multiple Data
**MIMO** Multiple Input Multiple Output
**MMIC** Monolithic Microwave Integrated Circuit
**MMSE** Minimum Mean Square Error filtering
**MQW** Multiple Quantum Well
**mITF** Mobile IT Forum
**MgO** Magnesium Oxide
**ML** Markup Language
**Moletronics** Molecular Electronics
**MOSFET** Metal Oxide Semiconductor Field Effect Transistor
**MP3** MPEG-1 Audio Layer-3
**MPLS** Multi-Protocol Label Switching
**MRAM** Magnetoresistive Random Assess Memory
**MRI** Magnetic Resonance Imaging
**MSB** Most Significant Bit
**MSDCT** Modified Symmetric Discrete Cosine Transform
**MSRC** Modular Software-Programmable Radio Consortium
**MSK** Minimum Shift Keying
**MTBF** Mean Time Between Failure
**mW** milli-watt
**MWNT** Multiwall Carbon Nanotube
**NASA** National Aeronautics and Space Administration
**NAT** Network Address Translation
**NED** Nano Emission Display
**NEMS** Nano Electro Mechanical System
**NGN** Next Generation Network
**Ni-Cad** Nickel Cadmium
**NiMHd** Nickel Metal Hydride
**NIST** National Institute of Standards & Technology
**NRAM** Nanotube Random Access Memory
**ns** nanosecond
**OADM** Optical Add-Drop Multiplexer
**OCDM** Optical Code Division Multiplexing
**ODCT**Odd Discrete CosineTransform
**OECD** Organisation for Economic Co-operation & Development
**OEO** Optical Electrical Optical
**OFCDM** Orthogonal Frequency Code Division Multiplexing
**OFDM** Orthogonal Frequency Division Multiplexing
**OFET** Organic Field Effect Transistor
**OLED** Organic Light Emitting Diode
**OLT** Optical Line Terminating
**ONU** Optical Network Unit
**OOO** All Optical
**OPSK** Offset Phase Shift Keying
**OFDM** Orthogonal Frequency Division Multiplexing
**OS** Operating System
**OSC** Organic Solar Cell
**OSI** Open Systems Interconnection
**OTDM** Optical Time Division Multiplexing
**OUM** Ovonics Unified Memory
**OVSF** Orthogonal Variable Spreading Factor
**OXC** Optical Cross Connect
**P2P** Peer-to-Peer
**PAM** Pulse Amplitude Modulation
**PAN** Personal Area Network
**PBG** Photonic Band Gap
**PC** Personal Computer
**PCF** Photonic Crystal Fibre
**PDP** Plasma Display Panel
**PE** Processing Element

**PEDGUI** Printed Embedded Data Graphical User Interface
**PET** Privacy Enhancing Technology
**PHI** Public Health for the Internet
**PHY** Physical Layer
**PIM** Passive Intermodulation
**PKI** Public Key Infrastructure
**PLC** Power-line Communications
**PLD** Programmable Logic Device
**PLED** Polymer Light Emitting Diode
**PLL** Phase Locked Loop
**PM** Phase Modulation
**PMR** Perpendicular Magnetic Record.
**PON** Passive Optical Network
**PRAM** Phase Change Random Access Memory
**PSTN** Public Switched Telephone Net
**PV** Photovoltaics
**PZT** Lead-Zirconium-Titanate
**QAM** Quadrature Amplitude Modul.
**QD** Quantum Dot
**QoS** Quality of Service
**QRNG_MFQ** Quantum Random Number Generator on Magnetic Flux Qubits
**QW** Quantum Well
**R&D** Research & Development
**RAM** Random Access Memory
**RANT** Random Access Nanotube Test
**RAW** Reconfigurable Architecture Workstation
**RBA** Role-Based Architecture
**RDF** ResourceDescriptionFramework
**redox** Reduced or Oxidised
**RF** Radio Frequency
**RFIC** Radio Frequency Integrated Circuit
**RFID** Radio Frequency Identification
**RISC** ReducedInstructionSetComputer
**RMS** Recursive Least Squares adaptive algorithm object
**RSA** Rivest, Shamir, & Adleman public key encryption technol.
**RSFQ** Rapid Single Flux Logic
**RSVP** Resource ReSerVation Protocol
**RTD** Resonant Tunnelling Diode
**SAIC** Scientific Applications Industrial Corporation
**SAR** Synthetic Aperture Radar
**SAW** Surface Acoustic Wave filter
**S&T** Science & Technology
**SBO** Services-Based Operators
**SBT** Strontium-Bismuth-Tantalum oxide
**SCA** Software Communications Architecture
**SCTP** Stream Control Transmission Protocol
**SDCT** Symmetric Discrete Cosine Transform
**SDMA** Spatial Division Multiple Access
**SDR** Software Defined Radio
**SED** Surface-conduction Emission Display
**SEM** Scanning Electron Microscopy
**SET** Single Electron Transistor
**SHDSL** Single-Pair High-Speed Digital Subscriber Line
**SHG** Second Harmonic Generation
**SIMD** SingleInstructionMultipleData
**SIS** Superconductor Insulator Superconductor
**SISD** Single Instruction Single Data
**$Si/SiO_2$** Silicon / Silicon Dioxide
**SiGe** Silicon Germanium
**SiC** Silicon Carbide
**SME** Small Medium Enterprise
**SMOLED** Small Molecule Organic Light Emitting Diode
**SMP** Symmetric multiprocessing
**SOA** Semiconductor Optical Amplifier
**SOC** System On the Chip
**SOFC** Solid Oxide Fuel Cell
**SOI** Silicon On Insulator
**SOMA** Self Organized Magnetic Arrays
**SONET** Synchronous Optical Net.
**SPRING** Standards, Productivity and Innovation Board
**SQUID** Superconducting Quantum Interference Device
**SRAM** Static Random AccessMemory
**SNR** Signal to Noise Ratio

**SSB** Single Side Band modulation
**SSH** Secure Shell
**SSL** Secure Socket Layer
**ST-FFT** Sande-Tukey FFT
**STN** Super Twisted Nematic
**SWNT** Single Wall Carbon Nanotube
**TB** Terra Byte
**$Tb/in^2$** Terabit per square inch
**TC** Turbo-Codes
**TCP** Transmission Control Protocol
**TDD** Time Division Duplex
**TDDB** Time Dependant Dielectric Breakdown
**TDM** Time Division Multiplexing
**TDMA** Time Division Multiple Access
**TD-SCDMA** Time Division Synchronous CDMA
**TFT** Thin Film Transistor
**THz** Terahertz
**$TiO_2$** Titanium Dioxide
**TLD** Top Level Domain Names
**TOLED** Top-emitting Organic Light Emitting Diode
**TTL** Transistor-Transistor Logic
**TVM** Track via Missile
**UAV** Unmanned Aerial Vehicle
**UDP** User Datagram Protocol
**UHF** Ultra High Frequency
**ULSI** Ultra Large Scale Integration
**UMTS** Universal Mobile Telephone System
**URSI** International Union Radio Science
**USB** Universal Serial Bus
**UTRA** Universal Terrestrial Radio Access
**UWB** Ultra-Wideband
**VCO** Voltage Controlled Oscillator
**VCSEL** Vertical Cavity Surface Emitting Laser
**VD** Viterbi decoding
**VDSL** Very High-Speed Digital Subscriber Line
**VHF** Very High Frequency
**VLSI** Very Large Scale Integration
**VoIP** Voice over Internet Protocol
**VRD** Virtual Retinal Display
**VSF** Variable Spreading Factor
**W3C** World Wide Web Consortium
**WAN** Wide Area Network
**WCDMA** Wideband Code Division Multiple Access
**WDF** Wave Digital Filter
**WDM** Wavelength Division Multiplexing
**WFTA** Winograd's fast Fourier Transform algorithm
**WGN** White Gaussian Noise
**Wi-Fi** Wireless Fidelity IEEE802.11b
**WiMAX** Worldwide Interoperability for Microwave Access
**WLAN** Wireless Local Area Network
**WMAN**WirelessMetropolitanAreaNet
**WORM** Write Once Read Many
**WPAN** Wireless Personal Area Netw.
**WRAN**Wireless Regional Area Netw.
**WWiSE** World Wide Spectrum Effic.
**WWRF** Wireless World Research Forum
**XML** Extensible Markup Language
**ZnMnTe** Zinc Manganese Tellurium

# ACKNOWLEDGMENT

The successful completion of dissertation with innovative research results on the possible origin of nonlinearities in microwave superconductivity, nature of nonlinear properties of *HTS* thin films at microwaves, and modeling techniques to predict the nonlinear characteristics of *HTS* thin films at microwaves, is a result of more than 11 years of intensive research work by the authors of book at *Department of Electrical and Computer Engineering* at *James Cook University* in Australia in close international collaboration with many scientists from different countries around the World. Authors would like to recognize indeed significant contributions to our book by the great scientists, we had a privilege to make our research with over the years.

D.L. wish especially to express my personal deep feelings of endless gratitude to my dear supervisor, Prof. (Personal Chair) Janina E. Mazierska, for providing me with the clear sense of research strategy, numerous scientific discussions and advices, tremendous support and guidance toward the development of creative integrative imperative co-lateral thinking on the nonlinearities in microwave superconductivity during many years of intensive theoretical and experimental research work at *Microwave and Electronic Material Research Group, Department of Electrical and Computer Engineering, James Cook University* (*JCU*) in Townsville, Australia. I give my candid thanks to my co-supervisor, Prof. Cornelis J. Kikkert, *Department of Electrical and Computer Engineering, JCU* in Townsville, Australia for thoughtful research advises and discussions on the design of microwave resonators. I wish to thank Prof. Jerzy Krupka from the *Institute of Microelectronics and Optoelectronics, Warsaw University of Technology*, Poland for the design and fabrication of *Hakki-Coleman type dielectric resonator*. My colleagues from the *Theva GmbH* in Germany are appreciated for the synthesis of high quality $YBa_2Cu_3O_{7-\delta}$ thin films for microwave applications. The Japanese industrial partners from *Tateho Chemical Industries Co. Ltd.* deserve a lot of thanks for the fabrication of a number of batches of high quality *MgO* substrates. My research collaborators at *Commonwealth Scientific Industrial Research Organisation* (*CSIRO*) in Canberra, Australia have completed a wonderful work, when designing the microstrip resonators, using the *Theva's* $YBa_2Cu_3O_{7-\delta}$ thin films deposited on the *Tateho's MgO* substrates. My most profound thanks go to Dr. Graham Woods of the *Department of Electrical and Computer Engineering, JCU*, Townsville, Australia; Prof. Ian Whittingham, *Department of Mathematical and Physical Sciences, JCU*, Townsville, Australia; A/Prof. Mohan V. Jacob, Drs. A. Knack, R. Grabovickic, *Department of Electrical and Computer Engineering*,

*JCU*, Townsville, Australia; Dr. K. Leong, *National Institute of Standards and Technology*, Boulder, Colorado, U.S.A.; Dr. S. Takeuchi, *Tokyo Denki University*, Tokyo, Japan; Dr. T. Takken, *IBM Corporation*, U.S.A.; Prof. J. Wosik, *Texas Center for Superconductivity, Electrical and Engineering Department*, University of Houston, Texas, U.S.A., A/Prof. Greg Allen, *Department of Information Technology, JCU*, Townsville, Australia, for their valuable professional advices, stimulating discussions and recommendations in course of research. I thank all staff of the *School of Engineering* at *JCU*, Townsville, Australia for always being positive and supportive. *The Federal Government of Australia*, *The Government of Queensland*, *The Government of Victoria*, and all my Australian Friends in Townsville, Cairns, Brisbane, Gold Coast, Canberra, Sydney, Melbourne, who have been always in support for me inside and outside the academic environment over the years, deserve my sincere gratitude.

We appreciate Dr. Harold Weinstock, *U.S. Air Force Office of Scientific Research* and Dr. Martin Nisenoff, *U.S. Naval Research Laboratory* for their encouragements through learning at memorable *NATO ASI* meeting on microwave superconductivity in Millau, France in 1999. Prof. H.O. Everitt, *U.S. Army Research Office, Duke University* was generous, expressing strong interest in our research since 2001. Prof. M. J. Lancaster, *Department of Electronic, Electrical & Computing Engineering*, *University of Birmingham*, U.K. was gracious to organize the *Marconi seminar* with my invited talk in 2000. I would like to acknowledge Prof. P.H. Kes for inviting me to visit the *Kamerlingh Onnes Laboratory, Department of Physics, Leiden University*, The Netherlands in 1999. Many thanks to Prof. D. E. Oates, *Massachusetts Institute of Technology* (*MIT*), U.S.A., and Prof. A. Porch, Cardiff University, U.K. for careful reading of manuscript and wise research opinions on nonlinearities origin in *HTS* at microwaves. Profs. Drs. K. Kitazawa, *University of Tokyo;* H. Takayanagi, *NTT BRL;* K. Yamanaka, A. Akasegawa, *Fujitsu Ltd.*; Sh. Futatsumori, *Hokkaido University*; Sh. Ohshima, *Yamagata University*; N. Sekiya, *Yamanashi University*; H. Hashimoto, M. Yamazaki, *Toshiba Corp.*, H. Nakakita, *NHK*, Japan are very much appreciated.

D.L. is grateful to the *IEEE* for the *2002 IEEE Microwave Theory and Techniques Society Award*, signed by John I. Barr, *University of Virginia*, U.S.A.

We wish to convey my warmest thanks to our family in the City of Kharkov, Ukraine, our father, Oleg P. Ledenyov and mother, Tamara V. Ledenyova, who have given us encouragements to complete the research on the nonlinearities in microwave superconductivity.

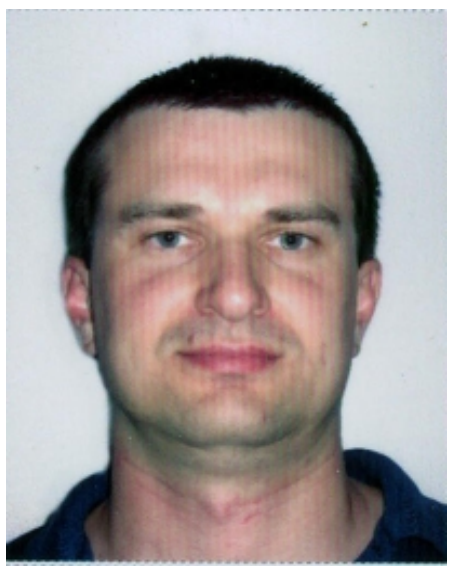

**Dimitri O. Ledenyov obtained M.Sc. degree in Radiophysics and Electronics with specialization in Microwave Engineering from Department of Radiophysics and Electronics at Kharkov National University named after V. N. Karazin (founded: 1805) in Kharkov, Ukraine in 1999. Three scientists from Kharkov National University: Landau (physics), Mechnikov (medicine), Kuznets (economics) were awarded by Nobel prizes in XX century. Mathematician Lyapunov (L. exponent) worked at KhNU too. In 1999-2000, Dimitri conducted his research on: superconductors at ultrasonic frequencies; Abricosov magnetic vortices in Type II superconductors at Cryogenic Lab (established in 1931) at Solid State Physics Institute at National Scientific Center Kharkov Institute of Physics and Technology (founded: 1928) in Ukraine, where Lifshits, Landau, Shubnikov, Kikoin, Akhiezer, Kapitza, Gamow, Kompaneyets, Lazarev, Leipunski, Bohr, Koestler, Dirac, Ruhemann, Pauli, Weisskopf, Tisza, had a privilege to make research in physics in different years. In 2000, Dimitri presented an invited talk: "Quantum Knots of Abricosov Magnetic Vortices in Superconductors" at Marconi Seminar, chaired by Prof. Lancaster at Birmingham University in the U.K. In 2000–2011, Dimitri worked on Ph.D. research on nonlinearities in microwave superconductivity under Prof. J.E. Mazierska supervision, at Microwave and Electronic Material Research Group, Electrical and Computer Engineering, Department, James Cook University, Queensland, Australia, where he is recognized as professional engineer. Dimitri was awarded by the IEEE Microwave Theory and Techniques Society Award, signed by John I. Barr, in the USA in 2002. Dimitri's research is focused on:**

**Many tens of innovative research publications were authored and co-authored by Dimitri over the last 15 years.**

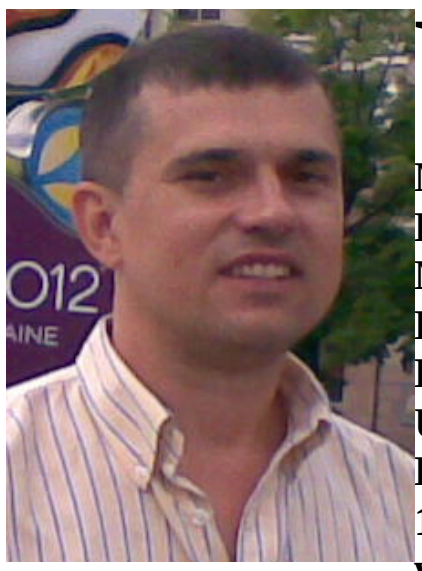

**Viktor O. Ledenyov obtained M.Sc. degree in Radiophysics and Electronics with specialization in Microwave Engineering from Department of Radiophysics and Electronics at Kharkov National University named after V. N Karazin in Kharkov, Ukraine in 1993. Viktor's university research was focused on design of microwave signal generators at frequency of 10GHz, using the Gunn diode in 1992; and design of RF resonators for high energy particles accelerators in 1993. In 1993-1999, Viktor conducted innovative research on LT/HT superconductors at ultrasonic frequencies and established the superconducting quantum computing scientific field at National Scientific Center Kharkov Institute of Physics and Technology in Ukraine. In 1996-1997, Viktor joined the Department of Physics, Technical University of Denmark, Lyngby, Denmark to continue his innovative research on the measurements of magnetic flux qubit by the dc-SQUID; nonlinear dynamics of complex systems in quantum computing. In 1997, Viktor's research on N-S boundary was presented at 4th Nordic Conference on Surface Science & Surface and Interface Optics in Ålesund, Norway. In 1997, Viktor attended the NATO ASI on Technical Applications of Superconductors in Loen, Norway. In 1998, Viktor participated at Leonardo da Vinci IAS on superconducting materials in Bologna, Italy. In 1998, Viktor as an inventor of quantum random number generator on magnetic flux qubits chipset, presented invited lecture with innovative scientific report: "Measurements of magnetic flux qubit by dc-SQUID" at Leiden University, The Netherlands. In 1998 – 1999, Viktor conducted his advanced research in the fields of microwave superconductivity and quantum computing at Departments of Physics and Engineering, University of Toronto in Canada. In 1999, Viktor attended the URSI General Assembly in Toronto, Canada. In 2000, Viktor made the research on the measurement of energy of single chip in Direct Sequence Spread Spectrum (DSSS), using the Rodhe & Schwartz and HP spectrum analyzers, at WaveRider in Calgary, Canada. In 2000-2001, Viktor performed R&D work on the WCDMA/UMTS at Nortel Networks in Calgary, Canada. In 2000-2002, Viktor participated in three wireless conferences and many seminars by TR Lab in Calgary, Alberta, Canada. In 1998-2002, Viktor completed technology assessment assignments at: JDS Uniphase, Nortel Semiconductors, Nortel Microwaves, TeleSat Canada, Computing Devices Canada, Ericsson, Philips, Siemens, SR Telecom, TR Lab in Canada. Viktor conducted his research on econophysics at Danish Technical University, Rotman School of Management, Haskayne Business School; Fuqua Business School. Since 2008, Viktor participated in the lectures and debates in the finances, economics, science with more than 100 CEOs, professors, deans, directors, ambassadors, from multinationals, universities, thin tanks, and governments, facilitated by the Financial Times, The Economist, Forbes. Viktor frequently writes in the blogs at Harvard Business School, MIT Sloan School of Management. Viktor made significant contribution to the fundamental and applied sciences, and is well known for the origination of superconducting quantum computing; creation of Ledenyov theory on nonlinearities in microwave superconductivity, creation of Ledenyov economic waves (cycles) modulation theory in econophysics. Viktor continues to make his innovative research in the fields of condensed matter, microwave superconductivity, quantum computing and econophysics.**

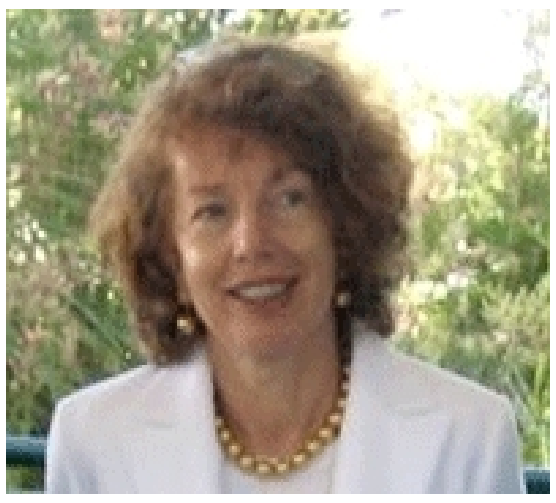

**Janina E. Mazierska obta-inned M.E. Eng. degree in Electro-nic Engineering and Ph.D. degree in Telecommunication" Analysis of Tran-sient Processes in p-n-n+ Step Recovery Diodes and in Pulse Circuit with SRDs" from Warsaw University of Technology, Warsaw, Poland, where she worked at the Institute of Electronic Fundamentals from 1972 to 1987. In the period from 1983 till 1987, Janina was on secondment to University of Jos in Nigeria, West Africa to assist in development of the Bachelor degree in Electronics. Janina E. Mazierska established Microwave and Electronic Material Research Group at James Cook University in Australia, serving in various positions, including Deputy and Acting Dean of the Faculty of Engineering. She was also a Visiting Scholar at Stanford University in the USA in 1991 and 1996. From 2004 to 2008, she was a Head of the Institute of Information Sciences and Technology at Massey University in New Zealand.**

**Prof. Janina E. Mazierska's research expertise is in microwave characterisation of low energy loss materials, including the High Temperature Superconductors (HTS), novel dielectric materials and semiconductors, as well as HTS thin films applications in wireless communication systems, with over 125 journal and conference proceedings papers published. For her contributions to knowledge in this area, she was elected an IEEE Fellow (class 2005). In the period 1998-2003, Janina was awarded four ARC Discovery Grants as the first Chief Investigator for over $M1.3, and was a CI on four other ARC grants for over $M2.1. Janina gave many keynote and invited expert level lectures around the world.**

**Janina E. Mazierska has been an active member of the IEEE, the biggest non-profit, technical professional association of more than 375,000 members in 160 countries, since 1987. Janina served the organisation in many roles and was instrumental in creation the IEEE North Queensland (now Northern Australia Section). In the period 2007-2008, Janina was on the IEEE Board of Directors and a Director of Region 10 (Asia & Pacifics). Prof. Janina E. Mazierska has been awarded the Order of Poland for service to the Polish Australian Community.**

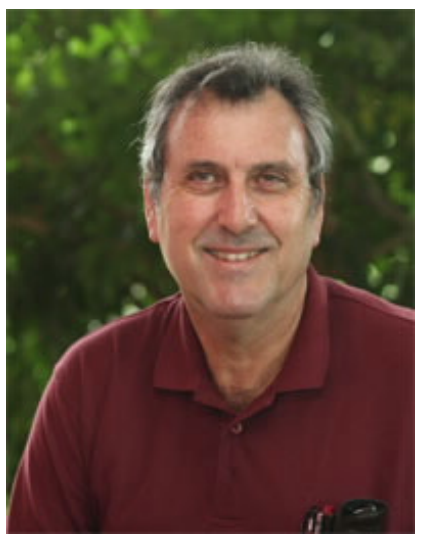

**Cornelis Jan Kikkert was born in Amsterdam, The Netherlands, and received his primary and High School education there. Cornelis Jan Kikkert obtained a BE with first class Honors and a Ph.D. from Adelaide University and lectured there for 3 years before coming to James Cook University. Cornelis Jan Kikkert is an active Christian. On 1/1/ 2010, after teaching 37 years at JCU, Cornelis Jan Kikkert retired from working fulltime at James Cook University and moved to Mt Barker, South Australia. Since that time Cornelis Jan Kikkert has been active in research, as well as teaching at different Universities. Cornelis Jan Kikkert is an Adjunct Associate Professor at James Cook University and Adelaide University.**

**Prof. Cornelis Jan Kikkert lectured on the EE3700 - Communication Systems Principles and Analogue, Digital; and CC2510 - Digital Logic and Computing Methods , EE3300 - Electronics II. He has the research expertise in many aspects of Electronic Design for Communication Systems, including Computer simulation and Hardware Design and Measurement Techniques. Selected examples are as follows:**

- **Power Line Carrier (PLC) on SWER Power lines.**
- **RF models of Power Distribution Transformers.**
- **Filters for 3G Mobile Radio Systems.**
- **RF and Microwave Filters.**
- **Predistortion for RF amplifiers.**
- **Satellite Beacon Receivers.**
- **Satellite Transmission Rain Attenuation.**
- **Spread Spectrum Techniques such as OFDM and Network Analyser Techniques.**
- **Analogue to Digital Converters.**
- **Low Cost Detection of Landmines.**
- **BPSK Demodulation and Phase Detectors using FPGAs.**
- **Ad-Hoc Radio Networks.**

**Prof. Cornelis Jan Kikkert wrote a book: "RF Electronics," and a book chapter: "Calculating Radiation from Power Lines for Power Line Communications" in the book: "MATLAB for Engineers – Applications in Control, Electrical Engineering, IT and Robotics."**

# Modeling of Nonlinear Properties of High Temperature Superconducting (HTS) Thin Films, Using Bardeen, Cooper, Schrieffer and Lumped Element Circuit Theories, with Purpose to Enhance Microwave Power Handling Capabilities of Microwave Filters and Optimize Design of Microwave Circuits in Micro- and Nano- Electronics.

MS Power Point Presentation Based on Ph. D. Thesis submitted by

Dimitri Olegovych LEDENYOV B.E., M.Sc.

April, 2012

Supervisor: Prof. Janina E. Mazierska

**Abstract of Thesis:** The aim of the thesis is to research the nonlinear properties of high temperature superconducting (HTS) thin films for application in electronic devices in information communication technologies at microwaves.

**Theoretical research:**

**1.** Analysis of present state of theoretical research on the superconductivity with focus on the electromagnetic and nonlinear properties of HTS materials at microwaves in Chapters 1, 2 and 3.

**2.** Creation of both a lumped element model of a superconductor and a lumped element model of a microwave resonator for accurate characterisation of nonlinear properties of HTS thin films in microwave resonators in Chapters 4. The consideration of limitations of r- parameter application for nonlinear models analysis in HTS microwave resonators is also described in Chapter 4.

**3.** Chapter 5 deals with modeling of nonlinear microwave responses of system with three types of R and L element dependences (linear, quadratic & exponential). The proposed lumped element model adequately describes the nonlinear properties of researched HTS thin films in Hakki-Coleman dielectric resonator (HCDR).

**4.** Chapter 6 focuses on the Rs(T) and Rs(P) dependences and theoretical consideration of measurement accuracy issues during accurate characterisation of HTS thin films in the HCDR.

**5.** Chapter 7 concentrates on the creation of a simplified microwave model, based on the Bardeen, Cooper, Schrieffer (BCS) theory, to determine nonlinear behaviour of Rs(P) of HTS thin films near $Hc_1$ and $Hc_2$ in microstrip resonator. The simulated S-type dependences Rs(P) closely approximate the measured experimental results.

**Abstract of Thesis:** The aim of the thesis is to research the nonlinear properties of high temperature superconducting (HTS) thin films for application in electronic devices in information communication technologies at microwaves.

**Experimental research:**

**1.** Research on the microwave properties of MgO substrates in split post dielectric resonator (SPDR) at f=10.48GHz. It is understood that the MgO substrate is not a source of nonlinearities in HTS thin films at microwaves in operational temperatures range in Chapter 6.

**2.** Research on the nonlinear surface resistance of $YBa_2Cu_3O_{7-\delta}$ thin films on MgO substrates in Hakki-Coleman dielectric resonator (HCDR) at f=25GHz. It was observed that $YBa_2Cu_3O_{7-\delta}$ thin films have nonlinear characteristics in form of S-type dependence with increase of Rs(P) at elevated power levels, when Hrf is higher than Hc1 in Chapter 6.

**3.** Research on nonlinear surface resistance of $YBa_2Cu_3O_{7-\delta}$ superconductor thin films on MgO substrates in microstrip resonators at f0=1.985GHz. Microstrip resonators expressed S-type nonlinearity at the same power levels, when Hrf is higher than $Hc_1$ in Chapter 7.

**Thesis's conclusion:** the magnitude of critical magnetic field $Hc_1$ of a superconductor needs to be firstly measured to predict the nonlinear behaviour of any HTS thin films, because the $Hc_1$ is the level of magnitude of magnetic fields at which the nonlinear effects arise. The developed software program in Matlab can accurately model the nonlinearities in HTS thin films at microwaves, using the measured critical magnetic fields $Hc_1$ and $Hc_2$.

# Thesis Contents:

**Chapter 1.** INTRODUCTION.

**Chapter 2**. THE SUPERCONDUCTIVITY: THEORIES AND EXPERIMENTS TOWARD PRECISE CHARACTERISATION OF PHYSICAL PROPERTIES OF SUPERCONDUCTORS AND OVERVIEW OF SOME TECHNICAL APPLICATIONS OF SUPERCONDUCTORS.

**Chapter 3.** MICROWAVE SUPERCONDUCTIVITY: ACCURATE CHARACTERISATION AND APPLICATIONS OF SUPERCONDUCTORS AT MICROWAVES.

**Chapter 4.** LUMPED ELEMENT MODELING OF NONLINEAR PROPERTIES OF HIGH TEMPERATURE SUPERCONDUCTORS IN MICROWAVE RESONANT CIRCUITS.

**Chapter 5.** DEVELOPMENT OF A LUMPED ELEMENT MODEL FOR ACCURATE MICROWAVE CHARACTERISATION OF SUPERCONDUCTORS.

**Chapter 6.** EXPERIMENTAL AND THEORETICAL RESEARCHES ON MICROWAVE PROPERTIES OF MgO SUBSTRATES IN A SPLIT POST DIELECTRIC RESONATOR AND NONLINEAR SURFACE RESISTANCE OF $YBa_2Cu_3O_{7-\delta}$ THIN FILMS ON MgO SUBSTRATES IN A DIELECTRIC RESONATOR AT ULTRA HIGH FREQUENCIES.

**Chapter 7.** EXPERIMENTAL AND THEORETICAL RESEARCHES ON NONLINEAR SURFACE RESISTANCE OF $YBa_2Cu_3O_{7-\delta}$ THIN FILMS ON MgO SUBSTRATES IN SUPERCONDUCTING MICROSTRIP RESONATORS AT ULTRA HIGH FREQUENCIES.

**Appendix I.**

**Conclusion.**

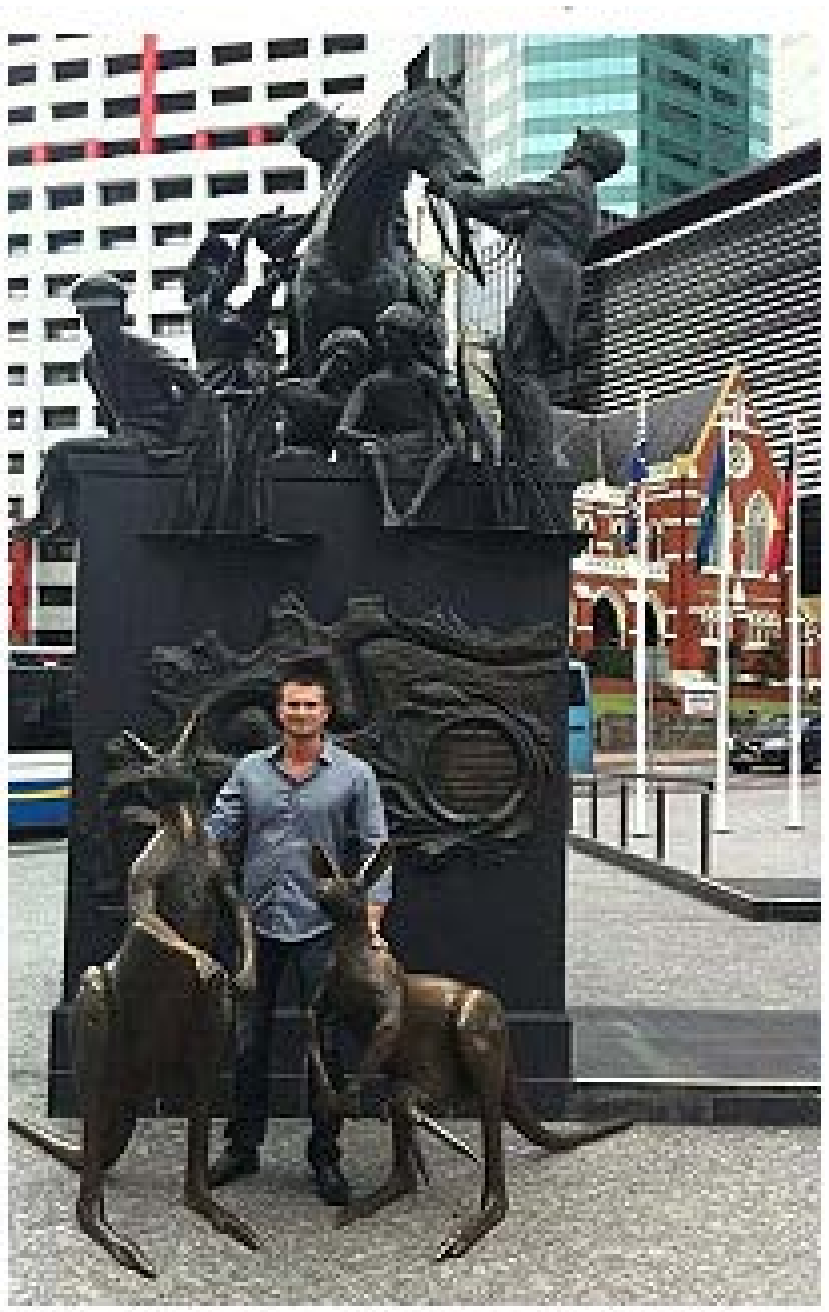

# Discovery of Superconductivity.

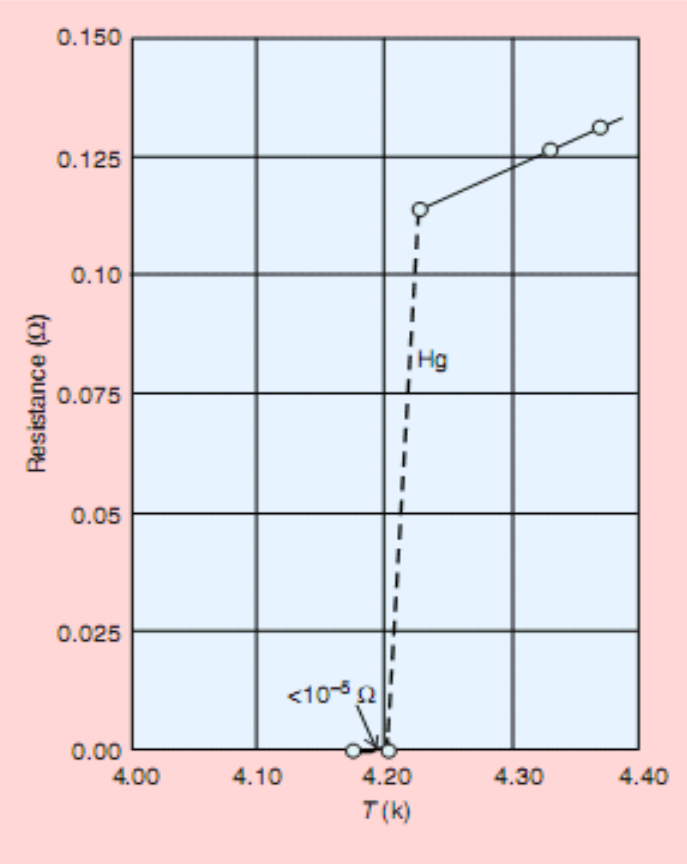

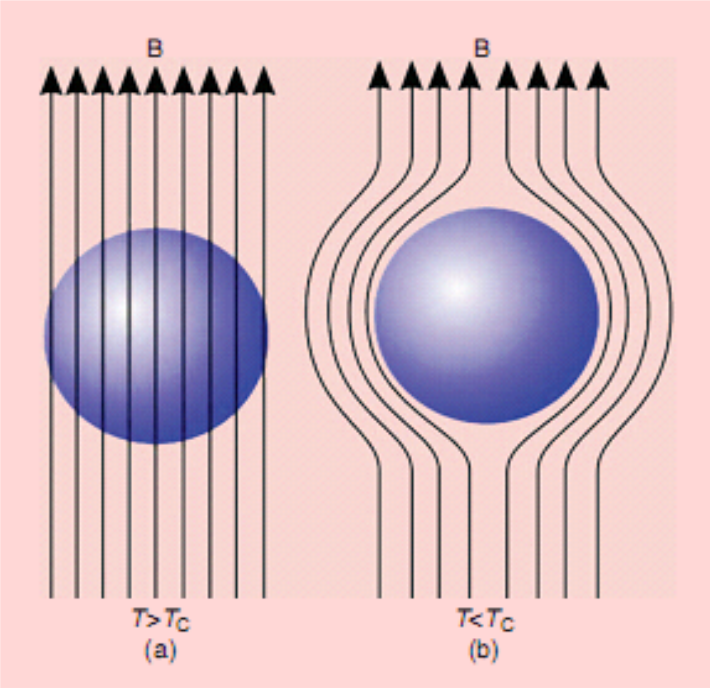

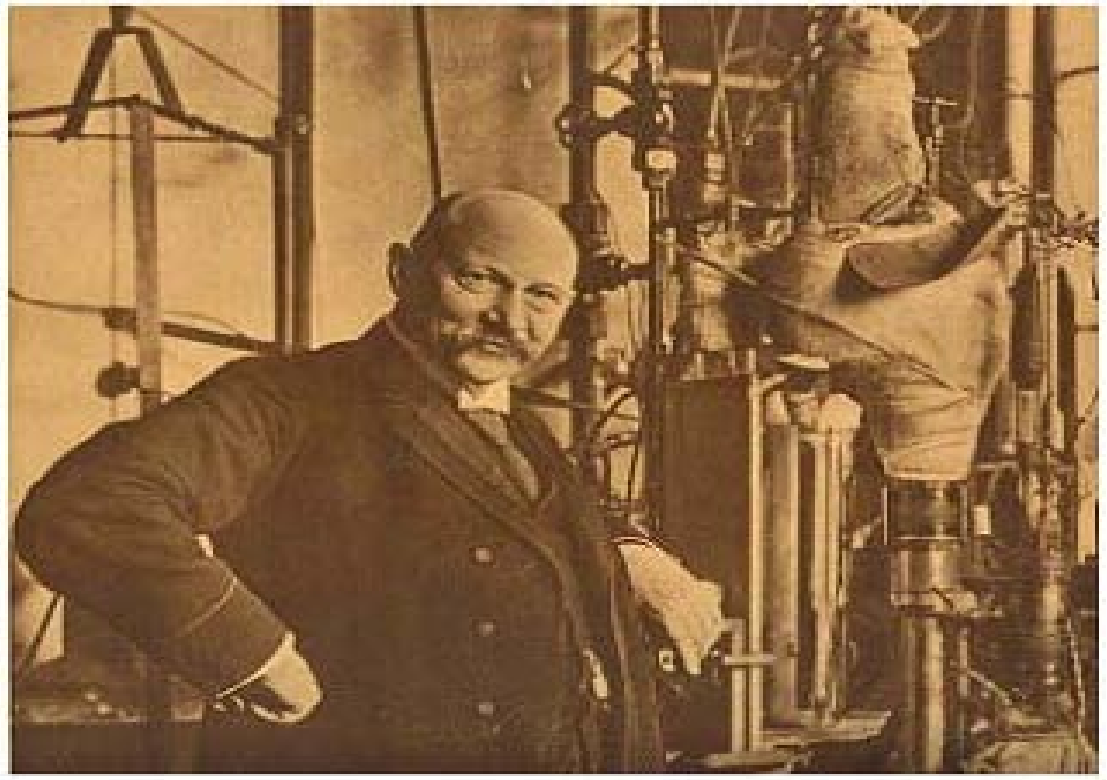

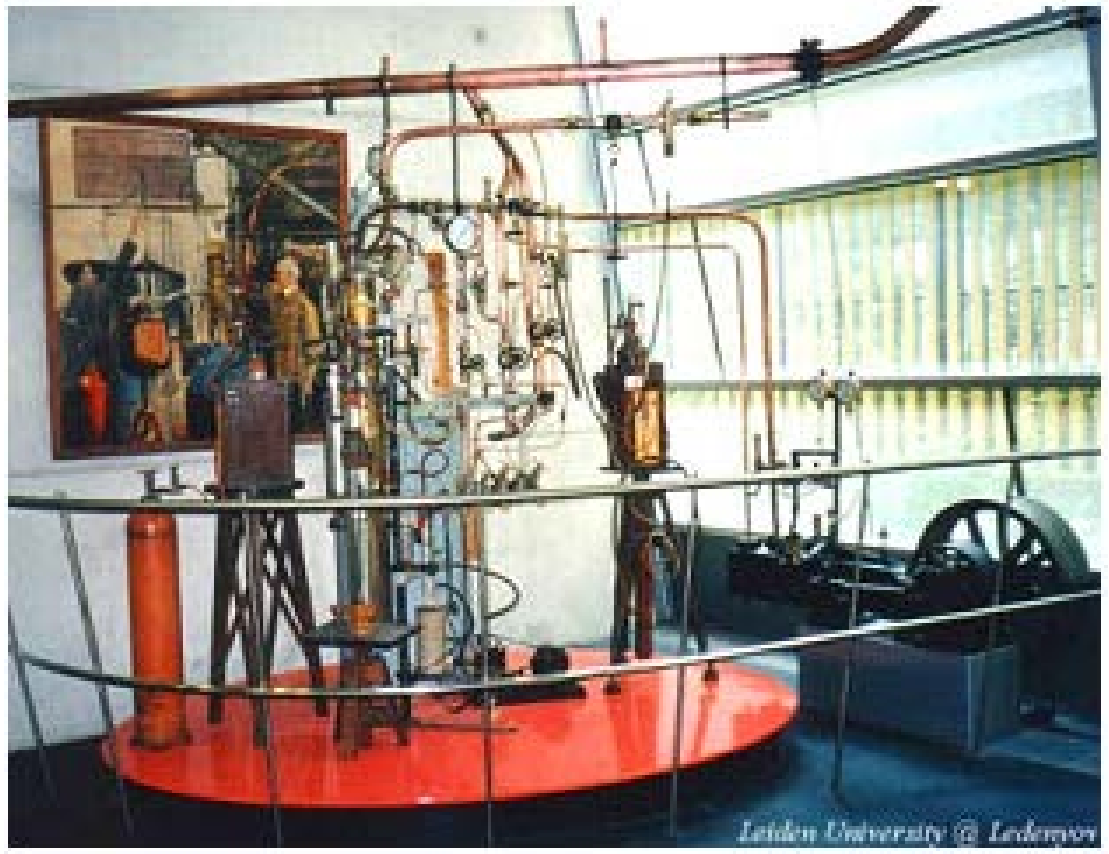

**Fig. 1**. Discovery of superconductivity in mercury by Professor Heike Kamerlingh Onnes and his research collaborators, Cornelis Dorsman, Gerrit Jan Flim, and Gilles Holst [1] at Leiden University in Leiden in The Netherlands on April 8, 1911 (after[3]).
**Fig. 2.** Schematic representation of the Meissner effect [2]in a superconductor (after [3]).
**Fig. 3.** Prof. Heike Kamerlingh Onnes (1853–1926), standing in front of the helium liquefier (Photograph is reproduced with permission, archives Leiden Institute of Physics, Leiden, The Netherlands) (after [R. de Bruyn Ouboter, Journal Physics.: Condensed Matter, vol. 21, pp. 1-8, IOP Publishing Ltd., UK, 2009.]).
**Fig. 4.** Experimental equipment set up used by Professor Heike Kamerlingh Onnes and his research collaborators, Cornelis Dorsman, Gerrit Jan Flim, and Gilles Holst to discover the superconductivity phenomena at Leiden University, Leiden, The Netherlands on April 8, 1911. The photo was taken by the author of thesis during his invited visit to the New Building at Department of Physics, Leiden University, Leiden, The Netherlands in 1999.

# The Three Eras of Superconductors Discovery.

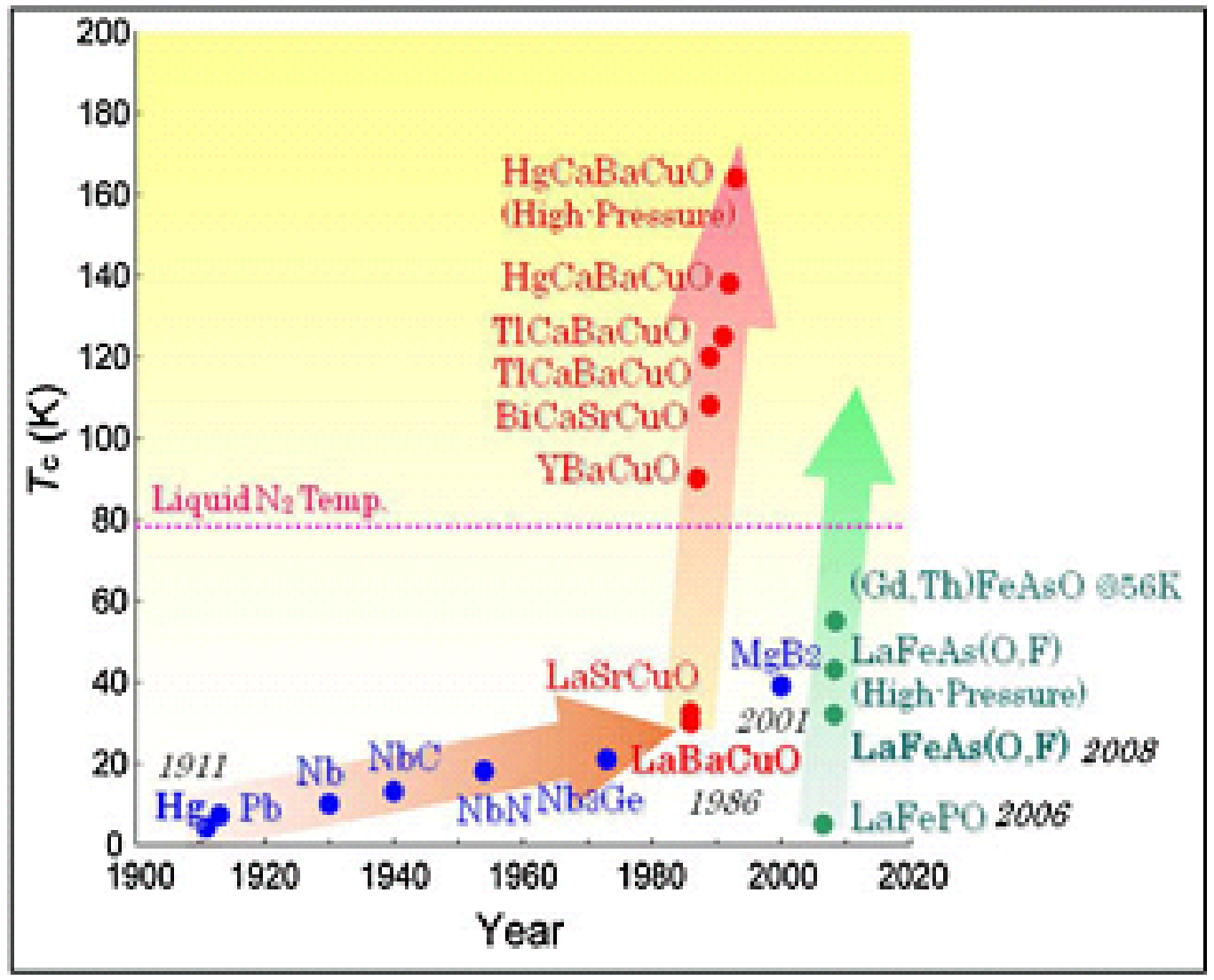

**1.** Kamerlingh Onnes first discovered superconductivity in mercury in 1911 [1];
**2.** Matthias, Geballe, Geller, Corenzwit discovered the superconductivity of a compound $Nb_3Sn$ in 1954 [2];
**3.** Bednorz and Muller discovered a cuprate superconductor in 1986 [3];
**4.** Wu, Ashburn, Torng, Hor, Meng, Gao, Huang, Wang, Chu discovered superconductivity with a critical temperature above the boiling point of liquid nitrogen for the first time in $YBa_2Cu_3O_{7-\delta}$ in 1987 [4];
**5.** Maeda, Tanaka, Fukutomi, Asano discovered superconductivity in $Bi_2Sr_2Ca_2Cu_3O_{10}$ at the highest critical temperature ever reported for potentially practically applicable superconductors in 1988 [5];
**6.** Nagamatsu, Nakagawa, Muranaka, Zenitani, Akimitsu discovered superconductivity in $MgB_2$ at a new highest reported critical temperature for metal superconductors in 2001 [6];
**7.** Kamihara, Watanabe, Hirano, Hosono discovered the Fe-based superconductor LaFeAsO in 2008 [7].

**Fig. 5.** Developing trend of superconducting critical temperature over last 100 years showing the three eras of discovery: the era of metal and intermetallic compound superconductors from 1911 to 1986, the era of cuprate superconductors from 1986 to 2000, and the subsequent era of superconductors based on new materials such as borides and Fe pnictides (after [K. Kitazawa, Superconductivity: 100th Anniversary of Its Discovery and Its Future, Japanese Journal of Applied Physics, vol. 51, pp. 1-14, 2012.]).

# Large Scale Applications of Superconductivity.

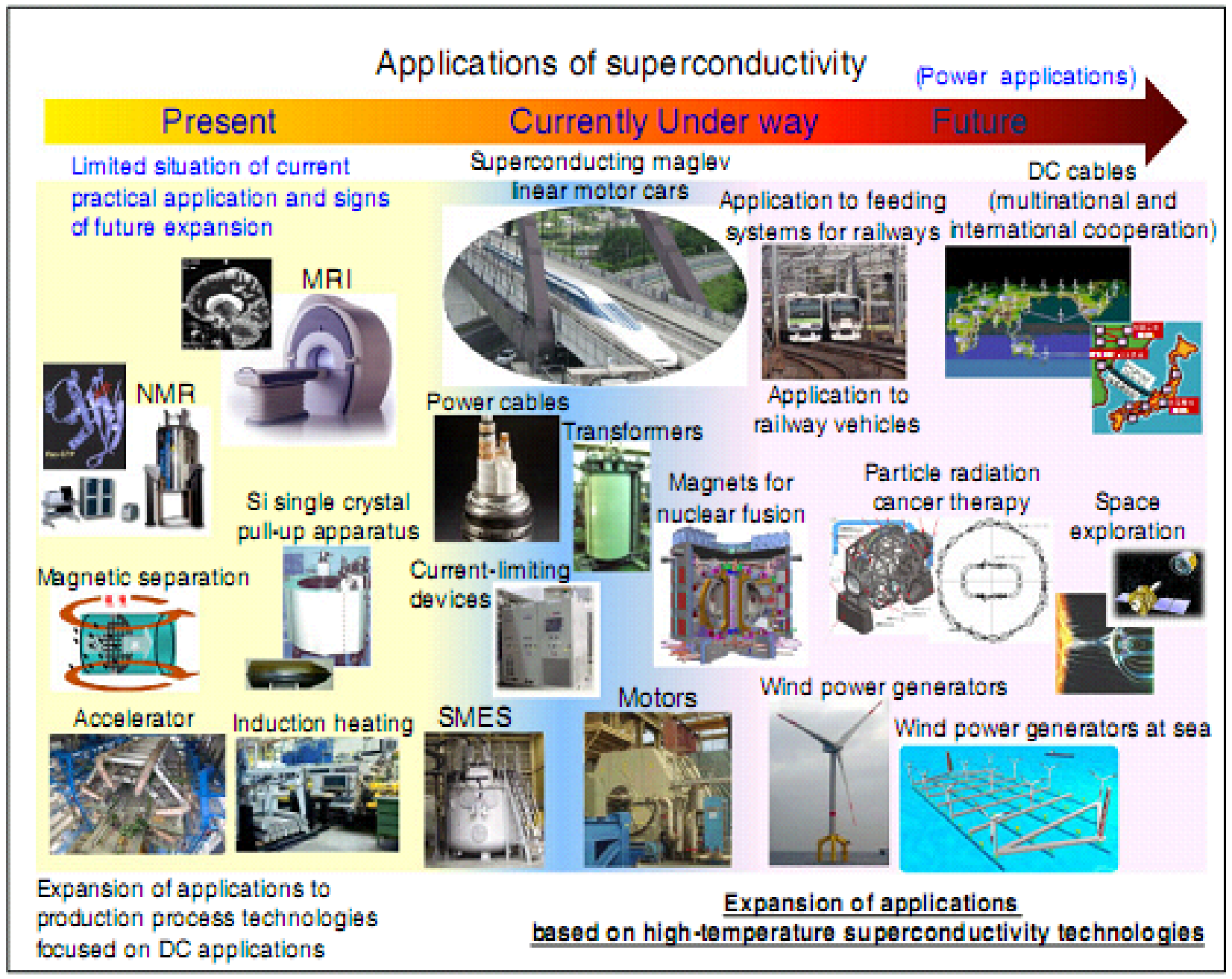

**Fig. 6.** Main applications of superconducting power and magnetic technologies (provided by Professor Hiroyuki Ohsaki of School of Engineering, The University of Tokyo) (after[8]).

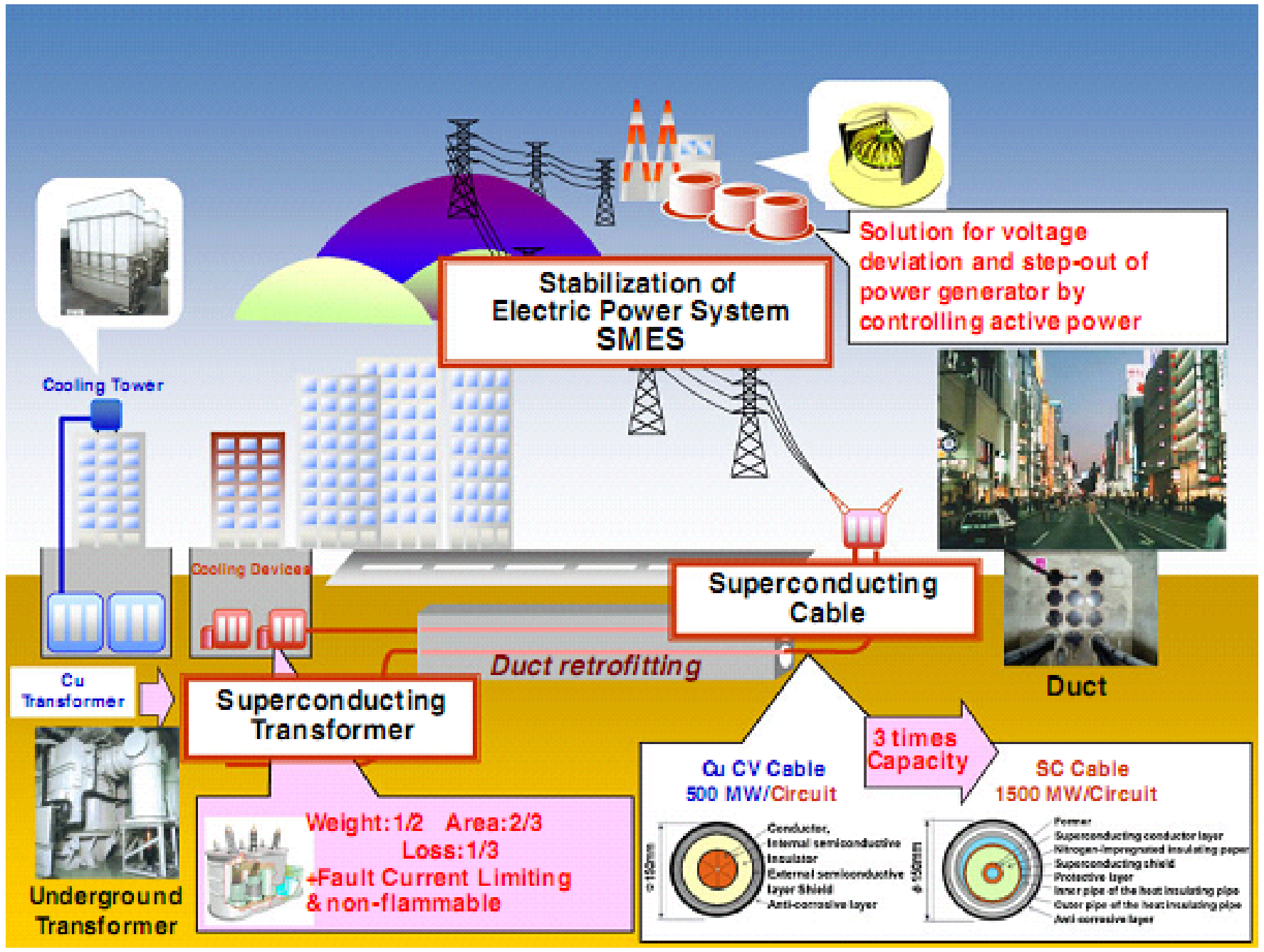

**Fig. 7.** Schematics of stable, large-capacity power supply system using superconducting electric power devices in urban area (after[9]).

# Small Scale Applications of Superconductivity.

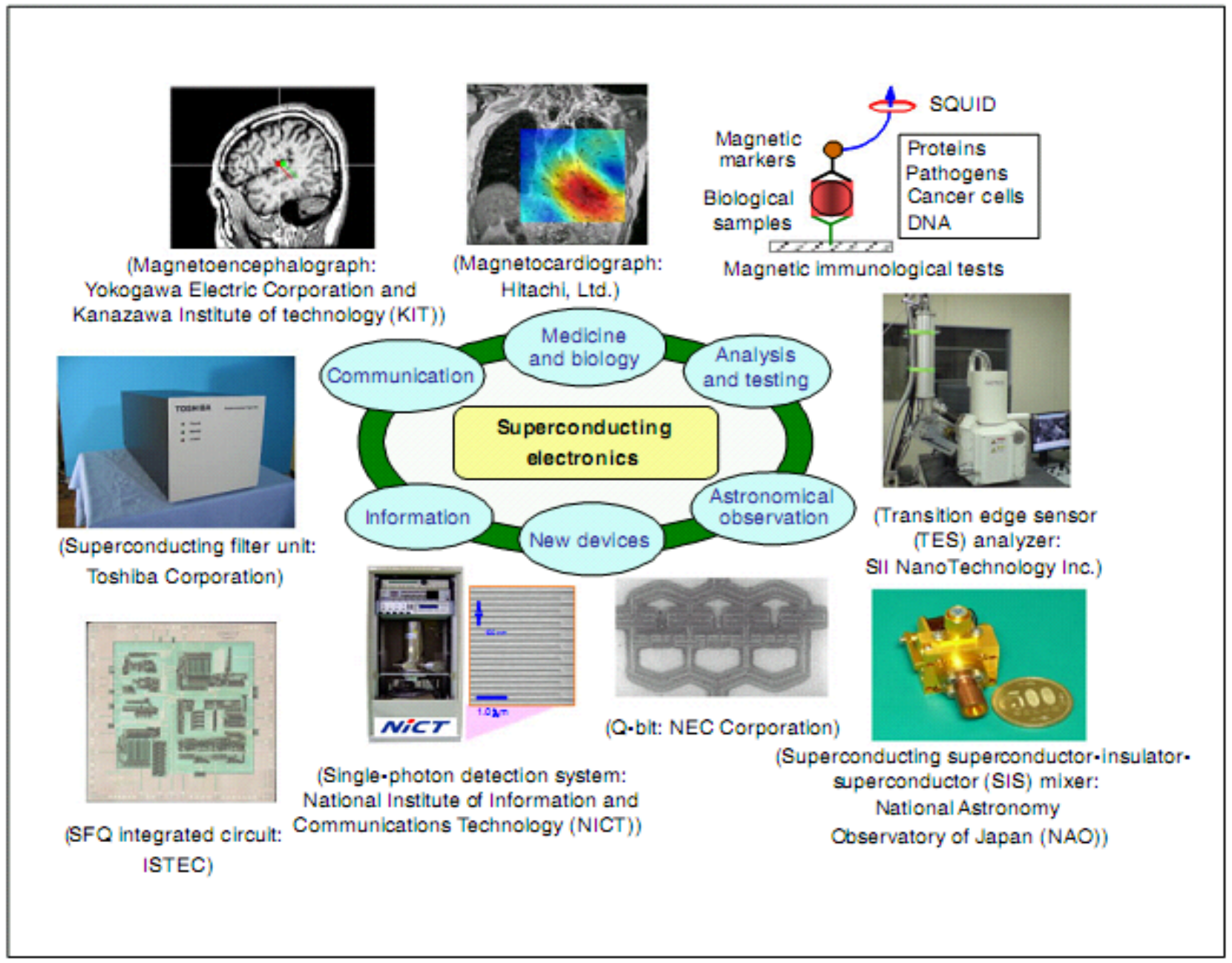

**Fig. 8.** Main applications of superconducting electronics (provided by Professor Keiji Enpuku of Faculty of Information Science and Electrical Engineering, Kyushu University) (after[8]).

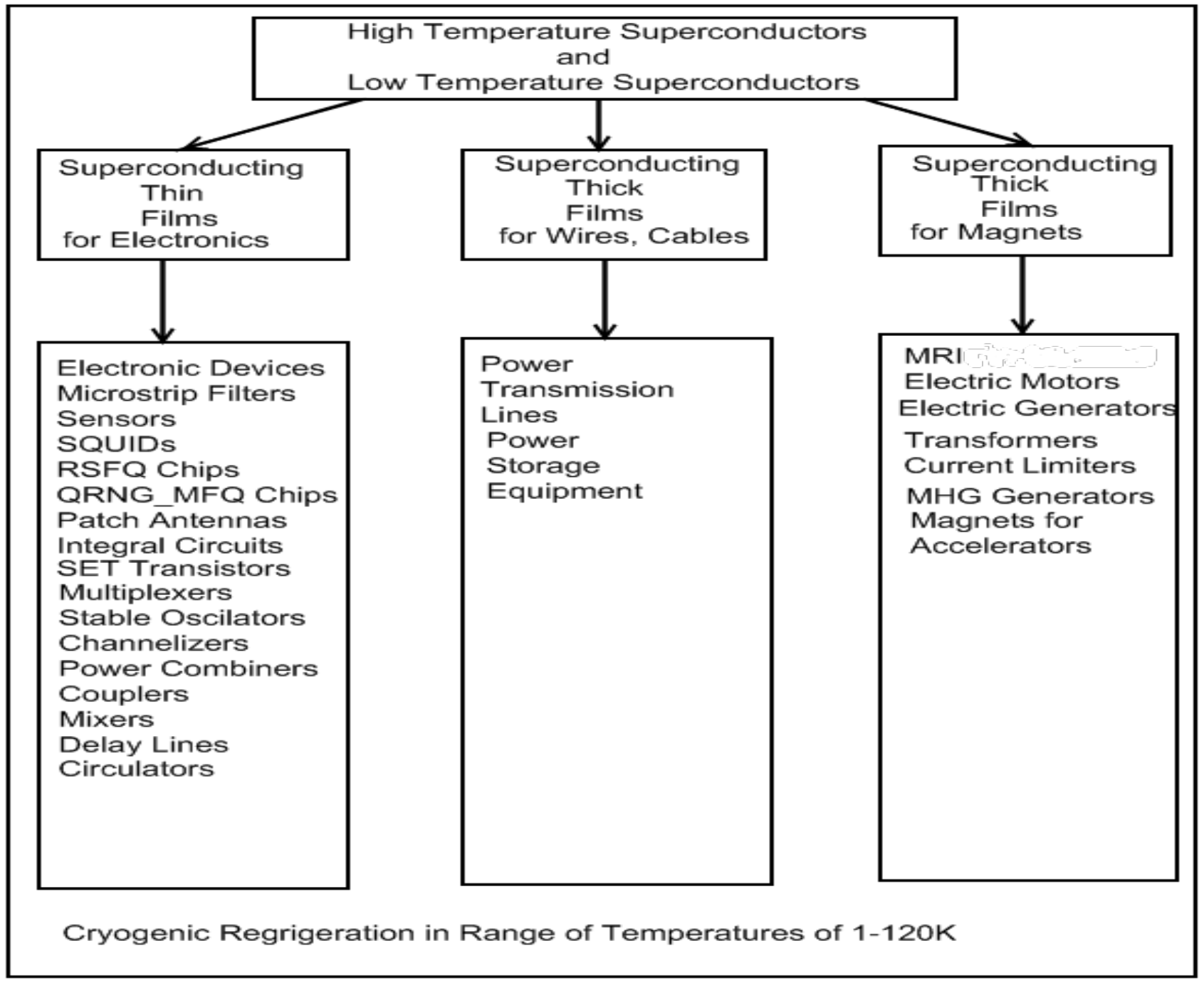

**Fig. 9.** General technical applications of High Temperature Superconductors (*HTS*) and Low Temperature Superconductors (*LTS*) summarized by author of thesis (after[10]).

# Microwave Applications of Superconductivity.

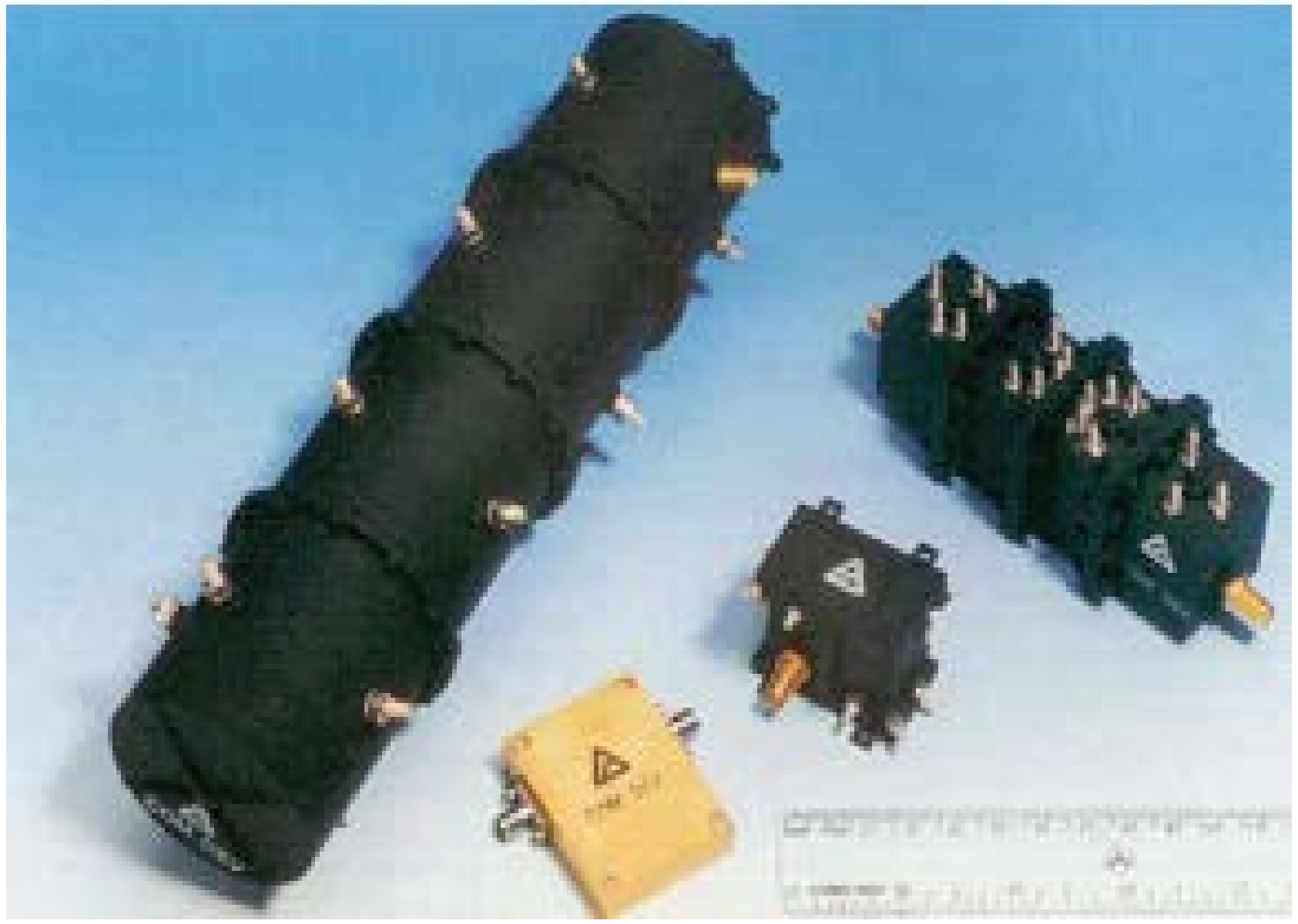

**Fig. 10.** A photograph of microwave filters fabricated using various filter technologies. The reduction in volume resulting from the utilization of superconducting filter technologies is obvious. (after[3]).

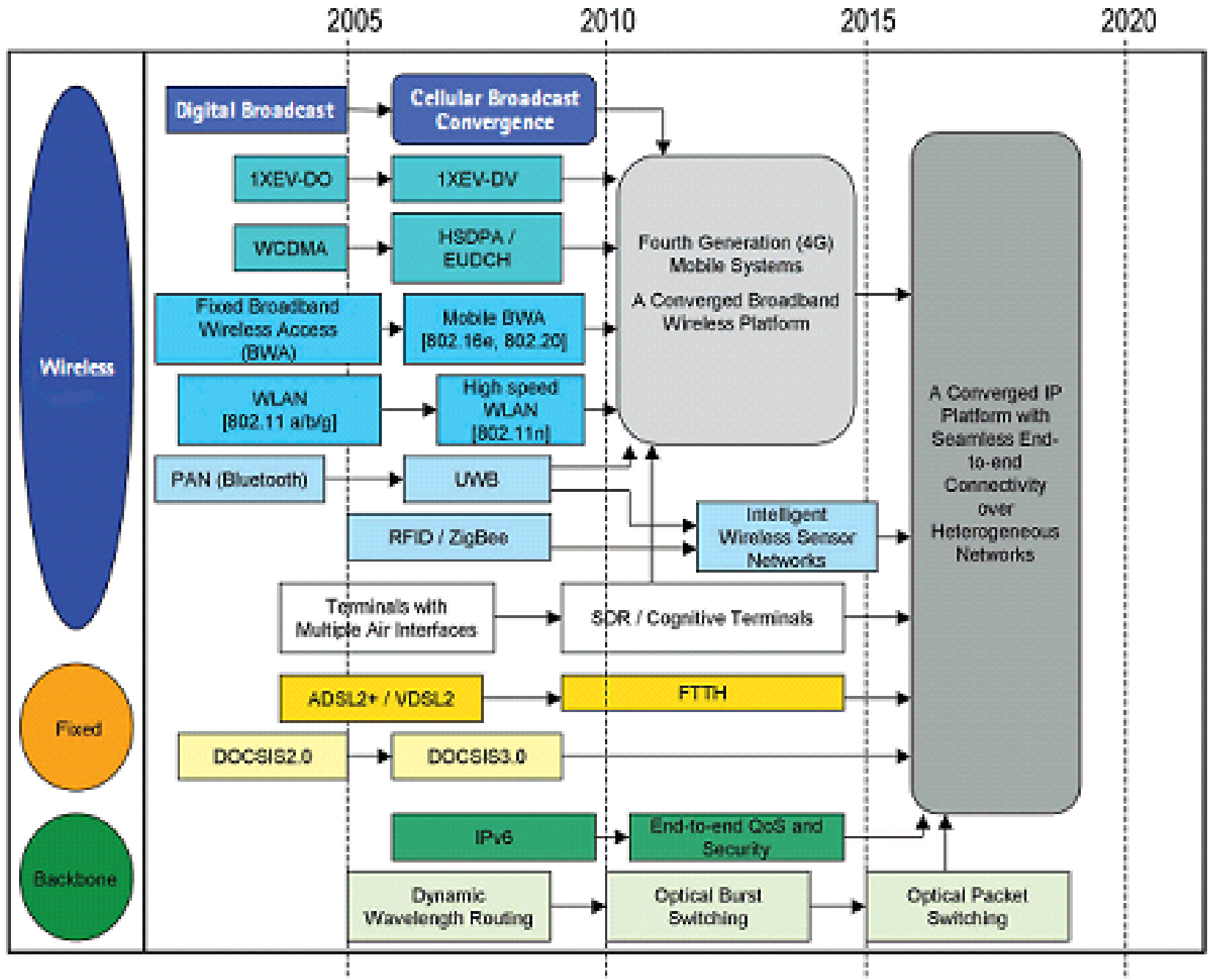

**Fig. 11.** Wireless Communication Technology Roadmap. (after[11]).

# Microwave Filters.

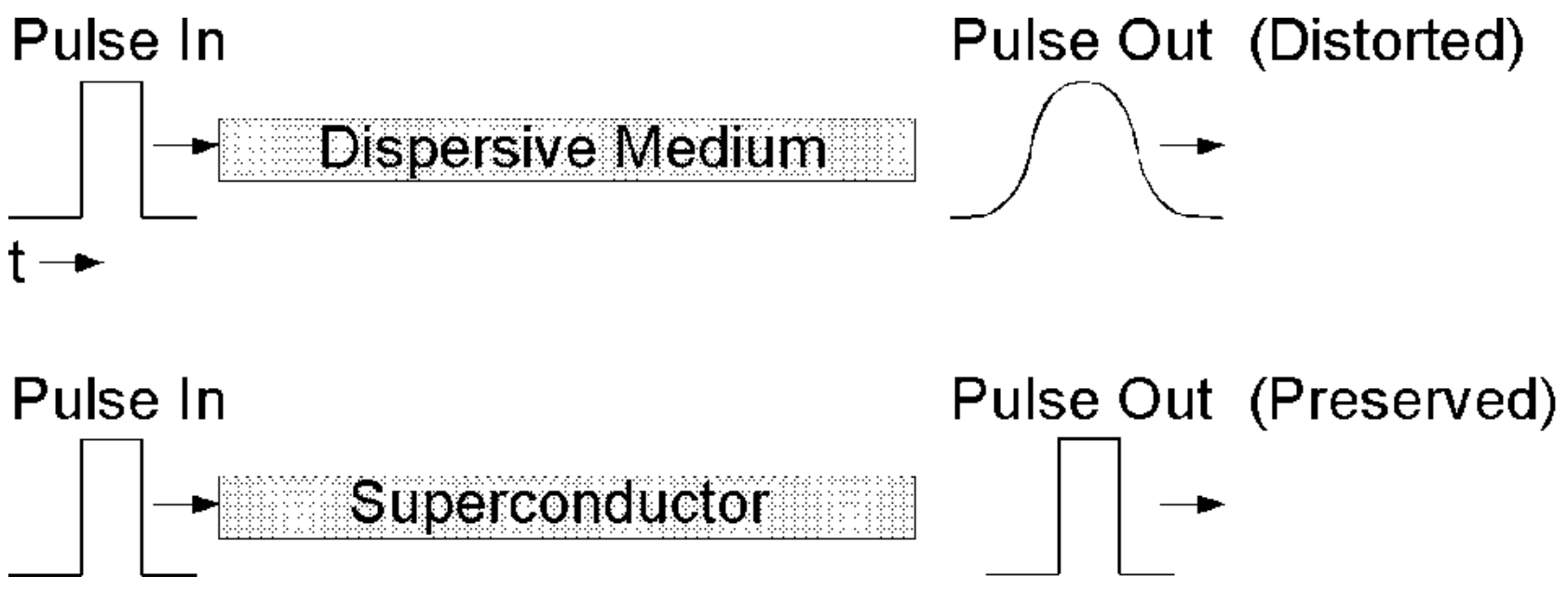

**Important Characteristic of Superconductors:**
**Superconductors have the zero energy dispersion property, which is used in microwave filters.**

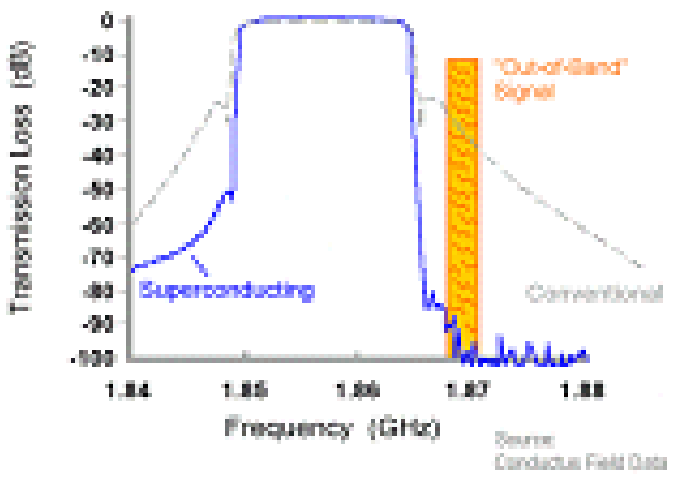

**Fig. 12.** Measured microwave characteristics of conventional metal microwave filters and High Temperature Superconducting (HTS) microwave filters in conditions of strong adjacent channel interference (after [12]).

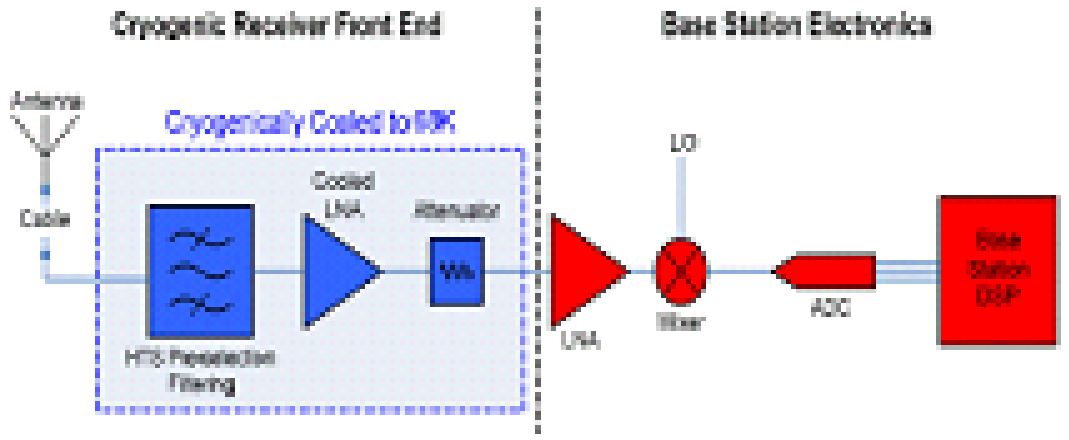

**Fig. 13.** Cryogenic Receiver Front Ends (CRFE) in base station (after [13, 14]).

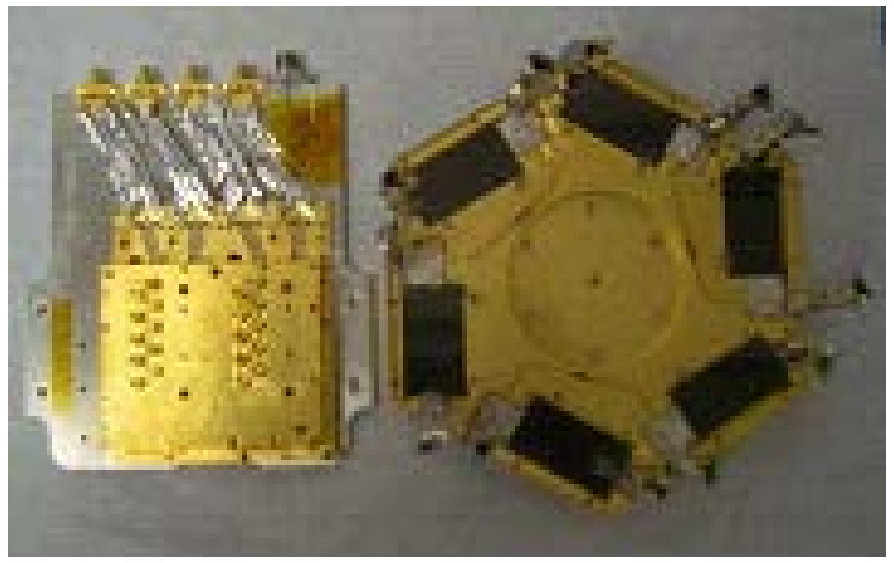

**Fig. 14.** Expanded views of one of the three microenclosures in a 2004 dewar (left) and a 2008 microenclosure (after[15]).

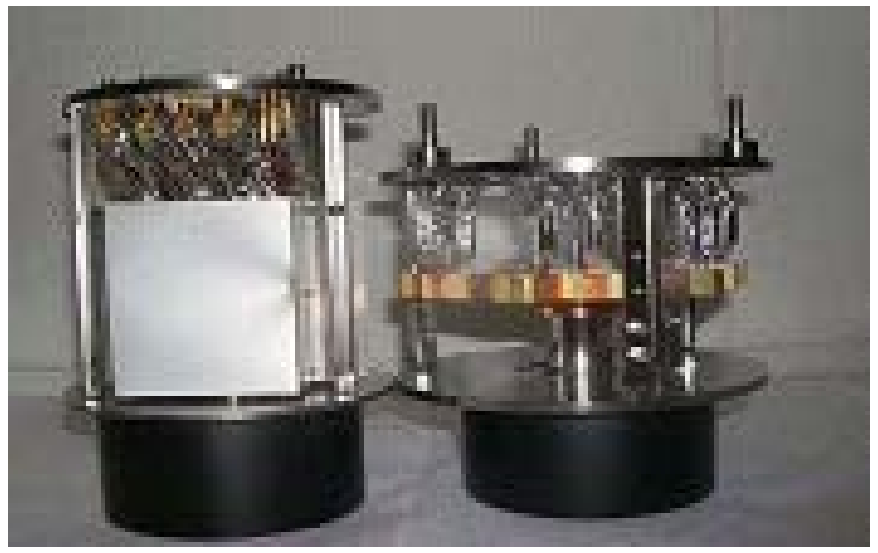

**Fig. 15.** Inside views of the cryogenic RF housings (microenclosures) inside 2004 dewar (left) and 2008 dewar (right). The 2008 dewar has been optimized to eliminate parts as compared to the 2004 design (after [15]).

# Nonlinearities in Microwave Superconductivity.

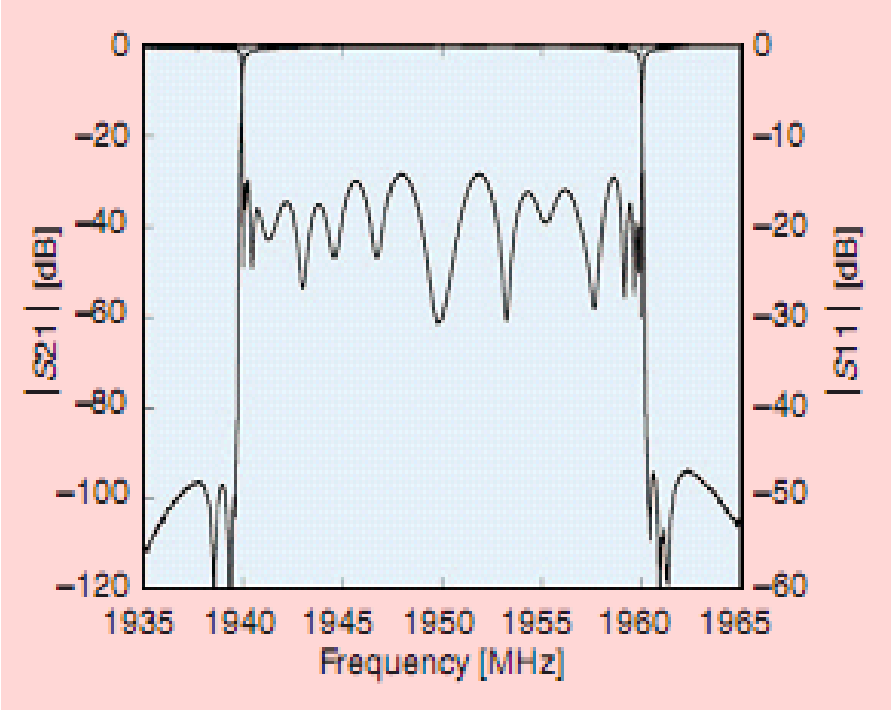

**Fig. 16.** Response of a 22-pole, ten-transmission-zero superconducting filter with performance exceeding a 50-pole Chebyhshev response (after[3]).

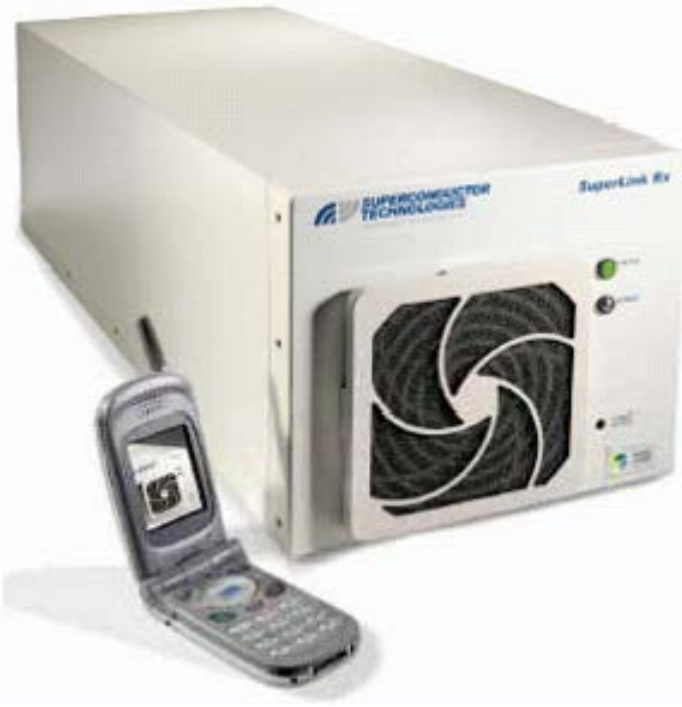

**Fig. 17.** HTS filter system for wireless base station consisting of six HTS filters (each integrated with a cryogenically cooled LNA), a cryogenic refrigerator, and associated control electronics (after[3]).

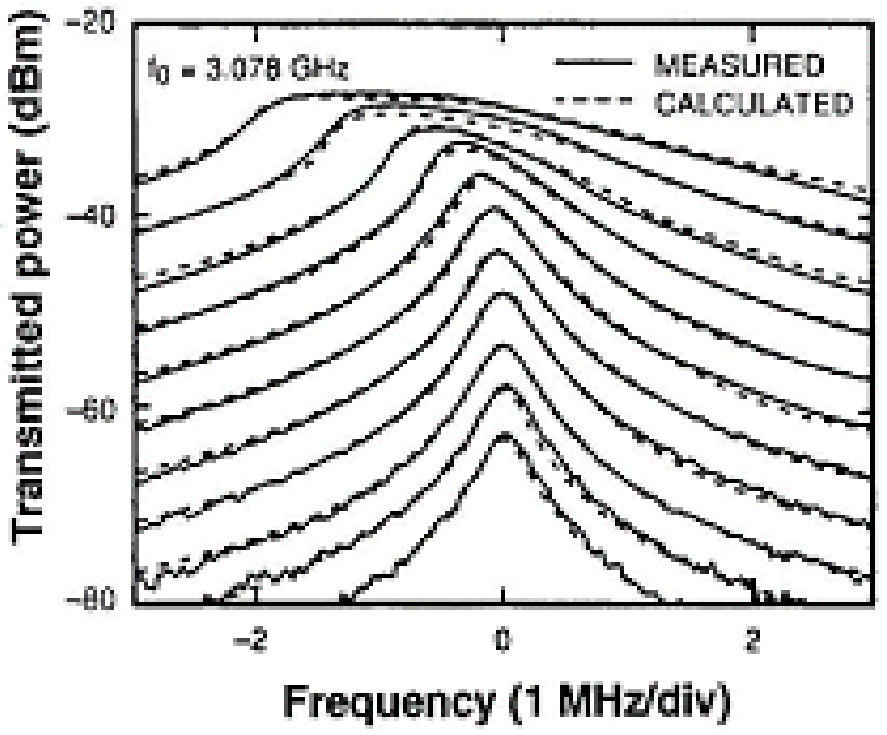

**Fig. 18.** Transmitted microwave power vs. frequency in $YBa_2Cu_3O_{7-\delta}$ microwave resonator as a function of input microwave power. The maximum power is +30dBm, and the curves are in 5dBm steps. The frequency is 3GHz, and the temperature is 77K (after [16]).

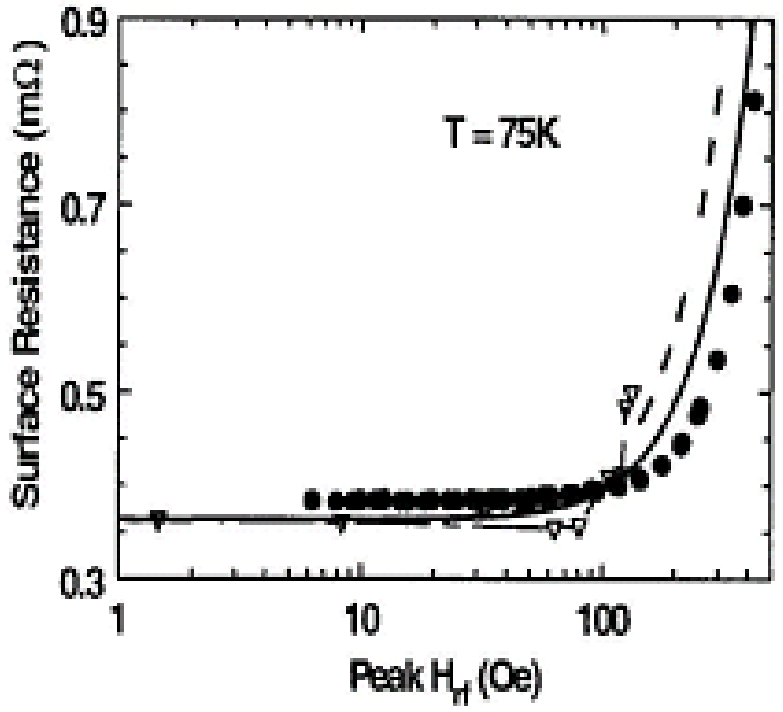

**Fig. 19**. Surface resistance $R_S$ vs. peak magnetic field Hrf in $YBa_2Cu_3O_{7-\delta}$ films at 1.5GHz at 75K. Points are measured data. Solid and dashed lines are calculated from models created for the stripline and dielectric resonators respectively (after [16]).

## Desirable characteristics of HTS microwave filters [17]:

**1. Higher power handling capability (PHC),**
**2. More compact dimensions, and**
**3. Wider tunability of resonance central frequency fc.**

## Nonlinear increase of surface resistance in HTS at high levels of applied microwave power results in [18]:

**1. Increased insertion loss of microwave filter,**
**2. Unwanted harmonics / sub-harmonics generation, and**
**3. Intermodulation distortions (IMD).**

# Theories Toward Understanding of Superconductivity.

The understanding of nature of superconductivity phenomena is based on the theories, which make it possible to better understand the nature of nonlinearities in microwave superconductivity:

1. **Becker, Sauter, Heller Macroscopic Theory of a Perfect Conductor [20] in 1933.**
2. **Gorter - Casimir Two-Fluid Phenomenological Theory [19] in 1934.**
3. **London Theory of Electrodynamics of Superconductors in AC Electromagnetic Fields [21] in 1935.**
4. **Richard A. Ogg Theory of Electron Pairs and their Bose-Einstein Condensation [22] in 1946.**
5. **Ginzburg-Landau (GL) Theory of Superconductivity [23] in 1950.**
6. **Fröhlich Theory of Interaction between the Electrons and Phonons [24] in 1950.**
7. **Pippard Nonlocal Electrodynamics [25] in 1953.**
8. **Bardeen, Cooper and Schrieffer (BCS) Theory of Superconductivity [26] in 1957.**
9. **Abrikosov–Gor'kov (AG) Theory [27] in 1959, linking the macroscopic GL theory and the microscopic BCS theory.**

# Gorter - Casimir Two-Fluid Phenomenological Theory.

In 1934, Gorter and Casimir [19] proposed the **Two-Fluid phenomenological theory** to understand the electrical losses in superconductors, introducing the Two-Fluid model, and assuming that the electrons in a superconductor may occupy either of two sets of states. In the Two-Fluid model, the current is considered to be flown by the two types of charge carriers in a superconductor:

**1.** The superconductive current is flown by superconducting electrons with carrier density $N_S$, and

**2.** The normal current is flown by normal electrons with carrier density $N_n$.

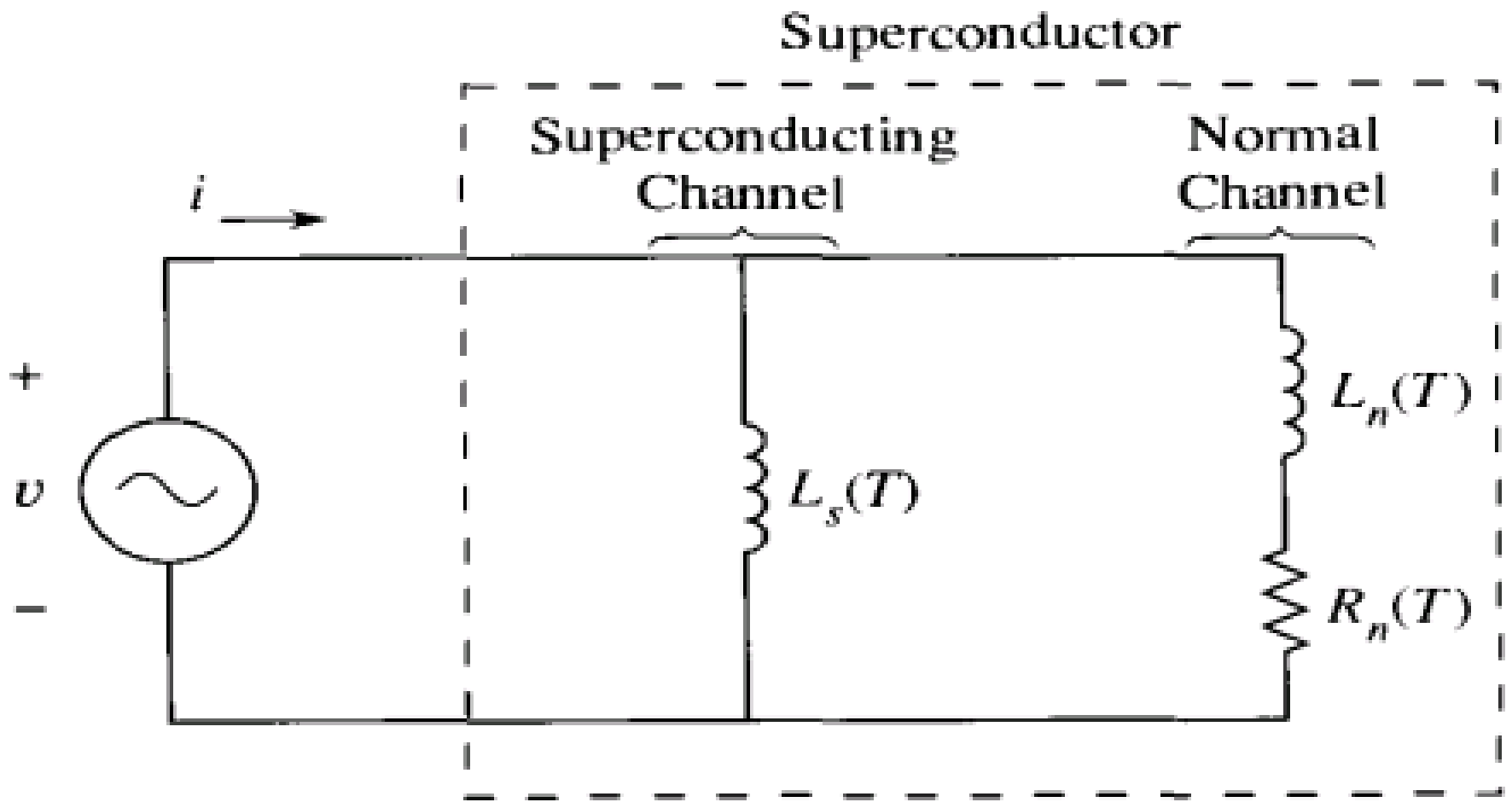

**Fig. 20.** Two Fluid lumped element model of a superconductor. In practice, the inductor $L_n$ is often neglected, which is similar to modeling the normal channel as nondispersive or independent of frequency (after [38]).

**The total electron density N is the sum of the densities of the superelectrons Ns and the normal electrons Nn [19]:**

$$N = N_s + N_n$$

**At temperatures below $T_C$, the equilibrium fractions of normal and superconducting electrons Nn/N and Ns/N vary with absolute temperature T as described in eq. (2.1) [19]:**

$$\frac{N_n}{N} = \left(\frac{T}{T_C}\right)^4, \qquad \frac{N_s}{N} = 1 - \left(\frac{T}{T_C}\right)^4 \qquad (2.1)$$

**The total current density J in a superconductor, is the sum of superconducting and normal currents in eq. (2.2) [19]:**

$$\boldsymbol{J} = \boldsymbol{J}_n + \boldsymbol{J}_s = (\sigma_1 - i\sigma_2)\boldsymbol{E} \qquad (2.2)$$

**where σ1 and σ2 are the real and imaginary parts of the complex conductivity in eq. (2.3):**

$$\sigma_1 = \frac{N_n e^2 \tau}{m(1+\omega^2\tau^2)} \quad \text{and} \quad \sigma_2 = \frac{N_s e^2}{m\omega} + \frac{N_n e^2 (\omega\tau)^2}{m\omega(1+\omega^2\tau^2)}, \qquad (2.3)$$

# Becker, Sauter, Heller Macroscopic Theory of a Perfect Conductor.
# London Theory of Electrodynamics of Superconductors in ac Electromagnetic Fields .

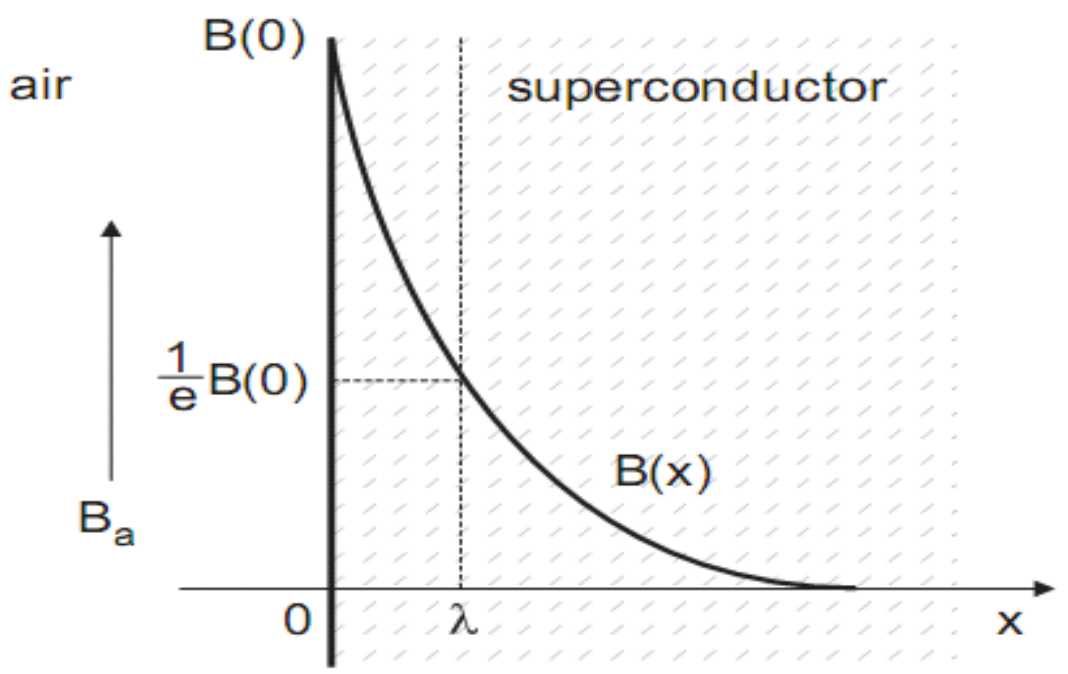

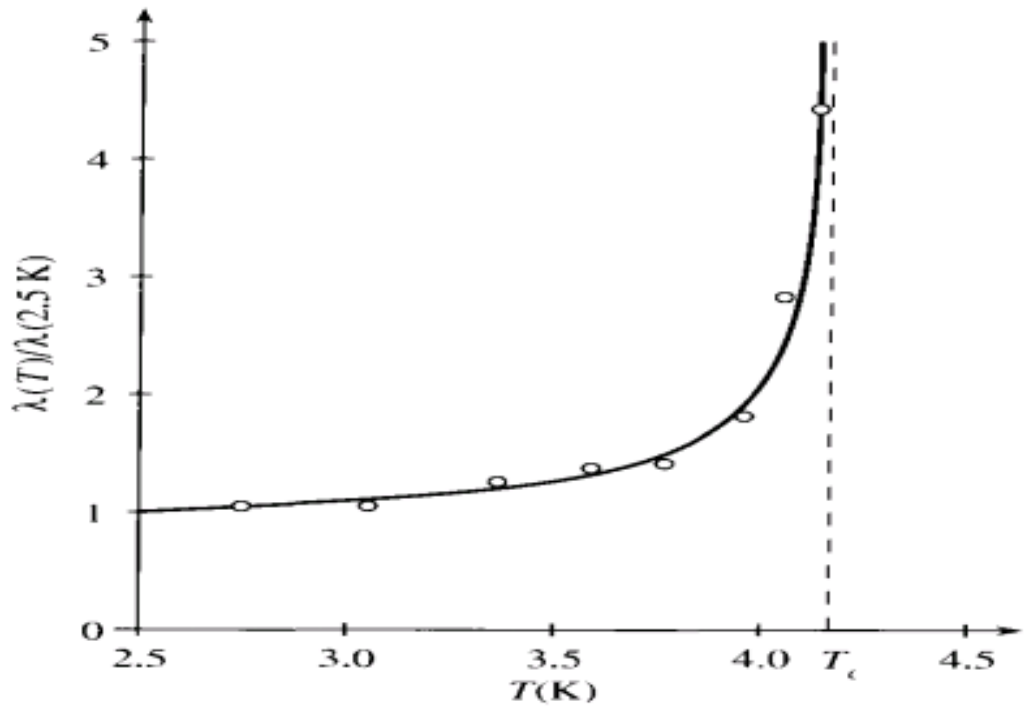

**Fig. 21. a)** Magnetic flux density as a function of distance B(x) inside the superconductor in eq. (2.7). **b)** Temperature dependence of the London penetration depth, λ(T), in mercury in eq. (2.8). Source: F. London, Superfluids: Macroscopic theory of superconductivity, John Willey & Sons, Inc., vol. 1, pp. 1-161, 1950 (after [21]).

In 1933, Becker, Sauter, Heller [20] used a simple free electron model to analyze the electrodynamic behaviors of superconductors as perfect normal conductors, in which the electrons accelerate without any resistance under the exertion of an electric field E. Becker, Sauter, Heller [20] argued that, if the electrons encountered no resistance, an applied electric field E would accelerate the electrons steadily. In 1935, the brothers H. and F. London [21] proposed that, since the Becker, Sauter, Heller [20] macroscopic theory of a perfect conductor makes correct predictions about superconductors for the special case Bo = 0, it might be reasonable To suppose that the magnetic behavior of a superconductor may be correctly described, taking to the account the Meissner effect, and derived their famous equations [21].

**The London equations [23] describe the electrodynamic behaviour of superconductors on a macroscopic scale in weak electromagnetic fields and relate the microscopic strength E of electric field and induction B of magnetic field to the supercurrent density JS**

$$\frac{\partial}{\partial t}\boldsymbol{J}_S = \frac{1}{\mu_0\lambda_L^2}\boldsymbol{E} \qquad (2.4)$$

$$\nabla\times\boldsymbol{J}_S = -\frac{1}{\mu_0\lambda_L^2}\boldsymbol{B} \qquad (2.5)$$

**where λL is the London penetration depth in eq. (2.6) [21]:**

$$\lambda_L = \sqrt{\frac{m_s}{\mu_0 n_s e^2}} \qquad (2.6)$$

**where $m_S$ and $n_S$ are the superconducting electron mass and volume density respectively, e is the electron charge.**

$$\boldsymbol{B}(x) = \boldsymbol{B}_0\exp\left(-\frac{x}{\alpha^{1/2}}\right) \qquad (2.7)$$

$$\lambda(T) = \lambda(0)/\left[1-(T/T_C)^4\right]^{1/2} \qquad (2.8)$$

# Pippard Nonlocal Electrodynamics.

In 1953, Pippard [25] conducted the experimental and theoretical study of the relation between magnetic field H and current I in a superconductor, considered the coherence concept in superconductivity, and introduced the coherence length $\xi_0$ while proposing a nonlocal generalization of the London equation (2.4). Pippard found that the penetration depth was noticeably dependent upon the impurity content, which could not be explained by the local theory of London since the density of superelectron and its effective mass could only be weak functions of the impurity concentration.

| Superconductor | $\lambda_0$ (nm) | $\xi_0$ (nm) | $T_c$ (K) |
|---|---|---|---|
| Al | 16 | 1500 | 1.2 |
| In | 25 | 400 | 3.3 |
| Sn | 28 | 300 | 3.7 |
| Pb | 28 | 110 | 7.2 |
| Nb | 32 | 39 | 8.95–9.3 |
| $Nb_3Sn$ | 50 | 6 | 18 |
| $YBa_2Cu_3O_x$ | 140 | 1.5 | 90 |

**Fig. 22.** Material parameters such as penetration depth λ0 and coherence length ξ for some LTS and HTS superconductors (after [38]).

**Pippard [25] considered Reuter and Sondheimer [25] and Chambers [25] nonlocal generalization of Ohm's law J(r) = σE(r) for explaining anomalous skin effect:**

$$J_n(r) = \frac{3\sigma}{4\pi\ell}\int \frac{R[R\bullet E(r')]e^{-R/\ell}}{R^4}dr'$$

**where R = r – r'; this formula takes into account the fact that the current at a point r depends on E(r') throughout a volume of radius ~ℓ about r.**

**Pippard proposed that a local relation should be replaced by a nonlocal relation of the form [25]:**

$$J_s(r) = -\frac{3}{4\pi\xi_0\Lambda c}\int \frac{\mathbf{R}[\mathbf{R}\bullet\mathbf{A}(\mathbf{r}')]}{\mathbf{R}^4}e^{-R/\xi}d\mathbf{r}'$$

**where R = r — r' and the effective coherence length ξ in the presence of scattering was assumed to be related to that of pure material characteristic length ξ0 [25]:**

$$\frac{1}{\xi} = \frac{1}{\xi_0} + \frac{1}{\ell}.$$

# Ginzburg-Landau (GL) Theory of Superconductivity.

In 1950, Ginzburg and Landau (GL) [23] intuited a remarkable phenomenological theory of superconductivity that integrated the electrodynamics, quantum mechanical and thermodynamic properties of superconductors.

| Material | $T_c$ (K) | ξ (nm) | λ (nm) | κ (= λ/ξ) |
|---|---|---|---|---|
| Nb | 9.25 | 39 | 50 | 1.28 |
| Nb-Ti | 9.5 | 4 | 300 | 75 |
| $Nb_3Ge$ (A15) | 23.2 | 3 | 90 | 30 |
| $YBa_2Cu_3O_7$ | 89 | 1.8 | 170 | 95 |

**Fig. 23.** Data for critical temperatures Tc, coherence length ξ, penetration depth λ and GL parameter κ for some LTS and HTS superconductors (after [38]).

**The GL coherence length $\xi_{GL}$ characterizes the distance over which the wave function Ψ can change without the undue energy increase in eq. (2.21) [23]:**

$$\xi_{GL}(T) = \frac{\hbar}{\left|2m^{*}\alpha(T)\right|^{1/2}}, \qquad (2.21)$$

**The ratio between two characteristic lengths: λ(T) and ξ(T) defines the GL parameter k in eq. (2.25) [23]:**

$$k = \frac{\lambda(T)}{\xi_{GL}(T)}, \qquad (2.25)$$

**The Ginzburg-Landau (GL) theory [23] states that the physical behaviour of superconducting electrons may be described by the effective order parameter or the wave function:**

$$\Psi(\mathbf{r}) = \Psi_0 \cdot \exp(i\theta(\mathbf{r})), \qquad (2.16)$$

**The density of superconducting electrons is [23]**

$$N_S = |\Psi|^2, \qquad (2.17)$$

**The superconducting current density is [23]**

$$\boldsymbol{J}_S \propto \nabla\theta(\mathbf{r}),$$

**where θ is the phase of the wave function, and ∇=d/dr, and r is the distance.**

**Minimizing expression for the free energy on Ψ and vector potential A, Ginzburg and Landau derived the two famous GL equations (2.19, 2.20) [23]**

$$\alpha\Psi + \beta|\Psi|^2\Psi + \frac{1}{2m^*}\left(i\hbar\nabla + e^*A\right)^2\Psi = 0 \qquad (2.19)$$

$$J_s = -i\frac{e^*\hbar}{2m^*}\left(\Psi^*\nabla\Psi - \Psi\nabla\Psi^*\right) + \frac{e^{*2}}{m^*}|\Psi|^2 A \qquad (2.20)$$

**where $J_S$ is the superconducting current density, α and β are the expansion coefficients:**

$$\alpha = -\mu_0 H_C^2/|\Psi_0|^2$$

$$\beta = \mu_0 H_C^2/|\Psi_0|^4,$$

**in which $H_C$ is the thermodynamic critical field, and $Ns=|\Psi 0|^2$ is the density of superconducting electrons in low magnetic field.**

# Richard A. Ogg Theory of Electron Pairs and their Bose-Einstein Condensation. Fröhlich Theory of Interaction between the Electrons and Phonons. Bardeen, Cooper and Schrieffer (BCS) Microscopic Theory of Superconductivity.

In 1946, Richard A. Ogg [22] made a first ever suggestion of electron pairs and their Bose-Einstein condensation, when describing his experiments on very dilute solutions of alkali metals in liquid ammonia.

In 1950, Fröhlich [24] from University of Liverpool made his original proposition about the possible existence of weak attraction between the electrons as a result of electron-phonon interaction in crystal lattice of a superconductor.

In 1957, Bardeen, Cooper and Schrieffer [26] published the microscopic theory of superconductivity, known as the BCS theory.

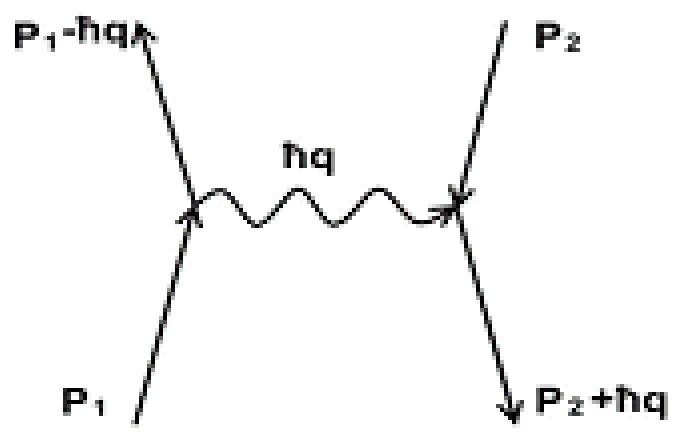

**Fig. 24.** Phonon exchange between two electrons with momentums $p_1 = -p_2$ and phonon quasi momentum ħq .

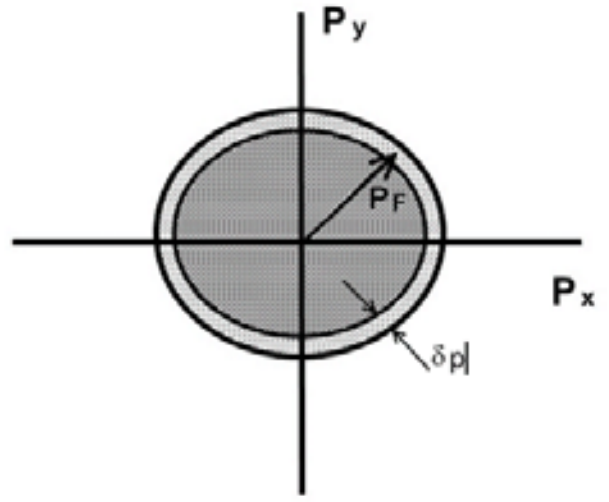

**Fig. 26.** Fermi surface and energy gap layer of superconducting electrons with thickness δq in conventional superconductors (singlet s-wave wave function).

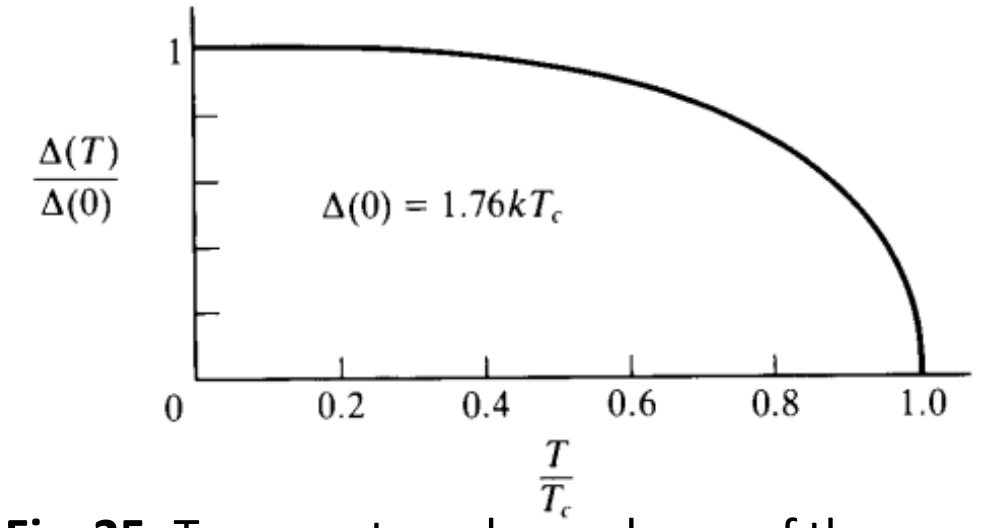

**Fig. 25.** Temperature dependence of the energy gap in the BCS theory.

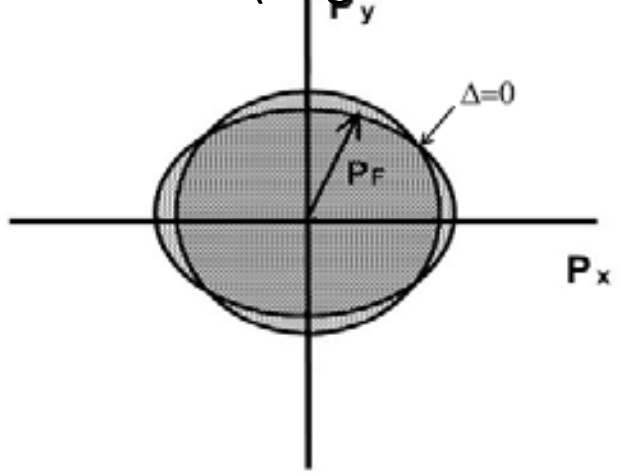

**Fig. 27.** Fermi surface and energy gap in d-wave high Tc superconductors ($d_{x2-y2}$ wave function).

**The BCS theory [15] is based on the paired Hamiltonian**

$$H = \sum_{k\sigma} \varepsilon_k n_{k\sigma} + \sum_{kl} V_{kl} c^*_{k\uparrow} c^*_{-k\downarrow} c_{-l\downarrow} c_{l\uparrow}$$

**In the BCS theory, the origin of ac energy losses in a superconductor is explained as a co-existence of two types of charge carriers within the superconductor: the Cooper superconducting electron pairs and the normal electron excitations near the Fermi surface, which depend on the temperature T. The surface resistance of superconductors Rs in the BCS theory is in eq. (2.27) [*]:**

$$R_s \approx \frac{\Delta(T)}{k_B T} \omega^{3/2} \exp\left[-\Delta(T)/k_B T\right]. \quad (2.27)$$

**In the application of the BCS theory to d-pairing superconductors in case ξ<<λ (local approximation), the surface resistance Rs can be expressed as an integral in eq. (2.28)**

$$R_s \propto \frac{\omega^{3/2}}{k_B T} \int_S \Delta(p_x, p_y) \exp\left[-\Delta(p_x, p_y)/k_B T\right] dp_x dp_y , \quad (2.28)$$

# Critical Parameters of Superconductors.

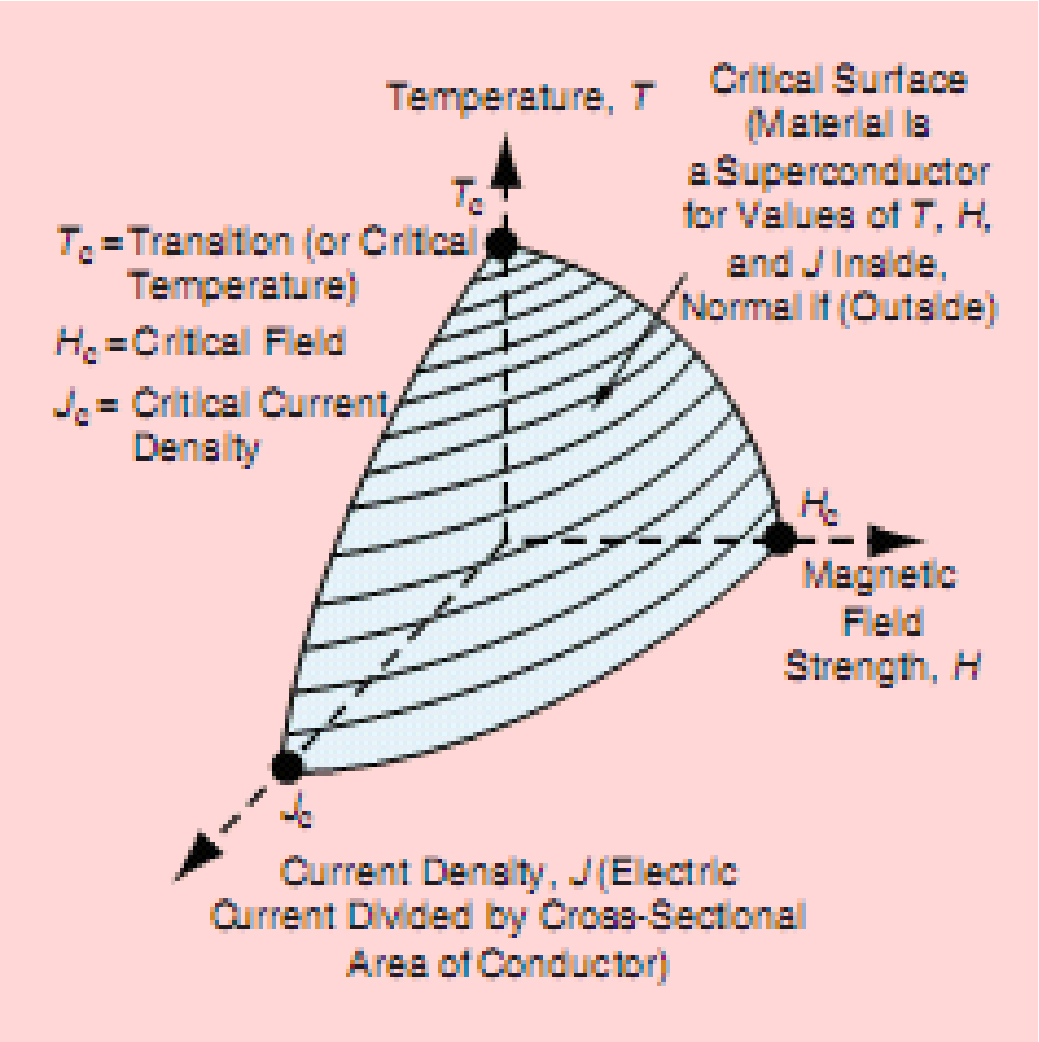

**Fig. 28.** A three-dimensional (3-D) plot of **Tc**, **Hc**, and **Jc** surface that defines superconductivity. Materials are Superconducting, if all of these values are below the surface shown, whereas they are in normal state if any of the three parameters fall outside the surface (after [3]).

**Critical parameters of superconductors:**

1. **Critical Temperature:** the highest temperature at which the superconductor can maintain the superconducting state. The critical temperature Tc for $YBa_2Cu_3O_{7-\delta}$ is 91-93K.
2. **Critical Frequency:** the frequency at which the photons of electromagnetic waves excite the superconducting electrons with enough energy to drive them to the normal state. The magnitudes of the frequency ωC , which force a material to lose its superconducting properties at T=0, are within the low optical frequency range of $10^{12}$ Hz for low temperature superconductors (LTS), and in frequency range of $10^{13}$ Hz for high temperature superconductors (HTS) approximately. The equation for critical frequency: $\hbar\omega c(T)=2\Delta(T)$, where ħ is the reduced Planck's constant, and Δ(T) is superconducting energy gap in the BCS theory.
3. **Critical Magnetic Fields:** the highest value of external magnetic field strength applied to a superconductor without causing superconductivity to disappear. The equation for the critical magnetic field: $H_c=H_0\,[1-(T/Tc)^2]$, where $H_0$ is the critical field at absolute zero and Tc is the critical temperature.
4. **Critical Current Density:** the maximum current that can be transported through a superconductor without causing resistance to appear is called the critical current Jc (T) = Bc (T)/(μ0λ(T)), where λ is the magnetic field penetration depth.

# Classification of Superconductors in Type I and Type II.

A significant contribution to the theory of superconductivity was made by Abrikosov [28] in 1957, who introduced the classification of superconductors as the Type I and Type II superconductors, showing the different physical behaviors of superconductors, when they are exposed to the external magnetic fields.

1) 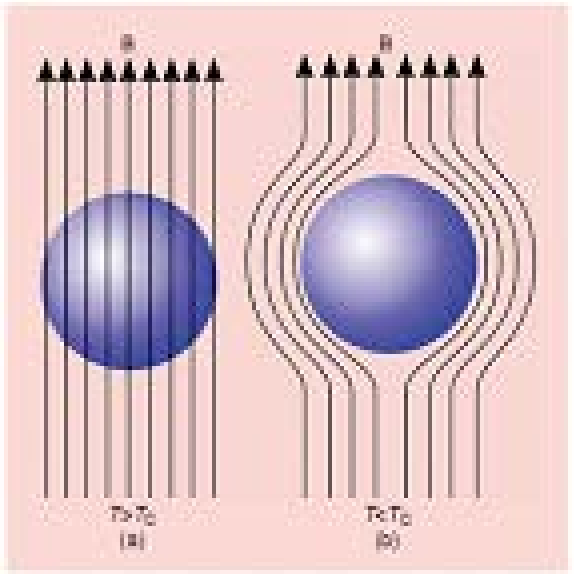

**Fig. 29.** 1) Schematic representation of the Meissner effect [2] in a super-conductor: (a) The magnetic behavior when the specimen is in the normal state (i.e., at temperatures above Tc). (b) The magnetic behavior when the specimen is in the superconducting state (i.e., at temperatures below Tc) (after[3]).

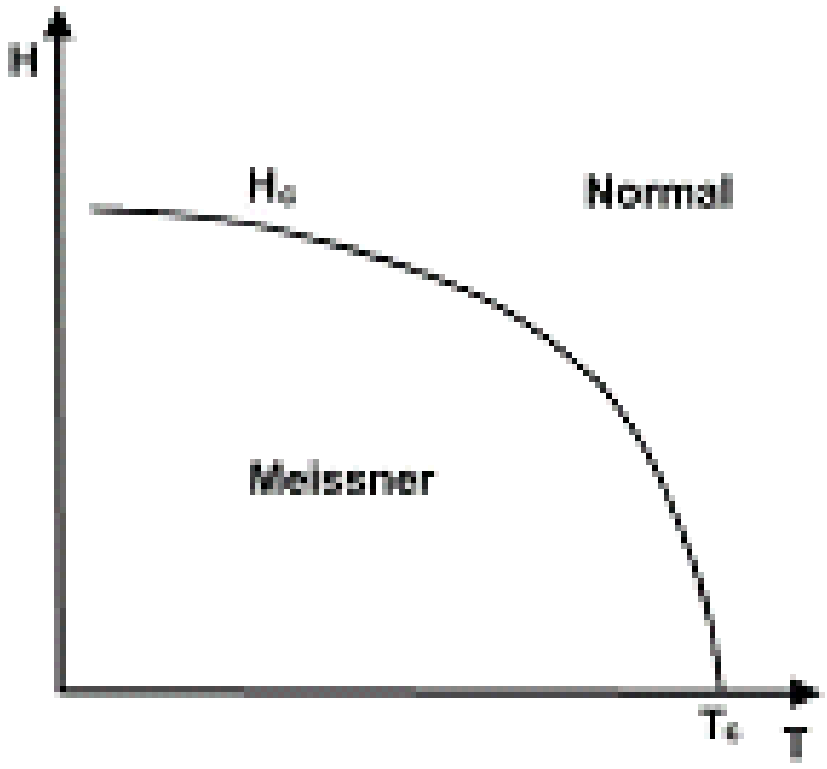

2)

**Fig. 30.** Meissner Effect in Type I superconductors (after [4]).

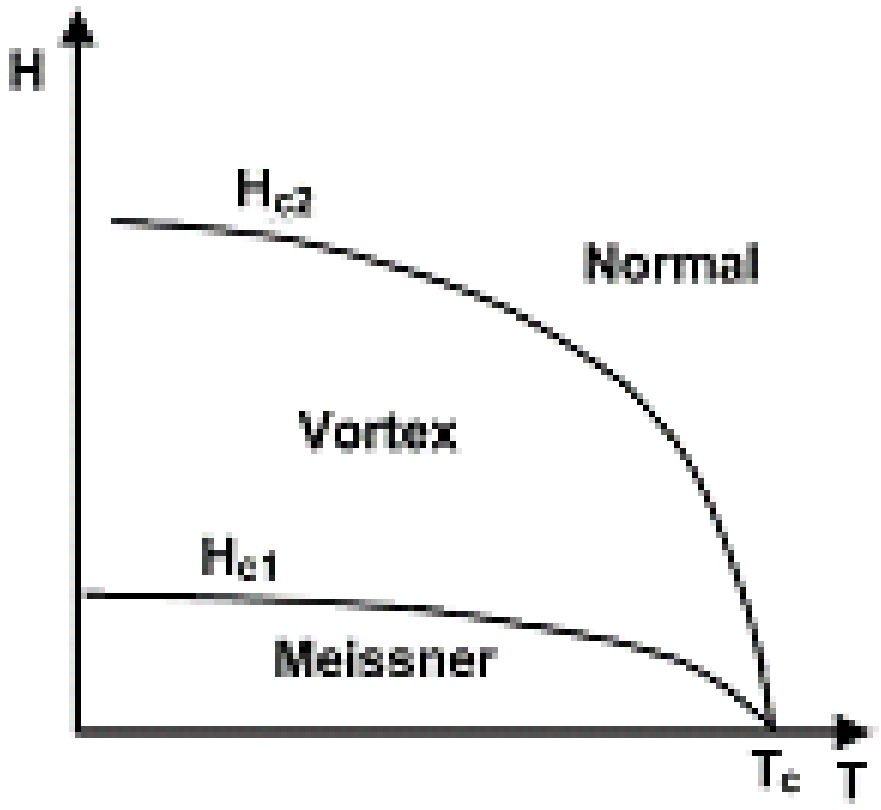

3)

**Fig. 31.** Meissner Effect in Type II superconductors (after [4]).

$$H_{C1} \approx \frac{H_C \ln k}{k\sqrt{2}},$$

$$H_{C2} = H_C k\sqrt{2},$$

where k = ξ/λ is the GL parameter, $H_C$ is the thermodynamic critical magnetic field.

# Abricosov Magnetic Vortices in Type II Superconductor.

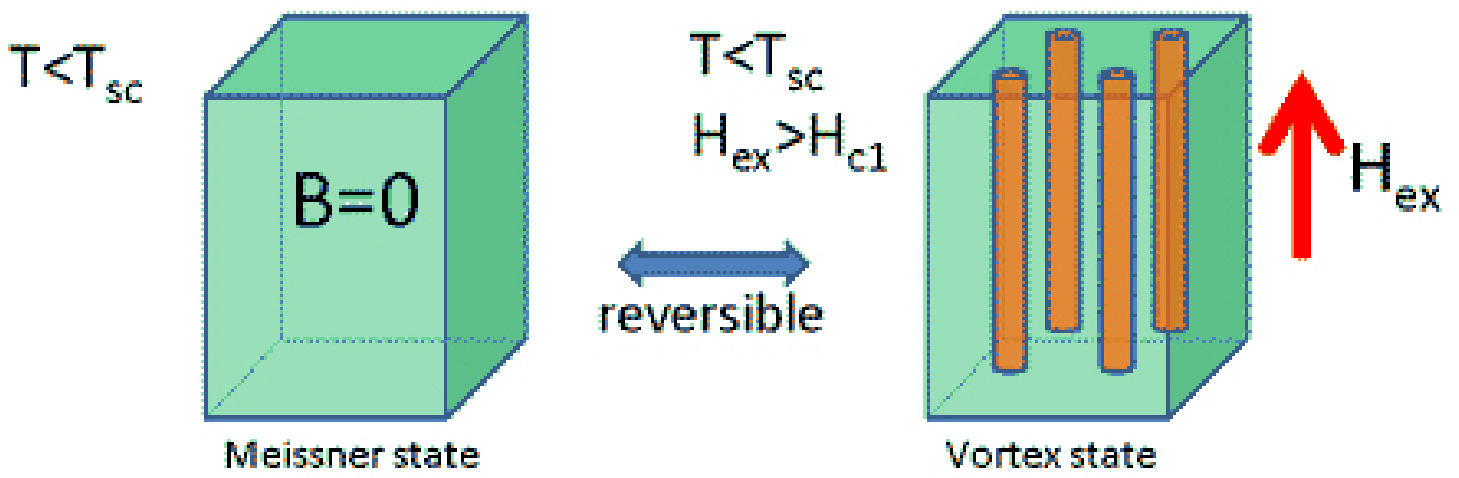

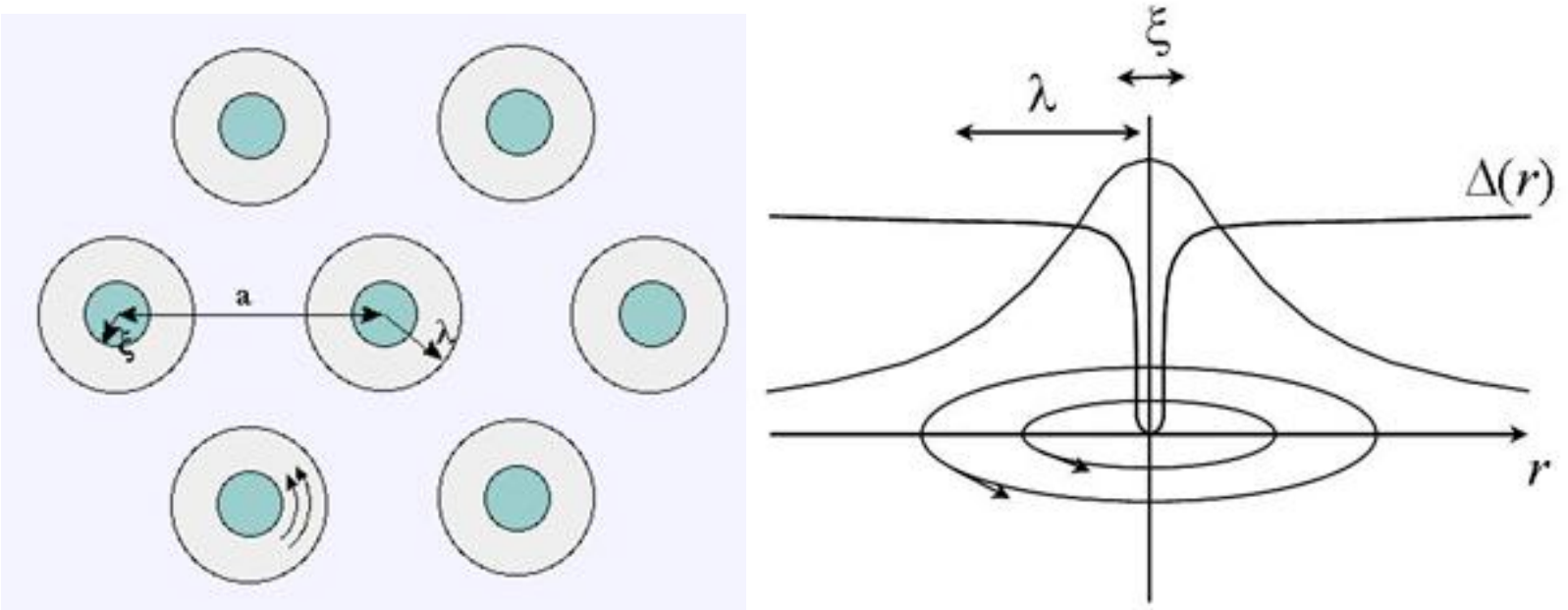

**Fig. 32.** Abricosov magnetic vortices generation in Type II superconductors at external magnetic field $H_{ex}$ action. Superconducting state exists in the region outside of Abricosov magnetic vortex core. In the case **$\xi < \lambda$ ,** the BCS superconducting gap vanishes and the normal state appears, where **$\lambda$** is the radius of Abricosov magnetic flux line, **$\xi$** is the radius of Abricosov magnetic vortex core.

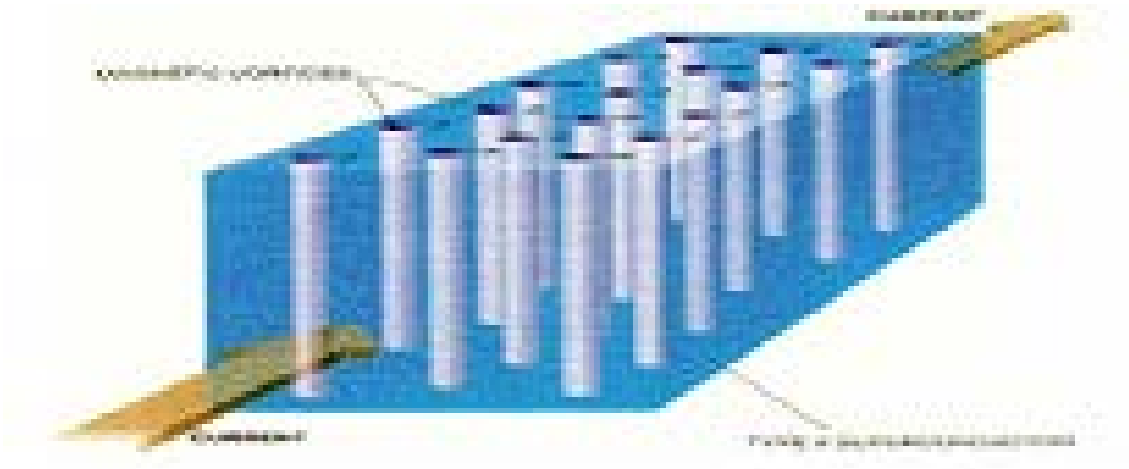

**Fig. 33.** Type II superconductor in mixed state with Abricosov magnetic vortices and transport current direction shown (after [29]).

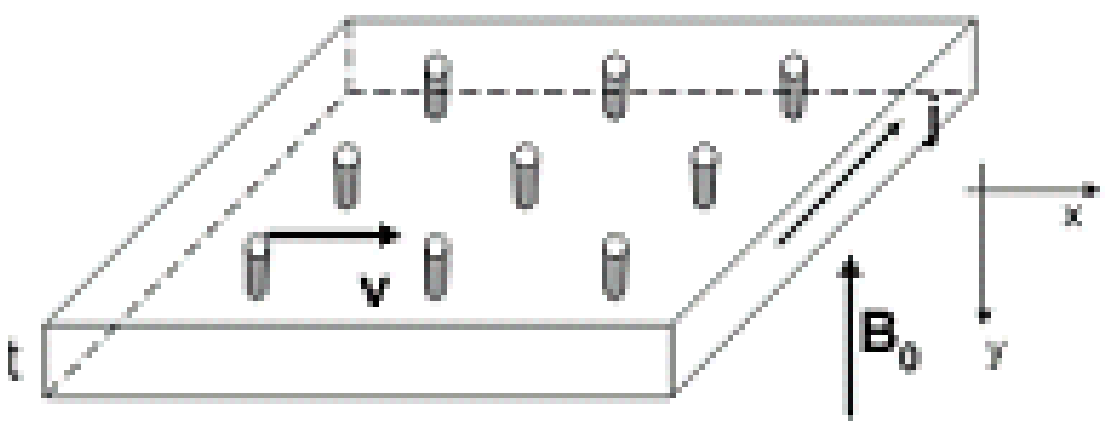

**Fig. 34.** Abricosov magnetic vortex lattice generation in Type II HTS thin film at application of external magnetic field B (after [30]).

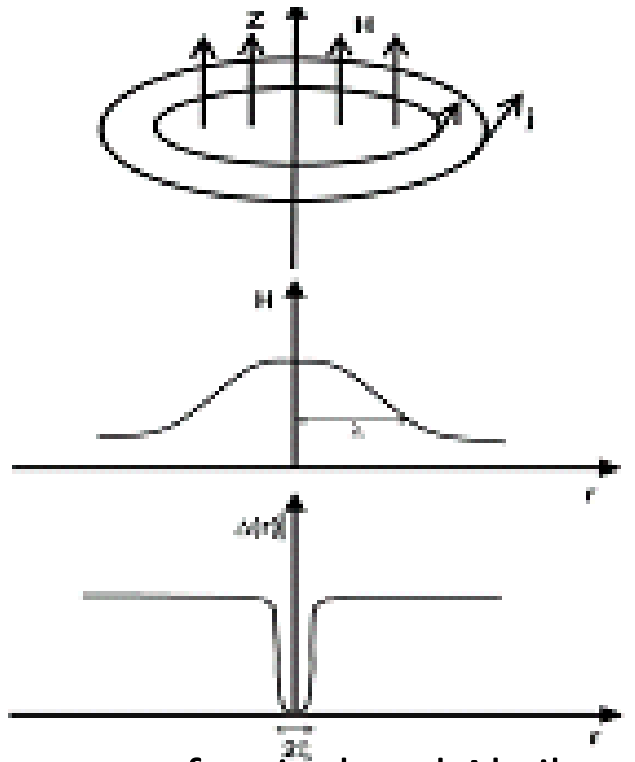

**Fig. 35.** Structure of an isolated Abrikosov vortex line in a Type II superconductor (after [31, 32]).

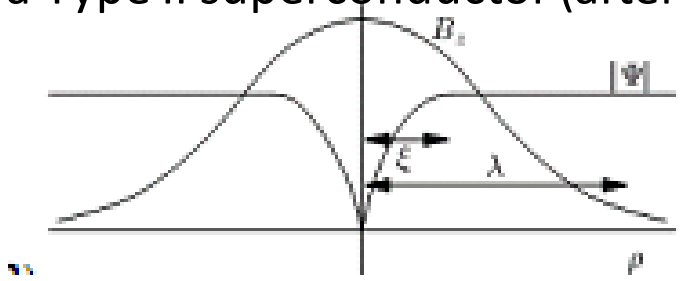

**Fig. 36.** Structure of an isolated Abrikosov vortex line in a Type II superconductor(after[33]).

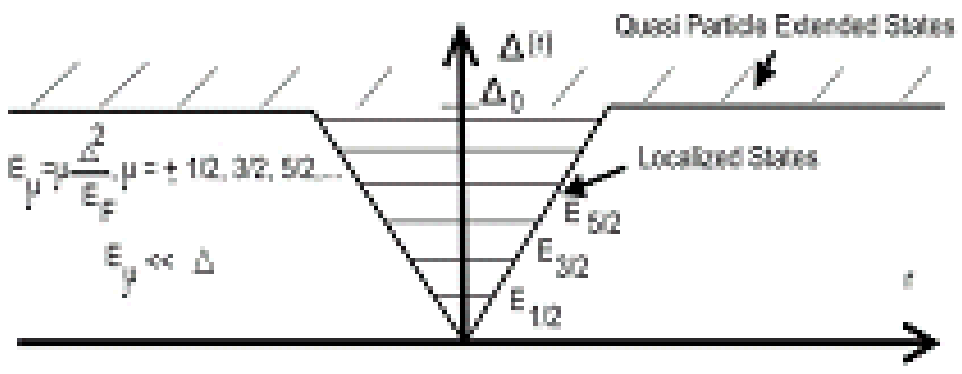

**Fig. 37.** Electronic structure of an Abricosov magnetic vortex core (after [34]).

# Synthesis of HTS Thin Films for Application in Microwave Devices and Circuits.

**The HTS thin films with thickness 400-600nm can be fabricated by deposition of high-Tc superconductor on dielectric substrates with application of different methods [34, 35, 36]:**

**1.** Evaporation of superconductors onto heated substrate in presence of oxygen.

**2.** Reactive co-evaporation by cyclic deposition - reaction.

**3.** Pulsed laser deposition.

**4.**Metal-organic chemical vapor deposition (MOCVD).

**5.** Sputtering method (DC and RF plasma discharge).

**6.** Liquid phase epitaxy.

**Substrates for HTS thin film deposition and their loss tangent magnitude at specified frequencies [37]:**

| Substrate | Frequency | tanδ |
|---|---|---|
| *Sapphire α-$Al_2O_3$* | 9 GHz | $1.5x10^{-8}$ |
| *Lanthanum Aluminate $LaAl_2O_3$* | 10 GHz | $7.6x10^{-6}$ |
| *Magnesium Oxide MgO* | 8 GHz | $2x10^{-6}$ |
| *Strontium Lanthanum Aluminate $LaSrAlO_4$* | 12 GHz | $2x10^{-6}$ |
| *Alumina $Al_2O_3$* | 7.7 GHz | $2x10^{-5}$ |

**Fig. 38.** Substrates for HTS thin film deposition and their loss tangent magnitude at specified frequencies (after [37]).

**The microwave device and circuit requirements for the fabrication of HTS thin films and multilayers on epitaxial substrates [35]:**

**1.** Growth at a lowest possible temperature, Ts, insuring the exact stoichiometry of all constituents, formation of the desired single phase and full crystalline order.

**2.** Adhesion to, but no inter-diffusion or reaction with, the substrate.

**3.** Film thickness from monocell to >> $\lambda_L$ with narrow thickness tolerances (goal:<< +-10%); $\lambda_L$ is the London penetration depth.

**4.** Thermal expansion match over a broad temperature range.

**5.** Smooth surfaces and interfaces (goal: roughness << 1 nm).

**6.** Large, uniform area (goal: at least 10 cm in diameter).

**7.** Epitaxial, i.e., structural and lattice parameter match.

**8.** Control of orientation.

**9.** Epitaxial growth on ex-situ processed underlayer.

**10.** No grain boundaries, except those controllably nucleated.

**11.** High crystalline quality, low defect number density.

**12.** No pinholes in insulators, barriers.

# Technical Characteristics of Synthesized High Temperature Superconducting (HTS) Thin Films for Microwave Applications.

**Technical parameters of Type II superconductors [38]:**

| Material Properties | YBaCuO | TlBaCaCuO | Nb |
|---|---|---|---|
| Transition temperature $T_c$ | up to 95 K | up to 130 K | 9.2 K |
| Coherence length $\xi_{ab}$ | 1.5 nm | 3.0 nm | 39 nm |
| Coherence length $\xi_c$ | 0.2 nm | 0.1 nm | — |
| Penetration depth $\lambda_{ab}$ | 150 nm | 200 nm | 50 nm |
| Lower critical field $B_{c1}$ | 10–100 mT | 10 mT | 0.13 T |
| Upper critical field $B_{c2}$ | 100–200 T | 60 T | 0.3 T |

**Fig. 39.** Technical parameters of Type II superconductors (after [38]).

**Substrates for HTS thin film deposition and their loss tangent magnitude at specified frequencies [37]:**

| Substrate | Frequency | tanδ |
|---|---|---|
| *Sapphire α-$Al_2O_3$* | 9 GHz | $1.5x10^{-8}$ |
| *Lanthanum Aluminate $LaAl_2O_3$* | 10 GHz | $7.6x10^{-6}$ |
| *Magnesium Oxide MgO* | 8 GHz | $2x10^{-6}$ |
| *Strontium Lanthanum Aluminate $LaSrAlO_4$* | 12 GHz | $2x10^{-5}$ |
| *Alumina $Al_2O_3$* | 7.7 GHz | $2x10^{-5}$ |

**Fig. 38.** Substrates for HTS thin film deposition and their loss tangent magnitude at specified frequencies (after [37]).

**Present achievements in synthesis of HTS thin films [39]:**

The synthesis of $YBa_2Cu_3O_{7-\delta}$ thin films is quite mature. Large-sized (≥8″) and double-sided $YBa_2Cu_3O_{7-\delta}$ thin films are commercialized in the developed countries, such as Theva GmbH in Germany, the films for microwave usage have $T_{C0}$ of 88 K, the critical current density $J_C \approx 3\times10^6$ A/cm$^2$ (77 K, 0 T), and $R_S \approx 500$ μΩ (77 K, 10 GHz). By employing these $YBa_2Cu_3O_{7-\delta}$ thin films to make the microstrip filters, the filters can operate under liquid nitrogen temperature. Alternatively, much research work has also been carried out on the synthesis of $Tl_2Ba_2CaCu_2O_{8-\delta}$ thin films. It was found that two phases of $Tl_2Ba_2CaCu_2O_8$ thin films are suitable for passive microwave devices: $Tl_2Ba_2CaCu_2O_8$ (Tl-2212) and $Tl_2Ba_2CaCu_2O_{10}$ (Tl-2223) [40]. The former has a $T_{C0}$ of 110 K, $J_C \approx 2\times10^6$ A/cm$^2$ and $R_S \approx 130$μΩ (77K,10GHz), and the latter has a $T_{C0}$ of 125 K, $J_C$ (77K)≈$10^5$A/cm$^2$ ,and $R_S \approx 86$ μΩ (77K,10GHz) [40]. The synthesis of Tl-2212 is easier than Tl-2223, and because of the intergrowth of Tl-2212 with Tl-2223, it is not possible to grow pure Tl-2223 phase without doping, there is always some Tl-2212 left in the Tl-2223 film. Therefore for microwave applications it is better to use Tl-2212. The microstrip filter made by Tl-2212 can be operated at 82 K [39].

# Synthesis of First $YBa_2Cu_3O_{7-\delta}$ Thin Films on Various Substrates.

**1.** $YBa_2Cu_3O_{7-\delta}$ thin films on $SrTiO_3$(100) by Laibowitz, Koch, Chaudhari, Gambino [41];
**2.** $YBa_2Cu_3O_{7-\delta}$ thin films on $SrTiO_3$(100) by co-evaporation by Mankiewich, Scofield, Skocpol, Howard, Dayem, Good [42];
**3.** $YBa_2Cu_3O_{7-\delta}$ thin films by Yoshizako, Tonouchi, Kobayashi [43];
**4.** $YBa_2Cu_3O_{7-\delta}$ thin films on $SrTiO_3$(100) and MgO(100) by activated reactive evaporation by Terashima, Iijima, Yamamoto, Hirata, Bando, Takada [44];
**5.** $YBa_2Cu_3O_{7-\delta}$ thin films on $SrTiO_3$(100) by laser ablation by Koren, Gupta, Baseman [45];
**6.** $YBa_2Cu_3O_{7-\delta}$ thin films on $SrTiO_3$(100) by pulse laser deposition by Koren, Gupta, Beserman, Lutwyche, Laibowitz [46];
**7.** $YBa_2Cu_3O_{7-\delta}$ thin films on SrTiO3(100) by laser ablation by Koren, Gupta, Segmuller [47];
**8.** $YBa_2Cu_3O_{7-\delta}$ epitaxially grown thin films by Klein et al. [48];
**9.** $YBa_2Cu_3O_{7-\delta}$ thin films on MgO(100) by vapour deposition by Humphreys, Satchell, Chew, Edwards, Goodyear, Blenkinsop, Dosser, Cullis[49];
**10.** $YBa_2Cu_3O_{7-\delta}$ thin films on MgO(100) by laser ablation by Moeckly, Russek, Lathrop, Buhrman, Jian Li, Mayer [50];
**11.** $YBa_2Cu_3O_{7-\delta}$ thin films by off-axis magnetron sputtering by Newman, Char, Garisson, Barton, Taber, Eom, Geballe, Wilkens [51];
**12.** $YBa_2Cu_3O_{7-\delta}$ /Pr thin films by inverted cylindrical magnetron sputtering and pulsed laser deposition by Xi [52].

# Main Parameters for Accurate Characterization of High Temperature Superconducting (HTS) Thin Films at Microwaves.

**Surface Impedance $Z_S$**

$$Z_S = \left(\frac{E_x}{H_y}\right)_{surf} = R_S + jX_S = \left(\frac{j\omega\mu_0}{\sigma_S}\right)^{1/2}$$

**Surface Resistance R**

$$R_S = \frac{\sigma_1}{2\sigma_2}\left(\frac{\omega\mu}{\sigma_2}\right)^{1/2}$$

**Surface Reactance Xs**

$$X_S = \left(\frac{\omega\mu}{\sigma_2}\right)^{1/2}$$

**Complex Surface Conductivity $\sigma_S$**

$$\sigma_S = \sigma_1 - j\sigma_2 = \frac{2\omega\mu_0 R_S X_S}{\left(R_S^2 + X_S^2\right)^2} - j\frac{\omega\mu_0\left(X_S^2 - R_S^2\right)}{\left(R_S^2 + X_S^2\right)^2}$$

**Surface Impedance $Z_S$ :**

In 1940, the first accurate measurements of surface impedance $Z_S$ of superconductors by London [53], using the microwave methods, provided the important information about the electromagnetic properties of superconductors at microwaves [54].

It was understood that the surface impedance $Z_S$ is a measure of energy losses by an electromagnetic wave in superconductor at microwaves [54]. The improved methods for the accurate measurements of surface impedance $Z_S$ were developed by Pippard [55, 56]. Blatt [57] writes that study of the surface impedance $Z_S$ of superconductors at microwaves led to the creation of the coherence length concept of Pippard [55, 56].

**Nature of ac Loss [3]:**

The origin of the energy loss at any nonzero frequency, when the temperature of superconductor is above 0 K follows from the existence of two types of charge carriers within the superconductor. Although, the Cooper pairs move without resistance, the carriers in the normal state, those above the energy gap, behave as electrons in a normal conductor. The reason the Cooper pairs don't short out the normal electrons is that they possess mass, and hence, kinetic energy.

# Experimental Methods and Types of Microwave Resonators for Precise Microwave Characterisation of Superconductors.

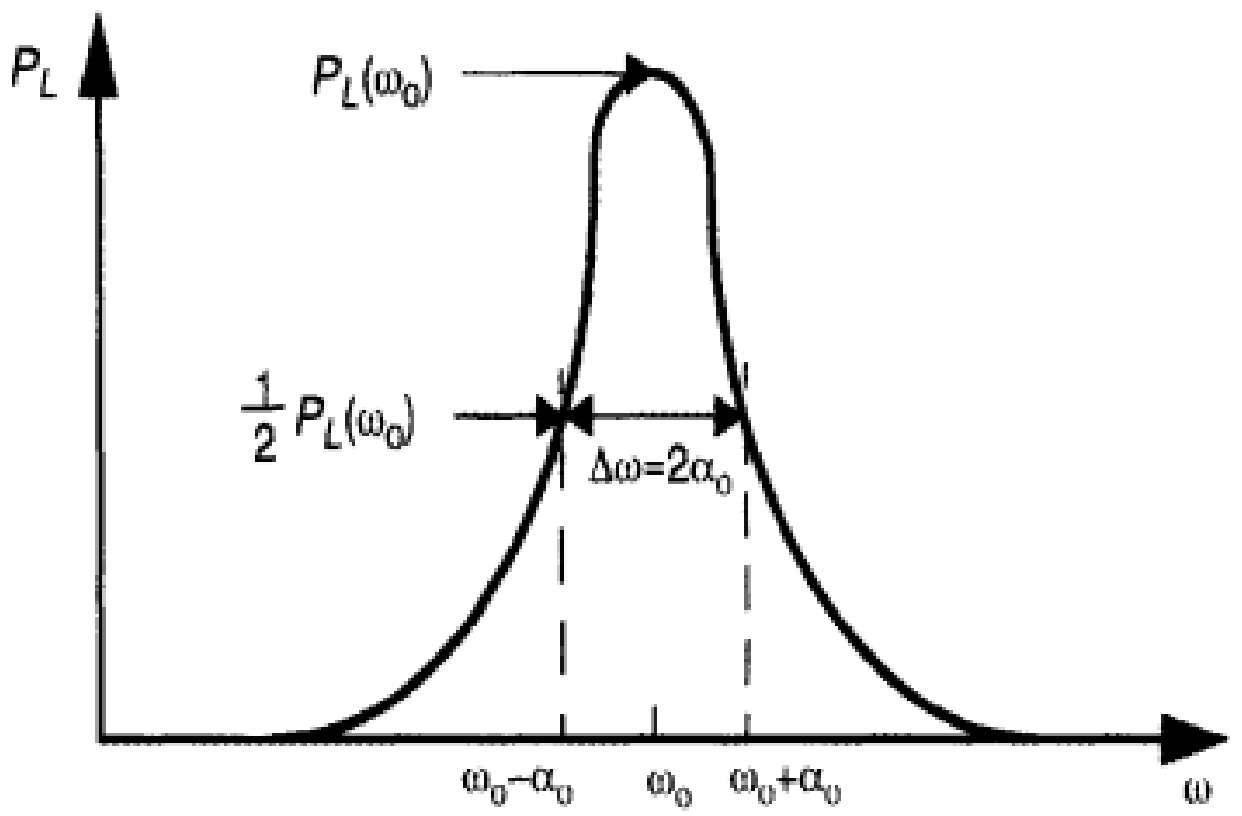

**Fig. 40.** Resonance curve of a simple RLC circuit (after [38]).

**Quality Factor Q** $Q = f_0 / \Delta f$

**Interconnection between Q and Rs**

$$\frac{1}{Q} = \frac{R_S}{A_S} + \frac{1}{Q_{par}} + \frac{\tan\delta}{A_{diel}}$$

**Geometric Factor A** $A_S \approx V_{res} / V_s$.

| Type of resonator | Advantages | Disadvantages |
|---|---|---|
| Cavity | (a) Geometric factor can be calculated analytically; (b) if the cavity is made of low $T_c$ superconductor, it can have a very high quality factor $Q$ and therefore the sensitivity of the technique can be quite high; (c) easy sample mounting; (d) quite reproducible; (e) easy to realize a fixed temperature operation which allows the measurement of both $R_s$ and λ (using a thermally isolated platform within the cavity) | (a) At typical microwave frequencies of interest (~10 GHz) the cavities are quite big (≳6 cm in diameter) and bulky; (b) they require large area films (several inches in diameter) or use diaphragms in the cavity end plate, which however bear potential problems due to current path interruption at the boundary between the plate and the film |
| Dielectric | (a) Quite sensitive (geometric factor is typically a few hundred ohms); (b) easy sample mounting; (c) does not require sample preparation or modification; (d) sufficiently reproducible; (e) relatively small size (at typical microwave frequencies of ~10 GHz for dielectric permittivity $\varepsilon \gtrsim 10$ the pucks are usually $\lesssim$ 10 mm in diameter, which is acceptable for measuring both small (10 x 10 mm$^2$) and large (2″ in diameter or more) area HTS films) | (a) Penetration depth measurements are greatly encumbered because the change in frequency is mostly due to change in the dielectric permittivity of the puck, not change in $\lambda(T)$ of the film. The problem may be overcome by keeping the puck at fixed temperature, which is however quite a technical challenge. More details on the DR technique can be found in [6] |
| Parallel plate | (a) Low geometric factor and therefore high sensitivity; (b) many easily excitable modes | (a) Parasitic modes interfering; (b) two films required; (c) reproducibility is not great |
| Microstrip | (a) High sensitivity (low geometric factor); (b) large enhanced current crowding at the edges of the strip, which is useful for nonlinear measurements | (a) Possible influence of patterned edges; (b) loss of the substrate may noticeably contribute to the measurements; (c) requires double-sided films and measured $R_s$ will be weighted by the contributions of the two; (d) finite edge roughness due to patterning may enhance nonlinear effects and promote vortex penetration |
| Stripline | (a) Extremely low geometric factor; (b) possibly highest current crowding at the edges | (a) Possible influence of patterning; (b) three films required, although the major contribution ~90–95% comes from the film with the resonant centre strip; (c) impossible to perform DC field measurements due to screening of the superconducting groundplanes; (d) same as for the microstrip; (e) hard to analyse and requires numerical modelling |
| Coplanar | (a) Geometric factor comparable to stripline; (b) current crowding comparable to stripline; (c) only one single-sided film is required; (d) losses are nearly insensitive to the substrate; (e) allows application of DC magnetic field | (a) Same as for the stripline; (b) shape of resonance (especially higher order modes) highly sensitive to grounding; (c) finite edge roughness; (d) requires numerical modelling |

**Fig. 41.** Review of advantages and disadvantages of various types of microwave resonators employed for surface impedance measurements (after [58]).

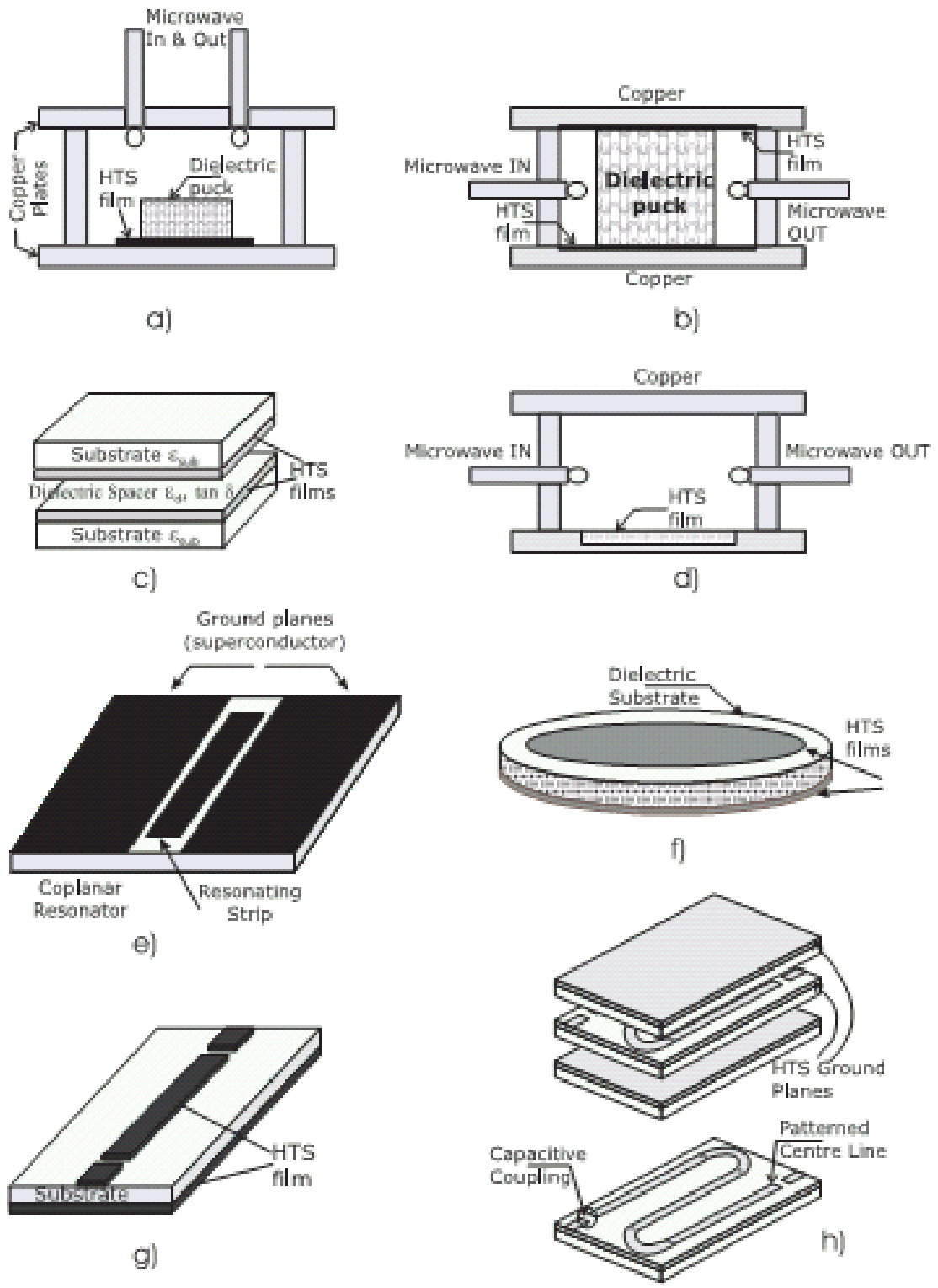

**Fig. 42.** Eight main types of microwave resonators used for thin film characterization: (a) open-ended dielectric resonator;(b) Hakki–Coleman dielectric resonator; (c) parallel plate resonator; (d) cavity resonator; (e) coplanar resonator; (f) disc resonator; (g) microstrip resonator; (h) stripline resonator (after [58]).

**Microwave resonators used in the research:**

**1.** Hakki-Colleman type dielectric resonator.

**2.** Split post dielectric resonator.

**3.** Microstrip resonator.

# Nonlinear Surface Resistance Rs in Low Temperature Superconductors at Ultra High Frequencies.

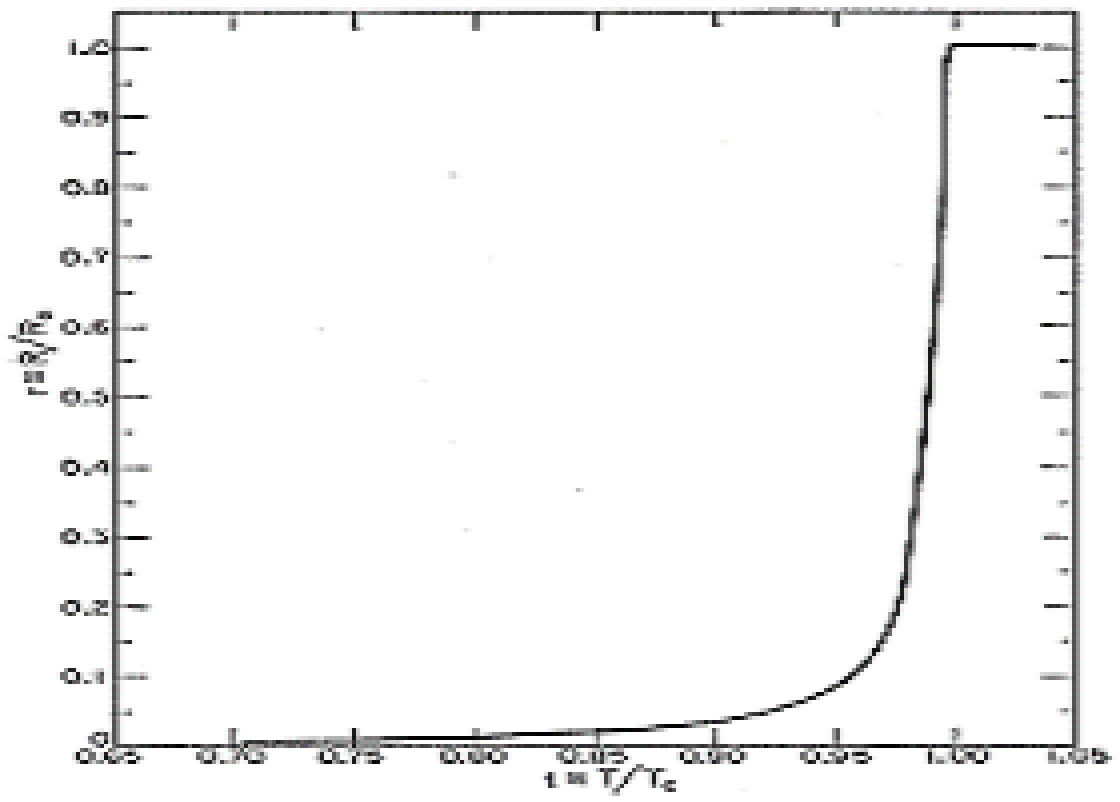

**Fig. 43.** Surface resistance ratio r=Rs/Rn as a function of reduced temperature t=T/Tc for superconducting aluminum at wavelength of 25cm (after [59]).

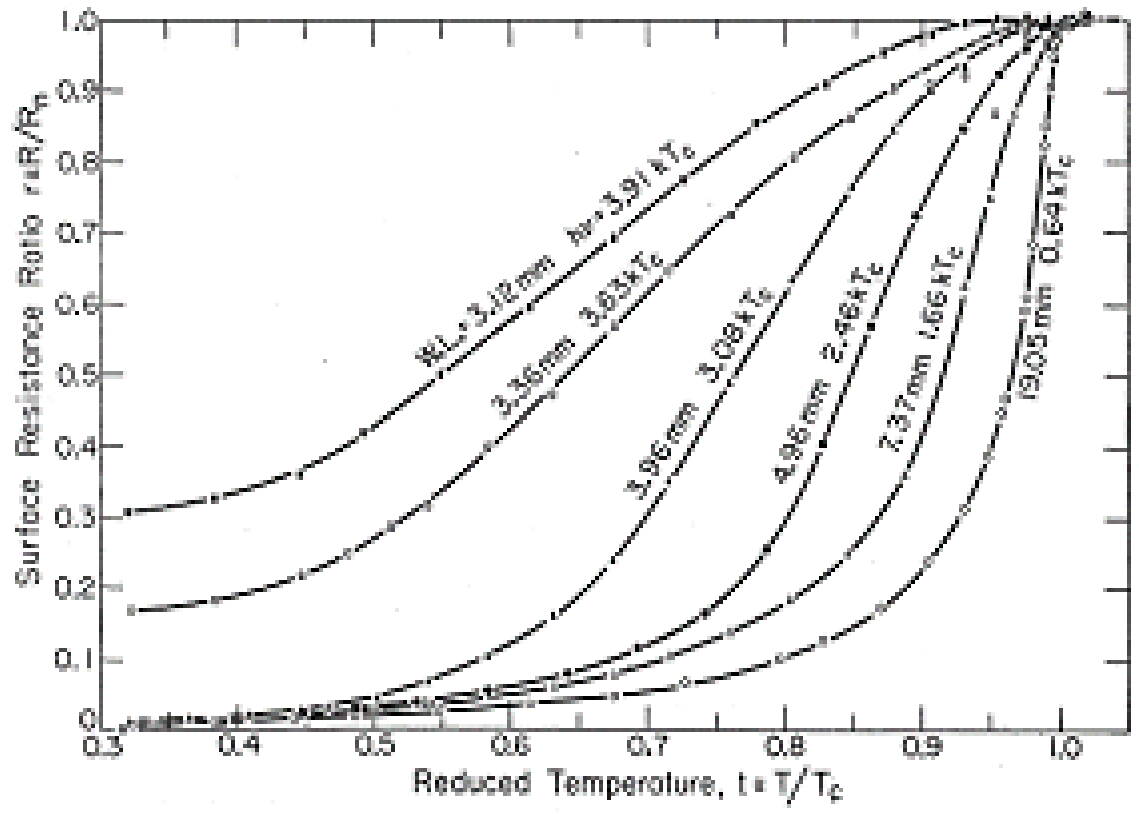

**Fig. 44.** Measured values of the surface resistance ratio r of superconducting aluminum as a function of the reduced temperature t at several representative wavelengths. The wavelengths and corresponding photon energies are indicated on curves (after [60]).

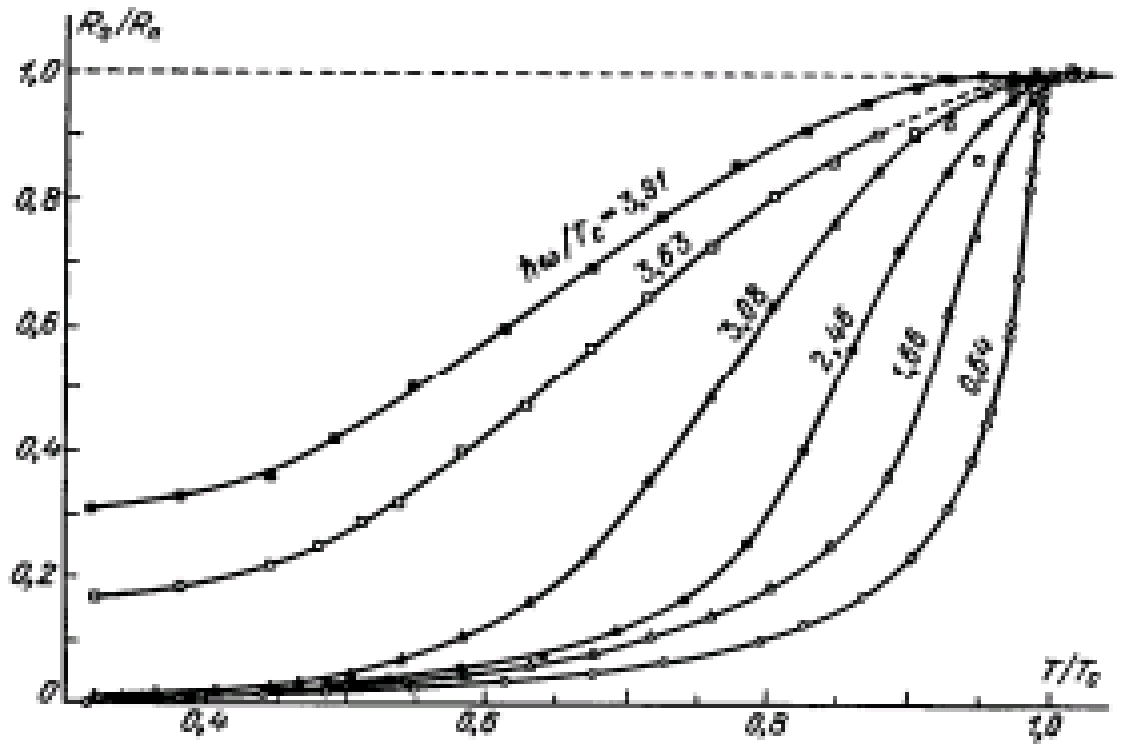

**Fig. 45.** Dependence surface resistance ratio r=Rs/Rn as a function of reduced temperature t=T/Tc for superconducting aluminum at several wavelengths in Abrikosov's interpretation (after [60, 61]).

The initial research on nonlinear effects in conventional superconductors was started in the beginning of the 1950s. Faber, Pippard [59] obtained the experimental dependence of surface resistance ratio r=Rs/Rn as a function of reduced temperature t=T/Tc for superconducting aluminum at wavelength of 25cm in London, U.K. in 1955 in Fig. 32. Biondi, Garfunkel [60] measured the temperature dependence of energy absorbed by superconducting aluminum, and found the surface resistance $R_S$ of superconducting aluminum in the range of wavelengths of 20-3mm in Pittsburgh, U.S.A. in 1959 in Fig 33.

# Nonlinear Microwave Properties of High Temperature Superconductors.

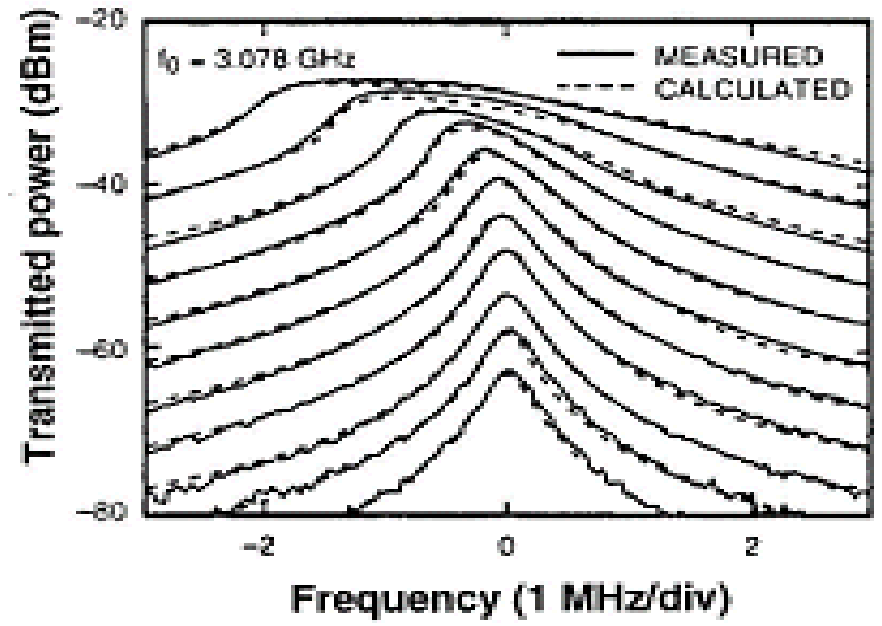

**Fig. 46.** Transmitted microwave power vs. frequency in $YBa_2Cu_3O_{7-\delta}$ microwave resonator as a function of input microwave power. The maximum power is +30dBm, and the curves are in 5dBm steps. The frequency is 3GHz, and the temperature is 77K (after[16]).

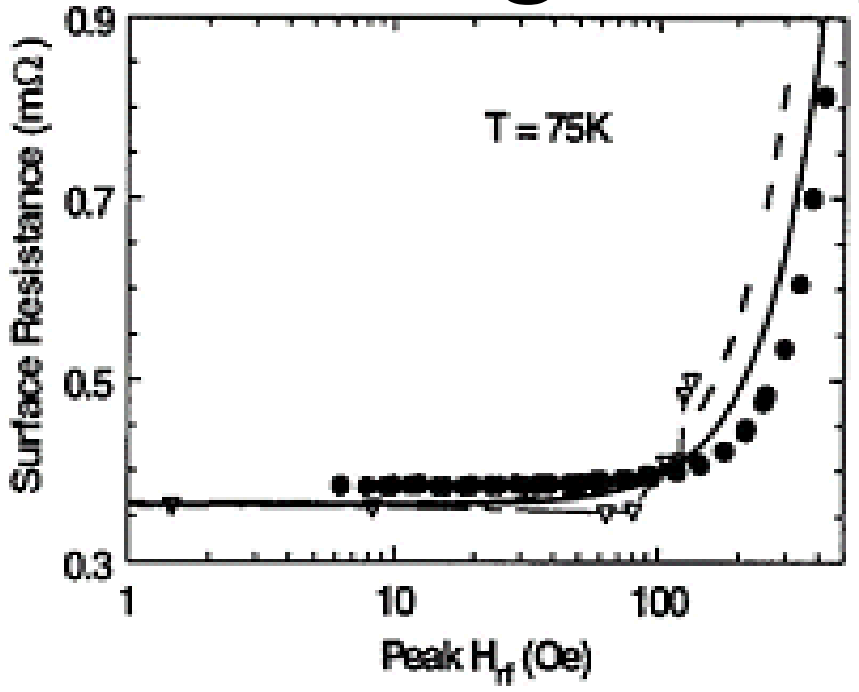

**Fig. 47.** Surface resistance $R_S$ vs. peak magnetic field Hrf in $YBa_2Cu_3O_{7-\delta}$ films at 1.5GHz at 75K. Points are measured data. Solid and dashed lines are calculated from models created for the stripline and dielectric resonators respectively (after [16]).

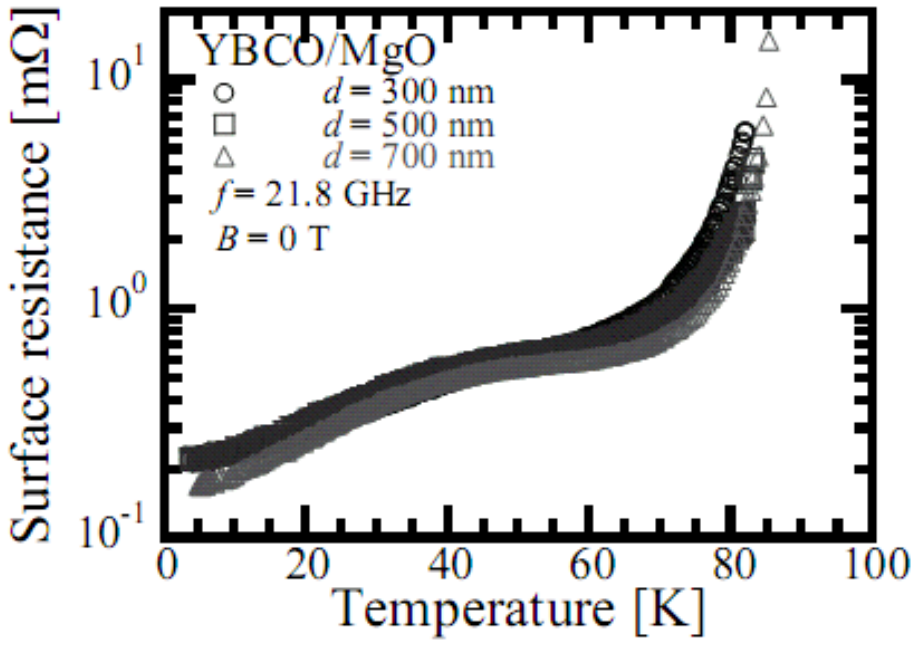

**Fig. 48.** Temperature dependence of Rs of YBCO thin films with thicknesses of 300, 500 and 700 nm in a zero magnetic field at 21.8GHz (after[62]).

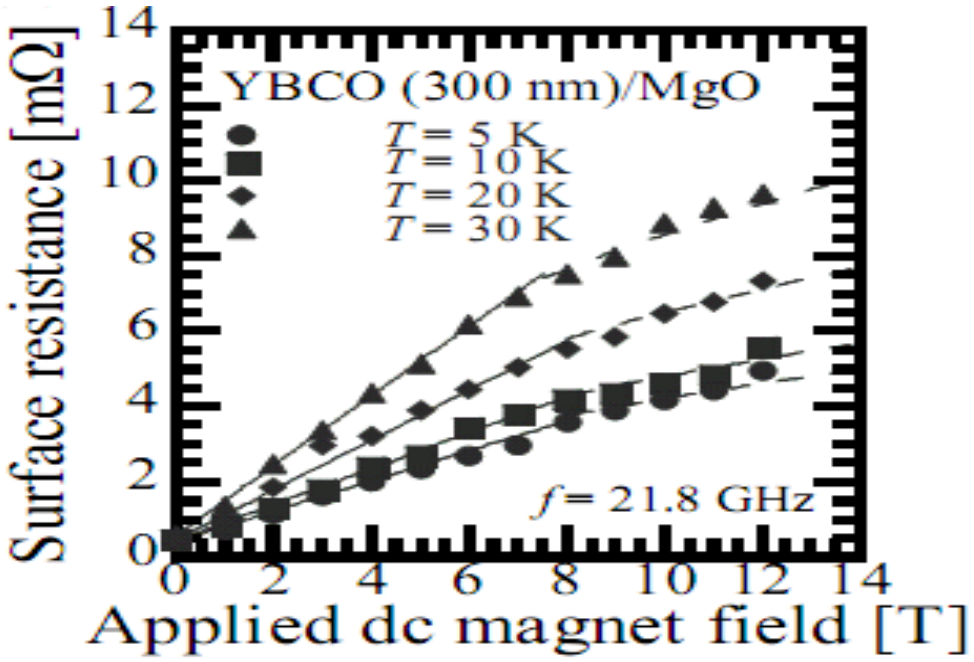

**Fig. 49.** The dc magnetic field dependence of Rs at 5, 10, 20 and 30 K for 300 nm-thick YBCO thin film (after [62]).

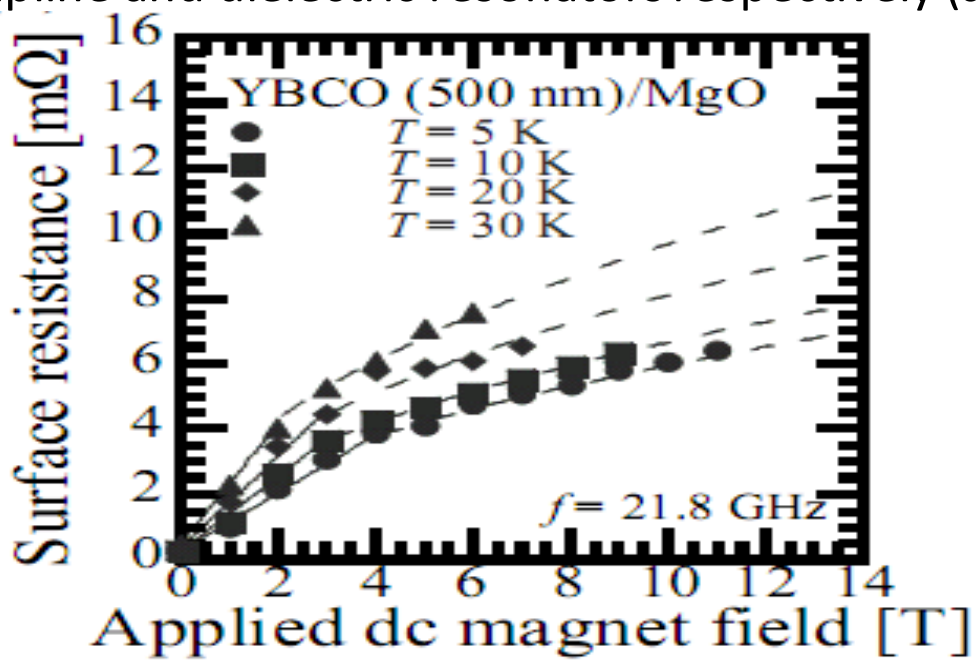

**Fig. 50.** The dc magnetic field dependence of Rs at 5, 10, 20 and 30 K for 500 nm-thick YBCO thin film (after [62]).

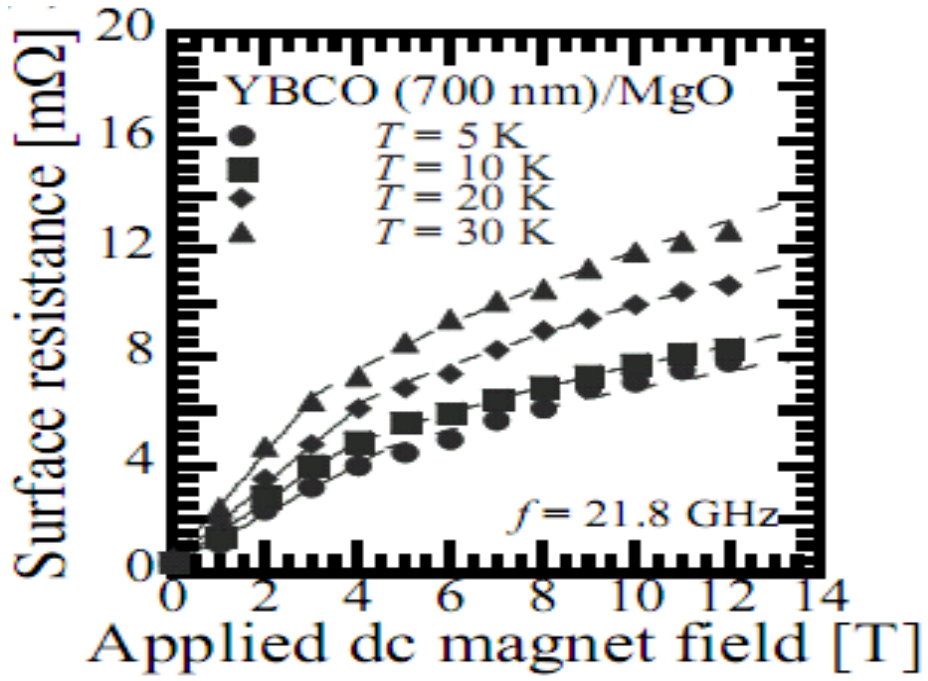

**Fig. 51.** The dc magnetic field dependence of Rs at 5, 10, 20 and 30 K for 700 nm-thick YBCO thin film (after [62]).

# Origins of Nonlinearities in HTS Thin Films at Microwaves.

**The microwave nonlinear properties of HTS films are usually studied by measuring [58]:**

1. Microwave power dependence of surface impedance $Z_S$ = Rs +i Xs;
2. Harmonic generation;
2. Intermodulation distortion (IMD) using both resonant and non-resonant transmission line techniques.

**Main sources of nonlinearities in HTS thin films at microwaves [16]:**

**1. Extrinsic nonlinearities, originated by extrinsic effects in superconductors at microwaves [16]:**

a) grain boundaries nonlinearity;
b) twin boundaries nonlinearity;
c) dislocation nonlinearity;
d) normal phase inclusion nonlinearity;
e) impurities nonlinearity;
f) oxygen doping nonlinearity;
g) possible patterning nonlinearity.

**2. Intrinsic nonlinearities, originated by intrinsic effects in superconductors at microwaves [16]:**

a) Cooper electron pair breaking nonlinearity.

**Main sources of nonlinearities in HTS thin films at microwaves [63]:**

1. Intrinsic nonlinearity due to pair breaking;
2. Weakly coupled grain;
3. Josephson magnetic vortices in weak links;
4. Abricosov magnetic vortices penetration into the grains;
5. Non-Uniform heating;
6. Heating of weak links.

# S-Shape Dependences of Surface Resistance on Magnetic Field $R_s(H)$ in HTS Thin Films at Microwaves.

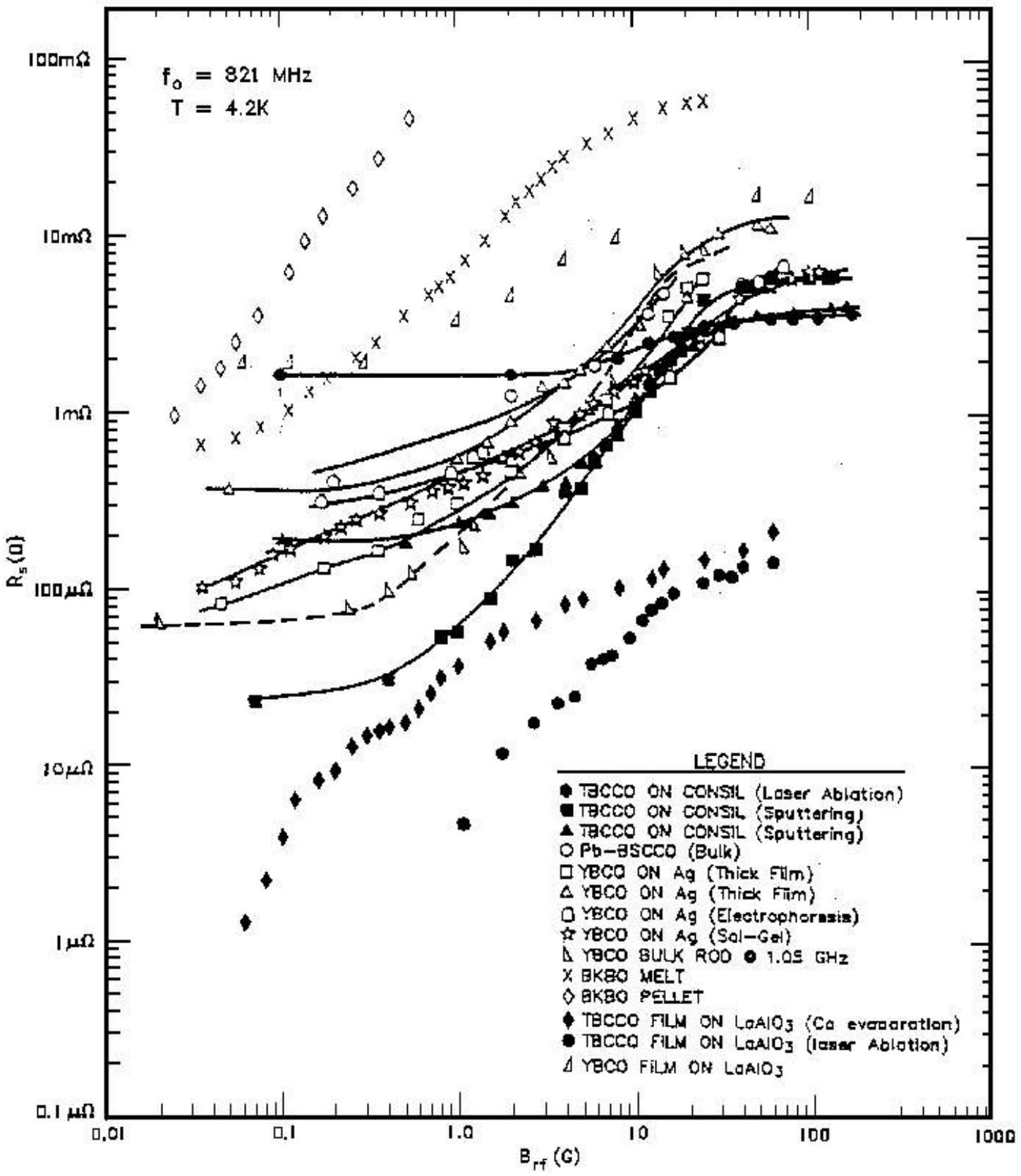

**Fig. 52.** Surface resistance $R_S$ vs. RF magnetic field $B_{rf}$ at the frequency of 821MHz and the temperature of 4.2K for a variety of HTS samples (after [94]).

**The dependence of surface resistance of superconducting film on the electromagnetic field $R_S(H_{rf})$ at ultra high frequencies can be characterised by the four regimes:**

**1. Linear dependence in small magnetic fields:** it is well described by the coupled-grain model [95];

**2. Weak nonlinear dependence in medium magnetic fields:** it originates because of the presence of defects such as the weak links at grain boundaries in crystal lattice of superconductor. This regime can be described by the extended coupled-grain model, which takes into the account the nonlinear inductance of the weak links [96];

**3. Strong nonlinear dependence in strong magnetic fields:** it originates because of the Abricosov magnetic vortices generation phenomena in strong magnetic field [28];

**4. Breakdown regime in highest saturated magnetic fields:** it originates because of the heating effect and the formation of normal-state domains in a superconductor [94].

**NOTE: It is assumed that, in the high quality HTS films, the nonlinearities, originated near the higher critical field $H_{c1}$, will mainly contribute to the nonlinear behaviour of a superconductor at microwaves.**

# Lumped Element Modeling of Nonlinear Properties of High Temperature Superconductors in Microwave Resonant Circuits.

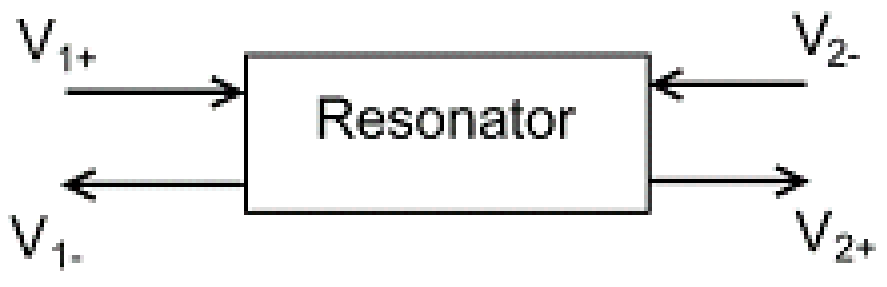

**Fig. 53**. Electrical scheme of two ports network for S-parameters definition (after [64]).

$$\begin{bmatrix} V_{1-} \\ V_{2-} \end{bmatrix} = \begin{bmatrix} S_{11} & S_{12} \\ S_{21} & S_{22} \end{bmatrix} \begin{bmatrix} V_{1+} \\ V_{2+} \end{bmatrix},$$

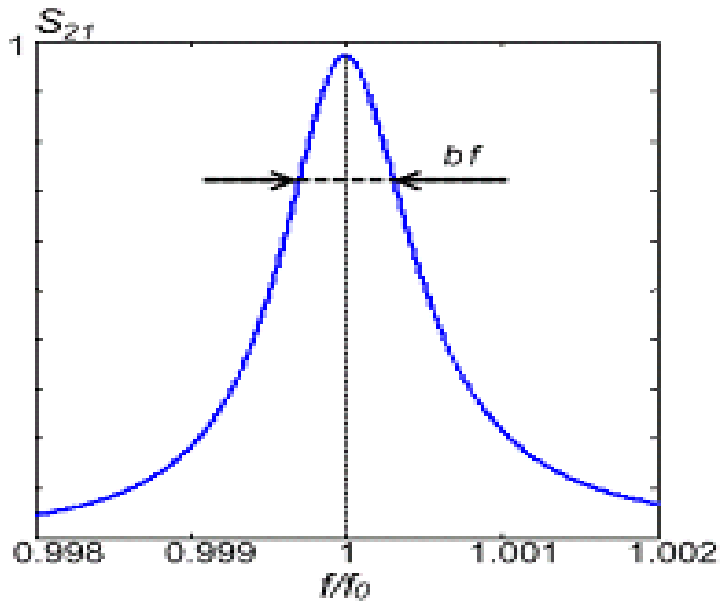

**Fig. 54**. Linear Lorenztian resonance curve with resonance frequency $f_0$ and band width bf.

**Lumped Element Equivalent Circuits for Representation of Superconductors at Microwaves.**

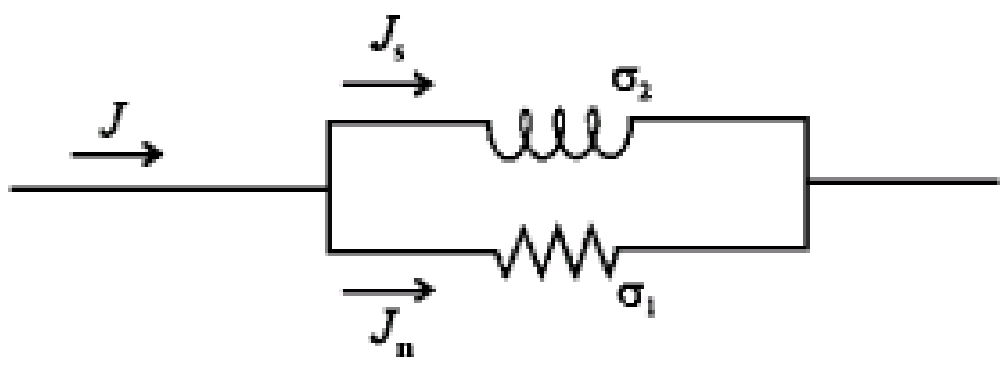

$$\sigma_S = \sigma_1 - j\sigma_2,$$

**Fig. 55**. Equivalent circuit representation of superconductor conductivity σs in two fluid model (after [65]).

**Lumped Element Equivalent RLC Circuits for Modeling of Microwave Resonators.**

The theorem of microwave theory: Any resonator in proximity to its resonant frequency can be represented as a lumped elements circuit with appropriate values of R, L and C.

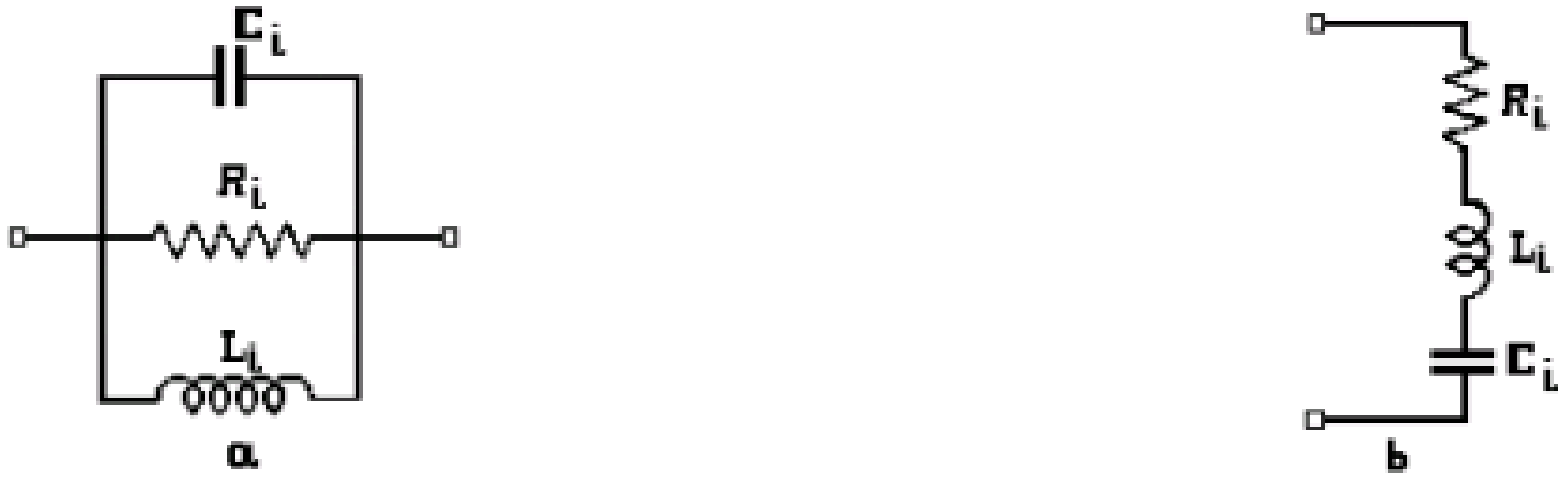

**Fig. 56** . Parallel (**a**) and series (**b**) equivalent lumped element networks for representation of microwave resonators taking into account dissipation effects (after [66]).

# r- Parameter for Nonlinear Lumped Element Models Analysis.

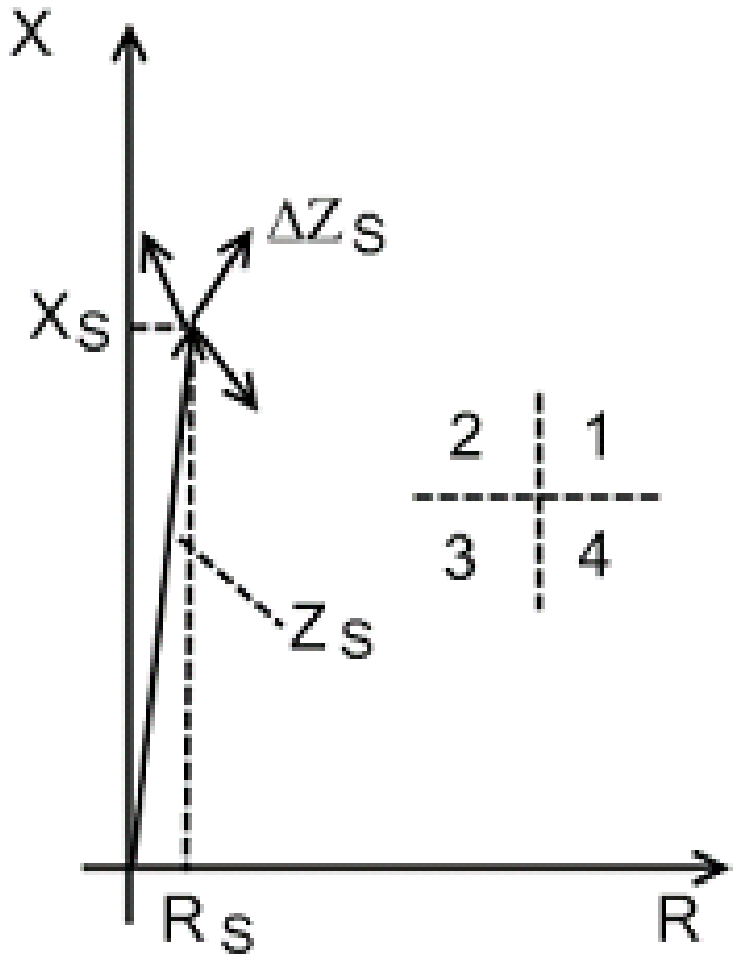

**Fig. 57.** Change of vector of impedance $Z_S$ on magnitude $\Delta Z_S$ at action of external parameters (temperature T, microwave power P) in superconductor (after [68]).

**The dimensionless parameter $r_H$ is defined as the ratio [66, 67]:**

$$r_H = \Delta X_S(P)/\Delta R_S(P).$$

**The parameter $r_H$ is equal to the differential quality factor $Q_{Sdif}$ on superconductor surface:**

$$r_H = Q_{Sdif}.$$

**The inverted parameter $r_G$ was introduced as the ratio [67]:**

$$r_G = \Delta R_S(P)/\Delta X_S(P).$$

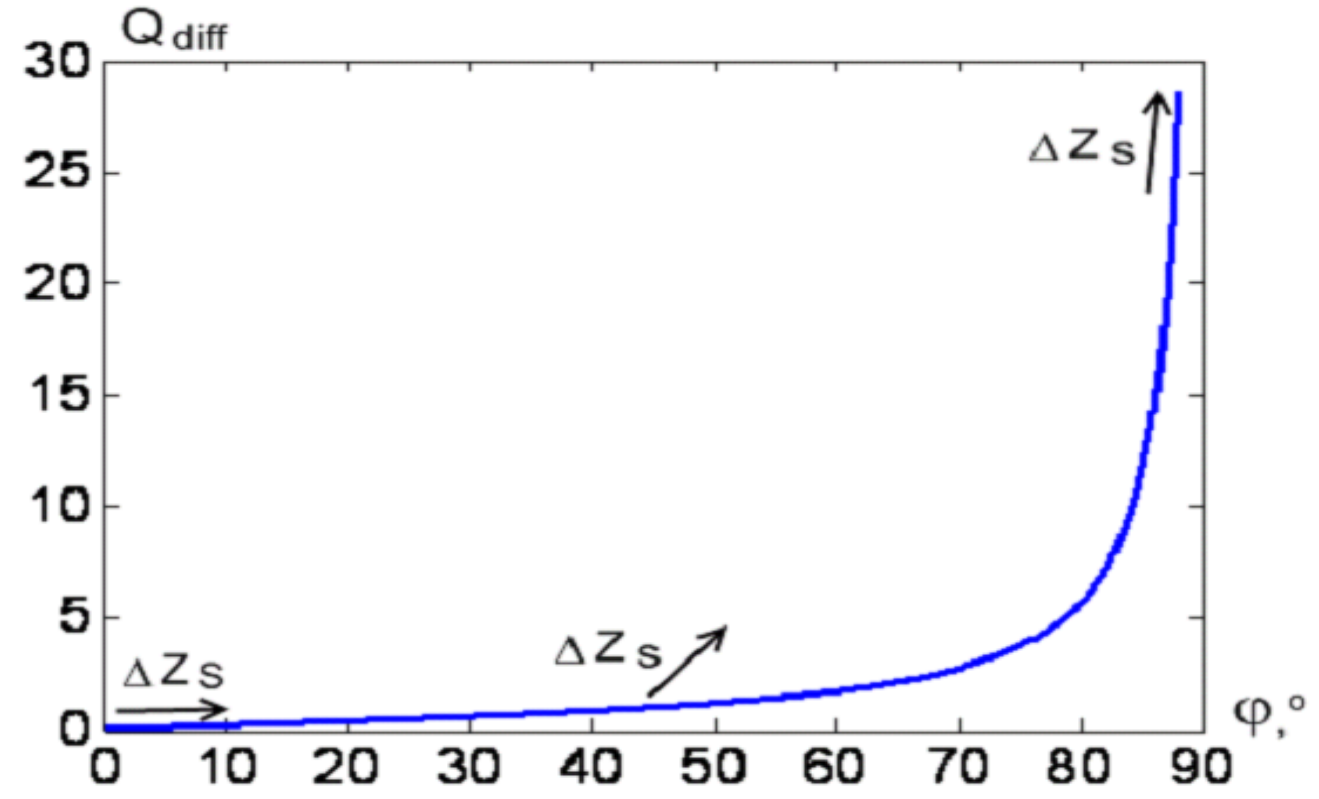

**Fig. 58**. Dependence of magnitude of differential quality factor $Q_{Sdiff}$ on orientation of vector $\Delta Z_S$ in R-X space ($\phi$ is the angle between vector $\Delta Z_S$ and axe R) (after [68]).

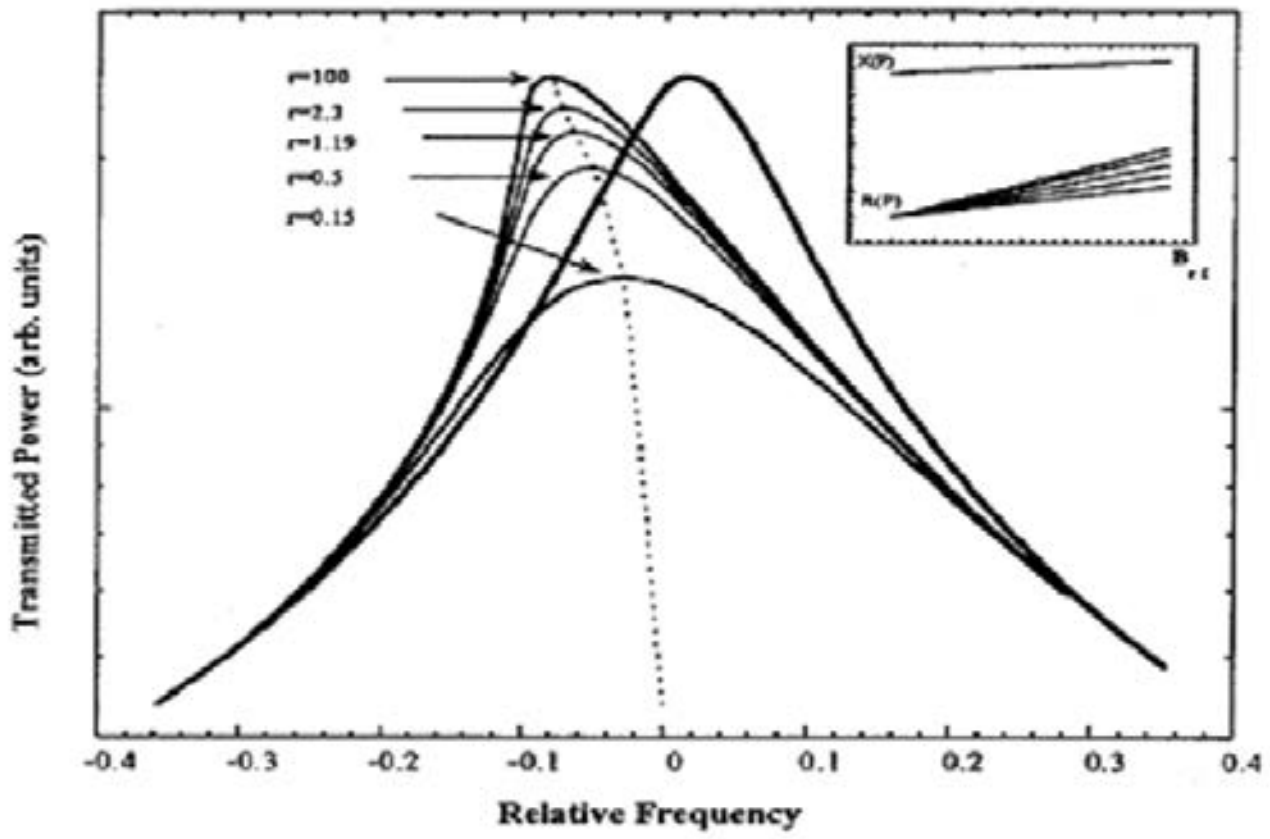

**Fig. 59**. Simulation of transmitted power P vs. frequency f for different r-parameters in microwave resonator (after [69]).

**Note: The parameter r is not an universal parameter for description of nonlinear phenomena, because the parameter r is only a differential quality factor, which formally describes the direction of movement of vector of impedance $Z_S$ in R-X space, and it is not directly connected with the physical mechanisms, which are the root causes of nonlinearities in superconductors.**

# Superconductor in a Hakki-Coleman Dielectric Resonator at Microwaves.

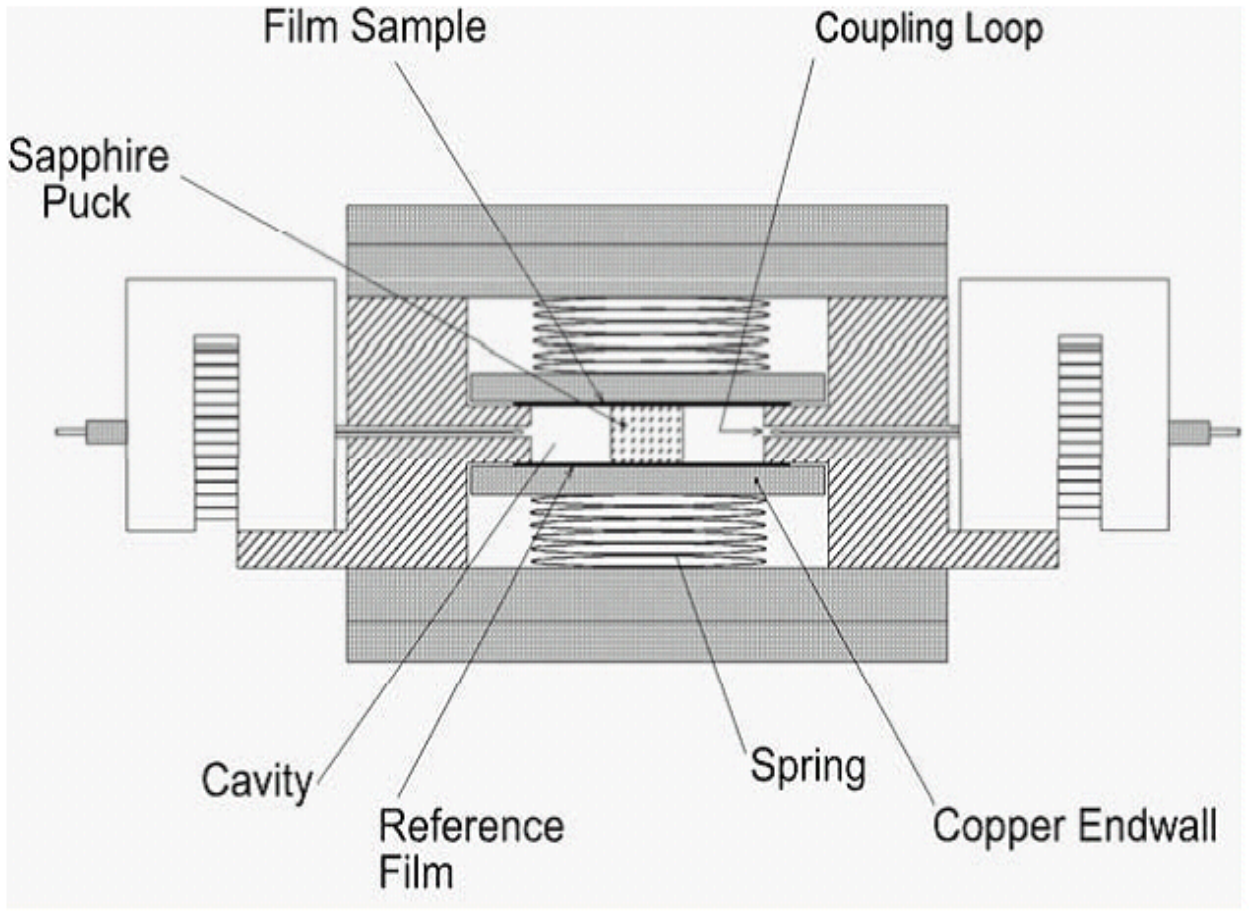

**Fig. 60.** Hakki-Coleman dielectric resonator [70] with HTS thin films, where the reference and sample films are superconductors (after [71]).

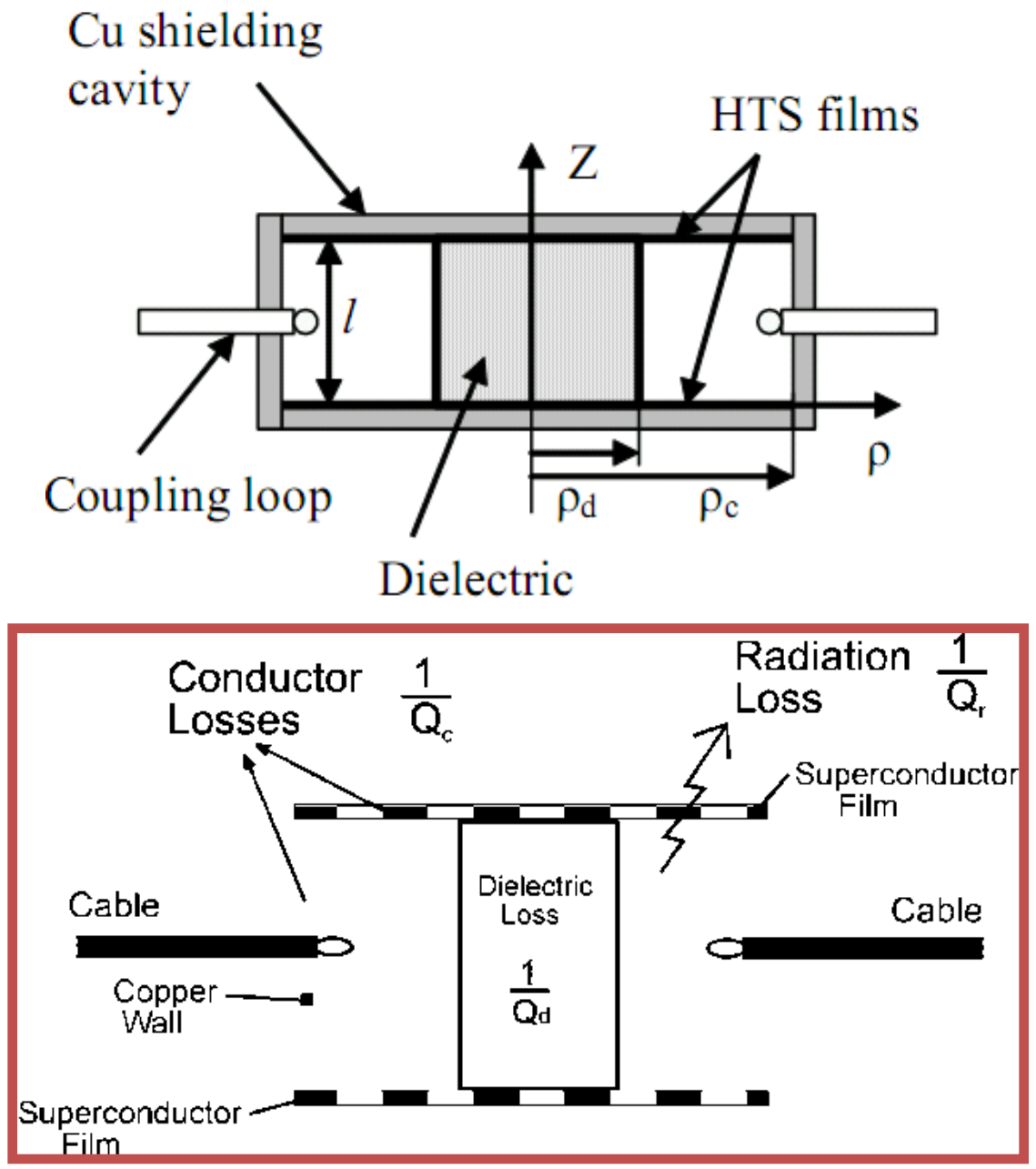

**Fig. 61.** Dielectric dimensions and orientation of axes in dielectric resonators (after [71]).

**The surface resistance Rs of HTS thin film:**

$$R_S = A_S\left\{\frac{1}{Q_0} - \frac{R_m}{A_m} - p_e \tan\delta\right\}$$

**where As and Am are the geometrical factors, Rm is the surface resistance of metal, $p_e$ is the constant for dielectric resonators in trapped state is approximately equal to one, $Q_0$ is the unloaded quality factor, which can be determined from the measured loaded $Q_L$-factor and two port coupling coefficients β1 and β2.**

$$Q_0 = Q_L(1 + \beta_1 + \beta_2)$$

# Electromagnetic Waves in a Hakki-Coleman Dielectric Resonator.

**These formulas represent a complete set of equations for accurate characterisation of electromagnetic properties of HTS thin films in a dielectric resonator at microwaves:**

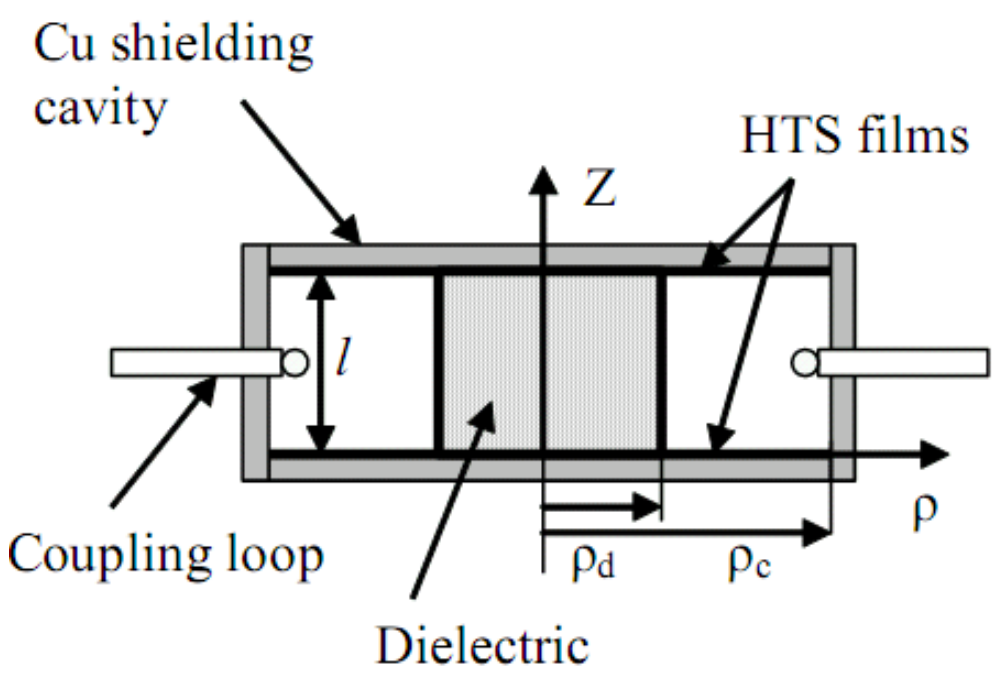

**Fig. 62.** Dielectric dimensions and orientation of axes in dielectric resonators (after [71]).

For $\rho < \rho_d$ in eq. (5.12)

$$E_{\rho 1} = 0$$
$$E_{\varphi 1}(\rho, z) = -jA\left(\frac{\omega\mu_0}{k_1}\right)J_1(k_1\rho)\sin(\beta z)$$
$$E_{z1} = 0 \qquad (5.12)$$
$$H_{\rho 1}(\rho, z) = -A\left(\frac{\beta}{k_1}\right)J_1(k_1\rho)\cos(\beta z)$$
$$H_{\varphi 1} = 0$$
$$H_{z1}(\rho, z) = AJ_0(k_1\rho)\sin(\beta z)$$

For $\rho > \rho_d$ in eq. (5.13)

$$E_{\rho 2} = 0$$
$$E_{\varphi 2}(\rho, z) = jB\left(\frac{\omega\mu_0}{k_2}\right)F_1(k_2\rho)\sin(\beta z)$$
$$E_{z2} = 0 \qquad (5.13)$$
$$H_{\rho 2}(\rho, z) = B\left(\frac{\beta}{k_2}\right)F_1(k_2\rho)\cos(\beta z)$$
$$H_{\varphi 2} = 0$$
$$H_{z2}(\rho, z) = BF_0(k_2\rho)\sin(\beta z)$$

where in eq. 5.14)

$$F_0(k_2\rho) = I_0(k_2\rho) + K_0(k_2\rho)\frac{I_1(k_2\rho_C)}{K_1(k_2\rho_C)}$$
$$F_1(k_2\rho) = -I_1(k_2\rho) + K_1(k_2\rho)\frac{I_1(k_2\rho_C)}{K_1(k_2\rho_C)}, \qquad (5.14)$$

where, $J_m$ is the $m$-th order *Bessel function* of the first kind, $I_m$ and $K_m$ are the $m$-th order modified *Bessel and Hankel functions* respectively [17], $\beta = \pi/l$, $l$ is the height of the cavity.

The wave number in the dielectric can be expressed as

$$k_1^2 = k_0^2 - \beta^2$$

and outside of the cavity as

$$k_2^2 = \beta^2 - k_0^2,$$

where $k_0 = \omega / c$, $c$ is the velocity of electromagnetic wave in free space.

Components of the electric and magnetic fields depend on time as $e^{j\omega t}$ and are shifted in phase by $\pi/2$. The fields in the cylindrical resonator are related by the *Maxwell equations* in eq. (5.15). Taking into the account eqs. (5.12) and (5.13)

$$rotE_\varphi = -\frac{\partial H_{(\rho,z)}}{\partial t} \qquad (5.15)$$

The density of shifting currents can be expressed as in eq. (5.16)

$$J_{\varphi,disp} = -\frac{\partial D_\varphi}{\partial t} \qquad (5.16)$$

where, $D_\varphi = \varepsilon_d E_\varphi$.

# Parallel and Series Equivalent Lumped Elements Models for Representation of a Superconductor in Hakki-Coleman Dielectric Resonator at Microwaves.

## Parallel and Series Equivalent Lumped Elements Models:

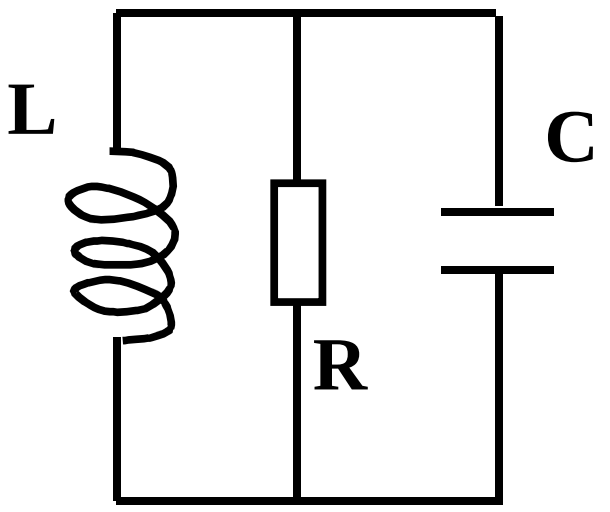

$$Q_{par} = \frac{R}{\omega L}$$

Fig. 63. Parallel lumped elements model.

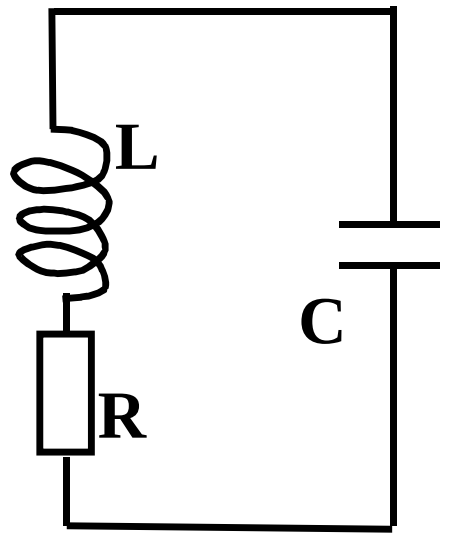

$$Q_{ser} = \frac{\omega L}{R}$$

Fig. 64. Series lumped elements model.

## Density of Current in Two-Fluid Analogy:

$$J = J_n + J_S$$

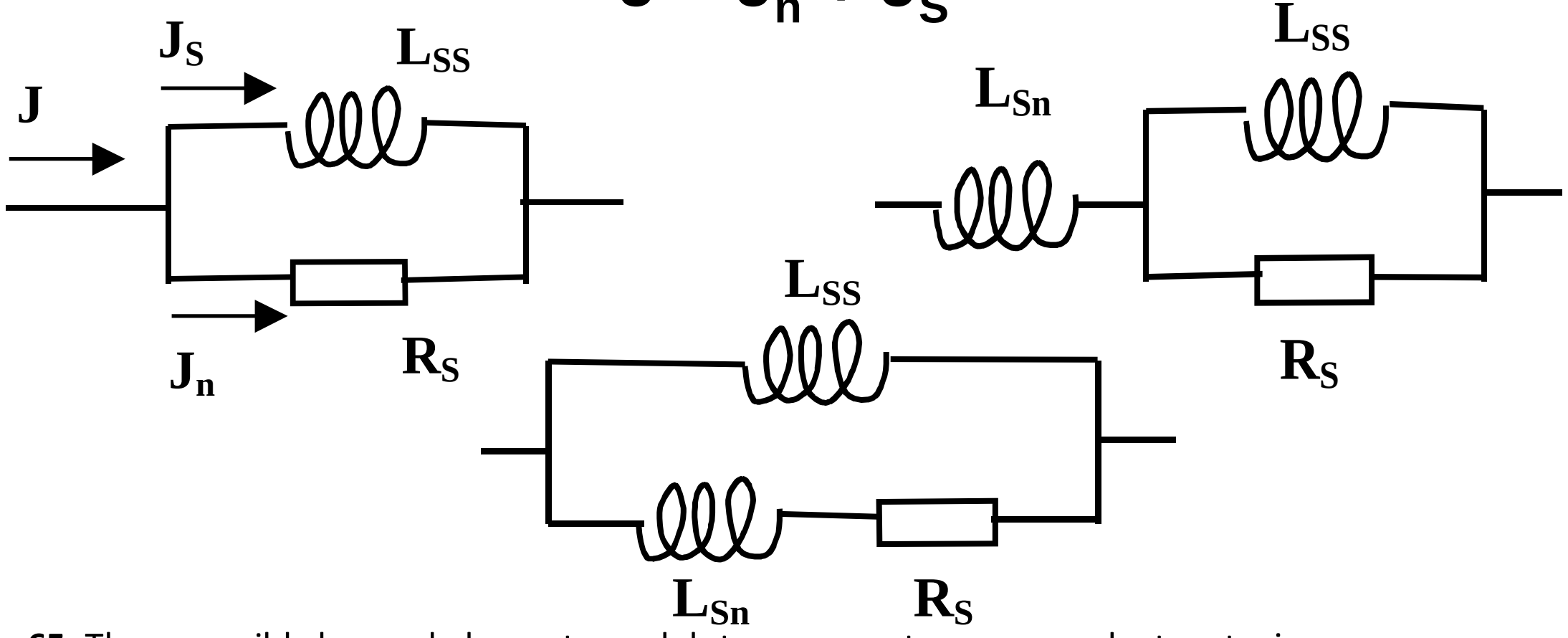

**Fig. 65.** Three possible lumped elements models to represent a superconductor at microwaves, where $L_{SS}$ is the magnetic induction to model the superconducting current.

**Note: The superconducting materials have very small values of surface resistance $R_S$, and as a result, the superconductors need to be placed in the position, where the strength of magnetic field reaches its maximum in a microwave resonator. In the case of superconductors, it is more convenient to represent a microwave resonator-sample system in the form of series resonant equivalent circuit with the small total resistance value. The quality factor Q of the microwave resonator-superconductor system will be higher, if the surface resistance $R_S$ of a superconducting sample is lower [94, 95].**

# Proposed Equivalent Lumped Element Models for Representation of a Superconductor in Hakki-Coleman Dielectric Resonator at Microwaves.

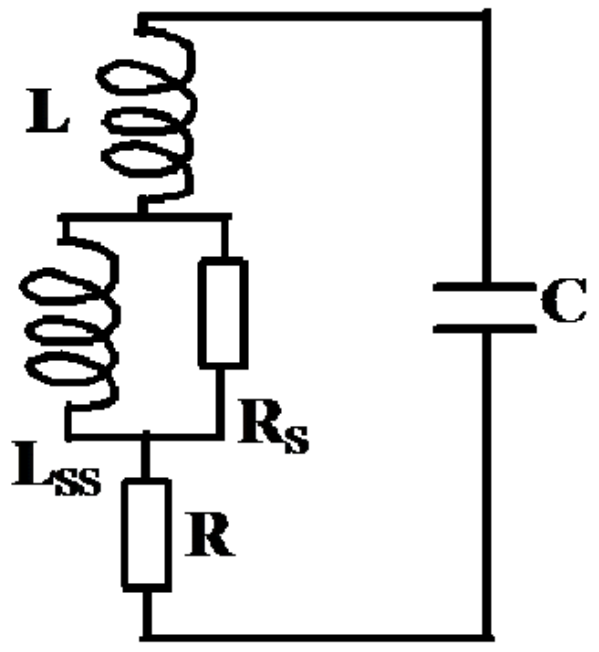

**Fig. 66.** Equivalent lumped element model of resonator circuit, which represents a superconductor in a dielectric resonator.

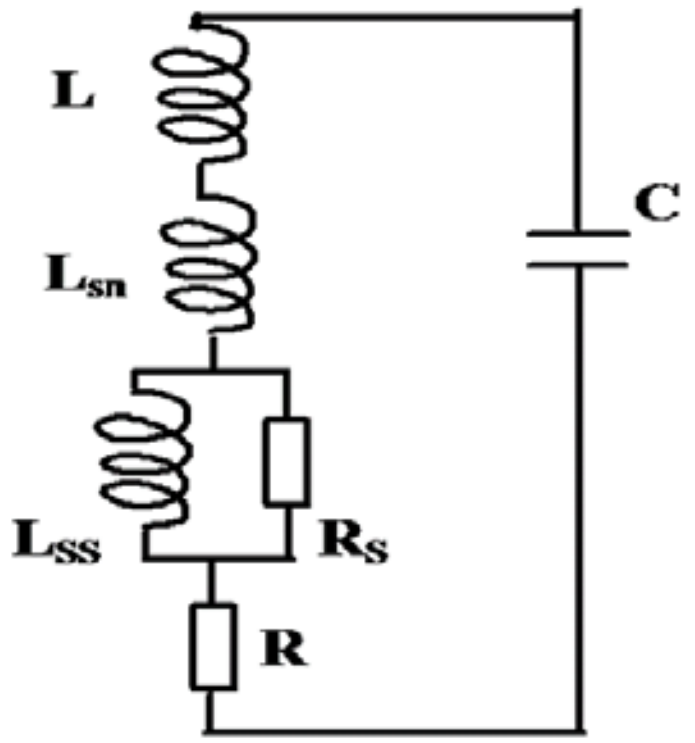

**Fig. 67.** Equivalent lumped element model of resonator circuit, which represents a superconductor in series with its inductance L in a dielectric resonator (**L-σ model**).

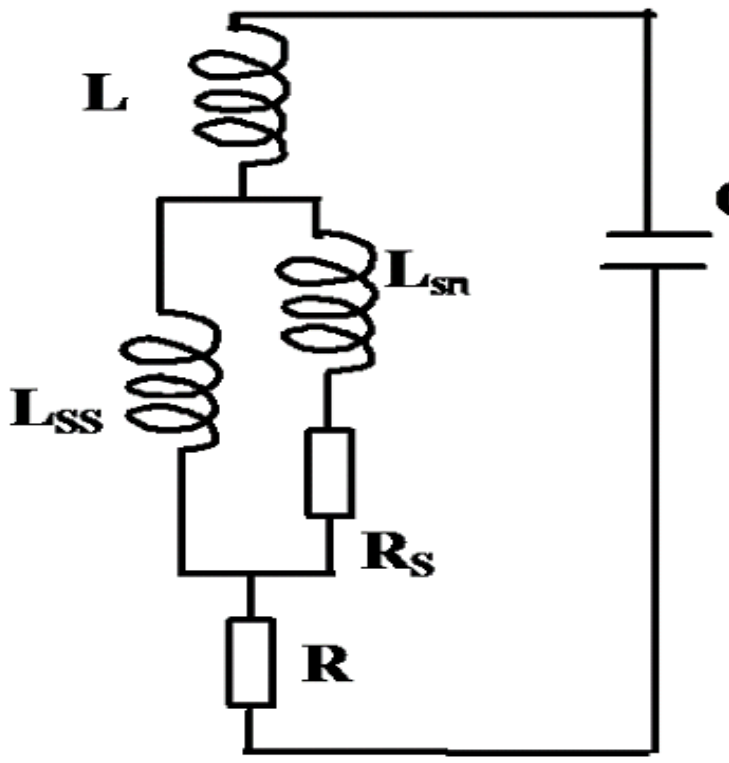

**Fig. 68.** Equivalent lumped element model of resonator circuit, which represents a superconductor in a normal part of its impedance Z in series with its inductance L in a dielectric resonator (**L-Z model**).

**Transmitted RF power for the L-σ model for the circuit in Fig. 67:**

$$P=\frac{1}{2}\frac{V^2\omega^2C^2(R\omega^2L_{SS}^2+RR_S^2+L_{SS}^2\omega^2R_S)}{R_S^2-2\omega^2(L+L_{Sn})CR_S^2-2R_S^2\omega^2L_{SS}C+R^2\omega^4C^2L_{SS}^2+\omega^4(L+L_{Sn})^2C^2R_S^2+R_S^2\omega^4L_{SS}^2C^2-2\omega^4L_{SS}^2LC+\omega^2L_{SS}^2+2R\omega^4C^2L_{SS}^2R_S+R^2\omega^2C^2R_S^2+\omega^6(L+L_{Sn})^2C^2L_{SS}^2+2\omega^4(L+L_{Sn})C^2R_S^2L_{SS}} \quad (5.19)$$

**Transmitted RF power for the L-Z model for the circuit in Fig. 68:**

$$P=\frac{1}{2}\frac{\omega^2C^2V^2(L_{SS}^2\omega^2R_S+RR_S^2+R\omega^2L_{SS}^2+2R\omega^2L_{SS}L_{Sn}+R\omega^2L_{Sn}^2)}{2\omega^2L_{SS}L_{Sn}+R^2\omega^4C^2L_{SS}^2+R_S^2+\omega^2L_{SS}+\omega^2L_{Sn}^2+2\omega^6LC^2L_{Sn}^2L_{SS}+\omega^4L_{SS}^2C^2R_S^2+2\omega^4LC^2R_S^2L_{SS}+2\omega^6L^2C^2L_{SS}L_{Sn}+2\omega^6LC^2L_{SS}^2L_{Sn}+2R\omega^4C^2L_{SS}^2R_S+R^2\omega^4C^2L_{Sn}^2+\omega^4L^2C^2R_S^2-4\omega^4L_{SS}L_{Sn}CL_{Sn}+2R^2\omega^4C^2L_{SS}L_{Sn}-2\omega^4L_{SS}^2LC-2\omega^4L_{SS}^2CL_{Sn}-2\omega^4L_{Sn}^2LC-2\omega^4L_{Sn}^2L_{SS}C+R^2\omega^2C^2R_S^2+\omega^6L^2C^2L_{SS}^2+\omega^6L^2C^2L_{Sn}^2+\omega^6L_{SS}^2C^2L_{Sn}^2-2R_S^2\omega^2LC-2R_S^2\omega^2L_{SS}C} \quad (5.20)$$

# Microwave Power vs. Electromagnetic Field Dependence of Lumped Element Model Parameters.

**Microwave Power vs. Electromagnetic Field Dependence of Lumped Element Model Parameters:**

In order to model the nonlinear behaviour of HTS materials, the following elements $R_{ss}$, $L_{ss}$, and $L_{sn}$ of equivalent lumped element model circuits need to be expressed as a function of RF magnetic field $H_{rf}$, or microwave power P, where $P \propto H_{rf2}$. There have been various mathematical dependencies proposed to describe the magnetic field dependence in terms of:

**linear** [75, 76],

**quadratic** [75, 76] and,

**exponential** representations [77, 78].

In this research, all three types of RF field dependences are used and analyzed to define $R_s$, $L_{ss}$, and $L_{sn}$. The expressions for each of the dependences are presented in eqs. (5.21 – 5.23).

***Linear dependence*** in eq. (5.21)

$$R_S = R_{S0}(1 + \rho H(i-1));$$
$$L_{SS} = L_{SS0}(1 + lH(i-1));$$
$$L_{Sn} = L_{Sn0}(1 + lH(i-1));\quad (5.21)$$

***Quadratic dependence*** in eq. (5.22)

$$R_S = R_{S0}(1 + \rho_1 H(i-1) + \rho_2 H^2(i-1));$$
$$L_{SS} = L_{SS0}(1 + l_1 H(i-1) + l_2 H^2(i-1));$$
$$L_{Sn} = L_{Sn0}(1 + l_1 H(i-1) + l_2 H^2(i-1));\quad (5.22)$$

***Exponential dependence*** in eq. (5.23)

$$R_S = R_{S0}(1 + a\exp(bH(i-1)));$$
$$L_{SS} = L_{SS0}(1 + c\exp(dH(i-1)));$$
$$L_{Sn} = L_{Sn0}(1 + c\exp(dH(i-1)));\quad (5.23)$$

where $\rho$, $\rho 1$, $\rho 2$, $l$, $l1$, $l2$, $a$, $b$, $c$, $d$ are fitting constants, $R_{S0}$ is the initial value of surface resistance, $L_{SS0}$ is the initial values of superconducting surface inductance and $L_{Sn0}$ is the initial values of normal metal surface inductance.

# Modeling and Identification of Lumped Element Model Parameters for a Superconducting Hakki-Coleman Dielectric Resonator.

The following system parameters have been applied:

1) Dielectric resonator (based on the *Hakki-Coleman sapphire dielectric design* of 10GHz): $R=3\times10^{-4}\,\Omega$, $C=10^{-12}F$, $L=2.5\times10^{-10}H$.

2) *HTS* film – $YBa_2Cu_3O_7$: $L_{ss0}=L_{sn0}=1.8\times10^{-13}H$ ($\lambda=1.4\times10^{-7}m$), $R_{s0}=5\times10^{-4}\Omega$, $\rho=1.6\times10^{-2}$, $l=1.1\times10^{-3}$, $\rho_1=4.1\times10^{-2}$, $l_1=1.8\times10^{-3}$, $\rho_2=3.2\times10^{-7}$, $l_2=2\times10^{-8}$, $a=9.5\times10^{-1}$, $b=2\times10^{-2}$, $c=3.2\times10^{-2}$, $d=2.5\times10^{-2}$, $f_0=10GHz$.

The ideal curve reflects the case, where components of $R_s$, $L_{ss}$ and $L_{sn}$ do not depend on *RF* magnetic field.

**Dielectric resonator:**

- **R** – the surface resistance of resonator walls at the given frequency and involved geometrical factor.
- **C**, **L** – the parameters calculated from the resonator cavity dimensions, dielectric geometry and resonance frequency.

**Superconductor:**

- $L_{ss0}=L_{sn0}$ (related to $\lambda$), - the values calculated from the characteristic $\lambda$ (penetration depth) for the given superconductor. Practically, it can be found from the resonance frequency change.
- $R_{s0}$ – the surface resistance of the superconductor.

**Fitting Parameters:**

- $\rho$ - proportional coefficient between P and $R_S$, meaning $dR_S/dP$;
- $l$ - proportional coefficient between P and $L_S$, meaning $dL_S/dP$;
- $\rho_1$, $l_1$ - similar coefficients ($dR_S/dP$, $dL_S/dP$), involving quadratic field dependence;
- $\rho_2$, $l_2$- second order - $d^2R_S/dP^2$, $d^2L_S/dP^2$;
- ***a***, ***c*** - amplitudes of exponential dependence;
- ***b***, ***d*** - exponential coefficients;
- $f_0$ - resonance frequency.

Electrical and Computer Engineering JCU

JAMES COOK UNIVERSITY AUSTRALIA

# Modeling and Identification of Lumped Element Model Parameters for a Superconducting Hakki-Coleman Dielectric Resonator.

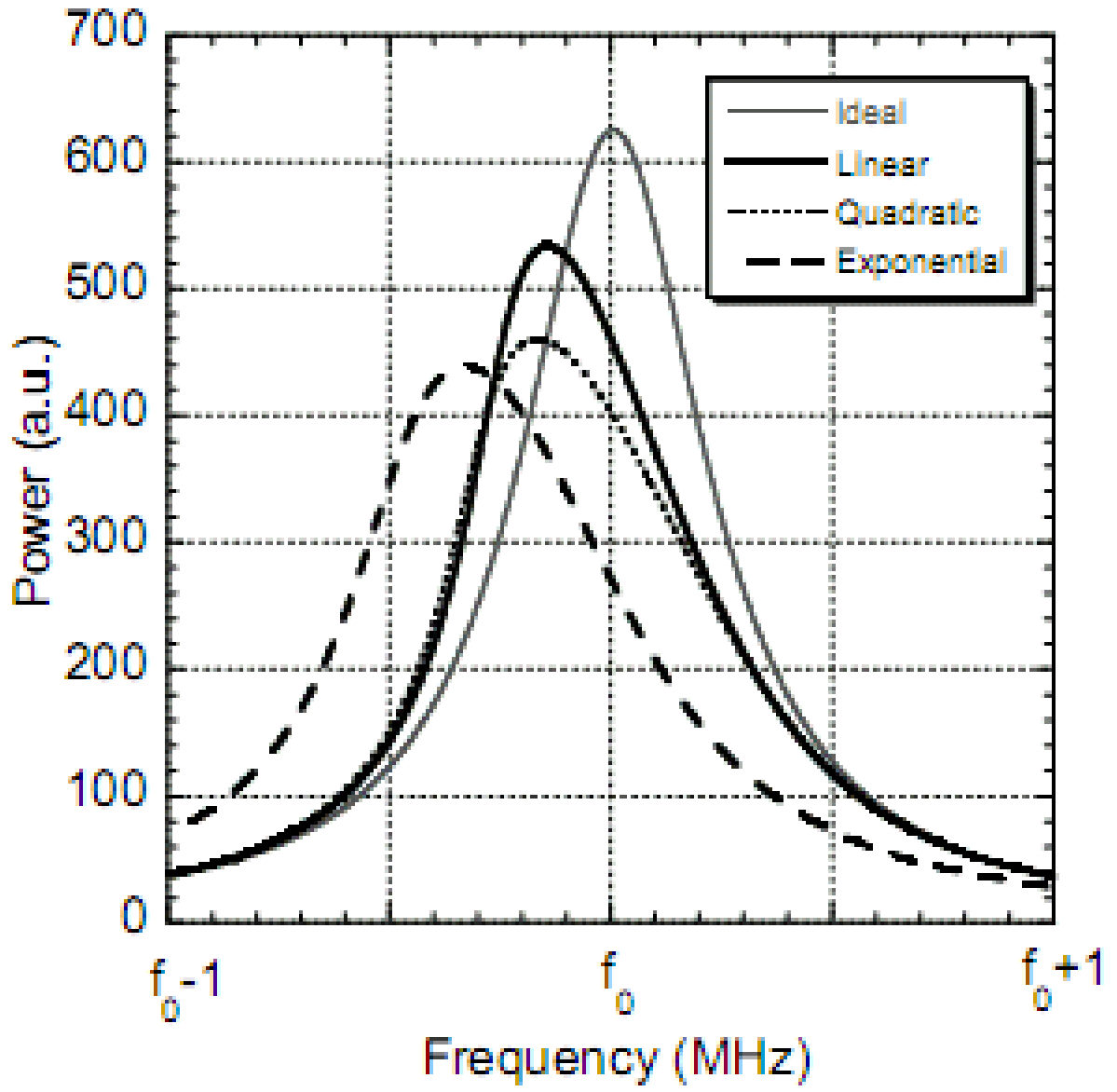

**Fig. 69.** Simulated dependence of microwave power on frequency P(f) for **L-σ model** of network in Fig. 67 ($f_0$ = 10 GHz).

**L-σ model: Ideal, Linear Quadratic, Exponential.**

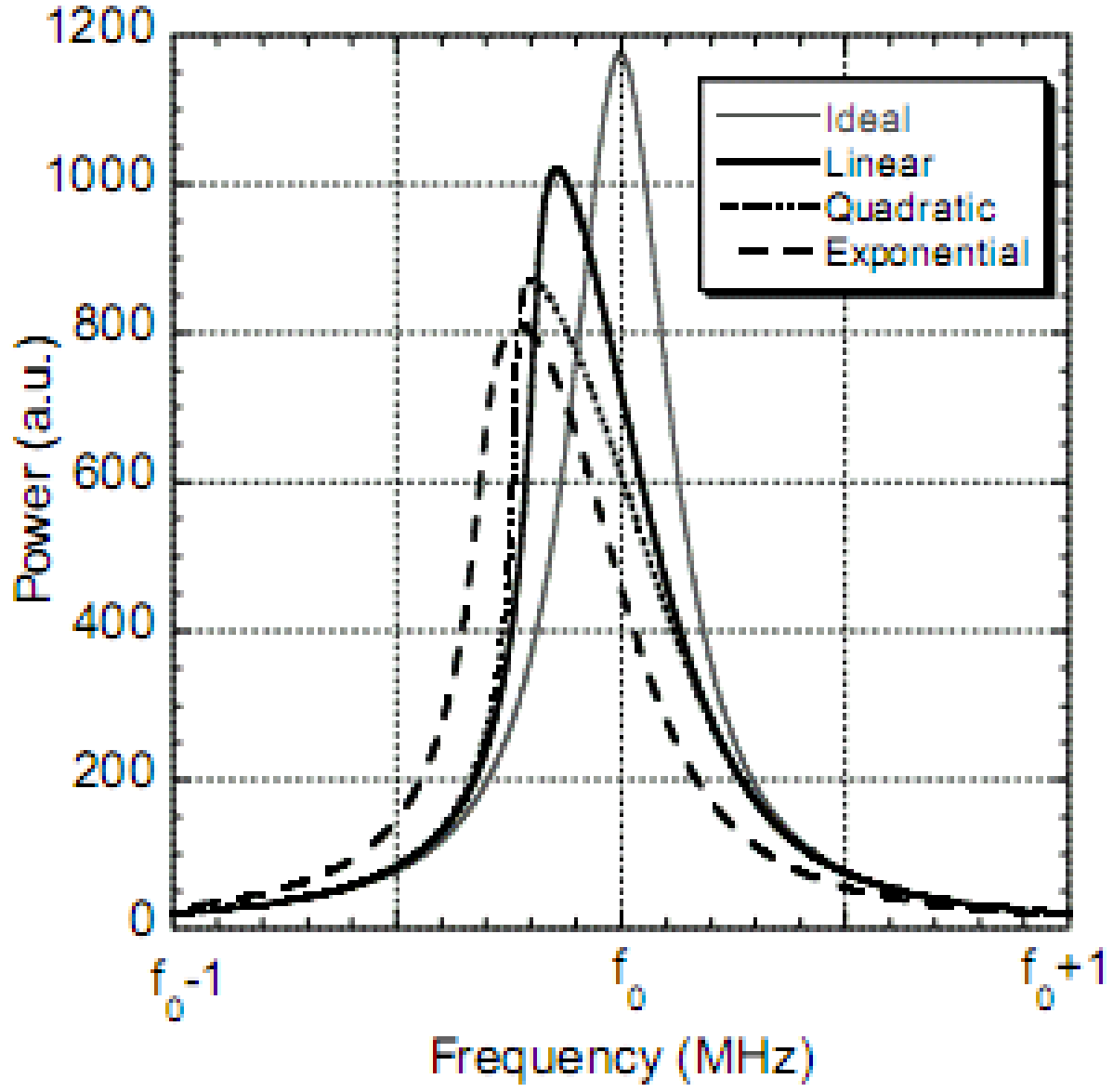

**Fig. 70.** Simulated dependence of microwave power on frequency P(f) for **L-Z model** of network in Fig. 68 ($f_0$ = 10 GHz).

**L-Z model: Ideal, Linear Quadratic, Exponential.**

# Modeling and Identification of Lumped Element Model Parameters for a Superconducting Hakki-Coleman Dielectric Resonator.

In order to verify the fitting parameters of the models in terms of their physical correlations with the simulated responses and to investigate their influence on the obtained results, the simulations have been performed for the two models with varying values of the following components: ρ, ρ1, ρ2, l, $l_1$, $l_2$, a and c. Each parameter was changed at a time for the linear, quadratic and exponential complex dependences.

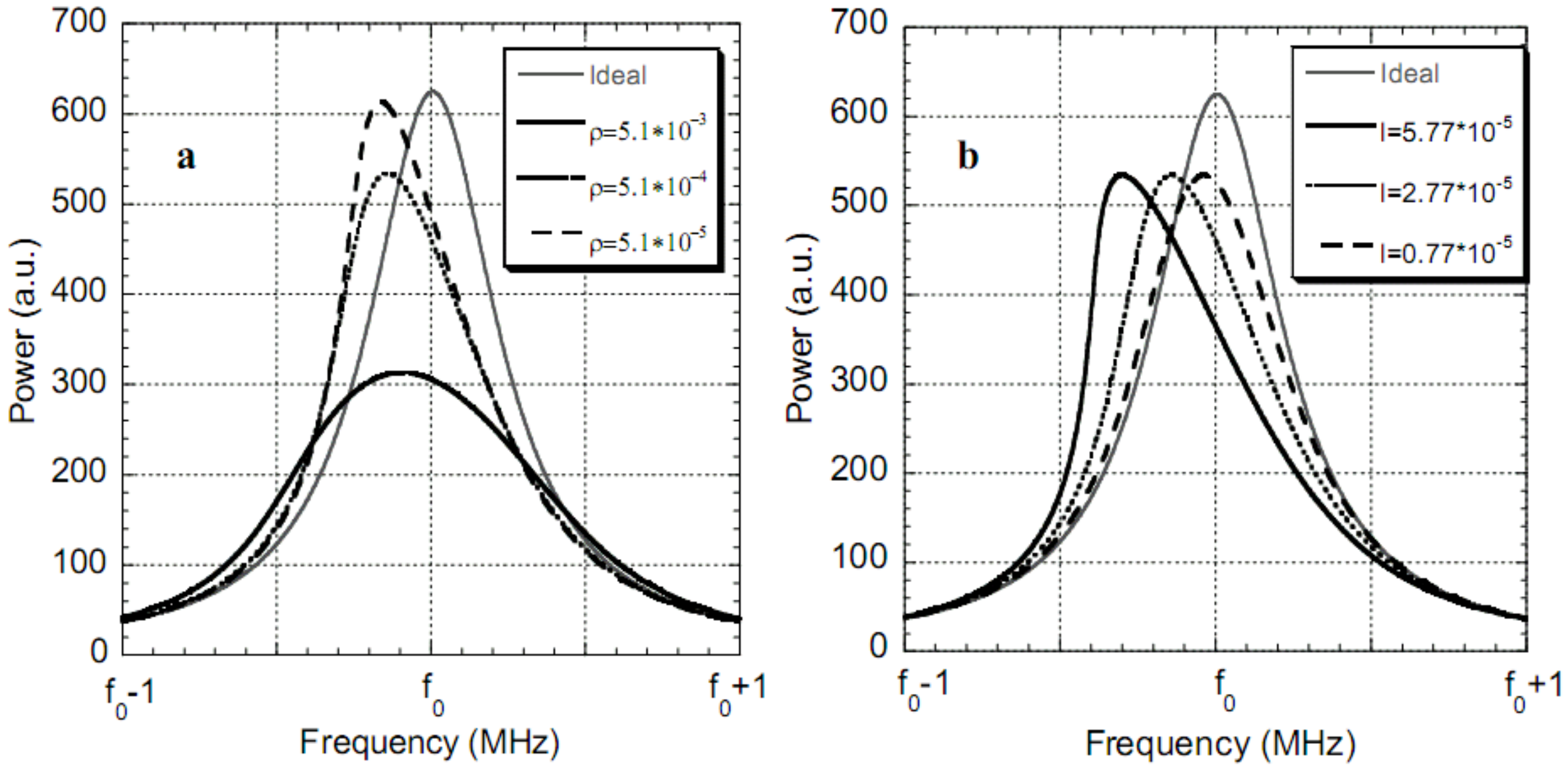

**Fig. 71.** Simulated dependences of microwave power on frequency P(f) for varying fitting parameters for linear dependence for **L-σ model** of network in Fig. 67 ($f_0$=10GHz).

**L-σ model with various magnitudes of parameters ρ and l**

# Modeling and Identification of Lumped Element Model Parameters for a Superconducting Hakki-Coleman Dielectric Resonator.

In order to verify the fitting parameters of the models in terms of their physical correlations with the simulated responses and to investigate their influence on the obtained results, the simulations have been performed for the two models with varying values of the following components: ρ, ρ1, ρ2, l, $l_1$, $l_2$, a and c. Each parameter was changed at a time for the linear, quadratic and exponential complex dependences.

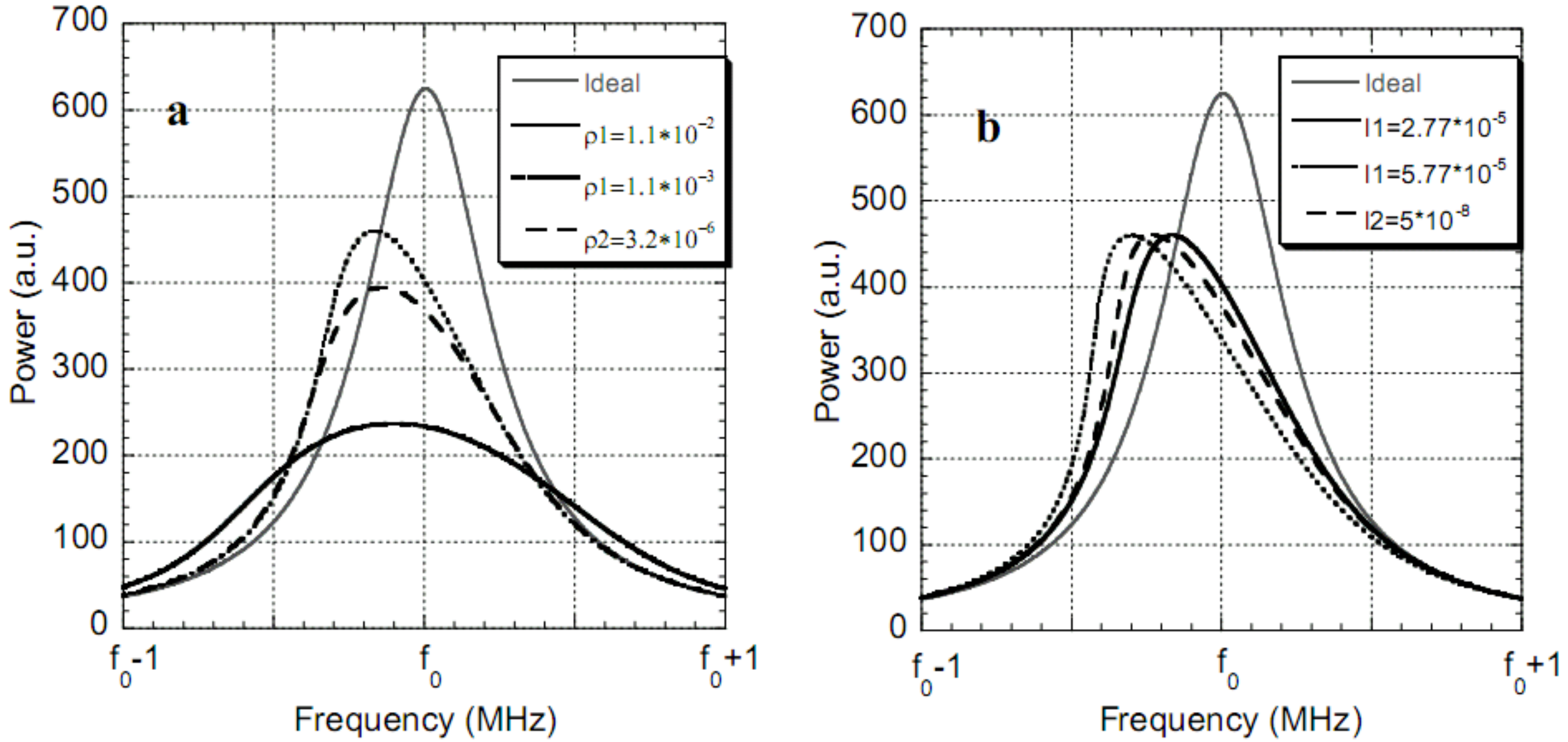

**Fig. 72.** Simulated dependences of RF power on frequency P(f) for varying fitting parameters for quadratic dependence for **L-σ model** of network in Fig. 67 ($f_0$=10GHz).

**L-σ model with various magnitudes of parameters $\rho_1$, $\rho_2$ and $l_1$, $l_2$**

# Modeling and Identification of Lumped Element Model Parameters for a Superconducting Hakki-Coleman Dielectric Resonator.

In order to verify the fitting parameters of the models in terms of their physical correlations with the simulated responses and to investigate their influence on the obtained results, the simulations have been performed for the two models with varying values of the following components: ρ, ρ1, ρ2, l, $l_1$, $l_2$, a and c. Each parameter was changed at a time for the linear, quadratic and exponential complex dependences.

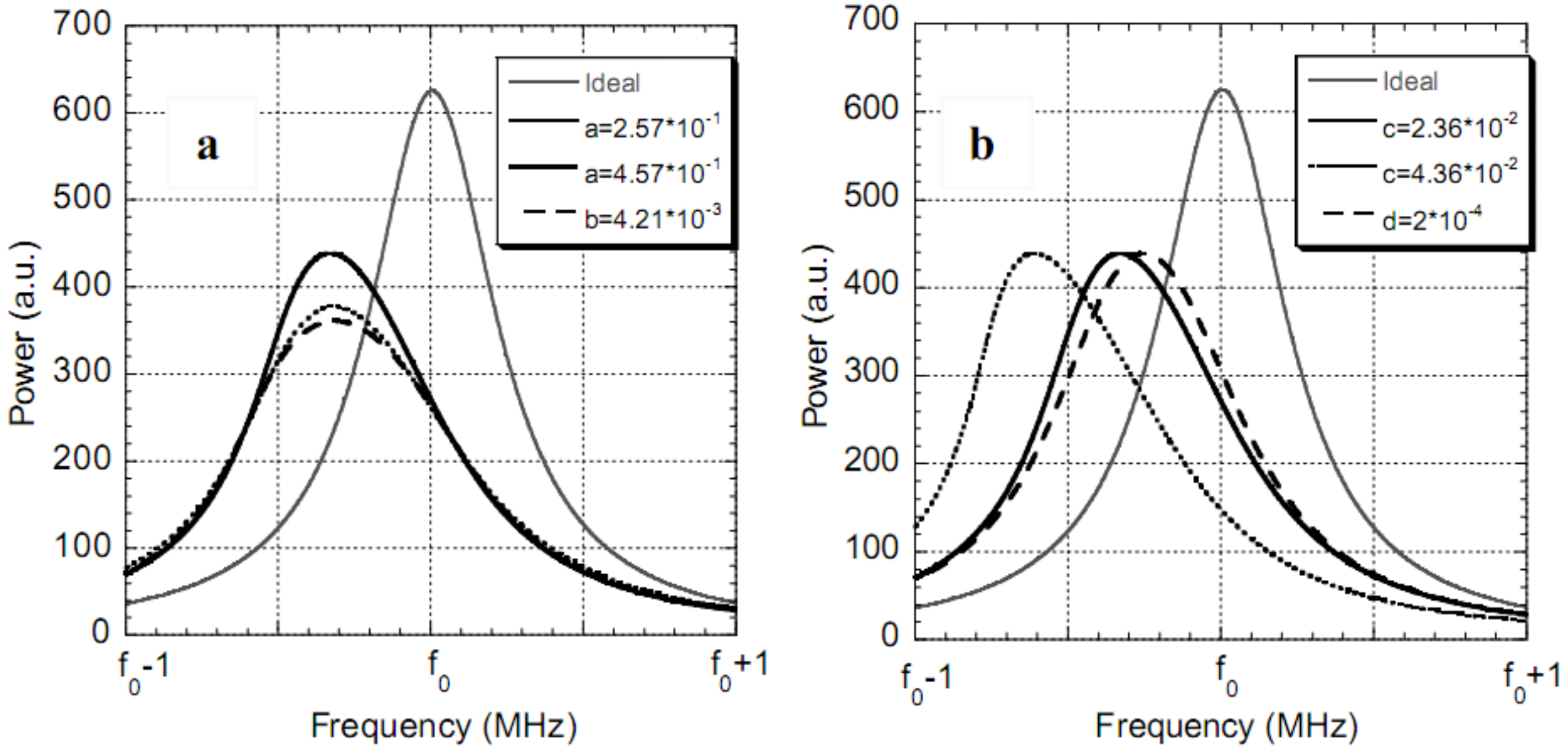

**Fig. 73.** Simulated dependences of RF power on frequency P(f) for varying fitting parameters for exponential dependence for **L-σ model** of network in Fig. 67 ($f_0$=10GHz).

**L-σ model with various magnitudes of parameters a, b, c and d**

# Modeling and Identification of Lumped Element Model Parameters for a Superconducting Hakki-Coleman Dielectric Resonator.

In order to verify the fitting parameters of the models in terms of their physical correlations with the simulated responses and to investigate their influence on the obtained results, the simulations have been performed for the two models with varying values of the following components: ρ, ρ1, ρ2, l, $l_1$, $l_2$, a and c. Each parameter was changed at a time for the linear, quadratic and exponential complex dependences.

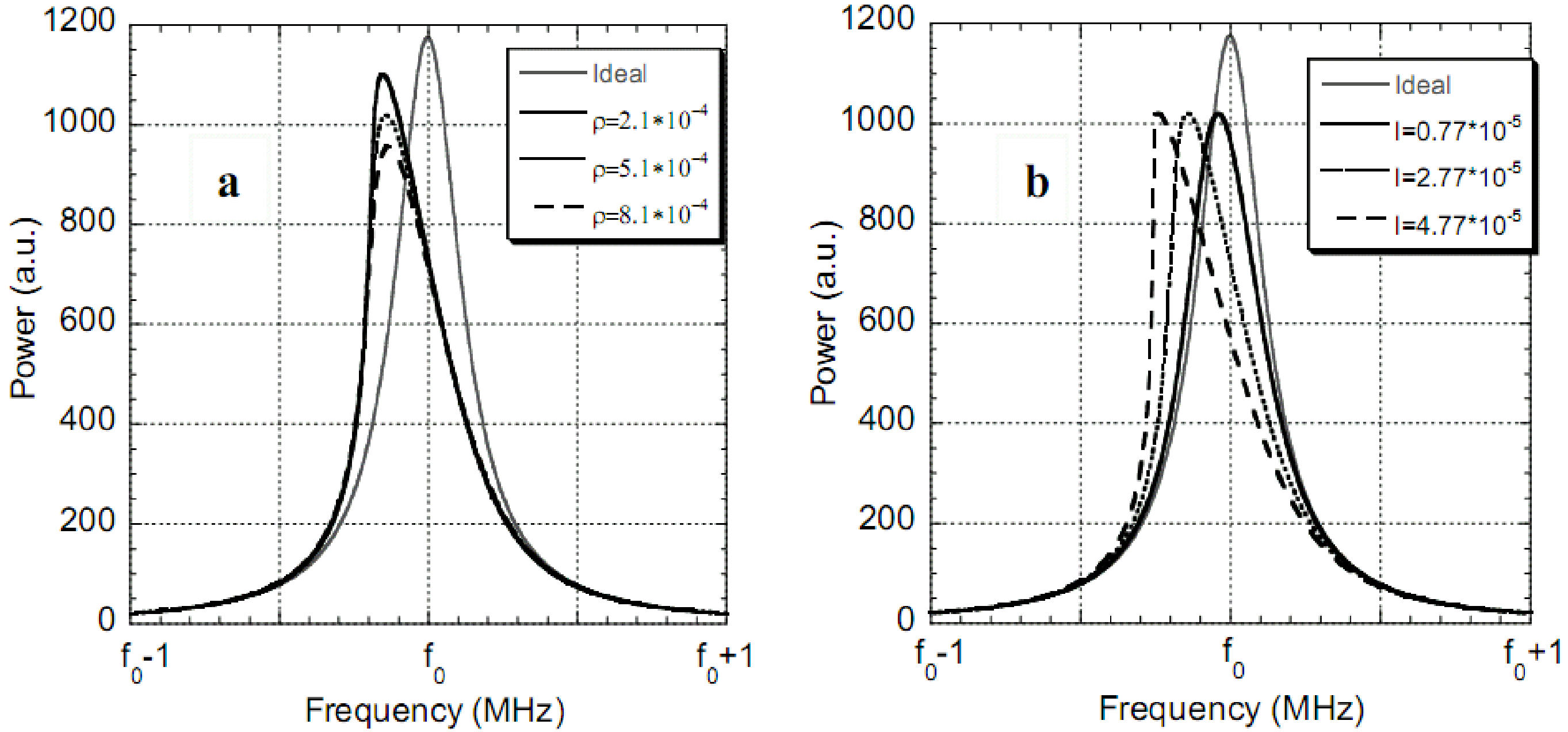

**Fig. 74.** Simulated dependences of RF power on frequency P(f) for varying fitting parameters for linear dependence for **L-Z model** of network in Fig. 68 ($f_0$=10GHz).

**L-Z model with various magnitudes of parameters ρ and l**

# Modeling and Identification of Lumped Element Model Parameters for a Superconducting Hakki-Coleman Dielectric Resonator.

In order to verify the fitting parameters of the models in terms of their physical correlations with the simulated responses and to investigate their influence on the obtained results, the simulations have been performed for the two models with varying values of the following components: ρ, ρ1, ρ2, l, $l_1$, $l_2$, a and c. Each parameter was changed at a time for the linear, quadratic and exponential complex dependences.

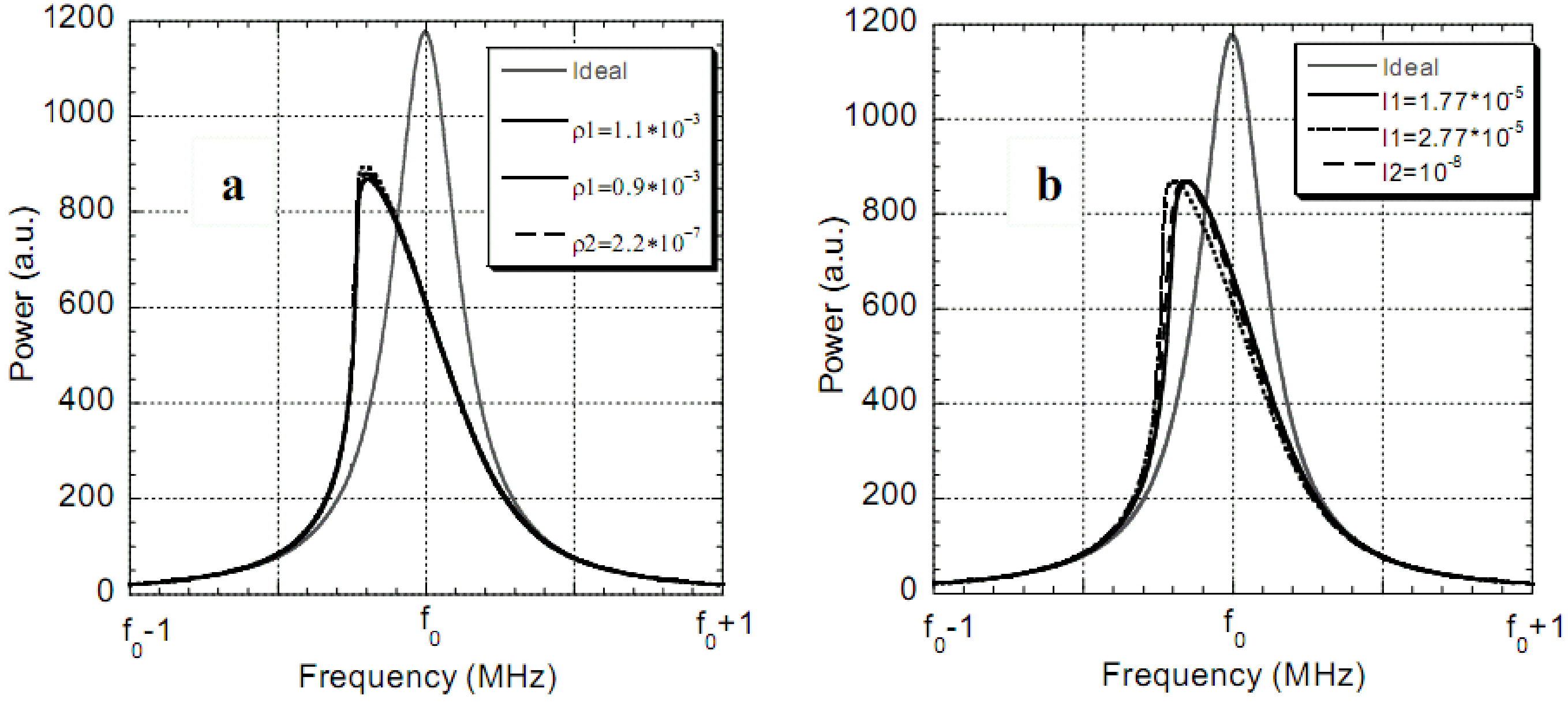

**Fig. 75.** Simulated dependences of microwave power on frequency P(f) for varying fitting parameters for quadratic dependence for **L-Z model** of network in Fig. 68 ($f_0$=10GHz).

**L-Z model with various magnitudes of parameters $\rho_1$, $\rho_2$ and $l_1$, $l_2$**

# Modeling and Identification of Lumped Element Model Parameters for a Superconducting Hakki-Coleman Dielectric Resonator.

In order to verify the fitting parameters of the models in terms of their physical correlations with the simulated responses and to investigate their influence on the obtained results, the simulations have been performed for the two models with varying values of the following components: ρ, ρ1, ρ2, l, $l_1$, $l_2$, a and c. Each parameter was changed at a time for the linear, quadratic and exponential complex dependences.

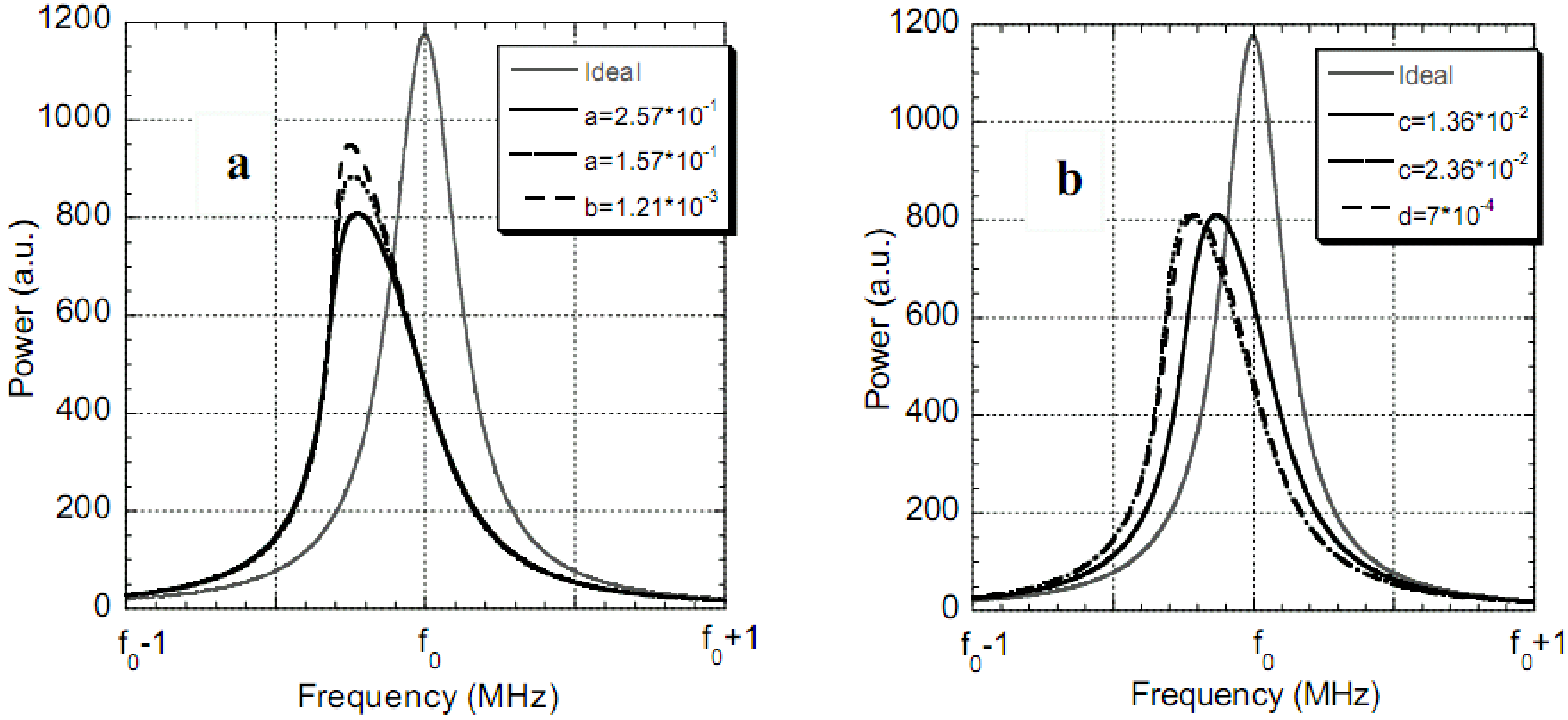

**Fig. 76.** Simulated dependences of microwave power on frequency P(f) for varying fitting coefficients for exponential nonlinear dependence for **L-Z model** of network in Fig. 68 (f0=10GHz).

**L-Z model with various magnitudes of parameters a, b, c and d**

# Calculation of Parameter r = ΔX/ΔR for Exponential Nonlinear Dependences.

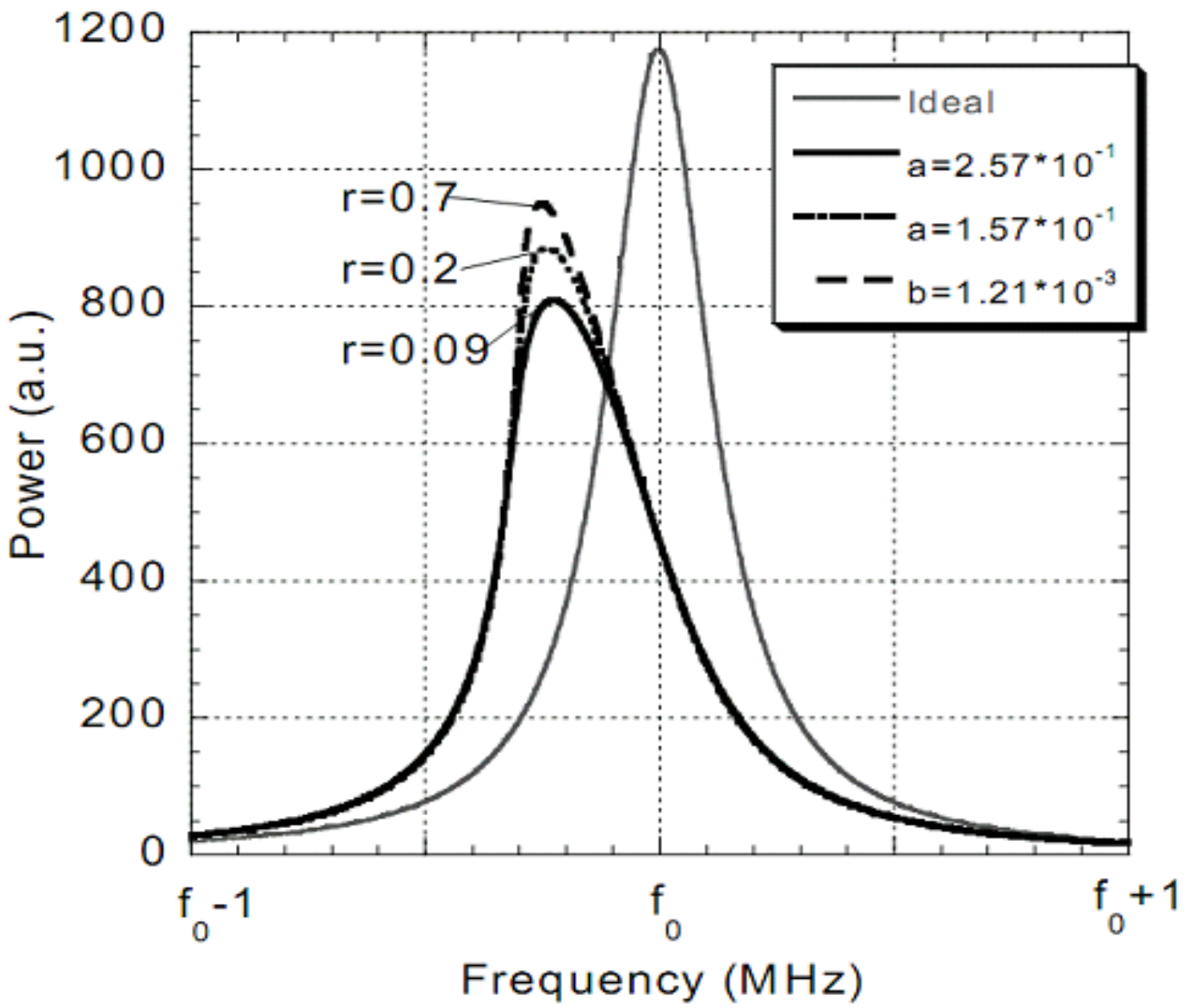

**Fig. 77.** Simulated change of value of r-parameters on frequency for exponential nonlinear dependences of microwave power vs. frequency P(f) for L-Z model of network in Fig. 68 (f0=10GHz).

All the numerical values of fitting coefficients are picked up in a way that allows to observe the changes in shapes of resonance curves in simulation at deviation frequency limits Δf = ±1 MHz at resonance frequency $f_0$ = 10 GHz. The increase of the magnitudes of fitting coefficients in nonlinear dependencies leads to the type of nonlinear dependencies, which was not observed in HTS thin films in author's researches. Therefore, the selected fitting coefficients, and hence the shapes of resonance curves in simulations in Matlab, closely approximate the experimental results, obtained by the author of dissertation during the measurements toward the accurate microwave characterization of HTS thin films at James Cook University in Australia.

# Experimental and Theoretical Researches on Microwave Properties of MgO Substrates in a Split Post Dielectric Resonator and Nonlinear Surface Resistance of $YBa_2Cu_3O_{7-\Delta}$ Thin Films on MgO Substrates in a Dielectric Resonator at Ultra High Frequencies.

The author of thesis decided to research the nonlinear phenomena in $YBa_2Cu_3O_{7-\delta}$ thin films on MgO substrate at high microwave powers up to 30 dBm at temperatures T = 25K and 50K. The results of experimental researches on microwave properties of pure MgO substrates in the split post dielectric resonator (SPDR) at ultra high frequency f=10.48GHz are presented in the beginning of presentation. Then, the results of experimental researches on nonlinear resonance response of $YBa_2Cu_3O_{7-\delta}$ thin films on MgO substrates in the Hakki-Coleman dielectric resonator (HCDR) at f = 25GHz are described.

# Ultra High Frequency Measurement Setup and Procedure.

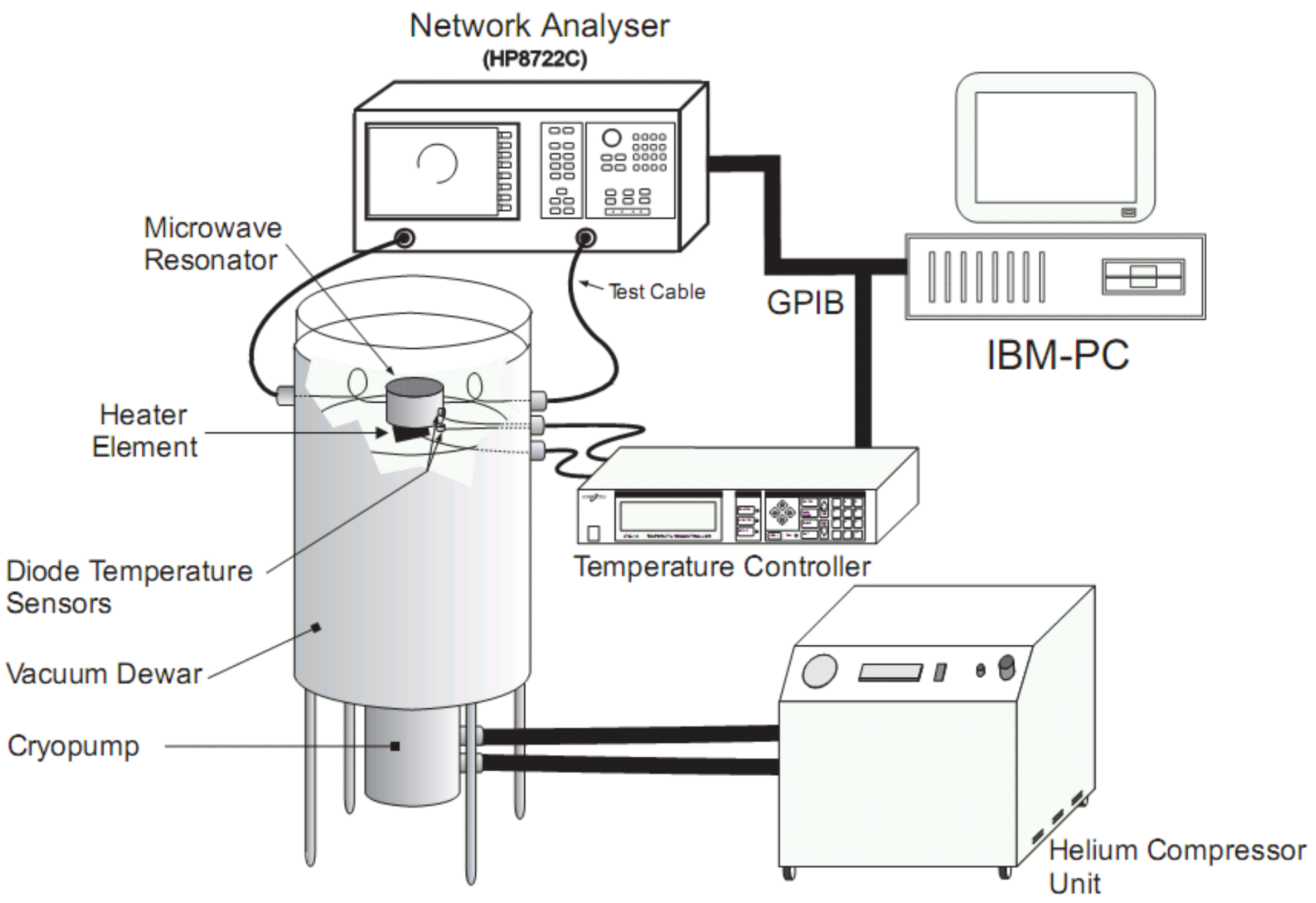

**Fig. 78**. Cryogenic measurement setup for research on nonlinear resonance response of $YBa_2Cu_3O_{7-\delta}$ thin films on MgO substrates in a dielectric resonator at microwaves.

The ultra high frequency experimental measurement system for research on nonlinear resonance response of $YBa_2Cu_3O_{7-\delta}$ thin films on MgO substrates in a dielectric resonator at various levels of microwave power consisted of the following equipment:

- **Vector Network Analyser** (HP 8722C).
- **Temperature Controller** (Conductus LTC-10) fitted with a resistive heating element and two silicon temperature diode sensors.
- **Vacuum Dewar.**
- **Close Cycle Cryogenic Laboratory System** (APC-HC4) suitable for measurements in a wide range of temperatures (10 K– 300 K).
- **Computer System** (IBM-PC) fitted with a GPIB card utilized for the computer control of temperature controller and network analyser, and S-parameter measurement data transfer from network analyser to computer.

The measurement set up was precisely calibrated by measuring the voltage standing wave ratio VSWR=1+ IΓI/1-IΓI, where Γ is the reflection coefficient, before accurate experimental measurements of S-parameters as schematically shown in Fig. 62.

# Surface Resistance Rs of $YBa_2Cu_3O_{7-\delta}$ Thin Films on MgO Substrates Measured with the Use of Hakki-Coleman Dielectric Resonator.

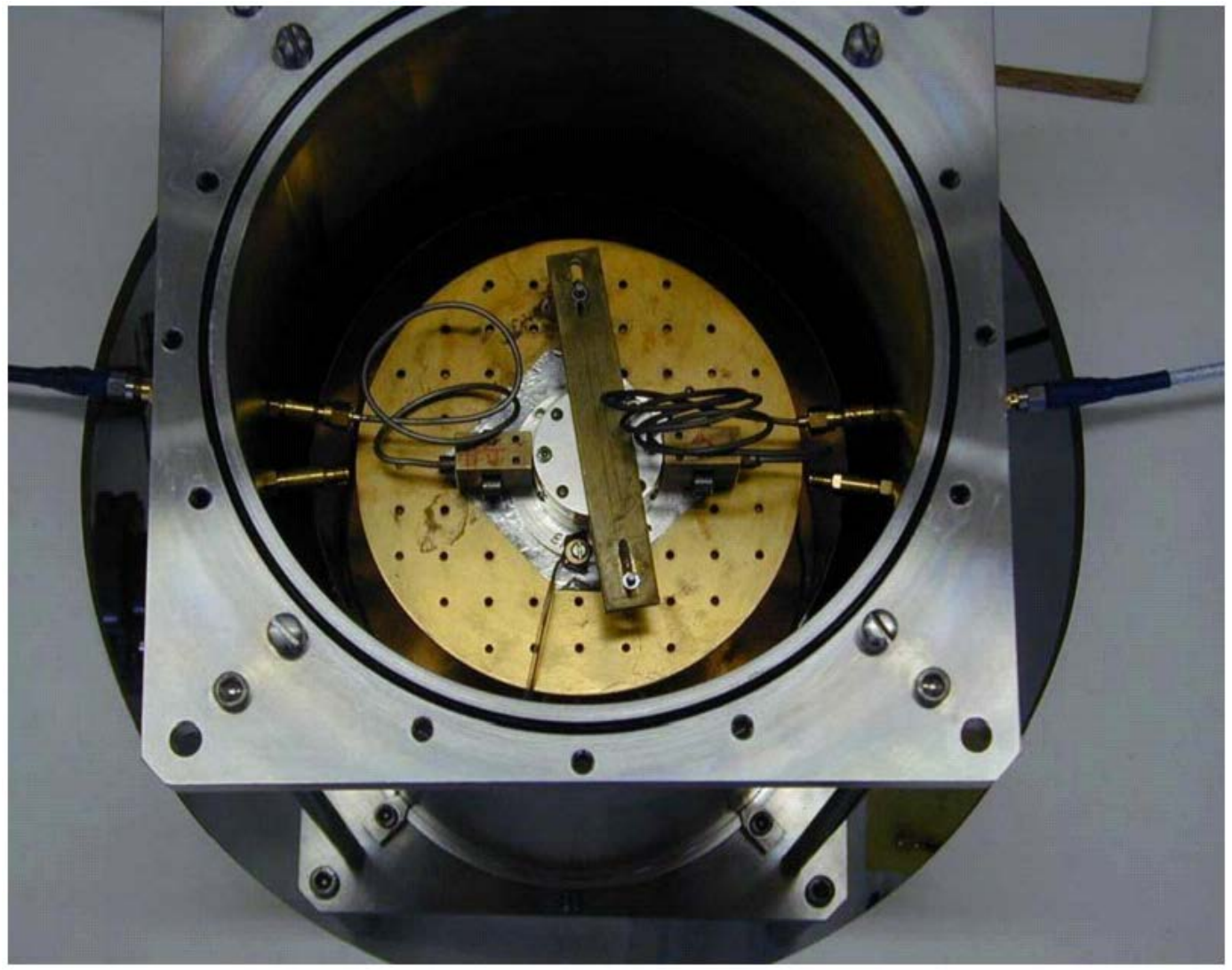

**Fig. 79**. Hakki-Coleman dielectric resonator mounted inside vacuum dewar with RF cable interconnections.

The loaded $Q_L$-factor and coupling coefficients $\beta_1$ and $\beta_2$ of resonator were obtained from multi-frequency measurements of $S_{21}$, $S_{11}$ and $S_{22}$ parameters measured around the resonance using the Transmission Mode Q-Factor (TMQF) Technique [79, 80]. The TMQF method enables to obtain accurate values of surface resistance accounting for factors such as noise, delays due to uncompensated transmission lines, and crosstalks occurring in measured data. The unloaded $Q_0$-factor was calculated from the exact equation

$$Q_0 = Q_L(1+\beta_1+\beta_2)$$

using the TMQF method for all temperatures. Dielectric resonator with embedded superconducting samples, mounted inside the vacuum dewar, was cooled down to temperature T around 12 K, and the S parameters were measured at resonance frequency up to temperature T = 85 K. The RF power of source signal was -5 dBm and the number of points was 1601.

Surface resistance Rs of $YBa_2Cu_3O_{7-\delta}$ thin films on MgO substrates has been computed using the software SUPER [81], based on the equation (6.2)

$$R_S = A_S\left\{\frac{1}{Q_0} - \frac{R_m}{A_m} - p_e \tan\delta\right\}$$

The geometric factors As, Am, and $p_e$ were computed using the incremental frequency rules as follows in eq. (6.3) [82]

$$A_s = \frac{\omega^2\mu_0}{4}\bigg/\frac{\partial\omega}{\partial L}$$

$$A_m = \frac{\omega^2\mu_0}{2}\bigg/\frac{\partial\omega}{\partial a}$$

$$p_e = 2\left|\frac{\partial\omega}{\partial\varepsilon}\right|\frac{\varepsilon_r}{\omega}$$

# Precise Microwave Characterization of MgO Substrates for HTS Circuits with Split Post Dielectric Resonator.

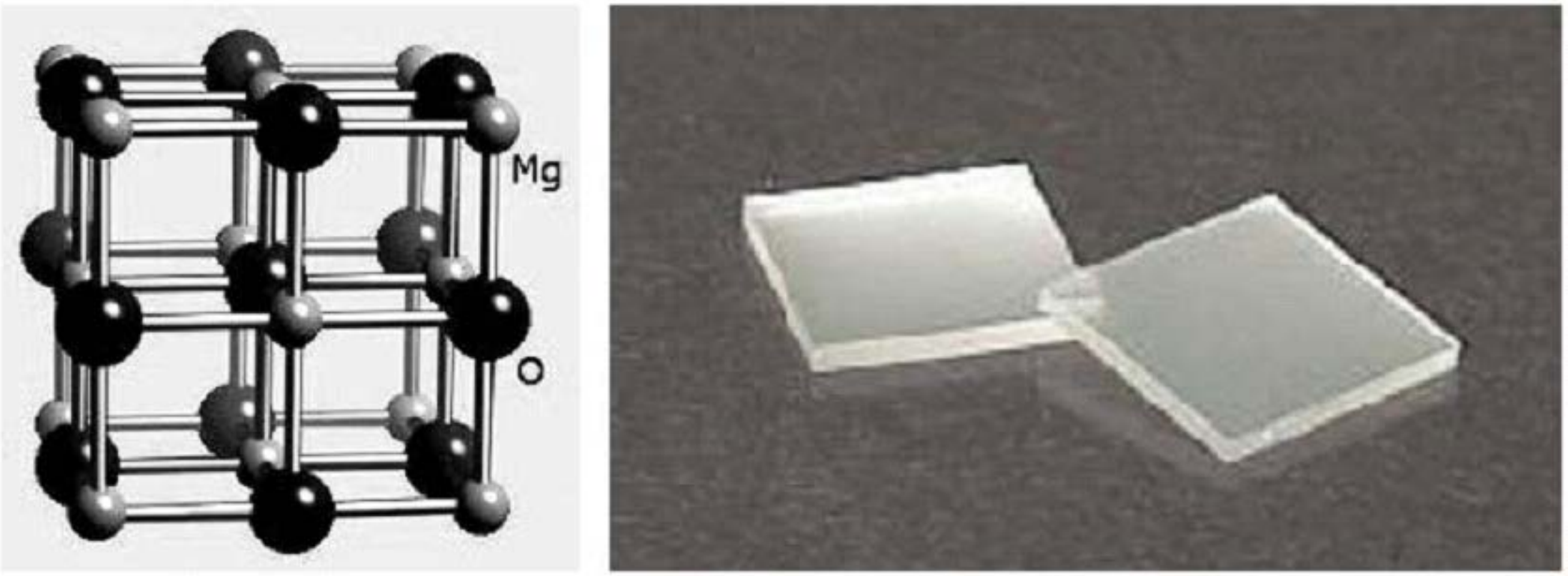

**Fig. 80**. Crystal structure of magnesium oxide and view of its substrates.

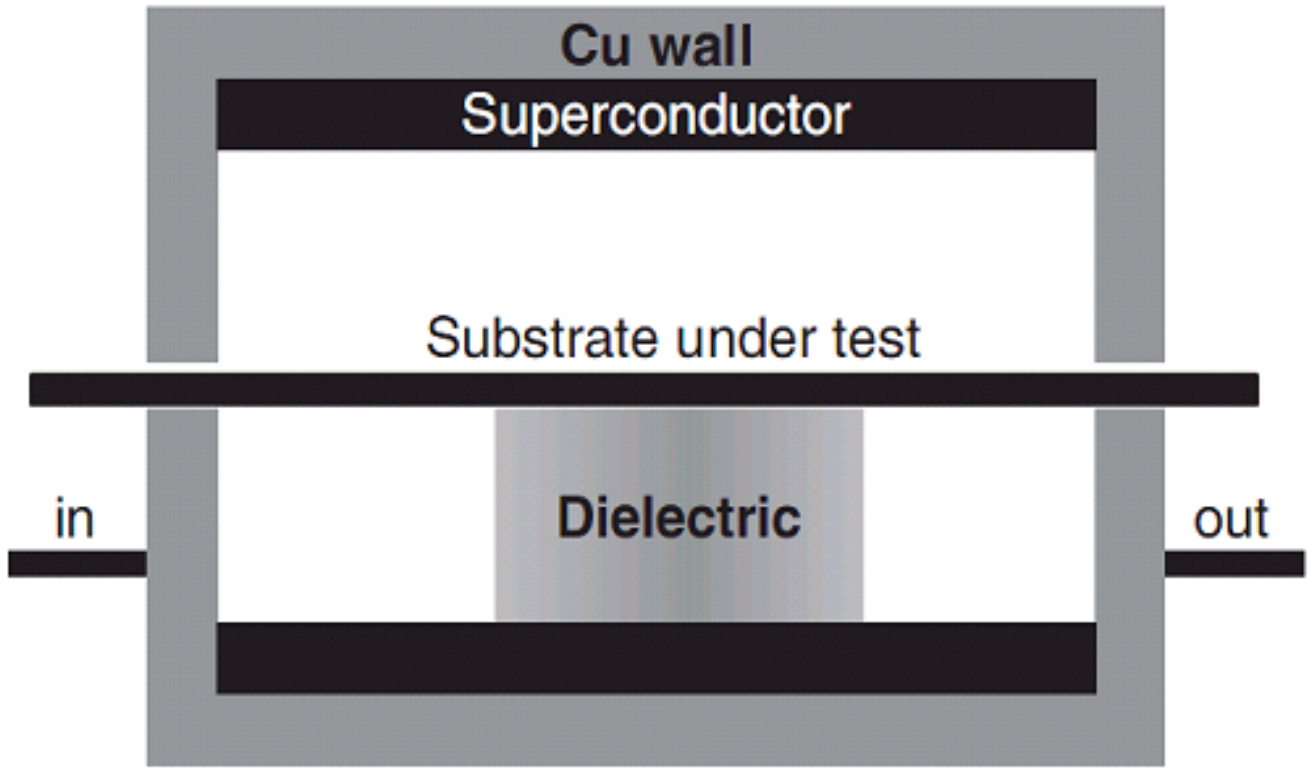

**Fig. 81**. Diagram of cryogenic split post dielectric resonator (after [19]).

Accurate data on complex permittivity of dielectric substrates are needed for efficient design of HTS microwave planar circuits. The author of thesis has researched MgO substrates from three different manufacturing batches, using a dielectric resonator with superconducting parts recently developed for precise microwave characterization of laminar dielectrics at cryogenic temperatures.

The measurement fixture has been fabricated using a $SrLaAlO_3$ post dielectric resonator with $DyBa_2Cu_3O_7$ endplates and silver-plated copper sidewalls to achieve the resolution of loss tangent measurements of $2 \cdot 10^{-6}$. The MgO substrates are essentially free of twinning, strain defects and air bubbles; they do not require buffer layers for HTS films; their orientation is typically along the (100) planes.

The split post dielectric resonator (SPDR) contains one $SrLaAlO_3$ dielectric rod resonating at frequency of 10.48GHz on which the dielectric substrate under test is placed. The split post dielectric resonator was designed to measure dielectric properties of substrates of thickness of 0.7 mm or smaller, and diameters in the range from 35 mm to 55 mm.

# Experimental Measurements Results on MgO Substrates for HTS Circuits with the Use of Split Post Dielectric Resonator.

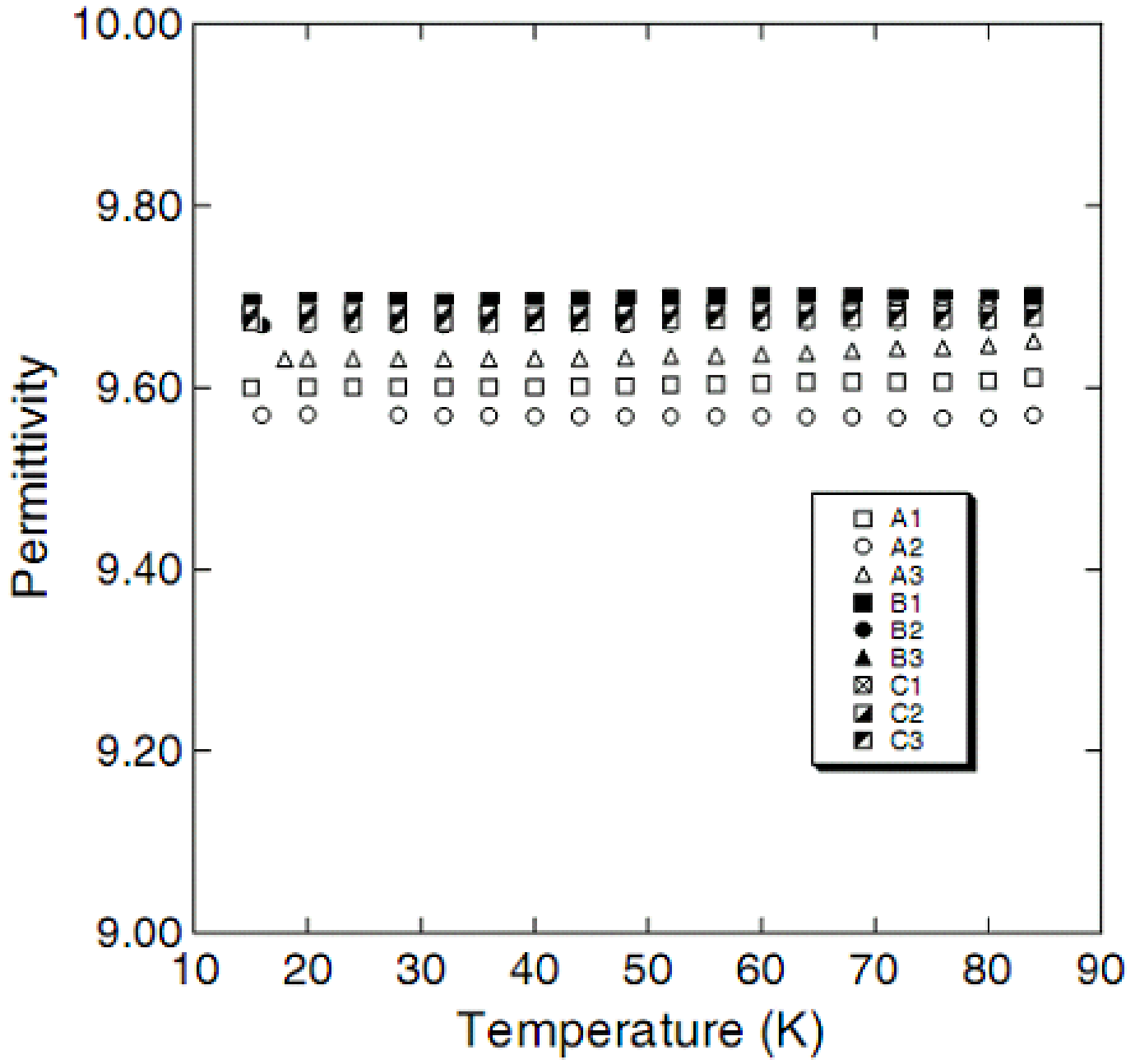

**Fig. 82**. Dependence of measured permittivity $\varepsilon'_r$ on temperature of MgO substrates.

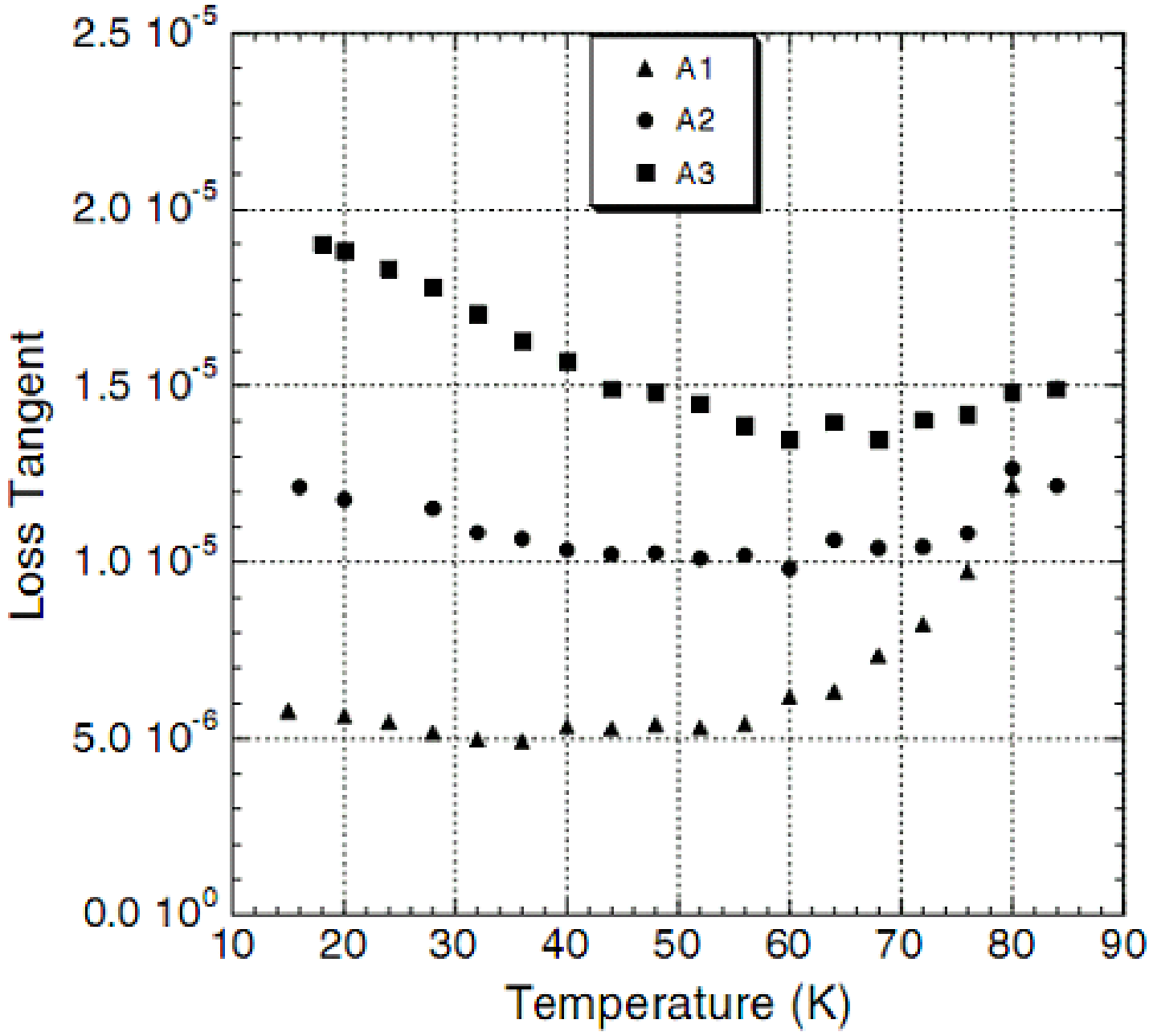

**Fig. 83.** Dependence of measured loss tan δ on temperature of MgO substrates of batch A.

# Experimental Measurements Results on MgO Substrates for HTS Circuits with the Use of Split Post Dielectric Resonator.

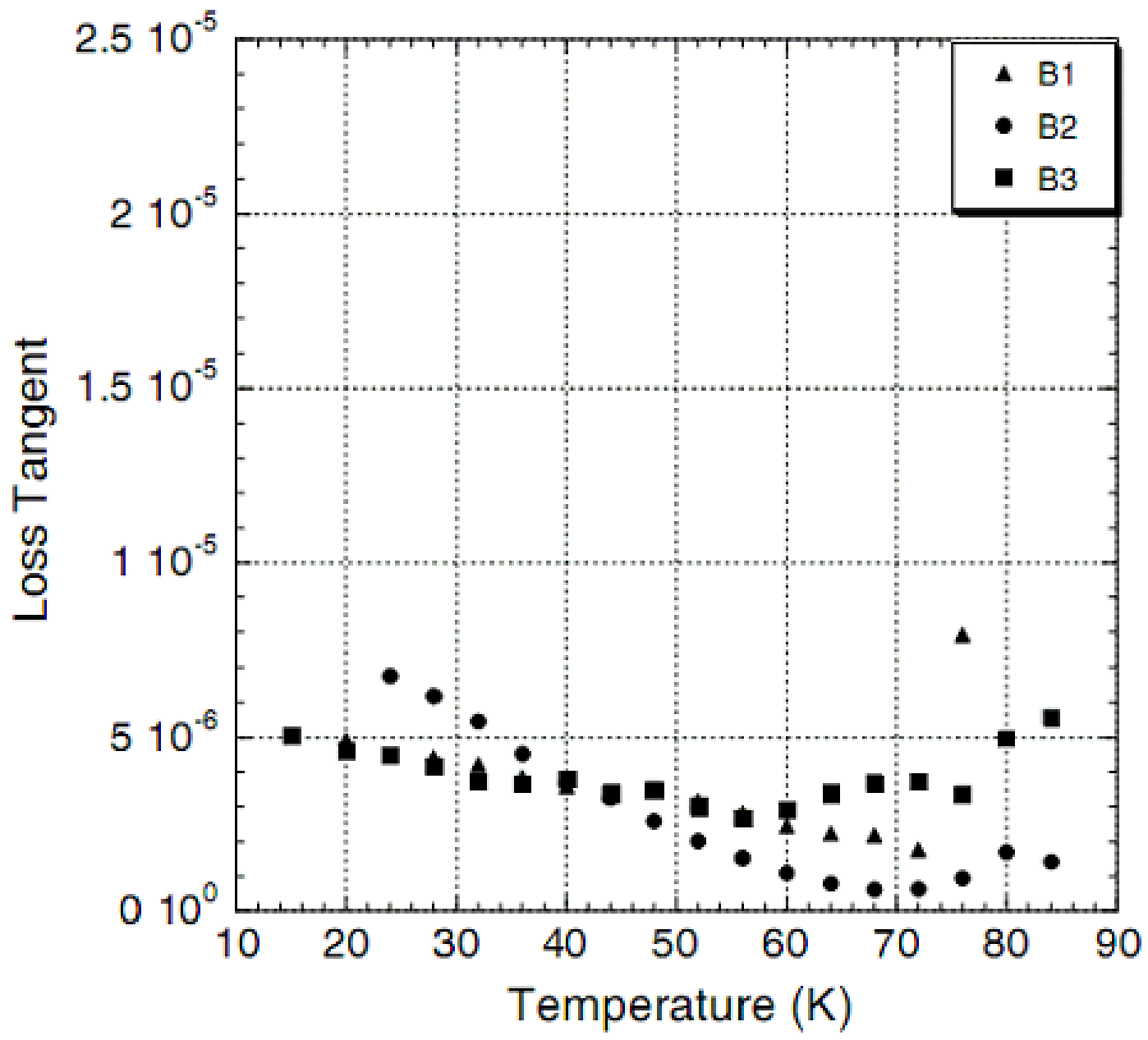

**Fig. 84.** Dependence of measured loss tan δ on temperature of MgO substrates of batch B.

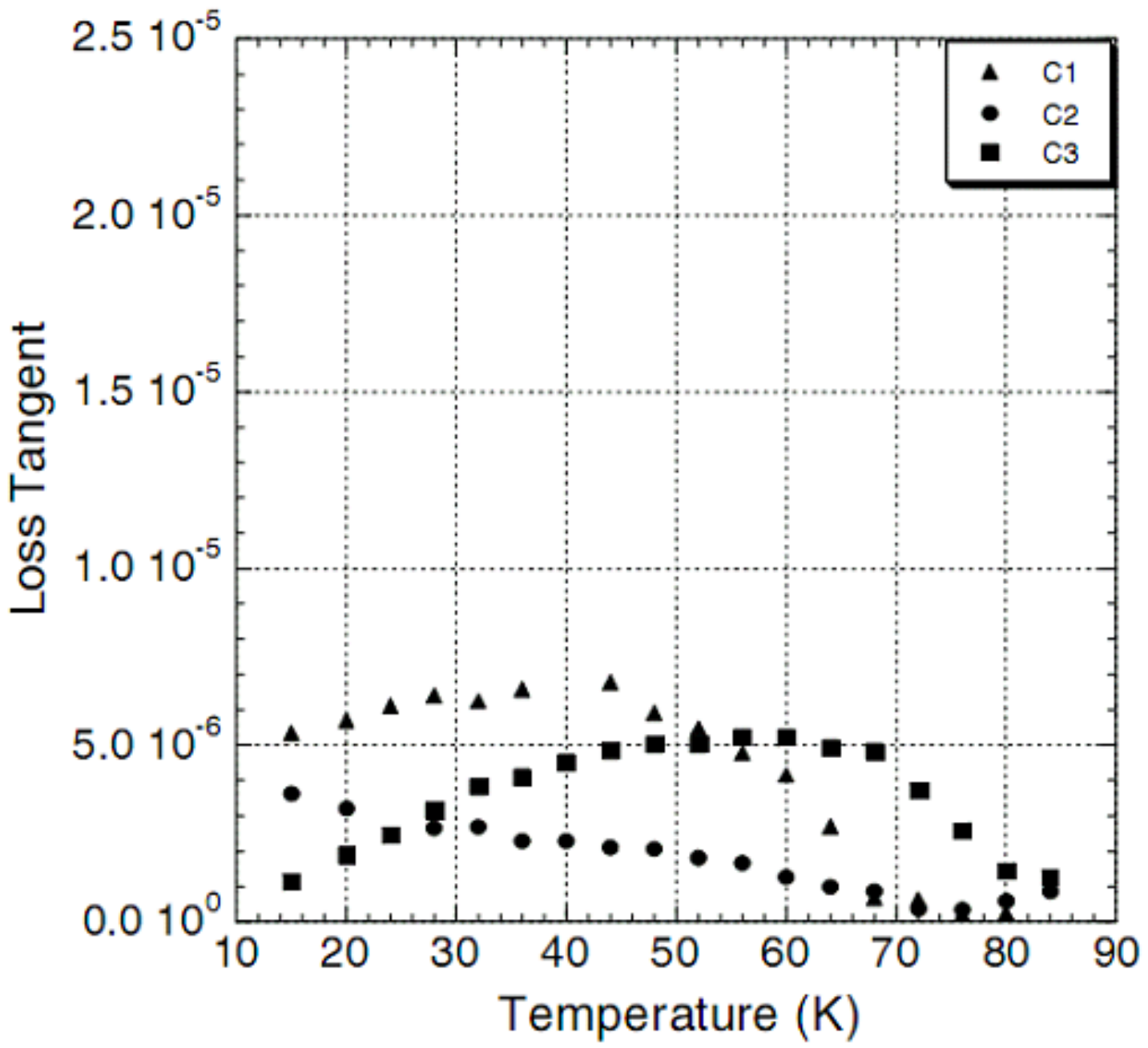

**Fig. 85.** Dependence of measured loss tan δ on temperature of MgO substrates of batch C.

# Experimental Measurements Results on MgO Substrates for HTS Circuits with the Use of Split Post Dielectric Resonator.

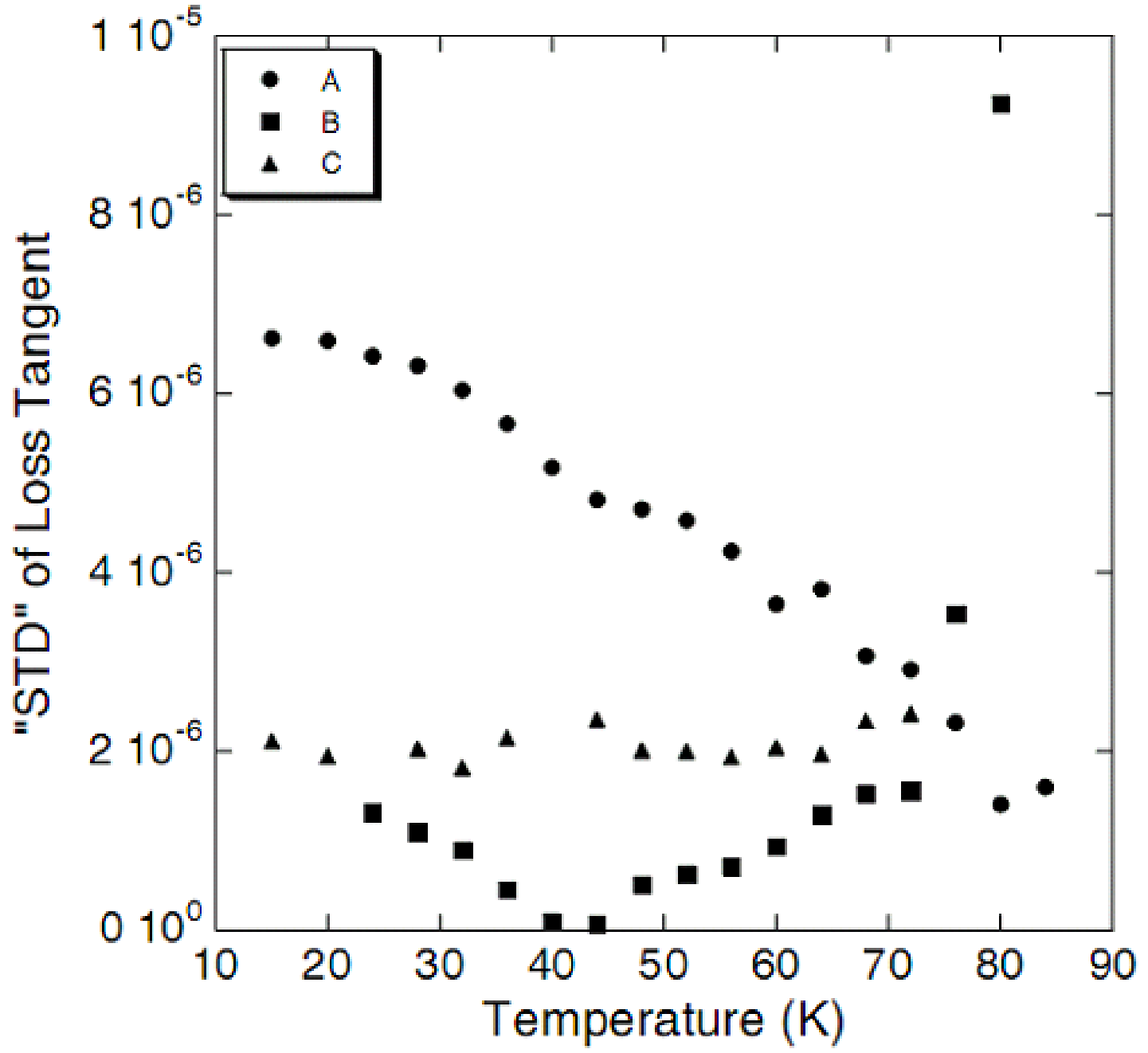

**Fig. 86.** Dependence of 'Standard Deviation' in tan δ values on temperature for three MgO batches.

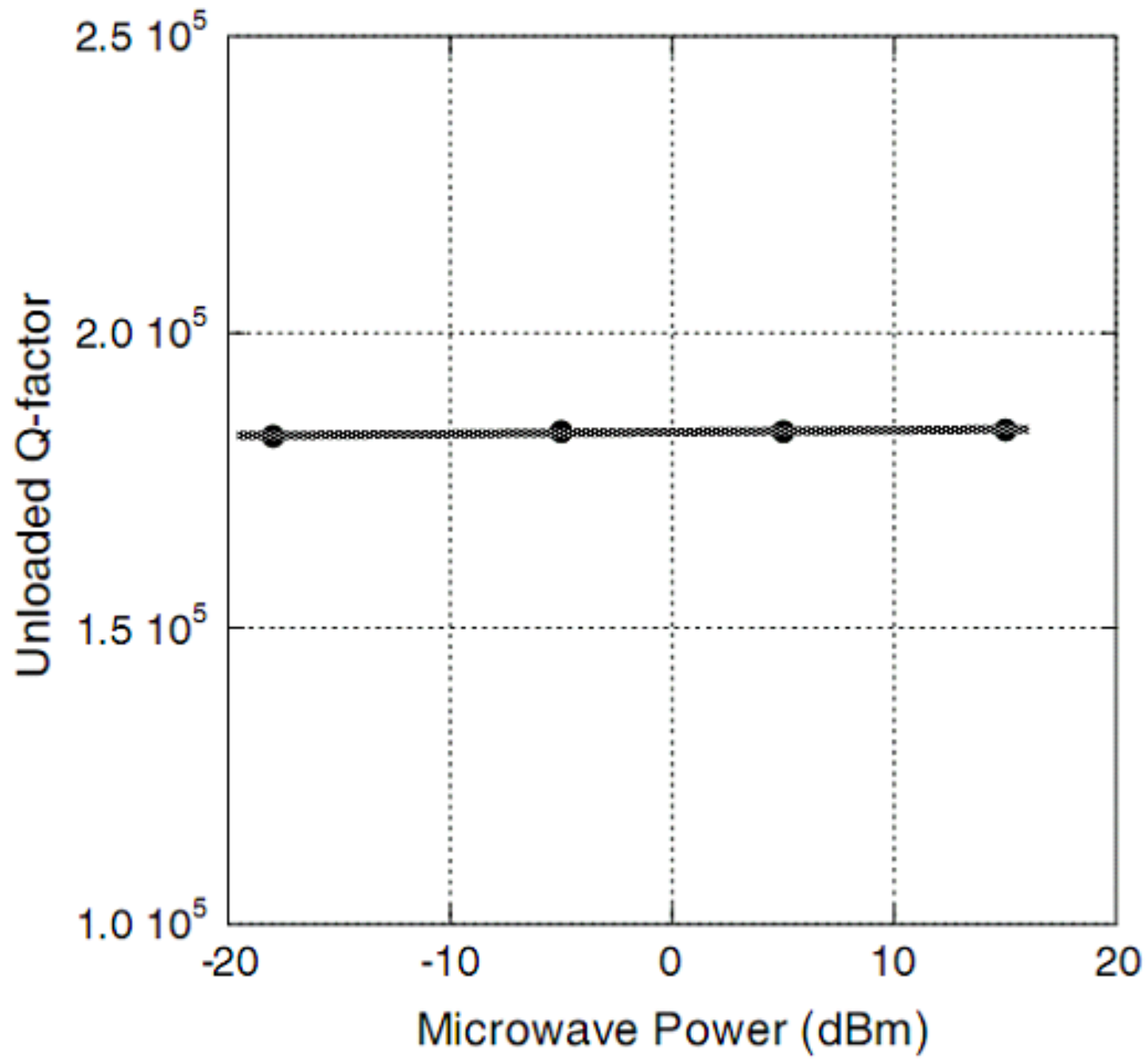

**Fig. 87.** Dependence of $Q_0$-factor versus microwave input power of split post dielectric resonator at frequency f = 10.4 GHz and temperature T = 24 K.

# Experimental Measurements Results on MgO Substrates for HTS Circuits with the Use of Split Post Dielectric Resonator.

| | Substrate | 15 K | 52 K | 72 K |
|---|---|---|---|---|
| Real relative permittivity | A1 | $9.60 \pm 0.5\%$ | $9.60 \pm 0.5\%$ | $9.61 \pm 0.5\%$ |
| | A2 | $9.57 \pm 0.5\%$ | $9.57 \pm 0.5\%$ | $9.57 \pm 0.5\%$ |
| | A3 | $9.63 \pm 0.5\%$ | $9.63 \pm 0.5\%$ | $9.64 \pm 0.5\%$ |
| | B1 | $9.69 \pm 0.5\%$ | $9.70 \pm 0.5\%$ | $9.70 \pm 0.5\%$ |
| | B2 | $9.67 \pm 0.5\%$ | $9.67 \pm 0.5\%$ | $9.67 \pm 0.5\%$ |
| | B3 | $9.68 \pm 0.5\%$ | $9.68 \pm 0.5\%$ | $9.69 \pm 0.5\%$ |
| | C1 | $9.68 \pm 0.5\%$ | $9.68 \pm 0.5\%$ | $9.69 \pm 0.5\%$ |
| | C2 | $9.68 \pm 0.5\%$ | $9.68 \pm 0.5\%$ | $9.68 \pm 0.5\%$ |
| | C3 | $9.67 \pm 0.5\%$ | $9.67 \pm 0.5\%$ | $9.68 \pm 0.5\%$ |
| $\tan\delta$ | A1 | $5.79 \times 10^{-6} \pm 27\%$ | $5.32 \times 10^{-6} \pm 35\%$ | $8.24 \times 10^{-6} \pm 30\%$ |
| | A2 | $1.20 \times 10^{-5} \pm 14\%$ | $1.01 \times 10^{-5} \pm 19\%$ | $1.04 \times 10^{-5} \pm 24\%$ |
| | A3 | $1.90 \times 10^{-5} \pm 9\%$ | $1.45 \times 10^{-5} \pm 14\%$ | $1.40 \times 10^{-5} \pm 18\%$ |
| | B1 | $5.07 \times 10^{-6} \pm 30\%$ | $3.17 \times 10^{-6} \pm 58\%$ | $1.77 \times 10^{-6} \pm 134\%$ |
| | B2 | | $2.02 \times 10^{-6} \pm 91\%$ | $6.5 \times 10^{-7} \pm 360\%$ |
| | B3 | $5.03 \times 10^{-6} \pm 30\%$ | $3.02 \times 10^{-6} \pm 61\%$ | $3.74 \times 10^{-6} \pm 64\%$ |
| | C1 | $5.33 \times 10^{-6} \pm 29\%$ | $5.48 \times 10^{-6} \pm 34\%$ | $6.5 \times 10^{-7} \pm 331\%$ |
| | C2 | $3.63 \times 10^{-6} \pm 42\%$ | $1.8 \times 10^{-6} \pm 101\%$ | $3.7 \times 10^{-7} \pm 564\%$ |
| | C3 | $1.13 \times 10^{-6} \pm 131\%$ | $5.0 \times 10^{-6} \pm 37\%$ | $3.6 \times 10^{-6} \pm 60\%$ |

**Fig. 88.** Measured parameters of MgO substrates at temperatures of 15, 52, and 72 K.

# Temperature Dependence of Surface Resistance $R_S(T)$ of $YBa_2Cu_3O_{7-\delta}$ Superconducting Thin Films on MgO Substrates in a Dielectric Resonator at Ultra High Frequencies.

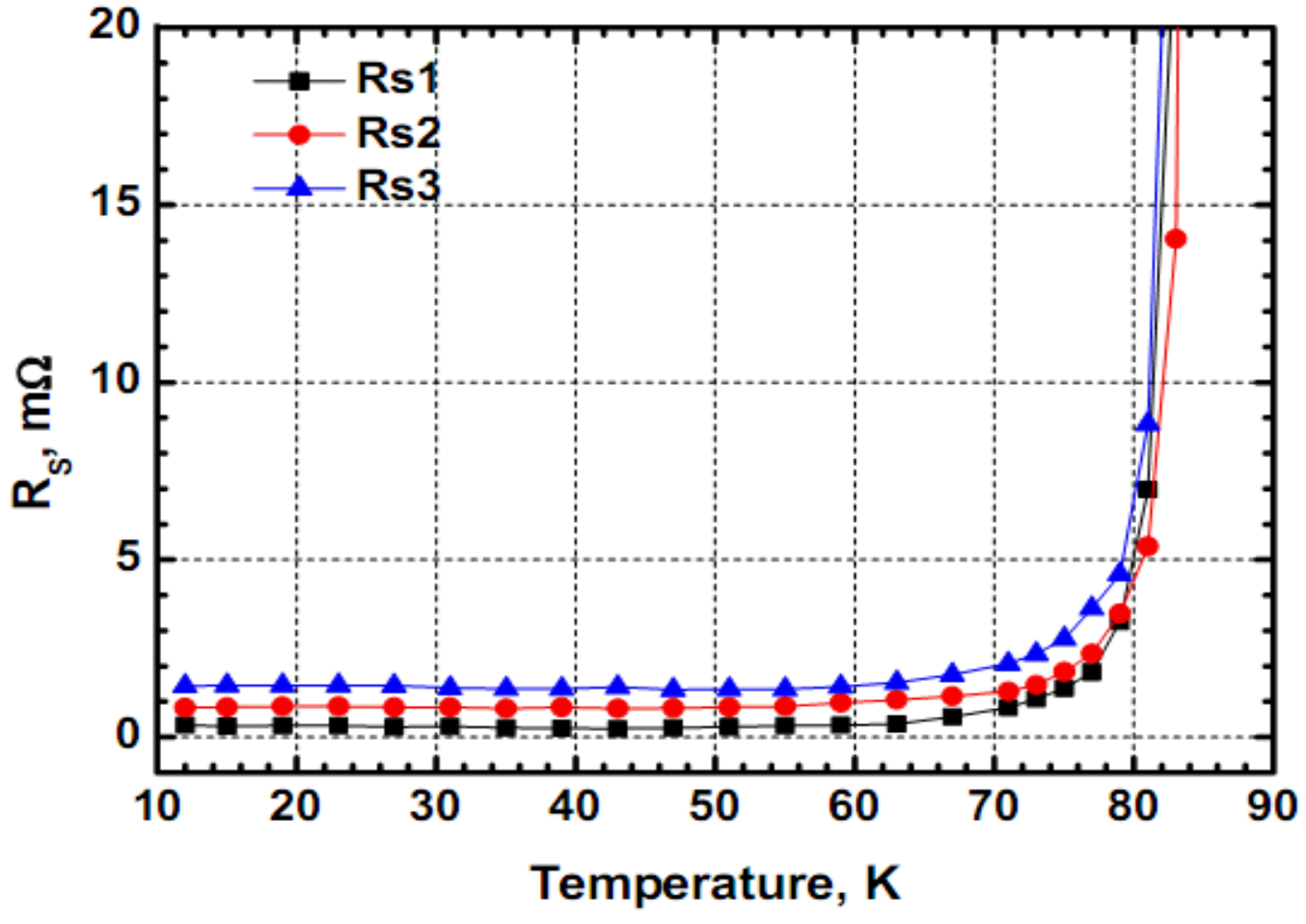

**Fig. 89.** Temperature dependence of surface resistance Rs(T) of $YBa_2Cu_3O_7$ thin films on MgO substrate (1-3) at 25GHz at -5dBm [83].

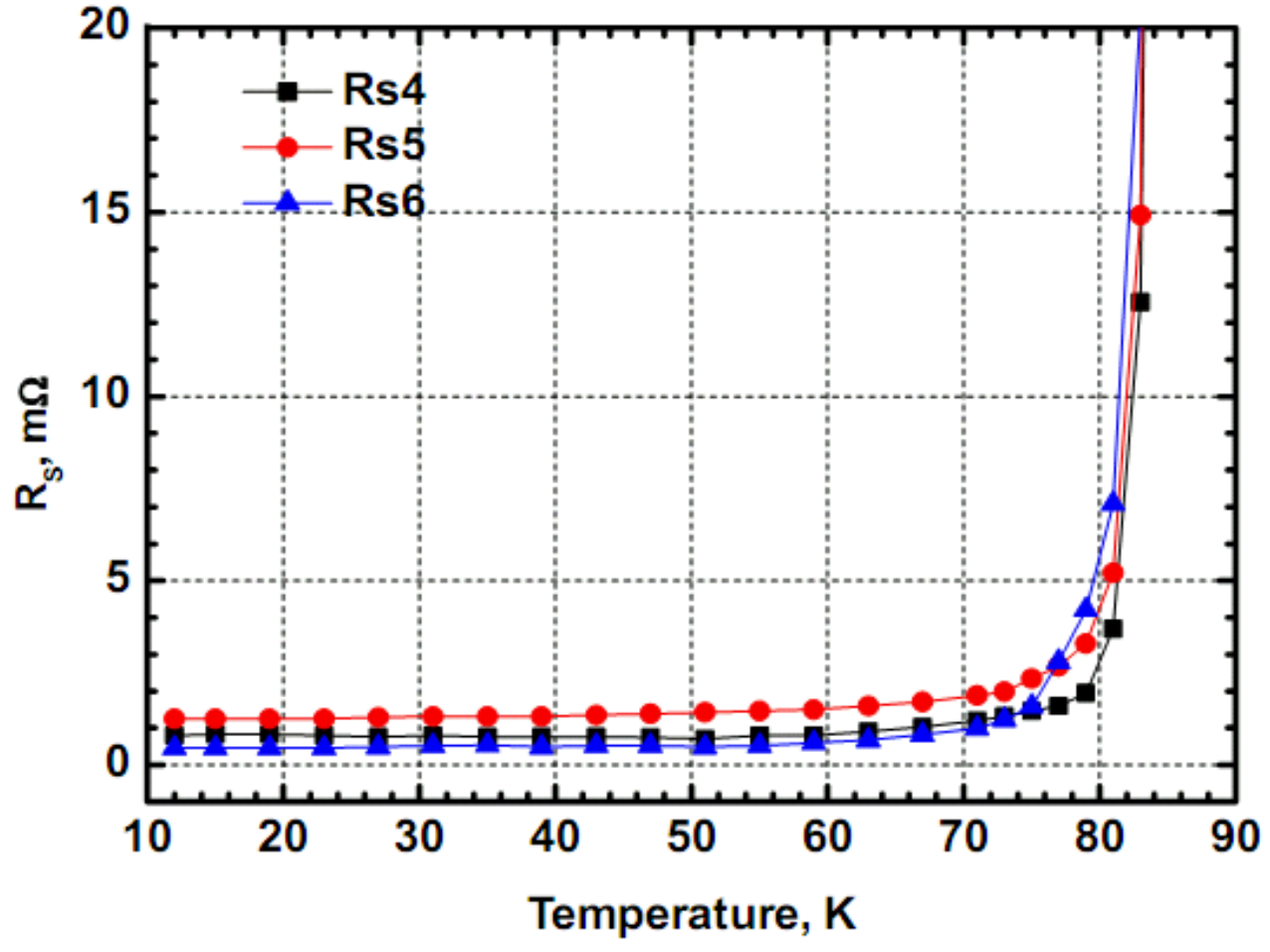

**Fig. 90.** Temperature dependence of surface resistance Rs(T) of $YBa_2Cu_3O_7$ thin films on MgO substrate (4-6) at 25GHz at -5dBm [83].

Six $YBa_2Cu_3O_{7-\delta}$ on MgO thin films have been utilized for microwave measurements. All the thin films were divided into two groups, and each group was measured on the "round robin" rotation basis by pairs to enable determination of microwave parameters for each sample. For the first group of three samples, the temperature dependences of sums of surface resistances $Rs_1+Rs_2$, $Rs_1+Rs_3$, $Rs_2+Rs_3$ were obtained in the process of measurements in temperature range from 12K to 85K. Dependences of surface resistances on temperature: $Rs_1(T)$, $Rs_2(T)$ and $Rs_3(T)$ were derived from the above data. The same measurements were conducted with the second group of samples. The obtained data for dependences of surface resistances on temperature Rs(T) for each tested sample are shown in Figs. 89 and 90.

# Microwave Power Dependence of Surface Resistance $R_S(P)$ of $YBa_2Cu_3O_{7-\delta}$ Superconducting Thin Films on MgO Substrates in a Dielectric Resonator at Ultra High Frequencies.

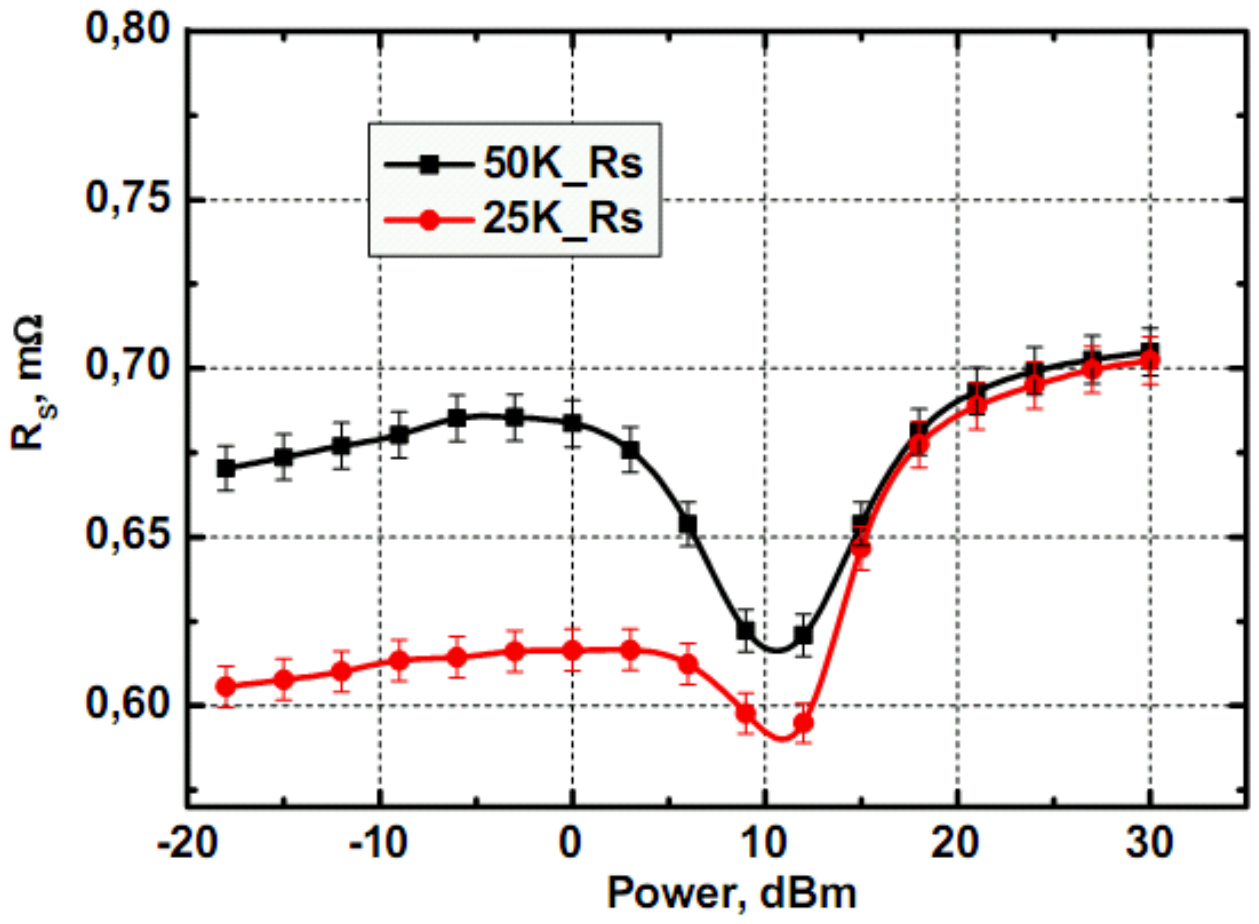

**Fig. 91.** Microwave power dependence of surface resistance Rs(P) of $YBa_2Cu_3O_{7-\delta}$ thin films on MgO substrate at 25GHz at T = 25 K, 50 K (samples 1-2) [84].

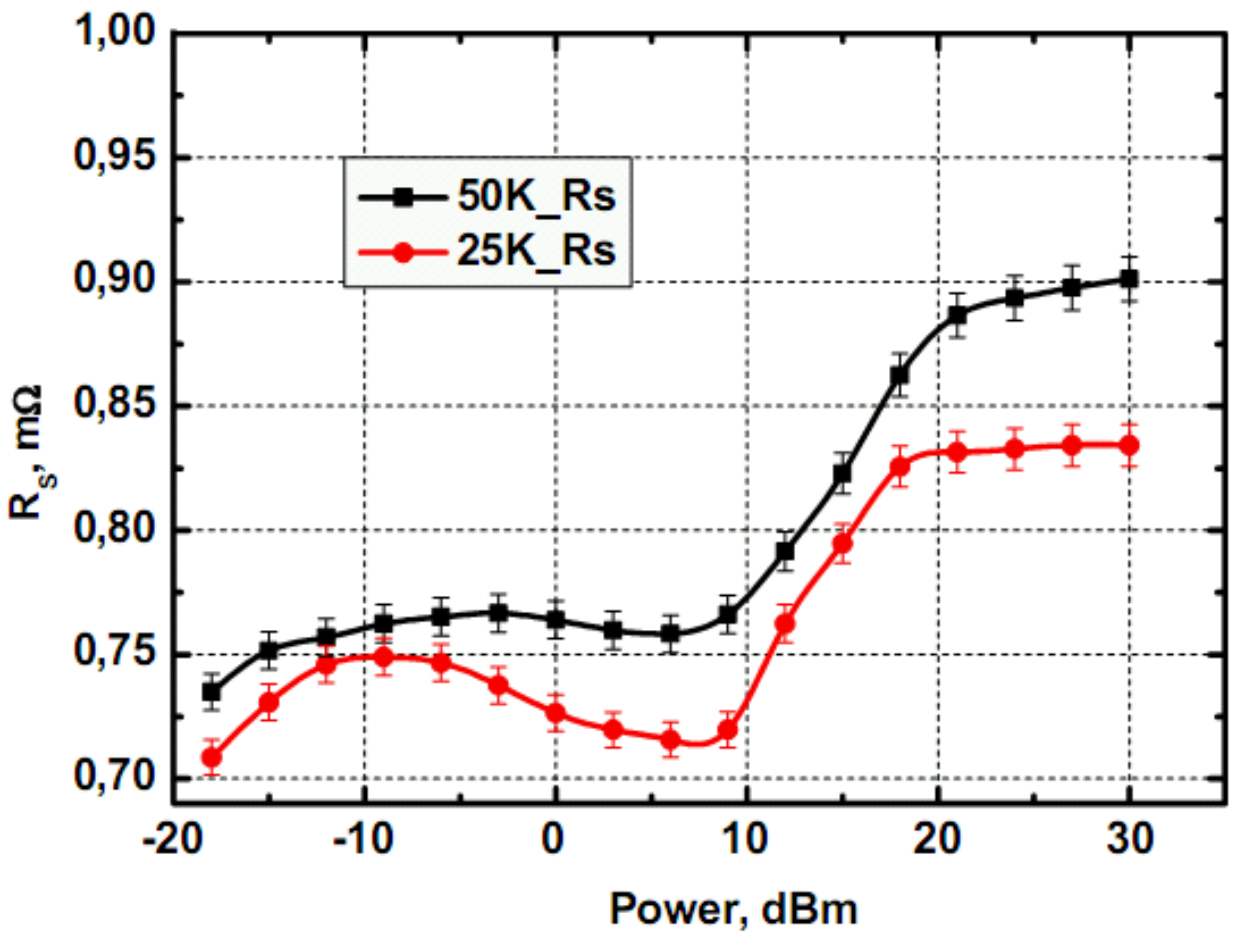

**Fig. 92.** Microwave power dependence of surface resistance Rs(P) of $YBa_2Cu_3O_{7-\delta}$ thin films on MgO substrate at 25GHz at T = 25 K, 50 K (samples 1-3) [84].

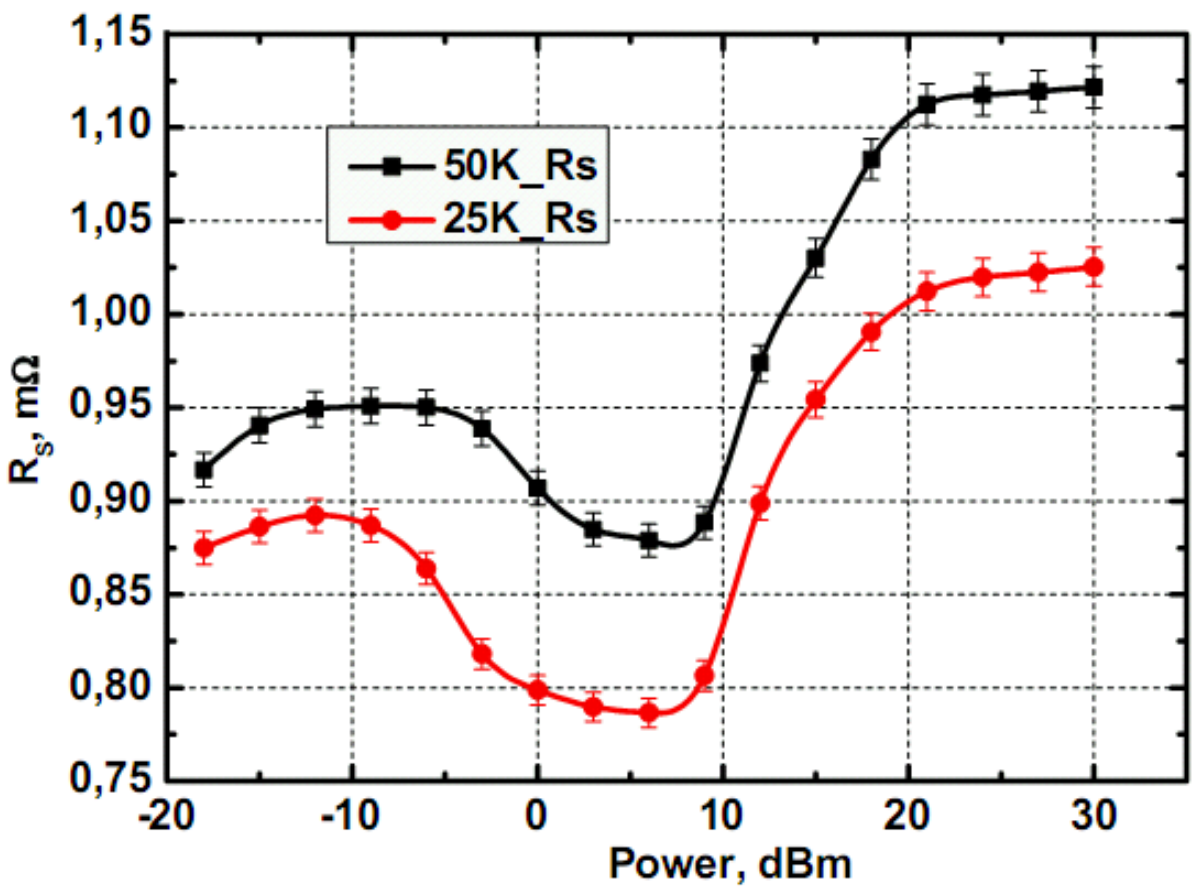

**Fig. 93.** Microwave power dependence of surface resistance Rs(P) of $YBa_2Cu_3O_{7-\delta}$ thin films on MgO substrate at 25GHz at T = 25 K, 50 K (samples 2-3) [84].

**Accurate measurements of surface resistance Rs of $YBa_2Cu_3O_{7-\delta}$ superconductor thin films on MgO substrate as a function of microwave power Rs(P) were conducted utilizing the measurement system depicted in Fig. 62. The additional microwave power amplifier, connected to the cable, leading to the input port of microwave resonator under test inside the vacuum dewar, enabled measurements from - 18dBm to + 30dBm microwave signal amplitude level. It should be mentioned that at higher microwave power levels, the attenuator was placed at output port of microwave resonator to prevent the high microwave power signals entering the input port of vector network analyser system. The attenuator had 20dBm attenuation characteristics. The amplifier had a maximum of +20dBm to +25dBm microwave signal amplification range, and enabled the measurements from 0dBm to the maximum range in our experiments +30 dBm. Appropriate scale modifications were made to the output signal in order to numerically compensate the attenuation effect.**

# Microwave Power Dependence of Surface Resistance $R_S(P)$ of $YBa_2Cu_3O_{7-\delta}$ Superconducting Thin Films on MgO Substrates in a Dielectric Resonator at Ultra High Frequencies.

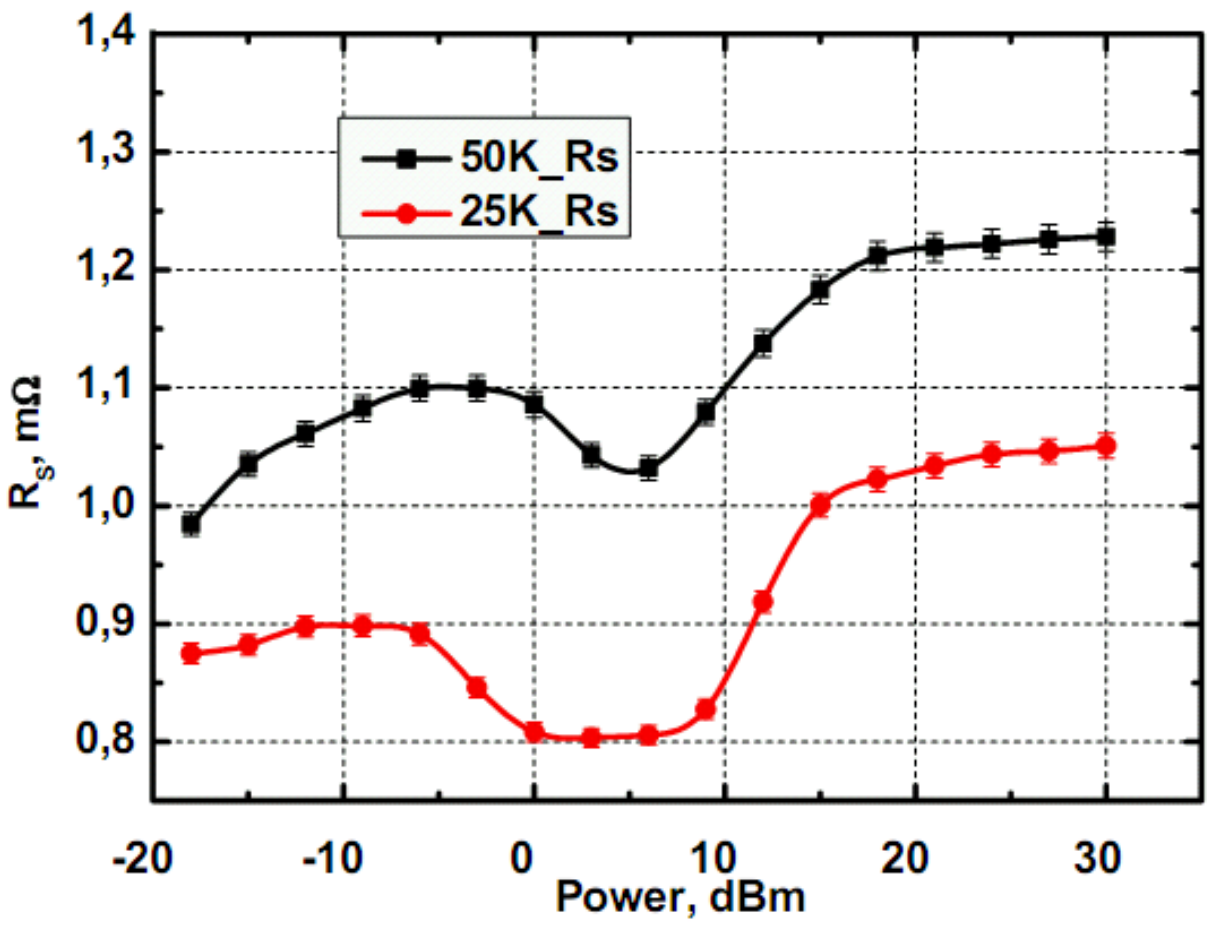

**Fig. 94.** Microwave power dependence of surface resistance Rs(P) of $YBa_2Cu_3O_{7-\delta}$ thin films on MgO substrate at 25GHz at T = 25 K, 50 K (samples 4-5) [84].

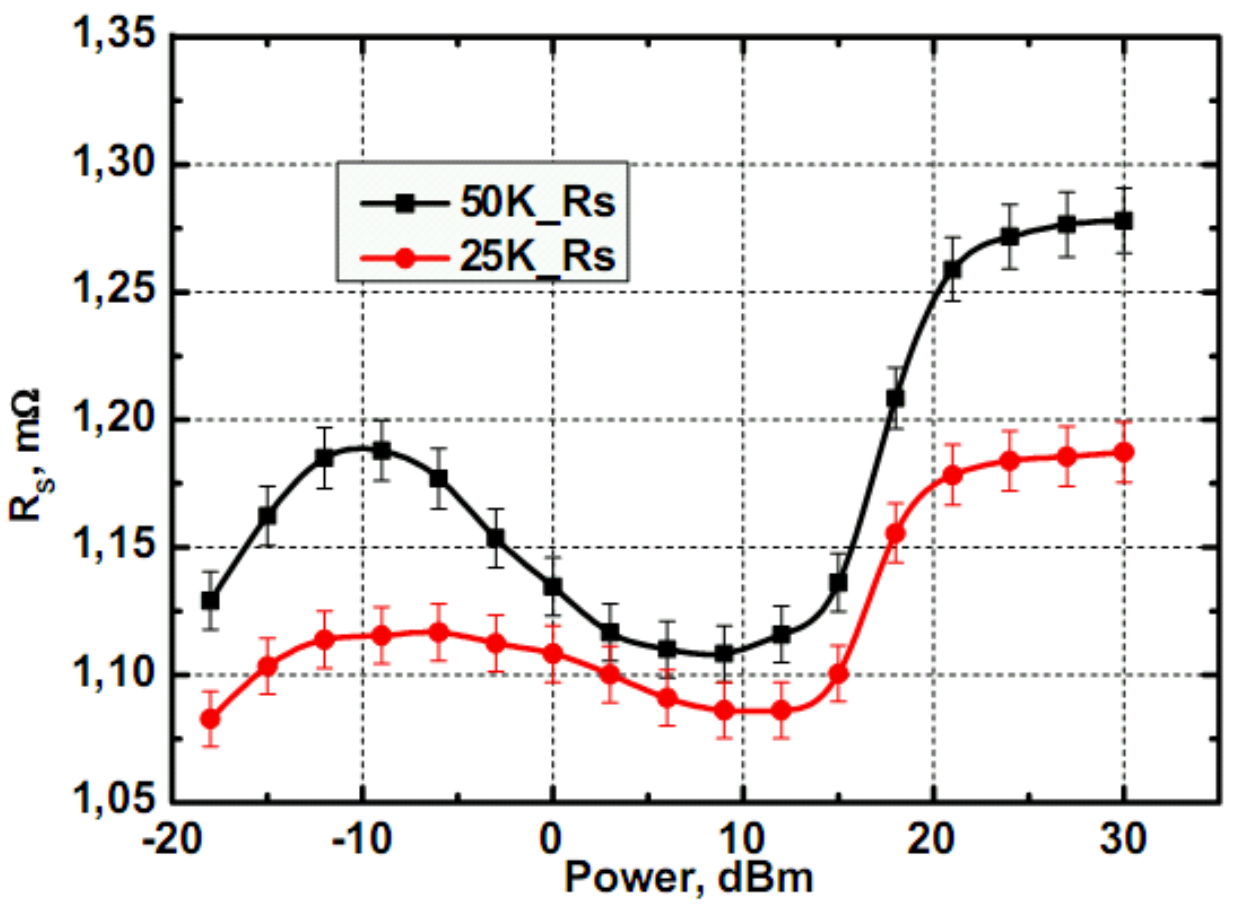

**Fig. 95.** Microwave power dependence of surface resistance Rs(P) of $YBa_2Cu_3O_{7-\delta}$ thin films on MgO substrate at 25GHz at T = 25 K, 50 K (samples 4-6) [84].

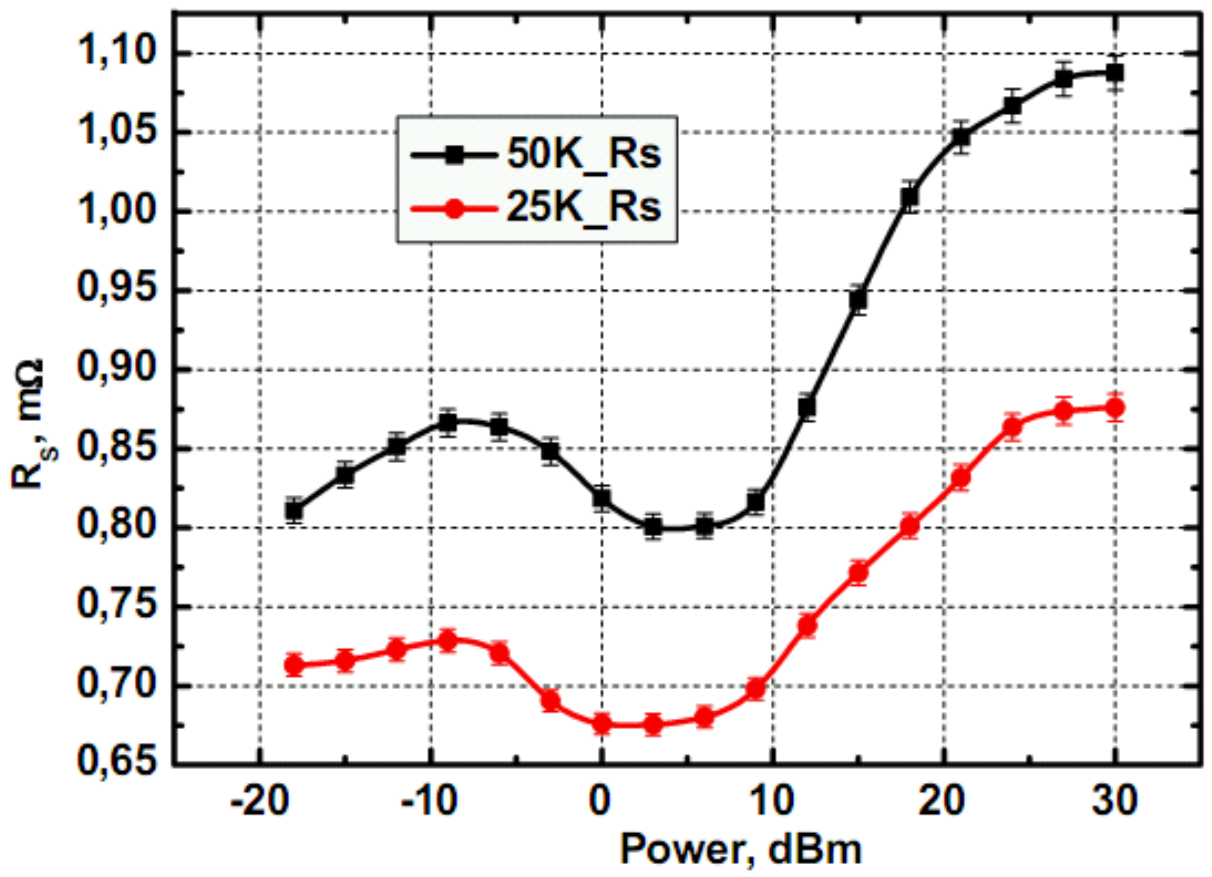

**Fig. 96.** Microwave power dependence of surface resistance Rs(P) of $YBa_2Cu_3O_{7-\delta}$ thin films on MgO substrate at 25GHz at T = 25 K, 50 K (samples 5-6) [84].

**Accurate measurements of surface resistance Rs of $YBa_2Cu_3O_{7-\delta}$ superconductor thin films on MgO substrate as a function of microwave power Rs(P) were conducted utilizing the measurement system depicted in Fig. 62. The additional microwave power amplifier, connected to the cable, leading to the input port of microwave resonator under test inside the vacuum dewar, enabled measurements from - 18dBm to + 30dBm microwave signal amplitude level. It should be mentioned that at higher microwave power levels, the attenuator was placed at output port of microwave resonator to prevent the high microwave power signals entering the input port of vector network analyser system. The attenuator had 20dBm attenuation characteristics. The amplifier had a maximum of +20dBm to +25dBm microwave signal amplification range, and enabled the measurements from 0dBm to the maximum range in our experiments +30 dBm. Appropriate scale modifications were made to the output signal in order to numerically compensate the attenuation effect.**

# Microwave Measurement Accuracy.

**The accuracy of measured results is an important issue to consider, when conducting novel research projects. In order to identify the range of acceptable variations in the obtained data, the theoretical analysis is made in reference to the equations and parameters related to the obtained experimental results. The following final formula was derived during the theoretical analysis (see the thesis):**

$$\left|\frac{\Delta R_S}{R_S}\right| \approx \left[\left(\frac{A_S}{R_S}\right)^2\left(\frac{R_m}{A_m}\right)^2\left|\frac{\Delta A_S}{A_S}\right|^2 + \left(\frac{R_m A_S}{R_S A_m}\right)^2\left(\left|\frac{\Delta R_m}{R_m}\right|^2 + \left|\frac{\Delta A_S}{A_S}\right|^2\right)\right]^{1/2}$$
$$= \left[\left(\frac{R_m A_S}{R_S A_m}\right)^2\left(\left|\frac{\Delta A_S}{A_S}\right|^2 + \left|\frac{\Delta R_m}{R_m}\right|^2 + \left|\frac{\Delta A_S}{A_S}\right|^2\right)\right]^{1/2} \qquad (6.14)$$

**The biggest error will be in the case, when all the elements values are added, having taken them by modulo. It is clear that the main error and uncertainty in the Rs magnitude depends on the ΔAm and ΔAs (their values are expressed in ohms – same as for the R), i.e. on the geometrical factors of normal metal and superconductor, which are determined by the geometry of both the sample and the resonator; and it does not depend on the signal level at first approximation. It is assumed that the geometrical factors do not depend on the microwave power or other parameters of the formula at certain temperatures. Then, the absolute value of the correction does not depend on microwave power either and is equal to the total value of the absolute corrections. From the analysis, it is evident that the error values will not have practical grounds to exceed a certain range that is approximately equal to around 1% in our case. The error band limits at 1% are shown in Figs. 91 - 95.**

# Experimental and Theoretical Researches on Nonlinear Surface Resistance of $YBa_2Cu_3O_{7-\delta}$ Thin Films on MgO Substrates in Superconducting Microstrip Resonators at Ultra High Frequencies.

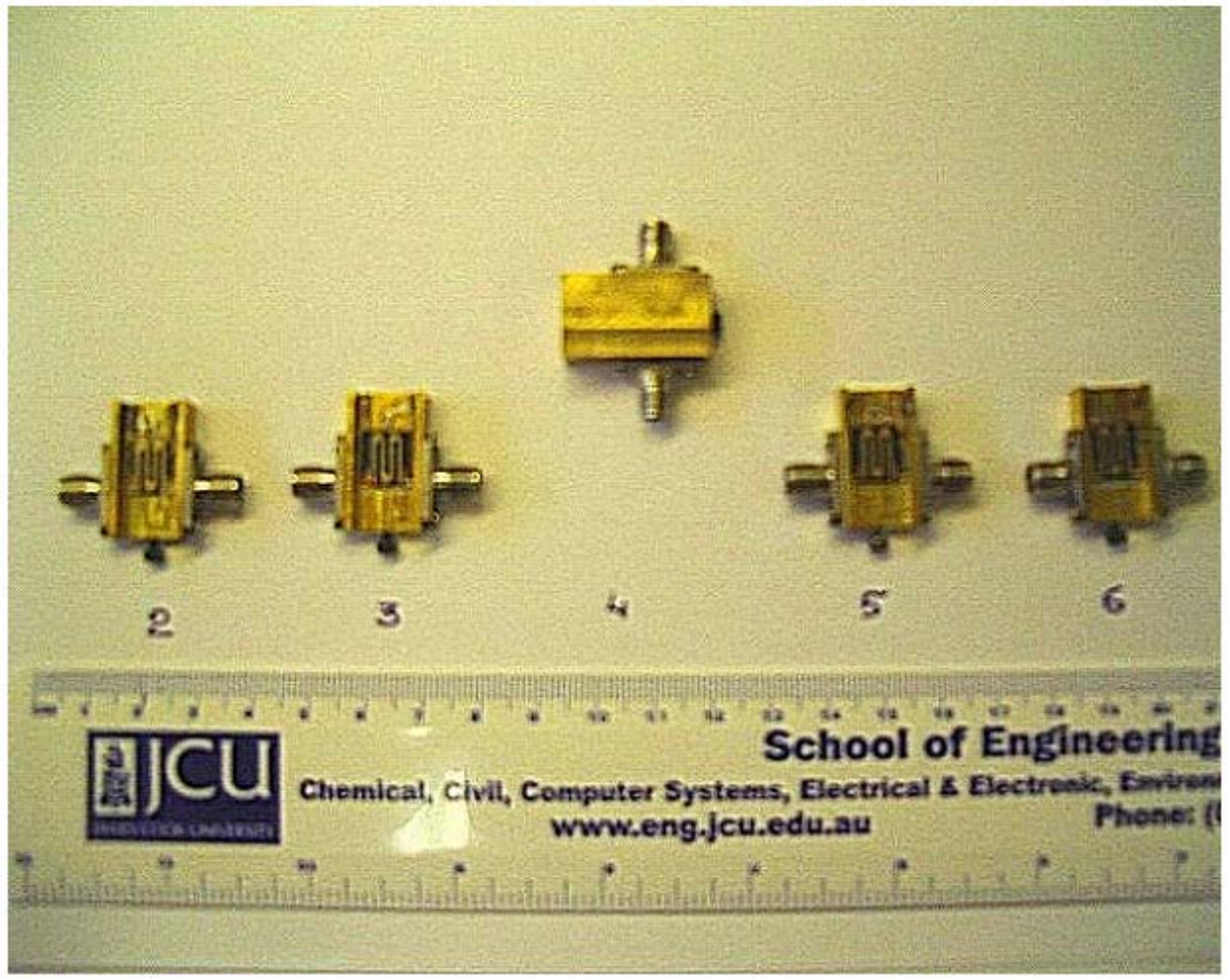

**Fig. 97.** Researched microstrip resonators with resonance frequency $f_0$ = 1.985GHz made from $YBa_2Cu_3O_{7-\delta}$ thin films on MgO substrate.

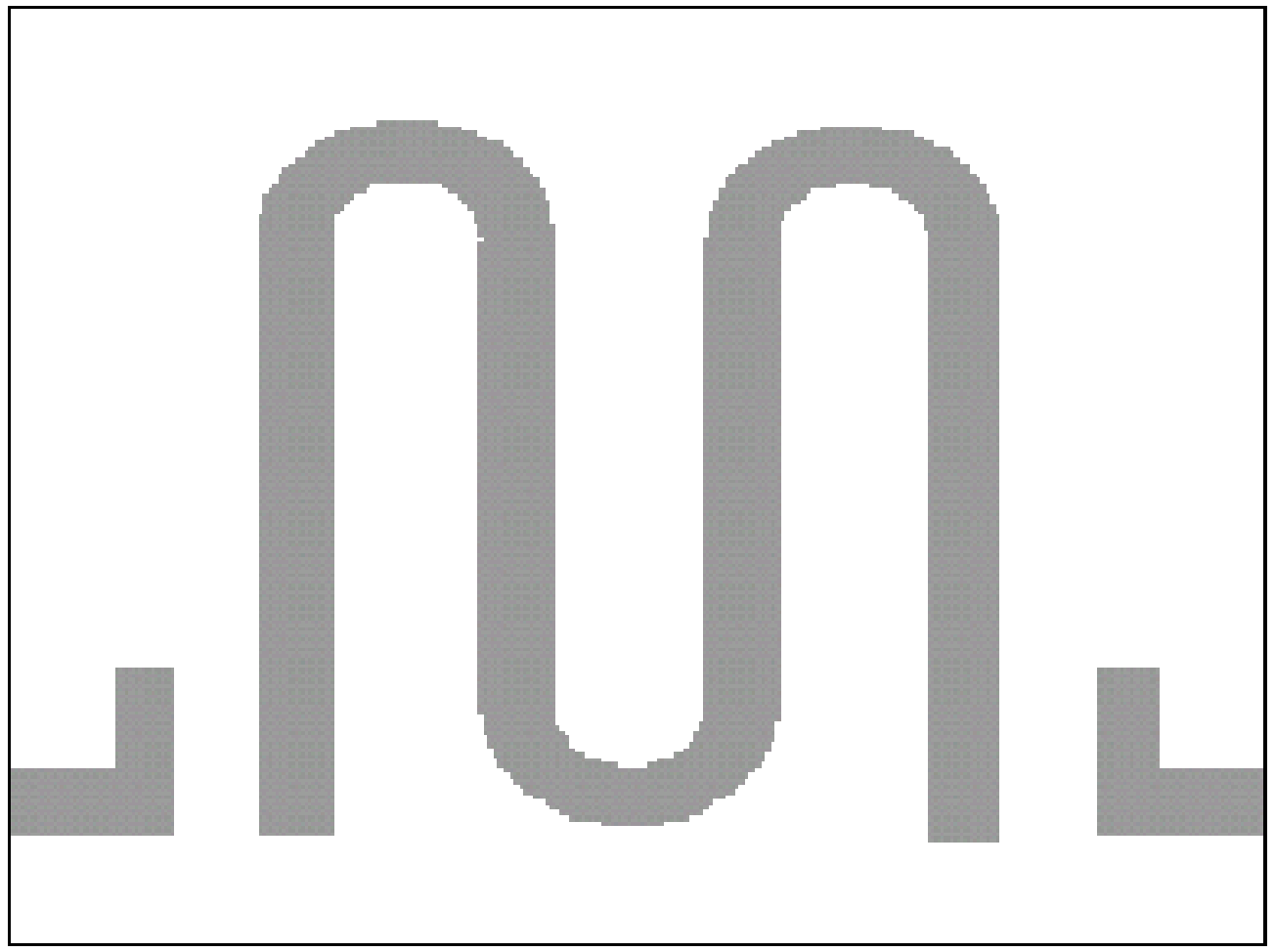

**Fig. 98.** Geometric form of microstrip resonators with resonance frequency $f_0$ = 1.985GHz made of $YBa_2Cu_3O_{7-\delta}$ thin films on MgO substrate researched at Department of Electrical and Computer Engineering at James Cook University in Australia.

**The geometrical form of microstrip resonators with the resonance frequency $f_0$ =1.985GHz is shown in Figs 97 and 98. The microstrip resonators were manufactured from $YBa_2Cu_3O_{7-\delta}$ high-temperature superconductor thin films by Theva Gmbh in Germany, and researched by the author of dissertation at Department of Electrical and Computer Engineering, James Cook University in Australia. The thickness of $YBa_2Cu_3O_{7-\delta}$ thin film was equal to 700 nm and its width was 0.49 mm. The film was deposited on a substrate with the width 0.5 mm made from MgO. The size of a microstrip resonator was approximately 10x8.5 $mm^2$.**

# Elementary Standing Half-Wave Microwave Resonator.

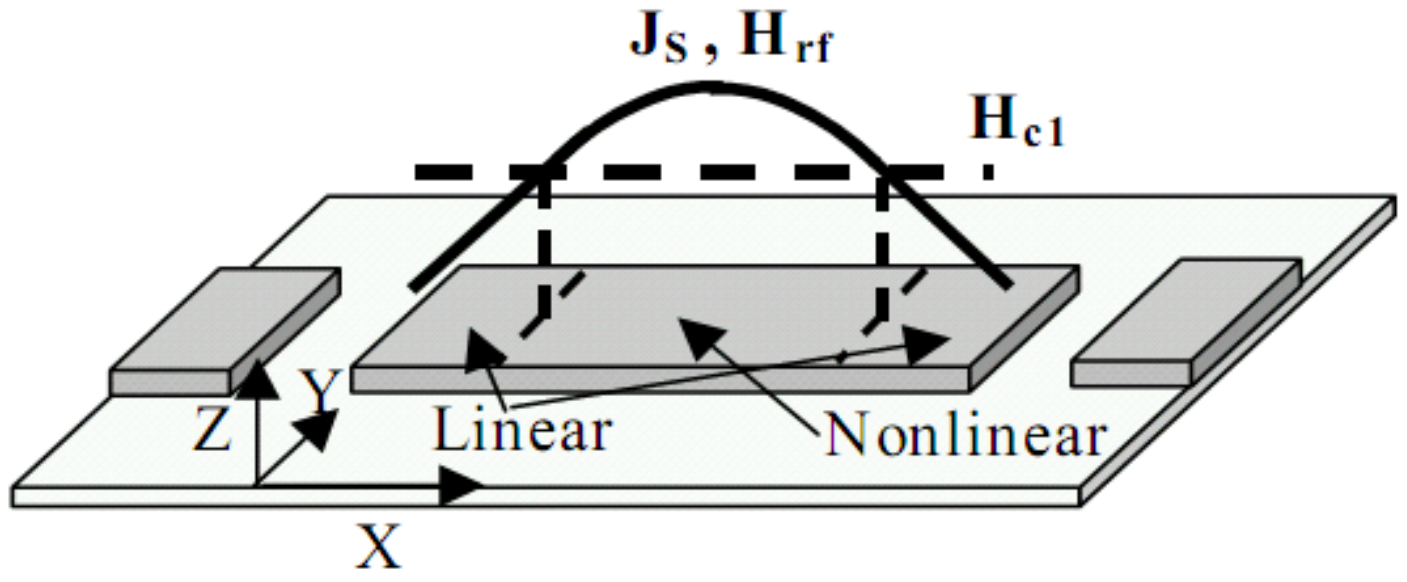

**Fig. 99.** Standing half-wave microstrip resonator and distribution of current density along resonating element.

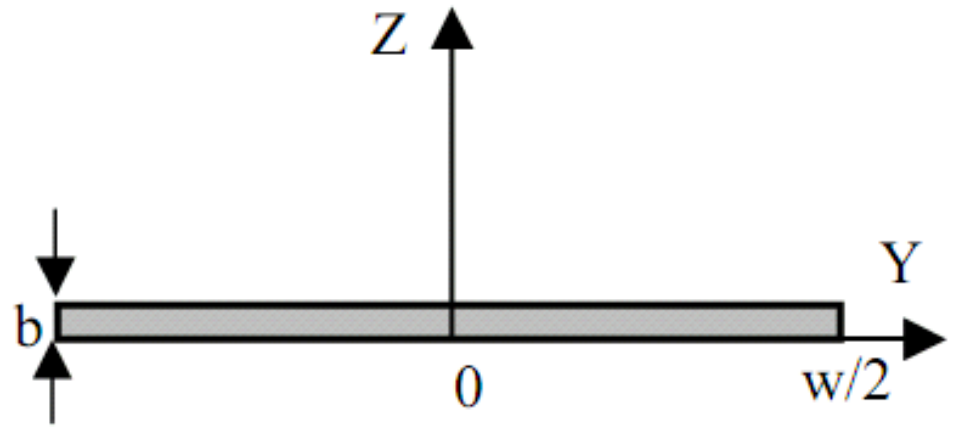

**Fig. 100.** Cross section of HTS thin film used in microstrip resonator. Superconducting current flows toward the direction of X axis.

**The current density distribution along microstrip resonator:**

$$J_X = J_{max} \sin(2\pi x / \lambda) \sin(\omega t), \qquad (7.1)$$

**The expression for the average magnitude of surface resistance $R_S$:**

$$\langle R_S(H_{rf}) \rangle = \frac{\int_0^{\pi/2}\int_0^{\lambda/4} R_S(t,x)\,dt\,d\ell}{\pi/2} = \frac{\int_0^{\pi/2}\int_0^{\lambda/4} \left(R_{SLin}(x,t) + R_{SNlin}(x,t)\right) d\ell\, dt}{\pi/2}, \qquad (7.2)$$

where $R_{SNlin}$ is the nonlinear surface resistance.

**The current density Js near to the center of HTS thin film:**

$$J_S(y) = \frac{J_S(0)}{\sqrt{1-(2y/w)^2}}, \qquad (7.3)$$

where y is counted from the centre of the film, w is the width of HTS thin film, the thickness of a film is equal to b.

**The current density Js near to the edges of HTS thin film:**

$$J_S(y) = J_S(0)\left(\frac{1.165}{\lambda}\right)\left(\frac{wb}{a}\right)^{1/2} \exp\left(-\frac{\left(\frac{w}{2}-|y|\right)b}{a\lambda^2}\right), \qquad (7.4)$$

where "a" is the constant ~ 1, λ is the penetration depth of magnetic field.

**In the computer simulation and modeling of nonlinear effects in superconducting microstrip resonators, the microstrip resonators are usually considered as the elementary standing half-wave microwave resonators, in which the superconducting sample looks like a linear plane unit. The microstrip resonator made of HTS thin film with composite form can easily be transformed into this relatively simple half-wave straight lines microwave resonator and modeled by an equivalent circuit during the mathematical modeling [85].**

# Modeling of Current Density Distribution on Cross Section of Superconducting Microstrip Resonator with Resonance Frequency $f_0$=1.985GHz Made of $YBa_2Cu_3O_{7-\delta}$ Thin Films on MgO substrate.

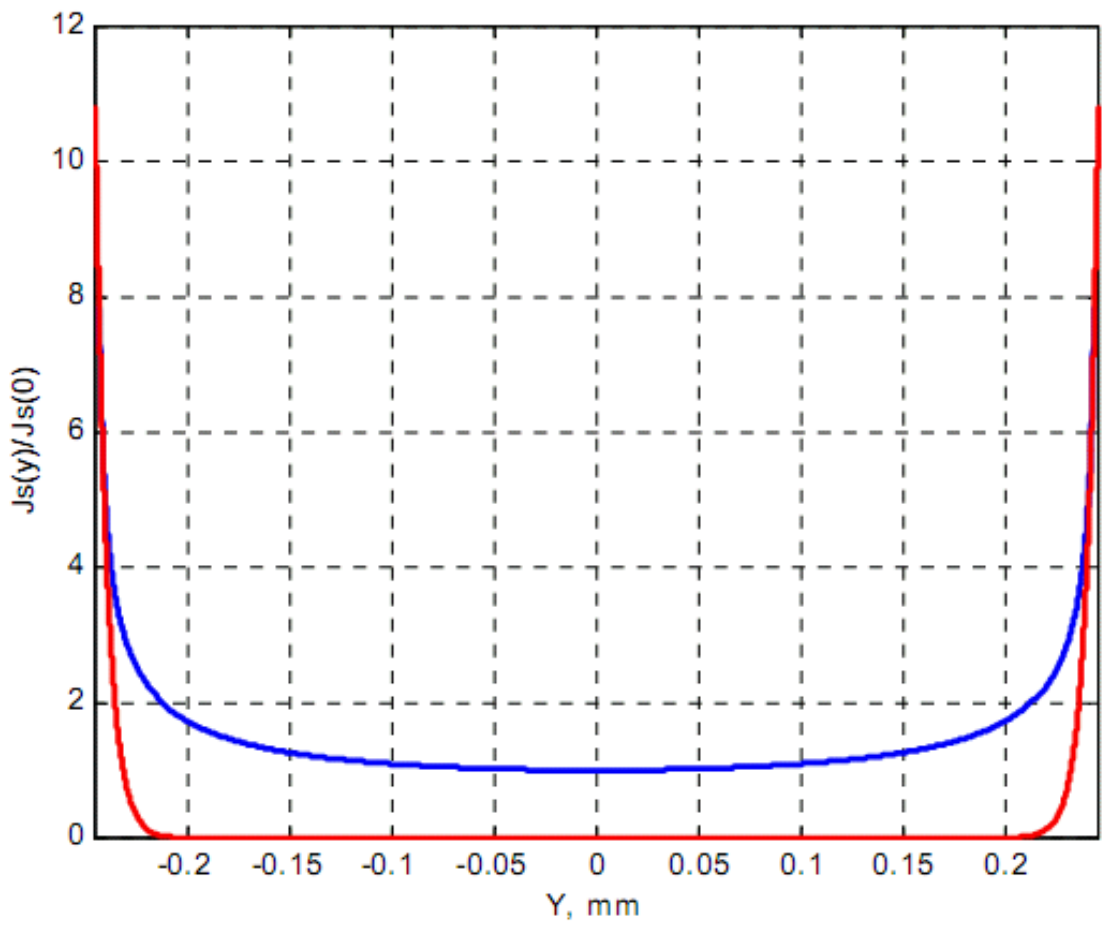

**Fig. 101.** Modeling of current density distribution on cross section of superconducting microstrip resonator with resonance frequency f0=1.985GHz made of $YBa_2Cu_3O_{7-\delta}$ thin films on MgO substrate [85]. The upper curve corresponds to the expression (7.3). The lower curve concerns the formula (7.4) and features the allocation of superconducting current near to the edges of HTS thin film, when $y \approx \pm w/2$. The first curve passes to the second curve in the points:

$$y = \pm\left(\frac{w}{2} - \frac{a\lambda^2}{2b}\right).$$

**The example of Js modeling in stripline done by D.E. Oates et al. [86]:**

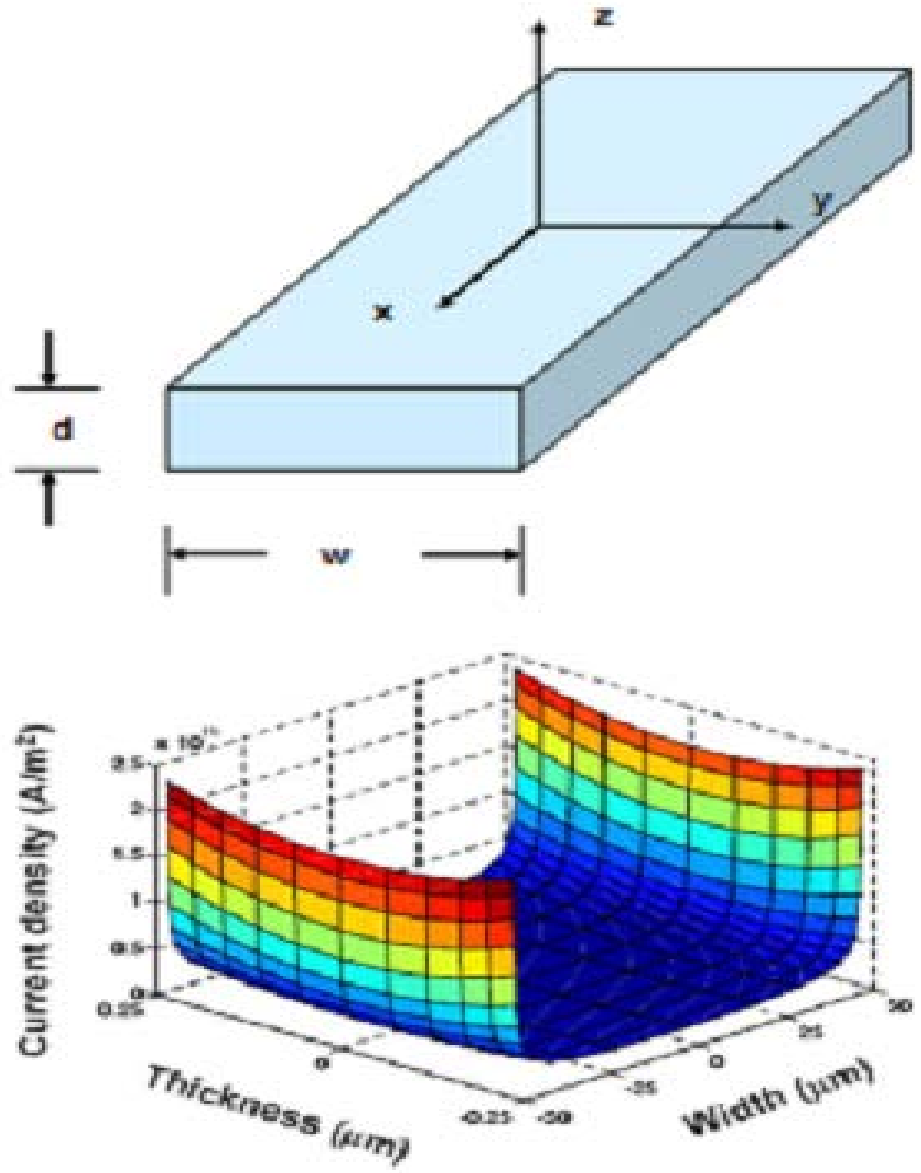

**Fig. 102. (a)** The choice of coordinate system and definition of the width w and thickness d parameter in a generic thin film strip w>>d. **(b)** An example of a numerically exact calculation of the pair-current density Js(y, z) for a typical strip, where w=100 μm, d=500nm, and $\lambda_0$=250nm. The circulating current is 0.026mA (after [86]).

# Ultra High Frequency and Microwave Power Dependences of Transmission Coefficient $S_{21}(f, P)$ in $YBa_2Cu_3O_{7-\delta}$ Superconducting Microstrip Resonator at Microwaves at Different Temperatures.

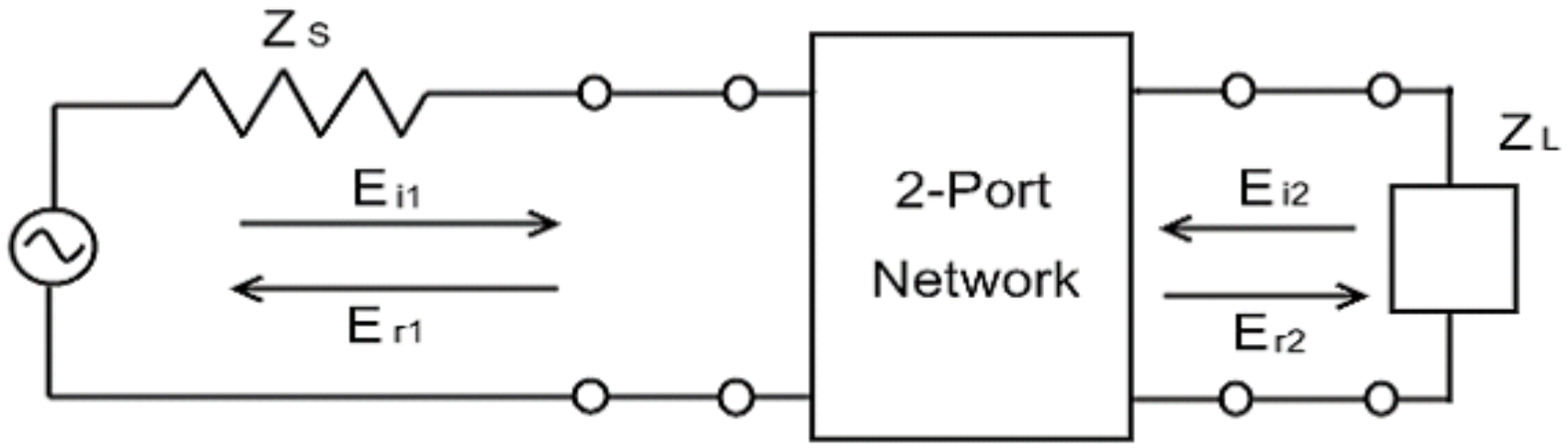

$$S_{21} = \frac{b_2}{a_1}\Big|_{a_2=0} = \frac{E_{r2}}{E_{i1}}.$$

**Fig. 103.** Model of two-port network inserted into transmission line with incident and reflected electromagnetic waves schematically shown at microwaves.

The **S-parameters** are widely used for the characterization of microwave devices in microwave circuits in micro- and nano-electronics [87]:
$S_{11}$ = **input reflection coefficient** with the output matched.
$S_{21}$ = **forward transmission coefficient** with the output matched.
$S_{22}$ = **output reflection coefficient** with the input matched.
$S_{12}$ = **reverse transmission coefficient** with the input matched.
The S-parameters are important, because of the several reasons [29]:
**1.** S-parameters are determined with resistive terminations. This obviates the difficulties involved in obtaining the broadband open and short circuit conditions required for the H, Y, and Z-parameters.
**2.** Parasitic oscillations in active devices are minimized, when these devices are terminated in resistive loads.
**3.** Equipment is available for determining S-parameters since only incident and reflected voltages need to be measured.

# Ultra High Frequency and Microwave Power Dependences of Transmission Coefficient $S_{21}(f, P)$ in $YBa_2Cu_3O_{7-\delta}$ Superconducting Microstrip Resonator at Microwaves at Different Temperatures.

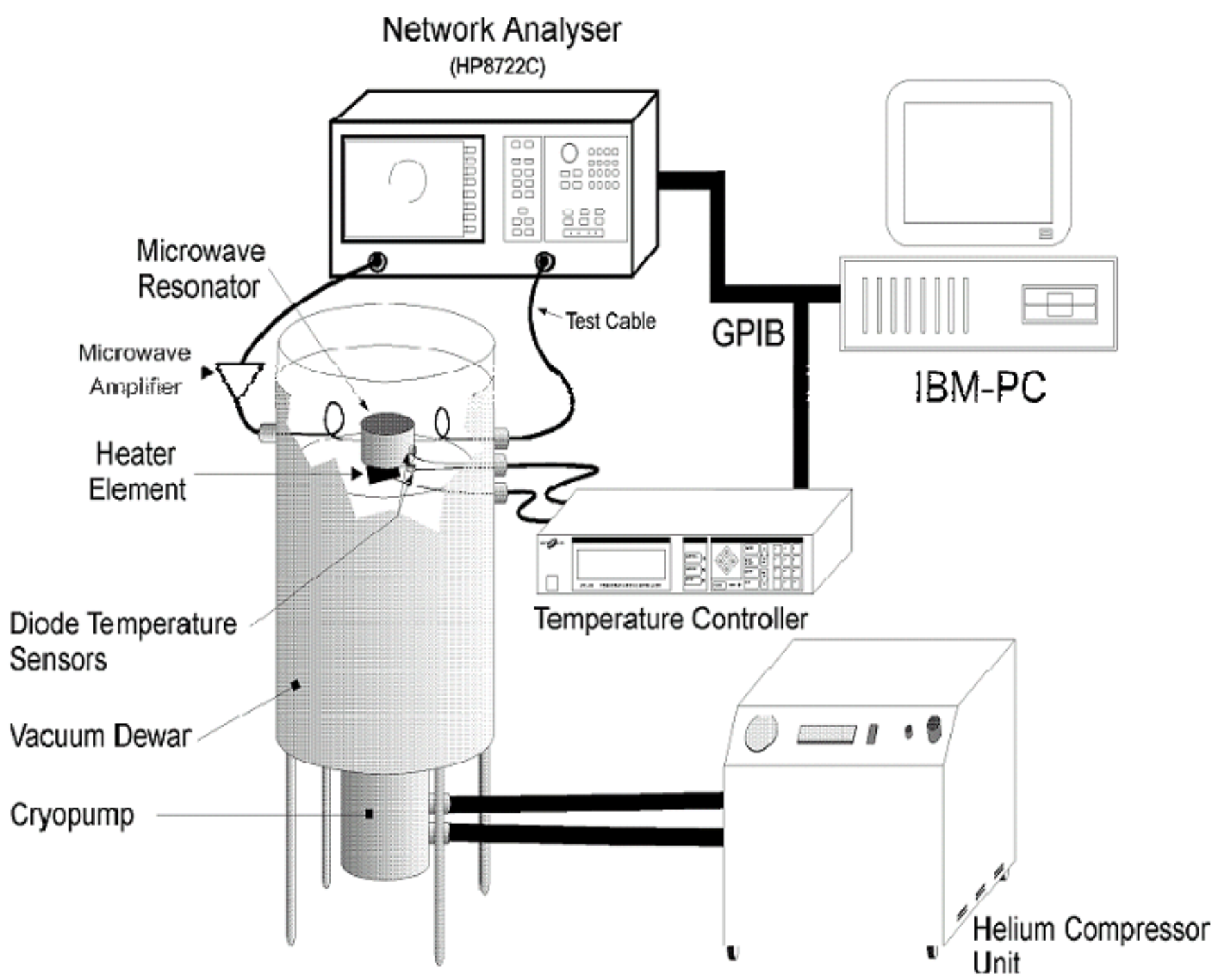

**Fig. 104**. Cryogenic measurement setup for research on nonlinear resonance response of $YBa_2Cu_3O_{7-\delta}$ thin films on MgO substrates in microstrip resonators at microwaves.

The ultra high frequency experimental measurement system for research on nonlinear resonance response of $YBa_2Cu_3O_{7-\delta}$ thin films on MgO substrates in a dielectric resonator at various levels of microwave power consisted of the following equipment:

- **Vector Network Analyser** (HP 8722C).
- **Microwave Amplifier.**
- **Temperature Controller** (Conductus LTC-10) fitted with a resistive heating element and two silicon temperature diode sensors.
- **Vacuum Dewar.**
- **Close Cycle Cryogenic Laboratory System** (APC-HC4) suitable for measurements in a wide range of temperatures (10 K–300 K).
- **Computer System** (IBM-PC) fitted with a GPIB card utilized for the computer control of temperature controller and network analyser, and S-parameter measurement data transfer from network analyser to computer.

The measurement set up was precisely calibrated by measuring the voltage standing wave ratio VSWR=1+ IΓI/1-IΓI, where Γ is the reflection coefficient, before accurate experimental measurements of S-parameters as schematically shown in Fig. 104.

# Experimental Dependences of Forward Transmission Coefficient on Ultra High Frequency $S_{21}(f)$ in $YBa_2Cu_3O_{7-\delta}$ Superconducting Microstrip Resonator at Different Microwave Power Levels and Temperatures.

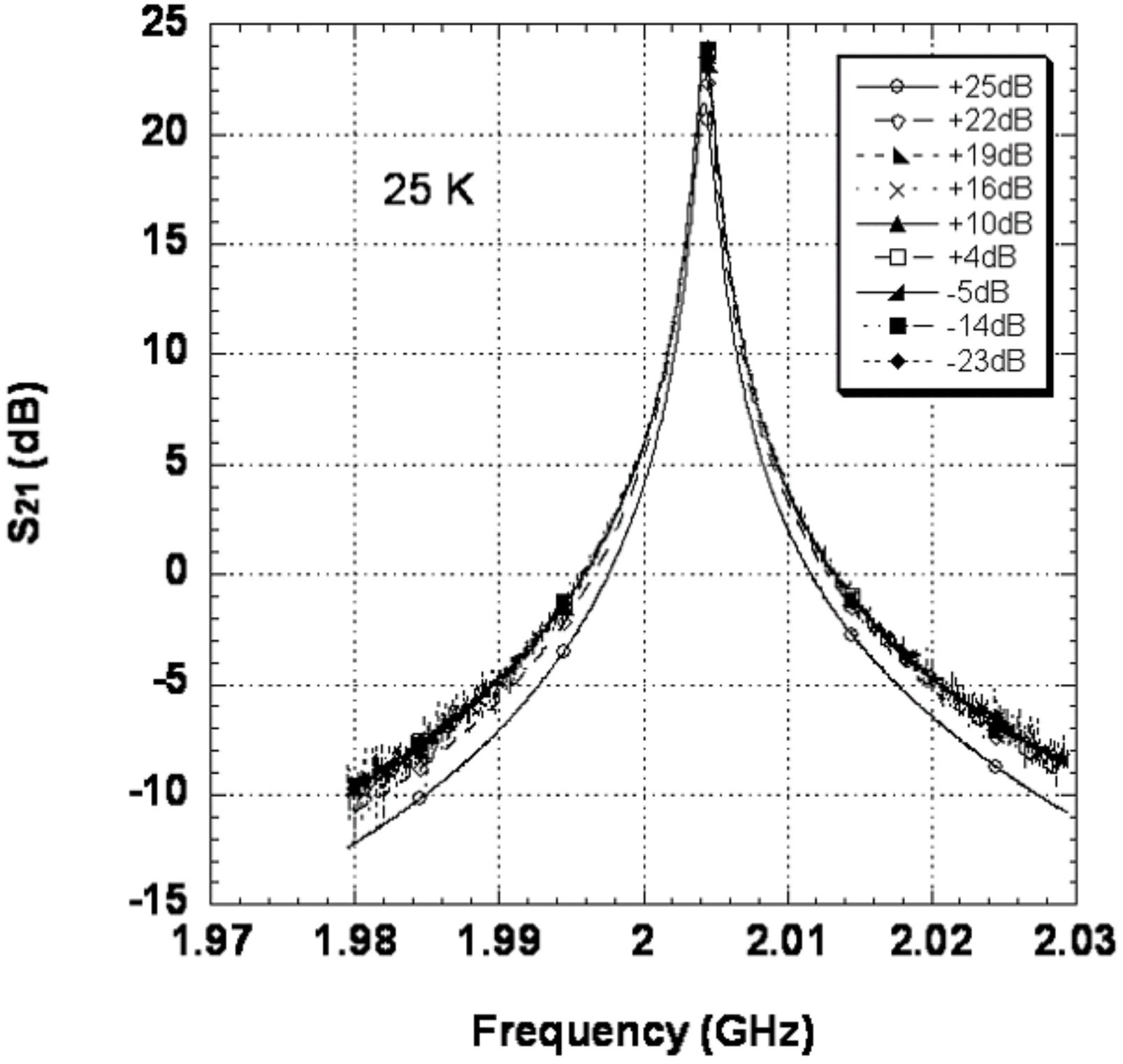

**Fig. 105**. Experimental dependence of forward transmission coefficient on ultra high frequency $S_{21}(f)$ in $YBa_2Cu_3O_{7-\delta}$ superconducting microstrip resonator at different microwave power levels at temperature of 25 K.

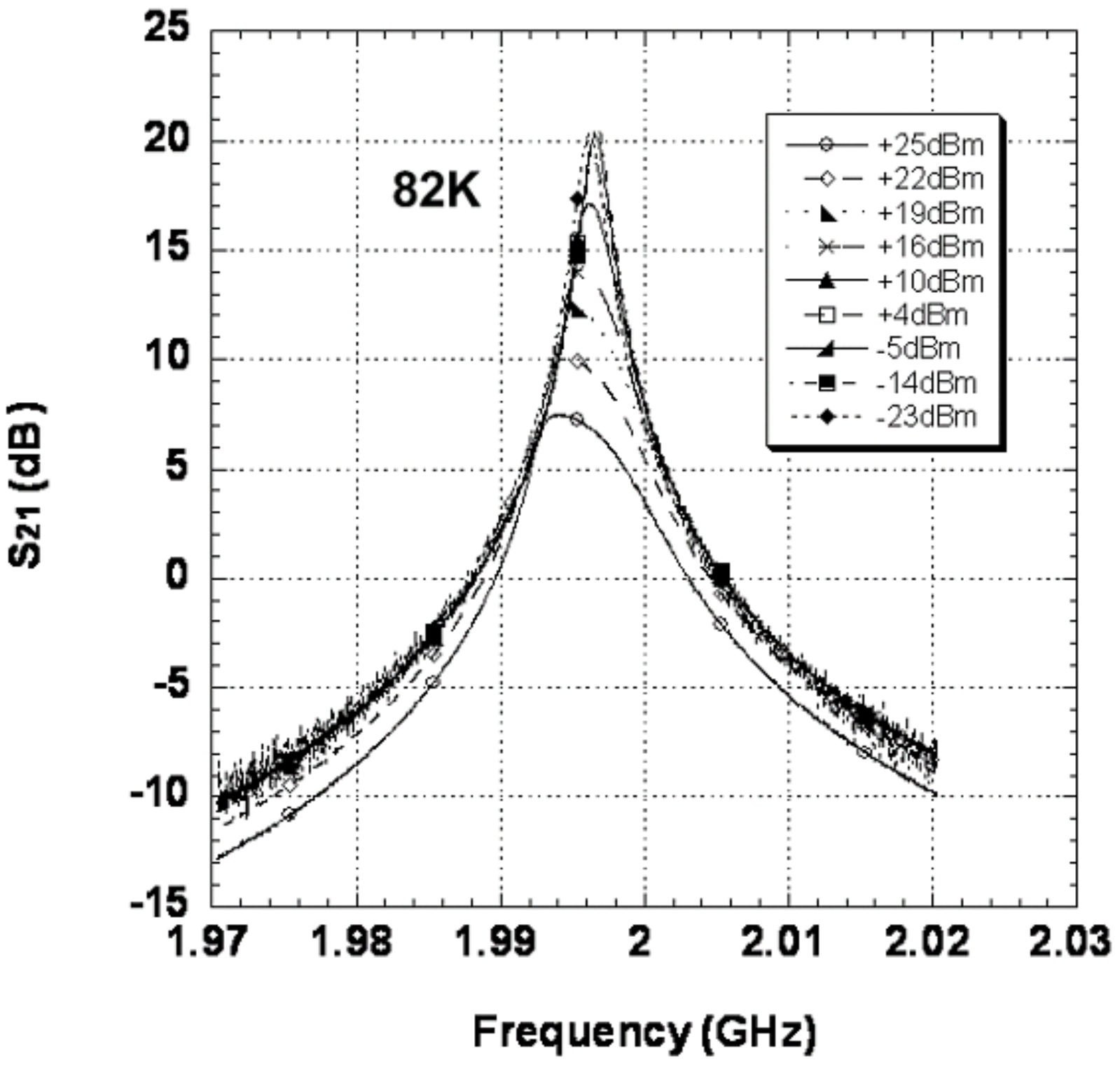

**Fig. 106**. Experimental dependence of forward transmission coefficient on ultra high frequency $S_{21}(f)$ in $YBa_2Cu_3O_{7-\delta}$ superconducting microstrip resonator at different microwave power levels at temperature of 82 K.

# Shift of Resonant Frequency $f_0$, Change of Quality Factor Q, and Change of Surface Resistance $R_s$, Depending on Applied Microwave Power $H_{rf}$, in $YBa_2Cu_3O_{7-\delta}$ Superconducting Microstrip Resonators at Ultra High Frequencies.

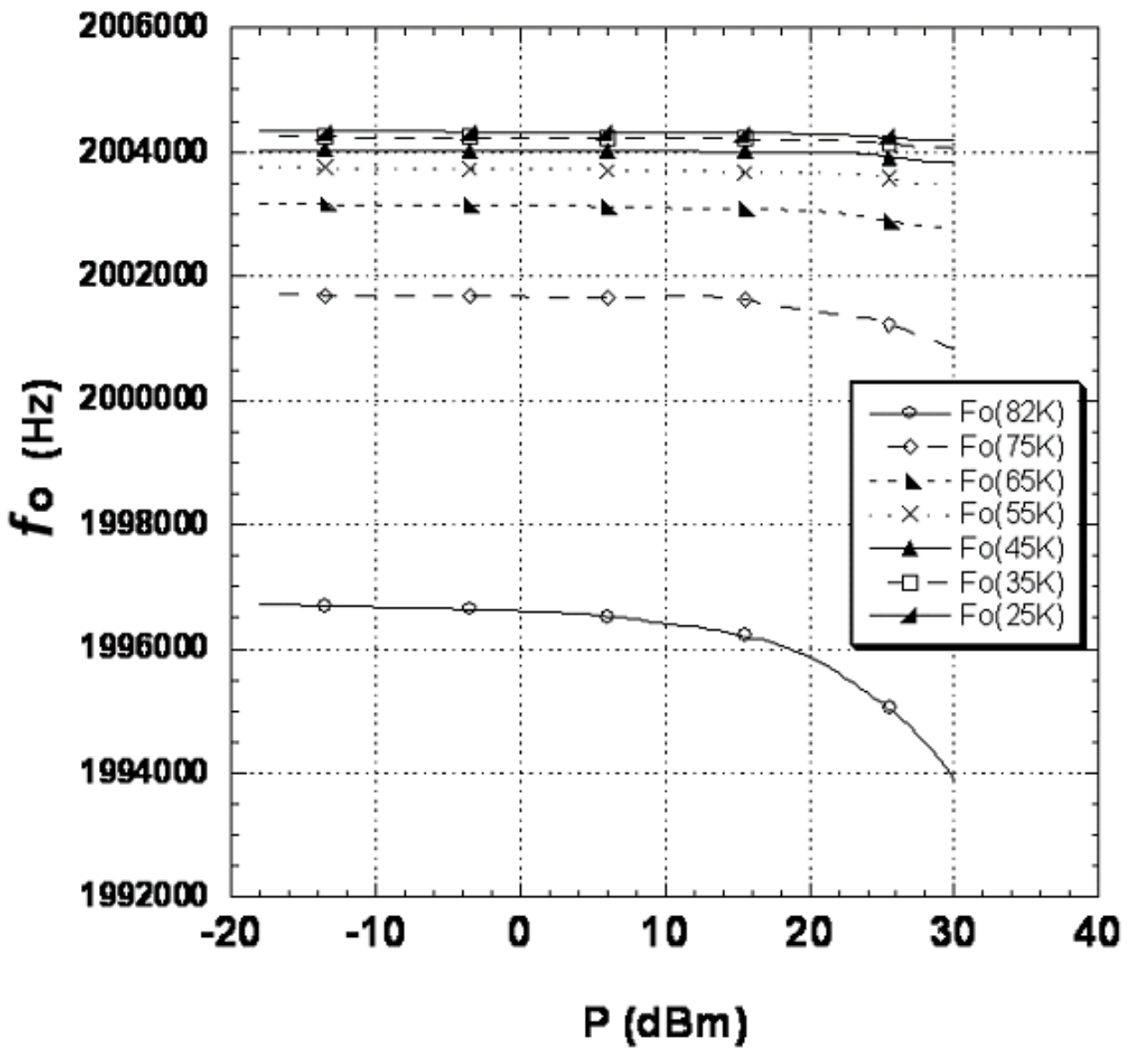

**Fig. 107**. Experimental dependence of resonance frequency on microwave signal power $f_0$ (P) in $YBa_2Cu_3O_{7-\delta}$ superconducting microstrip resonator at microwaves at different temperatures. Shift of resonance frequency is observed.

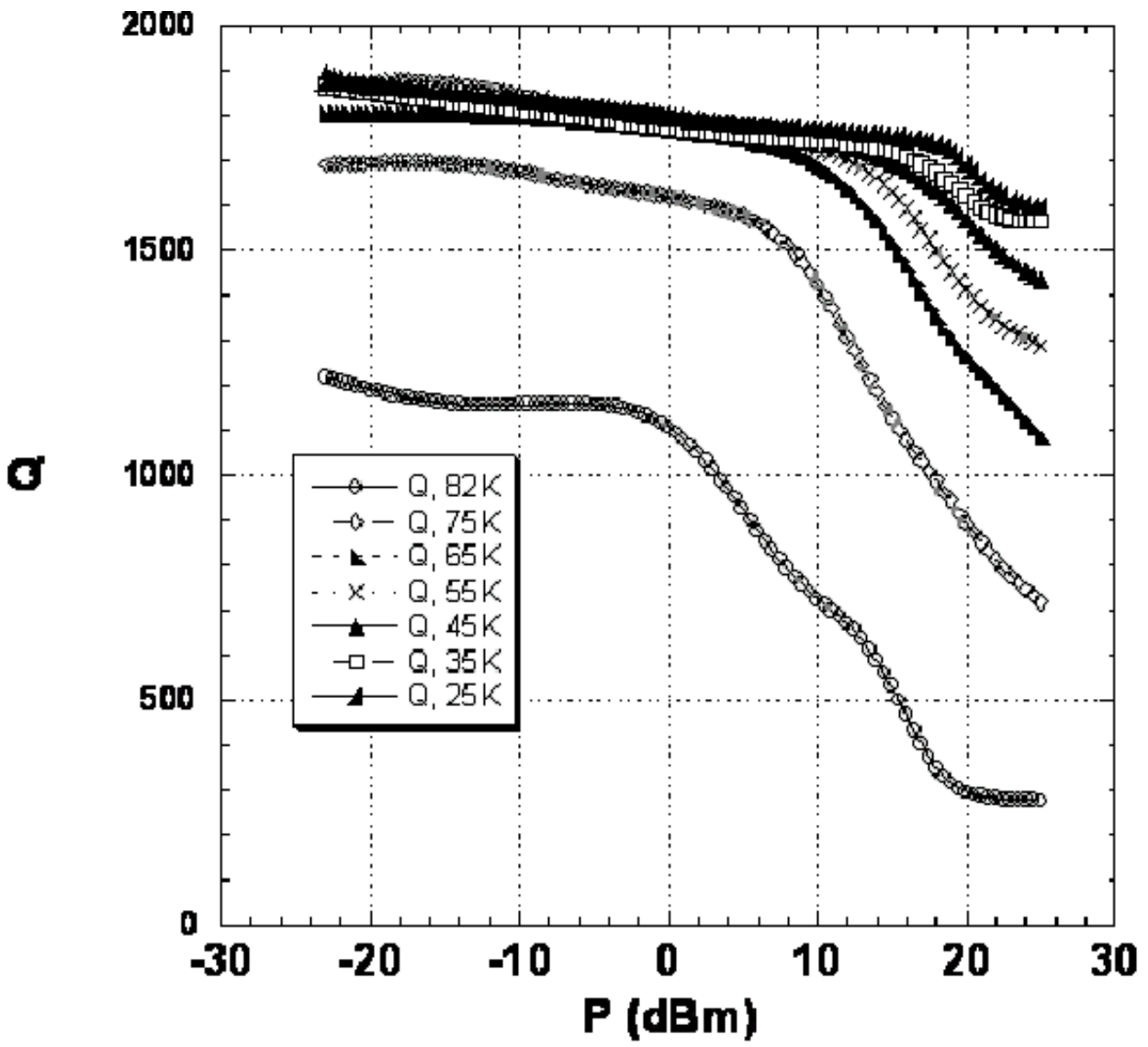

**Fig. 108**. Experimental dependence of quality factor on microwave signal power Q(P) in $YBa_2Cu_3O_{7-\delta}$ superconducting microstrip resonator at frequency f ≈ 2GHz at different temperatures.

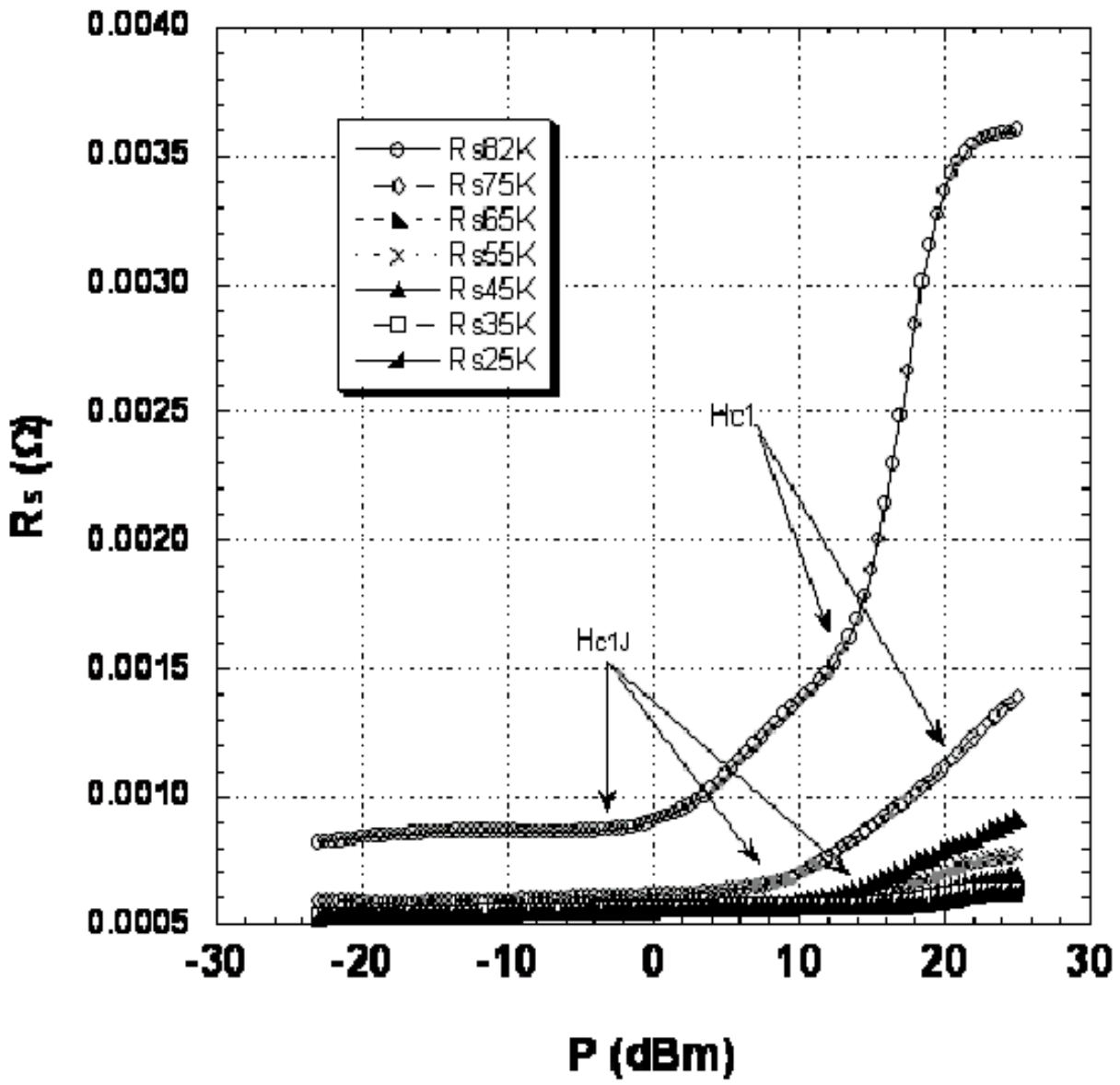

**Fig. 109**. Experimental dependence of surface resistance on microwave signal power $R_S$(P) in $YBa_2Cu_3O_{7-\delta}$ superconducting microstrip resonator at frequency f ≈ 2GHz at different temperatures.

**The nonlinearities of surface resistance Rs in $YBa_2Cu_3O_{7-\delta}$ superconducting microstrip resonator at microwaves are somehow interlinked with some unusual features of physical properties of superconductors appearing near to the critical magnetic fields $Hc_1$ and $Hc_2$ in HTS thin films at microwaves.**

# Shift of Resonant Frequency $f_0$, Change of Quality Factor Q, and Change of Surface Resistance $R_s$, Depending on Applied Microwave Power $H_{rf}$, in $YBa_2Cu_3O_{7-\delta}$ Superconducting Microstrip Resonators at Ultra High Frequencies.

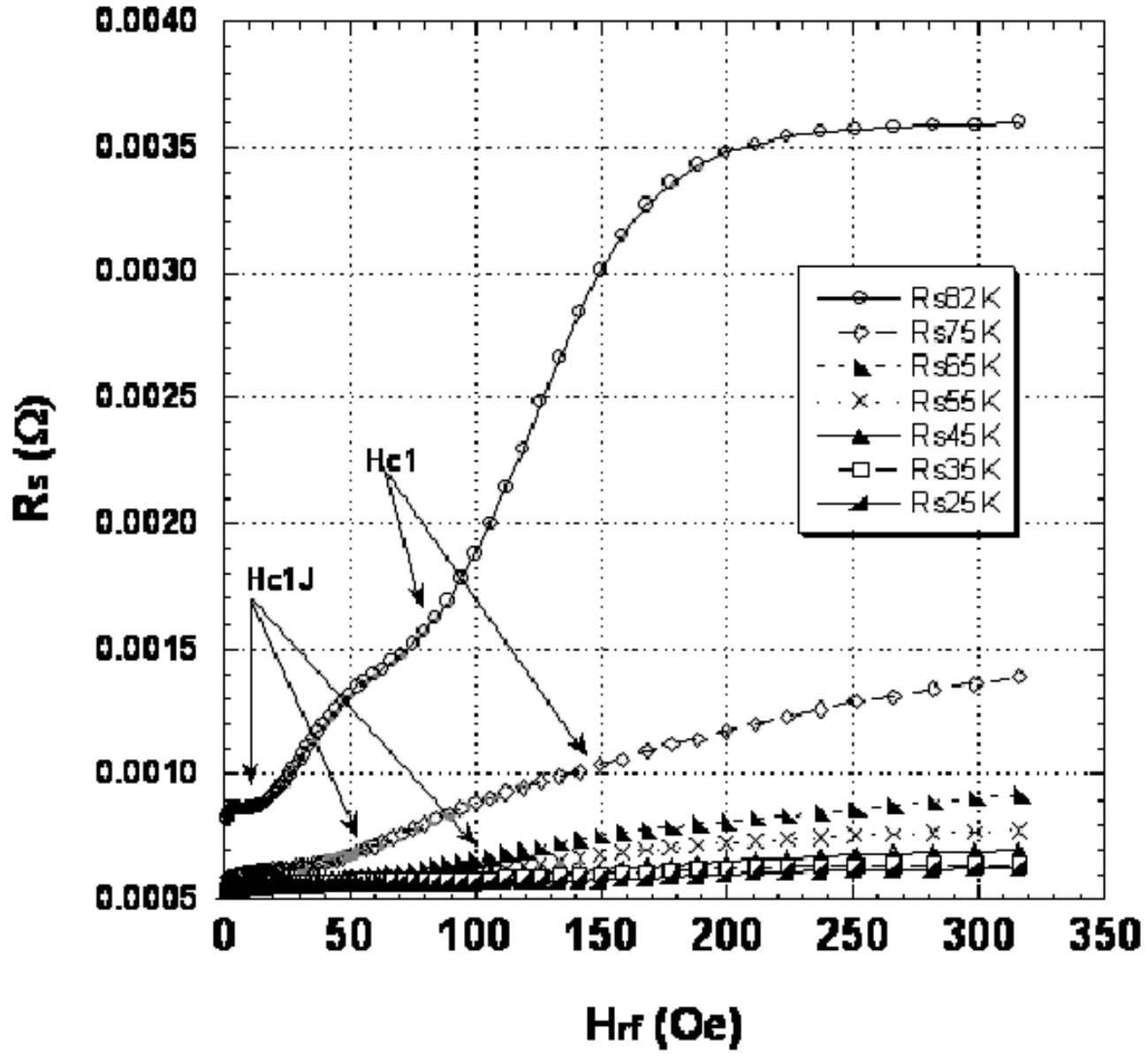

**Fig. 110**. Experimental dependence of surface resistance on magnetic field Rs(Hrf) in $YBa_2Cu_3O_{7-\delta}$ superconducting microstrip resonator at frequency f ≈ 2GHz at different temperatures.

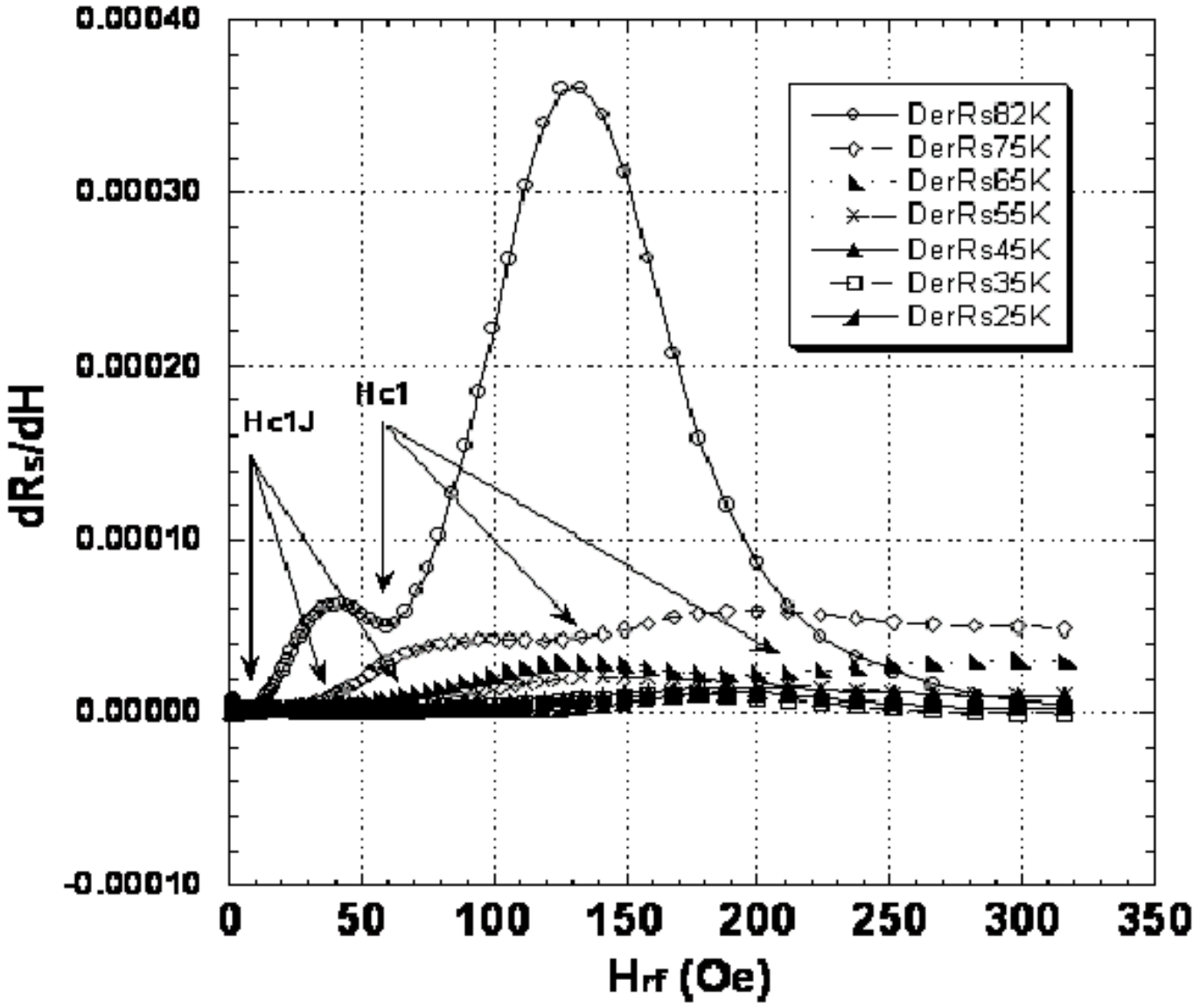

**Fig. 111**. Experimental dependences dRs/dHrf (Hrf), representing the ratio of dRs/dHrf derivatives as a function of magnetic field Hrf, in $YBa_2Cu_3O_{7-\delta}$ superconducting microstrip resonator at frequency f≈2GHz at different temperatures.

**The obtained graphics of experimental dependencies dRs/dHrf (Hrf) in $YBa_2Cu_3O_{7-\delta}$ superconducting microstrip resonator at frequency f≈2GHz at different temperatures confirm a conclusion that the nonlinearities of surface resistance Rs in $YBa_2Cu_3O_{7-\delta}$ superconducting microstrip resonator at microwaves are somehow interlinked with some unusual features of physical properties of superconductors appearing near to the critical magnetic fields $Hc_1$ and $Hc_2$ in HTS thin films at microwaves.**

# Modeling of Nonlinear Dependence of Surface Resistance on External Magnetic Field $R_S(H_e)$ in Proximity to Critical Magnetic Fields $H_{c1}$ and $H_{c2}$ in HTS Microstrip Resonators at Ultra High Frequencies.

The nonlinear change of surface resistance Rs under the operation of external magnetic field He, in close proximity to the magnetic fields $H_{c1}$ and $H_{c2}$, is stipulated by the fact that there are phase transitions of second type at critical magnetic fields $H_{c1}$ and $H_{c2}$ in a superconductor, and the superconductor is in a mixed state in the range between the critical magnetic fields from $H_{c1}$ up to $H_{c2}$. At an increase of the external magnetic field He, when it becomes higher than the critical magnetic field $H_{c1}$, the quantum magnetic lines (Abricosov magnetic vortices) start to appear and penetrate into the superconductor. The area of Abricosov magnetic vortex core, which has a radius $r \sim \xi$, is similar to the normal metal by its properties ($\xi$ – the correlation length of a superconductor). The total number of Abricosov magnetic vortices and the amount of normal phase, connected with the presence of the normal cores of the Abricosov magnetic vortices, have a nonlinear nature of change depending on the applied external magnetic field He or magnetic field of an electromagnetic wave $H_{rf}$, leading to the nonlinear increase of a surface resistance Rs of superconductor, because the normal cores of Abricosov magnetic vortices have the essentially bigger surface resistance $R_{sn}$ than the surface resistance Rs of superconductor. To estimate the effect of these phase transitions on the surface resistance Rs, it is necessary to consider the average model of a superconductor, in which he author of thesis supposes that the increasing external magnetic field will penetrate into a superconductor and suppress its superconducting energy gap <Δ> close to the low critical magnetic field $H_{c1}$. It is assumed that the similar suppression of superconducting properties of a sample by an external magnetic field has place at the phase transition from mixed state to normal state in close proximity to the upper critical magnetic field $H_{c2}$. Considering the interaction between the magnetic fields in a superconductor, the change of magnitude of external magnetic field He leads to the proportional change of a number of Abricosov magnetic vortices, which has smoothly varying character. In both cases, the suppression of superconductivity phenomena happens as a measure of change of average free energy of a superconductor in the external magnetic field.

The change of surface resistance $\Delta R_S$ as a function of the temperature T and magnetic field H:

$$\Delta R_S(T, H) = R_{S0} + A \exp\{-n_S(T, H)\cdot\Delta(T)/\, n_S(T,H)\cdot kT\}. \quad (7.6)$$

The energy of condensation of electrons during the sample's transition to a superconducting state:

$$n_S(T,H)\cdot\Delta(T) = F_n(T,H) - F_S(T,H).$$

# Modeling of Nonlinear Dependence of Surface Resistance on External Magnetic Field $R_S(H_e)$ in Proximity to Critical Magnetic Fields $H_{c1}$ and $H_{c2}$ in HTS Microstrip Resonators at Ultra High Frequencies.

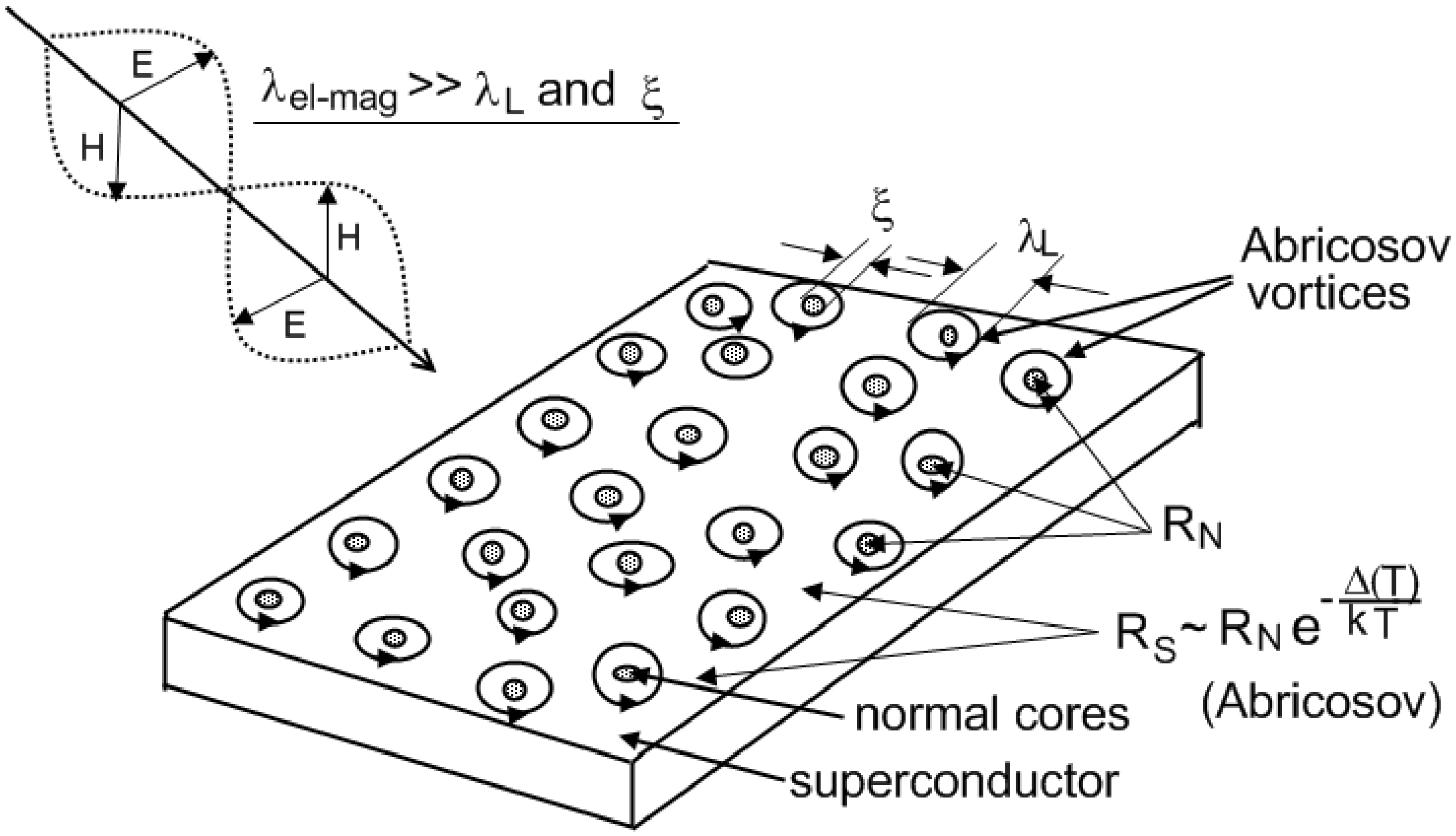

after averaging :

near $H_{c1}$ and $H_{c2}$

$$R_S(T, H) = R_{S0} + A \exp\{- n_S(T, H) \cdot \Delta(T) / n_S(T,H) \cdot kT\}$$

$$n_S(T,H) \cdot \Delta(T) = F_n(T,H) - F_S(T,H).$$

(F is free energy)

**Fig. 112**. Effect of Abricosov magnetic vortices generation on nonlinear surface resistance Rs(He) in close proximity to critical magnetic fields $H_{c1}$ and $H_{c2}$.

# Modeling of Nonlinear Dependence of Surface Resistance on External Magnetic Field $R_S(H_e)$ in Proximity to Critical Magnetic Fields $H_{c1}$ and $H_{c2}$ in HTS Microstrip Resonators at Ultra High Frequencies.

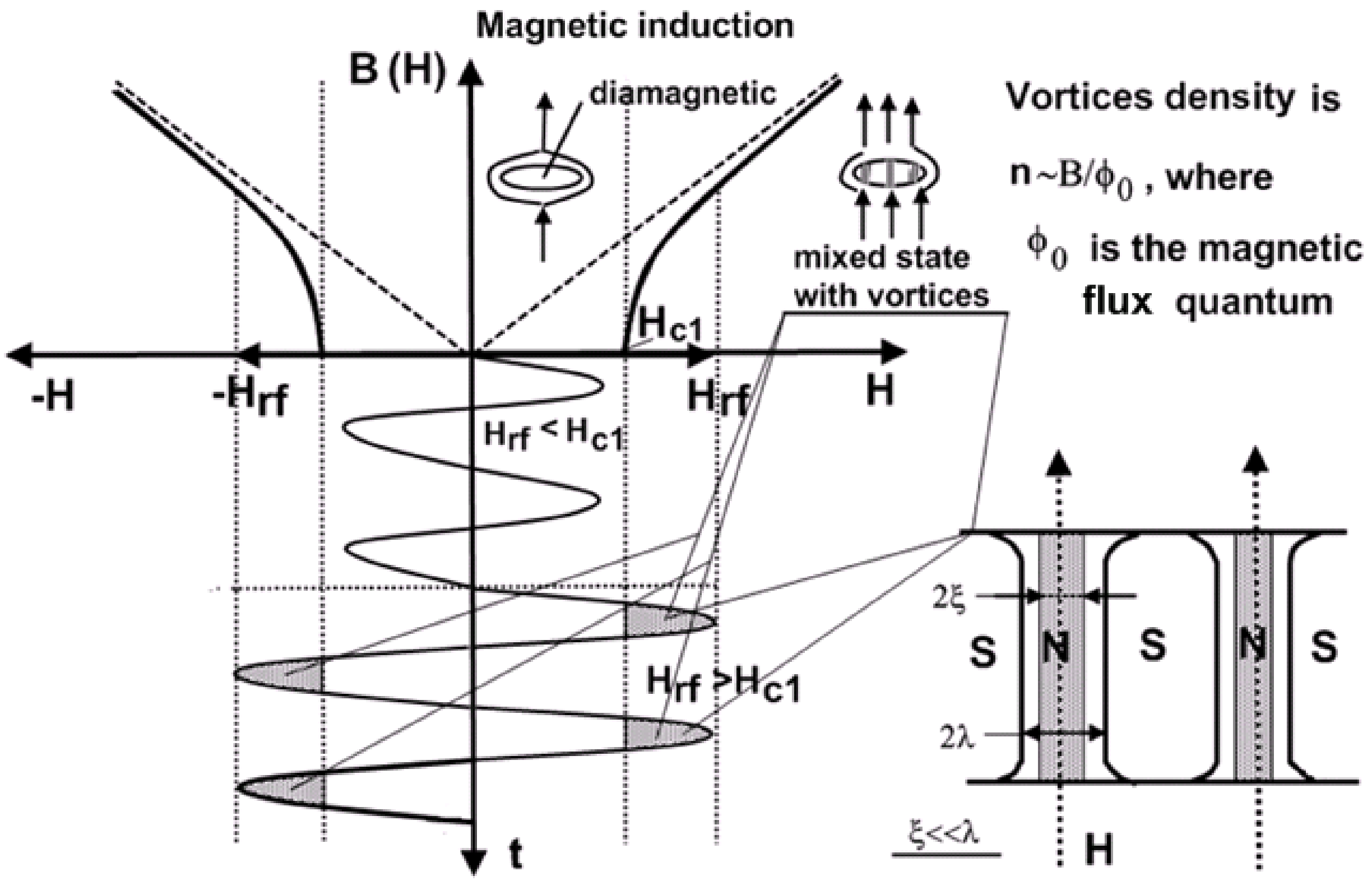

**Fig. 113**. Nature of nonlinear surface resistance Rs(He) appearance due to Abricosov magnetic vortex generation in close proximity to critical magnetic fields $H_{c1}$ and $H_{c2}$. In the case $H_{rf} > H_{C1}$ , the Hrf penetrates into superconductor, generating Abricosov magnetic vortices and resulting in nonlinear surface resistance appearance, because of normal metal cores of Abricosov magnetic vortices.

# Modeling of Nonlinear Dependence of Surface Resistance on External Magnetic Field $R_S(H_e)$ in Close Proximity to Low Critical Magnetic Field $H_{c1}$ in HTS Microstrip Resonators at Ultra High Frequencies.

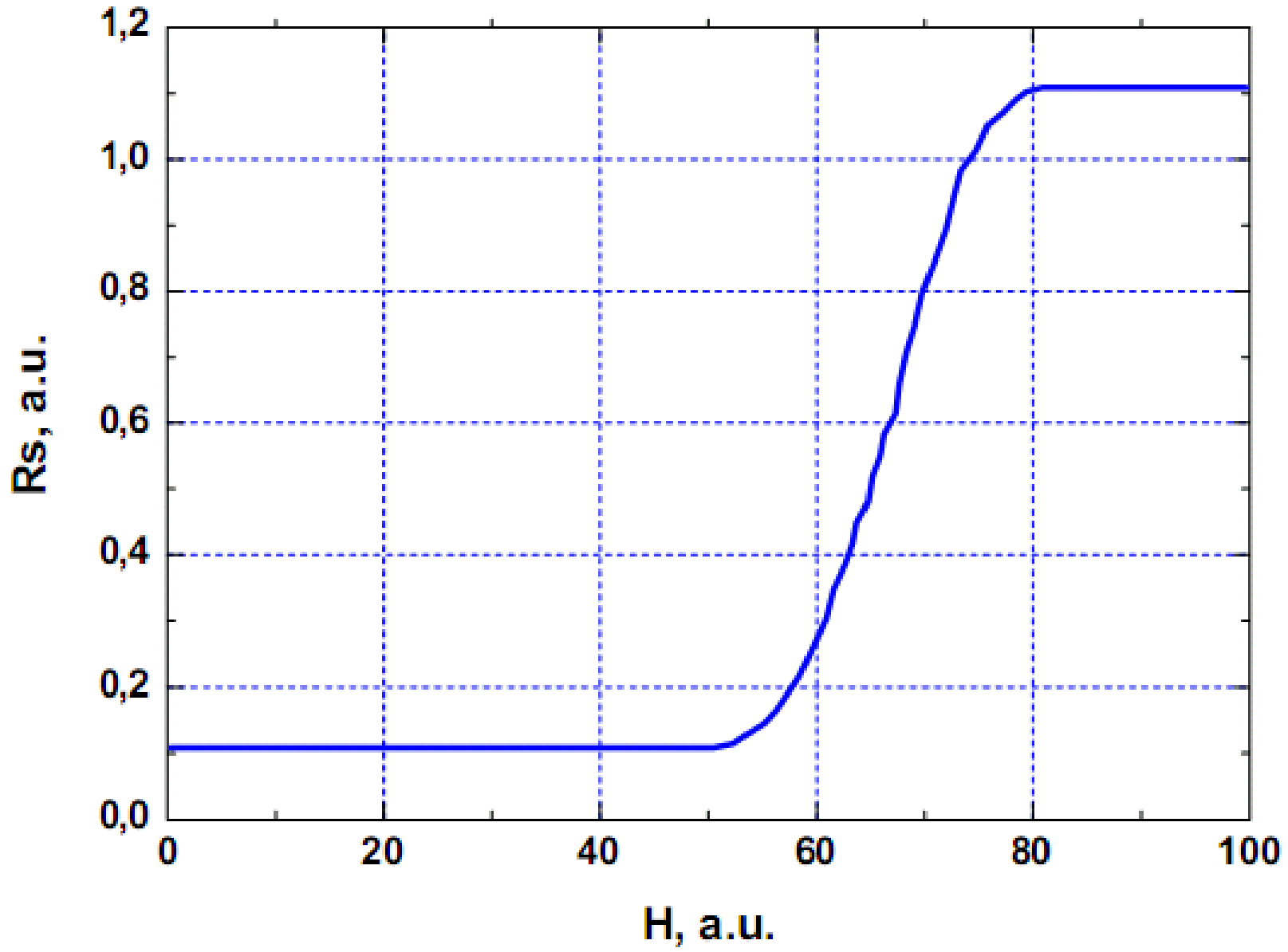

**Fig. 114**. Modeling of dependence of surface resistance on magnetic field Rs(H) in $YBa_2Cu_3O_{7-\delta}$ superconducting microstrip resonator at frequency f≈2GHz at different temperatures in close proximity to critical magnetic field $H_{C1}$=50a.u. with residual resistance $Rs_0$=0.1a.u. according to equation (7.6) [85].

**Near to the low critical field Hc1, it is convenient to represent the free energy in the magnetic field as the Gibbs free energy Gs . The Abricosov magnetic lines contribute to this energy. This contribution will reduce the density of superconducting electrons $n_s$ by the $G_{M1}$ .:**

$$G_{M1} = \frac{2\phi_0}{\sqrt{3}\lambda^2}\,\frac{(H_e - H_{C1})}{4\pi\left\{\ln\left[\frac{3\phi_0}{4\pi\lambda^2(H_e - H_{C1})}\right]\right\}^2}, \quad (7.7)$$

**The surface resistance $R_S$, which is close to the low critical magnetic field $H_{C1}$:**

$$R_S(T,H) \propto A\cdot\exp[-(n_S(T,H)\cdot\Delta(T)-G_{M1}) / n\cdot kT] \propto B\exp(G_{M1} / n\cdot kT), \quad (7.8)$$

**Let's emphasize that each mechanism of dissipation gives a contribution to the total resistance, and since the surface impedance of a superconductor is a sum of the resistances, hence the additional term of sort Rs (H) will correspond to this mechanism. Let's write :**

$$R_s(H) \sim B\exp(g_{M1} / kT) \quad (7.9)$$

**It is possible to shift this function on –1 on the axis of ordinates:**

$$R_s(H) \sim B\{\exp(g_{M1} / kT) - 1\} \quad (7.10)$$

**Let's conduct normalization of this function on 1, we will divide it on $\exp(g_{M1}/kT)$. Then, it is possible to obtain the expression for the transition close to $H_{C1}$ at $H \geq H_{C1}$ in eq. (7.11) :**

$$R_{S1}(H) \approx B\frac{\exp(g_{M1}/kT)-1}{\exp(g_{M1}/kT)} = B\left\{1-\exp\left(-\frac{a_1\cdot g_{M1}}{kT}\right)\right\} \quad (7.11)$$

# Modeling of Nonlinear Dependence of Surface Resistance on External Magnetic Field $R_S(H_e)$ in Close Proximity to High Critical Magnetic Field $H_{c2}$ in HTS Microstrip Resonators at Ultra High Frequencies.

**Let's consider, now, the variation of the surface resistance Rs near to the high critical magnetic field Hc2. In this case, the change of free energy of a superconductor is featured by the expression:**

$$G_{M2} = \frac{(H_{C2} - H_e)^2}{8\pi(2\kappa^2 - 1)\beta_A}, \qquad (7.12)$$

where He ≤ Hc2 , βA ≥ 1 and κ is the Ginzburg-Landau parameter, which is equal to 70 for the high-temperature superconductors.

**The expression from the thermodynamics theory of superconductors:**

$$n_s(T)\,\Delta(T) - H_c^2(T)/8\pi = 0$$

where $n_s$ is the density of superconducting electrons.

**The surface resistance $R_S$, which is close to the low critical magnetic field $H_{C2}$:**

$$R_s(H) \propto A_1 \cdot \exp[-(n_s \cdot \Delta + G_{M2}) / n \cdot kT] \propto B_1 \exp(-G_{M2} / n \cdot kT). \qquad (7.13)$$

**The surface resistance $R_S$, which is close to the low critical magnetic field $H_{C2}$:**

$$R_{S2}(H) \approx B_1 \exp\left(-\frac{a_2 \cdot g_{M2}}{kT}\right), \qquad (7.14)$$

where B1 is the amplitude of change for surface resistance Rs, a2 is the fitting coefficient defining the speed of change Rs(H).

**The fitting coefficient, which defines the speed of change Rs(H):**

$$a_2 \approx b / (50^2)$$

where b is the coefficient, which matches the dimensions of the multi-pair numerator and denominator.

# Modeling of Nonlinear Dependence of Surface Resistance on External Magnetic Field $R_S(H_e)$ in Close Proximity to High Critical Magnetic Field $H_{c2}$ in HTS Microstrip Resonators at Ultra High Frequencies.

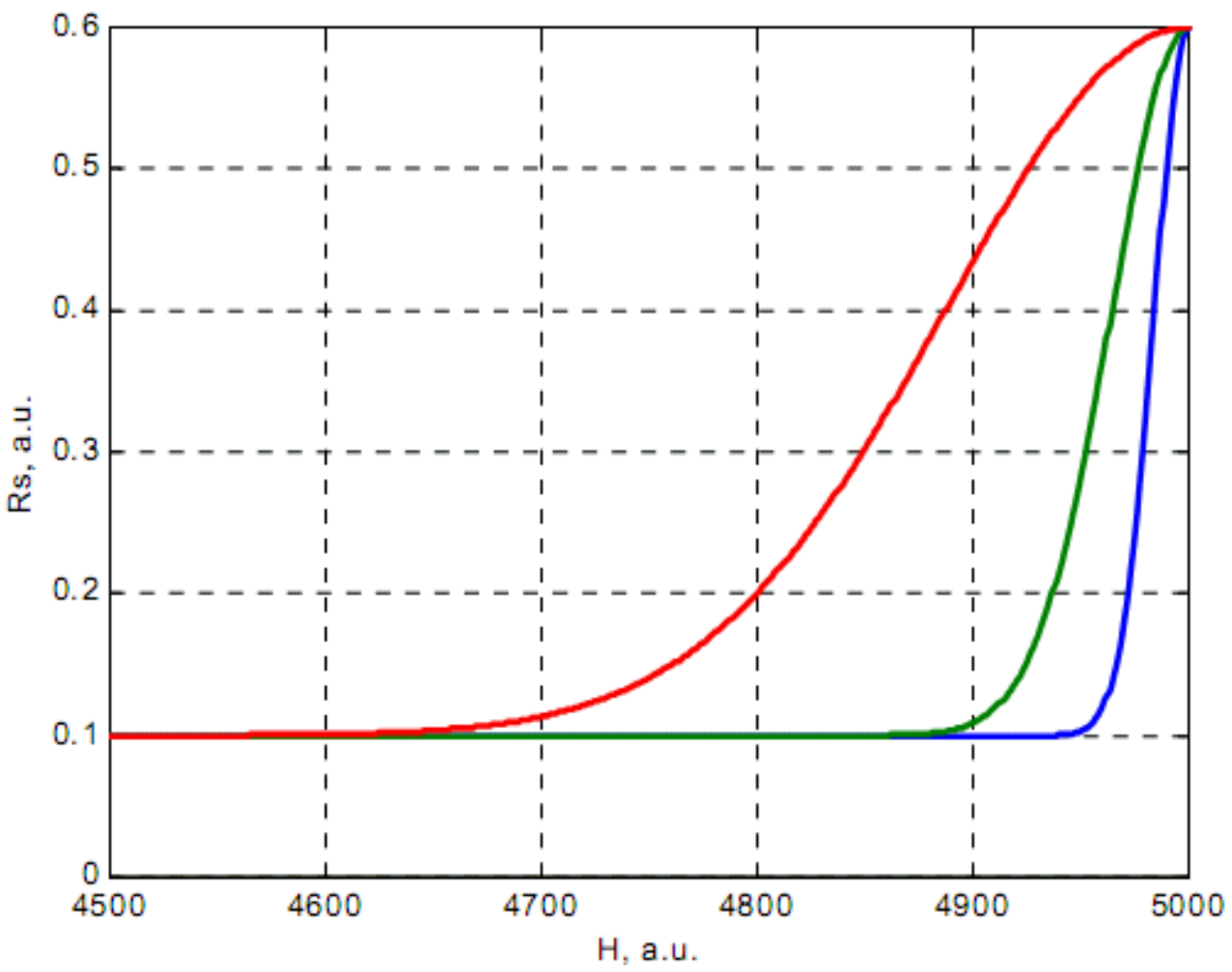

**Fig. 115.** Dependence of surface resistance on magnetic field Rs(H) near to high critical magnetic field $Hc_2$ =5000 a.u. at different coefficients a1 = 0.001 (red curve), 0.01 (green curve), 0.05 (blue curve); B1 = 0.5, $Rs_0$ =0.1 [85].

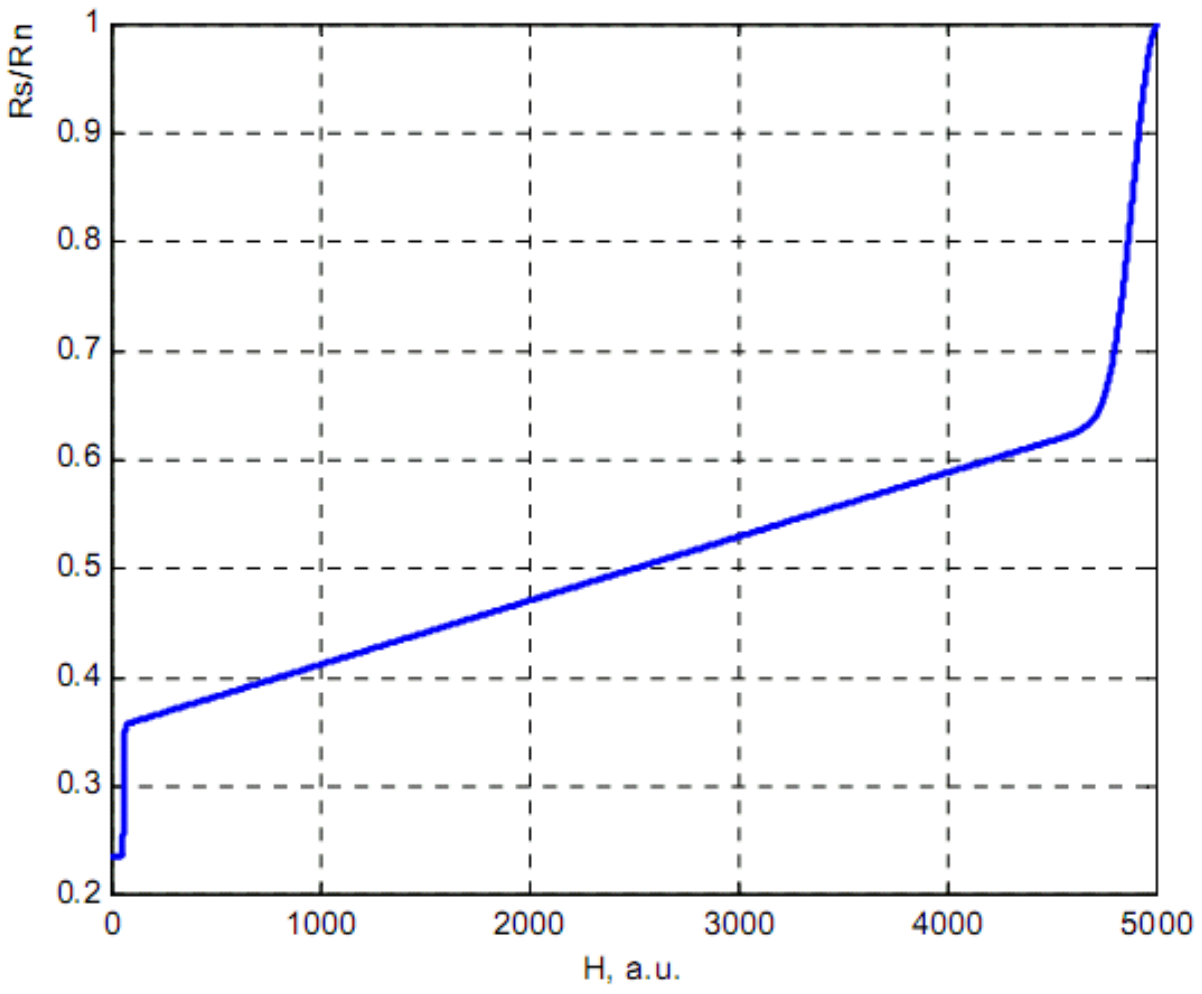

**Fig. 116.** Modeling of conditional dependence of superconductor surface resistance to normal metal resistance ratio on magnetic field Rs/Rn(H) with linear sort of dependence Rs/Rn(H) between critical magnetic fields Hc1÷Hc and nonlinear dependence Rs/Rn(H) close to low critical magnetic field $Hc_1$=50 and high critical magnetic field $Hc_2$=5000. Dependence Rs/Rn(H) represents symmetric function between critical magnetic fields $Hc_1$ and $Hc_2$ [85].

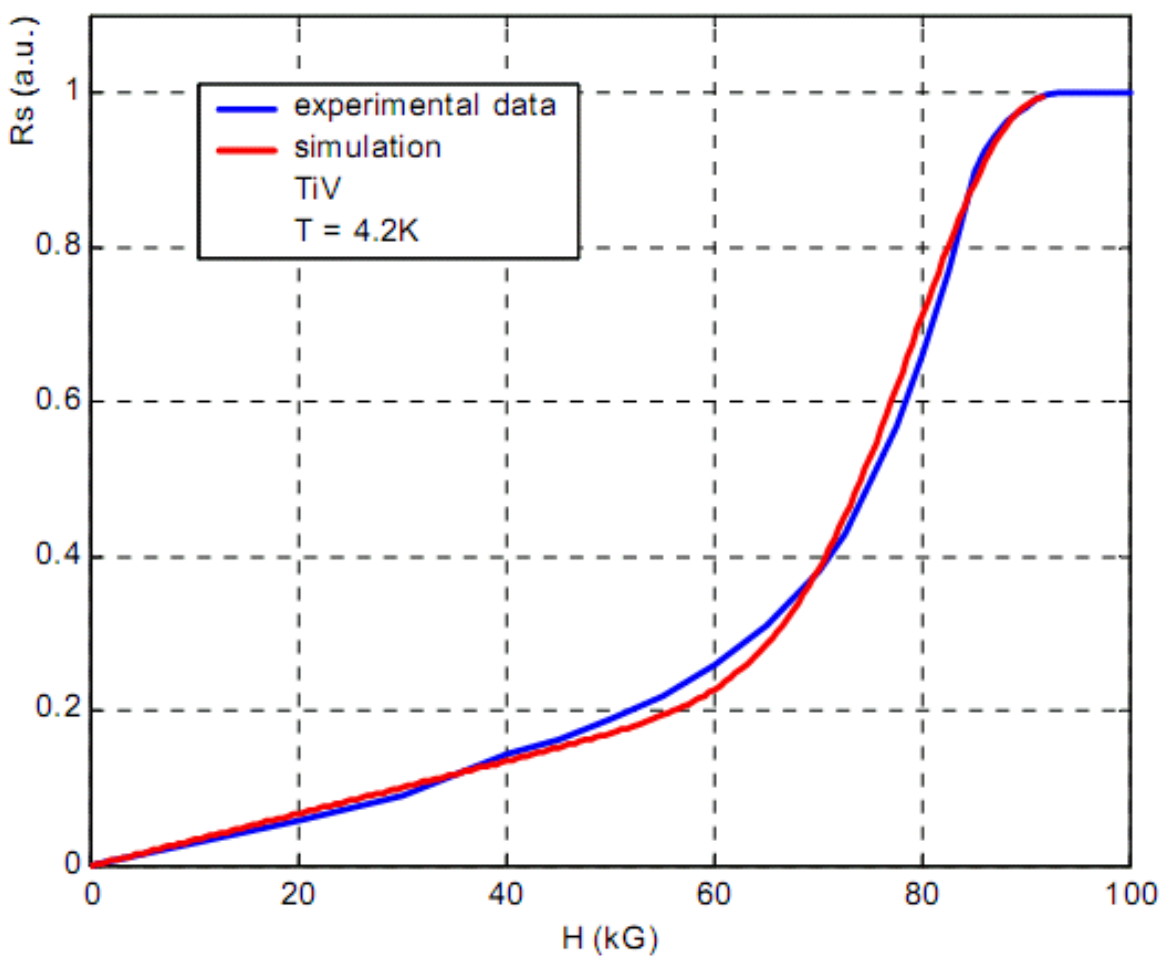

**Fig. 117.** Experimental dependence of surface resistance on magnetic field Rs(H) in $Ti_{0.6}V_{0.4}$ at frequency of 14.4GHz at temperature T = 4.2K is shown by blue curve and simulation of experimental dependence of surface resistance on magnetic field Rs(H) in $Ti_{0.6}V_{0.4}$ at frequency of 14.4GHz at temperature T = 4.2K is approximated by red curve. Linear relation at He < $Hc_2$ and formula (6.14) at He≤$Hc_2$ are utilized to simulate Rs(H) dependence in Ti0.6V0.4 at frequency of 14.4GHz at temperature T = 4.2K [85].

# Modeling of Nonlinear Dependence of Surface Resistance on External Magnetic Field $R_S(H_e)$ in Close Proximity to High Critical Magnetic Field $H_{c2}$ in HTS Microstrip Resonators at Ultra High Frequencies.

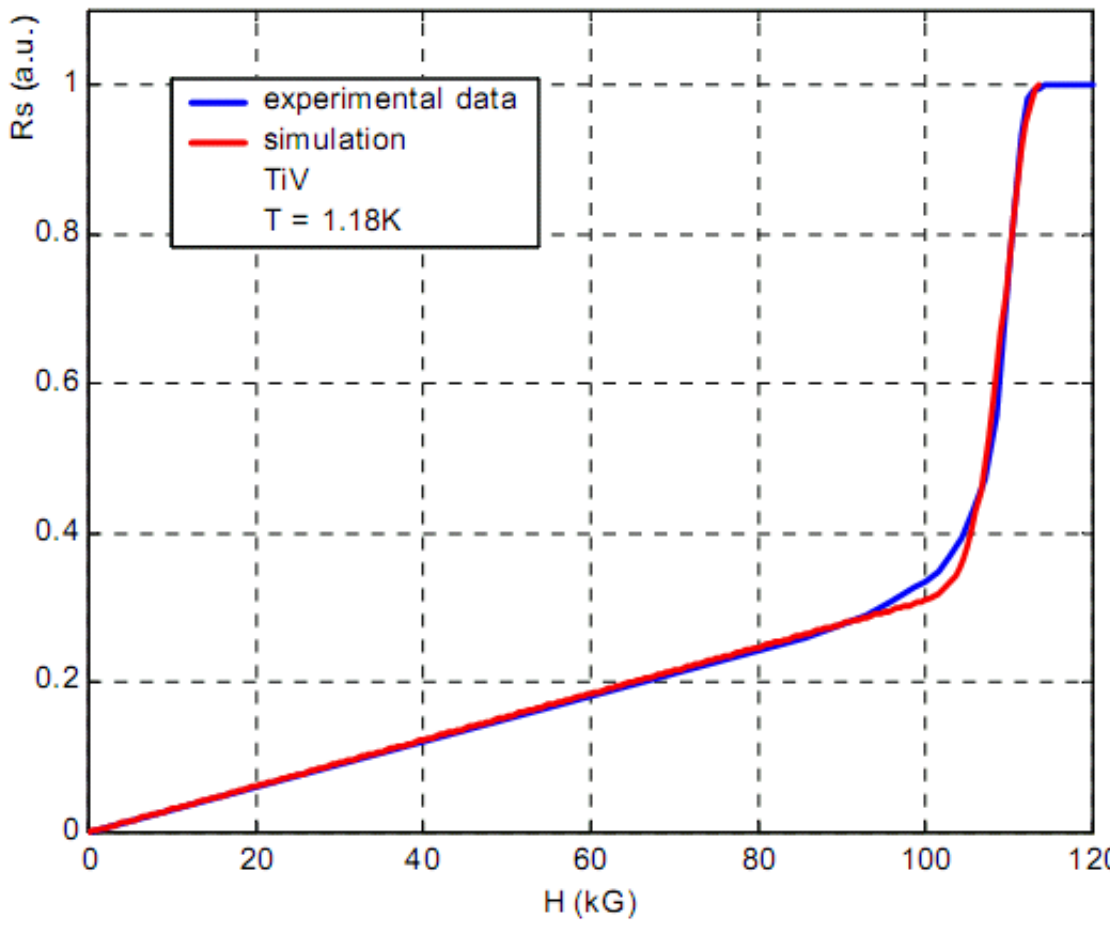

**Fig. 118.** Experimental dependence of surface resistance on magnetic field Rs(H) in $Ti_{0.6}V_{0.4}$ at frequency of 14.4GHz at temperature T = 1.18K is shown by blue curve [6], and simulation of experimental dependence of surface resistance on magnetic field Rs(H) in $Ti_{0.6}V_{0.4}$ at frequency of 14.4GHz at temperature T = 1.18K is approximated by red curve. Linear relation at He < $Hc_2$ and formula (7.14) at $H_e \leq H_{c2}$ are utilized to simulate Rs(H) dependence in $Ti_{0.6}V_{0.4}$ at frequency of 14.4GHz at temperature T = 1.18K [85].

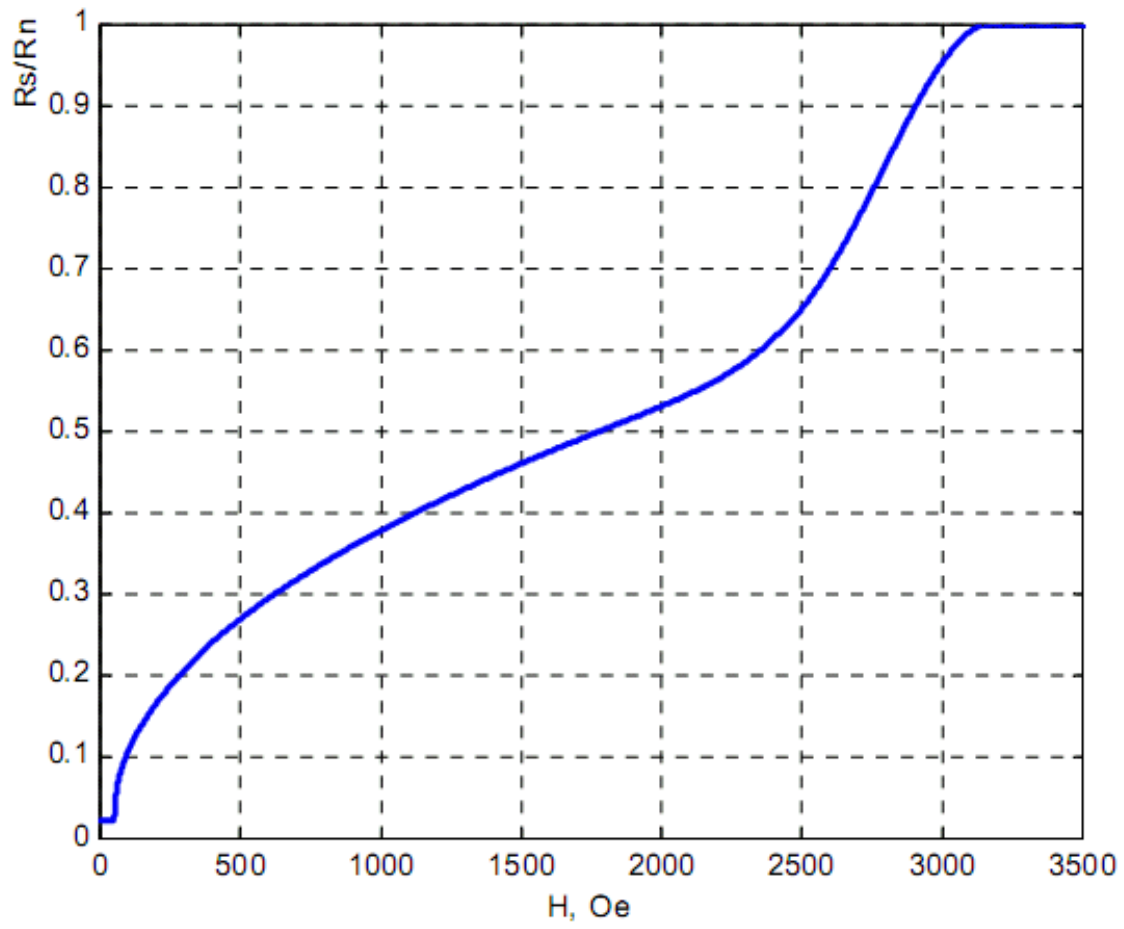

**Fig. 119.** Modeling of conditional dependence of superconductor surface resistance to normal metal resistance ratio on magnetic field Rs/Rn(H) in HTS thin film at microwaves with dependence Rs/Rn(H) ~ $(H)_{1/2}$ between critical magnetic fields $Hc_1 \div Hc$ and nonlinear dependence Rs/Rn(H) close to low critical magnetic field $Hc_1$=50Oe and high critical magnetic field $Hc_2$=3500 Oe in HTS thin film at microwaves. Dependence Rs/Rn(H) in HTS thin film at microwaves represents almost symmetric function between critical magnetic fields $Hc_1$ and $Hc_2$. [85].

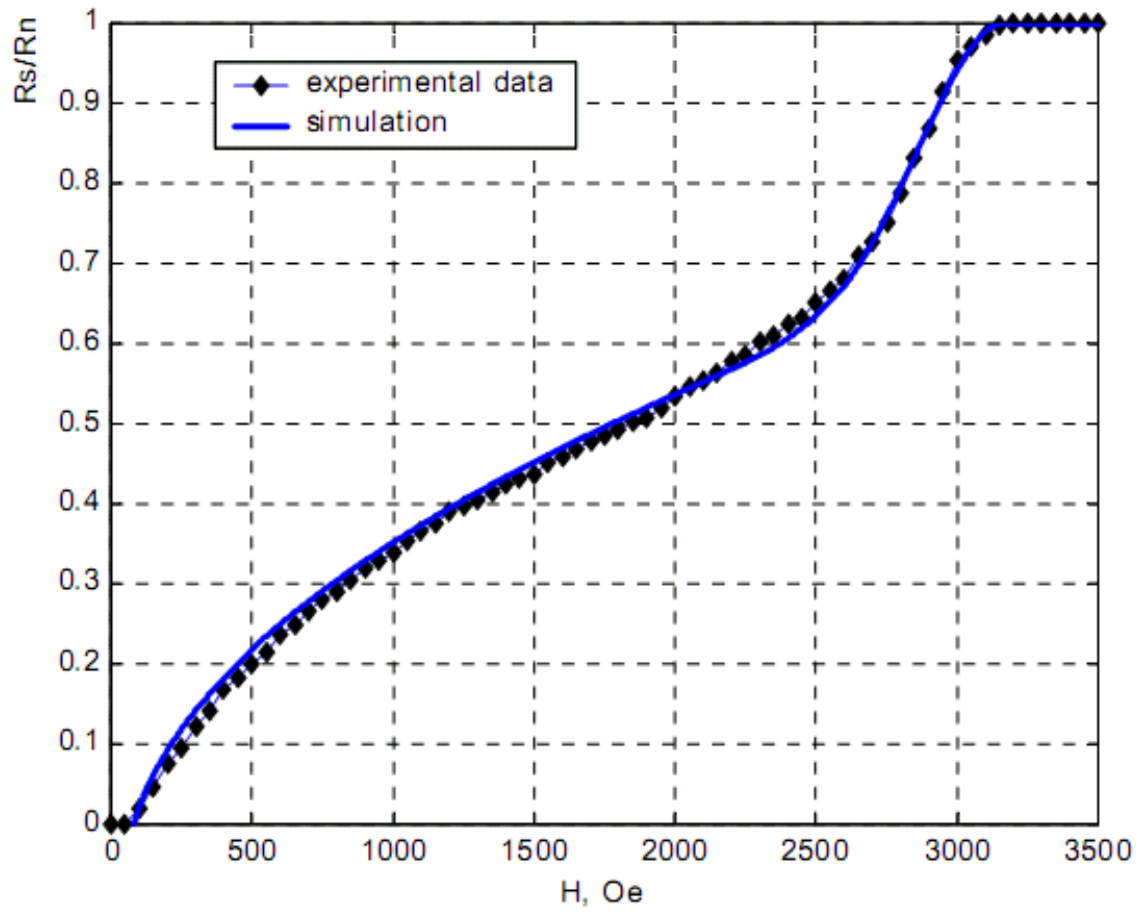

**Fig. 120.** Experimental dependence of surface resistance on magnetic field Rs/Rn(H) $In_{0.19}Pb_{0.8}$ at frequency 170 MHz at temperature of 4.2K is shown by black curve and simulation of dependence of superconductor surface resistance to normal metal resistance ratio on magnetic field Rs/Rn(H) with dependence Rs/Rn(H) ~ $(H)^{1/2}$ between critical magnetic fields Hc1 ÷ Hc and the nonlinear dependence Rs/Rn(H) close to the low critical magnetic field Hc1=50 Oe and high critical magnetic field $Hc_2$=3000 Oe in case of $In_{0.19}Pb_{0.8}$ at frequency 170 MHz at temperature of 4.2K is shown by blue curve [19].

# On Influence of Magnetic Field $H_{rf}$ on Nonlinear Surface Resistance $R_S$ of $YBa_2Cu_3O_{7-\delta}$ Thin Films on MgO Substrates in Close Proximity to Low Critical Magnetic Field $H_{c1}$ in Superconducting Microstrip Resonators at Ultra High Frequencies.

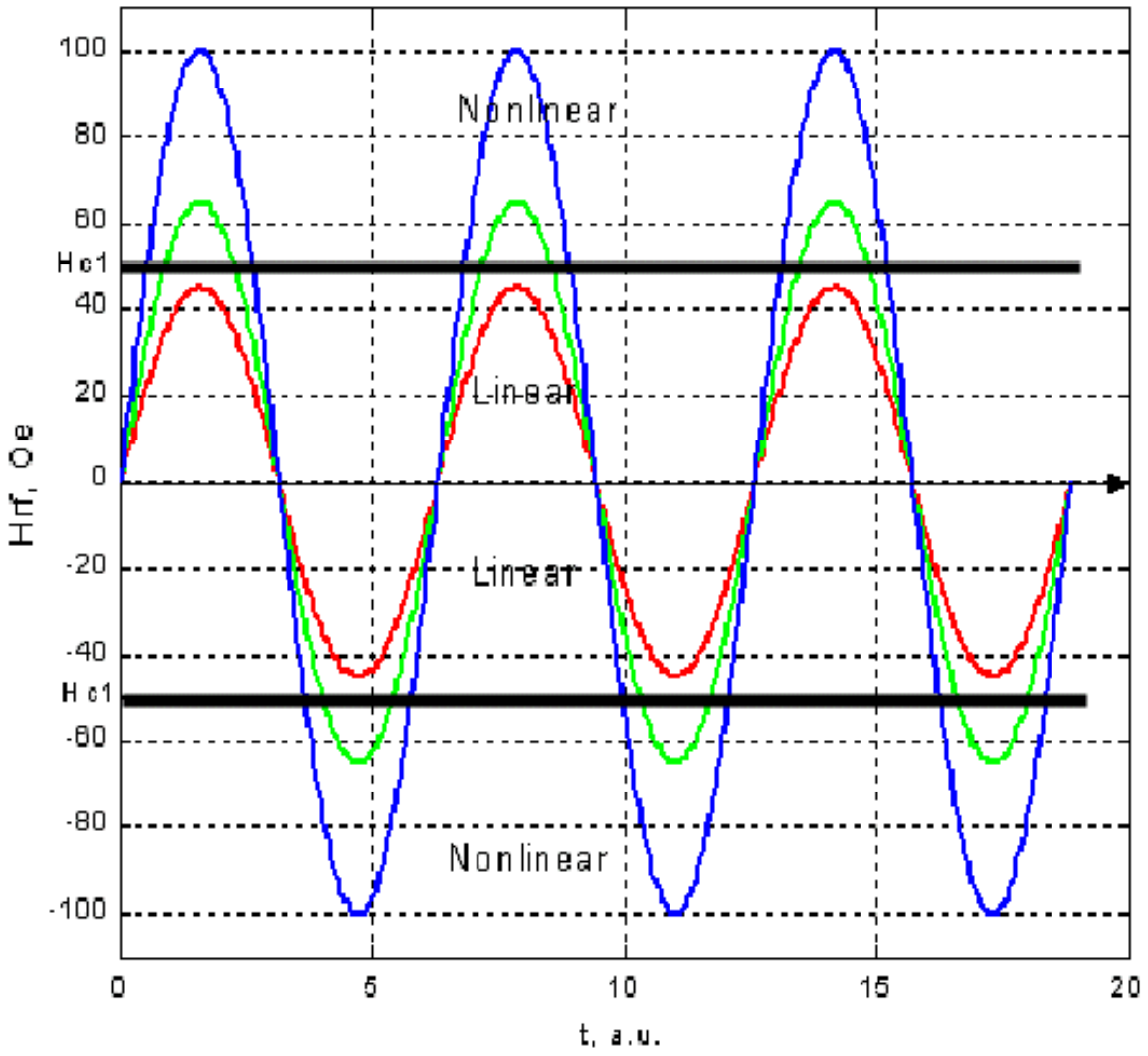

**Fig. 121.** Modeling of time dependence of magnetic fields of electromagnetic waves Hrf with amplitudes 45a.u. (red curve), 65a.u. (green curve), 100a.u. (blue curve) during their propagation in region with nonlinear surface resistance Rs in HTS thin films in microstrip resonator in proximity to low critical magnetic field $Hc_1$=50a.u. at microwaves [85].

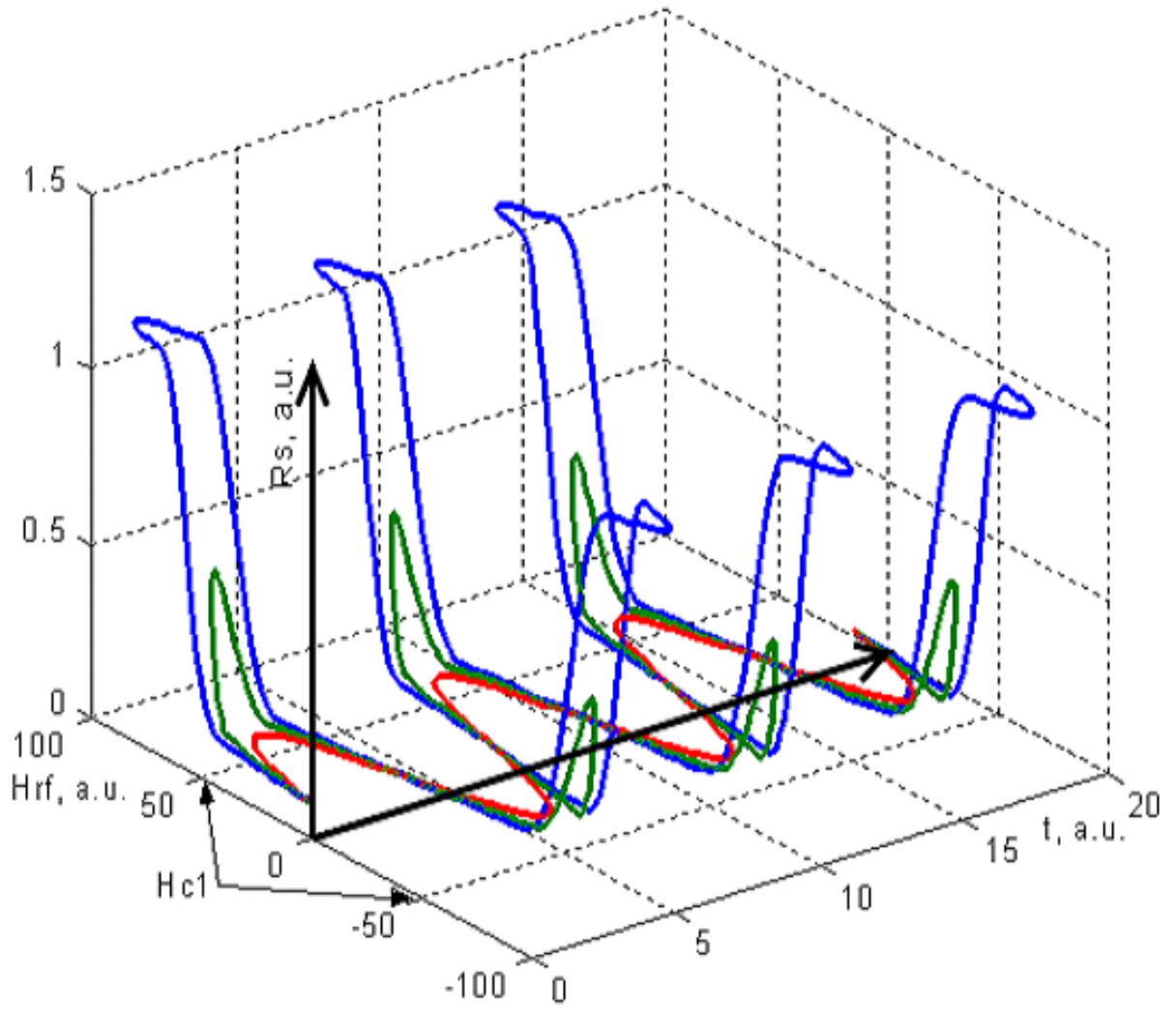

**Fig. 122.** Modeling of surface resistance dependence on time and on magnetic field Rs(t, Hrf) for magnetic fields Hrf with amplitudes 45a.u. (red curve), 65a.u. (green curve), 100a.u. (blue curve) during their propagation in region with the nonlinear surface resistance Rs in HTS thin films in microstrip resonator in proximity to low critical magnetic field $Hc_1$=50a.u. at microwaves [85].

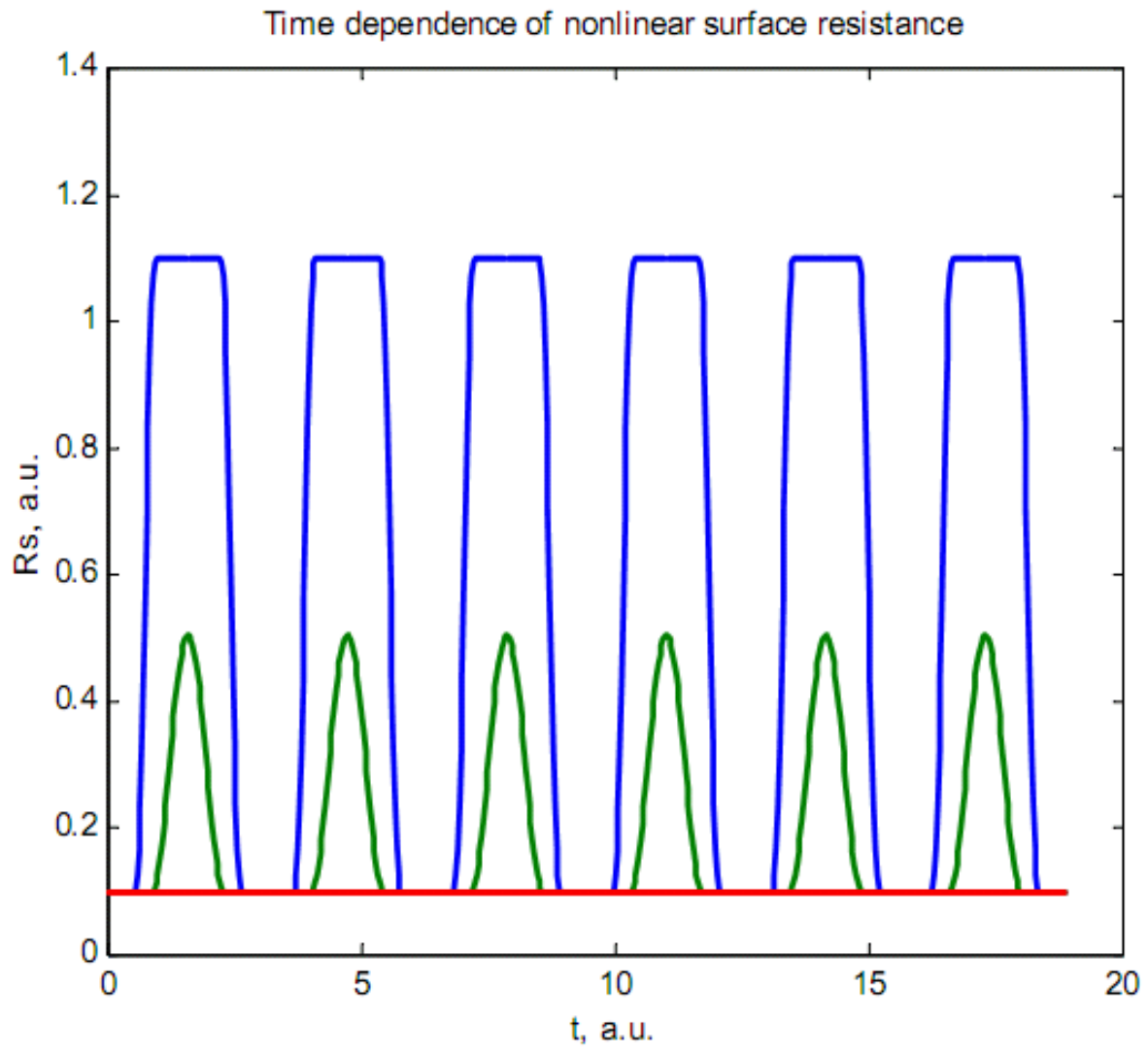

**Fig. 123.** Modeling of time dependence of surface resistance Rs(t) in linear (red curve) and nonlinear (green and blue curves) cases for magnetic fields Hrf with amplitudes 45a.u. (red curve), 65a.u. (green curve), 100a.u. (blue curve) during their propagation in region with nonlinear surface resistance Rs in HTS thin films in microstrip resonator in proximity to low critical magnetic field Hc1=50a.u. at microwaves. Residual resistance $R_0$=0.1 [85].

# On Influence of Magnetic Field $H_{rf}$ on Nonlinear Surface Resistance $R_S$ of $YBa_2Cu_3O_{7-\delta}$ Thin Films on MgO Substrates in Close Proximity to Low Critical Magnetic Field $H_{c1}$ in Superconducting Microstrip Resonators at Ultra High Frequencies.

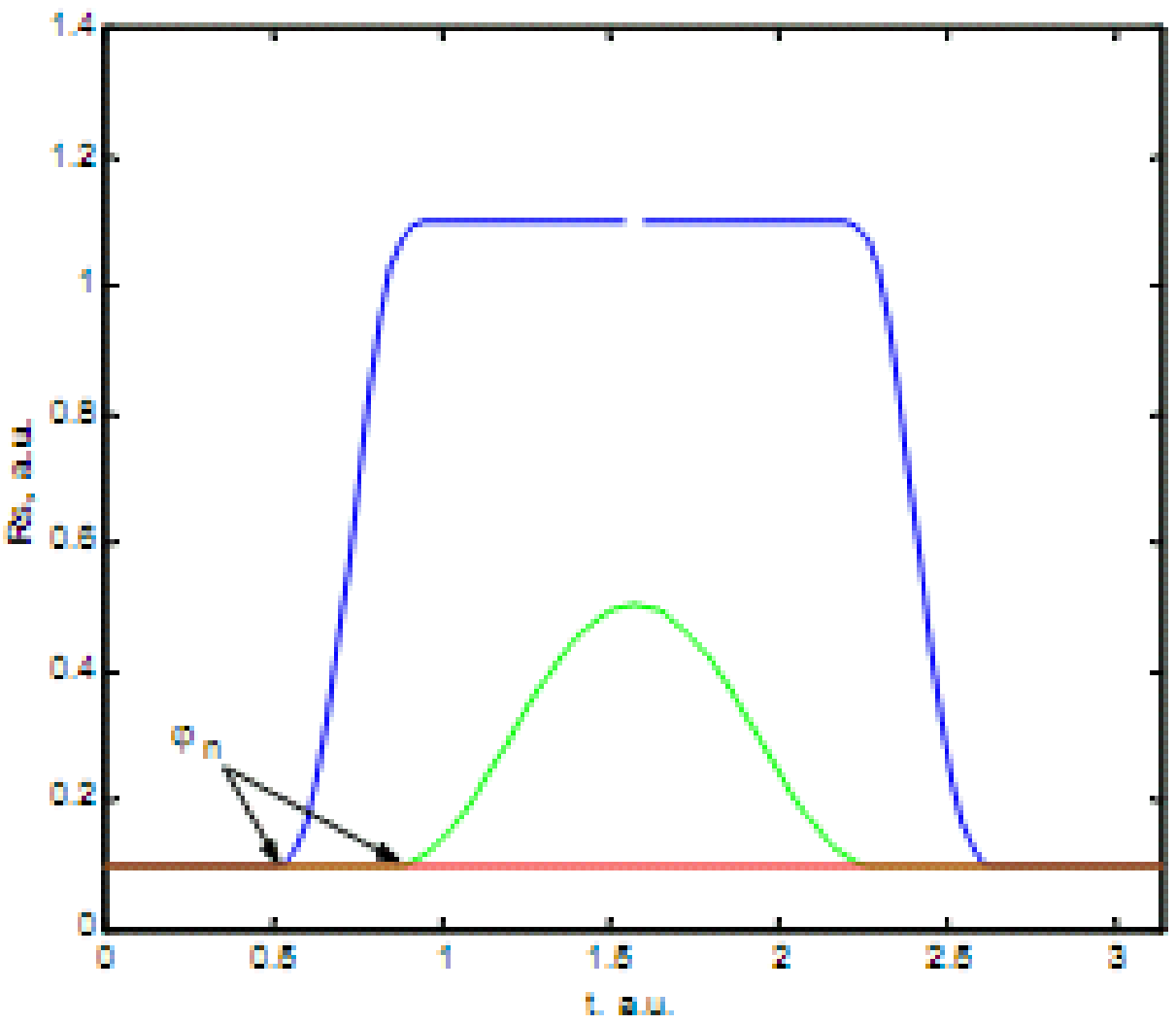

**Fig. 124.** Modeling of time dependence of surface resistance Rs(t) in linear (red curve) and nonlinear (green and blue curves) cases for magnetic fields Hrf with amplitudes 45a.u. (red curve), 65a.u. (green curve), 100a.u. (blue curve) during their propagation in region with nonlinear surface resistance Rs in $YBa_2Cu_3O_{7-\delta}$ superconducting microstrip resonator in proximity to low critical magnetic field $Hc_1$=50a.u. at microwaves is shown in oscillation period phase limits from 0 up to π. Residual resistance $R_0$ = 0.1 [85].

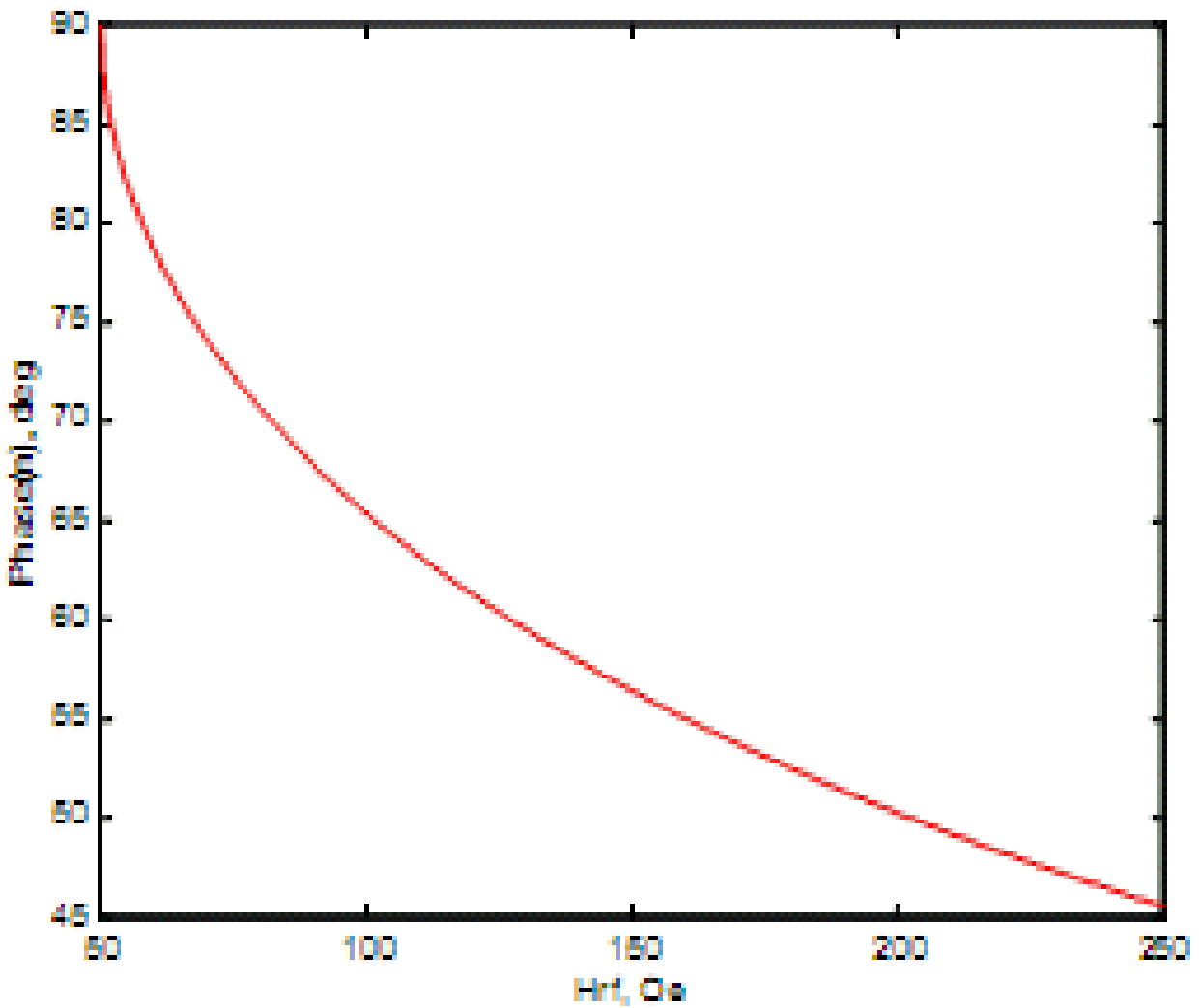

**Fig. 125.** Modeling of dependence of phase angle of electromagnetic wave on magnetic field of electromagnetic wave φn(Hrf) in $YBa_2Cu_3O_{7-\delta}$ superconducting microstrip resonator for case, when magnetic field is above or equal to low critical magnetic field Hrf ≥ $Hc_1$ =50Oe at microwaves [85].

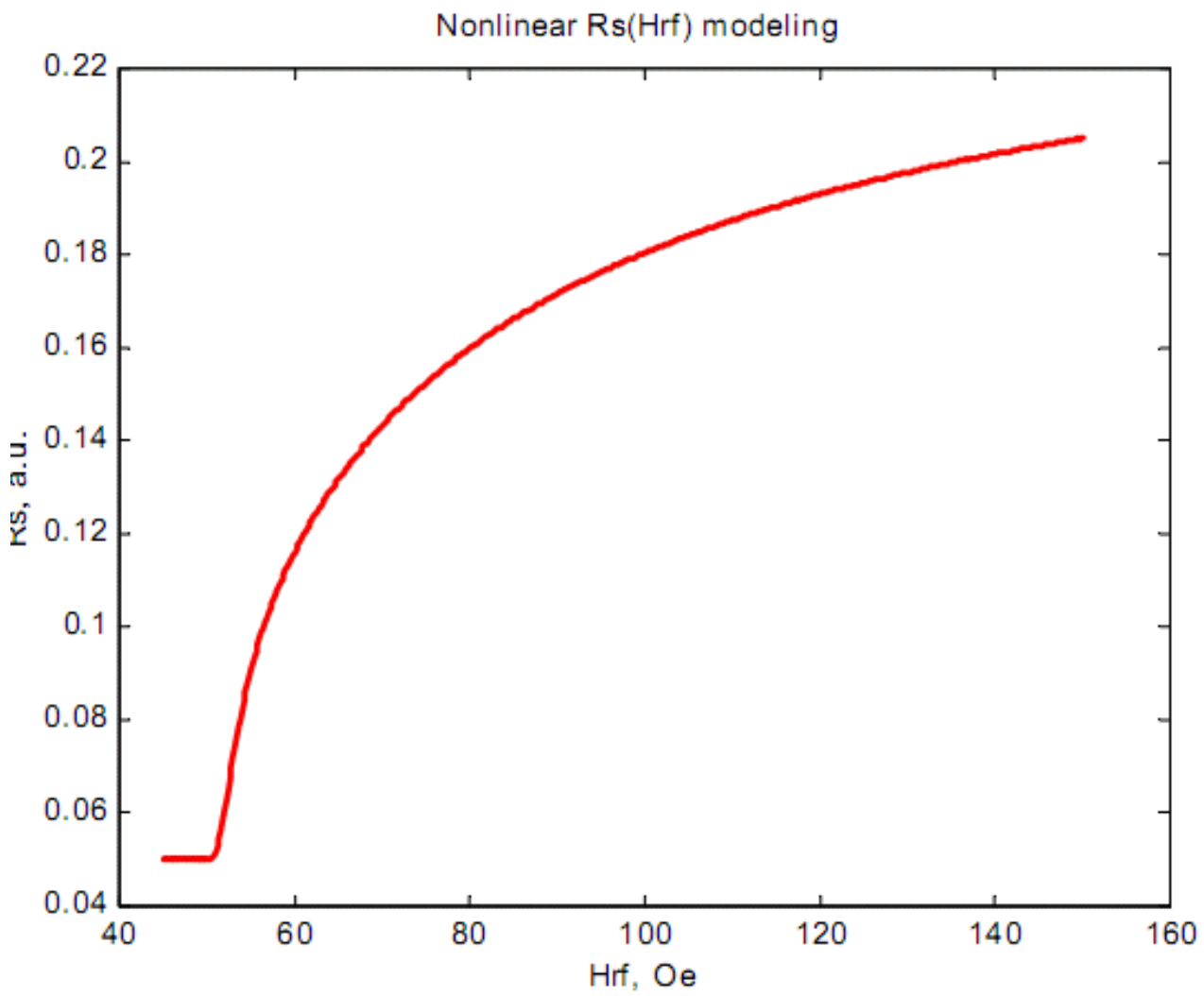

**Fig. 126.** Modeling of dependence of average surface resistance on magnetic field Rs(Hrf) in $YBa_2Cu_3O_{7-\delta}$ superconducting microstrip resonator with low critical magnetic field $Hc_1$ = 50Oe at microwaves [85].

# On Influence of Magnetic Field $H_{rf}$ on Nonlinear Surface Resistance $R_S$ of $YBa_2Cu_3O_{7-\delta}$ Thin Films on MgO Substrates in Close Proximity to Low Critical Magnetic Field $H_{c1}$ in Superconducting Microstrip Resonators at Ultra High Frequencies.

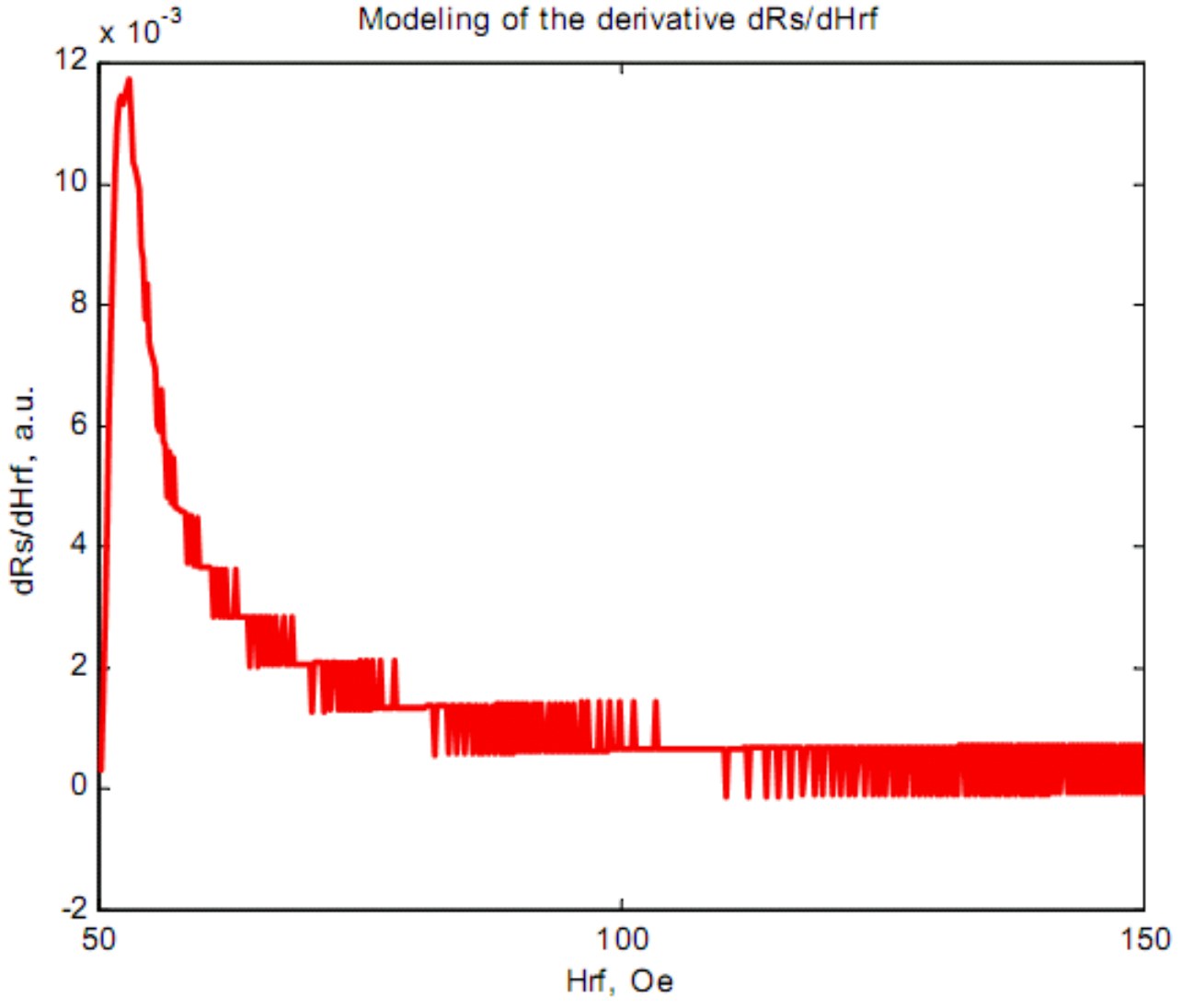

**Fig. 127.** Modeling of dependences dRs/dHrf (Hrf), representing ratio of dRs/dHrf derivatives as function of magnetic field Hrf , in $YBa_2Cu_3O_{7-\delta}$ superconducting microstrip resonator at frequency f≈2GHz at selected temperature of 82K. Modeling of dependence dRs/dHrf (Hrf) is performed to simulate experimental dependence dRs/dHrf (Hrf) in Fig. 126 [85].

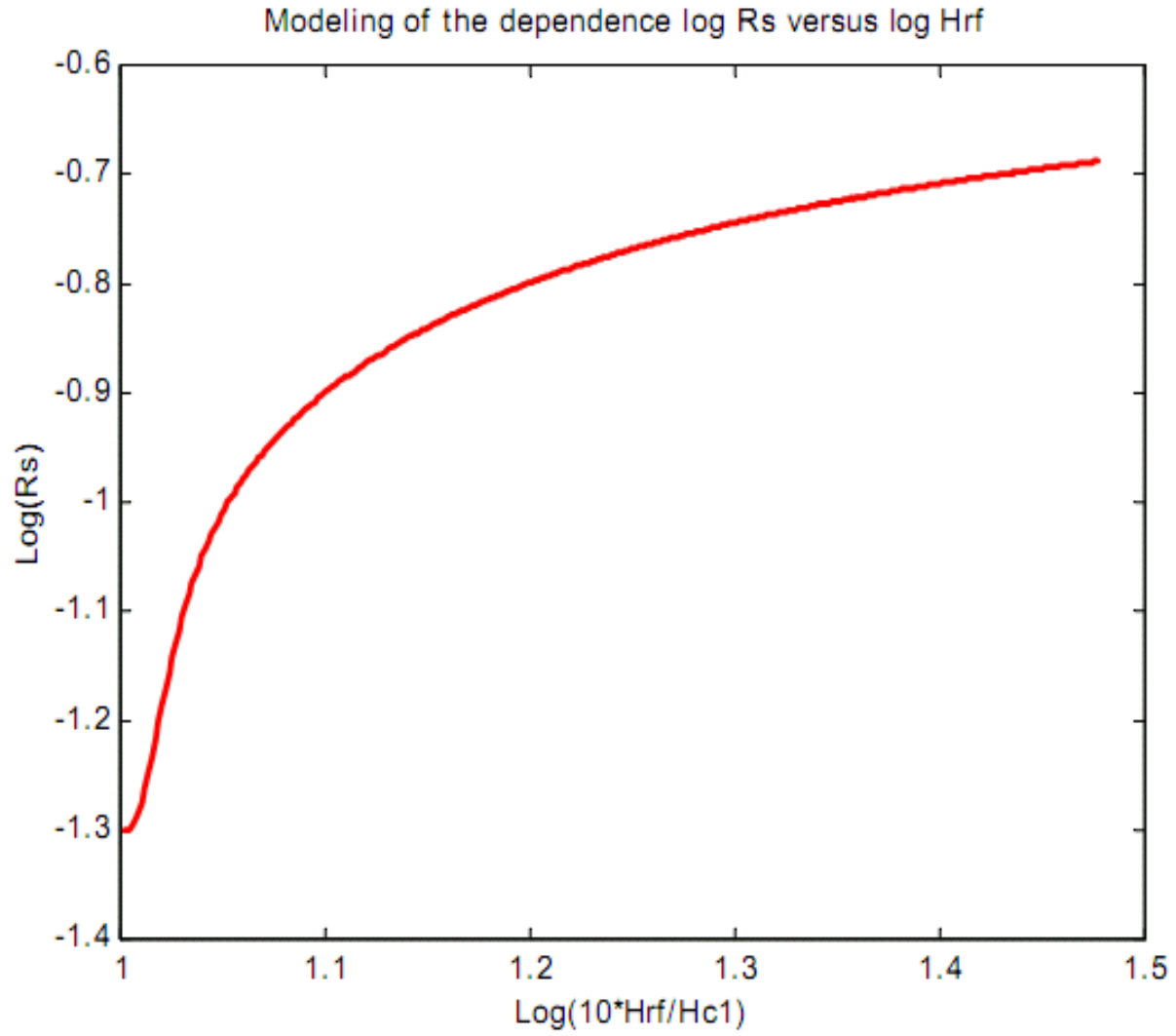

**Fig. 128.** Modeling of dependence of log (Rs) vs. log (10*Hrf/Hc1) in $YBa_2Cu_3O_{7-\delta}$ superconducting microstrip resonator with low critical magnetic field $Hc_1$ = 50Oe at microwaves. Modeling of dependence of log (Rs) vs. log (10*Hrf/Hc1) is performed for Rs(Hrf) curve in Fig. 126 [85].

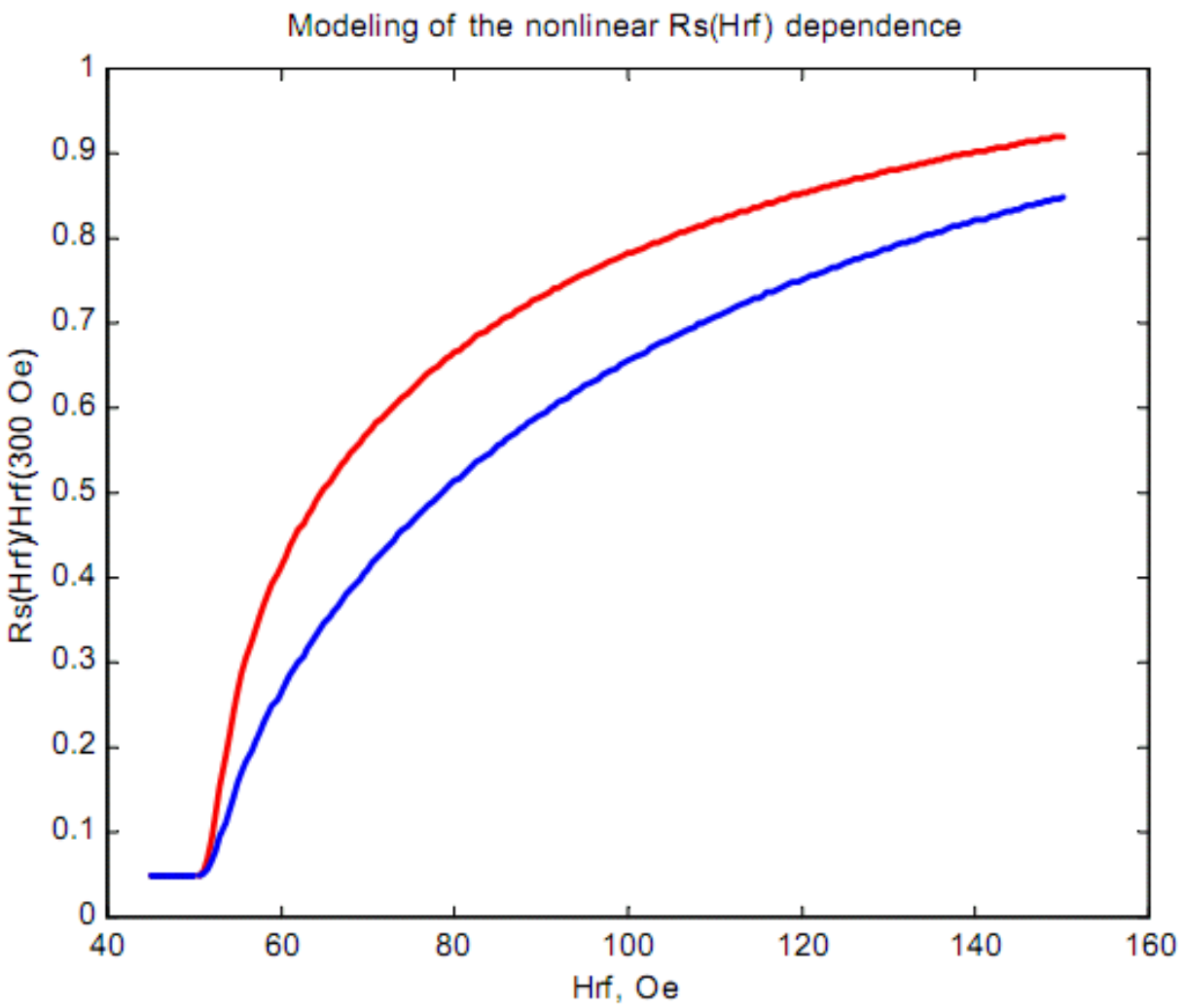

**Fig. 129.** Modeling of nonlinear dependence of relative surface resistance on magnetic field Rs(Hrf)/Hrf(300 Oe) vs. Hrf in $YBa_2Cu_3O_{7-\delta}$ superconducting dielectric and microstrip resonators with low critical magnetic field $Hc_1$=50Oe at microwaves: red curve corresponds to dependence Rs(Hrf)/Hrf(300 Oe) vs. Hrf in the case of dielectric resonator; blue curve corresponds to dependence Rs(Hrf)/Hrf(300Oe) vs. Hrf in the case of microstrip resonator [85].

# On Influence of Magnetic Field $H_{rf}$ on Nonlinear Surface Resistance $R_S$ of $YBa_2Cu_3O_{7-\delta}$ Thin Films on MgO Substrates in Close Proximity to Low Critical Magnetic Field $H_{c1}$ in Superconducting Microstrip Resonators at Ultra High Frequencies.

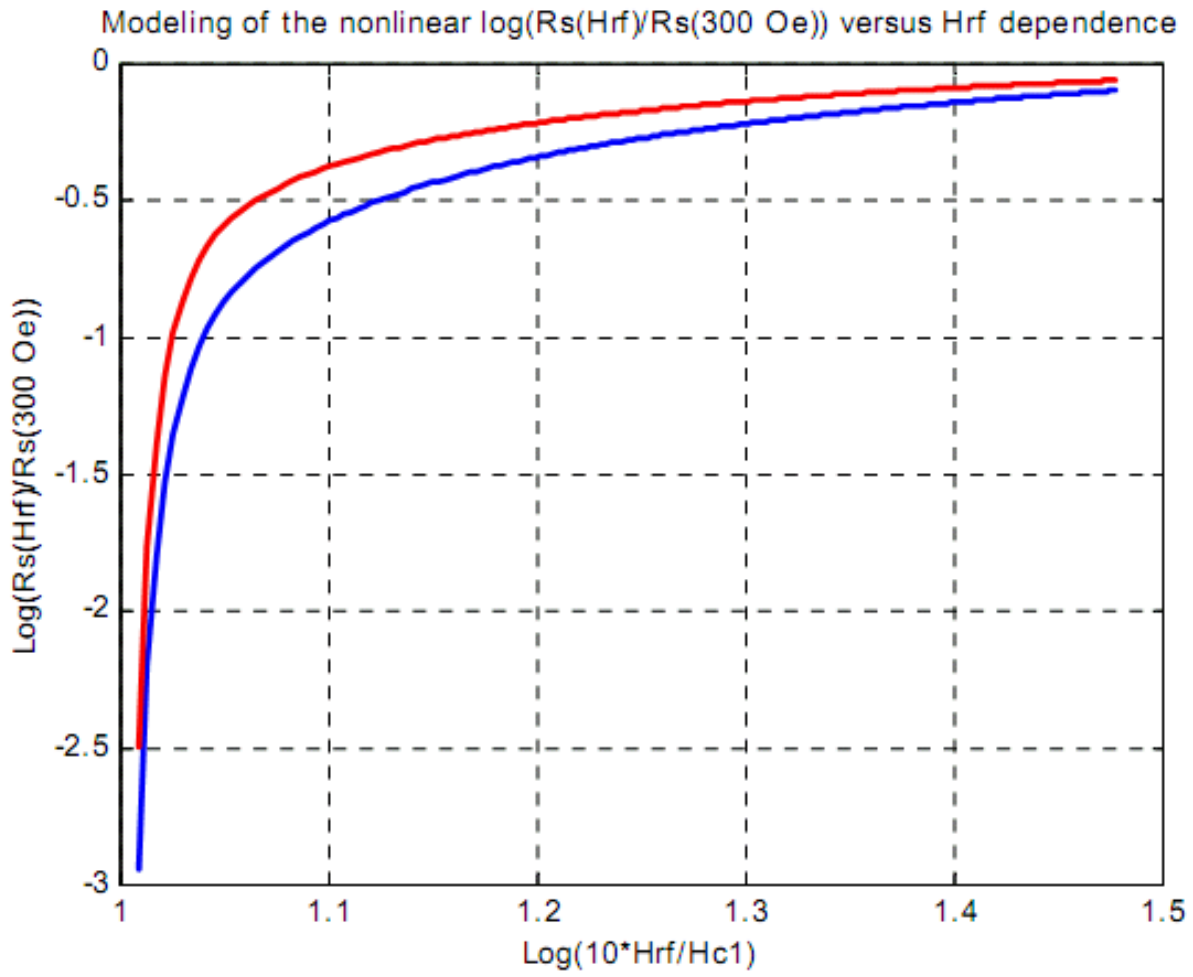

**Fig. 130.** Modeling of nonlinear dependences of relative surface resistance on magnetic field log(Rs(Hrf)/Hrf(300 Oe)) vs. log (10*Hrf/$Hc_1$) in $YBa_2Cu_3O_{7-\delta}$ superconducting dielectric and microstrip resonators with low critical magnetic field Hc1=50Oe at microwaves shown at a logarithmic scale: red curve corresponds to dependence log(Rs(Hrf)/Hrf(300 Oe)) vs. log (10*Hrf/Hc1) in the case of dielectric resonator; blue curve corresponds to dependence log(Rs(Hrf)/Hrf(300 Oe)) vs. log (10*Hrf/Hc1) in case of microstrip resonator. Modeling of nonlinear dependences log(Rs(Hrf)/Hrf(300 Oe)) vs. log (10*Hrf/Hc1) is performed for Rs(Hrf)/Hrf(300 Oe) vs. Hrf curves in Fig. 129 [85].

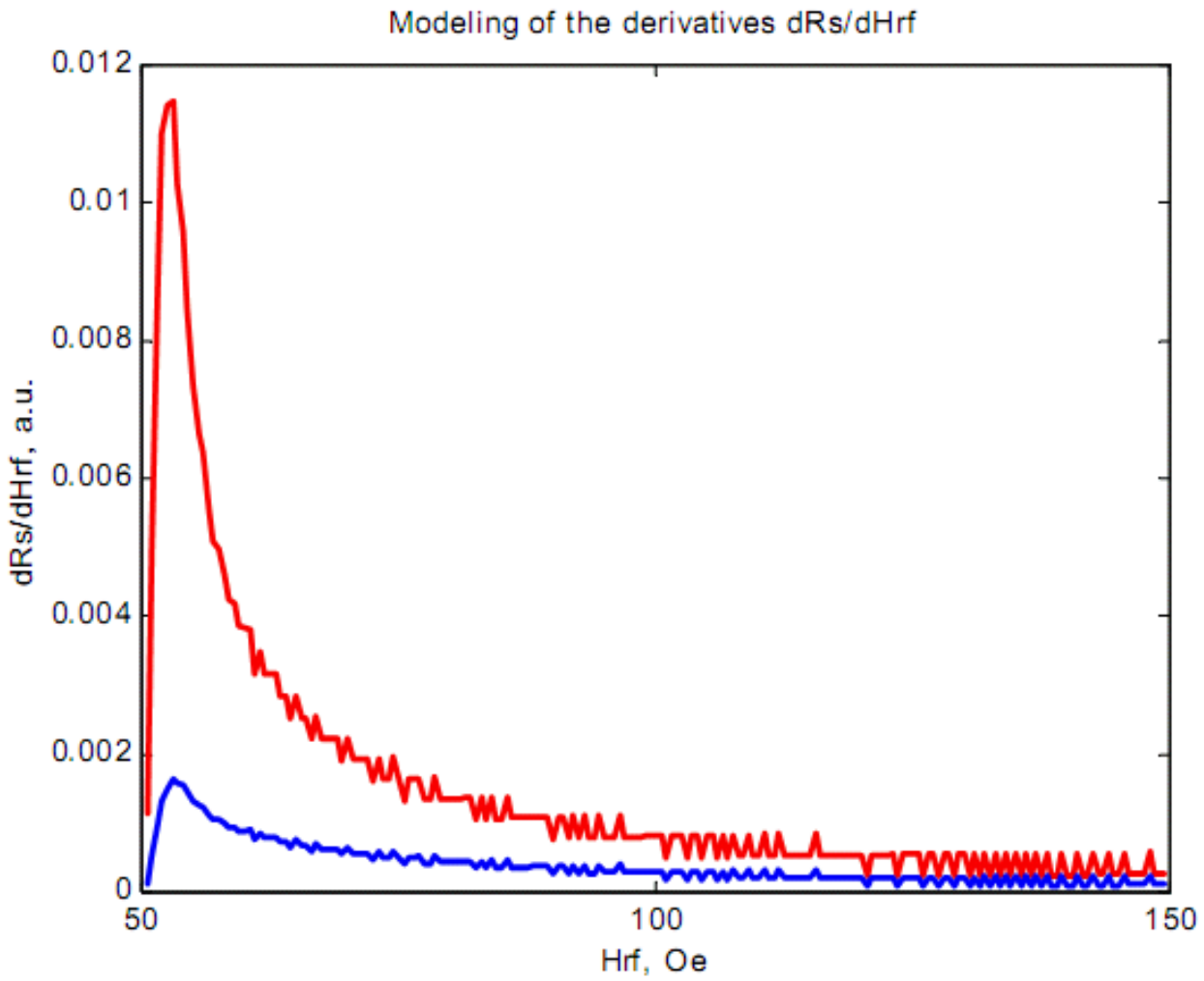

**Fig. 131.** Modeling of dependences dRs/dHrf(Hrf), representing ratio of dRs/dHrf derivatives as function of magnetic field Hrf, in $YBa_2Cu_3O_{7-\delta}$ superconducting dielectric and microstrip resonators with low critical magnetic field Hc1 = 50Oe at microwaves: red curve corresponds to dependence dRs/dHrf(Hrf) in case of dielectric resonator; blue curve corresponds to dependence dRs/dHrf(Hrf) in the case of microstrip resonator. Modeling of dependences dRs/dHrf(Hrf) is performed for Rs(Hrf)/Hrf(300 Oe) vs. Hrf curves shown in Fig. 129 [85].

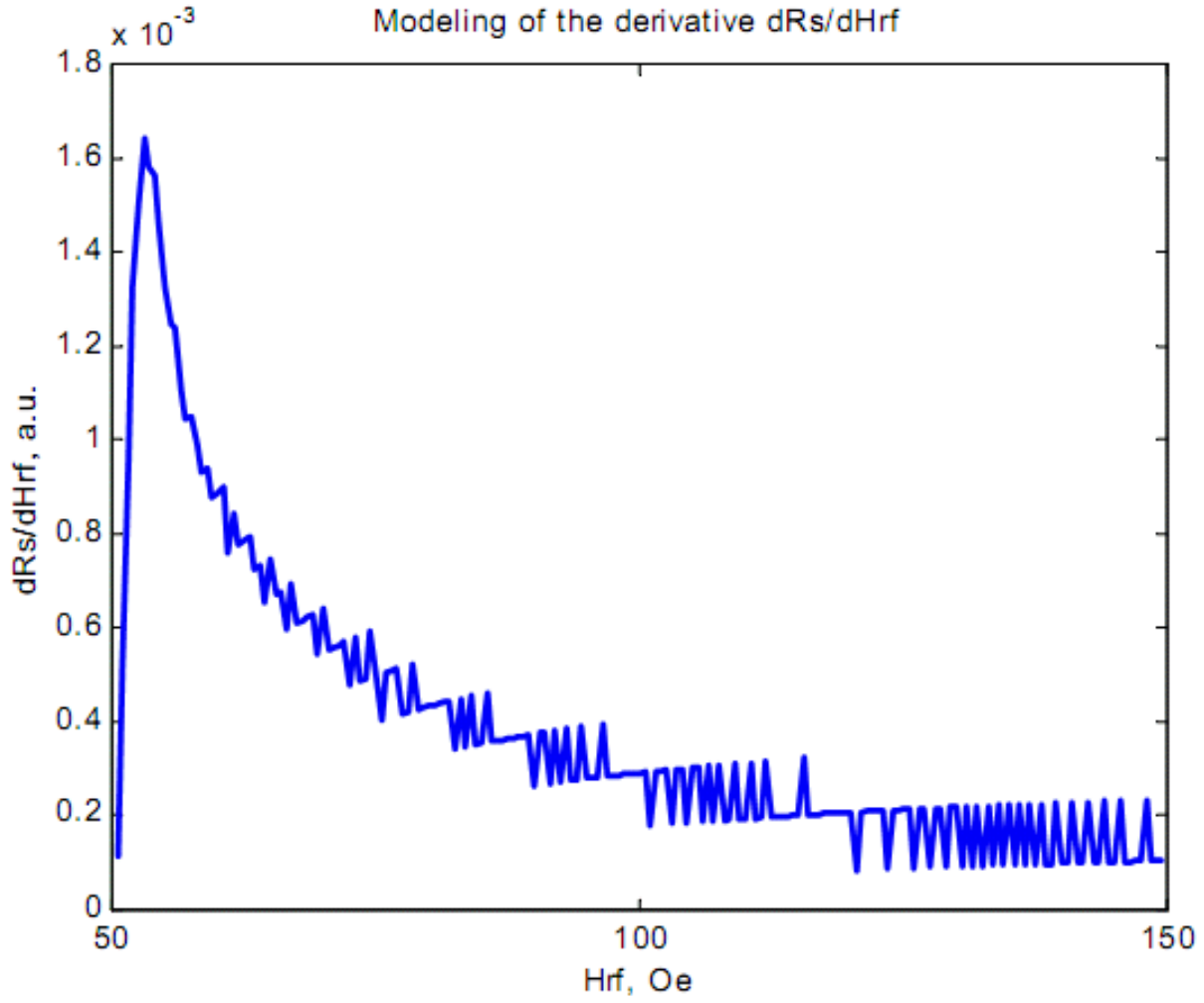

**Fig. 132.** Modeling of dependences dRs/dHrf(Hrf), representing ratio of dRs/dHrf derivatives as function of magnetic field Hrf , in $YBa_2Cu_3O_{7-\delta}$ superconducting microstrip resonator with low critical magnetic field $Hc_1$=50Oe at microwaves constructed on a different scale. Modeling of dependences dRs/dHrf(Hrf) is performed for Rs(Hrf)/Hrf(300 Oe) vs. Hrf curve for microstrip resonator in Fig. 129 [85].

# On Influence of Magnetic Field $H_{rf}$ on Nonlinear Surface Resistance $R_S$ of $YBa_2Cu_3O_{7-\delta}$ Thin Films on MgO Substrates in Close Proximity to Low Critical Magnetic Field $H_{c1}$ in Superconducting Microstrip Resonators at Ultra High Frequencies.

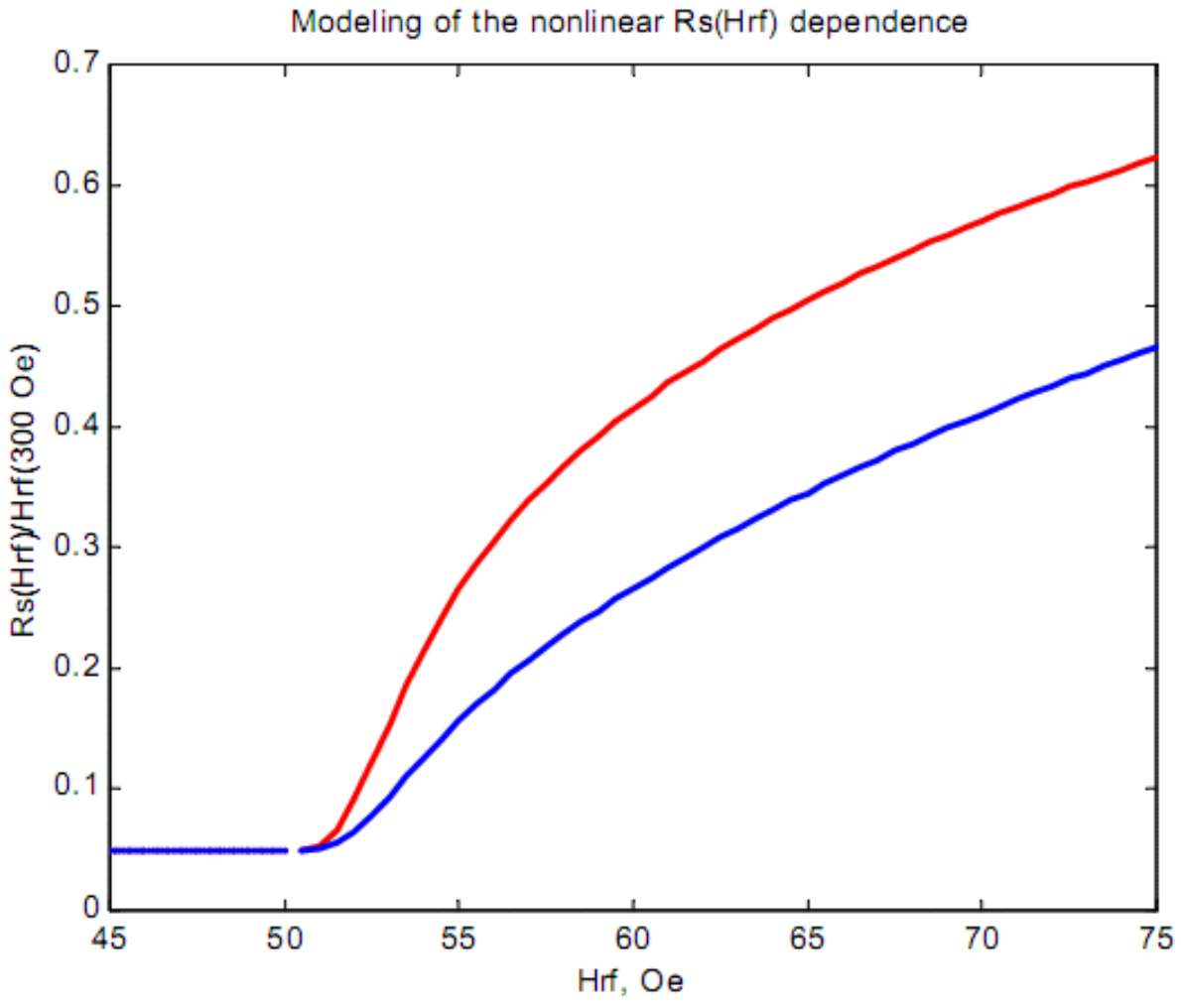

**Fig. 133.** Modeling of nonlinear dependence of relative surface resistance on magnetic field Rs(Hrf)/Hrf(300 Oe) vs. Hrf in $YBa_2Cu_3O_{7-\delta}$ superconducting dielectric and microstrip resonators in close proximity to low critical magnetic field Hc1=50Oe at microwaves: red curve corresponds to dependence Rs(Hrf)/Hrf(300 Oe) vs. Hrf in case of dielectric resonator; blue curve corresponds to dependence Rs(Hrf)/Hrf(300 Oe) vs. Hrf in case of microstrip resonator [85].

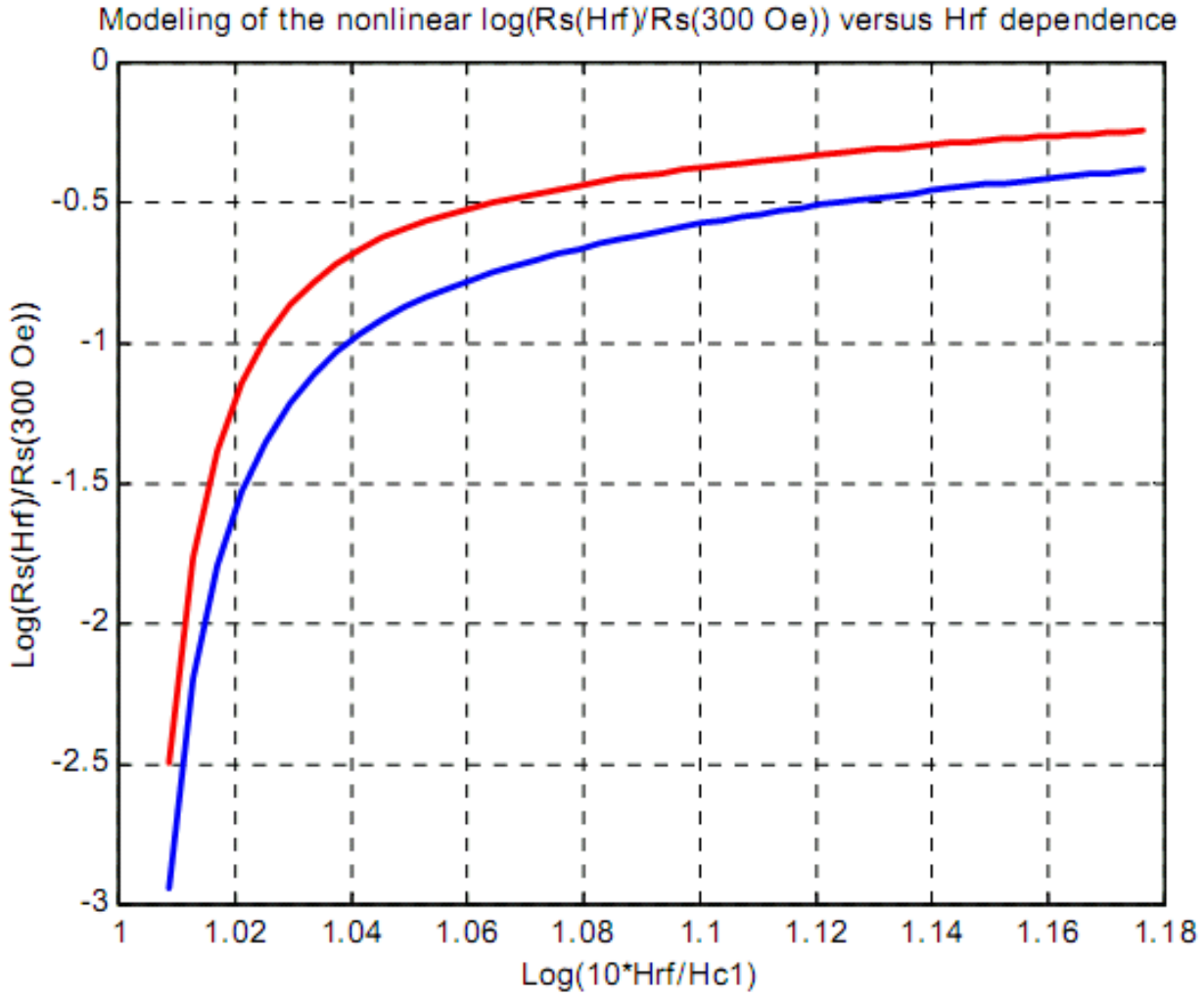

**Fig. 134.** Modeling of nonlinear dependences of relative surface resistance on magnetic field log(Rs(Hrf)/Hrf(300 Oe)) vs. log (10*Hrf/Hc1) in $YBa_2Cu_3O_{7-\delta}$ superconducting dielectric and microstrip resonators in close proximity to low critical magnetic field Hc1=50Oe at microwaves shown at a logarithmic scale: red curve corresponds to dependence log(Rs(Hrf)/Hrf(300 Oe)) vs. log (10*Hrf/Hc1) in the case of dielectric resonator; blue curve corresponds to dependence log(Rs(Hrf)/Hrf(300 Oe)) vs. log (10*Hrf/Hc1) in case of microstrip resonator. Modeling of nonlinear dependences log(Rs(Hrf)/Hrf(300 Oe)) vs. log (10*Hrf/Hc1) is performed for Rs(Hrf)/Hrf(300 Oe) vs. Hrf curves in Fig. 133 [85].

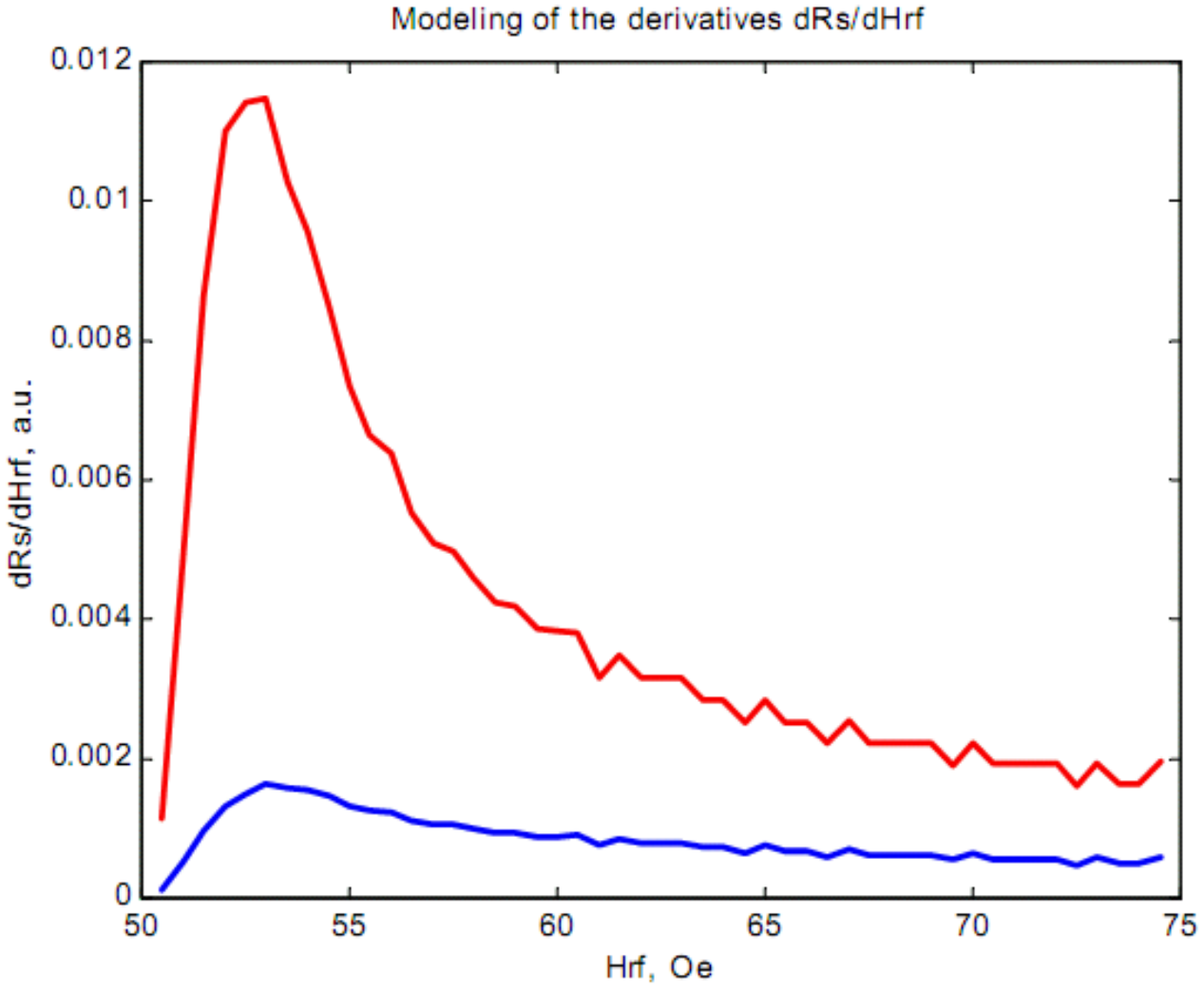

**Fig. 135.** Modeling of dependences dRs/dHrf(Hrf), representing the ratio of dRs/dHrf derivatives as a function of magnetic field Hrf , in $YBa_2Cu_3O_{7-\delta}$ superconducting dielectric and microstrip resonators in close proximity to low critical magnetic field Hc1 = 50Oe at microwaves: red curve corresponds to dependence dRs/dHrf(Hrf) in the case of dielectric resonator; blue curve corresponds to dependence dRs/dHrf(Hrf) in case of microstrip resonator. Modeling of dependences dRs/dHrf(Hrf) is performed for Rs(Hrf)/Hrf(300 Oe) vs. Hrf curves in Fig. 133 [85].

# On Influence of Magnetic Field $H_{rf}$ on Nonlinear Surface Resistance $R_S$ of $YBa_2Cu_3O_{7-\delta}$ Thin Films on MgO Substrates in Close Proximity to Low Critical Magnetic Field $H_{c1}$ in Superconducting Microstrip Resonators at Ultra High Frequencies.

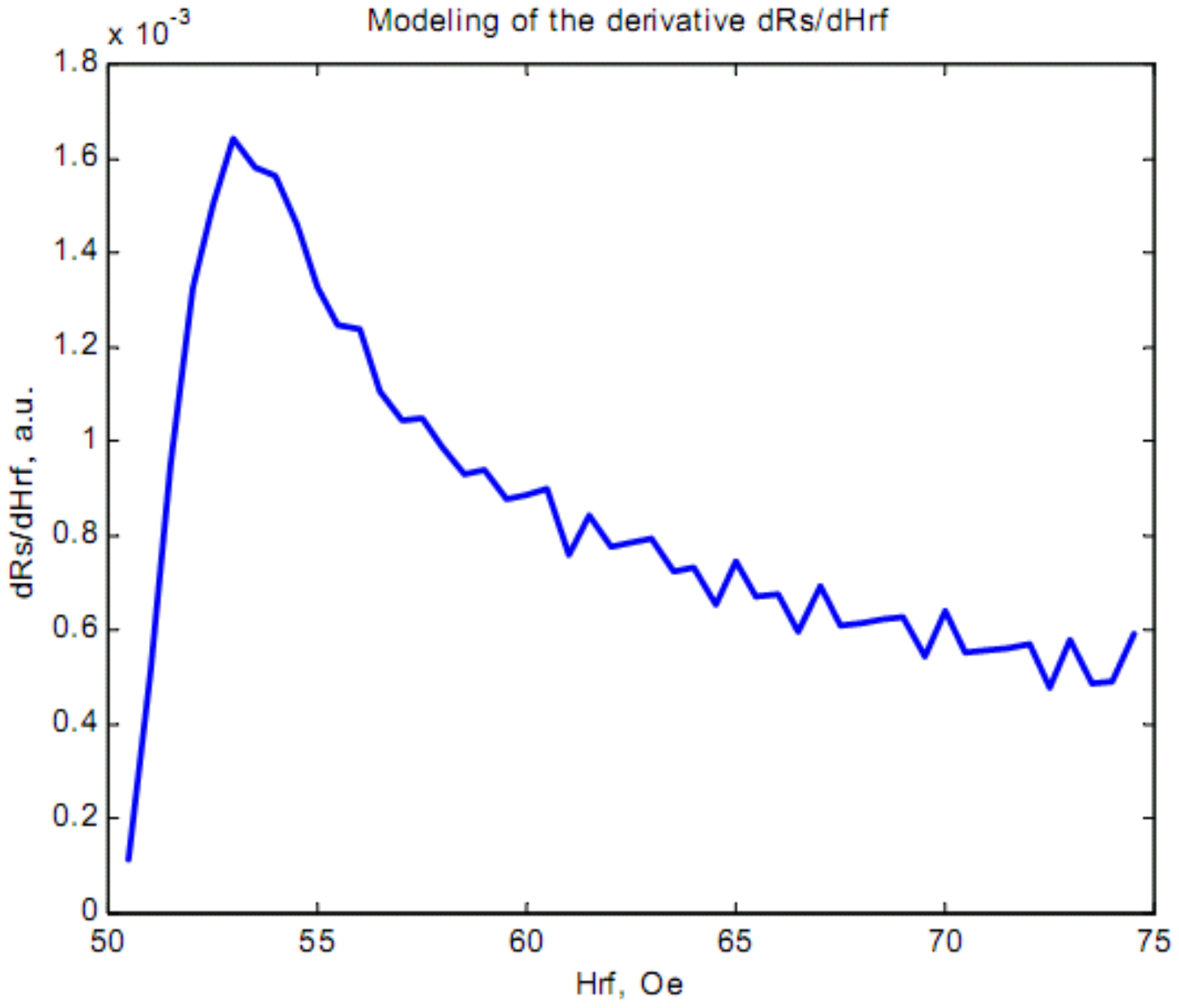

**Fig. 136**. Modeling of dependences dRs/dHrf(Hrf), representing ratio of dRs/dHrf derivatives as function of magnetic field Hrf , in $YBa_2Cu_3O_{7-\delta}$ superconducting microstrip resonator in close proximity to low critical magnetic field $Hc_1$ = 50Oe at microwaves constructed on a different scale. Modeling of dependences dRs/dHrf(Hrf) is performed for Rs(Hrf)/Hrf(300 Oe) vs. Hrf curve for microstrip resonator in Fig. 133 [85].

**In this research, the author conducted a number of computer modelings [85], and made a comparative analysis of the simulated results with the experimental results with the purpose to analyze the obtained research results and understand the nature of nonlinear surface resistance Rs in $YBa_2Cu_3O_{7-\delta}$ thin films on MgO substrate in a microstrip resonator. The characteristic experimental S-shape dependences of surface resistance on applied magnetic field Rs(Hrf) in $YBa_2Cu_3O_{7-\delta}$ thin films on MgO substrate in a microstrip resonator were observed by different researchers in a number of research papers. The similar Rs(Hrf) curves were measured by the author of dissertation. The typical S-dependences of the surface resistance on the magnetic field Rs(Hrf) in $YBa_2Cu_3O_{7-\delta}$ thin films on MgO substrates in the microstrip and dielectric resonators were modeled by the author of dissertation in [85]. The circumstance that, the typical S-dependences Rs(Hrf) are observed in rather small fields, can be interlinked to the fact that the plane HTS thin film samples could have a major demagnetization factor and consequently, the effective magnetic fields Hrf* on their surface have higher magnitude than the magnetic field Hrf of an electromagnetic wave. Therefore, these magnetic fields can reach the magnitudes comparable to the magnitude of low critical magnetic field $H_{rf}$~$Hc_1$, at which the transitions to the nonlinear mode in rather small applied magnetic fields are registered in $YBa_2Cu_3O_{7-\delta}$ thin films on MgO substrate in a microstrip resonators at microwaves. As it is visible from the presented research, the computer modeling results are in good agreement with the experimental data. The author of thesis performed research with the aim to clarify the nature of S-type dependences of nonlinear surface resistance on magnetic field Rs(Hrf) in $YBa_2Cu_3O_{7-\delta}$ thin films on MgO substrates in superconducting microstrip resonators at microwaves.**

# Conclusion.

**Research main conclusion: the magnitude of critical magnetic field $Hc_1$ of a superconductor needs to be firstly measured to predict the nonlinear behaviour of any HTS thin films, because the $Hc_1$ is the level of magnitude of magnetic fields at which the nonlinear effects begin to arise. The developed software program in Matlab can accurately model the nonlinearities in HTS thin films at microwaves, using the magnitudes of measured critical magnetic fields $Hc_1$ and $Hc_2$.**

- **Proposed Three New Equivalent Lumped Elements Models for Representation of a Superconductor in a Hakki-Coleman Dielectric Resonator at Microwaves.**
- **Performed Modeling of Microwave Power on Frequency Dependence P(f) with Identification of Lumped Elements Model Parameters for a Superconductor in a Hakki-Coleman Dielectric Resonator at Microwaves.**
- **Completed Precise Microwave Characterization of Microwave Properties of MgO Substrates in a Split Post Dielectric Resonator.**
- **Measured Temperature Dependences of Surface Resistance $R_S(T)$ of $YBa_2Cu_3O_{7-\delta}$ Superconducting Thin Films on MgO Substrates in a Hakki-Coleman Dielectric Resonator at Ultra High Frequencies.**
- **Measured Microwave Power Dependences of Surface Resistance $R_S(P)$ of $YBa_2Cu_3O_{7-\delta}$ Superconducting Thin Films on MgO Substrates in a Hakki-Coleman Dielectric Resonator at Ultra High Frequencies.**
- **Evaluated Microwave Measurement Accuracy by Deriving Formula, and Set the Error Band Limits at 1%.**
- **Proposed New Theory on Nature of Nonlinearities in HTS at Microwaves due to Abricosov Magnetic Vortices Generation, and Derived New Formula of Surface Resistance $R_S$ as a Function of Magnetic Field $H_e$ in Proximity to Low Critical Magnetic Field $H_{c1}$ and High Critical Magnetic Field $H_{c2}$ in the Case of High Magnitudes of Microwave Power Levels.**
- **Measured Dependences of Forward Transmission Coefficient on Ultra High Frequency $S_{21}(f)$ in $YBa_2Cu_3O_{7-\delta}$ Superconducting Microstrip Resonator at Different Microwave Power Levels and Temperatures.**
- **Completed Modeling of Nonlinear Dependence of Surface Resistance on External Magnetic Field $R_S(H_e)$ in Close Proximity to Low Critical Magnetic Field $H_{c1}$ in HTS Microstrip Resonators at Ultra High Frequencies.**
- **Completed Modeling of Nonlinear Dependence of Surface Resistance on External Magnetic Field $R_S(H_e)$ in Close Proximity to High Critical Magnetic Field $H_{c2}$ in HTS Microstrip Resonators at Ultra High Frequencies.**
- **Compared Experimental and Simulation Results, which Showed a Good Agreement at Low, Medium and High Microwave Power Levels.**
- **Engineered Models and Wrote Software, which can be Used to Predict EM Responses of HTS Planar Microwave Devices Accurately.**
- **Created Bibliography in the Field of Microwave Superconductivity.**

# New Innovative Strategies to Design Novel Highly Linear HTS Microstrip Filters.

**New innovative strategies to design novel highly linear HTS microstrip filters with enhanced microwave power handling capabilities, reduced microwave power level of intermodulation distortion (IMD), high interception point three (IP3), low insertion loss characteristics and compact design are:**

**1. Improvement of material properties of High Temperature Superconductors (HTS) through the synthesis of high quality HTS thin films without crystal defects or imperfections and with big magnitude of critical magnetic field $Hc_1$, leading to small non-linear response of microwave resonator to high level of microwave power.**

**2. Improvement of material properties of High Temperature Superconductors (HTS) through the use of composite HTS thin films with green phase nano-clusters, oxide nano-clusters, nano-impurities to improve the Abricosov magnetic vortices pinning, and hence increase the magnitude of critical current Ic, leading to small non-linear response of microwave resonator to high microwave power at $Hrf>Hc_1$.**

**3. Optimization of geometrical parameters of HTS thin films by increasing the thickness of HTS thin films, using of stacks with many HTS thin films layers, or by increasing the width of microstrip line in HTS resonator in HTS microstrip filter.**

**4. Optimization of geometric structure of HTS thin films, deposited on substrate, by creation of sliced HTS microstrip lines. For example, the use of sliced HTS microstrip lines for the design of microwave resonators in a transmitting filter.**

**5. Optimization of geometric structure of HTS thin films in HTS microstrip filter by the use of HTS microstrip line with slots instead of solid microstrip line to create the split resonators, aiming to reduce the peak current density at outer edges of HTS microstrip and avoid high current density HTS microstrip discontinuities.**

**6. Optimization of geometry of HTS microstrip filter by rounding all the edges and corners of HTS thin films in microstrip filters in receivers and in transmitters. For instance, the use of HTS split open-ring resonators (SRR) in microstrip filters in a transmitter.**

**7. Miniaturization of high pole HTS microwave filters with increased microwave power handling capability, by using the stripline (SL) resonators with shorter distance between resonators, resulting from a weak coupling. The modified SL filter with increased spacing between resonators to reduce maximum surface current has smaller size in comparison with the HTS microstrip line (MSL), dual-mode or bulk resonators.**

**8. Optimization of HTS microstrip filter design, using the different design techniques toward the intermodulation distortion reduction in HTS microstrip filters. The ferroelectrics may be used for maximum IMD reduction with a minimum effect on losses in HTS/FER microstrip filters. For example, the suppression of IMD, generated by HTS materials, with the use of a nonlinear ferroelectric segment for nonlinear pre-distortion in HTS band-pass filter.**

**9. Introduction of quality control during microwave filters fabrication to avoid the edge defects in HTS thin films, which can lead to the decrease of critical magnetic field $Hc_1$. The absence of edge defects in HTS thin films will significantly improve the characteristics of HTS microstrip filter.**

**10. Maintenance of thermal stability of HTS microwave filters by constantly adjusting the real operational temperature to normal operational temperature, which is equal to 0,8Tc, where Tc is the critical temperature of HTS.**

**11. Improvement of the packaging of HTS microwave filters to avoid the vibration related problems with the tuning in space and airborne applications.**

# Example of Cryogenic Transceiver Front End (CTFE) with HTS Transmitting and Receiving Microwave Filters in Wireless Communication Systems.

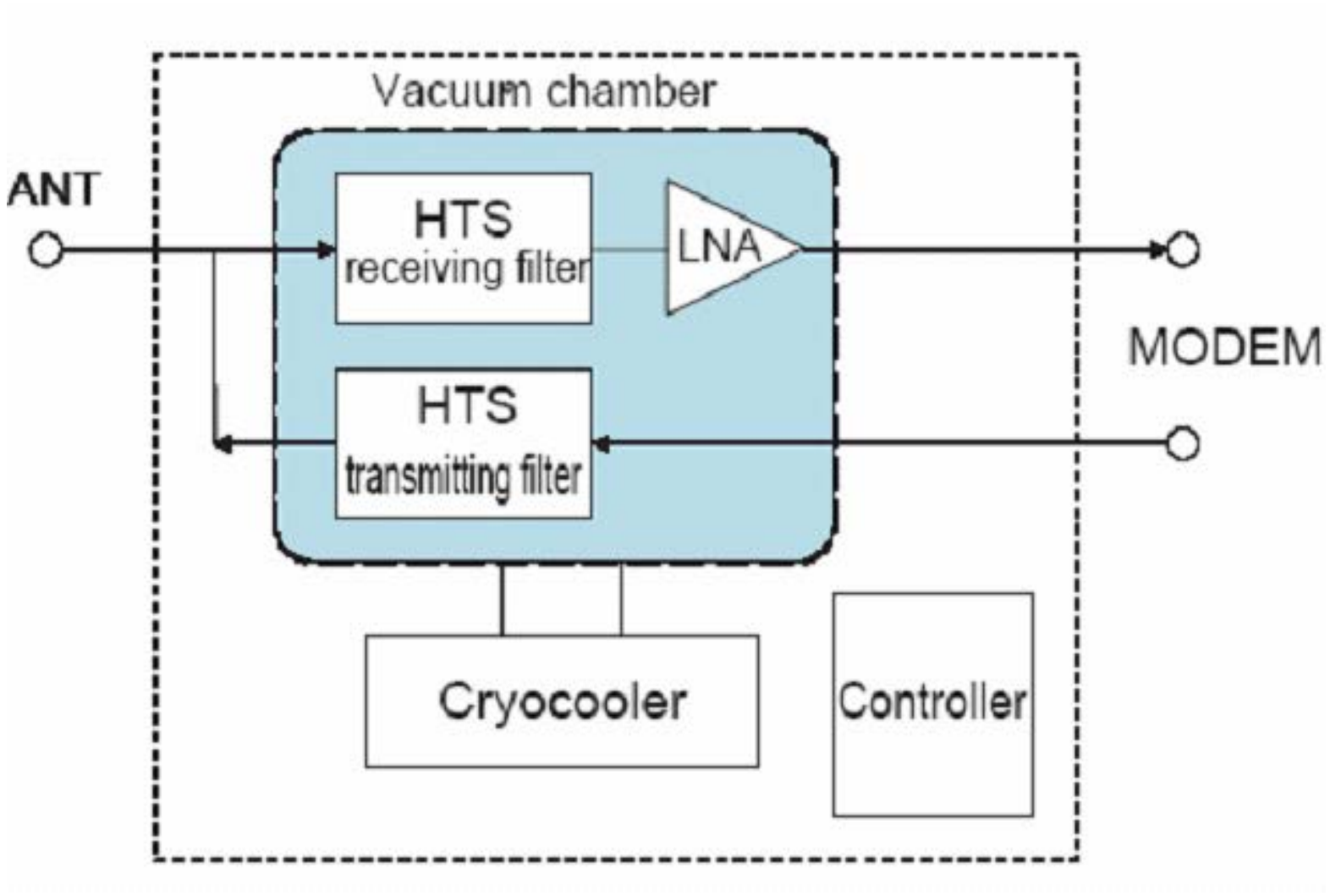

**Fig. 129**. Cryogenic Transceiver Front End (CTFE) block diagram that implements HTS microwave transmitting and receiving filters (after [90]).

**The superconducting filters in combination with cryogenically cooled LNAs have the following technical advantages [3]:**

**1.** Decreased noise figure for the system within the bandpass of HTS filter, and

**2.** Much steeper frequency roll-off on the edges of bandpass of HTS filter due to the higher order filter that can be used and the higher Q values of HTS resonators.

**The superconducting filters in combination with cryogenically cooled LNAs have the following technical advantages [90]:**

**1**. Better signal-to-noise ratio,

**2.** Greater input signal selectivity,

**3.** Small insertion losses,

**4.** Better adjacent channel leakage power ratio (in transmitter), and

**5.** Compact design.

# Examples of Tuning and Trimming Techniques for High Temperature Superconducting (HTS) Microstrip Filters Characteristics Optimization.

**The HTS microwave filters tuning and trimming techniques may include the use of specially designed trimmers, which are small dielectric screws, used to change the distance between the screw and the strip line, varying the couplings between the filter components , tuning the resonant frequencies, and adjusting the microstrip filter characteristics. The electric pads can also be used for center frequency tuning.**

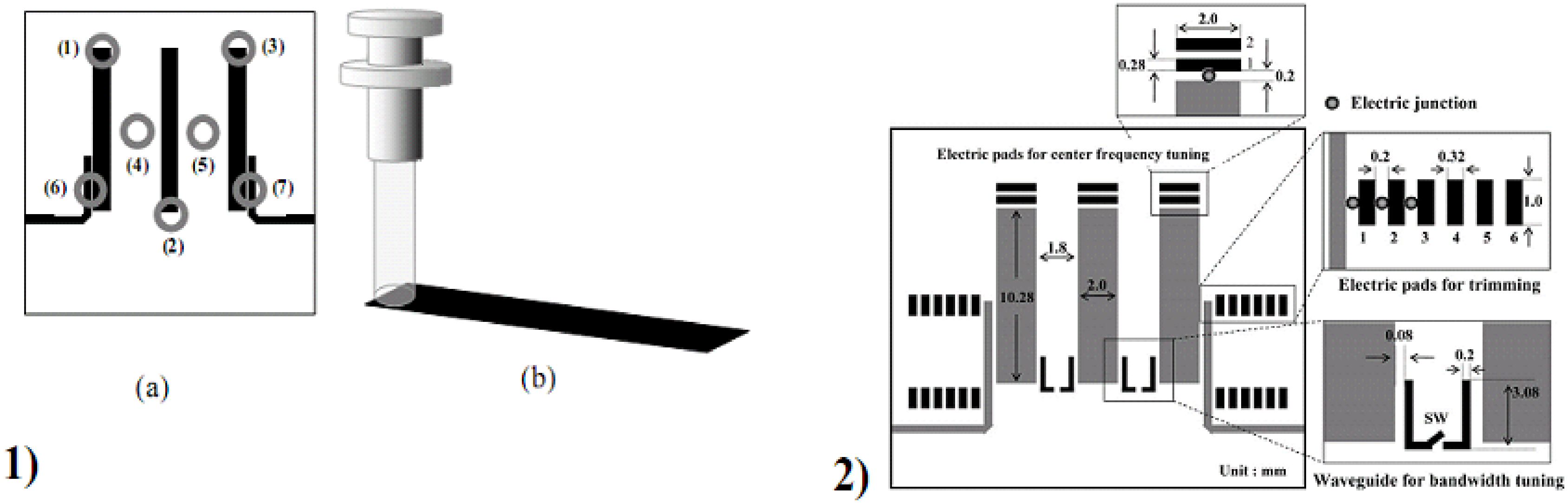

**Fig. 130**. **1**) Schematic drawing of filter configuration of a forward-coupled filter (**a**) and sapphire trimming rod (**b**): The circles in (**a**) indicate the positions of the rod trimmer (after [91]).
**2**) Layout of center frequency and bandwidth tunable filter with trimming and tuning elements (after [92]).

# Example of Schematic of Layout of Split Open-Ring Resonators in Multi Pole Microstrip Reaction Type Transmit Filter with its Frequency Response for Application in WCDMA Wireless Communications.

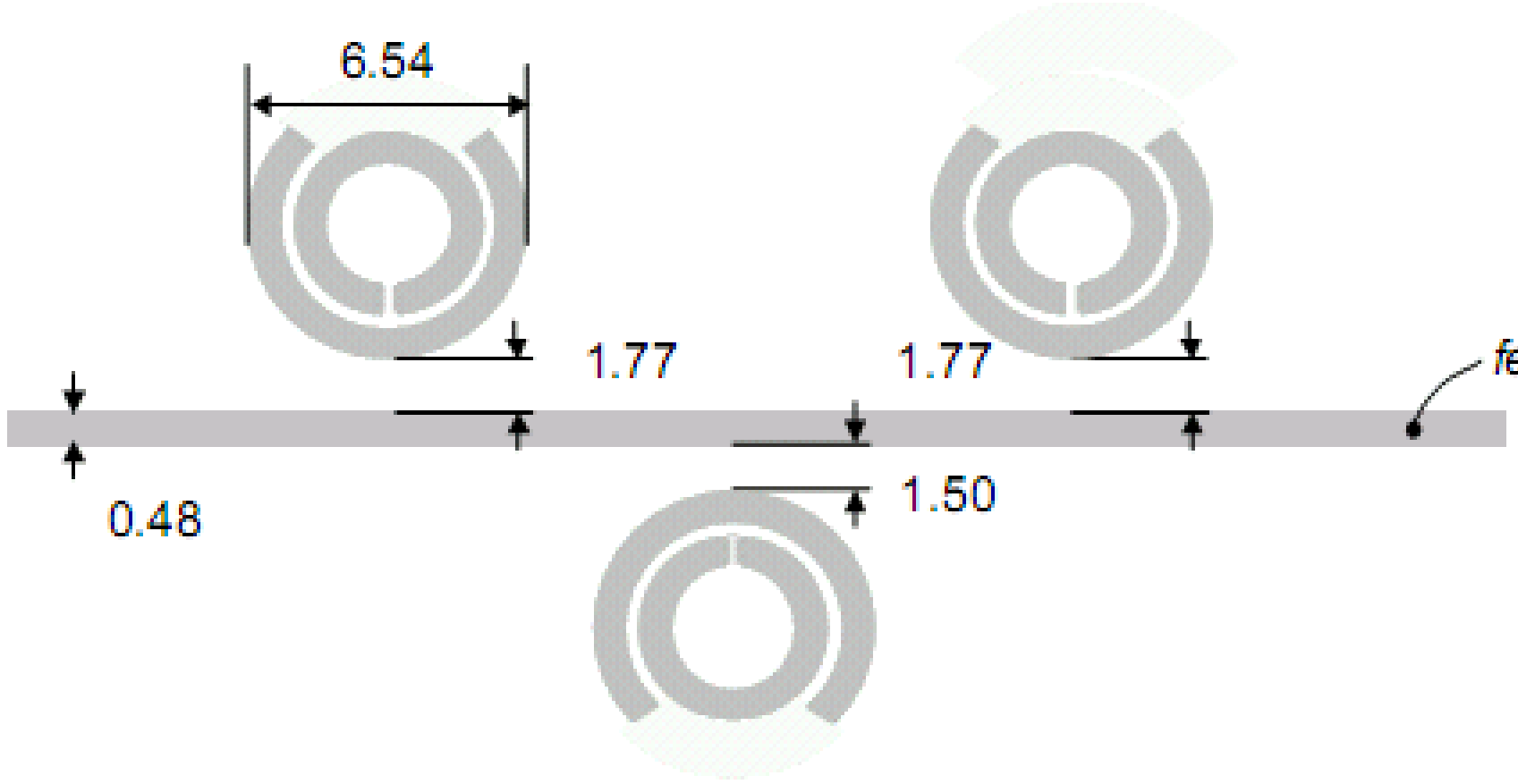

**Fig. 131**. The geometry of the 5-GHz HTS-RTF using the Split Open Ring Resonator (SORR) with w=0.8mm and g=0.4mm (after [93]).

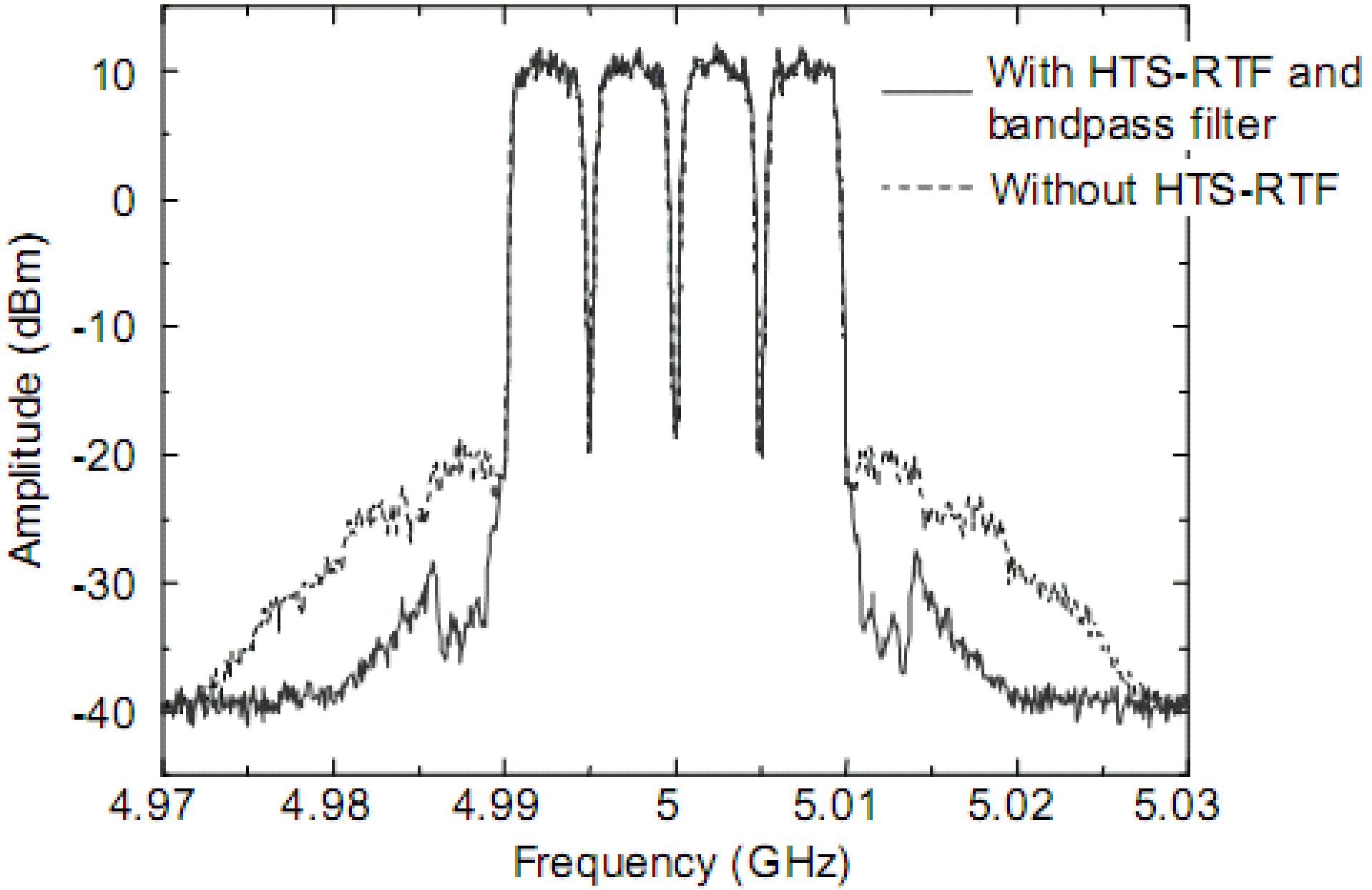

**Fig. 132**. The typical analyzed output spectrum with and without the designed HTS-RTFs with additional band pass filter (after [93]).

# Thank you very much!

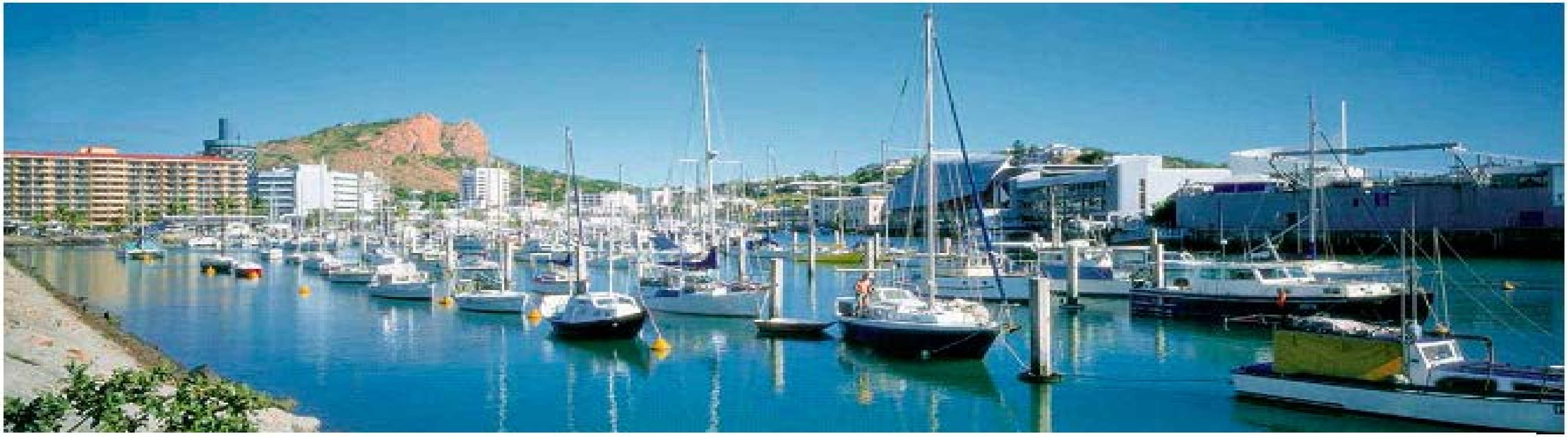

***This book describes the nature of nonlinearities in the field of microwave superconductivity, aiming to significantly improve our understanding on the sources of nonlinearities, effects originated by the nonlinearities, limitations imposed by the nonlinearities and application of knowledge gained about the nonlinearities to optimize the designs of microwave circuits.***

## Key Features:

- **Theories and experiments toward understanding of nonlinearities in microwave superconductivity.**
- **Microwave resonance techniques for complex permittivity and conductivity measurements of ceramics and superconductors.**
- **Precise characterization of nonlinear surface resistance of High Temperature Superconducting (HTS) thin films and semiconductors in dielectric resonators at microwaves.**
- **Precise characterization of nonlinear surface resistance of High Temperature Superconducting (HTS) thin films in superconducting microstrip resonators at microwaves.**
- **Ledenyov theory to describe the nature of nonlinearities in Type II superconductors at microwaves.**
- **Ledenyov theory to describe the noise origination by magnetic dipole two-level systems (MTLS) in microwave superconductivity.**
- **Ledenyov quantum theory to describe the nature of 1/f noise in microwave superconductivity.**
- **Ledenyov theory to describe the origin of differential noise in microwave superconductivity.**
- **Accurate characterization of Duffing like nonlinearities in nonlinear dynamic resonant microwave systems.**
- **New discovery of extreme quantum limit: quantum knot of Abricosov or Josephson magnetic vortex.**
- **Design of High Temperature Superconducting (HTS) microwave filters with exceptional microwave power handling capabilities and low Intermodulation Distortions (IMD) for application in Cryogenic Transceiver Front End (CTFE) in wireless communication systems.**

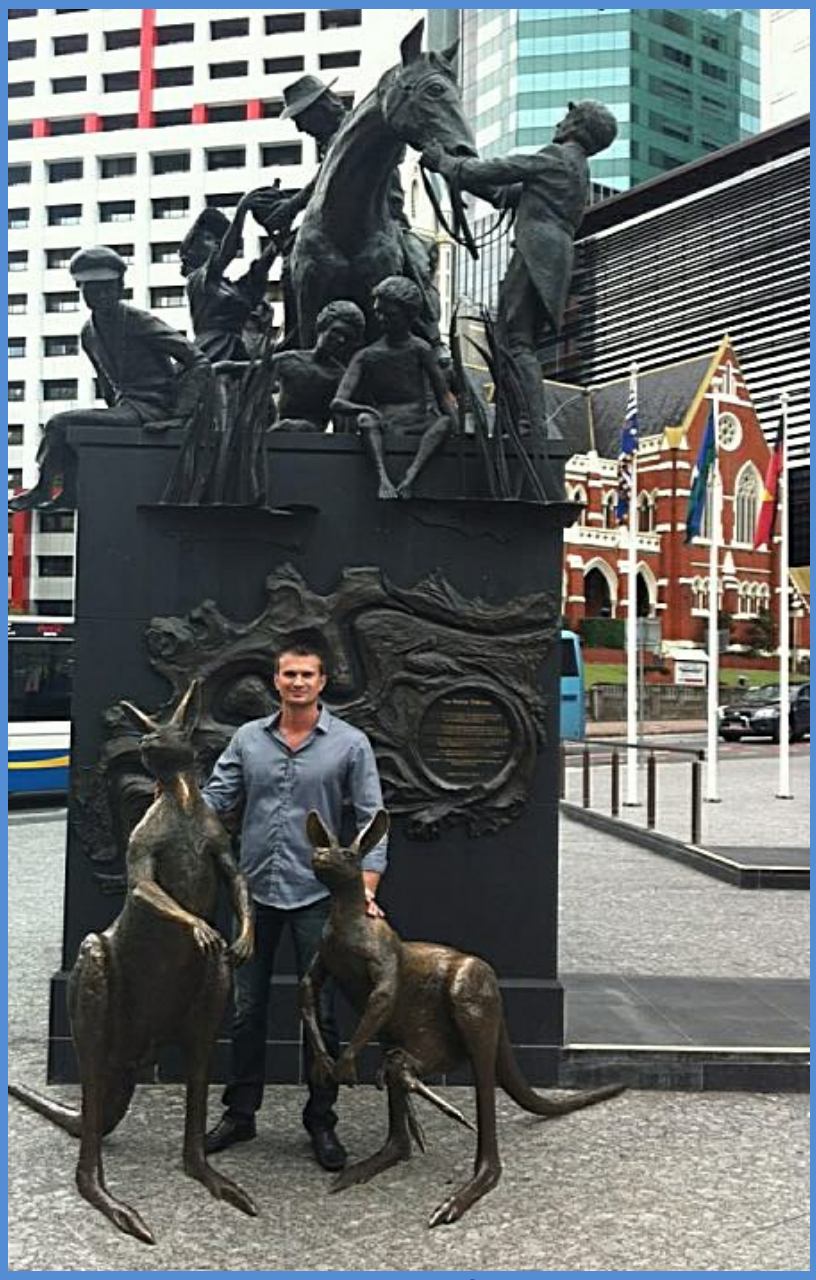

Dimitri O. Ledenyov

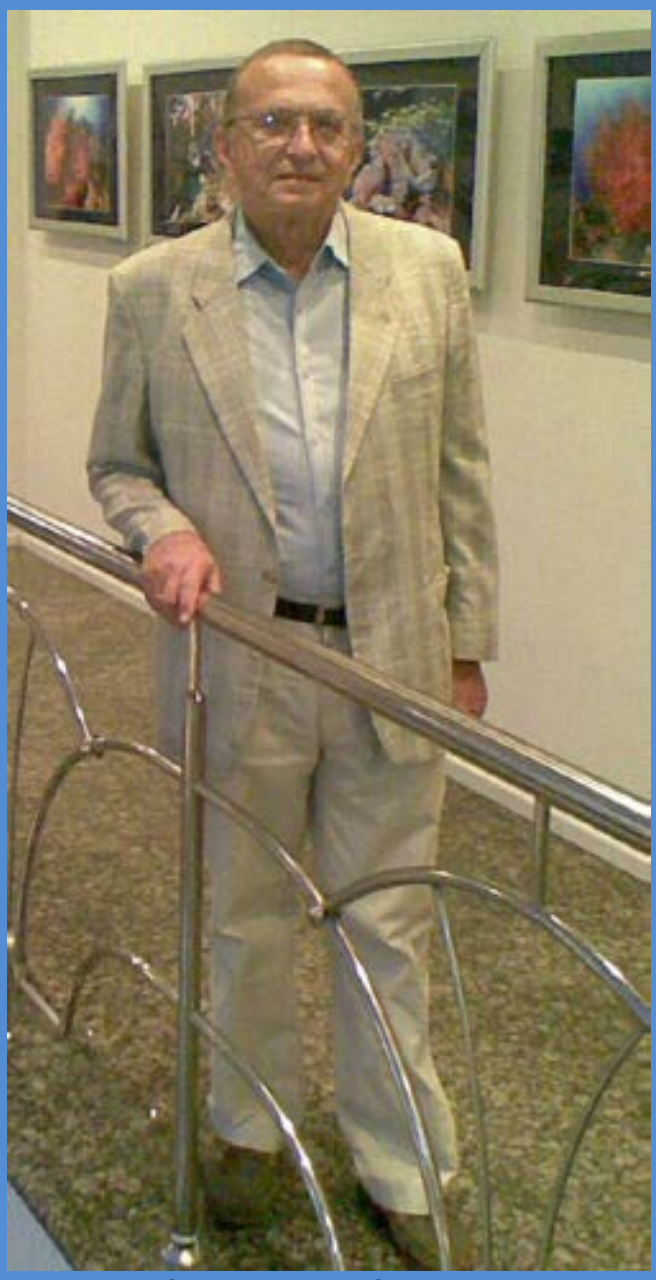

Oleg P. Ledenyov

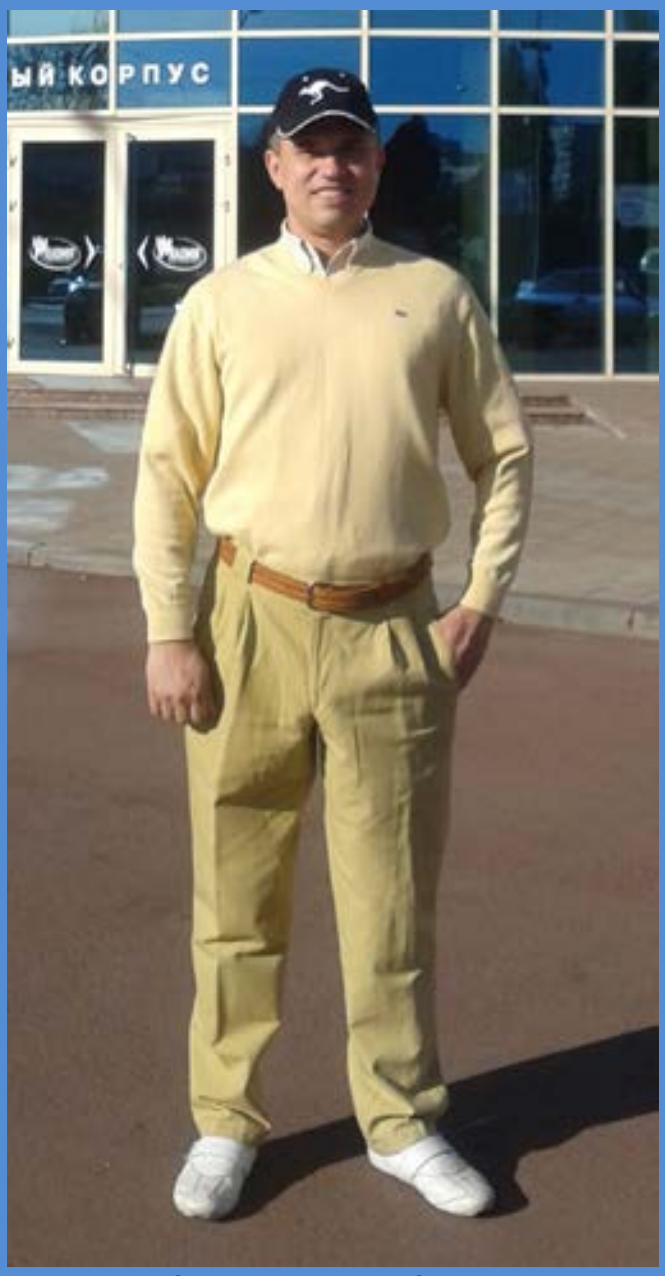

Viktor O. Ledenyov

**Engineering professionals, scientists, students, policymakers will greatly benefit from this insightful research on the nonlinearities in microwave superconductivity, learning to apply the fundamental knowledge to the applied engineering design tasks in microwave science and engineering.**

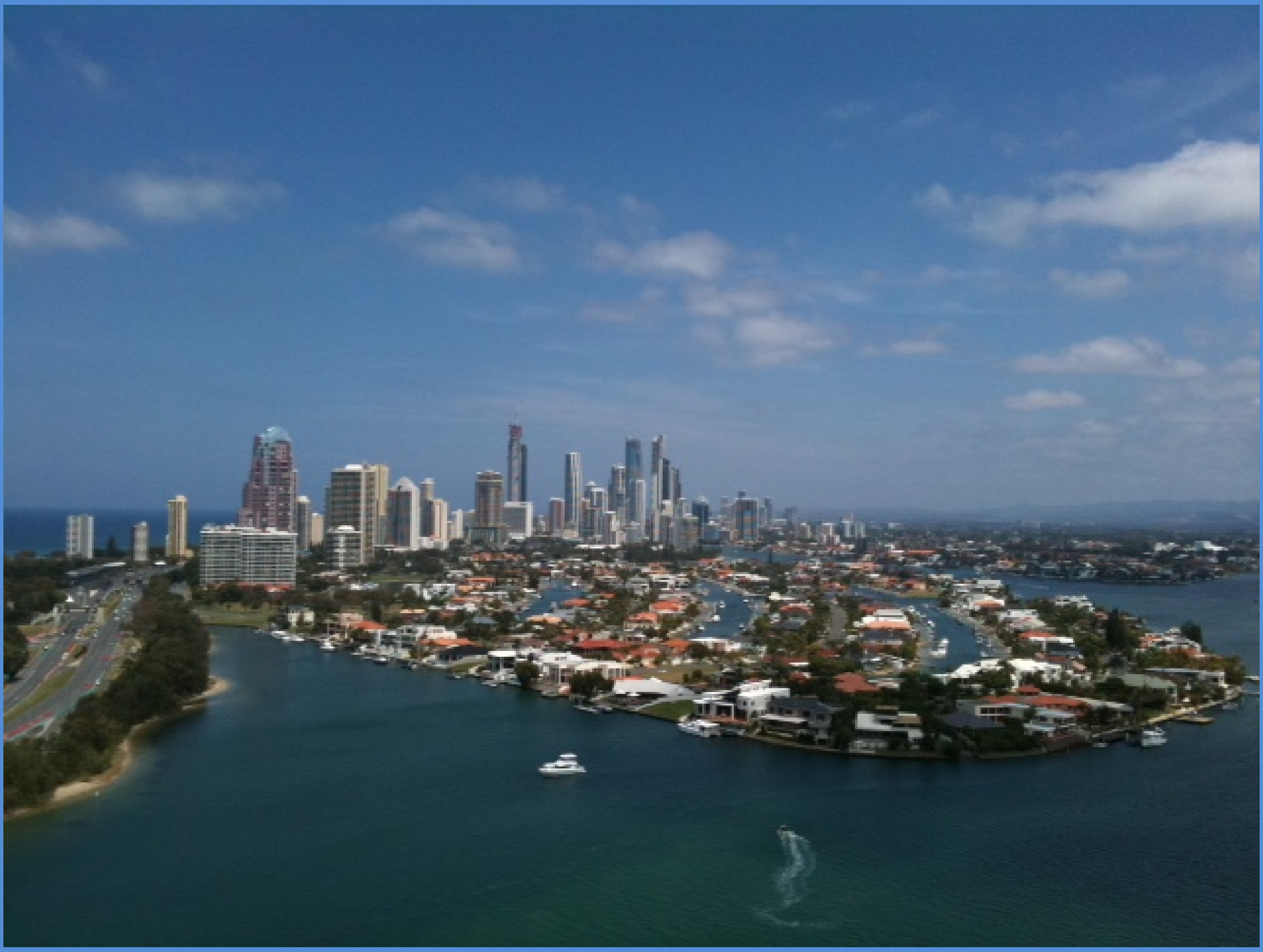

Published in Australia